**Fran De Aquino**

# TOE

## Theory of Everything

November
2012

**Fran De Aquino**

# T O E

## Theory of Everything

November
2012

**ABSTRACT**

This is a set of 25 articles, developed starting from the *Relativistic Theory of Quantum Gravity* (first article). Together they form the Theory of Everything.

# CONTENTS

(421 pages)

*

*   *

# Mathematical Foundations of the Relativistic Theory of Quantum Gravity

**Fran De Aquino**
Maranhao State University, Physics Department, S.Luis/MA, Brazil.

**Abstract:** Starting from the *action function*, we have derived a theoretical background that leads to the *quantization of gravity* and the deduction of a correlation between the gravitational and the inertial masses, which depends on the *kinetic momentum* of the particle. We show that the strong equivalence principle is reaffirmed and, consequently, Einstein's equations are preserved. In fact, such equations are deduced here directly from this new approach to Gravitation. Moreover, we have obtained a generalized equation for the inertial forces, which incorporates the Mach's principle into Gravitation. Also, we have deduced the equation of Entropy; the Hamiltonian for a particle in an electromagnetic field and the reciprocal fine structure constant directly from this new approach. It was also possible to deduce the expression of the *Casimir force* and to explain the *Inflation Period* and the *Missing Matter,* without assuming existence of *vacuum fluctuations*. This new approach to Gravitation will allow us to understand some crucial matters in Cosmology.

Key words: Quantum Gravity, Quantum Cosmology, Unified Field.
PACs: 04.60.-m; 98.80.Qc; 04.50. +h

**Contents**

## 1. INTRODUCTION

Quantum Gravity was originally studied, by Dirac and others, as the problem of quantizing General Relativity. This approach presents many difficulties, detailed by Isham [1]. In the 1970's, physicists tried an even more conventional approach: simplifying Einstein's equations by assuming that they are *almost linear*, and then applying the standard methods of quantum field theory to the thus oversimplified equations. But this method, too, failed. In the 1980's a very different approach, known as string theory, became popular. Thus far, there are many enthusiasts of string theory. But the mathematical difficulties in string theory are formidable, and it is far from clear that they will be resolved any time soon. At the end of 1997, Isham [2] pointed out several "Structural Problems Facing Quantum Gravity Theory". At the beginning of this new century, the problem of quantizing the gravitational field was still open.

In this work, we propose a new approach to Quantum Gravity. Starting from the generalization of the *action function* we have derived a theoretical background that leads to the quantization of gravity. Einstein's General Relativity equations are deduced directly from this theory of Quantum Gravity. Also, this theory leads to a complete description of the Electromagnetic Field, providing a consistent unification of gravity with electromagnetism.

## 2. THEORY

We start with the *action* for a free-particle that, as we know, is given by

$$S = -\alpha\int_a^b ds$$

where $\alpha$ is a quantity which characterizes the particle.

In Relativistic Mechanics, the action can be written in the following form [3]:

$$S = \int_{t_1}^{t_2} L dt = -\int_{t_1}^{t_2} \alpha c\sqrt{1 - V^2/c^2}\, dt$$

where

$$L = -\alpha c\sqrt{1 - V^2/c^2}$$

is the Lagrange's function.

In Classical Mechanics, the Lagrange's function for a free-particle is, as we know, given by: $L = aV^2$ where $V$ is the speed of the particle and $a$ is a quantity *hypothetically* [4] given by:

$$a = m/2$$

where $m$ is the mass of the particle. However, there is no distinction about the kind of mass (if *gravitational mass*, $m_g$, or *inertial mass* $m_i$) neither about its sign $(\pm)$.

The correlation between $a$ and $\alpha$ can be established based on the fact that, on the limit $c \to \infty$, the relativistic expression for $L$ must be reduced to the classic expression $L = aV^2$. The result [5] is: $L = \alpha V^2/2c$. Therefore, if $\alpha = 2ac = mc$, we obtain $L = aV^2$. Now, we must decide if $m = m_g$ or $m = m_i$. We will see in this work that the definition of $m_g$ includes $m_i$. Thus, the right option is $m_g$, i.e.,

$$a = m_g/2.$$

Consequently, $\alpha = m_g c$ and the generalized expression for the action of a free-particle will have the following form:

$$S = -m_g c\int_a^b ds \qquad (1)$$

or

$$S = -\int_{t_1}^{t_2} m_g c^2 \sqrt{1 - V^2/c^2}\, dt \qquad (2)$$

where the Lagrange's function is

$$L = -m_g c^2 \sqrt{1 - V^2/c^2}. \qquad (3)$$

The integral $S = \int_{t_1}^{t_2} m_g c^2 \sqrt{1 - V^2/c^2}\, dt$, preceded by the *plus* sign, cannot have a *minimum*. Thus, the integrand of Eq.(2) must be *always positive*. Therefore, if $m_g > 0$, then necessarily $t > 0$; if $m_g < 0$, then $t < 0$. The possibility of $t < 0$ is based on the well-known equation $t = \pm t_0 \big/ \sqrt{1 - V^2/c^2}$ of Einstein's Theory.

Thus if the *gravitational mass* of a particle is *positive,* then $t$ is also *positive* and, therefore, given by $t = +t_0 \big/ \sqrt{1 - V^2/c^2}$. This leads to the well-known relativistic prediction that the particle goes to the *future,* if $V \to c$. However, if the *gravitational mass* of the particle is *negative,* then $t$ is *negative* and given by $t = -t_0 \big/ \sqrt{1 - V^2/c^2}$. In this case, the prediction is that the particle goes to the *past,* if $V \to c$. Consequently, $m_g < 0$ is the necessary condition for the particle to go to the *past*. Further on, a correlation between the *gravitational* and the *inertial* masses will be derived, which contains the possibility of $m_g < 0$.

The Lorentz's transforms follow the same rule for $m_g > 0$ and $m_g < 0$, i.e., the sign before $\sqrt{1 - V^2/c^2}$ will be $(+)$ when $m_g > 0$ and $(-)$ if $m_g < 0$.

The *momentum*, as we know, is the vector $\vec{p} = \partial L / \partial \vec{V}$.Thus, from Eq.(3) we obtain

$$\vec{p} = \frac{m_g \vec{V}}{\pm\sqrt{1 - V^2/c^2}} = M_g \vec{V}$$

The $(+)$ sign in the equation above will be used when $m_g > 0$ and the $(-)$ sign if $m_g < 0$. Consequently, we will express the momentum $\vec{p}$ in the following form

$$\vec{p} = \frac{m_g \vec{V}}{\sqrt{1 - V^2/c^2}} = M_g \vec{V} \qquad (4)$$

The derivative $d\vec{p} / dt$ is the *inertial* force $F_i$ which acts on the particle. If the force is perpendicular to the speed, we have

$$\vec{F}_i = \frac{m_g}{\sqrt{1 - V^2/c^2}} \frac{d\vec{V}}{dt} \qquad (5)$$

However, if the force and the speed have the same direction, we find that

$$\vec{F}_i = \frac{m_g}{\left(1 - V^2/c^2\right)^{3/2}} \frac{d\vec{V}}{dt} \qquad (6)$$

From Mechanics [6], we know that $\vec{p} \cdot \vec{V} - L$ denotes the *energy* of the particle. Thus, we can write

$$E_g = \vec{p} \cdot \vec{V} - L = \frac{m_g c^2}{\sqrt{1 - V^2/c^2}} = M_g c^2 \qquad (7)$$

Note that $E_g$ is not null for$V$=0, but that it has the finite value

$$E_{g0} = m_{g0} c^2 \qquad (8)$$

Equation (7) can be rewritten in the following form:

$$E_g = m_g c^2 - \frac{m_g c^2}{\sqrt{1 - V^2/c^2}} - m_g c^2 =$$

$$= \frac{m_g}{m_i} \left[ m_i c^2 + \underbrace{\left( \frac{m_i c^2}{\sqrt{1 - V^2/c^2}} - m_i c^2 \right)}_{E_{Ki}} \right] =$$

$$= \frac{m_g}{m_i} \left( E_{i0} + E_{Ki} \right) = \frac{m_g}{m_i} E_i \qquad (9)$$

By analogy to Eq. (8), $E_{i0} = m_{i0} c^2$ into the equation above, is the inertial energy *at rest*. Thus, $E_i = E_{i0} + E_{Ki}$ is the *total* inertial energy, where $E_{Ki}$ is the *kinetic* *inertial energy*. From Eqs. (7) and (9) we thus obtain

$$E_i = \frac{m_{i0}c^2}{\sqrt{1-V^2/c^2}} = M_i c^2. \qquad (10)$$

For small velocities $(V<<c)$, we obtain

$$E_i \approx m_{i0}c^2 + \tfrac{1}{2} m_i V^2 \qquad (11)$$

where we recognize the classical expression for the *inertial kinetic energy* of the particle.

The expression for the *gravitational kinetic energy*, $E_{Kg}$, is easily deduced by comparing Eq.(7) with Eq.(9). The result is

$$E_{Kg} = \frac{m_g}{m_i} E_{Ki}. \qquad (12)$$

In the presented picture, we can say that the *gravity*, $\vec{g}$, in a gravitational field produced by a particle of gravitational mass $M_g$, depends on the particle's gravitational energy, $E_g$ (given by Eq.(7)), because we can write

$$g = -G\frac{E_g}{r^2c^2} = -G\frac{M_g c^2}{r^2c^2} \qquad (13)$$

Due to $g = \partial\Phi/\partial r$, the expression of the *relativistic gravitational potential*, $\Phi$, is given by

$$\Phi = -\frac{GM_g}{r} = -\frac{Gm_g}{r\sqrt{1-V^2/c^2}}$$

Then, it follows that

$$\Phi = -\frac{GM_g}{r} = -\frac{Gm_g}{r\sqrt{1-V^2/c^2}} = \frac{\phi}{\sqrt{1-V^2/c^2}}$$

where $\phi = -Gm_g/r$.

Then we get

$$\frac{\partial\Phi}{\partial r} = \frac{\partial\phi}{\partial r\sqrt{1-V^2/c^2}} = \frac{Gm_g}{r^2\sqrt{1-V^2/c^2}}$$

whence we conclude that

$$\frac{\partial\Phi}{\partial r} = \frac{Gm_g}{r^2\sqrt{1-V^2/c^2}}$$

By definition, *the gravitational potential energy per unit of gravitational mass* of a particle inside a gravitational field is equal to the *gravitational potential* $\Phi$ *of the field*. Thus, we can write that

$$\Phi = \frac{U(r)}{m'_g}$$

Then, it follows that

$$F_g = -\frac{\partial U(r)}{\partial r} = -m'_g\frac{\partial\Phi}{\partial r} = -G\frac{m_g m'_g}{r^2\sqrt{1-V^2/c^2}}$$

If $m_g > 0$ and $m'_g < 0$, or $m_g < 0$ and $m'_g > 0$ the force will be *repulsive*; *the force will never be null* due to the existence of a *minimum value* for $m_g$ (see Eq. (24)). However, if $m_g < 0$ and $m'_g < 0$, or $m_g > 0$ and $m'_g > 0$ the force will be *attractive*. Just for $m_g = m_i$ and $m'_g = m'_i$ we obtain the *Newton's attraction law*.

On the other hand, as we know, the gravitational force is *conservative*. Thus, gravitational energy, in agreement with the energy conservation law, can be expressed by the *decrease* of the inertial energy, i.e.,

$$\Delta E_g = -\Delta E_i \qquad (14)$$

This equation expresses the fact that a decrease of gravitational energy corresponds to an increase of the inertial energy.

Therefore, a variation $\Delta E_i$ in $E_i$ yields a variation $\Delta E_g = -\Delta E_i$ in $E_g$. Thus $E_i = E_{i0} + \Delta E_i$; $E_g = E_{g0} + \Delta E_g = E_{g0} - \Delta E_i$ and

$$E_g + E_i = E_{g0} + E_{i0} \qquad (15)$$

Comparison between (7) and (10) shows that $E_{g0} = E_{i0}$, i.e., $m_{g0} = m_{i0}$. Consequently, we have

$$E_g + E_i = E_{g0} + E_{i0} = 2E_{i0} \qquad (16)$$

However $E_i = E_{i0} + E_{Ki}$. Thus, (16) becomes

$$E_g = E_{i0} - E_{Ki}. \qquad (17)$$

Note the *symmetry* in the equations of $E_i$ and $E_g$ .Substitution of $E_{i0} = E_i - E_{Ki}$ into (17) yields

$$E_i - E_g = 2E_{Ki} \qquad (18)$$

Squaring the Eqs.(4) and (7) and comparing the result, we find the following correlation between gravitational energy and *momentum* :

$$\frac{E_g^2}{c^2} = p^2 + m_g{}^2 c^2. \qquad (19)$$

The energy expressed as a function of the *momentum* is, as we know, called *Hamiltonian* or Hamilton's function:

$$H_g = c\sqrt{p^2 + m_g{}^2 c^2}. \qquad (20)$$

Let us now consider the problem of quantization of gravity. Clearly there is something unsatisfactory about the whole notion of quantization. It is important to bear in mind that the quantization process is a series of rules-of-thumb rather than a well-defined algorithm, and contains many ambiguities. In fact, for electromagnetism we find that there are (at least) two different approaches to quantization and that while they appear to give the same theory they may lead us to very different quantum theories of gravity. Here we will follow a new theoretical strategy: It is known that starting from the Schrödinger equation we may obtain the well-known expression for the energy of a particle in periodic motion inside a cubical box of edge length $L$ [ 7 ]. The result now is

$$E_n = \frac{n^2 h^2}{8 m_g L^2} \qquad n = 1,2,3,... \qquad (21)$$

Note that the term $h^2/8m_g L^2$ (energy) will be minimum for $L = L_{max}$ where $L_{max}$ is the maximum edge length of a cubical box whose maximum diameter

$$d_{max} = L_{max}\sqrt{3} \qquad (22)$$

is equal to *the maximum* length scale *of the Universe*.

The minimum energy of a particle is obviously its inertial energy at rest $m_g c^2 = m_i c^2$. Therefore we can write

$$\frac{n^2 h^2}{8 m_g L_{max}^2} = m_g c^2$$

Then from the equation above it follows that

$$m_g = \pm \frac{nh}{cL_{max}\sqrt{8}} \qquad (23)$$

whence we see that there is a *minimum value* for $m_g$ given by

$$m_{g(min)} = \pm \frac{h}{cL_{max}\sqrt{8}} \qquad (24)$$

The *relativistic* gravitational mass $M_g = m_g\left(1 - V^2/c^2\right)^{-1/2}$, defined in the Eqs.(4), shows that

$$M_{g(\min)} = m_{g(\min)} \qquad (25)$$

The *box normalization* leads to the conclusion that the *propagation number* $k = |\vec{k}| = 2\pi/\lambda$ is restricted to the values $k = 2\pi n/L$ . This is deduced assuming an *arbitrarily large but finite* cubical box of volume $L^3$ [8]. Thus, we have

$$L = n\lambda$$

From this equation, we conclude that

$$n_{max} = \frac{L_{max}}{\lambda_{min}}$$

and

$$L_{min} = n_{min}\lambda_{min} = \lambda_{min}$$

Since $n_{min} = 1$. Therefore, we can write that

$$L_{max} = n_{max} L_{min} \qquad (26)$$

From this equation, we thus conclude that

$$L = nL_{min} \qquad (27)$$

or

$$L = \frac{L_{max}}{n} \qquad (28)$$

Multiplying (27) and (28) by $\sqrt{3}$ and reminding that $d = L\sqrt{3}$ , we obtain

$$d = nd_{min} \qquad or \qquad d = \frac{d_{max}}{n} \qquad (29)$$

Equations above show that the length (and therefore the *space*) is *quantized*.

By analogy to (23) we can also conclude that

$$M_{g(max)} = \frac{n_{max} h}{cL_{min}\sqrt{8}} \qquad (30)$$

since the relativistic gravitational mass, $M_g = m_g\left(1 - V^2/c^2\right)^{-½}$, is just a multiple of $m_g$.

Equation (26) tells us that $L_{min} = L_{max}/n_{max}$. Thus, Eq.(30) can be rewritten as follows

$$M_{g(max)} = \frac{n_{max}^2 h}{cL_{max}\sqrt{8}} \qquad (31)$$

Comparison of (31) with (24) shows that

$$M_{g(\max)} = n_{\max}^2 m_{g(\min)} \qquad (32)$$

which leads to following conclusion that

$$M_g = n^2 m_{g(\min)} \qquad (33)$$

This equation shows that *the gravitational mass is quantized*.

Substitution of (33) into (13) leads to *quantization of gravity*, i.e.,

$$g = \frac{GM_g}{r^2} = n^2\left(\frac{Gm_{g(\min)}}{\left(r_{\max}/n\right)^2}\right) = $$
$$= n^4 g_{\min} \qquad (34)$$

From the Hubble's law, it follows that

$$V_{max} = \tilde{H}l_{max} = \tilde{H}\left(d_{max}/2\right)$$
$$V_{min} = \tilde{H}l_{min} = \tilde{H}\left(d_{min}/2\right)$$

whence

$$\frac{V_{max}}{V_{min}} = \frac{d_{max}}{d_{min}}$$

Equations (29) tell us that $d_{max}/d_{min} = n_{max}$. Thus the equation above gives

$$V_{min} = \frac{V_{max}}{n_{max}} \qquad (35)$$

which leads to following conclusion

$$V = \frac{V_{max}}{n} \qquad (36)$$

this equation shows that *velocity* is also quantized.

From this equation one concludes that we can have $V = V_{max}$ or $V = V_{max}/2$, but there is nothing in between. This shows clearly that $V_{max}$ cannot be equal to $c$ (speed of light in vacuum). Thus, it follows that

$$
\begin{array}{lll}
n = 1 & V = V_{max} & \\
n = 2 & V = V_{max}/2 & \\
n = 3 & V = V_{max}/3 & Tachyons \\
\ldots\ldots & \ldots\ldots\ldots\ldots & \\
n = n_x - 1 & V = V_{max}/\left(n_x - 1\right) & \\
\hline
n = n_x & V = V_{max}/n_x = c \ \leftarrow & \\
n = n_x + 1 & V = V_{max}/\left(n_x + 1\right) & Tardyons \\
n = n_x + 2 & V = V_{max}/\left(n_x + 2\right) & \\
\ldots\ldots & \ldots\ldots\ldots\ldots &
\end{array}
$$

*where* $n_x$ *is a big number.*

Then $c$ is the speed *upper limit* of the *Tardyons* and also the speed *lower limit* of the *Tachyons*. Obviously, this limit is *always the same in all inertial frames*. Therefore $c$ can be used as a *reference speed*, to which we may compare any speed $V$, as occurs for the relativistic factor $\sqrt{1-V^2/c^2}$. Thus, in this factor, $c$ does not refer to maximum propagation speed of the interactions such as some authors suggest; $c$ is just a speed limit which remains the same in any inertial frame.

The temporal coordinate $x^0$ of space-time is now $x^0 = V_{max}t$ ( $x^0 = ct$ is then obtained when $V_{max} \rightarrow c$ ). Substitution of $V_{max} = nV = n\left(\tilde{H}l\right)$ into this equation yields $t = x^0/V_{max} = \left(1/n\tilde{H}\right)\left(x^0/l\right)$. On the other hand, since $V = \tilde{H}l$ and $V = V_{max}/n$ we can write that $l = V_{max}\tilde{H}^{-1}/n$. Thus $\left(x^0/l\right) = \tilde{H}\left(nt\right) = \tilde{H}t_{max}$. Therefore, we can finally write

$$t = \left(1/n\tilde{H}\right)\left(x^0/l\right) = t_{max}/n \qquad (37)$$

which shows the quantization of *time*.

From Eqs. (27) and (37) we can easily conclude that the *spacetime is not continuous* it is *quantized*.

Now, let us go back to Eq. (20) which will be called the *gravitational* Hamiltonian to distinguish it from the *inertial* Hamiltonian $H_i$:

$$H_i = c\sqrt{p^2 + m_{i0}{}^2 c^2}. \qquad (38)$$

Consequently, Eq. (18) can be rewritten in the following form:

$$H_i - H_g = 2\Delta H_i \qquad (39)$$

where $\Delta H_i$ is the *variation on the inertial Hamiltonian* or *inertial kinetic energy*. A *momentum* variation $\Delta p$ yields a variation $\Delta H_i$ given by:

$$\Delta H_i = \sqrt{(p+\Delta p)^2 c^2 + m_{i0}{}^2 c^4} - \sqrt{p^2 c^2 + m_{i0}{}^2 c^4} \qquad (40)$$

By considering that the particle is *initially at rest* $(p=0)$. Then, Eqs. (20), (38) and (39) give respectively: $H_g = m_g c^2$, $H_i = m_{i0} c^2$ and

$$\Delta H_i = \left[\sqrt{1+\left(\frac{\Delta p}{m_{i0}c}\right)^2} - 1\right] m_{i0} c^2$$

By substituting $H_g$, $H_i$ and $\Delta H_i$ into Eq.(39), we get

$$m_g = m_{i0} - 2\left[\sqrt{1+\left(\frac{\Delta p}{m_{i0}c}\right)^2} - 1\right] m_{i0}. \qquad (41)$$

This is the *general expression of the correlation between the gravitational and inertial mass*. Note that for $\Delta p > m_{i0}c\left(\sqrt{5}/2\right)$, the value of $m_g$ becomes *negative*.

Equation (41) shows that $m_g$ decreases of $\Delta m_g$ for an increase of $\Delta p$. Thus, starting from (4) we obtain

$$p + \Delta p = \frac{\left(m_g - \Delta m_g\right) V}{\sqrt{1-(V/c)^2}}$$

By considering that the particle is *initially at rest* $(p=0)$, the equation above gives

$$\Delta p = \frac{\left(m_g - \Delta m_g\right) V}{\sqrt{1-(V/c)^2}}$$

From the Eq.(16) we obtain: $E_g = 2E_{i0} - E_i = 2E_{i0} - \left(E_{i0} + \Delta E_i\right) = E_{i0} - \Delta E_i$

However, Eq.(14) tells us that $-\Delta E_i = \Delta E_g$; what leads to $E_g = E_{i0} + \Delta E_g$ or $m_g = m_{i0} + \Delta m_g$. Thus, in the expression of $\Delta p$ we can replace $\left(m_g - \Delta m_g\right)$ for $m_{i0}$, i.e,

$$\Delta p = \frac{m_{i0} V}{\sqrt{1-(V/c)^2}}$$

We can therefore write

$$\frac{\Delta p}{m_{i0}c} = \frac{V/c}{\sqrt{1-(V/c)^2}} \qquad (42)$$

By substitution of the expression above into Eq. (41), we thus obtain:

$$m_g = m_{i0} - 2\left[\left(1 - V^2/c^2\right)^{-\frac{1}{2}} - 1\right] m_{i0} \qquad (43)$$

For $V=0$ we obtain $m_g = m_{i0}$. Then,

$$m_{g(min)} = m_{i0(min)}$$

Substitution of $m_{g(min)}$ into the *quantized* expression of $M_g$ (Eq. (33)) gives

$$M_g = n^2 m_{i0(min)}$$

where $m_{i0(min)}$ is the *elementary quantum of inertial mass* to be determined.

For $V=0$, the *relativistic* expression $M_g = m_g \big/ \sqrt{1-V^2/c^2}$ becomes $M_g = M_{g0} = m_{g0}$. However, Eq. (43) shows that $m_{g0} = m_{i0}$. Thus, the *quantized* expression of $M_g$ reduces to

$$m_{i0} = n^2 m_{i0(min)}$$

In order to define the *inertial quantum number*, we will change *n* in the expression above for $n_i$. Thus we have

$$m_{i0} = n_i^2 m_{i0(\min)} \qquad (44)$$

which shows the quantization of *inertial mass*; $n_i$ is the *inertial quantum number*.

We will change $n$ in the quantized expression of $M_g$ for $n_g$ in order to define the *gravitational quantum number*. Thus, we have

$$M_g = n_g^2 m_{i0(min)} \qquad (44a)$$

Finally, by substituting $m_g$ given by Eq. (43) into the relativistic expression of $M_g$, we readily obtain

$$M_g = \frac{m_g}{\sqrt{1-V^2/c^2}} = $$
$$= M_i - 2\left[\left(1-V^2/c^2\right)^{-\frac{1}{2}} - 1\right]M_i \qquad (45)$$

By expanding in power series and neglecting infinitesimals, we arrive at:

$$M_g = \left(1 - \frac{V^2}{c^2}\right)M_i \qquad (46)$$

Thus, the well-known expression for the simple pendulum period, $T = 2\pi\sqrt{(M_i/M_g)(l/g)}$, can be rewritten in the following form:

$$T = 2\pi\sqrt{\frac{l}{g}}\left(1 + \frac{V^2}{2c^2}\right) \qquad for \quad V \ll c$$

Now, it is possible to learn why Newton's experiments using simple penduli do not find any difference between $M_g$ and $M_i$. The reason is due to the fact that, in the case of penduli, the ratio $V^2/2c^2$ is less than $10^{-17}$, which is much smaller than the accuracy of the mentioned experiments.

Newton's experiments have been improved upon (one part in 60,000) by Friedrich Wilhelm Bessel (1784–1846). In 1890, Eötvos confirmed Newton's results with accuracy of one part in $10^7$. Posteriorly, Eötvos experiment has been repeated with accuracy of one part in $10^9$. In 1963, the experiment was repeated with an even greater accuracy, one part in $10^{11}$. The result was the same previously obtained.

In all these experiments, the ratio $V^2/2c^2$ is less than $10^{-17}$, which is much smaller than the accuracy of $10^{-11}$ obtained in the previous more precise experiment.

Then, we arrive at the conclusion that all these experiments say nothing in regard to the relativistic behavior of masses in relative motion.

Let us now consider a planet in the Sun's gravitational field to which, in the absence of external forces, we apply Lagrange's equations. We arrive at the well-known equation:

$$\left(\frac{dr}{dt}\right)^2 + r^2\left(\frac{d\varphi}{dt}\right)^2 - \frac{2GM_i}{r} = \mathrm{E}$$

$$r^2\frac{d\varphi}{dt} = \mathrm{h}$$

where $M_i$ is the inertial mass of the Sun. The term $\mathrm{E} = -GM_i/a$, as we know, is called the *energy constant*; $a$ is the semiaxis major of the Kepler-ellipse described by the planet around the Sun.

By replacing $M_i$ into the differential equation above for the expression given by Eq. (46), and expanding in power series, neglecting infinitesimals, we arrive, at:

$$\left(\frac{dr}{dt}\right)^2 + r^2\left(\frac{d\varphi}{dt}\right)^2 - \frac{2GM_g}{r} = \mathrm{E} + \frac{2GM_g}{r}\left(\frac{V^2}{c^2}\right)$$

Since $V = \omega r = r(d\varphi/dt)$, we get

$$\left(\frac{dr}{dt}\right)^2 + r^2\left(\frac{d\varphi}{dt}\right)^2 - \frac{2GM_g}{r} = \mathrm{E} + \frac{2GM_g r}{c^2}\left(\frac{d\varphi}{dt}\right)^2$$

which is the *Einsteinian equation of the planetary motion*.

Multiplying this equation by $(dt/d\varphi)^2$ and remembering that $(dt/d\varphi)^2 = r^4/\mathrm{h}^2$, we obtain

$$\left(\frac{dr}{d\varphi}\right)^2 + r^2 = \mathrm{E}\left(\frac{r^4}{\mathrm{h}^2}\right) + \frac{2GM_g r^3}{\mathrm{h}^2} + \frac{2GM_g r}{c^2}$$

Making $r = 1/u$ and multiplying both members of the equation by $u^4$, we get

$$\left(\frac{du}{d\varphi}\right)^2+u^2=\frac{\mathrm{E}}{\mathrm{h}^2}+\frac{2GM_g u}{\mathrm{h}^2}+\frac{2GM_g u^3}{c^2}$$

This leads to the following expression

$$\frac{d^2u}{d\varphi^2}+u=\frac{GM_g}{\mathrm{h}^2}\left(1+\frac{3u^2\mathrm{h}^2}{c^2}\right)$$

In the absence of term $3\mathrm{h}^2u^2/c^2$, the integration of the equation should be immediate, leading to $2\pi$ period. In order to obtain the value of the perturbation we can use any of the well-known methods, which lead to an angle $\varphi$, for two successive perihelions, given by

$$2\pi+\frac{6G^2M_g^2}{c^2\mathrm{h}^2}$$

Calculating per century, in the case of Mercury, we arrive at an angle of 43” for the perihelion advance.

This result is the best theoretical proof of the accuracy of Eq. (45).

Now consider a relativistic particle inside a gravitational field. The condition for it to escape from the gravitational field is that its *inertial kinetic energy* becomes equal to the absolute value of the *gravitational energy of the field*, which is given by

$$U(r)=-\frac{Gm_g m'_g}{r\sqrt{1-V^2/c^2}}$$

By substituting $m_g$ and $m'_g$ given by Eq. (43) into this expression, and assuming that the velocity $V$ of the particle that creates the field is small $(V \ll c)$, we get

$$U(r)=-\frac{Gm_{i0}m'_{i0}}{r}\left[1-2\left(\frac{1}{\sqrt{1-V'^2/c^2}}-1\right)\right]$$

Due to the fact that the velocity $V'$ of particle inside the field is relativistic, then its *inertial kinetic energy* is given by

$$E'_{Ki}=(M'_{i0}-m'_{i0})c^2=\left(\frac{1}{\sqrt{1-V'^2/c^2}}-1\right)m'_{i0}c^2$$

Thus, by making $E'_{Ki}=|U(r)|$ we obtain

$$\left(\frac{1}{\sqrt{1-V'^2/c^2}}-1\right)=\left|-\frac{Gm_{i0}}{c^2r}\left[1-2\left(\frac{1}{\sqrt{1-V'^2/c^2}}-1\right)\right]\right|$$

By multiplying both members of this equation by $\sqrt{1-V'^2/c^2}$ the result is

$$\left(1-\sqrt{1-V'^2/c^2}\right)=\left|-\frac{Gm_{i0}}{c^2r}\left(3\sqrt{1-V'^2/c^2}-2\right)\right|$$

For $V'=c$ we obtain

$$1=\frac{2Gm_{i0}}{c^2r}$$

whence

$$r=\frac{2Gm_{i0}}{c^2}$$

Thus, we see that there is a limit at $r=2Gm_{i0}/c^2$. This is the called *Schwzarzschilds’ radius*, which defines the called *event horizon* of a Black-hole.

In the beginning of this article, it was shown that, besides the *inertial kinetic energy* $E_{Ki}$, there is also the *gravitational kinetic energy* $E_{Kg}$, which in agreement with Eq. (12), is expressed by

$$E_{Kg}=\frac{m_g}{m_i}E_{Ki}$$

Thus, we can write that

$$E'_{Kg}=\frac{m'_g}{m'_i}E'_{Ki}=\left(\frac{1}{\sqrt{1-V^2/c^2}}-1\right)m'_gc^2$$

By multiplying both members of this equation by $\phi=-Gm_{i0}/r$, we can write that

$$E'_{Kg}=-\frac{Gm_{i0}m'_g}{r\sqrt{1-V^2/c^2}}\left(-\frac{rc^2}{Gm_{i0}}\right)-m'_gc^2$$

As we have shown, in the case of *photons* $(V'=c)$ we have

$$1=\frac{2Gm_{i0}}{c^2r}$$

Thus, the expression of $E'_{Kg}$ becomes

$$E'_{Kg}=\frac{Gm_{i0}m'_g}{r\sqrt{1-V^2/c^2}}-m'_gc^2$$

If the energy of the photon $m'_gc^2$ is much less than the energy of the *relativistic* gravitational field, the equation above can be rewritten in the following form

$$E'_{Kg} = \frac{Gm_{i0} m'_g}{r\sqrt{1-V^2/c^2}}$$

Substitution of the expression of $E'_{Kg}$ into this expression gives

$$\left(\frac{1}{\sqrt{1-V^2/c^2}} - 1\right)c^2 = \frac{Gm_{i0}}{r\sqrt{1-V^2/c^2}}$$

which simplifies to

$$\sqrt{1-V^2/c^2} = 1 - \frac{Gm_{i0}}{rc^2} = 1 + \frac{\phi}{c^2}$$

For $V << c$ this expression gives

$$V^2 = \frac{2Gm_{i0}}{r}$$

By substituting this expression of $V^2$ into the equation of $g$, obtained at the beginning of this work, where we replace $m_g$ by $m_{i0}$ because $m_g \cong m_{i0}$ in the case of $V << c$, the result is

$$g = \frac{\partial \Phi}{\partial r} = \frac{Gm_{i0}}{r^2\sqrt{1-2Gm_{i0}/rc^2}}$$

Note that for $r = 2Gm_{i0}/c^2$ there occurs a singularity, $g \to \infty$.

Substitution of $\sqrt{1-V^2/c^2} = 1 + \phi/c^2$ into the well-known expression below

$$T = t\sqrt{1-V^2/c^2}$$

which expresses the relativistic correlation between *own time* ($T$) and *universal time* ($t$), gives

$$T = t\left(1 + \frac{\phi}{c^2}\right)$$

It is known from the Optics that the frequency of a wave, measured in units of *universal time,* remains constant during its propagation, and that it can be expressed by means of the following relation:

$$\omega_0 = \frac{\partial \psi}{\partial t}$$

where $d\psi/dt$ is the derivative of the *eikonal* $\psi$ with respect to the time.

On the other hand, the frequency of the wave measured in units of *own time* is given by

$$\omega = \frac{\partial \psi}{\partial T}$$

Thus, we conclude that

$$\frac{\omega}{\omega_0} = \frac{\partial t}{\partial T}$$

whence we obtain

$$\frac{\omega}{\omega_0} = \frac{t}{T} = \frac{1}{\left(1 + \frac{\phi}{c^2}\right)}$$

By expanding in power series, neglecting infinitesimals, we arrive at:

$$\omega = \omega_0\left(1 - \frac{\phi}{c^2}\right)$$

In this way, if a light ray with a frequency $\omega_0$ is emitted from a point where the gravitational potential is $\phi_1$, it will have a frequency $\omega_1$. Upon reaching a point where the gravitational potential is $\phi_2$ its frequency will be $\omega_2$. Then, according to equation above, it follows that

$$\omega_1 = \omega_0\left(1 - \frac{\phi_1}{c^2}\right) \quad and \quad \omega_2 = \omega_0\left(1 - \frac{\phi_2}{c^2}\right)$$

Thus, from point 1 to point 2 the frequency will be shifted in the interval $\Delta\omega = \omega_1 - \omega_2$, given by

$$\Delta\omega = \omega_0\left(\frac{\phi_2 - \phi_1}{c^2}\right)$$

If $\Delta\omega < 0$, $(\phi_1 > \phi_2)$, the shift occurs in the direction of the decreasing frequencies (*red-shift*). If $\Delta\omega > 0$, $(\phi_1 < \phi_2)$ the *blue-shift* occurs.

Let us now consider another consequence of the existence of correlation between $M_g$ and $M_i$.

*Lorentz's force* is usually written in the following form:

$$d\,\vec{p}/dt = q\vec{E} + q\vec{V} \times \vec{B}$$

where $\vec{p} = m_{i0}\vec{V}/\sqrt{1-V^2/c^2}$. However, Eq.(4) tells us that $\vec{p} = m_g V/\sqrt{1-V^2/c^2}$. Therefore, the expressions above must be corrected by multiplying its members by $m_g/m_{i0}$, i.e.,

$$\vec{p}\,\frac{m_g}{m_{i0}} = \frac{m_g}{m_{i0}}\,\frac{m_{i0}\vec{V}}{\sqrt{1-V^2/c^2}} = \frac{m_g\vec{V}}{\sqrt{1-V^2/c^2}} = \vec{p}$$

and

$$\frac{d\vec{p}}{dt}=\frac{d}{dt}\left(\vec{p}\frac{m_g}{m_{i0}}\right)=\left(q\vec{E}+q\vec{V}\times\vec{B}\right)\frac{m_g}{m_{i0}}$$

That is now the *general expression* for Lorentz's force. Note that it depends on $m_g$.

When the force is perpendicular to the speed, Eq. (5) gives $d\vec{p}/dt=m_g\left(d\vec{V}/dt\right)/\sqrt{1-V^2/c^2}$ .By comparing with Eq.(46), we thus obtain

$$\left(m_{i0}/\sqrt{1-V^2/c^2}\right)\left(d\vec{V}/dt\right)=q\vec{E}+q\vec{V}\times\vec{B}$$

Note that this equation is the expression of an *inertial* force.

Starting from this equation, well-known experiments have been carried out in order to verify the relativistic expression: $m_i/\sqrt{1-V^2/c^2}$ . In general, the *momentum* variation $\Delta p$ is expressed by $\Delta p=F\Delta t$ where $F$ is the applied force during a time interval $\Delta t$. Note that there is no restriction concerning the *nature* of the force $F$, i.e., it can be mechanical, electromagnetic, etc.

For example, we can look on the *momentum* variation $\Delta p$ as due to absorption or emission of *electromagnetic energy* by the particle (by means of *radiation* and/or by means of *Lorentz's force* upon the *charge* of the particle).

In the case of radiation (any type), $\Delta p$ can be obtained as follows. It is known that the *radiation pressure* , $dP$, upon an area $dA=dxdy$ of a volume $dV=dxdydz$ of a particle( the incident radiation normal to the surface $dA$ )is equal to the energy $dU$ absorbed per unit volume $\left(dU/dV\right)$.i.e.,

$$dP=\frac{dU}{dV}=\frac{dU}{dxdydz}=\frac{dU}{dAdz} \qquad (47)$$

Substitution of $dz=vdt$ ( $v$ is the speed of radiation) into the equation above gives

$$dP=\frac{dU}{dV}=\frac{\left(dU/dAdt\right)}{v}=\frac{dD}{v} \qquad (48)$$

Since $dPdA=dF$ we can write:

$$dFdt=\frac{dU}{v} \qquad (49)$$

However we know that $dF=dp/dt$, then

$$dp=\frac{dU}{v} \qquad (50)$$

From Eq. (48), it follows that

$$dU=dPdV=\frac{dVdD}{v} \qquad (51)$$

Substitution into (50) yields

$$dp=\frac{dVdD}{v^2} \qquad (52)$$

or

$$\int_0^{\Delta p}dp=\frac{1}{v^2}\int_0^D\int_0^V dVdD$$

whence

$$\Delta p=\frac{VD}{v^2} \qquad (53)$$

This expression is general for all types of waves including *non-electromagnetic waves* such as *sound waves.* In this case, $v$ in Eq.(53), will be the speed of sound in the medium and $D$ the *intensity* of the sound radiation.

In the case of *electromagnetic waves*, the Electrodynamics tells us that $v$ will be given by

$$v=\frac{dz}{dt}=\frac{\omega}{\kappa_r}=\frac{c}{\sqrt{\frac{\varepsilon_r\mu_r}{2}\left(\sqrt{1+\left(\sigma/\omega\varepsilon\right)^2}+1\right)}}$$

where $k_r$ is the real part of the *propagation vector* $\vec{k}$ ; $k=\left|\vec{k}\right|=k_r+ik_i$ ; $\varepsilon$ , $\mu$ and $\sigma$, are the electromagnetic characteristics of the medium in which the incident (or emitted) radiation is propagating ( $\varepsilon=\varepsilon_r\varepsilon_0$ where $\varepsilon_r$ is the *relative dielectric permittivity* and $\varepsilon_0=8.854\times10^{-12}F/m$ ; $\mu=\mu_r\mu_0$ where $\mu_r$ is the *relative magnetic permeability* and $\mu_0=4\pi\times10^{-7}H/m$; $\sigma$ is the *electrical conductivity*). For an *atom* inside a body, the incident (or emitted) radiation on this atom will be propagating inside the body, and consequently, $\sigma=\sigma_{body}$, $\varepsilon=\varepsilon_{body}$, $\mu=\mu_{body}$.

It is then evident that the *index of refraction* $n_r=c/v$ will be given by

$$n_r = \frac{c}{v} = \sqrt{\frac{\varepsilon_r \mu_r}{2}\left(\sqrt{1+(\sigma/\omega\varepsilon)^2}+1\right)} \qquad (54)$$

On the other hand, from Eq. (50) follows that

$$\Delta p = \frac{U}{v}\left(\frac{c}{c}\right) = \frac{U}{c} n_r$$

Substitution into Eq. (41) yields

$$m_g = \left\{1 - 2\left[\sqrt{1+\left(\frac{U}{m_{i0}c^2} n_r\right)^2} - 1\right]\right\} m_{i0} \qquad (55)$$

If the body is *also* rotating, with an angular speed $\omega$ around its central axis, then it acquires an additional energy equal to its rotational energy $\left(E_k = \frac{1}{2} I\omega^2\right)$. Since this is an increase in the internal energy of the body, and this energy is basically electromagnetic, we can assume that $E_k$, such as $U$, corresponds to an amount of electromagnetic energy absorbed by the body. Thus, we can consider $E_k$ as an increase $\Delta U = E_k$ in the electromagnetic energy $U$ absorbed by the body. Consequently, in this case, we must replace $U$ in Eq. (55) for $(U + \Delta U)$. If $U << \Delta U$, the Eq. (55) reduces to

$$m_g \cong \left\{1 - 2\left[\sqrt{1+\left(\frac{I\omega^2 n_r}{2m_{i0}c^2}\right)^2} - 1\right]\right\} m_{i0}$$

For $\sigma << \omega\varepsilon$, Eq. (54) shows that $n_r = c/v = \sqrt{\varepsilon_r \mu_r}$ and $n_r = \sqrt{\mu\sigma c^2/4\pi f}$ in the case of $\sigma >> \omega\varepsilon$. In this case, if the body is a *Mumetal* disk $\left(\mu_r = 105{,}000\, at\, 100 gauss; \sigma = 2.1\times10^7 S.m^{-1}\right)$ with radius $R$, $\left(I = \frac{1}{2} m_{i0} R^2\right)$, the equation above shows that the *gravitational mass* of the disk is

$$m_{g(disk)} \cong \left\{1 - 2\left[\sqrt{1 + 1.12\times10^{-13}\frac{R^4\omega^4}{f}} - 1\right]\right\} m_{i0(disk)}$$

Note that the effect of the electromagnetic radiation applied upon the disk is highly relevant, because in the absence of this radiation the index of refraction, present in equations above, becomes equal to 1. Under these circumstances, the possibility of strongly reducing the gravitational mass of the disk practically disappears. In addition, the equation above shows that, in practice, the frequency $f$ of the radiation cannot be high, and that *extremely-low frequencies* (ELF) are most appropriated. Thus, if the frequency of the electromagnetic radiation applied upon the disk is $f = 0.1Hz$ (See Fig. I (a)) and the radius of the disk is $R = 0.15m$, and its angular speed $\omega = 1.05\times10^4 rad/s (\sim 100{,}000\ rpm)$, the result is

$$m_{g(disk)} \cong -2.6 m_{i0(disk)}$$

This shows that the gravitational mass of a body can also be controlled by means of its *angular velocity*.

In order to satisfy the condition $U << \Delta U$, we must have $dU/dt << d\Delta U/dt$, where $P_r = dU/dt$ is the radiation power. By integrating this expression, we get $\overline{U} = P_r/2f$. Thus we can conclude that, for $U << \Delta U$, we must have $P_r/2f << \frac{1}{2} I\omega^2$, i.e.,

$$P_r << I\omega^2 f$$

By dividing both members of the expression above by the area $S = 4\pi r^2$, we obtain

$$D_r << \frac{I\omega^2 f}{4\pi r^2}$$

Therefore, this is the necessary condition in order to obtain $U << \Delta U$. In the case of the Mumetal disk, we must have

$$D_r << 10^5/r^2 \qquad \left(watts/m^2\right)$$

From Electrodynamics, we know that a radiation with frequency $f$ propagating within a material with electromagnetic characteristics $\varepsilon$, $\mu$ and $\sigma$ has the amplitudes of its waves attenuated by $e^{-1}$=0.37 (37%) when it penetrates a distance $z$, given by *

$$z = \frac{1}{\omega\sqrt{\frac{1}{2}\varepsilon\mu\left(\sqrt{1+(\sigma/\omega\varepsilon)^2} - 1\right)}}$$

---

* Quevedo, C. P. (1978) *Eletromagnetismo* McGraw-Hill, p.269-270.

For $\sigma >> \omega\varepsilon$, the equation above reduces to

$$z = \frac{1}{\sqrt{\pi\mu f \sigma}}$$

In the case of the *Mumetal* subjected to an ELF radiation with frequency $f = 0.1Hz$, the value is $z = 1.07mm$. Obviously, the thickness of the Mumetal disk must be less than this value.

Equation (55) is general for all types of electromagnetic fields including *gravitoelectromagnetic* fields (See Fig. I (b)).

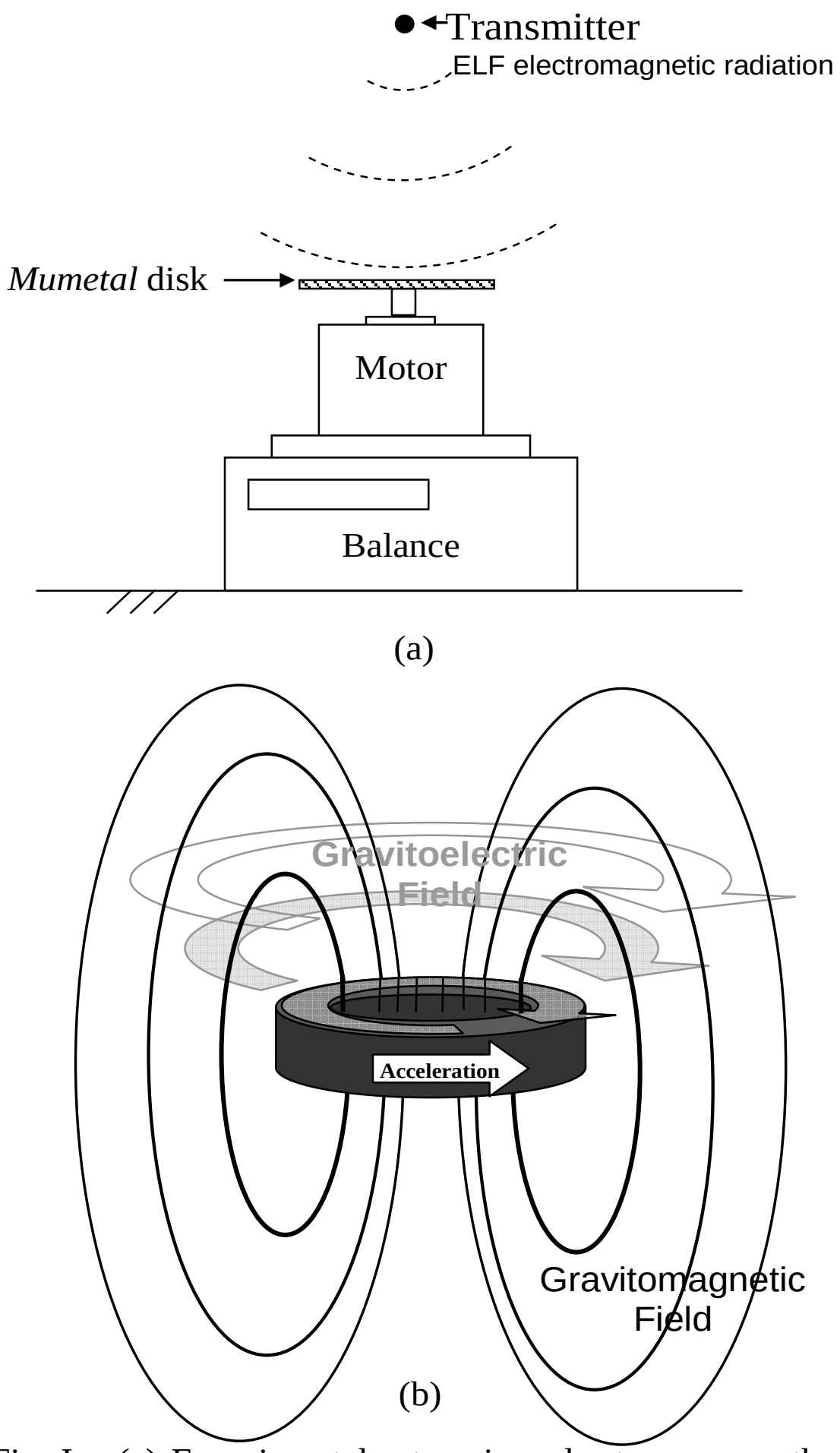

Fig. I – (a) Experimental set-up in order to measure the gravitational mass decreasing in the rotating Mumetal disk. A sample connected to a dynamometer can measure the decreasing of gravity above the disk. (b) Gravitoelectromagnetic Field.

The Maxwell-like equations for weak gravitational fields are [9]

$$\nabla . \boldsymbol{D_G} = -\rho$$

$$\nabla \times \boldsymbol{E_G} = -\frac{\partial \boldsymbol{B_G}}{\partial \boldsymbol{t}}$$

$$\nabla . \boldsymbol{B_G} = 0$$

$$\nabla \times \boldsymbol{H_G} = -\boldsymbol{j_G} + \frac{\partial \boldsymbol{D_G}}{\partial \boldsymbol{t}}$$

where $D_G = 4\varepsilon_{rG}\varepsilon_{0G}E_G$ is the *gravitodisplacement* field ( $\varepsilon_{rG}$ is the *gravitoelectric relative permittivity* of the medium; $\varepsilon_{0G}$ is the *gravitoelectric permittivity* for free space and $\boldsymbol{E_G} = \boldsymbol{g}$ is the *gravitoelectric* field intensity); $\rho$ is the density of local rest mass in the local rest frame of the matter; $B_G = \mu_{rG}\mu_{0G}H_G$ is the *gravitomagnetic* field ($\mu_{rG}$ is the *gravitomagnetic relative permeability*, $\mu_{0G}$ is the *gravitomagnetic permeability* for free space and $\boldsymbol{H_G}$ is the *gravitomagnetic* field intensity; $\boldsymbol{j_G} = -\sigma_G \boldsymbol{E_G}$ is the local rest-mass current density in this frame ($\sigma_G$ is the *gravitoelectric conductivity* of the medium).

Then, for *free space* we can write that

$$D_G = 4\varepsilon_{0G}E_G = 4\varepsilon_{0G}\,g = 4\varepsilon_{0G}\left(\frac{GM}{r^2}\right)$$

But from the electrodynamics we know that

$$D = \varepsilon E = \frac{q}{4\pi r^2}$$

By analogy we can write that

$$D_G = \frac{M_g}{4\pi r^2}$$

By comparing this expression with the previous expression of $D_G$ , we get

$$\varepsilon_{0G} = \frac{1}{16\pi G} = 2.98\times10^{8}\,kg^2.N^{-1}.m^{-2}$$

which is the expression of the *gravitoelectric permittivity* for free space.

The *gravitomagnetic permeability* for free space [10,11] is

$$\mu_{0G} = \frac{16\pi G}{c^2} = 3.73\times10^{-26}\,m/kg$$

We then convert Maxwell-like equations

for weak gravity into a wave equation for free space in the standard way. We conclude that *the speed* of *Gravitational Waves* in free space is

$$v = \frac{1}{\sqrt{\varepsilon_{0G}\mu_{0G}}} = c$$

This means that *both electromagnetic and gravitational plane waves propagate at the free space with the same speed*.

Thus, the impedance for free space is

$$Z_G = \frac{E_G}{H_G} = \sqrt{\mu_{0G}/\varepsilon_{0G}} = \mu_{0G}c = \frac{16\pi G}{c}$$

and the Poynting-like vector is

$$\vec{S} = \vec{E}_G \times \vec{H}_G$$

For a plane wave propagating in the vacuum, we have $|E_G| = Z_G|H_G|$. Then, it follows that

$$\left|\vec{S}\right| = \frac{1}{2Z_G}\left|\vec{E}_G\right|^2 = \frac{\omega^2}{2Z_G}\left|\vec{h}\right|^2 = \frac{c^2\omega^2}{32\pi G}\left|h_{0i}\right|^2$$

which is *the power per unit area* of a harmonic plane wave of angular frequency $\omega$.

In classical electrodynamics the density of energy in an *electromagnetic* field, $W_e$, has the following expression

$$W_e = \tfrac{1}{2}\varepsilon_r\varepsilon_0 E^2 + \tfrac{1}{2}\mu_r\mu_0 H^2$$

In analogy with this expression we define the energy density in a *gravitoelectromagnetic* field, $W_G$, as follows

$$W_G = \tfrac{1}{2}\varepsilon_{rG}\varepsilon_{0G}E_G^2 + \tfrac{1}{2}\mu_{rG}\mu_{0G}H_G^2$$

For free space we obtain

$$\mu_{rG} = \varepsilon_{rG} = 1$$

$$\varepsilon_{0G} = 1/\mu_{0G}\,c^2$$

$$E_G/H_G = \mu_{0G}c$$

and

$$\boldsymbol{B}_G = \mu_{0G}\,\boldsymbol{H}_G$$

Thus, we can rewrite the equation of $W_G$ as follows

$$W_G = \tfrac{1}{2}\left(\frac{1}{\mu_{0G}c^2}\right)c^2B_G^2 + \tfrac{1}{2}\mu_{0G}\left(\frac{B_G}{\mu_{0G}}\right)^2 = \frac{B_G^2}{\mu_{0G}}$$

Since $U_G = W_G V$, ($V$ is the *volume* of the particle) and $n_r = 1$ for free space we can write (55) in the following form

$$m_g = \left\{1 - 2\left[\sqrt{1 + \left(\frac{W_G}{\rho\, c^2}\right)^2} - 1\right]\right\}m_{i0}$$

$$= \left\{1 - 2\left[\sqrt{1 + \left(\frac{B_G^2}{\mu_{0G}\rho\, c^2}\right)^2} - 1\right]\right\}m_{i0} \quad (55a)$$

where $\rho = m_{i0}/\mathrm{V}$.

This equation shows how the gravitational mass of a particle is altered by a *gravitomagnetic* field.

A gravitomagnetic field, according to Einstein's theory of general relativity, arises from moving matter (matter current) just as an ordinary magnetic field arises from moving charges. The Earth rotation is the source of a very weak gravitomagnetic field given by

$$B_{G,Earth} = -\frac{\mu_{0G}}{16\pi}\left(\frac{M\omega}{r}\right)_{Earth} \approx 10^{-14}\, rad.s^{-1}$$

Perhaps ultra-fast rotating stars can generate very strong gravitomagnetic fields, which can make the gravitational mass of particles inside and near the star *negative*. According to (55a) this will occur if $B_G > 1.06c\sqrt{\mu_{0G}\rho}$. Usually, however, gravitomagnetic fields produced by *normal* matter are very weak.

Recently Tajmar, M. et al., [12] have proposed that in addition to the *London moment*, $B_L$, ($B_L = -\left(2m^*/e^*\right)\omega \cong 1.1\times10^{-11}\omega$ ; $m^*$ and $e^*$ are the Cooper-pair mass and charge respectively), a rotating superconductor should exhibit also a large *gravitomagnetic* field, $B_G$, to explain an apparent mass increase of Niobium Cooper-pairs discovered by Tate et al[13,14]. According to Tajmar and Matos [15], in the case of *coherent* matter, $B_G$ is given by: $B_G = -2\omega\rho_c\mu_{0G}\lambda_{gr}^2$ where $\rho_c$ is the mass density of *coherent* matter and $\lambda_{gr}$ is the *graviphoton* wavelength. By choosing $\lambda_{gr}$ proportional to the local density of *coherent* matter, $\rho_c$. i.e.,

$$\frac{1}{\lambda_{gr}^2}=\left(\frac{m_{gr}c}{\hbar}\right)=\mu_{0G}\rho_c$$

we obtain

$$B_G=-2\omega\rho_c\mu_{0G}\lambda_{gr}^2=-2\omega\rho_c\mu_{0G}\left(\frac{1}{\mu_{0G}\rho_c}\right)=$$
$$=-2\omega$$

and the graviphoton mass, $m_{gr}$, is

$$m_{gr}=\mu_{0G}\rho_c\hbar/c$$

Note that if we take the case of *no* local sources of *coherent* matter $(\rho_c=0)$, the graviphoton mass will be *zero*. However, graviphoton will have non-zero mass inside coherent matter $(\rho_c\neq 0)$. This can be interpreted as a consequence of the graviphoton gaining mass inside the superconductor via the Higgs mechanism due to the breaking of gauge symmetry.

It is important to note that the *minus* sign in the expression for $B_G$ can be understood as due to the change from the normal to the coherent state of matter, i.e., a switch between real and *imaginary* values for the particles inside the material when going from the normal to the coherent state of matter. Consequently, in this case the variable $U$ in (55) must be replaced by $iU_G$ and not by $U_G$ only. Thus we obtain

$$m_g=\left\{1-2\left[\sqrt{1-\left(\frac{U_G}{m_{i0}c^2}n_r\right)^2}-1\right]\right\}m_{i0}\qquad(55b)$$

Since $U_G=W_GV$, we can write (55b) for $n_r=1$, in the following form

$$m_g=\left\{1-2\left[\sqrt{1-\left(\frac{W_G}{\rho_c c^2}\right)^2}-1\right]\right\}m_{i0}$$
$$=\left\{1-2\left[\sqrt{1-\left(\frac{B_G^2}{\mu_{0G}\rho_c c^2}\right)^2}-1\right]\right\}m_{i0}\,(55c)$$

where $\rho_c=m_{i0}/\mathrm{V}$ is the local density of *coherent* matter.

Note the different sign (inside the square root) with respect to (55a).

By means of (55c) it is possible to check the changes in the gravitational mass of the *coherent part* of a given material (e.g. the Cooper-pair fluid). Thus for the *electrons* of the Cooper-pairs we have

$$m_{ge}=m_{ie}+2\left[1-\sqrt{1-\left(\frac{B_G^2}{\mu_{0G}\rho_e c^2}\right)^2}\right]m_{ie}=$$
$$=m_{ie}+2\left[1-\sqrt{1-\left(\frac{4\omega^2}{\mu_{0G}\rho_e c^2}\right)^2}\right]m_{ie}=$$
$$=m_{ie}+\chi_e m_{ie}$$

where $\rho_e$ is the mass density of the electrons.

In order to check the changes in the gravitational mass of *neutrons* and *protons* (non-coherent part) inside the superconductor, we must use Eq. (55a) and $B_G=-2\omega\rho\mu_{0G}\lambda_{gr}^2$ [Tajmar and Matos, op.cit.]. Due to $\mu_{0G}\rho_c\lambda_{gr}^2=1$, that expression of $B_G$ can be rewritten in the following form

$$B_G=-2\omega\rho\mu_{0G}\lambda_{gr}^2=-2\omega(\rho/\rho_c)$$

Thus we have

$$m_{gn}=m_{in}-2\left[\sqrt{1+\left(\frac{B_G^2}{\mu_{0G}\rho_n c^2}\right)^2}-1\right]m_{in}=$$
$$=m_{in}-2\left[\sqrt{1+\left(\frac{4\omega^2(\rho_n/\rho_c)^2}{\mu_{0G}\rho_n c^2}\right)^2}-1\right]m_{in}=$$
$$=m_{in}-\chi_n m_{in}$$

$$m_{gp}=m_{ip}-2\left[\sqrt{1+\left(\frac{B_G^2}{\mu_{0G}\rho_p c^2}\right)^2}-1\right]m_{ip}=$$
$$=m_{ip}-2\left[\sqrt{1+\left(\frac{4\omega^2(\rho_p/\rho_c)^2}{\mu_{0G}\rho_p c^2}\right)^2}-1\right]m_{ip}=$$
$$=m_{ip}-\chi_p m_{ip}$$

where $\rho_n$ and $\rho_p$ are the mass density of *neutrons* and *protons* respectively.

In Tajmar's experiment, induced accelerations fields outside the superconductor in the order of $100\mu g$, at angular velocities of about $500 rad.s^{-1}$ were observed.

Starting from $g = Gm_{g(initial)}/r$ we can write that $g + \Delta g = G(m_{g(initial)} + \Delta m_g)/r$. Then we get $\Delta g = G\Delta m_g/r$. For $\Delta g = \eta g = \eta G m_{g(initial)}/r$ it follows that $\Delta m_g = \eta m_{g(initial)} = \eta m_i$. Therefore a variation of $\Delta g = \eta g$ corresponds to a gravitational mass variation $\Delta m_g = \eta m_{i0}$. Thus $\Delta g \approx 100\mu g = 1\times10^{-4} g$ corresponds to

$$\Delta m_g \approx 1\times10^{-4} m_{i0}$$

On the other hand, the total gravitational mass of a particle can be expressed by

$$m_g = N_n m_{gn} + N_p m_{gp} + N_e m_{ge} + N_p \Delta E/c^2 =$$
$$N_n(m_{in} - \chi_n m_{in}) + N_p(m_{ip} - \chi_p m_{ip}) +$$
$$+ N_e(m_{ie} - \chi_e m_{ie}) + N_p \Delta E/c^2 =$$
$$= (N_n m_{in} + N_p m_{ip} + N_e m_{ie}) + N_p \Delta E/c^2 -$$
$$-(N_n \chi_n m_{in} + N_p \chi_p m_{ip} + N_e \chi_e m_{ie}) + N_p \Delta E/c^2 =$$
$$= m_i - (N_n \chi_n m_{in} + N_p \chi_p m_{ip} + N_e \chi_e m_{ie}) + N_p \Delta E/c^2$$

where $\Delta E$ is the interaction energy; $N_n$, $N_p$, $N_e$ are the number of neutrons, protons and electrons respectively. Since $m_{in} \cong m_{ip}$ and $\rho_n \cong \rho_p$ it follows that $\chi_n \cong \chi_p$ and consequently the expression of $m_g$ reduces to

$$m_g \cong m_{i0} - (2N_p \chi_p m_{ip} + N_e \chi_e m_{ie}) + N_p \Delta E/c^2 \quad (55d)$$

Assuming that $N_e \chi_e m_{ie} << 2N_p \chi_p m_{ip}$ and $N_p \Delta E/c^2 << 2N_p \chi_p m_{ip}$ Eq. (55d) reduces to

$$m_g \cong m_{i0} - 2N_p \chi_p m_{ip} = m_i - \chi_p m_i \quad (55e)$$

or

$$\Delta m_g = m_g - m_{i0} = -\chi_p m_{i0}$$

By comparing this expression with $\Delta m_g \approx 1\times10^{-4} m_i$ which has been obtained from Tajmar's experiment, we conclude that at angular velocities $\omega \approx 500 rad.s^{-1}$ we have

$$\chi_p \approx 1\times10^{-4}$$

From the expression of $m_{gp}$ we get

$$\chi_p = 2\left[\sqrt{1 + \left(\frac{B_G^2}{\mu_{0G}\rho_p c^2}\right)^2} - 1\right] =$$
$$= 2\left[\sqrt{1 + \left(\frac{4\omega^2(\rho_p/\rho_c)^2}{\mu_{0G}\rho_p c^2}\right)^2} - 1\right]$$

where $\rho_p = m_p/V_p$ is the mass density of the protons.

In order to calculate $V_p$ we need to know the type of space (metric) inside the proton. It is known that there are just 3 types of space: the space of *positive* curvature, the space of *negative* curvature and the space of *null* curvature. The negative type is obviously excluded since the volume of the proton is *finite*. On the other hand, the space of null curvature is also excluded since the space inside the proton is strongly curved by its enormous mass density. Thus we can conclude that inside the proton the space has *positive* curvature. Consequently, the volume of the proton, $V_p$, will be expressed by the 3-dimensional space that corresponds to a *hypersphere* in a 4-dimentional space, i.e., $V_p$ will be the space of positive curvature the volume of which is [16]

$$V_p = \int_0^{2\pi}\int_0^{\pi}\int_0^{\pi} r_p^3 \sin^2\chi \sin\theta \, d\chi \, d\theta \, d\phi = 2\pi^2 r_p^3$$

In the case of Earth, for example, $\rho_{Earth} << \rho_p$. Consequently the curvature of the space inside the Earth is approximately *null* (space approximately *flat)*. Then $V_{Earth} \cong \frac{4}{3}\pi r_{Earth}^3$.

For $r_p = 1.4\times10^{-15} m$ we then get

$$\rho_p = \frac{m_p}{V_p} \cong 3\times10^{16}\,kg/m^3$$

Starting from the London moment it is easy to see that by precisely measuring the magnetic field and the angular velocity of the superconductor, one can calculate the mass of the Cooper-pairs. This has been done for both classical and high-Tc superconductors [17-20]. In the experiment with the highest precision to date, Tate et al, op.cit., reported a disagreement between the theoretically predicted Cooper-pair mass in Niobium of $m^*/2m_e = 0.999992$ and its experimental value of $1.000084(21)$, where $m_e$ is the electron mass. This anomaly was actively discussed in the literature without any apparent solution [21-24].

If we consider that the apparent mass increase from Tate's measurements results from an *increase* in the gravitational mass $m_g^*$ of the Cooper-pairs due to $B_G$, then we can write

$$\frac{m_g^*}{2m_e} = \frac{m_g^*}{m_i^*} = 1.000084$$

$$\Delta m_g^* = m_g^* - m_{g(initial)}^* = m_g^* - m_i^* =$$

$$= 1.000084\,m_i^* - m_i^* =$$

$$= +0.84\times10^{-4}\,m_i^* = \chi^* m_i^*$$

where $\chi^* = 0.84\times10^{-4}$.

From (55c) we can write that

$$m_g^* = m_i^* + 2\left[1-\sqrt{1-\left(\frac{4\omega^2}{\mu_{0G}\rho^* c^2}\right)^2}\right] m_i^* =$$

$$= m_i^* + \chi^* m_i^*$$

where $\rho^*$ is the Cooper-pair mass density.

Consequently we can write

$$\chi^* = 2\left[1-\sqrt{1-\left(\frac{4\omega^2}{\mu_{0G}\rho^* c^2}\right)^2}\right] = 0.84\times10^{-4}$$

From this equation we then obtain

$$\rho^* \cong 3\times10^{16}\,kg/m^3$$

Note that $\rho_p \cong \rho^*$.

Now we can calculate the graviphoton mass, $m_{gr}$, inside the Cooper-pairs fluid (coherent part of the superconductor) as

$$m_{gr} = \mu_{0G}\rho^*\hbar/c \cong 4\times10^{-52}\,kg$$

Outside the coherent matter $(\rho_c = 0)$ the graviphoton mass will be *zero* $(m_{gr} = \mu_{0G}\rho_c\hbar/c = 0)$.

Substitution of $\rho_p, \rho_c = \rho^*$ and $\omega \approx 500\,rad.s^{-1}$ into the expression of $\chi_p$ gives

$$\chi_p \approx 1\times10^{-4}$$

Compare this value with that one obtained from the Tajmar experiment.

Therefore, the decrease in the gravitational mass of the superconductor, expressed by (55e), is

$$m_{g,SC} \cong m_{i,SC} - \chi_p m_{i,SC}$$

$$\cong m_{i,SC} - 10^{-4} m_{i,SC}$$

This corresponds to a decrease of the order of $10^{-2}\%$ in respect to the initial gravitational mass of the superconductor. However, we must also consider the *gravitational shielding effect,* produced by this decrease of $\approx 10^{-2}\%$ in the gravitational mass of the particles inside the superconductor (see Fig. II). Therefore, the *total* weight decrease in the superconductor will be much greater than $10^{-2}\%$. According to Podkletnov experiment [25] it can reach up to 1% of the total weight of the superconductor at $523.6\,rad.s^{-1}$ $(5000\,rpm)$. In this experiment a slight decrease (up to $\approx 1\%$) in the weight of samples hung above the disk (rotating at 5000rpm) was observed. A smaller effect on the order of $0.1\%$ has been observed when the disk is not rotating. The percentage of weight decrease is the same for samples of different masses and chemical compounds. The effect does not seem to diminish with increases in elevation above the disk. There appears to be a "shielding cylinder" over the disk that extends upwards for at least 3 meters. No weight reduction has been observed under the disk.

It is easy to see that the decrease in the weight of samples hung above the disk (inside the "shielding cylinder" over the disk) in the Podkletnov experiment, is also a consequence of the *Gravitational Shielding Effect* showed in Fig. II.

In order to explain the *Gravitational Shielding Effect*, we start with the gravitational field, $\vec{g} = -\frac{GM_g}{R^2}\hat{\mu}$, produced by a particle with gravitational mass, $M_g$. The gravitational flux, $\phi_g$, through a spherical surface, with area $S$ and radius $R$, concentric with the mass $M_g$, is given by

$$\phi_g = \oint_S \vec{g}d\vec{S} = g\oint_S dS = gS =$$

$$= \frac{GM_g}{R^2}\left(4\pi R^2\right) = 4\pi GM_g$$

Note that the flux $\phi_g$ does not depend on the radius $R$ of the surface $S$, i.e., it is the *same* through any surface concentric with the mass $M_g$.

Now consider a particle with gravitational mass, $m'_g$, placed into the gravitational field produced by $M_g$. According to Eq. (41), we can have $m'_g/m'_{i0} = -1$, $m'_g/m'_{i0} \cong 0$†, $m'_g/m'_{i0} = 1$, etc. In the first case, the gravity acceleration, $g'$, upon the particle $m'_g$, is $\vec{g}' = -g = +\frac{GM_g}{R^2}\hat{\mu}$. This means that in this case, the gravitational flux, $\phi'_g$, through the particle $m'_g$ will be given by $\phi'_g = g'S = -gS = -\phi_g$, i.e., it will be *symmetric* in respect to the flux when $m'_g = m'_{i0}$ (third case). In the second case $\left(m'_g \cong 0\right)$, the intensity of the gravitational force between $m'_g$ and $M_g$ will be very close to zero. This is equivalent to say that the gravity acceleration upon the particle with mass $m'_g$ will be $g' \cong 0$. Consequently we can write that $\phi'_g = g'S \cong 0$. It is easy to see that there is a correlation between $m'_g/m'_{i0}$ and $\phi'_g/\phi_g$, i.e.,

_ If $m'_g/m'_{i0} = -1 \quad \Rightarrow \quad \phi'_g/\phi_g = -1$

_ If $m'_g/m'_{i0} = 1 \quad \Rightarrow \quad \phi'_g/\phi_g = 1$

_ If $m'_g/m'_{i0} \cong 0 \quad \Rightarrow \quad \phi'_g/\phi_g \cong 0$

Just a simple algebraic form contains the requisites mentioned above, the correlation

$$\frac{\phi'_g}{\phi_g} = \frac{m'_g}{m'_{i0}}$$

By making $m'_g/m'_{i0} = \chi$ we get

$$\phi'_g = \chi\ \phi_g$$

This is the expression of the gravitational flux through $m'_g$. It explains the *Gravitational Shielding Effect* presented in Fig. II.

As $\phi_g = gS$ and $\phi'_g = g'S$, we obtain

$$g' = \chi\ g$$

This is the gravity acceleration inside $m'_g$. Figure II (b) shows the gravitational shielding effect produced by two particles at the same direction. In this case, the gravity acceleration inside and above the second particle will be $\chi^2 g$ if $m_{g2} = m_{i1}$.

† The quantization of the gravitational mass (Eq.(33)) shows that for $n = 1$ the gravitational mass is not *zero* but equal to $m_{g(min)}$. Although the gravitational mass of a particle is never null, Eq.(41) shows that it can be turned very close to zero.

These particles are representative of any material particles or material *substance* (solid, liquid, gas, plasma, electrons flux, etc.), whose gravitational mass have been reduced by the factor $\chi$. Thus, *above* the substance, the gravity acceleration $g'$ is reduced at the same proportion $\chi = m_g / m_{i0}$, and, consequently, $g' = \chi\, g$, where $g$ is the gravity acceleration *below* the substance.

Figure III shows an experimental set-up in order to check the factor $\chi$ above *a high-speed electrons flux*. As we have shown (Eq. 43), the *gravitational mass* of a particle decreases with the increase of the velocity $V$ of the particle.

Since the theory says that the factor $\chi$ is given by the correlation $m_g / m_{i0}$ then, in the case of an electrons flux, we will have that $\chi = m_{ge}/m_{ie}$ where $m_{ge}$ as function of the velocity $V$ is given by Eq. (43). Thus, we can write that

$$\chi = \frac{m_{ge}}{m_{ie}} = \left\{1 - 2\left[\frac{1}{\sqrt{1 - V^2/c^2}} - 1\right]\right\}$$

Therefore, if we know the velocity $V$ of the electrons we can calculate $\chi$. ($m_{ie}$ is the electron mass at rest).

When an electron penetrates the electric field $E_y$ (see Fig. III) an electric force, $\vec{F}_E = -e\vec{E}_y$, will act upon the electron. The direction of $\vec{F}_E$ will be contrary to the direction of $\vec{E}_y$. The magnetic force $\vec{F}_B$ which acts upon the electron, due to the magnetic field $\vec{B}$, is $\vec{F}_B = eVB\hat{\mu}$ and will be opposite to $\vec{F}_E$ because the electron charge is negative.

By adjusting conveniently $B$ we can make $\left|\vec{F}_B\right| = \left|\vec{F}_E\right|$. Under these circumstances in which the total force is zero, the spot produced by the electrons flux on the surface $\alpha$ returns from $O'$ to $O$ and is detected by the galvanometer $G$. That is, there is no deflection for the cathodic rays. Then it follows that $eVB = eE_y$ since $\left|\vec{F}_B\right| = \left|\vec{F}_E\right|$. Then, we get

$$V = \frac{E_y}{B}$$

This gives a measure of the velocity of the electrons.

Thus, by means of the experimental set-up, shown in Fig. III, we can easily obtain the velocity $V$ of the electrons below the body $\beta$, in order to calculate the *theoretical* value of $\chi$. The *experimental* value of $\chi$ can be obtained by dividing the weight, $P'_\beta = m_{g\beta} g'$ of the body $\beta$ for a voltage drop $\widetilde{V}$ across the anode and cathode, by its weight, $P_\beta = m_{g\beta} g$, when the voltage $\widetilde{V}$ is *zero*, i.e.,

$$\chi = \frac{P'_\beta}{P_\beta} = \frac{g'}{g}$$

According to Eq. (4), the gravitational mass, $M_g$, is defined by

$$M_g = \left|\frac{m_g}{\sqrt{1 - V^2/c^2}}\right|$$

While Eq. (43) defines $m_g$ by means of the following expression

$$m_g = \left\{1 - 2\left[\frac{1}{\sqrt{1 - V^2/c^2}} - 1\right]\right\} m_{i0}$$

In order to check the gravitational mass of the electrons it is necessary to know the pressure $P$ produced by the electrons flux. Thus, we have put a piezoelectric sensor in the bottom of the glass tube as shown in Fig. III. The electrons flux radiated from the cathode is accelerated by the anode1 and strikes on the piezoelectric sensor yielding a pressure $P$ which is measured by means of the sensor.

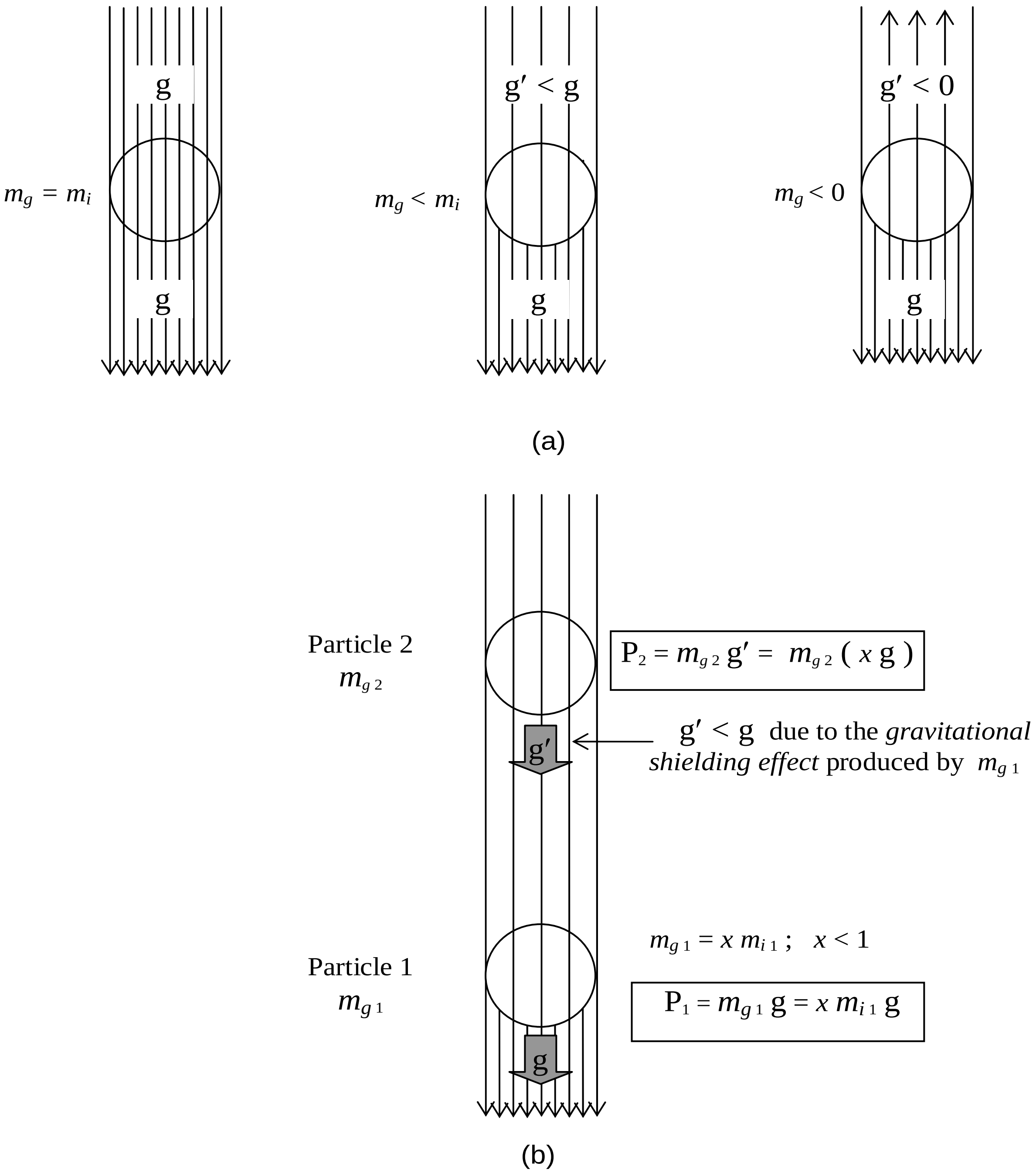

Fig. I I – The Gravitational Shielding effect.

Let us now deduce the correlation between $P$ and $M_{ge}$.

When the electrons flux strikes the sensor, the electrons transfer to it a *momentum* $Q = n_e q_e = n_e M_{ge} V$. Since $Q = F\Delta t = 2Fd/V$, we conclude that

$$M_{ge} = \frac{2d}{V^2}\left(\frac{F}{n_e}\right)$$

The amount of electrons, $n_e$, is given by $n_e = \rho S d$ where $\rho$ is the amount of electrons per unit of volume (electrons/m$^3$); $S$ is the cross-section of the electrons flux and $d$ the distance between cathode and anode.

In order to calculate $n_e$ we will start from the *Langmuir-Child law* and the *Ohm vectorial law*, respectively given by

$$J = \alpha\frac{\tilde{V}^{\frac{3}{2}}}{d} \text{ and } J = \rho_c V\ , \quad \left(\rho_c = \rho/e\right)$$

where $J$ is the thermoionic current density; $\alpha = 2.33\times10^{-6}\, A.m^{-1}.V^{-\frac{3}{2}}$ is the called *Child's constant*; $\tilde{V}$ is the voltage drop across the anode and cathode electrodes, and $V$ is the velocity of the electrons.

By comparing the Langmuir-Child law with the Ohm vectorial law we obtain

$$\rho = \frac{\alpha\tilde{V}^{\frac{3}{2}}}{ed^2 V}$$

Thus, we can write that

$$n_e = \frac{\alpha\tilde{V}^{\frac{3}{2}} S}{edV}$$

and

$$M_{ge} = \left(\frac{2ed^2}{\alpha V\tilde{V}^{\frac{3}{2}}}\right)P$$

Where $P = F/S$, is the pressure to be measured by the piezoelectric sensor.

In the experimental set-up the total force $F$ acting on the piezoelectric sensor is the resultant of all the forces $F_\phi$ produced by each electrons flux that passes through each hole of area $S_\phi$ in the grid of the anode 1, and is given by

$$F = nF_\phi = n\left(PS_\phi\right) = \left(\frac{\alpha n S_\phi}{2ed^2}\right)M_{ge}V\tilde{V}^{\frac{3}{2}}$$

where $n$ is the number of holes in the grid. By means of the piezoelectric sensor we can measure $F$ and consequently obtain $M_{ge}$.

We can use the equation above to evaluate the magnitude of the force $F$ to be measured by the piezoelectric sensor. First, we will find the expression of $V$ as a function of $\tilde{V}$ since the electrons speed $V$ depends on the voltage $\tilde{V}$.

We will start from Eq. (46) which is the general expression for Lorentz's force, i.e.,

$$\frac{d\vec{p}}{dt} = \left(q\vec{E} + q\vec{V}\times\vec{B}\right)\frac{m_g}{m_{i0}}$$

When the force and the speed have the same direction Eq. (6) gives

$$\frac{d\vec{p}}{dt} = \frac{m_g}{\left(1 - V^2/c^2\right)^{\frac{3}{2}}}\frac{d\vec{V}}{dt}$$

By comparing these expressions we obtain

$$\frac{m_{i0}}{\left(1 - V^2/c^2\right)^{\frac{3}{2}}}\frac{d\vec{V}}{dt} = q\vec{E} + q\vec{V}\times\vec{B}$$

In the case of electrons accelerated by a sole electric field $\left(B = 0\right)$, the equation above gives

$$\vec{a} = \frac{d\vec{V}}{dt} = \frac{e\vec{E}}{m_{ie}}\left(1 - V^2/c^2\right)\sqrt{\frac{2e\tilde{V}}{m_{ie}}}$$

Therefore, the velocity $V$ of the electrons in the experimental set-up is

$$V = \sqrt{2ad} = \left(1 - V^2/c^2\right)^{\frac{3}{4}}\sqrt{\frac{2e\tilde{V}}{m_{ie}}}$$

From Eq. (43) we conclude that

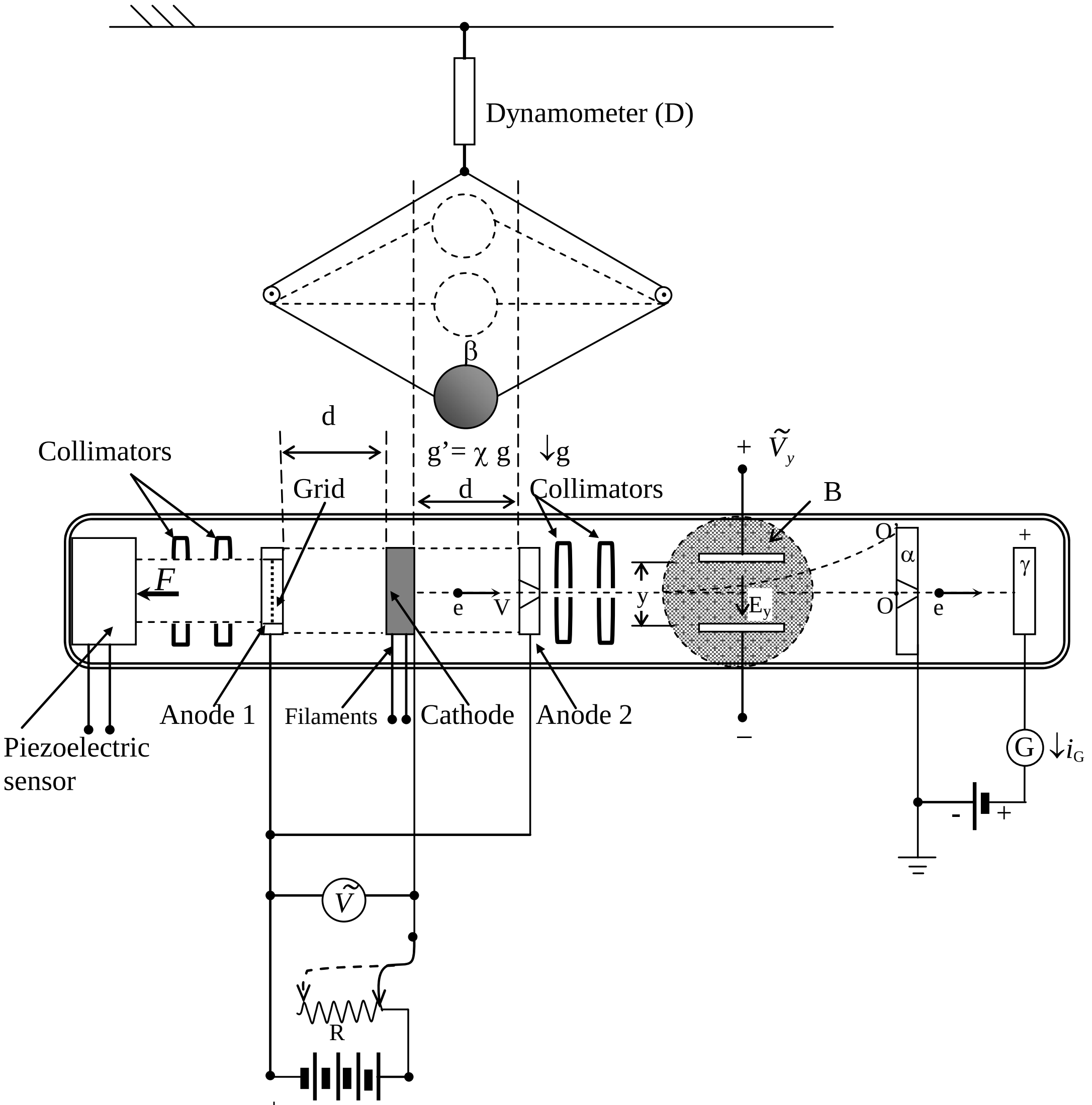

Fig. III – Experimental set-up in order to check the factor $\chi$ above a high-speed electrons flux. The set-up may also check the velocities and the gravitational masses of the electrons.

$m_{ge} \cong 0$ when $V \cong 0.745c$. Substitution of this value of $V$ into equation above gives $\widetilde{V} \cong 479.1KV$. This is the voltage drop necessary to be applied across the anode and cathode electrodes in order to obtain $m_{ge} \cong 0$.

Since the equation above can be used to evaluate the velocity $V$ of the electrons flux for a given $\widetilde{V}$, then we can use the obtained value of $V$ to evaluate the intensity of $\vec{B}$ in order to produce $eVB = eE_y$ in the experimental set-up. Then by adjusting $B$ we can check when the electrons flux is detected by the galvanometer $G$. In this case, as we have already seen, $eVB = eE_y$, and the velocity of the electrons flux is calculated by means of the expression $V = E_y / B$. Substitution of $V$ into the expressions of $m_{ge}$ and $M_{ge}$, respectively given by

$$m_{ge} = \left\{1 - 2\left[\frac{1}{\sqrt{1 - V^2/c^2}} - 1\right]\right\} m_{ie}$$

and

$$M_{ge} = \frac{m_{ge}}{\sqrt{1 - V^2/c^2}}$$

yields the corresponding values of $m_{ge}$ and $M_{ge}$ which can be compared with the values obtained in the experimental set-up:

$$m_{ge} = \chi m_{ie} = \left(P'_\beta / P_\beta\right) m_{ie}$$

$$M_{ge} = \frac{F}{V\widetilde{V}^{\frac{3}{2}}}\left(\frac{2ed^2}{\alpha n S_\phi}\right)$$

where $P'_\beta$ and $P_\beta$ are measured by the dynamometer $D$ and $F$ is measured by the piezoelectric sensor.

If we have $nS_\phi \cong 0.16m^2$ and $d = 0.08m$ in the experimental set-up then it follows that

$$F = 1.82 \times 10^{14} M_{ge} V \widetilde{V}^{\frac{3}{2}}$$

By varying $\widetilde{V}$ from 10KV up to 500KV we note that the maximum value for $F$ occurs when $\widetilde{V} \cong 344.7KV$. Under these circumstances, $V \cong 0.7c$ and $M_{ge} \cong 0.28 m_{ie}$. Thus the maximum value for $F$ is

$$F_{max} \cong 1.9N \cong 190gf$$

Consequently, for $\widetilde{V}_{max} = 500KV$, the piezoelectric sensor must satisfy the following characteristics:

– Capacity 200gf

– Readability 0.001gf

Let us now return to the explanation for the findings of Podkletnov's experiment. Next, we will explain the decrease of 0.1% in the weight of the superconductor when the disk is only levitating but not rotating.

Equation (55) shows how the gravitational mass is altered by *electromagnetic* fields.

The expression of $n_r$ for $\sigma >> \omega\varepsilon$ can be obtained from (54), in the form

$$n_r = \frac{c}{v} = \sqrt{\frac{\mu\sigma c^2}{4\pi f}} \qquad (56)$$

Substitution of (56) into (55) leads to

$$m_g = \left\{1 - 2\left[\sqrt{1 + \frac{\mu\sigma}{4\pi f}\left(\frac{U}{m_i c}\right)^2} - 1\right]\right\} m_{i0}$$

This equation shows that *atoms of ferromagnetic materials with very-high* $\mu$ can have gravitational masses strongly reduced by means of *Extremely Low Frequency* (ELF) electromagnetic radiation. It also shows that atoms of *superconducting* *materials* (due to *very-high* $\sigma$ ) can also have its gravitational masses strongly reduced by means of ELF electromagnetic radiation.

Alternatively, we may put Eq.(55) as a function of the *power density* ( or intensity ), $D$, of the radiation. The integration of (51) gives $U = VD/v$. Thus, we can write (55) in the following form:

$$m_g = \left\{1 - 2\left[\sqrt{1 + \left(\frac{n_r^2 D}{\rho c^3}\right)^2} - 1\right]\right\} m_{i0} \qquad (57)$$

where $\rho = m_{i0}/V$.

For $\sigma >> \omega\varepsilon$, $n_r$ will be given by (56) and consequently (57) becomes

$$m_g = \left\{1 - 2\left[\sqrt{1 + \left(\frac{\mu\sigma D}{4\pi f \rho c}\right)^2} - 1\right]\right\} m_{i0} \qquad (58)$$

In the case of *Thermal radiation*, it is common to relate the energy of photons to *temperature*, *T*, through the relation,

$$\langle hf \rangle \approx \kappa T$$

where $\kappa = 1.38 \times 10^{-23} J/°K$ is the *Boltzmann's constant*. On the other hand it is known that

$$D = \sigma_B T^4$$

where $\sigma_B = 5.67 \times 10^{-8} watts/m^2 °K^4$ is the *Stefan-Boltzmann's constant*. Thus we can rewrite (58) in the following form

$$m_g = \left\{1 - 2\left[\sqrt{1 + \left(\frac{\mu\sigma\sigma_B h T^3}{4\pi\kappa\rho c}\right)^2} - 1\right]\right\} m_{i0} \qquad (58a)$$

Starting from this equation, we can evaluate the effect of the *thermal radiation* upon the gravitational mass of the Copper-pair fluid, $m_{g,CPfluid}$. Below the *transition temperature*, $T_c$, $(T/T_c < 0.5)$ the conductivity of the superconducting materials is usually larger than $10^{22} S/m$ [26]. On the other hand the *transition temperature*, for high critical temperature (HTC) superconducting materials, is in the order of $10^2 K$. Thus (58a) gives

$$m_{g,CPfluid} = \left\{1 - 2\left[\sqrt{1 + \frac{\sim 10^{-9}}{\rho_{CPfluid}^2}} - 1\right]\right\} m_{i,CPfluid} \qquad (58b)$$

Assuming that the number of Copper-pairs per unit volume is $N \approx 10^{26} m^{-3}$ [27] we can write that

$$\rho_{CPfluid} = Nm^* \approx 10^{-4} kg/m^3$$

Substitution of this value into (58b) yields

$$m_{g,CPfluid} = m_{i,CPfluid} - 0.1\, m_{i,CPfluid}$$

This means that the gravitational masses of the electrons are decreased of ~10%. This corresponds to a decrease in the gravitational mass of the superconductor given by

$$\frac{m_{g,SC}}{m_{i,SC}} = \frac{N\left(m_{ge} + m_{gp} + m_{gn} + \Delta E/c^2\right)}{N\left(m_{ie} + m_{ip} + m_{in} + \Delta E/c^2\right)} =$$

$$= \left(\frac{m_{ge} + m_{gp} + m_{gn} + \Delta E/c^2}{m_{ie} + m_{ip} + m_{in} + \Delta E/c^2}\right) =$$

$$= \left(\frac{0.9 m_{ie} + m_{ip} + m_{in} + \Delta E/c^2}{m_{ie} + m_{ip} + m_{in} + \Delta E/c^2}\right) =$$

$$= 0.999976$$

Where $\Delta E$ is the interaction energy. Therefore, a decrease of $(1 - 0.999976) \approx 10^{-5}$, i.e., approximately $10^{-3}$*%* in respect to the initial gravitational mass of the superconductor, due to the local *thermal radiation* only. However, here we must also consider the *gravitational shielding effect* produced, in this case, by the decrease of $\approx 10^{-3}$*%* in the gravitational mass of the particles inside the superconductor (see Fig. II). Therefore the *total* weight decrease in the superconductor will be much greater than $\approx 10^{-3}\%$. This can explain the smaller effect on the order of $0.1\%$ observed in the Podkletnov measurements when the disk is not rotating.

Let us now consider an electric current $I$ through a conductor subjected to electromagnetic radiation with power density $D$ and frequency $f$.

Under these circumstances the *gravitational mass* $m_{ge}$ of the *electrons* of the conductor, according to Eq. (58), is given by

$$m_{ge} = \left\{1 - 2\left[\sqrt{1+\left(\frac{\mu\sigma D}{4\pi f \rho c}\right)^2} - 1\right]\right\} m_e$$

where $m_e = 9.11 \times 10^{-31} kg$.

Note that if the radiation upon the conductor has extremely-low frequency (ELF radiation) then $m_{ge}$ can be strongly reduced. For example, if $f \approx 10^{-6} Hz$, $D \approx 10^5 W/m^2$ and the conductor is made of *copper* ($\mu \cong \mu_0; \sigma = 5.8\times10^7 S/m$ and $\rho = 8900 kg/m^3$) then

$$\left(\frac{\mu\sigma D}{4\pi f \rho c}\right) \approx 1$$

and consequently $m_{ge} \approx 0.1 m_e$.

According to Eq. (6) the force upon each *free electron* is given by

$$\vec{F}_e = \frac{m_{ge}}{\left(1 - V^2/c^2\right)^{3/2}} \frac{d\vec{V}}{dt} = e\vec{E}$$

where $E$ is the applied electric field. Therefore, the decrease of $m_{ge}$ produces an increase in the velocity $V$ of the free electrons and consequently the *drift velocity* $V_d$ is also increased. It is known that the density of electric current $J$ through a conductor [28] is given by

$$\vec{J} = \Delta_e \vec{V}_d$$

where $\Delta_e$ is the density of the free electric charges ( For cooper conductors $\Delta_e = 1.3\times10^{10} C/m^3$ ). Therefore increasing $V_d$ produces an increase in the electric current $I$. Thus if $m_{ge}$ is reduced 10 times $\left(m_{ge} \approx 0.1 m_e\right)$ the drift velocity $V_d$ is increased 10 times as well as the electric current. Thus we conclude that strong fluxes of ELF radiation upon electric/electronic circuits can suddenly increase the electric currents and consequently damage these circuits.

Since the *orbital electrons* moment of inertia is given by $I_i = \Sigma(m_i)_j r_j^2$, where $m_i$ refers to *inertial mass* and not to gravitational mass, then the *momentum* $L = I_i\omega$ of the conductor *orbital electrons* are not affected by the ELF radiation. Consequently, this radiation just affects the conductor's *free electrons* velocities. Similarly, in the case of superconducting materials, the *momentum*, $L = I_i\omega$, of the *orbital electrons* are not affected by the gravitomagnetic fields.

The vector $\vec{D} = (U/V)\vec{v}$, which we may define from (48), has the same direction of the propagation vector $\vec{k}$ and evidently corresponds to the *Poynting vector*. Then $\vec{D}$ can be replaced by $\vec{E}\times\vec{H}$. Thus we can write $D = \frac{1}{2}EH = \frac{1}{2}E(B/\mu) = \frac{1}{2}E[(E/v)/\mu] = \frac{1}{2}(1/v\mu)E^2$.

For $\sigma >> \omega\varepsilon$ Eq. (54) tells us that $v = \sqrt{4\pi f/\mu\sigma}$. Consequently, we obtain

$$D = \tfrac{1}{2}E^2\sqrt{\frac{\sigma}{4\pi f \mu}}$$

This expression refers to the instantaneous values of $D$ and $E$. The average value for $E^2$ is equal to $\frac{1}{2}E_m^2$ because $E$ varies sinusoidaly

($E_m$ is the maximum value for $E$). Substitution of the expression of $D$ into (58) gives

$$m_g = \left\{1 - 2\left[\sqrt{1 + \frac{\mu}{4c^2}\left(\frac{\sigma}{4\pi f}\right)^3 \frac{E^2}{\rho^2}} - 1\right]\right\} m_{i0} \quad (59a)$$

Since $E_{rms} = E_m / \sqrt{2}$ and $E^2 = \frac{1}{2} E_m^2$ we can write the equation above in the following form

$$m_g = \left\{1 - 2\left[\sqrt{1 + \frac{\mu}{4c^2}\left(\frac{\sigma}{4\pi f}\right)^3 \frac{E_{rms}^2}{\rho^2}} - 1\right]\right\} m_{i0} \quad (59a)$$

Note that for *extremely-low frequencies* the value of $f^{-3}$ in this equation becomes highly expressive.

Since $E = vB$ equation (59a) can also be put as a function of $B$, i.e.,

$$m_g = \left\{1 - 2\left[\sqrt{1 + \left(\frac{\sigma}{4\pi f \mu c^2}\right)\frac{B^4}{\rho^2}} - 1\right]\right\} m_{i0} \quad (59b)$$

For *conducting* materials with $\sigma \approx 10^7 S/m$; $\mu_r = 1$; $\rho \approx 10^3 kg/m^3$ the expression (59b) gives

$$m_g = \left\{1 - 2\left[\sqrt{1 + \left(\frac{\approx 10^{-12}}{f}\right)B^4} - 1\right]\right\} m_{i0}$$

This equation shows that the decreasing in the *gravitational mass* of these conductors can become experimentally detectable for example, starting from 100Teslas at 10mHz.

One can then conclude that an interesting situation arises when a body penetrates a magnetic field in the direction of its center. The *gravitational mass* of the body decreases progressively. This is due to the intensity increase of the magnetic field upon the body while it penetrates the field. In order to understand this phenomenon we might, based on (43), think of the inertial mass as being formed by two parts: one *positive* and another *negative*. Thus, when the body penetrates the magnetic field, its negative inertial mass increases, but its total inertial mass decreases, i.e., although there is an increase of inertial mass, the total inertial mass (which is equivalent to *gravitational mass*) will be reduced.

On the other hand, Eq.(4) shows that the *velocity of the body must increase* as consequence of the gravitational mass decreasing since the *momentum* is conserved. Consider for example a spacecraft with velocity $V_s$ and gravitational mass $M_g$. If $M_g$ is reduced to $m_g$ then the velocity becomes

$$V_s' = \left(M_g / m_g\right) V_s$$

In addition, Eqs. 5 and 6 tell us that the *inertial forces* depend on $m_g$. Only in the particular case of $m_g = m_{i0}$ the expressions (5) and (6) reduce to the well-known Newtonian expression $F = m_{i0} a$. Consequently, one can conclude that the *inertial effects* on the spacecraft will also be reduced due to the *decreasing of its gravitational mass*. Obviously this leads to a new concept of aerospace flight.

Now consider an electric current $i = i_0 sin 2\pi ft$ through a conductor. Since the current density, $\vec{J}$, is expressed by $\vec{J} = di / d\vec{S} = \sigma \vec{E}$, then we can write that $E = i/\sigma S = (i_0/\sigma S) sin 2\pi ft$. Substitution of this equation into (59a) gives

$$m_g = \left\{1 - 2\left[\sqrt{1 + \frac{i_0^4 \mu}{64\pi^3 c^2 \rho^2 S^4 f^3 \sigma} sin^4 2\pi ft} - 1\right]\right\} m_{i0} \quad (59c)$$

If the conductor is a *supermalloy* rod $(1 \times 1 \times 400mm)$ then $\mu_r = 100{,}000$ (initial); $\rho = 8770 kg/m^3$; $\sigma = 1.6 \times 10^6 S/m$ and $S = 1 \times 10^{-6} m^2$. Substitution of these values into the equation above yields the following expression for the *gravitational mass* of the supermalloy rod

$$m_{g(sm)}=\left\{1-2\left[\sqrt{1+\left(5.71\times10^{-12}\,i_0^4/f^3\right)\sin^4 2\pi ft}-1\right]\right\}m_{i(sm)}$$

Some oscillators like the HP3325A (Op.002 High Voltage Output) can generate sinusoidal voltages with *extremely-low* frequencies down to $f=1\times10^{-6}Hz$ and amplitude up to 20V (into $50\Omega$ load). The maximum output current is $0.08A_{pp}$ .

Thus, for $i_0=0.04A\ \left(0.08A_{pp}\right)$ and $f<2.25\times10^{-6}Hz$ the equation above shows that the *gravitational mass* of the rod becomes *negative* at $2\pi ft=\pi/2$; for $f\cong1.7\times10^{-6}Hz$ at $t=1/4f=1.47\times10^5 s\cong40.8h$ it shows that $m_{g(sm)}\cong-m_{i(sm)}$.

This leads to the idea of the *Gravitational Motor*. See in Fig. IV a type of gravitational motor (Rotational Gravitational Motor) based on the possibility of gravity control on a ferromagnetic wire.

It is important to realize that this is not the unique way of decreasing the gravitational mass of a body. It was noted earlier that the expression (53) is general for all types of waves including non-electromagnetic waves like sound waves for example. In this case, the velocity $v$ in (53) will be the *speed of sound in the body* and $D$ the *intensity* of the sound radiation. Thus from (53) we can write that

$$\frac{\Delta p}{m_i c}=\frac{VD}{m_i c}=\frac{D}{\rho cv^2}$$

It can easily be shown that $D=2\pi^2\rho f^2A^2v$ where $A=\lambda P/2\pi\rho v^2$ ; $A$ and $P$ are respectively the amplitude and maximum pressure variation of the sound wave. Therefore we readily obtain

$$\frac{\Delta p}{m_{i0}c}=\frac{P^2}{2\rho^2cv^3}$$

Substitution of this expression into (41) gives

$$m_g=\left\{1-2\left[\sqrt{1+\left(\frac{P^2}{2\rho^2cv^3}\right)^2}-1\right]\right\}m_{i0} \quad (60)$$

This expression shows that in the case of sound waves the decreasing of gravitational mass is relevant for *very strong pressures* only.

It is known that in the nucleus of the Earth the pressure can reach values greater than $10^{13}N/m^2$. The equation above tells us that sound waves produced by pressure variations of this magnitude can cause strong decreasing of the *gravitational mass* at the surroundings of the point where the sound waves were generated. This obviously must cause an abrupt decreasing of the pressure at this place since pressure = weight /area = $m_g g$/area). Consequently a local instability will be produced due to the opposite internal pressure. The conclusion is that this effect may cause Earthquakes.

Consider a sphere of radius $r$ around the point where the sound waves were generated (at $\approx1{,}000km$ depth; the Earth's radius is $6{,}378km$). If the *maximum* pressure, at the explosion place ( sphere of radius $r_0$), is $P_{max}\approx10^{13}N/m^2$ and the pressure at the distance $r=10km$ is $P_{min}=\left(r_0/r\right)^2P_{max}\approx10^9N/m^2$ then we can consider that in the sphere $P=\sqrt{P_{max}P_{min}}\approx10^{11}N/m^2$.Thus assuming $v\approx10^3m/s$ and $\rho\approx10^3kg/m^3$ we can calculate the variation of gravitational mass in the sphere by means of the equation of $m_g$ , i.e.,

$$\alpha = \left\{ 1 - 2\left[ \sqrt{1 + \frac{i^4 \mu}{64\pi^3 c^2 \rho^2 S^4 f^3 \sigma}} - 1 \right] \right\}$$

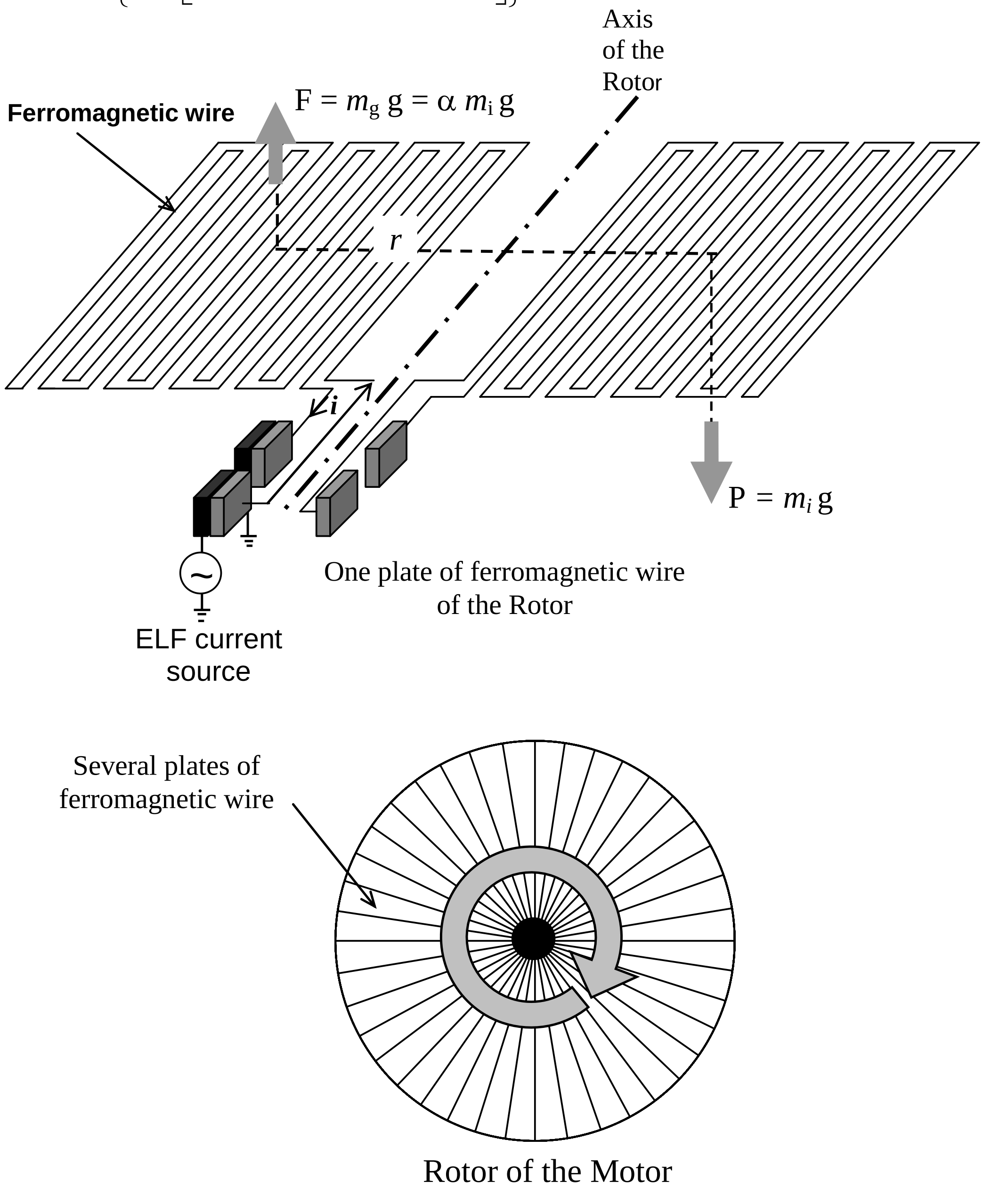

Fig. IV - Rotational Gravitational Motor

$$\Delta m_g = m_{g(initial)} - m_g =$$

$$= m_{i0} - \left\{1 - 2\left[\sqrt{1 + \left(\frac{P^2}{2\rho^2 cv^3}\right)^2} - 1\right]\right\} m_{i0} =$$

$$= 2\left[\sqrt{1 + \left(\frac{P^2}{2\rho^2 cv^3}\right)^2} - 1\right]\rho V \approx 10^{11} kg$$

The *transitory* loss of this great amount of gravitational mass may evidently produce a strong pressure variation and consequently a strong Earthquake.

Finally, we can evaluate the energy necessary to generate those sound waves. From (48) we can write $D_{max} = P_{max} v \approx 10^{16} W/m^2$. Thus, the released power is $P_0 = D_{max}\left(4\pi r_0^2\right) \approx 10^{21} W$ and the energy $\Delta E$ released at the time interval $\Delta t$ must be $\Delta E = P_0 \Delta t$. Assuming $\Delta t \approx 10^{-3} s$ we readily obtain

$$\Delta E = P_0 \Delta t \approx 10^{18}\, joules \approx 10^4\, Megatons$$

This is the amount of energy released by an earthquake of magnitude 9 $(M_s = 9)$, i.e., $E = 1.74 \times 10^{(5+1.44 M_s)} \cong 10^{18}\, joules$. The maximum magnitude in the *Richter* scale is 12. Note that the sole releasing of this energy at 1000km depth (without the effect of gravitational mass decreasing) cannot produce an Earthquake, since the sound waves reach 1km depth with pressures less than 10N/cm$^2$.

Let us now return to the Theory.

The equivalence between frames of non-inertial reference and gravitational fields assumed $m_g \equiv m_i$ because the inertial forces were given by $\vec{F}_i = m_i \vec{a}$, while the equivalent gravitational forces, by $\vec{F}_g = m_g \vec{g}$. Thus, to satisfy the equivalence ($\vec{a} \equiv \vec{g}$ and $\vec{F}_i \equiv \vec{F}_g$) it was *necessary that* $m_g \equiv m_i$. Now, the inertial force, $\vec{F}_i$, is given by Eq.(6), and from Eq.(13) we can obtain the gravitational force, $\vec{F}_g$. Thus, $\vec{F}_i \equiv \vec{F}_g$ leads to

$$\frac{m_g}{\left(1 - V^2/c^2\right)^{3/2}}\vec{a} \equiv G\frac{m_g'}{\left(r'\sqrt{1 - V^2/c^2}\right)^2}\frac{m_g}{\sqrt{1 - V^2/c^2}} \equiv$$

$$\equiv \left(G\frac{m_g'}{r'^2}\right)\frac{m_g}{\left(1 - V^2/c^2\right)^{3/2}} \equiv \vec{g}\frac{m_g}{\left(1 - V^2/c^2\right)^{3/2}} \qquad (61)$$

whence results

$$\vec{a} \equiv \vec{g} \qquad (62)$$

Consequently, the equivalence is evident, and therefore Einstein's equations from the General Relativity continue obviously valid.

The new expression for $F_i$ (Eqs. (5) and (6)) shows that the inertial forces are proportional to the *gravitational mass*, $m_g$. This means that these forces result from the gravitational interaction between the particle and the other gravitational masses of the Universe, just as *Mach's principle* predicts. Therefore the new expression for the inertial forces incorporates the Mach's principle into Gravitation Theory, and furthermore reveals that the inertial effects upon a particle can be reduced because, as we have seen, the gravitational mass may be reduced. When $m_g = m_{i0}$ the *nonrelativistic* equation for inertial forces, $\vec{F}_i = m_g \vec{a}$, reduces to $\vec{F}_i = m_{i0}\vec{a}$. This is the well-known *Newton's second law* for motion.

In Einstein's Special Relativity Theory the motion of a free-particle is described by means of $\delta S = 0$ [29]. Now based on Eq. (1), $\delta S = 0$ will be given by the following expression

$$\delta S = -m_g c \delta \int ds = 0. \qquad (63)$$

which also describes the motion of the particle inside the gravitational field. Thus, Einstein's equations from the General Relativity can be derived starting from $\delta(S_m + S_g) = 0$, where $S_g$ and $S_m$ refer to the *action* of the gravitational field and the action of the matter, respectively [30].

The variations $\delta S_g$ and $\delta S_m$ can be written as follows [31]:

$$\delta S_g = \frac{c^3}{16\pi G}\int\left(R_{ik} - \tfrac{1}{2}g_{ik}R\right)\delta g^{ik}\sqrt{-g}\,d\Omega \qquad (64)$$

$$\delta S_m = -\frac{1}{2c}\int T_{ik}\,\delta g^{ik}\sqrt{-g}\,d\Omega \qquad (65)$$

where $R_{ik}$ is the Ricci's tensor; $g_{ik}$ the metric tensor and $T_{ik}$ the matter's energy-momentum tensor:

$$T_{ik} = \left(P + \varepsilon_g\right)\mu_i\mu_k + Pg_{ik} \qquad (66)$$

where $P$ is the pressure and $\varepsilon_g = \rho_g c^2$ is now, the density of gravitational energy, $E_g$, of the particle; $\rho_g$ is then the density of gravitational mass of the particle, i.e., $M_g$ at the volume unit.

Substitution of (64) and (65) into $\delta S_m + \delta S_g = 0$ yields

$$\frac{c^3}{16\pi G}\int\left(R_{ik} - \tfrac{1}{2}g_{ik}R - \tfrac{8\pi G}{c^4}T_{ik}\right)\delta g^{ik}\sqrt{-g}\,d\Omega = 0$$

whence,

$$\left(R_{ik} - \tfrac{1}{2}g_{ik}R - \tfrac{8\pi G}{c^4}T_{ik}\right) = 0 \qquad (67)$$

because the $\delta g_{ik}$ are arbitrary.

Equations (67) in the following form

$$R_{ik} - \tfrac{1}{2}g_{ik}R = \tfrac{8\pi G}{c^4}T_{ik} \qquad (68)$$

or

$$R_i^k - \tfrac{1}{2}g\delta_i^k R = \tfrac{8\pi G}{c^4}T_i^k . \qquad (69)$$

are the Einstein's equations from the General Relativity.

It is known that these equations are *only valid* if the spacetime is *continuous*. We have shown at the beginning of this work that the spacetime *is not continuous* it is *quantized*. However, the spacetime can be considered approximately "*continuous*" when the *quantum number* $n$ is very large (Classical limit). Therefore, just under these circumstances the Einstein's equations from the General Relativity can be used in order to "*classicalize*" the quantum theory by means of approximated description of the spacetime.

Later on we will show that the length $d_{min}$ of Eq. (29) is given by

$$d_{min} = \tilde{k}\,l_{planck} = \tilde{k}\left(G\hbar/c^3\right)^{\frac{1}{2}} \approx 10^{-34}m \qquad (70)$$

(See Eq. (100)). On the other hand, we will find in the Eq. (129) the length scale of the *initial Universe*, i.e., $d_{initial} \approx 10^{-14}m$. Thus, from the Eq. (29) we get: $n = d_{initial}/d_{min} = 10^{-14}/10^{-34} \approx 10^{20}$ this is the quantum number of the spacetime at *initial instant*. That quantum number is sufficiently large for the spacetime to be considered approximately "*continuous*" starting from the beginning of the Universe. Therefore Einstein's equations can be used even at the Initial Universe.

Now, it is easy to conclude why the attempt to quantize gravity starting from the General Relativity was a bad theoretical strategy.

Since the gravitational interaction can be repulsive, besides attractive, such as the electromagnetic interaction, then the *graviton* must have spin 1 (called *graviphoton*) and not 2. Consequently, the gravitational forces are also *gauge* forces because they are yielded by the exchange of the so-called "virtual" *quanta* of spin 1, such as the electromagnetic forces and the weak and strong nuclear forces.

Let us now deduce the *Entropy Differential Equation* starting from Eq. (55). Comparison of Eqs. (55) and (41) shows that $Un_r = \Delta pc$. For small velocities, i.e., $(V << c)$, we have $Un_r << m_{i0}c^2$. Under these circumstances, the development of Eq. (55) in power of $\left(Un_r/m_{i0}c^2\right)$ gives

$$m_g = m_{i0} - \left(\frac{Un_r}{m_{i0}c^2}\right)^2 m_{i0} \qquad (71)$$

In the particular case of *thermal radiation,* it is usual to relate the energy of the photons to the temperature, through the relationship $\langle h\nu\rangle \approx kT$ where $k = 1.38\times10^{-23}\,J/K$ is the Boltzmann's constant. Thus, in that case, the energy absorbed by the particle will be $U = \eta\langle h\nu\rangle \approx \eta kT$, where $\eta$ is a particle-dependent absorption/emission coefficient. Therefore, Eq.(71) may be rewritten in the following form:

$$m_g = m_{i0} - \left[\left(\frac{n_r\eta k}{c^2}\right)^2 \frac{T^2}{m_{i0}^{\ 2}}\right] m_{i0} \qquad (72)$$

For electrons at T=300K, we have

$$\left(\frac{n_r\eta k}{c^2}\right)^2 \frac{T^2}{m_e^{\ 2}} \approx 10^{-17}$$

Comparing (72) with (18), we obtain

$$E_{Ki} = \frac{1}{2}\left(\frac{n_r\eta k}{c}\right)^2 \frac{T^2}{m_{i0}}. \qquad (73)$$

The derivative of $E_{Ki}$ with respect to temperature $T$ is

$$\frac{\partial E_{Ki}}{\partial T} = \left(n_r\eta k/c\right)^2\left(T/m_{i0}\right) \qquad (74)$$

Thus,

$$T\frac{\partial E_{Ki}}{\partial T} = \frac{\left(n_r\eta kT\right)^2}{m_{i0}c^2} \qquad (75)$$

Substitution of $E_{Ki} = E_i - E_{i0}$ into (75) gives

$$T\left(\frac{\partial E_i}{\partial T} + \frac{\partial E_{i0}}{\partial T}\right) = \frac{\left(n_r\eta kT\right)^2}{m_{i0}c^2} \qquad (76)$$

By comparing the Eqs.(76) and (73) and considering that $\partial E_{i0}/\partial T = 0$ because $E_{i0}$ does not depend on $T$, the Eq.(76) reduces to

$$T\left(\partial E_i/\partial T\right) = 2E_{Ki} \qquad (77)$$

However, Eq.(18) shows that $2E_{Ki} = E_i - E_g$. Therefore Eq.(77) becomes

$$E_g = E_i - T\left(\partial E_i/\partial T\right) \qquad (78)$$

Here, we can identify the energy $E_i$ with the *free-energy* of the system-F and $E_g$ with the *internal energy* of the system-U. Thus we can write the Eq.(78) in the following form:

$$U = F - T\left(\partial F/\partial T\right) \qquad (79)$$

This is the well-known equation of Thermodynamics. On the other hand, remembering that $\partial Q = \partial\tau + \partial U$ (1st principle of Thermodynamics) and

$$F = U - TS \qquad (80)$$

(Helmholtz's function), we can easily obtain from (79), the following equation

$$\partial Q = \partial\tau + T\partial S. \qquad (81)$$

For *isolated systems,* $\partial\tau = 0$, we have

$$\partial Q = T\partial S \qquad (82)$$

which is the well-known *Entropy Differential Equation*.

Let us now consider the Eq.(55) in the *ultra-relativistic case* where the inertial energy of the particle $E_i = M_ic^2$ is much larger than its inertial energy at rest $m_{i0}c^2$. Comparison of (4) and (10) leads to $\Delta p = E_iV/c^2$ which, in the ultra-relativistic case, gives $\Delta p = E_iV/c^2 \cong E_i/c \cong M_ic$. On the other hand, comparison of (55) and (41) shows that $Un_r = \Delta pc$. Thus $Un_r = \Delta pc \cong M_ic^2 >> m_{i0}c^2$. Consequently, Eq.(55) reduces to

$$m_g = m_{i0} - 2Un_r/c^2 \qquad (83)$$

Therefore, the *action* for such particle, in agreement with the Eq.(2), is

$$S=-\int_{t_1}^{t_2} m_g c^2\sqrt{1-V^2/c^2}\,dt=$$
$$=\int_{t_1}^{t_2}\left(-m_i+2Un_r/c^2\right)c^2\sqrt{1-V^2/c^2}\,dt=$$
$$=\int_{t_1}^{t_2}\left[-m_i c^2\sqrt{1-V^2/c^2}+2Un_r\sqrt{1-V^2/c^2}\right]dt. \quad (84)$$

The integrant function is the *Lagrangean*, i.e.,

$$L=-m_{i0}c^2\sqrt{1-V^2/c^2}+2Un_r\sqrt{1-V^2/c^2} \qquad (85)$$

Starting from the Lagrangean we can find the Hamiltonian of the particle, by means of the well-known general formula:

$$H=V(\partial L/\partial V)-L.$$

The result is

$$H=\frac{m_{i0}c^2}{\sqrt{1-V^2/c^2}}+Un_r\left[\frac{\left(4V^2/c^2-2\right)}{\sqrt{1-V^2/c^2}}\right]. \qquad (86)$$

The second term on the right hand side of Eq.(86) results from the particle's interaction with the *electromagnetic field*. Note the similarity between the obtained Hamiltonian and the well-known Hamiltonian for the particle in an electromagnetic field [32]:

$$H=m_{i0}c^2\Big/\sqrt{1-V^2/c^2}+Q\varphi. \qquad (87)$$

in which $Q$ is the electric charge and $\varphi$, the field's *scalar* potential. The quantity $Q\varphi$ expresses, as we know, the particle's interaction with the electromagnetic field in the same way as the second term on the right hand side of the Eq. (86).

It is therefore evident that it is the same quantity, expressed by different variables.

Thus, we can conclude that, in ultra-high energy conditions $\left(Un_r\cong M_i c^2>m_{i0}c^2\right)$, the gravitational and electromagnetic fields can be described by the *same* Hamiltonian, i.e., in these circumstances they are *unified* !

It is known that starting from that Hamiltonian we may obtain a complete description of the electromagnetic field. This means that from the present theory for gravity we can also derive *the equations of the electromagnetic field*.

Due to $Un_r=\Delta pc\cong M_i c^2$ the second term on the right hand side of Eq.(86) can be written as follows

$$\Delta pc\left[\frac{\left(4V^2/c^2-2\right)}{\sqrt{1-V^2/c^2}}\right]=$$
$$=\left[\frac{\left(4V^2/c^2-2\right)}{\sqrt{1-V^2/c^2}}\right]M_i c^2=$$
$$=Q\varphi=\frac{QQ'}{4\pi\varepsilon_0 R}=\frac{QQ'}{4\pi\varepsilon_0 r\sqrt{1-V^2/c^2}}$$

whence

$$\left(4V^2/c^2-2\right)M_i c^2=\frac{QQ'}{4\pi\varepsilon_0 r}$$

The factor $\left(4V^2/c^2-2\right)$ becomes equal to 2 in the ultra-relativistic case, then it follows that

$$2M_i c^2=\frac{QQ'}{4\pi\varepsilon_0 r} \qquad (88)$$

From (44), we know that there is a minimum value for $M_i$ given by $M_{i(min)}=m_{i(min)}$. Eq.(43) shows that $m_{g(min)}=m_{i0(min)}$ and Eq.(23) gives $m_{g(min)}=\pm h/cL_{max}\sqrt{8}=\pm h\sqrt{3/8}/cd_{max}$.

Thus we can write

$$M_{i(min)}=m_{i0(min)}=\pm h\sqrt{3/8}/cd_{max} \qquad (89)$$

According to (88) the value $2M_{i(min)}c^2$ is correlated to $\left(QQ'/4\pi\varepsilon_0 r\right)_{min}=Q^2_{min}/4\pi\varepsilon_0 r_{max}$, i.e.,

$$\frac{Q^2_{min}}{4\pi\varepsilon_0 r_{max}}=2M_{i(min)}c^2 \qquad (90)$$

where $Q_{min}$ is the *minimum electric charge* in the Universe ( therefore equal to minimum electric charge of the quarks, i.e., $\frac{1}{3}e$); $r_{max}$ is the *maximum distance* between $Q$ and $Q'$, which should be equal to the so-called "diameter", $d_c$, of the *visible* Universe ($d_c = 2l_c$ where $l_c$ is obtained from the Hubble's law for $V = c$, i.e., $l_c = c\widetilde{H}^{-1}$). Thus, from (90) we readily obtain

$$Q_{min} = \sqrt{\pi\varepsilon_0 hc\sqrt{24}\left(d_c/d_{max}\right)} = \\ = \sqrt{\left(\pi\varepsilon_0 hc^2\sqrt{96}\widetilde{H}^{-1}/d_{max}\right)} = \\ = \tfrac{1}{3}e \qquad (91)$$

whence we find

$$d_{max} = 3.4\times10^{30}\,m$$

This will be the maximum "diameter" that the Universe will reach. Consequently, Eq.(89) tells us that the *elementary quantum* of matter is

$$m_{i0(min)} = \pm h\sqrt{3/8}/cd_{max} = \pm 3.9\times10^{-73}kg$$

This is, therefore, the *smallest indivisible particle of matter*.

Considering that, the inertial mass of the *Observable Universe* is $M_U = c^3/2H_0G \cong 10^{53}kg$ and that its volume is $V_U = \frac{4}{3}\pi R_U^3 = \frac{4}{3}\pi\left(c/H_0\right)^3 \cong 10^{79}\,m^3$, where $H_0 = 1.75\times10^{-18}s^{-1}$ is the *Hubble constant*, we can conclude that the *number of these particles in the Observable Universe* is

$$n_U = \frac{M_U}{m_{i0(\min)}} \cong 10^{125}\ particles$$

By dividing this number by $V_U$, we get

$$\frac{n_U}{V_U} \cong 10^{46}\ particles/m^3$$

Obviously, the dimensions of the smallest indivisible particle of matter depend on its state of compression. In free space, for example, its volume is $V_U/n_U$. Consequently, its "radius" is $R_U/\sqrt[3]{n_U} \cong 10^{-15}m$.

If $N$ particles with diameter $\phi$ fill all space of $1m^3$ then $N\phi^3 = 1$. Thus, if $\phi \cong 10^{-15}m$ then the number of particles, with this diameter, necessary to fill all $1m^3$ is $N \cong 10^{45}\ particles$. Since the number of *smallest indivisible particles of matter* in the Universe is $n_U/V_U \cong 10^{46} particles/m^3$ we can conclude that these particles *fill all space* in the Universe, by forming a *Continuous*[3] *Universal Medium* or *Continuous Universal Fluid* (CUF), the density of which is

$$\rho_{CUF} = \frac{n_U\, m_{i0(\min)}}{V_U} \cong 10^{-27}\,kg/m^3$$

Note that this density is much smaller than the density of the *Intergalactic Medium* $\left(\rho_{IGM} \cong 10^{-26}\,kg/m^3\right)$.

The extremely-low density of the *Continuous Universal Fluid* shows that its *local gravitational mass* can be strongly affected by electromagnetic fields (including gravitoelectromagnetic fields), pressure, etc. (See Eqs. 57, 58, 59a, 59b, 55a, 55c and 60). The density of this fluid is clearly *not uniform* along the Universe, since it can be strongly compressed in several regions (galaxies, stars, blackholes, planets, etc). At the normal state (free space), the mentioned fluid is *invisible*. However, at *super compressed* state it can become *visible by giving origin* to the *known matter* since matter, as we have seen, is *quantized* and consequently, formed by an *integer number* of elementary quantum of matter with mass $m_{i0(\min)}$. Inside the proton, for example, there are $n_p = m_p/m_{i0(\min)} \cong 10^{45}$ *elementary quanta of matter* at supercompressed state, with volume $V_{proton}/n_p$ and "radius" $R_p/\sqrt[3]{n_p} \cong 10^{-30}\,m$.

Therefore, the solidification of the matter is just a *transitory state* of this Universal Fluid, which can back to the primitive state when the cohesion conditions disappear.

Let us now study another aspect of the present theory. By combination of gravity and the *uncertainty principle* we will derive the expression for the *Casimir force*.

An uncertainty $\Delta m_i$ in $m_i$ produces an uncertainty $\Delta p$ in $p$ and

---

[3] A*t very small scale.*

therefore an uncertainty $\Delta m_g$ in $m_g$, which according to Eq.(41) , is given by

$$\Delta m_g = \Delta m_i - 2\left[\sqrt{1+\left(\frac{\Delta p}{\Delta m_i c}\right)^2} - 1\right]\Delta m_i \quad (92)$$

From the uncertainty principle for position and momentum, we know that the product of the uncertainties of the simultaneously measurable values of the corresponding position and momentum components is at least of the magnitude order of $\hbar$, i.e.,

$$\Delta p \Delta r \sim \hbar$$

Substitution of $\Delta p \sim \hbar/\Delta r$ into (92) yields

$$\Delta m_g = \Delta m_i - 2\left[\sqrt{1+\left(\frac{\hbar/\Delta m_i c}{\Delta r}\right)^2} - 1\right]\Delta m_i \quad (93)$$

Therefore if

$$\Delta r << \frac{\hbar}{\Delta m_i c} \quad (94)$$

then the expression (93) reduces to:

$$\Delta m_g \cong -\frac{2\hbar}{\Delta r c} \quad (95)$$

Note that $\Delta m_g$ does not depend on $m_g$.

Consequently, the uncertainty $\Delta F$ in the gravitational force $F = -G m_g m'_g / r^2$, will be given by

$$\Delta F = -G\frac{\Delta m_g \Delta m'_g}{(\Delta r)^2} = -\left[\frac{2}{\pi(\Delta r)^2}\right]\frac{hc}{(\Delta r)^2}\left(\frac{G\hbar}{c^3}\right) \quad (96)$$

The amount $\left(G\hbar/c^3\right)^{1/2} = 1.61 \times 10^{-35} m$ is called the *Planck length*, $l_{planck}$ ,( the length scale on which quantum fluctuations of the metric of the space time are expected to be of order unity). Thus, we can write the expression of $\Delta F$ as follows

$$\Delta F = -\left(\frac{2}{\pi}\right)\frac{hc}{(\Delta r)^4} l^2_{planck} = -\left(\frac{\pi}{480}\right)\frac{hc}{(\Delta r)^4}\left[\left(\frac{960}{\pi^2}\right) l^2_{planck}\right] = -\left(\frac{\pi A_0}{480}\right)\frac{hc}{(\Delta r)^4} \quad (97)$$

or

$$F_0 = -\left(\frac{\pi A_0}{480}\right)\frac{hc}{r^4} \quad (98)$$

which is the expression of the *Casimir force* for $A = A_0 = \left(960/\pi^2\right) l^2_{planck}$ .

This suggests that $A_0$ is an *elementary area* related to the existence of a *minimum length* $d_{min} = \tilde{k}\, l_{planck}$ what is in accordance with the *quantization of space* (29) and which points out to the existence of $d_{min}$ .

It can be easily shown that the *minimum area* related to $d_{min}$ is the area of an *equilateral triangle* of side length $d_{min}$ ,i.e.,

$$A_{min} = \left(\tfrac{\sqrt{3}}{4}\right) d^2_{min} = \left(\tfrac{\sqrt{3}}{4}\right)\tilde{k}^2 l^2_{planck}$$

On the other hand, the *maximum area* related to $d_{min}$ is the area of a *sphere* of radius $d_{min}$ ,i.e.,

$$A_{max} = \pi d^2_{min} = \pi \tilde{k}^2 l^2_{planck}$$

Thus, the elementary area

$$A_0 = \delta_A d^2_{min} = \delta_A \tilde{k}^2 l^2_{planck} \quad (99)$$

must have a value between $A_{min}$ and $A_{max}$, i.e.,

$$\tfrac{\sqrt{3}}{4} < \delta_A < \pi$$

The previous assumption that $A_0 = \left(960/\pi^2\right) l^2_{planck}$ shows that $\delta_A \tilde{k}^2 = 960/\pi^2$ what means that

$$5.6 < \tilde{k} < 14.9$$

Therefore we conclude that

$$d_{min} = \tilde{k}\, l_{planck} \approx 10^{-34} m. \quad (100)$$

The $n-esimal$ area after $A_0$ is

$$A = \delta_A (nd_{min})^2 = n^2 A_0 \qquad (101)$$

It can also be easily shown that the *minimum volume* related to $d_{min}$ is the volume of a *regular tetrahedron* of edge length $d_{min}$, i.e.,

$$\Omega_{min} = \left(\tfrac{\sqrt{2}}{12}\right) d_{min}^3 = \left(\tfrac{\sqrt{2}}{12}\right) \tilde{k}^3 l_{planck}^3$$

The *maximum volume* is the volume of a *sphere* of radius $d_{min}$, i.e.,

$$\Omega_{max} = \left(\tfrac{4\pi}{3}\right) d_{min}^3 = \left(\tfrac{4\pi}{3}\right) \tilde{k}^3 l_{planck}^3$$

Thus, the elementary volume $\Omega_0 = \delta_V d_{min}^3 = \delta_V \tilde{k}^3 l_{planck}^3$ must have a value between $\Omega_{min}$ and $\Omega_{max}$, i.e.,

$$\left(\tfrac{\sqrt{2}}{12}\right) < \delta_V < \tfrac{4\pi}{3}$$

On the other hand, the $n-esimal$ volume after $\Omega_0$ is

$$\Omega = \delta_V (nd_{min})^3 = n^3 \Omega_0 \qquad n = 1,2,3,\ldots,n_{max}.$$

The existence of $n_{max}$ given by (26), i.e.,

$$n_{max} = L_{max}/L_{min} = d_{max}/d_{min} = $$
$$= \left(3.4 \times 10^{30}\right) / \tilde{k} l_{planck} \approx 10^{64}$$

shows that the Universe must have a *finite volume* whose value at the present stage is

$$\Omega_{Up} = n_{Up}^3 \Omega_0 = \left(d_p / d_{min}\right)^3 \delta_V d_{min}^3 = \delta_V d_p^3$$

where $d_p$ is the present length scale of the Universe. In addition as $\left(\tfrac{\sqrt{2}}{12}\right) < \delta_V < \tfrac{4\pi}{3}$ we conclude that the Universe must have a *polyhedral* space topology with volume between the volume of a *regular tetrahedron* of edge length $d_p$ and the volume of the *sphere of* diameter $d_p$.

A recent analysis of astronomical data suggests not only that the Universe is *finite*, but also that it has a *dodecahedral* space topology [33,34], what is in strong accordance with the previous theoretical predictions.

From (22) and (26) we have that $L_{max} = d_{max}/\sqrt{3} = n_{max} d_{min}/\sqrt{3}$. Since (100) gives $d_{min} \cong 10^{-34} m$ and $n_{max} \cong 10^{64}$ we conclude that $L_{max} \cong 10^{30} m$. From the *Hubble's law* and (22) we have that

$$V_{max} = \tilde{H} l_{max} = \tilde{H}(d_{max}/2) = \left(\sqrt{3}/2\right) \tilde{H} L_{max}$$

where $\tilde{H} = 1.7 \times 10^{-18} s^{-1}$. Therefore we obtain

$$V_{max} \cong 10^{12} m/s \qquad .$$

This is the speed upper limit imposed by the *quantization of velocity* (Eq. 36). It is known that the speed upper limit for *real* particles is equal to $c$. However, also it is known that *imaginary* particles can have velocities greater than $c$ (*Tachyons*). Thus, we conclude that $V_{max}$ is the speed upper limit for *imaginary particles in our ordinary space-time*. Later on, we will see that also exists a speed upper limit to the *imaginary* particles in the *imaginary* space-time.

Now, multiplying Eq. (98) (the expression of $F_0$) by $n^2$ we obtain

$$F = n^2 F_0 = -\left(\frac{\pi n^2 A_0}{480}\right)\frac{hc}{r^4} = -\left(\frac{\pi A}{480}\right)\frac{hc}{r^4} \qquad (102)$$

This is the general expression of the *Casimir force*.

Thus, we conclude that *the Casimir effect* is just a gravitational effect related to the *uncertainty principle*.

Note that Eq. (102) arises only when $\Delta m_i$ and $\Delta m_i'$ satisfy Eq.(94). If only $\Delta m_i$ satisfies Eq.(94), i.e., $\Delta m_i << \hbar/\Delta rc$ but $\Delta m_i' >> \hbar/\Delta rc$ then $\Delta m_g$ and $\Delta m_g'$ will be respectively given by

$$\Delta m_g \cong -2\hbar/\Delta rc \quad and \quad \Delta m_g' \cong \Delta m_i$$

Consequently, the expression (96) becomes

$$\Delta F = \frac{hc}{(\Delta r)^3}\left(\frac{G\Delta m_i'}{\pi c^2}\right) = \frac{hc}{(\Delta r)^3}\left(\frac{G\Delta m_i' c^2}{\pi c^4}\right) = $$

$$= \frac{hc}{(\Delta r)^3}\left(\frac{G\Delta E'}{\pi c^4}\right) \qquad (103)$$

However, from the uncertainty principle for *energy* and *time* we know that

$$\Delta E \sim \hbar/\Delta t \qquad (104)$$

Therefore, we can write the expression (103) in the following form:

$$\Delta F = \frac{hc}{(\Delta r)^3}\left(\frac{G\hbar}{c^3}\right)\left(\frac{1}{\pi \Delta t' c}\right) = $$

$$= \frac{hc}{(\Delta r)^3} l^2_{planck}\left(\frac{1}{\pi \Delta t' c}\right) \qquad (105)$$

From the General Relativity Theory we know that $dr = cdt/\sqrt{-g_{00}}$ . If the field is *weak* then $g_{00} = -1 - 2\phi/c^2$ and $dr = cdt/(1+\phi/c^2) = cdt/(1 - Gm/r^2c^2)$. For $Gm/r^2 c^2 << 1$ we obtain $dr \cong cdt$. Thus, if $dr = dr'$ then $dt = dt'$. This means that we can change $(\Delta t' c)$ by $(\Delta r)$ into (105). The result is

$$\Delta F = \frac{hc}{(\Delta r)^4}\left(\frac{1}{\pi} l^2_{planck}\right) = $$

$$= \left(\frac{\pi}{480}\right)\frac{hc}{(\Delta r)^4}\underbrace{\left(\frac{480}{\pi^2} l^2_{planck}\right)}_{\frac{1}{2}A_0} = $$

$$= \left(\frac{\pi A_0}{960}\right)\frac{hc}{(\Delta r)^4}$$

or

$$F_0 = \left(\frac{\pi A_0}{960}\right)\frac{hc}{r^4}$$

whence

$$F = \left(\frac{\pi A}{960}\right)\frac{hc}{r^4} \qquad (106)$$

Now, the Casimir force is *repulsive*, and its intensity is half of the intensity previously obtained (102).

Consider the case when both $\Delta m_i$ and $\Delta m_i'$ do not satisfy Eq.(94), and

$$\Delta m_i >> \hbar/\Delta rc$$

$$\Delta m_i' >> \hbar/\Delta rc$$

In this case, $\Delta m_g \cong \Delta m_i$ and $\Delta m_g' \cong \Delta m_i'$. Thus,

$$\Delta F = -G\frac{\Delta m_i \Delta m_i'}{(\Delta r)^2} = -G\frac{(\Delta E/c^2)(\Delta E'/c^2)}{(\Delta r)^2} = $$

$$= -\left(\frac{G}{c^4}\right)\frac{(\hbar/\Delta t)^2}{(\Delta r)^2} = -\left(\frac{G\hbar}{c^3}\right)\frac{hc}{(\Delta r)^2}\left(\frac{1}{c^2\Delta t^2}\right) = $$

$$= -\left(\frac{1}{2\pi}\right)\frac{hc}{(\Delta r)^4} l^2_{planck} = $$

$$= -\left(\frac{\pi}{1920}\right)\frac{hc}{(\Delta r)^4}\left(\frac{960}{\pi^2} l^2_{planck}\right) = -\left(\frac{\pi A_0}{1920}\right)\frac{hc}{(\Delta r)^4}$$

whence

$$F = -\left(\frac{\pi A}{1920}\right)\frac{hc}{r^4} \qquad (107)$$

The force will be *attractive* and its intensity will be the *fourth part* of the intensity given by the first expression (102) for the Casimir force.

We can also use this theory to explain some relevant cosmological phenomena. For example, the recent discovery that the cosmic expansion of the Universe may be *accelerating*, and not decelerating as many cosmologists had anticipated [35].

We start from Eq. (6) which shows that the *inertial forces*, $\vec{F}_i$, whose action on a particle, in the case of force and speed with *same direction*, is given by

$$\vec{F}_i = \frac{m_g}{(1 - V^2/c^2)^{3/2}}\vec{a}$$

Substitution of $m_g$ given by (43) into the expression above gives

$$\vec{F}_i = \left(\frac{3}{(1 - V^2/c^2)^{3/2}} - \frac{2}{(1 - V^2/c^2)^2}\right) m_{i0}\vec{a}$$

whence we conclude that a particle with rest inertial mass, $m_{i0}$, subjected to a force, $\vec{F}_i$, acquires an acceleration $\vec{a}$ given by

$$\vec{a} = \frac{\vec{F}_i}{\left(\frac{3}{\left(1 - V^2/c^2\right)^{3/2}} - \frac{2}{\left(1 - V^2/c^2\right)^2}\right) m_{i0}}$$

By substituting the well-known expression of Hubble's law for velocity, $V = \tilde{H}l$, ($\tilde{H} = 1.7 \times 10^{-18} s^{-1}$ is the Hubble constant) into the expression of $\vec{a}$, we get *the acceleration for any particle in the expanding Universe*, i.e.,

$$\vec{a} = \frac{\vec{F}_i}{\left(\frac{3}{\left(1 - \tilde{H}^2 l^2/c^2\right)^{3/2}} - \frac{2}{\left(1 - \tilde{H}^2 l^2/c^2\right)^2}\right) m_{i0}}$$

Obviously, the distance $l$ increases with the expansion of the Universe. Under these circumstances, it is easy to see that the term

$$\left(\frac{3}{\left(1 - \tilde{H}^2 l^2/c^2\right)^{3/2}} - \frac{2}{\left(1 - \tilde{H}^2 l^2/c^2\right)^2}\right)$$

decreases, *increasing the acceleration* of the expanding Universe.

Let us now consider the phenomenon of gravitational deflection of light.

A distant star's light ray, under the Sun's gravitational force field describes the usual central force hyperbolic orbit. The deflection of the light ray is illustrated in Fig. V, with the bending greatly exaggerated for a better view of the angle of deflection.

The distance CS is the distance $d$ of closest approach. The angle of deflection of the light ray, $\delta$, is shown in the Figure V and is

$$\delta = \pi - 2\beta.$$

where $\beta$ is the angle of the asymptote to the hyperbole. Then, it follows that

$$\tan\delta = \tan(\pi - 2\beta) = -\tan 2\beta$$

From the Figure V we obtain

$$\tan\beta = \frac{V_y}{c}.$$

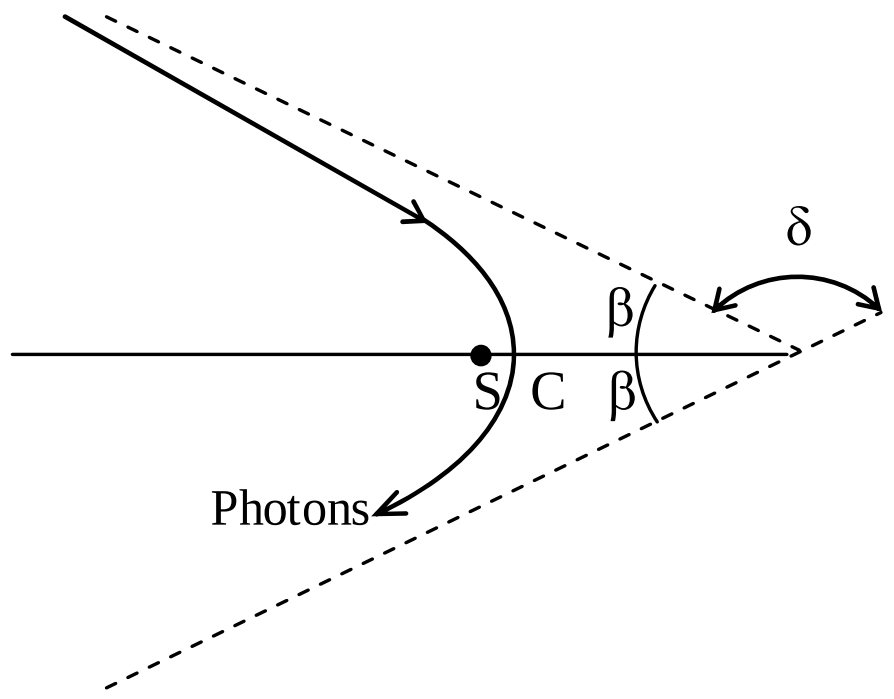

Fig. V – Gravitational deflection of light about the Sun.

Since $\delta$ and $\beta$ are very small we can write that

$$\delta = 2\beta \quad \text{and} \quad \beta = \frac{V_y}{c}$$

Then

$$\delta = \frac{2V_y}{c}$$

Consider the motion of the photons at some time $t$ after it has passed the point of closest approach. We impose Cartesian Co-ordinates with the origin at the point of closest approach, the x axis pointing along its path and the y axis towards the Sun. The gravitational pull of the Sun is

$$P = -G\frac{M_{gS}M_{gp}}{r^2}$$

where $M_{gp}$ is the relativistic *gravitational* mass of the photon and $M_{gS}$ the relativistic gravitational mass of the Sun. Thus, the component in a perpendicular direction is

$$F_y = -G\frac{M_{gS}M_{gp}}{r^2}\sin\beta =$$

$$= -G\frac{M_{gS}M_{gp}}{d^2 + c^2t^2}\frac{d}{\sqrt{d^2 + c^2t^2}}$$

According to Eq. (6) the expression of the force $F_y$ is

$$F_y = \left(\frac{m_{gp}}{\left(1 - V_y^2/c^2\right)^{\frac{3}{2}}}\right)\frac{dV_y}{dt}$$

By substituting Eq. (43) into this expression, we get

$$F_y = \left( \frac{3}{\left(1 - V_y^2/c^2\right)} - \frac{2}{\left(1 - V_y^2/c^2\right)^{\frac{3}{2}}} \right) M_{ip} \frac{dV_y}{dt}$$

For $V_y << c$, we can write this expression in the following form $F_y = M_{ip}\left(dV_y/dt\right)$. This force acts on the photons for a time $t$ causing an increase in the transverse velocity

$$dV_y = \frac{F_y}{M_{ip}} dt$$

Thus the component of transverse velocity acquired after passing the point of closest approach is

$$V_y = \frac{M_{gp}}{M_{ip}} \int \frac{d\left(-GM_{gS}\right)}{\left(d^2 + c^2 t^2\right)^{\frac{3}{2}}} dt =$$

$$= \frac{-GM_{gS}}{dc} \left( \frac{M_{gp}}{M_{ip}} \right) = \frac{-GM_{gS}}{dc} \left( \frac{m_{gp}}{m_{ip}} \right)$$

Since the angle of deflection $\delta$ is given by

$$\delta = 2\beta = \frac{2V_y}{c}$$

we readily obtain

$$\delta = \frac{2V_y}{c} = \frac{-2GM_{gS}}{c^2 d} \left( \frac{m_{gp}}{m_{ip}} \right)$$

If $m_{gp}/m_{ip} = 2$, the expression above gives

$$\delta = -\frac{4GM_{gS}}{c^2 d}$$

As we know, this is the correct formula indicated by the experimental results.

Equation (4) says that

$$m_{gp} = \left\{ 1 - 2 \left[ \sqrt{1 + \left( \frac{\Delta p}{m_{ip} c} \right)^2} - 1 \right] \right\} m_{ip}$$

Since $m_{gp}/m_{ip} = 2$ then, by making $\Delta p = h/\lambda$ into the equation above we get

$$m_{ip} = +\frac{2}{\sqrt{3}} \left( \frac{hf}{c^2} \right) i$$

Due to $m_{gp}/m_{ip} = 2$ we get

$$m_{gp} = +\frac{4}{\sqrt{3}} \left( \frac{hf}{c^2} \right) i$$

This means that the gravitational and inertial masses of the *photon* are *imaginaries*, and *invariants* with respect to speed of photon, i.e. $M_{ip} = m_{ip}$ and $M_{gp} = m_{gp}$. On the other hand, we can write that

$$m_{ip} = m_{ip(real)} + m_{ip(imaginary)} = \frac{2}{\sqrt{3}} \left( \frac{hf}{c^2} \right) i$$

and

$$m_{gp} = m_{gp(real)} + m_{gp(imaginary)} = \frac{4}{\sqrt{3}} \left( \frac{hf}{c^2} \right) i$$

This means that we must have

$$m_{ip(real)} = m_{gp(real)} = 0$$

The phenomenon of *gravitational deflection of light about the Sun* shows that *the gravitational interaction* between the Sun and the photons is *attractive*. Thus, due to the gravitational force between the Sun and a photon can be expressed by $F = -G M_{g(Sun)} m_{gp(imaginary)}/r^2$, where $m_{gp(imaginary)}$ is a quantity *positive* and *imaginary*, we conclude that the force $F$ will only be *attractive* if *the matter* $\left(M_{g(Sun)}\right)$ *has negative imaginary gravitational mass*.

The Eq. (41) shows that if the *inertial mass* of a particle is *null* then its *gravitational mass* is given by

$$m_g = \pm 2\Delta p/c$$

where $\Delta p$ is the *momentum* variation due to the energy absorbed by the particle. If the energy of the particle is *invariant*, then $\Delta p = 0$ and, consequently, its *gravitational mass* will also be null. This is the case of the photons, i.e., they have an invariant energy $hf$ and a *momentum* $h/\lambda$. As they cannot absorb additional energy, the variation in the *momentum* will be null $(\Delta p = 0)$ and, therefore, their *gravitational masses* will also be null.

However, if the energy of the particle is not invariant (it is able to absorb energy) then the absorbed energy will transfer the *amount of motion* (*momentum*) to the particle, and consequently its *gravitational mass* will be increased. This means that the *motion* generates gravitational mass*.*

On the other hand, if the *gravitational mass* of a particle is null then its *inertial mass*, according to Eq. (41), will be given by

$$m_i = \pm \frac{2}{\sqrt{5}}\frac{\Delta p}{c}$$

From Eqs. (4) and (7) we get

$$\Delta p = \left(\frac{E_g}{c^2}\right)\Delta V = \left(\frac{p_0}{c}\right)\Delta V$$

Thus we have

$$m_g = \pm\left(\frac{2p_0}{c^2}\right)\Delta V \text{ and } m_i = \pm\frac{2}{\sqrt{5}}\left(\frac{p_0}{c^2}\right)\Delta V$$

Note that, like the gravitational mass, the inertial mass is also directly related to the motion, i.e., it is also generated by the motion.

Thus, we can conclude that is the motion, or rather, the *velocity* is what makes the two types of mass.

In this picture, the fundamental particles can be considered as *immaterial vortex of velocity*; it is the velocity of these vortexes that causes the fundamental particles to have masses. That is, there exists not matter in the usual sense; but just *motion*. Thus, the difference between matter and energy just consists of the diversity of the motion direction; *rotating*, closed in itself, in the matter; *ondulatory*, with open cycle, in the energy (See Fig. VI).

Under this context, the *Higgs mechanism*† appears as a process, by which the velocity of an immaterial vortex can be increased or decreased by making the vortex (particle) *gain* or *lose mass*. If *real motion* is what makes *real mass* then, by analogy, we can say that *imaginary mass* is made by *imaginary motion*. This is not only a simple generalization of the process based on the theory of the *imaginary functions*, but also a fundamental conclusion related to the concept of *imaginary mass* that, as it will be shown, provides a coherent explanation for the *materialization* of the fundamental particles, in the beginning of the Universe.

It is known that the simultaneous disappearance of a pair (electron/positron) liberates an amount of energy, $2m_{i0e(real)}c^2$, under the form of two photons with frequency $f$, in such a way that

$$2m_{i0e(real)}c^2 = 2hf$$

Since the photon has *imaginary* masses associated to it, the phenomenon of transformation of the energy $2m_{i0e(real)}c^2$ into $2hf$ suggests that the imaginary energy of the photon, $m_{ip(imaginary)}c^2$, comes from the transformation of imaginary energy of the electron, $m_{i0e(imaginary)}c^2$, just as the real energy of the photon, $hf$, results from the transformation of real energy of the electron, i.e.,

$$2m_{i0e(imaginary)}c^2 + 2m_{i0e(real)}c^2 = \\ = 2m_{i0p(imaginary)}c^2 + 2hf$$

Then, it follows that

$$m_{i0e(imaginary)} = -m_{ip(imaginary)}$$

The sign (-) in the equation above, is due to the imaginary mass of the *photon* to be *positive*, on the contrary of the imaginary gravitational mass of the *matter*, which is *negative*, as we have already seen.

† The Standard Model is the name given to the current theory of fundamental particles and how they interact. This theory includes: *Strong interaction* and a combined theory of weak and electromagnetic interaction, known as *electroweak* theory. One part of the Standard Model is not yet well established. *What causes the fundamental particles to have masses*? The simplest idea is called the *Higgs mechanism*. This mechanism involves one additional particle, called the Higgs boson, and one additional force type, mediated by exchanges of this boson.

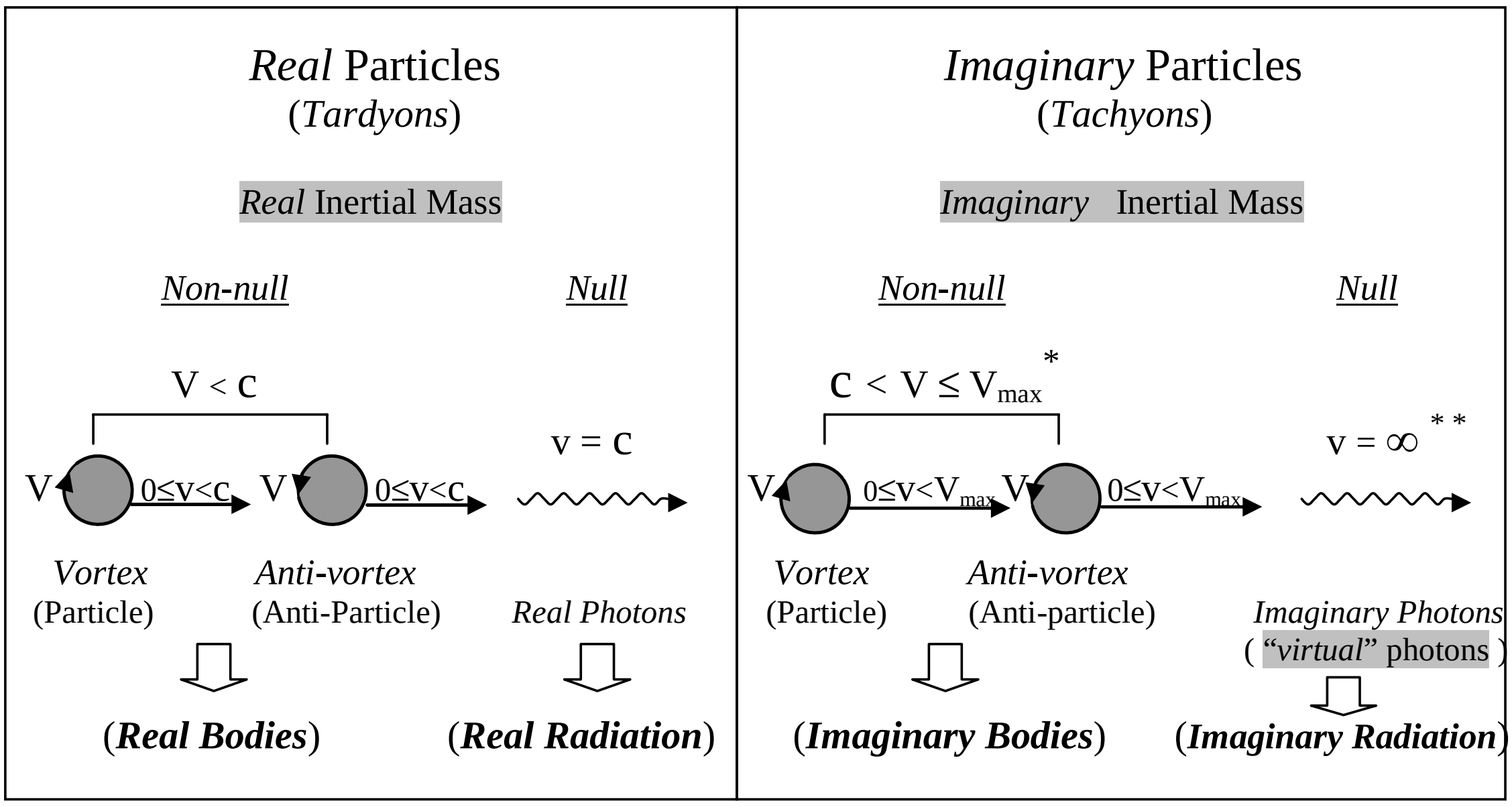

* $V_{max}$ is the *speed upper limit* for Tachyons with *non-null* imaginary inertial mass. It has been previously obtained starting from the *Hubble's law* and Eq.(22). The result is: $V_{max} = \left(\sqrt{3}/2\right)\tilde{H}L_{max} \cong 10^{12}\, m.s^{-1}$.

** In order to communicate instantaneously the *interactions* at infinite distance the velocity of the *quanta* ("virtual" photons) must be *infinity* and consequently their imaginary masses must be *null* .

Fig. VI - Real and Imaginary Particles.

Thus, we then conclude that

$$m_{i0e(imaginary)} = -m_{ip(imaginary)} =$$
$$= -\tfrac{2}{\sqrt{3}}\left(hf_e/c^2\right)\, i =$$
$$= -\tfrac{2}{\sqrt{3}}\left(h/\lambda_e c\right)\, i = -\tfrac{2}{\sqrt{3}}\, m_{i0e(real)}\, i$$

where $\lambda_e = h/m_{i0e(real)}\,c$ is the *Broglies' wavelength* for the electron.

By analogy, we can write for the *neutron* and the *proton* the following masses:

$$m_{i0neutron(imaginary)} = -\tfrac{2}{\sqrt{3}}\, m_{i0neutron(real)}\ i$$
$$m_{i0proton(imaginary)} = +\tfrac{2}{\sqrt{3}}\, m_{i0proton(real)}\ i$$

The sign (+) in the expression of $m_{i0proton(imaginary)}$ is due to the fact that $m_{i0neutron(imaginary)}$ and $m_{i0proton(imaginary)}$ must have contrary signs, as will be shown later on.

Thus, the *electron*, the *neutron* and the *proton* have respectively, the following masses:

*Electron*

$$m_{i0e(real)} = 9.11\times 10^{-31} kg$$

$$m_{i0e(im)} = -\tfrac{2}{\sqrt{3}}\, m_{i0e(real)}\, i$$

$$m_{ge(real)} = \left\{1 - 2\left[\sqrt{1+\left(\frac{U_{e(real)}}{m_{i0e(real)}c^2}\right)^2} - 1\right]\right\} m_{i0e(real)} =$$
$$= \chi_e m_{i0e(real)}$$

$$m_{ge(im)} = \left\{1 - 2\left[\sqrt{1+\left(\frac{U_{e(im)}}{m_{i0e(im)}c^2}\right)^2} - 1\right]\right\} m_{i0e(im)} =$$
$$= \chi_e m_{i0e(im)}$$

## Neutron

$m_{i0n(real)} = 1.6747 \times 10^{-27} kg$

$m_{i0n(im)} = -\frac{2}{\sqrt{3}} m_{i0n(real)}\, i$

$$m_{gn(real)} = \left\{1 - 2\left[\sqrt{1 + \left(\frac{U_{n(real)}}{m_{i0n(real)}c^2}\right)^2} - 1\right]\right\} m_{i0n(real)} = $$
$$= \chi_n m_{i0n(real)}$$

$$m_{gn(im)} = \left\{1 - 2\left[\sqrt{1 + \left(\frac{U_{n(im)}}{m_{i0n(im)}c^2}\right)^2} - 1\right]\right\} m_{i0n(im)} = $$
$$= \chi_n m_{i0n(im)}$$

## Proton

$m_{i0pr(real)} = 1.6723 \times 10^{-27} kg$

$m_{i0pr(im)} = +\frac{2}{\sqrt{3}} m_{i0pr(real)}\, i$

$$m_{gpr(real)} = \left\{1 - 2\left[\sqrt{1 + \left(\frac{U_{pr(real)}}{m_{i0pr(real)}c^2}\right)^2} - 1\right]\right\} m_{i0pr(real)} = $$
$$= \chi_{pr} m_{i0pr(real)}$$

$$m_{gpr(im)} = \left\{1 - 2\left[\sqrt{1 + \left(\frac{U_{pr(im)}}{m_{i0pr(im)}c^2}\right)^2} - 1\right]\right\} m_{i0pr(im)} = $$
$$= \chi_{pr} m_{i0pr(im)}$$

where $U_{(real)}$ and $U_{(im)}$ are respectively, the *real* and *imaginary* energies absorbed by the particles.

When *neutrons*, *protons* and *electrons* were created after the Big-bang, they absorbed quantities of electromagnetic energy, respectively given by

$$U_{n(real)} = \eta_n kT_n \qquad U_{n(imaginary)} = \eta_n kT_n\, i$$
$$U_{pr(real)} = \eta_{pr} kT_{pr} \qquad U_{pr(imaginary)} = \eta_{pr} kT_{pr}\, i$$
$$U_{e(real)} = \eta_e kT_e \qquad U_{e(imaginary)} = \eta_e kT_e\, i$$

where $\eta_n$, $\eta_{pr}$ and $\eta_e$ are the *absorption factors* respectively, for the neutrons, protons and electrons; $k = 1.38 \times 10^{-23} J/^{o}K$ is the *Boltzmann constant*; $T_n$, $T_{pr}$ and $T_e$ are the temperatures of the Universe, respectively when neutrons, protons and electrons were created.

In the case of the electrons, it was previously shown that $\eta_e \cong 0.1$. Thus, by considering that $T_e \cong 6.2 \times 10^{31} K$, we get

$$U_{e(im)} = \eta_e kT_e\, i = 8.5 \times 10^7\, i$$

It is known that the protons were created at the same epoch. Thus, we will assume that

$$U_{pr(im)} = \eta_{pr} kT_{pr}\, i = 8.5 \times 10^7\, i$$

Then, it follows that

$$\chi_e = -1.8 \times 10^{21}$$
$$\chi_{pr} = -9.7 \times 10^{17}$$

Now, consider the gravitational forces, due to the *imaginary masses* of *two electrons,* $F_{ee}$, *two protons,* $F_{prpr}$, and *one electron and one proton,* $F_{epr}$, all at rest.

$$F_{ee} = -G\frac{m^2_{ge(im)}}{r^2} = -G\chi_e^2 \frac{\left(-\frac{2}{\sqrt{3}} m_{i0e(real)} i\right)^2}{r^2} = $$
$$= +\frac{4}{3}G\chi_e^2 \frac{m^2_{i0e(real)}}{r^2} = \frac{+2.3 \times 10^{-28}}{r^2} \quad (repulsion)$$

$$F_{prpr} = -G\frac{m^2_{gpr(im)}}{r^2} = -G\chi_{pr}^2 \frac{\left(+\frac{2}{\sqrt{3}} m_{i0pr(real)} i\right)^2}{r^2} = $$
$$= +\frac{4}{3}G\chi_{pr}^2 \frac{m^2_{i0pr(real)}}{r^2} = \frac{+2.3 \times 10^{-28}}{r^2} \quad (repulsion)$$

$$F_{epr} = -G\frac{m_{ge(im)} m_{gpr(im)}}{r^2} = $$
$$= -G\chi_e\chi_{pr} \frac{\left(-\frac{2}{\sqrt{3}} m_{i0e(real)} i\right)\left(+\frac{2}{\sqrt{3}} m_{i0pr(real)} i\right)}{r^2} = $$
$$= -\frac{4}{3}G\chi_e\chi_{pr} \frac{m_{i0e(real)} m_{i0pr(real)}}{r^2} = \frac{-2.3 \times 10^{-28}}{r^2}$$
$$(atraction)$$

Note that

$$F_{electric}=\frac{e^2}{4\pi\varepsilon_0 r^2}=\frac{2.3\times10^{-28}}{r^2}$$

Therefore, we can conclude that

$$F_{ee}=F_{prpr}\equiv F_{electric}=+\frac{e^2}{4\pi\varepsilon_0 r^2}\quad (repulsion)$$

and

$$F_{ep}\equiv F_{electric}=-\frac{e^2}{4\pi\varepsilon_0 r^2}\quad (atraction)$$

These correlations permit to define the *electric charge* by means of the following relation:

$$q=\sqrt{4\pi\varepsilon_0 G}\;\; m_{g(imaginary)}\; i$$

For example, in the case of the *electron*, we have

$$q_e=\sqrt{4\pi\varepsilon_0 G}\;\; m_{ge(imaginary)}\; i=$$
$$=\sqrt{4\pi\varepsilon_0 G}\left(\chi_e m_{i0e(imaginary)} i\right)=$$
$$=\sqrt{4\pi\varepsilon_0 G}\left(-\chi_e \tfrac{2}{\sqrt{3}} m_{i0e(real)} i^2\right)=$$
$$=\sqrt{4\pi\varepsilon_0 G}\left(\chi_e \tfrac{2}{\sqrt{3}} m_{i0e(real)}\right)=-1.6\times10^{-19}C$$

In the case of the *proton*, we get

$$q_{pr}=\sqrt{4\pi\varepsilon_0 G}\;\; m_{gpr(imaginary)}\; i=$$
$$=\sqrt{4\pi\varepsilon_0 G}\left(\chi_{pr} m_{i0pr(imaginary)} i\right)=$$
$$=\sqrt{4\pi\varepsilon_0 G}\left(+\chi_{pr} \tfrac{2}{\sqrt{3}} m_{i0pr(real)} i^2\right)=$$
$$=\sqrt{4\pi\varepsilon_0 G}\left(-\chi_{pr} \tfrac{2}{\sqrt{3}} m_{i0pr(real)}\right)=+1.6\times10^{-19}C$$

For the *neutron*, it follows that

$$q_n=\sqrt{4\pi\varepsilon_0 G}\;\; m_{gn(imaginary)}\; i=$$
$$=\sqrt{4\pi\varepsilon_0 G}\left(\chi_n m_{i0n(imaginary)} i\right)=$$
$$=\sqrt{4\pi\varepsilon_0 G}\left(-\chi_n \tfrac{2}{\sqrt{3}} m_{i0n(real)} i^2\right)=$$
$$=\sqrt{4\pi\varepsilon_0 G}\left(\chi_n \tfrac{2}{\sqrt{3}} m_{i0n(real)}\right)$$

However, based on the *quantization of the mass* (Eq. 44), we can write that

$$\chi_n \tfrac{2}{\sqrt{3}} m_{i0n(real)}=n^2 m_{i0(\min)}\qquad n\neq 0$$

Since $n$ can have only discrete values *different of zero* (See Appendix B), we conclude that $\chi_n$ cannot be null. However, it is known that the electric charge of the neutron is *null*. Thus, it is necessary to assume that

$$q_n=q_n^+ + q_n^- =\sqrt{4\pi\varepsilon_0 G}\;\; m^+_{gn(imaginary)}\; i+$$
$$+\sqrt{4\pi\varepsilon_0 G}\;\; m^-_{gn(imaginary)}\; i=$$
$$=\sqrt{4\pi\varepsilon_0 G}\left(\chi_n m^+_{i0n(imaginary)}\; i\right)+$$
$$+\sqrt{4\pi\varepsilon_0 G}\left(\chi_n m^-_{i0n(imaginary)}\; i\right)=$$
$$=\sqrt{4\pi\varepsilon_0 G}\left[\chi_n\left(+\tfrac{2}{\sqrt{3}} m_{i0n} i^2\right)+\chi_n\left(-\tfrac{2}{\sqrt{3}} m_{i0n} i^2\right)\right]=0$$

We then conclude that in the neutron, *half* of the total amount of *elementary quanta of electric charge*, $q_{\min}$, is *negative*, while the other *half* is *positive*.

In order to obtain the value of the *elementary quantum of electric charge*, $q_{\min}$, we start with the expression obtained here for the electric charge, where we change $m_{g(imaginary)}$ by its quantized expression $m_{g(imaginary)}=n^2 m_{i0(imaginary)(\min)}$, derived from Eq. (44a). Thus, we get

$$q=\sqrt{4\pi\varepsilon_0 G}\;\; m_{g(imaginary)}\; i=$$
$$=\sqrt{4\pi\varepsilon_0 G}\;\; n^2 m_{i0(imaginary)(\min)} i=$$
$$=\sqrt{4\pi\varepsilon_0 G}\;\; \left[n^2\left(\pm\tfrac{2}{\sqrt{3}} m_{i0(\min)} i\right)\right] i=$$
$$=\mp\tfrac{2}{\sqrt{3}}\sqrt{4\pi\varepsilon_0 G}\;\; n^2 m_{i0(\min)}$$

This is the *quantized expression of the electric charge*.

For $n=1$ we obtain the value of the *elementary quantum of electric charge*, $q_{\min}$, i.e.,

$$q_{\min}=\mp\tfrac{2}{\sqrt{3}}\sqrt{4\pi\varepsilon_0 G}\;\; m_{i0(\min)}=\mp3.8\times10^{-83}C$$

where $m_{i0(\min)}$ is the *elementary quantum* of matter, whose value previously calculated, is $m_{i0(\min)}=\pm3.9\times10^{-73}kg$.

The existence of *imaginary* mass associated to a *real* particle suggests the possible existence of *imaginary*

*particles* with imaginary masses in Nature.

In this case, the concept of *wave associated* to a particle (De Broglie's waves) would also be applied to the imaginary particles. Then, by analogy, the imaginary wave associated to an imaginary particle with imaginary masses $m_{i\psi}$ and $m_{g\psi}$ would be described by the following expressions

$$\vec{p}_\psi = \hbar \vec{k}_\psi$$

$$E_\psi = \hbar \omega_\psi$$

Henceforth, for the sake of simplicity, we will use the Greek letter $\psi$ to stand for the word *imaginary*; $\vec{p}_\psi$ is the *momentum* carried by the $\psi$ *wave* and $E_\psi$ its energy; $\left|\vec{k}_\psi\right| = 2\pi/\lambda_\psi$ is the propagation number and $\lambda_\psi$ the wavelength of the $\psi$ *wave;* $\omega_\psi = 2\pi f_\psi$ is the cyclical frequency.

According to Eq. (4), the *momentum* $\vec{p}_\psi$ is

$$\vec{p}_\psi = M_{g\psi}\vec{V}$$

where $V$ is the velocity of the $\psi$ particle.

By comparing the expressions of $\vec{p}_\psi$ we get

$$\lambda_\psi = \frac{h}{M_{g\psi}V}$$

It is known that the variable quantity which characterizes the De Broglie's waves is called *wave function,* usually indicated by symbol $\Psi$. The wave function associated with a material particle describes the dynamic state of the particle: its value at a particular point x, y, z, t is related to the probability of finding the particle in that place and instant. Although $\Psi$ does not have a physical interpretation, its square $\Psi^2$ (or $\Psi\ \Psi^*$) calculated for a particular point x, y, z, t *is proportional to the probability of finding the particle in that place and instant*.

Since $\Psi^2$ is proportional to the probability $P$ of finding the particle described by $\Psi$, the integral of $\Psi^2$ on the *whole space* must be finite – inasmuch as the particle is somewhere.

On the other hand, if

$$\int_{-\infty}^{+\infty} \Psi^2 dV = 0$$

the interpretation is that the particle will not exist. However, if

$$\int_{-\infty}^{+\infty} \Psi^2 dV = \infty \qquad (108)$$

*The particle will be everywhere simultaneously*.

In Quantum Mechanics, the wave function $\Psi$ corresponds, as we know, to the displacement $y$ of the undulatory motion of a rope. However, $\Psi$, as opposed to $y$, is not a measurable quantity and can, hence, be a complex quantity. For this reason, it is assumed that $\Psi$ is described in the $x-direction$ by

$$\Psi = \Psi_0 e^{-(2\pi\, i/h)(Et-px)}$$

This is the expression of the wave function for a *free* particle, with total energy $E$ and *momentum* $\vec{p}$, moving in the direction $+x$.

As to the imaginary particle, the *imaginary particle wave function* will be denoted by $\Psi_\psi$ and, by analogy the expression of $\Psi$, will be expressed by:

$$\Psi_\psi = \Psi_{0\psi} e^{-(2\pi\, i/h)(E_\psi t - p_\psi x)}$$

Therefore, the *general expression* of the wave function for a *free* particle can be written in the following form

$$\Psi = \Psi_{0(real)} e^{-(2\pi\, i/h)(E_{(real)}t - p_{(real)}x)} + \\ + \Psi_{0\psi} e^{-(2\pi\, i/h)(E_\psi t - p_\psi x)}$$

It is known that the *uncertainty principle* can also be written as a function of $\Delta E$ (uncertainty in the energy) and $\Delta t$ (uncertainty in the time), i.e.,

$$\Delta E . \Delta t \geq \hbar$$

This expression shows that a variation of energy $\Delta E$, during a time interval $\Delta t$, can only be detected if $\Delta t \geq \hbar/\Delta E$. Consequently, a variation of energy $\Delta E$, during a time interval $\Delta t < \hbar/\Delta E$, cannot be experimentally detected. This is a limitation imposed by Nature and not by our equipments.

Thus, a *quantum* of energy $\Delta E = hf$ that varies during a time interval $\Delta t = 1/f = \lambda/c < \hbar/\Delta E$ (wave period) cannot be experimentally detected. This is an *imaginary* photon or a "*virtual*" photon.

Now, consider a particle with energy $M_g c^2$. The DeBroglie's gravitational and inertial wavelengths are respectively $\lambda_g = h/M_g c$ and $\lambda_i = h/M_i c$. In Quantum Mechanics, particles of matter and quanta of radiation are described by means of *wave packet* (DeBroglie's waves) with average wavelength $\lambda_i$. Therefore, we can say that during a time interval $\Delta t = \lambda_i/c$, a *quantum* of energy $\Delta E = M_g c^2$ varies. According to the uncertainty principle, the particle will be detected if $\Delta t \geq \hbar/\Delta E$, i.e., if $\lambda_i/c \geq \hbar/M_g c^2$ or $\lambda_i \geq \lambda_g/2\pi$. This condition is usually satisfied when $M_g = M_i$. In this case, $\lambda_g = \lambda_i$ and obviously, $\lambda_i > \lambda_i/2\pi$. However, when $M_g$ decreases $\lambda_g$ increases and $\lambda_g/2\pi$ can become bigger than $\lambda_i$, making the particle *non-detectable* or *imaginary*.

According to Eqs. (7) and (41) we can write $M_g$ in the following form:

$$M_g = \frac{m_g}{\sqrt{1 - V^2/c^2}} = \frac{\chi\ m_i}{\sqrt{1 - V^2/c^2}} = \chi M_i$$

where

$$\chi = \left\{1 - 2\left[\sqrt{1 + \left(\Delta p/m_{i0}c\right)^2} - 1\right]\right\}$$

Since the condition to make the particle *imaginary* is

$$\lambda_i < \frac{\lambda_g}{2\pi}$$

and

$$\frac{\lambda_g}{2\pi} = \frac{\hbar}{M_g c} = \frac{\hbar}{\chi M_i c} = \frac{\lambda_i}{2\pi\chi}$$

Then we get

$$\chi < \frac{1}{2\pi} = 0.159$$

However, $\chi$ can be *positive* or *negative* ($\chi < +0.159$ or $\chi > -0.159$).This means that when

$$-0.159 < \chi < +0.159$$

the particle becomes *imaginary*. Under these circumstances, we can say that the particle made a transition to the *imaginary space-time*.

Note that, when a particle becomes imaginary, its gravitational and inertial masses also become imaginary. However, the factor $\chi = M_{g(imaginary)}/M_{i(imaginary)}$ remains *real* because

$$\chi = \frac{M_{g(imaginary)}}{M_{i(imaginary)}} = \frac{M_g i}{M_i i} = \frac{M_g}{M_i} = real$$

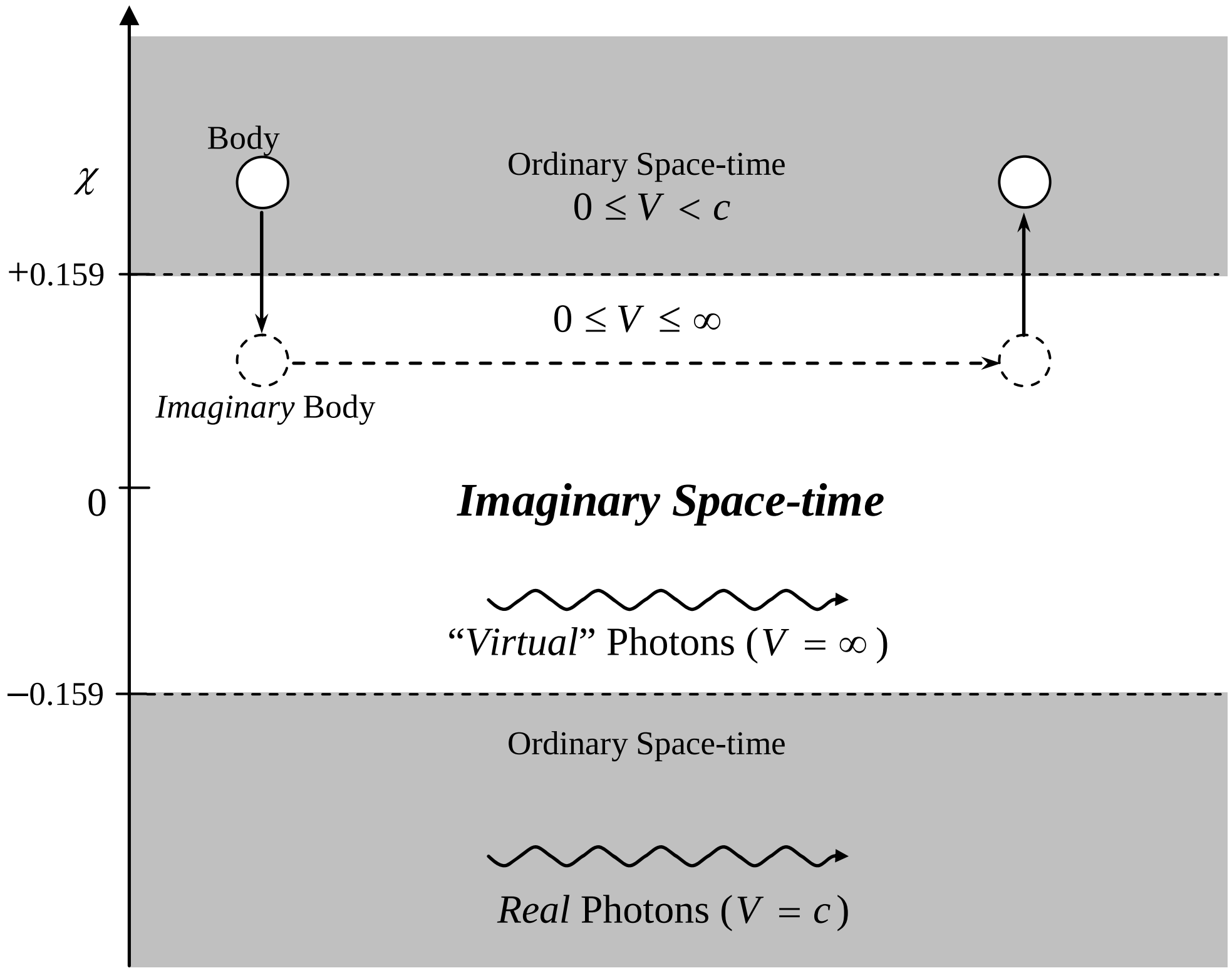

Fig. VII – *Travel in the imaginary space-time*. Similarly to the "*virtual*" photons, *imaginary* bodies can have *infinite speed* in the *imaginary space-time*.

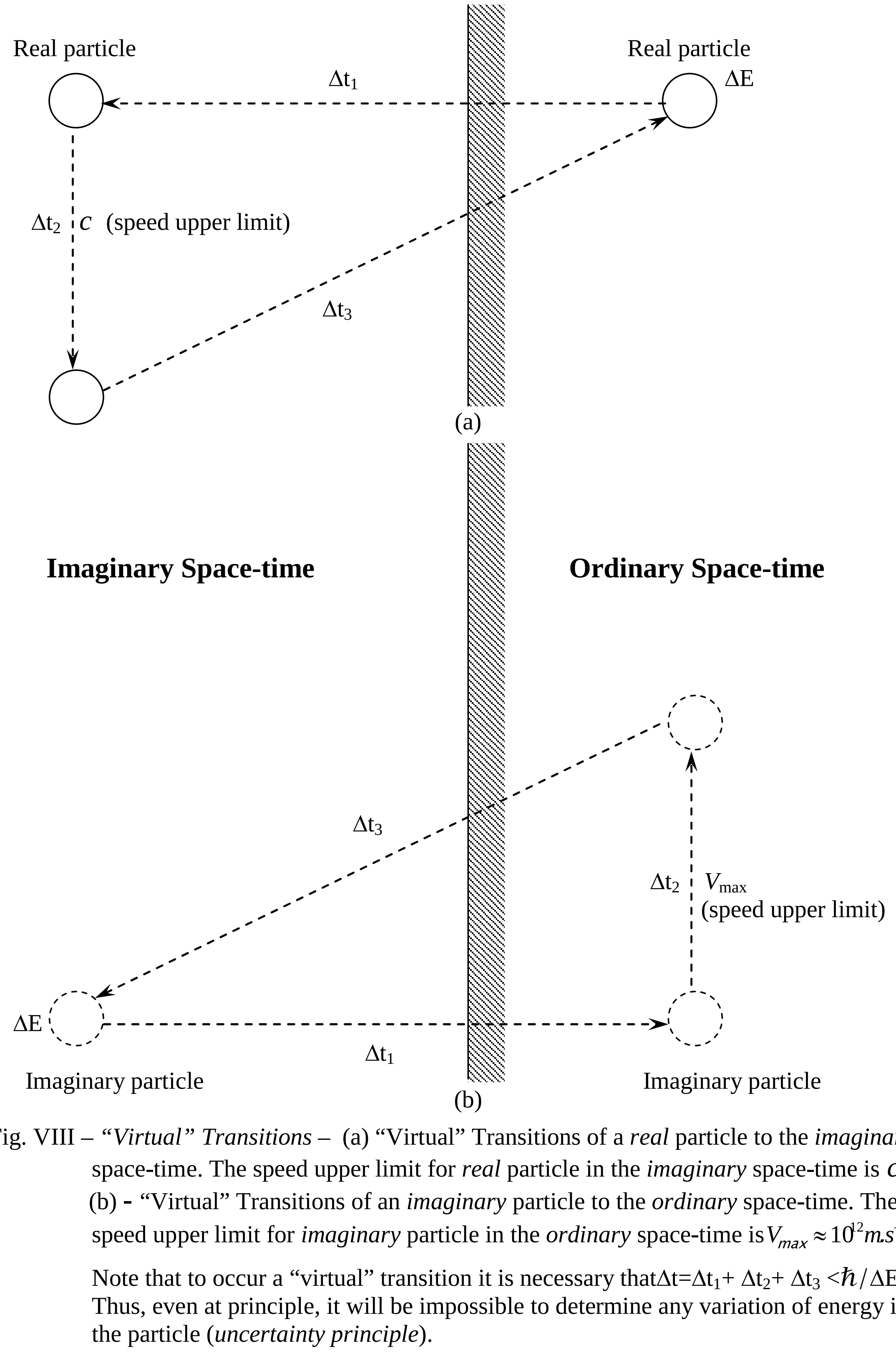

Fig. VIII – *"Virtual" Transitions* – (a) "Virtual" Transitions of a *real* particle to the *imaginary* space-time. The speed upper limit for *real* particle in the *imaginary* space-time is $c$.
(b) - "Virtual" Transitions of an *imaginary* particle to the *ordinary* space-time. The speed upper limit for *imaginary* particle in the *ordinary* space-time is $V_{max} \approx 10^{12} m.s^{-1}$

Note that to occur a "virtual" transition it is necessary that $\Delta t = \Delta t_1 + \Delta t_2 + \Delta t_3 < \hbar / \Delta E$
Thus, even at principle, it will be impossible to determine any variation of energy in the particle (*uncertainty principle*).

Thus, if the gravitational mass of the particle is reduced by means of the absorption of an amount of electromagnetic energy $U$, for example, we have

$$\chi = \frac{M_g}{M_i} = \left\{1 - 2\left[\sqrt{1 + \left(U/m_{i0}c^2\right)^2} - 1\right]\right\}$$

This shows that the energy $U$ of the electromagnetic field *remains acting on* the imaginary particle. In practice, this means that *electromagnetic fields act on imaginary particles*.

The gravity acceleration on a *imaginary* particle (due to the rest of the imaginary Universe) are given by

$$g'_j = \chi \ g_j \qquad j = 1,2,3,\ldots,n.$$

where $\chi = M_{g(imaginary)}/M_{i(imaginary)}$ and $g_j = -Gm_{gj(imaginary)}/r_j^2$. Thus, the gravitational forces acting on the particle are given by

$$F_{gj} = M_{g(imaginary)}g'_j = $$
$$= M_{g(imaginary)}\left(-\chi G m_{gj(imaginary)}/r_j^2\right) =$$
$$= M_g i\left(-\chi G m_{gj} i/r_j^2\right) = +\chi G M_g m_{gj}/r_j^2 \ .$$

Note that these forces are *real*. Remind that, the Mach's principle says that the *inertial effects* upon a particle are consequence of the gravitational interaction of the particle with the rest of the Universe. Then we can conclude that the *inertial forces* upon an *imaginary* particle are also real.

Equation (7) shows that , in the case of imaginary particles, the relativistic mass is

$$M_{g(imaginary)} = \frac{m_{g(imaginary)}}{\sqrt{1 - V^2/c^2}} =$$
$$= \frac{m_g i}{i\sqrt{V^2/c^2 - 1}} = \frac{m_g i}{\sqrt{V^2/c^2 - 1}}$$

This expression shows that *imaginary* particles can have velocities $V$ greater than $c$ in our ordinary space-time (Tachyons). The *quantization of velocity* (Eq. 36) shows that there is a speed upper limit $V_{max} > c$. As we have already calculated previously, $V_{max} \approx 10^{12} m.s^{-1}$, (Eq.102).

Note that this is the speed upper limit for *imaginary* particles *in our ordinary space-time* not in the *imaginary* space-time (Fig.7) because the *infinite* speed of the "virtual" *quanta* of the interactions shows that *imaginary* particles can have *infinite speed* in the *imaginary* space-time.

While the speed upper limit for imaginary particles in the ordinary space-time is $V_{max} \approx 10^{12} m.s^{-1}$, the speed upper limit for *real* particles in the *imaginary* space-time is $c$, because the relativistic expression of the mass shows that the velocity of *real* particles cannot be larger than $c$ in *any space-time*. The uncertainty principle permits that particles make "*virtual*" transitions, during a time interval $\Delta t$, if $\Delta t < \hbar/\Delta E$. The "*virtual*" transition of *mesons* emitted from nucleons that do not change of mass, during a time interval $\Delta t < \hbar/m_\pi c^2$, is a well-known example of "*virtual*" transition of particles. During a "virtual" transition of a *real* particle, the speed upper limit in the *imaginary* space-time is $c$, while the speed upper limit for an *imaginary* particle in the our ordinary space-time is $V_{max} \approx 10^{12} m.s^{-1}$. (See Fig. 8).

There is a crucial cosmological problem to be solved: the problem of the *hidden mass*. Most theories predict that the amount of known matter, detectable and available in the universe, is only about 1/100 to 1/10 of the amount needed to close the universe. That is, to achieve the density sufficient to close-up the universe by maintaining the gravitational curvature (escape velocity equal to the speed of light) at the outer boundary.

Eq. (43) may solve this problem. We will start by substituting the expression of *Hubble's* law for velocity, $V = \tilde{H}l$, into Eq.(43). The expression obtained shows that particles which are at distances $l = l_0 = \left(\sqrt{5}/3\right)\left(c/\tilde{H}\right) = 1.3\times10^{26} m$ have *quasi null* gravitational mass $m_g = m_{g(min)}$; beyond this distance, the particles have *negative* gravitational mass. Therefore, there are two well-defined regions in the Universe; the region of the bodies with *positive* gravitational masses and the region of the bodies with *negative* gravitational mass. The total gravitational mass of the first region, in accordance with Eq.(45), will be given by

$$M_{g1} \cong M_{i1} = \frac{m_{i1}}{\sqrt{1-\bar{V}_1^2/c^2}} \cong m_{i1}$$

where $m_{i1}$ is the total *inertial mass* of the bodies of the mentioned region; $\bar{V}_1 << c$ is the average velocity of the bodies at region 1. The total gravitational mass of the second region is

$$M_{g2} = \left|1 - 2\left(\frac{1}{\sqrt{1-\bar{V}_2^2/c^2}} - 1\right)\right| M_{i2}$$

where $\bar{V}_2$ is the average velocity of the bodies ; $M_{i2} = m_{i2}\Big/\sqrt{1-\bar{V}_2^2/c^2}$ and $m_{i2}$ is the total *inertial mass* of the bodies of region 2.

Now consider that from Eq.(7), we can write

$$\xi = \frac{E_g}{V} = \frac{M_g c^2}{V} = \rho_g c^2$$

where $\xi$ is the *energy density* of matter.

Note that the expression of $\xi$ only reduces to the well-known expression $\rho c^2$, where $\rho$ is the sum of the inertial masses per volume unit, when $m_g = m_i$. Therefore, in the derivation of the well-known difference

$$\frac{8\pi G\rho_U}{3} - \tilde{H}^2$$

which gives the *sign of the curvature* of the Universe [36], we must use $\xi = \rho_{gU} c^2$ instead of $\xi = \rho_U c^2$.The result obviously is

$$\frac{8\pi G\rho_{gU}}{3} - \tilde{H}^2 \qquad (109)$$

where

$$\rho_{gU} = \frac{M_{gU}}{V_U} = \frac{M_{g1} + M_{g2}}{V_U} \qquad (110)$$

$M_{gU}$ and $V_U$ are respectively the total gravitational mass and the volume of the Universe.

Substitution of $M_{g1}$ and $M_{g2}$ into expression (110) gives

$$\rho_{gU} = \frac{m_{iU} + \left[\left|\frac{3}{\sqrt{1-\bar{V}_2^2/c^2}} - \frac{2}{1-\bar{V}_2^2/c^2}\right| m_{i2} - m_{i2}\right]}{V_U}$$

where $m_{iU} = m_{i1} + m_{i2}$ is the total inertial mass of the Universe.

The volume $V_1$ of the region 1 and the volume $V_2$ of the region 2, are respectively given by

$$V_1 = 2\pi^2 l_0^3 \quad \text{and} \quad V_2 = 2\pi^2 l_c^3 - V_1$$

where $l_c = c/\tilde{H} = 1.8\times10^{26}\,m$ is the so-called "radius" of the visible Universe. Moreover, $\rho_{i1} = m_{i1}/V_1$ and $\rho_{i2} = m_{i2}/V_2$. Due to the *hypothesis of the uniform distribution of matter in the space*, it follows that $\rho_{i1} = \rho_{i2}$.Thus, we can write

$$\frac{m_{i1}}{m_{i2}} = \frac{V_1}{V_2} = \left(\frac{l_0}{l_c}\right)^3 = 0.38$$

Similarly,

$$\frac{m_{iU}}{V_U} = \frac{m_{i2}}{V_2} = \frac{m_{i1}}{V_1}$$

Therefore,

$$m_{i2} = \frac{V_2}{V_U} m_{iU} = \left[1 - \left(\frac{l_0}{l_c}\right)^3\right] m_{iU} = 0.62 m_{iU}$$

and $m_{i1} = 0.38 m_{iU}$.

Substitution of $m_{i2}$ into the expression of $\rho_{gU}$ yields

$$\rho_{gU} = \frac{m_{iU} + \left|\frac{1.86}{\sqrt{1-\overline{V}_2^2/c^2}} - \frac{1.24}{1-\overline{V}_2^2/c^2} - 0.62\right| m_{iU}}{V_U}$$

Due to $\overline{V}_2 \cong c$, we conclude that the term between bracket *is much larger than* $10 m_{iU}$. The amount $m_{iU}$ is the mass of matter in the universe (1/10 to 1/100 of the amount needed to close the Universe).

Consequently, the total mass

$$m_{iU} + \left|\frac{1.86}{\sqrt{1-\overline{V}_2^2/c^2}} - \frac{1.24}{1-\overline{V}_2^2/c^2} - 0.62\right| m_{iU}$$

must be sufficient to close the Universe.

There is another cosmological problem to be solved: the problem of the *anomalies* in the spectral red-shift of certain galaxies and stars.

Several observers have noticed red-shift values that cannot be explained by the Doppler-Fizeau effect or by the Einstein effect (the gravitational spectrum shift, supplied by Einstein's theory).

This is the case of the so-called *Stefan's quintet* (a set of five galaxies which were discovered in 1877), whose galaxies are located at approximately the same distance from the Earth, according to very reliable and precise measuring methods. But, when the velocities of the galaxies are measured by its red-shifts, the velocity of one of them is much larger than the velocity of the others.

Similar observations have been made on the *Virgo constellation* and spiral galaxies. Also the Sun presents a red-shift greater than the predicted value by the Einstein effect.

It seems that some of these anomalies can be explained if we consider the Eq.(45) in the calculation of the *gravitational mass* of the point of emission.

The expression of the gravitational spectrum shift was previously obtained in this work. It is the same supplied by Einstein's theory [37], and is given by

$$\Delta\omega = \omega_1 - \omega_2 = \frac{\phi_2 - \phi_1}{c^2}\omega_0 = \frac{-Gm_{g2}/r_2 + Gm_{g1}/r_1}{c^2}\omega_0 \qquad (111)$$

where $\omega_1$ is the frequency of the light at the point of emission ; $\omega_2$ is the frequency at the point of observation; $\phi_1$ and $\phi_2$ are respectively, the Newtonian gravitational potentials at the point of emission and at the point of observation.

In Einstein theory, this expression has been deduced from $T = t\sqrt{-g_{00}}$ [38] which correlates *own time* (real time), $t$, with the temporal coordinate $x^0$ of the space-time ( $t = x^0/c$ ).

When the gravitational field is *weak*, the temporal component $g_{00}$ of the metric tensor is given by $g_{oo} = -1 - 2\phi/c^2$[39].Thus, we readily obtain

$$T = t\sqrt{1 - 2Gm_g/rc^2} \tag{112}$$

This is the same equation that we have obtained previously in this work.

Curiously, this equation tell us that we can have $T < t$ when $m_g > 0$ ; and $T > t$ for $m_g < 0$. In addition, if $m_g = c^2 r/2G$, i.e., if $r = 2Gm_g/c^2$ (*Schwarzschild radius*) we obtain $T=0$.

Let us now consider the well-known process of stars' *gravitational contraction*. It is known that the destination of the star is directly correlated to its mass. If the star's mass is less than $1.4M_{\odot}$ (Schemberg-Chandrasekhar's limit), it becomes a *white dwarf*. If its mass exceeds that limit, the pressure produced by the degenerate state of the matter no longer counterbalances the gravitational pressure, and the star's contraction continues. Afterwards there occurs the reactions between protons and electrons (capture of electrons), where *neutrons* and anti-neutrinos are produced.

The contraction continues until the system regains stability (when the pressure produced by the neutrons is sufficient to stop the gravitational collapse). Such systems are called *neutron stars*.

There is also a critical mass for the stable configuration of neutron stars. This limit has not been fully defined as yet, but it is known that it is located between $1.8M_{\odot}$ and $2.4M_{\odot}$. Thus, if the mass of the star exceeds $2.4M_{\odot}$ , the contraction will continue.

According to Hawking [40] collapsed objects cannot have mass less than $\sqrt{\hbar c/4G} = 1.1\times10^{-8}kg$. This means that, with the progressing of the compression, the neutrons cluster must become a cluster of *superparticles* where the *minimal inertial mass* of the superparticle is

$$m_{i(sp)} = 1.1\times10^{-8}kg. \tag{113}$$

*Symmetry* is a fundamental attribute of the Universe that enables an investigator to study particular aspects of physical systems by themselves. For example, the assumption that space is homogeneous and isotropic is based on *Symmetry Principle*. Also here, by symmetry, we can assume that there are only *superparticles* with mass $m_{i(sp)} = 1.1\times10^{-8}kg$ in the cluster of *superparticles*.

Based on the mass-energy of the superparticles ( ~$10^{18}$ GeV ) we can say that they belong to a putative class of particles with mass-energy beyond the *supermassive* Higgs bosons ( the so-called X bosons). It is known that the GUT's theories predict an entirely new force mediated by a new type of boson, called simply X (or X boson ). The X bosons carry both electromagnetic and color charge, in order to ensure proper conservation of those charges in any interactions. The X bosons must be extremely massive, with mass-energy in the unification range of about $10^{16}$ GeV.

If we assume the superparticles *are not hyper*massive Higgs bosons then the possibility of the *neutrons cluster* become a *Higgs bosons cluster* before becoming a *superparticles cluster* must be considered. On the other hand, the fact that superparticles must be so massive also means that it is not possible to create them in any conceivable particle accelerator that could be built. They can exist as free particles only at a very early stage of the Big Bang from which the universe emerged.

Let us now imagine the Universe coming back to the past. There will be an instant in which it will be similar to a *neutrons cluster*, such as the stars at the final state of gravitational contraction. Thus, with the progressing of the compression, the *neutrons* cluster becomes a superparticles cluster. Obviously, this only can occur before $10^{-23}$s (after the Big-Bang).

The temperature T of the Universe at the $10^{-43}$s< t < $10^{-23}$s period can be calculated by means of the well-known expression[41]:

$$T \approx 10^{22}\left(t/10^{-23}\right)^{-1/2} \qquad (114)$$

Thus at $t \cong 10^{-43}s$ (at the *first* spontaneous breaking of symmetry) the temperature was $T \approx 10^{32}K$ (~$10^{19}$GeV).Therefore, we can assume that the absorbed electromagnetic energy by each *superparticle*, before $t \cong 10^{-43}s$, was $U = \eta kT > 1\times10^{9}J$ (see Eqs.(71) and (72)). By comparing with $m_{i(sp)}c^2 \cong 9\times10^{8}J$, we conclude that $U > m_{i(sp)}c$. Therefore, *the unification condition* $\left(U\eta_r \cong M_i c^2 > m_i c^2\right)$ is satisfied. This means that, *before* $t \cong 10^{-43}s$ *,the gravitational and electromagnetic interactions were unified*.

From the *unification condition* $\left(U\eta_r \cong M_i c^2\right)$, we may conclude that the superparticles' *relativistic inertial mass* $M_{i(sp)}$ is

$$M_{i(sp)} \cong \frac{U\eta_r}{c^2} = \frac{\eta n_r kT}{c^2} \approx 10^{-8}kg \qquad (115)$$

Comparing with the superparticles' *inertial mass at rest* (113), we conclude that

$$M_{i(sp)} \approx m_{i(sp)} = 1.1\times10^{-8}kg \qquad (116)$$

From Eqs.(83) and (115), we obtain the superparticle's *gravitational mass at rest*:

$$m_{g(sp)} = m_{i(sp)} - 2M_{i(sp)} \cong$$
$$\cong -M_{i(sp)} \cong -\frac{\eta n_r kT}{c^2} \qquad (117)$$

Consequently, the superparticle's *relativistic gravitational mass*, is

$$M_{g(sp)} = \frac{m_{g(sp)}}{\sqrt{1 - V^2/c^2}} =$$
$$= \frac{\eta n_r kT}{c^2\sqrt{1 - V^2/c^2}} \qquad (118)$$

Thus, the gravitational forces between two *superparticles* , according to (13), is given by:

$$\vec{F}_{12} = -\vec{F}_{21} = -G\frac{M_{g(sp)}M'_{g(sp)}}{r^2}\hat{\mu}_{21} =$$
$$= \left[\left(\frac{M_{i(sp)}}{m_{i(sp)}}\right)^2\left(\frac{G}{c^5\hbar}\right)\left(\eta n_r kT\right)^2\right]\frac{\hbar c}{r^2}\hat{\mu}_{21} \qquad (119)$$

Due to the *unification* of the gravitational and electromagnetic interactions at that period, we have

$$\vec{F}_{12} = -\vec{F}_{21} = G\frac{M_{g(sp)}M'_{g(sp)}}{r^2}\hat{\mu}_{21} =$$

$$= \left[\left(\frac{M_{i(sp)}}{m_{i(sp)}}\right)^2\left(\frac{G}{c^5\hbar}\right)(\eta\kappa T)^2\right]\frac{\hbar c}{r^2}\hat{\mu}_{21} =$$

$$= \frac{e^2}{4\pi\varepsilon_0 r^2} \qquad (120)$$

From the equation above we can write

$$\left[\left(\frac{M_{i(sp)}}{m_{i(sp)}}\right)^2\left(\frac{G}{c^5\hbar}\right)(\eta\kappa T)^2\hbar c\right] = \frac{e^2}{4\pi\varepsilon_0} \qquad (121)$$

Now assuming that

$$\left(\frac{M_{i(sp)}}{m_{i(sp)}}\right)^2\left(\frac{G}{c^5\hbar}\right)(\eta\kappa T)^2 = \psi \qquad (122)$$

the Eq. (121) can be rewritten in the following form:

$$\psi = \frac{e^2}{4\pi\varepsilon_0\hbar c} = \frac{1}{137} \qquad (123)$$

which is the well-known *reciprocal fine structure constant.*

For $T = 10^{32}K$ the Eq.(122) gives

$$\psi = \left(\frac{M_{i(sp)}}{m_{i(sp)}}\right)^2\left(\frac{G}{c^5\hbar}\right)(\eta n_r \kappa T)^2 \approx \frac{1}{100} \qquad (124)$$

This value has the same order of magnitude as the exact value(1/137) of the *reciprocal fine structure constant*.

From equation (120) we can write:

$$\left(G\frac{M_{g(sp)}M'_{g(sp)}}{\psi c\vec{r}}\right)\vec{r} = \hbar \qquad (125)$$

The term between parentheses has the same dimensions as the *linear momentum* $\vec{p}$. Thus, (125) tells us that

$$\vec{p}\cdot\vec{r} = \hbar. \qquad (126)$$

A component of the momentum of a particle cannot be precisely specified without loss of all knowledge of the corresponding component of its position at that time ,i.e., a particle cannot be precisely located in a particular direction without loss of all knowledge of its momentum component in that direction . This means that in intermediate cases the product of the uncertainties of the simultaneously measurable values of corresponding position and momentum components is *at least of the magnitude order* of $\hbar$,

$$\Delta p.\Delta r \geq \hbar \qquad (127)$$

This relation, *directly obtained here from the Unified Theory*, is the well-known relation of the *Uncertainty Principle* for position and momentum.

According to Eq.(83), the gravitational mass of the superparticles at the *center* of the cluster becomes *negative* when $2\eta n_r kT/c^2 > m_{i(sp)}$, i.e., when

$$T > T_{critical} = \frac{m_{i(sp)}c^2}{2\eta n_r k} \approx 10^{32}K.$$

According to Eq. (114) this temperature corresponds to $t_c \approx 10^{43}s$. With the progressing of the compression, more superparticles into the center will have *negative* gravitational mass. Consequently, there will be a critical point in which the *repulsive* gravitational forces between the superparticles with negative gravitational masses and the superparticles with positive gravitational masses will be so strong that an explosion will occur. This is the event that we call the Big Bang.

Now, starting from the Big Bang to the present time. Immediately after the Big Bang, the superparticles' *decompression*

begins. The gravitational mass of the most central superparticle will only be positive when the temperature becomes smaller than the critical temperature, $T_{critical} \approx 10^{32} K$. At the maximum state of compression (exactly at the Big Bang) the volumes of the superparticles were equal to the elementary volume $\Omega_0 = \delta_V d_{min}^3$ and the volume of the Universe was $\Omega = \delta_V (n d_{min})^3 = \delta_V d_{initial}^3$ where $d_{initial}$ was the *initial* length scale of the Universe. At this very moment the *average* density of the Universe was equal to the *average* density of the superparticles, thus we can write

$$\left(\frac{d_{initial}}{d_{min}}\right)^3 = \frac{M_{i(U)}}{m_{i(sp)}} \qquad (128)$$

where $M_{i(U)} \approx 10^{53} kg$ is the inertial mass of the Universe. It has already been shown that $d_{min} = \tilde{k} l_{planck} \approx 10^{-34} m.$ Then, from Eq.(128), we obtain:

$$d_{initial} \approx 10^{-14} m \qquad (129)$$

After the Big Bang the Universe expands itself from $d_{initial}$ up to $d_{cr}$ (when the temperature decrease reaches the critical temperature $T_{critical} \approx 10^{32} K$, and the gravity becomes *attractive*). Thus, it expands by $d_{cr} - d_{initial}$, under effect of the *repulsive* gravity

$$\bar{g} = \sqrt{g_{max} g_{min}} =$$

$$= \sqrt{\left[G\left(\tfrac{1}{2} M_{g(U)}\right) / \left(\tfrac{1}{2} d_{initial}\right)^2\right] \left[G \tfrac{1}{2} M_{i(U)} / \left(\tfrac{1}{2} d_{cr}\right)^2\right]} =$$

$$= \frac{2G\sqrt{M_{g(U)} M_{i(U)}}}{d_{cr} d_{initial}} = \frac{2G\sqrt{\sum m_{g(sp)} M_{i(U)}}}{d_{cr} d_{initial}} =$$

$$= \frac{2G\sqrt{\chi \sum m_{i(sp)} M_{i(U)}}}{d_{cr} d_{initial}} = \frac{2G M_{i(U)} \sqrt{\chi}}{d_{cr} d_{initial}}$$

during a period of time $t_c \approx 10^{-43} s$ .Thus,

$$d_{cr} - d_{initial} = \tfrac{1}{2} \bar{g} (t_c)^2 = \left(\sqrt{\chi}\right) \frac{G M_{i(U)}}{d_{cr} d_{initial}} (t_c)^2 \qquad (130)$$

The Eq.(83), gives

$$\chi = \frac{m_{g(sp)}}{m_{i(sp)}} = 1 - \frac{2Un_r}{m_{i(sp)} c^2} = 1 - \frac{2\eta n_r kT}{m_{i(sp)} c^2}$$

Calculations by Carr, B.J [41], indicate that it would seem reasonable to suppose that the fraction of initial *primordial black hole* mass ultimately converted into *photons* is about $0.11$. This means that we can take

$$\eta = 0.11$$

Thus, the amount $\eta M_{iU} c^2$, where $M_{iU}$ is the total inertial mass of the Universe, expresses the total amount of inertial energy converted into photons at the initial instant of the Universe(*Primordial Photons*).

It was previously shown that photons and also the matter have *imaginary* gravitational masses associated to them. The matter has *negative* imaginary gravitational mass, while the photons have *positive* imaginary gravitational mass, given by

$$M_{gp(imaginary)} = 2M_{ip(imaginary)} = +\frac{4}{\sqrt{3}}\left(\frac{hf}{c^2}\right) i$$

where $M_{ip(imaginary)}$ is the *imaginary inertial* mass of the photons.

Then, from the above we can conclude that, at the initial instant of the Universe, an amount of *imaginary* gravitational mass, $M_{gm(imaginary)}^{total}$, which was associated to the fraction of the *matter* transformed into photons, has been converted into *imaginary* gravitational mass of the primordial photons, $M^{total}_{gp(imaginary)}$, while an amount of *real inertial* mass of the matter, $M^{total}_{im(real)} = \eta\, M_{iU} c^2$, has been converted into *real* energy of the primordial photons, $E_p = \sum_{j=1}^{N} hf_j$ , i.e.,

$$M^{total}_{gm(imaginary)} + M^{total}_{im(real)} =$$
$$= M^{total}_{gp(imaginary)} + \underbrace{M^{total}_{ip(real)}}_{\frac{E_p}{c^2}}$$

where $M^{total}_{gm(imaginary)} = M^{total}_{gp(imaginary)}$ and

$$E_p / c^2 \equiv M^{total}_{ip(real)} = M^{total}_{im(real)} = \eta M_{iU} \cong 0.11 M_{iU}$$

It was previously shown that, for the *photons* equation: $M_{gp} = 2M_{ip}$, is valid. This means that

$$\underbrace{M_{gp(imagimary)} + M_{gp(real)}}_{M_{gp}} =$$
$$= 2\underbrace{\left(M_{ip(imagimary)} + M_{ip(real)}\right)}_{M_{ip}}$$

By substituting $M_{gp(imaginary)} = 2M_{ip(imaginary)}$ into the equation above, we get

$$M_{gp(real)} = 2M_{ip(real)}$$

Therefore we can write that

$$M^{total}_{gp(real)} = 2M^{total}_{ip(real)} = 0.22 M_{iU}$$

The phenomenon of *gravitational deflection of light about the Sun* shows that *the gravitational interaction* between the Sun and the photons is *attractive*. This is due to the gravitational force between the Sun and a photon, which is given by

$$F = -G\, M_{gSun(imaginary)} m_{gp(imaginary)} / r^2 \ ,$$

where $m_{gp(imaginary)}$ (the imaginary gravitational mass of the photon) is a quantity *positive* and *imaginary*, and $M_{gSun(imaginary)}$ (the imaginary gravitational mass associated to the matter of the Sun) is a quantity *negative* and imaginary.

The fact of the gravitational interaction between the imaginary gravitational masses of the primordial photons and the imaginary gravitational mass of the matter be *attractive* is highly relevant, because it shows that it is necessary to consider the effect of this gravitational interaction, which is equivalent to the gravitational effect produced by the amount of *real* gravitational mass, $M^{total}_{gp(real)} \cong 0.22 M_{iU}$, sprayed by all the Universe.

This means that this amount, which corresponds to 22% of the total inertial mass of the Universe, must be added to the overall computation of the *total mass of the matter* (stars, galaxies, etc., gas and dust of interstellar and intergalactic media). Therefore, this additional portion corresponds to what has been called *Dark Matter* (See Fig. IX).

On the other hand, the *total* amount of gravitational mass at the initial instant, $M^{total}_{g}$, according to Eq.(41), can be expressed by

$$M^{total}_{g} = \chi\ M_{iU}$$

This mass includes the total negative gravitational mass of the matter, $M^{total}_{gm(-)}$, plus the total gravitational mass, $M^{total}_{gp(real)}$, converted into primordial photons. This tells us that we can put

$$M^{total}_{g} = M^{total}_{gm(-)} + M^{total}_{gp(real)} = \chi\ M_{iU}$$

whence

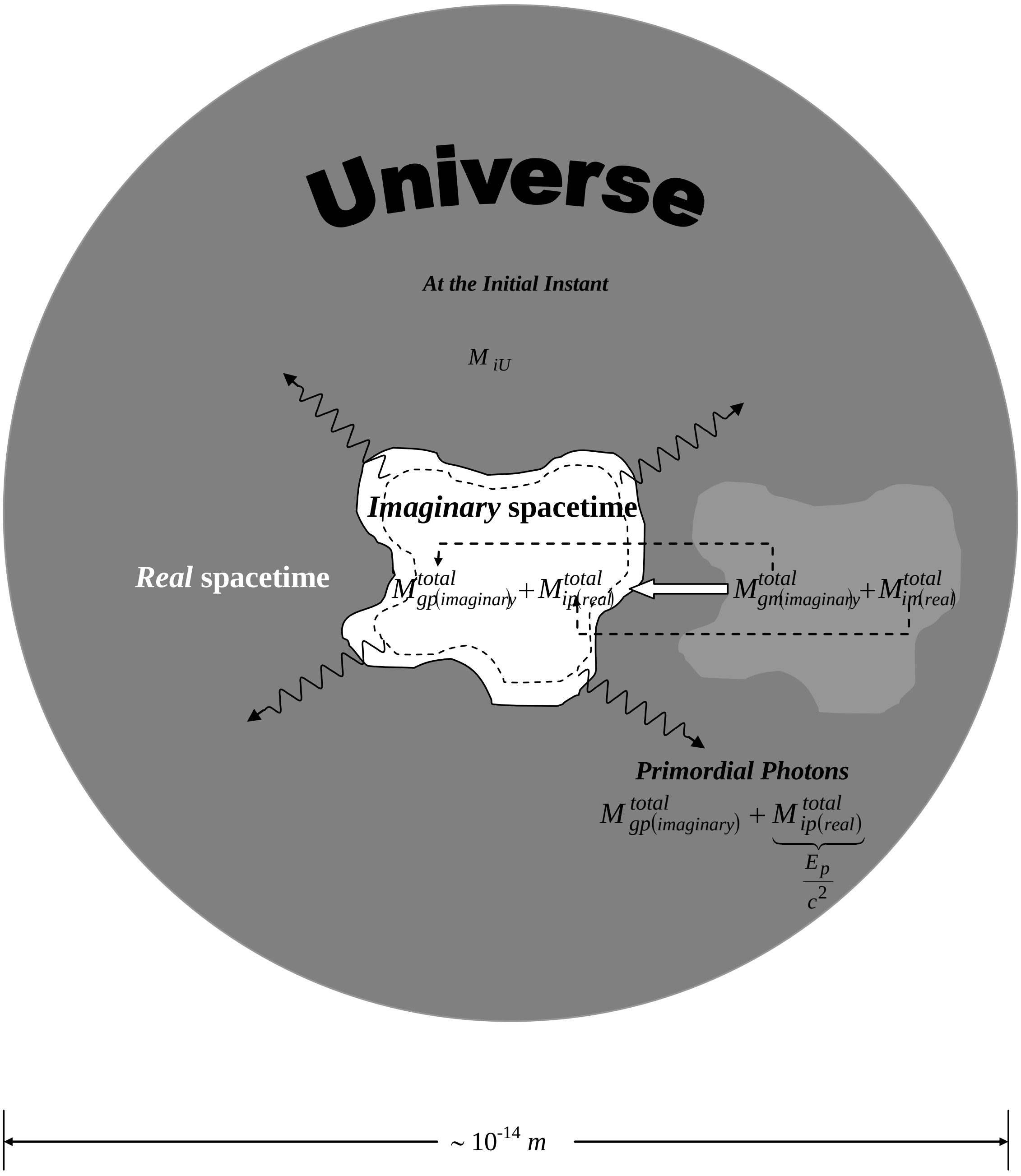

Fig. IX – Conversion of part of the *Real* Gravitational Mass of the Primordial Universe into *Primordial Photons*. The gravitational effect caused by the gravitational interaction of *imaginary* gravitational masses of the primordial photons with the *imaginary* gravitational mass associated to the matter is equivalent to the effect produced by the amount of *real* gravitational mass, $M^{total}_{gp(real)} \cong 0.22 M_{iU}$, sprayed by all Universe. This additional portion of mass corresponds to what has been called *Dark Matter*.

$$M_{gm(-)}^{total} = \chi \ M_{iU} - 0.22 M_{iU}$$

In order to calculate the value of $\chi$ we can start from the expression previously obtained for $\chi$, i.e.,

$$\chi = \frac{m_{g(sp)}}{m_{i(sp)}} = 1 - \frac{2\eta \ n_r kT}{m_{i(sp)} c^2} = 1 - \frac{T}{T_{critical}}$$

where

$$T_{critical} = \frac{m_{i(sp)} c^2}{2\eta n_r k} = 3.3 \times 10^{32} \, K$$

and

$$T = \frac{M_{i(sp)} c^2}{2\eta n_r k} = \frac{\left( \frac{m_{i(sp)}}{\sqrt{1 - V^2/c^2}} \right) c^2}{2\eta n_r k} = \frac{T_{critical}}{\sqrt{1 - V^2/c^2}}$$

We thus obtain

$$\chi = 1 - \frac{1}{\sqrt{1 - V^2/c^2}}$$

By substitution of this expression into the equation of $M_{gm(-)}^{total}$, we get

$$M_{gm(-)}^{total} = \left( 0.78 - \frac{1}{\sqrt{1 - V^2/c^2}} \right) M_{iU}$$

On the other hand, the *Unification condition* ($U\eta_r \cong \Delta pc = M_{iU} c^2$) previously shown and Eq. (41) show that at the initial instant of the Universe, $M_{g(sp)}$ has the following value:

$$M_{g(sp)} = \left\{ 1 - 2 \left[ \sqrt{1 + \left( \frac{U\eta_r}{M_{i(sp)}} \right)} - 1 \right] \right\} M_{i(sp)} \cong 0.1 M_{i(sp)}$$

Similarly, Eq.(45) tells us that

$$M_{g(sp)} = \left( 1 - 2 \left[ \left( 1 - V^2/c^2 \right)^{-\frac{1}{2}} \right] \right) M_{i(sp)}$$

By comparing this expression with the equation above, we obtain

$$\frac{1}{\sqrt{1 - V^2/c^2}} \cong 1.5$$

Substitution of this value into the expressions of $\chi$ and $M_{gm(-)}^{total}$ results in

$$\chi = -0.5$$

and

$$M_{gm(-)}^{total} \cong -0.72 M_{iU}$$

This means that 72% of the total energy of the Universe ($M_{iU} c^2$) is due to *negative gravitational mass* of the matter created at the initial instant.

Since the gravitational mass is correlated to the inertial mass (Eq. (41)), the energy related to the negative gravitational mass is where there is inertial energy (inertial mass). In this way, this negative gravitational energy permeates all space and tends to increase the rate of expansion of the Universe due to produce a strong gravitational repulsion between the material particles. Thus, this energy corresponds to what has been called *Dark Energy* (See Fig. X).

The value of $\chi = -0.5$ at the initial instant of the Universe shows that the gravitational interaction was *repulsive* at the Big-Bang. It remains repulsive until the temperature of the Universe is reduced down to the critical limit, $T_{critical}$. Below this temperature limit, the *attractive* component of the gravitational interaction became greater than the *repulsive* component, making *attractive* the resultant gravitational interaction. Therefore, at the beginning of the Universe – before the temperature decreased down to $T_{critical}$, there occurred an expansion of the Universe that was exponential in time rather than a normal power-law expansion. Thus, there was an evident *Inflation Period* during the beginning of the expansion of the Universe (See Fig. XI).

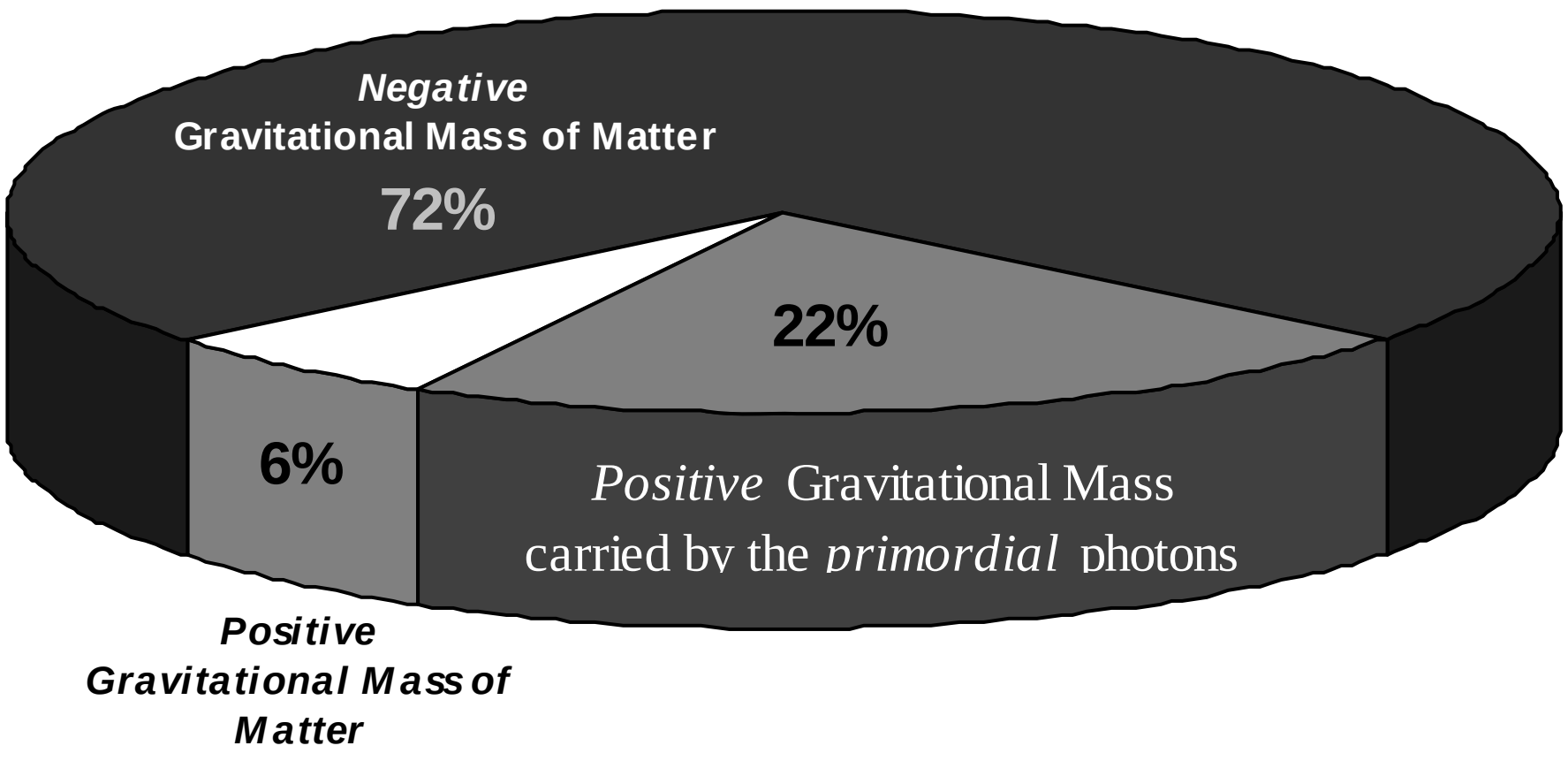

Fig. X - Distribution of Gravitational Masses in the Universe. The *total energy* related to *negative* gravitational mass of all the matter in the Universe corresponds to what has been called *Dark Energy*. While the *Dark Matter* corresponds to the *total* gravitational mass carried by the *primordial* photons, which is manifested in the *interaction of the imaginary* gravitational masses of the *primordial* photons with the imaginary mass of matter.

With the progressing of the *decompression* the *superparticles* cluster becomes a neutrons cluster. This means that the neutrons are created *without its antiparticle*, the antineutron. Thus, this solves the matter/antimatter dilemma that is unresolved in many cosmologies.

Now a question: How did the *primordial superparticles* appear at the beginning of the Universe?

It is a proven quantum fact that a wave function may collapse, and that, at this moment, all the possibilities that it describes are suddenly expressed in *reality*. This means that, through this process, particles can be suddenly *materialized*.

The materialization of the *primordial superparticles* into a critical volume denotes *knowledge* of what would happen with the Universe starting from that *initial condition*, a fact that points towards the *existence* of a Creator.

It was shown previously the possible existence of *imaginary particles* with imaginary masses in Nature. These particles can be associated with real particles, such as in case of the *photons* and *electrons*, as we have shown, or they can be associated with others imaginary particles by constituting the imaginary bodies. Just as the real particles constitute the real bodies.

The idea that we make about a *consciousness* is basically that of an *imaginary body* containing *psychic energy* and *intrinsic knowledge*. We can relate psychic energy with *psychic mass* (psychic mass= psychic energy/c$^2$). Thus, by analogy with the real bodies the psychic bodies would be constituted by psychic particles with psychic mass. Consequently, the psychic particles that constitute a consciousness would be equivalent to imaginary particles, and the *psychic mass* , $m_\Psi$ ,of the psychic particles would be equivalent to the *imaginary mass*, i.e.,

$$m_\Psi = m_{i(imaginary)} \qquad (131)$$

Thus, the imaginary masses associated to the *photons* and *electrons* would be *elementary psyche* actually, i.e.,

$$m_{\Psi photon} = m_{i(imaginary)photon} = $$
$$= \frac{2}{\sqrt{3}}\left(\frac{hf}{c^2}\right) i \qquad (132)$$

$$m_{\Psi electron} = m_{i(imaginary)electron} =$$
$$= -\frac{2}{\sqrt{3}}\left(\frac{hf_{electron}}{c^2}\right) i =$$
$$= -\frac{2}{\sqrt{3}} m_{i0(real)electron}\, i \qquad (133)$$

The idea that electrons have elementary psyche associated to themselves is not new. It comes from the pre-Socratic period.

By proposing the existence of psyche associated with matter, we are adopting what is called *panpsychic* posture. Panpsychism dates back to the pre-Socratic period;

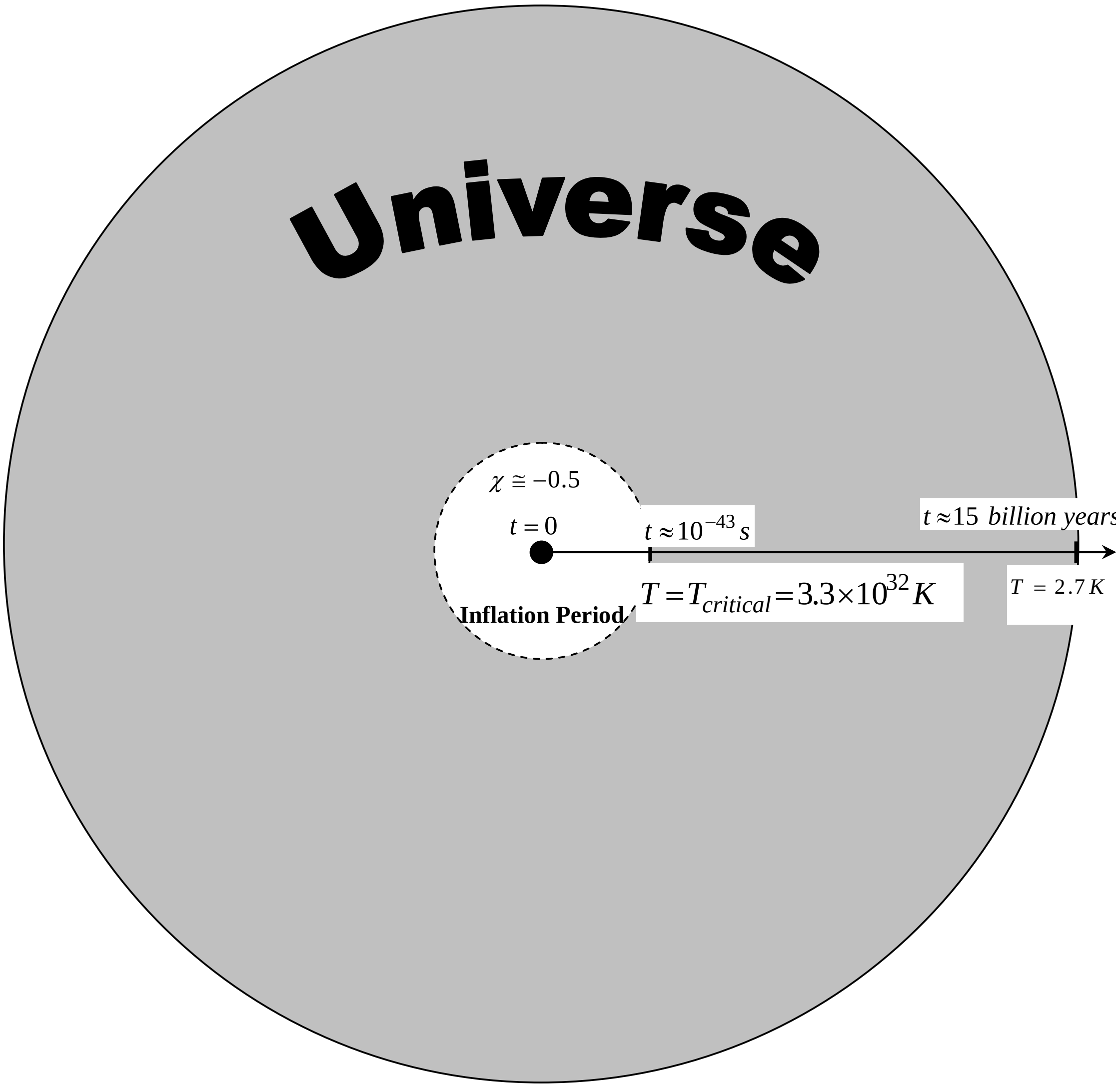

Fig. XI – *Inflation Period*. The value of $\chi \cong -0.5$ at the *Initial* Instant of the Universe shows that the gravitational interaction was repulsive at the Big-Bang. It remains repulsive until the temperature of the Universe is reduced down to the critical limit, $T_{critical}$. Below this temperature limit, the attractive component of the gravitational interaction became greater than the repulsive component, making attractive the resultant gravitational interaction. Therefore, at beginning of the Universe – before the temperature to be decreased down to $T_{critical}$, there occurred an expansion of the Universe that was exponential in time rather than a normal power-law expansion. Thus, there was an evident inflation period during the beginning of the expansion of the Universe.

remnants of organized panpsychism may be found in the Uno of Parmenides or in Heracleitus's Divine Flow. The scholars of Miletus's school were called *hylozoists*, that is, "those who believe that matter is alive". More recently, we will find the panpsychistic thought in Spinoza, Whitehead and Teilhard de Chardin, among others. The latter one admitted the existence of proto-conscious properties at the elementary particles' level.

We can find experimental evidences of the existence of psyche associated to electron in an experiment similar to that commonly used to show the wave duality of light. (Fig. XII). One merely substitutes an electron ray (fine electron beam) for the light ray. Just as in the experiment mentioned above, the ray which goes through the holes is detected as a wave if a *wave detector* is used (it is then observed that the interference pattern left on the detector screen is analogous with that produced by the light ray), and as a particle if a *particle detector* is used.

Since the electrons are detected on the other side of the metal sheet, it becomes obvious then that they passed through the holes. On the other hand, it is also evident that when they approached the holes, they had to decide which one of them to go through.

How can an electron "decide" which hole to go through? Where there is "choice", isn't there also *psyche*, by definition?

The existence of *psyche mass* associated to material particles can solve the black hole *information paradox* ‡ if we assume that any *informatio*n placed into a black hole is saved into the psyche mass associated to the black hole and after - during the evaporation of the black hole, carried by the psyche masses associated to the evaporated particles (Hawking radiation). Thus, if the material entering the black hole is *a pure quantum state*, the transformation of that state into the mixed state of Hawking radiation does *not* destroy information about the original quantum state because the psyche mass does not lose any structure of the original quantum state in the transformation. In this way, when information goes into a black hole it is not destroyed and might escape from the black hole during its evaporation.

The idea of psyche mass associated to a black hole leads to conclusion that there has been a psyche mass associated to the Initial Universe.

---

‡ In 1975, Stephen Hawking showed that black holes eventually evaporate releasing particles containing no information (Hawking radiation). It was soon realized that this prediction created an *information loss*. From the *no hair theorem* one would expect the Hawking radiation to be completely independent of the material entering the black hole. However, if the material entering the black hole were *a pure quantum state*, the transformation of that state into the mixed state of Hawking radiation would *destroy information* about the original quantum state. This violates Liouville's theorem and presents a *physical paradox*. It is a contentious subject for science since it violated a commonly assumed tenet of science—that *information cannot be destroyed.*

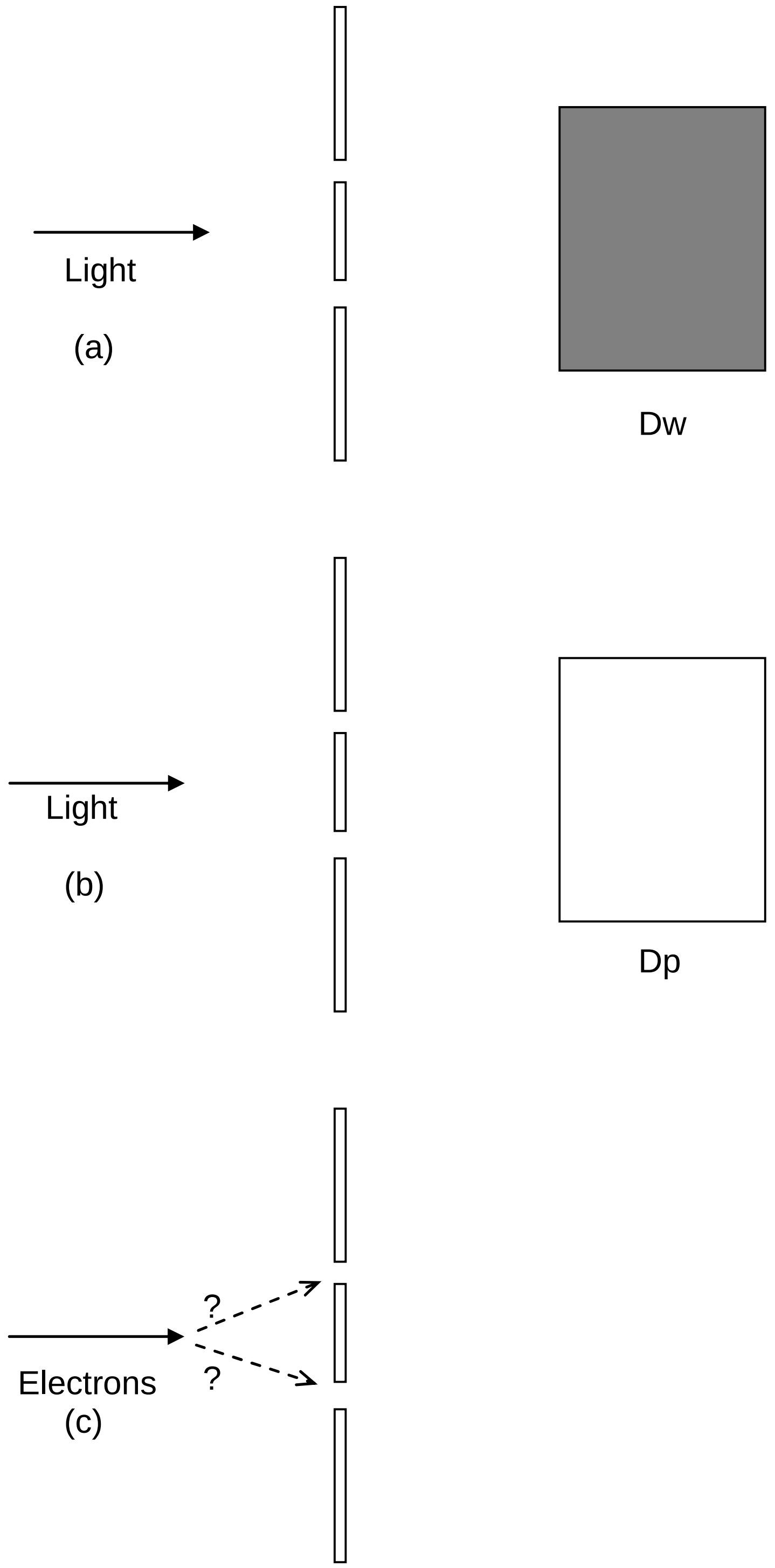

Fig. XII – A light ray, after going through the holes in the metal sheet, will be detected as a wave(a) by a wave detector Dw or as a particle if the wave detector is substituted for the wave detector Dp. Electron ray (c) has similar behavior as that of a light ray. However, before going through the holes, the electrons must “decide” which one to go through.

If the primordial superparticles that have been materialized at the beginning of the Universe came from the collapse of a primordial wave function, then the psychic form described by this wave function must have been generated in a consciousness with a psychic mass much greater than that needed to materialize the Universe (material and psychic).

This giant consciousness, in its turn, would not only be the greatest of all consciences in the Universe but also the *substratum* of everything that exists and, obviously, everything that exists would be entirely contained within it, including *all the spacetime*.

Thus, if the consciousness we refer to contains all the space, its volume is necessarily infinite, consequently having an *infinite psychic mass*.

This means that it contains all the existing psychic mass and, therefore, any other consciousness that exists will be contained in it. Hence, we may conclude that It is the *Supreme Consciousness* and that there is no other equal to It: It is *unique*.

Since the Supreme Consciousness also contains *all* time; past, present and future, then, for It the time does not flow as it flows for us.

Within this framework, when we talk about the Creation of the Universe, the use of the verb "to create" means that "something that was not" came into being, thus presupposing the concept of *time flow*. For the Supreme Consciousness, however, the instant of Creation is mixed up with all other times, consequently there being no "before" or "after" the Creation and, thus, the following question is not justifiable: "What did the Supreme Consciousness do before Creation?"

On the other hand, we may also infer, from the above that the existence of the Supreme Consciousness has no defined limit (beginning and end), what confers upon It the unique characteristic of *uncreated* and *eternal*.

If the Supreme Consciousness is eternal, Its wave function $\Psi_{SC}$ shall never collapse (will never be null). Thus, for having an infinite psychic mass, the value of $\Psi_{SC}^2$ will be always infinite and, hence, we may write that

$$\int_{-\infty}^{+\infty} \Psi_{SC}^2 dV = \infty$$

By comparing this equation with Eq. (108) derived from Quantum Mechanics, we conclude that the Supreme Consciousness is simultaneously everywhere, i.e., It is *omnipresent*.

Since the Supreme Consciousness contains all consciences, it is expected that It also contain *all the knowledge.* Therefore, It is also *omniscient*. Consequently, It *knows* how to formulate well-defined mental images with psychic masses sufficient for its contents to *materialize*. In this way, It can materialize *everything* It wishes (*omnipotence*).

All these characteristics of the Supreme Consciousness (infinite, unique, uncreated, eternal, omnipresent, omniscient and omnipotent) coincide with those traditionally ascribed to *God* by most religions.

It was shown in this work that the "virtual" *quanta* of the *gravitational interaction* must have spin 1 and not 2, and that they are "virtual" photons (*graviphotons*) with *zero mass* outside the *coherent* matter. Inside the coherent matter the graviphoton mass is *non-zero*. Therefore, the gravitational forces are also *gauge* forces, because they are yielded by the exchange of "virtual" *quanta* of spin 1, such as the electromagnetic forces and the weak and strong nuclear forces.

Thus, the gravitational forces are produced by the exchange of “virtual” photons. Consequently, this is precisely the *origin of the gravity*.

Newton’s theory of gravity does not explain *why* objects attract one another; it simply models this observation. Also Einstein’s theory does not explain the origin of gravity. Einstein’s theory of gravity only describes gravity with more precision than Newton’s theory does.

Besides, there is nothing in both theories explaining the *origin of the energy* that produces the gravitational forces. Earth’s gravity attracts all objects on the surface of our planet. This has been going on for well over 4.5 billions years, yet no known energy source is being converted to support this tremendous ongoing energy expenditure. Also is the enormous continuous energy expended by Earth’s gravitational field for maintaining the Moon in its orbit - millennium after millennium. In spite of the ongoing energy expended by Earth’s gravitational field to hold objects down on surface and the Moon in orbit, why the energy of the field never diminishes in strength or drains its energy source? Is this energy expenditure balanced by a conversion of energy from an unknown energy source?

The energy $W$ necessary to support the effort expended by the gravitational forces $F$ is well-known and given by

$$W = \int_{\infty}^{r} F dr = -G \frac{M_g m_g}{r}$$

According to the *principle of energy conservation*, this energy expenditure must be balanced by a conversion of energy from another energy type.

The Uncertainty Principle tells us that, due to the occurrence of exchange of *graviphotons* in a time interval $\Delta t < \hbar/\Delta E$ (where $\Delta E$ is the energy of the graviphoton), the energy variation $\Delta E$ cannot be detected in the system $M_g - m_g$. Since the total energy $W$ is the sum of the energy of the $n$ graviphotons, i.e., $W = \Delta E_1 + \Delta E_2 + \ldots + \Delta E_n$, then the energy $W$ *cannot be detected as well*. However, as we know it can be converted into another type of energy, for example, in rotational kinetic energy, as in the hydroelectric plants, or in the *Gravitational Motor*, as shown in this work.

It is known that a *quantum* of energy $\Delta E = hf$ which varies during a time interval $\Delta t = 1/f = \lambda/c < \hbar/\Delta E$ (wave period) cannot be experimentally detected. This is an *imaginary* photon or a “*virtual*” photon. Thus, the graviphotons are *imaginary* photons, i.e., the energies $\Delta E_i$ of the graviphotons are imaginaries energies and therefore the energy $W = \Delta E_1 + \Delta E_2 + \ldots + \Delta E_n$ is also an *imaginary* energy. Consequently, it belongs to the *imaginary space-time*.

According to Eq. (131), imaginary energy is equal to *psychic energy*. Consequently, the *imaginary space-time* is, in fact, the *psychic space-time*, which contains the Supreme Consciousness. Since the Supreme Consciousness has infinite psychic mass, then the *psychic space-time* has *infinite psychic energy*. This is highly relevant, because it confers to the *psychic space-time* the characteristic of *unlimited source of energy*.

This can be easily confirmed by the fact that, in spite of the enormous amount of energy expended by Earth’s gravitational field to hold objects down on the surface of the planet and maintain the Moon in its orbit, the energy of Earth’s gravitational field never diminishes in strength or drains its energy source.

If an experiment involves a large number of identical particles, all described by the same wave function $\Psi$, *real* density of mass $\rho$ of these particles in x, y, z, t is proportional to the corresponding value $\Psi^2$ ($\Psi^2$ is known as *density of probability*. If $\Psi$ is *complex* then $\Psi^2 = \Psi\Psi^*$. Thus, $\rho \propto \Psi^2 = \Psi.\Psi^*$). Similarly, in the case of psychic particles, the *density of psychic mass*, $\rho_\Psi$, in x, y, z, will be expressed by $\rho_\Psi \propto \Psi_\Psi^2 = \Psi_\Psi \Psi_\Psi^*$. It is known that $\Psi_\Psi^2$ is always *real* and *positive* while $\rho_\Psi = m_\Psi / V$ is an *imaginary* quantity. Thus, as the *modulus* of an imaginary number is always real and positive, we can transform the proportion $\rho_\Psi \propto \Psi_\Psi^2$, in equality in the following form:

$$\Psi_\Psi^2 = k\left|\rho_\Psi\right| \qquad (134)$$

where $k$ is a *proportionality constant* (real and positive) to be determined.

In Quantum Mechanics we have studied the *Superpositon Principle*, which affirms that, if a particle (or system of particles) is in a *dynamic state* represented by a wave function $\Psi_1$ and may also be in another dynamic state described by $\Psi_2$ then, the general dynamic state of the particle may be described by $\Psi$, where $\Psi$ is a linear combination(superposition)of $\Psi_1$ and $\Psi_2$, i.e.,

$$\Psi = c_1\Psi_1 + c_2\Psi_2 \qquad (135)$$

*Complex constants* $c_1$ and $c_2$ respectively indicates the percentage of dynamic state, represented by $\Psi_1$ and $\Psi_2$ in the formation of the general dynamic state described by $\Psi$.

In the case of psychic particles (psychic bodies, consciousness, etc.), by analogy, if $\Psi_{\Psi 1}$, $\Psi_{\Psi 2}$,..., $\Psi_{\Psi n}$ refer to the different dynamic states the psychic particle assume, then its general dynamic state may be described by the wave function $\Psi_\Psi$, given by:

$$\Psi_\Psi = c_1\Psi_{\Psi 1} + c_2\Psi_{\Psi 2} + \ldots + c_n\Psi_{\Psi n} \qquad (136)$$

The state of superposition of wave functions is, therefore, common for both psychic and material particles. In the case of material particles, it can be verified, for instance, when an electron changes from one orbit to another. Before effecting the transition to another energy level, the electron carries out "virtual transitions" [42]. A kind of *relationship* with other electrons before performing the real transition. During this relationship period, its wave function remains "*scattered*" by *a wide region of the space* [43] thus superposing the wave functions of the other electrons. In this relationship the electrons *mutually* influence one another, with the possibility of *intertwining* their wave functions[††]. When this happens, there occurs the so-called *Phase Relationship* according to quantum-mechanics concept.

In the electrons "virtual" transition mentioned before, the "listing" of all the possibilities of the electrons is described, as we know, by *Schrödinger's wave equation.* Otherwise, it is general for material particles. By analogy, in the case of psychic particles, we may say that the "listing" of all the possibilities of the psyches involved in the relationship will be described by *Schrödinger's equation* – for psychic case, i.e.,

[††] Since the electrons are simultaneously waves and particles, their wave aspects will interfere with each other; besides superposition, there is also the possibility of occurrence of *intertwining* of their wave functions.

$$\nabla^2\Psi_\Psi + \frac{p_\Psi^2}{\hbar^2}\Psi_\Psi = 0 \qquad (137)$$

Because the wave functions are capable of intertwining themselves, the quantum systems may "penetrate" each other, thus establishing an internal relationship where all of them are affected by the relationship, no longer being isolated systems but becoming an integrated part of a larger system. This type of internal relationship, which exists only in quantum systems, was called *Relational Holism* [44].

The equation of *quantization of mass* (33), in the generalized form, leads us to the following expression:

$$m_{i(imaginary)} = n^2 m_{i0(imagynary)(\min)}$$

Thus, we can also conclude that the *psychic mass is also quantized*, due to $m_\Psi = m_{i(imaginary)}$ (Eq. 131), i.e.,

$$m_\Psi = n^2 m_{\Psi(min)} \qquad (138)$$

where

$$m_{\Psi(\min)} = -\tfrac{2}{\sqrt{3}}\left(hf_{\min}/c^2\right) i = $$
$$= -\tfrac{2}{\sqrt{3}} m_{i0(real)\min}\; i \qquad (139)$$

It was shown that the *minimum quantum* of real inertial mass in the Universe, $m_{i0(real)\min}$, is given by:

$$m_{i0(real)\min} = \pm h\sqrt{3/8}\big/ cd_{\max} = $$
$$= \pm 3.9 \times 10^{-73}\,kg \qquad (140)$$

By analogy to Eqs. (132) and (133), the expressions of the psychic masses associated to the *proton* and the *neutron* are respectively given by:

$$m_{\Psi proton} = m_{i(imaginary)proton} = $$
$$= +\tfrac{2}{\sqrt{3}}\left(hf_{proton}/c^2\right) i = $$
$$= +\tfrac{2}{\sqrt{3}} m_{i0(real)proton}\; i \qquad (141)$$

$$m_{\Psi neutron} = m_{i(imaginary)neutron} = $$
$$= -\tfrac{2}{\sqrt{3}}\left(hf_{neutron}/c^2\right) i = $$
$$= -\tfrac{2}{\sqrt{3}} m_{i0(real)neutron}\; i \qquad (142)$$

The *imaginary* gravitational masses of the atoms must be *much smaller* than their *real* gravitational masses. On the contrary, the weight of the bodies would be very different of the observed values. This fact shows that $m_{i(imaginary)proton}$ and $m_{i(imaginary)neutron}$ must have *contrary* signs. In this way, the *imaginary* gravitational mass of an atom can be expressed by means of the following expression

$$m_{i(imaginary)atom} = N\left(m_e \pm \left(m_n - m_p\right) + \frac{\Delta E}{c^2}\right)i$$

where, $\Delta E$, is the interaction energy. By comparing this expression with the following expression

$$m_{i(real)atom} = N\left(m_e + m_p + m_n + \frac{\Delta E}{c^2}\right)$$

Thus,

$$\left|m_{i(imaginary)atom}\right| << m_{i(real)atom}$$

Now consider a monatomic body with *real* mass $M_{i(real)}$ and *imaginary* mass $M_{i(imaginary)}$. Then we have

$$\frac{M_{i(imaginary)}}{M_{i(real)}} = \frac{\Sigma\left(m_{i(imaginary)atom} + \frac{\Delta E_a i}{c^2}\right)}{\Sigma\left(m_{i(real)atom} + \frac{\Delta E_a}{c^2}\right)} = $$
$$= \frac{n\left(m_e \pm \left(m_n - m_p\right) + \frac{\Delta E}{c^2} + \frac{\Delta E_a}{c^2}\right)i}{n\left(m_e + m_p + m_n + \frac{\Delta E}{c^2} + \frac{\Delta E_a}{c^2}\right)} \cong$$
$$\cong \left(\frac{m_e \pm \left(m_n - m_p\right) + \frac{\Delta E}{c^2}}{m_e + m_p + m_n + \frac{\Delta E}{c^2}}\right)i$$

Since $\Delta E_a << \Delta E$.

The intensity of the gravitational forces between $M_{g(imaginary)}$ and an imaginary particle with mass $m_{g(imaginary)}$, *both at rest*, is given by

$$F = G M_{i(imaginary)} m_{i(imaginary)} / r^2 =$$

$$\cong \left( \frac{m_e \pm (m_n - m_p) + \frac{\Delta E}{c^2}}{m_e + m_p + m_n + \frac{\Delta E}{c^2}} \right) G \frac{M_{i(real)} i \ m_{i(real)} i}{r^2}$$

Therefore, the *total* gravity is

$$g_{real} + \Delta g_{(imaginary)} = -G \frac{M_{i(real)}}{r^2} -$$

$$- \left( \frac{m_e \pm (m_n - m_p) + \frac{\Delta E}{c^2}}{m_e + m_p + m_n + \frac{\Delta E}{c^2}} \right) G \frac{M_{i(real)}}{r^2}$$

Thus, the *imaginary* gravitational mass of a body produces an *excess* of gravity acceleration, $\Delta g$, given by

$$\Delta g \cong \left( \frac{m_e \pm (m_n - m_p) + \frac{\Delta E}{c^2}}{m_e + m_p + m_n + \frac{\Delta E}{c^2}} \right) G \frac{M_{i(real)}}{r^2}$$

In the case of *soft* atoms we can consider $\Delta E \cong 2 \times 10^{-13}\ joules$. Thus, in this case we obtain

$$\Delta g \cong 6 \times 10^{-4} G \frac{M_i}{r^2} \qquad (143)$$

In the case of the *Sun*, for example, there is an *excess* of gravity acceleration, due to its *imaginary* gravitational mass, given by

$$\Delta g \cong (6 \times 10^{-4}) G \frac{M_{iS}}{r^2}$$

At a distance from the Sun of $r = 1.0 \times 10^{13} m$ the value of $\Delta g$ is

$$\Delta g \cong 8 \times 10^{-10} m.s^{-2}$$

Experiments in the pioneer 10 spacecraft, at a distance from the Sun of about 67 AU or $r = 1.0 \times 10^{13} m$ [45], measured an excess acceleration towards the Sun of

$$\Delta g = 8.74 \pm 1.33 \times 10^{-10} m.s^{-2}$$

Note that the general expression for the gravity acceleration of the Sun is

$$g = (1 + \approx 6 \times 10^{-4}) G \frac{M_{iS}}{r^2}$$

Therefore, in the case of the *gravitational deflection of light about the Sun*, the new expression for the deflection of the light is

$$\delta = (1 + \approx 6 \times 10^{-4}) \frac{4GM_{iS}}{c^2 d} \qquad (144)$$

Thus, the increase in $\delta$ due to the excess acceleration towards the Sun can be considered *negligible*.

Similarly to the collapse of the real wave function, the collapse of the psychic wave function must suddenly also express in reality all the possibilities described by it. This is, therefore, *a point of decision* in which there occurs the compelling need of realization of the *psychic form*. Thus, this is moment in which the content of the psychic form realizes itself in the space-time. For an observer in space-time, something is *real* when it is in the form of matter or radiation. Therefore, the content of the psychic form may realize itself in space-time exclusively under the form of radiation, that is, it does not materialize. This must occur when the *Materialization Condition* is not satisfied, i.e., when the content of the psychic form is undefined (impossible to be defined by its own psychic) or it does not contain enough psychic mass to *materialize*[‡‡] the respective psychic contents.

Nevertheless, in both cases, there must always be a production of "virtual" photons to convey the psychic interaction to the other psychic particles, according to the quantum field theory, only through this type of quanta will interaction be conveyed, since it has an infinite reach and may be either attractive or repulsive, just as electromagnetic interaction which, as we know, is conveyed by the exchange of “virtual” photons.

[‡‡] By this we mean not only materialization proper but also the movement of matter to realize its psychic content (including radiation).

If electrons, protons and neutrons have psychic mass, then we can infer that the psychic mass of the atoms are *Phase Condensates*[§§]*.* In the case of the molecules the situation is similar. More molecular mass means more atoms and consequently, more psychic mass. In this case the phase condensate also becomes more structured because the great amount of elementary psyches inside the condensate requires, by stability reasons, a better distribution of them. Thus, in the case of molecules with very large molecular masses (*macromolecules*) it is possible that their psychic masses already constitute the most organized shape of a Phase Condensate, called Bose-Einstein Condensate[***].

The fundamental characteristic of a Bose-Einstein condensate is, as we know, that the various parts making up the condensed system not only behave as a whole but also *become a whole*, i.e., in the psychic case, the various consciousnesses of the system become a *single consciousness* with psychic mass equal to the sum of the psychic masses of all the consciousness of the condensate. This obviously, increases the available knowledge in the system since it is proportional to the psychic mass of the consciousness. This unity confers an *individual* character to this type of consciousness. For this reason, from now on they will be called *Individual Material Consciousness.*

We can derive from the above that most bodies do not possess individual material consciousness. In an iron rod, for instance, the cluster of elementary psyches in the iron molecules does not constitute Bose-Einstein condensate; therefore, the iron rod does not have an individual consciousness. Its consciousness is consequently, much more simple and constitutes just a phase condensate imprecisely structured made by the consciousness of the iron atoms.

The existence of consciousnesses in the atoms is revealed in the molecular formation, where atoms with strong mutual affinity (their consciousnesses) combine to form molecules. It is the case, for instance of the water molecules, in which two Hydrogen atoms join an Oxygen atom. Well, how come the combination between these atoms is always the same: the same grouping and the same invariable proportion? In the case of molecular combinations the phenomenon repeats itself. Thus, the chemical substances either mutually attract or repel themselves, carrying out specific motions for this reason. It is the so-called *Chemical Affinity*. This phenomenon certainly results from a specific interaction between the consciousnesses. From now on, it will be called *Psychic Interaction*.

*Mutual Affinity* is a dimensionless psychic quantity with which we are familiar and of which we have perfect understanding as to its meaning. The degree of *Mutual Affinity*, $A$, in the case of two consciousnesses, respectively described by $\Psi_{\Psi 1}$ and $\Psi_{\Psi 2}$, must be correlated to $\Psi_{\Psi1}^2$ and $\Psi_{\Psi2}^2$ [†††]. Only a simple algebraic form fills the requirements of interchange of the indices, the product

$$\Psi_{\Psi1}^2.\Psi_{\Psi2}^2 = \Psi_{\Psi2}^2.\Psi_{\Psi1}^2 =$$
$$= \left|A_{1,2}\right| = \left|A_{2,1}\right| = \left|A\right| \qquad (145)$$

---

[§§] Ice and NaCl crystals are common examples of imprecisely-structured *phase condensates*. Lasers, super fluids, superconductors and magnets are examples of phase condensates more structured.

[***] Several authors have suggested the possibility of the Bose-Einstein condensate occurring in the brain, and that it might be the physical base of memory, although they have not been able to find a suitable mechanism to underpin such a hypothesis. Evidences of the existence of Bose-Einstein condensates in living tissues abound (Popp, F.A Experientia, Vol. 44, p.576-585; Inaba, H., New Scientist, May89, p.41; Rattermeyer, M and Popp, F. A. Naturwissenschaften, Vol.68, Nº5, p.577.)

In the above expression, $|A|$ is due to the product $\Psi_{\Psi1}^2.\Psi_{\Psi2}^2$ will be always positive. From equations (143) and (134) we get

$$|A| = \Psi_{\Psi1}^2.\Psi_{\Psi2}^2 = k^2\left|\rho_{\Psi1}\right|\left|\rho_{\Psi2}\right| =$$
$$= k^2\frac{\left|m_{\Psi1}\right|}{V_1}\frac{\left|m_{\Psi2}\right|}{V_2} \qquad (146)$$

The psychic interaction can be described starting from the psychic mass because the psychic mass is the source of the psychic field. Basically, *the psychic mass is gravitational mass*, $m_\Psi \equiv m_{g(imaginary)}$. In this way, the equations of the gravitational interaction are also applied to the Psychic Interaction. However, due to the psychic mass, $m_\Psi$, to be an *imaginary* quantity, it is necessary to put $\left|m_\Psi\right|$ into the mentioned equations in order to homogenize them, because as we know, the module of an imaginary number is always real and positive.

Thus, based on gravity theory, we can write the equation of the *psychic field* in *nonrelativistic* Mechanics.

$$\Delta\Phi = 4\pi G\left|\rho_\Psi\right| \qquad (147)$$

It is similar to the equation of the gravitational field, with the difference that now instead of the density of gravitational mass we have the density of *psychic mass*. Then, we can write the general solution of Eq. (147), in the following form:

$$\Phi = -G\int\frac{\left|\rho_\Psi\right|dV}{r^2} \qquad (148)$$

This equation expresses, with nonrelativistic approximation, the potential of the psychic field of any distribution of psychic mass.

Particularly, for the potential of the field of only one particle with psychic mass $m_{\Psi1}$, we get:

$$\Phi = -\frac{G\left|m_{\Psi1}\right|}{r} \qquad (149)$$

Then the force produced by this field upon another particle with psychic mass $m_{\Psi2}$ is

$$\left|\vec{F}_{\Psi12}\right| = \left|-\vec{F}_{\Psi21}\right| = -\left|m_{\Psi2}\right|\frac{\partial\Phi}{\partial r} =$$
$$= -G\frac{\left|m_{\Psi1}\right|\left|m_{\Psi2}\right|}{r^2} \qquad (150)$$

By comparing equations (150) and (146) we obtain

$$\left|\vec{F}_{\Psi12}\right| = \left|-\vec{F}_{\Psi21}\right| = -G\left|A\right|\frac{V_1V_2}{k^2r^2} \qquad (151)$$

In the *vectorial* form the above equation is written as follows

$$\vec{F}_{\Psi12} = -\vec{F}_{\Psi21} = -GA\frac{V_1V_2}{k^2r^2}\hat{\mu} \qquad (152)$$

V*ersor* $\hat{\mu}$ has the direction of the line connecting the mass centers (psychic mass) of both particles and oriented from $m_{\Psi1}$ to $m_{\Psi2}$.

In general, we may distinguish and quantify two types of mutual affinity: *positive* and *negative* (*aversion*). The occurrence of the first type is synonym of psychic *attraction*, (as in the case of the atoms in the water molecule) while the aversion is synonym of *repulsion*. In fact, Eq. (152) shows that the forces $\vec{F}_{\Psi12}$ and $\vec{F}_{\Psi21}$ are attractive, if $A$ is *positive* (expressing *positive* mutual affinity between the two *psychic bodies*), and repulsive if $A$ is *negative* (expressing *negative* mutual affinity between the two *psychic bodies*). Contrary to the interaction of the matter, where the opposites attract themselves here, the *opposites repel themselves*.

[†††] Quantum Mechanics tells us that $\Psi$ do not have a physical interpretation or a simple meaning and also it cannot be experimentally observed. However such restriction does not apply to $\Psi^2$, which is known as *density of probability* and represents the probability of finding the body, described by the wave function $\Psi$, in the point x, y, z at the moment t. A large value of $\Psi^2$ means a strong possibility to find the body, while a small value of $\Psi^2$ means a weak possibility to find the body.

A method and device to obtain images of *psychic bodies* have been previously proposed [46]. By means of this device, whose operation is based on the gravitational interaction and the piezoelectric effect, it will be possible to observe psychic bodies.

Expression (146) can be rewritten in the following form:

$$A = k^2 \frac{m_{\Psi 1}}{V_1} \frac{m_{\Psi 2}}{V_2} \qquad (153)$$

The psychic masses $m_{\Psi 1}$ and $m_{\Psi 2}$ are *imaginary* quantities. However, the product $m_{\Psi 1}.m_{\Psi 2}$ is a *real* quantity. One can then conclude from the previous expression that the degree of mutual affinity between two consciousnesses depends basically on the densities of their psychic masses, and that:

1) If $m_{\Psi 1} > 0$ and $m_{\Psi 2} > 0$ then $A > 0$ (positive mutual affinity between them)
2) If $m_{\Psi 1} < 0$ and $m_{\Psi 2} < 0$ then $A > 0$ (positive mutual affinity between them)
3) If $m_{\Psi 1} > 0$ and $m_{\Psi 2} < 0$ then $A < 0$ (negative mutual affinity between them)
4) If $m_{\Psi 1} < 0$ and $m_{\Psi 2} > 0$ then $A < 0$ (negative mutual affinity between them)

In this relationship, as occurs in the case of material particles ("virtual" transition of the electrons previously mentioned), the consciousnesses interact mutually, *intertwining* or not their wave functions. When this happens, there occurs the so-called *Phase Relationship* according to quantum-mechanics concept. Otherwise a *Trivial Relationship* takes place.

The psychic forces such as the gravitational forces, must be very weak when we consider the interaction between two particles. However, in spite of the subtleties, those forces stimulate the relationship of the consciousnesses with themselves and with the Universe (Eq.152).

From all the preceding, we perceive that Psychic Interaction – unified with matter interactions, constitutes a single *Law* which links things and beings together and, in a network of continuous relations and exchanges, governs the Universe both in its material and psychic aspects. We can also observe that in the interactions the same principle reappears always identical. This *unity of principle* is the most evident expression of *monism* in the Universe.

## APPENDIX A: Allais effect explained

A Foucault-type pendulum slightly increases its period of oscillation at sites experiencing a *solar eclipse*, as compared with any other time. This effect was first observed by Allais [47] over 40 years ago. Also Saxl and Allen [48], using a torsion pendulum, have observed the phenomenon. Recently, an anomalous eclipse effect on gravimeters has become well-established [49], while some of the pendulum experiments have not. Here, we will show that the Allais gravity and pendulum effects during solar eclipses result from a *shielding effect of the Sun's gravity when the Moon is between the Sun and the Earth*.

The *interplanetary medium* includes interplanetary dust, cosmic rays and hot plasma from the solar wind. Its *density* is inversely proportional to the squared distance from the Sun, decreasing as this distance increases. Near the Earth-Moon system, this *density* is very low, with values about $5\,protons/cm^3\left(8.3\times10^{-21}kg/m^3\right)$. However, this density is *highly variable*. It can be increased up to $\sim 100\,protons/cm^3\left(1.7\times10^{-19}kg/m^3\right)$ [50].

The *atmosphere of the Moon* is very tenuous and insignificant in comparison with that of the Earth. The *average* daytime abundances of the elements known to be present in the lunar atmosphere, in atoms per cubic centimeter, are as follows: H <17, He 2-40x10$^3$, Na 70,K 17, Air 4x10$^4$, yielding ~8x10$^4$ total atoms per cubic centimeter $\left(\cong 10^{-16}kg.m^{-3}\right)$[51]. According to Öpik [52], near the Moon surface, the density of the lunar atmosphere can reach values up to $10^{-12}\,kg.m^{-3}$.The *minimum* possible density of the lunar atmosphere is in the top of the atmosphere and is essentially very close to the value of the *interplanetary medium*.

Since the density of the interplanetary medium is very small it cannot work as gravitational shielding. However, there is a top layer in the lunar atmosphere with density $\cong 1.3\times10^{-18}\,kg.m^{-3}$ that can work as a gravitational shielding and explain the Allais and pendulum effects. Below this layer, the density of the lunar atmosphere increases, making the effect of gravitational shielding negligible.

During the solar eclipses, when the Moon is between the Sun and the Earth, two *gravitational shieldings* $Sh1$ and $Sh2$, are established in the top layer of the lunar atmosphere (See Fig. 1A). In order to understand how these gravitational shieldings work (the gravitational shielding *effect*) see Fig. II. Thus, right after $Sh1$ (inside the system Moon-Lunar atmosphere), the *Sun's gravity acceleration*, $\vec{g}_S$, becomes $\chi\,\vec{g}_S$ where, according to Eq. (57) $\chi$ is given by

$$\chi=\left\{1-2\left[\sqrt{1+\left(\frac{n_r^2 D}{\rho c^3}\right)^2}-1\right]\right\} \qquad (1A)$$

The total density of *solar* radiation $D$ arriving at the top layer of the lunar atmosphere is given by

$$D=\sigma T^4=6.32\times10^7\,W/m^2$$

Since the temperature of the surface of the Sun is $T=5.778\times10^3\,K$ and $\sigma=5.67\times10^{-8}\,W.m^{-2}.K^{-4}$. The density of the top layer is $\rho\cong1.3\times10^{-18}\,kg.m^{-3}$ then Eq. (1A) gives[‡‡‡]

$$\chi=-1.1$$

The *negative* sign of $\chi$ shows that $\chi\vec{g}_S$, has *opposite* direction to $\vec{g}_S$. As previously showed (see Fig. II), after the second gravitational shielding $(Sh2)$ the gravity acceleration $\chi\vec{g}_S$ becomes $\chi^2 \vec{g}_S$. This means that $\chi^2 \vec{g}_S$ has the *same direction* of $\vec{g}_S$. In addition, right after $(Sh2)$ the lunar gravity becomes $\chi \vec{g}_{moon}$. Therefore, the *total gravity acceleration in the Earth* will be given by

$$\vec{g}' = \vec{g}_{\oplus} - \chi^2 \vec{g}_S - \chi \vec{g}_{moon} \qquad (2A)$$

[‡‡‡] The text in red in wrong. But the value of $\chi=-1.1$ is correct. It is not the solar radiation that produces the phenomenon. The exact description of the phenomenon starting from the same equation (1*A*) is presented in the end of my paper: "*Scattering of Sunlight in Lunar Exosphere Caused by Gravitational Microclusters of Lunar Dust*" (2013).

Since $g_S \cong 5.9 \times 10^{-3} m/s^2$ and $g_{moon} \cong 3.3 \times 10^{-5} m/s^2$ Eq. (2A), gives

$$\begin{aligned} g' &= g_{\oplus} - (-1.1)^2 g_S - (-1.1) g_{Moon} = \\ &\cong g_{\oplus} - 7.1 \times 10^{-3} m.s^{-2} = \\ &= (1 - 7.3 \times 10^{-4}) g_{\oplus} \qquad (3A) \end{aligned}$$

This decrease in $g$ increases the period $T = 2\pi\sqrt{l/g}$ of a *paraconical pendulum* (Allais effect) in about

$$T' = T\sqrt{\frac{g_{\oplus}}{(1 - 7.3 \times 10^{-4}) g_{\oplus}}} = 1.00037 \ T$$

This corresponds to $0.037\%$ increase in the period, and is roughly the value (0.0372%) obtained by Saxl and Allen during the total solar eclipse in March 1970 [48].

As we have seen, the density of the interplanetary medium near the Moon is *highly variable* and can reach values up to $\sim 100 \ protons/cm^3$ $(1.7 \times 10^{-19} kg/m^3)$.

When the density of the interplanetary medium increases, the top layer of the lunar atmosphere can also increase its density, by absorbing particles from the interplanetary medium due to the lunar gravitational attraction. In the case of a density increase of roughly 30% $(1.7 \times 10^{-18} kg/m^3)$, the value for $\chi$ becomes

$$\chi = -0.4$$

Consequently, we get

$$\begin{aligned} g' &= g_{\oplus} - (-0.4)^2 g_S - (-0.4) g_{Moon} = \\ &\cong g_{\oplus} - 9.6 \times 10^{-4} m.s^{-2} = \\ &= (1 - 9.7 \times 10^{-5}) g_{\oplus} \qquad (4A) \end{aligned}$$

This decrease in $g$ increases the pendulum's period by about

$$T' = T\sqrt{\frac{g_{\oplus}}{(1 - 9.4 \times 10^{-5}) g_{\oplus}}} = 1.000048 \ T$$

This corresponds to $0.0048\%$ increase in the pendulum's period. Jun's abstract [53] tells us of a relative change less than 0.005% in the pendulum's period associated with the 1990 solar eclipse.

For example, if the density of the top layer of the lunar atmosphere increase up to $2.0917 \times 10^{-18} kg/m^3$, the value for $\chi$ becomes

$$\chi = -1.5 \times 10^{-3}$$

Thus, we obtain

$$\begin{aligned} g' &= g_{\oplus} - (-1.5 \times 10^{-3})^2 g_S - (-1.5 \times 10^{-3}) g_{Moon} = \\ &\cong g_{\oplus} - 6.3 \times 10^{-8} m.s^{-2} = \\ &= (1 - 6.4 \times 10^{-9}) g_{\oplus} \qquad (5A) \end{aligned}$$

So, the total gravity acceleration in the Earth will *decrease* during the solar eclipses by about

$$6.4 \times 10^{-9} g_{\oplus}$$

The size of the effect, as measured with a *gravimeter*, during the 1997 eclipse, was roughly $(5-7) \times 10^{-9} g_{\oplus}$ [54, 55].

The decrease will be even smaller for $\rho \gtrsim 2.0917 \times 10^{-18}$ kg.m$^{-3}$. The lower limit now is set by Lageos satellites, which suffer an anomalous acceleration of only about $3 \times 10^{-13} g_{\oplus}$, during "seasons" where the satellite experiences eclipses of the Sun by the Earth [56].

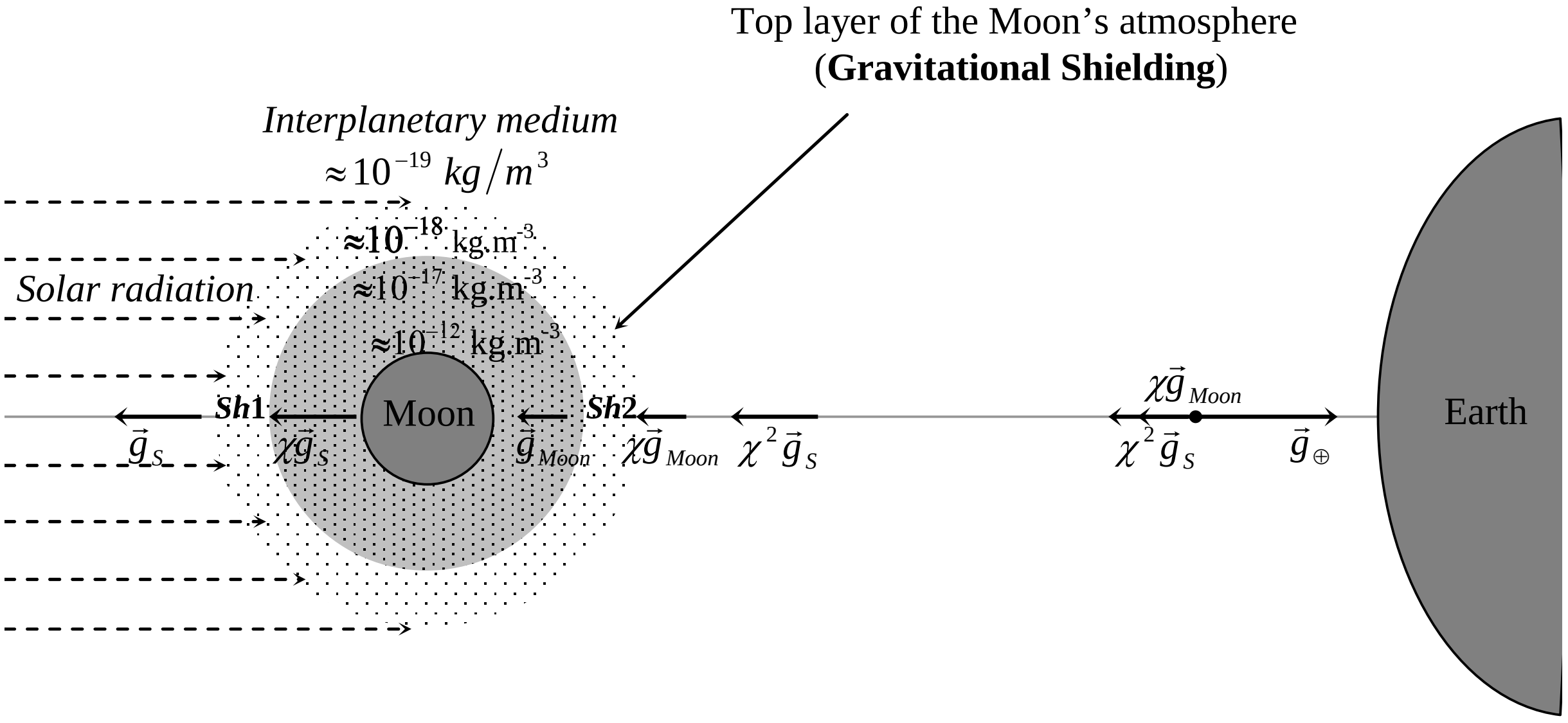

Fig. 1A – *Schematic diagram of the Gravitational Shielding around the Moon* – The top layer of the Moon's atmosphere with density of the order of $10^{-18}$ kg.m$^{-3}$, produces a gravitational shielding when subjected to the radiation from the Sun. Thus, the solar gravity $\vec{g}_S$ becomes $\chi\ \vec{g}_S$ after the first shielding $Sh1$ and $\chi^2\vec{g}_S$ after the second shielding $Sh2$. The Moon gravity becomes $\chi\ \vec{g}_{Moon}$ after $Sh2$. Therefore the *total gravity acceleration in the Earth* will be given by $\vec{g}' = \vec{g}_{\oplus} - \chi^2\vec{g}_S - \chi\vec{g}_{moon}$.

## APPENDIX B

In this appendix we will show why, in the *quantized gravity equation* (Eq.34), $n = 0$ is *excluded* from the sequence of possible values of $n$. Obviously, the exclusion of $n = 0$, means that the gravity can have only discrete values *different of zero*.

Equation (33) shows that the gravitational mass is *quantized* and given by

$$M_g = n^2 m_{g(\min)}$$

Since Eq. (43) leads to

$$m_{g(\min)} = m_{io(\min)}$$

where

$$m_{i0(\min)} = \pm h\sqrt{3/8}/cd_{\max} = \pm 3.9 \times 10^{-73} kg$$

is the *elementary quantum of inertial mass*. Then the equation for $M_g$ becomes

$$M_g = n^2 m_{g(\min)} = n^2 m_{i(\min)}$$

On the other hand, Eq. (44) shows that

$$M_i = n_i^2 m_{i0(\min)}$$

Thus, we can write that

$$\frac{M_g}{M_i} = \left(\frac{n}{n_i}\right)^2 \quad or \quad M_g = \eta^2 M_i \qquad (1B)$$

where $\eta = n/n_i$ is a *quantum number* different of $n$.

By multiplying both members of Eq. (1B) by $\sqrt{1 - V^2/c^2}$ we get

$$m_g = \eta^2 m_i \qquad (2B)$$

By substituting (2B) into Eq. (21) we get

$$E_n = \frac{n^2 h^2}{8 m_g L^2} = \frac{n^2 h^2}{8 \eta^2 m_i L^2} \qquad (3B)$$

From this equation we can easily conclude that $\eta$ cannot be *zero* $\left(E_n \to \infty \;\; or \;\; E_n \to \frac{0}{0}\right)$. On the other hand, the Eq. (2B) shows that the exclusion of $\eta = 0$ means the exclusion of $m_g = 0$ as a possible value for the gravitational mass. Obviously, this also means the exclusion of $M_g = 0$ (Relativistic mass). Equation (33) tells us that $M_g = n^2 m_{g(\min)}$, thus we can conclude that the exclusion of $M_g = 0$ implies in the exclusion of $n = 0$ since $m_{g(\min)} = m_{i0(\min)} = finite\ value$ (*elementary* quantum of mass). Therefore Eq. (3B) is only valid for values of $n$ and $\eta$ different of zero. Finally, from the *quantized gravity equation* (Eq. 34),

$$g = -\frac{GM_g}{r^2} = n^2\left(-\frac{Gm_{g(\min)}}{(r_{\max}/n)^2}\right) = \\ = n^4 g_{\min}$$

we conclude that the exclusion of $n = 0$ means that *the gravity* can have only discrete values *different of zero*.

# Gravity Control by means of *Electromagnetic Field* through *Gas* or *Plasma* at Ultra-Low Pressure

**Fran De Aquino**
Maranhao State University, Physics Department, S.Luis/MA, Brazil.

It is shown that the gravity acceleration just above a chamber filled with *gas* or *plasma* at ultra-low pressure can be strongly reduced by applying an Extra Low-Frequency (ELF) electromagnetic field across the gas or the plasma. This Gravitational Shielding Effect is related to recent discovery of *quantum correlation* between gravitational mass and inertial mass. According to the theory samples hung above the gas or the plasma should exhibit a weight decrease when the frequency of the electromagnetic field is decreased or when the intensity of the electromagnetic field is increased. This Gravitational Shielding Effect is unprecedented in the literature and can not be understood in the framework of the General Relativity. From the technical point of view, there are several applications for this discovery; possibly it will change the paradigms of *energy* generation, *transportation* and *telecommunications*.

**Key words:** Phenomenology of quantum gravity, Experimental Tests of Gravitational Theories, Vacuum Chambers, Plasmas devices. **PACs**: 04.60.Bc, 04.80.Cc, 07.30.Kf, 52.75.-d.

**CONTENTS**

## I. INTRODUCTION

It will be shown that the local gravity acceleration can be controlled by means of a device called Gravity Control Cell (GCC) which is basically a recipient filled with gas or plasma where is applied an *electromagnetic field*. According to the theory samples hung above the gas or plasma should exhibit a weight decrease when the frequency of the electromagnetic field is decreased or when the intensity of the electromagnetic field is increased. The electrical *conductivity* and the *density* of the gas or plasma are also highly relevant in this process.

With a GCC it is possible to convert the gravitational energy into rotational mechanical energy by means of the *Gravitational Motor*. In addition, a new concept of spacecraft (the *Gravitational Spacecraft*) and aerospace flight is presented here based on the possibility of gravity control. We will also see that the gravity control will be very important to *Telecommunication*.

## II. THEORY

It was shown [1] that the relativistic *gravitational mass* $M_g = m_g \big/ \sqrt{1 - V^2/c^2}$ and the relativistic *inertial mass* $M_i = m_{i0} \big/ \sqrt{1 - V^2/c^2}$ are *quantized*, and given by $M_g = n_g^2 m_{i0(min)}$ , $M_i = n_i^2 m_{i0(min)}$ where $n_g$ and $n_i$ are respectively, the *gravitational quantum number* and the *inertial quantum number* ; $m_{i0(min)} = \pm 3.9 \times 10^{-73} kg$ is the elementary *quantum* of inertial mass. The masses $m_g$ and $m_{i0}$ are correlated by means of the following expression:

$$m_g = m_{i0} - 2\left[\sqrt{1 + \left(\frac{\Delta p}{m_i c}\right)^2} - 1\right] m_{i0}. \qquad (1)$$

Where $\Delta p$ is the *momentum* variation on the particle and $m_{i0}$ is the inertial mass at rest.

In general, the *momentum* variation $\Delta p$ is expressed by $\Delta p = F\Delta t$ where $F$ is the applied force during a time interval $\Delta t$. Note that there is no restriction concerning the *nature* of the force $F$, i.e., it can be mechanical, electromagnetic, etc.

For example, we can look on the *momentum* variation $\Delta p$ as due to absorption or emission of *electromagnetic energy* by the particle.

In the case of radiation, $\Delta p$ can be obtained as follows: It is known that the *radiation pressure*, $dP$, upon an area $dA = dxdy$ of a volume $dV = dxdydz$ of a particle ( the incident radiation normal to the surface $dA$ )is equal to the energy $dU$ absorbed per unit volume $(dU/dV)$.i.e.,

$$dP = \frac{dU}{dV} = \frac{dU}{dxdydz} = \frac{dU}{dAdz} \qquad (2)$$

Substitution of $dz = vdt$ ($v$ is the speed of radiation) into the equation above gives

$$dP = \frac{dU}{dV} = \frac{(dU/dAdt)}{v} = \frac{dD}{v} \qquad (3)$$

Since $dPdA = dF$ we can write:

$$dFdt = \frac{dU}{v} \qquad (4)$$

However we know that $dF = dp/dt$, then

$$dp = \frac{dU}{v} \qquad (5)$$

From this equation it follows that

$$\Delta p = \frac{U}{v}\left(\frac{c}{c}\right) = \frac{U}{c} n_r$$

Substitution into Eq. (1) yields

$$m_g = \left\{1 - 2\left[\sqrt{1 + \left(\frac{U}{m_{i0}c^2} n_r\right)^2} - 1\right]\right\} m_{i0} \qquad (6)$$

Where $U$, is the electromagnetic energy absorbed by the particle; $n_r$ is the index of refraction.

Equation (6) can be rewritten in the following form

$$m_g = \left\{1 - 2\left[\sqrt{1 + \left(\frac{W}{\rho c^2} n_r\right)^2} - 1\right]\right\} m_{i0} \quad (7)$$

Where $W = U/V$ is the *density of electromagnetic energy* and $\rho = m_{i0}/V$ is the density of inertial mass.

The Eq. (7) is the expression of the *quantum correlation* between the *gravitational mass* and the *inertial mass* as a function of the *density of electromagnetic energy*. This is also the expression of correlation between gravitation and electromagnetism.

The density of electromagnetic energy in an *electromagnetic* field can be deduced from Maxwell's equations [2] and has the following expression

$$W = \tfrac{1}{2}\varepsilon E^2 + \tfrac{1}{2}\mu H^2 \quad (8)$$

It is known that $B = \mu H$, $E/B = \omega/k_r$ [3] and

$$v = \frac{dz}{dt} = \frac{\omega}{\kappa_r} = \frac{c}{\sqrt{\frac{\varepsilon_r \mu_r}{2}\left(\sqrt{1 + (\sigma/\omega\varepsilon)^2} + 1\right)}} \quad (9)$$

Where $k_r$ is the real part of the *propagation vector* $\vec{k}$ (also called *phase constant* [4]); $k = \left|\vec{k}\right| = k_r + ik_i$ ; $\varepsilon$, $\mu$ and $\sigma$, are the electromagnetic characteristics of the medium in which the incident (or emitted) radiation is propagating ($\varepsilon = \varepsilon_r \varepsilon_0$ where $\varepsilon_r$ is the *relative dielectric permittivity* and $\varepsilon_0 = 8.854 \times 10^{-12} F/m$ ; $\mu = \mu_r \mu_0$ where $\mu_r$ is the *relative magnetic permeability* and $\mu_0 = 4\pi \times 10^{-7} H/m$; $\sigma$ is the *electrical conductivity*). It is known that for *free-space* $\sigma = 0$ and $\varepsilon_r = \mu_r = 1$ then Eq. (9) gives

$$v = c \quad (10)$$

From (9) we see that the *index of refraction* $n_r = c/v$ will be given by

$$n_r = \frac{c}{v} = \sqrt{\frac{\varepsilon_r \mu_r}{2}\left(\sqrt{1 + (\sigma/\omega\varepsilon)^2} + 1\right)} \quad (11)$$

Equation (9) shows that $\omega/\kappa_r = v$. Thus, $E/B = \omega/k_r = v$, i.e., $E = vB = v\mu H$ . Then, Eq. (8) can be rewritten in the following form:

$$W = \tfrac{1}{2}\left(\varepsilon v^2 \mu\right)\mu H^2 + \tfrac{1}{2}\mu H^2 \quad (12)$$

For $\sigma << \omega\varepsilon$, Eq. (9) reduces to

$$v = \frac{c}{\sqrt{\varepsilon_r \mu_r}}$$

Then, Eq. (12) gives

$$W = \tfrac{1}{2}\left[\varepsilon\left(\frac{c^2}{\varepsilon_r \mu_r}\right)\mu\right]\mu H^2 + \tfrac{1}{2}\mu H^2 = \mu H^2 \quad (13)$$

This equation can be rewritten in the following forms:

$$W = \frac{B^2}{\mu} \quad (14)$$

or

$$W = \varepsilon E^2 \quad (15)$$

For $\sigma >> \omega\varepsilon$, Eq. (9) gives

$$v = \sqrt{\frac{2\omega}{\mu\sigma}} \quad (16)$$

Then, from Eq. (12) we get

$$W = \tfrac{1}{2}\left[\varepsilon\left(\frac{2\omega}{\mu\sigma}\right)\mu\right]\mu H^2 + \tfrac{1}{2}\mu H^2 = \left(\frac{\omega\varepsilon}{\sigma}\right)\mu H^2 + \tfrac{1}{2}\mu H^2 \cong$$

$$\cong \tfrac{1}{2}\mu H^2 \quad (17)$$

Since $E = vB = v\mu H$ , we can rewrite (17) in the following forms:

$$W \cong \frac{B^2}{2\mu} \quad (18)$$

or

$$W \cong \left(\frac{\sigma}{4\omega}\right)E^2 \quad (19)$$

By comparing equations (14) (15) (18) and (19) we see that Eq. (19) shows that the better way to obtain a strong value of $W$ *in practice* is by applying an *Extra Low-Frequency* (ELF) *electric field* $(w = 2\pi f << 1Hz)$ through a *mean with high electrical conductivity*.

Substitution of Eq. (19) into Eq. (7), gives

$$m_g = \left\{1 - 2\left[\sqrt{1 + \frac{\mu}{4c^2}\left(\frac{\sigma}{4\pi f}\right)^3 \frac{E^4}{\rho^2}} - 1\right]\right\} m_{i0} \quad (20)$$

This equation shows clearly that if an

electrical *conductor mean* has $\rho << 1\ Kg.m^{-3}$ and $\sigma >> 1$, then it is possible obtain strong changes in its gravitational mass, with a relatively small ELF *electric field*. An electrical *conductor mean* with $\rho << 1\ Kg.m^{-3}$ is obviously a *plasma*.

There is a very simple way to test Eq. (20). It is known that inside a fluorescent lamp *lit* there is *low-pressure Mercury plasma.* Consider a 20W T-12 *fluorescent lamp* (80044–F20T12/C50/ECO GE, Ecolux® T12), whose characteristics and dimensions are well-known [5]. At around $T \cong 318.15^0 K$, an optimum mercury vapor pressure of $P = 6\times10^{-3} Torr = 0.8 N.m^{-2}$ is obtained, which is required for maintenance of high luminous efficacy throughout life. Under these conditions, the mass density of the Hg plasma can be calculated by means of the well-known *Equation of State*

$$\rho = \frac{PM_0}{ZRT} \qquad (21)$$

Where $M_0 = 0.2006 kg.mol^{-1}$ is the molecular mass of the Hg; $Z \cong 1$ is the *compressibility factor* for the Hg plasma; $R = 8.314 joule.mol^{-1}.{}^0K^{-1}$ is the *gases universal constant*. Thus we get

$$\rho_{Hg\ plasma} \cong 6.067\times10^{-5} kg.m^{-3} \qquad (22)$$

The electrical conductivity of the Hg plasma can be deduced from the *continuum form of Ohm's Law* $\vec{j} = \sigma\vec{E}$, since the *operating current* through the lamp and the *current density* are well-known and respectively given by $i = 0.35 A$ [5] and $j_{lamp} = i/S = i/\frac{\pi}{4}\phi_{int}^2$, where $\phi_{int} = 36.1 mm$ is the inner diameter of the lamp. The voltage drop across the electrodes of the lamp is $57V$ [5] and the distance between them $l = 570 mm$. Then the electrical field *along* the lamp $E_{lamp}$ is given by $E_{lamp} = 57V/0.570m = 100\ V.m^{-1}$. Thus, we have

$$\sigma_{Hg\ plasma} = \frac{j_{lamp}}{E_{lamp}} = 3.419\ \ S.m^{-1} \qquad (23)$$

Substitution of (22) and (23) into (20) yields

$$\frac{m_{g(Hg\ plasma)}}{m_{i(Hg\ plasma)}} = \left\{1 - 2\left[\sqrt{1 + 1.909\times10^{-17}\frac{E^4}{f^3}} - 1\right]\right\} \qquad (24)$$

Thus, if an *Extra Low-Frequency electric field* $E_{ELF}$ with the following characteristics: $E_{ELF} \approx 100 V.m^{-1}$ and $f < 1 mHZ$ is applied through the Mercury plasma then a strong *decrease in the gravitational mass of the Hg plasma* will be produced.

It was shown [1] that there is an additional effect of *gravitational shielding* produced by a substance under these conditions. Above the substance the gravity acceleration $g_1$ is reduced at the same ratio $\chi = m_g/m_{i0}$, i.e., $g_1 = \chi\ g$, ($g$ is the gravity acceleration *under* the *substance*). Therefore, due to the *gravitational shielding effect* produced by the decrease of $m_{g(Hg\ plasma)}$ *in the region where the ELF electric field* $E_{ELF}$ *is applied*, the gravity acceleration just *above* this region will be given by

$$g_1 = \chi_{(Hg\ plasma)} g = \frac{m_{g(Hg\ plasma)}}{m_{i(Hg\ plasma)}} g =$$

$$= \left\{1 - 2\left[\sqrt{1 + 1.909\times10^{-17}\frac{E_{ELF}^4}{f_{ELF}^3}} - 1\right]\right\} g \qquad (25)$$

The trajectories of the electrons/ions through the lamp are determined by the electric field $E_{lamp}$ *along* the lamp. If the ELF electric field *across* the lamp $E_{ELF}$ is much greater than $E_{lamp}$, the current through the lamp can be interrupted. However, if $E_{ELF} << E_{lamp}$, these trajectories will be only slightly modified. Since here $E_{lamp} = 100\ V.m^{-1}$, then we can arbitrarily choose $E_{ELF}^{max} \cong 33\ \ V.m^{-1}$. This means that the *maximum* voltage drop, which can be applied across the metallic plates, placed at distance $d$, is equal to the outer diameter (max *) of the bulb $\phi_{lamp}^{max}$ of the 20W T-12 Fluorescent lamp, is given by

$$V_{max} = E_{ELF}^{max}\phi_{lamp}^{max} \cong 1.5\ V$$

Since $\phi_{lamp}^{max} = 40.3mm$ [5].

Substitution of $E_{ELF}^{max} \cong 33\ V.m^{-1}$ into (25) yields

$$g_1 = \chi_{(Hg\ plasma)}g = \frac{m_{g(Hg\ plasma)}}{m_{i(Hg\ plasma)}}g =$$

$$= \left\{1-2\left[\sqrt{1+\frac{2.264\times10^{-11}}{f_{ELF}^3}}-1\right]\right\}g \qquad (26)$$

Note that, for $f < 1mHz = 10^{-3}Hz$, the gravity acceleration can be strongly reduced. These conclusions show that the ELF Voltage Source of the set-up shown in Fig.1 should have the following characteristics:

- Voltage range: 0 – 1.5 V
- Frequency range: $10^{-4}$Hz – $10^{-3}$Hz

In the experimental arrangement shown in Fig.1, an ELF electric field with intensity $E_{ELF} = V/d$ crosses the fluorescent lamp; $V$ is the voltage drop across the metallic plates of the capacitor and $d = \phi_{lamp}^{max} = 40.3mm$. When the ELF electric field is applied, the gravity acceleration just above the lamp (inside the dotted box) decreases according to (25) and the changes can be measured by means of the system balance/sphere presented on the top of Figure 1.

In Fig. 2 is presented an experimental arrangement with *two* fluorescent lamps in order to test the gravity acceleration above the *second* lamp. Since gravity acceleration above the *first* lamp is given by $\vec{g}_1 = \chi_{1(Hg\ plasma)}\vec{g}$, where

$$\chi_{1(Hg\ plasma)} = \frac{m_{g1(Hg\ plasma)}}{m_{i1(Hg\ plasma)}} =$$

$$= \left\{1-2\left[\sqrt{1+1.909\times10^{-17}\frac{E_{ELF(1)}^4}{f_{ELF(1)}^3}}-1\right]\right\} \qquad (27)$$

Then, above the *second* lamp, the gravity acceleration becomes

$$\vec{g}_2 = \chi_{2(Hg\ plasma)}\vec{g}_1 = \chi_{2(Hg\ plasma)}\chi_{1(Hg\ plasma)}\vec{g} \qquad (28)$$

where

$$\chi_{2(Hg\ plasma)} = \frac{m_{g2(Hg\ plasma)}}{m_{i2(Hg\ plasma)}} =$$

$$= \left\{1-2\left[\sqrt{1+1.909\times10^{-17}\frac{E_{ELF(2)}^4}{f_{ELF(2)}^3}}-1\right]\right\} \qquad (29)$$

Then, results

$$\frac{g_2}{g} = \left\{1-2\left[\sqrt{1+1.909\times10^{-17}\frac{E_{ELF(1)}^4}{f_{ELF(1)}^3}}-1\right]\right\}\times$$

$$\times\left\{1-2\left[\sqrt{1+1.909\times10^{-17}\frac{E_{ELF(2)}^4}{f_{ELF(2)}^3}}-1\right]\right\} \qquad (30)$$

From Eq. (28), we then conclude that if $\chi_{1(Hg\ plasma)} < 0$ and also $\chi_{2(Hg\ plasma)} < 0$, then $g_2$ will have the *same direction* of $g$. This way it is possible to intensify several times the gravity in the direction of $\vec{g}$. On the other hand, if $\chi_{1(Hg\ plasma)} < 0$ and $\chi_{2(Hg\ plasma)} > 0$ the direction of $\vec{g}_2$ will be contrary to direction of $\vec{g}$. In this case will be possible to *intensify* and become $\vec{g}_2$ *repulsive* in respect to $\vec{g}$.

If we put a lamp above the *second* lamp, the gravity acceleration above the *third* lamp becomes

$$\vec{g}_3 = \chi_{3(Hg\ plasma)}\vec{g}_2 =$$

$$= \chi_{3(Hg\ plasma)}\chi_{2(Hg\ plasma)}\chi_{1(Hg\ plasma)}\vec{g} \qquad (31)$$

or

---

* After heating.

$$\frac{g_3}{g}=\left\{1-2\left[\sqrt{1+1.909\times10^{-17}\frac{E^4_{ELF(1)}}{f^3_{ELF(1)}}}-1\right]\right\}\times$$
$$\times\left\{1-2\left[\sqrt{1+1.909\times10^{-17}\frac{E^4_{ELF(2)}}{f^3_{ELF(2)}}}-1\right]\right\}\times$$
$$\times\left\{1-2\left[\sqrt{1+1.909\times10^{-17}\frac{E^4_{ELF(3)}}{f^3_{ELF(3)}}}-1\right]\right\} \quad (32)$$

If $f_{ELF(1)} = f_{ELF(2)} = f_{ELF(3)} = f$ and

$$E_{ELF(1)} = E_{ELF(2)} = E_{ELF(3)} = V/\phi =$$
$$= V_0 \sin \omega t / 40.3mm =$$
$$= 24.814 V_0 \sin 2\pi f t.$$

Then, for $t = T/4$ we get

$$E_{ELF(1)} = E_{ELF(2)} = E_{ELF(3)} = 24.814 V_0 .$$

Thus, Eq. (32) gives

$$\frac{g_3}{g}=\left\{1-2\left[\sqrt{1+7.237\times10^{-12}\frac{V_0^4}{f^3}}-1\right]\right\}^3 \quad (33)$$

For $V_0 = 1.5V$ and $f = 0.2mHz$ $(t = T/4 = 1250s = 20.83\min)$ the gravity acceleration $\vec{g}_3$ above the *third* lamp will be given by

$$\vec{g}_3 = -5.126\vec{g}$$

Above the *second* lamp, the gravity acceleration given by (30), is

$$\vec{g}_2 = +2.972\vec{g} \quad .$$

According to (27) the gravity acceleration above the *first* lamp is

$$\vec{g}_1 = \text{-}1{,}724\vec{g}$$

Note that, by this process an acceleration $\vec{g}$ can be increased several times in the direction of $\vec{g}$ or in the opposite direction.

In the experiment proposed in Fig. 1, we can start with ELF voltage sinusoidal wave of amplitude $V_0 = 1.0V$ and frequency $1mHz$. Next, the frequency will be progressively decreased down to $0.8mHz$, $0.6mHz$, $0.4mHz$ and $0.2mHz$. Afterwards, the amplitude of the voltage wave must be increased to $V_0 = 1.5V$ and the frequency decreased in the above mentioned sequence.

Table1 presents the *theoretical* values for $g_1$ and $g_2$, calculated respectively by means of (25) and (30).They are also plotted on Figures 5, 6 and 7 as a function of the frequency $f_{ELF}$.

Now consider a chamber filled with *Air* at $3\times10^{-12}torr$ and 300K as shown in Figure 8 (a). Under these circumstances, the mass density of the *air* inside the chamber, according to Eq. (21) is $\rho_{air} \cong 4.94\times10^{-15}kg.m^{-3}$.

If the frequency of the *magnetic* field, $B$, through the *air* is $f = 60Hz$ then $\omega\varepsilon = 2\pi f\varepsilon \cong 3\times10^{-9} S/m$. Assuming that the electric conductivity of the *air* inside the chamber, $\sigma_{(air)}$ is much less than $\omega\varepsilon$, i.e., $\sigma_{(air)} << \omega\varepsilon$ (The atmospheric air conductivity is of the order of $2-100\times10^{-15}S.m^{-1}$ [6, 7]) then we can rewritten the Eq. (11) as follows

$$n_{r(air)} \cong \sqrt{\varepsilon_r \mu_r} \cong 1 \quad (34)$$

From Eqs. (7), (14) and (34) we thus obtain

$$m_{g(air)} = \left\{1-2\left[\sqrt{1+\left(\frac{B^2}{\mu_{air}\rho_{air}c^2}n_{r(air)}\right)^2}-1\right]\right\}m_{i(air)} =$$
$$= \left\{1-2\left[\sqrt{1+3.2\times10^6 B^4}-1\right]\right\}m_{i(air)} \quad (35)$$

Therefore, due to the *gravitational shielding effect* produced by the decreasing of $m_{g(air)}$, the gravity acceleration *above* the *air* inside the chamber will be given by

$$g' = \chi_{air}\, g = \frac{m_{g(air)}}{m_{i(air)}} g =$$
$$= \left\{1-2\left[\sqrt{1+3.2\times10^6 B^4}-1\right]\right\}g$$

Note that the gravity acceleration above the *air* becomes *negative* for $B > 2.5\times10^{-2}T$.

For $B = 0.1T$ the gravity acceleration above the air becomes

$$g' \cong -32.8g$$

Therefore the ultra-low pressure air inside the chamber, such as the Hg plasma inside the fluorescent lamp, works like a Gravitational Shield that in practice, may be used to build *Gravity Control Cells* (GCC) for several practical applications.

Consider for example the GCCs of Plasma presented in Fig.3. The ionization of the plasma can be made of several manners. For example, by means of an electric field between the electrodes (Fig. 3(a)) or by means of a RF signal (Fig. 3(b)). In the first case the ELF electric field and the ionizing electric field can be the same.

Figure 3(c) shows a GCC filled with *air* (at ambient temperature and 1 atm) strongly ionized by means of alpha particles emitted from 36 radioactive ions sources (a very small quantity of *Americium* 241†). The radioactive element Americium has a half-life of 432 years, and emits *alpha particles* and low energy gamma rays $(\approx 60KeV)$. In order to shield the *alpha* particles and *gamma* rays emitted from the Americium 241 it is sufficient to encapsulate the GCC with *epoxy*. The alpha particles generated by the americium ionize the oxygen and nitrogen atoms of the air in the *ionization chamber* (See Fig. 3(c)) increasing the *electrical conductivity* of the air inside the chamber. The high-speed alpha particles hit molecules in the air and knock off electrons to form ions, according to the following expressions

$$O_2 + H_e^{++} \rightarrow O_2^{+} + e^{-} + H_e^{++}$$
$$N_2 + H_e^{++} \rightarrow N_2^{+} + e^{-} + H_e^{++}$$

It is known that the electrical conductivity is proportional to both the concentration and the mobility of the *ions* and the *free electrons*, and is expressed by

$$\sigma = \rho_e \mu_e + \rho_i \mu_i$$

Where $\rho_e$ and $\rho_i$ express respectively the concentrations $\left(C/m^3\right)$ of electrons and ions; $\mu_e$ and $\mu_i$ are respectively the mobilities of the electrons and the ions.

In order to calculate the electrical conductivity of the air inside the ionization chamber, we first need to calculate the concentrations $\rho_e$ and $\rho_i$. We start calculating the *disintegration constant*, $\lambda$, for the Am 241 :

$$\lambda = \frac{0.693}{T^{\frac{1}{2}}} = \frac{0.693}{432\left(3.15\times 10^{7}s\right)} = 5.1\times 10^{-11}s^{-1}$$

Where $T^{\frac{1}{2}} = 432 years$ is the *half-life* of the Am 241.

One $kmole$ of an isotope has mass equal to atomic mass of the isotope expressed in kilograms. Therefore, $1g$ of Am 241 has

$$\frac{10^{-3}kg}{241\ \ kg/kmole} = 4.15\times 10^{-6}kmoles$$

One $kmole$ of any isotope contains the Avogadro's number of atoms. Therefore $1g$ of Am 241 has

$$N = 4.15\times 10^{-6}kmoles \times$$
$$\times 6.025\times 10^{26}\ atoms/kmole = 2.50\times 10^{21} atoms$$

Thus, the *activity* [8] of the sample is

† The radioactive element *Americium* (Am-241) is widely used in *ionization smoke detectors*. This type of smoke detector is more common because it is inexpensive and better at detecting the smaller amounts of smoke produced by flaming fires. Inside an ionization detector there is a small amount (perhaps 1/5000th of a gram) of americium-241. The Americium is present in oxide form ($AmO_2$) in the detector. The cost of the $AmO_2$ is US$ 1,500 per gram. The amount of radiation in a smoke detector is extremely small. It is also predominantly alpha radiation. Alpha radiation cannot penetrate a sheet of paper, and it is blocked by several centimeters of air. The americium in the smoke detector could only pose a danger if inhaled.

$R = \lambda N = 1.3 \times 10^{11}$ disintegrations/s.

However, we will use 36 ionization sources each one with 1/5000th of a gram of Am 241. Therefore we will only use $7.2 \times 10^{-3}\,g$ of Am 241. Thus, $R$ reduces to:

$$R = \lambda N \cong 10^{9} \text{ disintegrations/s}$$

This means that at *one* second, about $10^{9}\,\alpha$ *particles* hit molecules in the air and knock off electrons to form ions $O_2^+$ and $N_2^+$ inside the ionization chamber. Assuming that *each* alpha particle yields *one* ion at each $1/10^{9}$ second then the total number of ions produced in one second will be $N_i \cong 10^{18}\,ions$. This corresponds to an ions concentration

$$\rho_i = eN_i / V \approx 0.1 / V \quad \left(C/m^3\right)$$

Where $V$ is the volume of the ionization chamber. Obviously, the concentration of electrons will be the same, i.e., $\rho_e = \rho_i$. For $d = 2cm$ and $\phi = 20cm$ (See Fig.3(c)) we obtain

$V = \frac{\pi}{4}(0.20)^2\left(2 \times 10^{-2}\right) = 6.28 \times 10^{-4}\,m^3$ Then we get:

$$\rho_e = \rho_i \approx 10^{2}\ C/m^3$$

This corresponds to the *minimum* concentration level in the case of *conducting materials*. For these materials, at temperature of 300K, the mobilities $\mu_e$ and $\mu_i$ vary from 10 up to $100\ m^2V^{-1}s^{-1}$ [9]. Then we can assume that $\mu_e = \mu_i \approx 10\ m^2V^{-1}s^{-1}$. (*minimum* mobility level for conducting materials). Under these conditions, the electrical conductivity of the air inside the ionization chamber is

$$\sigma_{air} = \rho_e\mu_e + \rho_i\mu_i \approx 10^{3}\,S.m^{-1}$$

At temperature of 300K, the air *density* inside the GCC, is $\rho_{air} = 1.1452\,kg.m^{-3}$. Thus, for $d = 2cm$, $\sigma_{air} \approx 10^{3}\,S.m^{-1}$ and $f = 60Hz$ Eq. (20) gives

$$\chi_{air} = \frac{m_{g(air)}}{m_{i(air)}} = \left\{1 - 2\left[\sqrt{1 + \frac{\mu}{4c^2}\left(\frac{\sigma_{air}}{4\pi f}\right)^3 \frac{V_{rms}^4}{d^4\rho_{air}^2}} - 1\right]\right\} = \left\{1 - 2\left[\sqrt{1 + 3.10 \times 10^{-16} V_{rms}^4} - 1\right]\right\}$$

Note that, for $V_{rms} \cong 7.96KV$, we obtain: $\chi_{(air)} \cong 0$. Therefore, if the voltages range of this GCC is: $0 - 10KV$ then it is possible to reach $\chi_{air} \cong -1$ when $V_{rms} \cong 10KV$.

It is interesting to note that $\sigma_{air}$ can be strongly increased by increasing the amount of Am 241. For example, by using $0.1g$ of Am 241 the value of $R$ increases to:

$$R = \lambda N \cong 10^{10} \text{ disintegrations/s}$$

This means $N_i \cong 10^{20}\,ions$ that yield

$$\rho_i = eN_i / V \approx 10 / V \quad \left(C/m^3\right)$$

Then, by reducing, $d$ and $\phi$ respectively, to 5mm and to 11.5cm, the volume of the ionization chamber reduces to:

$V = \frac{\pi}{4}(0.115)^2\left(5 \times 10^{-3}\right) = 5.19 \times 10^{-5}\,m^3$

Consequently, we get:

$$\rho_e = \rho_i \approx 10^{5}\ C/m^3$$

Assuming that $\mu_e = \mu_i \approx 10\ m^2V^{-1}s^{-1}$, then the electrical conductivity of the air inside the ionization chamber becomes

$$\sigma_{air} = \rho_e\mu_e + \rho_i\mu_i \approx 10^{6}\,S.m^{-1}$$

This reduces for $V_{rms} \cong 18.8V$ the voltage necessary to yield $\chi_{(air)} \cong 0$ and reduces to $V_{rms} \cong 23.5V$ the voltage necessary to reach $\chi_{air} \cong -1$.

If the outer surface of a metallic sphere with radius $a$ is covered with a radioactive element (for example Am 241), then the electrical conductivity of the air (very close to the sphere) can be strongly increased (for example up to $\sigma_{air} \cong 10^6 s.m^{-1}$). By applying a low-frequency electrical potential $V_{rms}$ to the sphere, in order to produce an electric field $E_{rms}$ starting from the outer surface of the sphere, then very close to the sphere the low-frequency electromagnetic field is $E_{rms} = V_{rms}/a$, and according to Eq. (20), the *gravitational mass* of the air in this region expressed by

$$m_{g(air)} = \left\{1 - 2\left[\sqrt{1 + \frac{\mu_0}{4c^2}\left(\frac{\sigma_{air}}{4\pi f}\right)^3 \frac{V_{rms}^4}{a^4 \rho_{air}^2}} - 1\right]\right\} m_{i0(air)},$$

can be easily reduced, making possible to produce a controlled *Gravitational Shielding* (similar to a GCC) surround the sphere.

This becomes possible to build a spacecraft to work with a gravitational shielding as shown in Fig. 4.

The *gravity accelerations* on the spacecraft (due to the rest of the Universe. See Fig.4) is given by

$$g_i' = \chi_{air} g_i \qquad i = 1, 2, 3 \ldots n$$

Where $\chi_{air} = m_{g(air)}/m_{i0(air)}$. Thus, the *gravitational forces* acting on the spacecraft are given by

$$F_{is} = M_g g_i' = M_g \left(\chi_{air} g_i\right)$$

By reducing the value of $\chi_{air}$, these forces can be reduced.

According to the *Mach's principle*;

"The *local inertial forces* are determined by the *gravitational interactions* of the local system with the distribution of the cosmic masses".

Thus, the local inertia is just the gravitational influence of the rest of matter existing in the Universe. Consequently, if we reduce the gravitational interactions between a spacecraft and the rest of the Universe, then *the inertial properties of the spacecraft* will be also reduced. This effect leads to a new concept of spacecraft and space flight.

Since $\chi_{air}$ is given by

$$\chi_{air} = \frac{m_{g(air)}}{m_{i0(air)}} = \left\{1 - 2\left[\sqrt{1 + \frac{\mu_0}{4c^2}\left(\frac{\sigma_{air}}{4\pi f}\right)^3 \frac{V_{rms}^4}{a^4 \rho_{air}^2}} - 1\right]\right\}$$

Then, for $\sigma_{air} \cong 10^6 s.m^{-1}$, $f = 6Hz$, $a = 5m$, $\rho_{air} \cong 1Kg.m^{-3}$ and $V_{rms} = 3.35KV$ we get

$$\chi_{air} \cong 0$$

Under these conditions, the gravitational forces upon the spacecraft become approximately nulls and consequently, the spacecraft practically *loses its inertial properties.*

Out of the terrestrial atmosphere, the gravity acceleration upon the spacecraft is negligible and therefore the gravitational shielding is not necessary. However, if the spacecraft is in the outer space and we want to use the gravitational shielding then, $\chi_{air}$ must be replaced by $\chi_{vac}$ where

$$\chi_{vac} = \frac{m_{g(vac)}}{m_{i0(vac)}} = \left\{1 - 2\left[\sqrt{1 + \frac{\mu_0}{4c^2}\left(\frac{\sigma_{vac}}{4\pi f}\right)^3 \frac{V_{rms}^4}{a^4 \rho_{vac}^2}} - 1\right]\right\}$$

The electrical conductivity of the ionized outer space (very close to the spacecraft) is small; however, its density is remarkably small $\left(<< 10^{-16} Kg.m^{-3}\right)$, in such a manner that the smaller value of the factor $\sigma_{vac}^3/\rho_{vac}^2$ can be easily compensated by the increase of $V_{rms}$.

It was shown that, when the gravitational mass of a particle is reduced to ranging between $+0.159M_i$ to $-0.159M_i$, it becomes *imaginary* [1], i.e., the gravitational and the inertial masses of the particle become *imaginary*. Consequently, the particle disappears from our ordinary space-time. However, the factor $\chi = M_{g(imaginary)}/M_{i(imaginary)}$ remains *real* because

$$\chi = \frac{M_{g(imaginary)}}{M_{i(imaginary)}} = \frac{M_g i}{M_i i} = \frac{M_g}{M_i} = real$$

Thus, if the gravitational mass of the particle is reduced by means of absorption of an amount of electromagnetic energy $U$, for example, we have

$$\chi = \frac{M_g}{M_i} = \left\{1-2\left[\sqrt{1+\left(U/m_{i0}c^2\right)^2}-1\right]\right\}$$

This shows that the energy $U$ of the electromagnetic field *remains acting on* the imaginary particle. In practice, this means that *electromagnetic fields act on imaginary particles*. Therefore, the electromagnetic field of a GCC remains acting on the particles inside the GCC even when their gravitational masses reach the gravitational mass ranging between $+0.159M_i$ to $-0.159M_i$ and they become imaginary particles. This is very important because it means that the GCCs of a gravitational spacecraft keep on working when the spacecraft becomes imaginary.

Under these conditions, the gravity accelerations on the *imaginary* spacecraft particle (due to the rest of the imaginary Universe) are given by

$$g'_j = \chi \ g_j \qquad j = 1,2,3,\ldots,n.$$

Where $\chi = M_{g(imaginary)}/M_{i(imaginary)}$ and $g_j = -Gm_{gj(imaginary)}/r_j^2$. Thus, the gravitational forces acting on the spacecraft are given by

$$F_{gj} = M_{g(imaginary)}g'_j =$$
$$= M_{g(imaginary)}\left(-\chi G m_{gj(imaginary)}/r_j^2\right) =$$
$$= M_g i\left(-\chi G m_{gj} i/r_j^2\right) = +\chi G M_g m_{gj}/r_j^2 \ .$$

Note that these forces are *real*. Remind that, the Mach's principle says that the *inertial effects* upon a particle are consequence of the gravitational interaction of the particle with the rest of the Universe. Then we can conclude that the *inertial forces* upon an *imaginary* spacecraft are also *real*. Consequently, it can travel in the imaginary space-time using its thrusters.

It was shown that, *imaginary particles* can have *infinite speed* in the *imaginary space-time* [1] . Therefore, this is also the speed upper limit for the spacecraft in the imaginary space-time.

Since the gravitational spacecraft can use its thrusters after to becoming an imaginary body, then if the thrusters produce a total thrust $F = 1000kN$ and the gravitational mass of the spacecraft is reduced from $M_g = M_i = 10^5 kg$ down to $M_g \cong 10^{-6} kg$, the acceleration of the spacecraft will be, $a = F/M_g \cong 10^{12} m.s^{-2}$. With this acceleration the spacecraft crosses the "visible" Universe ($diameter = d \approx 10^{26} m$) in a time interval $\Delta t = \sqrt{2d/a} \cong 1.4\times10^7 m.s^{-1} \cong 5.5 \ months$

Since the inertial effects upon the spacecraft are reduced by $M_g/M_i \cong 10^{-11}$ then, in spite of the effective spacecraft acceleration be $a = 10^{12} m.s^{-1}$, the effects for the crew and for the spacecraft will be equivalent to an acceleration $a'$ given by

$$a' = \frac{M_g}{M_i} a \approx 10 m.s^{-1}$$

This is the order of magnitude of the acceleration upon of a commercial jet aircraft.

On the other hand, the travel in the *imaginary* space-time can be very safe, because there won't any material body along the trajectory of the spacecraft.

Now consider the GCCs presented in Fig. 8 (a). Note that below and above the *air* are the bottom and the top of the chamber. Therefore the choice of the material of the chamber is highly relevant. If the chamber is made of steel, for example, and the gravity acceleration below the chamber is $g$ then at the bottom of the chamber, the gravity becomes $g' = \chi_{steel}\, g$ ; in the air, the gravity is $g'' = \chi_{air} g' = \chi_{air}\chi_{steel} g$ . At the top of the chamber, $g''' = \chi_{steel} g'' = (\chi_{steel})^2 \chi_{air} g$. Thus, out of the chamber (close to the top) the gravity acceleration becomes $g'''$. (See Fig. 8 (a)). However, for the steel at $B < 300T$ and $f = 1\times10^{-6}\,Hz$ , we have

$$\chi_{steel} = \frac{m_{g(steel)}}{m_{i(steel)}} = \left\{1 - 2\left[\sqrt{1 + \frac{\sigma_{(steel)} B^4}{4\pi f \mu \rho^2_{(steel)} c^2}} - 1\right]\right\} \cong 1$$

Since $\rho_{steel} = 1.1\times10^6\, S.m^{-1}$, $\mu_r = 300$ and $\rho_{(steel)} = 7800 k.m^{-3}$.

Thus, due to $\chi_{steel} \cong 1$ it follows that

$$g''' \cong g'' = \chi_{air}\, g' \cong \chi_{air}\, g$$

If instead of one GCC we have *three* GCC, all with steel box (Fig. 8(b)), then the gravity acceleration above the *second* GCC, $g_2$ will be given by

$$g_2 \cong \chi_{air}\, g_1 \cong \chi_{air}\chi_{air}\, g$$

and the gravity acceleration above the *third* GCC, $g_3$ will be expressed by

$$g_3 \cong \chi_{air}\, g'' \cong \chi_{air}^{\;3}\, g$$

## III. CONSEQUENCES

These results point to the possibility to convert gravitational energy into rotational mechanical energy. Consider for example the system presented in Fig. 9. Basically it is a motor with massive iron rotor and a box filled with gas or plasma at ultra-low pressure (Gravity Control Cell-GCC) as shown in Fig. 9. The GCC is placed below the rotor in order to become *negative* the acceleration of gravity inside *half* of the rotor $\left(g' = (\chi_{steel})^2 \chi_{air}\, g \cong \chi_{air}\, g = -ng\right)$. Obviously this causes a torque $T = (-F' + F)r$ and the rotor spins with angular velocity $\omega$. The average power, $P$ , of the motor is given by

$$P = T\omega = [(-F' + F)r]\omega \qquad (36)$$

Where

$$F' = \tfrac{1}{2} m_g\, g' \qquad F = \tfrac{1}{2} m_g\, g$$

and $m_g \cong m_i$ ( mass of the rotor ). Thus, Eq. (36) gives

$$P = (n+1)\frac{m_i g \omega\, r}{2} \qquad (37)$$

On the other hand, we have that

$$-g' + g = \omega^2 r \qquad (38)$$

Therefore the angular speed of the rotor is given by

$$\omega = \sqrt{\frac{(n+1)g}{r}} \qquad (39)$$

By substituting (39) into (37) we obtain the expression of the average power of the *gravitational motor*, i.e.,

$$P = \tfrac{1}{2} m_i \sqrt{(n+1)^3 g^3 r} \qquad (40)$$

Now consider an electric generator coupling to the gravitational motor in order to produce electric energy.

Since $\omega = 2\pi f$ then for $f = 60Hz$ we have $\omega = 120\pi rad.s^{-1} = 3600\, rpm$ .

Therefore for $\omega = 120\pi rad.s^{-1}$ and $n = 788$ $(B \cong 0.22T)$ the Eq. (40) tell us that we must have

$$r = \frac{(n+1)g}{\omega^2} = 0.0545m$$

Since $r = R/3$ and $m_i = \rho\pi R^2 h$ where $\rho$, $R$ and $h$ are respectively the mass density, the radius and the height of the rotor then for $h = 0.5m$ and $\rho = 7800 Kg.m^{-3}$ (iron) we obtain

$$m_i = 327.05kg$$

Then Eq. (40) gives

$$P \cong 2.19\times10^5\, watts \cong 219\ KW \cong 294HP \qquad (41)$$

This shows that the *gravitational motor* can be used to yield electric energy at large scale.

The possibility of gravity control leads to a new concept of spacecraft which is presented in Fig. 10. Due to the *Meissner effect*, the magnetic field *B* is expelled from the *superconducting shell*. The Eq. (35) shows that a magnetic field, $B$, through the *aluminum shell* of the spacecraft reduces its gravitational mass according to the following expression:

$$m_{g(Al)} = \left\{1 - 2\left[\sqrt{1+\left(\frac{B^2}{\mu c^2 \rho_{(Al)}} n_{r(Al)}\right)^2} - 1\right]\right\} m_{i(Al)} \qquad (42)$$

If the frequency of the magnetic field is $f = 10^{-4}\,Hz$ then we have that $\sigma_{(Al)} >> \omega\varepsilon$ since the electric conductivity of the aluminum is $\sigma_{(Al)} = 3.82\times10^7\, S.m^{-1}$. In this case, the Eq. (11) tell us that

$$n_{r(Al)} = \sqrt{\frac{\mu c^2 \sigma_{(Al)}}{4\pi f}} \qquad (43)$$

Substitution of (43) into (42) yields

$$m_{g(Al)} = \left\{1 - 2\left[\sqrt{1+\frac{\sigma_{(Al)} B^4}{4\pi f \mu \rho_{(Al)}^2 c^2}} - 1\right]\right\} m_{i(Al)} \qquad (44)$$

Since the mass density of the Aluminum is $\rho_{(Al)} = 2700\,kg.m^{-3}$ then the Eq. (44) can be rewritten in the following form:

$$\chi_{Al} = \frac{m_{g(Al)}}{m_{i(Al)}} = \left\{1 - 2\left[\sqrt{1+3.68\times10^{-8} B^4} - 1\right]\right\} \qquad (45)$$

In practice it is possible to adjust $B$ in order to become, for example, $\chi_{Al} \cong 10^{-9}$. This occurs to $B \cong 76.3T$. (Novel superconducting magnets are able to produce up to $14.7T$ [10, 11]).

Then the gravity acceleration in any direction *inside* the spacecraft, $g'_l$ , will be reduced and given by

$$g'_l = \frac{m_{g(Al)}}{m_{i(Al)}} g_l = \chi_{Al} g_l \cong -10^{-9} g_l \quad l = 1,2,..,n$$

Where $g_l$ is the *externa*l gravity in the direction $l$. We thus conclude that the gravity acceleration inside the spacecraft becomes negligible if $g_l << 10^9\, m.s^{-2}$. This means that the aluminum shell, under these conditions, works like a gravity shielding.

Consequently, the gravitational forces between anyone point inside the spacecraft with gravitational mass, $m_{gj}$, and another external to the spacecraft (gravitational mass $m_{gk}$) are given by

$$\vec{F}_j = -\vec{F}_k = -G\frac{m_{gj} m_{gk}}{r_{jk}^2}\hat{\mu}$$

where $m_{gk} \cong m_{ik}$ and $m_{gj} = \chi_{Al} m_{ij}$. Therefore we can rewrite equation above in the following form

$$\vec{F}_j = -\vec{F}_k = -\chi_{Al} G\frac{m_{ij} m_{ik}}{r_{jk}^2}\hat{\mu}$$

Note that when $B = 0$ the *initial gravitational forces* are

$$\vec{F}_j = -\vec{F}_k = -G\frac{m_{ij} m_{ik}}{r_{jk}^2}\hat{\mu}$$

Thus, if $\chi_{Al} \cong -10^{-9}$ then the initial gravitational forces are reduced from $10^9$ times and become repulsives.

According to the new expression for the *inertial forces* [1], $\vec{F} = m_g \vec{a}$, we see that these forces have origin in the *gravitational interaction* between a particle and the others of the Universe, just as *Mach's principle* predicts. Hence mentioned expression incorporates the Mach's principle into Gravitation Theory, and furthermore reveals that the inertial effects upon a body can be strongly reduced by means of the decreasing of its gravitational mass.

Consequently, we conclude that if the *gravitational forces* upon the spacecraft are reduced from $10^9$ times then also the *inertial forces* upon the spacecraft will be reduced from $10^9$ times when $\chi_{Al} \cong -10^{-9}$. Under these conditions, the inertial effects on the crew would be strongly decreased. Obviously this leads to a new concept of aerospace flight.

Inside the spacecraft the gravitational forces between the dielectric with gravitational mass, $M_g$ and the man (gravitational mass, $m_g$), when $B = 0$ are

$$\vec{F}_m = -\vec{F}_M = -G\frac{M_g m_g}{r^2}\hat{\mu} \qquad (46)$$

or

$$\vec{F}_m = -G\frac{M_g}{r^2}m_g\hat{\mu} = -m_g g_M\hat{\mu} \qquad (47)$$

$$\vec{F}_M = +G\frac{m_g}{r^2}M_g\hat{\mu} = +M_g g_m\hat{\mu} \qquad (48)$$

If the *superconducting box* under $M_g$ (Fig. 10) is filled with *air* at ultra-low pressure ($3\times10^{-12}$ torr, 300K for example) then, when $B \neq 0$, the gravitational mass of the *air* will be reduced according to (35). Consequently, we have

$$g'_M = (\chi_{steel})^2\chi_{air}g_M \cong \chi_{air}g_M \qquad (49)$$

$$g'_m = (\chi_{steel})^2\chi_{air}g_m \cong \chi_{air}g_m \qquad (50)$$

Then the forces $\vec{F}_m$ and $\vec{F}_M$ become

$$\vec{F}_m = -m_g(\chi_{air}g_M)\hat{\mu} \qquad (51)$$

$$\vec{F}_M = +M_g(\chi_{air}g_m)\hat{\mu} \qquad (52)$$

Therefore if $\chi_{air} = -n$ we will have

$$\vec{F}_m = +nm_g g_M\hat{\mu} \qquad (53)$$

$$\vec{F}_M = -nM_g g_m\hat{\mu} \qquad (54)$$

Thus, $\vec{F}_m$ and $\vec{F}_M$ become *repulsive*. Consequently, the man inside the spacecraft is subjected to a gravity acceleration given by

$$\vec{a}_{man} = ng_M\hat{\mu} = -\chi_{air}G\frac{M_g}{r^2}\hat{\mu} \qquad (55)$$

Inside the GCC we have,

$$\chi_{air} = \frac{m_{g(air)}}{m_{i(air)}} = \left\{1 - 2\left[\sqrt{1 + \frac{\sigma_{(air)}B^4}{4\pi f\mu\rho^2_{(air)}c^2}} - 1\right]\right\} \qquad (56)$$

By ionizing the air inside the GCC (Fig. 10), for example, by means of a *radioactive* material, it is possible to increase the *air conductivity* inside the GCC up to $\sigma_{(air)} \cong 10^6 S.m^{-1}$. Then for $f = 10\ Hz$; $\rho_{(air)} = 4.94\times10^{-15}kg.m^{-3}$ (Air at $3\times10^{-12}$ torr, 300K) and we obtain

$$\chi_{air} = \left\{2\left[\sqrt{1 + 2.8\times10^{21}B^4} - 1\right] - 1\right\} \qquad (57)$$

For $B = B_{GCC} = 0.1T$ (note that, due to the *Meissner effect*, the magnetic field $B_{GCC}$ stay confined inside the *superconducting box*) the Eq. (57) yields

$$\chi_{air} \cong -10^9$$

Since there is no magnetic field through the *dielectric* presented in Fig.10 then, $M_g \cong M_i$. Therefore if $M_g \cong M_i = 100Kg$ and $r = r_0 \cong 1m$ the gravity acceleration upon the man, according to Eq. (55), is

$$a_{man} \cong 10m.s^{-1}$$

Consequently it is easy to see that this system is ideal to yield artificial gravity inside the spacecraft in the case of *inter-stellar travel*, when the gravity acceleration out of the spacecraft - due to the Universe - becomes negligible.

The *vertical* displacement of the spacecraft can be produced by means of *Gravitational Thrusters*. A schematic diagram of a Gravitational Thruster is shown in Fig.11. The Gravitational Thrusters can also provide the *horizontal* displacement of the spacecraft.

The concept of Gravitational Thruster results from the theory of the *Gravity Control Battery*, showed in Fig. 8 (b). Note that the number of GCC increases the thrust of the thruster. For example, if the thruster has *three* GCCs then the gravity acceleration upon the gas sprayed inside the thruster will be *repulsive* in respect to $M_g$ (See Fig. 11(a)) and given by

$$a_{gas} = (\chi_{air})^3(\chi_{steel})^4 g \cong -(\chi_{air})^3 G\frac{M_g}{r_0^2}$$

Thus, if inside the GCCs, $\chi_{air} \cong -10^9$

(See Eq. 56 and 57) then the equation above gives

$$a_{gas} \cong +10^{27} G \frac{M_i}{r_0^2}$$

For $M_i \cong 10kg$, $r_0 \cong 1m$ and $m_{gas} \cong 10^{-12} kg$ the thrust is

$$F = m_{gas} a_{gas} \cong 10^5 N$$

Thus, the Gravitational Thrusters are able to produce strong thrusts.

Note that in the case of very strong $\chi_{air}$, for example $\chi_{air} \cong -10^9$, the gravity accelerations upon the boxes of the second and third GCCs become very strong (Fig.11 (a)). Obviously, the walls of the mentioned boxes cannot to stand the enormous pressures. However, it is possible to build a similar system with 3 or more GCCs, *without material boxes*. Consider for example, a surface with several radioactive sources (Am-241, for example). The *alpha* particles emitted from the Am-241 cannot reach besides 10cm of air. Due to the trajectory of the alpha particles, three or more successive layers of air, with different electrical conductivities $\sigma_1$, $\sigma_2$ and $\sigma_3$, will be established in the ionized region (See Fig.11 (b)). It is easy to see that the gravitational shielding effect produced by these three layers is similar to the effect produced by the 3 GCCs shown in Fig. 11 (a).

It is important to note that if $F$ is force produced by a thruster then the spacecraft acquires acceleration $a_{spacecraft}$ given by [1]

$$a_{spacecraft} = \frac{F}{M_{g(spacecraft)}} = \frac{F}{\chi_{Al} M_{i(inside)} + m_{i(Al)}}$$

Therefore if $\chi_{Al} \cong 10^{-9}$; $M_{i(inside)} = 10^4 Kg$ and $m_{i(Al)} = 100 Kg$ (inertial mass of the aluminum shell) then it will be necessary $F = 10kN$ to produce

$$a_{spacecraft} = 100 m.s^{-2}$$

Note that the concept of Gravitational Thrusters leads directly to the *Gravitational Turbo Motor* concept (See Fig. 12).

Let us now calculate the gravitational forces between two very close *thin* layers of the *air* around the spacecraft. (See Fig. 13).

The gravitational force $dF_{12}$ that $dm_{g1}$ exerts upon $dm_{g2}$, and the gravitational force $dF_{21}$ that $dm_{g2}$ exerts upon $dm_{g1}$ are given by

$$d\vec{F}_{12} = d\vec{F}_{21} = -G \frac{dm_{g2} dm_{g1}}{r^2} \hat{\mu} \qquad (58)$$

Thus, the gravitational forces between the *air layer* 1, gravitational mass $m_{g1}$, and the *air layer* 2, gravitational mass $m_{g2}$, around the spacecraft are

$$\vec{F}_{12} = -\vec{F}_{21} = -\frac{G}{r^2} \int_0^{m_{g1}} \int_0^{m_{g2}} dm_{g1} dm_{g2} \hat{\mu} =$$

$$= -G \frac{m_{g1} m_{g2}}{r^2} \hat{\mu} = -\chi_{air} \chi_{air} G \frac{m_{i1} m_{i2}}{r^2} \hat{\mu} \qquad (59)$$

At 100km altitude the air pressure is $5.691 \times 10^{-3} torr$ and $\rho_{(air)} = 5.998 \times 10^{-6} kg.m^{-3}$ [12]. By ionizing the air surround the spacecraft, for example, by means of an oscillating electric field, $E_{osc}$, starting from the surface of the spacecraft ( See Fig. 13) it is possible to increase the *air conductivity* near the spacecraft up to $\sigma_{(air)} \cong 10^6 S.m^{-1}$ . Since $f = 1Hz$ and, in this case $\sigma_{(air)} >> \omega\varepsilon$, then, according to Eq. (11), $n_r = \sqrt{\mu \sigma_{(air)} c^2 / 4\pi f}$ . From Eq.(56) we thus obtain

$$\chi_{air} = \frac{m_{g(air)}}{m_{i(air)}} = \left\{1 - 2\left[\sqrt{1 + \frac{\sigma_{(air)} B^4}{4\pi f \mu_0 \rho_{(air)}^2 c^2}} - 1\right]\right\} \qquad (60)$$

Then for $B = 763T$ the Eq. (60) gives

$$\chi_{air} = \left\{1 - 2\left[\sqrt{1 + \sim 10^4 B^4} - 1\right]\right\} \cong -10^8 \qquad (61)$$

By substitution of $\chi_{air} \cong -10^8$ into Eq., (59) we get

$$\vec{F}_{12} = -\vec{F}_{21} = -10^{16} G \frac{m_{i1} m_{i2}}{r^2} \hat{\mu} \qquad (62)$$

If $m_{i1} \cong m_{i2} = \rho_{air} \boldsymbol{V}_1 \cong \rho_{air} \boldsymbol{V}_2 \cong 10^{-8} kg$, and $r = 10^{-3} m$ we obtain

$$\vec{F}_{12} = -\vec{F}_{21} \cong -10^{-4} N \qquad (63)$$

These forces are much more intense than the *inter-atomic forces* (the forces which maintain joined atoms, and molecules that make the solids and liquids) whose intensities, according to the Coulomb's law, is of the order of 1-1000×10$^{-8}$N.

Consequently, the air around the spacecraft will be strongly compressed upon their surface, making an "*air shell*" that will accompany the spacecraft during its displacement and will protect the *aluminum shell* of the direct attrition with the Earth's atmosphere.

In this way, during the flight, the attrition would occur just between the "air shell" and the atmospheric air around her. Thus, the spacecraft would stay free of the thermal effects that would be produced by the direct attrition of the aluminum shell with the Earth's atmosphere.

Another interesting effect produced by the magnetic field $B$ of the spacecraft is the possibility of to lift a body from the surface of the Earth to the spacecraft as shown in Fig. 14. By ionizing the air surround the spacecraft, by means of an oscillating electric field, $E_{osc}$, the *air conductivity* near the spacecraft can reach, for example, $\sigma_{(air)} \cong 10^6 S.m^{-1}$. Then for $f = 1Hz$; $B = 40.8T$ and $\rho_{(air)} \cong 1.2kg.m^{-3}$ (300K and 1 atm) the Eq. (56) yields

$$\chi_{air} = \left\{1 - 2\left[\sqrt{1 + 4.9 \times 10^{-7} B^4} - 1\right]\right\} \cong -0.1$$

Thus, the weight of the body becomes

$$P_{body} = m_{g(body)} g = \chi_{air} m_{i(body)} g = m_{i(body)} g'$$

Consequently, the body will be lifted on the direction of the spacecraft with acceleration

$$g' = \chi_{air} g \cong +0.98 m.s^{-1}$$

Let us now consider an important aspect of the flight dynamics of a Gravitational Spacecraft.

Before starting the flight, the *gravitational mass of the spacecraft*, $M_g$, must be strongly reduced, by means of a gravity control system, in order to produce – with a weak thrust $\vec{F}$, a strong acceleration, $\vec{a}$, given by [1]

$$\vec{a} = \frac{\vec{F}}{M_g}$$

In this way, the spacecraft could be strongly accelerated and quickly to reach very high speeds near speed of light.

If the gravity control system of the spacecraft is *suddenly* turned off, the *gravitational mass* of the spacecraft becomes immediately equal to its *inertial mass*, $M_i$, $\left(M'_g = M_i\right)$ and the velocity $\vec{V}$ becomes equal to $\vec{V}'$. According to the *Momentum* Conservation Principle, we have that

$$M_g V = M'_g V'$$

Supposing that the spacecraft was traveling in space with speed $V \approx c$, and that its gravitational mass it was $M_g = 1Kg$ and $M_i = 10^4 Kg$ then the velocity of the spacecraft is reduced to

$$V' = \frac{M_g}{M'_g} V = \frac{M_g}{M_i} V \approx 10^{-4} c$$

Initially, when the velocity of the spacecraft is $\vec{V}$, its kinetic energy is $E_k = \left(M_g - m_g\right)c^2$. Where $M_g = m_g \big/ \sqrt{1 - V^2/c^2}$. At the instant in which the gravity control system of the spacecraft is turned off, the kinetic energy becomes $E'_k = \left(M'_g - m'_g\right)c^2$. Where $M'_g = m'_g \big/ \sqrt{1 - V'^2/c^2}$.

We can rewritten the expressions of $E_k$ and $E'_k$ in the following form

$$E_k = \left(M_g V - m_g V\right)\frac{c^2}{V}$$

$$E'_k = \left(M'_g V' - m'_g V'\right)\frac{c^2}{V'}$$

Substitution of $M_g V = M'_g V' = p$,

$m_g V = p\sqrt{1-V^2/c^2}$ and $m'_g V' = p\sqrt{1-V'^2/c^2}$ into the equations of $E_k$ and $E'_k$ gives

$$E_k = \left(1-\sqrt{1-V^2/c^2}\right)\frac{pc^2}{V}$$

$$E'_k = \left(1-\sqrt{1-V'^2/c^2}\right)\frac{pc^2}{V'}$$

Since $V \approx c$ then follows that

$$E_k \approx pc$$

On the other hand, since $V' << c$ we get

$$E'_k = \left(1-\sqrt{1-V'^2/c^2}\right)\frac{pc^2}{V'} =$$

$$\cong \left(1 - \frac{1}{1+\frac{V'^2}{2c^2}+\ldots}\right)\frac{pc^2}{V'} \cong \left(\frac{V'}{2c}\right)pc$$

Therefore we conclude that $E_k >> E'_k$. Consequently, when the gravity control system of the spacecraft is turned off, occurs an *abrupt* decrease in the kinetic energy of the spacecraft, $\Delta E_k$, given by

$$\Delta E_k = E_k - E'_k \approx pc \approx M_g c^2 \approx 10^{17} J$$

By comparing the energy $\Delta E_k$ with the *inertial energy* of the spacecraft, $E_i = M_i c^2$, we conclude that

$$\Delta E_k \approx \frac{M_g}{M_i} E_i \approx 10^{-4} M_i c^2$$

The energy $\Delta E_k$ (several *megatons*) must be released in very short time interval. It is approximately the same amount of energy that would be released in the case of collision of the spacecraft[‡]. However, the situation is very different of a collision ($M_g$ just becomes suddenly equal to $M_i$), and possibly the energy $\Delta E_k$ is converted into a *High Power Electromagnetic Pulse*.

[‡] In this case, the collision of the spacecraft would release $\approx 10^{17}$J (several megatons) and it would be similar to a powerful *kinetic weapon*.

Obviously this electromagnetic pulse (EMP) will induce heavy currents in all electronic equipment that mainly contains semiconducting and conducting materials. This produces immense heat that melts the circuitry inside. As such, *while not being directly responsible for the loss of lives*, these EMP are capable of disabling electric/electronic systems. Therefore, we possibly have a new type of *electromagnetic bomb*. An *electromagnetic bomb* or *E-bomb* is a well-known weapon designed to disable electric/electronic systems on a wide scale with an intense electromagnetic pulse.

Based on the theory of the GCC it is also possible to build a *Gravitational Press* of *ultra-high* pressure as shown in Fig.15.

The chamber 1 and 2 are GCCs with air at $1\times10^{-4}$torr, 300K $\left(\sigma_{(air)} \approx 10^6 S.m^{-1}; \rho_{(air)} = 5\times10^{-8} kg.m^{-3}\right)$. Thus, for $f = 10Hz$ and $B = 0.107T$ we have

$$\chi_{air} = \left\{1 - 2\left[\sqrt{1+\frac{\sigma_{(air)}B^4}{4\pi f \mu_0 \rho^2_{(air)} c^2}} - 1\right]\right\} \cong -118$$

The gravity acceleration above the air of the chamber 1 is

$$\vec{g}_1 = \chi_{stell}\chi_{air} g\hat{\mu} \cong +1.15\times10^3 \hat{\mu} \qquad (64)$$

Since, in this case, $\chi_{steel} \cong 1$; $\hat{\mu}$ is an *unitary vector* in the opposite direction of $\vec{g}$.

Above the air of the chamber 2 the gravity acceleration becomes

$$\vec{g}_2 = (\chi_{stell})^2(\chi_{air})^2 g\hat{\mu} \cong -1.4\times10^5 \hat{\mu} \qquad (65)$$

Therefore the *resultant* force $\vec{R}$ acting on $m_2$, $m_1$ and $m$ is

$$\vec{R} = \vec{F}_2 + \vec{F}_1 + \vec{F} = m_2\vec{g}_2 + m_1\vec{g}_1 + m\vec{g} =$$
$$= -1.4\times10^5 m_2\hat{\mu} + 1.15\times10^3 m_1\hat{\mu} - 9.81m\hat{\mu} =$$
$$\cong -1.4\times10^5 m_2\hat{\mu} \qquad (66)$$

where

$$m_2 = \rho_{steel}V_{disk\,2} = \rho_{steel}\left(\frac{\pi}{4}\phi_{inn}^2 H\right) \qquad (67)$$

Thus, for $\rho_{steel} \cong 10^4 kg.m^{-3}$ we can write that

$$F_2 \cong 10^9 \phi_{inn}^2 H$$

For the steel $\tau \cong 10^5 kg.cm^{-2} = 10^9 kg.m^{-2}$ consequently we must have $F_2/S_\tau < 10^9 kg.m^{-2}$ ($S_\tau = \pi\phi_{inn}H$ see Fig.15). This means that

$$\frac{10^9 \phi_{inn}^2 H}{\pi\phi_{inn} H} < 10^9 kg.m^{-2}$$

Then we conclude that

$$\phi_{inn} < 3.1m$$

For $\phi_{inn} = 2m$ and $H = 1m$ the Eq. (67) gives

$$m_2 \cong 3\times10^4 kg$$

Therefore from the Eq. (66) we obtain

$$R \cong 10^{10} N$$

Consequently, in the area $S = 10^{-4} m^2$ of the Gravitational Press, the pressure is

$$p = \frac{R}{S} \cong 10^{14} N.m^{-2}$$

This enormous pressure is much greater than the pressure in the center of the Earth ($3.617\times10^{11} N.m^{-2}$) [13]. It is near of the gas pressure in the *center of the sun* ($2\times10^{16} N.m^{-2}$). Under the action of such intensities new states of matter are created and astrophysical phenomena may be simulated in the lab for the first time, e.g. supernova explosions. Controlled thermonuclear fusion by inertial confinement, fast nuclear ignition for energy gain, novel collective acceleration schemes of particles and the numerous variants of material processing constitute examples of progressive applications of such *Gravitational Press* of ultra-high pressure.

The GCCs can also be applied on generation and detection of *Gravitational Radiation*.

Consider a cylindrical GCC (GCC antenna) as shown in Fig.16 (a). The *gravitational mass* of the *air* inside the GCC is

$$m_{g(air)} = \left\{1 - 2\left[\sqrt{1 + \frac{\sigma_{(air)}B^4}{4\pi f\mu\rho_{(air)}^2 c^2}} - 1\right]\right\} m_{i(air)} \qquad (68)$$

By varying $B$ one can varies $m_{g(air)}$ and consequently to vary the gravitational field generated by $m_{g(air)}$, producing then gravitational radiation. Then a GCC can work like a *Gravitational Antenna*.

Apparently, Newton's theory of gravity had no gravitational waves because, if a gravitational field changed in some way, that change took place *instantaneously* everywhere in space, and one can think that there is not a wave in this case. However, we have already seen that the gravitational interaction can be repulsive, besides attractive. Thus, as with electromagnetic interaction, the gravitational interaction must be produced by the exchange of "virtual" *quanta of* spin 1 and mass null, i.e., the *gravitational* "virtual" *quanta* (*graviphoton*) must have spin 1 and not 2. Consequently, the fact of a change in a gravitational field reach *instantaneously* everywhere in space occurs simply due to the speed of the *graviphoton* to be *infinite*. It is known that there is no speed limit for "*virtual*" photons. On the contrary, the *electromagnetic quanta* ("virtual" photons) could not communicate the *electromagnetic interaction* an infinite distance.

Thus, there are *two types* of gravitational radiation: the *real* and *virtual*, which is constituted of graviphotons; the *real* gravitational waves are ripples in the space-time generated by *gravitational field* changes. According to Einstein's theory of gravity the velocity of propagation of these waves is equal to the speed of light (*c*).

Unlike the electromagnetic waves the *real* gravitational waves have low interaction with matter and consequently low scattering. Therefore *real* gravitational waves are suitable as a means of transmitting information. However, when the distance between transmitter and receiver is too large, for example of the order of magnitude of several light-years, the transmission of information by means of gravitational waves becomes impracticable due to the long time necessary to receive the information. On the other hand, there is no delay during the transmissions by means of *virtual* gravitational radiation. In addition the scattering of this radiation is null. Therefore the *virtual* gravitational radiation is very suitable as a means of transmitting information at any distances including astronomical distances.

As concerns detection of the *virtual* gravitational radiation from GCC antenna, there are many options. Due to *Resonance Principle* a similar GCC antenna (receiver) *tuned at the same frequency* can absorb energy from an incident *virtual* gravitational radiation (See Fig.16 (b)). Consequently, the gravitational mass of the air inside the GCC receiver will vary such as the gravitational mass of the air inside the GCC transmitter. This will induce a magnetic field similar to the magnetic field of the GCC transmitter and therefore the current through the coil inside the GCC receiver will have the same characteristics of the current through the coil inside the GCC transmitter. However, the *volume* and *pressure* of the air inside the two GCCs must be exactly the same; also the *type* and the *quantity of atoms* in the air inside the two GCCs must be exactly the same. Thus, the GCC antennas are simple but they are not easy to build.

Note that a GCC antenna radiates *graviphotons* and *gravitational waves* simultaneously (Fig. 16 (a)). Thus, it is not only a gravitational antenna: it is a *Quantum Gravitational Antenna* because it can also emit and detect gravitational "virtual" *quanta* (graviphotons), which, in turn, can transmit information *instantaneously* from any distance in the Universe *without* scattering.

Due to the difficulty to build two similar GCC antennas and, considering that the electric current in the receiver antenna can be detectable even if the gravitational mass of the nuclei of the antennas are not *strongly* reduced, then we propose to replace the gas at the nuclei of the antennas by a thin *dielectric lamina*. The dielectric lamina with exactly $10^8$ atoms ($10^3$atoms $\times$ $10^3$atoms $\times$ $10^2$atoms) is placed between the plates (electrodes) as shown in Fig. 17. When the *virtual* gravitational radiation strikes upon the dielectric lamina, its gravitational mass varies similarly to the gravitational mass of the dielectric lamina of the transmitter antenna, inducing an electromagnetic field ($E$, $B$) similar to the transmitter antenna. Thus, the electric current in the receiver antenna will have the same characteristics of the current in the transmitter antenna. In this way, it is then possible to build two similar antennas whose nuclei have the same volumes and the same types and quantities of atoms.

Note that the Quantum Gravitational Antennas can also be used to transmit *electric power*. It is easy to see that the Transmitter and Receiver (Fig. 17(a)) can work with strong voltages and electric currents. This means that strong electric power can be transmitted among Quantum Gravitational Antennas. This obviously solves the problem of *wireless* electric power transmission.

The existence of *imaginary masses* has been predicted in a previous work [1]. Here we will propose a method and a device using GCCs for obtaining *images* of *imaginary bodies*.

It was shown that the *inertial* imaginary mass associated to an *electron* is given by

$$m_{ie(ima)} = \frac{2}{\sqrt{3}}\left(\frac{hf}{c^2}\right)i = \frac{2}{\sqrt{3}}m_{ie(real)}\ i \qquad (69)$$

Assuming that the correlation between the gravitational mass and the inertial mass (Eq.6) is the same for both imaginary and real masses then follows that the *gravitational* imaginary mass associated to an *electron* can be written in the following form:

$$m_{ge(ima)} = \left\{1 - 2\left[\sqrt{1+\left(\frac{U}{m_i c^2}n_r\right)^2} - 1\right]\right\} m_{ie(ima)} \qquad (70)$$

Thus, the gravitational *imaginary* mass *associated to matter* can be *reduced*, made *negative* and *increased,* just as the gravitational *real* mass.

It was shown that also *photons* have imaginary mass. Therefore, the imaginary mass can be associated or *not* to the matter.

In a general way, the gravitational forces between two gravitational imaginary masses are then given by

$$\vec{F} = -\vec{F} = -G\frac{(iM_g)(im_g)}{r^2}\hat{\mu} = +G\frac{M_g m_g}{r^2}\hat{\mu} \qquad (71)$$

Note that these forces are *real* and *repulsive*.

Now consider a gravitational imaginary mass, $m_{g(ima)} = im_g$, *not associated with matter* (like the gravitational imaginary mass associated to the photons) and another gravitational imaginary mass $M_{g(ima)} = iM_g$ *associated to* a *material* body.

*Any material body has an imaginary mass associated to it, d*ue to the existence of imaginary masses associated to the electrons*.* We will choose a *quartz crystal* (for the material body with gravitational imaginary mass $M_{g(ima)} = iM_g$) because quartz crystals are widely used to detect forces (piezoelectric effect).

By using GCCs as shown in Fig. 18(b) and Fig.18(c), we can increase the gravitational acceleration, $\vec{a}$, produced by the imaginary mass $im_g$ upon the crystals. Then it becomes

$$a = -\chi_{air}^3 G\frac{m_g}{r^2} \qquad (72)$$

As we have seen, the value of $\chi_{air}$ can be increased up to $\chi_{air} \cong -10^9$ (See Eq.57). Note that in this case, the gravitational forces become *attractive*. In addition, if $m_g$ is not small, the gravitational forces between the imaginary body of mass $im_g$ and the crystals can become sufficiently intense to be easily detectable.

Due to the piezoelectric effect, the gravitational force acting on the crystal will produce a voltage proportional to its intensity. Then consider a board with hundreds micro-crystals behind a set of GCCs, as shown in Fig.18(c). By amplifying the voltages generated in each micro-crystal and sending to an appropriated data acquisition system, it will be thus possible to obtain an image of the imaginary body of mass $m_{g(ima)}$ placed in front of the board.

In order to decrease strongly the gravitational effects produced by bodies placed behind the imaginary body of mass $im_g$, one can put five GCCs making a *Gravitational Shielding* as shown in Fig.18(c). If the GCCs are filled with air at 300K and $3\times10^{-12} torr$. Then $\rho_{air} = 4.94\times10^{-15} kg.m^{-3}$ and $\sigma_{air} \cong 1\times10^{-14} S.m^{-1}$. Thus, for $f = 60Hz$ and $B \cong 0.7T$ the Eq. (56) gives

$$\chi_{air} = \frac{m_{g(air)}}{m_{i(air)}} = \left\{1 - 2\left[\sqrt{1+5B^4} - 1\right]\right\} \cong -10^{-2} \qquad (73)$$

For $\chi_{air} \cong 10^{-2}$ the gravitational shielding presented in Fig.18(c) will reduce any value of $g$ to $\chi_{air}^5 g \cong 10^{-10} g$. This will be sufficiently to reduce strongly the gravitational effects proceeding from both sides of the gravitational shielding.

Another important consequence of the correlation between gravitational mass and inertial mass expressed by Eq. (1) is the possibility of building *Energy Shieldings* around objects in order to protect them from *high-energy particles* and *ultra-intense fluxes of radiation*.

In order to explain that possibility, we start from the new expression [1] for the *momentum* $q$ of a particle with gravitational mass $M_g$ and velocity $V$, which is given by

$$q = M_g V \qquad (74)$$

where $M_g = m_g \big/ \sqrt{1 - V^2/c^2}$ and $m_g = \chi\, m_i$ [1]. Thus, we can write

$$\frac{m_g}{\sqrt{1-V^2/c^2}} = \frac{\chi\, m_i}{\sqrt{1-V^2/c^2}} \qquad (75)$$

Therefore, we get

$$M_g = \chi M_i \qquad (76)$$

It is known from the Relativistic Mechanics that

$$q = \frac{UV}{c^2} \qquad (77)$$

where $U$ is the *total* energy of the particle. This expression is valid for *any* velocity $V$ of the particle, including $V = c$.

By comparing Eq. (77) with Eq. (74) we obtain

$$U = M_g c^2 \qquad (78)$$

It is a well-known experimental fact that

$$M_i c^2 = hf \qquad (79)$$

Therefore, by substituting Eq. (79) and Eq. (76) into Eq. (74), gives

$$q = \frac{V}{c}\chi\frac{h}{\lambda} \qquad (80)$$

Note that this expression is valid for *any* velocity $V$ of the particle. In the particular case of $V = c$, it reduces to

$$q = \chi\frac{h}{\lambda} \qquad (81)$$

By comparing Eq. (80) with Eq. (77), we obtain

$$U = \chi hf \qquad (82)$$

Note that only for $\chi = 1$ the Eq. (81) and Eq. (82) are reduced to the well=known expressions of DeBroglie $(q = h/\lambda)$ and Einstein $(U = hf)$.

Equations (80) and (82) show for example, that *any* real particle (material particles, real photons, etc) that penetrates a region (with density $\rho$ and electrical conductivity $\sigma$), where there is an ELF electric field $E$, will have its *momentum* $q$ and its energy $U$ reduced by the factor $\chi$, given by

$$\chi = \frac{m_g}{m_i} = \left\{1 - 2\left[\sqrt{1 + \frac{\mu}{4c^2}\left(\frac{\sigma}{4\pi f}\right)^3 \frac{E^4}{\rho^2}} - 1\right]\right\} \qquad (83)$$

The remaining amount of *momentum* and *energy*, respectively given by $(1-\chi)\left(\frac{V}{c}\right)\frac{h}{\lambda}$ and $(1-\chi)\ hf$, are *transferred to* the *imaginary* particle associated to the *real* particle[§] (material particles or real photons) that penetrated the mentioned region.

It was previously shown that, when the *gravitational mass* of a particle is reduced to ranging between $+0.159M_i$ to $-0.159M_i$, i.e., when $\chi < 0.159$, it becomes *imaginary* [1], i.e., the gravitational and the inertial masses of the particle become *imaginary*. Consequently, the particle disappears from our ordinary space-time. It goes to the Imaginary Universe. On the other hand, when the gravitational mass of the particle becomes greater than $+0.159M_i$, or less than $-0.159M_i$, i.e., when $\chi > 0.159$, the particle return to our Universe.

Figure 19 (a) clarifies the phenomenon of reduction of the *momentum* for $\chi > 0.159$, and Figure 19 (b) shows the effect in the case of $\chi < 0.159$. In this case, the particles become imaginary and consequently, they go to the *imaginary space-time* when they penetrate the electric field $E$. However, the electric field $E$ stays at the *real* space-time. Consequently, the particles return immediately to the real space-time in order to return soon after to the *imaginary* space-time, due to the action of the electric field $E$. Since the particles are moving at a direction, they *appear* and *disappear* while they are crossing the region, up to collide with the plate (See Fig.19) with a *momentum,* $q_m = \chi\left(\frac{V}{c}\right)\frac{h}{\lambda}$, in the case of the *material particle*, and $q_r = \chi\frac{h}{\lambda}$ in the case of the *photon*. Note that by making $\chi \cong 0$, it is possible to block high-energy particles and ultra-intense fluxes of radiation. These *Energy Shieldings* can be built around objects in order to protect them from such particles and radiation.

It is also important to note that the gravity control process described here points to the possibility of obtaining *Controlled Nuclear Fusion* by means of increasing of the intensity of the gravitational interaction between the nuclei. When the gravitational forces $F_G = Gm_g m_g'/r^2$ become greater than the electrical forces $F_E = qq'/4\pi\varepsilon_0 r^2$ between the nuclei, then nuclear fusion reactions can occur.

Note that, according to Eq. (83), the gravitational mass can be strongly increased. Thus, if $E = E_m \sin\omega t$, then the average value for $E^2$ is equal to $\frac{1}{2}E_m^2$, because $E$ varies sinusoidaly ($E_m$ is the maximum value for $E$). On the other hand, $E_{rms} = E_m/\sqrt{2}$. Consequently, we can replace

§ As previously shown, there are *imaginary particles* associated to each *real particle* [1].

$E^4$ for $E^4_{rms}$. In addition, as $j = \sigma E$ (*Ohm's vectorial Law*), then Eq. (83) can be rewritten as follows

$$\chi = \frac{m_g}{m_{i0}} = \left\{1 - 2\left[\sqrt{1 + K\frac{\mu_r j^4_{rms}}{\sigma\rho^2 f^3}} - 1\right]\right\} \qquad (84)$$

where $K = 1.758\times10^{-27}$ and $j_{rms} = j/\sqrt{2}$.

Thus, the gravitational force equation can be expressed by

$$F_G = Gm_g m'_g/r^2 = \chi^2 Gm_{i0}m'_{i0}/r^2 =$$

$$= \left\{1 - 2\left[\sqrt{1 + K\frac{\mu_r j^4_{rms}}{\sigma\rho^2 f^3}} - 1\right]\right\}^2 Gm_{i0}m'_{i0}/r^2 \qquad (85)$$

In order to obtain $F_G > F_E$ we must have

$$\left\{1 - 2\left[\sqrt{1 + K\frac{\mu_r j^4_{rms}}{\sigma\rho^2 f^3}} - 1\right]\right\} > \sqrt{\frac{qq'/4\pi\varepsilon_0}{Gm_{i0}m'_{i0}}} \qquad (86)$$

The *carbon fusion* is a set of nuclear fusion reactions that take place in massive stars (at least $8M_{sun}$ at birth). It requires high temperatures ($>5\times10^8 K$) and densities ($>3\times10^9\,kg.m^{-3}$). The principal reactions are:

$$^{12}C + {}^{12}C \rightarrow \begin{cases} {}^{23}Na + p + 2.24\ MeV \\ {}^{20}Ne + \alpha + 4.62\ MeV \\ {}^{24}Mg + \gamma + 13.93\ MeV \end{cases}$$

In the case of Carbon nuclei ($^{12}C$) of a *thin carbon wire* ($\sigma \cong 4\times10^4 S.m^{-1}; \rho = 2.2\times10^3 S.m^{-1}$) Eq. (86) becomes

$$\left\{1 - 2\left[\sqrt{1 + 9.08\times10^{-39}\frac{j^4_{rms}}{f^3}} - 1\right]\right\} > \sqrt{\frac{e^2}{16\pi\varepsilon_0 Gm_p^2}}$$

whence we conclude that the condition for the $^{12}C$ + $^{12}C$ fusion reactions occur is

$$j_{rms} > 1.7\times10^{18} f^{\frac{3}{4}} \qquad (87)$$

If the electric current through the carbon wire has Extremely-Low Frequency (ELF), for example, if $f = 1\mu Hz$, then the current density, $j_{rms}$, must have the following value:

$$j_{rms} > 5.4\times10^{13} A.m^{-2} \qquad (88)$$

Since $j_{rms} = i_{rms}/S$ where $S = \pi\phi^2/4$ is the area of the cross section of the wire, we can conclude that, for an *ultra-thin carbon* wire with $10\mu m$-diameter, it is necessary that the current through the wire, $i_{rms}$, have the following intensity

$$i_{rms} > 4.24\ kA$$

Obviously, this current will *explode* the carbon wire. However, this explosion becomes negligible in comparison with the very strong *gravitational implosion*, which occurs simultaneously due to the enormous increase in intensities of the gravitational forces among the carbon nuclei produced by means of the ELF current through the carbon wire as predicted by Eq. (85). Since, in this case, the gravitational forces among the carbon nuclei become greater than the repulsive electric forces among them the result is the production of $^{12}C$ + $^{12}C$ fusion reactions.

Similar reactions can occur by using a *lithium* wire. In addition, it is important to note that $j_{rms}$ is directly proportional to $f^{\frac{3}{4}}$ (Eq. 87). Thus, for example, if $f = 10^{-8} Hz$, the current necessary to produce the nuclear reactions will be $i_{rms} = 130A$.

## IV.CONCLUSION

The process described here is clearly the better way in order to control the gravity. This is because the *Gravity Control Cell* in this case is very easy to be built, the cost is low and it works at ambient temperature. The Gravity Control is the starting point for the generation of and detection of *Virtual Gravitational Radiation* (Quantum Gravitational *Transceiver*) also for the construction of the *Gravitational Motor* and the *Gravitational Spacecraft* which includes the system for generation of *artificial gravity* presented in Fig.10 and the *Gravitational Thruster* (Fig.11). While the *Gravitational Transceiver* leads to a new concept in *Telecommunication*, the Gravitational Motor changes the paradigm of *energy conversion* and the Gravitational Spacecraft points to a new concept in *aerospace flight*.

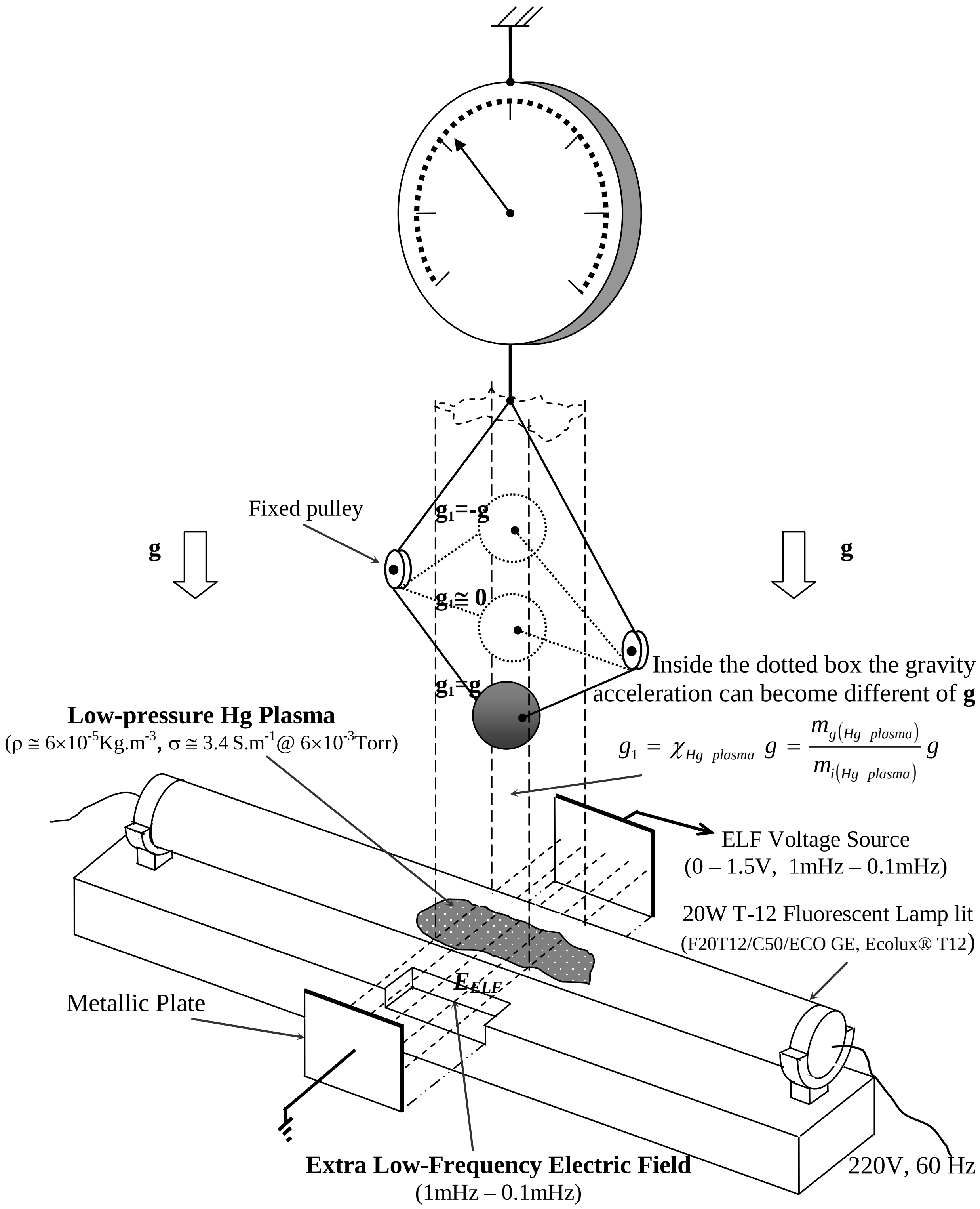

Fig. 1 – Gravitational Shielding Effect by means of an ELF electric field through low- pressure Hg Plasma.

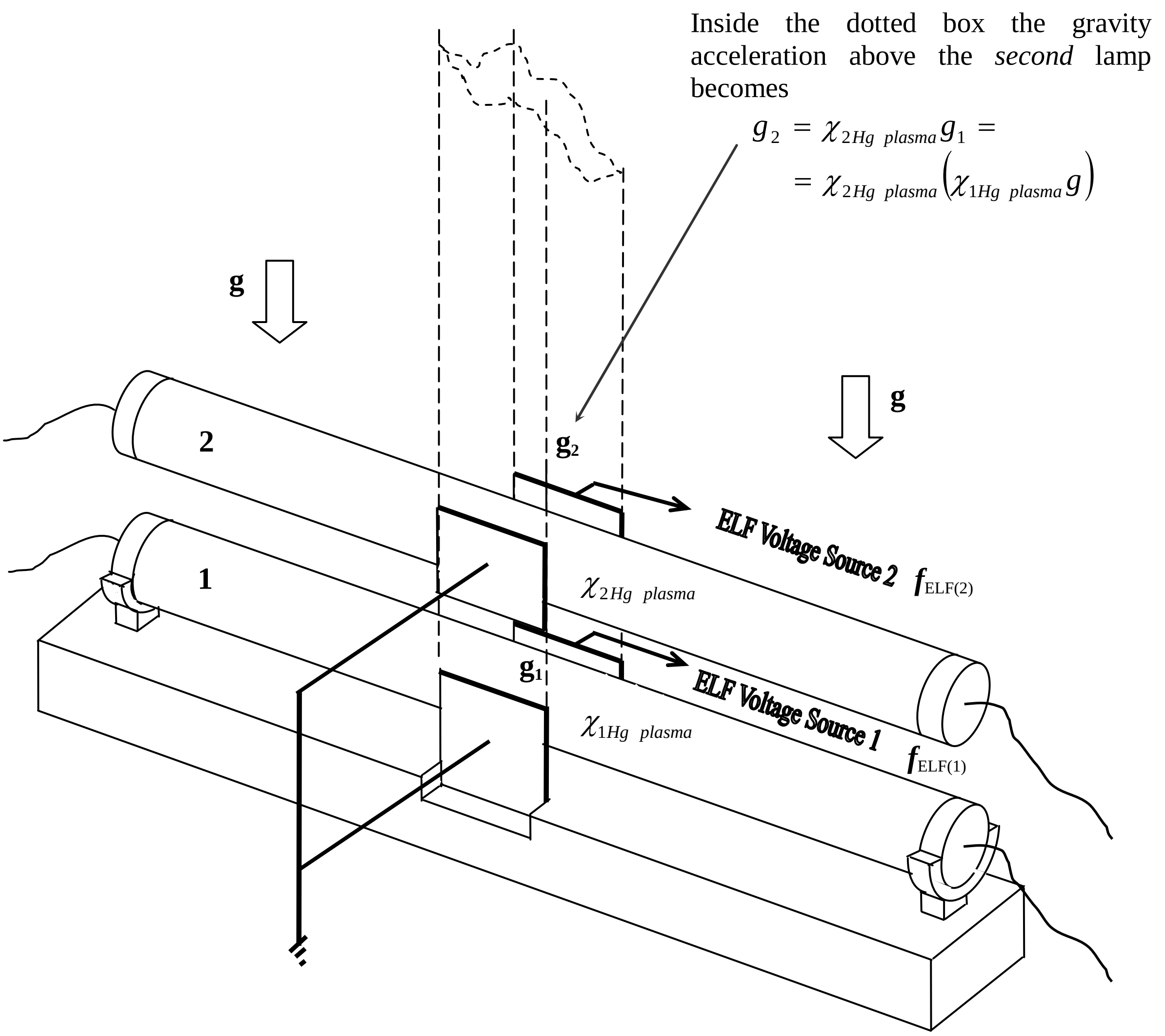

Fig. 2 – Gravity acceleration above a *second* fluorescent lamp.

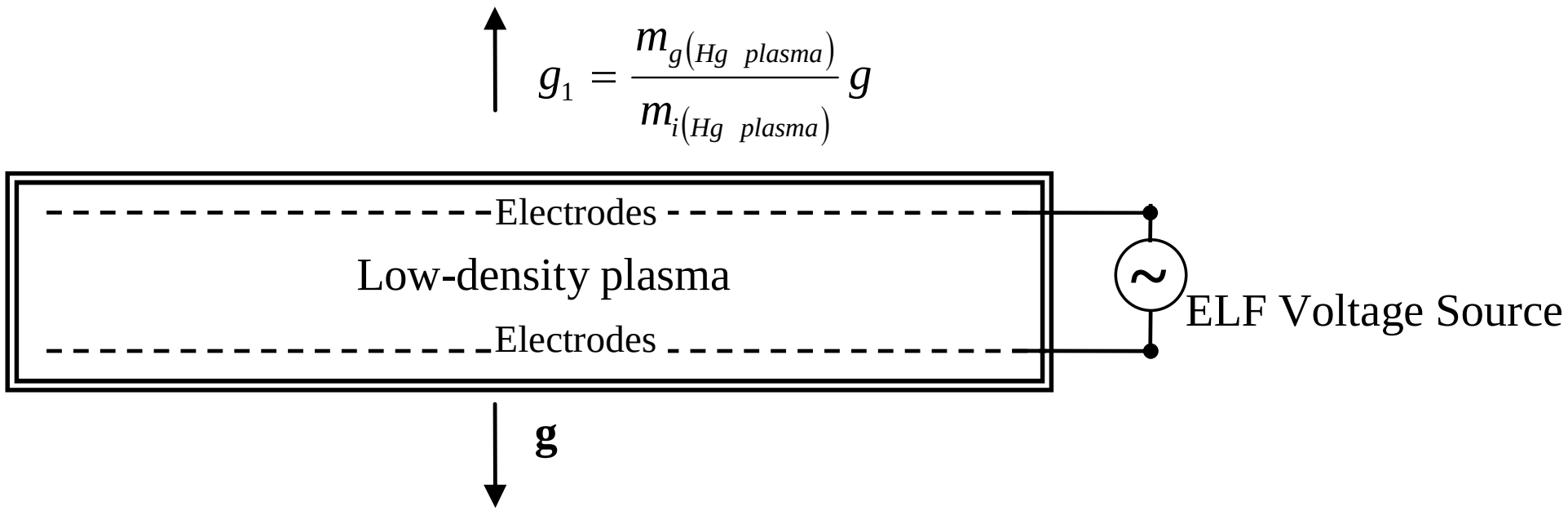

(a)

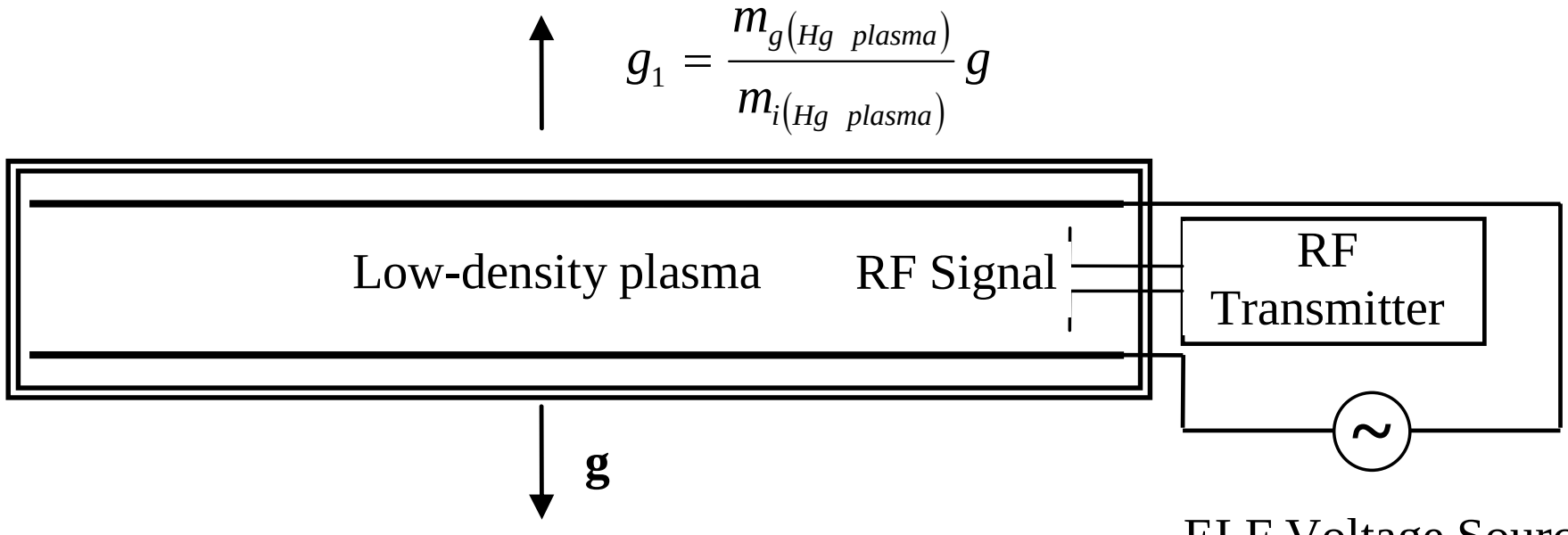

(b)

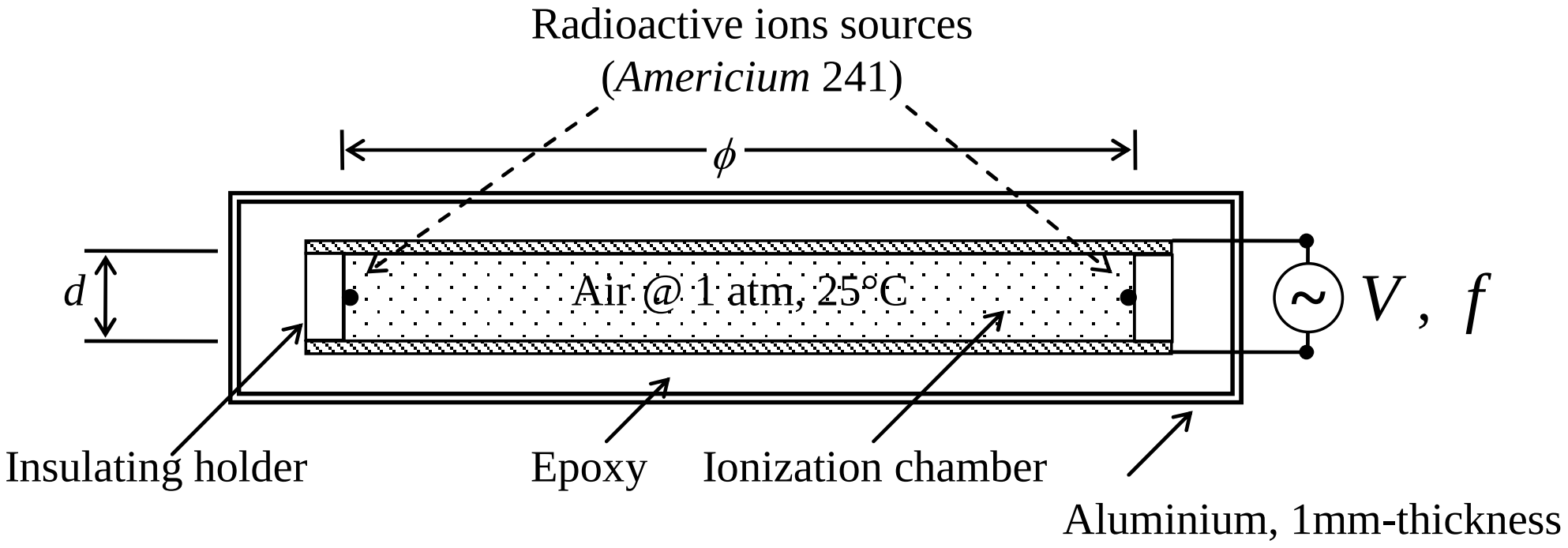

(c)

Fig. 3 – *Schematic diagram of Gravity Control Cells (GCCs).*
(a) GCC where the ELF electric field and the ionizing electric field can be the same. (b) GCC where the plasma is ionized by means of a RF signal. (c) GCC filled with *air* (at ambient temperature and 1 atm) strongly ionized by means of alpha particles emitted from radioactive ions sources (Am 241, *half-life* 432 years). Since the electrical conductivity of the ionized air depends on the amount of ions then it can be strongly increased by increasing the amount of Am 241 in the GCC. This GCC has 36 radioactive ions sources each one with 1/5000$^{th}$ of gram of Am 241, conveniently positioned around the ionization chamber, in order to obtain $\sigma_{air} \cong 10^3 S.m^{-1}$.

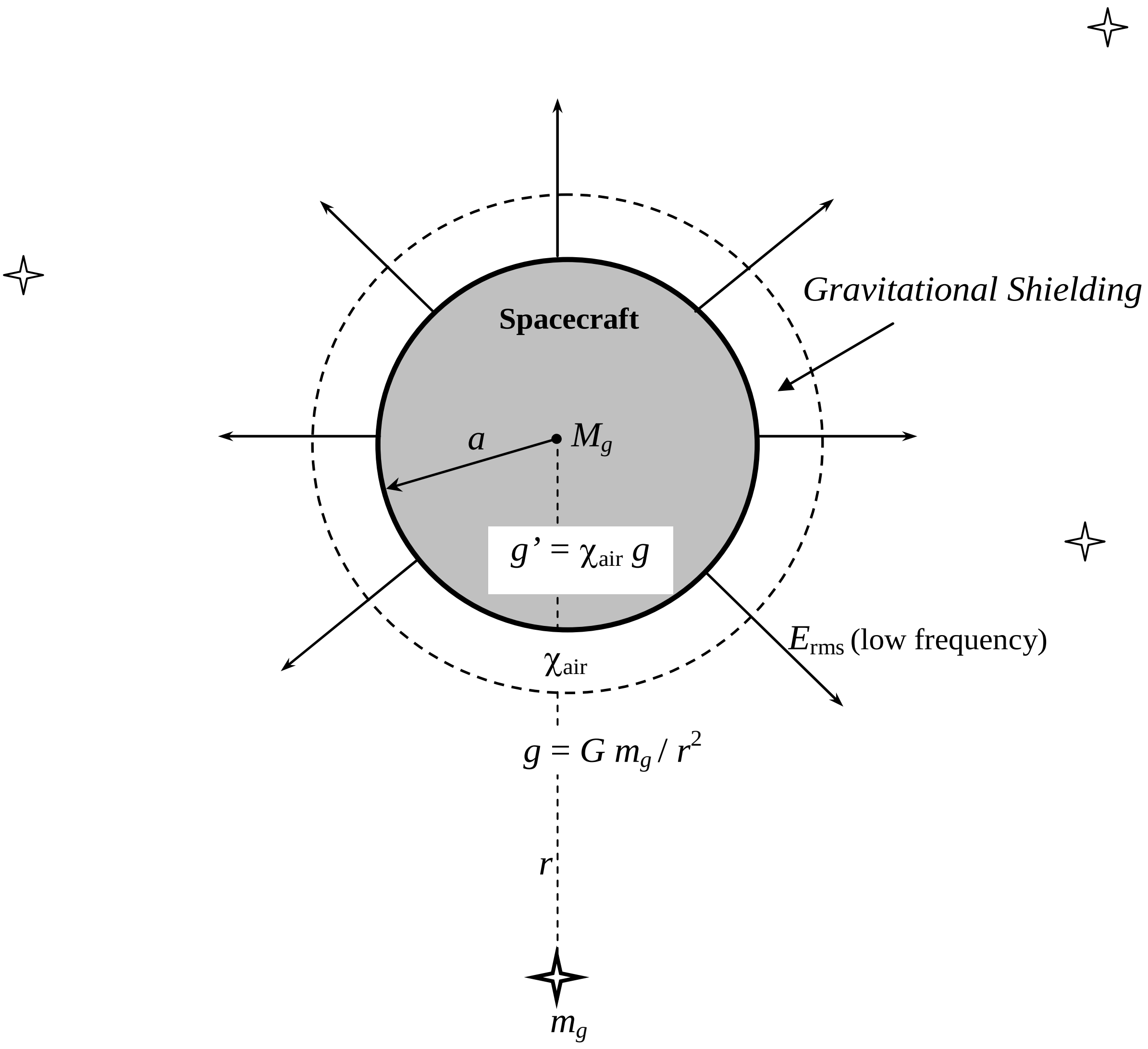

The *gravity accelerations* on the spacecraft (due to the rest of the Universe) can be controlled by means of the *gravitational shielding,* i.e.,

$$g'_i = \chi_{air}\, g_i \qquad i = 1, 2, 3 \ldots n$$

Thus,

$$F_{is} = F_{si} = M_g\, g'_i = M_g\, (\chi_{air}\, g_i)$$

Then the inertial forces acting on the spacecraft (s) can be strongly reduced. According to the *Mach's principle* this effect can reduce *the inertial properties of the spacecraft* and consequently, leads to a new concept of spacecraft and aerospace flight.

Fig. 4 – Gravitational Shielding surround a Spherical Spacecraft.

| V = $V_0$ (Volts) | t = T /4 (s) | ( min) | $E_{ELF\,(1)}$ (V/m) | $f_{ELF\,(1)}$ (mHz) | $g_1/g$ Exp. | $g_1/g$ Teo. | $E_{ELF\,(2)}$ (V/m) | $f_{ELF\,(2)}$ (mHz) | $g_2/g$ Exp. | $g_2/g$ Teo. |
|---|---|---|---|---|---|---|---|---|---|---|
| **1.0** V | 250 | **4.17** | 24.81 | 1 | - | **0.993** | 24.81 | 1 | - | **0.986** |
| | 312.5 | **5.21** | 24.81 | 0.8 | - | **0.986** | 24.81 | 0.8 | - | **0.972** |
| | 416.6 | **6.94** | 24.81 | 0.6 | - | **0.967** | 24.81 | 0.6 | - | **0.935** |
| | 625 | **10.42** | 24.81 | 0.4 | - | **0.890** | 24.81 | 0.4 | - | **0.792** |
| | 1250 | **20.83** | 24.81 | 0.2 | - | **0.240** | 24.81 | 0.2 | - | **0.058** |
| **1.5**V | 250 | **4.17** | 37.22 | 1 | - | **0.964** | 37.22 | 1 | - | **0.929** |
| | 312.5 | **5.21** | 37.22 | 0.8 | - | **0.930** | 37.22 | 0.8 | - | **0.865** |
| | 416.6 | **6.94** | 37.22 | 0.6 | - | **0.837** | 37.22 | 0.6 | - | **0.700** |
| | 625 | **10.42** | 37.22 | 0.4 | - | **0,492** | 37.22 | 0.4 | - | **0.242** |
| | 1250 | **20.83** | 37.22 | 0.2 | - | **-1,724** | 37.22 | 0.2 | - | **2.972** |

Table 1 – Theoretical Results.

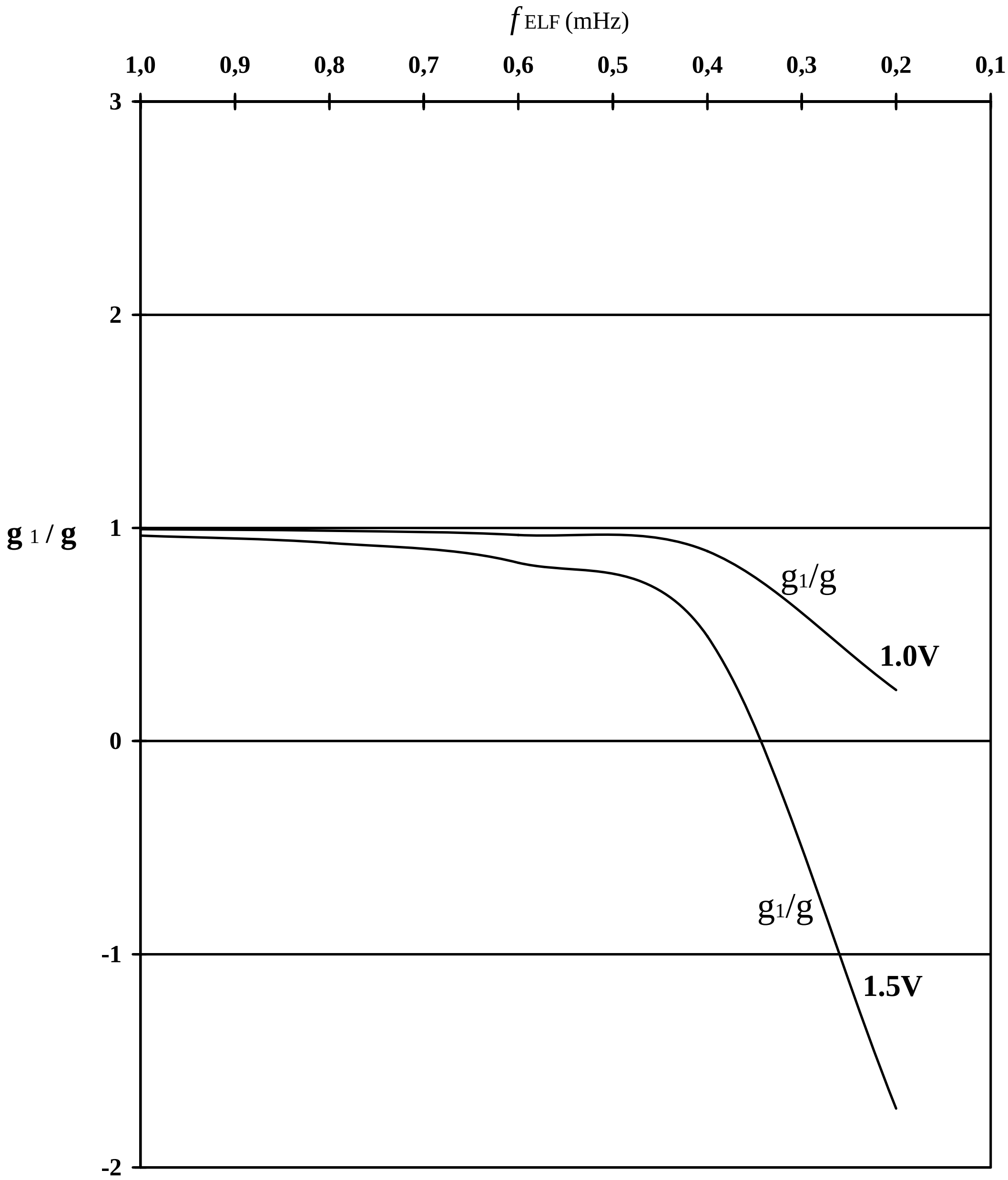

Fig. 5- Distribution of the correlation **$g_1$/ g** as a function of $f_{ELF}$

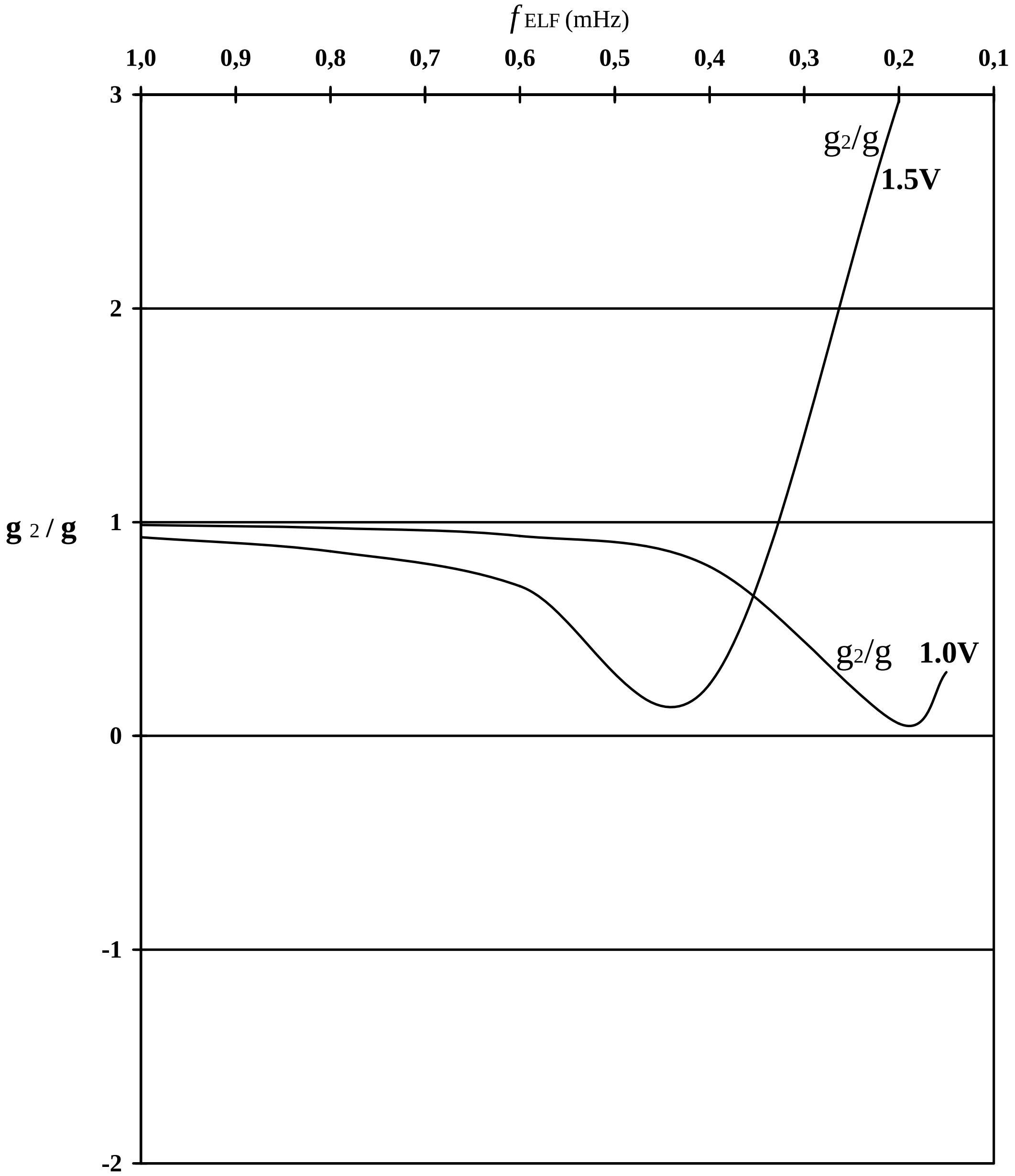

Fig. 6- Distribution of the correlation **g**2 / **g** as a function of $f_{ELF}$

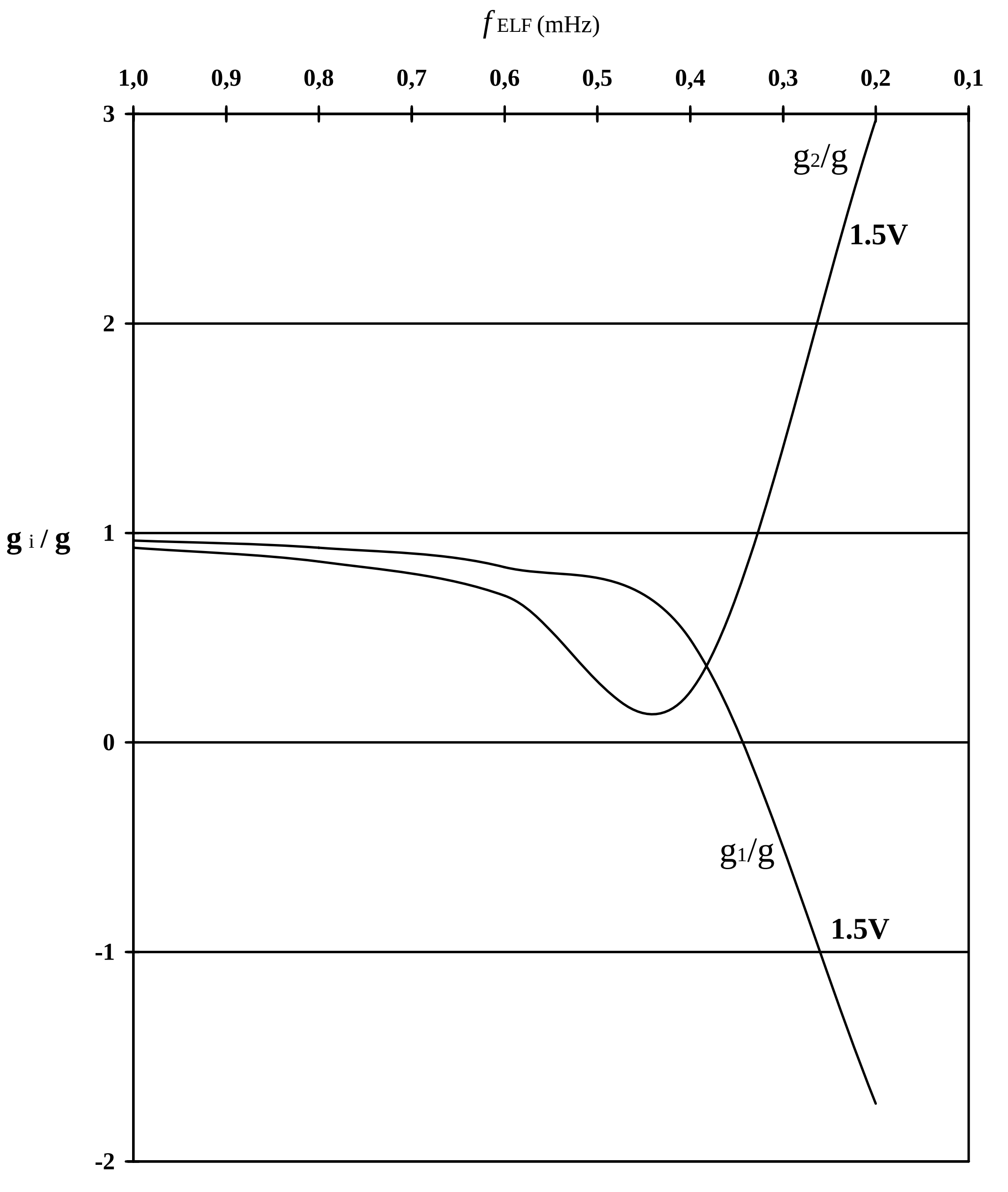

Fig. 7- Distribution of the correlations $\mathbf{g_i / g}$ as a function of $f_{ELF}$

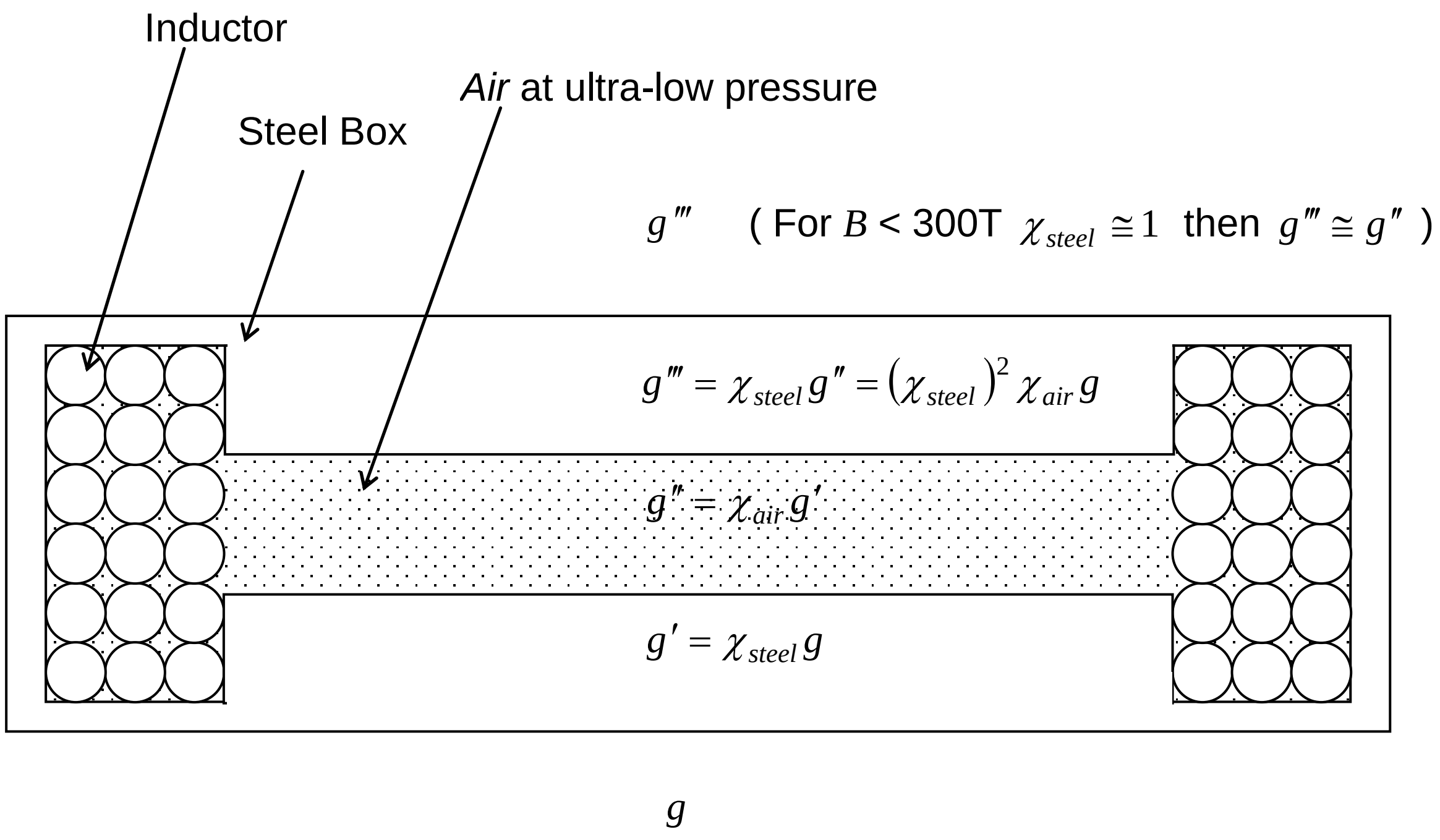

(a)

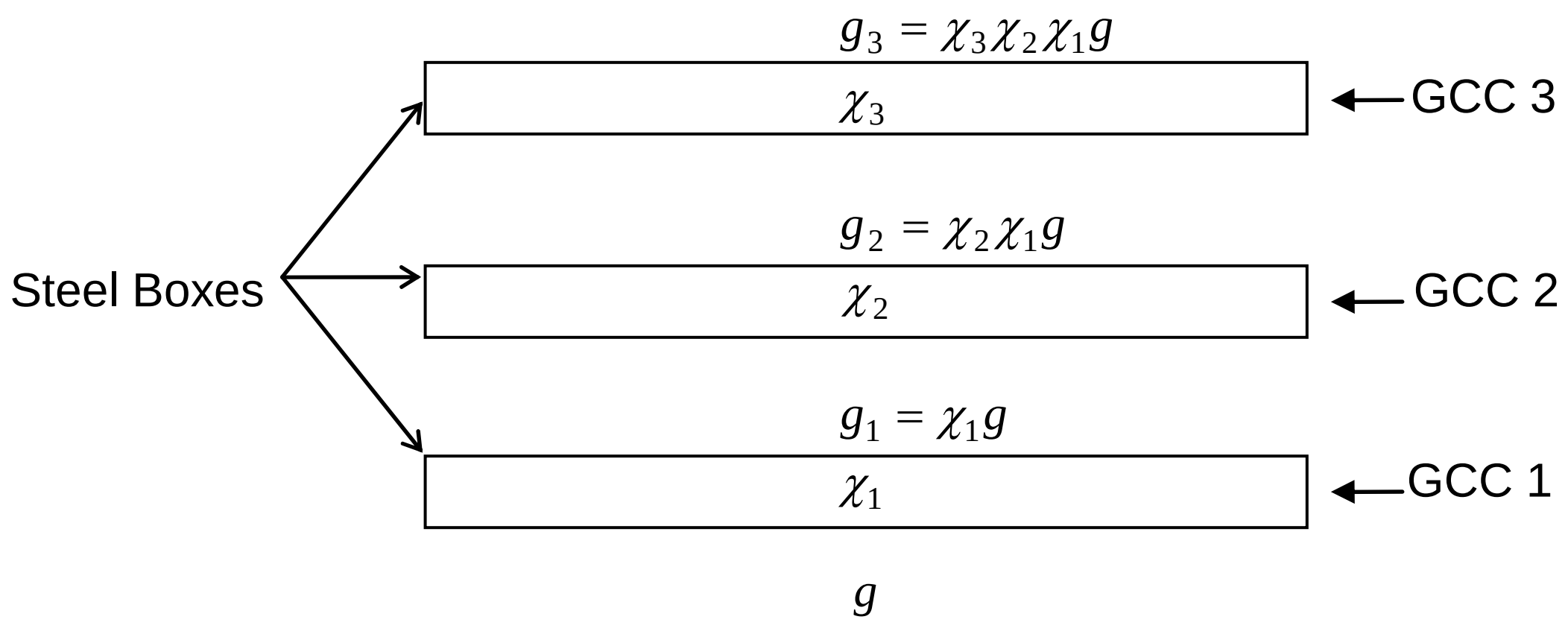

(b)

Fig. 8 – (a) Gravity Control Cell (GCC) filled with *air* at ultra-low pressure.
(b) Gravity Control Battery (Note that if $\chi_1 = \chi_2^{-1} = -1$ then $g'' = g$ )

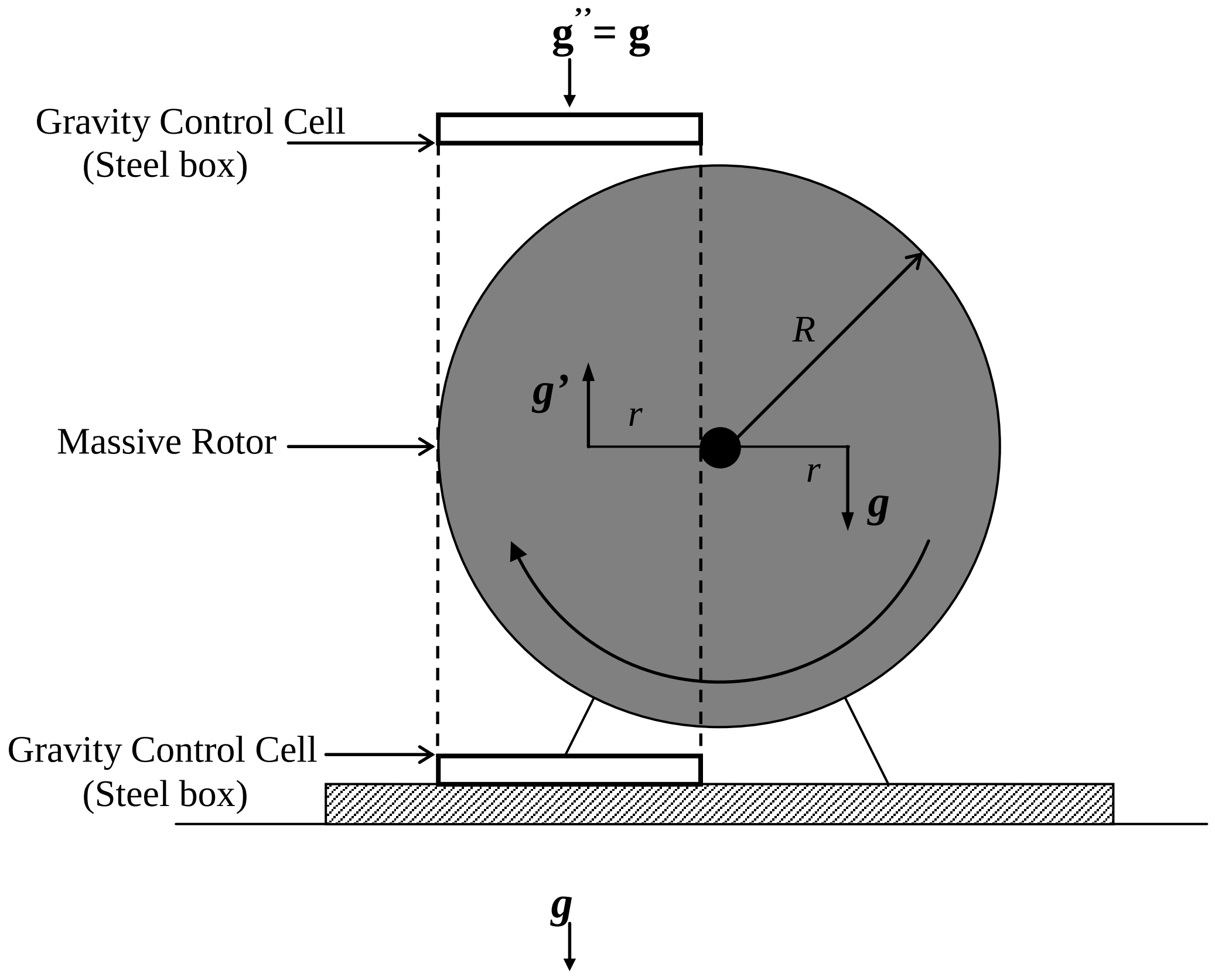

Note that $g' = (\chi_{steel})^2 \chi_{air} g$ and $g'' = (\chi_{steel})^4 (\chi_{air})^2 g$ therefore for $\chi_{steel} \cong 1$ and $\chi_{air(1)} = \chi_{air(2)}^{-1} = -n$ we get $g' \cong -ng$ and $g'' = g$

Fig. 9 – The Gravitational Motor

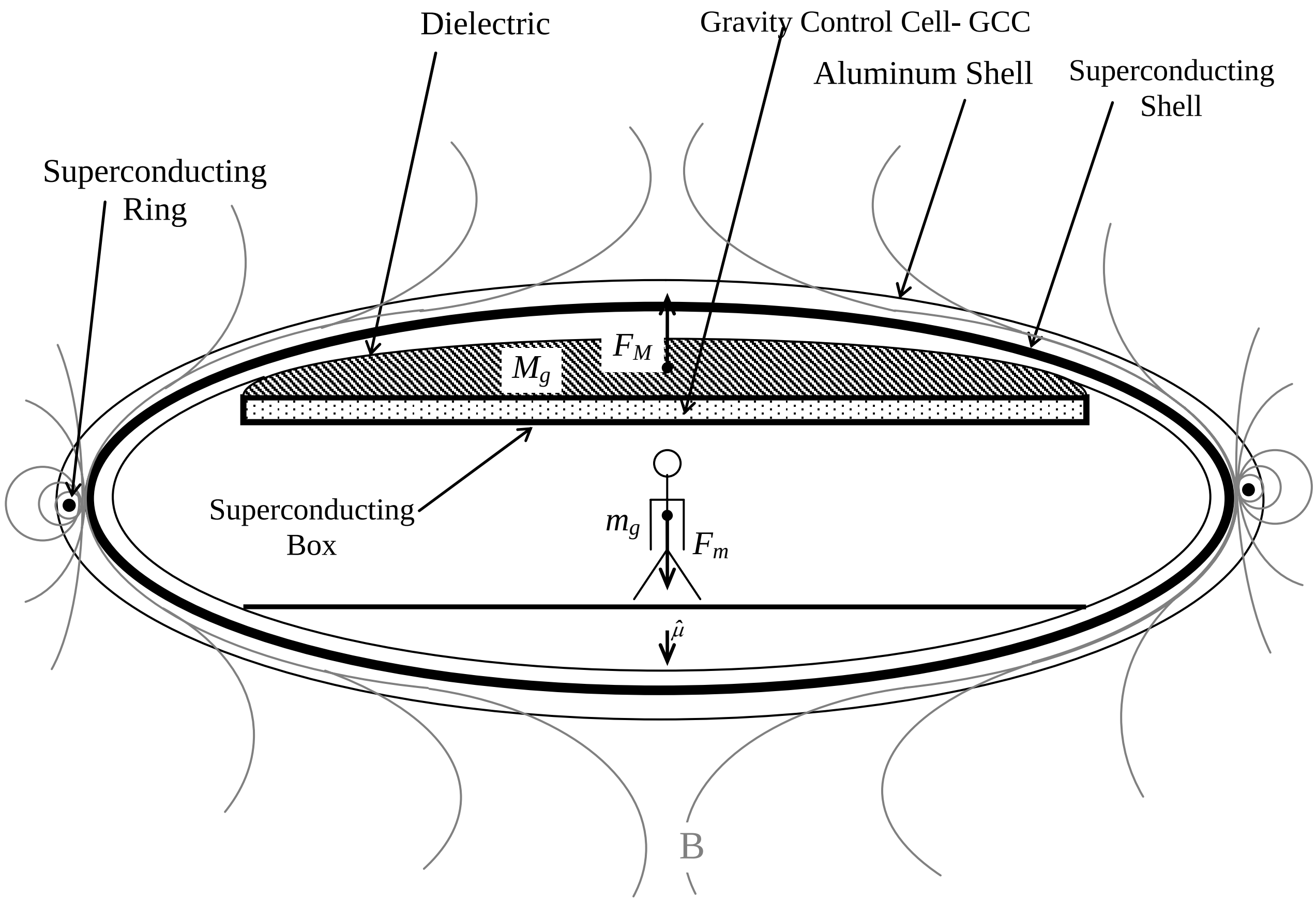

Fig. 10 – The Gravitational Spacecraft – Due to the *Meissner effect*, the magnetic field $B$ is expelled from the *superconducting shell*. Similarly, the magnetic field $B_{GCC}$, of the GCC stay confined inside the *superconducting box*.

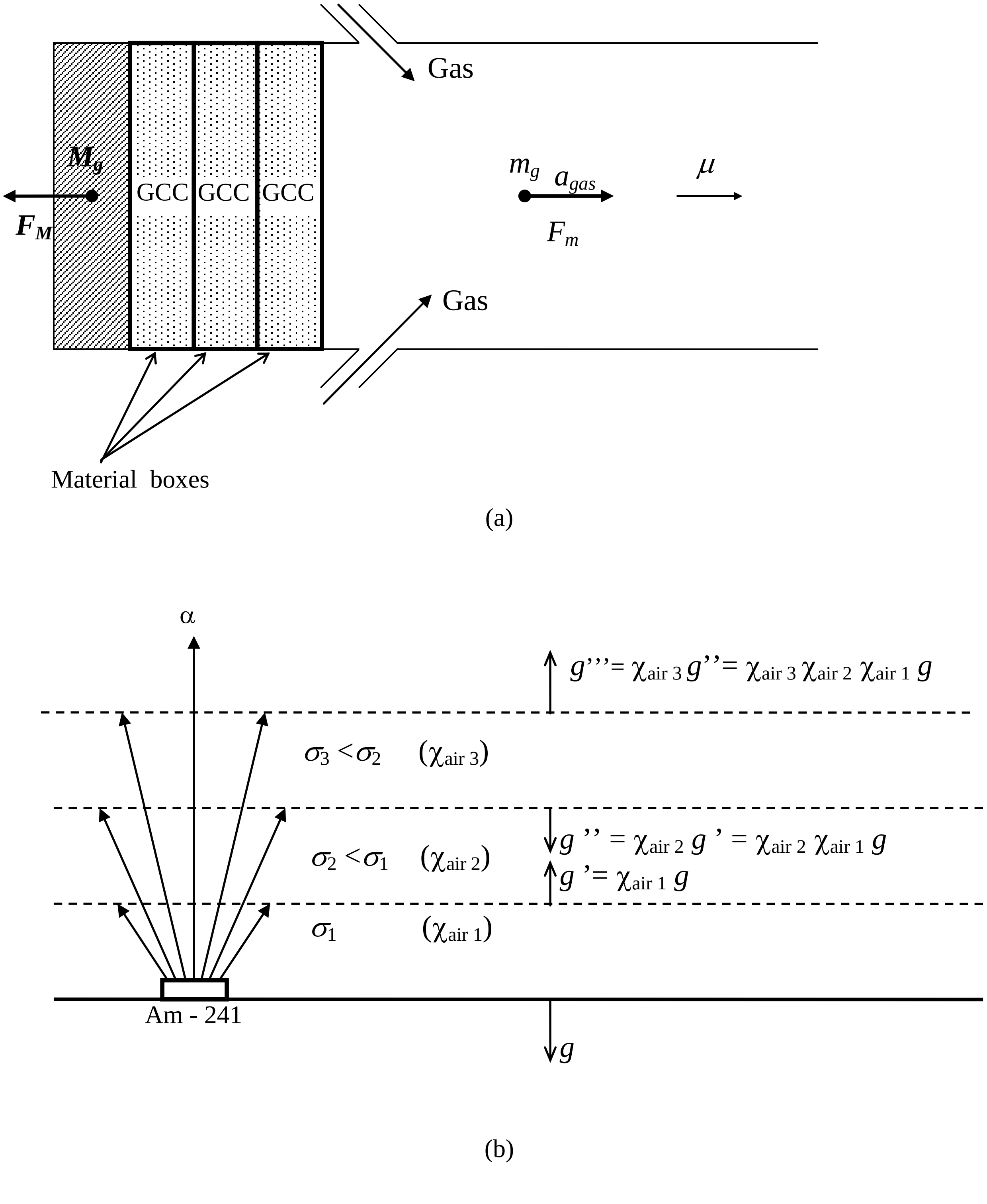

Fig. 11 – The Gravitational Thruster .
(a) Using material boxes. (b) Without material boxes

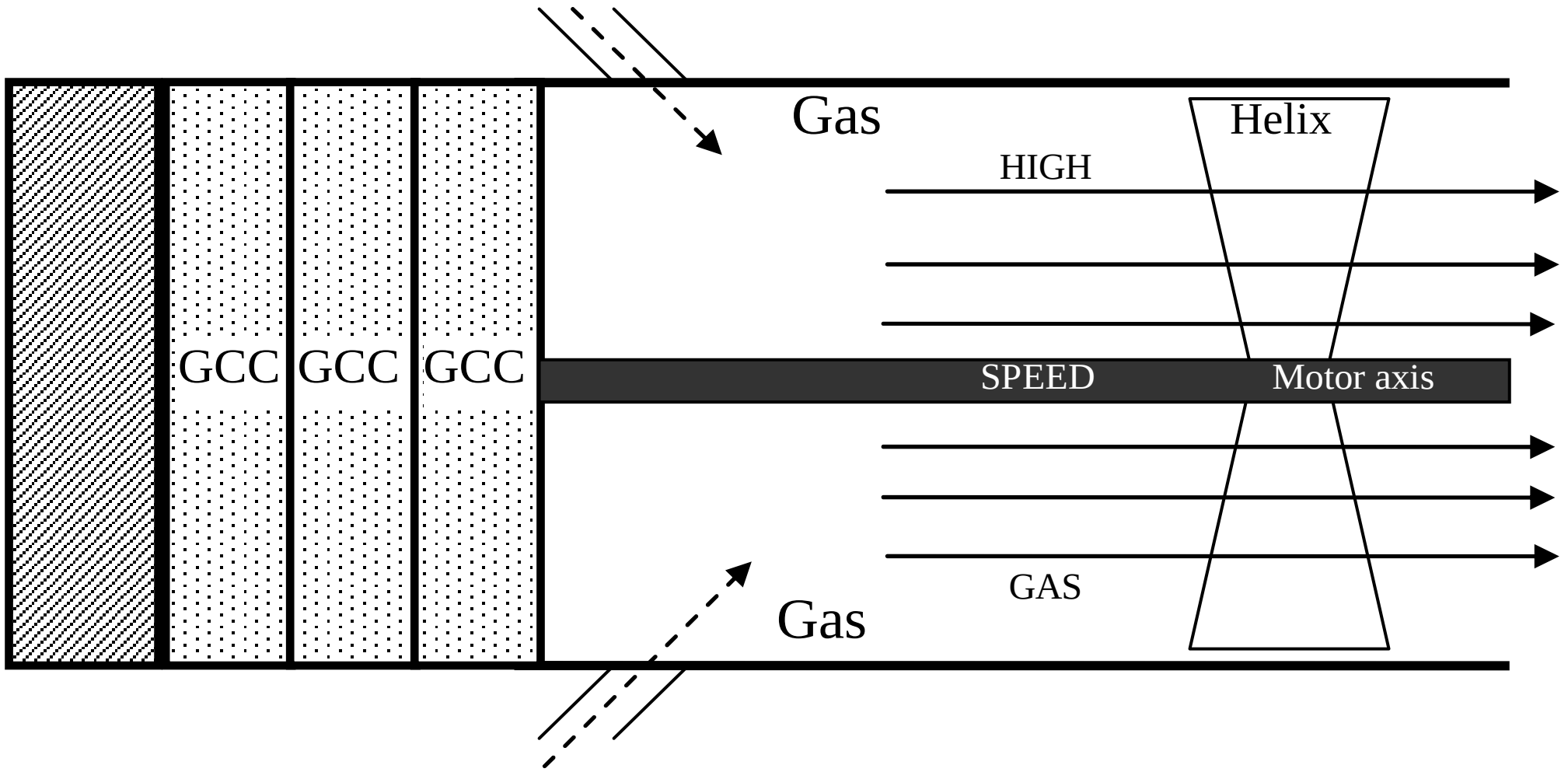

Fig. 12 - The Gravitational Turbo Motor – The gravitationally accelerated gas, by means of the GCCs, propels the helix which movies the motor axis.

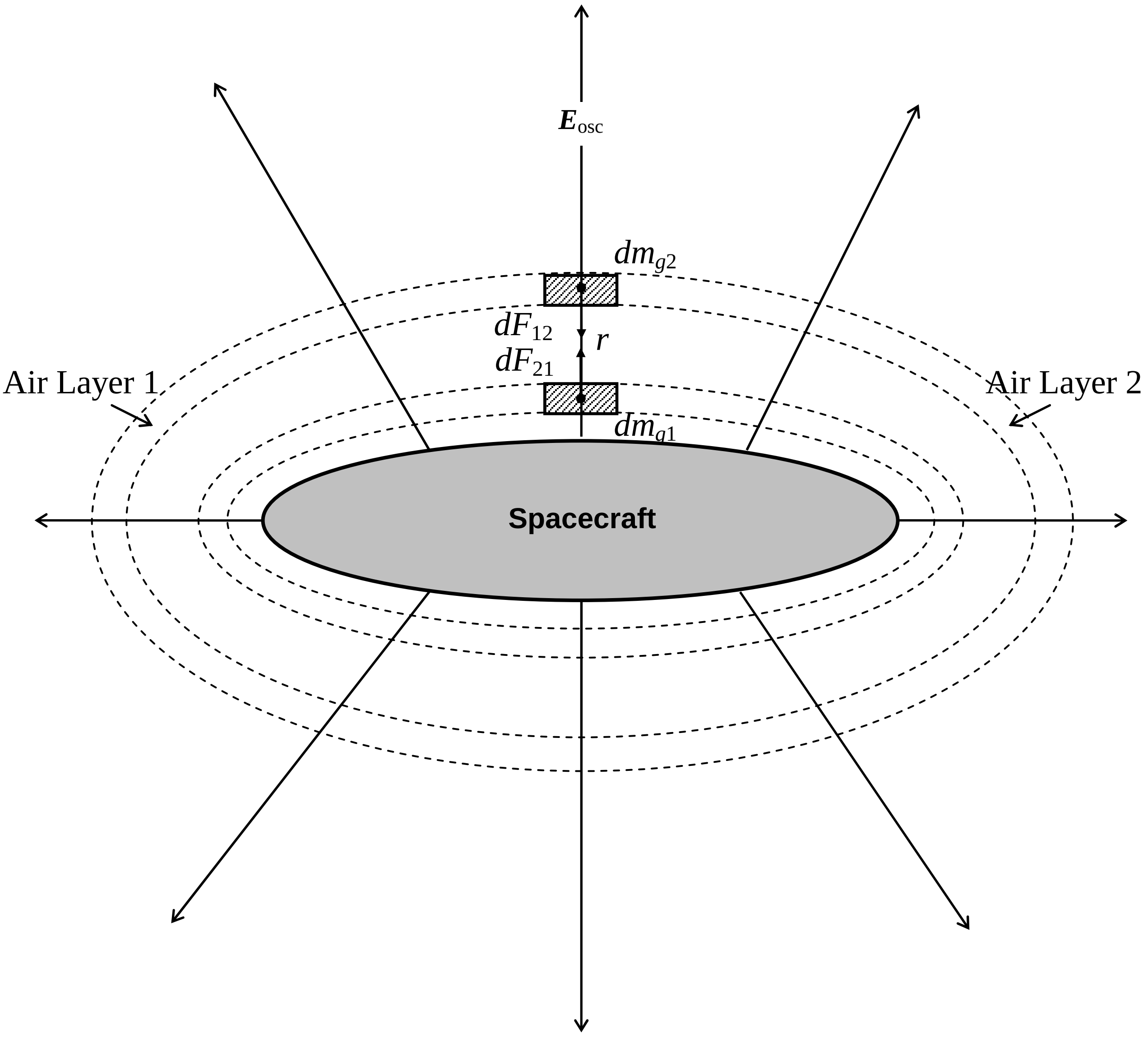

Fig. 13 – Gravitational forces between two layers of the “*air shell*”. The electric field $\boldsymbol{E}_{osc}$ provides the *ionization* of the *air*.

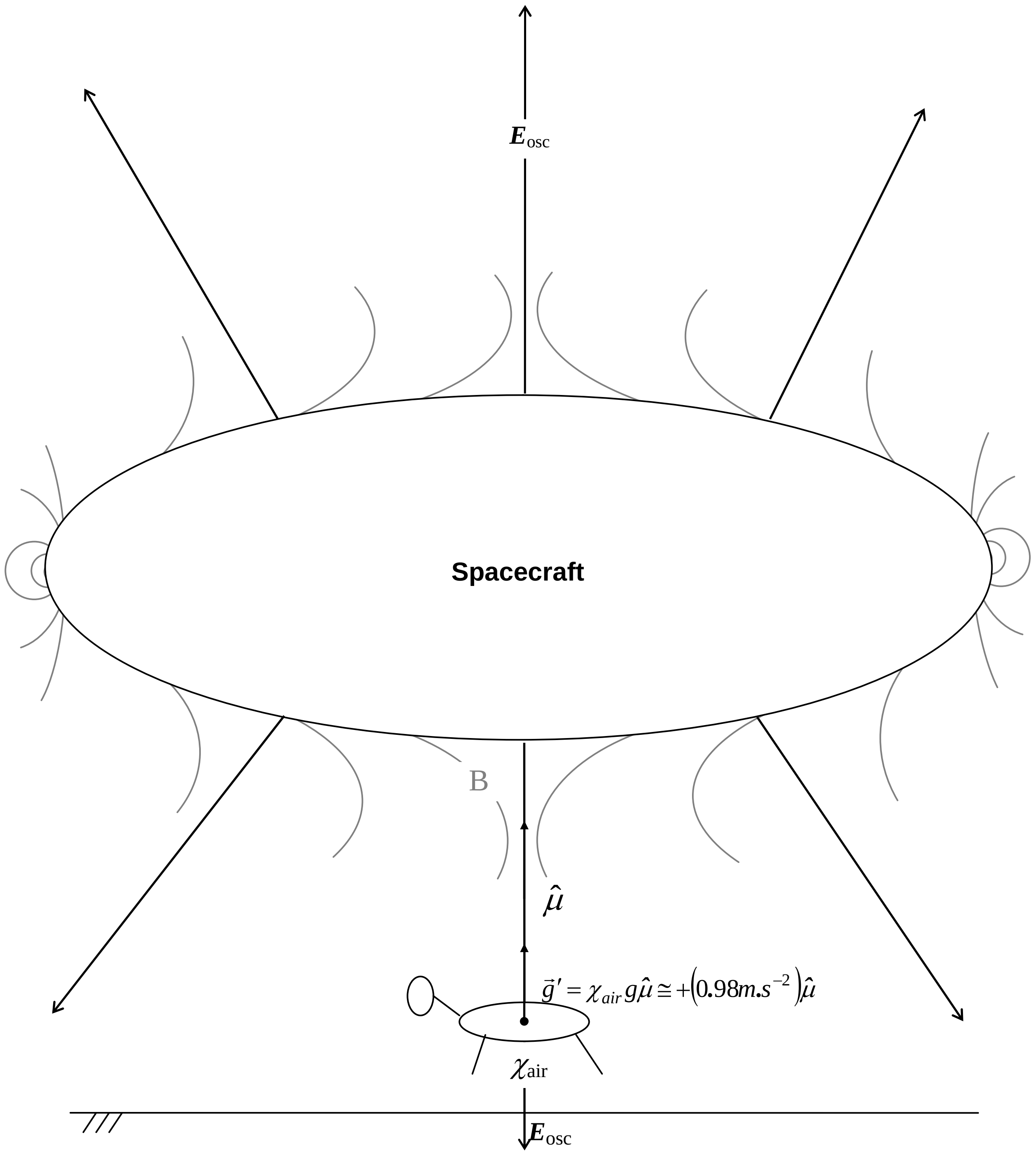

Fig. 14 – The Gravitational Lifter

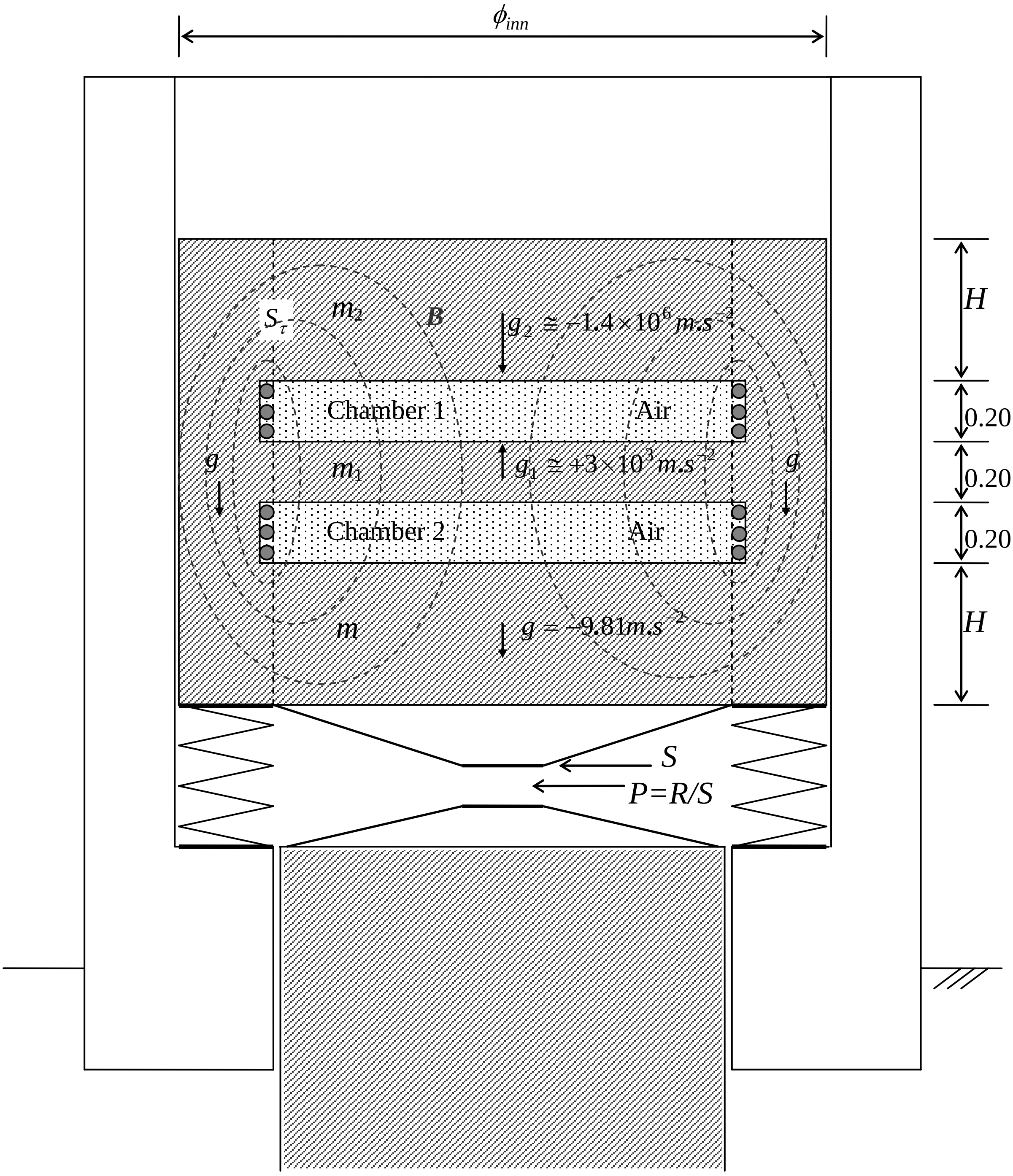

Fig. 15 – *Gravitational Press*

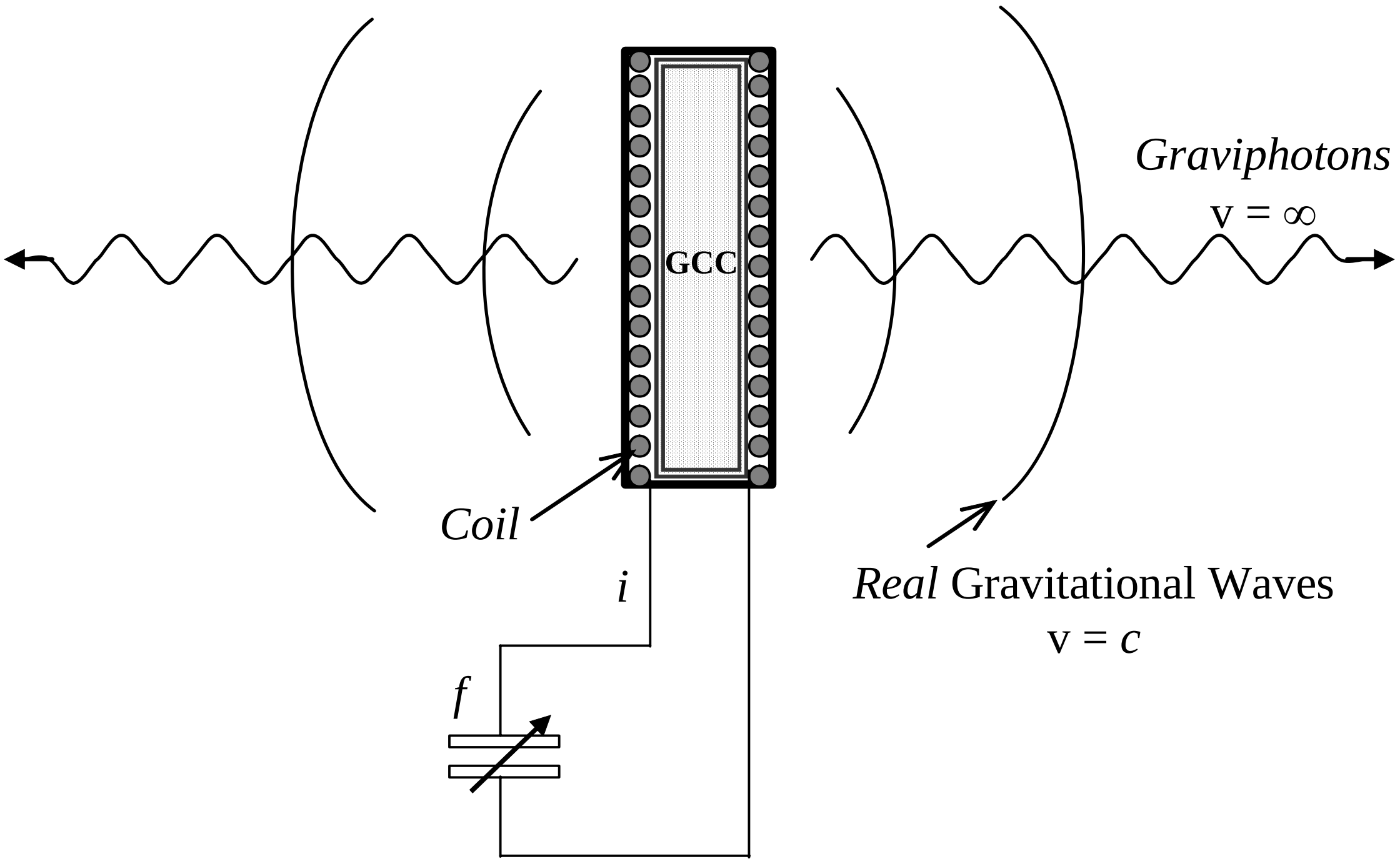

(a) GCC Antenna

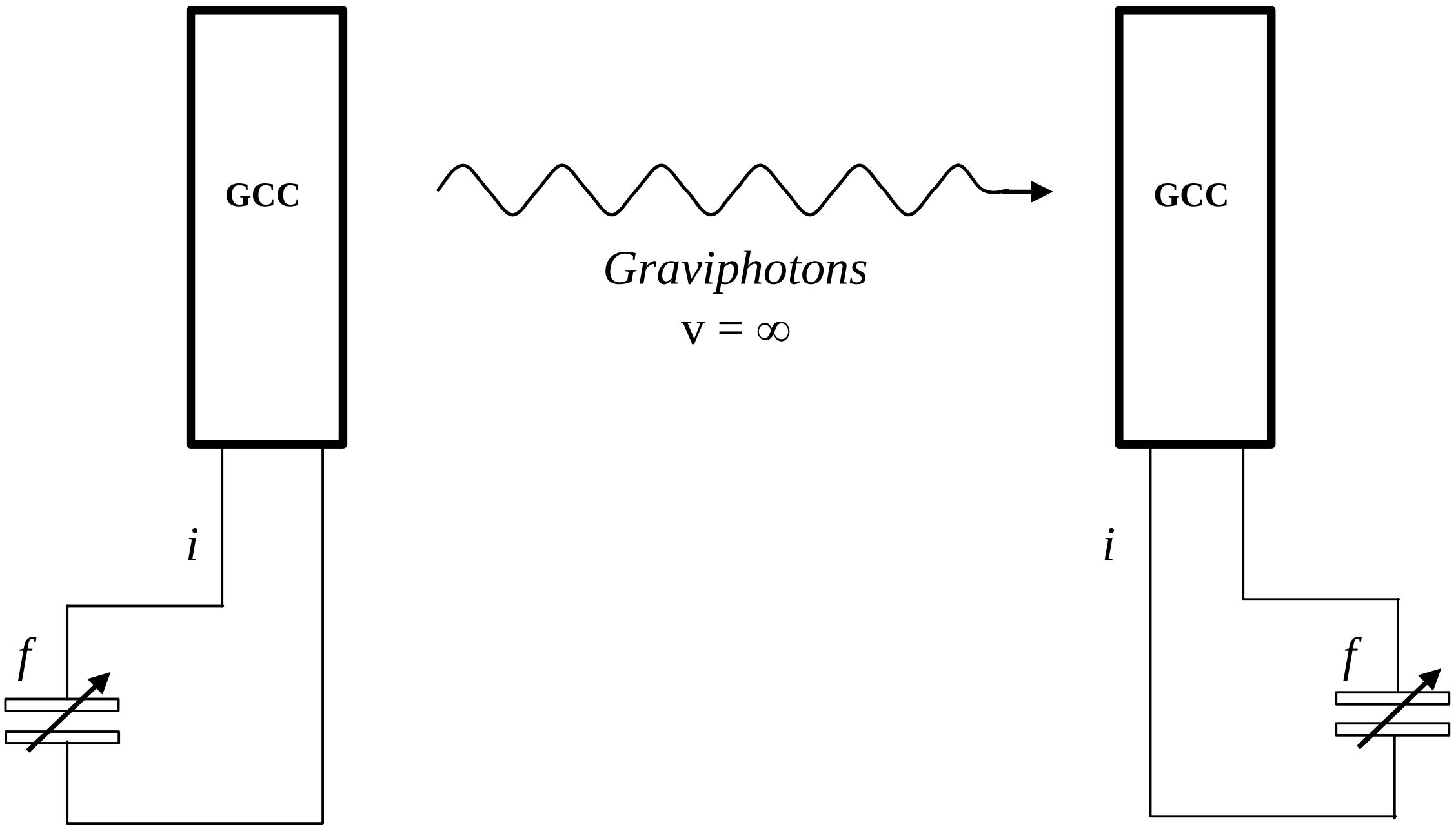

(b)

Fig. 16 - Transmitter and Receiver of *Virtual* Gravitational Radiation.

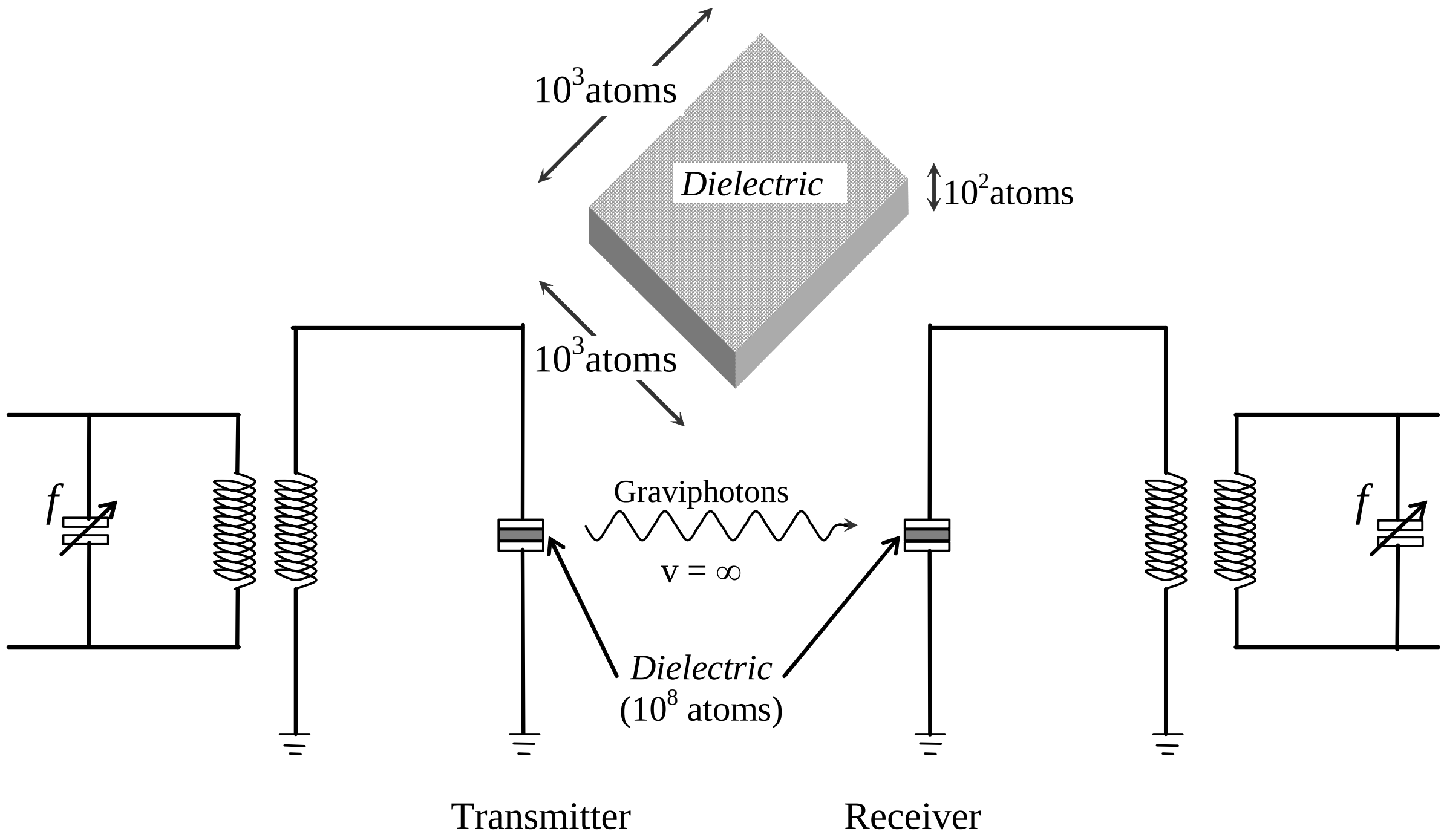

(a)

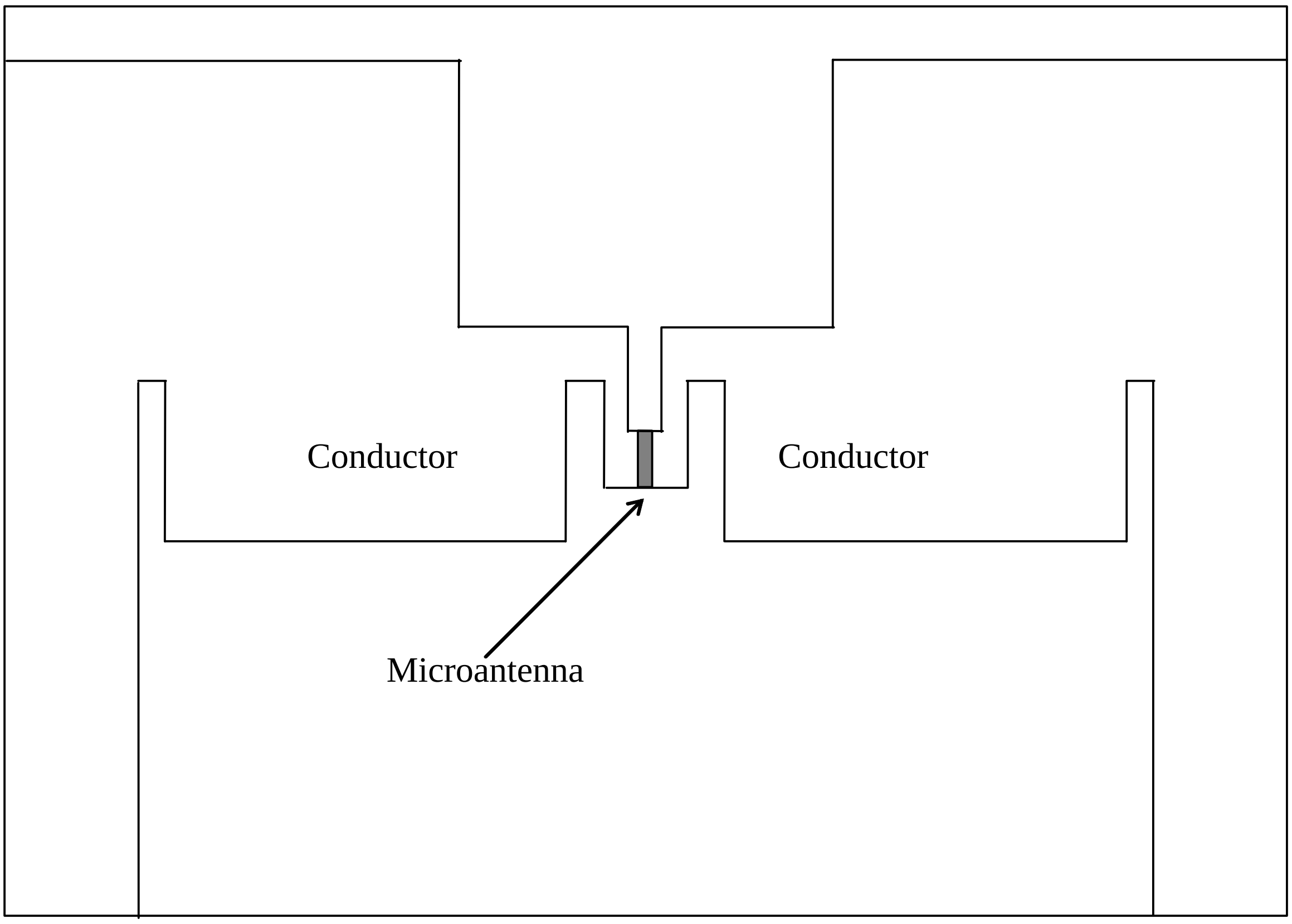

(b)

Fig. 17 – Quantum Gravitational Microantenna

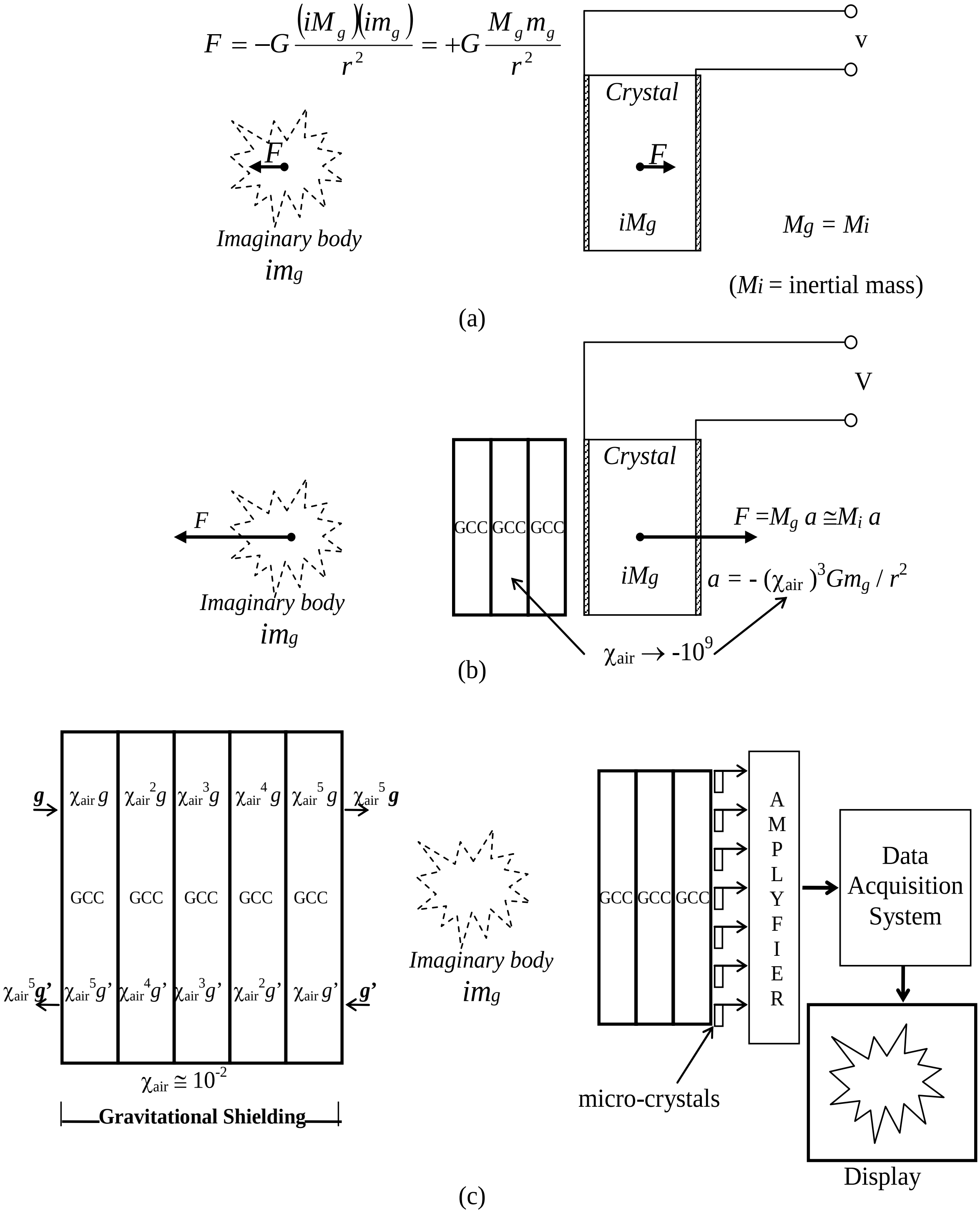

Fig.18 – Method and device using GCCs for obtaining *images* of *imaginary bodies*.

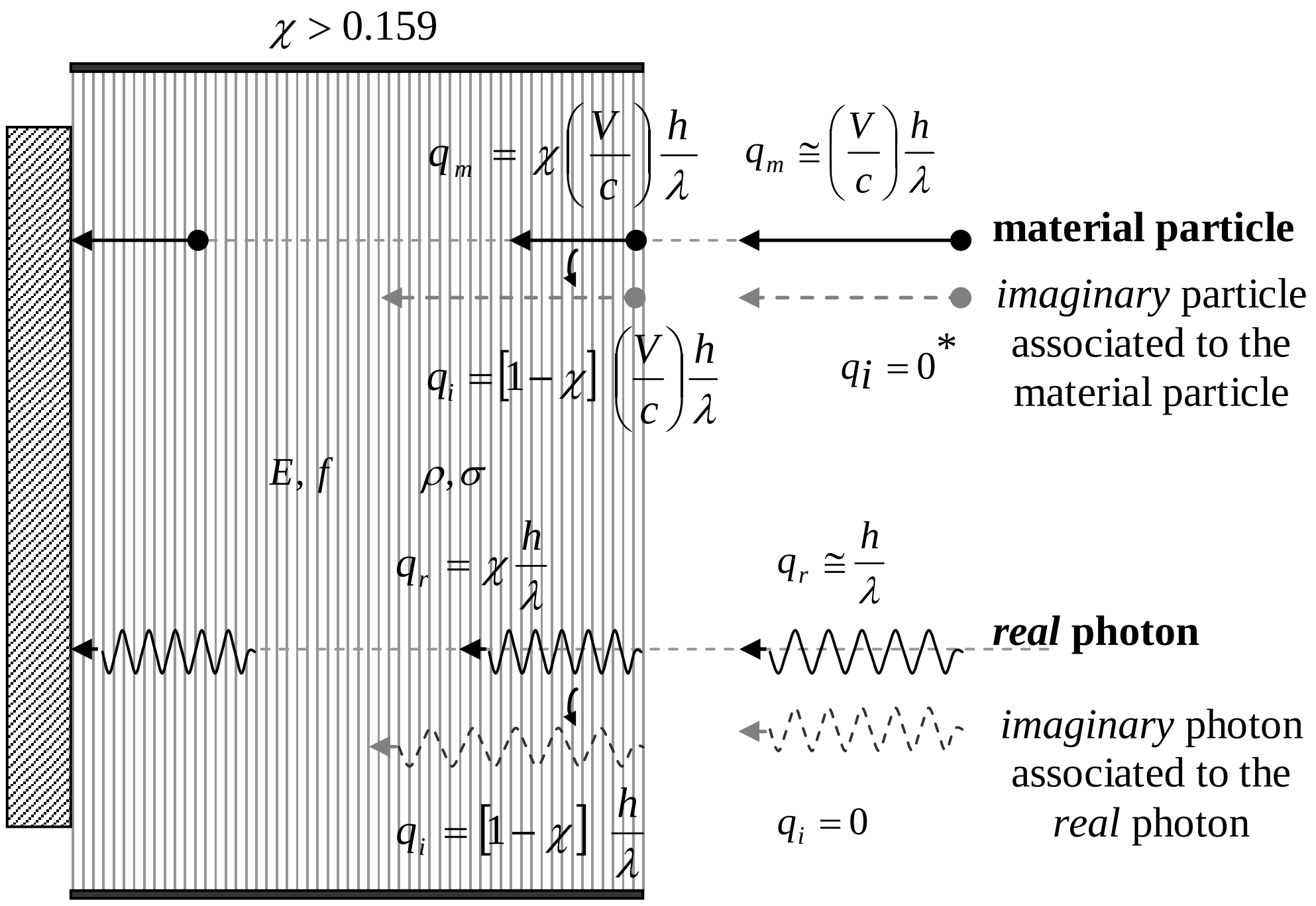

(a)

* There are a type of neutrino, called "ghost" neutrino, predicted by General Relativity, with *zero mass* and *zero momentum*. In spite its *momentum be zero*, it is known that there are wave functions that describe these neutrinos and that prove that really they exist.

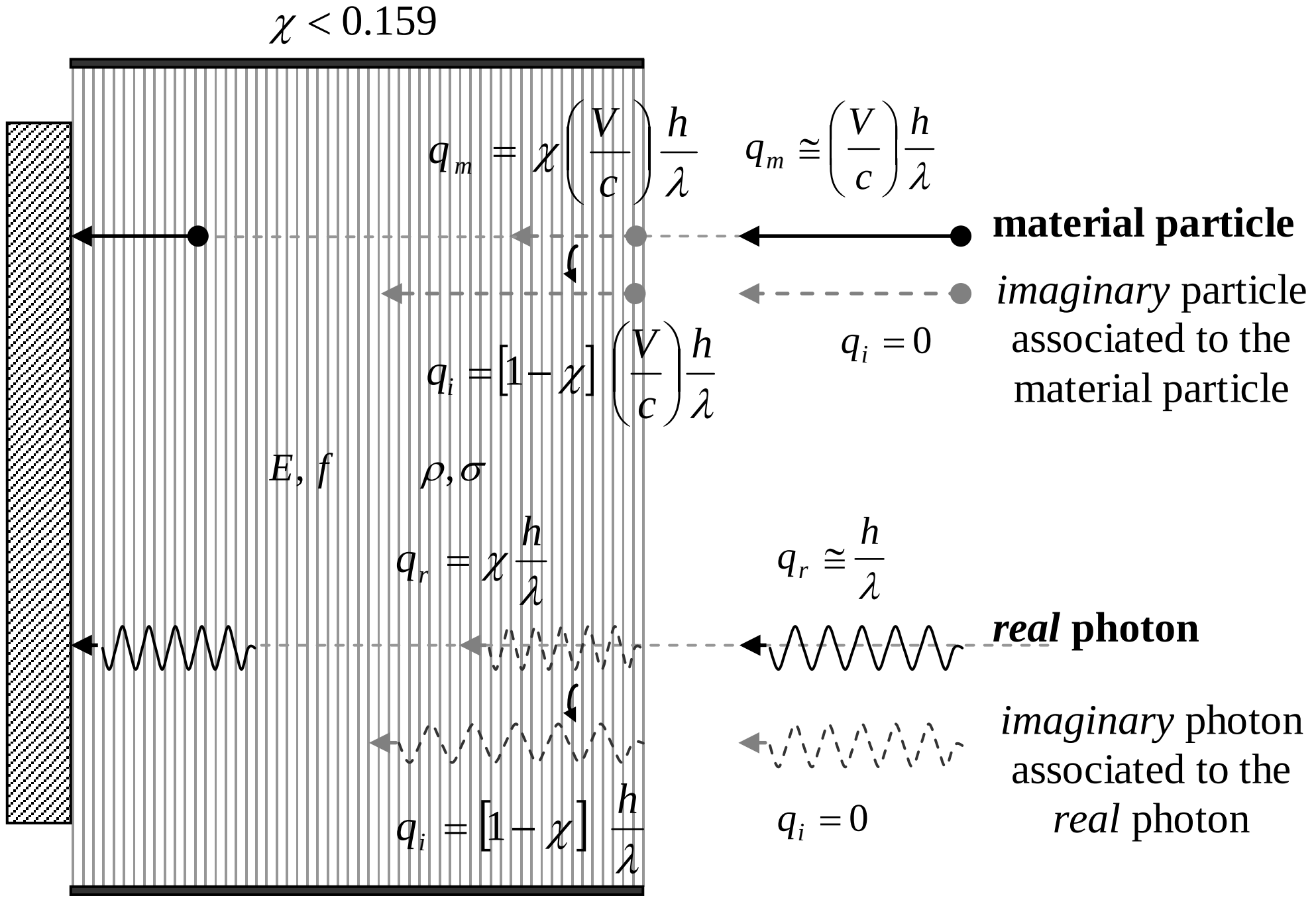

(b)

Fig. 19 – *The phenomenon of reduction of the momentum*. (a) Shows the reduction of *momentum* for $\chi > 0.159$. (b) Shows the effect when $\chi < 0.159$. Note that in both cases, the *material* particles collide with the cowl with the *momentum* $q_m = \chi(V/c)(h/\lambda)$, and the photons with $q_r = \chi\frac{h}{\lambda}$ . Therefore, that by making $\chi \cong 0$, it is possible to block high-energy particles and ultra-intense fluxes of radiation.

## APPENDIX A: THE SIMPLEST METHOD TO CONTROL THE GRAVITY

In this Appendix we show the simplest method to control the gravity.

Consider a body with mass density $\rho$ and the following electric characteristics: $\mu_r, \varepsilon_r, \sigma$ (relative permeability, relative permittivity and electric conductivity, respectively). Through this body, passes an electric current $I$, which is the sum of a sinusoidal current $i_{osc} = i_0 \sin \omega t$ and the DC current $I_{DC}$, i.e., $I = I_{DC} + i_0 \sin \omega t$; $\omega = 2\pi f$. If $i_0 << I_{DC}$ then $I \cong I_{DC}$. Thus, the current $I$ varies with the frequency $f$, but the variation of its intensity is quite small in comparison with $I_{DC}$, i.e., $I$ will be practically constant (Fig. 1A). This is of fundamental importance for maintaining the value of the gravitational mass of the body, $m_g$, sufficiently stable during all the time.

The *gravitational mass* of the *body* is given by [1]

$$m_g = \left\{1 - 2\left[\sqrt{1 + \left(\frac{n_r U}{m_{i0} c^2}\right)^2} - 1\right]\right\} m_{i0} \qquad (A1)$$

where $U$, is the electromagnetic energy absorbed by the body and $n_r$ is the index of refraction of the body.

Equation (A1) can also be rewritten in the following form

$$\frac{m_g}{m_{i0}} = \left\{1 - 2\left[\sqrt{1 + \left(\frac{n_r W}{\rho c^2}\right)^2} - 1\right]\right\} \qquad (A2)$$

where, $W = U/V$ is the *density of electromagnetic energy* and $\rho = m_{i0}/V$ is the density of inertial mass.

The *instantaneous values* of the density of electromagnetic energy in an *electromagnetic* field can be deduced from Maxwell's equations and has the following expression

$$W = \tfrac{1}{2}\varepsilon E^2 + \tfrac{1}{2}\mu H^2 \qquad (A3)$$

where $E = E_m \sin \omega t$ and $H = H \sin \omega t$ are the *instantaneous values* of the electric field and the magnetic field respectively.

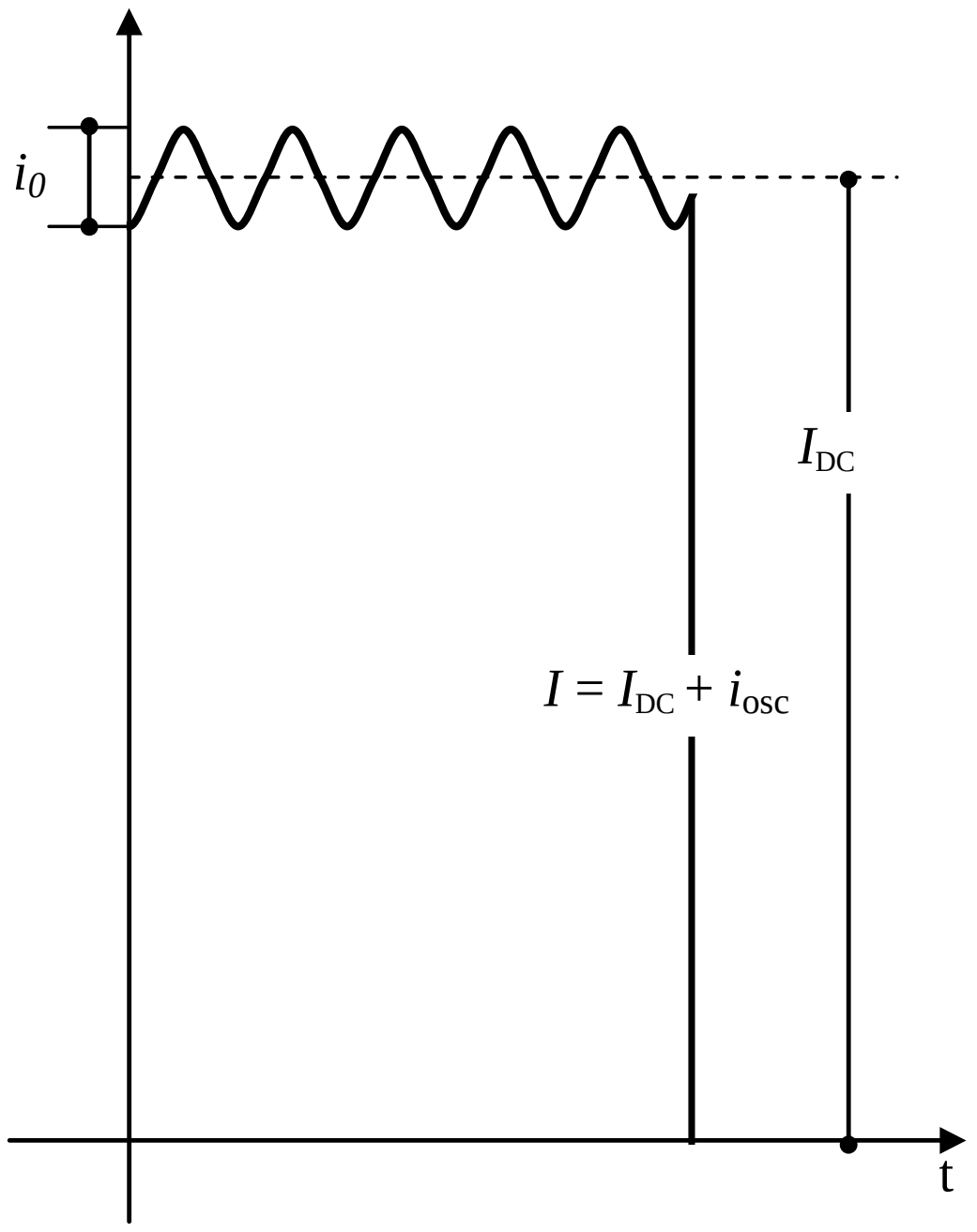

Fig. A1 - The electric current $I$ varies with frequency $f$. But the variation of $I$ is quite small in comparison with $I_{DC}$ due to $i_o << I_{DC}$. In this way, we can consider $I \cong I_{DC}$.

It is known that $B = \mu H$, $E/B = \omega/k_r$ [11] and

$$v = \frac{dz}{dt} = \frac{\omega}{\kappa_r} = \frac{c}{\sqrt{\frac{\varepsilon_r \mu_r}{2}\left(\sqrt{1 + (\sigma/\omega\varepsilon)^2} + 1\right)}} \qquad (A4)$$

where $k_r$ is the real part of the *propagation vector* $\vec{k}$ (also called *phase constant*); $k = |\vec{k}| = k_r + ik_i$; $\varepsilon$, $\mu$ and $\sigma$, are the electromagnetic characteristics of the medium in which the incident (or emitted) radiation is propagating($\varepsilon = \varepsilon_r \varepsilon_0$; $\varepsilon_0 = 8.854 \times 10^{-12} F/m$; $\mu = \mu_r \mu_0$ where $\mu_0 = 4\pi \times 10^{-7} H/m$). It is known that for *free-space* $\sigma = 0$ and $\varepsilon_r = \mu_r = 1$. Then Eq. (A4) gives

$$v = c$$

From (A4), we see that the *index of refraction* $n_r = c/v$ is given by

$$n_r = \frac{c}{v} = \sqrt{\frac{\varepsilon_r \mu_r}{2}\left(\sqrt{1 + (\sigma/\omega\varepsilon)^2} + 1\right)} \qquad (A5)$$

Equation (A4) shows that $\omega/\kappa_r = v$. Thus, $E/B = \omega/k_r = v$, i.e.,

$$E = vB = v\mu H \qquad (A6)$$

Then, Eq. (A3) can be rewritten in the following form:

$$W = \tfrac{1}{2}\left(\varepsilon v^2 \mu\right)\mu H^2 + \tfrac{1}{2}\mu H^2 \qquad (A7)$$

For $\sigma << \omega\varepsilon$, Eq. (A4) reduces to

$$v = \frac{c}{\sqrt{\varepsilon_r \mu_r}}$$

Then, Eq. (A7) gives

$$W = \tfrac{1}{2}\left[\varepsilon\left(\frac{c^2}{\varepsilon_r \mu_r}\right)\mu\right]\mu H^2 + \tfrac{1}{2}\mu H^2 = \mu H^2$$

This equation can be rewritten in the following forms:

$$W = \frac{B^2}{\mu} \qquad (A8)$$

or

$$W = \varepsilon\ E^2 \qquad (A9)$$

For $\sigma >> \omega\varepsilon$, Eq. (A4) gives

$$v = \sqrt{\frac{2\omega}{\mu\sigma}} \qquad (A10)$$

Then, from Eq. (A7) we get

$$W = \tfrac{1}{2}\left[\varepsilon\left(\frac{2\omega}{\mu\sigma}\right)\mu\right]\mu H^2 + \tfrac{1}{2}\mu H^2 = \left(\frac{\omega\varepsilon}{\sigma}\right)\mu H^2 + \tfrac{1}{2}\mu H^2 \cong$$

$$\cong \tfrac{1}{2}\mu H^2 \qquad (A11)$$

Since $E = vB = v\mu H$, we can rewrite (A11) in the following forms:

$$W \cong \frac{B^2}{2\mu} \qquad (A12)$$

or

$$W \cong \left(\frac{\sigma}{4\omega}\right)E^2 \qquad (A13)$$

By comparing equations (A8) (A9) (A12) and (A13), we can see that Eq. (A13) shows that the best way to obtain a strong value of $W$ *in practice* is by applying an *Extra Low-Frequency* (ELF) *electric field* $(w = 2\pi f << 1Hz)$ through a *medium with high electrical conductivity*.

Substitution of Eq. (A13) into Eq. (A2), gives

$$m_g = \left\{1 - 2\left[\sqrt{1 + \frac{\mu}{4c^2}\left(\frac{\sigma}{4\pi f}\right)^3 \frac{E^4}{\rho^2}} - 1\right]\right\} m_{i0} =$$

$$= \left\{1 - 2\left[\sqrt{1 + \left(\frac{\mu_0}{256\pi^3 c^2}\right)\left(\frac{\mu_r \sigma^3}{\rho^2 f^3}\right)E^4} - 1\right]\right\} m_{i0} =$$

$$= \left\{1 - 2\left[\sqrt{1 + 1.758 \times 10^{-27}\left(\frac{\mu_r \sigma^3}{\rho^2 f^3}\right)E^4} - 1\right]\right\} m_{i0}$$

$$(A14)$$

Note that $E = E_m \sin \omega t$. The average value for $E^2$ is equal to $\tfrac{1}{2}E_m^2$ because $E$ varies sinusoidaly ($E_m$ is the maximum value for $E$). On the other hand, $E_{rms} = E_m/\sqrt{2}$. Consequently, we can change $E^4$ by $E_{rms}^4$, and the equation above can be rewritten as follows

$$m_g = \left\{1 - 2\left[\sqrt{1 + 1.758 \times 10^{-27}\left(\frac{\mu_r \sigma^3}{\rho^2 f^3}\right)E_{rms}^4} - 1\right]\right\} m_{i0}$$

Substitution of the well-known equation of the *Ohm's vectorial Law*: $j = \sigma E$ into (A14), we get

$$m_g = \left\{1 - 2\left[\sqrt{1 + 1.758 \times 10^{-27}\frac{\mu_r j_{rms}^4}{\sigma \rho^2 f^3}} - 1\right]\right\} m_{i0} \qquad (A15)$$

where $j_{rms} = j/\sqrt{2}$.

Consider a 15 cm square *Aluminum thin foil* of *10.5 microns thickness* with the following characteristics: $\mu_r = 1$ ; $\sigma = 3.82 \times 10^7 S.m^{-1}$; $\rho = 2700\,Kg.m^{-3}$. Then, (A15) gives

$$m_g = \left\{1 - 2\left[\sqrt{1 + 6.313 \times 10^{-42}\frac{j_{rms}^4}{f^3}} - 1\right]\right\} m_{i0} \qquad (A16)$$

Now, consider that the ELF electric current $I = I_{DC} + i_0 \sin \omega t$, $(i_0 << I_{DC})$ passes through that Aluminum foil. Then, the current density is

$$j_{rms} = \frac{I_{rms}}{S} \cong \frac{I_{DC}}{S} \qquad (A17)$$

where

$$S = 0.15m\left(10.5 \times 10^{-6} m\right) = 1.57 \times 10^{-6} m^2$$

If the ELF electric current has frequency $f = 2\mu Hz = 2 \times 10^{-6} Hz$, then, the gravitational mass of the aluminum foil, given by (A16), is expressed by

$$m_g = \left\{1 - 2\left[\sqrt{1 + 7.89\times10^{-25}\frac{I_{DC}^4}{S^4}} - 1\right]\right\} m_{i0} =$$

$$= \left\{1 - 2\left[\sqrt{1 + 0.13 I_{DC}^4} - 1\right]\right\} m_{i0} \qquad (A18)$$

Then,

$$\chi = \frac{m_g}{m_{i0}} \cong \left\{1 - 2\left[\sqrt{1 + 0.13 I_{DC}^4} - 1\right]\right\} \qquad (A19)$$

For $I_{DC} = 2.2A$, the equation above gives

$$\chi = \left(\frac{m_g}{m_{i0}}\right) \cong -1 \qquad (A20)$$

This means that *the gravitational shielding* produced by the aluminum foil can change the gravity acceleration *above* the foil down to

$$g' = \chi\ g \cong -1g \qquad (A21)$$

Under these conditions, the Aluminum foil works basically as a Gravity Control Cell (GCC).

In order to check these theoretical predictions, we suggest an experimental set-up shown in Fig.A2.

A 15cm square Aluminum foil of *10.5 microns thickness* with the following composition: Al 98.02%; Fe 0.80%; Si 0.70%; Mn 0.10%; Cu 0.10%; Zn 0.10%; Ti 0.08%; Mg 0.05%; Cr 0.05%, and with the following characteristics: $\mu_r = 1$; $\sigma = 3.82\times10^7 S.m^{-1}$; $\rho = 2700 Kg.m^{-3}$, is fixed on a 17 cm square *Foam Board* ** plate of 6mm thickness as shown in Fig.A3. This device (the simplest Gravity Control Cell GCC) is placed on a pan balance shown in Fig.A2.

Above the Aluminum foil, a *sample* (any type of material, any mass) connected to a dynamometer will check the decrease of the *local gravity acceleration* upon the sample $(g' = \chi\ g)$, due to the gravitational shielding produced by the decreasing of gravitational mass of the Aluminum foil $(\chi = m_g / m_{i0})$. Initially, the sample lies 5 cm above the Aluminum foil. As shown in Fig.A2, the board with the dynamometer can be displaced up to few meters in height. Thus, the initial distance between the Aluminum foil and the sample can be increased in order to check the reach of the gravitational shielding produced by the Aluminum foil.

In order to generate the ELF electric current of $f = 2\mu Hz$, we can use the widely-known Function Generator HP3325A (Op.002 High Voltage Output) that can generate sinusoidal voltages with *extremely-low* frequencies down to $f = 1\times10^{-6} Hz$ and amplitude up to 20V (40$V_{pp}$ into $500\Omega$ load). The maximum output current is $0.08A_{pp}$; output impedance <2Ω at ELF.

Figure A4 shows the equivalent electric circuit for the experimental set-up. The electromotive forces are: $\varepsilon_1$ (HP3325A) and $\varepsilon_2$ (*12V* DC Battery).The values of the *resistors* are : $R_1 = 500\Omega - 2W$ ; $r_{i1} < 2\Omega$; $R_2 = 4\Omega - 40W$; $r_{i2} < 0.1\Omega$; $R_p = 2.5\times10^{-3}\Omega$; *Rheostat* (0≤ *R* ≤10Ω - 90W). The *coupling transformer* has the following characteristics: air core with diameter $\phi = 10mm$; area $S = \pi\phi^2/4 = 7.8\times10^{-5} m^2$; wire#12AWG; $N_1 = N_2 = N = 20$; $l = 42mm$; $L_1 = L_2 = L = \mu_0 N^2 (S/l) = 9.3\times10^{-7} H$ .Thus, we get

$$Z_1 = \sqrt{(R_1 + r_{i1})^2 + (\omega L)^2} \cong 501\Omega$$

and

$$Z_2 = \sqrt{(R_2 + r_{i2} + R_p + R)^2 + (\omega L)^2}$$

For $R = 0$ we get $Z_2 = Z_2^{min} \cong 4\Omega$; for $R = 10\Omega$ the result is $Z_2 = Z_2^{max} \cong 14\Omega$. Thus,

$$Z_{1,total}^{min} = Z_1 + Z_{1,reflected}^{min} = Z_1 + Z_2^{min}\left(\frac{N_1}{N_2}\right)^2 \cong 505\Omega$$

$$Z_{1,total}^{max} = Z_1 + Z_{1,reflected}^{max} = Z_1 + Z_2^{max}\left(\frac{N_1}{N_2}\right)^2 \cong 515\Omega$$

The maxima *rms* currents have the following values:

$$I_1^{max} = \tfrac{1}{\sqrt{2}} 40V_{pp} / Z_{1,total}^{min} = 56mA$$

(The maximum output current of the Function Generator HP3325A (Op.002 High Voltage Output) is $80mA_{pp} \cong 56.5mA_{rms}$);

$$I_2^{max} = \frac{\varepsilon_2}{Z_2^{min}} = 3A$$

and

$$I_3^{max} = I_2^{max} + I_1^{max} \cong 3A$$

The new expression for the *inertial forces*, (Eq.5) $\vec{F}_i = M_g \vec{a}$ , shows that the inertial forces are proportional to *gravitational mass*. Only in the particular case of $m_g = m_{i0}$, the expression above reduces to the well-known Newtonian expression $\vec{F}_i = m_{i0}\vec{a}$ . The equivalence

** *Foam board* is a very strong, *lightweigh*t (density: 24.03 kg.m$^{-3}$) and easily cut material used for the mounting of photographic prints, as backing in picture framing, in 3D design, and in painting. It consists of three layers — an inner layer of polystyrene clad with outer facing of either white clay coated paper or brown Kraft paper.

between gravitational and inertial forces $\left(\vec{F}_i \equiv \vec{F}_g\right)$ [1] shows then that a balance measures the *gravitational mass* subjected to acceleration $a = g$. Here, the decrease in the *gravitational mass* of the Aluminum foil will be measured by a pan balance with the following characteristics: range 0-200g; readability 0.01g.

The mass of the Foam Board plate is: $\cong 4.17g$, the mass of the Aluminum foil is: $\cong 0.64g$, the total mass of the ends and the electric wires of connection is $\cong 5g$. Thus, *initially* the balance will show $\cong 9.81g$. According to (A18), when the electric current through the Aluminum foil (resistance $r_p^* = l/\sigma S = 2.5\times10^{-3}\Omega$) reaches the value: $I_3 \cong 2.2A$, we will get $m_{g(Al)} \cong -m_{i0(Al)}$. Under these circumstances, the balance will show:

$$9.81g - 0.64g - 0.64g \cong 8.53g$$

and the gravity acceleration $g'$ *above* the Aluminum foil, becomes $g' = \chi\ g \cong -1g$.

It was shown [1] that, when the gravitational mass of a particle is reduced to the gravitational mass ranging between $+0.159M_i$ to $-0.159M_i$, it becomes *imaginary*, i.e., the gravitational and the inertial masses of the particle become *imaginary*. Consequently, the particle *disappears* from our ordinary space-time. This phenomenon can be observed in the proposed experiment, i.e., *the Aluminum foil will disappear* when its gravitational mass becomes smaller than $+0.159M_i$. It will become visible again, only when its gravitational mass becomes smaller than $-0.159M_i$, or when it becomes greater than $+0.159M_i$.

Equation (A18) shows that the gravitational mass of the Aluminum foil, $m_{g(Al)}$, goes *close to zero* when $I_3 \cong 1.76A$. Consequently, the gravity acceleration *above* the Aluminum foil also goes close to zero since $g' = \chi\ g = m_{g(Al)}/m_{i0(Al)}$. Under these circumstances, the Aluminum foil remains *invisible*.

Now consider a rigid Aluminum wire # 14 AWG. The area of its cross section is

$$S = \pi\left(1.628\times10^{-3}m\right)^2/4 = 2.08\times10^{-6}m^2$$

If an ELF electric current with frequency $f = 2\mu Hz = 2\times10^{-6}Hz$ passes through this wire, its gravitational mass, given by (A16), will be expressed by

$$m_g = \left\{1-2\left[\sqrt{1+6.313\times10^{-42}\frac{j_{rms}^4}{f^3}}-1\right]\right\}m_{i0} =$$

$$= \left\{1-2\left[\sqrt{1+7.89\times10^{-25}\frac{I_{DC}^4}{S^4}}-1\right]\right\}m_{i0} =$$

$$= \left\{1-2\left[\sqrt{1+0.13I_{DC}^4}-1\right]\right\}m_{i0} \qquad (A22)$$

For $I_{DC} \cong 3A$ the equation above gives

$$m_g \cong -3.8m_{i0}$$

Note that we can replace the Aluminum foil for this wire in the experimental set-up shown in Fig.A2. It is important also to note that an ELF electric current that passes through a wire - which makes a spherical form, as shown in Fig A5 - reduces the gravitational mass of the wire (Eq. A22), and the gravity *inside sphere* at the same proportion, $\chi = m_g/m_{i0}$, (Gravitational Shielding Effect). In this case, that effect can be checked by means of the Experimental set-up 2 (Fig.A6). Note that the spherical form can be transformed into an ellipsoidal form or a disc in order to coat, for example, a Gravitational Spacecraft. It is also possible to coat with a wire several forms, such as cylinders, cones, cubes, etc.

The circuit shown in Fig.A4 (a) can be modified in order to produce a new type of Gravitational Shielding, as shown in Fig.A4 (b). In this case, the Gravitational Shielding will be produced in the Aluminum plate, with thickness $h$, of the parallel plate capacitor connected in the point $P$ of the circuit (See Fig.A4 (b)). Note that, in this circuit, the Aluminum foil (resistance $R_p$) (Fig.A4(a)) has been replaced by a Copper wire # 14 AWG with *1cm* length ($l = 1cm$) in order to produce a resistance $R_\phi = 5.21\times10^{-5}\Omega$. Thus, the voltage in the point $P$ of the circuit will have the maximum value $V_p^{max} = 1.1\times10^{-4}V$ when the resistance of the rheostat is null $(R = 0)$ and the minimum value $V_p^{min} = 4.03\times10^{-5}V$ when $R = 10\Omega$. In this way, the voltage $V_p$ (with frequency $f = 2\mu Hz$) applied on the capacitor will produce an electric field $E_p$ with intensity $E_p = V_p/h$ through the Aluminum plate of thickness $h = 3mm$. It is important to note that *this plate cannot be connected to ground* (earth), in other words, cannot be grounded, because, in this case, the electric field through it will be *null* [††].

According to Eq. A14, when $E_p^{max} = V_p^{max}/h = 0.036\,V/m$, $f = 2\mu Hz$ and $\sigma_{Al} = 3.82 \times 10^7\ S/m$, $\rho_{Al} = 2700 kg/m^3$ (Aluminum), we get

$$\chi = \frac{m_{(Al)}}{m_{i(Al)}} \cong -0.9$$

Under these conditions, the maximum *current density* through the plate with thickness $h$ will be given by $j^{max} = \sigma_{Al} E_p^{max} = 1.4 \times 10^6 A/m^2$ (It is well-known that the maximum current density supported by the Aluminum is $\approx 10^8 A/m^2$ ).

Since the area of the plate is $A = (0.2)^2 = 4 \times 10^{-2} m^2$, then the maximum current is $i^{max} = j^{max} A = 56kA$. Despite this enormous current, the maximum dissipated power will be just $P^{max} = \left(i^{max}\right)^2 R_{plate} = 6.2W$, because the resistance of the plate is very small, i.e., $R_{plate} = h/\sigma_{Al} A \cong 2 \times 10^{-9} \Omega$.

Note that the area $A$ of the plate (where the Gravitational Shielding takes place) can have several geometrical configurations. For example, it can be the area of the external surface of an ellipsoid, sphere, etc. Thus, it can be the area of the external surface of a Gravitational Spacecraft. In this case, if $A \cong 100m^2$, for example, the maximum dissipated power will be $P^{max} \cong 15.4kW$, i.e., approximately $154W/m^2$.

All of these systems work with Extra-Low Frequencies $\left(f << 10^{-3} Hz\right)$. Now, we show that, by simply changing *the geometry of the surface of the Aluminum foil*, it is possible to increase the working frequency $f$ up to more than *1Hz*.

Consider the Aluminum foil, now with several semi-spheres stamped on its surface, as shown in Fig. A7 . The semi-spheres have radius $r_0 = 0.9\ mm$, and are joined one to another. The Aluminum foil is now coated by an insulation layer with relative permittivity $\varepsilon_r$ and dielectric strength $k$. A voltage source is connected to the Aluminum foil in order to provide a voltage $V_0$ (rms) with frequency $f$. Thus, the electric potential $V$ at a distance $r$, in the interval from $r_0$ to $a$, is given by

$$V = \frac{1}{4\pi\varepsilon_r\varepsilon_0}\frac{q}{r} \qquad (A23)$$

In the interval $a < r \leq b$ the electric potential is

$$V = \frac{1}{4\pi\varepsilon_0}\frac{q}{r} \qquad (A24)$$

since for the air we have $\varepsilon_r \cong 1$.

Thus, on the surface of the metallic spheres $(r = r_0)$ we get

$$V_0 = \frac{1}{4\pi\varepsilon_r\varepsilon_0}\frac{q}{r_0} \qquad (A25)$$

Consequently, the electric field is

$$E_0 = \frac{1}{4\pi\varepsilon_r\varepsilon_0}\frac{q}{r_0^2} \qquad (A26)$$

By comparing (A26) with (A25), we obtain

$$E_0 = \frac{V_0}{r_0} \qquad (A27)$$

The electric potential $V_b$ at $r = b$ is

$$V_b = \frac{1}{4\pi\varepsilon_0}\frac{q}{b} = \frac{\varepsilon_r V_0 r_0}{b} \qquad (A28)$$

Consequently, the electric field $E_b$ is given by

$$E_b = \frac{1}{4\pi\varepsilon_0}\frac{q}{b^2} = \frac{\varepsilon_r V_0 r_0}{b^2} \qquad (A29)$$

From $r = r_0$ up to $r = b = a + d$ *the electric field is approximately constant* (See Fig. A7). Along the distance $d$ it will be called $E_{air}$. For $r > a + d$, the electric field stops being constant. Thus, the intensity of the electric field at $r = b = a + d$ is approximately equal to $E_0$, i.e., $E_b \cong E_0$. Then, we can write that

$$\frac{\varepsilon_r V_0 r_0}{b^2} \cong \frac{V_0}{r_0} \qquad (A30)$$

whence we get

$$b \cong r_0\sqrt{\varepsilon_r} \qquad (A31)$$

Since the intensity of the electric field through the air, $E_{air}$, is $E_{air} \cong E_b \cong E_0$, then, we can write that

$$E_{air} = \frac{1}{4\pi\varepsilon_0}\frac{q}{b^2} = \frac{\varepsilon_r V_0 r_0}{b^2} \qquad (A32)$$

Note that $\varepsilon_r$ refers to the *relative permittivity of*

[††] When the voltage $V_p$ is applied on the capacitor, the charge distribution in the dielectric induces positive and negative charges, respectively on opposite sides of the Aluminum plate with thickness *h*. If the plate is not connected to the ground (Earth) this charge distribution produces an electric field $E_p = V_p/h$ through the plate. However, if the plate is connected to the ground, the negative charges (electrons) escapes for the ground and the positive charges are redistributed along the entire surface of the Aluminum plate making *null* the electric field through it.

*the insulation layer, which is covering the Aluminum foil*.

If the intensity of this field is greater than the dielectric strength of the air $\left(3\times 10^6 V/m\right)$ there will occur the well-known *Corona effect*. Here, this effect is necessary in order to increase the electric conductivity of the air at this region (layer with thickness *d*). Thus, we will assume

$$E_{air}^{min} = \frac{\varepsilon_r V_0^{min} r_0}{b^2} = \frac{V_0^{min}}{r_0} = 3\times 10^6 V/m$$

and

$$E_{air}^{max} = \frac{\varepsilon_r V_0^{max} r_0}{b^2} = \frac{V_0^{max}}{r_0} = 1\times 10^7 V/m \qquad (A33)$$

The electric field $E_{air}^{min} \le E_{air} \le E_{air}^{max}$ will produce an *electrons flux* in a direction and an *ions flux* in an opposite direction. From the viewpoint of electric current, the ions flux can be considered as an "electrons" flux at the same direction of the real electrons flux. Thus, the current density through the air, $j_{air}$, will be the *double* of the current density expressed by the well-known equation of Langmuir-Child

$$j = \frac{4}{9}\varepsilon_r \varepsilon_0 \sqrt{\frac{2e}{m_e}}\frac{V^{\frac{3}{2}}}{d^2} = \alpha \frac{V^{\frac{3}{2}}}{d^2} = 2.33\times 10^{-6}\frac{V^{\frac{3}{2}}}{d^2} \qquad (A34)$$

where $\varepsilon_r \cong 1$ for the *air*; $\alpha = 2.33\times 10^{-6}$ is the called *Child's constant*.

Thus, we have

$$j_{air} = 2\alpha \frac{V^{\frac{3}{2}}}{d^2} \qquad (A35)$$

where $d$, in this case, is the thickness of the air layer where the electric field is approximately constant and $V$ is the voltage drop given by

$$V = V_a - V_b = \frac{1}{4\pi\varepsilon_0}\frac{q}{a} - \frac{1}{4\pi\varepsilon_0}\frac{q}{b} =$$

$$= V_0 r_0 \varepsilon_r \left(\frac{b-a}{ab}\right) = \left(\frac{\varepsilon_r r_0 d}{ab}\right) V_0 \qquad (A36)$$

By substituting (A36) into (A35), we get

$$j_{air} = \frac{2\alpha}{d^2}\left(\frac{\varepsilon_r r_0 d V_0}{ab}\right)^{\frac{3}{2}} = \frac{2\alpha}{d^{\frac{1}{2}}}\left(\frac{\varepsilon_r r_0 V_0}{b^2}\right)^{\frac{3}{2}}\left(\frac{b}{a}\right)^{\frac{3}{2}} =$$

$$= \frac{2\alpha}{d^{\frac{1}{2}}} E_{air}^{\frac{3}{2}}\left(\frac{b}{a}\right)^{\frac{3}{2}} \qquad (A37)$$

According to the equation of the *Ohm's vectorial Law*: $j = \sigma E$, we can write that

$$\sigma_{air} = \frac{j_{air}}{E_{air}} \qquad (A38)$$

Substitution of (A37) into (A38) yields

$$\sigma_{air} = 2\alpha \left(\frac{E_{air}}{d}\right)^{\frac{1}{2}}\left(\frac{b}{a}\right)^{\frac{3}{2}} \qquad (A39)$$

If the insulation layer has thickness $\Delta = 0.6\ mm$, $\varepsilon_r \cong 3.5$ (1- 60Hz), $k = 17kV/mm$ (Acrylic sheet 1.5mm thickness), and the semi-spheres stamped on the metallic surface have $r_0 = 0.9\ mm$ (See Fig.A7) then $a = r_0 + \Delta = 1.5\ mm$. Thus, we obtain from Eq. (A33) that

$$V_0^{min} = 2.7kV$$

$$V_0^{max} = 9kV \qquad (A40)$$

From equation (A31), we obtain the following value for $b$:

$$b = r_0\sqrt{\varepsilon_r} = 1.68\times 10^{-3} m \qquad (A41)$$

Since $b = a + d$ we get

$$d = 1.8\times 10^{-4} m$$

Substitution of $a$, $b$, $d$ and A(32) into (A39) produces

$$\sigma_{air} = 4.117\times 10^{-4} E_{air}^{\frac{1}{2}} = 1.375\times 10^{-2} V_0^{\frac{1}{2}}$$

Substitution of $\sigma_{air}$, $E_{air}(rms)$ and $\rho_{air} = 1.2\ kg.m^{-3}$ into (A14) gives

$$\frac{m_{g(air)}}{m_{i0(air)}} = \left\{1 - 2\left[\sqrt{1 + 1.758\times 10^{-27}\frac{\sigma_{air}^3 E_{air}^4}{\rho_{air}^2 f^3}} - 1\right]\right\} =$$

$$= \left\{1 - 2\left[\sqrt{1 + 4.923\times 10^{-21}\frac{V_0^{5.5}}{f^3}} - 1\right]\right\} \qquad (A42)$$

For $V_0 = V_0^{max} = 9kV$ and $f = 2Hz$, the result is

$$\frac{m_{g(air)}}{m_{i0(air)}} \cong -1.2$$

Note that, by increasing $V_0$, the values of $E_{air}$ and $\sigma_{air}$ are increased. Thus, as show (A42), there are two ways for decrease the value of $m_{g(air)}$: increasing the value of $V_0$ or decreasing the value of $f$.

Since $E_0^{max} = 10^7 V/m = 10kV/mm$ and $\Delta = 0.6\ mm$ then the dielectric strength of the insulation must be $\ge 16.7kV/mm$. As mentioned above, the dielectric strength of the acrylic is $17kV/mm$.

It is important to note that, due to the strong value of $E_{air}$ (Eq. A37) the *drift velocity* $v_d$, $\left(v_d = j_{air}/ne = \sigma_{air}E_{air}/ne\right)$ of the free charges inside the ionized air put them at a distance $x = v_d / t = 2 f v_d \cong 0.4 m$, which is much greater than the distance $d = 1.8 \times 10^{-4} m$. Consequently, the number $n$ of free charges decreases strongly inside the air layer of thickness $d$ [‡‡], except, obviously, in a thin layer, very close to the dielectric, where the number of free charges remains sufficiently increased, to maintain the air conductivity with $\sigma_{air} \cong 1.1 S/m$ (Eq. A39).

The thickness $h$ of this thin air layer close to the dielectric can be easily evaluated starting from the charge distribution in the neighborhood of the dielectric, and of the repulsion forces established among them. The result is $h = \sqrt{0.06 e / 4\pi\varepsilon_0 E} \cong 4 \times 10^{-9} m$. This is, therefore, the thickness of the *Air* Gravitational Shielding. If the area of this Gravitational Shielding is equal to the area of a format A4 sheet of paper, i.e., $A = 0.20 \times 0.291 = 0.0582 m^2$, we obtain the following value for the resistance $R_{air}$ of the Gravitational Shielding: $R_{air} = h / \sigma_{air} A \cong 6 \times 10^{-8} \Omega$. Since the maximum electrical current through this air layer is $i^{max} = j^{max} A \cong 400 kA$, then the maximum power radiated from the Gravitational Shielding is $P_{air}^{max} = R_{air} \left( i_{air}^{max} \right)^2 \cong 10 kW$. This means that a very strong light will be radiated from this type of Gravitational Shielding. Note that this device can also be used as a lamp, which will be much more efficient than conventional lamps.

Coating a ceiling with this lighting system enables the entire area of ceiling to produce light. This is a form of lighting very different from those usually known.

Note that the value $P_{air}^{max} \cong 10 kW$, defines the power of the transformer shown in Fig.A10. Thus, the maximum current in the secondary is $i_s^{max} = 9kV / 10kW = 0.9A$.

Above the Gravitational Shielding, $\sigma_{air}$ is reduced to the normal value of conductivity of the atmospheric air $\left( \approx 10^{-14} S/m \right)$. Thus, the power radiated from this region is

$$P_{air}^{max} = (d - h)\left( i_{air}^{max} \right)^2 / \sigma_{air} A =$$

$$= (d - h) A \sigma_{air} \left( E_{air}^{max} \right)^2 \cong 10^{-4} W$$

Now, we will describe a method to coat the Aluminum semi-spheres with acrylic in the necessary dimensions $(\Delta = a - r_0)$, we propose the following method. First, take an Aluminum plate with $21cm \times 29.1cm$ (A4 format). By means of a convenient process, several semi-spheres can be stamped on its surface. The semi-spheres have radius $r_0 = 0.9\ mm$, and are joined one to another. Next, take an acrylic sheet (A4 format) with 1.5mm thickness (See Fig.A8 (a)). Put a heater below the Aluminum plate in order to heat the Aluminum (Fig.A8 (b)). When the Aluminum is sufficiently heated up, the acrylic sheet and the Aluminum plate are pressed, one against the other, as shown in Fig. A8 (c). The two D devices shown in this figure are used in order to impede that the press compresses the acrylic and the aluminum to a distance shorter than $y + a$. After some seconds, remove the press and the heater. The device is ready to be subjected to a voltage $V_0$ with frequency $f$, as shown in Fig.A9. Note that, in this case, the balance is not necessary, because *the substance that produces the gravitational shielding* is an *air layer* with thickness $d$ *above* the acrylic sheet. This is, therefore, more a type of Gravity Control Cell (GCC) with *external gravitational shielding*.

It is important to note that this GCC can be made very thin and as flexible as a fabric. Thus, it can be used to produce *anti- gravity clothes*. These clothes can be extremely useful, for example, to walk on the surface of high gravity planets.

Figure A11 shows some geometrical forms that can be stamped on a metallic surface in order to produce a Gravitational Shielding effect, similar to the produced by the *semi-spherical form*.

An obvious evolution from the semi-spherical form is the *semi-cylindrical* form shown in Fig. A11 (b); Fig.A11(c) shows *concentric metallic rings* stamped on the metallic surface, an evolution from Fig.A11 (b). These geometrical forms produce the same effect as the semi-spherical form, shown in Fig.A11 (a). By using concentric metallic rings, it is possible to build *Gravitational Shieldings* around bodies or spacecrafts with several formats (spheres, ellipsoids, etc); Fig. A11 (d) shows a Gravitational Shielding around a Spacecraft with *ellipsoidal form*.

The previously mentioned Gravitational Shielding, produced on a thin layer of ionized air, has a *behavior different from* the Gravitational Shielding produced on a *rigid substance*. When the gravitational masses of the air molecules, inside the shielding, are reduced to within the range $+0.159 m_i < m_g < -0.159 m_i$, they go to the *imaginary space-time*, as previously shown in this article. However, the electric field $E_{air}$ stays at the real space-time. Consequently, the molecules return immediately to the real space-time in order to return soon after to the *imaginary* space-time, due to the action of the electric field $E_{air}$.

[‡‡] Reducing therefore, the conductivity $\sigma_{air}$, to the normal value of conductivity of the atmospheric air.

In the case of the Gravitational Shielding produced on a *solid substance*, when the molecules of the substance go to the *imaginary* space-time, *the electric field that produces the effect, also goes to the imaginary space-time together with them*, since in this case, the substance of the Gravitational Shielding is rigidly connected to the metal that produces the electric field. (See Fig. A12 (b)). This is the fundamental difference between the *non-solid* and *solid* Gravitational Shieldings.

Now, consider a Gravitational Spacecraft that is able to produce an *Air* Gravitational Shielding and also a *Solid* Gravitational Shielding, as shown in Fig. A13 (a) [§§]. Assuming that the intensity of the electric field, $E_{air}$, necessary to reduce the gravitational mass of the *air molecules* to within the range $+0.159m_i < m_g < -0.159m_i$, *is much smaller* than the intensity of the electric field, $E_{rs}$, necessary to reduce the gravitational mass of the *solid substance* to within the range $+0.159m_i < m_g < -0.159m_i$, then we conclude that the Gravitational Shielding made of ionized air goes to the imaginary space-time *before* the Gravitational Shielding made of *solid substance*. When this occurs the spacecraft does not go to the imaginary space-time together with the Gravitational Shielding of air, because the air molecules are not rigidly connected to the spacecraft. Thus, while the air molecules go into the imaginary space-time, the spacecraft stays in the *real space-time*, and remains subjected to the effects of the Gravitational Shielding around it, since the shielding does not stop to work, due to its extremely short permanence at the imaginary space-time. Under these circumstances, the gravitational mass of the Gravitational Shielding can be reduced to $m_g \cong 0$. For example, $m_g \cong 10^{-4}kg$. Thus, if the *inertial mass* of the Gravitational Shielding is $m_{i0} \cong 1kg$, then $\chi = m_g / m_{i0} \cong 10^{-4}$. As we have seen, this means that *the inertial effects on the spacecraft* will be reduced by $\chi \cong 10^{-4}$. Then, in spite of the effective acceleration of the spacecraft be, for example, $a = 10^5 m.s^{-2}$, the effects on the crew of the spacecraft will be equivalent to an acceleration of only

$$a' = \frac{m_g}{m_{i0}} a = \chi \ a \approx 10 m.s^{-1}$$

This is the magnitude of the acceleration upon the passengers in a contemporary commercial jet.

Then, it is noticed that Gravitational Spacecrafts can be subjected to enormous *accelerations* (or *decelerations*) without imposing any harmful impacts whatsoever on the spacecrafts or its crew.

Now, imagine that the intensity of the electric field that produces the Gravitational Shielding around the spacecraft is *increased* up to reaching the value $E_{rs}$ that reduces the gravitational mass of the *solid* Gravitational Shielding to within the range $+0.159m_i < m_g < -0.159m_i$. Under these circumstances, the *solid* Gravitational Shielding goes to the imaginary space-time and, since it is rigidly connected to the spacecraft, also the spacecraft goes to the imaginary space-time together with the Gravitational Shielding. Thus, the spacecraft can travel within the imaginary space-time and make use of the Gravitational Shielding around it.

As we have already seen, the maximum velocity of propagation of the interactions in the imaginary space-time is *infinite* (in the real space-time this limit is equal to the light velocity $c$). This means that *there are no limits for the velocity of the spacecraft in the imaginary space-time*. Thus, the acceleration of the spacecraft can reach, for example, $a = 10^9 m.s^{-2}$, which leads the spacecraft to attain velocities $V \approx 10^{14} m.s^{-1}$ (about 1 million times the speed of light) after one day of trip. With this velocity, after 1 month of trip the spacecraft would have traveled about $10^{21} m$. In order to have idea of this distance, it is enough to remind that the diameter of our Universe (visible Universe) is of the order of $10^{26} m$.

---

[§§] The *solid* Gravitational Shielding can also be obtained by means of *an ELF electric current through a metallic lamina* placed *between the semi-spheres and the Gravitational Shielding of Air* (See Fig.A13 (a)). The gravitational mass of the solid Gravitational Shielding will be controlled just by means of the intensity of the ELF electric current. Recently, it was discovered that Carbon nanotubes (CNTs) can be added to *Alumina* ($Al_2O_3$) to convert it into a good electrical conductor. It was found that the electrical conductivity increased up to 3375 S/m at 77°C in samples that were 15% nanotubes by volume [12]. It is known that the density of α-Alumina is 3.98kg.m$^{-3}$ and that it can withstand 10-20 KV/mm. Thus, these values show that the Alumina-CNT can be used to make a *solid* Gravitational Shielding. In this case, the electric field produced by means of the semi-spheres will be used to control the gravitational mass of the Alumina-CNT.

Due to the extremely low density of the *imaginary* bodies, the collision between them cannot have the same consequences of the collision between the real bodies.

Thus, *for a Gravitational Spacecraft in imaginary state, the problem of the collision in high-speed doesn't exist.* Consequently, the Gravitational Spacecraft can transit freely in the imaginary Universe and, in this way, reach easily any point of our real Universe once they can make the transition back to our Universe by only increasing the gravitational mass of the Gravitational Shielding of the spacecraft in such way that it leaves the range of $+0.159M_i$ to $-0.159M_i$.

The return trip would be done in similar way. That is to say, the spacecraft would transit in the imaginary Universe back to the departure place where would reappear in our Universe. Thus, trips through our Universe that would delay millions of years, at speeds close to the speed of light, could be done in just a few *months* in the imaginary Universe.

In order to produce the acceleration of $a \approx 10^9 m.s^{-2}$ upon the spacecraft we propose a Gravitational Thruster with 10 GCCs (10 Gravitational Shieldings) of the type with several semi-spheres stamped on the metallic surface, as previously shown, or with the *semi-cylindrical* form shown in Figs. A11 (b) and (c). The 10 GCCs are filled with air at 1 atm and 300K. If the insulation layer is made with *Mica* $(\varepsilon_r \cong 5.4)$ and has thickness $\Delta = 0.1\ mm$, and the semi-spheres stamped on the metallic surface have $r_0 = 0.4\ mm$ (See Fig.A7) then $a = r_0 + \Delta = 0.5\ mm$. Thus, we get

$$b = r_0\sqrt{\varepsilon_r} = 9.295\times10^{-4} m$$

and

$$d = b - a = 4.295\times10^{-4} m$$

Then, from Eq. A42 we obtain

$$\chi_{air} = \frac{m_{g(air)}}{m_{i0(air)}} = \left\{1 - 2\left[\sqrt{1 + 1.758\times10^{-27}\frac{\sigma_{air}^3 E_{air}^4}{\rho_{air}^2 f^3}} - 1\right]\right\} =$$

$$= \left\{1 - 2\left[\sqrt{1 + 1.0\times10^{-18}\frac{V_0^{5.5}}{f^3}} - 1\right]\right\}$$

For $V_0 = V_0^{max} = 15.6kV$ and $f = 0.12Hz$, the result is

$$\chi_{air} = \frac{m_{g(air)}}{m_{i0(air)}} \cong -1.6\times10^4$$

Since $E_0^{max} = V_0^{max}/r_0$ is now given by $E_0^{max} = 15.6kV/0.9mm = 17.3kV/mm$ and $\Delta = 0.1\ mm$ then the dielectric strength of the insulation must be $\geq 173 kV/mm$. As shown in the table below***, *0.1mm - thickness of* Mica *can withstand 17.6 kV* (that is greater than $V_0^{max} = 15.6kV$), in such way that the dielectric strength is *176 kV/mm*.

The Gravitational Thrusters are positioned at the spacecraft, as shown in Fig. A13 (b). Then, when the spacecraft is in the *intergalactic space*, the gravity acceleration upon the gravitational mass $m_{gt}$ of the bottom of the thruster (See Fig.A13 (c)), is given by [2]

$$\vec{a} \cong (\chi_{air})^{10}\vec{g}_M \cong -(\chi_{air})^{10} G\frac{M_g}{r^2}\hat{\mu}$$

where $M_g$ is the gravitational mass in front of the spacecraft.

For simplicity, let us consider just the effect of a hypothetical volume $V = 10\times10^3\times10^3 = 10^7 m^3$ of intergalactic matter in front of the spacecraft $(r \cong 30m)$. The average density of matter in the *intergalactic medium (IGM)* is $\rho_{ig} \approx 10^{-26} kg.m^{-3}$)†††. Thus, for $\chi_{air} \cong -1.6\times10^4$ we get

$$a = -(-1.6\times10^4)^{10}(6.67\times10^{-11})\left(\frac{10^{-19}}{30^2}\right) =$$

$$= -10^9 m.s^{-2}$$

In spite of this gigantic acceleration, the inertial effects for the crew of the spacecraft can be strongly reduced if, for example, the gravitational mass of the Gravitational Shielding is reduced

---

*** The *dielectric strength* of some dielectrics can have different values in lower thicknesses. This is, for example, the case of the *Mica*.

| *Dielectric* | *Thickness* (mm) | *Dielectric Strength* (kV/mm) |
|---|---|---|
| Mica | 0.01 mm | 200 |
| **Mica** | **0.1 mm** | **176** |
| Mica | 1 mm | 61 |

††† Some theories put the average density of the Universe as the equivalent of *one hydrogen atom per cubic meter* [13,14]. The density of the universe, however, is clearly not uniform. Surrounding and stretching between galaxies, there is a rarefied plasma [15] that is thought to possess a cosmic filamentary structure [16] and that is slightly denser than the average density in the universe. This material is called the *intergalactic medium (IGM)* and is mostly ionized hydrogen; i.e. a plasma consisting of equal numbers of electrons and protons. The IGM is thought to exist at a density of 10 to 100 times the average density of the Universe (10 to 100 hydrogen atoms per cubic meter, i.e., $\approx 10^{-26} kg.m^{-3}$).

down to $m_g \cong 10^{-6} kg$ and its inertial mass is $m_{i0} \cong 100kg$. Then, we get $\chi = m_g / m_{i0} \cong 10^{-8}$. Therefore, *the inertial effects on the spacecraft* will be reduced by $\chi \cong 10^{-8}$, and consequently, the inertial effects on the crew of the spacecraft would be *equivalent to* an acceleration $a'$ of only

$$a' = \frac{m_g}{m_{i0}} a = \left(10^{-8}\right)\left(10^{9}\right) \approx 10 m.s^{-2}$$

Note that the Gravitational Thrusters in the spacecraft must have a very small diameter (of the order of *millimeters*) since, obviously, the hole through the Gravitational Shielding cannot be large. Thus, these thrusters are in fact, *Micro-Gravitational Thrusters*. As shown in Fig. A13 (b), it is possible to place several micro-gravitational thrusters in the spacecraft. This gives to the Gravitational Spacecraft, several degrees of freedom and shows the enormous superiority of this spacecraft in relation to the contemporaries spacecrafts.

The density of matter in the *intergalactic medium (IGM)* is about *10 $^{-26}$ kg.m$^{-3}$* , which is very less than the density of matter in the *interstellar medium* (*~10$^{-21}$ kg.m$^{-3}$*) that is less than the density of matter in the *interplanetary medium* (*~10$^{-20}$ kg.m$^{-3}$*). The density of matter is enormously increased inside the Earth's atmosphere (1.2*kg.m$^{-3}$* near to Earth's surface). Figure A14 shows the gravitational acceleration acquired by a Gravitational Spacecraft, in these media, using Micro-Gravitational thrusters.

In relation to the *Interstellar* and *Interplanetary medium*, the *Intergalactic medium* requires the greatest value of $\chi_{air}$ ($\chi$ inside the *Micro-Gravitational Thrusters*), i.e., $\chi_{air} \cong -1.6\times 10^{4}$. This value strongly decreases when the spacecraft is within the Earth's atmosphere. In this case, it is sufficient only[‡‡‡] $\chi_{air} \cong -10$ in order to obtain:

$$a = -\left(\chi_{air}\right)^{10} G \frac{\rho_{atm} V}{r^2} \cong$$

$$\cong -\left(-10\right)^{10}\left(6.67\times 10^{-11}\right)\frac{1.2\left(10^{7}\right)}{\left(20\right)^{2}} \cong 10^{4} m.s^{-2}$$

With this acceleration the Gravitational Spacecraft can reach about *50000 km/h* in a few seconds. Obviously, the Gravitational Shielding of the spacecraft will reduce strongly *the inertial effects upon the crew* of the spacecraft, in such way that the inertial effects of this strong acceleration will not be felt. In addition, the *artificial atmosphere*, which is possible to build around the spacecraft, by means of gravity control technologies shown in this article (See Fig.6) and [2], will protect it from the *heating* produced by the friction with the Earth's atmosphere. Also, the gravity can be controlled inside of the Gravitational Spacecraft in order to maintain a value close to the Earth's gravity as shown in Fig.3.

Finally, it is important to note that a Micro-Gravitational Thruster does not work *outside* a Gravitational Shielding, because, in this case, *the resultant upon the thruster is null* due to the symmetry (See Fig. A15 (a)). Figure A15 (b) shows a micro-gravitational thruster inside a Gravitational Shielding. This thruster has 10 Gravitational Shieldings, in such way that the gravitational acceleration upon the *bottom* of the thruster, due to a gravitational mass $M_g$ *in front* of the thruster, is $a_{10} = \chi_{air}^{10} a_0$ where $a_0 = -G M_g / r^2$ is the gravitational acceleration acting on the front of the micro-gravitational thruster. *In the opposite direction*, the gravitational acceleration upon the bottom of the thruster, produced by a gravitational mass $M_g$, is

$$a'_0 = \chi_s \left(-G M_g / r'^2\right) \cong 0$$

since $\chi_s \cong 0$ due to the Gravitational Shielding around the micro-thruster (See Fig. A15 (b)). Similarly, the acceleration in front of the thruster is

$$a'_{10} = \chi_{air}^{10} a'_0 = \left[\chi_{air}^{10}\left(-G M_g / r'^2\right)\right] \chi_s$$

where $\left[\chi_{air}^{10}\left(-G M_g / r'^2\right)\right] < a_{10}$, since $r' > r$. Thus, for $a_{10} \cong 10^{9} m.s^{-2}$ and $\chi_s \approx 10^{-8}$ we conclude that $a'_{10} < 10 m.s^{-2}$. This means that $a'_{10} << a_{10}$. Therefore, we can write that the resultant on the micro-thruster can be expressed by means of the following relation

$$R \cong F_{10} = \chi_{air}^{10} F_0$$

Figure A15 (c) shows a Micro-Gravitational Thruster with *10 Air Gravitational Shieldings* (10 GCCs). Thin Metallic laminas are placed after each *Air* Gravitational Shielding in order to retain the electric field $E_b = V_0/x$, produced by metallic *surface behind* the semi-spheres. The laminas with semi-spheres stamped on its surfaces are connected to the ELF voltage source $V_0$ and the thin laminas in front of the Air Gravitational Shieldings are grounded. The air inside this Micro-Gravitational Thruster is at 300K, 1atm.

---

[‡‡‡] This value is within the range of values of $\chi$ $\left(\chi < -10^{3}.\ \textit{See Eq.A15}\right)$, which can be produced by means of *ELF electric currents* through metals as *Aluminum*, etc. This means that, in this case, if convenient, we can replace *air* inside the GCCs of the Gravitational Micro-thrusters by metal laminas with *ELF electric currents* through them.

We have seen that the insulation layer of a GCC can be made up of Acrylic, Mica, etc. Now, we will design a GCC using *Water* (*distilled water*, $\varepsilon_{r(H_2O)} = 80$) and Aluminum *semi-cylinders* with radius $r_0 = 1.3mm$. Thus, for $\Delta = 0.6mm$, the new value of $a$ is $a = 1.9mm$. Then, we get

$$b = r_0\sqrt{\varepsilon_{r(H_2O)}} = 11.63\times10^{-3}m \qquad (A43)$$

$$d = b - a = 9.73\times10^{-3}m \qquad (A44)$$

and

$$E_{air} = \frac{1}{4\pi\varepsilon_{r(air)}\varepsilon_0}\frac{q}{b^2} =$$

$$= \varepsilon_{r(H_2O)}\frac{V_0 r_0}{\varepsilon_{r(air)}b^2} =$$

$$= \frac{V_0/r_0}{\varepsilon_{r(air)}} \cong \frac{V_0}{r_0} = 1111.1\ V_0 \qquad (A45)$$

Note that

$$E_{(H_2O)} = \frac{V_0/r_0}{\varepsilon_{r(H_2O)}}$$

and

$$E_{(acrylic)} = \frac{V_0/r_0}{\varepsilon_{r(acrylic)}}$$

Therefore, $E_{(H_2O)}$ and $E_{(acrylic)}$ are much smaller than $E_{air}$. Note that for $V_0 \le 9kV$ the intensities of $E_{(H_2O)}$ and $E_{(acrylic)}$ are not sufficient to produce the ionization effect, which increases the electrical conductivity. Consequently, the conductivities of the water and the acrylic remain $<< 1\,S.m^{-1}$. In this way, with $E_{(H_2O)}$ and $E_{(acrylic)}$ much smaller than $E_{air}$, and $\sigma_{(H2O)} << 1$, $\sigma_{(acrylic)} << 1$, the decrease in both the gravitational mass of the acrylic and the gravitational mass of water, according to Eq.A14, is negligible. This means that only in the air layer the decrease in the gravitational mass will be relevant.

Equation A39 gives the electrical conductivity of the air layer, i.e.,

$$\sigma_{air} = 2\alpha\left(\frac{E_{air}}{d}\right)^{\frac{1}{2}}\left(\frac{b}{a}\right)^{\frac{3}{2}} = 0.029V_0^{\frac{1}{2}} \qquad (A46)$$

Note that $b = r_0\sqrt{\varepsilon_{r(H_2O)}}$. Therefore, here the value of $b$ is larger than in the case of the acrylic. Consequently, *the electrical conductivity of the air layer will be larger here than in the case of acrylic.*

Substitution of $\sigma_{(air)}$, $E_{air}$ (*rms*) and $\rho_{air} = 1.2kg.m^{-3}$ into Eq. A14, gives

$$\frac{m_{g(air)}}{m_{i0(air)}} = \left\{1 - 2\left[\sqrt{1 + 4.54\times10^{-20}\frac{V_0^{5}}{f^3}} - 1\right]\right\} \qquad (A47)$$

For $V_0 = V_0^{max} = 9kV$ and $f = 2Hz$, the result is

$$\frac{m_{g(air)}}{m_{i0(air)}} \cong -8.4$$

This shows that, by using *water* instead of acrylic, the result is much better.

In order to build the GCC based on the calculations above (See Fig. A16), take an Acrylic plate with *885mm* X *885m* and *2mm* thickness, then paste on it an Aluminum sheet with *895.2mm* X *885mm* and *0.5mm* thickness(note that two edges of the Aluminum sheet are bent as shown in Figure A16 (b)). Next, take *342* Aluminum yarns with *884mm* length and *2.588mm* diameter (wire # 10 AWG) and insert them side by side on the Aluminum sheet. See in Fig. A16 (b) the detail of fixing of the yarns on the Aluminum sheet. Now, paste acrylic strips (with *13.43mm* height and *2mm* thickness) around the Aluminum/Acrylic, making a box. Put *distilled water* (approximately *1 litter*) inside this box, up to a height of exactly *3.7mm* from the edge of the acrylic base. Afterwards, paste an Acrylic lid (*889mm* X *889mm* and *2mm* thickness) on the box. Note that above the water there is an *air* layer with *885mm* X *885mm* and *7.73mm* thickness (See Fig. A16). This thickness plus the acrylic lid thickness (*2mm*) is equal to $d = b - a = 9.73mm$ where $b = r_0\sqrt{\varepsilon_{r(H_2O)}} = 11.63mm$ and $a = r_0 + \Delta = 1.99mm$, since $r_0 = 1.3mm$, $\varepsilon_{r(H_2O)} = 80$ and $\Delta = 0.6mm$.

Note that the gravitational action of the electric field $E_{air}$, extends itself only up to the distance $d$, which, in this GCC, is given by the sum of the Air layer thickness (*7.73mm*) plus the thickness of the Acrylic lid (*2mm*).

Thus, it is ensured the gravitational effect on the air layer while it is practically nullified in the acrylic sheet above the air layer, since $E_{(acrylic)} << E_{air}$ and $\sigma_{(acrylic)} << 1$.

With this GCC, we can carry out an experiment where the *gravitational mass of the air layer* is progressively reduced when the voltage applied to the GCC is increased (or when the frequency is decreased). A precision balance is placed below the GCC in order to measure the mentioned mass decrease for comparison with the values predicted by Eq. A(47). In total, this GCC weighs about *6kg*; the *air layer 7.3grams*. The balance has the following characteristics: *range 0-6kg; readability 0.1g*. Also, in order to prove the *Gravitational Shielding Effect*, we can put a *sample* (connected to a dynamometer) above the GCC in order to check the gravity acceleration in this region.

In order to prove *the exponential effect* produced by the superposition of the Gravitational Shieldings, we can take three similar GCCs and put them one above the other, in such way that above the GCC 1 the gravity acceleration will be $g' = \chi g$; above the GCC2 $g'' = \chi^2 g$, and above the GCC3 $g''' = \chi^3 g$. Where $\chi$ is given by Eq. (A47).

It is important to note that the intensity of the electric field through the air *below* the GCC is *much smaller* than the intensity of the electric field through the air layer inside the GCC. In addition, the electrical conductivity of the air below the GCC is much smaller than the conductivity of the air layer inside the GCC. Consequently, the decrease of the gravitational mass of the air below the GCC, according to Eq.A14, is negligible. This means that the GCC1, GCC2 and GCC3 can be simply overlaid, on the experiment proposed above. However, since it is necessary to put samples among them in order to measure the gravity above each GCC, we suggest a spacing of 30cm or more among them.

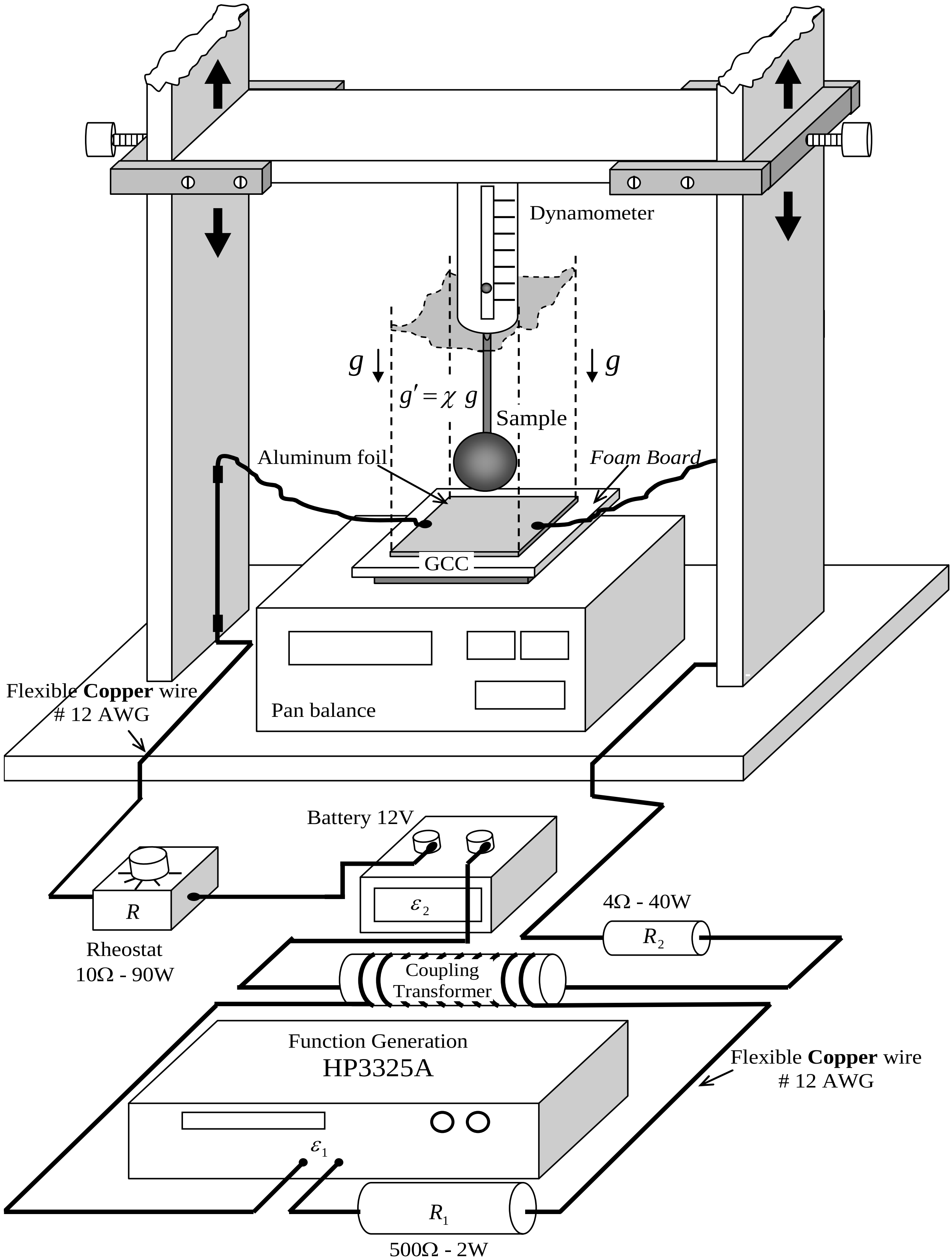

Figure A2 – Experimental Set-up 1.

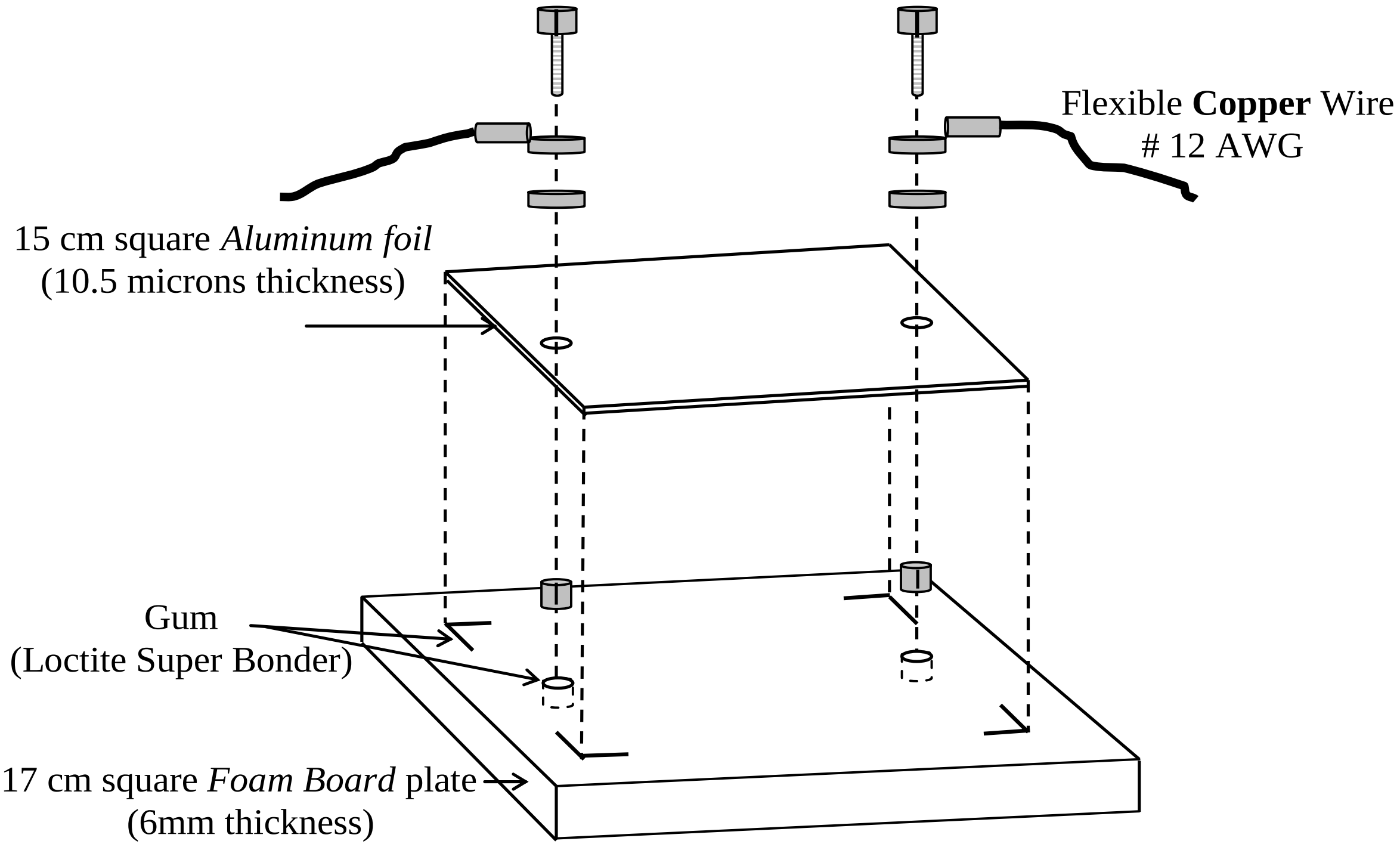

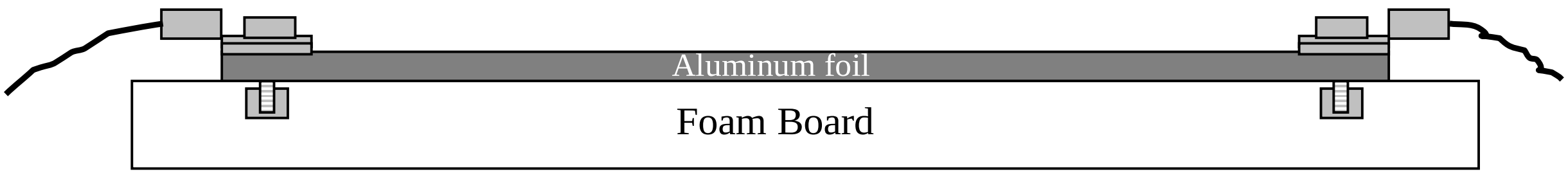

Figure A3 – The Simplest *Gravity Control Cell* (GCC).

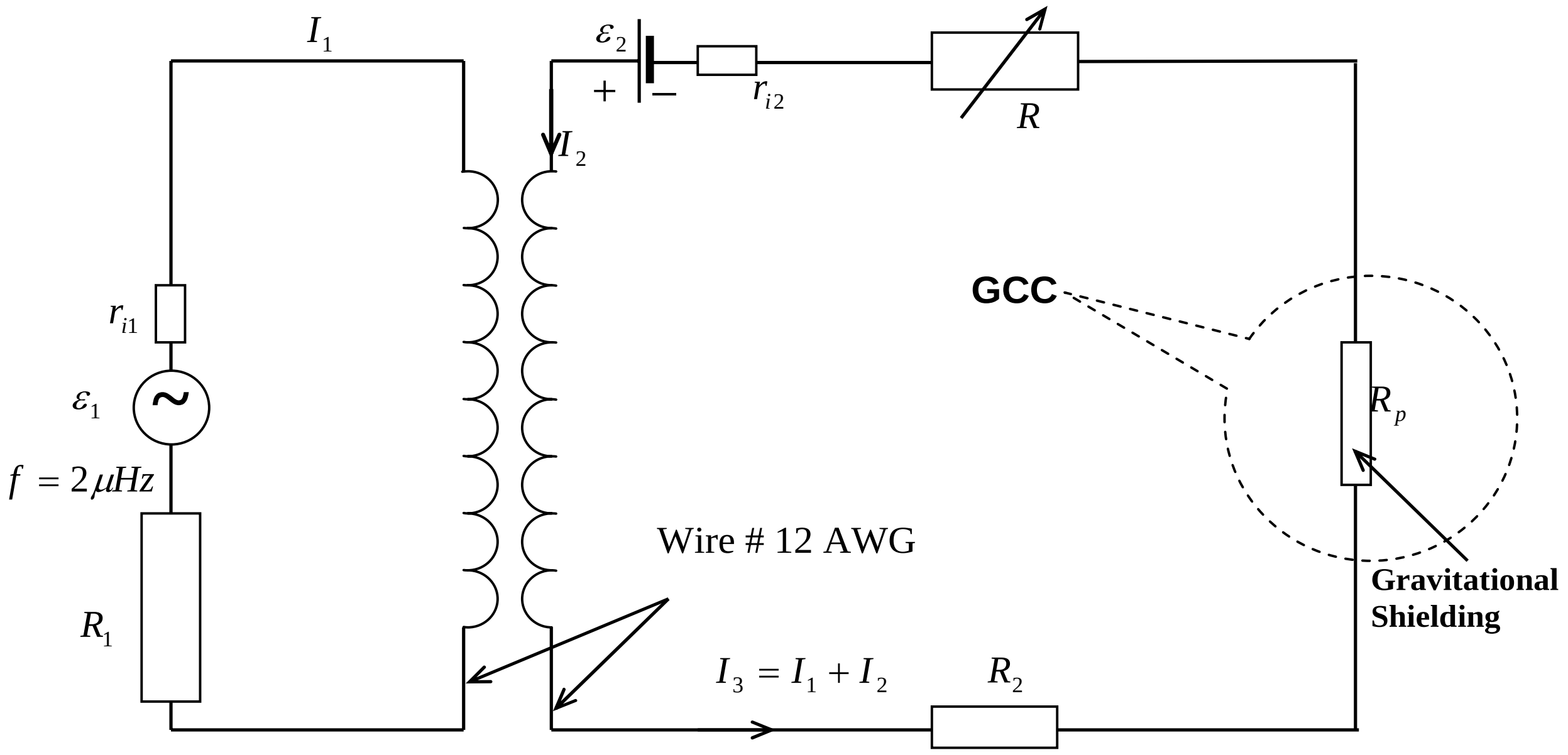

(a)

$\varepsilon_1$ = Function Generator HP3325A(*Option* 002 High Voltage Output)

$r_{i1} < 2\Omega$; $R_1 = 500\Omega - 2\ W$; $\varepsilon_2 = 12V\ DC$; $r_{i2} < 0.1\Omega$ $(Battery)$;

$R_2 = 4\Omega - 40W$; $R_p = 2.5\times10^{-3}\Omega$; $Reostat = 0 \le R \le 10\Omega - 90W$

$I_1^{max} = 56mA$ $(rms)$; $I_2^{max} = 3A$ ; $I_3^{max} \cong 3A$ $(rms)$

*Coupling Transformer* to isolate the *Function Generator* from the Battery

- Air core 10 - mm diameter; wire #12 AWG; $N_1 = N_2 = 20; l = 42mm$

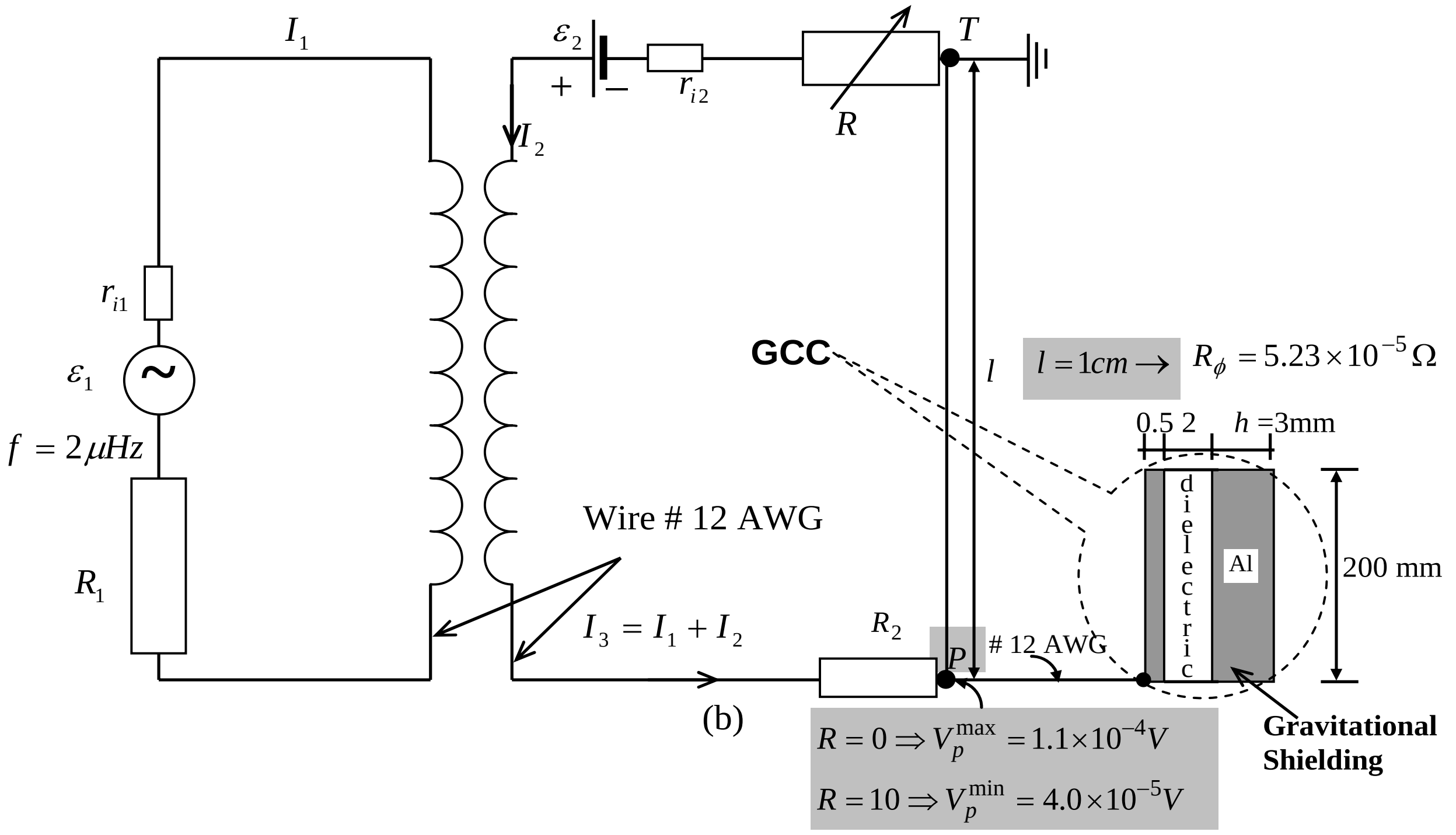

(b)

Fig. A4 – Equivalent Electric Circuits

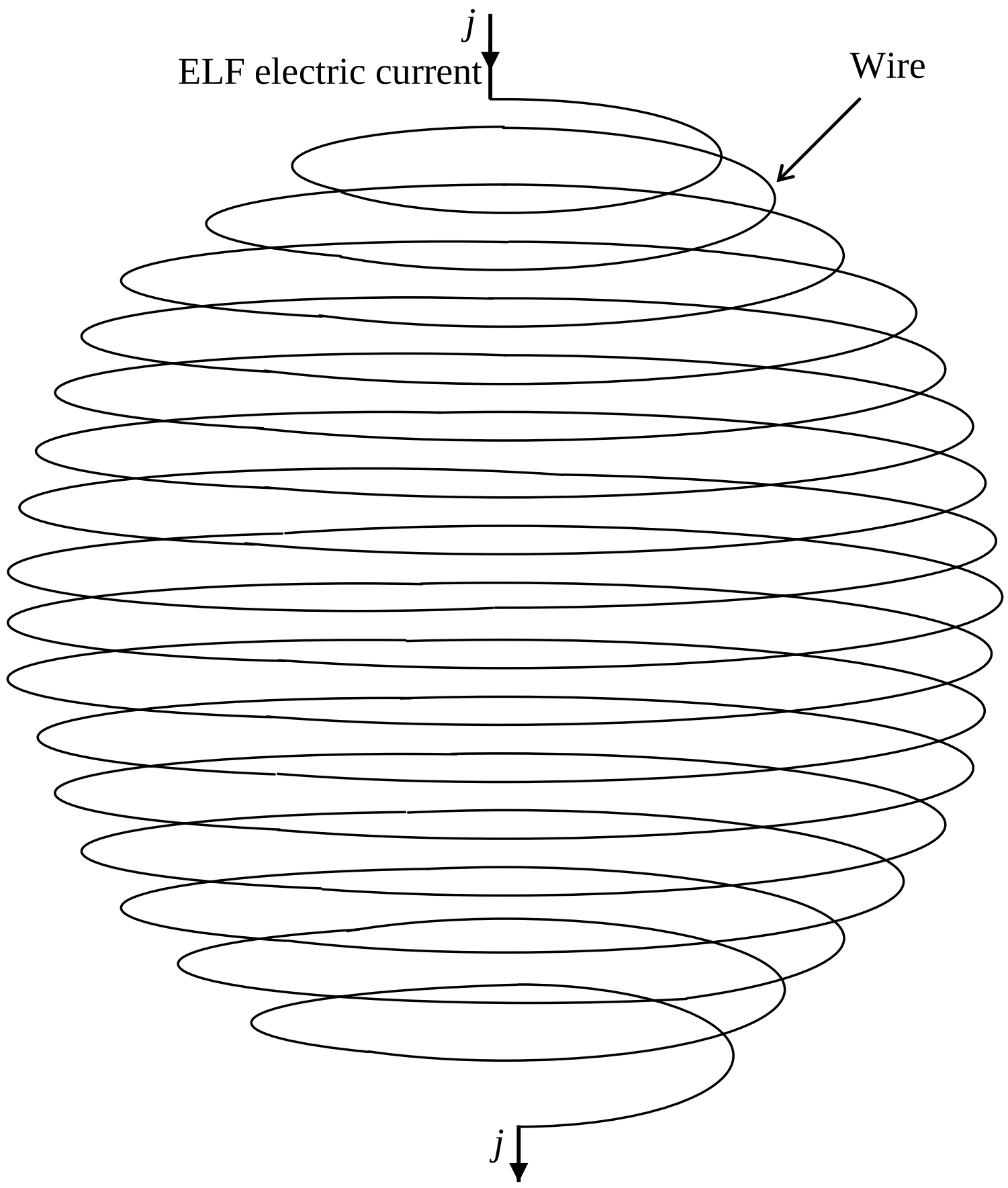

$$m_g = \left\{1 - 2\left[\sqrt{1 + 1.758\times10^{-27}\frac{\mu_r j^4}{\sigma\rho^2 f^3}} - 1\right]\right\}m_{i0}$$

Figure A5 – An ELF electric current through a wire, that makes a spherical form as shown above, reduces the gravitational mass of the wire and the gravity inside sphere at the same proportion $\chi = m_g / m_{i0}$ (Gravitational Shielding Effect). Note that this spherical form can be transformed into an ellipsoidal form or a disc in order to coat, for example, a Gravitational Spacecraft**.** It is also possible to coat with a wire several forms, such as cylinders, cones, cubes, etc. The characteristics of the wire are expressed by: $\mu_r$, $\sigma$, $\rho$; $j$ is the electric current density and $f$ is the frequency.

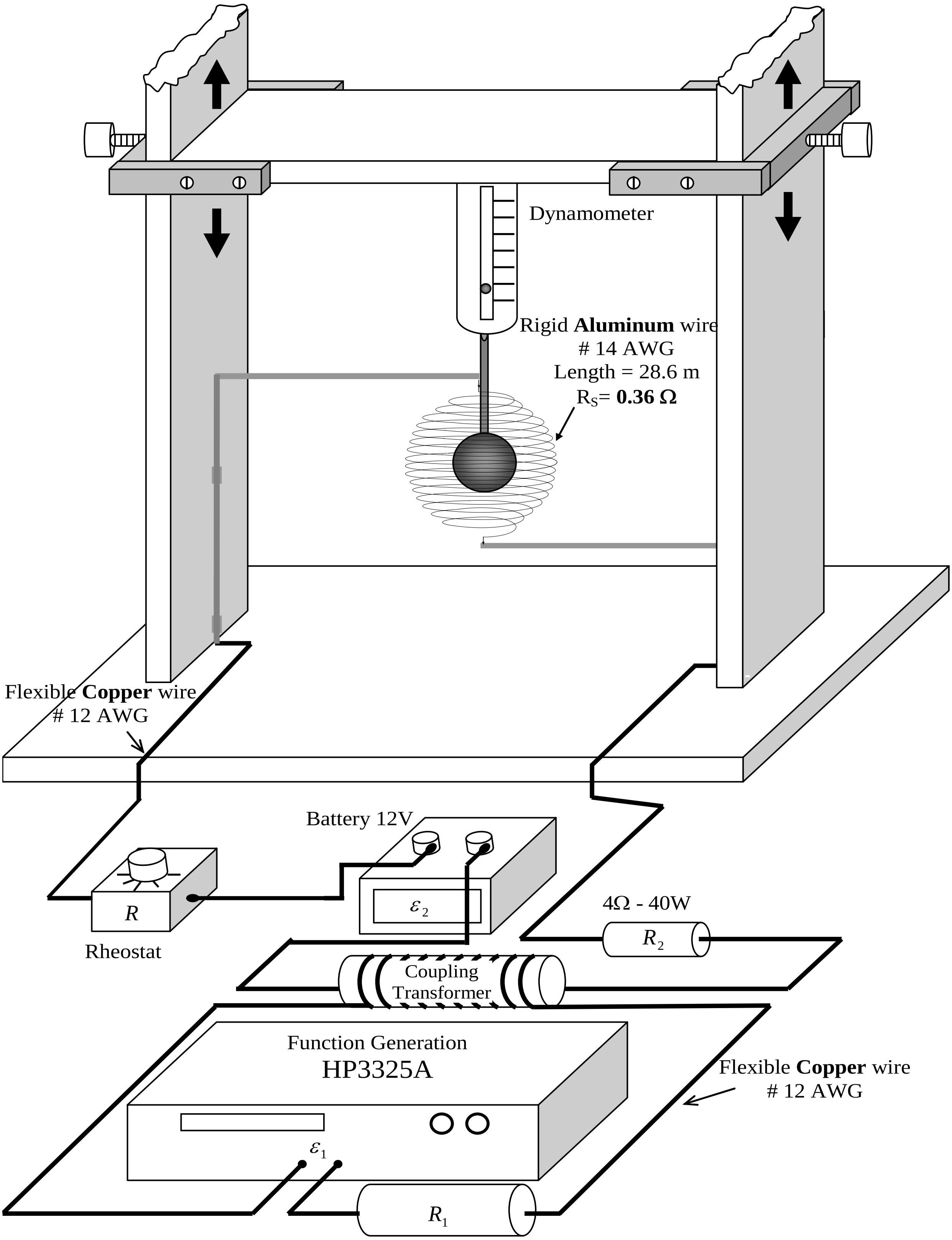

Figure A6 – Experimental set-up 2.

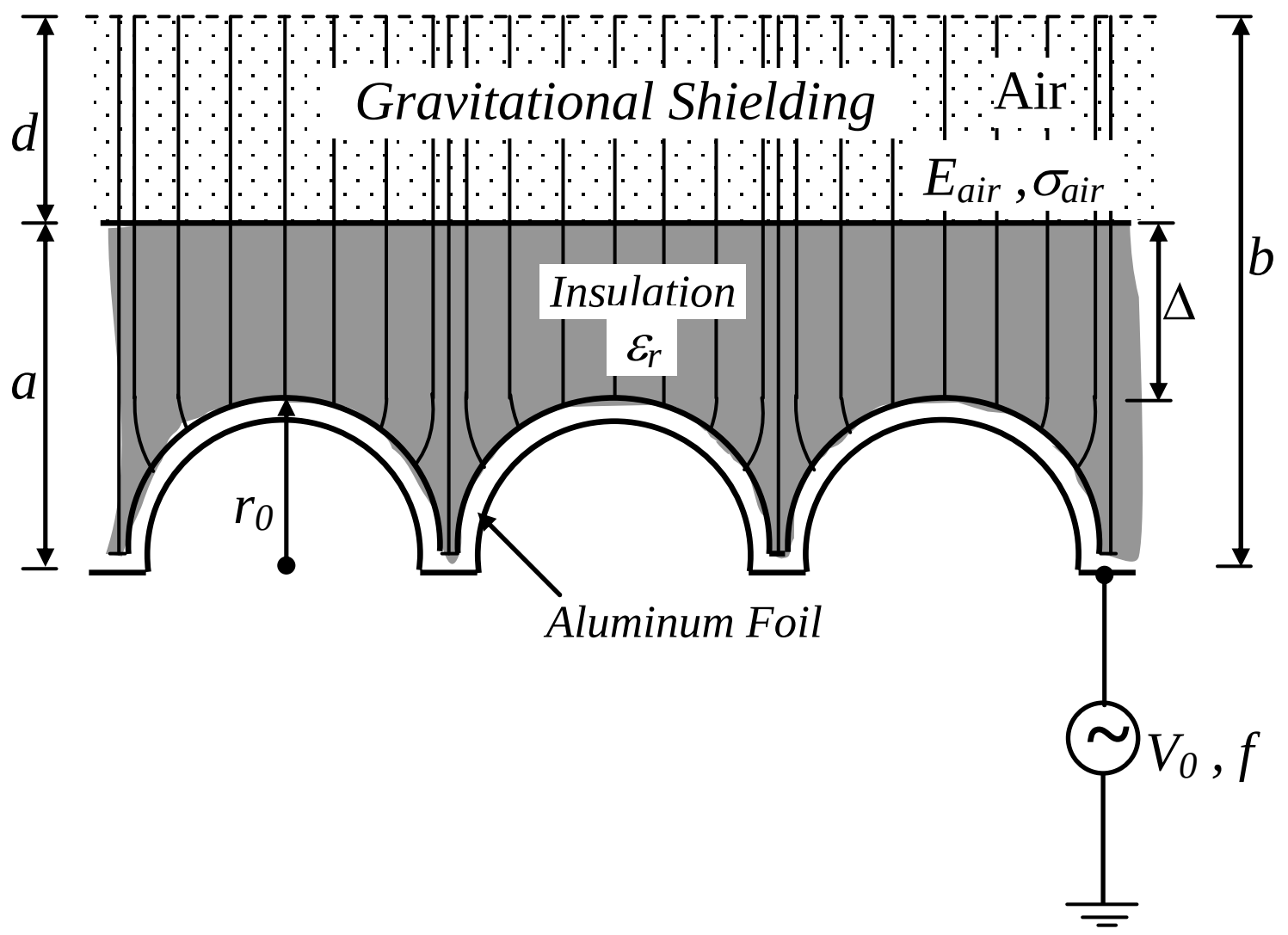

Figure A7 – *Gravitational shielding produced by semi-spheres stamped on the Aluminum foil* - By simply changing the geometry of the surface *of* the Aluminum foil it is possible to increase the working frequency $f$ up to more than *1Hz*.

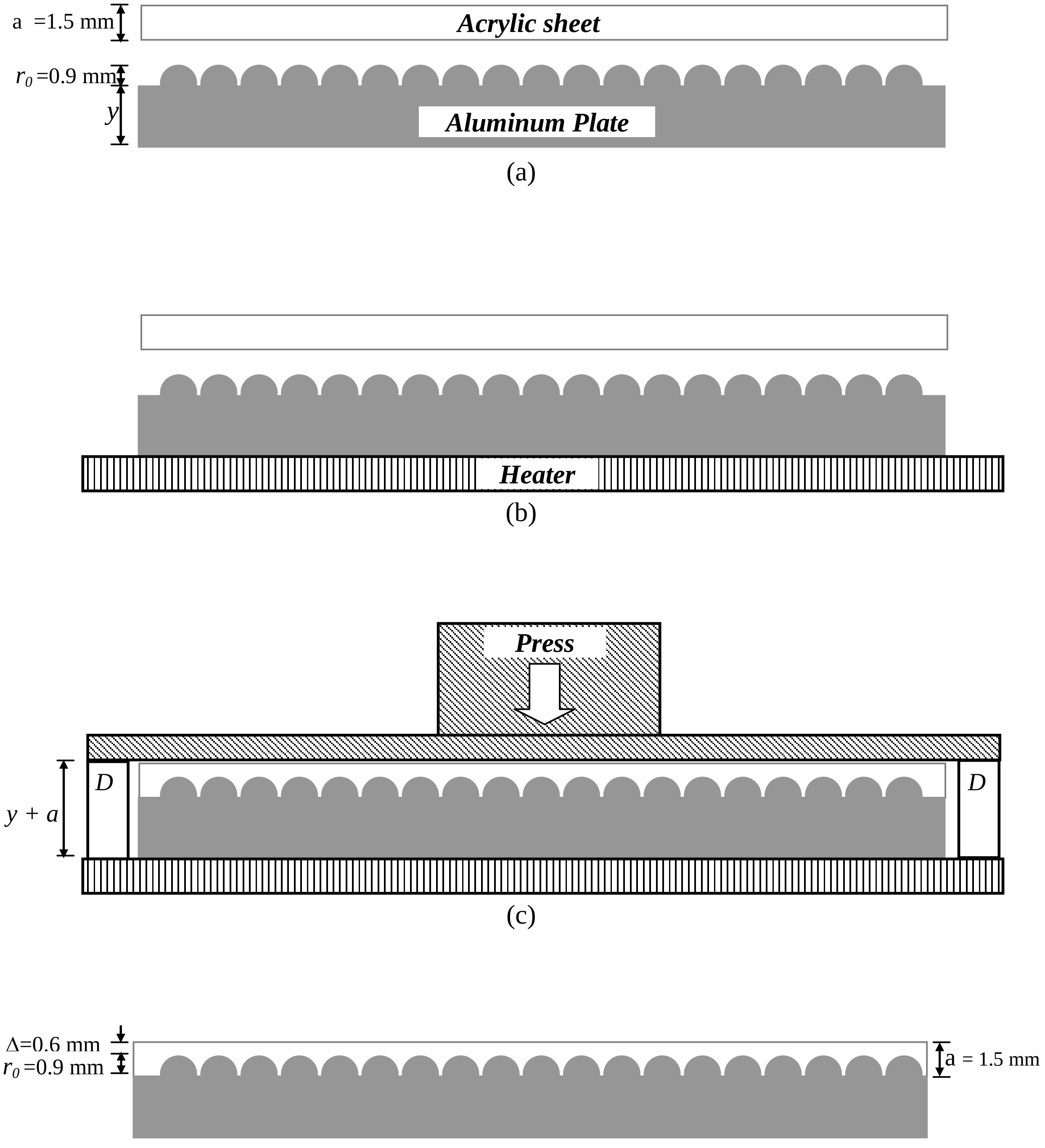

Figure A8 – *Method to coat the Aluminum semi-spheres with acrylic* $(\Delta = a - r_0 = 0.6mm)$. (a)Acrylic sheet (A4 format) with 1.5mm thickness and an Aluminum plate (A4) with several semi-spheres (radius $r_0 = 0.9\ mm$) stamped on its surface. (b)A heater is placed below the Aluminum plate in order to heat the Aluminum. (c)When the Aluminum is sufficiently heated up, the acrylic sheet and the Aluminum plate are pressed, one against the other (The two D devices shown in this figure are used in order to impede that the press compresses the acrylic and the aluminum besides distance $y + a$). (d)After some seconds, the press and the heater are removed, and the device is ready to be used.

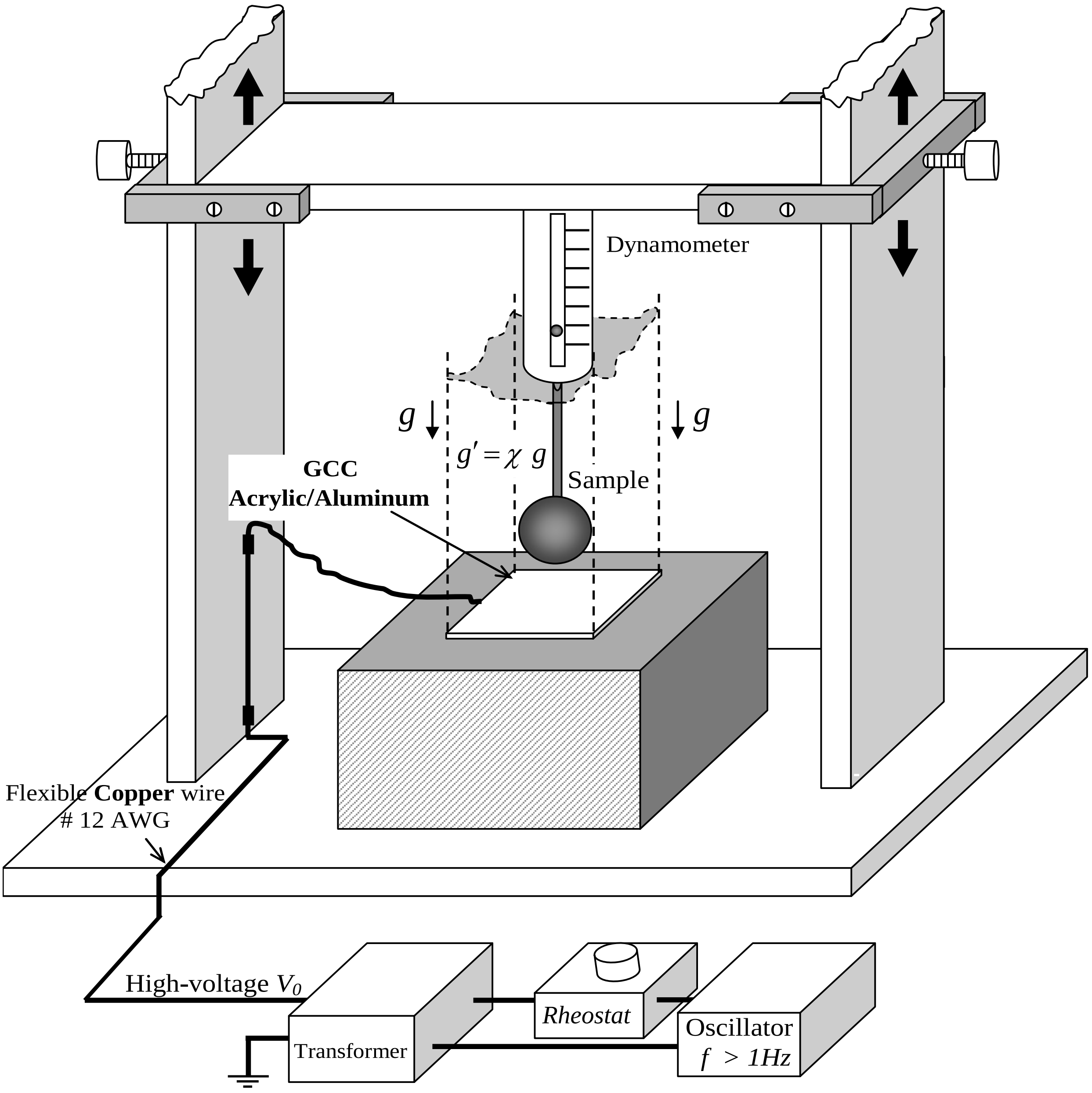

Figure A9 – *Experimental Set-up using a GCC subjected to high-voltage* $V_0$ *with frequency* $f > 1Hz$ . Note that in this case, the pan balance is not necessary because the substance of the Gravitational Shielding is an *air layer* with thickness $d$ *above* the acrylic sheet. This is therefore, more a type of Gravity Control Cell (GCC) with *external gravitational shielding.*

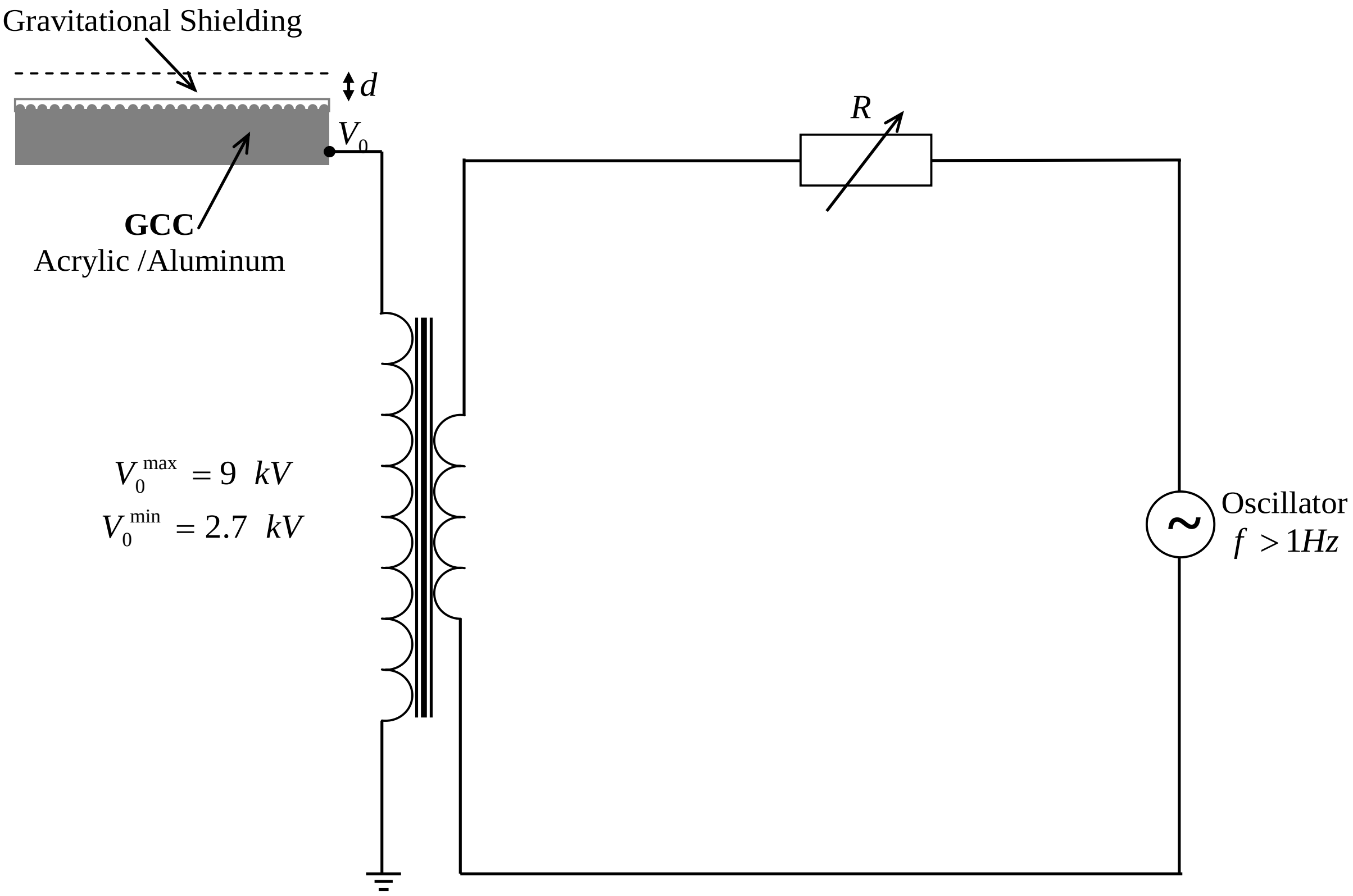

(a)

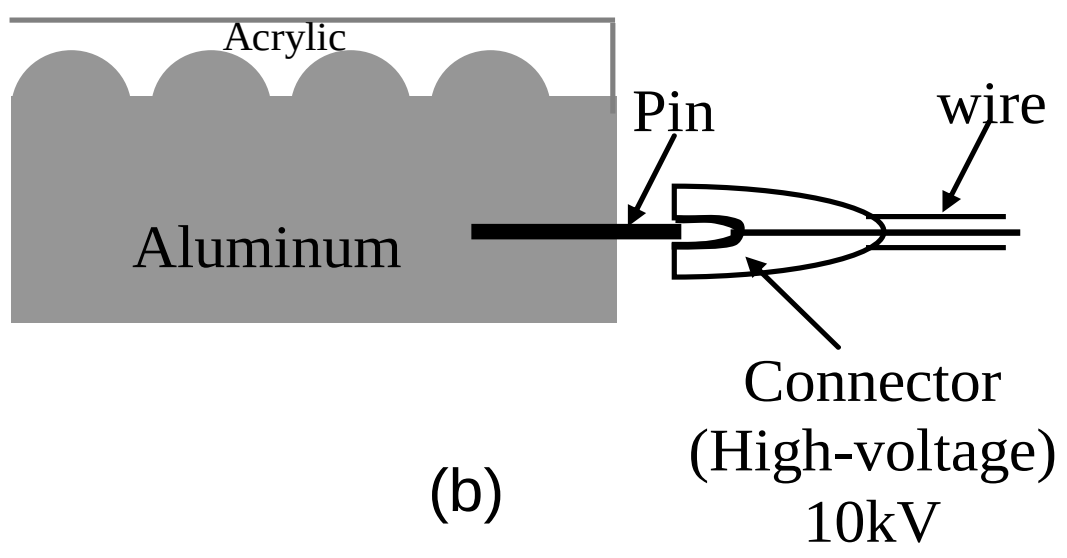

(b)

Figure A10 – (a) *Equivalent Electric Circuit*. (b) Details of the electrical connection with the Aluminum plate. Note that others connection modes (by the top of the device) can produce destructible interference on the electric lines of the $E_{air}$ field.

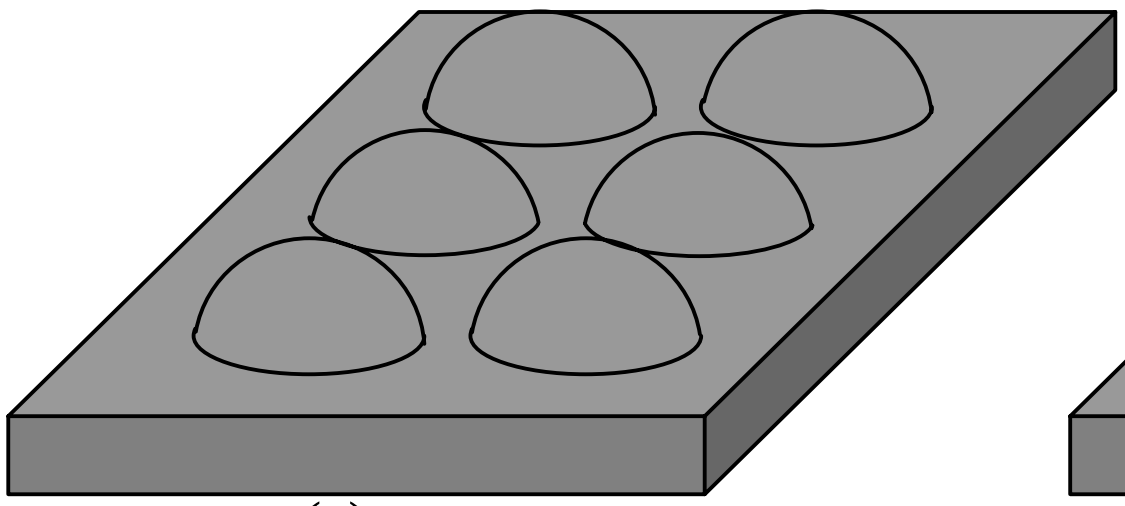

(a)

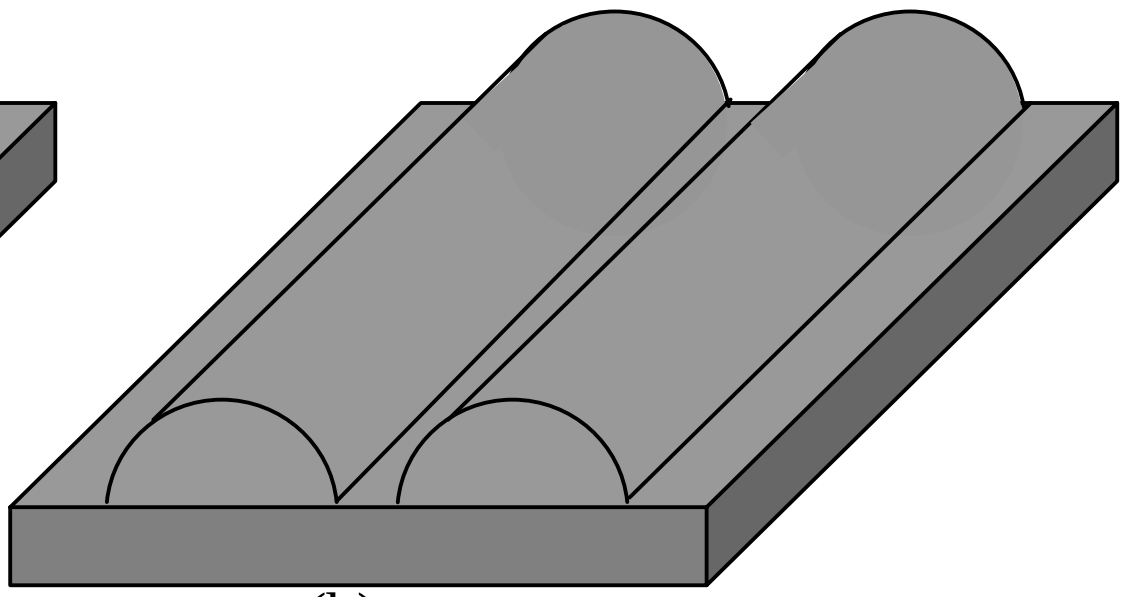

(b)

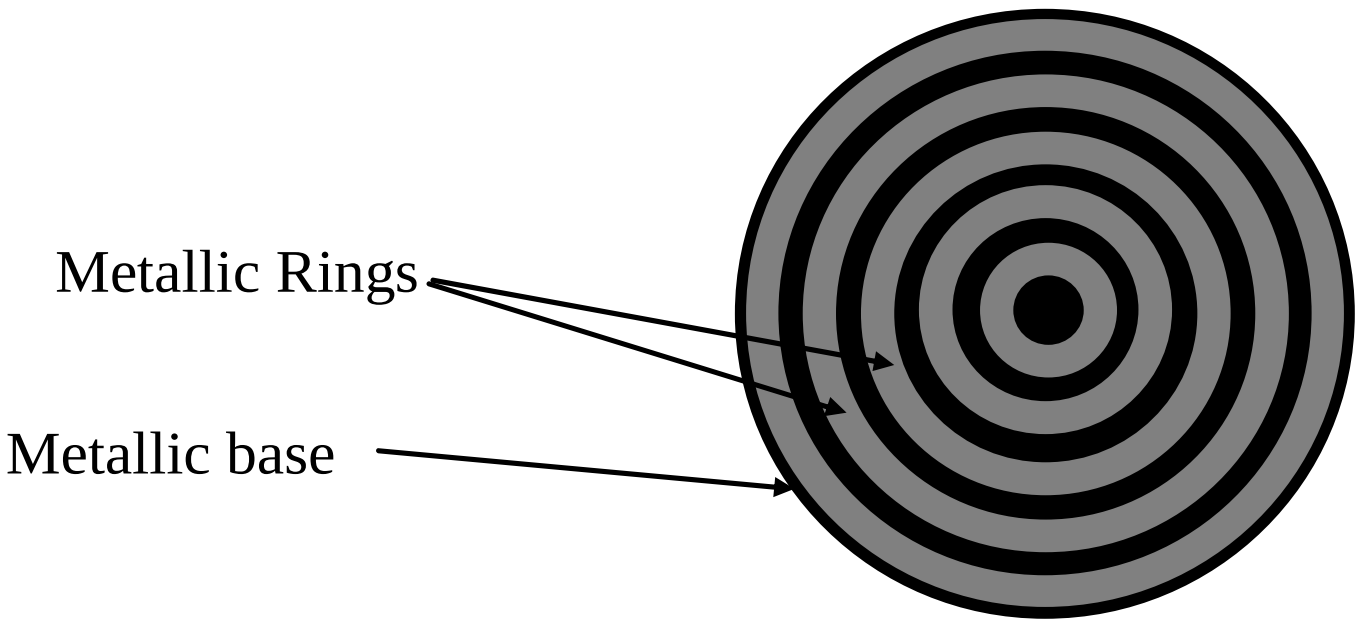

(c)

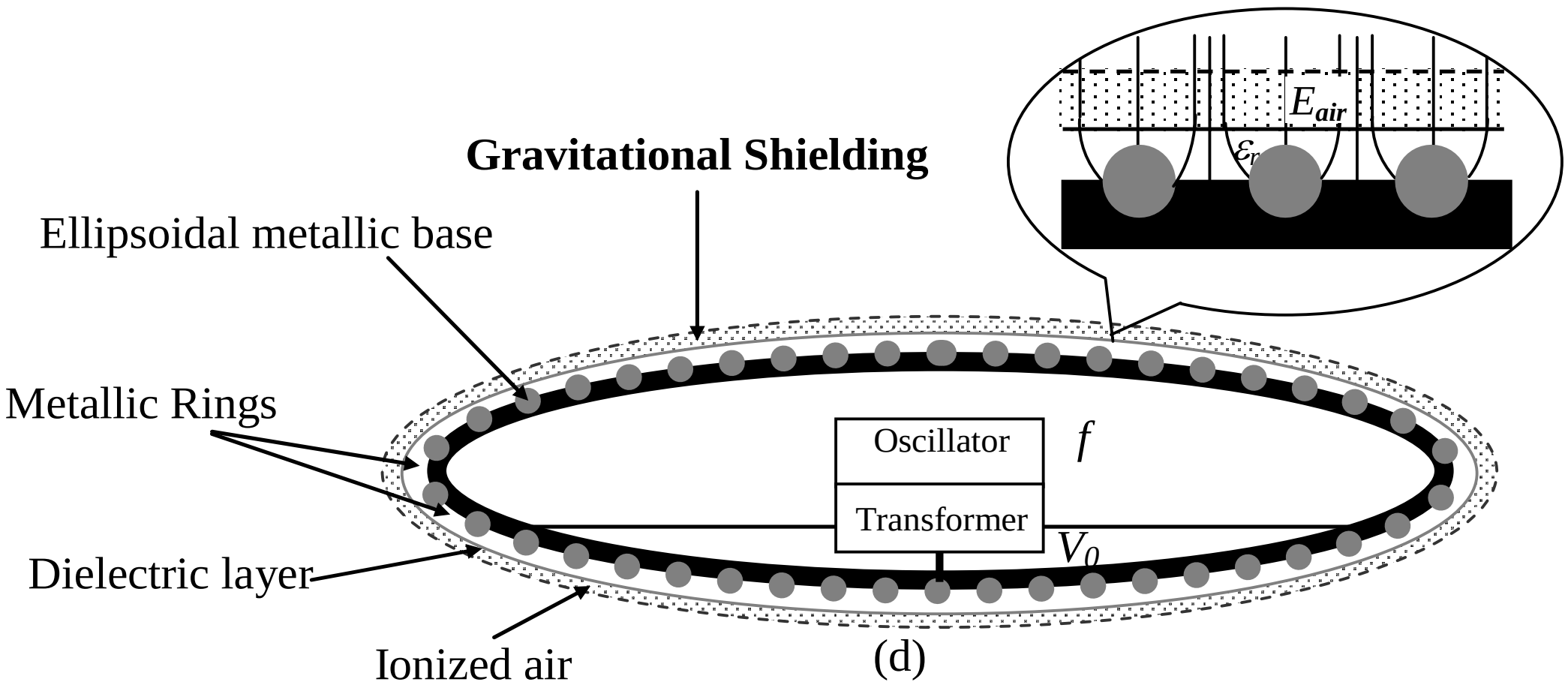

(d)

Figure A11 – *Geometrical forms with similar effects as those produced by the semi-spherical form* – (a) shows the semi-spherical form stamped on the metallic surface; (b) shows the *semi-cylindrical* form (an obvious evolution from the semi-spherical form); (c) shows *concentric metallic rings* stamped on the metallic surface, an evolution from semi-cylindrical form. These geometrical forms produce the same effect as that of the semi-spherical form, shown in Fig.A11 (a). By using concentric metallic rings, it is possible to build *Gravitational Shieldings* around bodies or spacecrafts with several formats (spheres, ellipsoids, etc); (d) shows a Gravitational Shielding around a Spacecraft with *ellipsoidal form*.

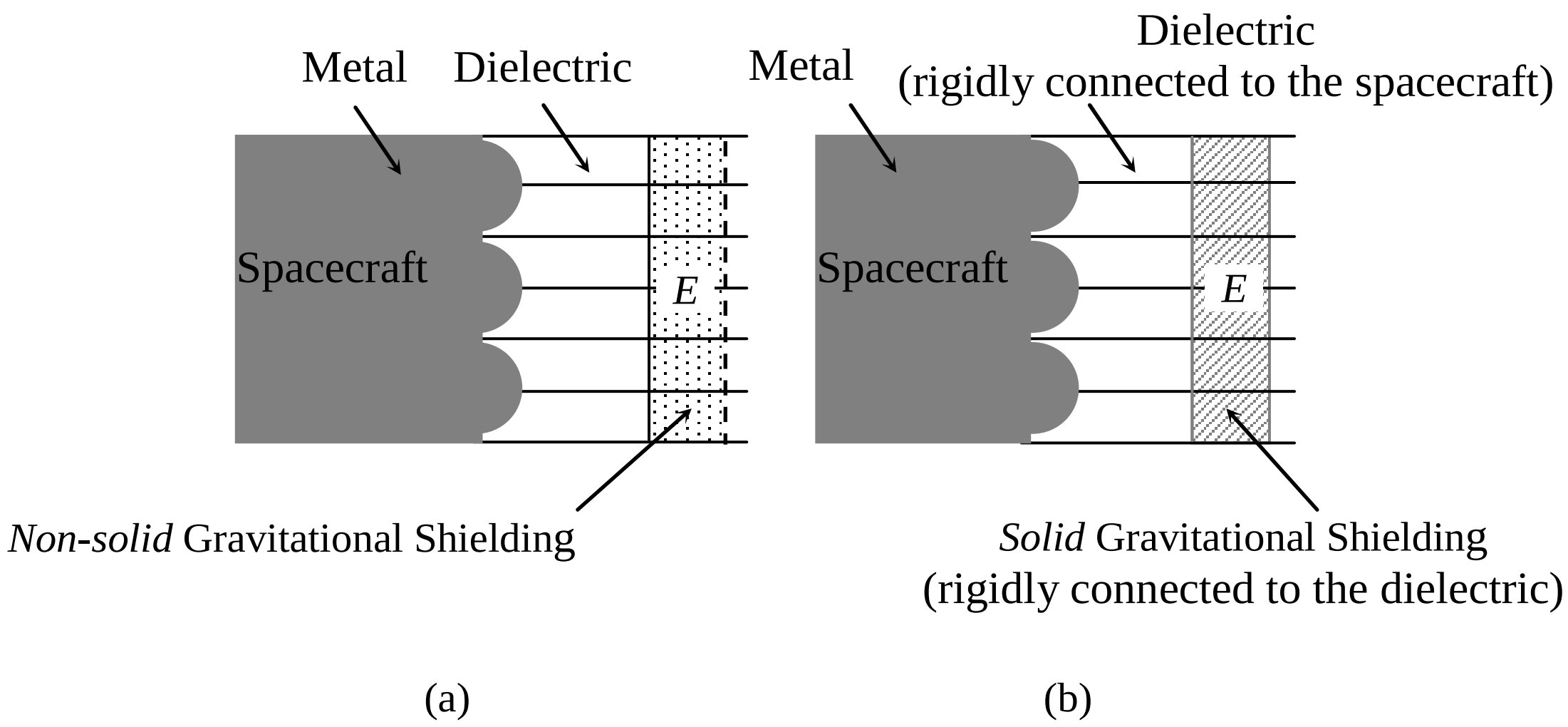

Figure A12 – *Non-solid and Solid Gravitational Shieldings* - In the case of the Gravitational Shielding produced on a *solid substance* (b), when its molecules go to the *imaginary* space-time, *the electric field that produces the effect also goes to the imaginary space-time together with them*, because in this case, the substance of the Gravitational Shielding is *rigidly connected (by means of the dielectric) to the metal* that produces the electric field. This does not occur in the case of *Air* Gravitational Shielding.

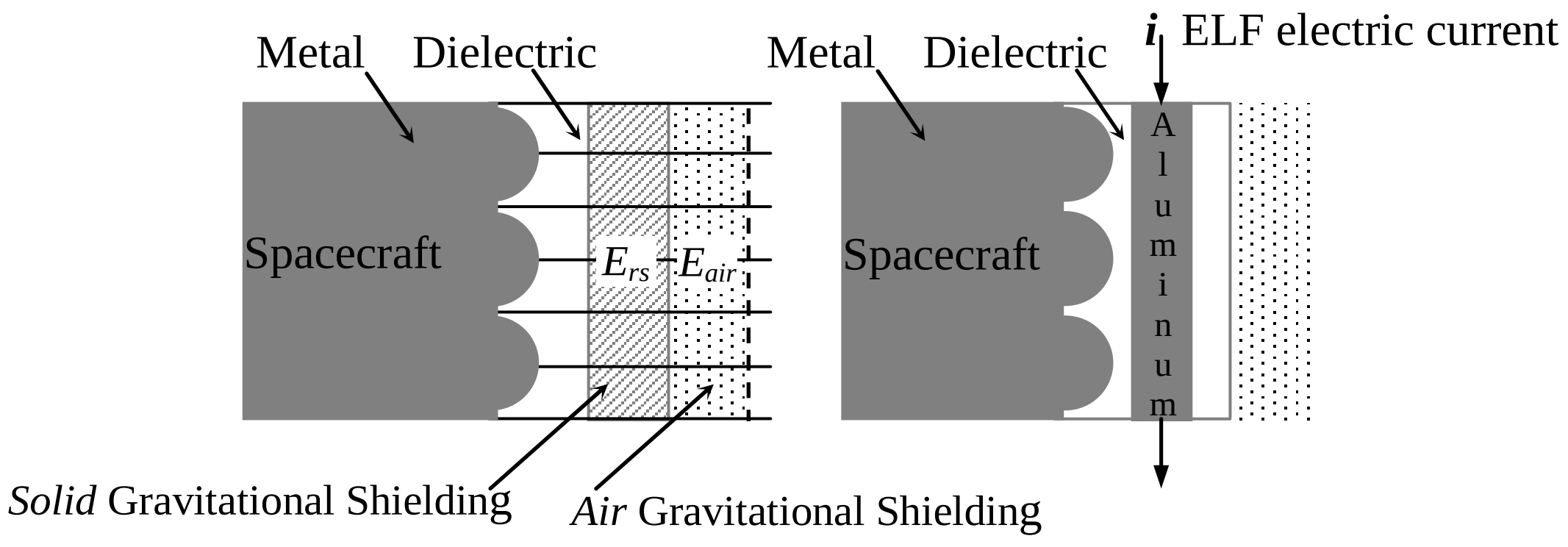

(a)

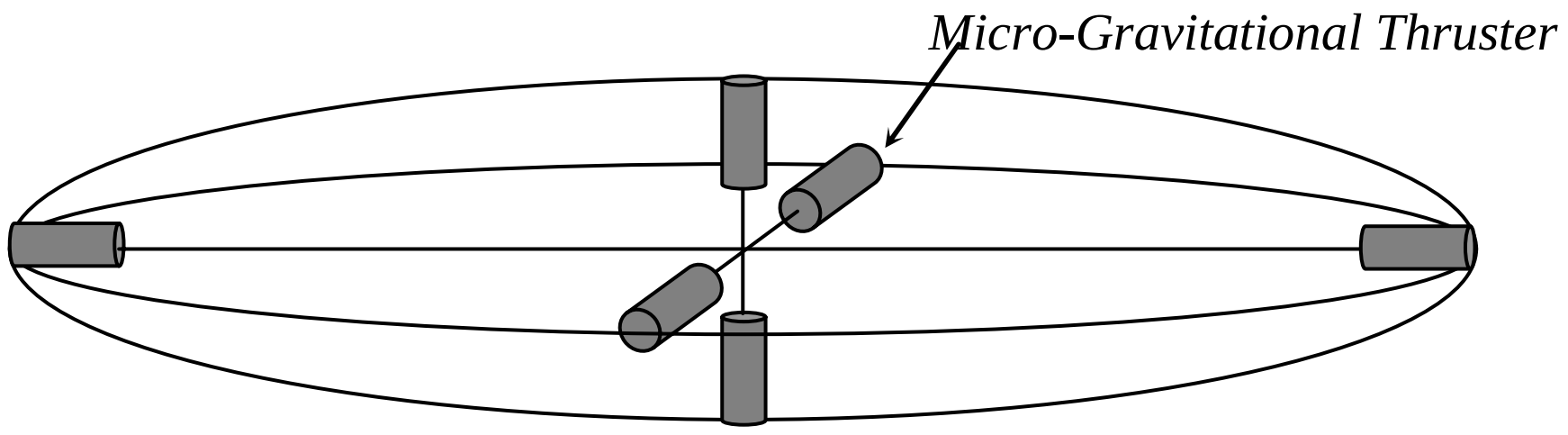

(b)

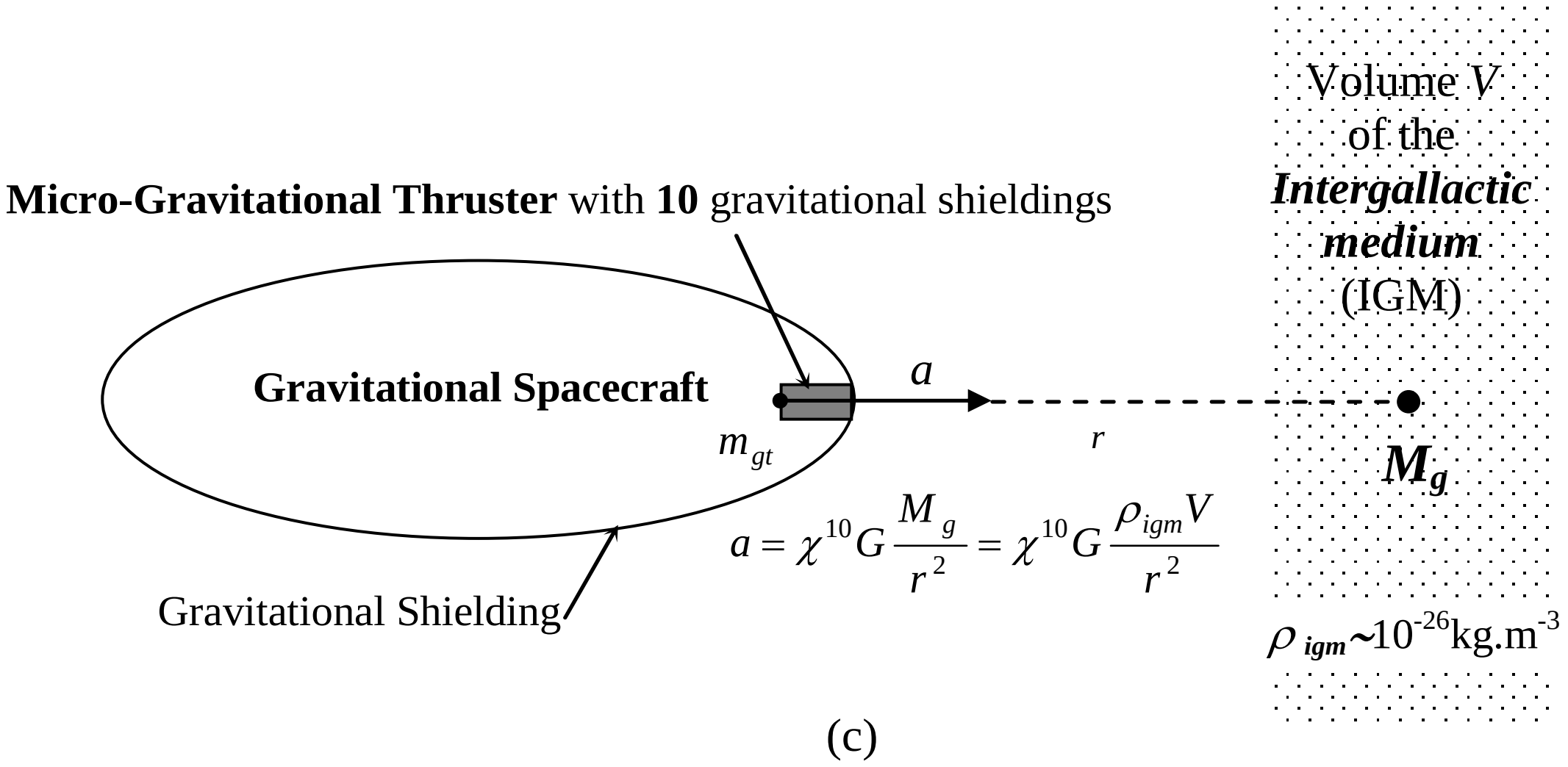

(c)

Figure A13 – *Double Gravitational Shielding and Micro-thrusters* – (a) Shows a double gravitational shielding that makes possible to decrease the *inertial effects* upon the spacecraft when it is traveling both in the *imaginary* space-time and in the *real* space-time. The *solid* Gravitational Shielding also can be obtained by means of *an ELF electric current through a metallic lamina* placed *between the semi-spheres and the Gravitational Shielding of Air* as shown above. (b) Shows 6 *micro-thrusters* placed inside a Gravitational Spacecraft, in order to propel the spacecraft in the directions x, y and z. Note that the Gravitational Thrusters in the spacecraft must have a very small diameter (of the order of *millimeters*) because the hole through the Gravitational Shielding of the spacecraft cannot be large. Thus, these thrusters are in fact *Micro-thrusters*. (c) Shows a micro-thruster inside a spacecraft, and in front of a volume *V* of the intergalactic medium (IGM). Under these conditions, the spacecraft acquires an acceleration *a* in the direction of the volume *V*.

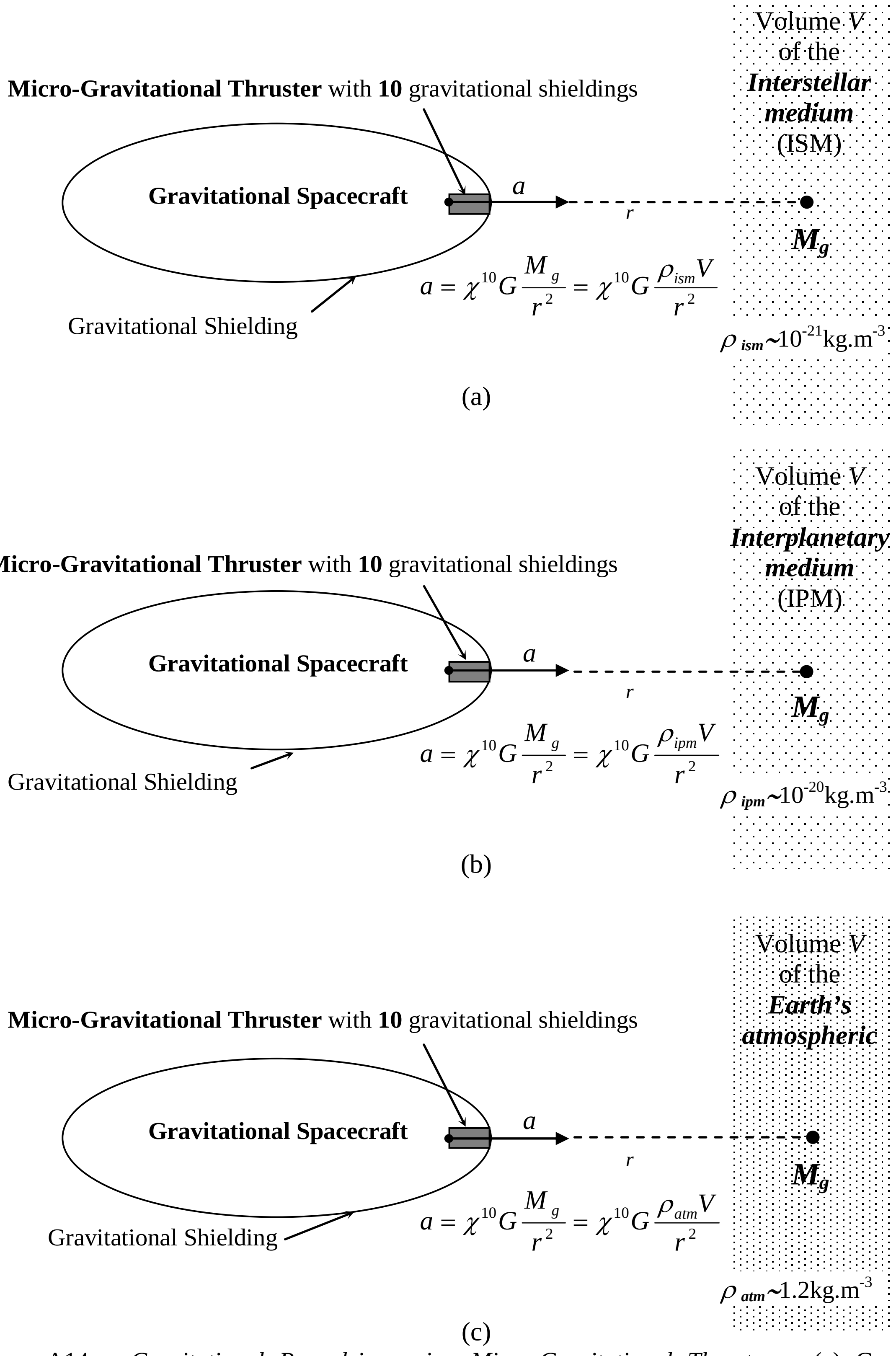

Figure A14 – *Gravitational Propulsion using Micro-Gravitational Thruster* – (a) Gravitational acceleration produced by a gravitational mass $M_g$ of the *Interstellar Medium*. The density of the Interstellar Medium is about $10^5$ times greater than the density of the *Intergalactic Medium* (b) Gravitational acceleration produced in the *Interplanetary Medium*. (c) Gravitational acceleration produced in the *Earth's atmosphere*. Note that, in this case, $\rho_{atm}$ (*near to the Earth's surface*)is about $10^{26}$ times greater than the density of the *Intergalactic Medium*.

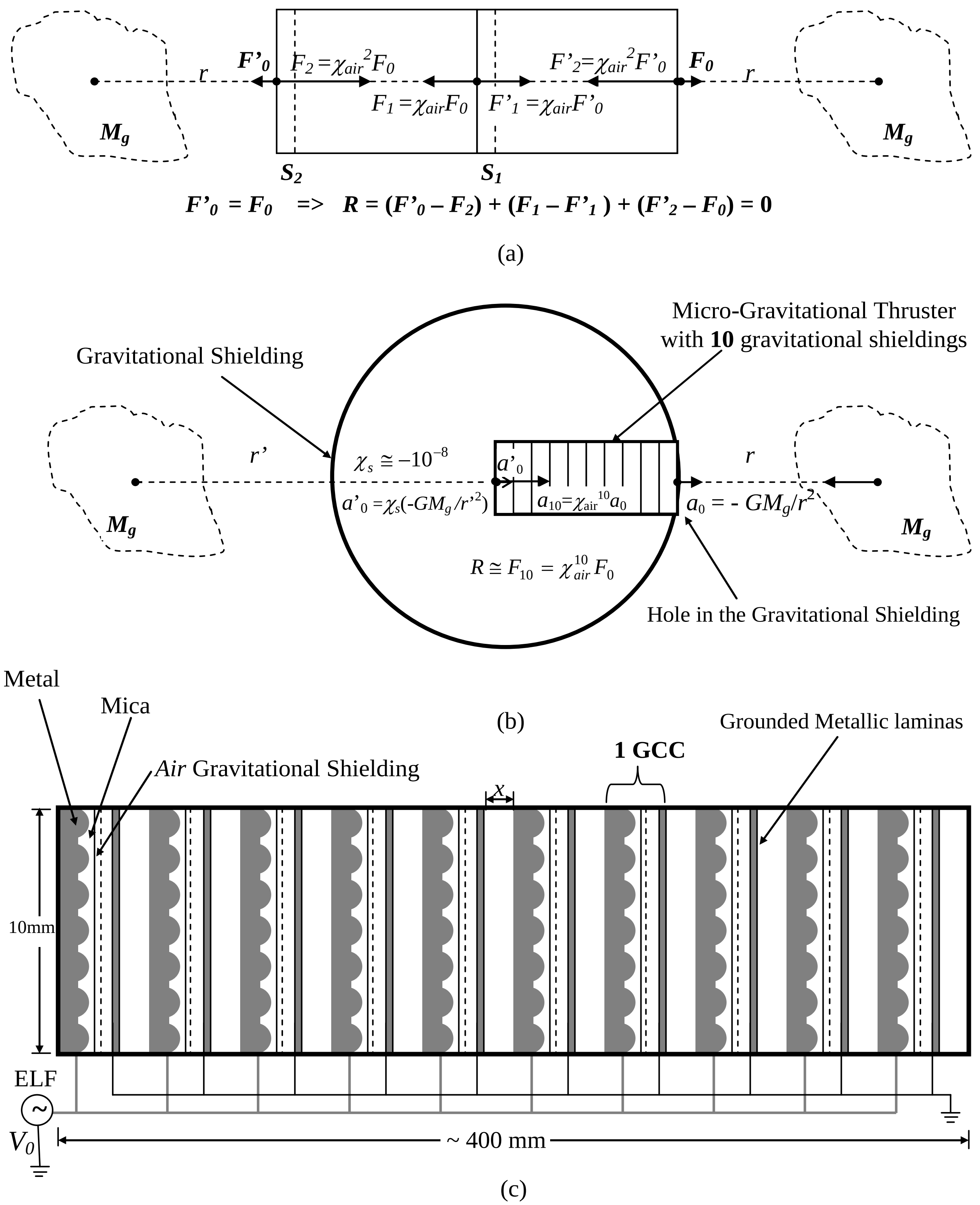

Figure A15 – *Dynamics and Structure of the Micro-Gravitational Thrusters* - (a) The Micro-Gravitational Thrusters do not work *outside* the Gravitational Shielding, because, in this case, *the resultant upon the thruster is null* due to the symmetry. (b) The Gravitational Shielding $\left(\chi_s \cong 10^{-8}\right)$ reduces strongly the intensities of the gravitational forces acting on the micro-gravitational thruster, except obviously, through the hole in the gravitational shielding. (c) Micro-Gravitational Thruster with *10 Air Gravitational Shieldings* (10GCCs). The grounded metallic laminas are placed so as to retain the electric field produced by metallic surface behind the semi-spheres.

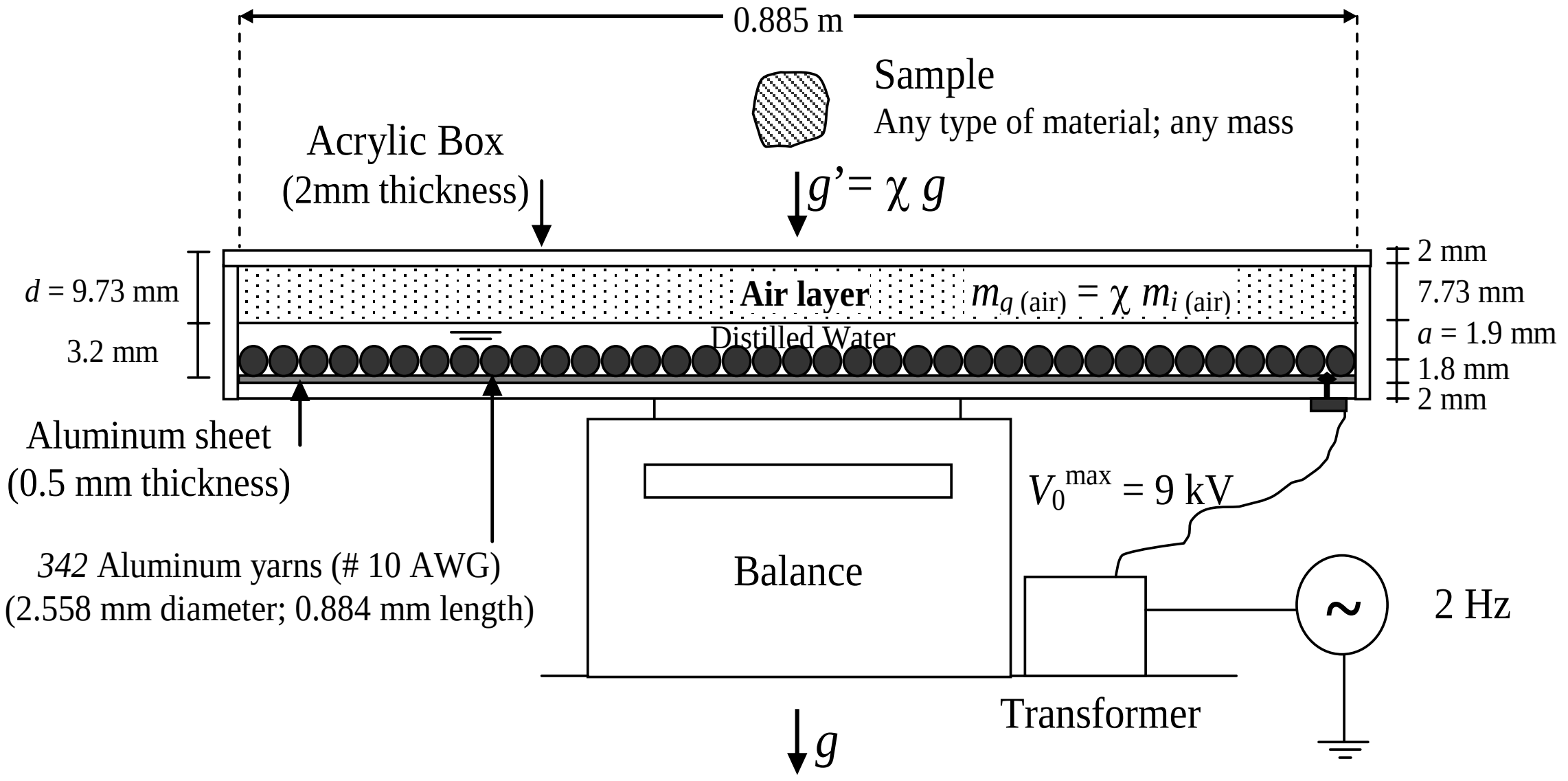

**GCC Cross-section Front view**

(a)

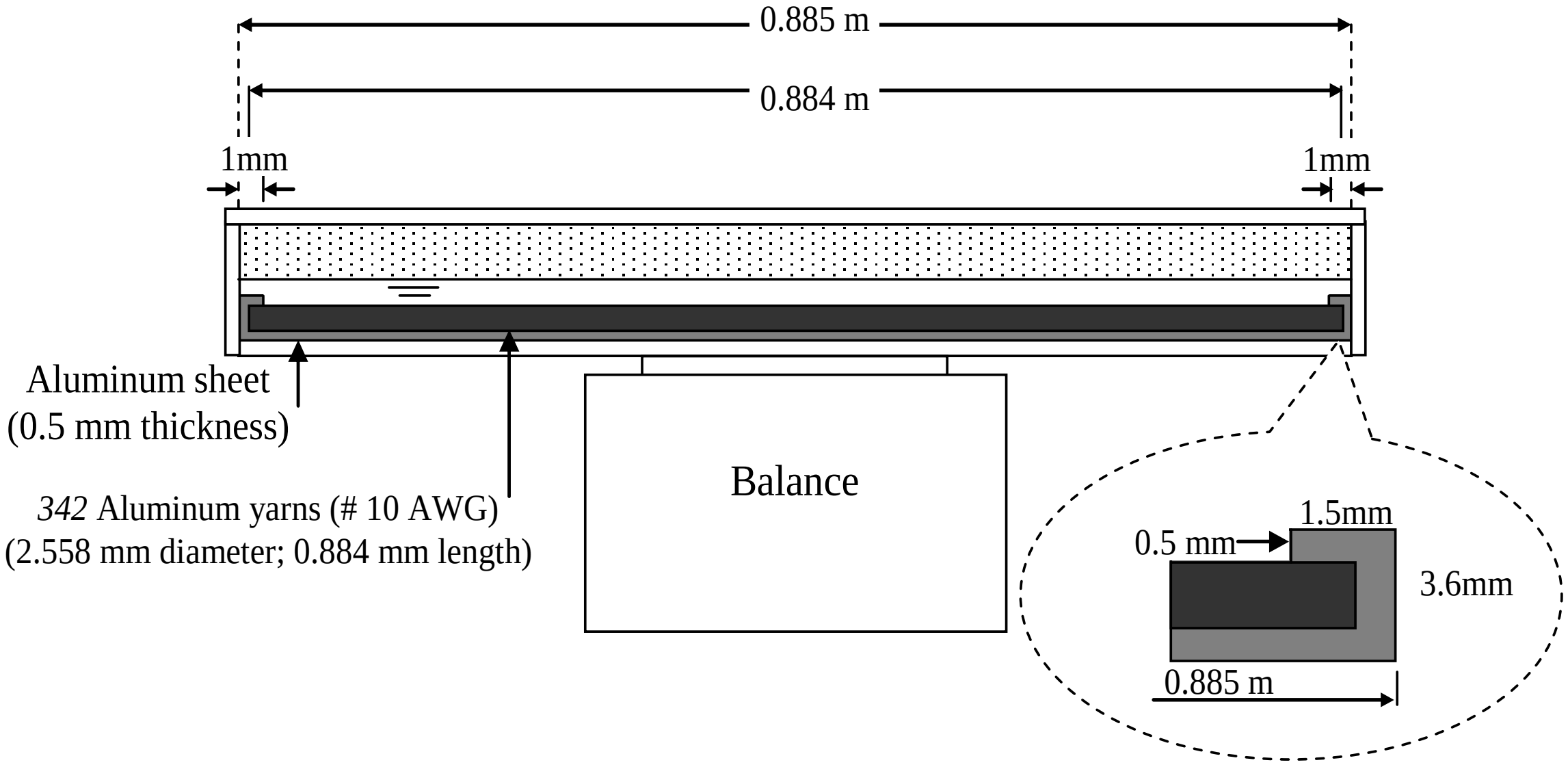

**GCC Cross-section Side View**

(b)

Fig. A16 – *A GCC using distilled Water*.
In total this GCC weighs about 6kg; the air layer 7.3 grams. The balance has the following characteristics: Range 0 – 6kg; readability 0.1g. The yarns are inserted side by side on the Aluminum sheet. Note the detail of fixing of the yarns on the Aluminum sheet.

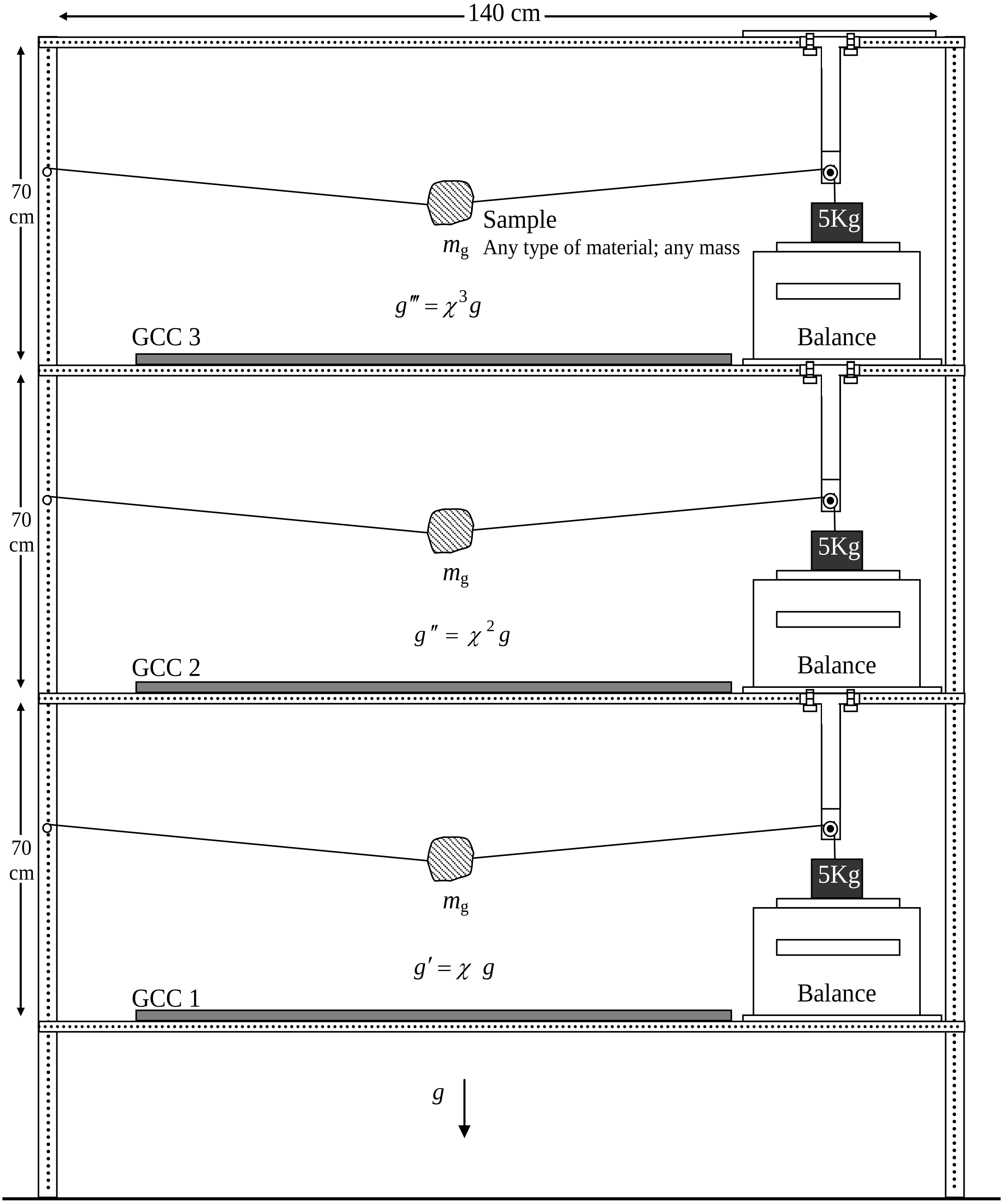

Fig. A17 – *Experimental set-up.* In order to prove *the exponential effect* produced by the superposition of the Gravitational Shieldings, we can take three similar GCCs and put them one above the other, in such way that above the GCC 1 the gravity acceleration will be $g' = \chi g$; above the GCC2 $g'' = \chi^2 g$, and above the GCC3 $g''' = \chi^3 g$. Where $\chi$ is given by Eq. (A47). The arrangement above has been designed for values of $m_g < 13g$ and $\chi$ up to -9 or $m_g < 1kg$ and $\chi$ up to -2 .

**APPENDIX B: A DIDACTIC *GCC* USING A BATTERY OF CAPACITORS**

Let us now show a new type of GCC - easy to be built with materials and equipments that also can be obtained with easiness.

Consider a battery of $n$ parallel plate capacitors with capacitances $C_1, C_2, C_3, \ldots, C_n$, connected in parallel. The voltage applied is $V$; $A$ is the area of each plate of the capacitors and $d$ is the distance between the plates; $\varepsilon_{r(water)}$ is the relative permittivity of the dielectric (water). Then the electric charge $q$ on the plates of the capacitors is given by

$$q = (C_1 + C_2 + C_3 + \ldots + C_n)V = n(\varepsilon_{r(water)}\varepsilon_0)\frac{A}{d}V \qquad (B1)$$

In Fig. I we show a GCC with *two* capacitors connected in parallel. It is easy to see that the electric charge density $\sigma_0$ on each area $A_0 = az$ of the edges B of the thin laminas (*z* is the thickness of the edges B and *a* is the length of them, see Fig.B2) is given by

$$\sigma_0 = \frac{q}{A_0} = n(\varepsilon_{r(water)}\varepsilon_0)\frac{A}{azd}V \qquad (B2)$$

Thus, the electric field $E$ between the edges B is

$$E = \frac{2\sigma_0}{\varepsilon_{r(air)}\varepsilon_0} = 2n\left(\frac{\varepsilon_{r(water)}}{\varepsilon_{r(air)}}\right)\frac{A}{azd}V \qquad (B3)$$

Since $A = L_x L_y$, we can write that

$$E = 2n\left(\frac{\varepsilon_{r(water)}}{\varepsilon_{r(air)}}\right)\frac{L_x L_y}{azd}V \qquad (B4)$$

Assuming $\varepsilon_{r(water)} = 81$ **** (bidistilled water); $\varepsilon_{r(air)} \cong 1$ (vacuum $10^{-4}$ Torr; 300K); $n = 2$; $L_x = L_y = 0.30m$; $a = 0.12m$; $z = 0.1mm$ and $d = 10mm$ we obtain

$$E = 2.43 \times 10^8 V$$

For $V_{max} = 220V$, the electric field is

$$E_{max} = 5.3 \times 10^{10} V/m$$

Therefore, if the frequency of the wave voltage is $f = 60Hz$, $(\omega = 2\pi f)$, we have that $\omega\varepsilon_{air} = 3.3 \times 10^{-9} S.m^{-1}$. It is known that the electric conductivity of the air, $\sigma_{air}$, at $10^{-4}$ Torr and 300K, is much smaller than this value, i.e.,

$$\sigma_{air} << \omega\varepsilon_{air}$$

Under this circumstance $(\sigma << \omega\varepsilon)$, we can substitute Eq. 15 and 34 into Eq. 7. Thus, we get

$$m_{g(air)} = \left\{1 - 2\left[\sqrt{1 + \frac{\mu_{air}\varepsilon_{air}^3}{c^2}\frac{E^4}{\rho_{air}^2}} - 1\right]\right\}m_{i0(air)}$$
$$= \left\{1 - 2\left[\sqrt{1 + 9.68 \times 10^{-57}\frac{E^4}{\rho_{air}^2}} - 1\right]\right\}m_{i0(air)} \qquad (B5)$$

The density of the air at $10^{-4}$ Torr and 300K is

$$\rho_{air} = 1.5 \times 10^{-7} kg.m^{-3}$$

Thus, we can write

$$\chi = \frac{m_{g(air)}}{m_{i(air)}} =$$
$$= \left\{1 - 2\left[\sqrt{1 + 4.3 \times 10^{-43} E^4} - 1\right]\right\} \qquad (B6)$$

Substitution of $E$ for $E_{max} = 5.3 \times 10^{10} V/m$ into this equation gives

$$\chi_{max} \cong -1.2$$

This means that, in this case, the *gravitational shielding* produced in the vacuum between the edges B of the thin laminas can reduce the local gravitational acceleration $g$ down to

$$g_1 \cong -1.2g$$

Under these circumstances, the weight, $P = +m_g g$, of any body just *above* the gravitational shielding becomes

$$P = m_g g_1 = -1.2 m_g g$$

---

****It is easy to see that by substituting the water for Barium Titanate ($BaTiO_3$) the dimensions $L_x, L_y$ of the capacitors can be strongly reduced due to $\varepsilon_{r(BaTiO3)} = 1200$.

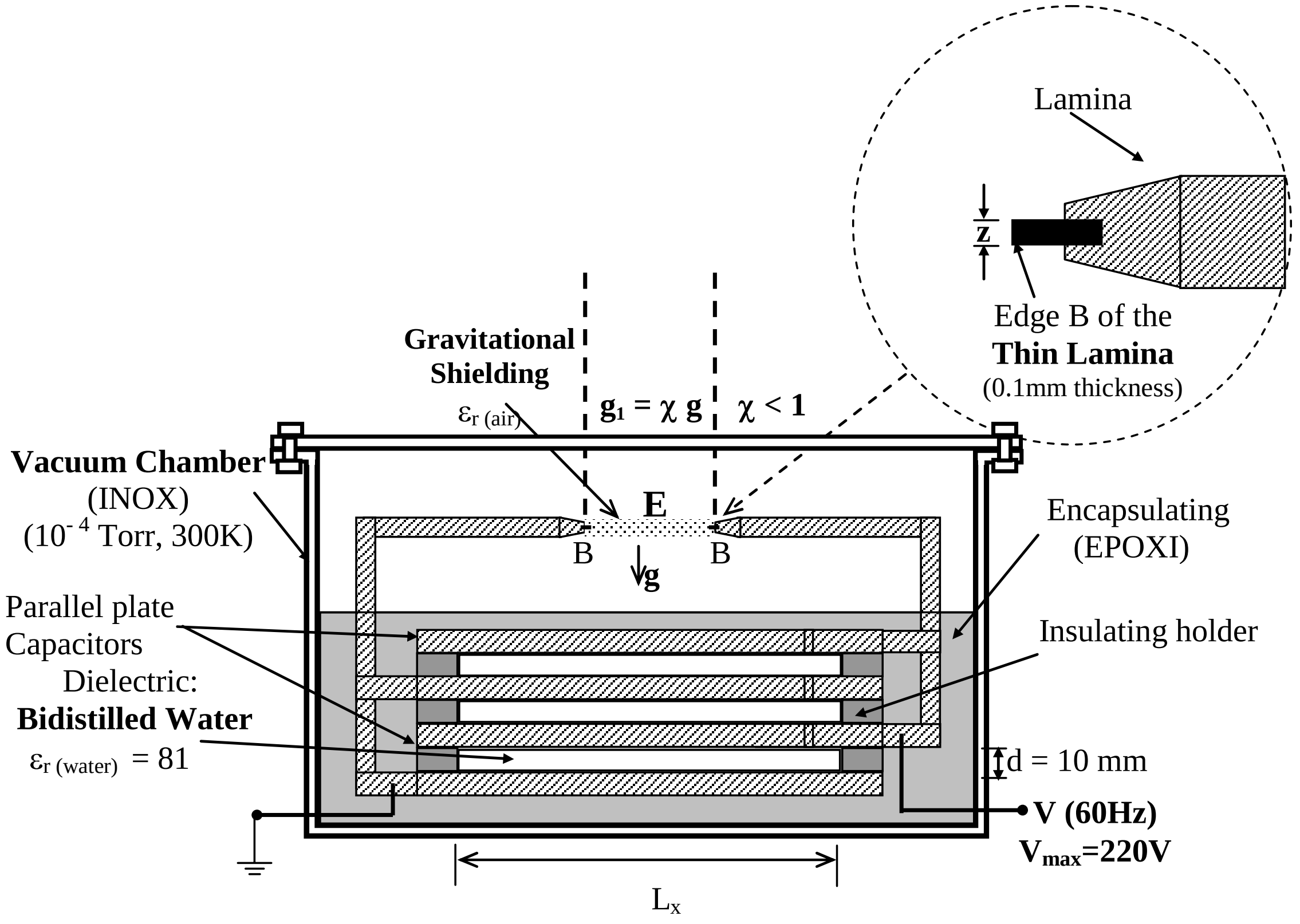

**Vertical Cross Section**

$$q = (C_1+C_2+...+C_n)\,V =$$
$$= n\,[\varepsilon_{r\,(water)} / \varepsilon_{r\,(air)}]\,[A/A_0]\,V/d$$

$$\varepsilon_{r\,(water)} = 81 \;\; ; \;\; \varepsilon_{r\,(air)} \cong 1$$

$$\mathbf{E} = [q/A_0] / \varepsilon_{r\,(air)}\,\varepsilon_0 = n\,[\varepsilon_{r\,(water)} / \varepsilon_{r\,(air)}]\,[A/A_0]\,V/d$$

A is the area of the plates of the capacitors and $A_0$ the cross section area of the edges B of the thin laminas (z is the thickness of the edges).

Figure B1 – **Gravity Control Cell** (GCC) using a *battery of capacitors*. According to Eq. 7 , the electric field, **E**, through the air at $10^{-4}$ Torr; 300K, in the vacuum chamber, produces a gravitational shielding effect. The gravity acceleration above this gravitational shielding is reduced to $\chi\mathbf{g}$ where $\chi < 1$.

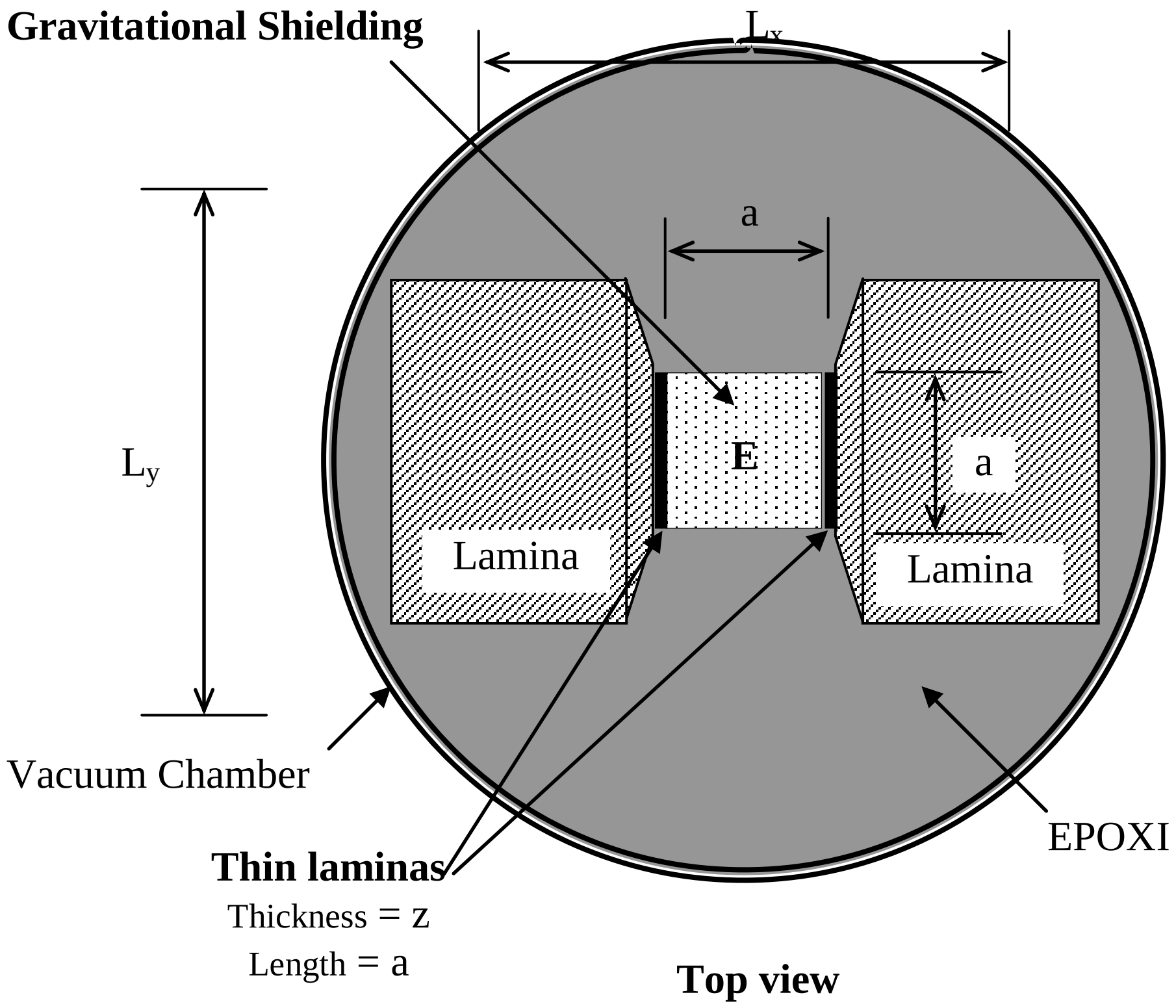

$A_0 = a\ z$ ; $A = L_x L_y$

Figure B2 – The gravitational shielding produced between the thin laminas.

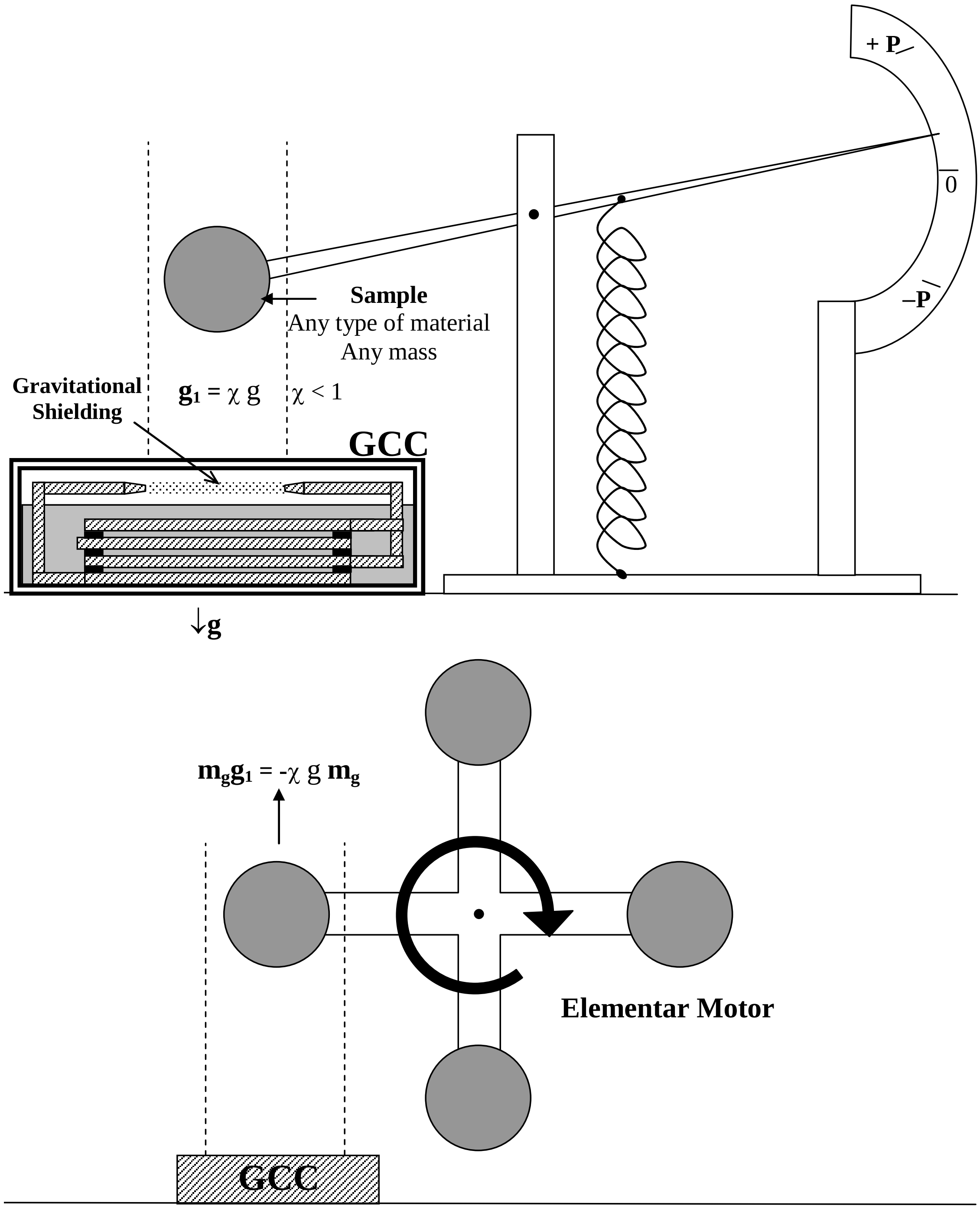

Figure B3 – Experimental arrangement with a GCC using battery of capacitors. By means of this set-up it is possible to check the weight of the sample even when it becomes *negative*.

# Physical Foundations of Quantum Psychology

**Fran De Aquino**
Maranhao State University, Physics Department, S.Luis/MA, Brazil.

**Abstract:** The existence of imaginary mass associated to the neutrino is already well-known. Although its imaginary mass is not physically observable, its square is. This amount is found experimentally to be negative. Recently, it was shown that *quanta* of imaginary mass exist associated to the *electron* and the *photon* too. These imaginary masses have unusual properties that violate the Parity Conservation Principle. The non-conservation of the parity is also found in the weak interactions, and possibly can be explained by means of the existence of the imaginary masses. Also protons and neutrons would have imaginary masses associated to them and, in this way, atoms and molecules would also have imaginary masses directly proportional to their atomic and molecular masses. The Parity Conservation Principle holds that the material particles are not able to distinguish their right from their left. The non-conservation of the parity would necessarily imply capability of "choice". Thus, as the particles with imaginary mass don't conserve the parity, they would have the elementary capability of "choosing between their right or left". Where there is "choice" isn't there also psychism, by definition? This fundamental discovery shows that, in some way, the consciousnesses are related to the imaginary masses. This fact, make it possible to redefine Psychology on a Quantum Physics basis.

*Key words*: Quantum Psychology, Quantized Fields, Unification and Mass Relations, Quantum Mechanics, Bose-Einstein Condensation, Origin of the Universe.
*PACs*: 03.70.+k; 12.10.Kt; 14.80.Cp; 03.65.-w; 03.75.Nt; 98.80.Bp.

## 1. INTRODUCTION

In the last decades it has become evident that the theoretical foundations of Natural Sciences are based on Physics. Today's Chemistry is completely based on Quantum Mechanics, Quantum Statistics, Thermodynamics and Kinetic Physics. Also Biology becomes progressively based on Physics, as more and more biological phenomena are being described on the basis of Quantum Physics. Modern Biophysics is now considered a branch of Physics and no longer a secondary part of Biology and Physiology. As regards Psychology, there are recently several authors making use of Quantum Physics in order to explain psychic phenomena [1,2].

The idea of psyche associated with matter dates back to the pre-Socratic period and is usually called *panpsychism*. Remnants of organized panpsychism may be found in the *Uno* of Parmenides or in Heracleitus's *Divine Flux*. Scholars of Miletus's school were called *hylozoist*s, that is, "those who believe that matter is alive". More recently, we will find the panpsychistic thought in Spinoza, Whitehead and Teilhard de Chardin, among others. The latter one admitted the existence of proto-conscious properties at level of elementary particles.

Generally, the people believe that there is some type of psyche associated to the animals, and some biologists agree that even very simple animals like the ameba and the sea anemone are endowed with psychism. This led several authors to consider the possibility of the psychic phenomena to be described in a theory based on Physics [3,4,5,6].

This work presents a possible theoretical foundation for Psychology based on Quantum Physics, starting from discoveries published in a recent article [7], where it is shown that there is a *quantum* of *imaginary mass* associated to the electron, which would be equivalent to an elementary particle that does not conserves the *parity*. Thus, besides its *inertial mass* the electron would have an *imaginary mass* that would have elementary capacity of "choice". The theory here presented describes the structures and the interaction between these imaginary particles and also explain their relations with the matter on all levels, from the atom to man. In addition, it gives us a better understanding of life and a more complete cosmological view, which lead us to understand our relationship with ourselves, with others, with the Universe and with God.

## 2. THEORY

It was shown [7] that *quanta* of imaginary mass exist associated to the *electron* and the *photon* and that these imaginary masses would have psychic properties (elementary capacity of "choice"). Thus, we can say that, besides its inertial mass, the electron would have a psychic mass, given by

$$m_{\Psi electron} = m_{g(imaginary)electron} = \tfrac{4}{\sqrt{3}}\left(\frac{hf_{electron}}{c^2}\right) i = \tfrac{4}{\sqrt{3}} m_{i(real)electron}\, i \qquad (01)$$

Where $m_{i(real)electron} = 9.11\times10^{-31}kg$ is the *real* inertial mass of the electron. In the case of the photons, it was shown that the *imaginary gravitational* mass of the photon is: $m_{g(imaginary)photon} = \frac{4}{\sqrt{3}}\left(\frac{hf}{c^2}\right) i$. Therefore, the psychic mass associated to a photon with frequency $f$ is expressed by the following equation:

$$m_{\Psi photon} = m_{g(imaginary)photon} = \frac{4}{\sqrt{3}}\left(\frac{hf}{c^2}\right) i \qquad (02)$$

The equation of *quantization of mass* [7], in the generalized form is expressed by: $m_{g(imaginary)} = n^2 m_{g(imagynary)(min)}$. Thus, we can also conclude that the psychic *mass is also quantized*, due to $m_\Psi = m_{g(imaginary)}$, i.e.,

$$m_\Psi = n^2 m_{\Psi(min)} \qquad (03)$$

Where

$$m_{\Psi(min)} = \tfrac{4}{\sqrt{3}}\left(hf_{min}/c^2\right) i = \tfrac{4}{\sqrt{3}} m_{i(real)min}\, i \qquad (04)$$

The *minimum quantum* of real inertial mass in the Universe, $m_{i(real)min}$, is given by [7]:

$$m_{i(real)min} = \pm h\sqrt{3/8}/cd_{max} = \pm 3.9\times10^{-73}kg \qquad (05)$$

By analogy to Eq. (01), the expressions of the psychic masses associated to the *proton* and the *neutron* are respectively given by:

$$m_{\Psi proton} = m_{g(imaginary)proton} = \tfrac{4}{\sqrt{3}}\left(hf_{proton}/c^2\right) i = \tfrac{4}{\sqrt{3}} m_{i(real)proton}\, i \qquad (06)$$

$$m_{\Psi neutron} = m_{g(imaginary)neutron} = \tfrac{4}{\sqrt{3}}\left(hf_{neutron}/c^2\right) i = \tfrac{4}{\sqrt{3}} m_{i(real)neutron}\, i \qquad (07)$$

Where $f_{proton}$ and $f_{neutron}$ are respectively the frequencies of the DeBroglie's waves associated to the proton and the neutron.

Thus, from a quantum viewpoint, the psychic particles are similar to the material particles, so that we can use the Quantum Mechanics to describe the psychic particles. In this case, by analogy to the material particles, a particle with psychic mass $m_\Psi$ will be described by the following expressions:

$$\vec{p}_\psi = \hbar \vec{k}_\psi \qquad (08)$$

$$E_\psi = \hbar \omega_\psi \qquad (09)$$

Where $\vec{p}_\psi = m_\Psi \vec{V}$ is the *momentum* carried by the wave and $E_\psi$ its energy; $\left|\vec{k}_\psi\right| = 2\pi/\lambda_\psi$ is the *propagation number* and $\lambda_\psi = h/m_\Psi V$ the *wavelength* and $\omega_\psi = 2\pi f_\psi$ its cyclic *frequency*.

The variable quantity that characterizes DeBroglie's waves is called *Wave Function*, usually indicated by $\Psi$. The wave function associated to a material particle describes the dynamic state of the particle: its value at a particular point x, y, z, t is related to the probability of finding the particle in that place and instant. Although $\Psi$ does not have a physical interpretation, its square $\Psi^2$ or $\Psi\ \Psi^*$) calculated for a particular point x, y, z, t is *proportional to the probability of experimentally finding the particle in that place and instan*t.

Since $\Psi^2$ is proportional to the probability $P$ of finding the particle described by $\Psi$, the integral of $\Psi^2$ on the *whole space* must be finite – inasmuch as the particle is someplace. Therefore, if

$$\int_{-\infty}^{+\infty} \Psi^2 dV = 0 \qquad (10)$$

The interpretation is that the particle does not exist. However, if

$$\int_{-\infty}^{+\infty} \Psi^2 dV = \infty \qquad (11)$$

*the particle will be everywhere simultaneously* (Omnipresence).

The wave function $\Psi$ corresponds, as we know, to the displacement $y$ of the undulatory motion of a rope. However, $\Psi$ as opposed to $y$, is not a measurable quantity and can, hence, be a *complex* quantity. For this reason, it is admitted that $\Psi$ is described in the $x$-direction by

$$\Psi = Be^{-(2\pi\, i/h)(Et-px)} \qquad (12)$$

This equation is the mathematical description of the wave associated with a free material particle, with total energy $E$ and *momentum* $p$, moving in the direction $+x$.

As concerns the psychic particle, the variable quantity characterizing psyche waves will also be called wave function, denoted by $\Psi_\Psi$ ( to differentiate it from the material particle wave function), and, by analogy with equation Eq. (12), expressed by:

$$\Psi_\Psi = \Psi_0 e^{-(2\pi i/h)(E_\Psi t - p_\Psi x)} \qquad (13)$$

If an experiment involves a large number of identical particles, all described by the same wave function $\Psi$ , *real* density of mass $\rho$ of these particles in x, y, z, t is proportional to the corresponding value $\Psi^2$ ($\Psi^2$ is known as *density of probability*. If $\Psi$ is *complex* then $\Psi^2 = \Psi\Psi^*$. Thus, $\rho \propto \Psi^2 = \Psi.\Psi^*$). Similarly, in the case of psychic particles, the *density of psychic mass*, $\rho_\Psi$, in x, y, z, will be expressed by $\rho_\Psi \propto \Psi_\Psi^2 = \Psi_\Psi \Psi_\Psi^*$. It is known that $\Psi_\Psi^2$ is always *real* and *positive* while $\rho_\Psi = m_\Psi / V$ is an *imaginary* quantity. Thus, as the *modulus* of an imaginary number is always real and positive, we can transform the proportion $\rho_\Psi \propto \Psi_\Psi^2$, in equality in the following form:

$$\Psi_\Psi^2 = k|\rho_\Psi| \qquad (14)$$

Where $k$ is a *proportionality constant* (real and positive) to be determined.

In Quantum Mechanics we have studied the *Superpositon Principle*, which affirms that, if a particle (or system of particles) is in a *dynamic state* represented by a wave function $\Psi_1$ and may also be in another dynamic state described by $\Psi_2$ then, the general dynamic state of the particle may be described by $\Psi$, where $\Psi$ is a linear combination (superposition) of $\Psi_1$ and $\Psi_2$, i.e.,

$$\Psi = c_1\Psi_1 + c_2\Psi_2 \qquad (15)$$

C*omplex constants* $c_1$ e $c_2$ respectively indicate the percentage of dynamic state, represented by $\Psi_1$ e $\Psi_2$ in the formation of the general dynamic state described by $\Psi$.

In the case of psychic particles (psychic bodies, consciousness, etc.), by analogy, if $\Psi_{\Psi 1}$, $\Psi_{\Psi 2}$,..., $\Psi_{\Psi n}$ refer to the different dynamic states the psychic particle assume, then its general dynamic state may be described by the wave function $\Psi_\Psi$ , given by:

$$\Psi_\Psi = c_1\Psi_{\Psi 1} + c_2\Psi_{\Psi 2} + ... + c_n\Psi_{\Psi n} \qquad (16)$$

The state of superposition of wave functions is, therefore, common for both psychic and material particles. In the case of material particles, it can be verified, for instance, when an electron changes from one orbit to another. Before effecting the transition to another energy level, the electron carries out "virtual transitions" [8]. A kind of *relationship* with other electrons before performing the real transition. During this relationship period, its wave function remains "*scattered*" by *a wide region of the space* [9] thus superposing the wave functions of the other electrons. In this relationship the electrons *mutually* influence each other, with the possibility of *intertwining* their wave functions[1]. When this happens, there occurs the so-called *Phase Relationship* according to quantum-mechanics concept.

[1] Since the electrons are simultaneously waves and particles, their wave aspects will interfere with each other; besides superposition, there is also the possibility of occurrence of *intertwining* of their wave functions.

In the electrons "virtual" transition mentioned before, the "listing" of all the possibilities of the electrons is described, as we know, by *Schrödinger's wave equation.* Otherwise, it is general for material particles. By analogy, in the case of psychic particles, we may say that the "listing" of all the possibilities of the psyches involved in the relationship will be described by *Schrödinger's equation* – for psychic case, i.e.,

$$\nabla^2 \Psi_\Psi + \frac{p_\Psi^2}{\hbar^2} \Psi_\Psi = 0$$

Because the wave functions are capable of intertwining themselves, the quantum systems may "penetrate" each other, thus establishing an internal relationship where all of them are affected by the relationship, no longer being isolated systems but becoming an integrated part of a larger system. This type of internal relationship, which exists only in quantum systems, was called *Relational Holism* [10].

It is a proven quantum fact that a wave function may *collapse*, and that, at this moment, all the possibilities that it describes are suddenly expressed in *reality*. This means that, through this process, particles can be suddenly *materialized*. Similarly, the collapse of the psychic wave function must suddenly also express in reality all the possibilities described by it. This is, therefore, *a point of decision* in which there occurs the compelling need of realization of the *psychic form*. Thus, this is moment in which the content of the psychic form realizes itself in the space-time. For an observer in space-time, something is *real* when it is under a matter or radiation form. Therefore, the content of the psychic form may realize itself in space-time exclusively under the form of radiation, that is, it does not materialize. This must occur when the *Materialization Condition* is not satisfied, i.e., when the content of the psychic form is undefined (impossible to be defined by its own psychic) or it does not contain enough psychic mass to *materialize*[2] the respective psychic contents.

Nevertheless, in both cases, there must always be a production of "virtual" photons to convey the psychic interaction to the other psychic particles, according to the quantum field theory, only through this type of quanta will interaction be conveyed, since it has an infinite reach and may be either attractive or repulsive, just as electromagnetic interaction which, as we know, is conveyed by the exchange of "virtual" photons.

If electrons, protons and neutrons have psychic mass, then we can infer that the psychic mass of the atoms are *Phase Condensates*[3]. In the case of the molecules the situation is similar. More molecular mass means more atoms and consequently, more psychic mass. In this case the phase condensate also becomes more structured because the great amount of elementary psyches inside the condensate requires, by stability reasons, a better distribution of them. Thus, in the case of molecules with very large molecular masses (*macromolecules*) it is possible that their psychic masses already constitute the most organized shape of a Phase Condensate, called Bose-Einstein Condensate[4].

---

[2] By this we mean not only materialization proper but also the movement of matter to realize its psychic content (including radiation).

[3] Ice and NaCl crystals are common examples of imprecisely-structured *phase condensates*. Lasers, super fluids, superconductors and magnets are examples of phase condensates more structured.

[4] Several authors have suggested the possibility of the Bose-Einstein condensate occurring in the brain, and that it might be the physical base of memory, although they have not been able to find a suitable mechanism to underpin such a hypothesis. Evidences of the existence of Bose-Einstein condensates in living tissues abound (Popp, F.A Experientia, Vol. 44, p.576-585; Inaba, H., New Scientist, May89, p.41; Rattermeyer, M and Popp, F. A. Naturwissenschaften, Vol.68, Nº5, p.577.)

The fundamental characteristic of a Bose-Einstein condensate is, as we know, that the various parts making up the condensed system not only behave as a whole but also *become a whole*, i.e., in the psychic case, the various consciousnesses of the system become a *single consciousness* with psychic mass equal to the sum of the psychic masses of all the consciousness of the condensate. This obviously, increases the available knowledge in the system since it is proportional to the psychic mass of the consciousness. This unity confers an *individual* character to this type of consciousness. For this reason, from now on they will be called *Individual Material Consciousness.*

It derives from the above that most bodies do not possess individual material consciousness. In an iron rod, for instance, the cluster of elementary psyches in the iron molecules does not constitute Bose-Einstein condensate; therefore, the iron rod does not have an individual consciousness. Its consciousness is consequently, much more simple and constitutes just a phase condensate imprecisely structured made by the consciousness of the iron atoms.

The existence of consciousnesses in the atoms is revealed in the molecular formation, where atoms with strong mutual affinity (their consciousnesses) combine to form molecules. It is the case, for instance of the water molecules, in which two Hydrogen atoms join an Oxygen atom. Well, how come the combination between these atoms is always the same: the same grouping and the same invariable proportion? In the case of molecular combinations the phenomenon repeats itself. Thus, the chemical substances either mutually attract or repel themselves, carrying out specific motions for this reason. It is the so-called *Chemical Affinity*. This phenomenon certainly results from a specific interaction between the consciousnesses. From now on, it will be called *Psychic Interaction*.

*Mutual Affinity* is a dimensionless psychic quantity with which we are familiar and of which we have perfect understanding as to its meaning. The degree of *Mutual Affinity*, $A$, in the case of two consciousnesses, respectively described by $\Psi_{\Psi 1}$ e $\Psi_{\Psi 2}$, must be correlated to $\Psi^2_{\Psi 1}$ e $\Psi^2_{\Psi 2}$ [5]. Only a simple algebraic form fills the requirements of interchange of the indices, the product

$$\Psi^2_{\Psi 1} . \Psi^2_{\Psi 2} = \Psi^2_{\Psi 2} . \Psi^2_{\Psi 1} = \left|A_{1,2}\right| = \left|A_{2,1}\right| = \left|A\right| \tag{17}$$

In the above expression, $|A|$ is due to the product $\Psi^2_{\Psi 1} . \Psi^2_{\Psi 2}$ will be always positive. From equations (17) and (14) we get

$$\left|A\right| = \Psi^2_{\Psi 1} . \Psi^2_{\Psi 2} = k^2 \left|\rho_{\Psi 1}\right|\left|\rho_{\Psi 2}\right| = k^2 \frac{\left|m_{\Psi 1}\right|}{V_1} \frac{\left|m_{\Psi 2}\right|}{V_2} \tag{18}$$

---

[5] Quantum Mechanics tells us that $\Psi$ does not have a physical interpretation nor a simple meaning and also it cannot be experimentally observed. However such restriction does not apply to $\Psi^2$ , which is known as *density of probability* and represents the probability of finding the body, described by the wave function $\Psi$ , in the point x, y, z at the moment t. A large value of $\Psi^2$ means a strong possibility to find the body, while a small value of $\Psi^2$ means a weak possibility to find the body.

The psychic interaction can be described starting from the psychic mass because the psychic mass is the source of the psychic field. Basically, *the psychic mass is gravitational mass*, since $m_\Psi = m_{g(imaginary)}$. In this way, the equations of the gravitational interaction are also applied to the Psychic Interaction. That is, we can use Einstein's General Relativity equations, given by:

$$R_i^k = \frac{8\pi G}{c^4}\left(T_i^k - \tfrac{1}{2}\delta_i^k T\right) \tag{19}$$

in order to describe the Psychic Interaction. In this case, the expression of the energy-momentum tensor, $T_i^k$, must have the following form [11]:

$$T_i^k = |\rho_\Psi| c^2 \mu_i \mu^k \tag{20}$$

The psychic mass density, $\rho_\Psi$, is a imaginary quantity. Thus, in order to homogenize the above equation it is necessary to put $|\rho_\Psi|$ because, as we know, the module of an imaginary number is always real and positive.

Making on the transition to Classical Mechanics [12] one can verify that Eqs. (19) are reduced to:

$$\Delta\Phi = 4\pi G|\rho_\Psi| \tag{21}$$

This is, therefore, the equation of the psychic field in *nonrelativistic* Mechanics. With respect to its form, it is similar to the equation of the gravitational field, with the difference that now, instead of the density of gravitational mass we have the density of *psychic mass*. Then, we can write the general solution of Eq. (21), in the following form:

$$\Phi = -G\int \frac{|\rho_\Psi| dV}{r^2} \tag{22}$$

This equation expresses, with nonrelativistic approximation, the potential of the psychic field of any distribution of psychic mass.

Particularly, for the potential of the field of only one particle with psychic mass $m_{\Psi 1}$, we get:

$$\Phi = -\frac{G|m_{\Psi 1}|}{r} \tag{23}$$

Then the force produced by this field upon another particle with psychic mass $m_{\Psi 2}$ is

$$\left|\vec{F}_{\Psi 12}\right| = \left|-\vec{F}_{\Psi 21}\right| = -|m_{\Psi 2}|\frac{\partial\Phi}{\partial r} = -G\frac{|m_{\Psi 1}||m_{\Psi 2}|}{r^2} \tag{24}$$

By comparing equations (24) and (18) we obtain

$$\left|\vec{F}_{\Psi 12}\right| = \left|-\vec{F}_{\Psi 21}\right| = -G|A|\frac{V_1 V_2}{k^2 r^2} \tag{25}$$

In the *vectorial* form the above equation is written as follows

$$\vec{F}_{\Psi 12} = -\vec{F}_{\Psi 21} = -GA\frac{V_1 V_2}{k^2 r^2}\hat{\mu} \tag{26}$$

V*ersor* $\hat{\mu}$ has the direction of the line connecting the mass centers (psychic mass) of both particles and oriented from $m_{\Psi 1}$ to $m_{\Psi 2}$.

In general, we may distinguish and quantify two types of mutual affinity: *positive* and *negative* (*aversion*). The occurrence of the first type is synonym of psychic *attraction*, (as in the case of the atoms in the water molecule) while the aversion is synonym of *repulsion*. In fact, Eq. (26) shows that the forces $\vec{F}_{\Psi 12}$ and $\vec{F}_{\Psi 21}$ are attractive, if $A$ is *positive* (expressing *positive* mutual affinity between the two *psychic bodies*), and repulsive if $A$ is *negative* (expressing *negative* mutual affinity between the two *psychic bodies*). Contrary to the interaction of the matter, where the opposites attract themselves here, the *opposites repel themselves*.

A method and device to obtain images of *psychic bodies* have been previously proposed [13]. By means of this device, whose operation is based on the gravitational interaction and the piezoelectric effect, it will be possible to observe psychic bodies.

Expression (18) can be rewritten in the following form:

$$A = k^2 \frac{m_{\Psi 1}}{V_1} \frac{m_{\Psi 2}}{V_2} \qquad (27)$$

The psychic masses $m_{\Psi 1}$ and $m_{\Psi 2}$ are *imaginary* quantities. However, the product $m_{\Psi 1} . m_{\Psi 2}$ is a *real* quantity. One can then conclude from the previous expression that the degree of mutual affinity between two consciousnesses depends basically on the densities of their psychic masses, and that:

1) If $m_{\Psi 1} > 0$ and $m_{\Psi 2} > 0$ then $A > 0$ (positive mutual affinity between them)
2) If $m_{\Psi 1} < 0$ and $m_{\Psi 2} < 0$ then $A > 0$ (positive mutual affinity between them)
3) If $m_{\Psi 1} > 0$ and $m_{\Psi 2} < 0$ then $A < 0$ (negative mutual affinity between them)
4) If $m_{\Psi 1} < 0$ and $m_{\Psi 2} > 0$ then $A < 0$ (negative mutual affinity between them)

In this relationship, such as occurs in the case of material particles (“virtual” transition of the electrons previously mentioned), the consciousnesses interact mutually, *intertwining* or not their wave functions. When this happens, there occurs the so-called *Phase Relationship* according to quantum-mechanics concept. Otherwise a *Trivial Relationship* takes place.

The psychic forces such as the gravitational forces, must be very weak when we consider the interaction between two particles. However, in spite of the subtleties, those forces stimulate the relationship of the consciousnesses with themselves and with the Universe (Eq.26).

From all the preceding, we perceive that Psychic Interaction – unified with matter interactions, constitutes a single *Law* which links things and beings together and, in a network of continuous relations and exchanges, governs the Universe both in its material and psychic aspects. We can also observe that in the interactions the same principle reappears always identical. This *unity of principle* is the most evident expression of *monism* in the Universe.

## 3. UNIFIED COSMOLOGY

In traditional Cosmology, the Universe arises from a great explosion where everything that exists would be initially concentrated in a minuscule particle with the size of a proton and with a gigantic mass equal to the mass of the Universe. However, the origin this tiny particle is not explained, nor is the reason for its critical volume.

This critical volume denotes *knowledge* of what would happen with the Universe starting from that *initial condition*, a fact that points towards the *existence* of a Creator.

It was shown that a wave function may *collapse* and, at this moment, all the possibilities that it describes are suddenly expressed in *reality*. This means that, through this process, particles can be suddenly *materialized*. This is a materialization process which can explain the materialization of the Universe. That is, the Primordial Universe would have arisen at the exact moment in which the *Primordial Wave Function* collapsed (Initial Instant) realizing the content of the psychic form generated at the consciousness of the Creator when He *thought* to create the Universe.

The psychic form described by this primordial wave function must have been generated in a consciousness with a psychic mass much greater than that needed to materialize the Universe (material and psychic).

This giant consciousness, in its turn, would not only be the greatest of all consciousnesses in the Universe but also the *substratum* of everything that exists and, obviously, everything that exists would be entirely contained within it, including *all the spacetime*.

Based on General Theory of Relativity and recent cosmological observations, it is known today that the Universe occupies a space of positive curvature. This space, as we known, is “closed in itself”, its volume is finite but, clearly understood, the space has no frontiers, it is *unlimited*. Thus, if the consciousness we refer to contains *all* the space, its volume is necessarily infinite, consequently having an *infinite* psychic mass.

This means that It contains *all* the existing psychic mass and, therefore, any other consciousness that may exist will contained in It. Hence, we may conclude that It is the *Supreme Consciousness* and that there no other equal to It: It is *unique*.

The manifestation of the knowledge or *auto-accessible knowledge* in a consciousness should be related to its quantity of psychic mass. In the Supreme Consciousness, whose psychic mass is infinite, the manifestation of the knowledge is *total*, and as such, necessarily, It should be *omniscient*. In the *elementary psyche* $\left(m_{\Psi(min)}\right)$ most of the knowledge should be in latent state. Being omniscient, the Supreme Conscience knows evidently, how to formulate well-defined mental images and with sufficiently psychic masses in order to materialize their contents (Materialization Condition). Consequently, It can materialize everything which It wants (Omnipotence).

Since the Supreme Consciousness occupies *all the space*, we can conclude that It cannot be displaced by another consciousness, not even by Itself. Therefore, the Supreme Consciousness is *immovable*.

As Augustine says (Gen. Ad lit viii, 20), "The Creator Spirit moves Himself neither by time, nor by place."

Thomas Aquinas also had already considered Creator's *immobility* as necessary:
"From this we infer that it is necessary that the God that moves everything is *immovable*." (Summa Theologica).

On the other hand, since the Supreme Consciousness contains all the space-time, It should contain obviously, *all* the time. More explicit, for the Supreme Consciousness, past, present and future are an eternal present, and the time does not flow as it flows for us.

Within this framework, when we talk about the Creation of the Universe, the use of the verb "to create" means that something that was not came into being, thus presupposing the concept of *time flow*. For the Supreme Consciousness, however, the instant of Creation is mixed up with all other times, consequently there being no "*before*" or "*after*" the Creation and, thus, the following questions like "What did the Supreme Consciousness do *before* Creation?

We can also infer from the above that the existence of the Supreme Consciousness has no defined limit (beginning and end), which confers upon It the unique characteristic of *uncreated* and *eterna*l.

Being eternal, Its wave function $\Psi_{SC}$ shall never collapse. On the other hand, for having an infinite psychic mass, the value of $\Psi_{SC}$ will always be infinite and, hence, in agreement with Eq. (11), the Supreme Consciousness *is simultaneously everywhere*, that is, It is *omnipresent*.

All these characteristics of the Supreme Consciousness (*infinite, unique, uncreated, eternal, omnipresent, omniscient and omnipotent*) coincide with those traditionally ascribed to God by most religions.

The option of the Supreme Consciousness to materialize the primordial Universe into a critical volume denotes the knowledge of what was would happen in the Universe starting from that initial condition. Therefore, It knew how the Universe would behave under already *existing laws*. Consequently, the laws were not created *for the Universe* and, hence, are not "Nature's laws" or "laws placed on Nature by God", as written by Descartes. They already existed as an intrinsic part of the Supreme Consciousness; Thomas Aquinas had a very clear understanding about this. He talks about the Eternal Law "…which *exists* in God's mind and governs the whole Universe".

The Supreme Consciousness had all freedom to choose the initial conditions of the Universe, but opted for the concentration in a critical volume so that the evolution of the Universe would proceed in the most convenient form for the purpose It had in mind and in accordance with the laws inherent in Its own nature. This reasoning then answers Einstein's famous question: "What level of choice would God have had when building the Universe?"

Apparently, Newton was the first one to notice the Divine option. In his book *Optiks*, he gives us a perfect view of how he imagined the creation of the Universe:
" *It seems possible to me that God, in the beginning, gave form to matter in solid, compacted particles[…] in the best manner possible to contribute to the purpose He had in mind*…"

With what purpose did the Supreme Consciousness create the Universe? This question seems to be difficult to answer. Nevertheless, if we admit the Supreme Consciousness's primordial desire *to procreate*, i.e., to generate individual consciousnesses from Itself so that the latter could evolve and manifest Its same creating attributes, then we can infer that, in order for them to evolve, such consciousness would need a Universe, and this might have been the main reason for its creation. Therefore, the origin of the Universe would be related to the generation of said consciousness and, consequently, the materialization of the primordial Universe must have taken place at the same epoch when the Supreme Consciousness decided to *individualize* the postulated consciousness, hereinafter called *Primordial Consciousness*.

For having been directly individualized from the Supreme Consciousness, the primordial consciousness certainly contained in themselves, although in a latent state, all the possibilities of the Supreme Consciousness, including the germ of independent will, which enables original starting points to be established. However, in spite of the similarity to Supreme Consciousness, the primordial consciousness could not have the understanding of themselves. This self-understanding only arises with the *creative mental state* that such consciousnesses can only reach by evolution.

Thus, in the first evolutionary period, the primordial consciousness must have remained in total unconscious state, this being then the beginning of an evolutionary pilgrimage from *unconsciousness* to *superconsciousness*.

The evolution of the primordial consciousness in this *unconsciousness* period takes place basically through psychic *relationship* among them (superposition of psychic wave functions, having or not *intertwining*). Thus, the speed at which they evolved was determined by what they obtained in these relationships.

After the origin of the first planets, some of them came to develop favorable conditions for the appearance of macromolecules. These macromolecules, as we have shown, may have a special type of consciousness formed by a Bose-Einstein condensate (Individual Material Consciousness). In this case, since the molecular masses of the macromolecules are very large, they will have individual material consciousness of large psychic mass and, therefore, access to a considerable amount of information in its own consciousness. Consequently, macromolecules with individual material consciousness are potentially very capable and some certainly already can carry out autonomous motions, thus being considered as “living” entities.

However, if we decompose one of these molecules so as to destroy its individual consciousness, its parts will no longer have access to the information which “instructed” said molecule and, hence, will not be able to carry out the autonomous motions it previously did. Thus, the “life” of the molecule disappears – as we can see, *Delbrück’s Paradox* is then solved[6].

The appearance of “living” molecules in a planet marks the beginning of the most important evolutionary stage for the psyche of matter, for it is from the combination of these molecules that there appear living beings with individual material consciousness with even larger psychic masses.

Biologists have shown that all living organisms existing on Earth come from two types of molecules – aminoacids and nucleotides – which make up the fundamental building blocks of living beings. That is, the nucleotides and aminoacids are identical in all living beings, whether they are bacteria, mollusks or men. There are twenty different species of aminoacids and five of nucleotides.

---

[6] This paradox ascribed to Max Delbrück (Delbrück, Max., (1978) *Mind from Matter*? American Scholar, **47**. pp.339-53.) remained unsolved and was posed as follows: How come the same matter studied by Physics, when incorporated into a living organism, assumes an unexpected behavior, although not contradicting physical laws?

In 1952, Stanley Miller and Harold Urey proved that aminoacids could be produced from inert chemical products present in the atmosphere and oceans in the first years of existence of the Earth. Later, in 1962, nucleotides were created in laboratory under similar conditions. Thus, it was proved that the molecular units making up the living beings could have formed during the Earth's primitive history.

Therefore, we can imagine what happened from the moment said molecules appeared. The concentration of aminoacids and nucleotides in the oceans gradually increased. After a long period of time, when the amount of nucleotides was already large enough, they began to group themselves by mutual psychic attraction, forming the molecules that in the future will become DNA molecules.

When the molecular masses of these molecules became large enough, the distribution of elementary psyches in their consciousnesses took the most orderly possible form of phase condensate (Bose-Einstein condensate) and such consciousnesses became the *individual material consciousness*.

Since the psychic mass of the consciousnesses of these molecules is very large (as compared with the psychic mass of the atoms), the amount of self-accessible knowledge became considerable in such consciousnesses and thus, they became apt to *instruct* the joining of aminoacids in the formation of the first proteins (origin of the *Genetic Code*). Consequently, the DNA's capability to serve as guide for the joining of aminoacids in the formation of proteins is fundamentally a result of their psychism.

In the psychic of DNA molecules, the formation of proteins certainly had a definite objective: *the construction of cells*.

During the cellular construction, the most important function played by the consciousnesses of the DNA molecules may have been that of organizing the distribution of the new molecules incorporated to the system so that the consciousnesses of these molecules jointly formed with the consciousness of the system a Bose-Einstein condensate. In this manner, more knowledge would be available to the system and, after the cell is completed, the latter would also have an individual material consciousness.

Afterwards, under the action of psychic interaction, the cells began to group themselves according to different degrees of positive mutual affinity, in an organized manner so that the distribution of their consciousnesses would also form Bose-Einstein condensates. Hence, collective cell units began to appear with individual consciousnesses of larger psychic masses and, therefore, with access to more knowledge. With greater knowledge available, these groups of cells began to perform specialized functions to obtain food, assimilation, etc. That is when the first multi-celled beings appeared.

Upon forming the tissues, the cells gather structurally together in an organized manner. Thus, the tissues and, hence, the organs and the organisms themselves also possess individual material consciousnesses.

The existence of the material consciousness of the organisms is proved in a well-known experiment by Karl Lashley, a pioneer in neurophysiology.

Lashley initially taught guinea pigs to run through a maze, an ability they remember and keep in their memories in the same way as we acquire new skills. He then systematically removed small portions of the brain tissue of said guinea pigs. He thought that, if the guinea pigs still remembered how to run through the maze, the memory centers would still be intact.

Little by little he removed the brain mass; the guinea pigs, curiously enough, kept remembering how to run through the maze. Finally, with more than90% of their cortex removed, the guinea pigs still kept remembering how to run through the maze. Well, as we have seen, the consciousness of an organism is formed by the concretion of all its cellular consciousnesses. Therefore, the removal of a portion of the organism cells does not make it disappear. Their cells, or better saying, the consciousnesses of their cells contribute to the formation of the consciousness of the organism just as the others, and it is exactly due to this that, even when we remove almost all of the guinea pigs' cortex, they were still able to remember from the memories of their individual material consciousnesses. In this manner, what Lashley's experiment proved was precisely the existence of individual material consciousnesses in the guinea pigs.

Another proof of the existence of the individual material consciousnesses in organisms is given by the *regeneration* phenomenon, so frequent in animals of simple structure: sponges, isolated coelenterates, worms of various groups, mollusks, echinoderms and tunicates. The arthropods regenerate their pods. Lizards may regenerate only their tail after autoctomy. Some starfish may regenerate so easily that a simple detached arm may, for example, give origin to a wholly new animal.

The organization of the psychic parts in the composition of an organism's individual material consciousness is directly related to the organization of the material parts of the organism, as we have already seen. Thus, due to this interrelationship between body and consciousness, any disturbance of a material (physiological) nature in the body of the being will affect its individual material consciousness, and any psychic disturbance imposed upon its consciousness affects the physiology of its body.

When a consciousness is strongly affected to the extent of unmaking the Bose-Einstein's condensate, which gives it the status of individual consciousness, there also occurs the simultaneous disappearance of the knowledge made accessible by said condensation. Therefore, when a cell's consciousness no longer constitutes a Bose-Einstein condensate, there is also the simultaneous disappearance of the knowledge that *instructs and maintains* the cellular metabolism. Consequently, the cell no longer functions thus initiating its decomposition (molecular disaggregation).

Similarly, when the consciousness of an animal (or plant) no longer constitutes a Bose-Einstein condensate, the knowledge that instructs and maintains its body functioning also disappears, and it dies. In this process, after the unmaking of the being's individual consciousness, there follows the unmaking of the individual consciousnesses of the organs; next will be the consciousnesses of their own cells which no longer exit. At the end there will remain the isolated psyches of the molecules and atoms. *Death, indeed, destroys nothing, neither what makes up matter nor what makes up psyche*.

As we have seen, all the information available in the consciousnesses of the beings is also accessible by the consciousnesses of their organs up to their molecules'. Thus, when an individual undergoes a certain experience, the information concerning it not only is recorded somewhere in this consciousness but also pervades all the individual consciousnesses that make up its total consciousness. Consequently, psychic disturbances imposed to a being reflect up to the level of their individual molecular consciousnesses, perhaps even structurally affecting said molecules, due to the interrelationship between body and consciousness already mentioned here.

Therefore, one can expect that there may occur modifications in the sequences of nucleotides of DNA molecules when the psychism of the organism to which they are incorporated is sufficiently affected.

It is known that such modifications in the structure of DNA molecules may also occur because of the chemical products in the blood stream ( as in the case of the mustard gas used in chemical warface) or by the action of radiation sufficiently energetic.

Modifications in the sequences of nucleotides in DNA molecules are called mutations. Mutations as we know, determine hereditary variations which make up the basis of Darwin's theory of evolution.

There may occur "favorable" and unfavorable" mutation to the individuals; the former enhances the individuals' possibility of survival, whereas the latter decrease such possibility.

The theory of evolution is established as a consequence of individuals' efforts to survive in the environment where they live. This means that their descendants may become different from their ancestors. This is the mechanism that leads to the frequent appearance of new species. Darwin believed that the mutation process was slow and gradual. Nevertheless, it is known today that this is not the general rule, for there are evidences of the appearance of new species in a relatively short period of time [14]. We also know that the characteristics are transmitted from parents to offsprings by means of genes and that the recombination of the parents' genes, when *genetic instructions* are transmitted by such genes.

However, it was shown that the genetic instructions are basically associated with the psychism of DNA molecules. Consequently, *the genes transmit not only physiological but also psychic differences.*

Thus, as a consequence of genetic transmission, besides the great physiological difference between individuals of the same species, there is also a great psychic dissimilarity.

Such psychic dissimilarity associated with the progressive enhancement of the individual's psychic quantities may have given rise, in immemorial time, to a variety of individuals (most probably among anthropoid primates) which unconsciously established a positive mutual affinity with *primordial consciousnesses* must have been attracted to the Earth. Thus, the relationship established among them and the consciousnesses of said individuals is enhanced.

In the course of evolutionary transformation, there was a time when the fetuses of said variety already presented such a high degree of mutual affinity with the primordial consciousnesses attracted to the Earth that, during pregnancy, the incorporation of primordial consciousnesses may have occurred in said fetuses.

In spite of absolute psychic mass of the fetus's material consciousness being much smaller than that of the mother's consciousness, the degree of positive mutual affinity between the fetus's consciousness and the primordial consciousness that is going to be incorporated is much greater than that between the latter and the mother's, which makes the psychic attraction between the fetus's consciousness and primordial consciousness much stronger than the attraction between the latter and the mother's. That is the reason why primordial consciousness incorporates the fetus. Thus, when these new individuals are born, they bring along, besides their individual material consciousness, an individualized consciousness of the Supreme Consciousness. This is how the first *hominids* were born.

Having been directly individualized from Supreme Consciousness, the primordial consciousnesses constitutes as perfect individualities and not as phase condensates as the consciousnesses of matter. In this manner, they do not dissociate upon the death of those that incorporated them. Afterwards, upon the action of psychic attraction, they were again able to incorporate into other fetuses to proceed with their evolution.

These consciousnesses (hereinafter called *human consciousness*) constitutes individualities and, therefore, the larger their psychic mass the more available knowledge they will have and, consequently, greater ability to evolve.

Just as the human race evolves biologically, human consciousnesses have also been evolving. When they are incorporated, the difficulties of the material world provide them with more and better opportunities to acquire psychic mass (later on we will see how said consciousnesses may gain or lose psychic mass). That is why they need to perform successive reincorporations. Each reincorporation arises as a new opportunity for said consciousnesses to increase their psychic mass and thus evolve.

The belief in the reincarnation is millenary and well known, although it has not yet been scientifically recognized, due to its *antecedent probability* being very small. In other words, there is small amount of data contributing to its confirmation. This, however, does not mean that the phenomenon is not true, but only that there is the need for a considerable amount of experiments to establish a significant degree of antecedent probability.

The rational acceptance of reincarnation entails deep modifications in the general philosophy of the human being. For instance, it frees him from negative feelings, such as nationalistic or racial prejudices and other response patterns based on the naive conception that we are simply what we appear to be.

Darwin's lucid perception upon affirming that not only the individual's corporeal qualities but also his psychic qualities tend to improve made implicit in his "natural selection" one of the most important rules of evolution: *the psychic selection*, which basically consists in the *survival of the most apt consciousnesses*. Psychic aptitude means, in the case of human consciousnesses, mental quality, i.e., *quality of thinking*.

Further on, we will see that the human consciousnesses may gain or lose psychic mass from the Supreme consciousness, respectively due to the mode of *resonance* (quality) of their thoughts. This means that the consciousnesses that cultivate a greater amount of bad-quality thoughts will have a *lesser* chance of psychic survival than the others. A human consciousness that permanently cultivates bad-quality thoughts progressively loses psychic mass and may even be extinguished.

With the progressive disappearance of psychically less apt consciousnesses, it will be increasingly easy for the more apt consciousnesses to increase their psychic masses during reincorporation periods. There will be a time when psychic selection will have produced consciousnesses of large psychic mass and, therefore, highly evolved. It may happen that such time will precede the critical time from which material life will no longer be possible in the Universe.

## 4. INTERACTION OF HUMAN CONSCIOUSNESSES

The thought originated in a consciousness (static thought) presupposes the individualization of a *quantum* of psychic mass $\Delta m_{\Psi}$ in the very consciousness where the thought originated. Consequently, the wave function $\Psi_{\Psi}$ associated with this psychic body must collapse after a time interval $\Delta t$, expressing in the space-time its psychic content when it contains sufficiently psychic mass for that, or otherwise transforming itself in radiation (*psychic radiation*). In both cases, there is also production of "virtual" photons ("virtual' psychic radiation) to convey the psychic interaction.

According to the Uncertainty Principle, "virtual" *quanta* cannot be observed experimentally. However, since they are interaction *quanta*, their effects may be verified in the very particles or bodies subjected to the interactions.

Obviously, only one specific type of interaction occurs between two particles if each one *absorbs* the *quanta* of said interaction emitted by the other; otherwise, the interaction will be null. Thus, the null interaction between psychic bodies particularly means that there is no mutual absorption of the "virtual" psychic photons (psychic interaction *quanta*) emitted by them. That is, the emission spectrum of each one of them does not coincide with the absorption spectrum of the other.

By analogy with material bodies, whose emission spectra are, as we know, identical with the absorption ones, also the psychic bodies must absorb within the spectrum they emit. Specifically, in the case of human consciousness, their thoughts cause them to become emitters of psychic radiation in certain frequency spectra and, consequently, receivers in the same spectra. Thus, when a human consciousness, by its thoughts, is receptive coming from a certain thought, said radiation will be absorbed by the consciousness (resonance absorption). Under these circumstances, the radiation absorbed must *stimulate* – through the *Resonance Principle* – said consciousness to emit in the same spectrum, just as it happens with matter.

Nevertheless, in order for that emission to occur in a human consciousness, it must be preceded by the individualization of thoughts identical with that which originated the radiation absorbed because obviously only identical thoughts will be able to reproduce, when they collapse, the spectrum of "virtual" psychic radiations absorbed.

These *induced thoughts* – such as the thoughts of consciousnesses themselves – must remain individualized for a period of time $\Delta t$ (lifetime of the thought) after which its wave function will collapse, thus producing the "virtual" psychic radiation in the same spectrum of frequencies absorbed.

The Supreme Consciousness, just as the other consciousnesses, has Its own spectrum of absorption determined by Its thoughts – which make up the standard of a good-quality thought is hereby established. That is, they are *resonant* thoughts in Supreme Consciousness. Thus, only thoughts of this kind, produced in human consciousnesses, may induce the individualization of similar thoughts in Supreme Consciousness.

In this context, a system of judgment is established in which the good and the evil are psychic values, with their origin in free thought. *The good is related to the good-quality thoughts, which are thoughts resonant in Supreme Consciousness. The evil, in turn, is related to the bad-quality thoughts, non-resonant in the Supreme Consciousness.*

Consequently, the moral derived thereof results from the Law itself, inherent in the Supreme Consciousness and, therefore, this psychic moral must be the *fundamental moral*. Thus, fundamental ethics is neither biological nor located in the aggressive action, as thought by Nietzsche. It is psychic and located in the good-quality thoughts. It has a theological basis and in it the creation of the Universe by a pre-existing God is of an essential nature, opposed, for instance, to Spinoza's "geometrical ethics", which eliminated the ideas of the Creation of the Universe by a pre-existing God the main underpinning of Christian theology and philosophy. However, it is very close to Aristotle's ethics, to the extent that, from it, we understand that we are what we repeatedly do (think) and that *excellence is not an act, but a habit* (Ethics, II, 4). According to Aristotle: '" *the goodness of a man is a work of the soul towards excellence in a complete lifetime*: … it is not a day or a short period that makes a man fortunate and happy. " (Ibid, I, 7).

The "virtual" psychic radiation coming from a thought may induce *several* similar thoughts in the consciousness absorbing it, because each photon of radiation absorbed carries in itself the electromagnetic expression of the thought which produced it and, consequently, each one of them stimulates the individualization of a similar thought. However, the amount of thoughts induced is, of course, limited by the amount of psychic mass of the consciousness proper.

In the specific case of the Supreme Consciousness, the "virtual" psychic radiation coming from a good-quality thought must induce many similar thoughts. On the other hand, since Supreme Consciousness involves human consciousness the induced thoughts appear in the surroundings of the very consciousness which induced them. These thoughts are then strongly attracted by said consciousness and fuse therewith, for just as the thoughts generated in a consciousness have a high degree of positive mutual affinity with it, they will also have the thoughts induced by it.

The fusion of these thoughts in the consciousness obviously determines an *increase* in its psychic mass. We then conclude that the cultivation of good-quality thoughts is highly beneficial to the individual. On the contrary, the cultivation of bad-quality thoughts makes consciousness lose psychic mass.

When bad-quality thoughts are generated in a consciousness, they do not induce identical thoughts in Supreme Consciousness, because the absorption spectrum of Supreme Consciousness excludes psychic radiations coming from bad-quality thoughts. Thus, such radiation directs itself to other consciousnesses; however, it will only induce identical thoughts in those that are receptive in the same frequency spectrum. When this happens and right after the wave functions corresponding to these induced thoughts collapse and *materialize* said thoughts or changing them into radiation, the receptive consciousness will lose psychic mass, similarly to what happens in the consciousness which first produced the thought. Consequently, both the consciousness which gave rise to the bad-quality thought and those receptive to the psychic radiations coming from this type of thoughts will lose psychic mass.

We must observe, however, that our thoughts are not limited only to harming or benefiting ourselves, since they also can, as we have already seen, induce similar thoughts in other consciousnesses, thus affecting them. In this case, it is important to observe that the psychic radiation produced by the induced thoughts may return to the consciousness which initially produced the bad-quality thought, inducing other similar thoughts in it, which evidently cause more loss of psychic mass in said consciousness.

The fact of our thoughts not being restricted to influencing ourselves is highly relevant because it leads us to understand we have a great responsibility towards the others as regards what we think.

Let us now approach the intensity of thoughts. If two thoughts have the same psychic form and equal psychic masses, they have the same psychic density and, consequently, the same intensity, from the psychic viewpoint. However, if one of them has more psychic mass than the other, it will evidently have a larger psychic density and, thus, will be more intense.

The same thought repeated with different intensities in a consciousness – in a time period much shorter than the lifetime of thought – has its psychic mass increased due to the fusion of the psychic masses corresponding to each repetition. The fusion is caused by a strong psychic attraction between them, because the inertial thought and the repeated ones have high degree of positive mutual affinity.

It is then possible by this process that the thought may appear with enough psychic mass to materialize when its wave function collapses.

If the process is jointly shared with other consciousnesses, the thoughts in these consciousnesses evidently correspond to different dynamics states in the same thought. Thus, if $\Psi_{\Psi 1}$, $\Psi_{\Psi 2}$,..., $\Psi_{\Psi n}$ refer to the different dynamic states that the same thought may assume, then its general dynamic state, according to the superposition principle, may be described by a single wave function $\Psi_{\Psi}$, given by:

$$\Psi_{\Psi} = c_1\Psi_{\Psi 1} + c_2\Psi_{\Psi 2} + \dots + c_n\Psi_{\Psi n}$$

Therefore, everything happens as if there were only a single thought described by $\Psi_{\Psi}$, with psychic mass determined by the set of psychic masses of all the similar thoughts repeated in the various consciousnesses. In this manner, it is possible that in this process the thought materializes even faster than in the case of a single consciousness.

It was shown that the consciousnesses may increase their psychic masses by cultivating good-quality thoughts and avoiding the bad-quality thoughts ones. However, both the cultivation of good thoughts and the ability to instantly perceive nature in our thoughts to quickly repel the bad-quality thoughts result in a slow and difficult process.

The fact of intense enough mental images being capable of materializing suggests that we must be careful with mental images of fear. Thus more than anything else, it is imperative to avoid their repetition in our consciousnesses, because at each repetition they acquire more psychic mass.

Great are the possibilities encompassed in the consciousnesses, just as many are the effects of psychic interaction. At cellular level, the intervention of psychic interaction in the formation of the embryo's organs is particularly interesting.

Despite the recent advances in Embryology, embryologists cannot understand how the cells of the *internal cellular mass* [7] migrate to defined places in the embryo in order to form the organs of the future child.

We will show that this is a typical biological phenomenon which is fundamentally derived from the psychic interaction between the cells' consciousnesses.

Just as the consciousnesses of the children have a high degree of positive mutual affinity with the consciousnesses of their parents, and among themselves (*principle of familiar formation*), the embryo cells, by having originated from cellular duplication, have a high degree of positive mutual affinity. The embryo cells result, as we know, from the cellular duplication of a single cell containing the paternal and maternal genes and, hence, have a high degree of positive mutual affinity.

Thus, under the action of psychic interaction the cells of the internal cellular mass start gathering into small groups, according to the different degrees of mutual affinity.

When there is a positive mutual affinity between two consciousnesses there occurs the *intertwining* between their wave functions, and a *Phase Relationship* is established among them. Consequently, since the degree of positive mutual affinity among the embryo cells is high, also the relationship among them will be intense, and it is exactly this what enables the construction of the organs of the future child. In other words, when a cell is attracted by certain group in the embryo, it is through the cell-group relationship that determines where the cell is to aggregate to the group. In this manner, each cell finds its correct place in the embryo; that is why observers frequently say that, "*the cells appear to know where to go*", when experimentally observed.

The cells of the internal cellular mass are capable of originating any organ, and are hence called *totipotents*; thus, the organs begin to appear. In the endoderm, there appear the urinary organs, the respiratory system, and part of the digestive system; in the mesoderm are formed the muscles, bones, cartilages, blood, vessels, heart, kidneys; in the ectoderm there appear the skin, the nervous system, etc.

Thus, it is the mutual affinity among the consciousnesses of the cells that determines the formation of the body organs and keeps their own physical integrity. For this reason, every body rejects cells from other bodies, unless the latter have positive mutual affinity with their own cells. The higher the degree of cellular positive mutual affinity, the faster the integration of the transplanted cells and, therefore, the less problematic the transplant. In the case of cells from identical twins, this integration takes place practically with no problems, since said degree of mutual affinity is very high.

---

[7] When a spermatozoon penetrates the ovum, an *egg* is formed. Roughly twelve to fourteen hours later, the egg divides into two identical cells. This is the beginning of the phase where the embryo is called *morula*. Six days later, in the *blastula* phase, the external cells fix the embryo to the uterus. The cells inside the blastula remain equal to each other and are known as *internal cellular mass*.

In eight weeks of life, all organs are practically formed in the embryo. From there on, it begins to be called *fetus*.

The embryo's material individual consciousness is formed by the consciousnesses of its cells united in a Bose-Einstein condensate. As more cells become incorporated into the embryo, its material consciousness acquires more psychic mass. This means that this type of consciousness will be greater in the fetus than in the embryo and even greater in the child.

Thus, the psychic mass of the mother-fetus consciousness progressively increases during pregnancy, consequently increasing the psychic attraction between this consciousness and that new one about to incorporate. In normal pregnancies, this psychic attraction also increases due to the habitual increase in the degree of positive mutual affinity between said consciousnesses.

Since the embryo's consciousness has greater degree of positive mutual affinity with the consciousness that is going to incorporate, then the embryo's consciousness becomes the center of psychic attraction to where the human consciousness destined to the fetus will go.

When the psychic attraction becomes intense enough, human consciousness penetrates the mother-fetus consciousness, forming with it a new Bose–Einstein condensate. From that instant on, the fetus begins to have two consciousnesses: *the individual material one and the human consciousness attracted to it.*

It easy to see that the psychic attraction upon this human consciousness tends to continue, being progressively *compressed* until effectively incorporating the fetus. When this takes place, it will be ready to be born.

It is probably due to this *psychic compression* process that the incorporated consciousness suffers amnesia of its preceding history. Upon death, after the psychic decompression that arises from the definitive disincorporation of the consciousness, the preceding memory must return.

It was shown that particles of matter perform transitions to the *imaginary* space-time when their gravitational masses reach the gravitational mass ranging between $+0.159M_i$ to $-0.159M_i$ [7]. Under these circumstances, the total energy of the particle becomes *imaginary* and consequently it disappears from our *ordinary space-time.* Since *imaginary mass is equal to psychic mass* we can infer that the particle makes a transition to the *psychic space-time.*

The consciousnesses are in the *psychic space-time*. Therefore, if material bodies can become psychic bodies and to interact with others psychic bodies in this space-time, then they reach a new part of the Universe where the consciousnesses live and from where they come in order to incorporate the human fetus, and to where they should return, after the death of the material bodies. Consequently, the transition to the *psychic space-time* is a door for us to visit the spiritual Universe.

# The Gravitational Spacecraft

**Fran De Aquino**
Maranhao State University, Physics Department, S.Luis/MA, Brazil.

There is an electromagnetic factor of correlation between gravitational mass and inertial mass, which in specific electromagnetic conditions, can be reduced, made negative and increased in numerical value. This means that gravitational forces can be reduced, inverted and intensified by means of electromagnetic fields. Such control of the gravitational interaction can have a lot of practical applications. For example, a new concept of spacecraft and aerospace flight arises from the possibility of the electromagnetic control of the gravitational mass. The novel spacecraft called Gravitational Spacecraft possibly will change the paradigm of space flight and transportation in general. Here, its operation principles and flight possibilities, it will be described. Also it will be shown that other devices based on gravity control, such as the Gravitational Motor and the Quantum Transceivers, can be used in the spacecraft, respectively, for Energy Generation and Telecommunications.

**Key words:** Gravity, Gravity Control, Quantum Devices.

**CONTENTS**

## 1. Introduction

The discovery of the correlation between gravitational mass and inertial mass [1] has shown that the gravity can be *reduced*, *nullified* and *inverted*. Starting from this discovery several ways were proposed in order to obtain experimentally the local gravity control [2]. Consequently, new concepts of spacecraft and aerospace flight have arisen. This novel spacecraft, called Gravitational Spacecraft, can be equipped with other devices also based on gravity control, such as the Gravitational Motor and the Quantum Transceiver that can be used, respectively, for energy generation and telecommunications. Based on the theoretical background which led to the gravity control, the operation principles of the Gravitational Spacecraft and of the devices above mentioned, will be described in this work.

## 2. Gravitational Shielding

The contemporary greatest challenge of the Theoretical Physics was to prove that, Gravity is a *quantum* phenomenon. Since the General Relativity describes gravity as related to the curvature of the space-time then, the quantization of the gravity implies the quantization of the proper space-time. Until the end of the century XX, several attempts to quantify gravity were accomplished. However, all of them resulted fruitless [3, 4].

In the beginning of this century, it has been clearly noticed that there was something unsatisfactory about the whole notion of quantization and that the quantization process had many ambiguities. Then, a new approach has been proposed starting from the generalization of the *action function**. The result has been the derivation of a theoretical background, which finally led to the so-sought quantization of the gravity and of the space-time. Published under the title: "*Mathematical Foundations of the Relativistic Theory of Quantum Gravity*"†, this theory predicts a consistent *unification* of Gravity with Electromagnetism. It shows that the *strong* equivalence principle is reaffirmed and, consequently Einstein's equations are preserved. In fact, Einstein's equations can be deduced directly from the *Relativistic Theory of Quantum Gravity*. This shows, therefore, that the General Relativity is a particularization of this new theory, just as the Newton's theory is a particular case from the General Relativity. Besides, it was deduced from the new theory an important correlation between the *gravitational mass* and the *inertial mass*, which shows that the gravitational mass of a particle can be *decreased* and even made *negative*, independently of its inertial mass, i.e., while the gravitational mass is

* The formulation of the *action* in Classical Mechanics extends to the Quantum Mechanics and it has been the basis for the development of the *Strings Theory*.

† http://arxiv.org/abs/physics/0212033

progressively reduced, the inertial mass does not vary. This is highly relevant because it means that the weight of a body can also be reduced and even inverted in certain circumstances, since Newton's gravity law defines the weight $P$ of a body as the product of its *gravitational mass* $m_g$ by the local gravity acceleration $g$, i.e.,

$$P = m_g g \tag{1}$$

It arises from the mentioned law that the gravity acceleration (or simply the gravity) produced by a body with gravitational mass $M_g$ is given by

$$g = \frac{GM_g}{r^2} \tag{2}$$

The physical property of mass has two distinct aspects: *gravitational mass* $m_g$ and *inertial mass* $m_i$. The gravitational mass produces and responds to gravitational fields. It supplies the mass factors in Newton's famous inverse-square law of gravity $\left(F = GM_g m_g / r^2\right)$. The inertial mass is the mass factor in *Newton's 2nd Law of Motion* $(F = m_i a)$. These two masses are not equivalent but correlated by means of the following factor [1]:

$$\frac{m_g}{m_{i0}} = \left\{1 - 2\left[\sqrt{1 + \left(\frac{\Delta p}{m_{i0} c}\right)^2} - 1\right]\right\} \tag{3}$$

Where $m_{i0}$ is the *rest* inertial mass and $\Delta p$ is the variation in the particle's *kinetic momentum*; $c$ is the speed of light.

This equation shows that only for $\Delta p = 0$ the gravitational mass is equal to the inertial mass. Instances in which $\Delta p$ is produced by *electromagnetic radiation*, Eq. (3) can be rewritten as follows:

$$\frac{m_g}{m_{i0}} = \left\{1 - 2\left[\sqrt{1 + \left(\frac{n_r^2 D}{\rho c^3}\right)^2} - 1\right]\right\} \tag{4}$$

Where $n_r$ is the *refraction index* of the particle; $D$ is the power density of the electromagnetic radiation absorbed by the particle; and $\rho$ its density of inertial mass.

It was shown [1] that there is an additional effect of *gravitational shielding* produced by a substance whose gravitational mass was reduced or made negative. This effect shows that just *above the substance* the gravity acceleration $g_1$ will be reduced at the same proportion $\chi = m_g / m_{i0}$, i.e., $g_1 = \chi g$, ($g$ is the gravity acceleration *bellow* the substance).

Equation (4) shows, for example, that, in the case of a gas at ultra-low pressure (*very low density of inertial mass*), the *gravitational mass* of the gas can be strongly reduced or made negative by means of the incidence of electromagnetic radiation with power density relatively low.

Thus, it is possible to use this effect in order to produce gravitational shieldings and, thus, *to control the local gravity*.

The *Gravity Control Cells* (GCC) shown in the article "*Gravity Control by means of Electromagnetic*

*Field through Gas or Plasma at Ultra-Low Pressure*"‡, are devices designed on the basis, of this effect, and usually are chambers containing gas or plasma at ultra-low pressure. Therefore, when an oscillating electromagnetic field is applied upon the gas its gravitational mass will be reduced and, consequently, the gravity *above* the mentioned GCC will also be reduced at the same proportion.

It was also shown that it is possible to make a gravitational shielding even with the chamber filled with *Air* at one atmosphere. In this case, the *electric conductivity of the air* must be strongly increased in order to reduce the intensity of the electromagnetic field or the power density of the applied radiation.

This is easily obtained by *ionizing the air* in the local where we want to build the gravitational shielding. There are several manners of ionizing the air. One of them is by means of ionizing radiation produced by a radioactive source of low intensity, for example, by using the radioactive element *Americium* (Am-241). The Americium is widely used as air ionizer in smoke detectors. Inside the detectors, there is just a little amount of americium 241 (about of 1/5000 grams) in the form of $AmO_2$. Its cost is very low (about of US$ 1500 per gram). The dominant radiation is composed of alpha particles. Alpha particles cannot cross a paper sheet and are also blocked by some centimeters of air. The Americium used in the smoke detectors can only be dangerous if inhaled.

The Relativistic Theory of Quantum Gravity also shows the existence of a *generalized equation for the inertial forces* which has the following form

$$F_i = M_g a \tag{5}$$

This expression means a *new law for the Inertia*. Further on, it will be shown that it *incorporates the Mach's principle* to Gravitation theory [5].

Equation (3) tell us that the gravitational mass is only equal to the inertial mass when $\Delta p = 0$. Therefore, we can easily conclude that only in this particular situation the new expression of $F_i$ reduces to $F_i = m_i a$, which is the expression for Newton's 2nd Law of Motion. Consequently, this Newton's law is just a particular case from the new law expressed by the Eq. (5), which clearly shows how *the local inertial forces are correlated to the gravitational interaction of the local system with the distribution of cosmic masses* (via $m_g$) and thus, *incorporates* definitively *the Mach's principle* to the Gravity theory.

The Mach's principle postulates that: "*The local inertial forces would be produced by the gravitational interaction of the local system with the distribution of cosmic masses*". However, in spite of the several attempts carried out, this principle had not yet been incorporated to the Gravitation theory. Also Einstein had carried out several attempts. The *ad hoc* introduction of the cosmological

‡ http://arxiv.org/abs/physics/0701091

term in his gravitation equations has been one of these attempts.

With the advent of equation (5), *the origin of the inertia* - that was considered the most obscure point of the particles' theory and field theory – becomes now evident.

In addition, this equation also reveals that, if the gravitational mass of a body is very close to zero or if there is around the body a *gravitational shielding* which reduces closely down to zero the *gravity accelerations due to the rest of the Universe*, then the intensities of the inertial forces that act on the body become also very close to zero.

This conclusion is highly relevant because it shows that, under these conditions, the spacecraft could describe, with great velocities, unusual trajectories (such as curves in right angles, abrupt inversion of direction, etc.) without inertial impacts on the occupants of the spacecraft. Obviously, out of the above-mentioned condition, the spacecraft and the crew would be destroyed due to the strong presence of the inertia.

When we make a sharp curve with our car we are pushed towards a direction contrary to that of the motion of the car. This happens due to existence of *the inertial forces*. However, if our car is involved by a *gravitational shielding*, which reduces strongly the gravitational interaction of the car (and everything that is inside the car) with the rest of the Universe, then in accordance with the Mach's principle, the local inertial forces would also be strongly reduced and, consequently, we would not feel anything during the maneuvers of the car.

## 3. Gravitational Motor: Free Energy

It is known that the energy of the gravitational field of the Earth can be converted into rotational kinetic energy and electric energy. In fact, this is exactly what takes place in hydroelectric plants. However, the construction these hydroelectric plants have a high cost of construction and can only be built, obviously, where there are rivers.

The gravity control by means of any of the processes mentioned in the article: "*Gravity Control by means of Electromagnetic Field through Gas or Plasma at Ultra-Low Pressure*" allows the inversion of the weight of any body, practically at any place. Consequently, the conversion of the gravitational energy into rotational mechanical energy can also be carried out at any place.

In Fig. (1), we show a schematic diagram of a *Gravitational Motor*. The first *Gravity Control Cell* (GCC1) changes the local gravity from $g$ to $g' = -ng$, propelling the left side of the rotor in a direction contrary to the motion of the right side. The second GCC changes the gravity back again to $g$ i.e., from $g' = -ng$ to $g$, in such a way that the gravitational change occurs just on the region indicated in Fig.1. Thus, a *torque* $T$ given by

$$T = (-F' + F)r = \left[-(m_g/2)g' + (m_g/2)g\right]r =$$
$$= (n+1)\tfrac{1}{2}m_g gr$$

Is applied on the rotor of gravitational mass $m_g$, making the rotor spin with angular velocity $\omega$.

The average *power*, $P$, of the motor is $P = T\omega$. However, $-g' + g = \omega^2 r$. Thus, we have

$$P = \tfrac{1}{2} m_i \sqrt{(n+1)^3 g^3 r} \qquad (6)$$

Consider a cylindrical rotor of iron $(\rho = 7800 Kg.m^{-3})$ with height $h = 0.5m$, radius $r = R/3 = 0.0545m$ and inertial mass $m_i = \rho\pi R^2 h = 327.05kg$. By adjusting the GCC 1 in order to obtain $\chi_{air(1)} = m_{g(air)}/m_{i(air)} = -n = -19$ and, since $g = 9.81m.s^{-2}$, then Eq. (6) gives

$$P \cong 2.19 \times 10^5 \, watts \cong 219 \;\; KW \cong 294HP$$

This shows that this small motor can be used, for example, to substitute the conventional motors used in the cars. It can also be coupled to an electric generator in order to produce *electric energy*. The conversion of the rotational mechanical energy into electric energy is not a problem since it is a problem technologically resolved several decades ago. Electric generators are usually produced by the industries and they are commercially available, so that it is enough to couple a gravitational motor to an electric generator for we obtaining electric energy. In this case, just a gravitational motor with the power above mentioned it would be enough to supply the need of electric energy of, for example, at least 20 residences. Finally, it can substitute the conventional motors of the same power, with the great advantage of *not needing of fuel for its opera*tion. What means that the gravitational motors can produce energy practically free.

It is easy to see that gravitational motors of this kind can be designed for powers needs of just some watts up to millions of kilowatts.

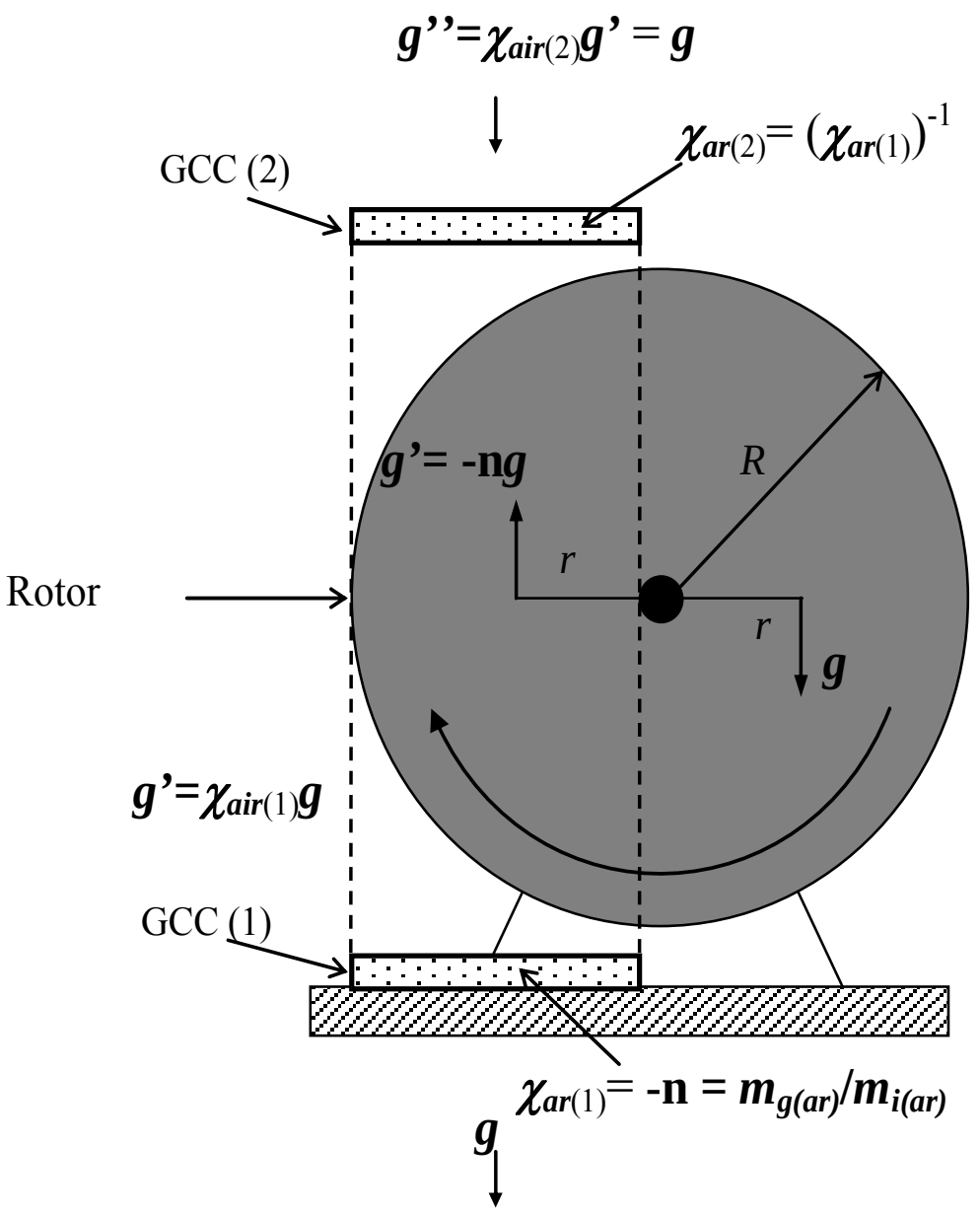

Fig. 1 – Gravitational Motor - The first *Gravity Control Cell* (GCC1) changes the local gravity from $g$ to $g' = -ng$, propelling the left side of the rotor in contrary direction to the motion of the right side. The second GCC changes the gravity back again to $g$ i.e., from $g' = -ng$ to $g$, in such a way that the gravitational change occurs just on the region shown in figure above.

## 4. The Gravitational Spacecraft

Consider a metallic sphere with radius $r_s$ in the terrestrial atmosphere. If the external surface of the sphere is recovered with a radioactive substance (for example, containing Americium 241) then the air in the space close to the surface of the sphere will be strongly ionized by the radiation emitted from the radioactive element and, consequently, the electric conductivity of the air close to sphere will become strongly increased.

By applying to the sphere an electric potential of low frequency $V_{rms}$, in order to produce an electric field $E_{rms}$ starting from the surface of the sphere, then very close to the surface, the intensity of the electric field will be $E_{rms} = V_{rms}/r_s$ and, in agreement with Eq. (4), the *gravitational mass* of the *Air* in this region will be expressed by

$$m_{g(air)} = \left\{1 - 2\left[\sqrt{1 + \frac{\mu_0}{4c^2}\left(\frac{\sigma_{air}}{4\pi f}\right)^3 \frac{V_{rms}^4}{r_s^4 \rho_{air}^2}} - 1\right]\right\} m_{i0(air)} \qquad (7)$$

Therefore we will have

$$\chi_{air} = \frac{m_{g(air)}}{m_{i0(air)}} = \left\{1 - 2\left[\sqrt{1 + \frac{\mu_0}{4c^2}\left(\frac{\sigma_{air}}{4\pi f}\right)^3 \frac{V_{rms}^4}{r_s^4 \rho_{air}^2}} - 1\right]\right\} \qquad (8)$$

The gravity accelerations acting on the sphere, due to the rest of the Universe (See Fig. 2), will be given by

$$g_i' = \chi_{air} g_i \qquad i = 1,2,...,n$$

Note that by varying $V_{rms}$ or the frequency $f$, we can easily to *reduce and control* $\chi_{air}$. Consequently, we can also control the intensities of the gravity accelerations $g_i'$ in order to produce a *controllable gravitational shielding* around the sphere.

Thus, the *gravitational forces* acting on the sphere, due to the rest of the Universe, will be given by

$$F_{gi} = M_g g_i' = M_g (\chi_{air} g_i)$$

where $M_g$ is the gravitational mass of the sphere.

The gravitational shielding around of the sphere reduces both the gravity accelerations acting on the sphere, due to the rest of the Universe, and the gravity acceleration produced by the gravitational mass $M_g$ of the own sphere. That is, if inside the shielding the gravity produced by the sphere is $g = -GM_g/r^2$, then, *out of* the shielding it becomes $g' = \chi_{air} g$. Thus, $g' = \chi_{air}(-GM_g/r^2) = -G(\chi_{air} M_g)/r^2 = -Gm_g/r^2$, where

$$m_g = \chi_{air} M_g$$

Therefore, *for the Universe out of the shielding* the gravitational mass of the sphere is $m_g$ and not $M_g$. In these circumstances, the *inertial forces* acting on the sphere, in agreement with the *new law for inertia*, expressed by Eq. (5), will be given by

$$F_{ii} = m_g a_i \qquad (9)$$

Thus, these forces will be almost null when $m_g$ becomes almost null by means of the action of the gravitational shielding. This means that, in these circumstances, the sphere practically loses its *inertial properties*. This effect leads to *a new concept of spacecraft and aerospatial flight*. The spherical form of the spacecraft is just *one* form that the Gravitational Spacecraft can have, since the gravitational shielding can also be obtained with other formats.

An important aspect to be observed is that it is possible to control the gravitational mass of the spacecraft, $M_{g(spacecraf)}$, simply by controlling the gravitational mass of a body *inside* the spacecraft. For instance, consider a parallel plate capacitor inside the spacecraft. The gravitational mass of the *dielectric* between the plates of the capacitor can be controlled by means of the ELF electromagnetic field through it. Under these circumstances, the *total* gravitational mass of the spacecraft will be given by

$$M_{g(spacecraf)}^{total} = M_{g(spacecraf)} + m_g = $$
$$= M_{i0} + \chi_{dielectric} m_{i0} \qquad (10)$$

where $M_{i0}$ is the rest inertial mass of the spacecraft(without the dielectric) and $m_{i0}$ is the rest inertial mass of the dielectric; $\chi_{dielectric} = m_g / m_{i0}$, where $m_g$ is the gravitational mass of the dielectric. By decreasing the value of $\chi_{dielectric}$, the gravitational mass of the spacecraft decreases. It was shown, that the value of $\chi$ can be negative. Thus, when $\chi_{dielectric} \cong -M_{i0}/m_{i0}$, the *gravitational mass of the spacecraft gets very close to zero*. When $\chi_{dielectric} < -M_{i0}/m_{i0}$, the gravitational mass of the spacecraft becomes negative.

Therefore, *for an observer out of the spacecraft*, the gravitational mass of the spacecraft is $M_{g(spacecraf)} = M_{i0} + \chi_{dielectric} m_{i0}$, and not $M_{i0} + m_{i0}$.

Another important aspect to be observed is that we can *control the gravity inside the spacecraft*, in order to produce, for example, a gravity acceleration equal to the Earth's gravity $(g = 9.81 m.s^{-2})$. This will be very useful in the case of space flight, and can be easily obtained by putting in the ceiling of the spacecraft the system shown in Fig. 3. This system has three GCC with nuclei of ionized air (or air at low pressure). Above these GCC there is a massive block with mass $M_g$.

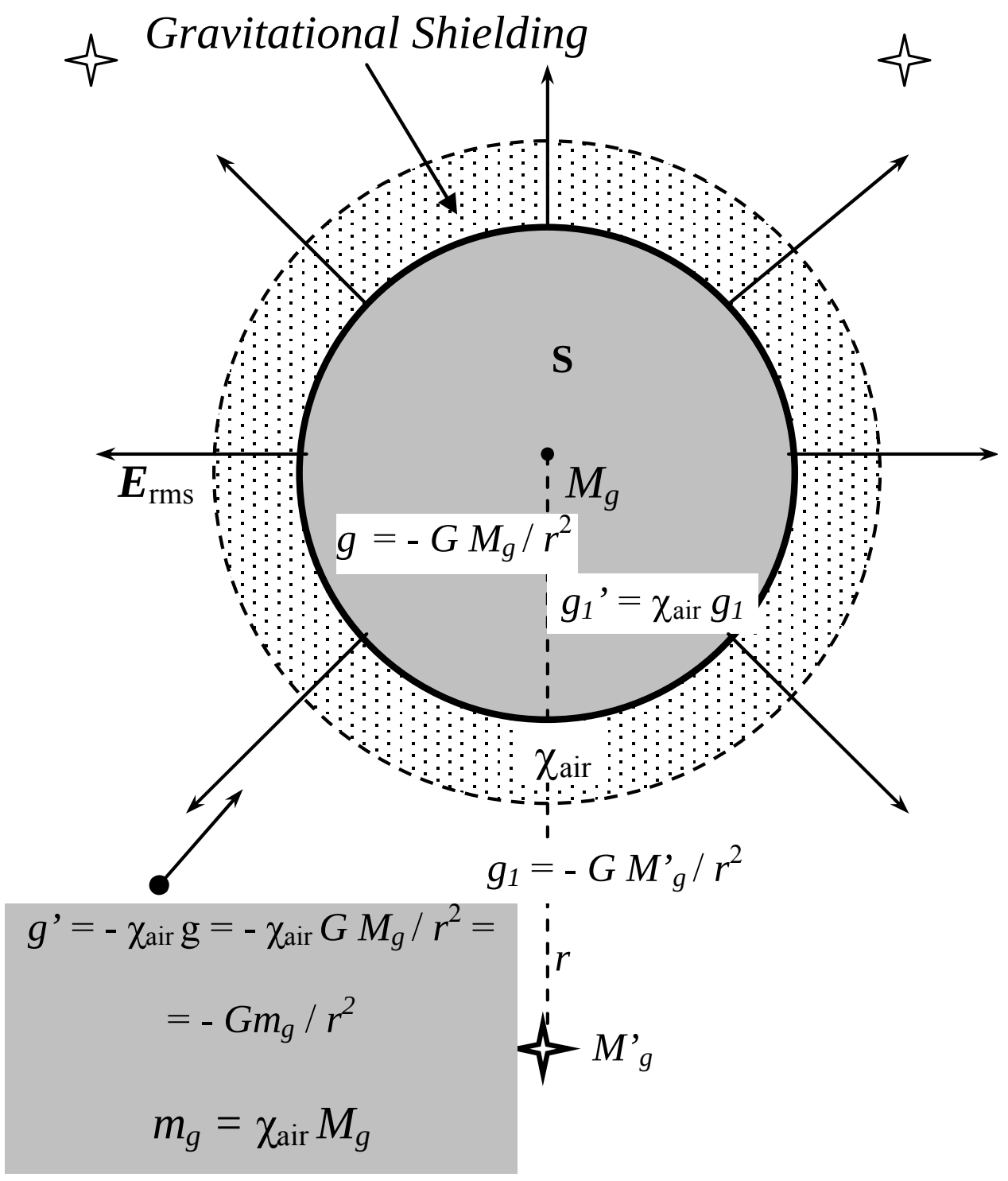

Fig.2- The gravitational shielding reduces the *gravity accelerations* ( $g_1'$) acting on the sphere (due to the rest of the Universe) and also reduces the gravity acceleration that the sphere produces upon all the particles of the Universe (**g'**). For the Universe, the gravitational mass of the sphere will be $m_g = \chi_{air} M_g$.

As we have shown [2], a gravitational *repulsion* is established between the mass $M_g$ and any positive gravitational mass below the mentioned system. This means that the particles in this region will stay subjected to a gravity acceleration $a_b$, given by

$$\vec{a}_b \cong (\chi_{air})^3 \vec{g}_M \cong -(\chi_{air})^3 G \frac{M_g}{r_0^2} \hat{\mu} \qquad (11)$$

If the Air inside the GCCs is sufficiently ionized, in such way that $\sigma_{air} \cong 10^3 S.m^{-1}$, and if $f = 1\ Hz$, $\rho_{air} \cong 1\ kg.m^{-3}$, $V_{rms} \cong 10\ KV$ and $d = 1\ cm$ then the Eq.8 shows that inside the GCCs we will have

$$\chi_{air} = \frac{m_{g(air)}}{m_{i0(air)}} = \left\{1 - 2\left[\sqrt{1 + \frac{\mu_0}{4c^2}\left(\frac{\sigma_{air}}{4\pi f}\right)^3 \frac{V_{rms}^4}{d^4 \rho_{air}^2}} - 1\right]\right\} \cong -10^3$$

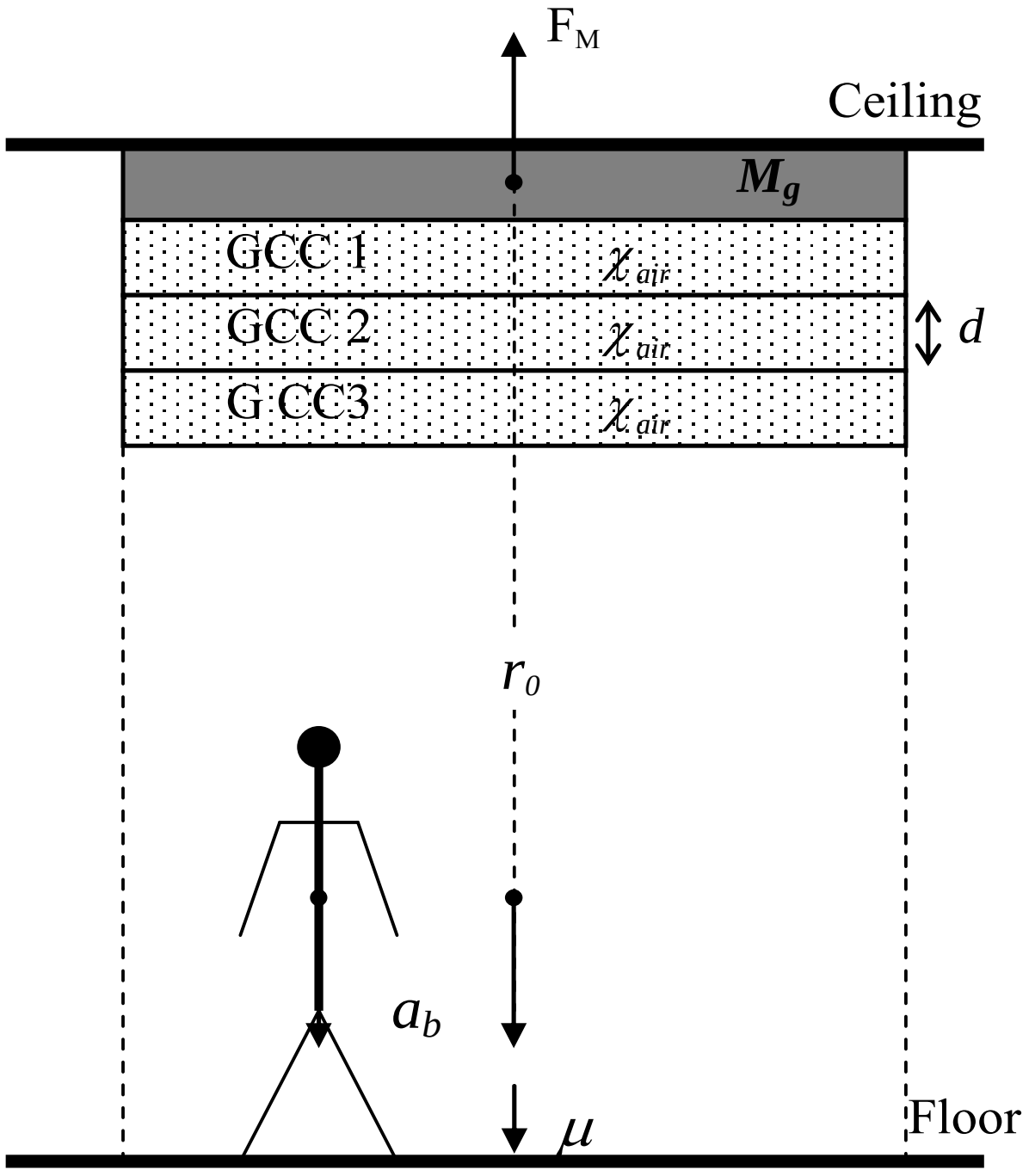

$$\vec{a}_b \cong (\chi_{air})^3 \vec{g}_M \cong -(\chi_{air})^3 G \frac{M_g}{r_0^2} \hat{\mu}$$

Fig.3 – If the Air inside the GCC is sufficiently ionized, in such way that $\sigma_{air} \cong 10^3 S.m^{-1}$ and if $f = 1\ Hz$; $d = 1cm$; $\rho_{air} \cong 1\ kg.m^{-3}$ and $V_{rms} \cong 10\ KV$ then Eq. 8 shows that inside the CCGs we will have $\chi_{air} \cong -10^3$. Therefore, for $M_g \cong M_i \cong 100kg$ and $r_o \cong 1m$ the gravity acceleration inside the spacecraft will be directed from the ceiling to the floor of the spacecraft and its intensity will be $a_b \approx 10 m.s^{-2}$.

Therefore the equation (11) gives

$$a_b \approx +10^9 G \frac{M_g}{r_0^2} \qquad (12)$$

For $M_g \cong M_i \cong 100kg$ and $r_0 \cong 1m$ (See Fig.3), the gravity inside the spacecraft will be directed from the ceiling to the floor and its intensity will have the following value

$$a_b \approx 10 m.s^{-2} \qquad (13)$$

Therefore, an interstellar travel in a gravitational spacecraft will be particularly comfortable, since we can travel during all the time subjected to the gravity which we are accustomed to here in the Earth.

We can also use the system shown in Fig. 3 as a thruster in order to propel the spacecraft. Note that the gravitational repulsion that occurs between the block with mass $M_g$ and any particle after the GCCs *does not depend on* of the place where the system is working. Thus, this *Gravitational Thruster* can propel the gravitational spacecraft in *any direction*. Moreover, it can work in the terrestrial atmosphere as well as in the cosmic space. In this case, the energy that produces the propulsion is obviously the *gravitational energy*, which is always present in any point of the Universe.

The schematic diagram in Fig. 4 shows in details the operation of the Gravitational Thruster. A gas of any type injected into the chamber beyond the GCCs acquires an acceleration $a_{gas}$, as shown in Fig.4, the intensity of which, as we have seen, is given by

$$a_{gas} = (\chi_{gas})^3 g_M \cong -(\chi_{gas})^3 G \frac{M_g}{r_0^2} \qquad (14)$$

Thus, if inside of the GCCs, $\chi_{gas} \cong -10^9$ then the equation above gives

$$a_{gas} \cong +10^{27} G \frac{M_g}{r_0^2} \qquad (15)$$

For $M_g \cong M_i \cong 10kg$, $r_0 \cong 1m$ we have $a_{gas} \cong 6.6 \times 10^{17} m.s^{-2}$. With this enormous acceleration the particles of the gas reach velocities close to the speed of the light in just a few nanoseconds. Thus, if the emission rate of the gas is $dm_{gas}/dt \cong 10^{-3} kg/s \cong 4000 litres/hour$, then the trust produced by the gravitational thruster will be

$$F = v_{gas}\frac{dm_{gas}}{dt} \cong c\frac{dm_{gas}}{dt} \cong 10^5 N \qquad (16)$$

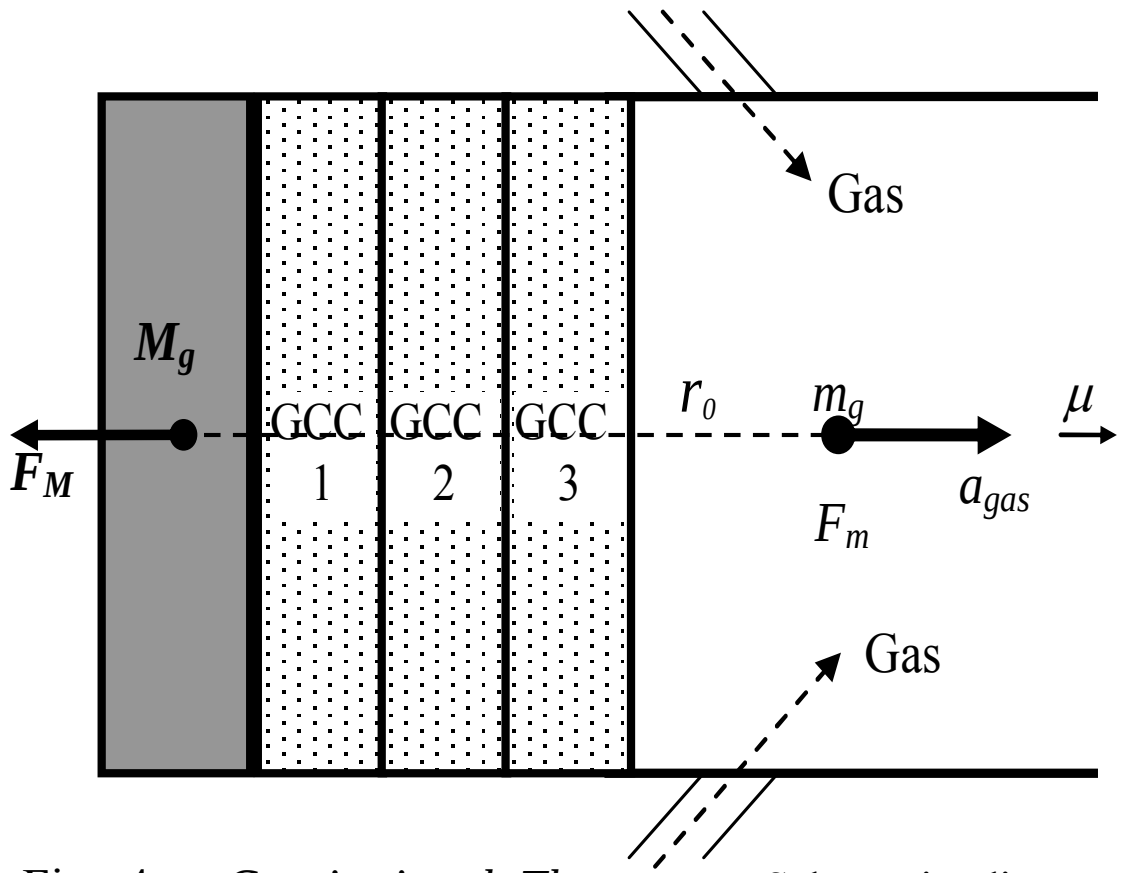

Fig. 4 – *Gravitational Thruster* – Schematic diagram showing the operation of the Gravitational Thruster. Note that in the case of very strong $\chi_{air}$, for example $\chi_{air} \cong -10^9$, the gravity accelerations upon the boxes of the second and third GCCs become very strong. Obviously, the walls of the mentioned boxes cannot to stand the enormous pressures. However, it is possible to build a similar system [2] with 3 or more GCCs, *without material boxes*. Consider for example, a surface with several radioactive sources (Am-241, for example). The *alpha* particles emitted from the Am-241 cannot reach besides 10cm of air. Due to the trajectory of the alpha particles, three or more successive layers of air, with different electrical conductivities $\sigma_1$, $\sigma_2$ and $\sigma_3$, will be established in the ionized region. It is easy to see that the gravitational shielding effect produced by these three layers is similar to the effect produced by the 3 GCCs above.

It is easy to see that the gravitational thrusters are able to produce strong trusts (similarly to the produced by the powerful thrusters of the modern aircrafts) *just by consuming the injected gas for its operation.*

It is important to note that, if $F$ is the thrust produced by the gravitational thruster then, in agreement with Eq. (5), the spacecraft acquires an acceleration $a_{spacecraft}$, expressed by the following equation

$$a_{spacecraft} = \frac{F}{M_{g(spacecraft)}} = \frac{F}{\chi_{out} M_{i(spacecraft)}} \qquad (17)$$

Where $\chi_{out}$, given by Eq. (8), is the factor of gravitational shielding which depends on the external medium where the spacecraft is placed. By adjusting the shielding for $\chi_{out} = 0.01$ and if $M_{spacecraft} = 10^4 Kg$ then for a thrust $F \cong 10^5 N$, the acceleration of the spacecraft will be

$$a_{spacecraft} = 1000 m.s^{-2} \qquad (18)$$

With this acceleration, in just at 1(one) day, the velocity of the spacecraft will be close to the speed of light. However it is easy to see that $\chi_{out}$ can still be much more reduced and, consequently, the thrust much more increased so that it is possible to increase up to 1 million times the acceleration of the spacecraft.

It is important to note that, the inertial effects upon the spacecraft will be reduced by $\chi_{out} = M_g / M_i \cong 0.01$. Then, in spite of its effective acceleration to be $a = 1000 m.s^{-2}$, the effects for the crew of the spacecraft will be equivalents to an acceleration of only

$$a' = \frac{M_g}{M_i} a \approx 10 m.s^{-1}$$

This is the magnitude of the acceleration on the passengers in a contemporary commercial jet.

Then, it is noticed that the gravitational spacecrafts can be subjected to enormous *accelerations* (or *decelerations*) without imposing any harmful impacts whatsoever on the spacecrafts or its crew.

We can also use the system shown in Fig. 3, as a *lifter*, inclusively within the spacecraft, in order to lift peoples or things into the spacecraft as shown in Fig. 5. Just using two GCCs, the gravitational acceleration produced below the GCCs will be

$$\vec{a}_g = (\chi_{air})^2 g_M \cong -(\chi_{air})^2 G M_g / r_0^2 \, \hat{\mu} \qquad (19)$$

Note that, in this case, if $\chi_{air}$ is *negative*, the acceleration $\vec{a}_g$ will have a direction *contrary* to the versor $\hat{\mu}$, i.e., the body will be *attracted* in the direction of the GCCs, as shown in Fig.5. In practice, this will occur when the air inside the GCCs is sufficiently ionized, in such a way that $\sigma_{air} \cong 10^3 S.m^{-1}$. Thus, if the internal thickness of the GCCs is now $d = 1\ mm$ and if $f = 1\ Hz$; $\rho_{air} \cong 1\ kg.m^{-3}$ and $V_{rms} \cong 10\ KV$, we will then have $\chi_{air} \cong -10^5$. Therefore, for $M_g \cong M_i \cong 100kg$ and, for example, $r_0 \cong 10m$ the gravitational acceleration acting on the body will be $a_b \approx 0.6 m.s^{-2}$. It is obvious that this value can be easily increased or decreased, simply by varying the voltage $V_{rms}$. Thus, by means of this *Gravitational Lifter,* we can lift or lower persons or materials with great versatility of operation.

It was shown [1] that, when the gravitational mass of a particle is reduced into the range, $+0.159 M_i$ to $-0.159 M_i$, it becomes imaginary, i.e., its masses (gravitational and inertial) becomes imaginary. Consequently, the particle disappears from our ordinary Universe, i.e., it becomes *invisible* for us. This is therefore a manner of to obtain the transitory invisibility of persons, animals, spacecraft, etc. However, the factor $\chi = M_{g(imaginary)} / M_{i(imaginary)}$ remains real because

$$\chi = \frac{M_{g(imaginary)}}{M_{i(imaginary)}} = \frac{M_g i}{M_i i} = \frac{M_g}{M_i} = real$$

Thus, if the gravitational mass of the particle is reduced by means of the absorption of an amount of electromagnetic energy $U$, for example, then we have

$$\chi = \frac{M_g}{M_i} = \left\{ 1 - 2\left[ \sqrt{1 + \left( U / m_{i0} c^2 \right)^2} - 1 \right] \right\}$$

This shows that the energy $U$ *continues acting* on the particle turned imaginary. In practice this means that *electromagnetic fields act on imaginary particles.* Therefore, the internal electromagnetic field of a GCC remains acting upon the particles inside the GCC even when their gravitational masses are in the range $+0.159 M_i$ to $-0.159 M_i$, turning them *imaginaries.* This is very important because it means that the GCCs of a gravitational spacecraft remain working even when the spacecraft becomes imaginary.

Under these conditions, the gravity accelerations acting on the imaginary spacecraft, due to the rest of the Universe will be, as we have see, given by

$$g_i' = \chi \ g_i \qquad i = 1,2,...,n$$

Where $\chi = M_{g(imaginary)} / M_{i(imaginary)}$ and $g_i = -G m_{gi(imaginary)} / r_i^2$. Thus, the gravitational forces acting on the spacecraft will be given by

$$F_{gi} = M_{g(imaginary)} g_i' =$$
$$= M_{g(imaginary)} \left( -\chi G m_{gj(imaginary)} / r_j^2 \right) =$$
$$= M_g i \left( -\chi G m_{gi} i / r_i^2 \right) = +\chi G M_g m_{gi} / r_i^2 . \qquad (20)$$

Note that these forces are *real*. By calling that, the *Mach's principle* says that the *inertial effects* upon a particle are consequence of the gravitational interaction of the particle with the rest of the Universe. Then we can conclude that the inertial forces acting on the spacecraft in imaginary state are also *real*. Therefore, it can travel in the imaginary space-time using the gravitational thrusters.

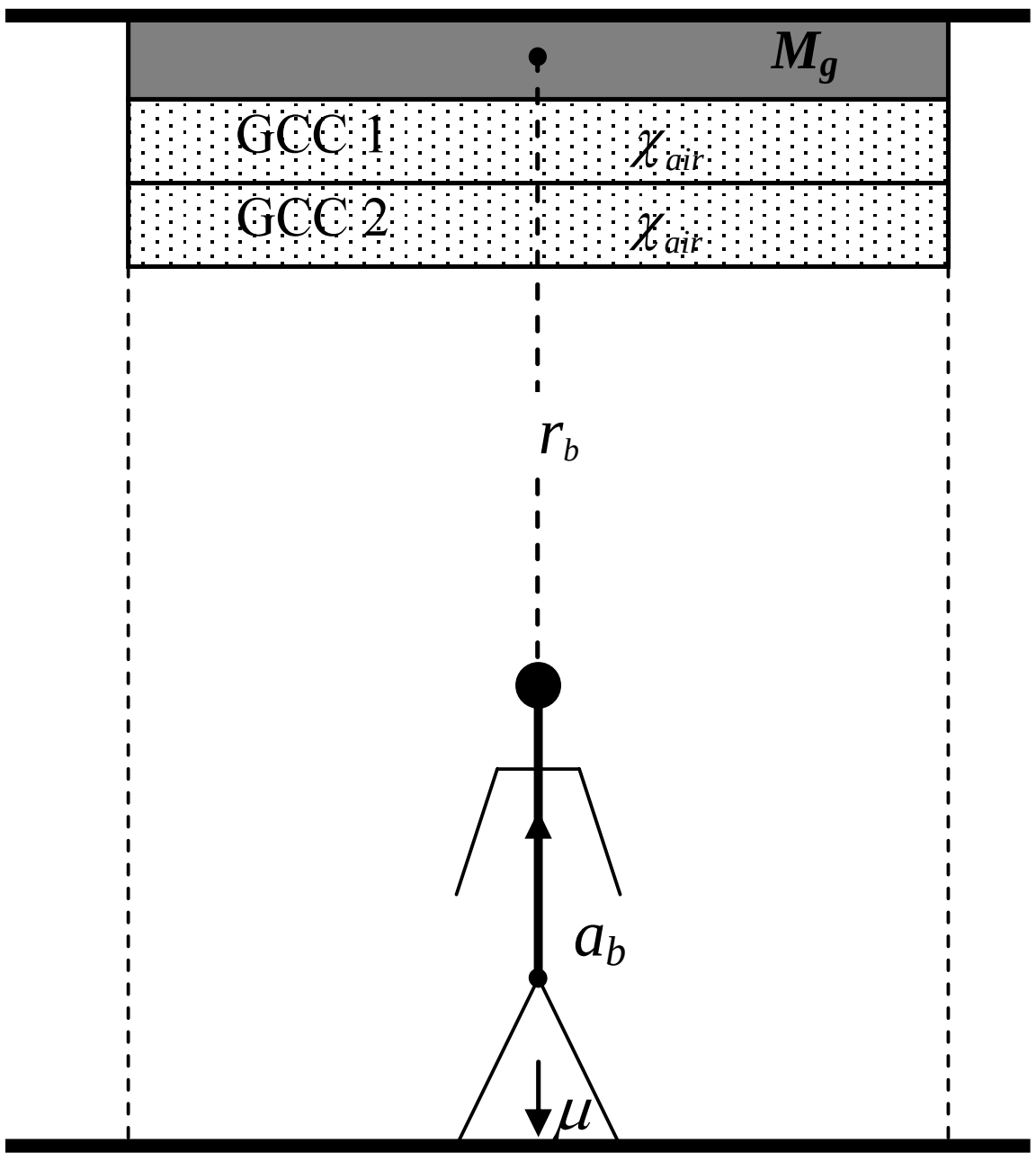

$$\vec{a}_b \cong (\chi_{air})^2 \vec{g}_M \cong -(\chi_{air})^2 G \frac{M_g}{r_b^2} \hat{\mu}$$

Fig.5 – *The Gravitational Lifter* – If the air inside the GCCs is sufficiently ionized, in such way that $\sigma_{air} \cong 10^3 S.m^{-1}$ and the internal thickness of the GCCs is now $d = 1\ mm$ then, if $f = 1\ Hz$; $\rho_{air} \cong 1\ kg.m^{-3}$ and $V_{rms} \cong 10\ KV$, we have $\chi_{air} \cong -10^5$. Therefore, for $M_g \cong M_i \cong 100 kg$ and $r_0 \cong 10 m$ the gravity acceleration acting on the body will be $a_b \approx 0.6 m.s^{-2}$.

It was also shown [1] that *imaginary particles* can have *infinity velocity* in the imaginary space-time. Therefore, this is also the upper limit of velocity for the gravitational spacecrafts traveling in the imaginary space-time. On the other hand, the travel in the imaginary space-time can be very safe, because there will not be any material body in the trajectory of the spacecraft.

It is easy to show that the gravitational forces between two thin layers of air (with masses $m_{g1}$ and $m_{g2}$ ) around the spacecraft , are expressed by

$$\vec{F}_{12} = -\vec{F}_{21} = -(\chi_{air})^2 G \frac{m_{i1} m_{i2}}{r^2} \hat{\mu} \qquad (21)$$

Note that these forces can be strongly increased by increasing the value of $\chi_{air}$. In these circumstances, the air around the spacecraft would be strongly compressed upon the external surface of the spacecraft creating an atmosphere around it. This can be particularly useful in order to minimize the *friction* between the spacecraft and the atmosphere of the planet in the case of very high speed movements of the spacecraft. With the atmosphere around the spacecraft the friction will occur between the atmosphere of the spacecraft and the atmosphere of the planet. In this way, the friction will be minimum and the spacecraft could travel at very high speeds without overheating.

However, in order for this to occur, it is necessary to put the gravitational shielding in another position as shown in Fig.2. Thus, the values of $\chi_{airB}$ and $\chi_{airA}$ will be independent (See Fig.6). Thus, while inside the gravitational shielding, the value of $\chi_{airB}$ is put close to zero, in order to strongly reduce the gravitational mass of the spacecraft (inner part of the shielding), the value of $\chi_{airA}$ must be reduced to about $-10^8$ in order to strongly increase the gravitational attraction between the air molecules around the spacecraft. Thus, by substituting $\chi_{airA} \cong -10^8$ into Eq.21, we get

$$\vec{F}_{12} = -\vec{F}_{21} = -10^{16} G \frac{m_{i1} m_{i2}}{r^2} \hat{\mu} \qquad (22)$$

If, $m_{i1} \cong m_{i2} = \rho_{air} V_1 \cong \rho_{air} V_2 \cong 10^{-8} kg$ and $r = 10^{-3} m$ then Eq. 22 gives

$$\vec{F}_{12} = -\vec{F}_{21} \cong -10^{4} N \qquad (23)$$

These forces are much more intense than the *inter-atomic forces* (the forces that unite the atoms and molecules) the intensities of which are of the order of $1 - 1000 \times 10^{-8} N$. Consequently, the air around the spacecraft will be strongly compressed upon the surface of the spacecraft and thus will produce a crust of air which will accompany the spacecraft during its displacement and will protect it from the friction with the atmosphere of the planet.

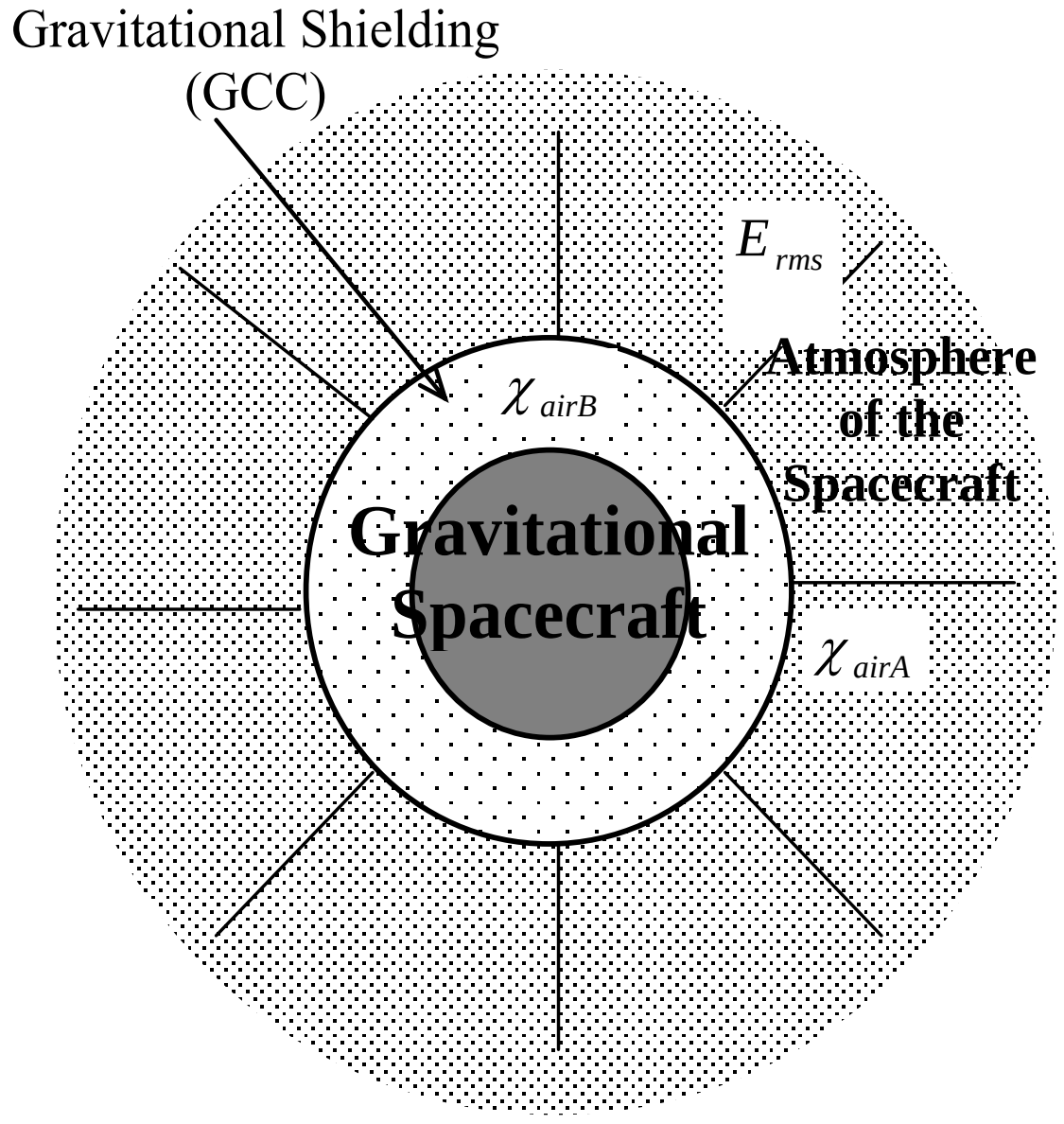

Fig. 6 – *Artificial atmosphere* around the gravitational spacecraft - while inside the gravitational shielding the value of $\chi_{airB}$ is putted close to zero, in order to strongly reduces the gravitational mass of the spacecraft (inner part of the shielding), the value of $\chi_{airA}$ must be reduced for about $-10^8$ in order to strongly increase the gravitational attraction between the air molecules around the spacecraft.

## 5. **The Imaginary Space-time**

The speed of light in free space is, as we know, about of 300.000 km/s. The speeds of the fastest modern airplanes of the present time do not reach 2 km/s and the speed of rockets do not surpass 20 km/s. This shows how much our aircraft and rockets are slow when compared with the speed of light.

The star nearest to the Earth (excluding the Sun obviously) is the Alpha of Centaur, which is about of *4 light-years* distant from the Earth (Approximately 37.8 trillions of kilometers). Traveling at a speed about 100 times greater than the maximum speed of our faster spacecrafts, we would take about 600 years to reach Alpha of Centaur. Then imagine how many years we would take to leave our own galaxy. In fact, it is not difficult to see that our spacecrafts are very slow, even for travels in our own solar system.

One of the fundamental characteristics of the gravitational spacecraft, as we already saw, is its capability to acquire enormous accelerations without submitting the crew to any discomfort.

Impelled by gravitational thrusters gravitational spacecrafts can acquire accelerations until $10^8 m.s^{-2}$ or more. This means that these spacecrafts can reach speeds very close to the speed of light in just a few seconds. These gigantic accelerations can be unconceivable for a layman, however they are common in our Universe. For example, when we submit an electron to an electric field of just $1\ Volt/m$ it acquires an acceleration $a$, given by

$$a = \frac{eE}{m_e} = \frac{(1.6\times10^{-19}C)(1\ V/m)}{9.11\times10^{-31}} \cong 10^{11} m.s^{-2}$$

As we see, this acceleration is about 100 times greater than that acquired by the gravitational spacecraft previously mentioned.

By using the gravitational shieldings it is possible to reduce the inertial effects upon the spacecraft. As we have shown, they are reduced by the factor $\chi_{out} = M_g / M_i$. Thus, if the inertial mass of the spacecraft is $M_i = 10.000kg$ and, by means of the gravitational shielding effect the gravitational mass of the spacecraft is reduced to $M_g \approx 10^{-8} M_i$ then , in spite of the effective acceleration to be gigantic, for example, $a \approx 10^9 m.s^{-2}$, the effects for the crew of the spacecraft would be *equivalents to* an acceleration $a'$ of only

$$a' = \frac{M_g}{M_i} a = (10^{-8})(10^9) \approx 10 m.s^{-2}$$

This acceleration is similar to that which the passengers of a contemporary commercial jet are subjected.

Therefore the crew of the gravitational spacecraft would be comfortable while the spacecraft would reach speeds close to the speed of light in few seconds. However to travel at such velocities in the Universe may note be practical. Take for example, Alpha of Centaur (4 light-years far from the Earth): a round trip to it would last about eight years. Trips beyond that star could take then several decades, and this obviously is impracticable. Besides, to travel at such a speed would be very dangerous, because a shock with other celestial bodies would be inevitable. However, as we showed [1] there is a possibility of a spacecraft travel *quickly* far beyond our galaxy without the risk of being destroyed by a sudden shock with some celestial body. The solution is the gravitational spacecraft travel through the *Imaginary* or *Complex Space-time.*

It was shown [1] that it is possible to carry out a transition to the *Imaginary space-time* or *Imaginary Universe*. It is enough that the body has its *gravitational mass* reduced to a value in the range of $+0.159M_i$ to $-0.159M_i$. In these circumstances, the masses of the body (gravitational and inertial) become *imaginaries* and, so does the body. (Fig.7). Consequently, the body disappears from our ordinary space-time and appears in the imaginary space-time. In other words, it becomes *invisible* for an observer at the real Universe. Therefore, this is a way to get temporary *invisibility* of human beings, animals, spacecrafts, etc.

Thus, a spacecraft can leave our Universe and appear in the Imaginary Universe, where it can travel at any speed since in the Imaginary Universe *there is no speed limit for the gravitational spacecraft*, as it occurs in our Universe, where the particles cannot surpass the light speed. In this way, as the gravitational spacecraft is propelled by the gravitational thrusters, it can attain accelerations up to $10^9 m.s^{-2}$, then after one day of trip with this acceleration, it can reach velocities $V \approx 10^{14} m.s^{-1}$(about 1 million times the speed of light). With this velocity, after 1 month of trip the spacecraft would have traveled about$10^{21} m$. In order to have idea of this distance, it is enough to remind that the diameter of our Universe (visible Universe) is of the order of$10^{26} m$.

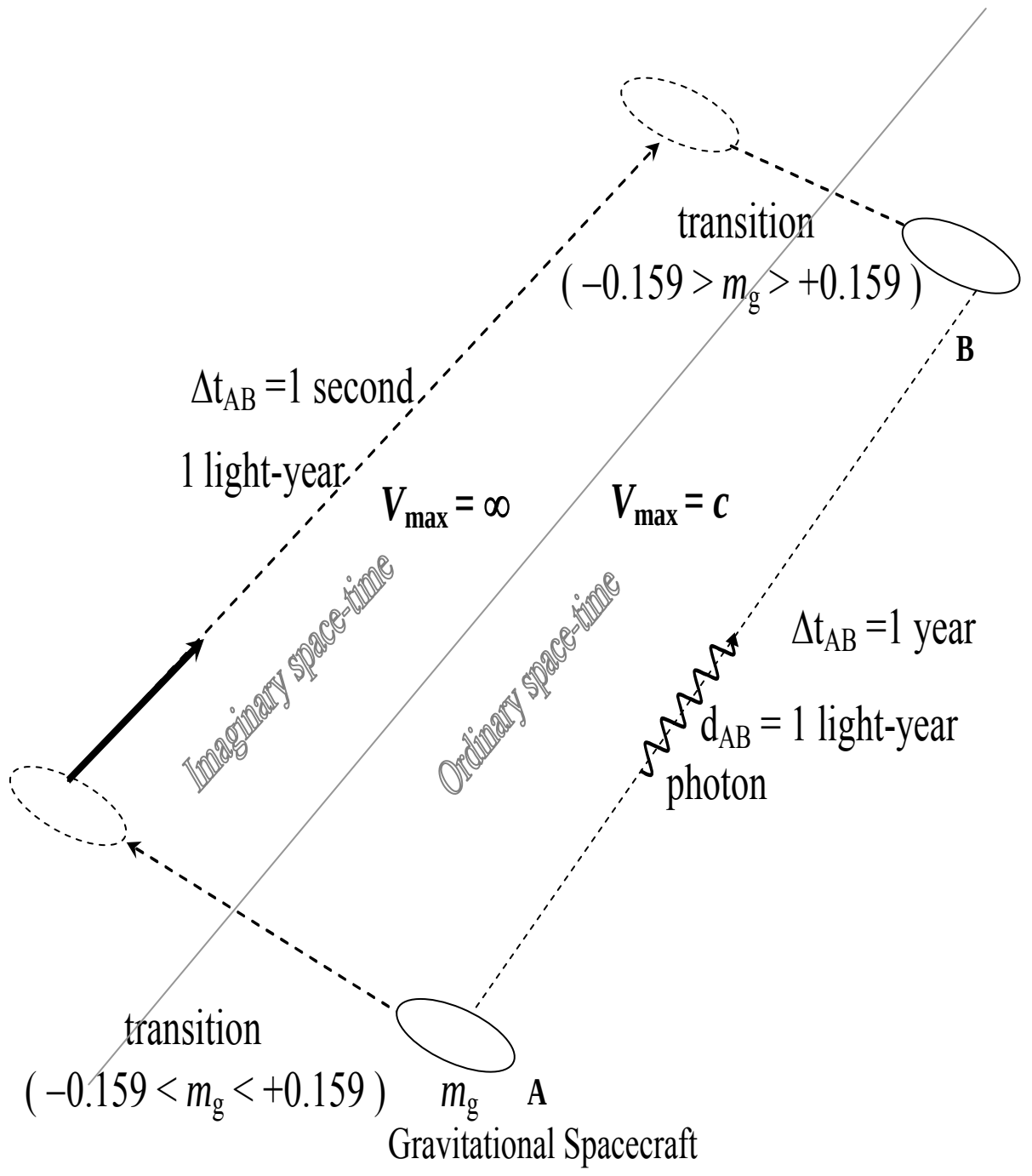

Fig. 7 – Travel in the *Imaginary* Space-time.

Due to the extremely low density of the imaginary bodies, the collision between them cannot have the same consequences of the collision between the dense real bodies.

Thus for a gravitational spacecraft in imaginary state the problem of the collision doesn't exist in high-speed. Consequently, the gravitational spacecraft can transit freely in the imaginary Universe and, in this way reach easily any point of our real Universe once they can make the transition back to our Universe by only increasing the gravitational mass of the spacecraft in such way that it leaves the range of $+0.159M_i$ to$-0.159M_i$. Thus the spacecraft can reappear in our Universe near its target.

The return trip would be done in similar way. That is to say, the spacecraft would transit in the imaginary Universe back to the departure place where would reappear in our Universe and it would make the approach flight to the wanted point. Thus, trips through our Universe that would delay millions of years, at speeds close to the speed of light, could be done in just a few *months* in the imaginary Universe.

What will an observer see when in the imaginary space-time? It will see light, bodies, planets, stars, etc., everything formed by imaginary photons, imaginary atoms, imaginary protons, imaginary neutrons and imaginary electrons. That is to say, the observer will find an Universe similar to ours, just formed by particles with imaginary masses. The term *imaginary* adopted from the Mathematics, as we already saw, gives the false impression that these masses do not exist. In order to avoid this misunderstanding we researched the true nature of that new mass type and matter.

The existence of imaginary mass associated to the *neutrino* is well-known. Although its imaginary mass is not physically observable, its square is. This amount is found experimentally to be negative. Recently, it was shown [1] that *quanta* of imaginary mass exist associated to the *photons, electrons, neutrons,* and *protons*, and that these imaginary masses would have psychic properties (elementary capability of "choice"). Thus, the true nature of this new kind of mass and matter shall be psychic and, therefore we should not use the term *imaginary* any longer. Consequently from the above exposed we can conclude that the gravitational spacecraft penetrates in the *Psychic Universe* and not in an "imaginary" Universe.

In this Universe, the matter would be, obviously composed by psychic molecules and psychic atoms formed by psychic neutrons, psychic protons and psychic electrons. i.e., the matter would have psychic mass and consequently it would be *subtle*, much less dense than the matter of our *real* Universe.

Thus, from a quantum viewpoint, the psychic particles are similar to the material particles, so that we can use the Quantum Mechanics to describe the psychic particles. In this case, by analogy to the material particles, a particle with psychic mass $m_\Psi$ will be described by the following expressions:

$$\vec{p}_\psi = \hbar\vec{k}_\psi$$

$$E_\psi = \hbar\omega_\psi$$

Where $\vec{p}_\psi = m_\Psi\vec{V}$ is the *momentum* carried by the wave and $E_\psi$ its energy; $\left|\vec{k}_\psi\right| = 2\pi/\lambda_\psi$ is the *propagation number* and $\lambda_\psi = h/m_\Psi V$ the *wavelength* and $\omega_\psi = 2\pi f_\psi$ its cyclic *frequency*.

The variable quantity that characterizes DeBroglie's waves is called *Wave Function*, usually indicated by $\Psi$. The wave function associated to a material particle describes the dynamic state of the particle: its value at a particular point x, y, z, t is related to the probability of finding the particle in that place and instant. Although $\Psi$ does not have a physical interpretation, its square $\Psi^2$ (or $\Psi\ \Psi^*$) calculated for a particular point x, y, z, t is *proportional to the probability of experimentally finding the particle in that place and instant*.

Since $\Psi^2$ is proportional to the probability $P$ of finding the particle described by $\Psi$, the integral of $\Psi^2$ on the *whole space* must be finite – inasmuch as the particle is someplace. Therefore, if

$$\int_{-\infty}^{+\infty}\Psi^2 dV = 0$$

The interpretation is that the particle does not exist. Conversely, if

$$\int_{-\infty}^{+\infty}\Psi^2 dV = \infty$$

*the particle will be everywhere simultaneously*.

The wave function $\Psi$ corresponds, as we know, to the displacement $y$ of the undulatory motion of a rope. However, $\Psi$ as opposed to $y$, is not a measurable quantity and can, hence, being a *complex* quantity. For this reason, it is admitted that $\Psi$ is described in the $x$-direction by

$$\Psi = Be^{-(2\pi i/h)(Et-px)}$$

This equation is the mathematical description of the wave associated with a free material particle, with total energy $E$ and *momentum* $p$, moving in the direction $+x$.

As concerns the psychic particle, the variable quantity characterizing psyche waves will also be called wave function, denoted by $\Psi_\Psi$ ( to distinguish it from the material particle wave function), and, by analogy with equation of $\Psi$, expressed by:

$$\Psi_\Psi = \Psi_0 e^{-(2\pi i/h)(E_\Psi t - p_\Psi x)}$$

If an experiment involves a large number of identical particles, all described by the same wave function $\Psi$ , the *real* density of mass $\rho$ of these particles in x, y, z, t is proportional to the corresponding value $\Psi^2$ ($\Psi^2$ is known as *density of probability*. If $\Psi$ is *complex* then $\Psi^2 = \Psi\Psi^*$. Thus, $\rho \propto \Psi^2 = \Psi.\Psi^*$). Similarly, in the case of psychic particles, the *density of psychic mass*, $\rho_\Psi$, in x, y, z, will be expressed by $\rho_\Psi \propto \Psi_\Psi^2 = \Psi_\Psi \Psi_\Psi^*$. It is known that $\Psi_\Psi^2$ is always *real* and *positive* while $\rho_\Psi = m_\Psi / V$ is an *imaginary* quantity. Thus, as the *modulus* of an imaginary number is always real and positive, we can transform the proportion $\rho_\Psi \propto \Psi_\Psi^2$, in equality in the following form:

$$\Psi_\Psi^2 = k|\rho_\Psi|$$

Where $k$ is a *proportionality constant* (real and positive) to be determined.

In Quantum Mechanics we have studied the *Superposition Principle*, which affirms that, if a particle (or system of particles) is in a *dynamic state* represented by a wave function $\Psi_1$ and may also be in another dynamic state described by $\Psi_2$ then, the general dynamic state of the particle may be described by $\Psi$, where $\Psi$ is a linear combination (superposition) of $\Psi_1$ and $\Psi_2$, i.e.,

$$\Psi = c_1\Psi_1 + c_2\Psi_2$$

The C*omplex constants* $c_1$ e $c_2$ respectively express the percentage of dynamic state, represented by $\Psi_1$ e $\Psi_2$ in the formation of the general dynamic state described by $\Psi$.

In the case of psychic particles (psychic bodies, consciousness, etc.), by analogy, if $\Psi_{\Psi 1}$, $\Psi_{\Psi 2}$,..., $\Psi_{\Psi n}$ refer to the different dynamic states the psychic particle takes, then its general dynamic state may be described by the wave function $\Psi_\Psi$ , given by:

$$\Psi_\Psi = c_1\Psi_{\Psi 1} + c_2\Psi_{\Psi 2} + \ldots + c_n\Psi_{\Psi n}$$

The state of superposition of wave functions is, therefore, common for both psychic and material particles. In the case of material particles, it can be verified, for instance, when an electron changes from one orbit to another. Before effecting the transition to another energy level, the electron carries out "virtual transitions" [6]. A kind of *relationship* with other electrons before performing the real transition. During this relationship period, its wave function remains "*scattered*" by *a wide region of the space* [7] thus superposing the wave functions of the other electrons. In this relationship the electrons *mutually* influence each other, with the possibility of *intertwining* their wave functions[§]. When this happens, there occurs the so-called *Phase Relationship* according to quantum-mechanics concept.

In the electrons "virtual" transition mentioned before, the "listing" of all the possibilities of the electrons is described, as we know, by *Schrödinger's wave equation.*

---

[§] Since the electrons are simultaneously waves and particles, their wave aspects will interfere with each other; besides superposition, there is also the possibility of occurrence of *intertwining* of their wave functions.

Otherwise, it is general for material particles. By analogy, in the case of psychic particles, we may say that the "listing" of all the possibilities of the psyches involved in the relationship will be described by *Schrödinger's equation* – for psychic case, i.e.,

$$\nabla^2 \Psi_\Psi + \frac{p_\Psi^2}{\hbar^2} \Psi_\Psi = 0$$

Because the wave functions are capable of intertwining themselves, the quantum systems may "penetrate" each other, thus establishing an internal relationship where all of them are affected by the relationship, no longer being isolated systems but becoming an integrated part of a larger system. This type of internal relationship, which exists only in quantum systems, was called *Relational Holism* [8].

We have used the Quantum Mechanics in order to describe the foundations of the Psychic Universe which the Gravitational Spacecrafts will find, and that influences us daily. These foundations recently discovered – particularly the *Psychic Interaction*, show us that a rigorous description of the Universe cannot to exclude the psychic energy and the psychic particles. This verification makes evident the need of to redefine the Psychology with basis on the quantum foundations recently discovered. This has been made in the article: "*Physical Foundations of Quantum Psychology*"** [9], recently published, where it is shown that the Psychic Interaction leads us to understand the Psychic Universe and the extraordinary relationship that the human consciousnesses establish among themselves and with the Ordinary Universe. Besides, we have shown that the Psychic Interaction postulates a new model for the evolution theory, in which the evolution is interpreted not only as a biological fact, but mainly as *psychic* fact. Therefore, is not only the mankind that evolves in the Earth's planet, but all the ecosystem of the Earth.

** http://htpprints.yorku.ca/archive/00000297

## 6. Past and Future

It was shown [1,9] that the *collapse* of the *psychic* wave function must suddenly also express in reality (*real* space-time) all the possibilities described by it. This is, therefore, *a point of decision* in which there occurs the compelling need of *realization* of the *psychic form*. We have seen that the *materialization* of the psychic form, in the real space-time, occurs when it contains enough *psychic mass* for the total materialization†† of the psychic form (*Materialization Condition*). When this happens, all the psychic energy contained in the psychic form is transformed in real energy in the real space-time. Thus, in the psychic space-time just the *holographic* register of the psychic form, which gives origin to that fact, survives, since the psychic energy deforms the *metric* of the psychic space-time‡‡, producing the holographic register. Thus, the past survive in the psychic space-time just in the form of holographic register. That is to say, all that have occurred in the past is holographically registered in the *psychic space-time*. Further ahead, it will be seen that this register can be accessed by an observer in the *psychic* space-time as well as by an observer in the *real* space-time.

†† By this we mean not only materialization proper but also the movement of matter to realize its psychic content (including radiation).

‡‡ As shown in *General Theory of Relativity* the energy modifies the metric of the space-time (deforming the space-time).

A psychic form is intensified by means of a continuous addition of psychic mass. Thus, when it acquires sufficiently psychic mass, its realization occurs in the real space-time. Thus the future is going being built in the present. By means of our current thoughts we shape the psychic forms that will go (or will not) take place in the future. Consequently, those psychic forms are continually being holographically registered in the *psychic space-time* and, just as the holographic registrations of the *past* these future registration can also be accessed by the *psychic* space-time as well as by the *real* space-time.

The access to the holographic registration of the past doesn't allow, obviously, the modification of the past. This is not possible because there would be a clear violation of the *principle of causality* that says that the causes should precede the effects. However, the psychic forms that are being shaped now in order to manifest themselves in the future, can be modified before they manifest themselves. Thus, the access to the registration of those psychic forms becomes highly relevant for our present life, since we can avoid the manifestation of many unpleasant facts in the future.

Since both registrations are in the *psychic* space-time, then the access to their information only occur by means of the interaction with another psychic body, for example, our *consciousness* or a *psychic observer* (body totally formed by psychic mass). We have seen that, if the gravitational mass of a body is reduced to within the range $+0.159M_i$ to $-0.159M_i$, its gravitational and inertial masses become *imaginaries* (*psychics*) and, therefore, the body becomes a *psychic body*. Thus, a *real observer* can also become in a *psychic observer*. In this way, a gravitational spacecraft can transform all its inertial mass into psychic mass, and thus carry out a transition to the psychic space-time and become a psychic spacecraft. In these circumstances, an observer inside the spacecraft also will have its mass transformed into psychic mass, and, therefore, the observer also will be transformed into a psychic observer. What will this observer see when it penetrates the psychic Universe? According to the *Correspondence principle*, all that exists in the real Universe must have the correspondent in the psychic Universe and vice-versa. This principle reminds us that we live in more than one world. At the present time, we live in the real Universe, but we can also live in the psychic Universe. Therefore, the psychic observer will see the psychic bodies and their correspondents in the real Universe. Thus, a pilot of a gravitational spacecraft, in travel through the psychic space-time, won't have difficulty to spot the spacecraft in its trips through the Universe.

The fact of the psychic forms manifest themselves in the real space-time exactly at its images and likeness, it indicates that real forms (forms in the real space-time) are prior to all reflective *images* of psychic forms of the past. Thus, the real space-time is a mirror of the psychic space-time. Consequently, any register in the psychic space-time will have a correspondent image in the real space-time. This means that it is possible that we find in the real space-time the *image* of the holographic register existing in the psychic space-time, corresponding to our *past*. Similarly, every psychic form that is being shaped in the psychic space-time will have reflective image in the real space-time. Thus, the *image* of the holographic register of our future (existing in the psychic space-time) can also be found in the *real* space-time.

Each image of the holographic register of our future will be obviously correlated to a future epoch in the temporal coordinate of the space-time. In the same way, each image of the holographic registration of our past will be correlated to a passed time in the temporal coordinate of the referred space-time. Thus, in order to access the mentioned registrations we should accomplish trips to the past or future in the real space-time. This is possible now, with the advent of the gravitational spacecrafts because they allow us to reach speeds close to the speed of light. Thus, by varying the gravitational mass of the spacecraft for *negative* or *positive* we can go respectively to the *past* or *future* [1].

If the gravitational mass of a particle is *positive*, then $t$ is always *positive* and given by

$$t = +t_0 \Big/ \sqrt{1 - V^2/c^2}$$

This leads to the well-known relativistic prediction that the particle goes to the *future* if $V \rightarrow c$. However, if the gravitational mass of the particle is negative, then $t$ is also *negative* and, therefore, given by

$$t = -t_0 \Big/ \sqrt{1 - V^2/c^2}$$

In this case, the prevision is that the particle goes to the *past* if $V \rightarrow c$. In this way, *negative gravitational mass* is the necessary condition to the particle to go to the *past*.

Since the acceleration of a spacecraft with gravitational mass $m_g$, is given by $a = F/m_g$, where $F$ is the thrust of its thrusters, then the more we reduce the value of $m_g$ the bigger the acceleration of the spacecraft will be. However, since the value of $m_g$ cannot be reduced to the range $+0.159M_i$ to $-0.159M_i$ because the spacecraft would become a psychic body, and it needs to remain in the real space-time in order to access the past or the future in the real space-time, then, the ideal values for the spacecraft to operate with safety would be $\pm 0.2m_i$. Let us consider a gravitational spacecraft whose inertial mass is $m_i = 10.000kg$. If its gravitational mass was made *negative* and equal to $m_g = -0.2m_i = -2000kg$ and, at this instant the thrust produced by the thrusters of the spacecraft was $F = 10^5 N$ then, the spacecraft would acquire acceleration $a = F/m_g = 50ms^{-2}$ and, after $t = 30 days = 2.5 \times 10^6 s$, the speed of the spacecraft would be $v = 1.2 \times 10^8 m.s^{-1} = 0.4c$. Therefore, right after that the spacecraft returned to the Earth, its crew would find the Earth in the *past* (due to the *negative* gravitational mass of the spacecraft) at a time $t = -t_0 / \sqrt{1 - V^2/c^2}$; $t_0$ is the time measured by an observer at rest on the Earth. Thus, if $t_0 = 2009$ AD, the time interval $\Delta t = t - t_0$ would be expressed by

$$\Delta t = t - t_0 = -t_0 \left( \frac{1}{\sqrt{1 - V^2/c^2}} - 1 \right) = -t_0 \left( \frac{1}{\sqrt{1 - 0.16}} - 1 \right) \cong$$

$$\cong -0.091 t_0 \cong -183 years$$

That is, the spacecraft would be in the year 1826 AD. On the other hand, if the gravitational mass of the spacecraft would have become positive $m_g = +0.2 m_i = +2000 kg$, instead of negative, then the spacecraft would be in the future at $\Delta t = +183$ *years* from 2009. That is, it would be in the year 2192 AD.

## 7. Instantaneous Interestelar Communications

Consider a cylindrical GCC (GCC antenna) as shown in Fig.8. The *gravitational mass* of the *air* inside the GCC is

$$m_{g(air)} = \left\{ 1 - 2 \left[ \sqrt{1 + \frac{\sigma_{(air)} B^4}{4\pi f \mu \rho^2_{(air)} c^2}} - 1 \right] \right\} m_{i(air)} \quad (24)$$

Where $\sigma_{(ar)}$ is the electric conductivity of the ionized air inside the GCC and $\rho_{(ar)}$ is its density; $f$ is the frequency of the magnetic field.

By varying $B$ one can vary $m_{g(air)}$ and consequently to vary the gravitational field generated by $m_{g(air)}$, producing then *Gravitational Radiation*. Then a GCC can work as a *Gravitational Antenna*.

Apparently, Newton's theory of gravity had no gravitational waves because, if a gravitational field changed in some way, that change would have taken place *instantaneously* everywhere in space, and one can think that there is not a wave in this case. However, we have already seen that the gravitational interaction can be repulsive, besides attractive. Thus, as with electromagnetic interaction, the gravitational interaction must be produced by the exchange of "virtual" *quanta of* spin 1 and mass null, i.e., the *gravitational* "virtual" *quanta* (*graviphoton*) must have spin 1 and not 2. Consequently, the fact that a change in a gravitational field reaches *instantaneously* every point in space occurs simply due to the speed of the *graviphoton* to be *infinite*. It is known that there is no speed limit for "*virtual*" photons. On the other hand, the *electromagnetic quanta* ("virtual" photons) can not communicate the *electromagnetic interaction* to an infinite distance.

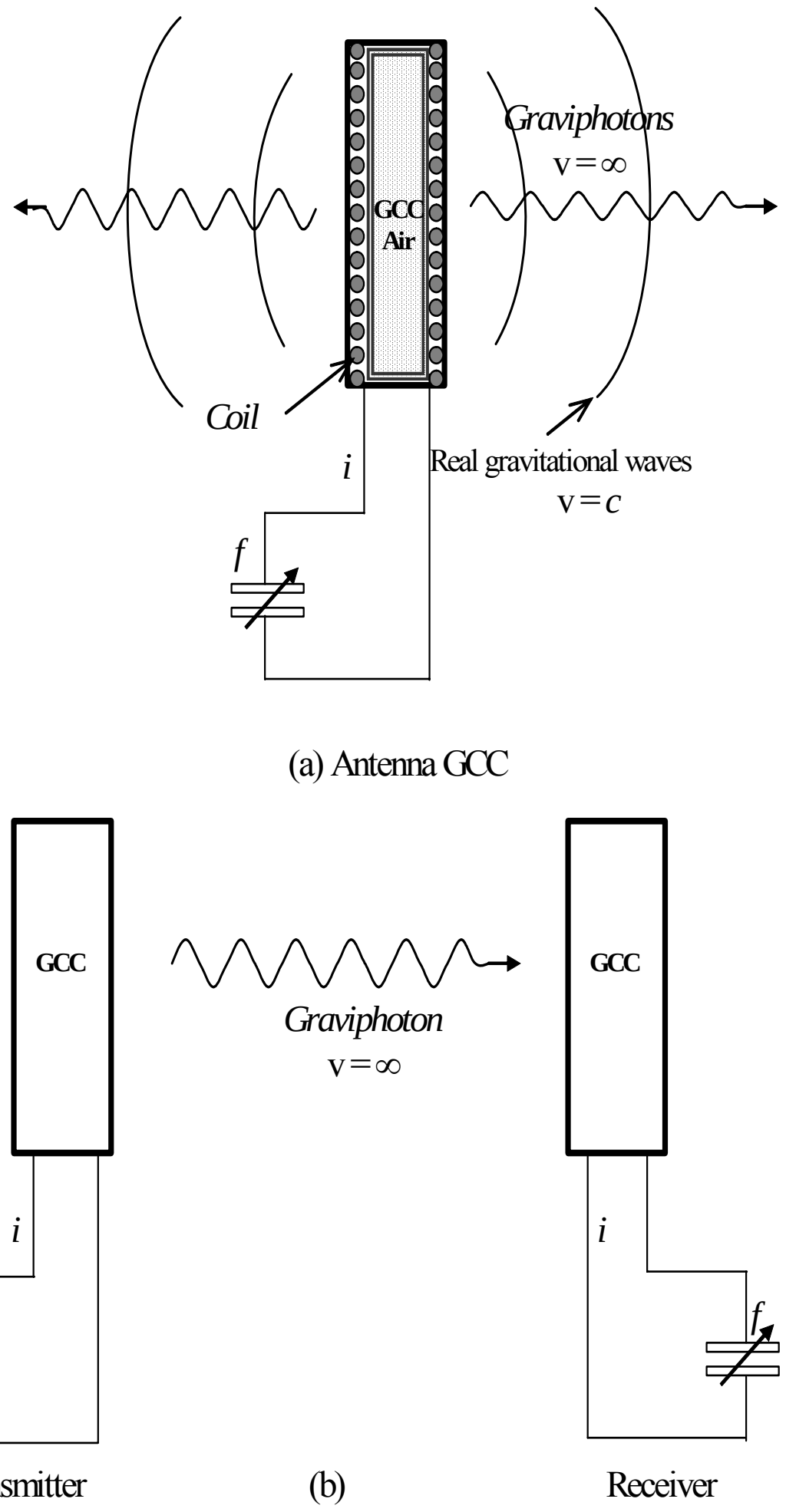

Fig. 8 – Transmitter and Receiver of *Virtual* Gravitational Radiation.

Thus, there are *two types* of gravitational radiation: the *real* and *virtual*, which is constituted of graviphotons; the *real* gravitational waves are ripples in the space-time generated by *gravitational field* changes. According to Einstein's theory of gravity the velocity of propagation of these waves is equal to the speed of light [10].

Unlike the electromagnetic waves the *real* gravitational waves have low interaction with matter and consequently low scattering. Therefore *real* gravitational waves are suitable as a means of transmitting information. However, when the distance between transmitter and receiver is too large, for example of the order of magnitude of several light-years, the transmission of information by means of gravitational waves becomes impracticable due to the long time necessary to receive the information. On the other hand, there is no delay during the transmissions by means of *virtual* gravitational radiation. In addition, the scattering of this radiation is null. Therefore the *virtual* gravitational radiation is very suitable as a means of transmitting information at any distances, including astronomical distances.

As concerns detection of the *virtual* gravitational radiation from GCC antenna, there are many options. Due to *Resonance Principle* a similar GCC antenna (receiver) *tuned at the same frequency* can absorb energy from an incident *virtual* gravitational radiation (See Fig.8 (b)). Consequently, the gravitational mass of the air inside the GCC receiver will vary such as the gravitational mass of the air inside the GCC transmitter. This will induce a magnetic field similar to the magnetic field of the GCC transmitter and therefore the current through the coil inside the GCC receiver will have the same characteristics of the current through the coil inside the GCC transmitter. However, the *volume* and *pressure* of the air inside the two GCCs must be exactly the same; also the *type* and the *quantity of atoms* in the air inside the two GCCs must be exactly the same. Thus, the GCC antennas are simple but they are not easy to build.

Note that a GCC antenna radiates *graviphotons* and *gravitational waves* simultaneously (Fig. 8 (a)). Thus, it is not only a gravitational antenna: it is a *Quantum Gravitational Antenna* because it can also emit and detect gravitational "virtual" *quanta* (graviphotons), which, in turn, can transmit information *instantaneously* from any distance in the Universe *without* scattering.

Due to the difficulty to build two similar GCC antennas and, considering that the electric current in the receiver antenna can be detectable even if the gravitational mass of the nuclei of the antennas are not *strongly* reduced, then we propose to replace the gas at the nuclei of the antennas by a thin *dielectric lamina*. When the *virtual* gravitational radiation strikes upon the dielectric lamina, its gravitational mass varies similarly to the gravitational mass of the dielectric lamina of the transmitter antenna, inducing an electromagnetic field $(E, B)$ similar to the transmitter antenna. Thus, the electric current in the receiver antenna will have the same characteristics of the current in the transmitter antenna. In this way, it is then possible to build two similar antennas whose nuclei have the same volumes and the same types and quantities of atoms.

Note that the Quantum Gravitational Antennas can also be used to transmit *electric power*. It is easy to see that the Transmitter and Receiver can work with strong voltages and electric currents. This means that strong electric power can be transmitted among Quantum Gravitational Antennas. This obviously solves the problem of *wireless* electric power transmission. Thus, we can conclude that the spacecrafts *do not necessarily need* to have a system for generation of electric energy inside them. Since the electric energy to be used in the spacecraft can be *instantaneously transmitted* from *any point of the Universe*, by means of the above mentioned systems of transmission and reception of "virtual" gravitational waves.

## 8. Origin of Gravity and Genesis of the Gravitational Energy

It was shown [1] that the "virtual" *quanta* of the *gravitational interaction* must have spin 1 and not 2, and that they are "virtual" photons (*graviphotons*) with *zero mass* outside the *coherent* matter. Inside the coherent matter the graviphotons mass is *non-zero*. Therefore, the gravitational forces are also *gauge* forces, because they are yielded by the exchange of "virtual" *quanta* of spin 1, such as the electromagnetic forces and the weak and strong nuclear forces.

Thus, the gravitational forces are produced by the exchanging of "virtual" photons (Fig.9). Consequently, this is precisely the *origin of the gravity*.

Newton's theory of gravity does not explain *why* objects attract one another; it simply models this observation. Also Einstein's theory does not explain the origin of gravity. Einstein's theory of gravity only describes gravity with more precision than Newton's theory does.

Besides, there is nothing in both theories explaining the *origin of the energy* that produces the gravitational forces. Earth's gravity attracts all objects on the surface of our planet. This has been going on for over 4.5 billions years, yet no known energy source is being converted to support this tremendous ongoing energy expenditure. Also is the enormous continuous energy expended by Earth's gravitational field for maintaining the Moon in its orbit - millennium after millennium. In spite of the ongoing energy expended by Earth's gravitational field to hold objects down on surface and the Moon in orbit, why the energy of the field never diminishes in strength or drains its energy source? Is this energy expenditure balanced by a conversion of energy from an unknown energy source?

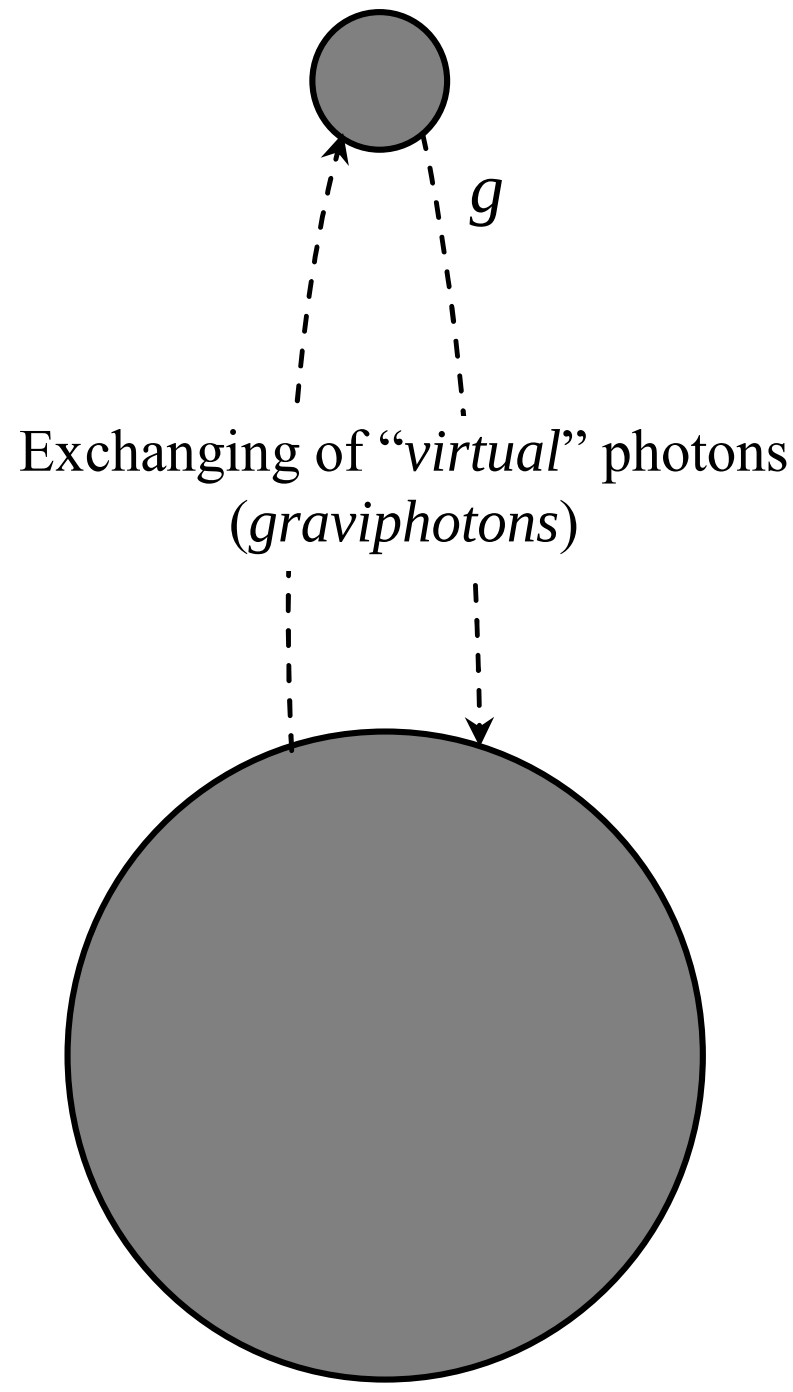

Fig. 9 – *Origin of Gravity*: The gravitational forces are produced by the exchanging of "virtual" photons (*graviphotons*).

The energy $W$ necessary to support the effort expended by the gravitational forces $F$ is well-known and given by

$$W = \int_{\infty}^{r} F dr = -G \frac{M_g m_g}{r}$$

According to the *Principle of Energy Conservation*, the spending of this energy must be compensated by a conversion of another type of energy.

The Uncertainty Principle tells us that, due to the occurrence of exchange of *graviphotons* in a time interval $\Delta t < \hbar/\Delta E$ (where $\Delta E$ is the energy of the graviphoton), the energy variation $\Delta E$ cannot be detected in the system $M_g - m_g$. Since the total energy $W$ is the sum of the energy of the $n$ graviphotons, i.e., $W = \Delta E_1 + \Delta E_2 + ... + \Delta E_n$, then the energy $W$ *cannot be detected as well*. However, as we know it can be converted into another type of energy, for example, in rotational kinetic energy, as in the hydroelectric plants, or in the *Gravitational Motor*, as shown in this work.

It is known that a *quantum* of energy $\Delta E = hf$, which varies during a time interval $\Delta t = 1/f = \lambda/c < \hbar/\Delta E$ (wave period) cannot be experimentally detected. This is an *imaginary* photon or a "*virtual*" photon. Thus, the graviphotons are *imaginary* photons, i.e., the energies $\Delta E_i$ of the graviphotons are imaginaries energies and therefore the energy $W = \Delta E_1 + \Delta E_2 + ... + \Delta E_n$ is also an *imaginary* energy. Consequently, it belongs to the *imaginary space-time*.

It was shown [1] that, imaginary energy is equal to *psychic energy*. Consequently, the *imaginary space-time* is, in fact, the *psychic space-time*, which contains the Supreme Consciousness. Since the Supreme Consciousness has *infinite* psychic mass [1], then the *psychic space-time* contains *infinite psychic energy*. This is highly relevant, because it confers to the *Psychic Universe* the characteristic of *unlimited source of energy*. Thus, as the origin of the gravitational energy it is correlated to the psychic energy, then the spending of gravitational energy can be supplied *indefinitely* by the Psychic Universe.

This can be easily confirmed by the fact that, in spite of the enormous amount of energy expended by Earth's gravitational field to hold objects down on the surface of the planet and maintain the Moon in its orbit, the energy of Earth's gravitational field never diminishes in strength or drains its energy source.

***Acknowledgement***

The author would like to thank Dr. *Getúlio Marques Martins* (COPPE – UFRJ, Rio de Janeiro-Brasil) for revising the manuscript.

# APPENDIX A: The Simplest Method to Control the Gravity

In this Appendix we show the simplest method to control the gravity.

Consider a body with mass density $\rho$ and the following electric characteristics: $\mu_r, \varepsilon_r, \sigma$ (relative permeability, relative permittivity and electric conductivity, respectively). Through this body, passes an electric current $I$, which is the sum of a sinusoidal current $i_{osc} = i_0 \sin \omega t$ and the DC current $I_{DC}$, i.e., $I = I_{DC} + i_0 \sin \omega t$ ; $\omega = 2\pi f$ . If $i_0 << I_{DC}$ then $I \cong I_{DC}$. Thus, the current $I$ varies with the frequency $f$, but the variation of its intensity is quite small in comparison with $I_{DC}$, i.e., $I$ will be practically constant (Fig. A1). This is of fundamental importance for maintaining the value of the gravitational mass of the body, $m_g$, sufficiently stable during all the time.

The *gravitational mass* of the *body* is given by [1]

$$m_g = \left\{1 - 2\left[\sqrt{1 + \left(\frac{n_r U}{m_{i0} c^2}\right)^2} - 1\right]\right\} m_{i0} \qquad (A1)$$

where $U$, is the electromagnetic energy absorbed by the body and $n_r$ is the index of refraction of the body.

Equation (A1) can also be rewritten in the following form

$$\frac{m_g}{m_{i0}} = \left\{1 - 2\left[\sqrt{1 + \left(\frac{n_r W}{\rho c^2}\right)^2} - 1\right]\right\} \qquad (A2)$$

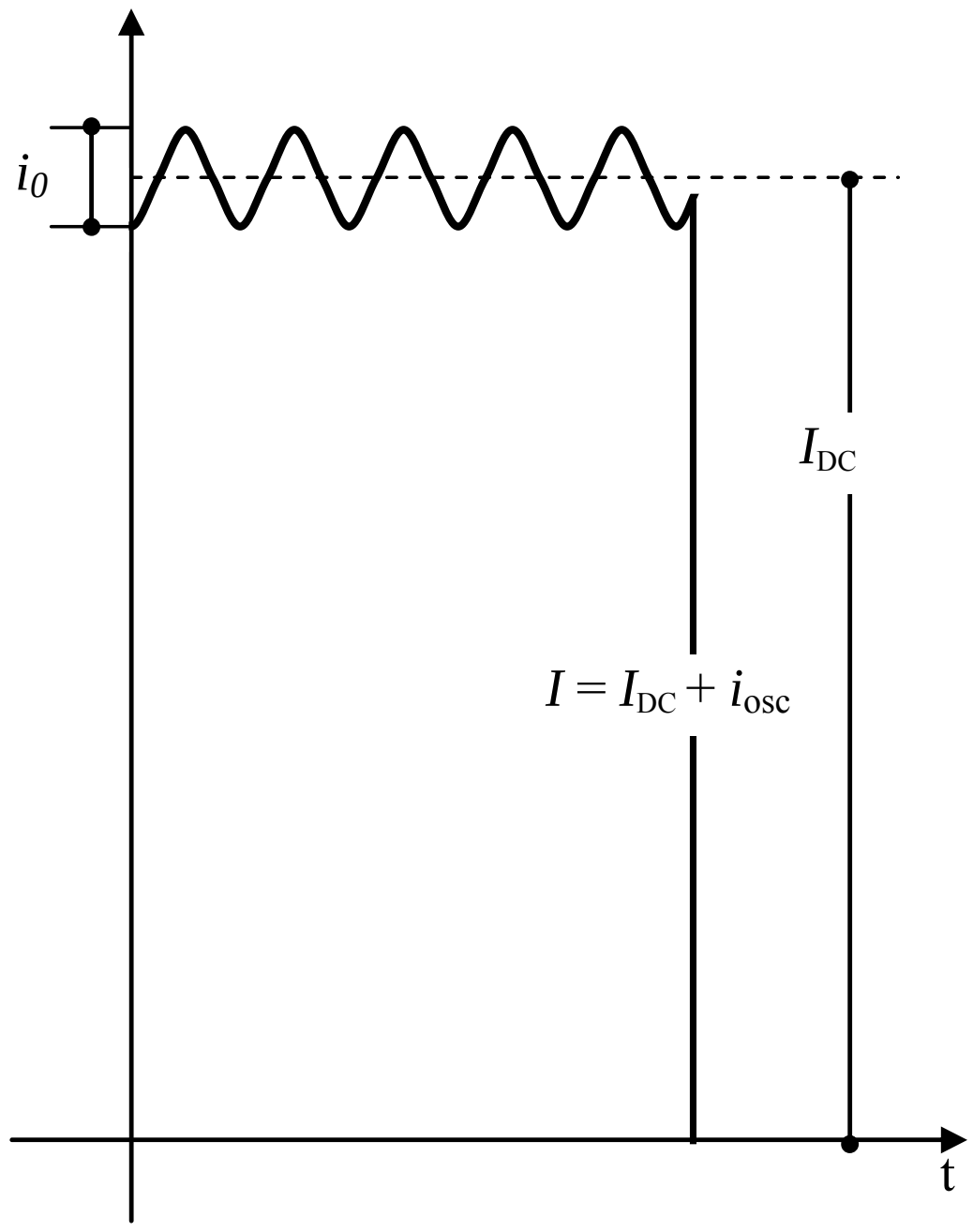

Fig. A1 - The electric current $I$ varies with frequency $f$. But the variation of $I$ is quite small in comparison with $I_{DC}$ due to $i_o << I_{DC}$. In this way, we can consider $I \cong I_{DC}$.

where, $W = U/V$ is the *density of electromagnetic energy* and $\rho = m_{i0}/V$ is the density of inertial mass.

The *instantaneous values* of the density of electromagnetic energy in an *electromagnetic* field can be deduced from Maxwell's equations and has the following expression

$$W = \tfrac{1}{2}\varepsilon E^2 + \tfrac{1}{2}\mu H^2 \qquad (A3)$$

where $E = E_m \sin \omega t$ and $H = H \sin \omega t$ are the *instantaneous values* of the electric field and the magnetic field respectively.

It is known that $B = \mu H$, $E/B = \omega/k_r$ [11] and

$$v = \frac{dz}{dt} = \frac{\omega}{\kappa_r} = \frac{c}{\sqrt{\frac{\varepsilon_r \mu_r}{2}\left(\sqrt{1+(\sigma/\omega\varepsilon)^2}+1\right)}} \quad (A4)$$

where $k_r$ is the real part of the *propagation vector* $\vec{k}$ (also called *phase constant* ); $k = |\vec{k}| = k_r + ik_i$ ; $\varepsilon$ , $\mu$ and $\sigma$, are the electromagnetic characteristics of the medium in which the incident (or emitted) radiation is propagating( $\varepsilon = \varepsilon_r \varepsilon_0$ ; $\varepsilon_0 = 8.854\times10^{-12} F/m$ ; $\mu = \mu_r \mu_0$ where $\mu_0 = 4\pi\times10^{-7} H/m$). It is known that for *free-space* $\sigma = 0$ and $\varepsilon_r = \mu_r = 1$. Then Eq. (A4) gives

$$v = c$$

From (A4), we see that the *index of refraction* $n_r = c/v$ is given by

$$n_r = \frac{c}{v} = \sqrt{\frac{\varepsilon_r \mu_r}{2}\left(\sqrt{1+(\sigma/\omega\varepsilon)^2}+1\right)} \quad (A5)$$

Equation (A4) shows that $\omega/\kappa_r = v$. Thus, $E/B = \omega/k_r = v$, i.e.,

$$E = vB = v\mu H \quad (A6)$$

Then, Eq. (A3) can be rewritten in the following form:

$$W = \tfrac{1}{2}(\varepsilon v^2 \mu)\mu H^2 + \tfrac{1}{2}\mu H^2 \quad (A7)$$

For $\sigma << \omega\varepsilon$, Eq. (A4) reduces to

$$v = \frac{c}{\sqrt{\varepsilon_r \mu_r}}$$

Then, Eq. (A7) gives

$$W = \tfrac{1}{2}\left[\varepsilon\left(\frac{c^2}{\varepsilon_r \mu_r}\right)\mu\right]\mu H^2 + \tfrac{1}{2}\mu H^2 = \mu H^2$$

This equation can be rewritten in the following forms:

$$W = \frac{B^2}{\mu} \quad (A8)$$

or

$$W = \varepsilon E^2 \quad (A9)$$

For $\sigma >> \omega\varepsilon$, Eq. (A4) gives

$$v = \sqrt{\frac{2\omega}{\mu\sigma}} \quad (A10)$$

Then, from Eq. (A7) we get

$$W = \tfrac{1}{2}\left[\varepsilon\left(\frac{2\omega}{\mu\sigma}\right)\mu\right]\mu H^2 + \tfrac{1}{2}\mu H^2 = \left(\frac{\omega\varepsilon}{\sigma}\right)\mu H^2 + \tfrac{1}{2}\mu H^2 \cong$$

$$\cong \tfrac{1}{2}\mu H^2 \quad (A11)$$

Since $E = vB = v\mu H$ , we can rewrite (A11) in the following forms:

$$W \cong \frac{B^2}{2\mu} \quad (A12)$$

or

$$W \cong \left(\frac{\sigma}{4\omega}\right)E^2 \quad (A13)$$

By comparing equations (A8) (A9) (A12) and (A13), we can see that Eq. (A13) shows that the best way to obtain a strong value of $W$ *in practice* is by applying an *Extra Low-Frequency* (ELF) *electric field* $(w = 2\pi f << 1Hz)$ through a *medium with high electrical conductivity.*

Substitution of Eq. (A13) into Eq. (A2), gives

$$m_g = \left\{1 - 2\left[\sqrt{1 + \frac{\mu}{4c^2}\left(\frac{\sigma}{4\pi f}\right)^3 \frac{E^4}{\rho^2}} - 1\right]\right\} m_{i0} =$$

$$= \left\{1 - 2\left[\sqrt{1 + \left(\frac{\mu_0}{256\pi^3 c^2}\right)\left(\frac{\mu_r \sigma^3}{\rho^2 f^3}\right)E^4} - 1\right]\right\} m_{i0} =$$

$$= \left\{1 - 2\left[\sqrt{1 + 1.758\times10^{-27}\left(\frac{\mu_r \sigma^3}{\rho^2 f^3}\right)E^4} - 1\right]\right\} m_{i0}$$

$$(A14)$$

Note that $E = E_m \sin\omega t$. The average value for $E^2$ is equal to $\tfrac{1}{2}E_m^2$ because $E$ varies sinusoidaly ($E_m$ is the maximum value for $E$). On the other hand, $E_{rms} = E_m/\sqrt{2}$. Consequently we can change $E^4$ by $E^4_{rms}$, and the equation above can be rewritten as follows

$$m_g = \left\{1 - 2\left[\sqrt{1 + 1.758\times10^{-27}\left(\frac{\mu_r\sigma^3}{\rho^2 f^3}\right)E^4_{rms}} - 1\right]\right\}m_{i0}$$

Substitution of the well-known equation of the *Ohm's vectorial Law*: $j = \sigma E$ into (A14), we get

$$m_g = \left\{1 - 2\left[\sqrt{1 + 1.758\times10^{-27}\frac{\mu_r j^4_{rms}}{\sigma\rho^2 f^3}} - 1\right]\right\}m_{i0} \qquad (A15)$$

where $j_{rms} = j/\sqrt{2}$.

Consider a 15 cm square *Aluminum thin foil* of *10.5 microns thickness* with the following characteristics: $\mu_r = 1$ ; $\sigma = 3.82\times10^7 S.m^{-1}$; $\rho = 2700\,Kg.m^{-3}$. Then, (A15) gives

$$m_g = \left\{1 - 2\left[\sqrt{1 + 6.313\times10^{-42}\frac{j^4_{rms}}{f^3}} - 1\right]\right\}m_{i0} \qquad (A16)$$

Now, consider that the ELF electric current $I = I_{DC} + i_0 \sin\omega t$, $(i_0 << I_{DC})$ passes through that Aluminum foil. Then, the current density is

$$j_{rms} = \frac{I_{rms}}{S} \cong \frac{I_{DC}}{S} \qquad (A17)$$

where

$$S = 0.15m\left(10.5\times10^{-6}m\right) = 1.57\times10^{-6}m^2$$

If the ELF electric current has frequency $f = 2\mu Hz = 2\times10^{-6}Hz$, then, the gravitational mass of the aluminum foil, given by (A16), is expressed by

$$m_g = \left\{1 - 2\left[\sqrt{1 + 7.89\times10^{-25}\frac{I^4_{DC}}{S^4}} - 1\right]\right\}m_{i0} =$$

$$= \left\{1 - 2\left[\sqrt{1 + 0.13 I^4_{DC}} - 1\right]\right\}m_{i0} \qquad (A18)$$

Then,

$$\chi = \frac{m_g}{m_{i0}} \cong \left\{1 - 2\left[\sqrt{1 + 0.13 I^4_{DC}} - 1\right]\right\} \qquad (A19)$$

For $I_{DC} = 2.2A$, the equation above gives

$$\chi = \left(\frac{m_g}{m_{i0}}\right) \cong -1 \qquad (A20)$$

This means that *the gravitational shielding* produced by the aluminum foil can change the gravity acceleration *above* the foil down to

$$g' = \chi\, g \cong -1g \qquad (A21)$$

Under these conditions, the Aluminum foil works basically as a Gravity Control Cell (GCC).

In order to check these theoretical predictions, we suggest an experimental set-up shown in Fig.A2.

A 15cm square Aluminum foil of *10.5 microns thickness* with the following composition: Al 98.02%; Fe 0.80%; Si 0.70%; Mn 0.10%; Cu 0.10%; Zn 0.10%; Ti 0.08%; Mg 0.05%; Cr 0.05%, and with the following characteristics: $\mu_r = 1$; $\sigma = 3.82\times10^7 S.m^{-1}$; $\rho = 2700 Kg.m^{-3}$, is fixed on a 17 cm square *Foam Board* §§ plate of 6mm thickness as shown in Fig.A3. This device (the simplest

§§ *Foam board* is a very strong, *lightweight* (density: 24.03 kg.m$^{-3}$) and easily cut material used for the mounting of photographic prints, as backing in picture framing, in 3D design, and in painting. It consists of three layers — an inner layer of polystyrene clad with outer facing of either white clay coated paper or brown Kraft paper.

Gravity Control Cell GCC) is placed on a pan balance shown in Fig.A2.

Above the Aluminum foil, a *sample* (any type of material, any mass) connected to a dynamometer will check the decrease of the *local gravity acceleration* upon the sample $(g' = \chi\ g)$, due to the gravitational shielding produced by the decreasing of gravitational mass of the Aluminum foil $(\chi = m_g / m_{i0})$. Initially, the sample lies 5 cm above the Aluminum foil. As shown in Fig.A2, the board with the dynamometer can be displaced up to few meters in height. Thus, the initial distance between the Aluminum foil and the sample can be increased in order to check the reach of the gravitational shielding produced by the Aluminum foil.

In order to generate the ELF electric current of $f = 2\mu Hz$, we can use the widely-known Function Generator HP3325A (Op.002 High Voltage Output) that can generate sinusoidal voltages with *extremely-low* frequencies down to $f = 1\times10^{-6}\,Hz$ and amplitude up to 20V ($40V_{pp}$ into $500\Omega$ load). The maximum output current is $0.08A_{pp}$; output impedance <2Ω at ELF .

Figure A4 (a) shows the equivalent electric circuit for the experimental set-up. The electromotive forces are: $\varepsilon_1$ (HP3325A) and $\varepsilon_2$ (*12V* DC Battery).The values of the *resistors* are : $R_1 = 500\Omega - 2W$ ; $r_{i1} < 2\Omega$ ; $R_2 = 4\Omega - 40W$ ; $r_{i2} < 0.1\Omega$ ; $R_p = 2.5\times10^{-3}\Omega$ ; *Rheostat* (0≤ $R$ ≤10Ω - 90W). The *coupling transformer* has the following characteristics: air core with diameter $\phi = 10mm$ ; area $S = \pi\phi^2/4 = 7.8\times10^{-5}m^2$ ; wire#12AWG; $N_1 = N_2 = N = 20$; $l = 42mm$; $L_1 = L_2 = L = \mu_0 N^2 (S/l) = 9.3\times10^{-7}H$ .Thus, we get

$$Z_1 = \sqrt{(R_1 + r_{i1})^2 + (\omega L)^2} \cong 501\Omega$$

and

$$Z_2 = \sqrt{(R_2 + r_{i2} + R_p + R)^2 + (\omega L)^2}$$

For $R = 0$ we get $Z_2 = Z_2^{min} \cong 4\Omega$; for $R = 10\Omega$ the result is $Z_2 = Z_2^{max} \cong 14\Omega$. Thus,

$$Z_{1,total}^{min} = Z_1 + Z_{1,reflected}^{min} = Z_1 + Z_2^{min}\left(\frac{N_1}{N_2}\right)^2 \cong 505\Omega$$

$$Z_{1,total}^{max} = Z_1 + Z_{1,reflected}^{max} = Z_1 + Z_2^{max}\left(\frac{N_1}{N_2}\right)^2 \cong 515\Omega$$

The maxima *rms* currents have the following values:

$$I_1^{max} = \tfrac{1}{\sqrt{2}} 40V_{pp} / Z_{1,total}^{min} = 56mA$$

(The maximum output current of the Function Generator HP3325A (Op.002 High Voltage Output) is $80mA_{pp} \cong 56.5mA_{rms}$);

$$I_2^{max} = \frac{\varepsilon_2}{Z_2^{min}} = 3A$$

and

$$I_3^{max} = I_2^{max} + I_1^{max} \cong 3A$$

The new expression for the *inertial forces*, (Eq.5) $\vec{F}_i = M_g \vec{a}$ , shows that the inertial forces are proportional to *gravitational mass*. Only in the particular case of $m_g = m_{i0}$, the expression above reduces to the well-known Newtonian expression $\vec{F}_i = m_{i0}\vec{a}$ . The equivalence between gravitational and inertial forces $(\vec{F}_i \equiv \vec{F}_g)$ [1] shows then that a balance measures the *gravitational mass* subjected to acceleration $a = g$. Here, the decrease in the *gravitational mass* of the Aluminum foil will be measured by a pan balance with the following characteristics: range 0-200g; readability 0.01g.

The mass of the Foam Board plate is: $\cong 4.17g$, the mass of the Aluminum foil is: $\cong 0.64g$, the total mass of the ends and the electric wires of connection is $\cong 5g$. Thus, *initially* the balance will show $\cong 9.81g$. According to (A18), when the electric current through the Aluminum foil (resistance $r_p^* = l/\sigma S = 2.5\times10^{-3}\Omega$) reaches the value $I_3 \cong 2.2A$, we will get $m_{g(Al)} \cong -m_{i0(Al)}$. Under these circumstances, the balance will show:

$$9.81g - 0.64g - 0.64g \cong 8.53g$$

and the gravity acceleration $g'$ *above* the Aluminum foil, becomes $g' = \chi\, g \cong -1g$.

It was shown [1] that, when the gravitational mass of a particle is reduced to the gravitational mass ranging between $+0.159M_i$ to $-0.159M_i$, it becomes *imaginary*, i.e., the gravitational and the inertial masses of the particle become *imaginary*. Consequently, the particle *disappears* from our ordinary space-time. This phenomenon can be observed in the proposed experiment, i.e., *the Aluminum foil will disappear* when its gravitational mass becomes smaller than $+0.159M_i$. It will become visible again, only when its gravitational mass becomes smaller than $-0.159M_i$, or when it becomes greater than $+0.159M_i$.

Equation (A18) shows that the gravitational mass of the Aluminum foil, $m_{g(Al)}$, goes *close to zero* when $I_3 \cong 1.76A$. Consequently, the gravity acceleration *above* the Aluminum foil also goes close to zero since $g' = \chi\, g = m_{g(Al)}/m_{i0(Al)}$. Under these circumstances, the Aluminum foil remains *invisible*.

Now consider a rigid Aluminum wire # 14 AWG. The area of its cross section is

$$S = \pi\left(1.628\times10^{-3}m\right)^2/4 = 2.08\times10^{-6}m^2$$

If an ELF electric current with frequency $f = 2\mu Hz = 2\times10^{-6}Hz$ passes through this wire, its gravitational mass, given by (A16), will be expressed by

$$m_g = \left\{1 - 2\left[\sqrt{1 + 6.313\times10^{-42}\frac{j_{rms}^4}{f^3}} - 1\right]\right\}m_{i0} =$$
$$= \left\{1 - 2\left[\sqrt{1 + 7.89\times10^{-25}\frac{I_{DC}^4}{S^4}} - 1\right]\right\}m_{i0} =$$
$$= \left\{1 - 2\left[\sqrt{1 + 0.13I_{DC}^4} - 1\right]\right\}m_{i0} \qquad (A22)$$

For $I_{DC} \cong 3A$ the equation above gives

$$m_g \cong -3.8m_{i0}$$

Note that we can replace the Aluminum foil for this wire in the experimental set-up shown in Fig.A2. It is important also to note that an ELF electric current that passes through a wire - which makes a spherical form, as shown in Fig A5 - reduces the gravitational mass of the wire (Eq.

A22), and the gravity *inside sphere* at the same proportion, $\chi = m_g/m_{i0}$, (Gravitational Shielding Effect). In this case, that effect can be checked by means of the Experimental set-up 2 (Fig.A6). Note that the spherical form can be transformed into an ellipsoidal form or a disc in order to coat, for example, a Gravitational Spacecraft. It is also possible to coat with a wire several forms, such as cylinders, cones, cubes, etc.

The circuit shown in Fig.A4 (a) can be modified in order to produce a new type of Gravitational Shielding, as shown in Fig.A4 (b). In this case, the Gravitational Shielding will be produced in the Aluminum plate, with thickness $h$, of the parallel plate capacitor connected in the point $P$ of the circuit (See Fig.A4 (b)). Note that, in this circuit, the Aluminum foil (resistance $R_p$) (Fig.A4(a)) has been replaced by a Copper wire # 14 AWG with *1cm* length ($l = 1cm$) in order to produce a resistance $R_\phi = 5.21 \times 10^{-5}\Omega$. Thus, the voltage in the point $P$ of the circuit will have the maximum value $V_p^{max} = 1.1 \times 10^{-4} V$ when the resistance of the rheostat is null$(R = 0)$ and the minimum value $V_p^{min} = 4.03 \times 10^{-5} V$ when $R = 10\Omega$. In this way, the voltage $V_p$(with frequency $f = 2\mu Hz$) applied on the capacitor will produce an electric field $E_p$ with intensity $E_p = V_p/h$ through the Aluminum plate of thickness $h = 3mm$. It is important to note that *this plate cannot be connected to ground* (earth), in other words, cannot be grounded, because, in this case, the electric field through it will be *null* ***.

According to Eq. A14, when $E_p^{max} = V_p^{max}/h = 0.036\, V/m$, $f = 2\mu Hz$ and $\sigma_{Al} = 3.82 \times 10^7 S/m$, $\rho_{Al} = 2700\, kg/m^3$ (Aluminum), we get

$$\chi = \frac{m_{(Al)}}{m_{i(Al)}} \cong -0.9$$

Under these conditions, the maximum *current density* through the plate with thickness $h$ will be given by $j^{max} = \sigma_{Al} E_p^{max} = 1.4 \times 10^6 A/m^2$ (It is well-known that the maximum current density supported by the Aluminum is $\approx 10^8 A/m^2$).

Since the area of the plate is $A = (0.2)^2 = 4 \times 10^{-2} m^2$, then the maximum current is $i^{max} = j^{max} A = 56kA$. Despite this enormous current, the maximum dissipated power will be just $P^{max} = (i^{max})^2 R_{plate} = 6.2W$, because the resistance of the plate is very small, i.e., $R_{plate} = h/\sigma_{Al} A \cong 2 \times 10^{-9}\Omega$.

Note that the area $A$ of the plate (where the Gravitational Shielding takes place) can have several geometrical configurations. For example, it can be the area of the external surface of an ellipsoid, sphere, etc. Thus, it can be the area of the external surface of a Gravitational Spacecraft. In this case, if $A \cong 100m^2$, for example, the maximum dissipated

*** When the voltage $V_p$ is applied on the capacitor, the charge distribution in the dielectric induces positive and negative charges, respectively on opposite sides of the Aluminum plate with thickness $h$. If the plate is not connected to the ground (Earth) this charge distribution produces an electric field $E_p = V_p/h$ through the plate. However, if the plate is connected to the ground, the negative charges (electrons) escapes for the ground and the positive charges are redistributed along the entire surface of the Aluminum plate making *null* the electric field through it.

power will be $P^{\max} \cong 15.4kW$, i.e., approximately $154W/m^2$.

All of these systems work with Extra-Low Frequencies $\left(f << 10^{-3}Hz\right)$. Now, we show that, by simply changing *the geometry of the surface of the Aluminum foil*, it is possible to increase the working frequency $f$ up to more than *1Hz*.

Consider the Aluminum foil, now with several semi-spheres stamped on its surface, as shown in Fig. A7 . The semi-spheres have radius $r_0 = 0.9\ mm$, and are joined one to another. The Aluminum foil is now coated by an insulation layer with relative permittivity $\varepsilon_r$ and dielectric strength $k$. A voltage source is connected to the Aluminum foil in order to provide a voltage $V_0$ (rms) with frequency $f$. Thus, the electric potential $V$ at a distance $r$, in the interval from $r_0$ to $a$, is given by

$$V = \frac{1}{4\pi\varepsilon_r\varepsilon_0}\frac{q}{r} \qquad (A23)$$

In the interval $a < r \leq b$ the electric potential is

$$V = \frac{1}{4\pi\varepsilon_0}\frac{q}{r} \qquad (A24)$$

since for the air we have $\varepsilon_r \cong 1$.

Thus, on the surface of the metallic spheres $(r = r_0)$ we get

$$V_0 = \frac{1}{4\pi\varepsilon_r\varepsilon_0}\frac{q}{r_0} \qquad (A25)$$

Consequently, the electric field is

$$E_0 = \frac{1}{4\pi\varepsilon_r\varepsilon_0}\frac{q}{r_0^2} \qquad (A26)$$

By comparing (A26) with (A25), we obtain

$$E_0 = \frac{V_0}{r_0} \qquad (A27)$$

The electric potential $V_b$ at $r = b$ is

$$V_b = \frac{1}{4\pi\varepsilon_0}\frac{q}{b} = \frac{\varepsilon_r V_0 r_0}{b} \qquad (A28)$$

Consequently, the electric field $E_b$ is given by

$$E_b = \frac{1}{4\pi\varepsilon_0}\frac{q}{b^2} = \frac{\varepsilon_r V_0 r_0}{b^2} \qquad (A29)$$

From $r = r_0$ up to $r = b = a + d$ *the electric field is approximately constant* (See Fig. A7). Along the distance $d$ it will be called $E_{air}$. For $r > a + d$, the electric field stops being constant. Thus, the intensity of the electric field at $r = b = a + d$ is approximately equal to $E_0$, i.e., $E_b \cong E_0$. Then, we can write that

$$\frac{\varepsilon_r V_0 r_0}{b^2} \cong \frac{V_0}{r_0} \qquad (A30)$$

whence we get

$$b \cong r_0\sqrt{\varepsilon_r} \qquad (A31)$$

Since the intensity of the electric field through the air, $E_{air}$, is $E_{air} \cong E_b \cong E_0$, then, we can write that

$$E_{air} = \frac{1}{4\pi\varepsilon_0}\frac{q}{b^2} = \frac{\varepsilon_r V_0 r_0}{b^2} \qquad (A32)$$

Note that, $\varepsilon_r$ refers to the *relative permittivity of the insulation layer, which is covering the Aluminum foil*.

If the intensity of this field is greater than the dielectric strength of the air $\left(3\times10^6 V/m\right)$ there will occur the well-known *Corona effect*. Here, this effect is necessary in order to increase the electric conductivity of the air at this region (layer with thickness *d*). Thus, we will assume

$$E_{air}^{\min} = \frac{\varepsilon_r V_0^{\min} r_0}{b^2} = \frac{V_0^{\min}}{r_0} = 3\times10^6 V/m$$

and

$$E_{air}^{\max} = \frac{\varepsilon_r V_0^{\max} r_0}{b^2} = \frac{V_0^{\max}}{r_0} = 1\times10^7 V/m \qquad (A33)$$

The electric field $E_{air}^{\min} \leq E_{air} \leq E_{air}^{\max}$ will produce an *electrons flux* in a direction and an *ions flux* in an opposite direction. From the viewpoint of electric current, the ions flux can be considered as an "electrons" flux at the same direction of the real electrons flux. Thus, the current density through the air, $j_{air}$, will be the *double* of the current density expressed by the well-known equation of Langmuir-Child

$$j=\frac{4}{9}\varepsilon_r\varepsilon_0\sqrt{\frac{2e}{m_e}}\frac{V^{\frac{3}{2}}}{d^2}=\alpha\frac{V^{\frac{3}{2}}}{d^2}=2.33\times10^{-6}\frac{V^{\frac{3}{2}}}{d^2} \qquad (A34)$$

where $\varepsilon_r \cong 1$ for the *air*; $\alpha = 2.33\times10^{-6}$ is the called *Child's constant*.

Thus, we have

$$j_{air}=2\alpha\frac{V^{\frac{3}{2}}}{d^2} \qquad (A35)$$

where $d$, in this case, is the thickness of the air layer where the electric field is approximately constant and $V$ is the voltage drop given by

$$V=V_a-V_b=\frac{1}{4\pi\varepsilon_0}\frac{q}{a}-\frac{1}{4\pi\varepsilon_0}\frac{q}{b}=$$

$$=V_0r_0\varepsilon_r\left(\frac{b-a}{ab}\right)=\left(\frac{\varepsilon_r r_0 d}{ab}\right)V_0 \qquad (A36)$$

By substituting (A36) into (A35), we get

$$j_{air}=\frac{2\alpha}{d^2}\left(\frac{\varepsilon_r r_0 dV_0}{ab}\right)^{\frac{3}{2}}=\frac{2\alpha}{d^{\frac{1}{2}}}\left(\frac{\varepsilon_r r_0 V_0}{b^2}\right)^{\frac{3}{2}}\left(\frac{b}{a}\right)^{\frac{3}{2}}=$$

$$=\frac{2\alpha}{d^{\frac{1}{2}}}E_{air}^{\frac{3}{2}}\left(\frac{b}{a}\right)^{\frac{3}{2}} \qquad (A37)$$

According to the equation of the *Ohm's vectorial Law*: $j=\sigma E$, we can write that

$$\sigma_{air}=\frac{j_{air}}{E_{air}} \qquad (A38)$$

Substitution of (A37) into (A38) yields

$$\sigma_{air}=2\alpha\left(\frac{E_{air}}{d}\right)^{\frac{1}{2}}\left(\frac{b}{a}\right)^{\frac{3}{2}} \qquad (A39)$$

If the insulation layer has thickness $\Delta = 0.6\ mm$, $\varepsilon_r \cong 3.5$ (1-60Hz), $k=17kV/mm$ (Acrylic sheet 1.5mm thickness), and the semi-spheres stamped on the metallic surface have $r_0 = 0.9\ mm$ (See Fig.A7) then $a=r_0+\Delta=1.5\ mm$. Thus, we obtain from Eq. (A33) that

$$V_0^{min}=2.7kV$$
$$V_0^{max}=9kV \qquad (A40)$$

From equation (A31), we obtain the following value for $b$:

$$b=r_0\sqrt{\varepsilon_r}=1.68\times10^{-3}m \qquad (A41)$$

Since $b=a+d$ we get

$$d=1.8\times10^{-4}m$$

Substitution of $a$, $b$, $d$ and A(32) into (A39) produces

$$\sigma_{air}=4.117\times10^{-4}E_{air}^{\frac{1}{2}}=1.375\times10^{-2}V_0^{\frac{1}{2}}$$

Substitution of $\sigma_{air}$, $E_{air}(rms)$ and $\rho_{air}=1.2\ kg.m^{-3}$ into (A14) gives

$$\frac{m_{g(air)}}{m_{i0(air)}}=\left\{1-2\left[\sqrt{1+1.758\times10^{-27}\frac{\sigma_{air}^3E_{air}^4}{\rho_{air}^2f^3}}-1\right]\right\}=$$

$$=\left\{1-2\left[\sqrt{1+4.923\times10^{-21}\frac{V_0^{5.5}}{f^3}}-1\right]\right\} \qquad (A42)$$

For $V_0=V_0^{max}=9kV$ and $f=2Hz$, the result is

$$\frac{m_{g(air)}}{m_{i0(air)}}\cong-1.2$$

Note that, by increasing $V_0$ the values of $E_{air}$ and $\sigma_{air}$ are increased. Thus, as show (A42), there are two ways for decrease the value of $m_{g(air)}$: increasing the value of $V_0$ or decreasing the value of $f$.

Since $E_0^{max} = 10^7 V/m = 10kV/mm$ and $\Delta = 0.6\ mm$ then the dielectric strength of the insulation must be $\geq 16.7kV/mm$. As mentioned above, the dielectric strength of the acrylic is $17kV/mm$.

It is important to note that, due to the strong value of $E_{air}$ (Eq. A37) the *drift velocity* $v_d$, $\left(v_d = j_{air}/ne = \sigma_{air}E_{air}/ne\right)$ of the free charges inside the ionized air put them at a distance $x = v_d/t = 2fv_d \cong 0.4m$, which is much greater than the distance $d = 1.8\times10^{-4}m$. Consequently, the number $n$ of free charges decreases strongly inside the air layer of thickness $d$ †††, except, obviously, in a thin layer, very close to the dielectric, where the number of free charges remains sufficiently increased, to maintain the air conductivity with $\sigma_{air} \cong 1.1S/m$ (Eq. A39).

The thickness $h$ of this thin air layer close to the dielectric can be easily evaluated starting from the charge distribution in the neighborhood of the dielectric, and of the repulsion forces established among them. The result is $h = \sqrt{0.06e/4\pi\varepsilon_0 E} \cong 4\times10^{-9}m$. This is, therefore, the thickness of the *Air* Gravitational Shielding. If the area of this Gravitational Shielding is equal to the area of a format A4 sheet of paper, i.e., $A = 0.20\times0.291 = 0.0582m^2$, we obtain the following value for the resistance $R_{air}$ of the Gravitational Shielding: $R_{air} = h/\sigma_{air}A \cong 6\times10^{-8}\Omega$. Since the maximum electrical current through this air layer is $i^{max} = j^{max}A \cong 400kA$, then the maximum power radiated from the Gravitational Shielding is $P_{air}^{max} = R_{air}\left(i_{air}^{max}\right)^2 \cong 10kW$. This means that a very strong light will be radiated from this type of Gravitational Shielding. Note that this device can also be used as a lamp, which will be much more efficient than conventional lamps.

Coating a ceiling with this lighting system enables the entire area of ceiling to produce light. This is a form of lighting very different from those usually known.

Note that the value $P_{air}^{max} \cong 10kW$, defines the power of the transformer shown in Fig.A10. Thus, the maximum current in the secondary is $i_s^{max} = 9kV/10kW = 0.9A$.

Above the Gravitational Shielding, $\sigma_{air}$ is reduced to the normal value of conductivity of the atmospheric air $\left(\approx 10^{-14}S/m\right)$. Thus, the power radiated from this region is

$$P_{air}^{max} = (d-h)\left(i_{air}^{max}\right)^2/\sigma_{air}A = \\ = (d-h)A\sigma_{air}\left(E_{air}^{max}\right)^2 \cong 10^{-4}W$$

Now, we will describe a method to coat the Aluminum semi-spheres with acrylic in the necessary dimension $(\Delta = a - r_0)$. First, take an Aluminum plate with $21cm\times29.1cm$ (A4 format). By means of a convenient process, several semi-spheres can be stamped on its surface. The semi-spheres have radius $r_0 = 0.9\ mm$, and are joined one to another. Next, take an acrylic sheet (A4 format) with 1.5mm thickness (See Fig.A8 (a)). Put a heater below the Aluminum plate in order to heat the Aluminum (Fig.A8 (b)). When the Aluminum is sufficiently heated up, the acrylic sheet and the Aluminum plate are pressed, one against the other, as shown in Fig. A8 (c). The two D devices shown in this figure are used in order to impede that the press compresses the acrylic and the aluminum to a distance shorter than $y+a$. After some seconds, remove the press and the heater. The device is ready to be subjected to a voltage $V_0$ with frequency $f$, as shown in Fig.A9. Note that, in this case, the balance is not necessary, because *the substance that produces the gravitational shielding* is an *air layer* with thickness $d$ *above* the acrylic sheet. This is, therefore, more a type of Gravity Control Cell (GCC) with *external gravitational shielding*.

††† Reducing therefore the conductivity, $\sigma_{air}$, to the normal value of the conductivity of atmospheric air.

It is important to note that this GCC can be made very thin and as flexible as a fabric. Thus, it can be used to produce *anti- gravity clothes*. These clothes can be extremely useful, for example, to walk on the surface of high gravity planets.

Figure A11 shows some geometrical forms that can be stamped on a metallic surface in order to produce a Gravitational Shielding effect, similar to the produced by the *semi-spherical form*.

An obvious evolution from the semi-spherical form is the *semi-cylindrical* form shown in Fig. A11 (b); Fig.A11(c) shows *concentric metallic rings* stamped on the metallic surface, an evolution from Fig.A11 (b). These geometrical forms produce the same effect as the semi-spherical form, shown in Fig.A11 (a). By using concentric metallic rings, it is possible to build *Gravitational Shieldings* around bodies or spacecrafts with several formats (spheres, ellipsoids, etc); Fig. A11 (d) shows a Gravitational Shielding around a Spacecraft with *ellipsoidal form*.

The previously mentioned Gravitational Shielding, produced on a thin layer of ionized air, has a *behavior different from* the Gravitational Shielding produced on a *rigid substance*. When the gravitational masses of the air molecules, inside the shielding, are reduced to within the range $+0.159m_i < m_g < -0.159m_i$, they go to the *imaginary space-time*, as previously shown in this article. However, the electric field $E_{air}$ stays at the real space-time. Consequently, the molecules return immediately to the real space-time in order to return soon after to the *imaginary* space-time, due to the action of the electric field $E_{air}$.

In the case of the Gravitational Shielding produced on a *solid substance*, when the molecules of the substance go to the *imaginary* space-time, *the electric field that produces the effect, also goes to the imaginary space-time together with them*, since in this case, the substance of the Gravitational Shielding is rigidly connected to the metal that produces the electric field. (See Fig. A12 (b)). This is the fundamental difference between the *non-solid* and *solid* Gravitational Shieldings.

Now, consider a Gravitational Spacecraft that is able to produce an *Air* Gravitational Shielding and also a *Solid* Gravitational Shielding, as shown in Fig. A13 (a) [‡‡‡]. Assuming that the intensity of the electric field, $E_{air}$, necessary to reduce the gravitational mass of the *air molecules* to within the range $+0.159m_i < m_g < -0.159m_i$, *is much smaller* than the intensity of the electric field, $E_{rs}$, necessary to reduce the gravitational mass of the *solid substance* to within the range $+0.159m_i < m_g < -0.159m_i$, then we conclude that the Gravitational Shielding made of ionized air goes to the imaginary space-time *before* the Gravitational Shielding made of *solid substance*. When this occurs the spacecraft does not go to the imaginary space-time together with the Gravitational Shielding of air, because the air molecules are not rigidly connected to the spacecraft. Thus, while the air molecules go into the imaginary space-time, the spacecraft stays in the *real space-time*, and remains subjected to the effects of the Gravitational Shielding around it, since the shielding does not stop to work, due to its extremely short permanence at the imaginary space-time. Under these circumstances, the gravitational mass of the Gravitational Shielding can be reduced to $m_g \cong 0$. For example, $m_g \cong 10^{-4} kg$. Thus, if the *inertial mass* of the Gravitational Shielding is $m_{i0} \cong 1kg$, then $\chi = m_g / m_{i0} \cong 10^{-4}$. As we have seen, this means that *the inertial effects on the spacecraft* will be reduced by $\chi \cong 10^{-4}$. Then, in spite of the effective acceleration of the spacecraft be, for example, $a = 10^5 m.s^{-2}$, the effects on the crew of the spacecraft will be equivalent to an acceleration of only

$$a' = \frac{m_g}{m_{i0}} a = \chi \; a \approx 10 m.s^{-1}$$

This is the magnitude of the acceleration upon the passengers in a contemporary commercial jet.

Then, it is noticed that Gravitational Spacecrafts can be subjected to enormous *accelerations* (or *decelerations*) without imposing any harmful impacts whatsoever on the spacecrafts or its crew.

Now, imagine that the intensity of the electric field that produces the Gravitational Shielding around the spacecraft is *increased* up to reaching the value $E_{rs}$ that reduces the gravitational mass of the *solid* Gravitational Shielding to within the range $+0.159m_i < m_g < -0.159m_i$. Under these circumstances, the *solid* Gravitational Shielding goes to the imaginary space-time and, since it is rigidly connected to the spacecraft, also the spacecraft goes to the imaginary space-time together with the Gravitational Shielding. Thus, the spacecraft can travel within the imaginary space-time and make use of the Gravitational Shielding around it.

‡‡‡ The *solid* Gravitational Shielding can also be obtained by means of *an ELF electric current through a metallic lamina* placed *between the semi-spheres and the Gravitational Shielding of Air* (See Fig.A13 (a)). The gravitational mass of the solid Gravitational Shielding will be controlled just by means of the intensity of the ELF electric current. Recently, it was discovered that Carbon nanotubes (CNTs) can be added to *Alumina* ($Al_2O_3$) to convert it into a good electrical conductor. It was found that the electrical conductivity increased up to 3375 S/m at 77°C in samples that were 15% nanotubes by volume [12]. It is known that the density of α-Alumina is $3.98\times10^3 kg.m^{-3}$ and that it can withstand 10-20 KV/mm. Thus, these values show that the Alumina-CNT can be used to make a *solid* Gravitational Shielding.

As we have already seen, the maximum velocity of propagation of the interactions in the imaginary space-time is *infinite* (in the real space-time this limit is equal to the light velocity $c$). This means that *there are no limits for the velocity of the spacecraft in the imaginary space-time*. Thus, the acceleration of the spacecraft can reach, for example, $a = 10^9 m.s^{-2}$, which leads the spacecraft to attain velocities $V \approx 10^{14} m.s^{-1}$ (about 1 million times the speed of light) after one day of trip. With this velocity, after 1 month of trip the spacecraft would have traveled about $10^{21} m$. In order to have idea of this distance, it is enough to remind that the diameter of our Universe (visible Universe) is of the order of $10^{26} m$.

Due to the extremely low density of the *imaginary* bodies, the collision between them cannot have the same consequences of the collision between the real bodies.

Thus, *for a Gravitational Spacecraft in imaginary state, the problem of the collision in high-speed doesn't exist.* Consequently, the Gravitational Spacecraft can transit freely in the imaginary Universe and, in this way, reach easily any point of our real Universe once they can make the transition back to our Universe by only increasing the gravitational mass of the Gravitational Shielding of the spacecraft in such way that it leaves the range of $+0.159M_i$ to $-0.159M_i$.

The return trip would be done in similar way. That is to say, the spacecraft would transit in the imaginary Universe back to the departure place where would reappear in our Universe. Thus, trips through our Universe that would delay millions of years, at speeds close to the speed of light, could be done in just a few *months* in the imaginary Universe.

In order to produce the acceleration of $a \approx 10^9 m.s^{-2}$ upon the spacecraft we propose a Gravitational Thruster with 10 GCCs (10 Gravitational Shieldings) of the type with several semi-spheres stamped on the metallic surface, as previously shown, or with the *semi-cylindrical* form shown in Figs. A11 (b) and (c). The 10 GCCs are filled with air at 1 atm and 300K. If the insulation layer is made with *Mica* $(\varepsilon_r \cong 5.4)$ and has thickness $\Delta = 0.1\ mm$, and the semi-spheres stamped on the metallic surface have $r_0 = 0.4\ mm$ (See Fig.A7) then $a = r_0 + \Delta = 0.5\ mm$. Thus, we get

$$b = r_0\sqrt{\varepsilon_r} = 9.295 \times 10^{-4} m$$

and

$$d = b - a = 4.295 \times 10^{-4} m$$

Then, from Eq. A42 we obtain

$$\chi_{air} = \frac{m_{g(air)}}{m_{i0(air)}} = \left\{1 - 2\left[\sqrt{1 + 1.758 \times 10^{-27}\frac{\sigma_{air}^3 E_{air}^4}{\rho_{air}^2 f^3}} - 1\right]\right\} =$$

$$= \left\{1 - 2\left[\sqrt{1 + 1.0 \times 10^{-18}\frac{V_0^{5.5}}{f^3}} - 1\right]\right\}$$

For $V_0 = V_0^{max} = 15.6kV$ and $f = 0.12Hz$, the result is

$$\chi_{air} = \frac{m_{g(air)}}{m_{i0(air)}} \cong -1.6 \times 10^4$$

Since $E_0^{max} = V_0^{max}/r_0$ is now given by $E_0^{max} = 15.6kV/0.9mm = 17.3kV/mm$ and $\Delta = 0.1\ mm$ then the dielectric strength of the insulation must be $\geq 173kV/mm$. As shown in the table below[§§§], *0.1mm - thickness of* Mica *can withstand 17.6 kV* (that is greater than $V_0^{max} = 15.6kV$), in such way that the dielectric strength is *176 kV/mm.*

The Gravitational Thrusters are positioned at the spacecraft, as shown in Fig. A13 (b). Then, when the spacecraft is in the *intergalactic space*, the gravity acceleration upon the gravitational mass $m_{gt}$ of the bottom of the thruster (See Fig.A13 (c)), is given by [2]

$$\vec{a} \cong (\chi_{air})^{10} \vec{g}_M \cong -(\chi_{air})^{10} G \frac{M_g}{r^2} \hat{\mu}$$

where $M_g$ is the gravitational mass in front of the spacecraft.

For simplicity, let us consider just the effect of a hypothetical volume $V = 10 \times 10^3 \times 10^3 = 10^7 m^3$ of intergalactic matter in front of the spacecraft $(r \cong 30m)$. The average density of matter in the *intergalactic medium (IGM)* is $\rho_{ig} \approx 10^{-26} kg.m^{-3})$[****]. Thus, for $\chi_{air} \cong -1.6 \times 10^4$ we get

$$a = -\left(-1.6 \times 10^4\right)^{10} \left(6.67 \times 10^{-11}\right)\left(\frac{10^{-19}}{30^2}\right) =$$

$$= -10^9 m.s^{-2}$$

In spite of this gigantic acceleration, the inertial effects for the crew of the spacecraft can be strongly reduced if, for example, the gravitational mass of the Gravitational Shielding is reduced down to $m_g \cong 10^{-6} kg$ and its inertial mass is $m_{i0} \cong 100kg$. Then, we get $\chi = m_g / m_{i0} \cong 10^{-8}$. Therefore, *the inertial effects on the spacecraft* will be reduced by $\chi \cong 10^{-8}$, and consequently, the inertial effects on the crew of the spacecraft would be *equivalent to* an acceleration $a'$ of only

$$a' = \frac{m_g}{m_{i0}} a = \left(10^{-8}\right)\left(10^9\right) \approx 10 m.s^{-2}$$

Note that the Gravitational Thrusters in the spacecraft must have a very small diameter (of the order of *millimeters*) since, obviously, the hole through the Gravitational Shielding cannot be large. Thus, these thrusters are in fact, *Micro-Gravitational Thrusters*. As shown in Fig. A13 (b), it is possible to place several micro-gravitational thrusters in the spacecraft. This gives to the Gravitational Spacecraft, several degrees of freedom and shows the enormous superiority of this spacecraft in relation to the contemporaries spacecrafts.

The density of matter in the *intergalactic medium (IGM)* is about *10 $^{-26}$ kg.m$^{-3}$* , which is very less than the density of matter in the *interstellar medium* (*~10$^{-21}$ kg.m$^{-3}$*) that is less than the density of matter in the *interplanetary medium* (*~10$^{-20}$ kg.m$^{-3}$*). The density of matter is enormously

---

[§§§] The *dielectric strength* of some dielectrics can have different values in lower thicknesses. This is, for example, the case of the *Mica*.

| *Dielectric* | *Thickness* (mm) | *Dielectric Strength* (kV/mm) |
|---|---|---|
| Mica | 0.01 mm | 200 |
| **Mica** | **0.1 mm** | **176** |
| Mica | 1 mm | 61 |

[****] Some theories put the average density of the Universe as the equivalent of *one hydrogen atom per cubic meter* [13,14]. The density of the universe, however, is clearly not uniform. Surrounding and stretching between galaxies, there is rarefied plasma [15] that is thought to possess a cosmic filamentary structure [16] and that is slightly denser than the average density in the universe. This material is called the *intergalactic medium (IGM)* and is mostly ionized hydrogen; i.e. a plasma consisting of equal numbers of electrons and protons. The IGM is thought to exist at a density of 10 to 100 times the average density of the Universe (10 to 100 hydrogen atoms per cubic meter, i.e., $\approx 10^{-26} kg.m^{-3}$).

increased inside the Earth's atmosphere ($1.2kg.m^{-3}$ near to Earth's surface). Figure A14 shows the gravitational acceleration acquired by a Gravitational Spacecraft, in these media, using Micro-Gravitational thrusters.

In relation to the *Interstellar* and *Interplanetary medium*, the *Intergalactic medium* requires the greatest value of $\chi_{air}$ ($\chi$ inside the *Micro-Gravitational Thrusters*), i.e., $\chi_{air} \cong -1.6\times10^4$. This value strongly decreases when the spacecraft is within the Earth's atmosphere. In this case, it is sufficient only[††††] $\chi_{air} \cong -10$ in order to obtain:

$$a = -(\chi_{air})^{10} G \frac{\rho_{atm} V}{r^2} \cong$$

$$\cong -(-10)^{10}\left(6.67\times10^{-11}\right)\frac{1.2\left(10^7\right)}{(20)^2} \cong 10^4\, m.s^{-2}$$

With this acceleration the Gravitational Spacecraft can reach about *50000 km/h* in a few seconds. Obviously, the Gravitational Shielding of the spacecraft will reduce strongly *the inertial effects upon the crew* of the spacecraft, in such way that the inertial effects of this strong acceleration will not be felt. In addition, the *artificial atmosphere*, which is possible to build around the spacecraft, by means of gravity control technologies shown in this article (See Fig.6) and [2], will protect it from the *heating* produced by the friction with the Earth's atmosphere. Also, the gravity can be controlled inside of the Gravitational Spacecraft in order to maintain a value close to the Earth's gravity as shown in Fig.3.

Finally, it is important to note that a Micro-Gravitational Thruster does not work *outside* a Gravitational Shielding, because, in this case, *the resultant upon the thruster is null* due to the symmetry (See Fig. A15 (a)). Figure A15 (b) shows a micro-gravitational thruster inside a Gravitational Shielding. This thruster has 10 Gravitational Shieldings, in such way that the gravitational acceleration upon the *bottom* of the thruster, due to a gravitational mass $M_g$ *in front* of the thruster, is $a_{10} = \chi_{air}^{10} a_0$ where $a_0 = -GM_g/r^2$ is the gravitational acceleration acting on the front of the micro-gravitational thruster. *In the opposite direction*, the gravitational acceleration upon the bottom of the thruster, produced by a gravitational mass $M_g$, is

$$a_0' = \chi_s\left(-GM_g/r'^2\right) \cong 0$$

since $\chi_s \cong 0$ due to the Gravitational Shielding around the micro-thruster (See Fig. A15 (b)). Similarly, the acceleration in front of the thruster is

$$a_{10}' = \chi_{air}^{10} a_0' = \left[\chi_{air}^{10}\left(-GM_g/r'^2\right)\right]\chi_s$$

where $\left[\chi_{air}^{10}\left(-GM_g/r'^2\right)\right] < a_{10}$, since $r' > r$. Thus, for $a_{10} \cong 10^9 m.s^{-2}$ and $\chi_s \approx 10^{-8}$ we conclude that $a_{10}' < 10 m.s^{-2}$. This means that $a_{10}' << a_{10}$. Therefore, we can write that the resultant on the micro-thruster can be expressed by means of the following relation

$$R \cong F_{10} = \chi_{air}^{10} F_0$$

[††††] This value is within the range of values of $\chi$ ($\chi < -10^3$. *See Eq.A15*), which can be produced by means of *ELF electric currents* through metals as *Aluminum*, etc. This means that, in this case, if convenient, we can replace *air* inside the GCCs of the Gravitational Micro-thrusters by metal laminas with *ELF electric currents* through them.

Figure A15 (c) shows a Micro-Gravitational Thruster with *10 Air Gravitational Shieldings* (10 GCCs). Thin Metallic laminas are placed after each *Air* Gravitational Shielding in order to retain the electric field $E_b = V_0/x$, produced by metallic *surface behind* the semi-spheres. The laminas with semi-spheres stamped on its surfaces are connected to the ELF voltage source $V_0$ and the thin laminas in front of the Air Gravitational Shieldings are grounded. The air inside this Micro-Gravitational Thruster is at 300K, 1atm.

We have seen that the insulation layer of a GCC can be made up of Acrylic, Mica, etc. Now, we will design a GCC using *Water* (*distilled water*, $\varepsilon_{r(H_2O)} = 80$) and Aluminum *semi-cylinders* with radius $r_0 = 1.3mm$. Thus, for $\Delta = 0.6mm$, the new value of $a$ is $a = 1.9mm$. Then, we get

$$b = r_0\sqrt{\varepsilon_{r(H_2O)}} = 11.63\times10^{-3}m \qquad (A43)$$

$$d = b - a = 9.73\times10^{-3}m \qquad (A44)$$

and

$$E_{air} = \frac{1}{4\pi\varepsilon_{r(air)}\varepsilon_0}\frac{q}{b^2} = $$

$$= \varepsilon_{r(H_2O)}\frac{V_0 r_0}{\varepsilon_{r(air)}b^2} =$$

$$= \frac{V_0/r_0}{\varepsilon_{r(air)}} \cong \frac{V_0}{r_0} = 1111.1\ V_0 \qquad (A45)$$

Note that

$$E_{(H_2O)} = \frac{V_0/r_0}{\varepsilon_{r(H_2O)}}$$

and

$$E_{(acrylic)} = \frac{V_0/r_0}{\varepsilon_{r(acrylic)}}$$

Therefore, $E_{(H_2O)}$ and $E_{(acrylic)}$ are much smaller than $E_{air}$. Note that for $V_0 \le 9kV$ the intensities of $E_{(H_2O)}$ and $E_{(acrylic)}$ are not sufficient to produce the ionization effect, which increases the electrical conductivity. Consequently, the conductivities of the water and the acrylic remain $<< 1\ S.m^{-1}$. In this way, with $E_{(H_2O)}$ and $E_{(acrylic)}$ much smaller than $E_{air}$, and $\sigma_{(H2O)} << 1$, $\sigma_{(acrylic)} << 1$, the decrease in both the gravitational mass of the acrylic and the gravitational mass of water, according to Eq.A14, is negligible. This means that only in the air layer the decrease in the gravitational mass will be relevant.

Equation A39 gives the electrical conductivity of the air layer, i.e.,

$$\sigma_{air} = 2\alpha\left(\frac{E_{air}}{d}\right)^{\frac{1}{2}}\left(\frac{b}{a}\right)^{\frac{3}{2}} = 0.029V_0^{\frac{1}{2}} \qquad (A46)$$

Note that $b = r_0\sqrt{\varepsilon_{r(H_2O)}}$. Therefore, here the value of $b$ is larger than in the case of the acrylic. Consequently, *the electrical conductivity of the air layer will be larger here than in the case of acrylic.*

Substitution of $\sigma_{(air)}$, $E_{air}$ (*rms*) and $\rho_{air} = 1.2kg.m^{-3}$ into Eq. A14, gives

$$\frac{m_{g(air)}}{m_{i0(air)}} = \left\{1 - 2\left[\sqrt{1 + 4.54\times10^{-20}\frac{V_0^{5.5}}{f^3}} - 1\right]\right\} \qquad (A47)$$

For $V_0 = V_0^{max} = 9kV$ and $f = 2Hz$, the result is

$$\frac{m_{g(air)}}{m_{i0(air)}} \cong -8.4$$

This shows that, by using *water* instead of acrylic, the result is much better.

In order to build the GCC based on the calculations above (See Fig. A16), take an Acrylic plate with *885mm* x *885m* and *2mm* thickness, then paste on it an Aluminum sheet with *895.2mm* x *885mm* and *0.5mm* thickness(note that two edges of the Aluminum sheet are bent as shown in Figure A16 (b)). Next, take *342* Aluminum yarns with *884mm* length and *2.588mm* diameter (wire # 10 AWG) and insert them side by side on the Aluminum sheet. See in Fig. A16 (b) the detail of fixing of the yarns on the Aluminum sheet. Now, paste acrylic strips (with *13.43mm* height and *2mm* thickness) around the Aluminum/Acrylic, making a box. Put *distilled water* (approximately *1 litter*) inside this box, up to a height of exactly *3.7mm* from the edge of the acrylic base. Afterwards, paste an Acrylic lid (*889mm* x *889mm* and *2mm* thickness) on the box. Note that above the water there is an *air* layer with *885mm* x *885mm* and *7.73mm* thickness (See Fig. A16). This thickness plus the acrylic lid thickness (*2mm*) is equal to $d = b - a = 9.73mm$ where $b = r_0\sqrt{\varepsilon_{r(H_2O)}} = 11.63mm$ and $a = r_0 + \Delta = 1.99mm$, since $r_0 = 1.3mm$, $\varepsilon_{r(H_2O)} = 80$ and $\Delta = 0.6mm$.

Note that the gravitational action of the electric field $E_{air}$, extends itself only up to the distance $d$, which, in this GCC, is given by the sum of the Air layer thickness (*7.73mm*) plus the thickness of the Acrylic lid (*2mm*).

Thus, it is ensured the gravitational effect on the air layer while it is practically nullified in the acrylic sheet above the air layer, since $E_{(acrylic)} << E_{air}$ and $\sigma_{(acrylic)} << 1$.

With this GCC, we can carry out an experiment where the *gravitational mass of the air layer* is progressively reduced when the voltage applied to the GCC is increased (or when the frequency is decreased). A precision balance is placed below the GCC in order to measure the mentioned mass decrease for comparison with the values predicted by Eq. A(47). In total, this GCC weighs about *6kg*; the *air layer 7.3grams*. The balance has the following characteristics: *range 0-6kg; readability 0.1g*. Also, in order to prove the *Gravitational Shielding Effect*, we can put a *sample* (connected to a dynamometer) above the GCC in order to check the gravity acceleration in this region.

In order to prove *the exponential effect* produced by the superposition of the Gravitational Shieldings, we can take three similar GCCs and put them one above the other, in such way that above the GCC 1 the gravity acceleration will be $g' = \chi g$; above the GCC2 $g'' = \chi^2 g$, and above the GCC3 $g''' = \chi^3 g$. Where $\chi$ is given by Eq. (A47).

It is important to note that the intensity of the electric field through the air *below* the GCC is *much smaller* than the intensity of the electric field through the air layer inside the GCC. In addition, the electrical conductivity of the air below the GCC is much smaller than the conductivity of the air layer inside the GCC. Consequently, the decrease of the gravitational mass of the air below the GCC, according to Eq.A14, is negligible. This means that the GCC1, GCC2 and GCC3 can be simply overlaid, on the experiment proposed above. However, since it is necessary to put samples among them in order to measure the gravity above each GCC, we suggest a spacing of 30cm or more among them.

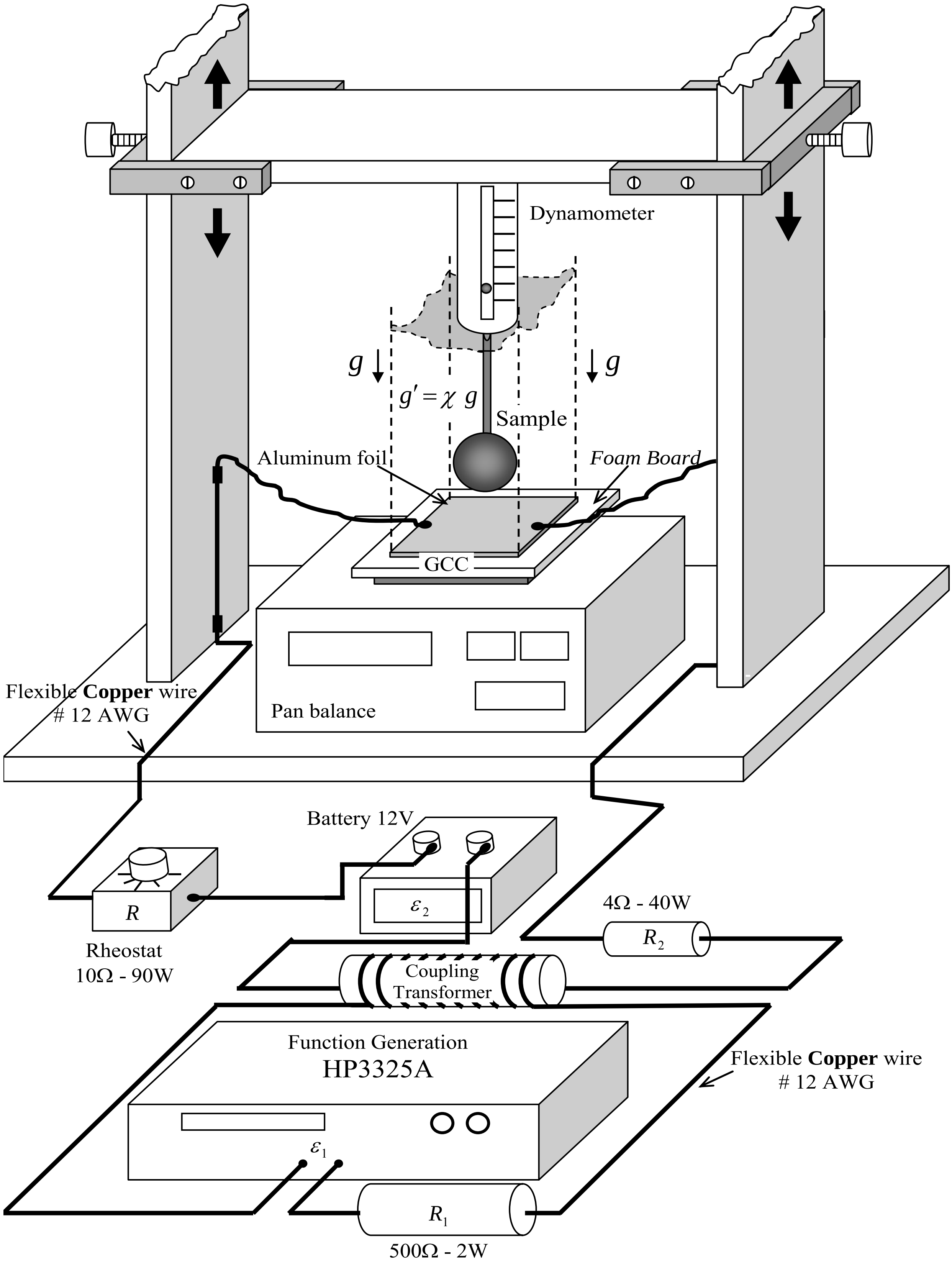

Fig. A2 – Experimental Set-up 1.

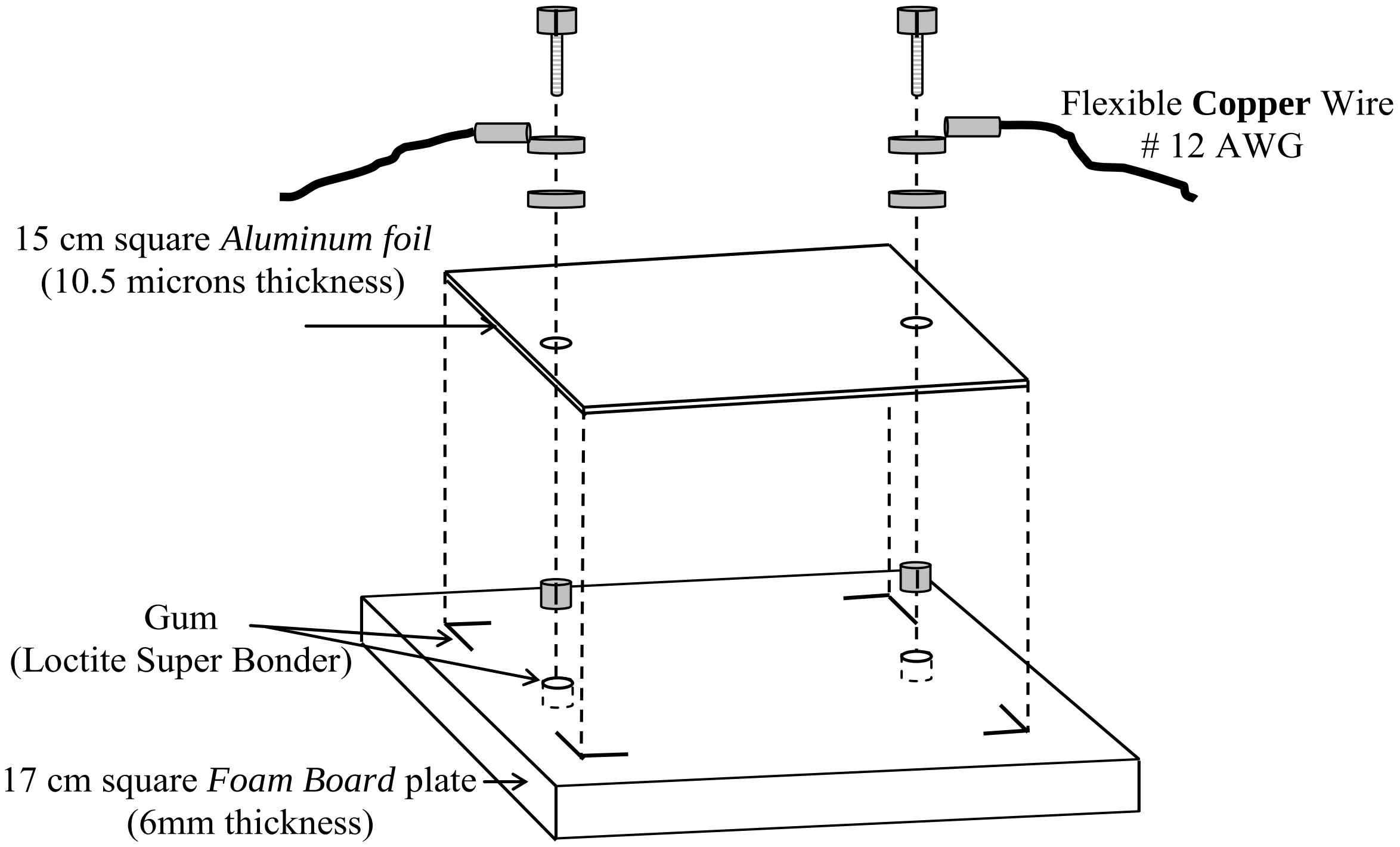

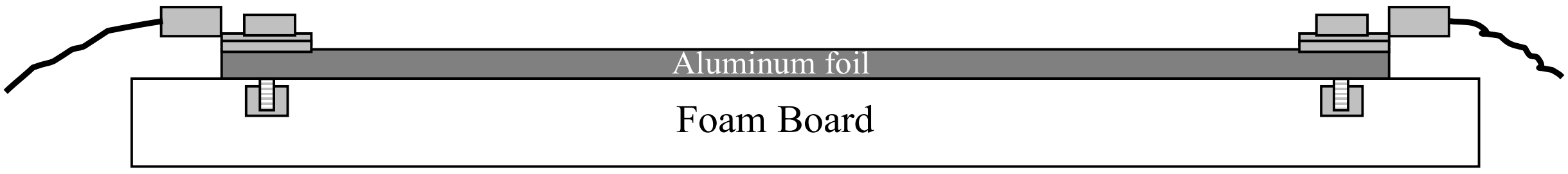

Fig. A3 – The Simplest *Gravity Control Cell* (GCC).

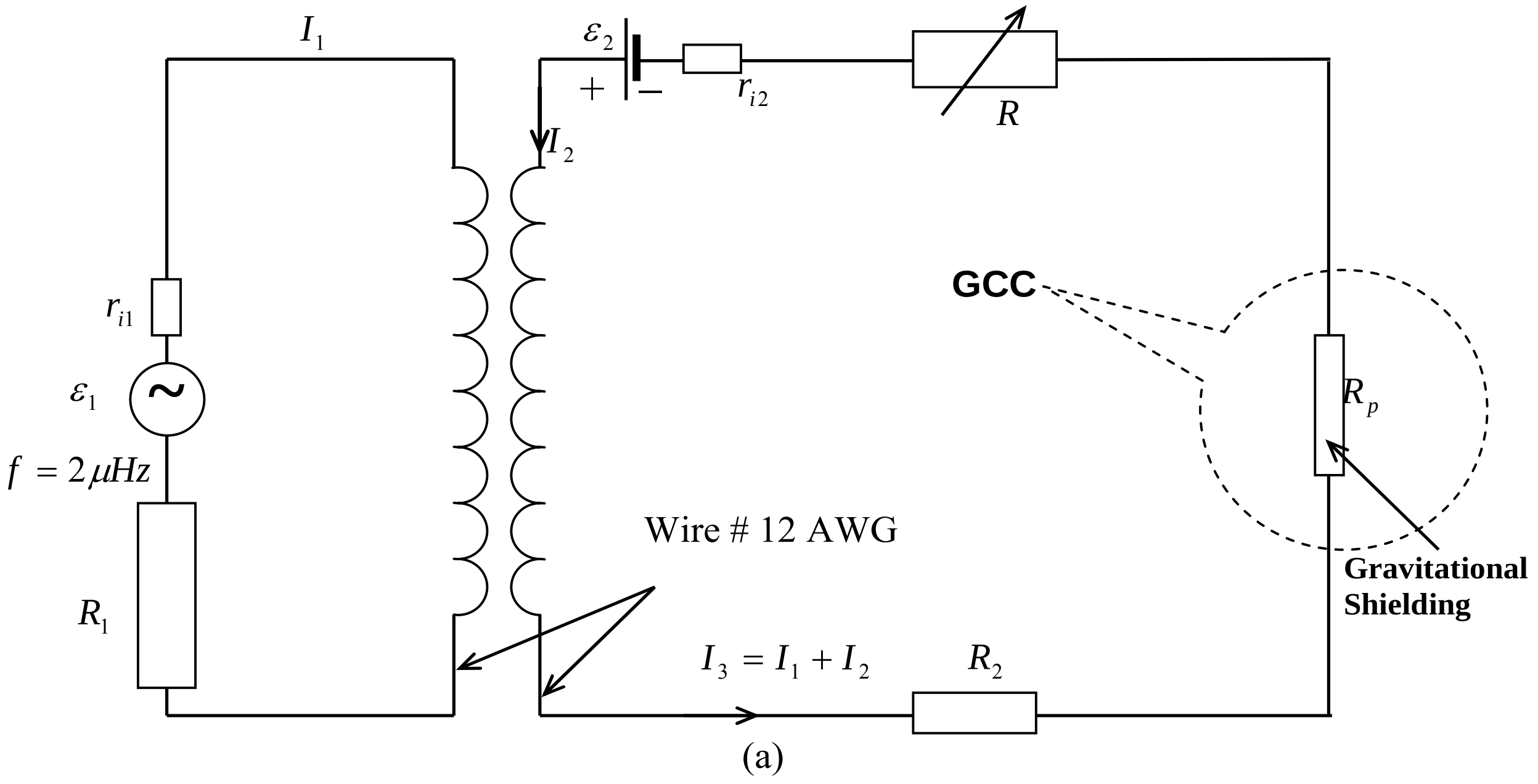

(a)

$\varepsilon_1 =$ Function Generator HP3325A(*Option* 002 High Voltage Output)

$r_{i1} < 2\Omega; \quad R_1 = 500\Omega - 2\ W; \quad \varepsilon_2 = 12V\ DC; \quad r_{i2} < 0.1\Omega\ (Battery);$

$R_2 = 4\Omega - 40W; \quad R_p = 2.5\times10^{-3}\Omega; \quad Reostat = 0 \le R \le 10\Omega - 90W$

$I_1^{max} = 56mA\ (rms); \quad I_2^{max} = 3A\ ; \quad I_3^{max} \cong 3A\ (rms)$

*Coupling Transformer* to isolate the *Function Generator* from the Battery

- Air core 10 - mm diameter; wire #12 AWG; $N_1 = N_2 = 20; l = 42mm$

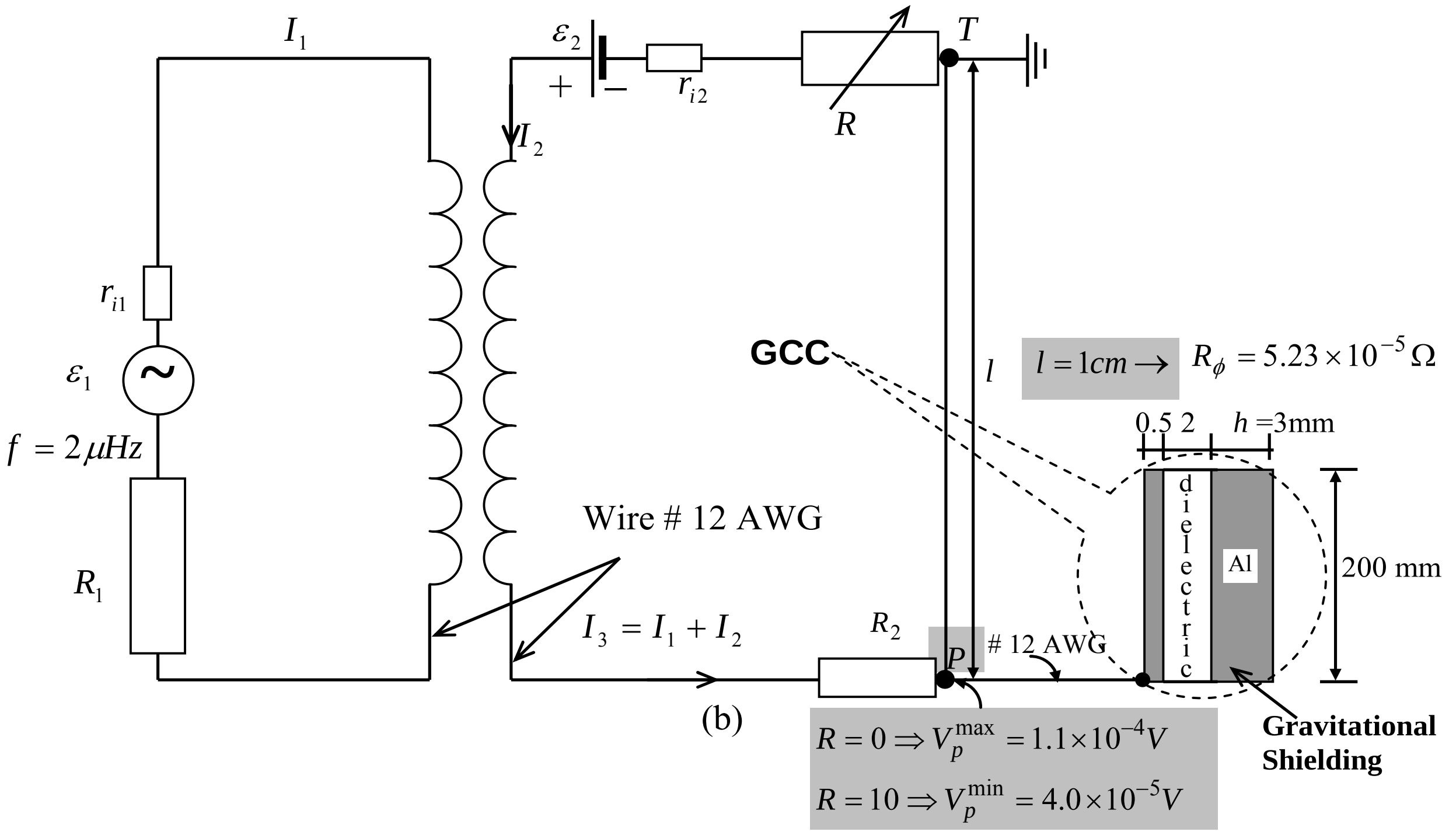

(b)

Fig. A4 – Equivalent Electric Circuits

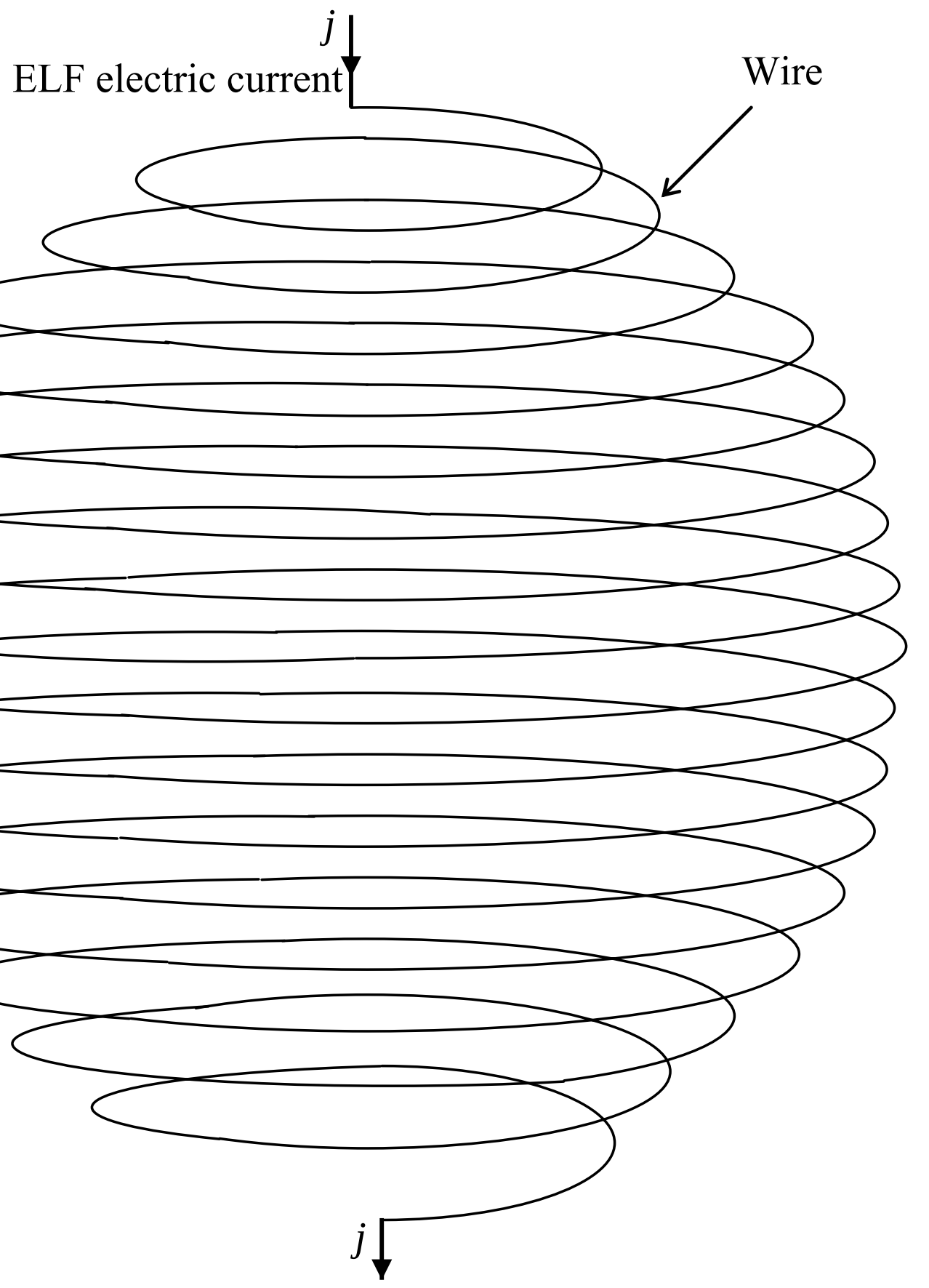

$$m_g = \left\{1 - 2\left[\sqrt{1 + 1.758\times10^{-27}\frac{\mu_r j^4}{\sigma\rho^2 f^3}} - 1\right]\right\} m_{i0}$$

Fig. A5 – An ELF electric current through a wire, that makes a spherical form as shown above, reduces the gravitational mass of the wire and the gravity inside sphere at the same proportion $\chi = m_g / m_{i0}$ (Gravitational Shielding Effect). Note that this spherical form can be transformed into an ellipsoidal form or a disc in order to coat, for example, a Gravitational Spacecraft. It is also possible to coat with a wire several forms, such as cylinders, cones, cubes, etc. The characteristics of the wire are expressed by: $\mu_r$, $\sigma$, $\rho$; $j$ is the electric current density and $f$ is the frequency.

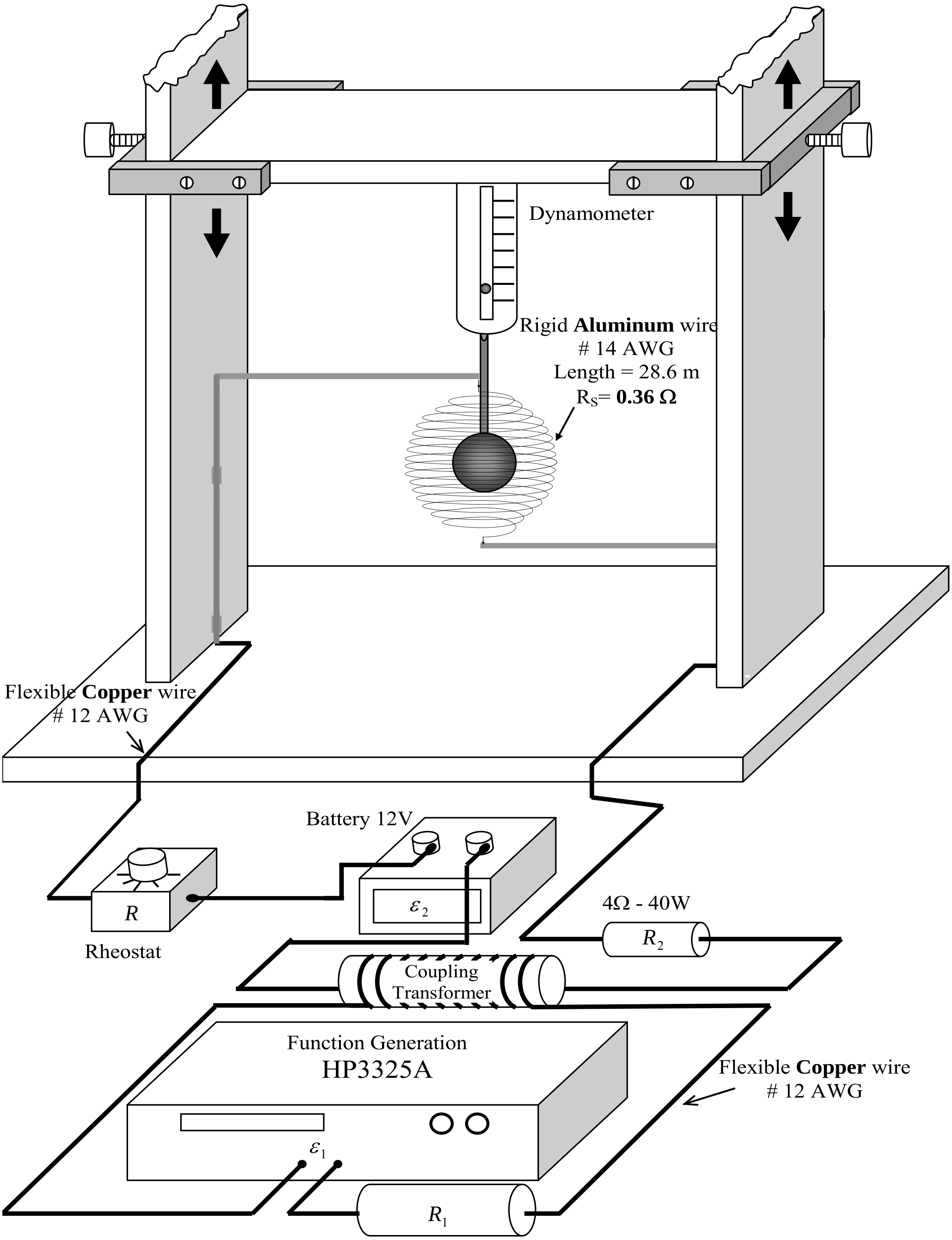

Fig. A6 – Experimental set-up 2.

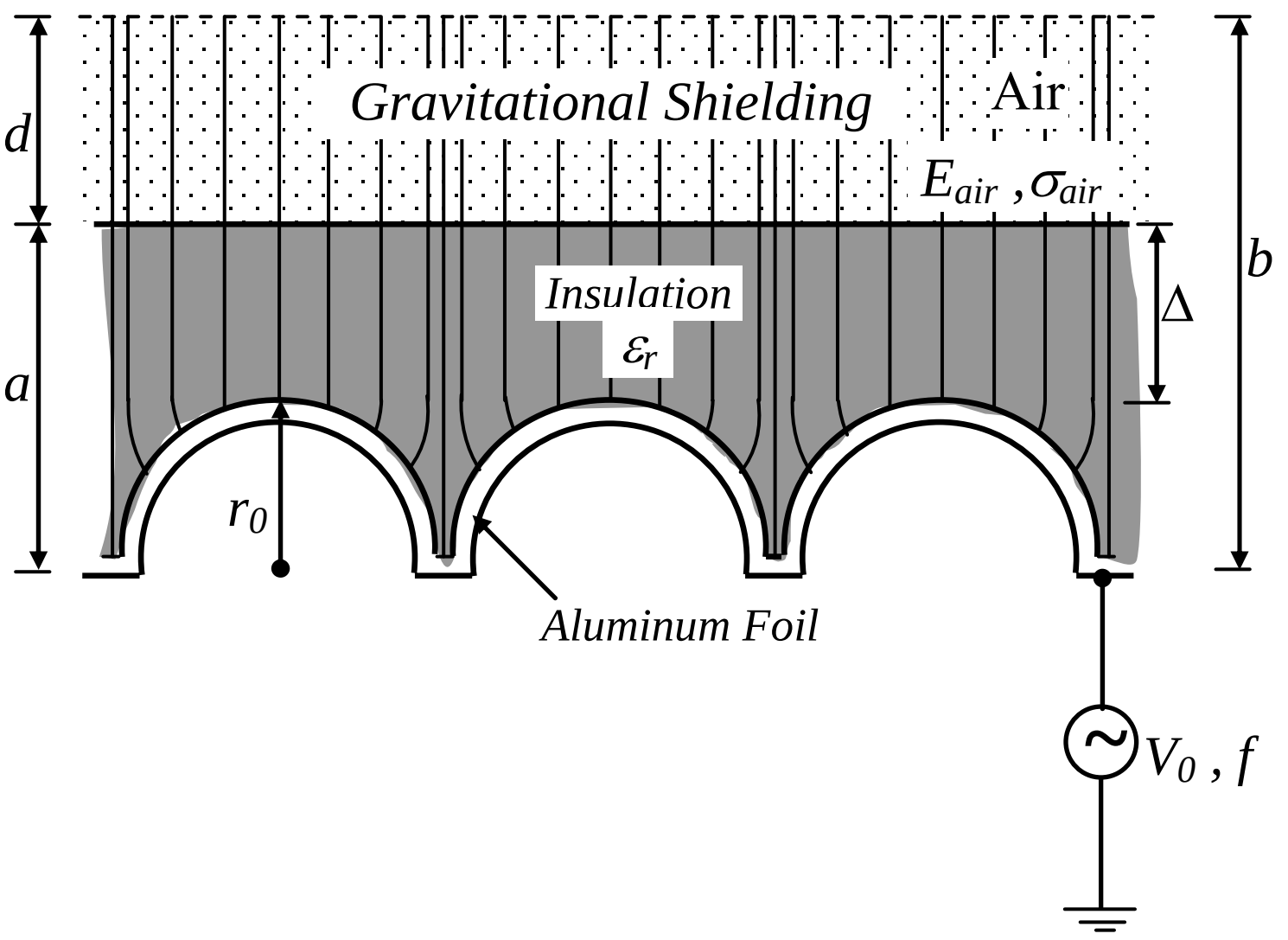

Fig A7 – *Gravitational shielding produced by semi-spheres stamped on the Aluminum foil* - By simply changing the geometry of the surface *of* the Aluminum foil it is possible to increase the working frequency $f$ up to more than *1Hz*.

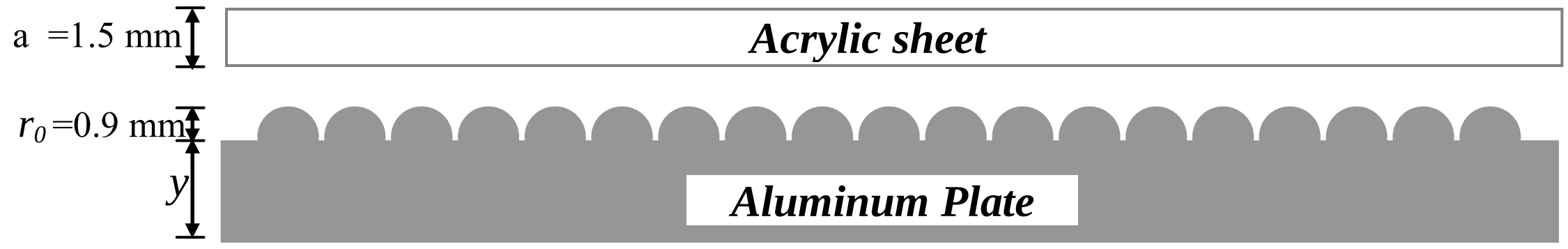

(a)

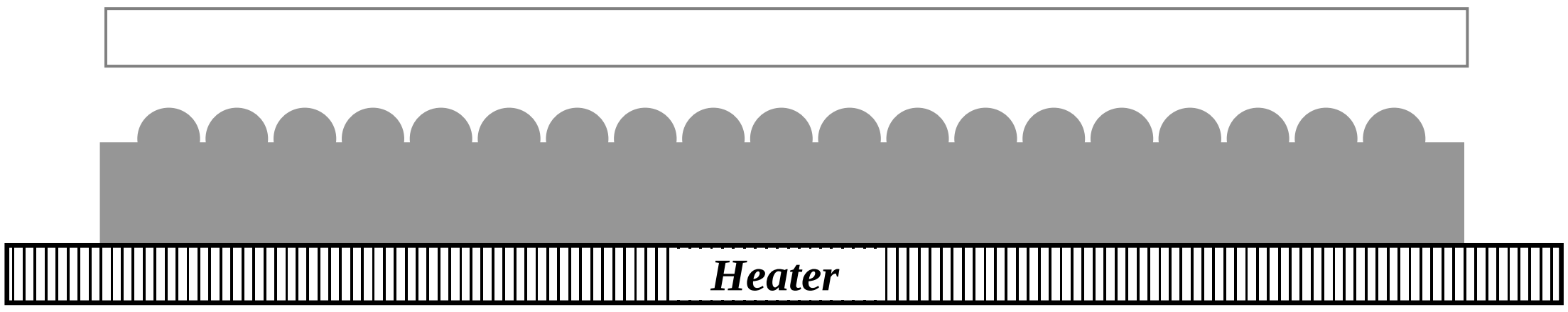

(b)

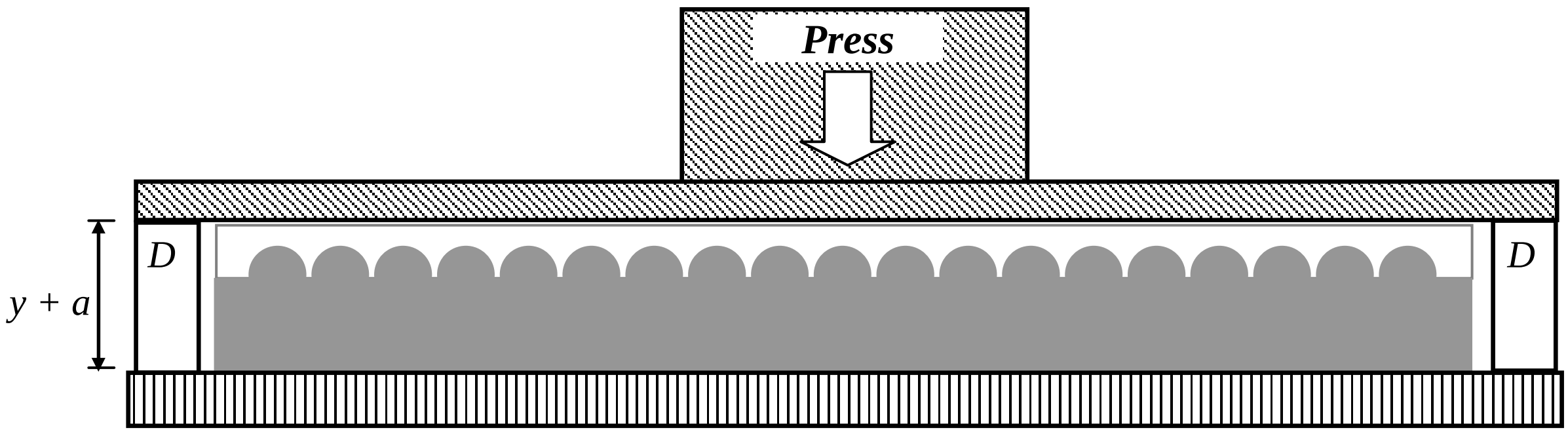

(c)

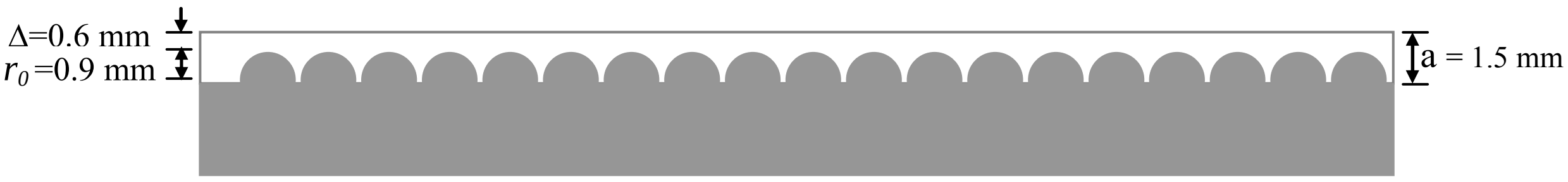

(d)

Fig A8 – *Method to coat the Aluminum semi-spheres with acrylic* $(\Delta = a - r_0 = 0.6mm)$. (a)Acrylic sheet (A4 format) with 1.5mm thickness and an Aluminum plate (A4) with several semi-spheres (radius $r_0 = 0.9\ mm$) stamped on its surface. (b)A heater is placed below the Aluminum plate in order to heat the Aluminum. (c)When the Aluminum is sufficiently heated up, the acrylic sheet and the Aluminum plate are pressed, one against the other (The two D devices shown in this figure are used in order to impede that the press compresses the acrylic and the aluminum besides distance $y + a$). (d)After some seconds, the press and the heater are removed, and the device is ready to be used.

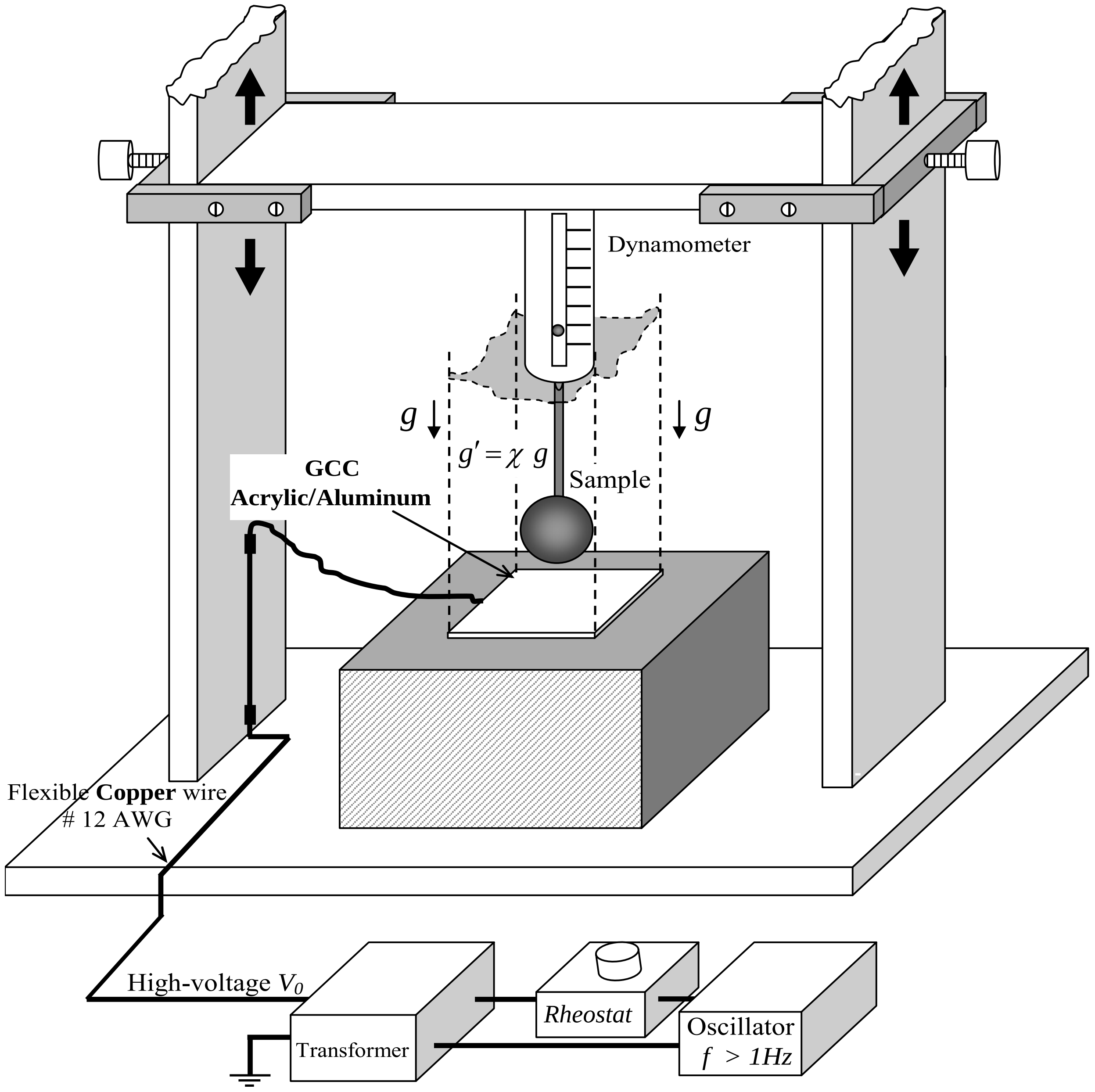

Fig. A9 – *Experimental Set-up using a GCC subjected to high-voltage* $V_0$ *with frequency* $f > 1Hz$ . Note that in this case, the pan balance is not necessary because the substance of the Gravitational Shielding is an *air layer* with thickness $d$ *above* the acrylic sheet. This is therefore, more a type of Gravity Control Cell (GCC) with *external gravitational shielding*.

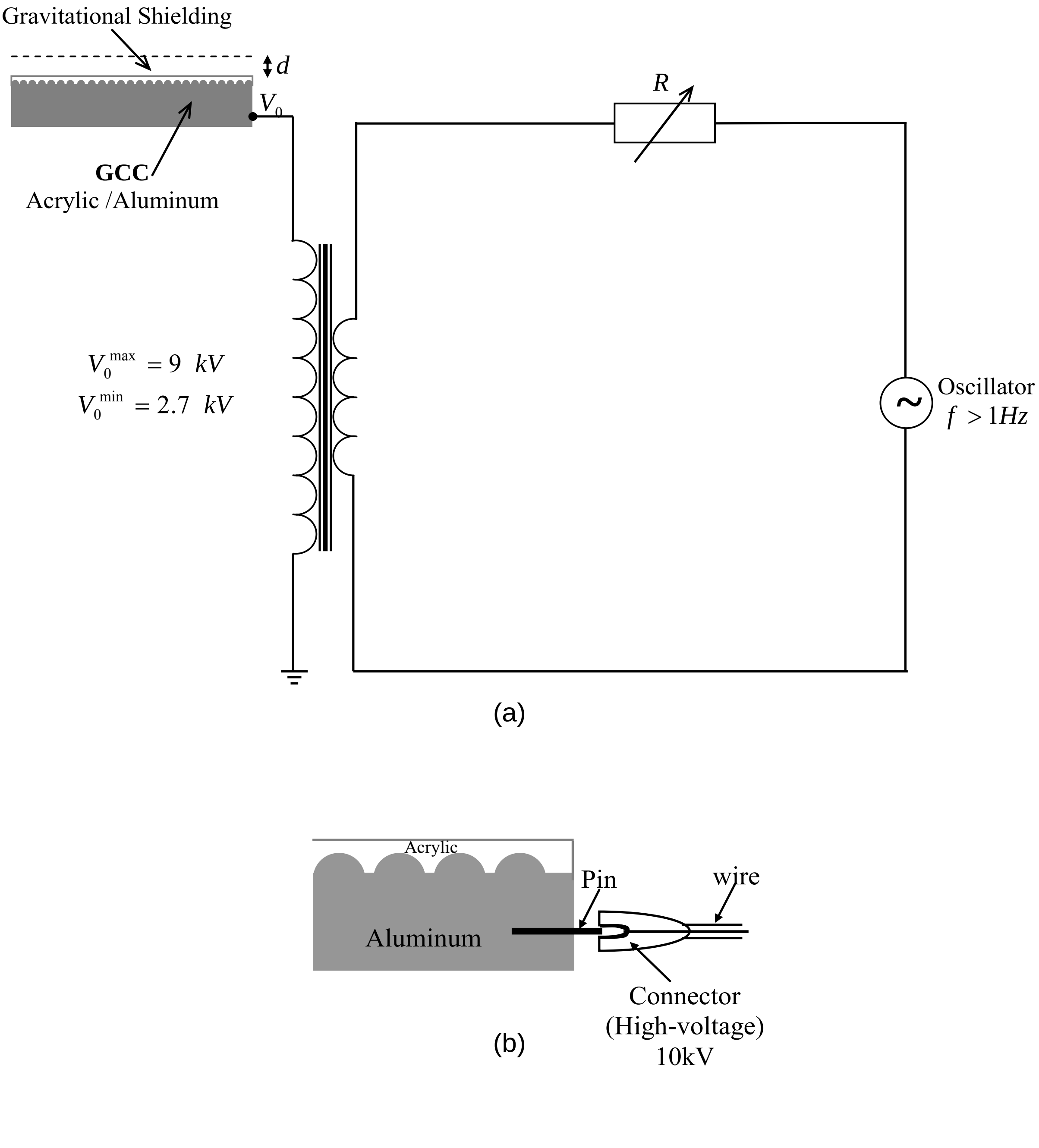

Fig. A10 – (a) *Equivalent Electric Circuit.* (b) Details of the electrical connection with the Aluminum plate. Note that others connection modes (by the top of the device) can produce destructible interference on the electric lines of the $E_{air}$ field.

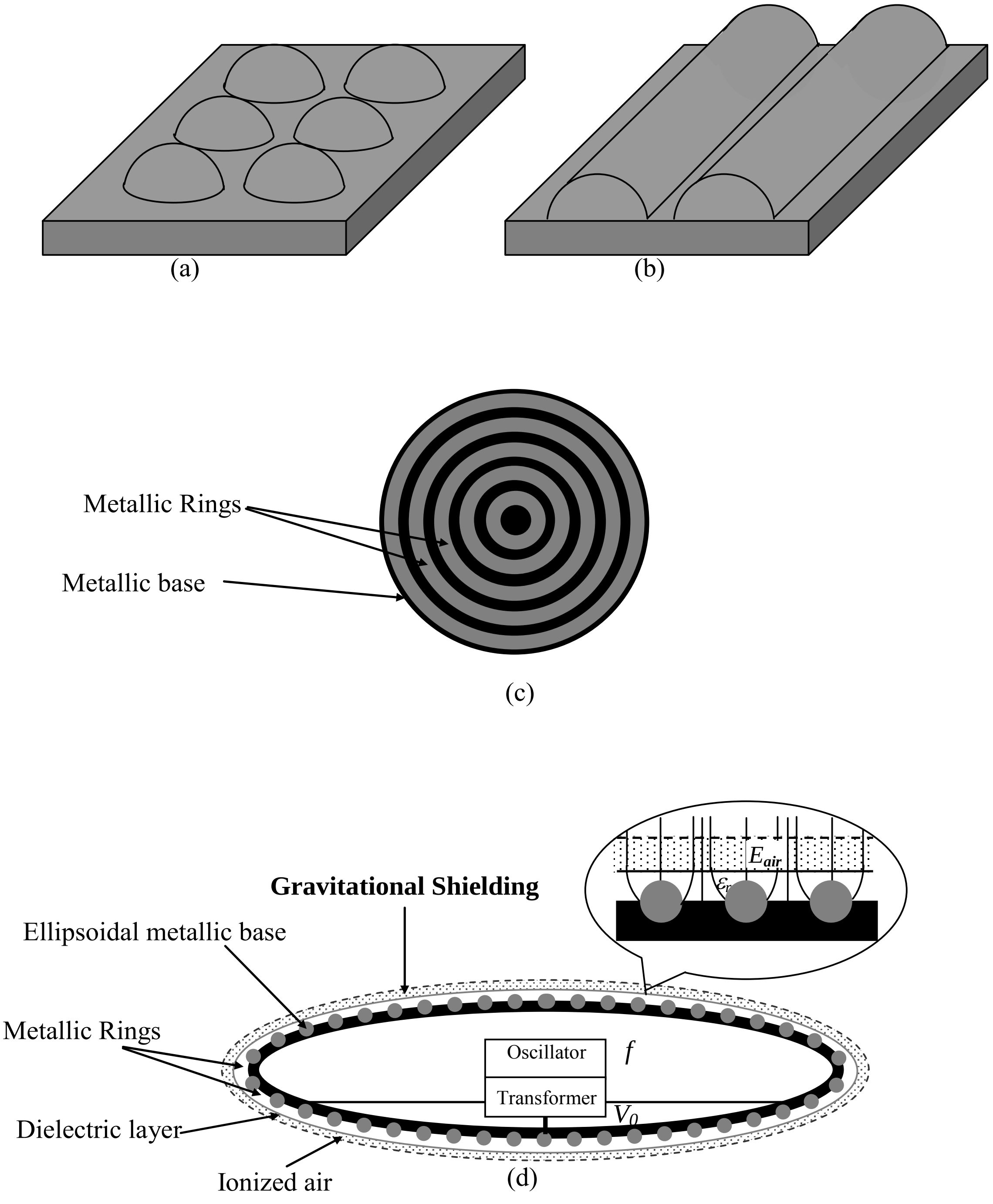

Fig. A11 – *Geometrical forms with similar effects as those produced by the semi-spherical form* – (a) shows the semi-spherical form stamped on the metallic surface; (b) shows the *semi-cylindrical* form (an obvious evolution from the semi-spherical form); (c) shows *concentric metallic rings* stamped on the metallic surface, an evolution from semi-cylindrical form. These geometrical forms produce the same effect as that of the semi-spherical form, shown in Fig.A11 (a). By using concentric metallic rings, it is possible to build *Gravitational Shieldings* around bodies or spacecrafts with several formats (spheres, ellipsoids, etc); (d) shows a Gravitational Shielding around a Spacecraft with *ellipsoidal form*.

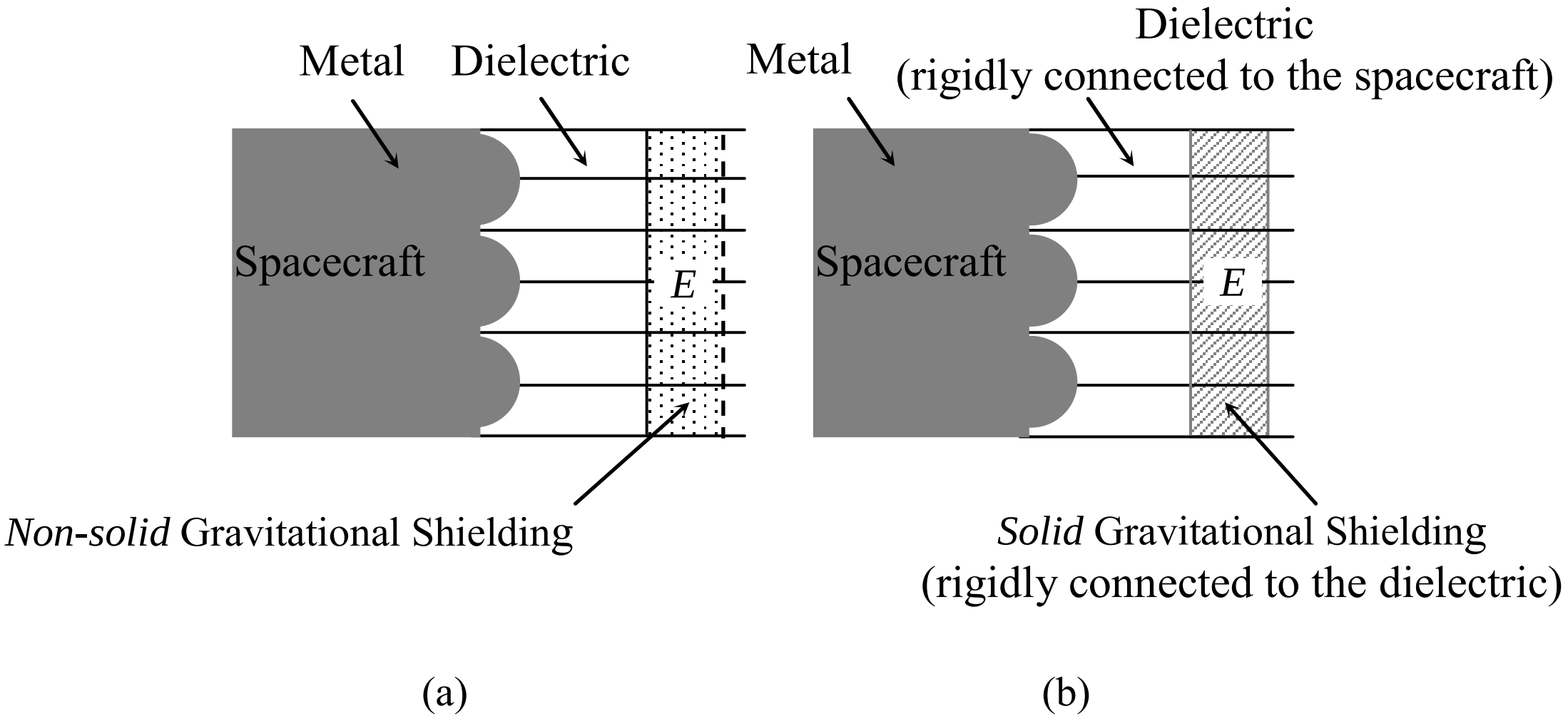

Fig. A12 – *Non-solid and Solid Gravitational Shieldings* - In the case of the Gravitational Shielding produced on a *solid substance* (b), when its molecules go to the *imaginary* space-time, *the electric field that produces the effect also goes to the imaginary space-time together with them*, because in this case, the substance of the Gravitational Shielding is *rigidly connected (by means of the dielectric) to the metal* that produces the electric field. This does not occur in the case of *Air* Gravitational Shielding.

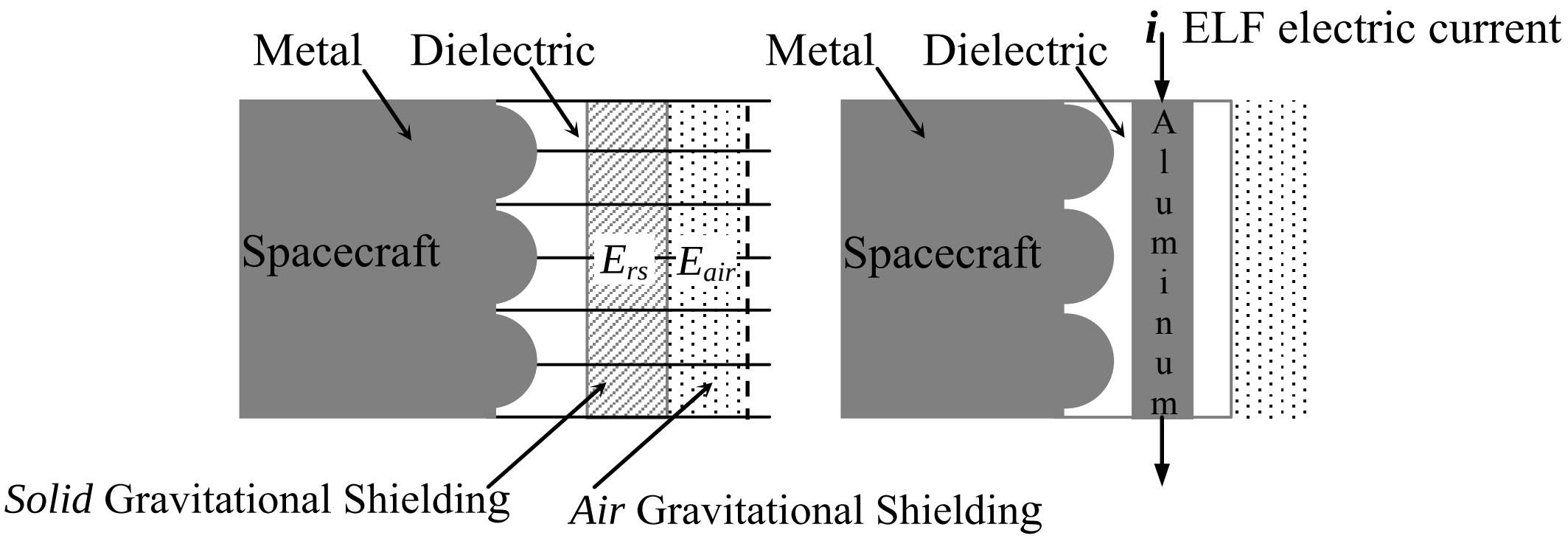

(a)

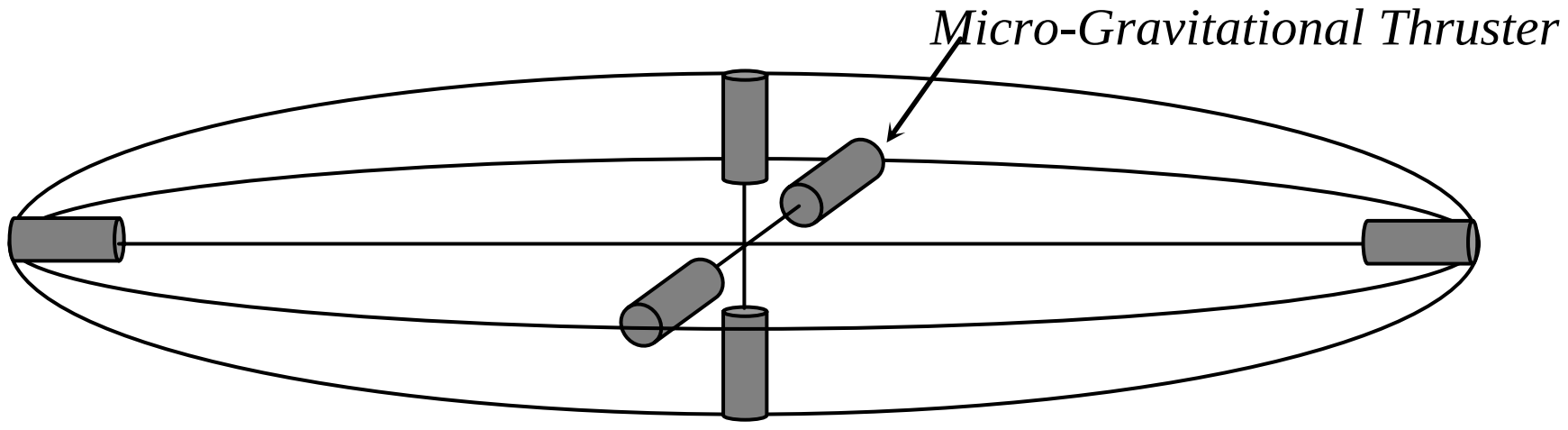

(b)

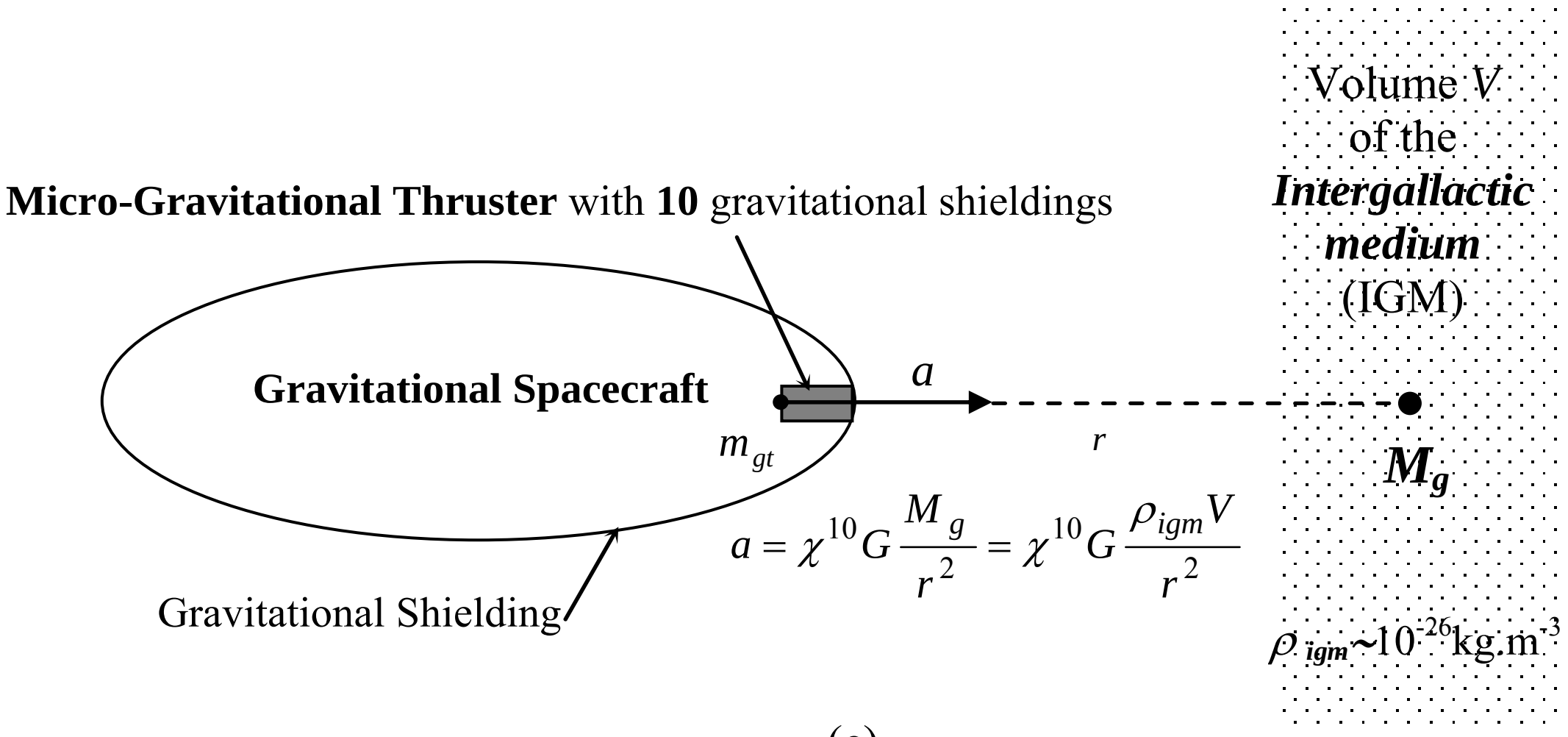

(c)

Fig. A13 – *Double Gravitational Shielding and Micro-thrusters* – (a) Shows a double gravitational shielding that makes possible to decrease the *inertial effects* upon the spacecraft when it is traveling both in the *imaginary* space-time and in the *real* space-time. The *solid* Gravitational Shielding also can be obtained by means of *an ELF electric current through a metallic lamina* placed *between the semi-spheres and the Gravitational Shielding of Air* as shown above. (b) Shows 6 *micro-thrusters* placed inside a Gravitational Spacecraft, in order to propel the spacecraft in the directions x, y and z. Note that the Gravitational Thrusters in the spacecraft must have a very small diameter (of the order of *millimeters*) because the hole through the Gravitational Shielding of the spacecraft cannot be large. Thus, these thrusters are in fact *Micro-thrusters*. (c) Shows a micro-thruster inside a spacecraft, and in front of a volume *V* of the intergalactic medium (IGM). Under these conditions, the spacecraft acquires an acceleration *a* in the direction of the volume *V*.

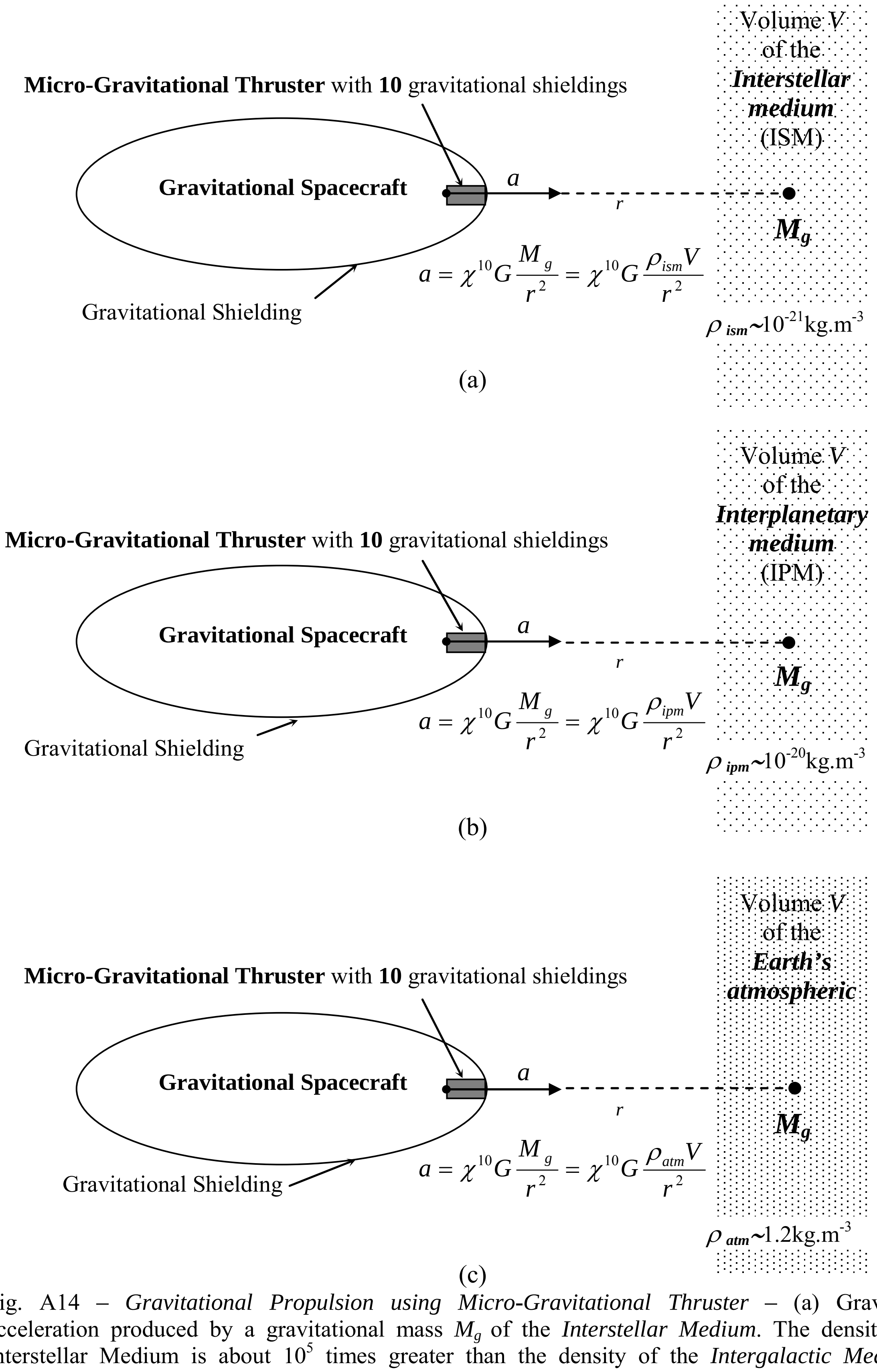

Fig. A14 – *Gravitational Propulsion using Micro-Gravitational Thruster* – (a) Gravitational acceleration produced by a gravitational mass $M_g$ of the *Interstellar Medium*. The density of the Interstellar Medium is about $10^5$ times greater than the density of the *Intergalactic Medium* (b) Gravitational acceleration produced in the *Interplanetary Medium*. (c) Gravitational acceleration produced in the *Earth's atmosphere*. Note that, in this case, $\rho_{atm}$ (*near to the Earth's surface*)is about $10^{26}$ times greater than the density of the *Intergalactic Medium.*

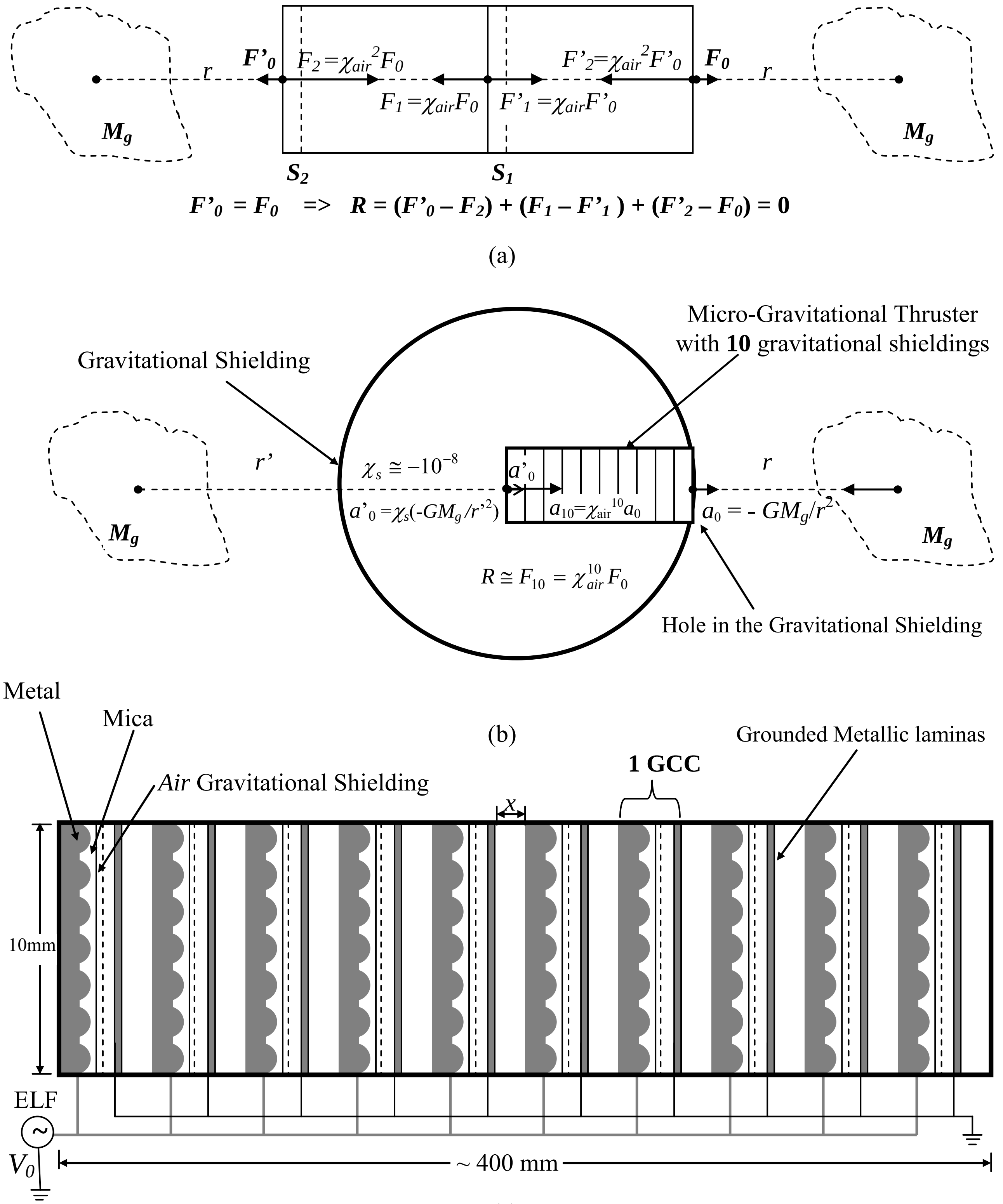

(c)

Fig. A15 – *Dynamics and Structure of the Micro-Gravitational Thrusters* - (a) The Micro-Gravitational Thrusters do not work *outside* the Gravitational Shielding, because, in this case, *the resultant upon the thruster is null* due to the symmetry. (b) The Gravitational Shielding $(\chi_s \cong 10^{-8})$ reduces strongly the intensities of the gravitational forces acting on the micro-gravitational thruster, except obviously, through the hole in the gravitational shielding. (c) Micro-Gravitational Thruster with *10 Air Gravitational Shieldings* (10GCCs). The grounded metallic laminas are placed so as to retain the electric field produced by metallic surface behind the semi-spheres.

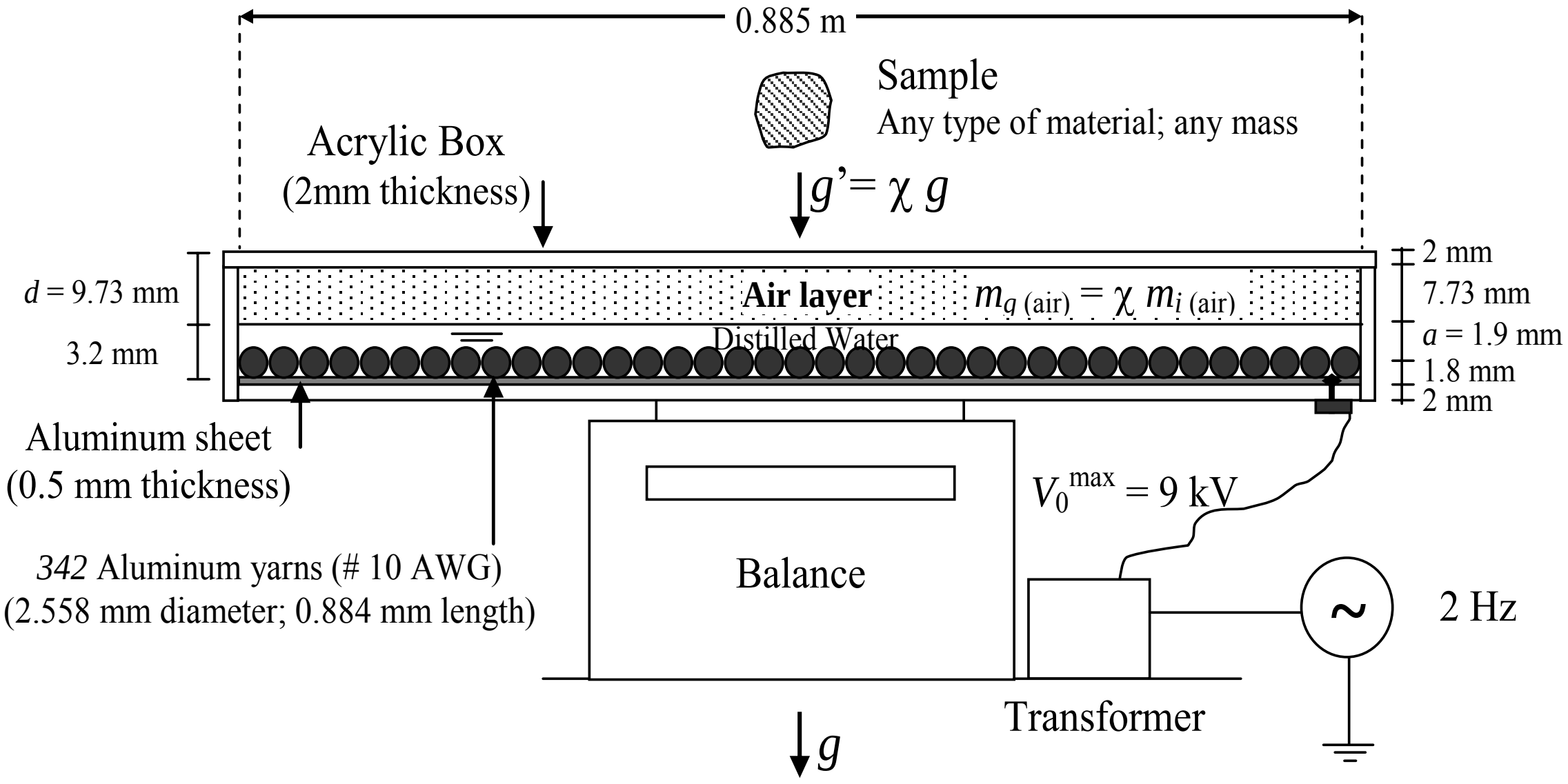

**GCC Cross-section Front view**

(a)

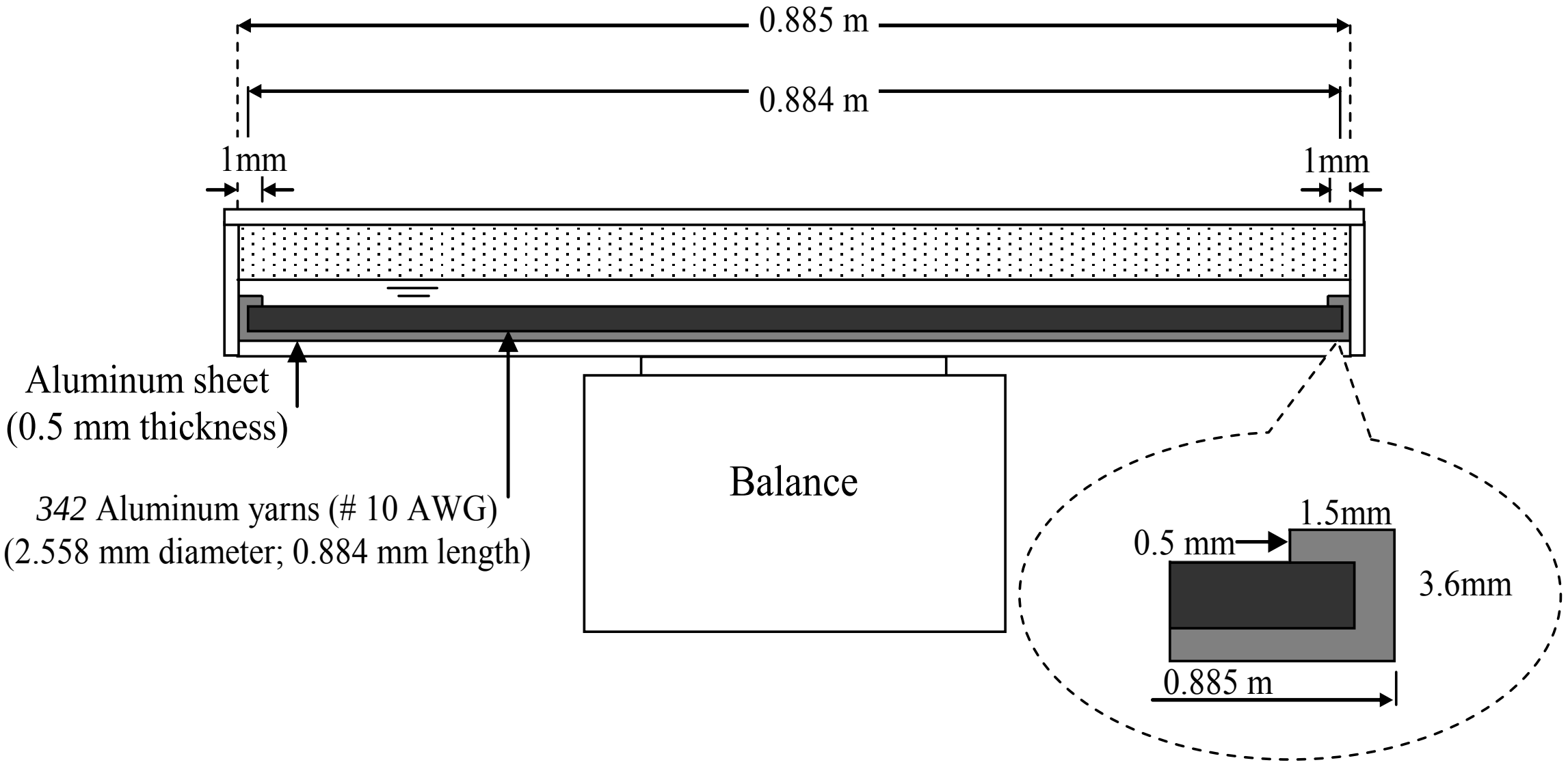

**GCC Cross-section Side View**

(b)

Fig. A16 – *A GCC using distilled Water.*
In total this GCC weighs about 6kg; the air layer 7.3 grams. The balance has the following characteristics: Range 0 – 6kg; readability 0.1g. The yarns are inserted side by side on the Aluminum sheet. Note the detail of fixing of the yarns on the Aluminum sheet.

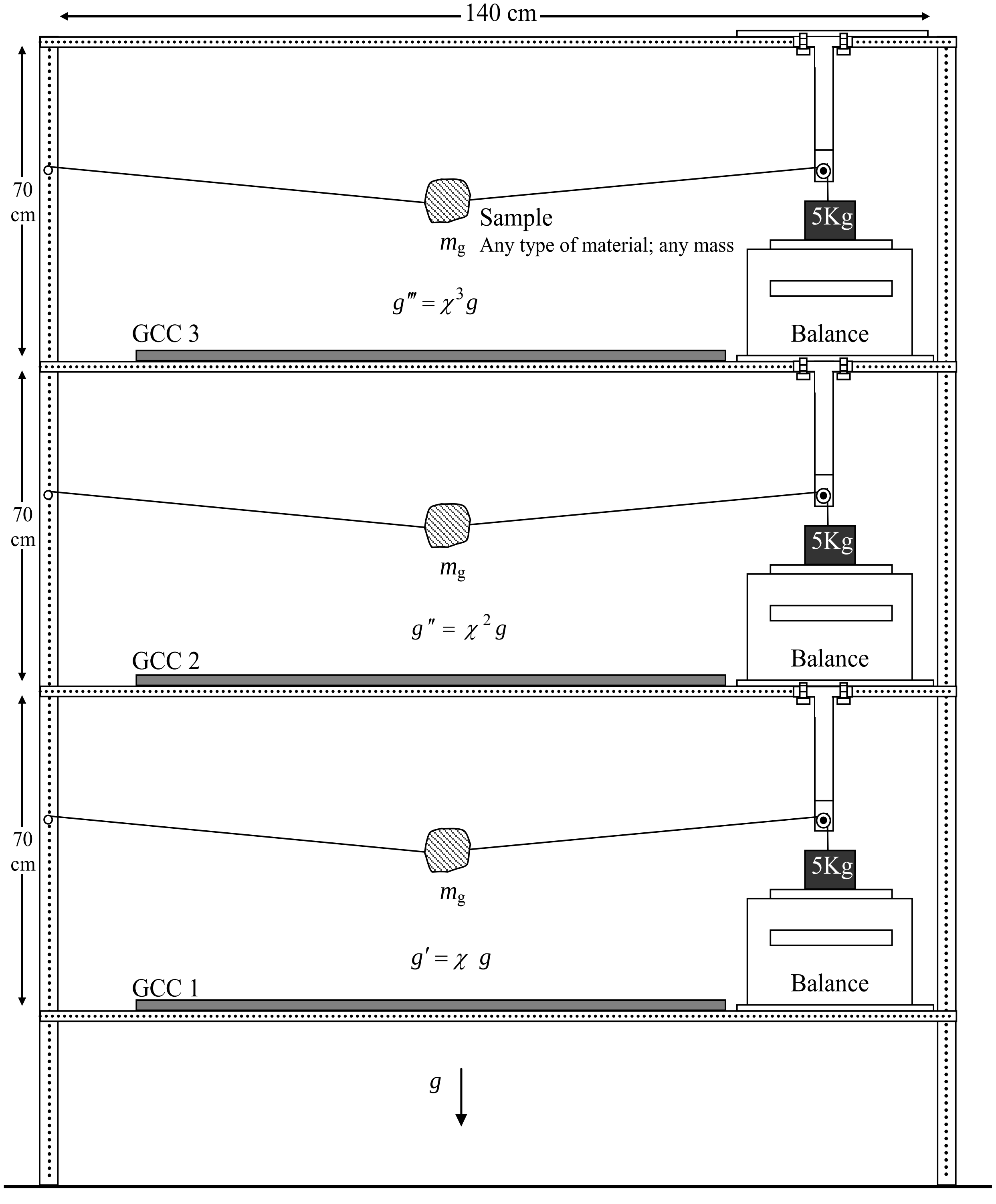

Fig. A17 – *Experimental set-up.* In order to prove *the exponential effect* produced by the superposition of the Gravitational Shieldings, we can take three similar GCCs and put them one above the other, in such way that above the GCC 1 the gravity acceleration will be $g' = \chi g$; above the GCC2 $g'' = \chi^2 g$, and above the GCC3 $g''' = \chi^3 g$. Where $\chi$ is given by Eq. (A47). The arrangement above has been designed for values of $m_g < 13g$ and $\chi$ up to -9 or $m_g < 1kg$ and $\chi$ up to -2 .

# APPENDIX B: Gravity Control Cells (GCCs) made from *Semiconductor Compounds*.

There are some semiconductors compounds with electrical conductivity between $10^4$S/m to 1 S/m, which can have their gravitational mass strongly decreased when subjected to ELF electromagnetic fields.

For instance, the polyvinyl chloride (PVC) compound, called Duracap™ 86103.

It has the following characteristics:

$$\mu_r = 1;\ \varepsilon_r = 3$$
$$\sigma = 3333.3 S/m$$
$$\rho = 1400 kg.m^{-3}$$
$$dieletric\ \ strength = 98 KV\ /\ mm$$

Then, according to the following equation below (derived from Eq.A14)

$$m_g = \left\{1 - 2\left[\sqrt{1 + 1.758 \times 10^{-27}\left(\frac{\mu_r \sigma^3}{\rho^2 f^3}\right) E_{rms}^4} - 1\right]\right\} m_{i0} \qquad (B1)$$

the *gravitational mass*, $m_g$, of the Duracap™ 86103, when subjected to an electromagnetic field of frequency $f$, is given by

$$m_g = \left\{1 - 2\left[\sqrt{1 + 3.3 \times 10^{-23}\frac{E_{rms}^4}{f^3}} - 1\right]\right\} m_{i0} \qquad (B2)$$

Note that, if the electromagnetic field through the Duracap has *extremely-low frequency*, for example, if $f = 2Hz$, and

$$E_{rms} = 9.4 \times 10^5 V/m \qquad (0.94 kV/mm)$$

Then, its *gravitational mass* will be reduced down to $m_g \cong -1.1 m_{i0}$, reducing in this way, the initial *weight* $(P_0 = m_g g = m_{i0} g)$ of the Duracap down to $-1.1 P_0$.

BACKGROUND FOR EXPERIMENTAL

The Duracap™ 86103 is sold under the form of small cubes. Its melting temperature varies from 177ºC to 188ºC. Thus, a 15cm square Duracap plate with 1 mm thickness can be shaped by using a suitable mold, as the shown in Fig.B1.

Figure B2(a) shows the Duracap plate between the Aluminum plates of a parallel plate capacitor. The plates have the following dimensions: 19cm x 15cm x 1mm. They are painted with an insulating varnish spray of high dielectric strength (ISOFILM). They are connected to the secondary of a transformer, which is connected to a Function Generator. The distance between the Aluminum plates is $d = 1mm$. Thus, the electric field through the Duracap is given by

$$E_{rms} = \frac{E_m}{\sqrt{2}} = \frac{V_0}{\varepsilon_r d \sqrt{2}} \qquad (B3)$$

where $\varepsilon_r$ is the relative permittivity of the dielectric (Duracap), and $V_0$ is the amplitude of the wave voltage applied on the capacitor.

In order to generate *ELF wave voltage* of $f = 2Hz$, we can use the widely-known Function Generator HP3325A (Op.002 High Voltage Output) that can generate sinusoidal voltages with *extremely-low* frequencies and amplitude up to *20V* ($40V_{pp}$ into $500\Omega$ load). The maximum output current is $0.08 A_{pp}$; output impedance <2Ω at ELF.

The turns ratio of the transformer (Bosch red coil) is $200:1$. Thus, since the maximum value of the amplitude of the voltage produced by the Function Generator is $V_p^{max} = 20\ V$, then the maximum secondary voltage will be $V_s^{max} = V_0^{max} = 4kV$. Consequently, Eq. (B3) gives

$$E_{rms}^{max} = 2.8 \times 10^6 V/m$$

Thus, for $f = 2Hz$, Eq. (B2) gives

$$m_g = -29.5 m_{i0}$$

The *variations on the gravitational mass* of the Duracap plate can be measured by a pan balance with the following characteristics: range 0 – 1.5kg ; readability 0.01g, using the set-up shown in Fig. B2(a).

Figure B2(b) shows the set-up to measure *the gravity acceleration variations above* the Duracap plate (Gravitational Shielding effect). The samples used in this case, can be of several types of material.

Since voltage waves with frequencies very below 1Hz have a very long period, we cannot consider, in practice, their *rms* values. However, we can add a sinusoidal voltage $V_{osc} = V_0 \sin \omega t$ with a DC voltage $V_{DC}$, by means of the circuit shown in Fig.B3. Thus, we obtain $V = V_{DC} + V_0 \sin \omega t$ ; $\omega = 2\pi f$. If $V_0 << V_{DC}$ then $V \cong V_{DC}$. Thus, the voltage $V$ varies with the frequency $f$, but its intensity is approximately equal to $V_{DC}$, i.e., $V$ will be practically constant. This is of fundamental importance for maintaining the value of the gravitational mass of the body, $m_g$, sufficiently stable during all the time, in the case of $f << Hz$.

We have shown in this paper that it is possible to control the gravitational mass of a spacecraft, simply by controlling the gravitational mass of a body *inside* the spacecraft (Eq.(10)). This body can be, for example, the *dielectric* between the plates of a capacitor, whose gravitational mass can be easily controlled by means of an ELF electromagnetic field produced between the plates of the capacitor. We will call this type of capacitor of *Capacitor of Gravitational Mass Control* (CGMC).

Figure B 4(a) shows a CGMC placed in the center of the spacecraft. Thus, the gravitational mass of the spacecraft can be controlled simply by varying the gravitational mass of the dielectric of the capacitor by means of an ELF electromagnetic field produced between the plates of the capacitor. Note that the Capacitor of Gravitational Mass Control can have the *spacecraft's own form* as shown in Fig. B 4(b). The dielectric can be, for example, a Duracap plate, as shown in this appendix. In this case, the gravitational mass of the dielectric is expressed by Eq. (B2). Under these circumstances, the *total* gravitational mass of the spacecraft will be given by Eq.(10):

$$M_{g(spacecraf)} = M_{i0} + \chi_{dielectric} m_{i0}$$

where $M_{i0}$ is the rest inertial mass of the spacecraft(without the dielectric) and $m_{i0}$ is the rest inertial mass of the dielectric; $\chi_{dielectric} = m_g / m_{i0}$, where $m_g$ is the gravitational mass of the dielectric. By decreasing the value of $\chi_{dielectric}$, the gravitational mass of the spacecraft decreases. It was shown, that the value of $\chi$ can be negative. Thus, for example, when $\chi_{dielectric} \cong -M_{i0}/m_{i0}$, the *gravitational mass of the spacecraft gets very close to zero.* When $\chi_{dielectric} < -M_{i0}/m_{i0}$, the gravitational mass of the spacecraft becomes negative.

Therefore, *for an observer out of the spacecraft* the gravitational mass of the spacecraft is $M_{g(spacecraft)} = M_{i0} + \chi_{dielectric} m_{i0}$, and not $M_{i0} + m_{i0}$.

Since the dielectric strength of the Duracap is $98kV/mm$, a Duracap plate with *1mm* thickness can withstand up to $98kV$. In this case, the value of $\chi_{dielectric}$ for $f = 2Hz$, according to Eq. (B2), is

$$\chi_{dielectric} = m_g / m_{i0} \cong -10^4$$

Thus, for example, if the inertial mass of the spacecraft is $M_{i0} \cong 10021.0014kg$ and, the inertial mass of the *dielectric* of the *Capacitor of Gravitational Mass Control* is $m_{i0} \cong 1.0021kg$, then the gravitational mass of the spacecraft becomes

$$M_{g(spacecraf)} = M_{i0} + \chi_{dielectric} m_{i0} \cong 10^{-3} kg$$

This value is much smaller than $+0.159 M_{i0}$.

It was shown [1] that, when the gravitational mass of a particle is reduced to values between $+0.159 M_i$ and $-0.159 M_i$, it becomes *imaginary*, i.e., the gravitational and the inertial masses of the particle become *imaginary*. Consequently, the particle *disappears* from our ordinary space-time.

This means that we cannot reduce the gravitational mass of *the spacecraft* below $+0.159 M_i$, *unless we want to make it imaginary*.

Obviously this limits the minimum value of $\chi_{dielectric}$, i.e. $\chi_{dielctric}^{\min} = 0.159$. Consequently, if the gravity acceleration *out of the spacecraft* (in a given direction) is $g$, then, according to the Gravitational Shielding Principle, the corresponding gravity acceleration upon the crew of the spacecraft can be reduced just down to $0.159g$. In addition, since the Mach's principle says that *the local inertial forces are produced by the gravitational interaction of the local system with the distribution of cosmic masses* then *the inertial effects* upon the crew would be reduced just by $\chi_{dielectric} = 0.159$.

However, there is a way to strongly reduce the inertial effects upon the crew of the spacecraft without making it *imaginary*. As shown in Fig. B4 (c), we can build an *inertial shielding*, with $n$ superimposed CGMCs. In this case, according to the Gravitational Shielding Principle, the gravity upon the crew will be given by $g_n = \chi_{dielectric}^n g$, where $g$ is the gravity acceleration out of the spacecraft (in a given direction) and $\chi_{dielectric} = m_g / m_{i0}$ ; $m_g$ and $m_{i0}$ are, respectively, the gravitational mass and the inertial mass of the dielectric. Under these conditions the *inertial effects upon the crew* will be reduced by $\chi_{dielectric}^n$.

Thus, for $n = 10$ (*ten* superimposed CGMCs), and $\chi_{dielectric} \cong 0.2$, the *inertial effects upon the crew* will be reduced by $\chi_{dielectric}^{10} \cong 1 \times 10^{-7}$. Therefore, if the maximum thrust produced by the thrusters of the spacecraft is $F = 10^5 N$, then the intensities of the inertial forces upon the crew will not exceed $0.01N$, i.e. they will be practically negligible.

Under these circumstances, the gravitational mass of the spacecraft, for an observer out of the spacecraft, will be just approximately equal to the gravitational mass of the *inertial shielding*, i.e. $M_{g(spacecraft)} \cong M_{g(inertial.shield)}$.

If $M_{g(inertial.shield)} \cong 10^3 kg$, and the thrusters of the spacecraft are able to produces up to $F = 3\times10^5 N$, the spacecraft will acquires an acceleration given by

$$a_{spacecraft} = \frac{F}{M_{g(spacecraft)}} \cong 3\times10^2 m.s^{-2}$$

With this acceleration it can reach velocities close to *Mach 10* in some seconds.

The velocity that the spacecraft can reach in the *imaginary* spacetime is much greater than this value, since $M_{g(spacecraft)}$, as we have seen, can be reduced down to $\cong 10^{-3} kg$ or less.

Thus, if the thrusters of the spacecraft are able to produces up to $F = 3\times10^5 N$, and $M_{g(spacecraft)} \cong 10^{-3} kg$, the spacecraft will acquires an acceleration given by

$$a_{spacecraft} = \frac{F}{M_{g(spacecraft)}} \cong 3\times10^8 m.s^{-2}$$

With this acceleration it can reach velocities close to the light speed in less than 1 second. After 1 month, the velocity of the spacecraft would be about $10^{15} m/s$ (remember that in the *imaginary* spacetime the maximum velocity of propagation of the interactions is *infinity* [1]).

OTHER SEMICONDUCTOR COMPOUNDS

A semiconductor compound, which can have its gravitational mass strongly decreased when subjected to ELF electromagnetic fields is the CoorsTek Pure SiC™ LR CVD *Silicon Carbide*, 99.9995% ‡‡‡‡. This Low-resistivity (LR) pure Silicon Carbide has electrical conductivity of *5000S/m* at room temperature; $\varepsilon_r = 10.8$; $\rho = 3210 kg.m^{-3}$; dielectric strength >10 KV/mm; maximum working temperature of 1600°C.

Another material is the *Alumina-CNT*, recently discovered §§§§. It has electrical conductivity of *3375 S/m* at 77°C in samples that were 15% nanotubes by volume [17]; $\varepsilon_r = 9.8$; $\rho = 3980 kg.m^{-3}$; dielectric strength 10-20KV/mm; maximum working temperature of 1750°C.

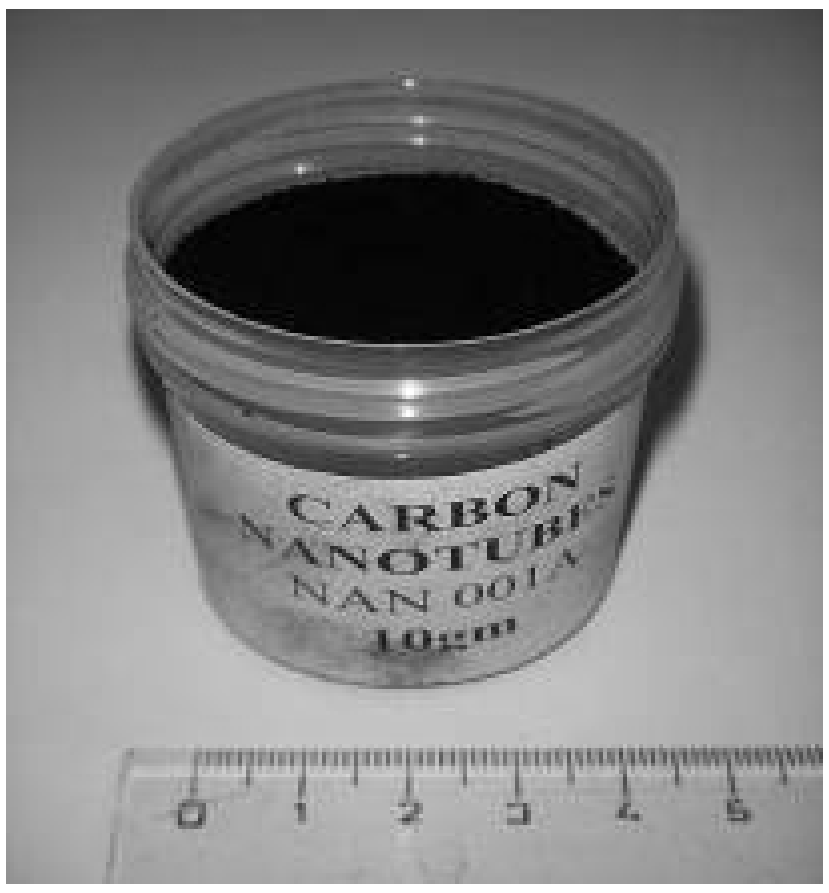

The novel *Carbon Nanotubes Aerogels* *****, called *CNT Aerogels* are also suitable to produce Gravitational Shieldings, mainly due to their very small densities. The electrical conductivity of the *CNT Aerogels* is *70.4S/m* for a density of $\rho = 7.5 kg.m^{-3}$ [18]; $\varepsilon_r \approx 10$. Recently (2010), it was announced the discovered of *Graphene Aerogel* with $\sigma = \sim 1\times10^2 S/m$ and $\rho = 10 kg.m^{-3}$ [19] (Aerogels exhibit higher dielectric strength than expected for porous materials).

‡‡‡‡ www.coorstek.com

§§§§ Recently, it was discovered that Carbon nanotubes (CNTs) can be added to *Alumina* ($Al_2O_3$) to convert it into a good electrical conductor.

***** In 2007, Mateusz Brying *et al.* working with Prof. Arjun Yodh at the University of Pennsylvania produced the first aerogels made entirely of carbon nanotubes (CNT Aerogels) [20] that, depending on the processing conditions, can have their electrical conductivity ranging as high as 100 S/m.

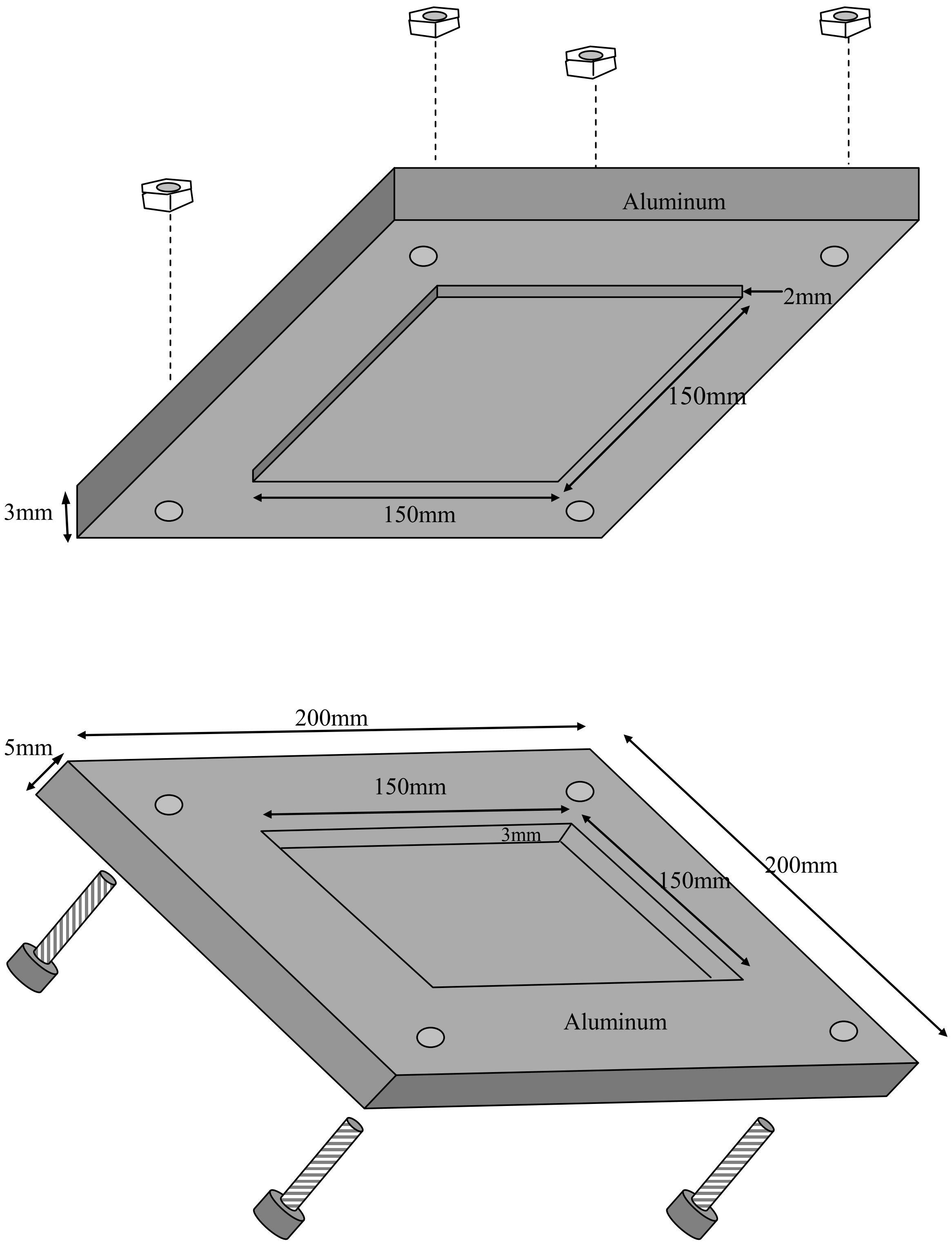

Fig.B1 – Mold design

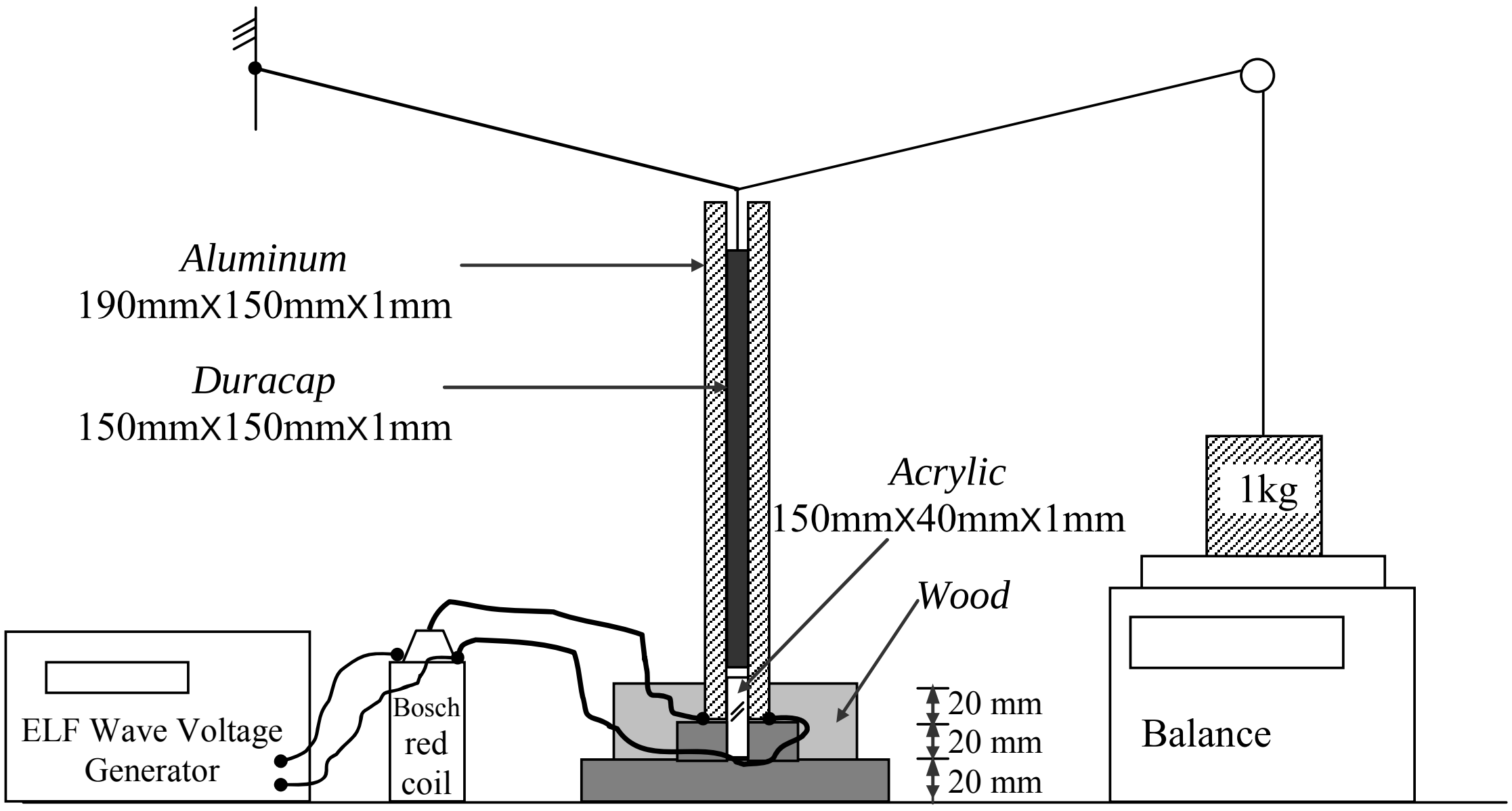

(a)

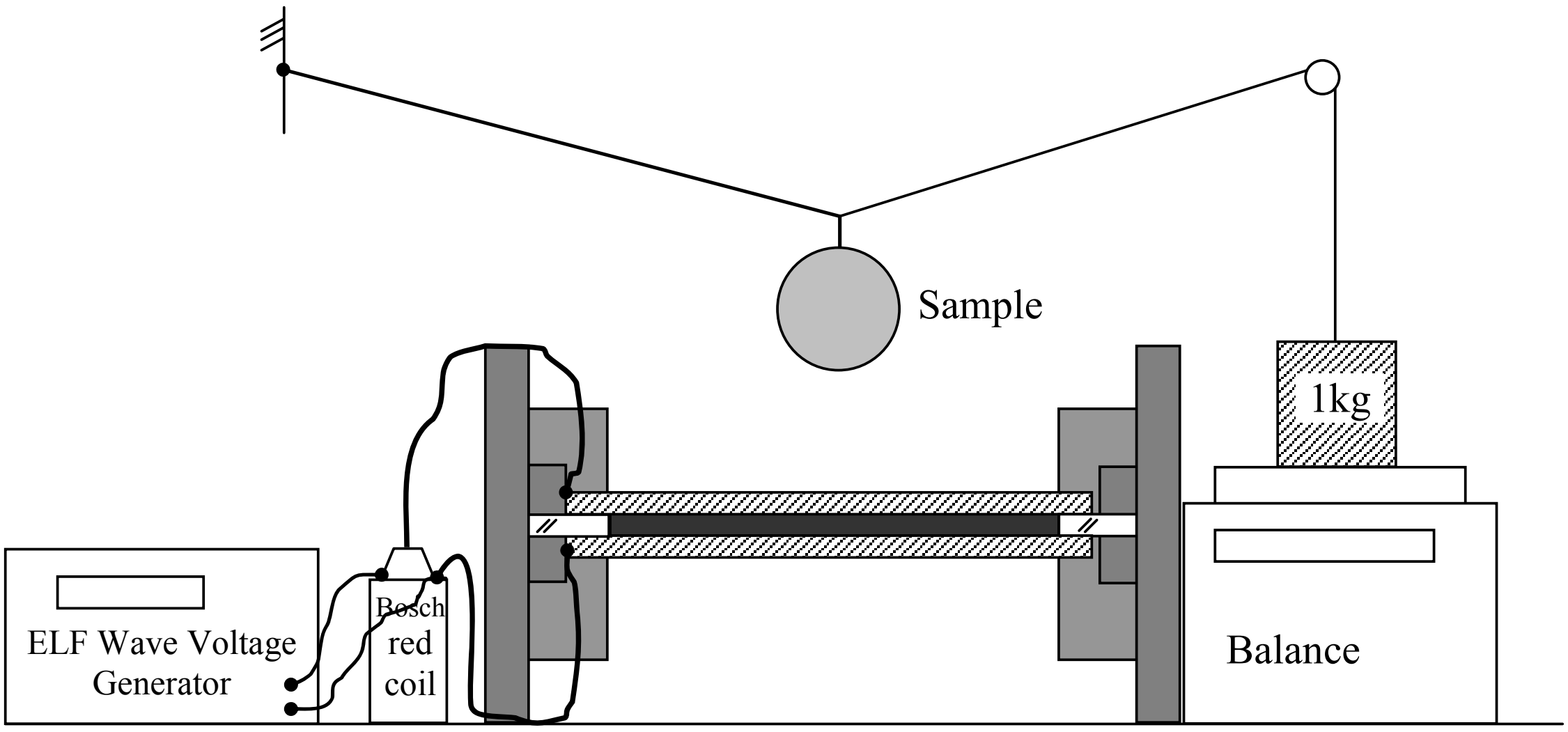

(b)

Fig.B2 – Schematic diagram of the experimental set-up

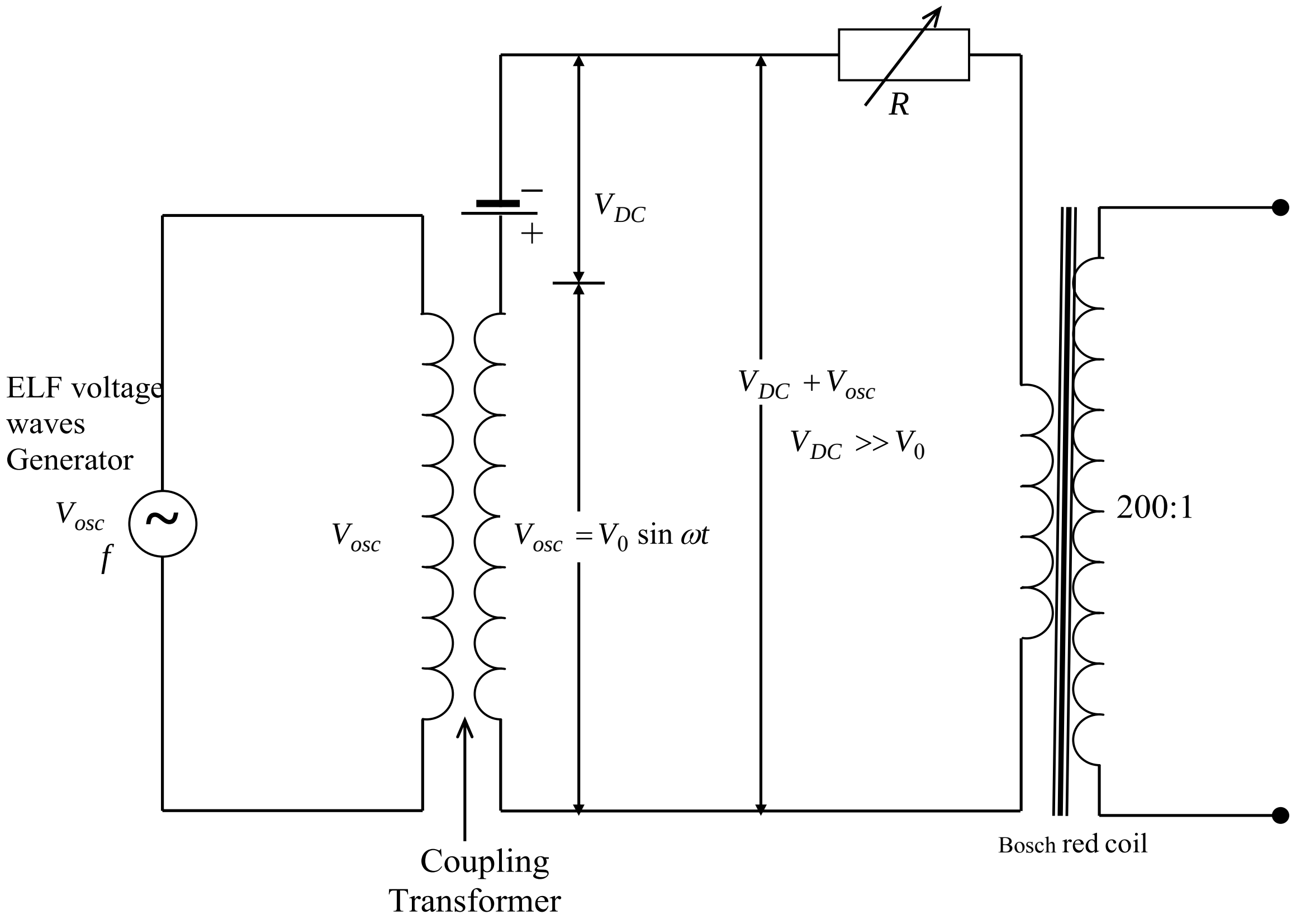

Fig. B3 – Equivalent Electric Circuit

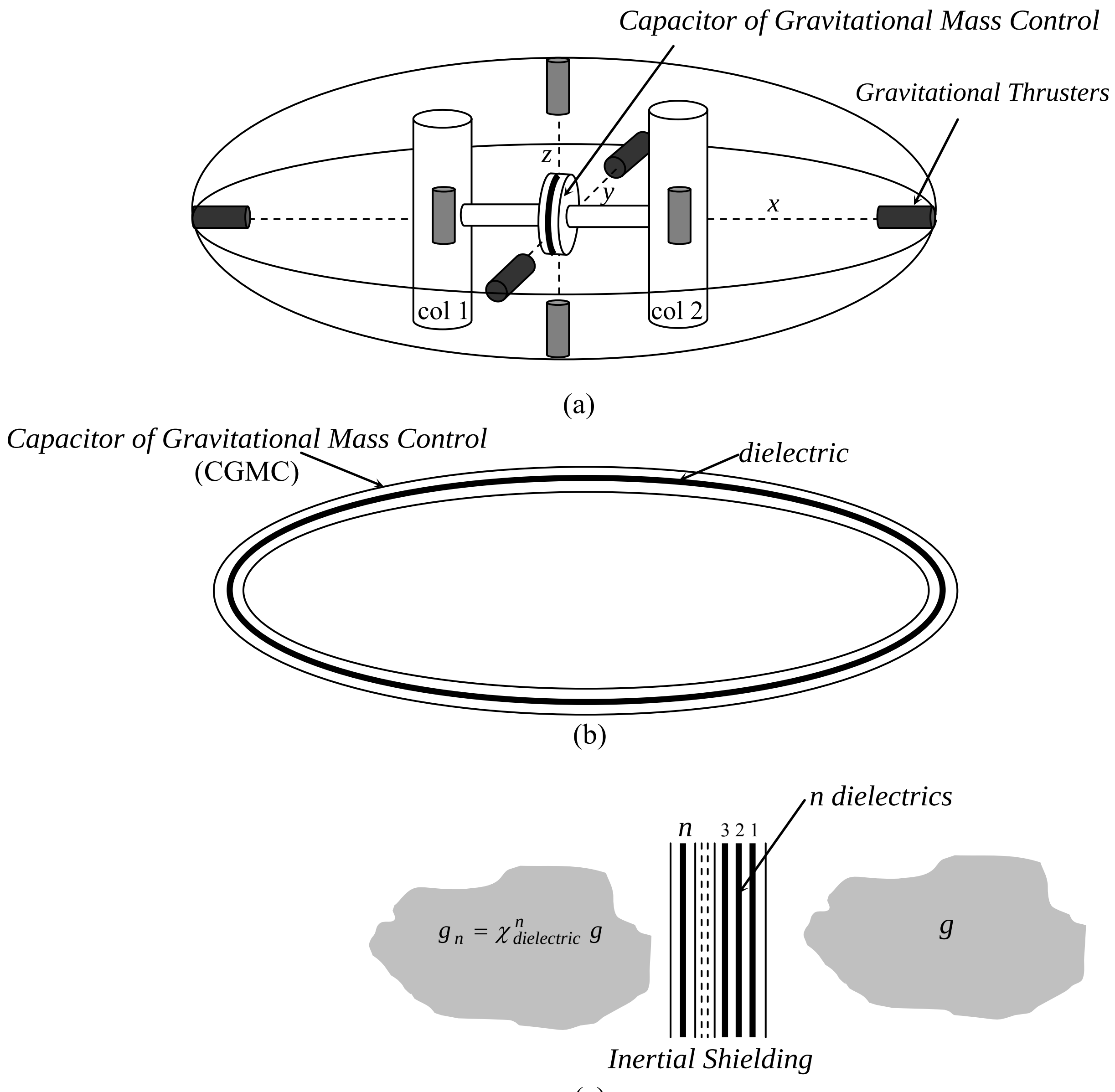

Fig.B4 – Gravitational Propulsion System and *Inertial Shielding* of the Gravitational Spacecraft – (a) eight *gravitational thrusters are* placed inside a Gravitational Spacecraft, in order to propel the spacecraft along the directions x, y and z. Two *gravitational thrusters* are inside the columns 1 and 2, in order to rotate the spacecraft around the y-axis. The functioning of the Gravitational Thrusters is shown in Fig.A14. *The gravitational mass of the spacecraft is controlled by the Capacitor of Gravitational Mass Control* (CGMC). Note that the CGMC can have the *spacecraft's own form*, as shown in (b). In order to strongly reduce the inertial effects upon the crew of the spacecraft, we can build an inertial shielding, with several CGMCs, as shown above (c). In this case, the gravity upon the crew will be given by $g_n = \chi^n_{dielectric}\, g$, where $g$ is the gravity acceleration out of the spacecraft (in a given direction) and $\chi_{dielectric} = m_g / m_{i0}$ ; $m_g$ and $m_{i0}$ are, respectively, the gravitational mass and the inertial mass of the dielectric. Under these conditions the *inertial effects upon the crew* will be reduced by $\chi^n_{dielectric}$. Thus, for example, if $n = 10$ and $\chi_{dielectric} \cong 0.2$, the inertial effects will be reduced by $\chi^{10}_{dielectric} \cong 1 \times 10^{-7}$. If the maximum thrust produced by the thrusters is $F = 10^5\,N$, then the intensities of the inertial forces upon the crew will not exceed $0.01N$.

# APPENDIX C: Longer-Duration Microgravity Environment Produced by *Gravity Control Cells* (GCCs).

The acceleration experienced by an object in a *microgravity* environment, by definition, is one-millionth ($10^{-6}$) of that experienced at Earth's surface (1g). Consequently, a *microgravity* environment is one where the acceleration induced by gravity has little or no measurable effect. The term *zero-gravity* is, obviously inappropriate since the *quantization of gravity* [1] shows that the gravity can have only discrete values *different of zero* [1, Appendix B].

Only three methods of creating a microgravity environment are currently known: to travel far enough into deep space so as to reduce the effect of gravity by *attenuation*, by *falling*, and by *orbiting* a planet.

The first method is the simplest in conception, but requires traveling an enormous distance, rendering it most impractical with the conventional spacecrafts. The second method, *falling*, is very common but approaches microgravity only when the fall is in a vacuum, as air resistance will provide some resistance to free fall acceleration. Also it is difficult to fall for long enough periods of time. There are also problems which involve avoiding too sudden of a stop at the end. The NASA Lewis Research Center has several drop facilities. One provides a 132 meter drop into a hole in the ground similar to a mine shaft. This drop creates a reduced gravity environment for *5.2 seconds*. The longest drop time currently available (about *10 seconds*) is at a 490 meter deep vertical mine shaft in Japan that has been converted to a drop facility.

Drop towers are used for experiments that only need a *short duration of microgravity*, or for an initial validation for experiments that will be carried out in longer duration of microgravity.

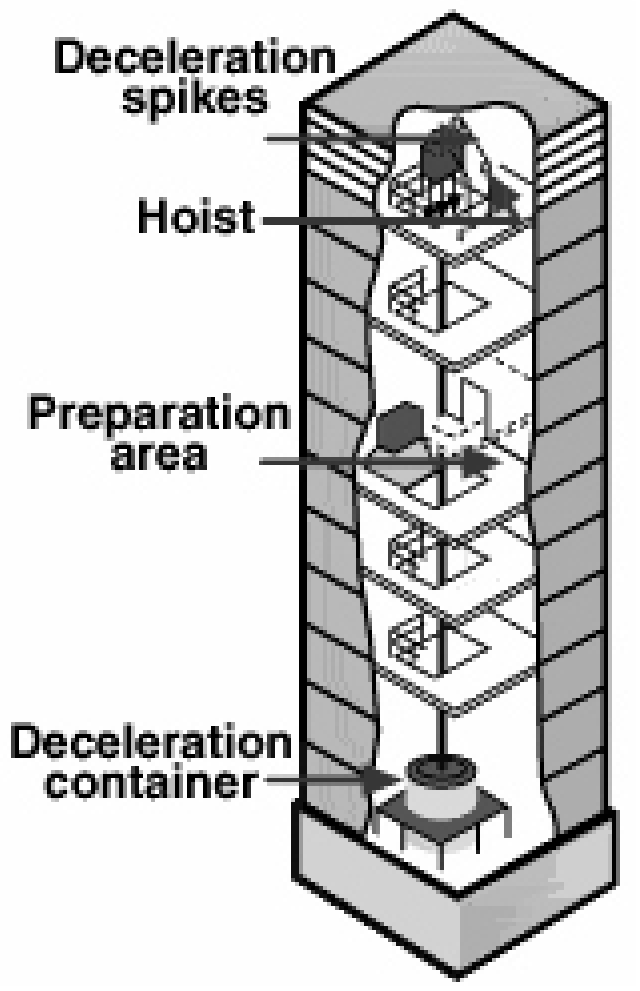

Aircraft can fly in parabolic arcs to achieve period of microgravity of 20 to 25 seconds with g-level of approximately 0.02 g. The airplane climbs rapidly until its nose is about 45-degree angle to the horizon then the engines are cut back. The airplane slows; the plane remains in free fall over the top of the parabola, then it nose-dives to complete the parabola, creating microgravity conditions.

Aircraft parabolic flights give the opportunity to perform medical experiments on human subjects in real microgravity environment. They also offer the possibility of direct intervention by investigators on board the aircraft during and between parabolas. In the mid-1980s, NASA KC-135, a modified Boeing 707, provided access to microgravity environment. A parabolic flight provided 15 to 20 seconds of 0.01 g or less, followed by a 2-g pull out. On a typical flight, up to 40 parabolic trajectories can be performed. The KC-135 can accommodate up to 21 passengers performing 12 different experiments. In 1993, the Falcon-20 performed its first parabolic flight with microgravity experiment on board. This jet can carry two experimenters and perform up to 3 experiments. Each flight can make up to 4 parabolic trajectories, with each parabola lasting 75 seconds, with 15 to 20 seconds of microgravity at 0.01g or less.

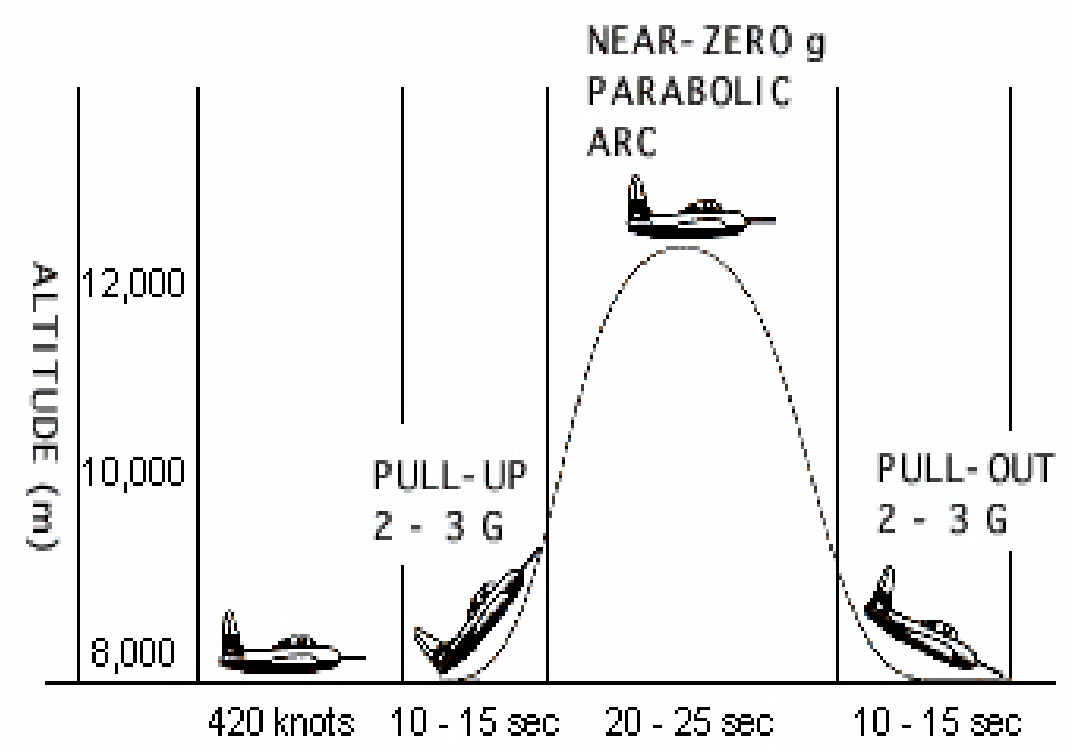

The third method of creating a microgravity environment is *orbiting* a planet. This is the environment commonly experienced in the space shuttle, International Space Station, Mir (no longer in orbit), etc. While this scenario is the most suitable for scientific experimentation and commercial exploitation, it is still quite expensive to operate in, mostly due to launch costs.

A space shuttle provides an ideal laboratory environment to conduct microgravity research. A large panoply of experiments can be carried out in microgravity conditions for up to *17 days*, and scientists can make adjustment to avoid experiment failure and potential loss of data. Unmanned capsules, platforms or satellites, such as the European retrievable carrier Eureka, DLR's retrievable carrier SPAS, or the Russian Photon capsules, the US Space Shuttle (in connection with the European Spacelab laboratory or the US Spacelab module), provide *weeks* or *months* of microgravity.

A space station, maintaining a low earth orbit for several decades, greatly improves access to microgravity environment for up to *several months.*

Thus, microgravity environment can be obtained via different means, providing different duration of microgravity. While *short-duration microgravity environments* can be achieved on Earth with relative easiness, *longer-duration microgravity environments* are too expensive to be obtained.

Here, we propose to use the Gravity Control Cells (GCCs), shown in this work, in order to create *longer-duration microgravity environments*. As we have seen, just above a GCC the gravity can be strongly reduced (*down to* $1\mu$g or less). In this way, the gravity above a GCC can remain at the microgravity ranging during a very long time (*several years*). Thus, GCCs can be used in order to create longer-duration microgravity environments on Earth. In addition, due to the cost of the GCCs to be relatively low, also the longer-duration microgravity environments will be produced with low costs.

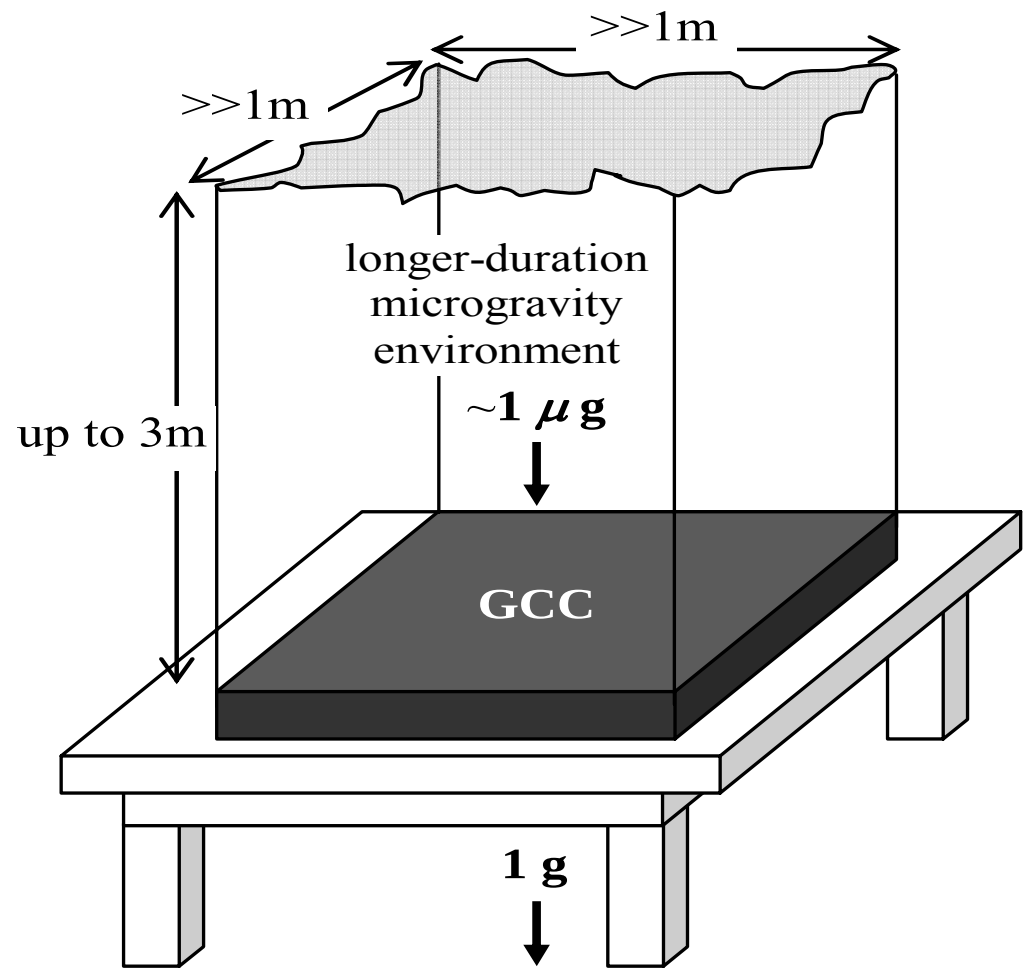

This possibility appears to be absolutely new and unprecedented in the literature since longer-duration microgravity environments are usually obtained via airplanes, sounding rockets, spacecraft and space station.

It is easy to see that the GCCs can be built with *width* and *length* of until some meters. On the other hand, as the effect of gravity reduction above the GCC can reach up to 3m, we can then conclude that the *longer-duration microgravity environments* produced above the GCCs can have sufficiently large volumes to perform any microgravity experiment on Earth.

The longer-duration microgravity environment produced by a GCC will be a special tool for microgravity research. It will allow to improve and to optimize physical, chemical and biological processes on Earth that are important in science, engineering and also medicine. The reduction of gravitational effects in a microgravity environment shows, for example, that temperature differences in a fluid do not produce convection, buoyancy or sedimentation. The changes in fluid behavior in microgravity lie at the heart of the studies in materials science, combustion and many aspects of space biology and life sciences. Microgravity research holds the promise to develop new materials which can not be made on Earth due to gravity. These new materials shall have properties that are superior to those made on Earth and may be used to:

-increase the speed of future computers,
-improve fiber optics,
-make feasible Room Temperature Superconductors,
-enable medical breakthroughs to cure several diseases (e.g., diabetes).

In a microgravity environment protein crystals can be grown larger and with a purity that is impossible to obtain under gravity of 1g. By analyzing the space-grown crystals it is possible to determine the structure and function of the thousands of proteins used in the human body and in valuable plants and animals. The determination of protein structure represents a huge opportunity for pharmaceutical companies to develop new drugs to fight diseases.

Crystal of HIV protease inhibitor grown in microgravity are significantly larger and of higher quality than any specimens grown under gravity of 1g. This will help in defining the structure of the protein crucial in fighting the AIDS virus.

Protein Crystal Isocitrate Lysase is an enzyme for fungicides. The isocitrate lysase crystals grown in microgravity environments are of larger sizes and fewer structural defects than crystals grown under gravity of 1g. They will lead to more powerful fungicides to treat serious crop diseases such as rice blast, and increase crop output.

Improved crystals of human insulin will help improve treatment for diabetes and *potentially create a cure*.

*Anchorage dependent cells* attached to a polymer and grown in a bioreactor in microgravity will lead to the production of a protein that is closer in structure and function to the three-dimensional protein living in the body.

Anchorage Dependant Cells
Attached to a Polymer

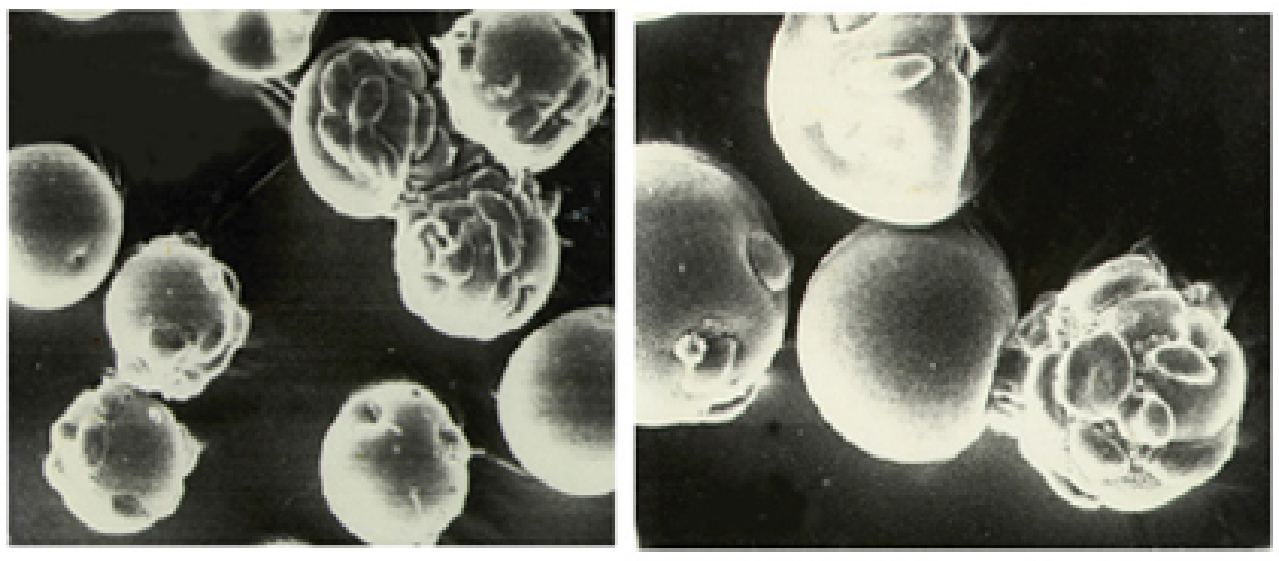

1g μg

This should help reduce or eliminate *transplant rejection* and is therefore critical for organ transplant and for the replacement of damaged bone and tissues. Cells grown on Earth are far from being three-dimensional due to the effect of 1g gravity.

The ZBLAN is a new substance with the potential to revolutionize fiber optics communications. A member of the heavy metal fluoride family of glasses, ZBLAN has promising applications in fiber optics. It can be used in a large array of industries, including manufacture of ultra high purity fiber optics, optical switches for computing, telecommunications, medical surgery and cauterization, temperature monitoring, infrared imaging, fiber-optic lasers, and optical power transmission. A ZBLAN fiber optic cable manufactured in a *microgravity environment* has the potential to carry 100 times the amount of data conveyed by conventional silica-based fibers.

ZBLAN Fiber

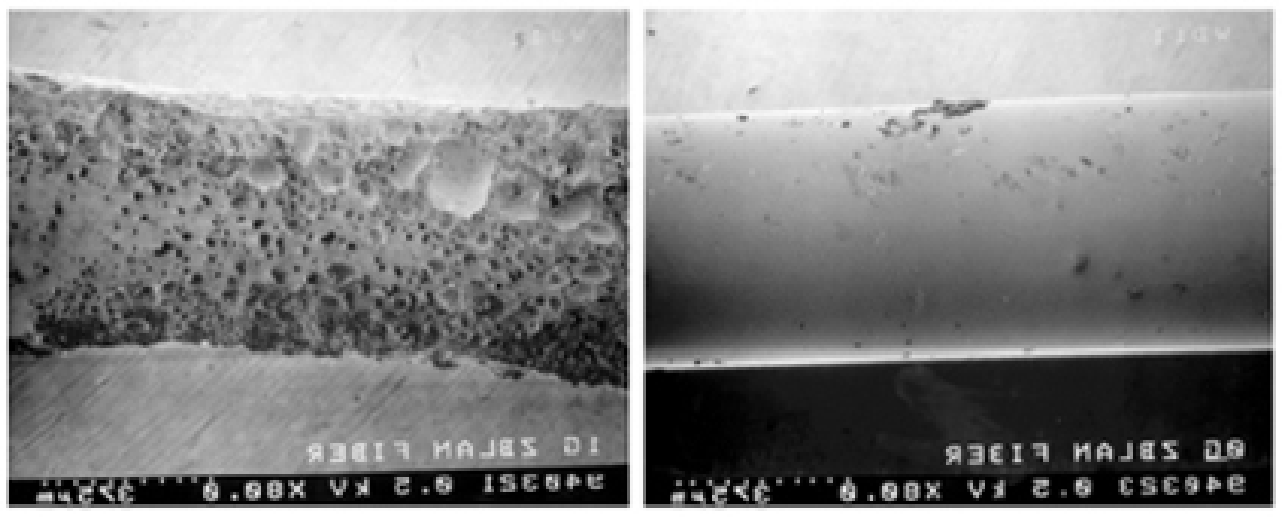

1g μg

In microgravity environment where complications of gravity-driven convection flows are eliminated, we can explore the fundamental processes in fluids of several types more easily and test fundamental theories of three-dimensional laminar, oscillatory and turbulent flow generated by various other forces.

By improving the basics for predicting and controlling the behavior of fluids, we open up possibilities for improving a whole range of industrial processes:

- Civil engineers can design safe buildings in earthquake-prone areas thanks to a better understanding of the fluid-like behavior of soils under stress.

- Materials engineers can benefit from a deeper knowledge of the determination of the structure and properties of a solid metal during its formation and can improve product quality and yield, and, in some cases, lead to the introduction of new products.

- Architects and engineers can design more stable and performing power

plants with the knowledge of the flow characteristics of vapor-liquid mixture.

- Combustion scientists can improve fire safety and fuel efficiency with the knowledge of fluid flow in microgravity.

In microgravity environment, medical researchers can observe the functional changes in cells when the effect of gravity is practically removed. It becomes possible to study fundamental life processes down to the cellular level.

Access to microgravity will provide better opportunities for research, offer repeated testing procedures, and enormously improve the test facilities available for life sciences investigations. This will provide valuable information for medical research and lead to improvements in the health and welfare of the six billion people, which live under the influence of 1g gravity on the Earth's surface.

The utilization of microgravity to develop new and innovative materials, pharmaceuticals and other products is waiting to be explored. Access to microgravity environments currently is limited. Better access, as the produced by GCCs, will help researchers accelerate the experimentation into these new products.

*Terrafoam* is a rigid, silicate based inorganic foam. It is *nonflammable* and does not five off noxious fumes when in the presence of fire. *It does not conduct heat to any measurable degree* and thus is an outstanding and possible unsurpassed thermal insulator. In addition, it appears to have unique *radiation shielding* capabilities, including an ability to block *alpha*, *beta*, gamma rays). Terrafoam can be constructed to be extremely lightweight. Altering the manufacturing process and the inclusion of other materials can vary the properties of Terrafoam. Properties such as cell structure, tensile strength, bulk density and temperature resistance can be varied to suit specific applications. It self-welds to concrete, aluminum and other metals. The useful variations on the base product are potentially in the thousands. Perhaps the most exciting potential applications for Terrafoam stem from its extraordinary capability as an ultra-lightweight thermal and radioactive shield.

Also, the formation of nanoscale *carbon structures* by electrical arc discharge plasma synthesis has already been investigated in microgravity experiments by NASA. Furthermore, complex plasmas are relevant for processes in which a particle formation is to be prevented, if possible, as, for example, within plasma etching processes for microchip production.

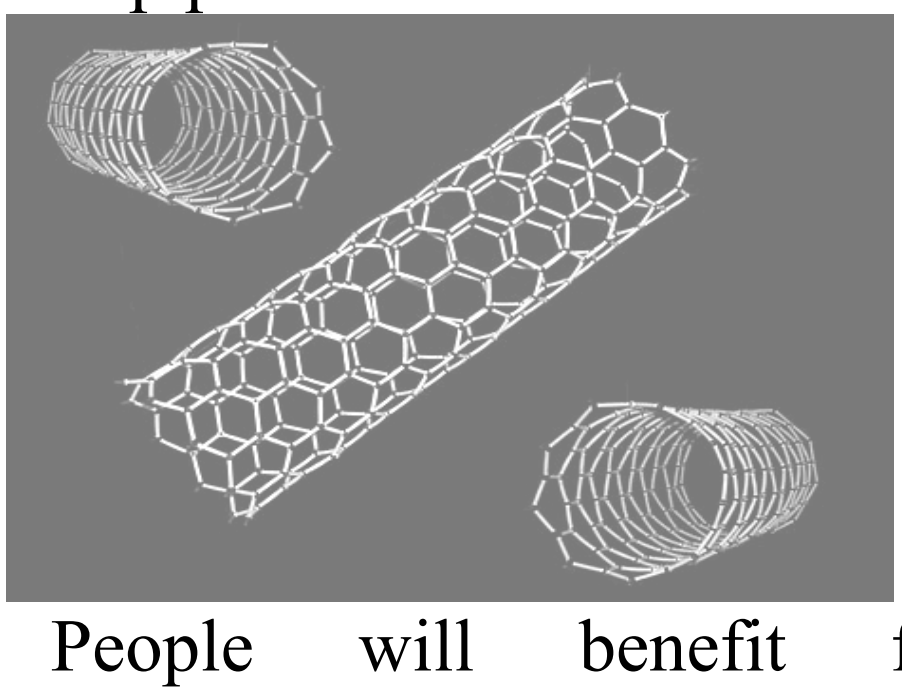

People will benefit from numerous microgravity experiments that can be conducted in Longer-Duration Microgravity Environment Produced by *Gravity Control Cells* (GCCs) on Earth.

# APPENDIX D: Antenna with Gravitational Transducer for *Instantaneous Communications* at any distance

It was previously shown in this article that *Quantum Gravitational Antennas* (GCC antennas, Fig.8) can emit and detect *virtual gravitational* radiation. The velocity of this radiation is *infinite*, as we have seen. This means that these quantum antennas can transmit and receive communications *instantaneously* to and from anywhere in the Universe. Here, it is shown how to transmit and receive communications *instantaneously* from any distance in the Universe by utilizing *virtual electromagnetic* (EM) radiation instead of *virtual gravitational* radiation. Starting from the principle that the antennas of usual transceivers (*real* antennas) radiate *real* EM radiation, then we can expect that *imaginary* antennas radiate *imaginary* EM radiation or *virtual* EM radiation. The velocity of this radiation is also *infinite*, in such a way that it can transmit communications *instantaneously* from any distance in the Universe.

It was shown [1] that when the *gravitational mass* of a body is decreased down to the range of $+0.159m_i$ to $-0.159m_i$ ($m_i$ is its inertial mass), the body becomes *imaginary* and goes to an *imaginary* Universe which contains our *real* Universe. Thus, we have the method to convert *real* antennas to *imaginary* antennas.

Now, consider a Gravitational Shielding $S$, whose gravitational mass is decreased down to the range of $+0.159m_{iS}$ to $-0.159m_{iS}$. By analogy, it becomes imaginary and goes to the imaginary Universe. It is easy to show that, in these circumstances, also a body inside the shielding $S$ becomes imaginary and goes to the imaginary Universe together with the gravitational shielding $S$. In order to prove it, consider, for example, Fig.D1 where we clearly see that the *Gravitational Shielding Effect* is equivalent to a decrease of $\chi = m_{gS}/m_{iS}$ in the gravitational masses of the bodies $A$ and $B$, since the *initial* gravitational masses: $m_{gA} \cong m_{iA}$ and $m_{gB} \cong m_{iB}$ become respectively $m_{gA} = \chi m_{iA}$ and $m_{gB} = \chi m_{iB}$, when the gravitational shielding is activated. Thus, when $\chi$ becomes less than $+0.159$, both the gravitational masses of $S$ and $A$ become respectively:

$$m_{gS} < +0.159m_{iS}$$

and

$$m_{gA} < +0.159m_{iA}$$

This proves, therefore, that when a Gravitational Shielding $S$ becomes *imaginary*, any particle (including *photons*[‡‡‡‡‡]) inside $S$, also becomes *imaginary* and goes to the *imaginary*

‡‡‡‡‡ As shown in the article "*Mathematical Foundations of the Relativistic Theory of Quantum Gravity*", *real* photons become *imaginary* photons or *virtual* photons.

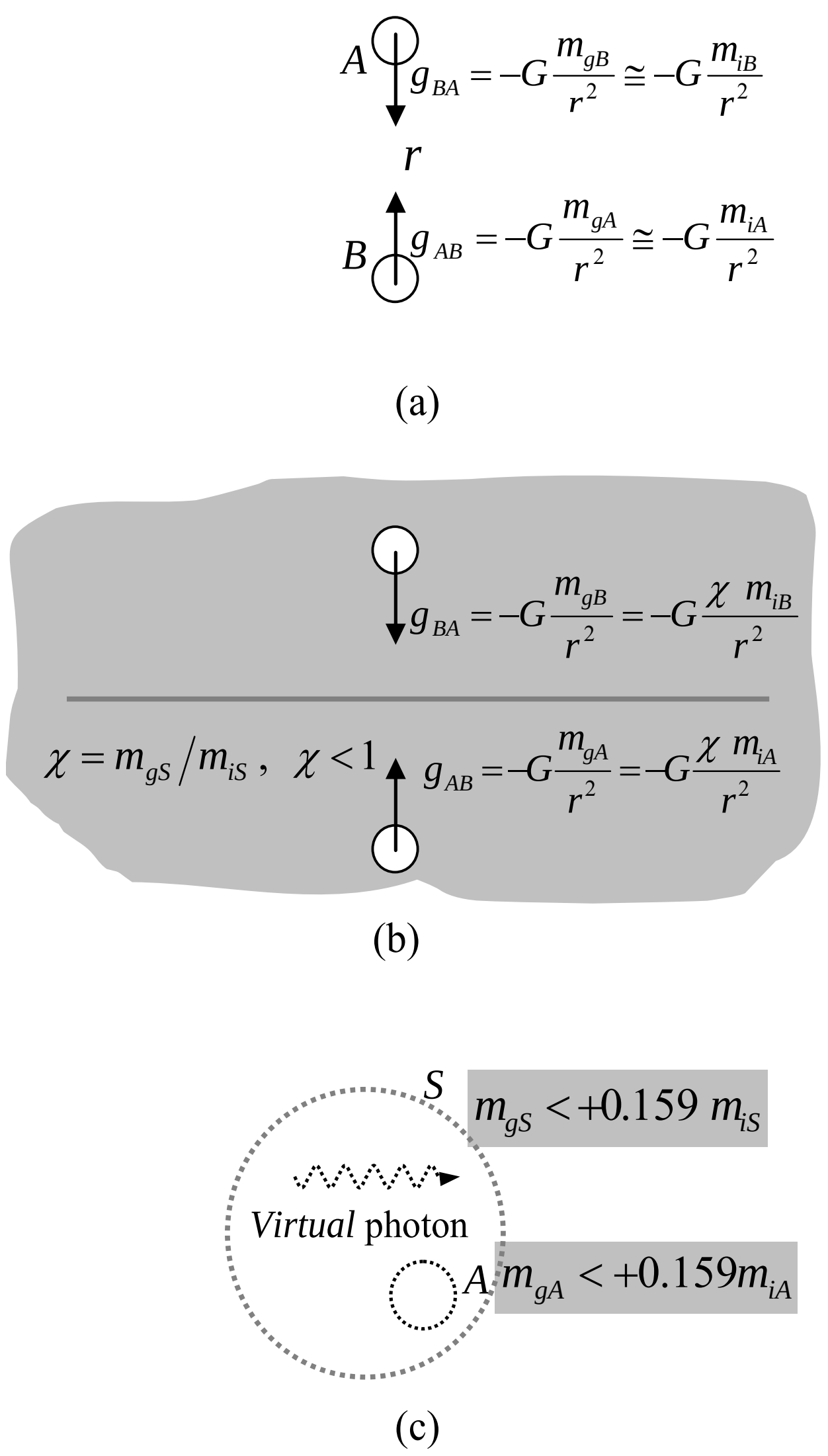

Fig.D1 – (a) (b) The *Gravitational Shielding Effect* is equivalent to a decrease of $\chi = m_{gS}/m_{iS}$ in the gravitational masses of the bodies $A$ and $B$. (c) When a Gravitational Shielding $S$ becomes *imaginary*, any particle (including *photons*) inside $S$, also becomes *imaginary*.

Universe together with the *Gravitational Shielding* $S$ [§§§§§].

Now, consider a transceiver antenna inside a Gravitational Shielding $S$. When the gravitational mass of $S$, $m_{gS}$, is reduced down to the range of $+0.159m_{iS}$ to $-0.159m_{iS}$, the antenna becomes imaginary, and, together with $S$, it goes to the imaginary Universe. In these circumstances, the *real* photons radiated from the antenna also become imaginary photons or *virtual* photons. Since the velocity of these photons is infinite, they can reach instantaneously the receiving antenna, *if it is also an imaginary antenna in the imaginary Universe*.

Therefore, we can say that the Gravitational Shielding around the antenna works as a *Gravitational* Transducer [******] converting *real* EM energy into *virtual* EM energy.

In practice, we can encapsulate antennas of transceivers with Aluminum cylinders, as shown in Fig.D2(a). By applying an appropriate ELF electric current through the Al cylinders, in order to put the gravitational masses of the cylinders within the range of $+0.159m_{iCyl}$ to $-0.159m_{iCyl}$, we can transform *real* antennas into *imaginary* antennas, making possible *instantaneously communications* at any distance, including astronomical distances.

Figure D2 (b) shows usual transceivers operating with *imaginary* antennas, i.e., real antennas turned into imaginary antennas. It is important to note that the communications between them occur through the *imaginary* space-time. At the end of transmissions, when the Gravitational Transducers are turned off, the antennas reappear in the *real* space-time, i.e., they become *real* antennas again.

[§§§§§] Similarly, the bodies inside a Gravitational Spacecraft become also imaginaries when the Gravitational Spacecraft becomes imaginary.

[******] A *Transducer* is substance or device that converts input energy of one form into output energy of another.

Imagine now cell phones using antennas with gravitational transducers. There will not be any more need of *cell phone signal transmission stations* because the reach of the *virtual* EM radiation is *infinite* (without *scattering*). The new cell phones will transmit and receive communications directly to and from one another. In addition, since the *virtual* EM radiation does not interact with matter, then there will not be any biological effects, as it happens in the case of usual cell phones.

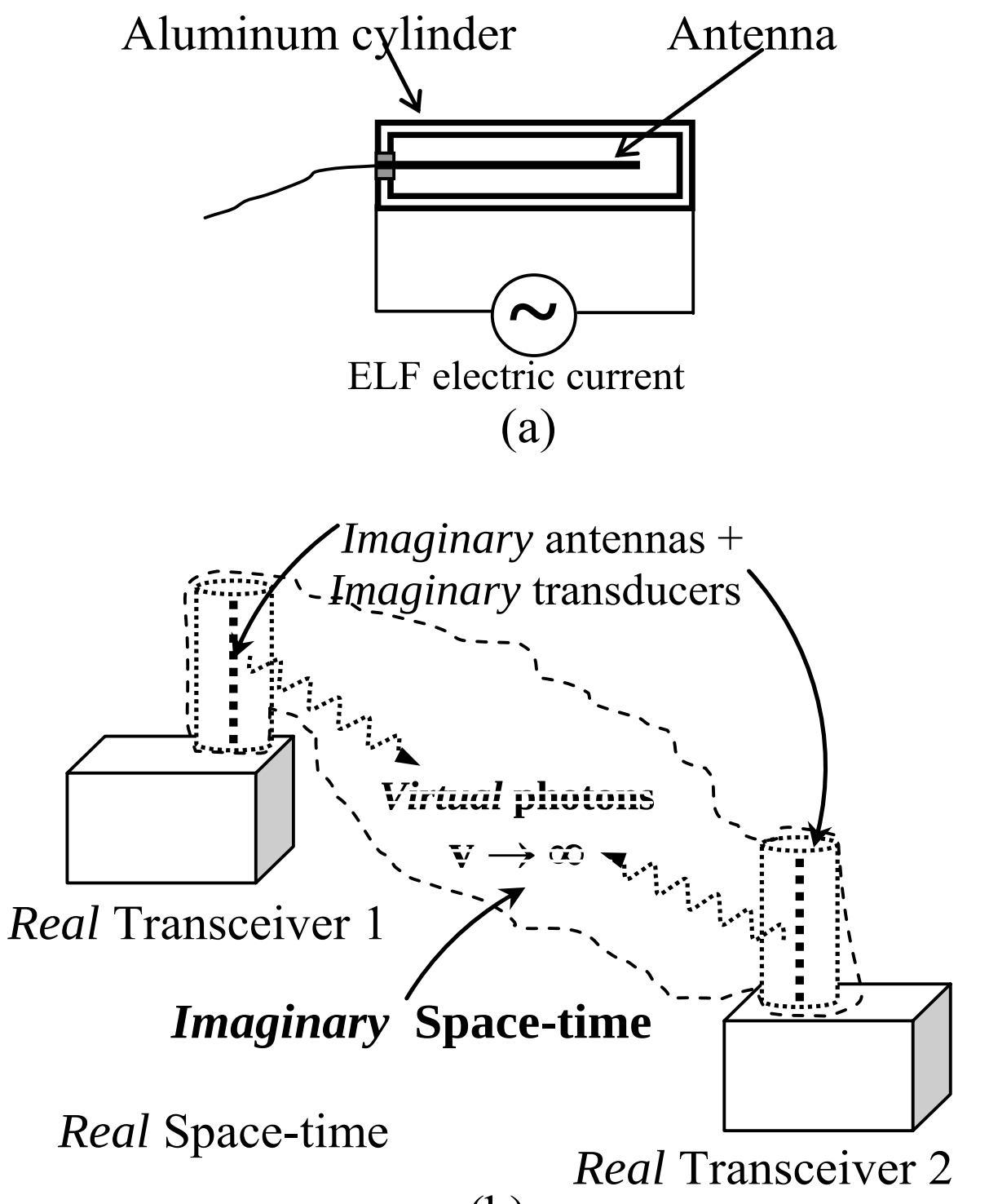

Fig. D2 – (a) Antenna with *Gravitational* Transducer. (b) Transceivers operating with *imaginary* antennas (*instantaneous* communications *at any distance*, including astronomical distances).

Let us now consider the case where a transceiver is *totally* turned into *imaginary* (Fig.D3). In order to convert real antennas into imaginary antennas, we have used the gravitational shielding effect, as we have already seen. Now, it is necessary to put the transceiver *totally* inside a Gravitational Shielding. Then, consider a transceiver $X$ inside the gravitational shielding of a Gravitational Spacecraft. When the spacecraft becomes imaginary, so does the transceiver $X$. Imagine then, another real transceiver $Y$ with *imaginary* antenna. With their antennas in the imaginary space-time, both transceivers $X$ and $Y$ are able to transmit and receive communications instantaneously between them, by means of *virtual* EM radiation (See Fig. D3(a)). Figure D3(b) shows another possibility: instantaneous communications between two transceivers at *virtual* state.

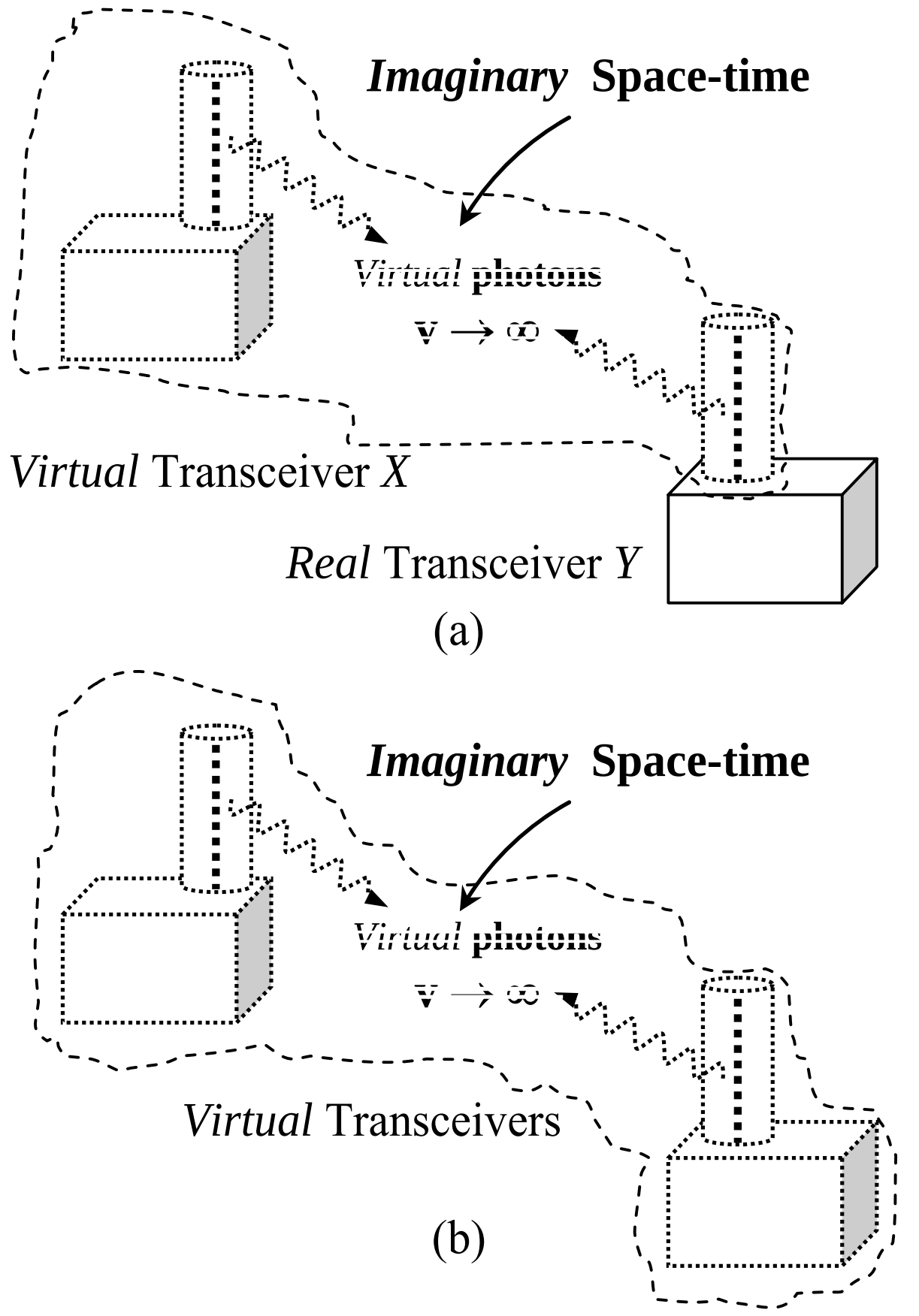

Fig. D3 – (a) Instantaneous communications between the *real* Universe and the *imaginary* Universe.(b) Instantaneous communications between two Virtual Transceivers in the *imaginary* Universe.

# Possibility of controlled nuclear fusion by means of Gravity Control

**Fran De Aquino**
Maranhao State University, Physics Department, S.Luis/MA, Brazil.

The gravity control process described in the articles *Mathematical Foundations of the Relativistic Theory of Quantum Gravity* [1] and *Gravity Control by means of Electromagnetic Field through Gas at Ultra-Low Pressure*, [2] points to the possibility of obtaining *Controlled Nuclear Fusion* by means of increasing of the intensity of the gravitational interaction between the nuclei. When the gravitational forces $F_G = Gm_g m'_g / r^2$ become greater than the electrical forces $F_E = qq'/4\pi\varepsilon_0 r^2$ between the nuclei, then nuclear fusion reactions can occur.

The equation of correlation between gravitational mass and inertial mass [1]

$$\chi = \frac{m_g}{m_i} = \left\{1 - 2\left[\sqrt{1 + \frac{\mu}{4c^2}\left(\frac{\sigma}{4\pi f}\right)^3 \frac{E^4}{\rho^2}} - 1\right]\right\} \qquad (1)$$

tells us that the gravitational mass can be strongly increased. Thus, if $E = E_m \sin \omega t$, then the average value for $E^2$ is equal to $\frac{1}{2}E_m^2$, because $E$ varies sinusoidaly ($E_m$ is the maximum value for $E$). On the other hand, $E_{rms} = E_m/\sqrt{2}$. Consequently, we can replace $E^4$ for $E^4_{rms}$. In addition, as $j = \sigma E$ (*Ohm's vectorial Law*), then Eq. (1) can be rewritten as follows

$$\chi = \frac{m_g}{m_{i0}} = \left\{1 - 2\left[\sqrt{1 + K\frac{\mu_r j^4_{rms}}{\sigma \rho^2 f^3}} - 1\right]\right\} \qquad (2)$$

where $K = 1.758 \times 10^{-27}$ and $j_{rms} = j/\sqrt{2}$.

Thus, the gravitational force equation can be expressed by

$$F_G = Gm_g m'_g / r^2 = \chi^2 Gm_{i0} m'_{i0} / r^2 =$$

$$= \left\{1 - 2\left[\sqrt{1 + K\frac{\mu_r j^4_{rms}}{\sigma \rho^2 f^3}} - 1\right]\right\}^2 Gm_{i0} m'_{i0} / r^2 \qquad (3)$$

In order to obtain $F_G > F_E$ we must have

$$\left\{1 - 2\left[\sqrt{1 + K\frac{\mu_r j^4_{rms}}{\sigma \rho^2 f^3}} - 1\right]\right\} > \sqrt{\frac{qq'/4\pi\varepsilon_0}{Gm_{i0} m'_{i0}}} \qquad (4)$$

The *carbon fusion* is a set of nuclear fusion reactions that take place in massive stars (at least $8M_{sun}$ at birth). It requires high temperatures ($> 5\times 10^8 K$) and densities ($> 3 \times 10^9\, kg.m^{-3}$). The principal reactions are:

$$^{12}C + {}^{12}C \rightarrow \begin{cases} ^{23}Na + p + 2.24\ MeV \\ ^{20}Ne + \alpha + 4.62\ MeV \\ ^{24}Mg + \gamma + 13.93\ MeV \end{cases}$$

In the case of Carbon nuclei ($^{12}C$) of a *thin carbon wire* (carbon fiber) ($\sigma \cong 4\times 10^4 S.m^{-1}$; $\rho = 2.2\times 10^3 S.m^{-1}$) Eq. (4) becomes

$$\left\{1 - 2\left[\sqrt{1 + 9.08\times 10^{-39}\frac{j^4_{rms}}{f^3}} - 1\right]\right\} > \sqrt{\frac{e^2}{16\pi\varepsilon_0 Gm_p^2}}$$

whence we conclude that the condition for the $^{12}C$ + $^{12}C$ fusion reactions occur is

$$j_{rms} > 1.7\times10^{18} f^{\frac{3}{4}} \qquad (5)$$

If the electric current through the carbon wire has Extremely-Low Frequency (ELF), for example, if $f = 1\mu Hz$, then the current density, $j_{rms}$, must have the following value:

$$j_{rms} > 5.4\times10^{13} Am^{-2} \qquad (6)$$

Since $j_{rms} = i_{rms}/S$ where $S = \pi\phi^2/4$ is the area of the cross section of the wire, we can conclude that, for an *ultra-thin carbon* wire with $10\mu m$-diameter, it is necessary that the current through the wire, $i_{rms}$, have the following intensity

$$i_{rms} > 4.24\ kA$$

Obviously, this current will *explode* the carbon wire. However, this explosion becomes negligible in comparison with the very strong *gravitational implosion*, which occurs simultaneously due to the enormous increase in intensities of the gravitational forces among the carbon nuclei produced by means of the ELF current through the carbon wire as predicted by Eq. (3). Since, in this case, the gravitational forces among the carbon nuclei become greater than the repulsive electric forces among them the result is the production of $^{12}C$ + $^{12}C$ fusion reactions.

Similar reactions can occur by using a *lithium* wire. In addition, it is important to note that $j_{rms}$ is directly proportional to $f^{\frac{3}{4}}$ (Eq.5). Thus, for example, if $f = 10^{-8} Hz$, the current necessary to produce the fusion reactions will be $i_{rms} = 130A$. However, it seems that in practice is better to reduce the diameter of the wire. For a diameter of $1\mu m$ $(10^{-6} m)$, the intensity of the current must have the following value

$$i_{rms} > 42.4\,A$$

In order to obtain an ELF current with these characteristics $\left(f = 10^{-6} Hz; i_{rms} = 42.4A\right)$ we can start from the following background: Consider an electric current $I$, which is the sum of a sinusoidal current $i_{osc} = i_0 \sin\omega t$ [1] and the DC current $I_{DC}$, i.e., $I = I_{DC} + i_0 \sin\omega t \quad ; \omega = 2\pi f$. If $i_0 << I_{DC}$ then $I \cong I_{DC}$. Thus, the current $I$ varies with the frequency $f$, but the variation of its intensity is quite small in comparison with $I_{DC}$, i.e., $I$ will be practically constant (Fig. 1). Thus, we obtain $i_{rms} \cong I_{DC}$ ( See Fig.2).

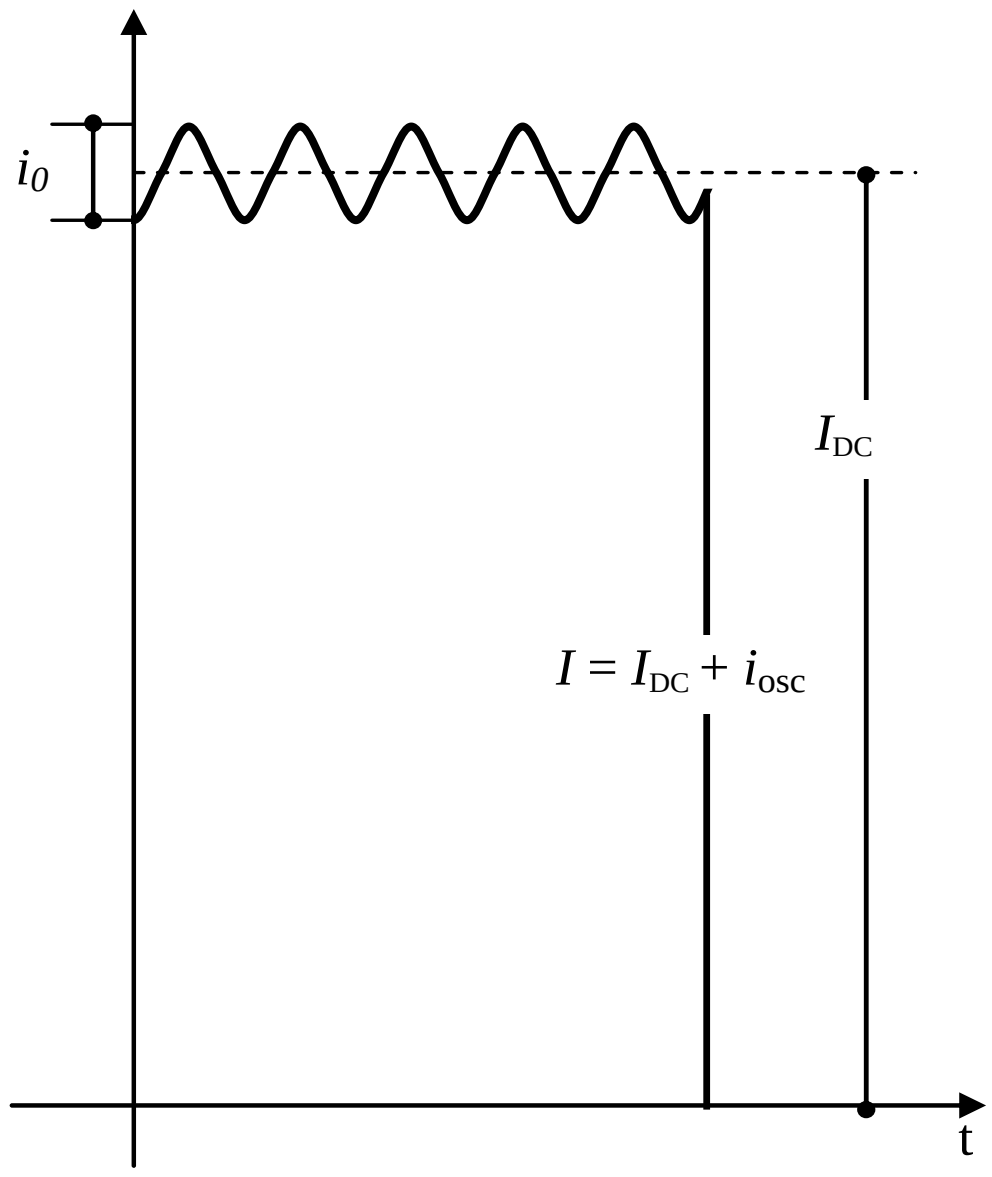

Fig. 1 - The electric current $I$ varies with frequency $f$. But the variation of $I$ is quite small in comparison with $I_{DC}$ due to $i_o << I_{DC}$. In this way, we can consider $I \cong I_{DC}$.

---

[1] In order to generate the ELF electric current $i_{osc}$ with $f = 10^{-6} Hz$, we can use the widely-known Function Generator HP3325A (Op.002 High Voltage Output) that can generate sinusoidal voltages with *extremely-low* frequencies down to $f = 1\times10^{-6} Hz$ and amplitude up to 20V ($40V_{pp}$ into $500\Omega$ load). The maximum output current is $0.08A_{pp}$; output impedance <2Ω at ELF .

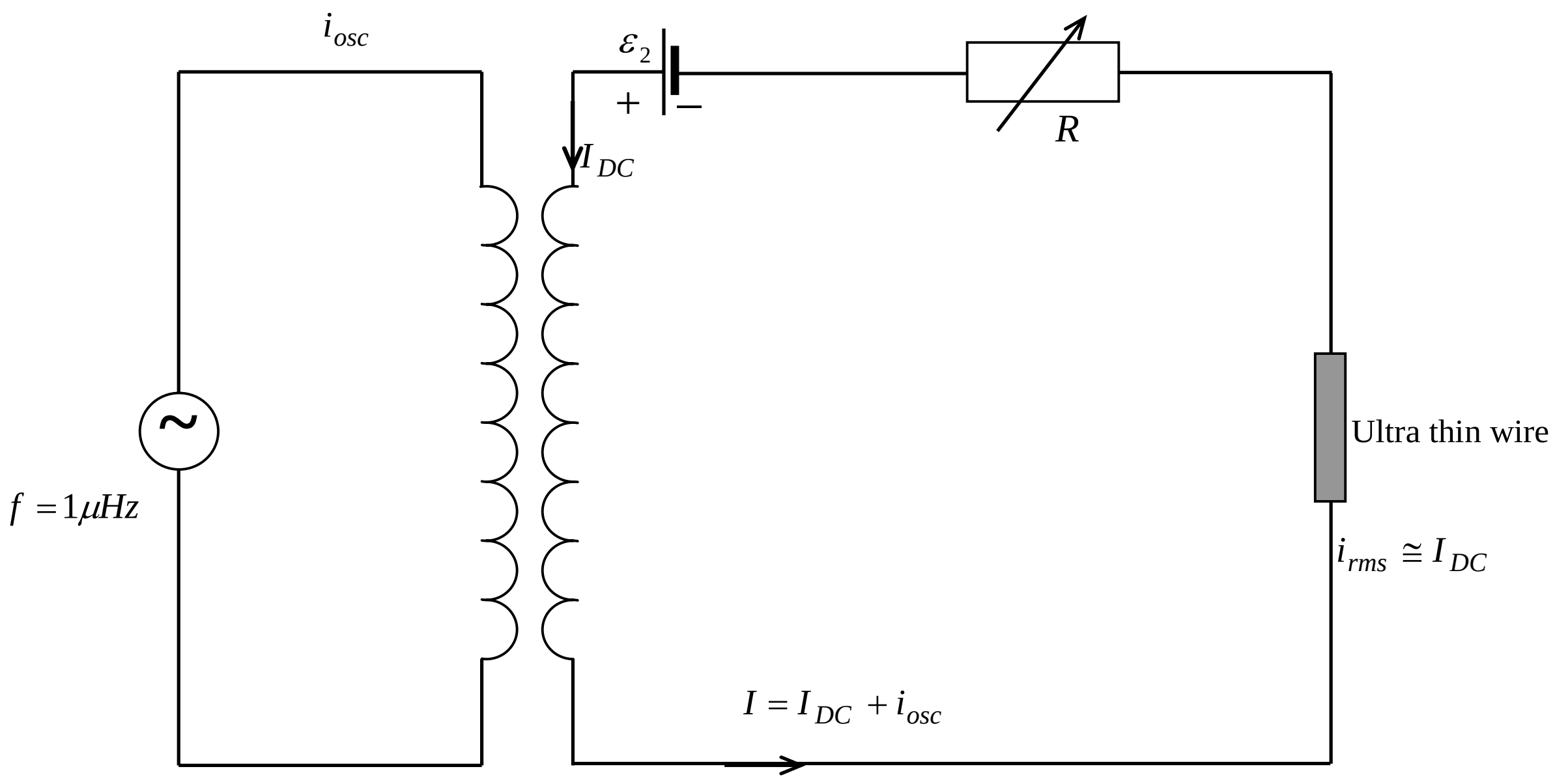

Fig. 2 – Electrical Circuit

# High-power ELF radiation generated by modulated HF heating of the ionosphere can cause Earthquakes, Cyclones and localized heating

**Fran De Aquino**
Maranhao State University, Physics Department, S.Luis/MA, Brazil.

The High Frequency Active Auroral Research Program (HAARP) is currently the most important facility used to generate extremely low frequency (ELF) electromagnetic radiation in the ionosphere. In order to produce this ELF radiation the HAARP transmitter radiates a strong beam of high-frequency (HF) waves modulated at ELF. This HF heating modulates the electrons' temperature in the D region ionosphere and leads to modulated conductivity and a time-varying current which then radiates at the modulation frequency. Recently, the HAARP HF transmitter operated with 3.6GW of effective radiated power modulated at frequency of 2.5Hz. It is shown that high-power ELF radiation generated by HF ionospheric heaters, such as the current HAARP heater, can cause Earthquakes, Cyclones and strong localized heating.
.

**Key words:** Physics of the ionosphere, radiation processes, Earthquakes, Tsunamis, Storms.
PACS: 94.20.-y ; 94.05.Dd ; 91.30.Px ; 91.30.Nw; 92.60.Qx

## 1. Introduction

Generating electromagnetic radiation at extremely-low frequencies is difficult because the long wavelengths require long antennas, extending for hundreds of kilometers. Natural ionospheric currents provide such an antenna if they can be modulated at the desired frequency [1-6]. The generation of ELF electromagnetic radiation by modulated heating of the ionosphere has been the subject matter of numerous papers [7-13].

In 1974, it was shown that ionospheric heater can generate ELF waves by heating the ionosphere with high-frequency (HF) radiation in the megahertz range [7]. This heating modulates the electron's temperature in the D region ionosphere, leading to modulated conductivity and a time-varying current, which then radiates at the modulation frequency.

Several HF ionospheric heaters have been built in the course of the latest decades in order to study the ELF waves produced by the heating of the ionosphere with HF radiation. Currently, the HAARP heater is the most powerful ionospheric heater, with 3.6GW of effective power using HF heating beam, modulated at ELF (2.5Hz) [14, 15]. This paper shows that high-power ELF radiation generated by modulated HF heating of the lower ionosphere, such as that produced by the current HAARP heater, can cause Earthquakes, Cyclones and strong localized heating.

## 2. Gravitational Shielding

The contemporary greatest challenge of the Theoretical Physics was to prove that, Gravity is a *quantum* phenomenon. Since General Relativity describes gravity as related to the curvature of space-time then, the quantization of the gravity implies the quantization of the proper space-time. Until the end of the century XX, several attempts to quantize gravity were made. However, all of them resulted fruitless [16, 17].

In the beginning of this century, it was clearly noticed that there was something unsatisfactory about the whole notion of quantization and that the quantization process had many ambiguities. Then, a new approach has been proposed starting from the generalization of the *action function**. The result has been the derivation of a theoretical background, which finally led to the so-sought quantization of the gravity and of the

* The formulation of the *action* in Classical Mechanics extends to Quantum Mechanics and has been the basis for the development of the *Strings Theory*.

space-time. Published with the title "*Mathematical Foundations of the Relativistic Theory of Quantum Gravity*"[18], this theory predicts a consistent *unification* of Gravity with Electromagnetism. It shows that the *strong* equivalence principle is reaffirmed and, consequently, Einstein's equations are preserved. In fact, Einstein's equations can be deduced directly from the mentioned theory. This shows, therefore, that the General Relativity is a particularization of this new theory, just as Newton's theory is a particular case of the General Relativity. Besides, it was deduced from the new theory an important correlation between the *gravitational mass* and the *inertial mass*, which shows that the gravitational mass of a particle can be *decreased* and even made *negative*, independently of its inertial mass, i.e., while the gravitational mass is progressively reduced, the inertial mass does not vary. This is highly relevant because it means that the weight of a body can also be reduced and even inverted in certain circumstances, since Newton's gravity law defines the weight $P$ of a body as the product of its *gravitational mass* $m_g$ by the local gravity acceleration $g$, i.e.,

$$P = m_g g \qquad (1)$$

It arises from the mentioned law that the gravity acceleration (or simply the gravity) produced by a body with gravitational mass $M_g$ is given by

$$g = \frac{GM_g}{r^2} \qquad (2)$$

The physical property of mass has two distinct aspects: *gravitational mass* $m_g$ and *inertial mass* $m_i$. The gravitational mass produces and responds to gravitational fields; it supplies the mass factor in Newton's famous inverse-square law of gravity $\left(F = GM_g m_g / r^2\right)$. The inertial mass is the mass factor in *Newton's 2nd Law of Motion* $\left(F = m_i a\right)$. These two masses are not equivalent but correlated by means of the following factor [18]:

$$\frac{m_g}{m_{i0}} = \left\{1 - 2\left[\sqrt{1 + \left(\frac{\Delta p}{m_{i0} c}\right)^2} - 1\right]\right\} \qquad (3)$$

Where $m_{i0}$ is the *rest* inertial mass and $\Delta p$ is the variation in the particle's *kinetic momentum*; $c$ is the speed of light.

This equation shows that only for $\Delta p = 0$ the gravitational mass is equal to the inertial mass. Instances in which $\Delta p$ is produced by *electromagnetic radiation*, Eq. (3) can be rewritten as follows [18]:

$$\frac{m_g}{m_{i0}} = \left\{1 - 2\left[\sqrt{1 + \left(\frac{n_r^2 D}{\rho c^3}\right)^2} - 1\right]\right\} \qquad (4)$$

Where $n_r$ is *the refraction index* of the particle; $D$ is the power density of the electromagnetic radiation absorbed by the particle; and $\rho$, its density of inertial mass.

From electrodynamics we know that

$$v = \frac{dz}{dt} = \frac{\omega}{\kappa_r} = \frac{c}{\sqrt{\frac{\varepsilon_r \mu_r}{2}\left(\sqrt{1 + (\sigma/\omega\varepsilon)^2} + 1\right)}} \qquad (5)$$

where $k_r$ is the real part of the *propagation vector* $\vec{k}$ (also called *phase constant*); $k = \left|\vec{k}\right| = k_r + ik_i$ ; $\varepsilon$, $\mu$ and $\sigma$, are the electromagnetic characteristics of the medium in which the incident radiation is propagating ($\varepsilon = \varepsilon_r \varepsilon_0$ ; $\varepsilon_0 = 8.854 \times 10^{-12} F/m$ ; $\mu = \mu_r \mu_0$, where $\mu_0 = 4\pi \times 10^{-7} H/m$).

From (5), we see that the *index of refraction* $n_r = c/v$, for $\sigma >> \omega\varepsilon$, is given by

$$n_r = \sqrt{\frac{\mu_r \sigma}{4\pi f \varepsilon_0}} \qquad (6)$$

Substitution of Eq. (6) into Eq. (4) yields

$$\chi = \frac{m_g}{m_{i0}} = \left\{1 - 2\left[\sqrt{1 + \left(\frac{\mu\sigma D}{4\pi\rho\, cf}\right)^2} - 1\right]\right\} \qquad (7)$$

It was shown that there is an additional effect - *Gravitational Shielding* effect - produced by a substance whose gravitational mass was reduced or made negative [18]. This effect shows that just *beyond* the substance the gravity acceleration $g_1$ will be reduced at the same proportion $\chi_1 = m_g/m_{i0}$, i.e., $g_1 = \chi_1 g$, ($g$ is the gravity acceleration *before* the substance). Consequently, *after a second gravitational shielding*, the gravity will be given by $g_2 = \chi_2 g_1 = \chi_1 \chi_2 g$, where $\chi_2$ is the value of the ratio $m_g/m_{i0}$ for the *second* gravitational shielding. In a generalized way, we can write that after the *nth* gravitational shielding the gravity, $g_n$, will be given by

$$g_n = \chi_1 \chi_2 \chi_3 \ldots \chi_n g$$

The dependence of the shielding effect on the height, at which the samples are placed above a superconducting disk with radius $r_D = 0.1375m$, has been recently measured up to a height of about 3m [19]. This means that the gravitational shielding effect extends, beyond the disk, for approximately *20 times* the disk *radius*.

## 3. Gravitational Shieldings in the Van Allen belts

The *Van Allen belts* are torus of plasma around Earth, which are held in place by Earth's magnetic field (See Fig.1). The existence of the belts was confirmed by the Explorer 1 and Explorer 3 missions in early 1958, under Dr James Van Allen at the University of Iowa. The term *Van Allen belts* refers specifically to the radiation belts surrounding Earth; however, similar radiation belts have been discovered around other planets.

Now consider the ionospheric heating with HF beam, modulated at ELF (See Fig. 2). The amplitude-modulated HF heating wave is absorbed by the ionospheric plasma, modulating the local conductivity $\sigma$. The current density $j = \sigma E_0$ radiates ELF electromagnetic waves that pass through the Van Allen belts producing two Gravitational Shieldings where the densities are *minima*, i.e., where they are approximately equal to density of the *interplanetary medium* near Earth. The quasi-vacuum of the interplanetary space might be thought of as beginning at an altitude of about 1000km above the Earth's surface [20]. Thus, we can assume that the densities $\rho_i$ and $\rho_o$ respectively, at the first gravitational shielding $S_i$ (at the *inner* Van Allen belt) and at $S_o$ (at the *outer* Van Allen belt) are $\rho_o \cong \rho_i \cong 0.8 \times 10^{-20} kg.m^{-3}$ (density of the *interplanetary medium* near the Earth [21]).

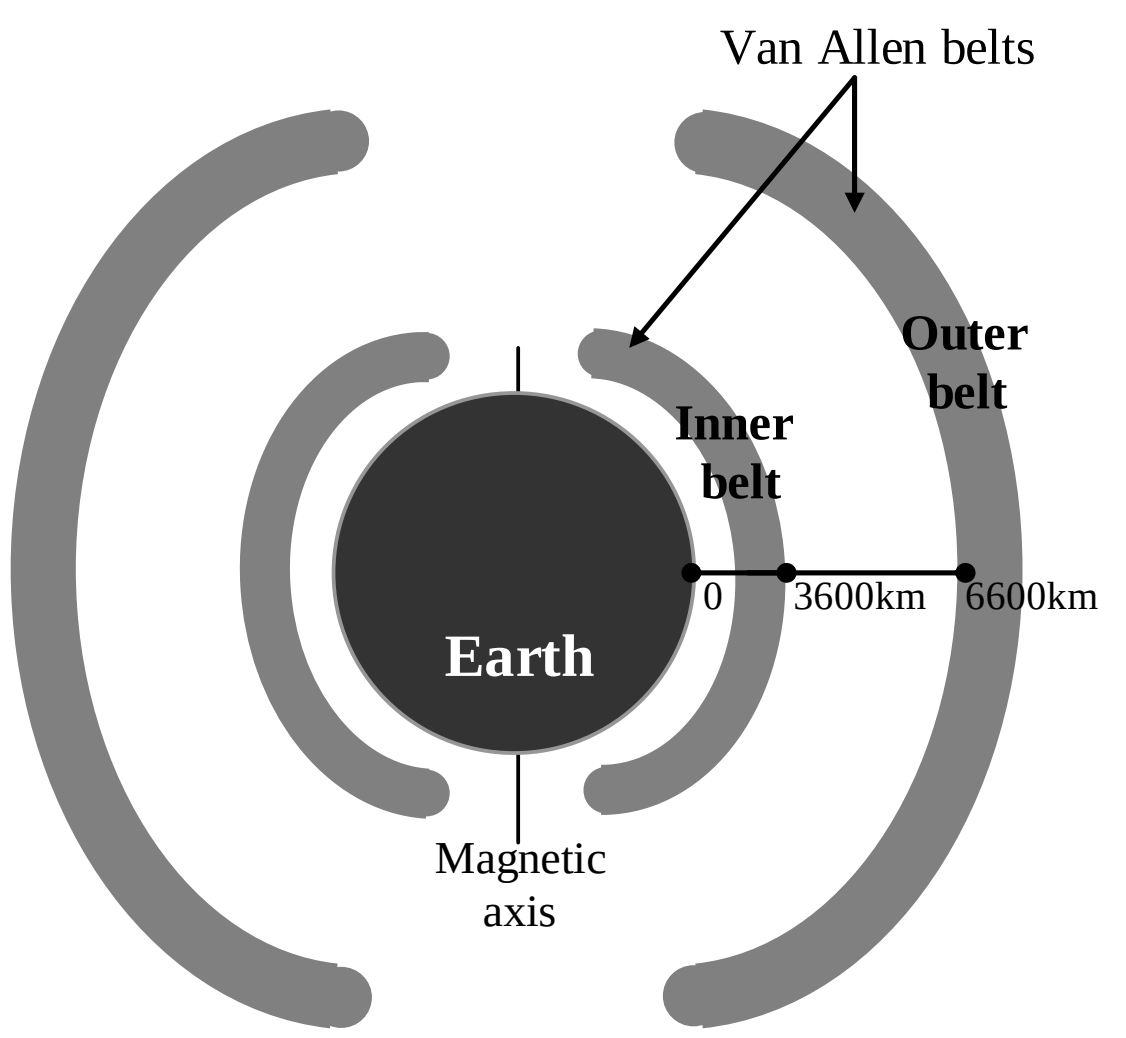

Fig.1 – Van Allen belts

The parallel conductivities,[†] $\sigma_{0i}$ and $\sigma_{0o}$, respectively at $S_i$ and $S_o$, present values which lie between those for metallic conductors and those for semiconductors [20], i.e., $\sigma_{0i} \cong \sigma_{0o} \sim 1S/m$. Thus, in these two Gravitational Shielding, according to Eq. (7), we have, respectively:

$$\chi_i = \left\{1 - 2\left[\sqrt{1 + \left(4.1 \times 10^4 \frac{D_i}{f}\right)^2} - 1\right]\right\} \qquad (8)$$

[†] Conductivity in presence of the Earth's magnetic field

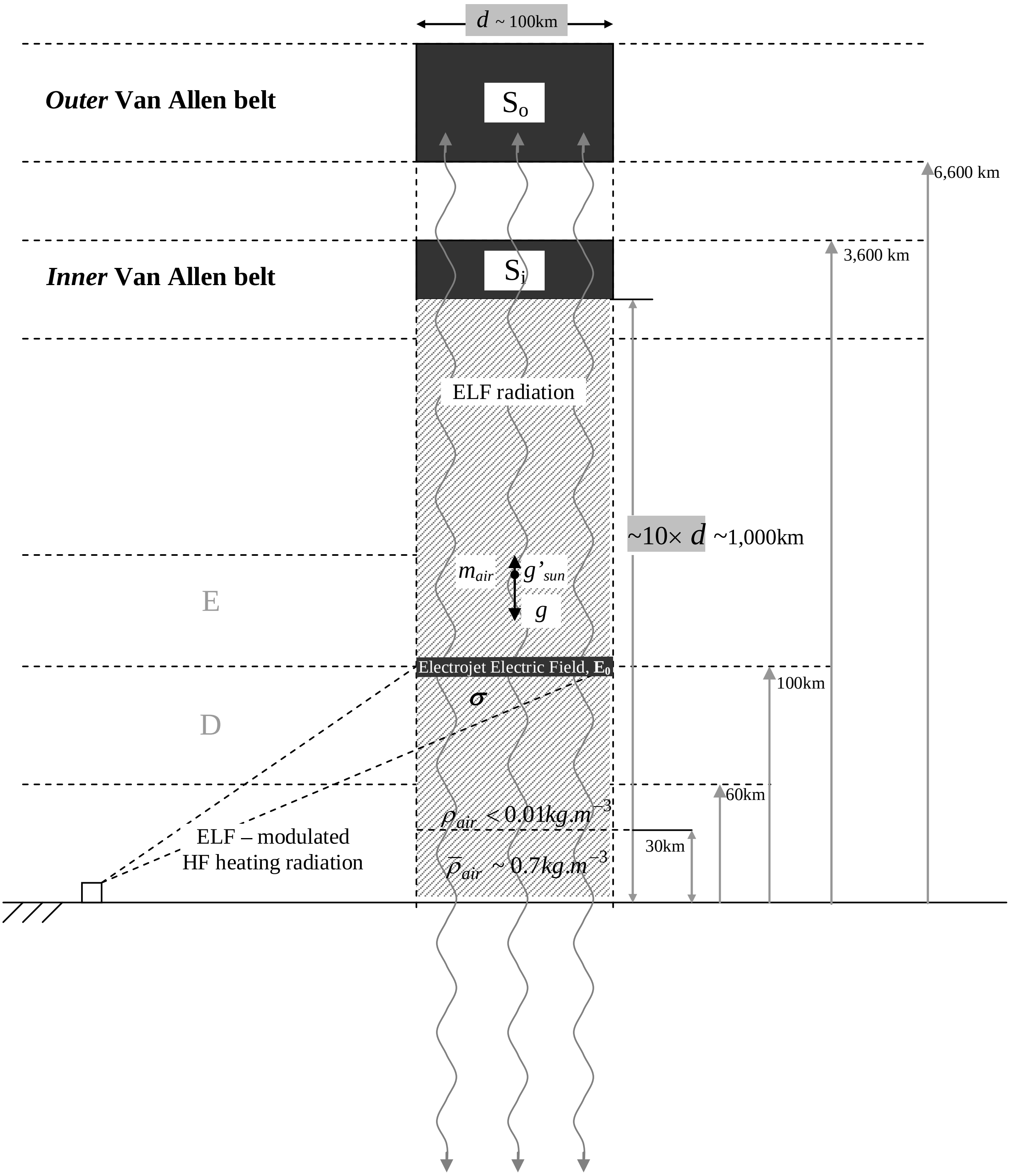

Fig. 2 – *Ionospheric Gravitational Shieldings* - The amplitude-modulated HF heating wave is absorbed by the ionospheric plasma, modulating the local conductivity $\sigma_0$. The current density $j = \sigma_0 E_0$ (*$E_0$* is the Electrojet Electric Field), radiates ELF electromagnetic waves ($d$ is the length of the ELF dipole). Two gravitational shieldings ($S_o$ and $S_i$) are formed at the Van Allen belts. Then, the gravity due to the Sun, after the shielding $S_i$, becomes $g'_{sun} = \chi_o \chi_i \, g_{sun}$. The effect of the gravitational shielding reaches $\sim 20 \times r_D = \sim 10 \times d \cong 1{,}000 km$.

and

$$\chi_o = \left\{ 1 - 2\left[ \sqrt{1 + \left( 4.1 \times 10^4 \frac{D_o}{f} \right)^2} - 1 \right] \right\} \qquad (9)$$

where

$$D_i \cong D_o \cong \frac{P_{ELF}}{S_a} \qquad (10)$$

$P_{ELF}$ is the ELF radiation power, radiated from the ELF ionospheric antenna; $S_a$ is the area of the antenna.

Substitution of (10) into (8) and (9) leads to

$$\chi_o \chi_i = \left\{ 1 - 2\left[ \sqrt{1 + \left( 4.1 \times 10^4 \frac{P_{ELF}}{S_a f} \right)^2} - 1 \right] \right\}^2 \qquad (11)$$

**4. Effect of the gravitational shieldings** $S_i$ **and** $S_o$ **on the Earth and its environment.**

Based on the Podkletnov experiment, previously mentioned, in which the effect of the Gravitational Shielding extends for approximately *20 times* the disk *radius* $(r_D)$, we can assume that the effect of the gravitational shielding $S_i$ extends for approximately 10 times the dipole length ($d$). For a dipole length of about 100km, we can conclude that the effect of the gravitational shielding reaches about 1,000Km below $S_i$ (See Fig.2), affecting therefore an air mass, $m_{air}$, given by ‡

$$m_{air} = \bar{\rho}_{air} V_{air} = = \left( \sim 0.7 kg.m^{-3} \right) (100{,}000m)^2 (30{,}000m) = \sim 10^{14} kg \qquad (12)$$

The gravitational potential energy related to $m_{air}$, with respect to the Sun's center, without the effects produced by the gravitational shieldings $S_o$ and $S_i$ is

$$E_{p0} = m_{air} r_{se} \left( g - g_{sun} \right) \qquad (13)$$

where, $r_{se} = 1.49 \times 10^{11} m$ (distance from the Sun to Earth, 1 AU), $g = 9.8 m/s^2$ and $g_{sun} = -GM_{sun}/r_{se}^2 = 5.92 \times 10^{-3} m/s^2$, is the gravity due to the Sun at the Earth.

The gravitational potential energy related to $m_{air}$, with respect to the Sun's center, considering the effects produced by the gravitational shieldings $S_o$ and $S_i$, is

$$E_p = m_{air} r_{se} \left( g - \chi_o \chi_i g_{sun} \right) \qquad (14)$$

Thus, the *decrease* in the gravitational potential energy is

$$\Delta E_p = E_p - E_{p0} = \left( 1 - \chi_o \chi_i \right) m_{air} r_{se} g_{sun} \qquad (15)$$

Substitution of (11) into (15) gives

$$\Delta E_p = \left\{ 1 - \left\{ 1 - 2\left[ \sqrt{1 + \left( 4.1 \times 10^4 \frac{P_{ELF}}{S_a f} \right)^2} - 1 \right] \right\}^2 \right\} m_{air} r_{se} g_{sun} \qquad (16)$$

The HF power produced by the HAARP transmitter is $P_{HF} = 3.6GW$ modulated at $f = 2.5Hz$. The ELF conversion efficiency at HAARP is estimated to be $\sim 10^{-4}\%$ for wave generated using sinusoidal amplitude modulation. This means that

$$P_{ELF} \sim 4kW$$

Substitution of $P_{ELF} \sim 4kW$, $f = 2.5Hz$ and $S_a = (100{,}000)^2 = 1 \times 10^{10} m^2$ into (16) yields

$$\Delta E_p \sim 10^{-4} m_{air} r_{se} g_{sun} \sim 10^{19} joules \qquad (17)$$

This decrease in the gravitational potential energy of the air column, $\Delta E_p$, produces a decrease $\Delta p$ in the local pressure $p$ ( *Bernoulli principle*). Then the pressure equilibrium between the Earth's mantle and the Earth's atmosphere, in the region corresponding to the air column, is broken. This is equivalent to an increase of pressure $\Delta p$ in the region of the mantle corresponding to the air column. This phenomenon is similar to an Earthquake, which liberates an energy equal to $\Delta E_p$ (see Fig.3).

‡ The mass of the air column above 30km height is negligible in comparison with the mass of the air column below 30km height, whose average density is ~0.7kg./m$^3$.

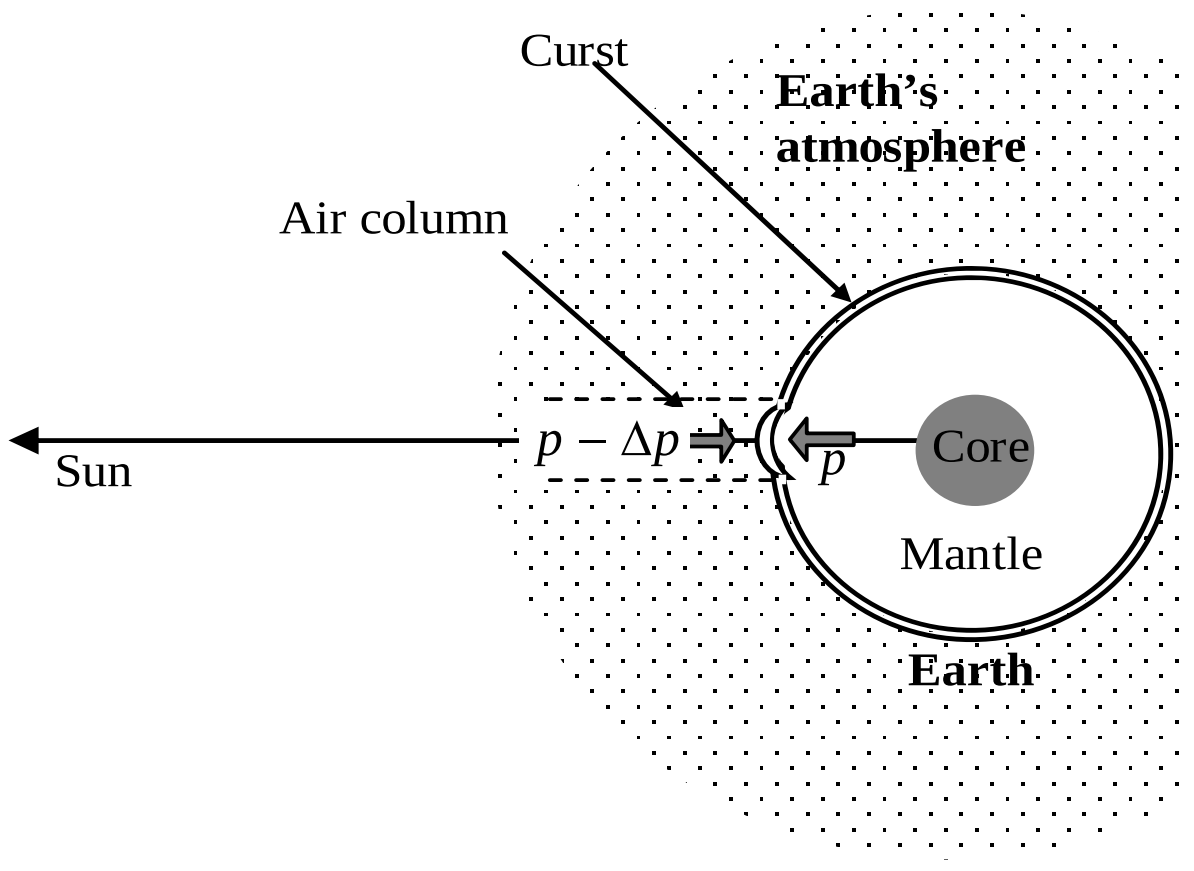

Fig. 3 - The decrease in the gravitational potential energy of the air column, $\Delta E_p$, produces a decrease $\Delta p$ in the local pressure $p$ (Principle of Bernoulli). Then the pressure equilibrium between the Earth's mantle and the Earth's atmosphere, in the region corresponding to the air column, is broken. This is equivalent to an increase of pressure $\Delta p$ in the region of the mantle corresponding to the air column. This phenomenon is similar to an Earthquake, which liberates an amount of energy equal to $\Delta E_p$.

The magnitude $M_s$ in the *Richter scales*, corresponding to liberation of an amount of energy, $\Delta E_p \sim 10^{19}\,joules$, is obtained by means of the well-known equation:

$$10^{19} = 10^{(5+1.44 M_s)} \qquad (18)$$

which gives $M_s = 9.1$. That is, an Earthquake with magnitude of about 9.1 in the *Richter scales*.

The decrease in the gravitational potential energy in the air column whose mass is $m_{air}$ gives to the air column an initial kinetic energy $E_k = \frac{1}{2} m_{air} V_{0air}^2 = \Delta E_p$, where $\Delta E_p$ is given by (15).

In the previously mentioned HAARP conditions, Eq.(11) gives $(1-\chi_o\chi_i) \sim 10^{-4}$. Thus, from (15), we obtain

$$\Delta E_p \sim 10^{-4} m_{air} r_{se} g_{sun} \qquad (19)$$

Thus, the initial air speed $V_{0air}$ is

$$V_{0air} \cong \sqrt{10^{-4} g_{sun} r_{se}} \sim 10^2 m/s \sim 400 km/h \qquad (20)$$

This velocity will strongly reduce the pressure in the air column (*Bernoulli principle*) and it is sufficient to produce a powerful *Cyclone* around the air column (*Coriolis Effect*).

Note that, by reducing the diameter of the HF beam radiation, it is possible to reduce dipole length ($d$) and consequently to reduce the reach of the Gravitational Shielding, since the effect of the gravitational shielding reaches approximately18 times the dipole length. By reducing $d$, we also reduce the area $S_a$, increasing consequently the value of $\chi_o\chi_i$ (See Eq. (18)). This can cause an increase in the velocity $V_{0air}$ (See Eq. (22)).

On the other hand, if the dipole length ($d$) is increased, the reach of the Gravitational Shielding will also be increased. For example, by increasing the value of $d$ for $d = 101 km$, the effect of the Gravitational Shielding reaches approximately $1010 km$, and can surpass the surface of the Earth or the Oceans (See Fig.2). In this case, the decrease in the gravitational potential energy at the local, by analogy to Eq.(15), is

$$\Delta E_p = (1-\chi_o\chi_i) m r_{se} g_{sun} \qquad (21)$$

where $m$ is the mass of the soil, or the mass of the ocean water, according to the case.

The decrease, $\Delta E_p$, in the gravitational potential energy increases the kinetic energy of the local at the same ratio, in such way that the mass $m$ acquires a kinetic energy $E_k = \Delta E_p$. If this energy is not enough to pluck the mass $m$ from the soil or the ocean, and launch it into space, then $E_k$ is converted into heat, raising the local temperature by $\Delta T$, the value of which can be obtained from the following expression:

$$\left\langle \frac{E_k}{N} \right\rangle \cong k\Delta T \qquad (22)$$

where $N$ is the number of atoms in the volume $V$ of the substance considered; $k = 1.38\times 10^{-23} J/K$ is the *Boltzmann constant*. Thus, we get

$$\Delta T \cong \frac{E_k}{Nk} = \frac{(1-\chi_o\chi_i) m\, r_{se} g_{sun}}{(nV)k} = $$

$$= \frac{(1-\chi_o\chi_i)\ \rho\ r_{se} g_{sun}}{nk} \qquad (23)$$

where $n$ is the number of atoms/m$^3$ in the substance considered.

In the previously mentioned HAARP conditions, Eq. (11) gives $(1-\chi_o\chi_i)\sim 10^{-4}$. Thus, from (23), we obtain

$$\Delta T \cong \frac{6.4\times 10^{27}}{n}\rho \qquad (24)$$

For most liquid and solid substances the value of $n$ is about $10^{28}\,atoms/m^3$, and $\rho\sim 10^3\,kg/m^3$. Therefore, in this case, Eq. (24) gives

$$\Delta T \cong 640K \cong 400°C$$

This means that, the region in the soil or in the ocean will have its temperature increased by approximately 400°C.

By increasing $P_{ELF}$ or decreasing the frequency, $f$, of the ELF radiation, it is possible to increase $\Delta T$ (See Eq.(16)). In this way, it is possible to produce *strong localized heating on Land or on the Oceans*.

This process suggests that, by means of two *small* Gravitational Shieldings built with *Gas* or *Plasma* at ultra-low pressure, as shown in the processes of gravity control [22], it is possible to produce the same heating effects. Thus, for example, the water inside a container can be strongly heated when the container is placed below the mentioned Gravitational Shieldings.

Let us now consider another source of ELF radiation, which can activate the Gravitational Shieldings $S_o$ and $S_i$.

It is known that the *Schumann resonances* [23] are global electromagnetic resonances (a set of spectrum peaks in the extremely low frequency ELF), excited by lightning discharges in the *spherical resonant cavity* formed by the Earth's surface and the inner edge of the ionosphere (60km from the Earth's surface). The Earth–ionosphere waveguide behaves like a resonator at ELF frequencies and amplifies the spectral signals from lightning at the resonance frequencies. In the normal mode descriptions of Schumann resonances, the fundamental mode $(n=1)$ is a standing wave in the Earth–ionosphere cavity with a wavelength equal to the circumference of the Earth. This lowest-frequency (and highest-intensity) mode of the Schumann resonance occurs at a frequency $f_1 = 7.83Hz$ [24].

It was experimentally observed that ELF radiation escapes from the *Earth–ionosphere waveguide* and reaches the Van Allen belts [25-28]. In the ionospheric spherical cavity, the ELF radiation power density, $D$, is related to the energy density inside the cavity, $W$, by means of the well-known expression:

$$D=\frac{c}{4}W \qquad (25)$$

where $c$ is the speed of light, and $W=\frac{1}{2}\varepsilon_0 E^2$. The electric field $E$, is given by

$$E=\frac{q}{4\pi\varepsilon_0 r_\oplus^2}$$

where $q=500{,}000C$ [24] and $r_\oplus = 6.371\times 10^6 m$. Therefore, we get

$$E = 110.7V/m,$$
$$W = 5.4\times 10^{-8} J/m^3,$$
$$D \cong 4.1\ W/m^2 \qquad (26)$$

The area, $S$, of the cross-section of the cavity is $S = 2\pi r_\oplus d = 2.4\times 10^{12}\ m^2$. Thus, the ELF radiation power is $P = DS \cong 9.8\times 10^{12} W$. The total power escaping from the Earth-ionosphere waveguide, $P_{esc}$, is only a fraction of this value and need to be determined.

When this ELF radiation crosses the Van Allen belts the Gravitational Shieldings $S_o$ and $S_i$ can be produced (See Fig.4).

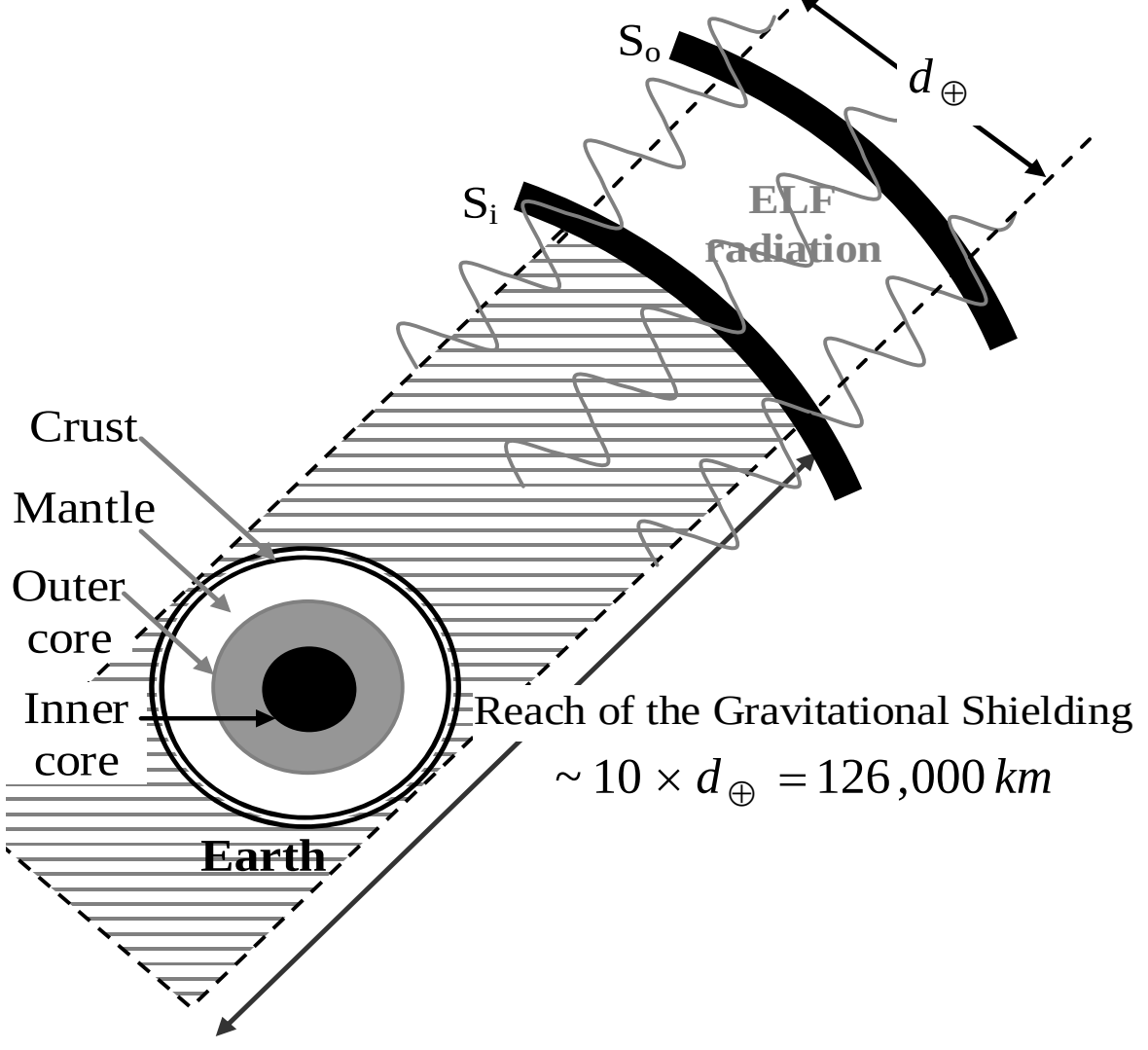

Fig.4 – ELF radiation escaping from the *Earth–ionosphere waveguide* can produce the Gravitational Shieldings $S_o$ and $S_i$ in the Van Allen belts.

The ELF radiation power densities $D_i$ and $D_o$, respectively in $S_i$ and $S_o$, are given by

$$D_i = \frac{P_{esc}}{4\pi r_i^2} \qquad (27)$$

and

$$D_o = \frac{P_{esc}}{4\pi r_o^2} \qquad (28)$$

where $r_i$ and $r_o$ are respectively, the distances from the Earth's center up to the Gravitational Shieldings $S_i$ and $S_o$ .

Under these circumstances, the *kinetic energy* related to the mass, $m_{oc}$, of the Earth's *outer core*[§], with respect to the Sun's center, considering the effects produced by the Gravitational Shieldings $S_o$ and $S_i$ [**] is

$$E_k = (1-\chi_o\chi_i)m_{oc}r_{se}g_{sun} = \tfrac{1}{2}m_{oc}\bar{V}_{oc}^2 \qquad (29)$$

Thus, we get

$$\bar{V}_{oc} = \sqrt{(1-\chi_o\chi_i)r_{se}g_{sun}} \qquad (30)$$

The average radius of the *outer core* is $\bar{r}_{oc} = 2.3\times10^6\,m$. Then, assuming that the average angular speed of the outer core, $\varpi_{oc}$, has the same order of magnitude of the average angular speed of the Earth's *crust*, $\varpi_{\oplus}$, i.e., $\varpi_{oc} \sim \varpi_{\oplus} = 7.29\times10^{-5}\,rad/s$, then we get $V_{oc} = \varpi_{oc}\bar{r}_{oc} \sim 10^2\,m/s$. Thus, Eq. (30) gives

$$(1-\chi_o\chi_i) \sim 10^{-5} \qquad (31)$$

This relationship shows that, *if the power of the ELF radiation escaping from the Earth-ionosphere waveguide is progressively increasing* (for example, by the increasing of the dimensions of the holes in the Earth-ionosphere waveguide[††]), then as soon as the value of $\chi_o\chi_i$ equals 1, and the speed $\bar{V}_{oc}$ will be null. After a time interval, the progressive increasing of the power density of the ELF radiation makes $\chi_o\chi_i$ greater than 1. Equation (29) shows that, at this moment, the velocity $V_{oc}$ resurges, but now in the *opposite direction*.

The Earth's magnetic field is generated by the outer core motion, i.e., the molten iron in the *outer core* is spinning with angular speed, $\varpi_{oc}$, and it's spinning inside the Sun's magnetic field, so a magnetic field is generated in the molten core. This process is called *dynamo effect*.

Since Eq. (31) tells us that the factor $(1-\chi_o\chi_i)$ is currently very close to zero, we can conclude that the moment of the *reversion* of the Earth's magnetic field is very close.

## 5. Device for moving very heavy loads.

Based on the phenomenon of reduction of local gravity related to the Gravitational Shieldings $S_o$ and $S_i$, it is possible to create a device for moving very heavy loads such as large monoliths, for example.

Imagine a large monolith on the Earth's surface. At noon the gravity acceleration upon the monolith is basically given by

$$g_R = g - g_{sun}$$

where $g_{sun} = -GM_{sun}/r_{se}^2 = 5.92\times10^{-3}\,m/s^2$ is the gravity due to the Sun at the monolith and $g = 9.8\,m/s^2$.

If we place upon the monolith a *mantle* with a set of $n$ Gravitational Shieldings inside, the value of $g_R$ becomes

$$g_R = g - \chi^n g_{sun}$$

This shows that, it is possible to reduce $g_R$ down to values very close to zero, and thus to transport very heavy loads (See Fig.5). We will call the mentioned mantle of *Gravitational Shielding Mantle*. Figure 5 shows one of these mantles with a set of 8 Gravitational Shieldings. Since the mantle thickness must be thin, the option is to use Gravitational Shieldings produced by layers of *high-dielectric strength* semiconductor [22]. When the Gravitational Shieldings are active the gravity due to the Sun is multiplied by the factor $\chi^8$, in such way that the gravity resultant upon the monolith becomes $g_R = g - \chi^8 g_{Sun}$. Thus, for example, if $\chi = -2.525$ the result is $g_R = 0.028 m/s^2$. Under these circumstances, the weight of the monolith becomes $2.9 \times 10^{-3}$ of the initial weight.

---

[§] The Earth is an oblate spheroid. It is composed of a number of different layers. An outer silicate solid *crust*, a highly viscous *mantle*, a liquid *outer core* that is much less viscous than the mantle, and a solid *inner core*. The *outer core* is made of *liquid* iron and nickel.

[**] Note that the *reach* of the Gravitational Shielding is $\sim 10\times d_{\oplus} = 126{,}000\,km$.

[††] The amount of ELF radiation that escapes from the Earth-ionosphere waveguide is directly proportional to the *number of holes* in inner edge of the ionosphere and the *dimensions* of these holes. Thus, if the amount of holes or its dimensions are increasing, then the *power of the ELF radiation escaping from the Earth-ionosphere waveguide* will also be increased.

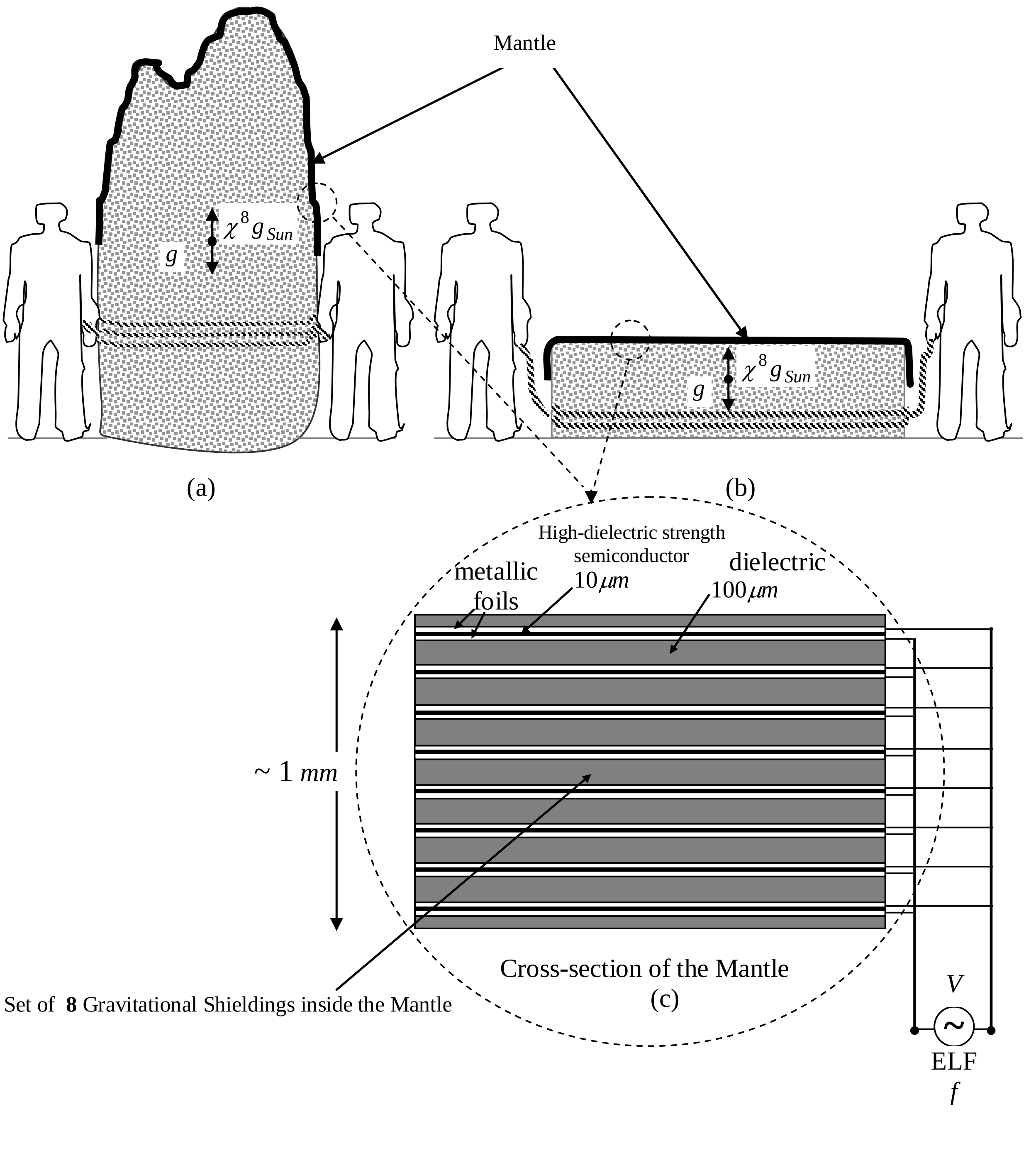

Fig. 5 – *Device for transporting very heavy loads.* It is possible to transport very heavy loads by using a Gravitational Shielding Mantle - A Mantle with a set of 8 semiconductor layers or more (each layer with $10\mu m$ thickness, sandwiched by two metallic foils with $10\mu m$ thickness). The total thickness of the mantle (including the insulation layers) is ~1*mm*. The metallic foils are connected to the ends of an ELF voltage source in order to generate ELF electromagnetic fields through the semiconductor layers. The objective is to create 8 Gravitational Shieldings as shown in (c). When the Gravitational Shieldings are active the gravity due to the Sun is multiplied by the factor $\chi^8$, in such way that the gravity resultant upon the monoliths (a) and (b) becomes $g_R = g - \chi^8 g_{Sun}$. Thus, for example, if $\chi = -2.525$ results $g_R = 0.028 m/s^2$. Under these circumstances, the weight of the monolith becomes $2.9 \times 10^{-3}$ of the initial weight.

## 6. Gates to the imaginary spacetime in the Earth-ionosphere waveguide.

It is known that strong densities of electric charges can occur in some regions of the upper boundary of the Earth-ionosphere waveguide, for example, as a result of the lightning discharges [29]. These anomalies increase strongly the electric field $E_w$ in the mentioned regions, and possibly can produce a tunneling effect to the imaginary spacetime.

The electric field $E_w$ will produce an *electrons flux* in a direction and an *ions flux* in an opposite direction. From the viewpoint of electric current, the ions flux can be considered as an "electrons" flux at the same direction of the real electrons flux. Thus, the current density through the air, $j_w$, will be the *double* of the current density expressed by the well-known equation of Langmuir-Child

$$j = \frac{4}{9}\varepsilon_r\varepsilon_0\sqrt{\frac{2e}{m_e}}\frac{V^{\frac{3}{2}}}{r^2} = \alpha\frac{V^{\frac{3}{2}}}{r^2} = 2.33\times10^{-6}\frac{V^{\frac{3}{2}}}{r^2} \qquad (32)$$

where $\varepsilon_r \cong 1$ for the *air*; $\alpha = 2.33\times10^{-6}$ is the called *Child's constant*; $r$, in this case, is the distance between the center of the charges and the Gravitational Shieldings $S_{w1}$ and $S_{w2}$ (see Fig.6) ($r = \frac{1}{2}\left(1.4\times10^{-15}m\right) = 7\times10^{-16}m$); $V$ is the voltage drop given by

$$V = E_w r = \frac{\sigma_Q}{2\varepsilon_0}r = \frac{Qr}{2\varepsilon_0 A} \qquad (33)$$

where $Q$ is the anomalous amount of charge in the region with area $A$, i.e., $\sigma_Q = Q/A = \eta\sigma_q$, $\eta$ is the ratio of proportionality, and $\sigma_q = q/4\pi R^2 \cong 9.8\times10^{-10} C/m^2$ is the normal charge density ; $q = 500{,}000C$ is the total charge[24], then $Q = \eta A\sigma_q = \eta q_n$ ($q_n = A\sigma_q$ is the normal amount of charge in the area $A$).

By substituting (33) into (32), we get

$$j_w = 2j = 2\alpha\frac{V^{\frac{3}{2}}}{r^2} = 2\alpha\frac{\left(\frac{Qr}{2\varepsilon_0 A}\right)^{\frac{3}{2}}}{r^2} = \frac{2\alpha}{\sqrt{r}}\left(\frac{Q}{2\varepsilon_0 A}\right)^{\frac{3}{2}} \qquad (34)$$

Since $E_w = \sigma_Q/2\varepsilon_0$ and $j_w = \sigma_w E_w$, we can write that

$$\sigma_w^3 E_w^4 = j_w^3 E_w = \left[\frac{2\alpha}{\sqrt{r}}\left(\frac{Q}{2\varepsilon_0 A}\right)^{\frac{3}{2}}\right]^3 \frac{Q}{2\varepsilon_0 A} =$$

$$= \frac{0.18\alpha^3 Q^{5.5}}{r^{1.5}\varepsilon_0^{5.5}A^{5.5}} = \frac{0.18\alpha^3\sigma_Q^{5.5}}{r^{1.5}\varepsilon_0^{5.5}} = \frac{0.18\alpha^3\left(\eta\sigma_q\right)^{5.5}}{r^{1.5}\varepsilon_0^{5.5}} =$$

$$= 2.14\times10^{16}\eta^{5.5} \qquad (35)$$

The electric field $E_w$ has an oscillating component, $E_{w1}$, with frequency, $f$, equal to the lowest Schumann resonance frequency $f_1 = 7.83Hz$. Then, by using Eq. (7), that can be rewritten in the following form [18]:

$$\chi = \frac{m_g}{m_i} = \left\{1 - 2\left[\sqrt{1 + 1.758\times10^{-27}\frac{\mu_r\sigma^3E^4}{\rho^2 f^3}} - 1\right]\right\} \qquad (36)$$

we can write that

$$\chi_w = \frac{m_g}{m_i} = \left\{1 - 2\left[\sqrt{1 + 1.758\times10^{-27}\frac{\mu_{rw}\sigma_w^3E_{w1}^4}{\rho_w^2 f_1^3}} - 1\right]\right\} \qquad (37)$$

By substitution of Eq. (35), $\mu_{rw} = 1$, $\rho_w = 1\times10^{-2} kg/m^3$ and $f_1 = 7.83Hz$ into the expression above, we obtain

$$\chi_w = \left\{1 - 2\left[\sqrt{1 + 7.84\times10^{-10}\eta^{5.5}} - 1\right]\right\} \qquad (38)$$

The gravity below $S_{w2}$ will be decreased by the effect of the Gravitational Shieldings $S_{w1}$ and $S_{w2}$, according to the following expression

$$\left(g - \chi_{w1}\chi_{w2}g_{sun}\right)$$

where $\chi_{w1} = \chi_{w2} = \chi_w$. Thus, we get

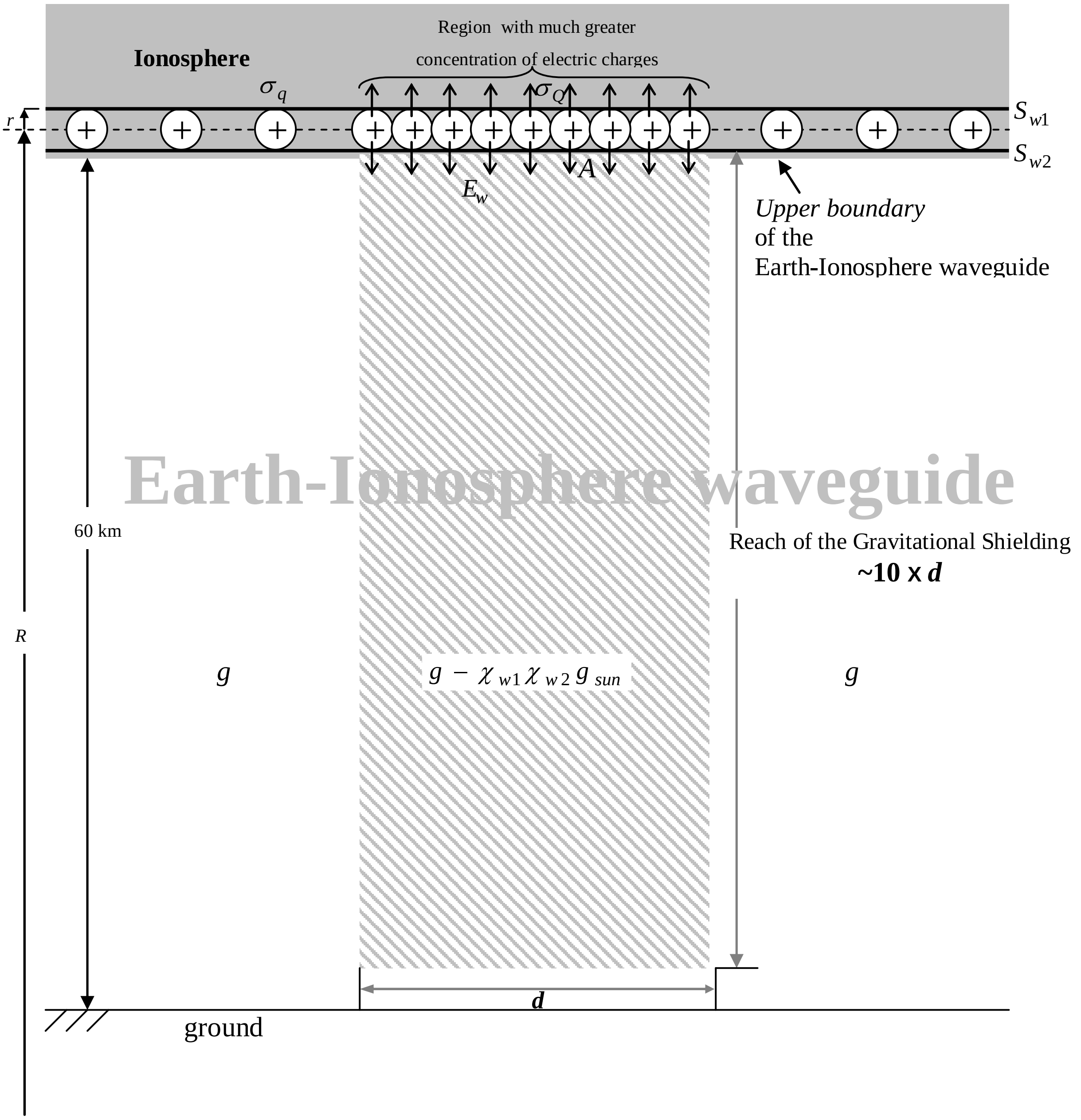

Fig. 6 - Gravitational Shieldings $S_{w1}$ and $S_{w2}$ produced by strong densities of electric charge in the upper boundary of the Earth-Ionosphere.

$$\left\{1-\left\{1-2\left[\sqrt{1+7.84\times10^{-10}\eta^{5.5}}-1\right]\right\}^2\frac{g_{sun}}{g}\right\}g=\chi g$$

where

$$\chi=\left\{1-\left\{1-2\left[\sqrt{1+7.84\times10^{-10}\eta^{5.5}}-1\right]\right\}^2\frac{g_{sun}}{g}\right\} \quad (39)$$

In a previous article [18], it was shown that, when the gravitational mass of a body is reduced to a value in the range of $+0.159m_i$ to $-0.159m_i$ *or the local gravity* $(g)$ *is reduced to a value in the range of* $+0.159g$ *to* $-0.159g$, the body performs a transition to the *imaginary spacetime*. This means that, if the value of $\chi$ given by Eq.(39) is in the range $0.159<\chi<-0.159$, then *any body* (aircrafts, ships, etc) that enters the region - defined by the volume $(A \times \sim 10d)$ *below* the Gravitational Shielding $S_{w2}$, will perform a transition to the imaginary spacetime. Consequently, it will disappear from our Real Universe and will appear in the Imaginary Universe. However, the electric field $E_{w1}$, which reduces the gravitational mass of the body (or the gravitational shieldings, which reduce the local gravity) does not accompany the body; they stay at the Real Universe. Consequently, the body returns immediately from the Imaginary Universe. Meanwhile, it is important to note that, in the case of *collapse* of the *wavefunction* $\Psi$ of the body, it will never more come back to the Real Universe.

Equation (39) shows that, in order to obtain $\chi$ in the range of $0.159<\chi<-0.159$ the value of $\eta$ must be in the following range:

$$127.1<\eta<135.4$$

Since the normal charge density is $\sigma_q \cong 9.8\times10^{-10}C/m^2$ then it must be increased by about 130 times in order to transform the region $(A \times \sim 10d)$, *below* the Gravitational Shielding $S_{w2}$, in a gate to the imaginary spacetime.

It is known that in the Earth's atmosphere occur transitorily large densities of electromagnetic energy across extensive areas. We have already seen how the density of electromagnetic energy affects the gravitational mass (Eq. (4)). Now, it will be shown that it also affects the length of an object. *Length contraction* or Lorentz contraction is the physical phenomenon of a decrease in length detected by an observer of objects that travel at any non-zero velocity relative to that observer. If $L_0$ is the length of the object in its rest frame, then the length $L$, observed by an observer in relative motion with respect to the object, is given by

$$L=\frac{L_0}{\gamma(V)}=L_0\sqrt{1-V^2/c^2} \quad (40)$$

where $V$ is the relative velocity between the observer and the moving object and $c$ the speed of light. The function $\gamma(V)$ is known as the *Lorentz factor*.

It was shown that Eq. (3) can be written in the following form [18]:

$$\frac{m_g}{m_{i0}}=\left\{1-2\left[\sqrt{1+\left(\frac{\Delta p}{m_{i0}c}\right)^2}-1\right]\right\}=\left\{1-2\left[\frac{1}{\sqrt{1-V^2/c^2}}-1\right]\right\}$$

This expression shows that

$$\sqrt{1+\left(\frac{\Delta p}{m_{i0}c}\right)^2}=\frac{1}{\sqrt{1-V^2/c^2}}=\gamma(V) \quad (41)$$

By substitution of Eq. (41) into Eq.(40) we get

$$L=\frac{L_0}{\gamma(V)}=\frac{L_0}{\sqrt{1+\left(\frac{\Delta p}{m_{i0}c}\right)^2}} \quad (42)$$

It was shown that, the term, $\Delta p/m_{i0}c$, in the equation above is equal to $Wn_r/\rho c^2$, where $W$ is the density of electromagnetic energy absorbed by the body and $n_r$ the index of refraction, given by

$$n_r=\frac{c}{v}=\sqrt{\frac{\varepsilon_r\mu_r}{2}\left(\sqrt{1+(\sigma/\omega\varepsilon)^2}+1\right)}$$

In the case of $\sigma>>2\pi f\varepsilon$, $W=(\sigma/8\pi f)E^2$ and $n_r=c/v=\sqrt{\mu\sigma c^2/4\pi f}$ [30]. Thus, in this case, Eq. (42) can be written as follows

$$L=\frac{L_0}{\sqrt{1+1.758\times10^{-27}\left(\frac{\mu_r\sigma^3}{\rho^2 f^3}\right)E^4}} \quad (43)$$

Note that $E = E_m \sin \omega t$. The average value for $E^2$ is equal to $\frac{1}{2}E_m^2$ because $E$ varies sinusoidaly ($E_m$ is the maximum value for $E$). On the other hand, $E_{rms} = E_m/\sqrt{2}$. Consequently we can change $E^4$ by $E_{rms}^4$, and the equation above can be rewritten as follows

$$L = \frac{L_0}{\sqrt{1+1.758\times10^{-27}\left(\frac{\mu_r\sigma^3}{\rho^2 f^3}\right)E_{rms}^4}} \qquad (44)$$

Now, consider an airplane traveling in a region of the atmosphere. Suddenly, along a distance $L_0$ of the trajectory of the airplane arises an ELF electric field with intensity $E_{rms} \sim 10^5 V.m^{-1}$ and frequency $f \sim 1Hz$. The Aluminum density is $\rho = 2.7\times10^3 kg.m^{-3}$ and its conductivity is $\sigma = 3.82\times10^7 S.m^{-1}$. According to Eq. (44), for the airplane the distance $L_0$ is shortened by $2.7\times10^{-5}$. Under these conditions, a distance $L_0$ of about 3000km will become just 0.08km.

*Time dilation* is an observed difference of elapsed time between two observers which are moving relative to each other, or being differently situated from nearby gravitational masses. This effect arises from the nature of space-time described by the theory of relativity. The expression for determining time dilation in special relativity is:

$$T = T_0\gamma(V) = \frac{T_0}{\sqrt{1-V^2/c^2}}$$

where $T_0$ is the interval time measured at the object in its rest frame (known as the *proper time*); $T$ is the time interval observed by an observer in relative motion with respect to the object.

Based on Eq. (41), we can write the expression of $T$ in the following form:

$$T = \frac{T_0}{\sqrt{1-V^2/c^2}} = T_0\sqrt{1+\left(\frac{\Delta p}{m_{i0}c}\right)^2}$$

For $V << c$, we can write that $\Delta p = m_{i0}V$ and

$$\tfrac{1}{2}m_{i0}V^2 = m_{i0}gr = m_{i0}\varphi \;\Rightarrow\; V^2 = 2\varphi$$

where $\varphi$ is the gravitational potential. Then, it follows that

$$\left(\frac{\Delta p}{m_{i0}}\right)^2 = V^2 = 2\varphi \quad and \quad \left(\frac{\Delta p}{m_{i0}c}\right)^2 = \frac{V^2}{c^2} = \frac{2\varphi}{c^2}$$

Consequently, the expression of $T$ becomes

$$T = \frac{T_0}{\sqrt{1-V^2/c^2}} = T_0\sqrt{1+\frac{2\varphi}{c^2}}$$

which is the well-known expression obtained in the General Relativity.

Based on Eq. (41) we can also write the expression of $T$ in the following form:

$$T = T_0\sqrt{1+\left(\frac{\Delta p}{m_{i0}c}\right)^2} = T_0\sqrt{1+1.758\times10^{-27}\left(\frac{\mu_r\sigma^3}{\rho^2 f^3}\right)E_{rms}^4} \qquad (45)$$

Now, consider a ship in the ocean. It is made of steel ($\mu_r = 300$; $\sigma = 1.1\times10^6 S.m^{-1}$; $\rho = 7.8\times10^3 kg.m^{-3}$). When subjected to a uniform ELF electromagnetic field, with intensity $E_{rms} = 1.36\times10^3 V.m^{-1}$ and frequency $f = 1Hz$, the ship will perform a transition in time to a time $T$ given by

$$T = T_0\sqrt{1+1.758\times10^{-27}\left(\frac{\mu_r\sigma^3}{\rho^2 f^3}\right)E_{rms}^4} = $$
$$= T_0(1.0195574) \qquad (46)$$

If $T_0 = January, 1\ 1943,\ 0h\ 0\min\ 0s$ then the ship performs a transition in time to $T = January, 1\ 1981,\ 0h\ 0\min\ 0s$. Note that the use of ELF $(f = 1Hz)$ is fundamental.

It is important to note that *the electromagnetic field $E_{rms}$, besides being uniform, must remain with the ship during the transition to the time $T$*. If it is not uniform, each part of the ship will perform transitions for different times in the future. On the other hand, the field must remain with the ship, because, if it stays at the time $T_0$, the transition is interrupted. In order to the electromagnetic field remains at the ship, it is necessary that all the parts, which are involved with the generation of the field, stay *inside* the ship. If persons are inside the ship they will perform transitions for different times in the future because their conductivities and densities are different. Since the conductivity and density of the ship and of the persons are different, they will perform transitions to different times. This means that the ship and the persons must have the same characteristics, in order to perform transitions to the same time. Thus, *in this way is unsuitable and highly dangerous to make transitions to the future with persons*. However, there is a way to solve this problem. If we can control the gravitational mass of a body, in such way that $m_g = \chi\ m_{i0}$, and we put this body inside a ship with gravitational mass $M_g \cong M_{i0}$, then the total gravitational mass of the ship will be given by‡‡

$$M_{g(total)} = M_g + m_g = M_{i0} + \chi\ m_{i0}$$

or

$$\chi_{ship} = \frac{M_{g(total)}}{M_{i0}} = 1 + \frac{\chi\ m_{i0}}{M_{i0}} \qquad (47)$$

Since

$$\chi_{ship} = \frac{M_g}{M_{i0}} = \left\{1 - 2\left[\sqrt{1+\left(\frac{\Delta p}{M_{i0}c}\right)^2} - 1\right]\right\}$$

we can write that

$$\sqrt{1+\left(\frac{\Delta p}{M_{i0}c}\right)^2} = \frac{3-\chi_{ship}}{2} \qquad (48)$$

Then it follows that

$$T = T_0\sqrt{1+\left(\frac{\Delta p}{M_{i0}c}\right)^2} = T_0\left(\frac{3-\chi_{ship}}{2}\right) \qquad (49)$$

Substitution of Eq. (47) into Eq. (49) gives

$$T = T_0\left(1 - \frac{\chi\, m_{i0}}{2M_{i0}}\right) \qquad (50)$$

Note that, if $\chi = -0.0391148(M_{i0}/m_{i0})$, Eq. (50) gives

$$T = T_0(1.0195574)$$

which is the same value given by Eq.(46).

‡‡ This idea was originally presented by the author in the paper: *The Gravitational Spacecraft* [30].

Other safe way to make transitions in the time is by means of flights with relativistic speeds, according to predicted by the equation:

$$T = \frac{T_0}{\pm\sqrt{1 - V^2/c^2}} \qquad (51)$$

With the advent of the Gravitational Spacecrafts [30], which could reach velocities close to the light speed, this possibility will become very promising.

It was shown in a previous paper [18] that by varying the gravitational mass of the spacecraft for *negative* or *positive* we can go respectively to the *past* or *future*.

If the gravitational mass of a particle is *positive*, then $t$ is always *positive* and given by

$$t = +t_0 \big/ \sqrt{1 - V^2/c^2} \qquad (52)$$

This leads to the well-known relativistic prediction that the particle goes to the *future* if $V \to c$. However, if the gravitational mass of the particle is negative, then $t$ is also *negative* and, therefore, given by

$$t = -t_0 \big/ \sqrt{1 - V^2/c^2} \qquad (53)$$

In this case, the prevision is that the particle goes to the *past* if $V \to c$. In this way, *negative gravitational mass* is the necessary condition to the particle to go to the *past*.

Now, consider a parallel plate capacitor, which has a *high-dielectric strength* semiconductor between its plates, with the following characteristics $\mu_r = 1$; $\sigma = 10^4 S.m^{-1}$; $\rho = 10^3 kg.m^{-3}$. According to Eq.(45), when the semiconductor is subjected to a uniform ELF electromagnetic field, with intensity $E_{rms} = 10^5 V.m^{-1}$ $(0.1KV/mm)$ and frequency $f = 1Hz$, *it should perform* a transition in time to a time $T$ given by

$$T = T_0\sqrt{1 + 1.758\times10^{-27}\left(\frac{\mu_r\sigma^3}{\rho^2 f^3}\right)E_{rms}^4} =$$

$$= T_0(1.08434) \qquad (54)$$

However, the transition *is not performed*, because the electromagnetic field is *external* to the semiconductor, and obviously would not accompany the semiconductor during the transition. In other words, the field stays at the time $T_0$, and the transition is not performed.

## 7. Detection of Earthquakes at the Very Early Stage

When an earthquake occurs, energy radiates outwards in all directions. The energy travels through and around the earth as three types of seismic waves called primary, secondary, and surface waves (P-wave, S-wave and Surface-waves). All various types of earthquakes follow this pattern. At a given distance from the epicenter, first the P-waves arrive, then the S-waves, both of which have such small energies that they are mostly not threatening. Finally, the surface waves arrive with all of their damaging energies. It is predominantly the surface waves that we would notice as the earthquake. This knowledge, that, preceding any destructive earthquake, there are telltales P-waves, are used by the earthquake warning systems to reliably initiate an alarm before the arrival of the destructive waves. Unfortunately, the warning time of these earthquake warning systems is less than 60 seconds.

Earthquakes are caused by the movement of tectonic plates. There are three types of motion: plates moving away from each other (at divergent boundaries); moving towards each other (at convergent boundaries) or sliding past one another (at transform boundaries). When these movements are interrupted by an obstacle (rocks, for example), an Earthquake occurs when the obstacle breaks (due to the sudden release of stored energy).

The pressure $P$ acting on the obstacle and the corresponding reaction modifies the *gravitational mass* of the matter along the pressing surfaces, according to the following expression [18]:

$$m_g = \left\{1 - 2\left[\sqrt{1 + \left(\frac{P^2}{2\rho^2 c v^3}\right)^2} - 1\right]\right\} m_{i0} \qquad (55)$$

where $\rho$ and $v$ are respectively, the *density* of matter and the *speed of the pressure waves* in the mentioned region.

*Hooke's* law tells us that $P = \rho v^2$, thus Eq. (55) can be rewritten as follows

$$m_g = \left\{1 - 2\left[\sqrt{1 + \frac{P}{4c^2\rho}} - 1\right]\right\} m_{i0} \qquad (56)$$

or

$$\chi = \frac{m_g}{m_{i0}} = \left\{1 - 2\left[\sqrt{1 + \frac{P}{4c^2\rho}} - 1\right]\right\} \qquad (57)$$

Thus, the matter subjected to the pressure $P$ works as a Gravitational Shielding. Consequently, if the gravity below it is $g_\oplus$, then the gravity *above* it is $\chi g_\oplus$, in such way that a gravimeter on the Earth surface (See Fig.7) shall detect a *gravity anomaly* $\Delta g$ given by

$$\Delta g = g_\oplus - \chi g_\oplus = (1 - \chi) g_\oplus \qquad (58)$$

Substitution of Eq. (57) into this Eq. (58) yields

$$\Delta g = 2\left[\sqrt{1 + \frac{P}{4c^2\rho}} - 1\right] g_\oplus \qquad (59)$$

Thus, when a gravity anomaly is detected, we can evaluate, by means of Eq. (59), the magnitude of the ratio $P/\rho$ in the compressing region. On the other hand, several experimental observations of the *time interval* between the appearing of gravity anomaly $\Delta g$ and the breaking of the obstacle (beginning of the Earthquake) will give us a statistical value for the mentioned time interval, which will warn us (earthquake warning system) when to initiate an alarm. Obviously, the earthquake warning time, in this case becomes much greater than 60 seconds.

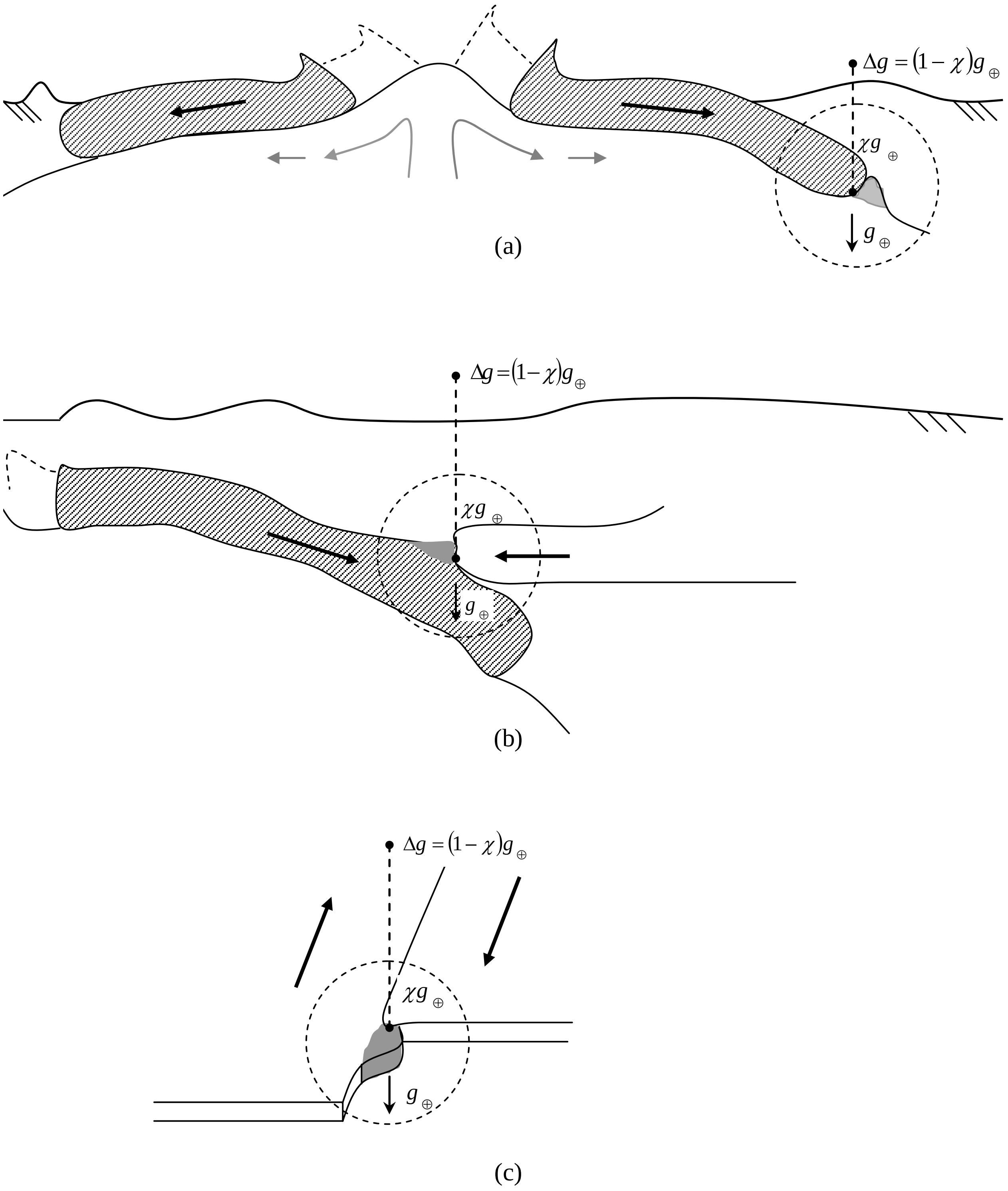

Fig. 7 – *Three main types of movements*: (a) Divergent (tectonic plates diverge). (b) Convergent (plates converge). (c) Transform (plates slide past each other). Earthquakes occur when the obstacle breaks (due to the sudden release of stored energy).

# The Universal Quantum Fluid

**Fran De Aquino**
Maranhao State University, Physics Department, S.Luis/MA, Brazil.

The quantization of gravity showed that the matter is also quantized, and that there is an *elementary quantum of matter*, indivisible, whose mass is $\pm 3.9\times10^{-73}kg$. This means that any body is formed by *a whole number* of these particles (quantization). It is shown here that these elementary *quanta* of matter should fill all the space in the Universe forming a *Quantum Fluid continuous and stationary*. In addition, it is also explained why the Michelson-Morley experiment was not able to detect this Universal Quantum Fluid.

**Key words:** Quantum Fluids, Quantum Gravity, Quantum Cosmology
PACS: 67.10.-j; 04.60.-m; 98.80.Qc

## 1. Introduction

Until the end of the century XX, several attempts to quantize gravity were made. However, all of them resulted fruitless [1, 2]. In the beginning of this century, it was clearly noticed that there was something unsatisfactory about the whole notion of quantization and that the quantization process had many ambiguities. Then, a new approach has been proposed starting from the generalization of the *action function**. The result has been the derivation of a theoretical background, which finally led to the so-sought quantization of gravity and of *matter* [3]. The quantization of matter shows that there is an *elementary quantum of matter* whose mass is $\pm 3.9\times10^{-73}kg$. This means that there are no particles in the Universe with masses smaller than this, and that any body is formed by *a whole number* of these particles. Here, it will be shown that these elementary *quanta* of matter should fill all the space in the Universe, forming a *quantum fluid continuous and stationary*. In addition, it is also explained why the Michelson-Morley experiment found no evidence of the existence of the universal fluid [4]. A modified Michelson-Morley experiment is proposed in order to observe the displacement of the interference bands.

## 2. The Universal Quantum Fluid

The quantization of gravity showed that the matter is also quantized, and that there is an *elementary quantum of matter*, indivisible, whose mass is $\pm 3.9\times10^{-73}kg$ [3].

Considering that the inertial mass of the *Observable Universe* is $M_U = c^3/2H_0G \cong 10^{53}kg$, and that its volume is $V_U = \frac{4}{3}\pi R_U^3 = \frac{4}{3}\pi\left(c/H_0\right)^3 \cong 10^{79}\,m^3$, where $H_0 = 1.75\times10^{-18}s^{-1}$ is the *Hubble constant*, we can conclude that the *number of these particles in the Observable Universe* is

$$n_U = \frac{M_U}{m_{i0(\min)}} \cong 10^{125}\,particles \qquad (1)$$

By dividing this number by $V_U$, we get

$$\frac{n_U}{V_U} \cong 10^{46}\,particles/m^3 \qquad (2)$$

Obviously, the dimensions of the *elementary quantum of matter* depend on its state of compression. In free space, for example, its volume is $V_U/n_U$. Consequently, its "radius" is $R_U/\sqrt[3]{n_U} \cong 10^{-15}m$.

If $N$ particles with diameter $\phi$ *fill all space* of $1m^3$ then $N\phi^3 = 1$. Thus, if $\phi \cong 10^{-15}m$ then the number of particles, with this diameter, necessary to fill all $1m^3$ is $N \cong 10^{45}\,particles$. Since the number of *elementary quantum of matter* in the Universe is $n_U/V_U \cong 10^{46}particles/m^3$ we can conclude that these particles *fill all space* in the Universe, forming a *Quantum Fluid continuous and stationary*, the density of which is

$$\rho_{CUF} = \frac{n_U\, m_{i0(\min)}}{V_U} \cong 10^{-27}\,kg/m^3 \qquad (3)$$

Note that this density is smaller than the density of the *Intergalactic Medium* $\left(\rho_{IGM} \cong 10^{-26} kg/m^3\right)$.

---

* The formulation of the *action* in Classical Mechanics extends to Quantum Mechanics and has been the basis for the development of the *Strings Theory*.

The density of the Universal Quantum Fluid is clearly *not uniform* along the Universe, since it can be strongly compressed in several regions (galaxies, stars, blackholes, planets, etc). At the normal state (free space), the mentioned fluid is *invisible*. However, at *supercompressed* state, it can become *visible by giving origin* to the *known matter*, since matter, as we have seen, is *quantized* and consequently, formed by an *integer number* of elementary quantum of matter with mass $m_{i0(\min)}$. Inside the proton, for example, there are $n_p = m_p / m_{i0(\min)} \cong 10^{45}$ *elementary quanta of matter* at supercompressed state, with volume $V_{proton}/n_p$ and "radius" $R_p / \sqrt[3]{n_p} \cong 10^{-30} m$.

Therefore, the solidification of the matter is just a *transitory state* of this Universal Quantum Fluid, which can turn back into the primitive state when the cohesion conditions disappear.

Due to the cohesion state of the *elementary quanta of matter* in the *Universal Quantum Fluid*, any amount of *linear momentum* transferred to any elementary *quantum* of matter propagates totally to the neighboring and so on, in such way that, during the propagation of the *momentum*, the elementary *quanta* of matter do not move, in the same way as the intermediate spheres in Newton's *pendulum* (the well-known device that demonstrates conservation of *momentum* and energy) [5, 6]. Thus, whether it is a photon that transfers its *momentum* to the elementary *quanta* of matter, then the momentum variation due to the incident photon is $\Delta p = h/\lambda$, where $\lambda$ is its wavelength. As we have seen, the diameter of the elementary *quantum* of matter is $\Delta x \approx 10^{-15} m$. According to the *Uncertainty Principle* the variation $\Delta p$ can only be detected if $\Delta p \Delta x \geq \hbar$. In order to satisfy this condition we must have $\lambda \leq 2\pi\Delta x \approx 10^{-14} m$. This means that *momentum* variations, in the elementary *quanta* of matter, *caused by photons with wavelength* $\lambda > 10^{-14} m$ *cannot be detected.* That is to say that the propagation of these photons through the Universal Quantum Fluid is equivalent to its propagation in the *free space*. In practice, it works as if *there was not the Universal Quantum Fluid*. This conclusion is highly important, because it can easily explain why in the historical Michelson-Morley experiment there was no displacement of the interference bands namely because the wavelength of the light used in the Michelson-Morley experiment was $\lambda = 5\times10^{-7} m$ fact that led Michelson to conclude that the hypothesis of a stationary ether was incorrect. Posteriorly, several experiments [7-13] have been carried out in order to check the Michelson-Morley experiment, but the results basically were the same obtained by Michelson.

Thus, actually there was no displacement of the interference bands in the Michelson-Morley experiment because the wavelength used in the experiment was $\lambda = 5\times10^{-7} m$, which is a value clearly much greater than $10^{-14} m$, and therefore, does not satisfy the condition $\lambda \leq 2\pi\Delta x \approx 10^{-14} m$ derived from the Uncertainty Principle. The substitution of light used in the Michelson-Morley experiment by radiation with $\lambda \leq 10^{-14} m$ is clearly impracticable. However, the Michelson-Morley experiment can be partially modified so as to yield the displacement of the interference bands. The idea is based on the generalized expression for the *momentum* obtained recently[3], which is given by

$$p = M_g V \tag{4}$$

where $M_g = \left| m_g / \sqrt{1 - V^2/c^2} \right|$ is the relativistic gravitational mass of the particle and $V$ its velocity; $m_g = \chi\, m_{i0}$ the *general expression of the correlation between the gravitational and inertial mass*; $\chi$ is the correlation factor[3].Thus, we can write

$$\left| \frac{m_g}{\sqrt{1 - V^2/c^2}} \right| = \left| \frac{\chi\ m_{i0}}{\sqrt{1 - V^2/c^2}} \right| \tag{5}$$

Therefore, we get

$$M_g = |\chi| M_i \tag{6}$$

The Relativistic Mechanics tells us that

$$p = \frac{UV}{c^2} \tag{7}$$

where $U$ is the *total* energy of the particle. This expression is valid for *any* velocity $V$ of the particle, including $V = c$.

By comparing Eq. (7) with Eq. (4) we obtain

$$U = M_g c^2 \tag{8}$$

It is a well-known experimental fact that

$$M_i c^2 = hf \tag{9}$$

Therefore, by substituting Eq. (9) and Eq. (6) into Eq. (4), gives

$$p = \frac{V}{c} |\chi| \frac{h}{\lambda} \tag{10}$$

Note that this expression is valid for *any* velocity $V$ of the particle. In the particular case of $V = c$, it reduces to

$$p = |\chi| \frac{h}{\lambda} \tag{11}$$

By comparing Eq. (10) with Eq. (7), we obtain

$$U = |\chi| hf \tag{12}$$

Note that only for $\chi = 1$ Eq. (11) and Eq. (12) are reduced to the well=known expressions of DeBroglie $(q = h/\lambda)$ and Einstein $(U = hf)$.

Equations (10) and (12) show, for example, that *any* real particle (material particles, real photons, etc) that penetrates a region (with density $\rho$, conductivity $\sigma$ and relative permeability $\mu_r$), where there is an electromagnetic field $(E, B)$, will have its *momentum* $p$ and its energy $U$ reduced by the factor $|\chi|$, where $\chi$ is given by[3]:

$$\chi = \frac{m_g}{m_{i0}} = \left\{1 - 2\left[\sqrt{1 + \left(\frac{\Delta p}{m_{i0}c}\right)^2} - 1\right]\right\} =$$

$$= \left\{1 - 2\left[\sqrt{1 + 1.758 \times 10^{-27}\left(\frac{\mu_r \sigma^3}{\rho^2 f^3}\right) c^4 B_{rms}^4} - 1\right]\right\} \tag{13}$$

where $B_{rms}$ is the *rms* value of the magnetic field $B$.

The remaining amount of *momentum* and *energy*, respectively given by $(1 - |\chi|)\left(\frac{V}{c}\right)\frac{h}{\lambda}$ and $(1 - |\chi|)\, hf$, are *transferred to* the *imaginary* particle associated to the *real* particle† (material particles or real photons) that penetrated the mentioned region.

It was previously shown that, when the *gravitational mass* of a particle is reduced to a range between $+0.159 M_i$ to $-0.159 M_i$, i.e., when $|\chi| < 0.159$, it becomes *imaginary*[3], i.e., the gravitational and the inertial masses of the particle becomes *imaginary*. Consequently, the particle disappears from our ordinary space-time. It goes to the Imaginary Universe. On the other hand, when the gravitational mass of the particle becomes greater than $+0.159 M_i$, or less than $-0.159 M_i$, i.e., when $|\chi| > 0.159$, the particle return to our Universe.

Figure 1 (a) clarifies the phenomenon of reduction of the *momentum* for $|\chi| > 0.159$, and Figure 1 (b) shows the effect in the case of $|\chi| < 0.159$. In this case, the particles become imaginary and, consequently, they go to the *imaginary space-time* when they penetrate the electric field $E$. However, the electric field $E$ stays in the *real* space-time. Consequently, the particles return immediately to the real space-time in order to return soon after to the *imaginary* space-time, due to the action of the electric field $E$. Since the particles are moving at a direction, they *appear* and *disappear* while they are crossing the region, up to collide with the plate (See Fig.1) with a *momentum*, $p_m = |\chi|\left(\frac{V}{c}\right)\frac{h}{\lambda}$, in the case of a *material particle*, and $p_r = |\chi|\frac{h}{\lambda}$ in the case of a *photon*.

If this photon transfers its *momentum* to elementary *quanta* of matter $(\Delta x \approx 10^{-15} m)$, then the momentum variation due to the incident photon is $\Delta p = |\chi| h/\lambda$. According to the *Uncertainty Principle* the variation $\Delta p$ can only be detected if $\Delta p \Delta x \geq \hbar$, i.e., if

$$\lambda \leq 2\pi |\chi| \Delta x \tag{14}$$

We conclude, then, that the *interaction* between the light used in the Michelson-

† As previously shown, there are *imaginary particles* associated to each *real particle*[3].

Morley experiment $(\lambda = 5\times10^{-7}m)$ and the Universal Quantum Fluid *just can be detected*, and to produce of the displacement of the interference bands, if

$$|\chi| \geq 8\times10^{7} \quad (15)$$

In order to satisfy this condition in the Michelson-Morley experiment, we must modify the medium where the experiment is performed (for example substituting the air by *low-pressure Mercury plasma*), and apply through it an electromagnetic field with frequency $f$. Under these conditions, according to Eq. (13), the value of $|\chi|$ will be given by

$$|\chi| = \left|\left\{1-2\left[\sqrt{1+1.758\times10^{-27}\left(\frac{\mu_r\sigma^3}{\rho^2 f^3}\right)c^4 B_{rms}^4}-1\right]\right\}\right| \quad (16)$$

If the *low-pressure Mercury plasma* is at $P = 6\times10^{-3}Torr = 0.8N.m^{-2}$ and $T \cong 318.15\ K$ [14], then the mass density, according to the well-known *Equation of State*, is

$$\rho = \frac{PM_0}{ZRT} \cong 6.067\times10^{-5}kg.m^{-3} \quad (17)$$

where $M_0 = 0.2006 kg.mol^{-1}$ is the molecular mass of the Hg; $Z \cong 1$ is the *compressibility factor* for the Hg plasma; $R = 8.314\, joule.mol^{-1}.^0K^{-1}$ is the *gases universal constant*.

The electrical conductivity of the Hg plasma, under the mentioned conditions, has already been calculated [15], and is given by

$$\sigma \cong 3.419\ \ S.m^{-1} \quad (18)$$

By substitution of the values of $\rho$ and $\sigma$ into Eq. (16) yields

$$|\chi| = \left|\left\{1-2\left[\sqrt{1+1.5471\times10^{17}\frac{B_{rms}^4}{f^3}}-1\right]\right\}\right| \quad (19)$$

By comparing with (15), we get

$$\frac{B_{rms}^4}{f^3} \geq 0.01037 \quad (20)$$

Thus, for $f = 1Hz$, the ELF magnetic field must have the following intensity:

$$B_{rms} \geq 0.32T \quad (21)$$

This means that, if in the Michelson-Morley experiment the air is substituted by *Hg plasma* at $6\times10^{-3}Torr$ and $318.15\ K$, and an ELF magnetic field with frequency $f = 1Hz$ and intensity $B_{rms} \geq 0.32T$ is applied through this plasma (Fig. 2), *then the displacement of the interference bands should appear*.

It is important to note that due to the *Gravitational Shielding effect* [3], the gravity above the magnetic field is given by $\chi\ g \geq 7.8\times10^{8}m.s^{-2}$. This value, extends above the vacuum chamber for approximately *10 times* its length. In order to eliminate this problem we can replace the ELF magnetic field, $B$, shown in Fig. 2, by two ELF magnetic fields, $B_1$ and $B_2$, sharing the same frequency, $f = 1Hz$. The field, $B_1$, is placed vertically through the region of the experimental set-up. The field, $B_2$, is also placed vertically, just above $B_1$ (See Fig. 3). Thus, the gravity above $B_2$ is given by $\chi_1\chi_2 g$ where $\chi_1 = m_{g1}/m_{i1}$ and $\chi_2 = m_{g2}/m_{i2}$ are respectively, the correlation factors in the Gravitational Shieldings 1 and 2, produced by the ELF magnetic fields $B_1$ and $B_2$, respectively. In order to become $\chi_1\chi_2 g = g$ we must make $\chi_2 = 1/\chi_1 = 1/-8\times10^{7}$. According to Eq. (19), this value can be obtained if $B_{rms(2)} = 5.331481522\times10^{-5}T$ and $B_{rms(1)} = 0.32T$. Note that the value of $B_{rms(2)}$ is less than the value of the Earth's magnetic field ($B_{\oplus} \cong 6\times10^{-5}T$). However, this is not a problem because the steel of the vacuum chamber works as a magnetic shielding, isolating the magnetic fields inside the vacuum chamber.

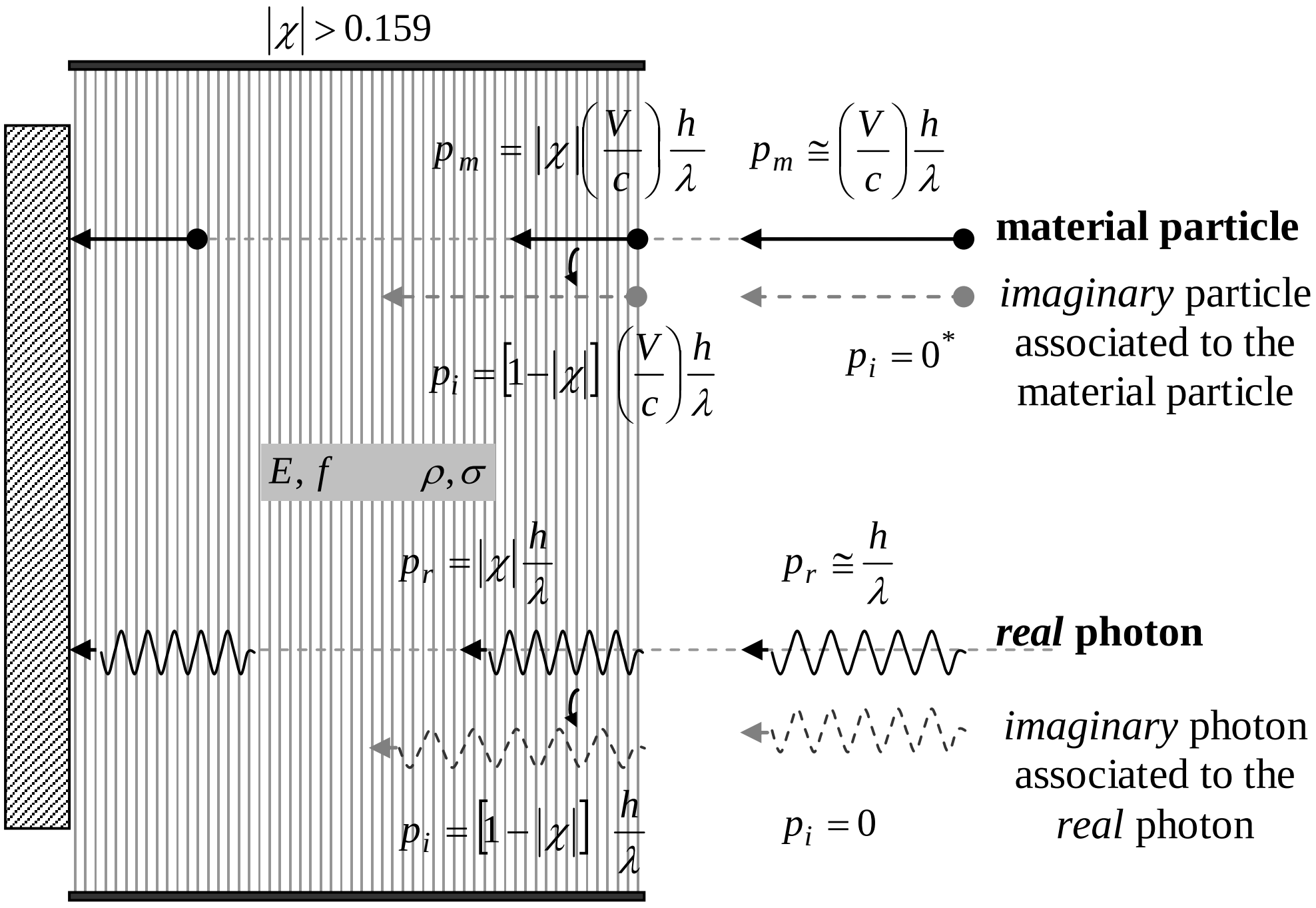

(a)

* There are a type of neutrino, called "ghost" neutrino, predicted by General Relativity, with *zero mass* and *zero momentum*. In spite its *momentum be zero*, it is known that there are wave functions that describe these neutrinos and that prove that really they exist.

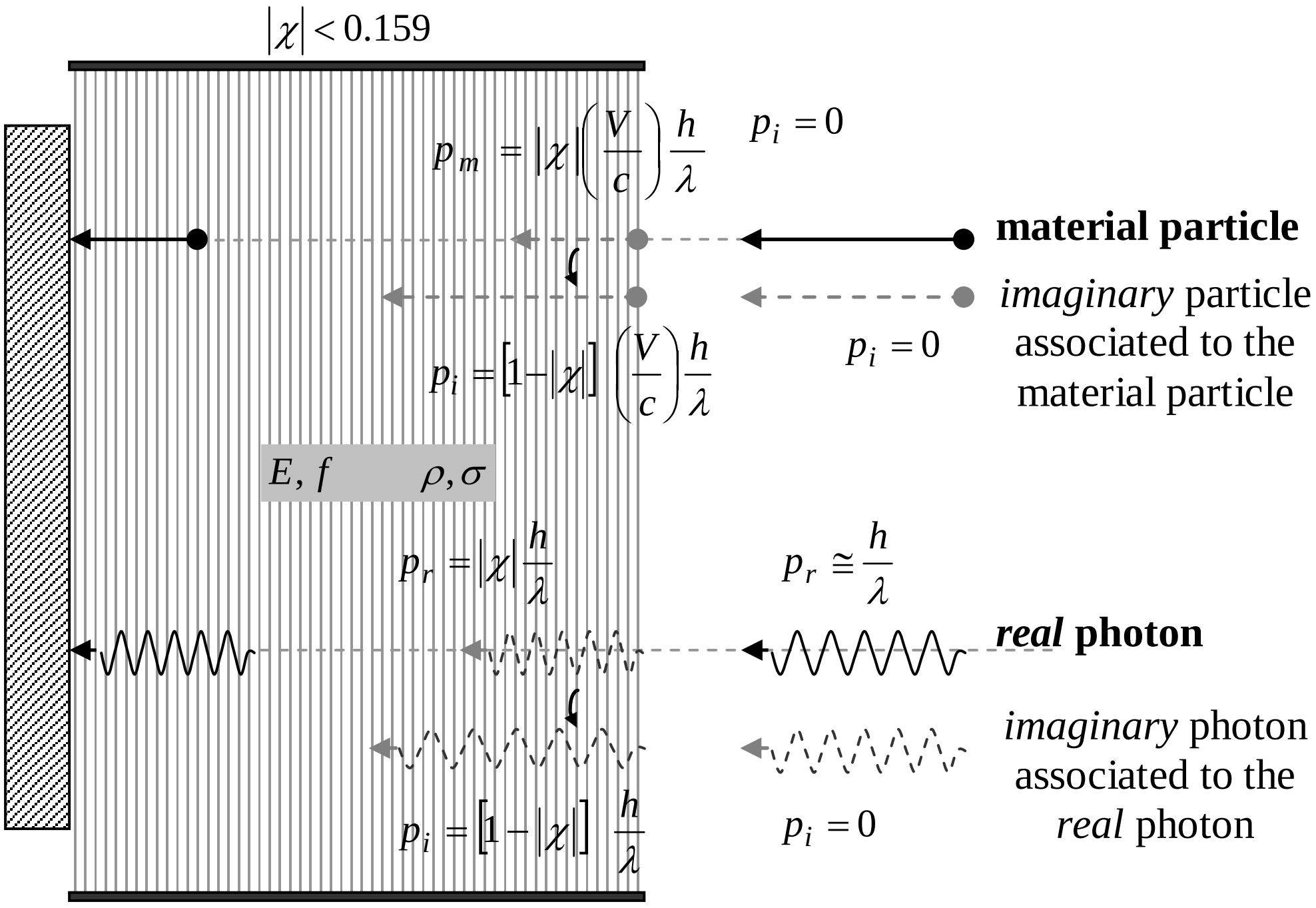

(b)

Fig. 1 –*The correlation factor in the expression of the Momentum*. (a) Shows the *momentum* for $|\chi| > 0.159$. (b) Shows the effect when $|\chi| < 0.159$. Note that in both cases, the *material* particles collide with the cowl with the *momentum* $p_m = |\chi|(V/c)\frac{h}{\lambda}$, and the photons with $p_r = |\chi|\frac{h}{\lambda}$ .

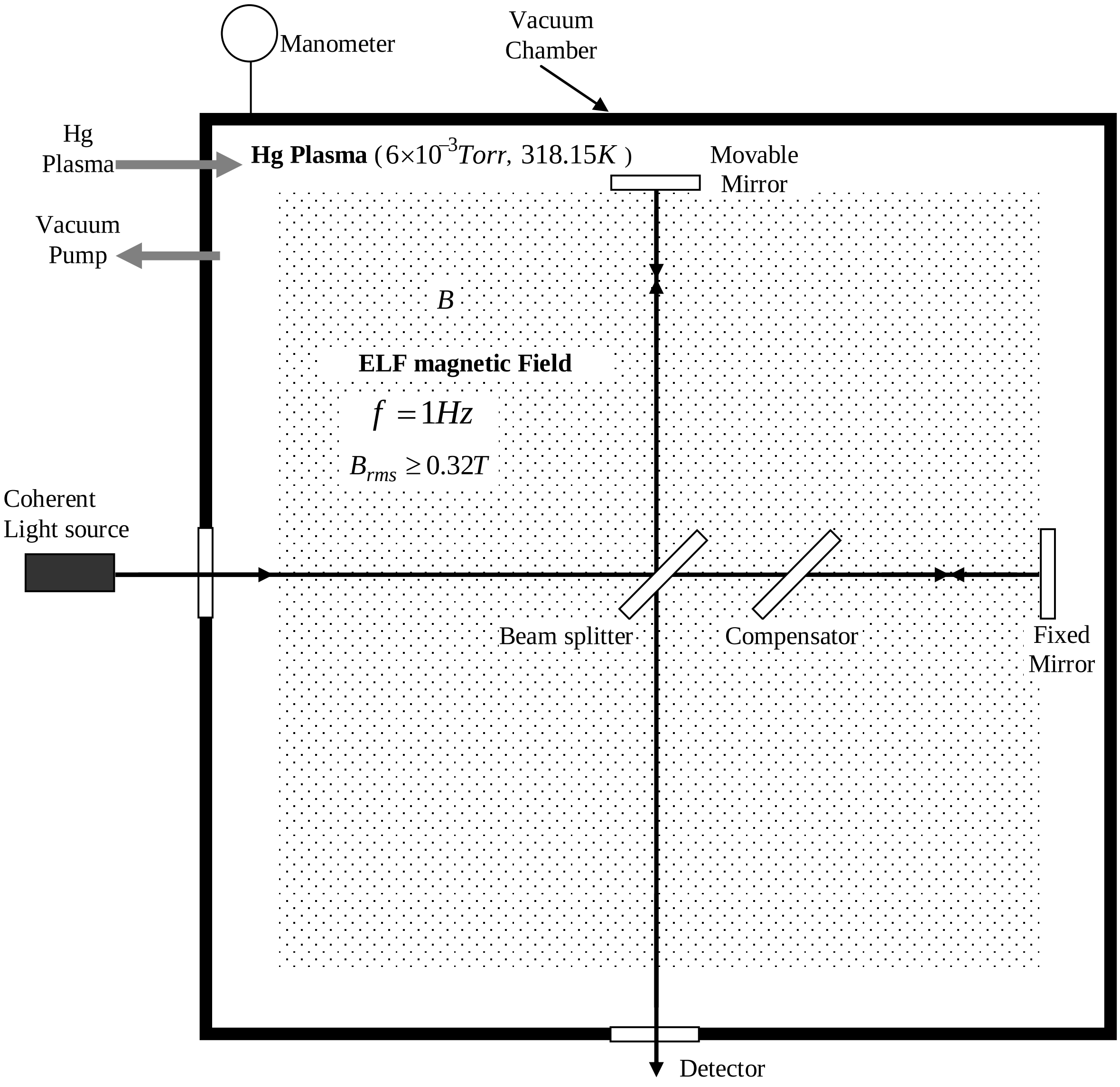

Fig. 2 - *The modified Michelson-Morley experiment*. The air is substituted by *Hg plasma* at $6\times10^{-3}Torr$ and $318.15\ K$, and an ELF magnetic field with frequency $f = 1Hz$ and intensity $B_{rms} \geq 0.32T$ is applied through this plasma, *then the displacement of the interference bands should appear.*

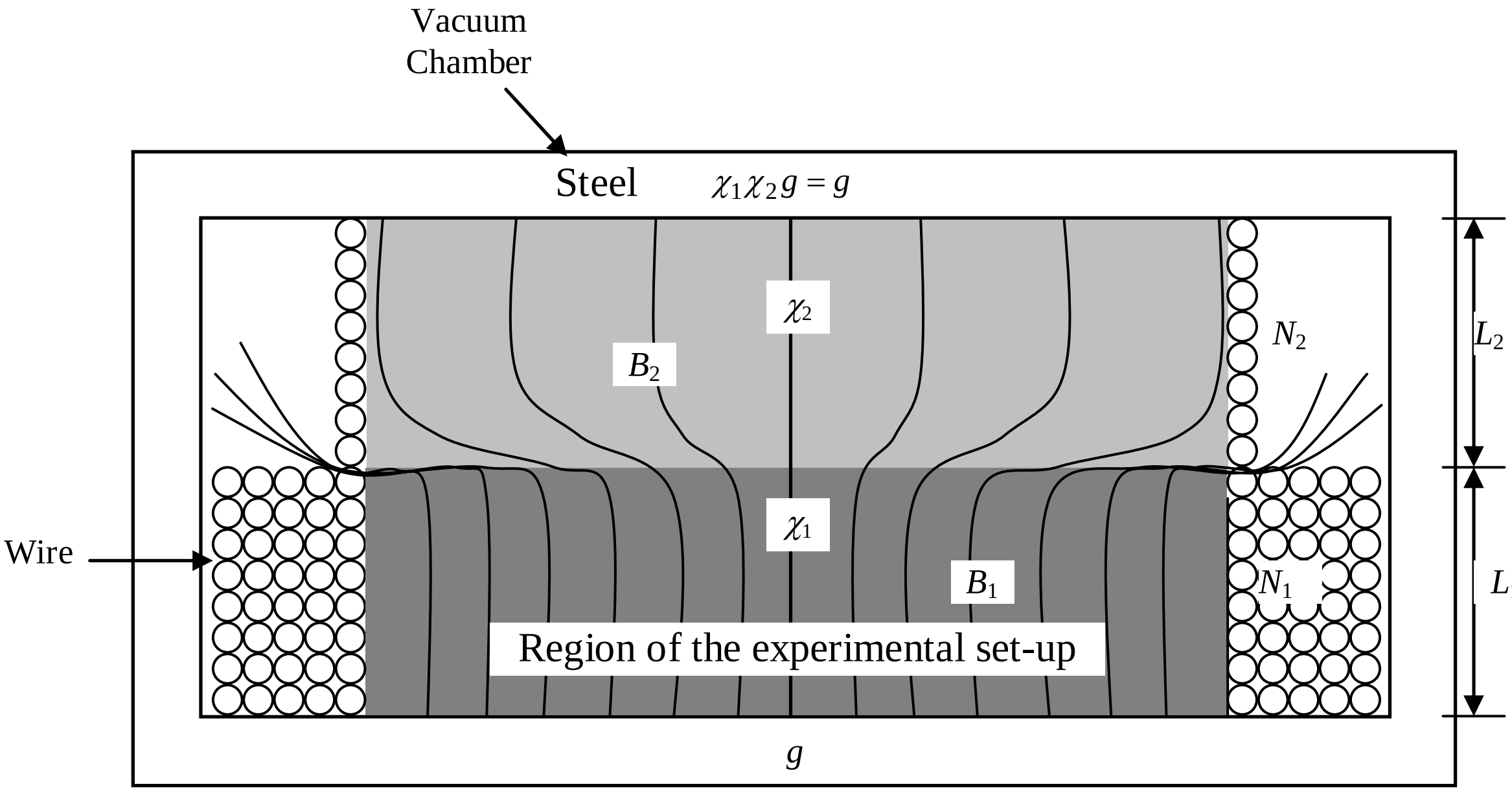

Fig. 3 – *Cross-section of the vacuum chamber showing the magnetic fields* $B_1$ and $B_2$.

$$B_1 = \mu_0 \left( \frac{N_1}{L_1} \right) i_1 \; ; \; B_2 = \mu_0 \left( \frac{N_2}{L_2} \right) i_2$$

# The Gravitational Mass of a Charged Supercapacitor

**Fran De Aquino**
Maranhao State University, Physics Department, S.Luis/MA, Brazil.

Electric double-layer capacitors (EDLCs), also known as supercapacitors, electrochemical double layer capacitors, or ultracapacitors, are electrochemical capacitors that have an unusually high energy density when compared to common capacitors, typically on the order of thousands of times greater than a high capacity electrolytic capacitor. It is shown here that when an EDLC is fully charged its *gravitational mass* is considerably greater than when it is discharged.

**Key words:** Supercapacitors, Energy storage systems, Experimental tests of gravitational theories
PACS: 88.80.fh; 84.60.Ve , 04.80.Cc

## 1. Introduction

The electric double-layer capacitor effect was first noticed in 1957 by General Electric engineers experimenting with devices using porous carbon electrode [1]. It was believed that the energy was stored in the carbon pores and it exhibited "exceptionally high capacitance", although the mechanism was unknown at that time.

General Electric did not immediately follow up on this work, and the modern version of the devices were eventually developed by researchers at Standard Oil of Ohio in 1966, after they accidentally re-discovered the effect while working on experimental fuel cell designs [2]. Their cell design used *two layers of activated carbon* separated by a thin porous insulator, and this basic mechanical design remains the basis of most electric double-layer capacitors to this day.

An electric double-layer capacitor (EDLC), is known as *supercapacitor*, or *ultracapacitor*. Their energy density is typically hundreds of times greater than conventional electrolytic capacitors. They also have a much higher power density than batteries or fuel cells. As of 2011 EDLCs had a maximum working voltage of 5 volts and capacities of up to 5,000 farads [3].

Currently, the EDLCs are used for energy storage rather than as general-purpose circuit components. The EDLCs also have two metal plates, but they are coated with *activated carbon* immersed in an electrolyte, and separated by an intervening insulator, forming in this manner, the double-layer of activated carbon inside the capacitor. During the charging process, ions from the electrolyte accumulate on the surface of each carbon-coated plate.

Here it is shown that when an EDLC is fully charged its *gravitational mass* is considerably greater than when it is discharged.

## 2. Theory

Consider the cross-section of an EDLC as shown in Fig. 1. The double-layer in the EDLCs is generally made of *activated carbon* immersed in an *electrolyte whose conductivity is much less than carbon conductivity* [4]. The result is that the conductivity of the double-layer becomes much less than the conductivity of the activated carbon and, in this way, the double-layer can withstand a low voltage, and *no significant current flows through the activated carbon layers* of an ELDC [3]. This means that they are similar to dielectrics with very low *dielectric strength*. Thus, due to the electrical charge stored in the activated carbon layers, each layer can be considered as a *non-conducting plane of charge,* with density of charge, $\sigma = q/S$ , where $S$ is the area of the plates of the capacitor, and $q = CV$ is the amount of electrical charge stored in the activated carbon layers; $C$ is the capacity of the EDLC. Thus, according to the well-known expression of the electric field produced by a *non-conducting plane of charge* [5], we can conclude that the electric field *through the layers* of activated carbon (See Fig.1) is given by

$$E_{(layer)} = E^- = E^+ = \frac{\sigma}{2\varepsilon_{r(layer)}\varepsilon_0} = \frac{CV}{2\varepsilon_{r(layer)}\varepsilon_0 S} \quad (1)$$

Consequently, the density of electromagnetic energy in the carbon layers is

$$W_{layer} = \tfrac{1}{2}\varepsilon_{r(layer)}\varepsilon_0 E_{layer}^2 = \frac{1}{8\varepsilon_{r(layer)}\varepsilon_0}\left(\frac{CV}{S}\right)^2 \quad (2)$$

It was shown that the *relativistic gravitational mass* $M_g$ is correlated with the *relativistic inertial mass* $M_i$ by means of the following factor [6]:

$$M_g = |\chi| M_i \quad (3)$$

where $\chi$ can be expressed by

$$\chi = \left\{1 - 2\left[\sqrt{1 + \left(\frac{n_r W}{\rho c^2}\right)^2} - 1\right]\right\} \quad (4)$$

where $n_r$ is the refraction index and $\rho$ the density of the material.

Substitution of Eq. (2) into Eq. (4), yields

$$\chi = \left\{1 - 2\left[\sqrt{1 + \left[\frac{n_{r(layer)}\mu_0}{8\varepsilon_{r(layer)}\rho_{layer}}\left(\frac{CV}{S}\right)^2\right]^2} - 1\right]\right\} \quad (5)$$

In the case of activated carbon layer: $n_{r(layer)} \cong 1$; $\varepsilon_{r(layer)} \cong 12$ and $\rho_{layer} \cong 800 kg.m^{-3}$. Thus, if the *supercapacitor* has $C = 3{,}000F$; $S = 0.08 \times 0.45 = 0.036m^2$ and is subjected to $V = 4Volts$ then Eq. (5) gives

$$\chi = -1.14 \quad (6)$$

Substitution of Eq.(6) into Eq. (3) yields

$$M_{g(layer)} = |-1.14| M_{i(layer)} = 1.14 M_{i(layer)} \quad (7)$$

This means an *increase* of 14% in the gravitational mass of the double-layer when the supercapacitor is fully charged. Since the mass of the double-layer is a significant part of the total mass of the supercapacitor, we can conclude that, *when fully charged the supercapacitor will display considerably more mass than when it is discharged.*

It is important to note that the gravitational mass of the double-layer can also be *reduced*, *decreasing the total mass* of the supercapacitor. This can occur, for example, when $1.5 Volts < V < 3.1 Volts$.

**Conclusion**

The theoretical results here obtained for the gravitational mass of an EDLC are general for *energy accumulator cells* which contain *non-conducting planes of charges* similar to the *activated carbon + electrolyte* layers of the EDLCs.

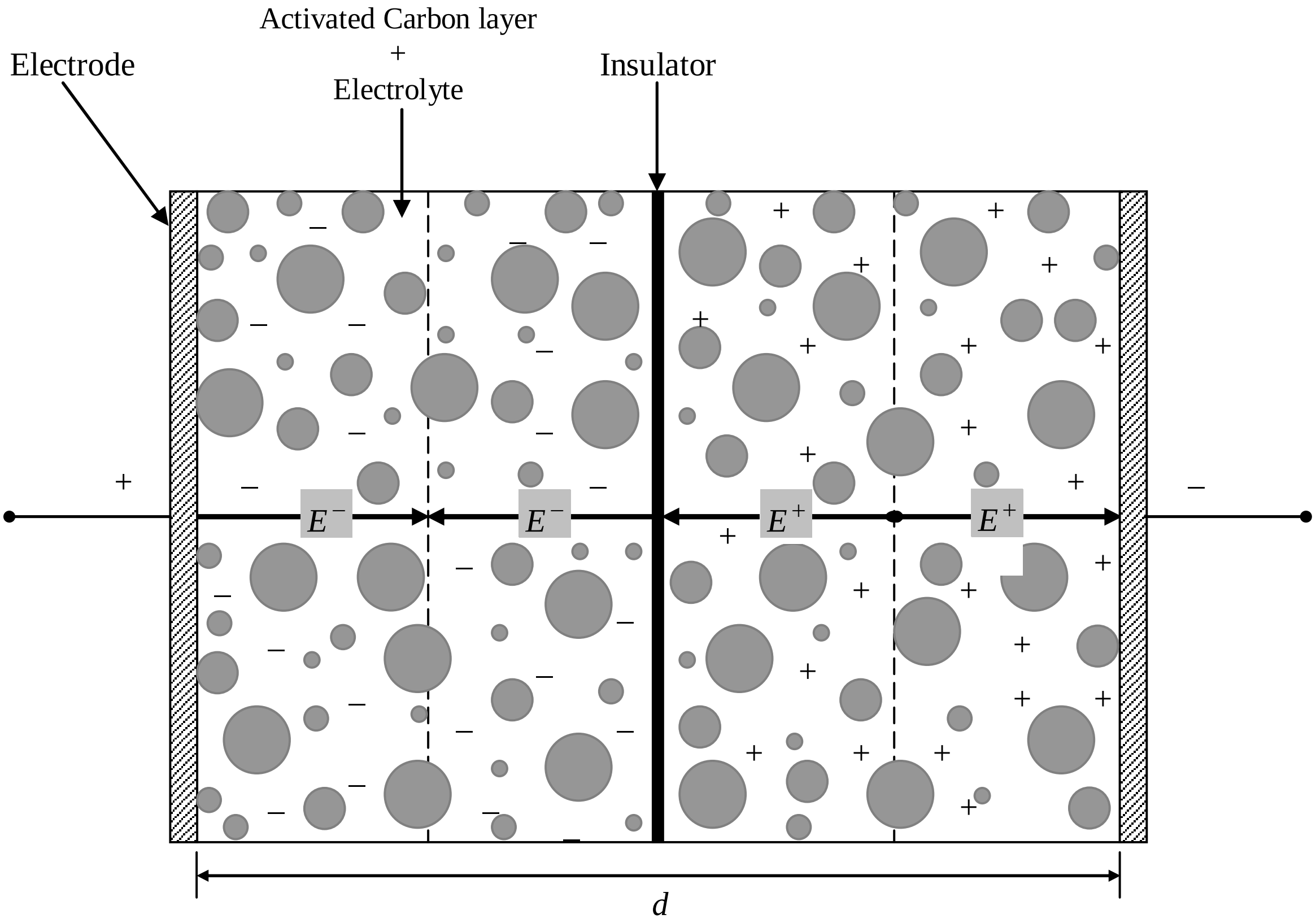

Fig. 1 – Cross-section of an *Electric Double-Layer Capacitor* (Supercapacitor) - Each *activated carbon + electrolyte* layer works as a *non-conducting plane of charge*, with density of charge $\sigma^- = q^-/S$ and $\sigma^+ = q^+/S$ respectively. The electric fields *through the layers*, due to these densities of charges $(E^-, E^+)$, are shown in the figure above.

# Beyond the Material Universe

**Fran De Aquino**
Maranhao State University, Physics Department, S.Luis/MA, Brazil.

Science and Religion have always observed the events from their exclusive viewpoint. It was necessary the arising of a bond that would make possible the *unification* of them. This bond was revealed in the last decades by Quantum Physics, which has shown us that some physical laws extend beyond the material world, pointing to the existence of the spiritual world. Thus, the spiritual world exists now no longer as supernatural one, but as world as real as our material world. This discovery marked the beginning of the understanding of the spiritual world *nature* and *its relationship with the material world*. Starting from this knowledge, here widely detailed, it is now possible to understand the eternal puzzles: *where have we come from, because here we are, and where do we go*.

**Key words:** Science, Religion, Spiritual World, Quantum Gravity, Quantum Cosmology, Quantum Consciousness.

## 1. Introduction

The Spiritual World has always been considered something supernatural. Only recently, with the advent of Quantum Physics, the first evidences of its existence arose. However, it was the theoretical background derived from Quantum Gravity [1] that has shown that our *Real* Universe is contained in an *Imaginary* Universe. Here, the terms real and imaginary are borrowed from mathematics (real and imaginary numbers). In addition, it has been possible to show that the Imaginary Energy has the same characteristics of the Psychic Energy, that is, both are equivalent. This means that the Imaginary Universe is in fact the Psychic Universe. This discovery was the starting point for understanding the *nature* of the Spiritual World and beings it contains. It also made possible the acquisition of a strong knowledge about the relationship of the beings of the Spiritual World with us, and with our Material World. This knowledge, which leads us beyond the Material Universe, is widely detailed in this work.

## 2. The Psychic Energy

It is known that the De Broglie waves are characterized by a variable quantity called the *wave function*, denoted by the symbol $\psi$. While the frequencies of the De Broglie waves are determined by a simple form, the value of $\psi$ is usually very complicated. The value of $\Psi^2$ (or $\Psi\Psi^*$) calculated for a particular point x, y, z, t is proportional to the probability of finding the particle experimentally in that place and time[1].

Thus, each particle has a particular *wave function*, which describes the particle fully. Roughly, it is similar to the "identity card" of the particle, containing all information about the particle or the body.

Since $\Psi^2$ is proportional to the probability $P$ of finding the particle described by $\Psi$, the integral of $\Psi^2$ on the *whole space* must be finite – inasmuch as the particle is someplace. Therefore, if

$$\int_{-\infty}^{+\infty} \Psi^2 dV = 0 \tag{01}$$

The interpretation is that the particle does not exist. However, if

[1] Interpretation by Max Born in 1926.

$$\int_{-\infty}^{+\infty} \Psi^{2} dV = \infty \qquad (02)$$

*the particle will be everywhere simultaneously.*

Despite the fact that the De Broglie waves are normally associated to material particles and, in general associated to material bodies, it is known that they are also associated to exotic particles that cannot even be detected, such as a type of neutrinos called "ghost" neutrinos, predicted by General Relativity. These are called "ghost" neutrinos, because they have *zero mass* and *zero momentum*, and cannot be detected. But even so, it is known that there are wave functions that describe them, which means that they exist and can be present in any place. As a rough analogy, it is like a person who, despite of existing and possessing an identity card, is never seen by anyone. The existence of a wave function associated with the "ghost" neutrino is very important, because, in this context, we can conclude that even a *thought* may have a wave function associated with it.

It is a proven quantum fact that the wave function $\Psi$ may "collapse" and, in that instant, the possibilities that it describes, suddenly express themselves in *reality*. The moment of the "collapse" of the wave function is then a decision point where the pressing need of *realization* of the possibilities described by the wave function occurs.

For an observer in our universe something is *real* when it is in the form of *matter* or *radiation*. Therefore, it can occur that the possibilities described by the wave function realize themselves in the form of radiation, i.e., they did not materialize. This obviously must occur when the energy that forms the content described by the wave function *is not equal* to the amount of energy needed for its materialization.

Then consider any *thought*. A thought is a psychic body, with a well-defined psychic energy, and, as such, with a wave function of its own as any other *psychic body*. When its wave function collapses, two possibilities may occur: (a) the psychic energy contained in the thought *is not sufficient* to materialize its content - in this case, the collapse of the wave function is *realized* in the form of radiation: (b) the psychic energy *is sufficient* for its materialization - in this case, the collapse of the wave function content will be fully materialized.

However, in both cases, there must always be production of "virtual" photons to communicate the psychic interaction to other bodies of the universe, because, according to the quantum theory, only through this type of *quanta*, can the interaction be communicated, since it has infinite range just in the same way as the electromagnetic interaction has, which we know is communicated by exchanging "virtual" photons. The term "virtual" derives from the *principle of uncertainty* due to the impossibility to detect these photons. This is a limitation imposed by Nature proper.

It can be easily seen that this *materialization process,* although theoretically possible, requires enormous amounts of psychic energy, because, according to Einstein's famous equation $E = mc^2$, even a tiny object contains an enormous amount of energy. Moreover, one can conclude that materializations of this type could only be produced by great consciousness with large psychic energy. In addition, it is evident, in this context, that the larger the amount of psychic energy of a consciousness the greater its chances of realization.

This is a materialization process that can explain the materialization of the Primordial Universe. In addition, it becomes evident that the Psychic Energy is a type of mother-energy or Universal Fluid, which can produce anything.

Since it is in the continuum 4-dimensional (space-time) that the realization of the wave functions which describe the psychic bodies occurs, then we can assume that they are generated in a continuum that, despite of containing all psychic forms, also interpenetrates the space-time continuum. Let us call it, hereafter, Psychic Continuum. By definition this continuum should also contain the Supreme Consciousness. Therefore, it should be infinite.

Then, from the above we can see that an accurate description of the universe cannot exclude the psychic energy, psychic particles and psychic bodies. That is, the situation calls for a cosmology that includes the psyche in the description of the Universe, thus complementing the traditional cosmology that is only a matter of cosmology. This idea is not new; it has existed for some time and seems to have arisen mainly in Princeton and Pasadena in the USA in the 70s [2] as a result of the joint effort of eminent physicists, biologists, psychologists and theologians as well. [3].

In traditional cosmology, the universe comes from a big bang where everything that exists would be concentrated in a tiny particle with the size of a proton and an enormous mass equal to the mass of the Universe at the instant before the Big Bang,. But its origin is not explained, nor the *why* of its *critical volume*.

The critical volume, in our opinion denotes *knowledge* of what would happen from these initial conditions, a fact which points to the *existence of a Creator*. In this case, the materialization process described above would explain the materialization of the Primordial Universe. That is, the Primordial Universe arose at the exact moment that *the primordial wave function* collapsed (initial instant) realizating the contents of the psychic form generated in the consciousness of the Creator when He *thought* of creating the universe.

The psychic form described by this primordial wave function must then have been generated in a consciousness with psychic energy much larger than that needed to materialize the Universe. This enormous consciousness in turn, not only would be the biggest of all consciences, but also would be the *substratum* of everything that exists, and obviously everything that exists would be wholly contained therein, including *all* space-time.

Based on the General Theory of Relativity and the recent cosmological observations, it is known that the Universe occupies a space of positive curvature. This space, as we know, is "closed in itself", its volume is *finite*, but rest well understood, the space has no boundaries, it is *unlimited*. Thus, if the consciousness which we refer to contains *all* the space, its volume is necessarily infinite, and therefore contains an infinite psychic energy.

This means that it contains *all the* psychic energy that exists and therefore, any other consciousness will be contained in it. Thus, we can conclude that it is the *Supreme Consciousness*, and there is no other like it: it is *unique*. In addition, since it contains all the psychic energy, it can accomplish *everything* that it wants, and therefore is *omnipotent*. Previously, we showed in the article "Physical Foundations of Quantum Psychology" the manifestation of knowledge, or *auto-accessible knowledge* in a consciousness, should be related to its quantity of psychic energy. In the Supreme Consciousness, whose psychic energy is infinite, the manifestation of the knowledge is *total*, thus, necessarily, it must be *omniscient*. Being omniscient, we cannot doubt its justice or goodness. Thus, God is supremely *just* and *good*. Moreover, as it also contains *all the time*, with past, present and future merging into it in an eternal present, so that the time will not flow as it does for us, in the four-dimensional continuum called space-time, as "we see" the future changing continuously into present and the present into past. Similarly, an observer in five-dimensional continuum would not have access to all time as the Supreme Consciousness, but his dimensional "vision" of the time would certainly be wider than that of the observer positioned in the four-dimensional continuum [3]. In this context, only the Supreme Consciousness would have a perfect "vision" of all dimensional levels.

When we speak of creation of the universe, the use of the verb to *create* means that anything that was not came to be; assuming therefore, the concept of time flow. For the Supreme Consciousness however, the instant of the creation coincides with all the other times, not existing neither *before* nor *after* the creation, and in this way, questions like "*What did the Supreme Consciousness do before creation*?” are not justified.

We can also infer from the above that the *existence* of the Supreme Consciousness has no defined limit (beginning and end), which gives It the peculiar qualities of *uncreated* and *eternal*.

Being eternal, its wave function $\psi$ never collapses. Moreover, as it has infinite psychic energy, the value of $\psi$ will also be infinite. In this way, according to Quantum Mechanics, it means that the Supreme Consciousness *is simultaneously everywhere*, or *omnipresent*.

All conclusions presented here about the Supreme Consciousness were mathematically demonstrated in the article "*Mathematical Foundations of the Relativistic Theory of Quantum Gravity*" [1], and represent nothing more than a formal finding of what was already accepted by most religions.

It is then justified the intuitive feeling that people have about the existence of God, and reveals that God is the Supreme Consciousness, *the first cause of all things.*

Although we can understand this and, thus, learn that God is psychic energy, we can say nothing about the nature of psychic energy. Likewise, we do not know the nature of electric charge, etc. This is a limitation imposed by the Creator Himself.

The option of the Supreme Consciousness in materializing the Primordial Universe into a critical volume, as we have seen, means that It knew what would happen from that initial condition. Therefore It also knew how the universe would behave under existing laws. Thus, the laws were not created *for* the Universe, and therefore are not "laws of Nature" or "laws that have been placed in Nature" as Descartes wrote. They already existed as an intrinsic part of the Supreme Consciousness; Thomas Aquinas had a very clear understanding about this. He speaks of the Eternal Law, "... those *exist* in the mind of God and rules the entire Universe."

The Supreme Consciousness had then complete freedom to choose the initial conditions of the Universe. But opted for the concentration of the early Universe in the critical volume in order that its development should be performed in the most convenient way for the purposes It had in mind, according to laws inherent in its very nature. This responds to Einstein's famous question: "What level of choice would God have had to build the universe?"

It seems that Newton was the first to realize the divine option. In his book *Opticks*, he gives us a perfect view of how he imagined the creation of the Universe:

"It seems to me that God, at the beginning, first gave way to matter in solid particles, compacted [...] in such way that best contribute to the purpose He had in mind ..."

For what purpose the Supreme Consciousness created the universe? This is a question that seems difficult to answer. However, if we assume the natural desire of the Supreme Consciousness of *procreating*, that is, of generating individual consciousness from itself so that they could evolve and express themselves creating attributes to Her, then, we can infer that in order to evolve, such consciousness would need a Universe, and this may have been the main reason for its creation. Thus, the origin of the Universe would be related to the generation of said consciousness and, consequently, the materialization of the Primordial Universe must have occurred at the same time at which the Supreme Consciousness decided *to individualize* the Primordial Consciousnesses.

As the Supreme Consciousness occupies *all* the space, it follows that it cannot be displaced by another consciousness, and not for himself. Therefore, the Supreme Consciousness is *immovable*.

As Augustine says (Gen. Ad lit vii, 20), "The Creator does not move either in time or space."

The immobility of God had been deemed necessary also by Thomas Aquinas, "we can infer be necessary that the God who puts into motion all the objects, it is immovable." (Summa Theologica).

Due to fact that they were individualized directly from the Supreme Consciousness, the primordial consciousnesses certainly contained in themselves - albeit in a latent state, all the possibilities of the Supreme Consciousness, including the germ of independent will that allows the establishment of original points of departure. However, although similar to the Supreme Consciousness, the primordial consciousnesses could have no understanding of themselves. This understanding comes only with the *creative mental state* that the consciousnesses can only achieve by evolution.

Thus, in the first evolutionary period, the primordial consciousnesses must have remained in complete unconsciousness. It was then, the beginning of a pilgrimage from *unconsciousness* to the *superconsciousness*.

## 3. The Good and the Evil

Basically, in the Universe there are two types of radiation: the *real* radiation constituted by of *real photon*, and the “virtual” radiation constituted by "virtual" photons. Previously, we talked about the "virtual" *quanta*, which are responsible for the interaction among the psychic particles. According to the Uncertainty Principle, “virtual” *quanta* cannot be observed experimentally. However, since they are interaction *quanta*, their effects may be verified in the very particles or bodies subjected to the interactions.

Obviously, only one specific type of interaction occurs between two particles if each one *absorbs* the *quanta* of said interaction emitted by the other; otherwise, the interaction will be null. Thus, the null interaction between psychic bodies particularly means that there is no mutual absorption of the “virtual” psychic photons (psychic interaction *quanta*) emitted by them. That is, the emission spectrum of each one of them does not coincide with the absorption spectrum of the other.

It was shown that, in *all interactions* (gravitational, electromagnetic, strong nuclear and weak nuclear), the "virtual" *quanta* are "virtual" *photons* [1].

It is obvious, then, that an interaction between two particles only occurs if each of them *absorbs* the *"virtual" photons* emitted by the other, otherwise the interaction will be zero. Thus, the null interaction means specifically that the emission spectrum of each particle does not coincide with the absorption spectrum of the other.

By analogy with material bodies, the emission spectra of which are, as we know, identical to the absorption spectra, also the psychic bodies must absorb radiation within the spectrum they emit. Specifically, in the case of human consciousness, their thoughts cause them to become emitters of psychic radiation in certain frequency spectra and, consequently, receivers in the same spectra. Thus, when a human consciousness, by its thoughts, is receptive to a radiation coming from a certain thought, said radiation will be absorbed by the consciousness (resonance absorption). Under these circumstances, the radiation absorbed must *stimulate* – through the *Resonance Principle* – said consciousness to emit in the same spectrum, just as it happens with matter.

Nevertheless, in order for that emission to occur in a human consciousness, it must be preceded by the individualization of thoughts identical with that which originated the radiation absorbed, because obviously only identical thoughts will be able to reproduce - when they collapse - the spectrum of “virtual” psychic radiations absorbed.

These *induced thoughts* – such as the thoughts of consciousnesses themselves – must remain individualized for a period of time $\Delta t$ (lifetime of the thought) after which its wave functions collapse, thus producing the “virtual” psychic radiation in the same spectrum of frequencies absorbed.

The Supreme Consciousness, just as other consciousnesses, has Its own spectrum of absorption determined by Its thoughts – *which make up the standard of a good-quality thought.*

Thus, the concept of good-quality thoughts is immediately established. That is, they are *resonant* thoughts in the Supreme Consciousness. Thus, only thoughts of this kind, produced in human consciousnesses, may induce the individualization of similar thoughts in the Supreme Consciousness.

In this context, a system of judgment is established in which the good and the evil are psychic values, with their origin in free thought. *The good is related to the good-quality thoughts, which are thoughts resonant in the Supreme Consciousness. The evil, in turn, is related to the bad-quality thoughts, non-resonant in the Supreme Consciousness.*

Consequently, the moral derived thereof results from the Law itself, inherent in the Supreme Consciousness and, therefore, this psychic moral must be the *fundamental moral*. Thus, *fundamental ethics* is neither biological nor located in the aggressive action, as thought by Nietzsche. It is psychic and located in the good-quality thoughts. It has a theological basis and, in it, the creation of the Universe by a pre-existing God is of an essential nature, opposed, for instance, to Spinoza's "geometrical ethics", which eliminated the ideas of Creation of the Universe by a pre-existing God, the main underpinning of Christian theology and philosophy. However, it is very close to Aristotle's ethics, to the extent that, from it, we understand that we are what we repeatedly do (think) and that *excellence is not an act, but a habit* (Ethics, II, 4). According to Aristotle: '" *the goodness of a man is a work of the soul towards excellence in a complete lifetime*: … it is not a day or a short period that makes a man fortunate and happy. " (Ibid, I, 7).

The "virtual" psychic radiation coming from a thought may induce *several* similar thoughts in the consciousness absorbing it, because each photon of radiation absorbed carries in itself the electromagnetic expression of the thought which produced it and, consequently, each one of them stimulates the individualization of a similar thought. However, the amount of thoughts induced is, of course, limited by the amount of psychic mass of the consciousness proper.

In the specific case of the Supreme Consciousness, the "virtual" psychic radiation coming from a good-quality thought must induce many similar thoughts. On the other hand, since Supreme Consciousness involves human consciousness, the induced thoughts appear in the surroundings of the very consciousness which induced them. These thoughts are then strongly attracted by said consciousness and fuse therewith, for, just as the thoughts generated in a consciousness have a high degree of positive mutual affinity [4] with it, they will also have the thoughts induced by it.

The fusion of these thoughts in the consciousness obviously determines an *increase* in its psychic mass. We then conclude that the cultivation of good-quality thoughts is highly beneficial to the individual. Reversally, the cultivation of bad-quality thoughts makes consciousness lose psychic mass.

When bad-quality thoughts are generated in a consciousness, they do not induce identical thoughts in Supreme Consciousness, because the absorption spectrum of Supreme Consciousness excludes psychic radiations coming from bad-quality thoughts. Thus, such radiation directs itself to other consciousnesses; however, it will only induce identical thoughts in those that are receptive in the same frequency spectrum. When this happens and right after the wave functions corresponding to these induced thoughts collapse and *materialize* said thoughts or change them into radiation, the receptive consciousness will lose psychic mass, similarly to what happens in the consciousness which first produced the thought. Consequently, both the consciousness which gave rise to the bad-quality thought and those receptive to the psychic radiations coming from this type of thoughts will lose psychic mass.

We must observe, however, that our thoughts are not limited only to harming or benefiting ourselves, since they also can, as we have already seen, induce similar thoughts in other consciousnesses, thus affecting them. In this case, it is important to observe that the psychic radiation produced by the induced thoughts may return to the consciousness which initially produced the bad-quality thought, inducing other similar thoughts in it, which evidently cause more loss of psychic mass in said consciousness.

The fact that our thoughts are not restricted to influencing ourselves is highly relevant, because it leads us to understand we have a great responsibility towards other persons as regards what we think.

## 4. The Psychic Universe

When we studied elementary Mathematics, we learned the called *Imaginary* Numbers. Just as there are the *real* numbers and *imaginary* numbers, there are also the *real* space-time and *imaginary* time. In the article "Mathematical Foundations of the Relativistic Theory of Quantum Gravity", we showed that the former contains our *Real* Universe, and the latter contains the *Imaginary Universe*. We also saw how a material body can make a transition to the Imaginary Universe. Simply reducing its gravitational mass to the range $+0.159M_i$ to $-0.159M_i$. Under these circumstances, its gravitational and inertial masses become imaginaries, and therefore, *the body becomes imaginary*. Consequently, *the body disappears* from our ordinary space-time and resurges in the imaginary space-time like an imaginary body. In other words, it becomes *invisible* for persons in the Real Universe.

What will an observer see when in the imaginary space-time? It will see light, bodies, planets, stars, etc., everything formed by imaginary photons, imaginary atoms, imaginary protons, imaginary neutrons and imaginary electrons. That is to say, the observer will find an Universe similar to ours, just formed by particles with imaginary masses. The term *imaginary* adopted from the Mathematics, as we already saw, gives the false impression that these masses do not exist. In order to avoid this misunderstanding we researched the true nature of that new mass type and matter.

The existence of imaginary mass associated to the *neutrino* is well-known. Although its imaginary mass is not physically observable, its square is. This amount is found experimentally to be negative. Recently, it was shown [1] that *quanta* of imaginary mass exist associated to the *photons*, *electrons*, *neutrons*, and *protons*, and that these imaginary masses would have psychic properties (elementary capability of “choice”). Thus, the true nature of this new kind of mass and matter shall be psychic and, therefore we should not use the term *imaginary* any longer. Consequently, from the previously described, we can conclude that the gravitational spacecraft penetrates in the *Psychic Universe* and not in an “imaginary” Universe.

In this Universe, the matter would be, obviously composed by psychic molecules and psychic atoms formed by psychic neutrons, psychic protons and psychic electrons. i.e., the matter would have psychic mass and consequently it would be *subtle*, much less dense than the matter of our *real* Universe.

From the quantum viewpoint, the psychic particles are similar to the material particles, so that we can use the Quantum Mechanics to describe the psychic particles. In this case, by analogy to the material particles, a particle with psychic mass $m_\Psi$ will be described by the following expressions:

$$\vec{p}_\psi = \hbar \vec{k}_\psi \qquad (02)$$

$$E_\psi = \hbar \omega_\psi \qquad (03)$$

where $\vec{p}_\psi = m_\Psi \vec{V}$ is the *momentum* carried by the wave and $E_\psi$ its energy; $\left|\vec{k}_\psi\right| = 2\pi/\lambda_\psi$ is the *propagation number* and $\lambda_\psi = h/m_\Psi V$ the *wavelength* and $\omega_\psi = 2\pi f_\psi$ its cyclic *frequency*.

As we already have seen, the variable quantity that characterizes DeBroglie’s waves is called *Wave Function*, usually indicated by $\Psi$ .

The wave function $\Psi$ corresponds, as we know, to the displacement $y$ of the undulatory motion of a rope. However, $\Psi$ as opposed to $y$, is not a measurable quantity and can, hence, be a *complex* quantity. For this reason, it is admitted that $\Psi$ is described in the $x$-direction by

$$\Psi = Be^{-(2\pi\ i/h)(Et-px)} \tag{04}$$

This equation is the mathematical description of the wave associated with a free material particle, with total energy $E$ and *momentum* $p$, moving in the direction $+x$.

As concerns the psychic particle, the variable quantity characterizing psyche waves will also be called wave function, denoted by $\Psi_\Psi$ ( to differentiate it from the material particle wave function), and, by analogy with equation Eq. (04), expressed by:

$$\Psi_\Psi = \Psi_0 e^{-(2\pi\ i/h)(E_\Psi t - p_\Psi x)} \tag{05}$$

If an experiment involves a large number of identical particles, all described by the same wave function $\Psi$ , *real* density of mass $\rho$ of these particles in x, y, z, t is proportional to the corresponding value $\Psi^2$ ($\Psi^2$ is known as *density of probability*. If $\Psi$ is *complex* then $\Psi^2 = \Psi\Psi^*$. Thus, $\rho \propto \Psi^2 = \Psi.\Psi^*$ ). Similarly, in the case of psychic particles, the *density of psychic mass*, $\rho_\Psi$, in x, y, z, will be expressed by $\rho_\Psi \propto \Psi_\Psi^2 = \Psi_\Psi\Psi_\Psi^*$. It is known that $\Psi_\Psi^2$ is always *real* and *positive* while $\rho_\Psi = m_\Psi/V$ is an *imaginary* quantity. Thus, as the *modulus* of an imaginary number is always real and positive, we can transform the proportion $\rho_\Psi \propto \Psi_\Psi^2$, in equality in the following form:

$$\Psi_\Psi^2 = k|\rho_\Psi| \tag{06}$$

Where $k$ is a *proportionality constant* (real and positive) to be determined.

In Quantum Mechanics we have studied the *Superposition Principle*, which affirms that, if a particle (or system of particles) is in a *dynamic state* represented by a wave function $\Psi_1$ and may also be in another dynamic state described by $\Psi_2$ then, the general dynamic state of the particle may be described by $\Psi$, where $\Psi$ is a linear combination (superposition) of $\Psi_1$ and $\Psi_2$, i.e.,

$$\Psi = c_1\Psi_1 + c_2\Psi_2 \tag{07}$$

C*omplex constants* $c_1$ e $c_2$ respectively indicate the percentage of dynamic state, represented by $\Psi_1$ e $\Psi_2$ in the formation of the general dynamic state described by $\Psi$ .

In the case of psychic particles (psychic bodies, consciousness, etc.), by analogy, if $\Psi_{\Psi 1}$, $\Psi_{\Psi 2}$,..., $\Psi_{\Psi n}$ refer to the different dynamic states the psychic particle assume, then its general dynamic state may be described by the wave function $\Psi_\Psi$ , given by:

$$\Psi_\Psi = c_1\Psi_{\Psi 1} + c_2\Psi_{\Psi 2} + ... + c_n\Psi_{\Psi n} \tag{08}$$

The state of superposition of wave functions is, therefore, common for both psychic and material particles. In the case of material particles, it can be verified, for instance, when an electron changes from one orbit to another. Before effecting the transition to another energy level, the electron carries out "virtual transitions" [5]. A kind of *relationship* with other electrons before performing the real transition. During this relationship period, its wave function remains "*scattered*" by *a wide region of the space* [6] thus superposing the wave functions of the other electrons. In this relationship the electrons *mutually* influence each other, with the possibility of *intertwining* their wave functions[2]. When this happens, there occurs the so-called *Phase Relationship* according to quantum-mechanics concept.

In the electrons "virtual" transition mentioned before, the "listing" of all the possibilities of the electrons is described, as we know, by *Schrödinger's wave equation.* Otherwise, it is general for material particles. By analogy, in the case of psychic particles, we may say that the "listing" of all the possibilities of the psyches involved in the relationship will be described by *Schrödinger's equation* – for psychic case, i.e.,

$$\nabla^2 \Psi_\Psi + \frac{p_\Psi^2}{\hbar^2} \Psi_\Psi = 0 \qquad (09)$$

Because the wave functions are capable of intertwining themselves, the quantum systems may "penetrate" each other, thus establishing an internal relationship where all of them are affected by the relationship, no longer being isolated systems but becoming an integrated part of a larger system. This type of internal relationship, which exists only in quantum systems, was called *Relational Holism* [7].

The idea of psyche associated with matter dates back to the pre-Socratic period and is usually called *panpsychism*. Remnants of organized panpsychism may be found in the *Uno* of Parmenides or in Heracleitus's *Divine Flux*. Scholars of Miletus's school were called *hylozoist*s, that is, "those who believe that matter is alive". More recently, we will find the panpsychistic thought in Spinoza, Whitehead and Teilhard de Chardin, among others. The latter one admitted the existence of proto-conscious properties at level of elementary particles.

Generally, the people believe that there is some type of psyche associated to the animals, and some biologists agree that even very simple animals like the ameba and the sea anemone are endowed with psychism. This led several authors to consider the possibility of the psychic phenomena to be described in a theory based on Physics [8-11].

The fact that an electron carries out "virtual" transitions to several energetic levels before performing the *real* transition [5] clearly shows a "choice" made by the electron. Where there is "choice" isn't there also *psyche,* by definition?

An *elementary psyche* associated to the electron would be an entity very similar to the *elementary electric charge* associated to the electron, whose existence was necessary to postulate for the establishment of electromagnetic theory. However, the elementary psyche has unique characteristics. Being a discrete quantity (quantum) of the Supreme Consciousness, which is omniscient, it must also contain within it *all* knowledge. In the Supreme Consciousness, whose psychic energy is infinite, the manifestation of this knowledge is total. In the case of the elementary psyche, would be minimal by definition, remaining the rest of the knowledge in a latent state.

But still this knowledge would be sufficient, for example, for electrons to define their orbital position (energy level) around the nuclei when they were electromagnetically attracted by the such nuclei.

How else could they have the knowledge of the exact orbit to stay? The electrosphere of atoms is a complex and accurate structure, and in no way could have been created randomly. Its construction undoubtedly involves knowledge.

Due to the fact that the formation of the electrosphere of the atoms is an organized process, the psyches of the electrons is also grouped in an organized manner, specifically in *phase condensates*, forming, what we can define as the *Individual Consciousnesses of the atoms*. Ice and NaCl crystals are common examples of imprecisely-structured phase condensates. Lasers, superfluids, superconductors, and magnets are examples of better-structured phase condensates.

---

[2] Since the electrons are simultaneously waves and particles, their wave aspects will interfere with each other. Besides superposition, there is also the possibility of occurrence of *intertwining* of their wave functions.

If electrons, protons and neutrons have psychic mass, then we can infer that the psychic mass of the atoms are *Phase Condensates.* In the case of the molecules the situation is similar. More molecular mass means more atoms and consequently, more psychic mass. In this case the phase condensate also becomes more structured because the great amount of elementary psyches inside the condensate requires, by stability reasons, a better distribution of them. Thus, in the case of molecules with very large molecular masses (*macromolecules*) it is possible that their psychic masses already constitute the most organized shape of a Phase Condensate, called Bose-Einstein Condensate[3].

The fundamental characteristic of a Bose-Einstein condensate is, as we know, that the various parts making up the condensed system not only behave as a whole but also *become a whole*, i.e., in the psychic case, the various consciousnesses of the system become a *single consciousness* with psychic mass equal to the sum of the psychic masses of all the consciousness of the condensate. This obviously, increases the available knowledge in the system since it is proportional to the psychic mass of the consciousness. This unity confers an *individual* character to this type of consciousness. For this reason, from now on they will be called *Individual Material Consciousness.*

It derives from the above that most bodies do not possess individual material consciousness. In an iron rod, for instance, the cluster of elementary psyches in the iron molecules does not constitute Bose-Einstein condensate; therefore, the iron rod does not have an individual consciousness. Its consciousness is consequently, much more simple and constitutes just a phase condensate imprecisely structured made by the consciousness of the iron atoms.

The existence of consciousnesses in the atoms is revealed in the molecular formation, where atoms with strong mutual affinity (their consciousnesses) combine to form molecules. It is the case, for instance of the water molecules, in which two Hydrogen atoms join an Oxygen atom. Well, how come the combination between these atoms is always the same: the same grouping and the same invariable proportion? In the case of molecular combinations the phenomenon repeats itself. Thus, the chemical substances either mutually attract or repel themselves, carrying out specific motions for this reason. It is the so-called *Chemical Affinity*. This phenomenon certainly results from a specific interaction between the consciousnesses. From now on, it will be called *Psychic Interaction*.

After the formation of the first planets, some of them came to develop favorable conditions for the appearance of macromolecules. These macromolecules, as we have shown, may have a special type of consciousness formed by a Bose-Einstein condensate (Individual Material Consciousness). In this case, since the molecular masses of the macromolecules are very large, they will have individual material consciousness of large psychic mass and, therefore, have access to a considerable amount of information in its own consciousness. Consequently, macromolecules with individual material consciousness are potentially very capable of, and some certainly already can carry out, autonomous motions, thus being considered as “living” entities.

However, if we decompose one of these molecules so as to destroy its individual consciousness, its parts will no longer have access to the information which “instructed” said molecule and, hence, will not be able to carry out the autonomous motions it previously did. Thus, the “life” of the molecule disappears – as we can see, *Delbrück’s Paradox* is then solved[4].

---

[3] Several authors have suggested the possibility of the Bose-Einstein condensate occur in the brain, and that it might be the physical base of memory, although they have not been able to find a suitable mechanism to underpin such a hypothesis. Evidences of the existence of Bose-Einstein condensates in living tissues abound (Popp, F.A Experientia, Vol. 44, p.576-585; Inaba, H., New Scientist, May89, p.41; Rattermeyer, M and Popp, F. A. Naturwissenschaften, Vol.68, Nº5, p.577.)

[4] This paradox ascribed to Max Delbrück (Delbrück, Max., (1978) *Mind from Matter*? American Scholar, **47**. pp.339-53.) remained unsolved and was posed as follows: How come the same matter studied by Physics, when incorporated into a living organism, assumes an unexpected behavior, although not contradicting physical laws?

The appearance of "living" molecules in a planet marks the beginning of the most important evolutionary stage for the psyche of matter, for it is from the combination of these molecules that there appear living beings with individual material consciousness with even larger psychic masses.

Biologists have shown that all living organisms existing on Earth come from two types of molecules – aminoacids and nucleotides – which make up the fundamental building blocks of living beings. That is, the nucleotides and aminoacids are identical in all living beings, whether they are bacteria, mollusks or men. There are twenty different species of aminoacids and five of nucleotides.

In 1952, Stanley Miller and Harold Urey proved that aminoacids could be produced from inert chemical products present in the atmosphere and oceans in the first years of existence of the Earth. Later, in 1962, nucleotides were created in laboratory under similar conditions. Thus, it was proved that the molecular units making up the living beings could have formed during the Earth's primitive history.

Therefore, we can imagine what happened from the moment said molecules appeared. The concentration of aminoacids and nucleotides in the oceans gradually increased. After a long period of time, when the amount of nucleotides was already large enough, they began to group themselves by mutual psychic attraction, forming the molecules that in the future will become DNA molecules.

When the molecular masses of these molecules became large enough, the distribution of elementary psyches in their consciousnesses took the most orderly possible form of phase condensate (Bose-Einstein condensate) and such consciousnesses became the *individual material consciousness*.

Since the psychic mass of the consciousnesses of these molecules is very large (as compared with the psychic mass of the atoms), the amount of self-accessible knowledge in such consciousnesses became considerable and, thus, they became apt to *instruct* the joining of aminoacids in the formation of the first proteins (origin of the *Genetic Code*). Consequently, the DNA's capability to serve as guide for the joining of aminoacids in the formation of proteins is fundamentally a result of their psychism.

In the psychic of DNA molecules, the formation of proteins certainly had a definite objective: *the construction of cells*.

During the cellular construction, the most important function played by the consciousnesses of the DNA molecules may have been that of organizing the distribution of the new molecules incorporated to the system so that the consciousnesses of these molecules jointly formed with the consciousness of the system a Bose-Einstein condensate. In this manner, more knowledge would be available to the system and, after the cell is completed, the latter would also have an individual material consciousness.

Afterwards, under the action of psychic interaction, the cells began to group themselves according to different degrees of positive mutual affinity, in an organized manner so that the distribution of their consciousnesses would also form Bose-Einstein condensates. Hence, collective cell units began to appear with individual consciousnesses of larger psychic masses and, therefore, with access to more knowledge. With greater knowledge available, these groups of cells began to perform specialized functions to obtain food, assimilation, etc. That is when the first multi-celled beings appeared.

Upon forming the tissues, the cells gather structurally together in an organized manner. Thus, the tissues and, hence, the organs and the organisms themselves also possess individual material consciousnesses.

The existence of the material consciousness of the organisms is proved in a well-known experiment by Karl Lashley, a pioneer in neurophysiology.

Lashley initially taught guinea pigs to run through a maze, an ability they remember and keep in their memories in the same way as we acquire new skills. He then systematically removed small portions of the brain tissue of said guinea pigs. He thought that, if the guinea pigs still remembered how to run through the maze, the memory centers would still be intact.

Little by little he removed the brain mass; the guinea pigs, curiously enough, kept remembering how to run through the maze. Finally, with more than90% of their cortex removed, the guinea pigs still kept remembering how to run through the maze. Well, as we have seen, the consciousness of an organism is formed by the concretion of all its cellular consciousnesses. Therefore, the removal of a portion of the organism cells does not make it disappear. Their cells, or better saying, the consciousnesses of their cells contribute to the formation of the consciousness of the organism just as the others, and it is exactly due to this fact that, even when we remove almost all of the guinea pigs' cortex, they were still able to remember from the memories of their individual material consciousnesses. In this manner, what Lashley's experiment proved was precisely the existence of individual material consciousnesses in the guinea pigs.

Another proof of the existence of the individual material consciousnesses in organisms is given by the *regeneration* phenomenon, so frequent in animals of simple structure: sponges, isolated coelenterates, worms of various groups, mollusks, echinoderms and tunicates. The arthropods regenerate their pods. Lizards may regenerate only their tail after autotomy. Some starfish may regenerate so easily that a simple detached arm may, for example, give origin to a wholly new animal.

The organization of the psychic parts in the composition of an organism's individual material consciousness is directly related to the organization of the material parts of the organism, as we have already seen. Thus, due to this interrelationship between body and consciousness, any disturbance of a material (physiological) nature in the body of the being will affect its individual material consciousness, and any psychic disturbance imposed upon its consciousness affects the physiology of its body.

When a consciousness is strongly affected to the extent of unmaking the Bose-Einstein's condensate, which gives it the status of individual consciousness, there also occurs the simultaneous disappearance of the knowledge made accessible by said condensation. Therefore, when a cell's consciousness no longer constitutes a Bose-Einstein condensate, there is also the simultaneous disappearance of the knowledge that *instructs and maintains* the cellular metabolism. Consequently, the cell no longer functions thus initiating its decomposition (molecular desegregation).

Similarly, when the consciousness of an animal (or vegetables) no longer constitutes a Bose-Einstein condensate, the knowledge that instructs and maintains its body functioning also disappears, and it dies. In this process, after the unmaking of the being's individual consciousness, there follows the unmaking of the individual consciousnesses of the organs; next will be the consciousnesses of their own cells which no longer exist. At the end there will remain the isolated psyches of the molecules and atoms. *Death, indeed, destroys nothing, neither what makes up matter nor what makes up psyche.*

As we have seen, all the information available in the consciousnesses of the beings is also accessible by the consciousnesses of their organs up to their molecules'. Thus, when an individual undergoes a certain experience, the information concerning it not only is recorded somewhere in this consciousness but also pervades all the individual consciousnesses that make up its total consciousness. Consequently, psychic disturbances imposed to a being reflect up to the level of their individual molecular consciousnesses, perhaps even structurally affecting said molecules, due to the interrelationship between body and consciousness already mentioned here.

Therefore, some modifications in the sequences of nucleotides of the DNA molecules can occur when the psychism of the organism in which the molecules are incorporated is sufficiently affected.

It is known that such modifications in the structure of DNA molecules may also occur as consequence of chemical products in the blood stream (as in the case of the mustard gas used in chemical warfare) or exposition to high-energy radiation.

Modifications in the sequences of nucleotides in DNA molecules are called mutations. Mutations, as we know, determine hereditary variations, which are the basis of Darwin's theory of evolution.

It is known that mutations of two types, "favorable" or "unfavorable", can occur. The former type enhances the individuals' possibility of survival, whereas the second reduces such possibility.

The theory of evolution is established as a consequence of individuals' efforts to survive in the environment where they live. This means that their descendants may become different from their ancestors. This is the mechanism that leads to frequent appearance of new species. Darwin believed that the mutation process was slow and gradual. Nevertheless, it is known today that this is not the general rule, for there are evidences of the appearance of new species in a relatively short period of time [12]. We also know that individual's characteristics are transmitted from parents to offsprings by means of genes and that the recombination of the parents' genes, when *genetic instructions* are transmitted, by such genes.

However, it was shown that the genetic instructions are basically associated with the psychism of DNA molecules. Consequently, *the genes transmit not only physiological but also psychic differences.*

Thus, as a consequence of genetic transmission, besides the great physiological difference between individuals of the same species, there is also a great psychic dissimilarity.

Such psychic dissimilarity associated with the progressive enhancement of the individual's psychic quantities may have given rise, in immemorial time, to a variety of individuals (most probably among anthropoid primates) which unconsciously established a positive mutual affinity with *primordial consciousnesses* that must have been attracted to Earth. Thus, the relationship established among them and the consciousnesses of said individuals is enhanced.

In the course of evolutionary transformation, there must have been a time when the fetuses of said variety already presented such a high degree of mutual affinity with the primordial consciousnesses attracted to Earth that, during pregnancy, the incorporation of primordial consciousnesses may have occurred in said fetuses.

In spite of absolute psychic mass of the fetus's material consciousness be much smaller than that of the mother's consciousness, the degree of positive mutual affinity between the fetus's consciousness and the primordial consciousness that is going to be incorporated is much greater than that between the latter and the mother's, which makes the psychic attraction between the fetus's consciousness and primordial consciousness much stronger than the attraction between the latter and the mother's. That is the reason why primordial consciousness incorporates the fetus. Thus, when these new individuals are born, they bring with them their individual material consciousness, an individualized consciousness of the Supreme Consciousness. In this way were the first *hominids* born.

Having been directly individualized from Supreme Consciousness, the primordial consciousnesses are perfect individualities and not phase condensates as the consciousnesses of the matter. In this manner, they do not dissociate after the death of those that incorporated them. Afterwards, upon the action of psychic attraction, they are again able to incorporate into other fetuses to proceed on their evolution.

These consciousnesses (hereinafter called *human consciousness*) constitutes individualities and, therefore, the larger their psychic mass the more available knowledge they will have and, consequently, greater ability to evolve.

Just as the human race evolves biologically, human consciousnesses have also been evolving. When they are incorporated, the difficulties of the material world provide them with more and better opportunities to acquire psychic mass (later on we will see how said consciousnesses may gain or lose psychic mass). That is why they need to perform successive reincorporations. Each reincorporation arises as a new opportunity for said consciousnesses to increase their psychic mass and thus evolve.

The belief in the reincarnation is millenary and well known, although it has not yet been scientifically recognized, due to its *antecedent probability* being very small. In other words, there is small amount of data contributing to its confirmation. This, however, does not mean that the phenomenon is not true, but only that there is the need for a considerable amount of experiments to establish a significant degree of antecedent probability.

The rational acceptance of reincarnation entails deep modifications in the general philosophy of the human being. For instance, it frees him from negative feelings, such as nationalistic or racial prejudices and other response patterns based on the naive conception that we are simply what we appear to be.

Darwin's lucid perception upon affirming that not only the individual's corporeal qualities but also his psychic qualities tend to improve made implicit in his "natural selection" one of the most important rules of evolution: *the psychic selection*, which basically consists in the *survival of the most apt consciousnesses*. Psychic aptitude means, in the case of human consciousnesses, mental quality, i.e., *quality of thinking*.

In this context, the human consciousnesses are equivalent to the called *Spirits*, mentioned in the Kardecist literature [13], where the reincarnation was strongly considered.

## 5. The Spirits

*Origin and Nature of the Spirits*

As we have already seen, the origin of the spirits is related to the natural desire of the Supreme Consciousness to procreate, that is, of generating individual consciousnesses in itself so that they could evolve and express the same creative attributes pertinent to Her. In this way, the nature of Spirits is the same of the Supreme Consciousness.

*Form and Ubiquity of the Spirits*

By definition the consciousnesses, the thought, etc., are psychic bodies, i.e., *psychic energy* locally concentrated. In the material world, we can not distinguish the *form* of thoughts probably because the density of concentrated psychic energy is so low that would be equivalent to a *fluid* with a density much lower than the densities of gases. We know that we can only see a body if the light emitted by it can be detected by our eyes. The solids and liquids generally reflect light well and this makes them visible. The gases, on the contrary, are only visible in a state of high density, as in the case of the clouds. In a state of low density, like the wind, become invisible, because, practically, do not reflect the rays of light. In the case of thoughts, whose density would be much lower than the density of the gases, we also cannot distinguish its shape. The same is true in the case of Spirits. Thus, it becomes very difficult for us to see the Spirits. However, as the concentration of energy in spirits is greater than the thoughts it is possible that we can see traces of its forms in certain circumstances. This would then correspond to the vision of figures, flashes, etc. Thus, the perfect vision of the forms of the spirits will probably only be possible for an observer in the Spiritual World.

As concerns the ubiquity of the Spirits, it is necessary to use the Quantum Physics in order to understand it. We start from the *Uncertainty Principle*, under the form obtained in 1927 by Werner Heisenberg, i.e.,

$$\Delta x \Delta p \geq h \qquad (10)$$

This expression shows that the product of the uncertainty $\Delta x$ in the position of a particle in a certain instant by the uncertainty $\Delta p$ in its *momentum* is equal or greater than the *Planck's constant h*. We cannot measure simultaneously both, position and *momentum*, with perfect accuracy. If we reduce $\Delta p$, then $\Delta x$ will be increased and vice-versa. Such uncertainties are not in our appliances, but in Nature.

A mathematical approach more accurate than the one proposed by Heisenberg presents to the uncertainty principle the following relationship:

$$\Delta x \Delta p \geq \frac{h}{2\pi} \qquad (11)$$

When we want to correlate the uncertainty $\Delta E$ in the energy with the uncertainty $\Delta t$ in the time interval it is customary to write the Uncertainty Principle in the following form:

$$\Delta E \Delta t \approx \hbar \qquad (12)$$

where $\hbar = h/2\pi$.

According to this expression, an event in which an amount of energy $\Delta E$ is not conserved is not prohibited, provided that the duration of the event does not exceed $\Delta t$. This means that it can occur variations of energy in a system, that *even in principle are impossible to determine them*. The emission of a meson by a nucleus that does not change its mass - clear violation of the principle of conservation of energy - can occur if the nucleon reabsorb the meson (or similar) in a time interval less than $\hbar/\Delta E = \hbar/m_\pi c^2$, ($m_\pi$ is the mass of the meson).

Therefore, it can also occur that a material particle moves temporarily to a certain position without actually leaving your starting position. In this case, it is said that the particle made a *Virtual Transition* to a certain position.

The designation virtual must not lead the reader to imagine that the transition was not made. It is effectively carried out: it is real. But, according to the uncertainty principle, it is impossible to be observed. This is a limitation imposed by Nature.

However, although we cannot observe the virtual transitions, their occurrence can often be detected by the produced effects.

The *psychic particles* can also perform virtual transitions, since the uncertainty principle also applies to them.

This means, therefore, that *quanta* of human consciousnesses (from the minds' conscious, subconscious and unconscious) can perform "temporary exits" *but without leaving them effectively*.

These transitions correspond to virtual transitions of the own minds where the *quanta* are originated, since these, when individualized, form Bose-Einstein condensates with the mind where they are originated, and therefore, share all the knowledge and attributes relevant to it.

During pseudo-medical deaths, projections, etc., people report later that they "saw" themselves out of the body, a clear indication of virtual transitions originating from the *conscious* and *subconscious*. In dreams, besides such transitions, there are also indications of transitions from the *unconscious*.

According to Feynman's Quantum Theory of Electromagnetic Interaction [14], no energy is spent in virtual transitions, which can occur *around at any distance*. Moreover, as is easily concluded from the uncertainty principle, one single quantum can perform several virtual transitions simultaneously. It all depends on how quickly it makes the transitions. Therefore, through this process, the *quanta* of human consciousnesses or *it all* may go to several places *simultaneously*. We conclude, therefore, that *a spirit can be in several places at a time*. But of course this is not a division of Spirit, but himself present simultaneously in several places.

*Incarnation of Spirits*

The great dissimilarity associated with the progressive enhancement of the individual's psychic quantities may have given rise, in immemorial time, to a variety of individuals (most probably among anthropoid primates) which unconsciously established a positive mutual affinity with *Primordial Spirits*, previously mentioned.

As this affinity was developed with the psychic enhancement, it is expected that natural selection has made it much higher in the offspring of this variety. Thus, due to the psychic interaction, several Primordial Spirits must have been attracted to the Earth. With this, the relationship established among them and the material consciousnesses of said individuals was intensified.

In the course of evolutionary transformation, there was a time when the fetuses of said variety already presented such a high degree of mutual affinity with the primordial consciousnesses attracted to the Earth that, during pregnancy, the incorporation of Primordial Spirits may have occurred in said fetuses[5].

This phenomenon should not have occurred only on Earth, may also have occurred in the same way on other planets with evolutionary conditions similar to Earth's. The belief that this phenomenon occurred only in the *incarnation* of Spirits on Earth would question the wisdom of God, favoring only the Earth and excluding thousands of other worlds.

As we have already seen, these Spirits constitute perfect individualities and, therefore, as greater their psychic energy greater auto-knowledge accessible and, consequently, they will have greater opportunities to evolve.

Thus, also the Spirits evolve as the human race evolved biologically.

*Return to Corporal Life*

Just as the consciousnesses of the children have a high degree of positive mutual affinity with the consciousnesses of their parents, and among themselves (*principle of familiar formation*), the embryo cells, by having originated from cellular duplication, have a high degree of positive mutual affinity. The embryo cells result, as we know, from the cellular duplication of a single cell containing the paternal and maternal genes and, hence, have a high degree of positive mutual affinity.

Thus, under the action of psychic interaction the cells of the internal cellular mass start gathering into small groups, according to the different degrees of mutual affinity.

When there is a positive mutual affinity between two consciousnesses there occurs the *intertwining* between their wave functions, and a *Phase Relationship* is established among them. Consequently, since the degree of positive mutual affinity among the embryo cells is high, also the relationship among them will be intense, and it is exactly this what enables the construction of the organs of the future child. In other words, when a cell is attracted by certain group in the embryo, it is through the cell-group relationship that determines where the cell is to aggregate to the group. In this manner, each cell finds its correct place in the embryo; that is why observers frequently say that, ”*the cells appear to know where to go*”, when experimentally observed.

---

[5] When incarnated, the Spirit is commonly called of *Soul*. However, considering that while incarnated the Spirit form a Bose-Einstein condensate with the material consciousness of the body, we can define the Soul as *the individual consciousness of being,* i.e., a Bose-Einstein condensate containing the incarnated *Spirit* and *material consciousness* of the body.

The cells of the internal cellular mass are capable of originating any organ, and are hence called *totipotents*; thus, the organs begin to appear. In the endoderm, there appear the urinary organs, the respiratory system, and part of the digestive system; in the mesoderm are formed the muscles, bones, cartilages, blood, vessels, heart, kidneys; in the ectoderm there appear the skin, the nervous system, etc.

Thus, it is the mutual affinity among the consciousnesses of the cells that determines the formation of the body organs and keeps their own physical integrity. For this reason, every body rejects cells from other bodies, unless the latter have positive mutual affinity with their own cells. The higher the degree of cellular positive mutual affinity, the faster the integration of the transplanted cells and, therefore, the less problematic the transplant. In the case of cells from identical twins, this integration takes place practically with no problems, since said degree of mutual affinity is very high.

In eight weeks of life, all organs are practically formed in the embryo. From there on, it begins to be called *fetus*.

The embryo’s material individual consciousness is formed by the consciousnesses of its cells united in a Bose-Einstein condensate. As more cells become incorporated into the embryo, its material consciousness acquires more psychic mass. This means that this type of consciousness *will be greater in the fetus than in the embryo* and even greater in the child.

Thus, the psychic mass of the mother-fetus consciousness progressively increases during pregnancy, consequently increasing the psychic attraction between this consciousness and that new one about to incorporate. In normal pregnancies, this psychic attraction also increases due to the habitual increase in the degree of positive mutual affinity between said consciousnesses.

Since the embryo's consciousness has greater degree of positive mutual affinity with the consciousness that is going to incorporate, then the embryo's consciousness becomes *the center of psychic attraction to where the human consciousness* (*Spirit*) *destined to the fetus will go.*

When the psychic attraction becomes intense enough, human consciousness penetrates the *mother consciousness*, forming with it a new Bose–Einstein condensate. From that instant on, the fetus begins to have two consciousnesses: *the individual material one and the human consciousness attracted to it.*

However, this should only occur after *eight weeks*, when all organs are practically formed in the embryo, and it is called *fetus*. This is a critical period in which *the imperfections of matter* can cause fetal death. Thus, the Spirit waits to complete formation of the fetus. If the fetus can not be structured conveniently, it will naturally be aborted and the Spirit will look for another body to reincarnate.

We conclude, therefore, that the initial eight weeks are a period imposed by Nature herself to finish the building of the fetus and test whether it will be able to be used by the Spirit that want to incarnate. Thus, in this period of "construction" of the fetus, both the Spirit and the mother, based on free will, also have the freedom to give up the process. In this case, breaks easily the Bose-Einstein condensate, and the Spirit and both the mother can restart on other basis, making sure they have fully exercised their rights and have not harmed or caused harm to either party involved in the process.

However, if the fetus is able, the process to continue the psychic attraction between material consciousness of the fetus and the Spirit that want to incarnate, accepted by the mother and by said Spirit, will progressively increase. In this way, with the psychic attraction, this human consciousness tends to continue, being progressively *compressed* until effectively incorporating the fetus. *When this takes place, it will be ready to be born.*

It is probably due to this *psychic compression* process that the incorporated consciousness suffers amnesia of its preceding history. Upon *death*, after the psychic decompression that arises from the definitive disincorporation of the consciousness, *the preceding memory must return*.

*Evolutionary Degree and Fate of the Spirit*

We have already seen that, when the *gravitational mass*, $M_g$, of a body is smaller than $-0.159M_i$ or larger than $+0.159\ M_i$ it is in the *Material World*. However, if its gravitational mass is reduced to the range between $-0.159M_i < M_g < +0.159\ M_i$, it performs a transition to the *Psychic World* or *Spiritual World*. [1] When this occurs, its *real gravitational mass,* $m_{g(real)}$*, is totally converted into imaginary gravitational mass,* $m_{g(imaginaria)}$, due to the *Principle of Conservation of Energy*.

On the other hand, it was shown that the *psychic mass,* $m_\Psi$, is equal to the *imaginary gravitational mass*[1] , i.e.,

$$m_\Psi = m_{g(imaginaria)} \tag{13}$$

Thus, when a body perform a transition to the Psychic World, its *real gravitational mass* $m_{g(real)}$*, is totally converted into psychic mass.*

$$m_\Psi = m_{g(imaginaria)} \equiv m_{g(real)}$$

Since the mass is quantized, the body performs a transition to a *quantum* level correspondent to its *psychic mass* (See Fig.1). Thus, the body goes to a region corresponding to the gravitational mass that it acquired in the Material World.

According to the new concepts of spacecraft and aerospatial flight shown in the book *Física dos UFOs*[6], the called *Gravitational Spacecrafts* must use the Psychic Universe in order to viabilize trips that would require much time in the Material Universe.

As the mentioned spacecrafts just perform transition to the Psychic Universe if and only if its gravitational masses are reduced to the range $-0.159M_i < M_g < +0.159\ M_i$, then, with *negative* gravitational mass in the range $-0.159M_i < M_g$, they perform the transition, and their gravitational masses would be transformed into *negative* gravitational mass. Thus, they will enter the Psychic Universe by the zone energetically located in the range $-\infty < M_\Psi < 0$ (See Fig. 1). In the case of the gravitational mass of the spacecraft be reduced to the *positive* range, i.e., $0 < M_g < +0.159\ M_i$, the spacecraft will enter the Psychic Universe by zone of *positive* energy of the psychic spectrum.

Only after the discovery of the correlation between the gravitational mass and the inertial mass could the finding of *negative* gravitational mass be achieved, making it possible to find ways to obtain it. It is clear, then, that the common material in the Material Universe is the existence of bodies with *positive* gravitational mass. Similarly, in the Psychic Universe, the common is to find psychic bodies with *positive* psychic mass. Thus, to find the World of Spirits, a Gravitational Spacecraft must enter the Psychic Universe with *positive* psychic mass.

When a spirit disincarnates, he does not makes a transition to the Spiritual Universe because, due to its own nature, *the Spirit is already in the Spiritual Universe*. Thus, the Spirit just *turn off* the material body which they lived. As it leaves the Material World with a given positive psychic mass[7], $m_\Psi$ – acquired during its evolution, and during its recent reincarnation in the Material World – *it should proceed to the region of the World of Spirits corresponding to its psychic mass*. Thus, as the evolutionary degree of each Spirit is defined by the amount of psychic mass contained in the Spirit[8], then the Spirits proceed precisely to the regions that correspond to their evolutionary degrees, and there, brought together by mutual affinity, they continue the evolutionary process and wait for the time of new reincarnation. Driven by the need for progress, this is therefore the destiny of Spirits.

[6] See at www.frandeaquino.org

[7] It is fact that the presence of *negative* psychic energy in the Spirits would cause a decrease in its total psychic energy, which would imply *involution* (since the addition of *positive* psychic energy implies in evolution) and, as we have seen, the Spirits do not involute.

Thus, in the World of Spirit there is a natural selection that brings together Spirits with the same level of evolution, and that does not allow the less evolved access to more evolved regions. The most evolved Spirits, however, can transit through the lower regions, making the already mentioned "virtual" transitions. In this way, they may intervene with less evolved regions.

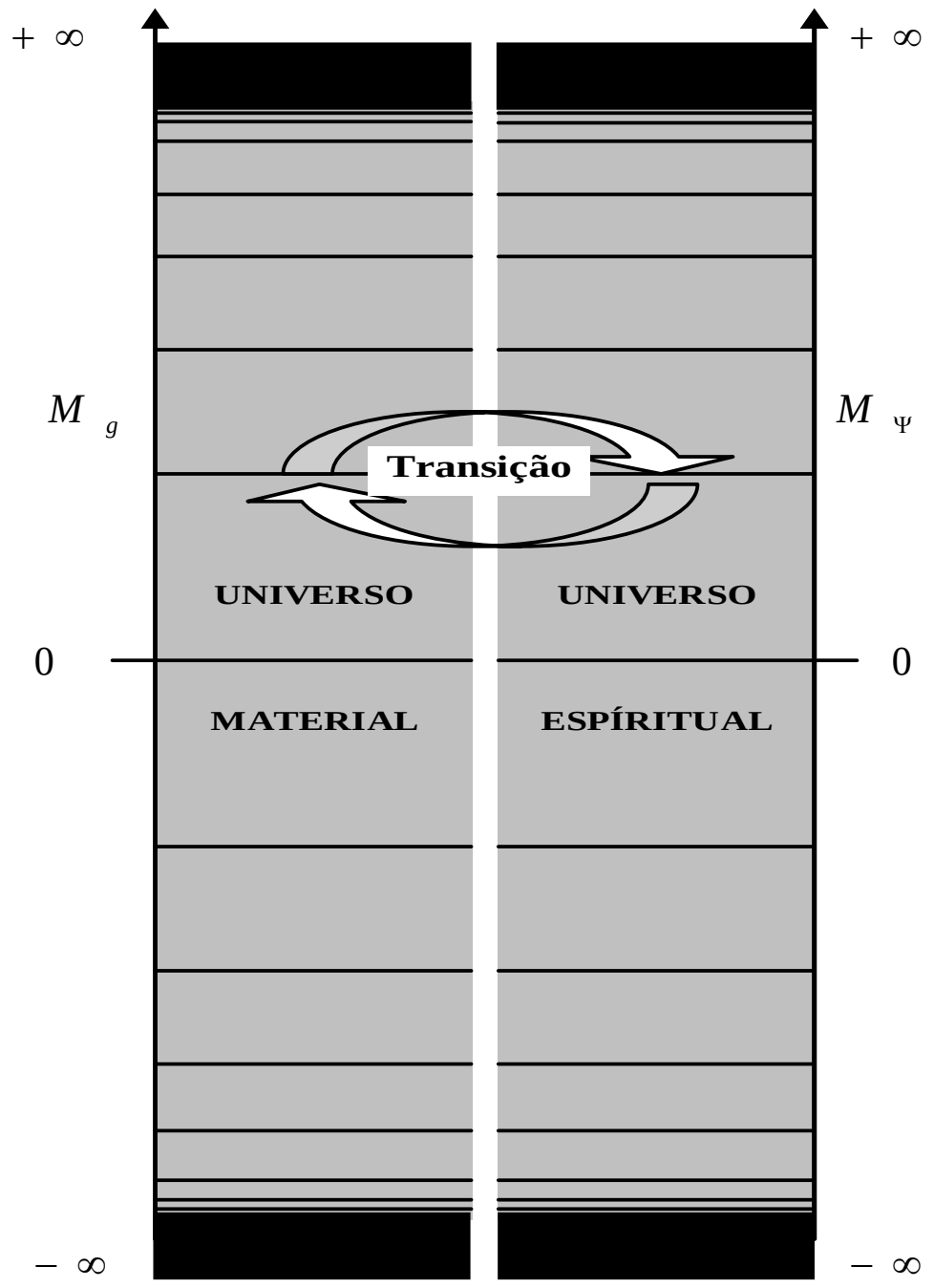

Fig. 1 – When the *gravitational mass*, $M_g$, of a body is in the range $-0.159M_i > M_g > +0.159\ M_i$ it is in the *Material Universe*. However, If $M_g$ is reduced to the range $-0.159M_i < M_g < +0.159\ M_i$, then it makes a transition to the *Psychic Universe* (or Spiritual) with its *gravitational mass* converted to in *psychic mass*. Since the mass is *quantized*, the body performs a transition to a quantum level correspondent to its psychic mass. Thus, it follows to a region correspondent to the gravitational mass, which it acquired in the Material Universe.

*Life in the World of Spirits*

By doing the good, spirits acquire more psychic mass, and thus, more latent powers are awaken, which facilitates their achievements, and makes them happier[9]. But they should not occupy themselves only with their personal improvement, since life in the Spirit World, such as life in the Material World, is a continuous occupation. We can also conclude from the above that even the spirits of the highest order, in having nothing more to improve, do not cease their activities, once the eternal idleness would also be an eternal punishment.

[8] As already seen, the Spirits were individualized in the Supreme Consciousness, and therefore brought with them, in a latent state, the same attributes pertinent to Her. With the progressive evolution of the Spirit, these attributes are being awakened, so that the degree of evolution of a Spirit is directly related to the quantity of attributes it aroused. On the other hand, the number of attributes in a Spirit is directly correlated to the amount of *mass psychic* of the Spirit. Thus, more psychic mass the Spirit has greater the amount of attributes awakened and, therefore, more evoluted is the Spirit.

[9]The happiness of the good Spirits certainly consists in *knowing* more and more; not having hate, jealousy, envy, or any of the passions that make the misery of men. They don't experience needs or suffering, or the anxieties of material life, and this in itself is synonymous of great happiness.

We have seen that the *realization* of what we want requires an expenditure of psychic energy proportional to the nature of desire. In other words, in order to have what we want realized through the collapse of its wave function, it is necessary an expenditure of psychic energy sufficient for its accomplishment. As the density of material bodies is much greater than the density of psychic bodies, the realization of our desires in the Material Universe usually requires much more the psychic energy than in the Psychic Universe. Thus, life in the World of Spirit becomes much easier and more enjoyable than in the Material World. But the difficulties of the Material World are what allow the Spirits to progress in their evolution, and that might have been the main reason for the creation of the Material Universe.

The possibility of transition to the Psychic Universe increased the likelihood of close encounters with beings from other planets in our ordinary Universe, and also with the people who live in planets of the Psychic Universe, since Gravitational Spacecraft trips can also be carried over in this Universe, as already shown. The characteristics associated to the subtle psychic mass indicate that the life of these beings should not be *finite* as the lives of the humans. This makes us think that maybe life in the Psychic Universe be the *real* life while our brief life in this Universe has only specific goals such as, for example, a learning period.

The Psychic Universe, by its very nature, it is constituted of photons, atoms and molecules *psychics*. This means that all types of photons, atoms and molecules that exist here may have its corresponding counterparts in the Psychic Universe. Therefore, all we have here can exist there with a similar form. However, considering the characteristics associated to the subtle psychic mass, we can conclude that life here may be an imperfect copy of the life there.

*Time in the World of Spirits*

We have already seen that the *Real Universe*, where we live, is contained in the *Psychic Universe* (Imaginary Universe), so the *real space-time* that corresponds to the Real Universe is within the Imaginary space-time, which forms the *Psychic Universe*, where the Spirits live. By definition, in the imaginary space-time both the *space* coordinates and the *time* coordinate are, obviously, imaginary. This means that *the time in the Universe of the Spirits (imaginary time) is different from the real time of our Universe*.

Only recently the concept of imaginary time was considered by physicists. Difficult to understand, but deemed essential to connect the Statistical Mechanics to Quantum Mechanics, the concept of imaginary time also became instrumental in Quantum Cosmology, where it was introduced in order to eliminate singularities (points where the curvature of space-time becomes infinite), which occur in the real time (See Hartle-Hawking state [15]). Twenty-two years ago, Hawking popularized the concept of imaginary time in his book: A Brief History of Time [16].

The imaginary time is not imaginary in the sense that it does not exist. Nor is it a mathematical artifice. No, it really exists, however, it has different characteristics of the time which we are used to.

The existence of the imaginary time is mathematically sustained by a mechanism called *Wick Rotation*[10], which transform the metrics of the *Minkowski* space-time

$$ds^2 = -\left(dt^2\right) + dx^2 + dy^2 + dz^2 \qquad (14)$$

into the metrics of the *Euclidean space-time*

$$ds^2 = d\tau^2 + dx^2 + dy^2 + dz^2 \qquad (15)$$

where $\tau = it$; $i = \sqrt{-1}$ is called *imaginary unit*, which defines the *imaginary numbers* in the form $z = a + bi$, where $a$ and $b$ are real numbers, called respectively, the *real* part of "*z*" and the *imaginary* part of "*z*".

---

[10] It is the called *rotation* because when we multiply an imaginary number by *i* the result, on the Cartesian plane, is equivalent to a rotation of $90^0$ of the vector that represents the number. Assim, $-dt^2 = -dt.\,dt = i^2\,dt.\,dt = i(i\,dt).\,dt = d\tau^2$.

From the definition of complex numbers follows that they can interpreted as points in the Cartesian plane (where conventionally we mark on the *x*-axis the *real part* and on the *y*-axis the *imaginary part* of a imaginary number "*z*") or, as *vectors* $\vec{OZ}$ whose origin "*O*" is at the origin of the Cartesian grid, and the point "*Z*" with the coordinates (a, b). (See Fig. 2).

Thus, imaginary numbers can be conceived as a new type of number *perpendicular* to the real numbers. This leads to the possibility of expressing mathematically *imaginary time* in a direction *perpendicular* to the common real time. In this model, the imaginary time is a function of real time and vice versa (See Fig.3). Thus, the imaginary time appears as a new dimension that makes a right angles to real time, and thereby, as Hawking showed [17], *it has much more possibilities than the real time*, which always flows from past to future, and only may have a beginning and an end.

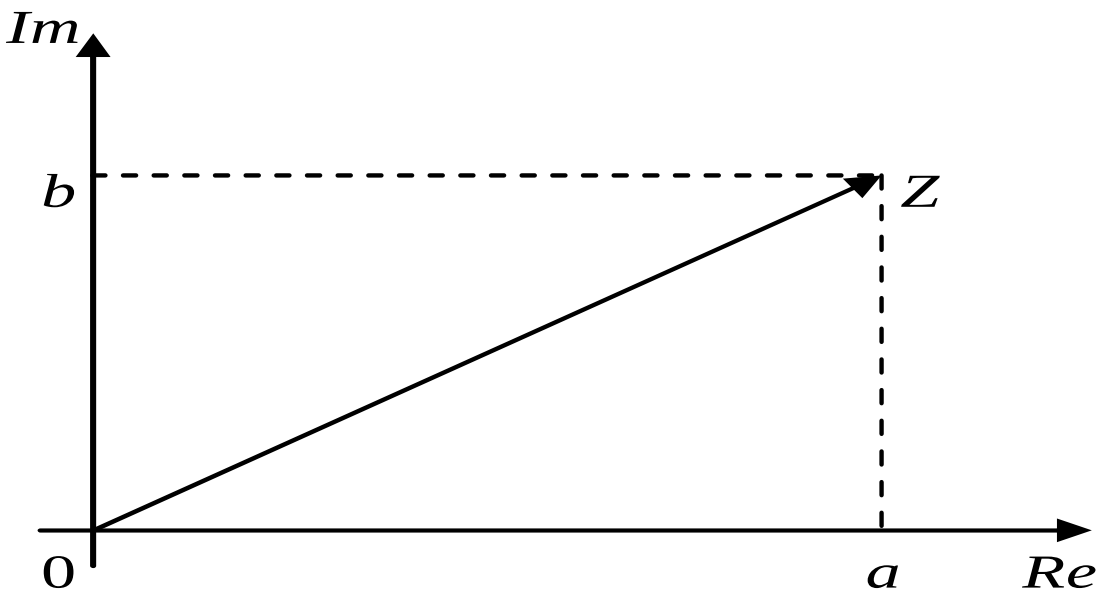

(a)

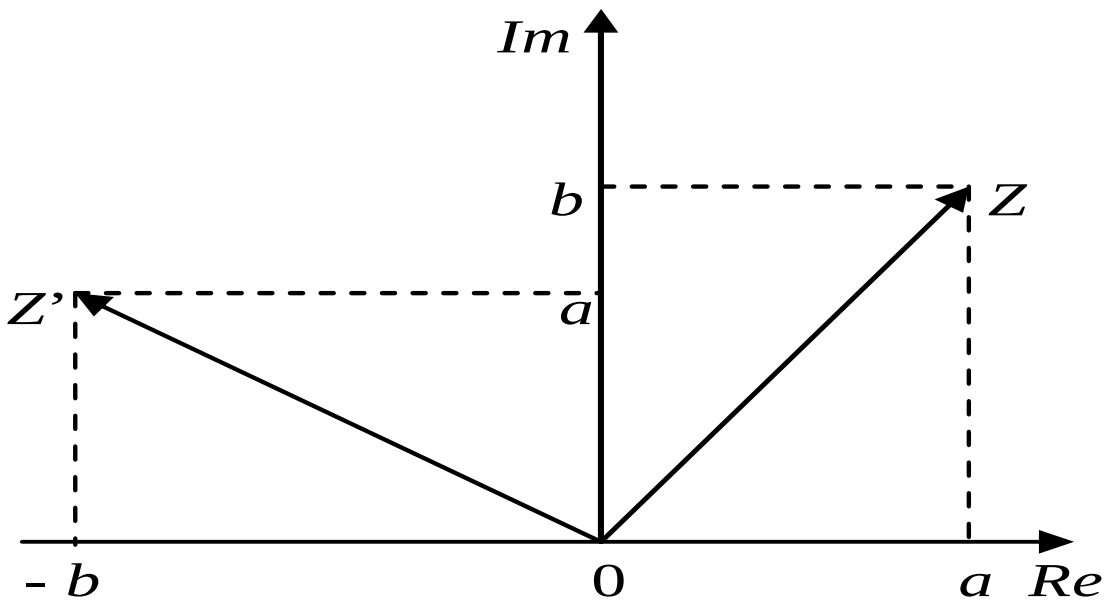

(b)

Fig. 2 — (a) *The Imaginary Plane* (or Argand-Gauss Plane) is a way to visualize the space of imaginary numbers. Can be understood as a modified Cartesian plane, where the *real part* is represented on the x-axis and the *imaginary part* on the y-axis. The x-axis is called real axis while the y-axis is called imaginary axis. (b) When we multiply a imaginary number $z = a+bi$ by $i$ ($iz = ai - b =$ z') the result on the Cartesian plane is equivalent to a *rotation* of 90° of the vector *OZ* that represents the number.

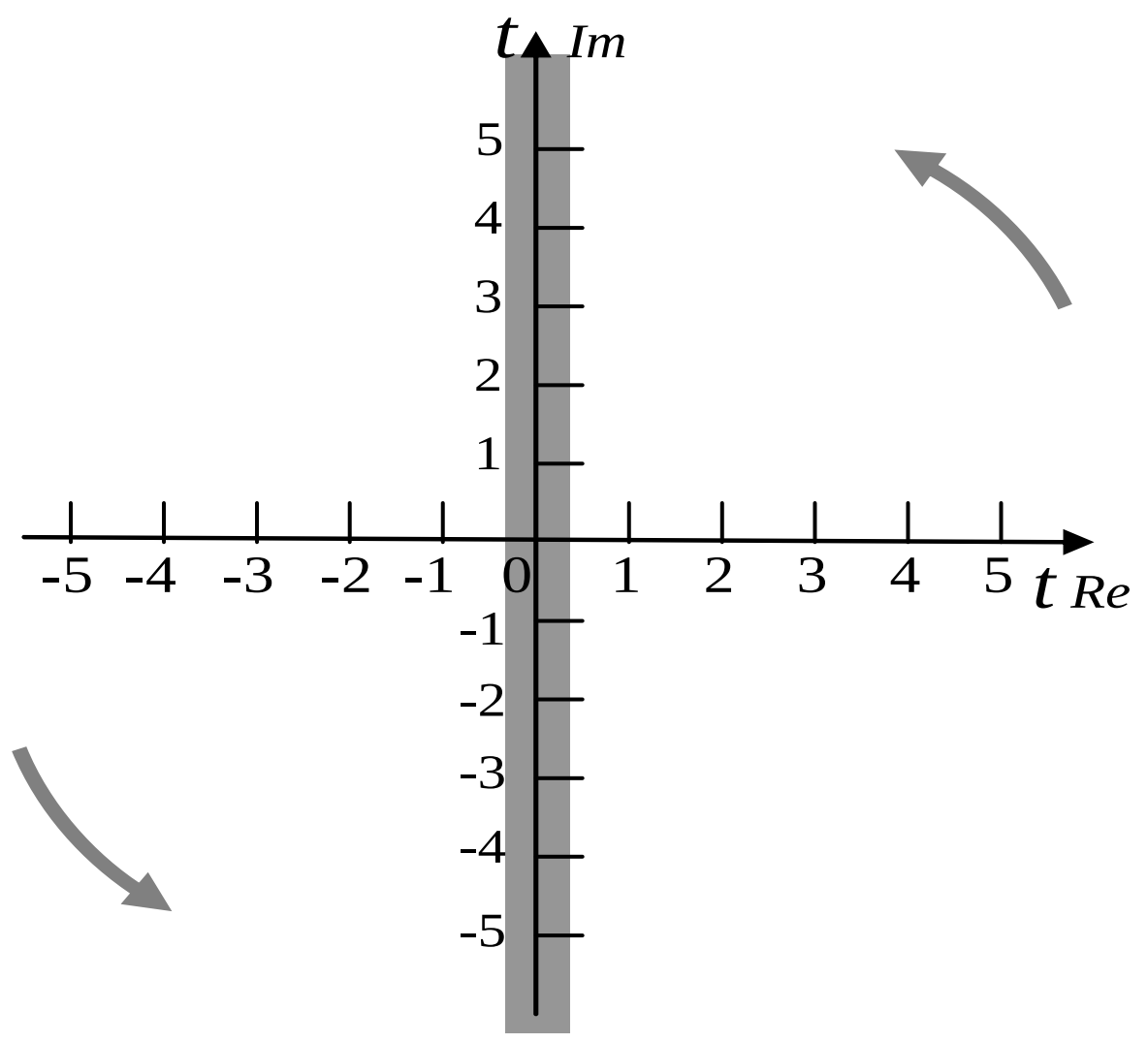

Fig.3 – Mathematically, it is possible to express the *imaginary time* in a direction *perpendicular* to the common real time. In this model, the imaginary time is a function of real time and vice versa.

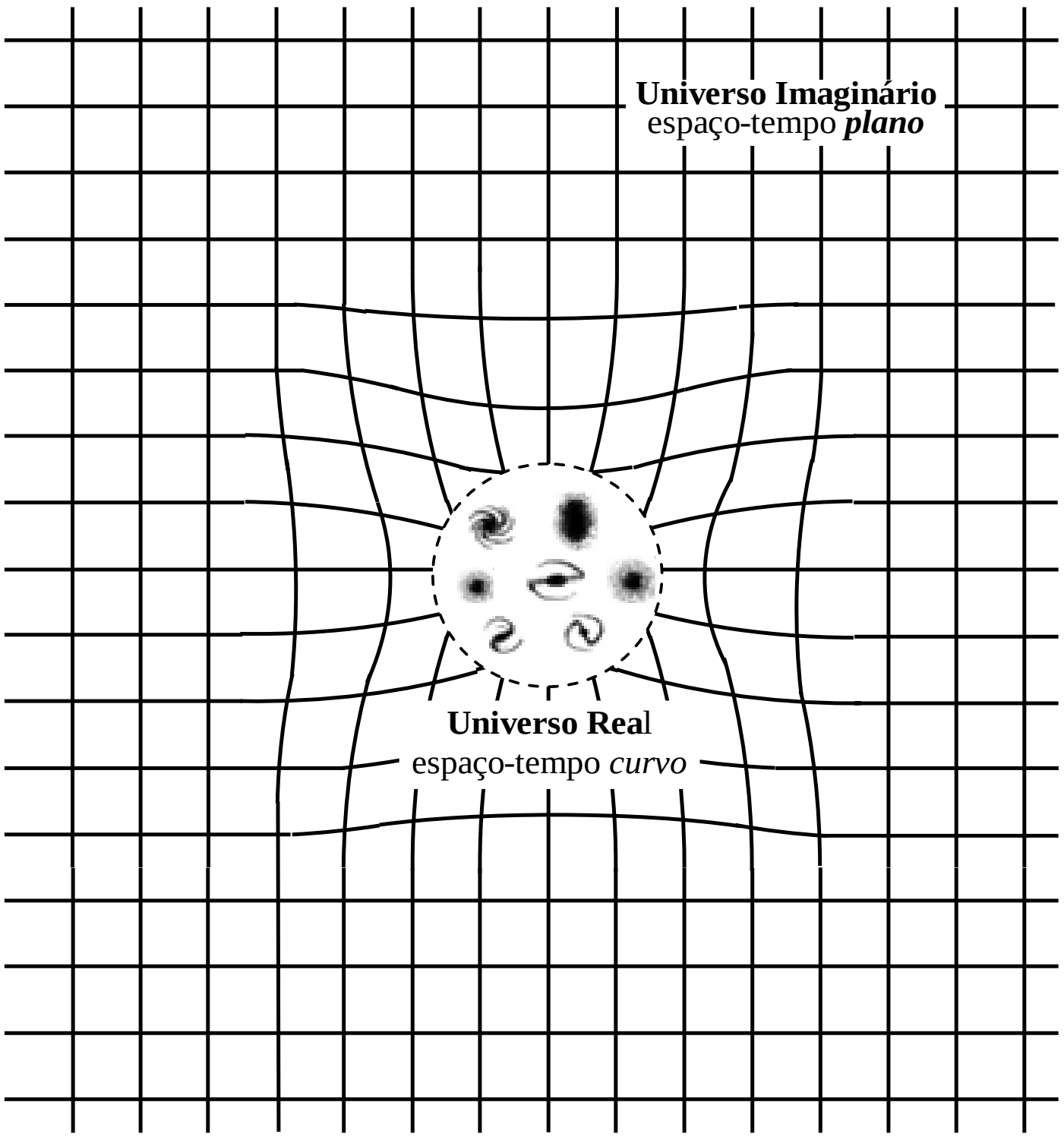

Fig. 4 – *The Model of the Imaginary Universe* (or Psychic Universe) *containing the Real Universe*. In the Imaginary Universe the space-time is *flat* (*Euclidean* metric), whereas in the Real Universe space-time is *curved* due to the presence of matter. In this model, the boundaries of the Real Universe confuse itself with the Imaginary Universe that is unlimited.

Imaginary time is measured in imaginary units (e.g., 2i seconds instead of 2 seconds). This imaginary unit of time may seem strange to us, such as our unit of real time measurement seems to the spirits, accustomed in their world measuring time in imaginary units. It all depends, of course, on the Universe where we are.

Based on the very definition of imaginary time, it is easy to see that we can interpret it as a vector $\vec{OZ}$ . Thus, being a vector, the imaginary time need not necessarily be always oriented in the same direction as the real time. It can freely change direction and intensity. This means that an observer in imaginary time can move in any direction for the *future* or *past*, such as we can move in any direction in space. This unique feature creates for us – accustomed to the flow of time always in the same direction – a horizon of events full of possibilities, hard to imagine because of the limitations imposed upon our consciousnesses by the Material Universe.

The existence of imaginary time derives from the very existence of the *imaginary space-time* contained in the Psychic Universe. As already shown, the speed of propagation of interactions in the Psychic Universe is *infinite*, which means that *the metrics of space-imaginary time is Euclidean* (or flat), while the metrics of the real universe is *non-Euclidean* (or curve). Since the Psychic Universe contains the Real Universe, we can conclude that the limits of the Real Universe mix itself up with the Psychic Universe that is unlimited (See Fig. 4).

The fact that the Psychic Universe have no limits implies that it has no *singularities* or boundaries in the imaginary time direction. With this condition, there is no beginning or end of the imaginary time.

**Conclusion**

Both the traditional physicists and most people recognize that there are phenomena where matter does not act alone, i.e., which involves also what we call psychism (consciousness, thought). These phenomena had hitherto been relegated to the professional affair of experts other than the traditional physicists. However, in recent decades, Quantum Physics has shown us that some physical laws are stretched beyond the Material World, revealing the existence of the Spiritual World. Thus, the Spiritual World arises not as a supernatural world[11], but as something as *real* as our Material World. On the other hand, this knowledge paved the way for Physics to study psychic phenomena using the same criteria adopted for the study of physical phenomena. In other words, it was evident that psychic phenomena could also be described by the laws of Physics. This unification is the basis for the *Grand Unification* of Science and Religion. Thereafter, both could no longer follow on separately. It was clear that Science could more accurately describe the truth postulated by Religion.

In this context, the Religion - absorbed by Science, must leave the scene just as the purely philosophical Cosmology gave way, in the past century, to Quantum Cosmology, when Quantum Physics discovered the laws that accurately describe the structures of the Universe.

The unification of Science and Religion is highly relevant because it will eliminate the spread of religious beliefs that have caused so much harm to Humanity in recent millennia. Now, the truth postulated by Religion will be transmitted by Science in schools and universities, and the human beings will understand it and use it, such as they use and understand, for example, the electric current, knowing that it can cause harm and also benefits for its users.

It would be too much presumptuous to believe that, due to the simple revelation of this new knowledge, the human nature could change suddenly. It will certainly take several decades for a complete assimilation of such truth.

It will then be taught to people from early stages of learning, the fundamental importance of the *quality* of our thoughts, since it is from them that the psychic interaction is defined and, consequently, the extraordinary relationship that is established among the human consciousnesses, the Universe and God.

[11] In the eyes of the general public, all phenomena of unknown cause become readily supernatural, wonderful and miraculous: the cause, once known, shows that the phenomenon, for more extraordinary it may seems, is nothing but a consequence of one or more natural laws. It is in this way that the set of supernatural facts is reduced with the Science progress.

Mankind then will begin to develop its psychic possibilities starting from the regular training at school.

There will come a time when, on Earth, the good will prevail over evil. The good spirits incarnated on Earth will become more numerous and, by the law of Psychic Interaction and Mutual Affinity, they will attract more and more the good spirits to Earth, warding off evil Spirits. The great transformation of Humanity then will be made by the progressive incarnation of better Spirits, which will give origin, on Earth, to a generation much more evolved than the current one.

# On the Cosmological Variation of the Fine Structure Constant

**Fran De Aquino**
Maranhao State University, Physics Department, S.Luis/MA, Brazil.

Recently, evidence indicating cosmological variations of the fine structure constant, α, has been reported. This result led to the conclusion that possibly the physical constants and the laws of physics vary throughout the universe. However, it will be shown here that variations in the value of the elementary electric charge, *e*, can occur under specific conditions, consequently producing variations in the value of α.
**Key words:** Fine Structure Constant, Elementary Electric Charge, Cosmology, Physics of Black-Holes.
PACS: 06.20. Jr, 98.80.-k, 98.80.Jk, 04.70.-s.

The well-known Fine Structure Constant determines the strength of the electromagnetic field and is expressed by the following equation (in SI units) [1]:

$$\alpha = \frac{e^2}{4\pi\varepsilon_0 \hbar c} = \frac{1}{137.035999 58(52)} \qquad (1)$$

However, recently, Webb, J.K *et al.*, [2] using data of the Very Large Telescope (VLT) and of the ESO Science archive, noticed small variation in the value of $\alpha$ in several distant galaxies. This led to the conclusion that $\alpha$ is not a constant [2- 4].

It will be shown here, that variations in the value of the elementary electric charge, *e*, can occur under specific conditions, consequently producing variations in the value of α. This effect may be explained starting from the expression recently obtained for the *electric charge* [5], i.e.,

$$q = \sqrt{4\pi\varepsilon_0 G}\; m_{g(im)}\, i \qquad (2)$$

where $m_{g(im)}$ are the *imaginary* gravitational mass of the elementary particle; $\varepsilon_0 = 8.854\times10^{-12} F/m$ is the permittivity of the free space and $G = 6.67\times10^{-11} N.m^2.kg^{-1}$ is the universal constant of gravitation.

For example, in the case of the *electron*, it was shown [5] that

$$m_{ge(im)} = \left\{1 - 2\left[\sqrt{1+\left(\frac{U_{e(im)}}{m_{i0e(im)}c^2}\right)^2} - 1\right]\right\} m_{i0e(im)} = $$
$$= \chi_e m_{i0e(im)} \qquad (3)$$

where $m_{i0e(im)} = -\frac{2}{\sqrt{3}} m_{i0e(real)}\, i$, $m_{i0e(real)} = 9.11\times10^{-31} kg$ and $U_{e(im)} = \eta_e k T_e i$. In this expression $\eta_e \cong 0.1$ is the *absorption factor* for the electron and $T_e \cong 6.2\times10^{31} K$ is its internal temperature (*temperature of the Universe when the electron was created*); $k = 1.38\times10^{-23} J/^oK$ is the *Boltzmann constant*.

Thus, according to Eq. (3), the value of $\chi_e$ is given by $\chi_e = -1.8\times10^{21}$. Then, according to Eq. (2), the *electric charge of the electron* is

$$q_e = \sqrt{4\pi\varepsilon_0 G}\; m_{ge(im)}\, i =$$
$$= \sqrt{4\pi\varepsilon_0 G}\left(\chi_e m_{i0e(im)} i\right) =$$
$$= \sqrt{4\pi\varepsilon_0 G}\left(-\chi_e \tfrac{2}{\sqrt{3}} m_{i0e(real)} i^2\right) =$$
$$= \sqrt{4\pi\varepsilon_0 G}\left(\chi_e \tfrac{2}{\sqrt{3}} m_{i0e(real)}\right) = -1.6\times10^{-19} C$$

As we know, the absolute value of this charge is called the *elementary electric charge*, $e$.

Since the internal temperature of the particle can vary, we then conclude that $\chi$ is not a constant, and consequently the value of $e$ also cannot be a constant in the Universe. Its value will depend on the local conditions that can vary the internal temperature of the particle. The *gravitational compression*, for example, can reduce the volume $V$ of the particles, diminishing their internal temperature $T$ to a temperature $T'$ according to the well-known equation: $T' = (V'/V)T$ [6]. This decreases the value of $U_{(im)}$, decreasing consequently the value of

$\chi$. Equation (2) shows that $e$ is proportional to $\chi$, i.e.,

$$\begin{aligned} e &= \sqrt{4\pi\varepsilon_0 G}\; m_{g(im)}\; i = \\ &= \sqrt{4\pi\varepsilon_0 G}\left(\chi\; m_{i0(im)} i\right) = \\ &= \sqrt{4\pi\varepsilon_0 G}\left(-\chi \tfrac{2}{\sqrt{3}} m_{i0(real)} i^2\right) = \\ &= \sqrt{4\pi\varepsilon_0 G}\left(\chi \tfrac{2}{\sqrt{3}} m_{i0(real)}\right) \end{aligned}$$

Therefore, when the volume of the particle decreases, the value of $e$ will be less than $1.6\times10^{-19}C$. Similarly, if the volume $V$ is *increased*, the temperature $T$ will be *increased* at the same ratio, increasing the value of $\chi$, and also the value of $e$. The *gravitational traction*, for example, can increase the volume $V$ of the particles, increasing their internal temperature $T$, and consequently increasing their electric charges (See Fig.1).

*Conclusions* – Our theoretical results show that variations in the value of the elementary electric charge, *e*, can occur under specific conditions, consequently producing variations of the *fine structure constant*, α, as shown in Fig.1. This excludes totally the erroneous hypothesis that the laws of physics vary throughout the universe.

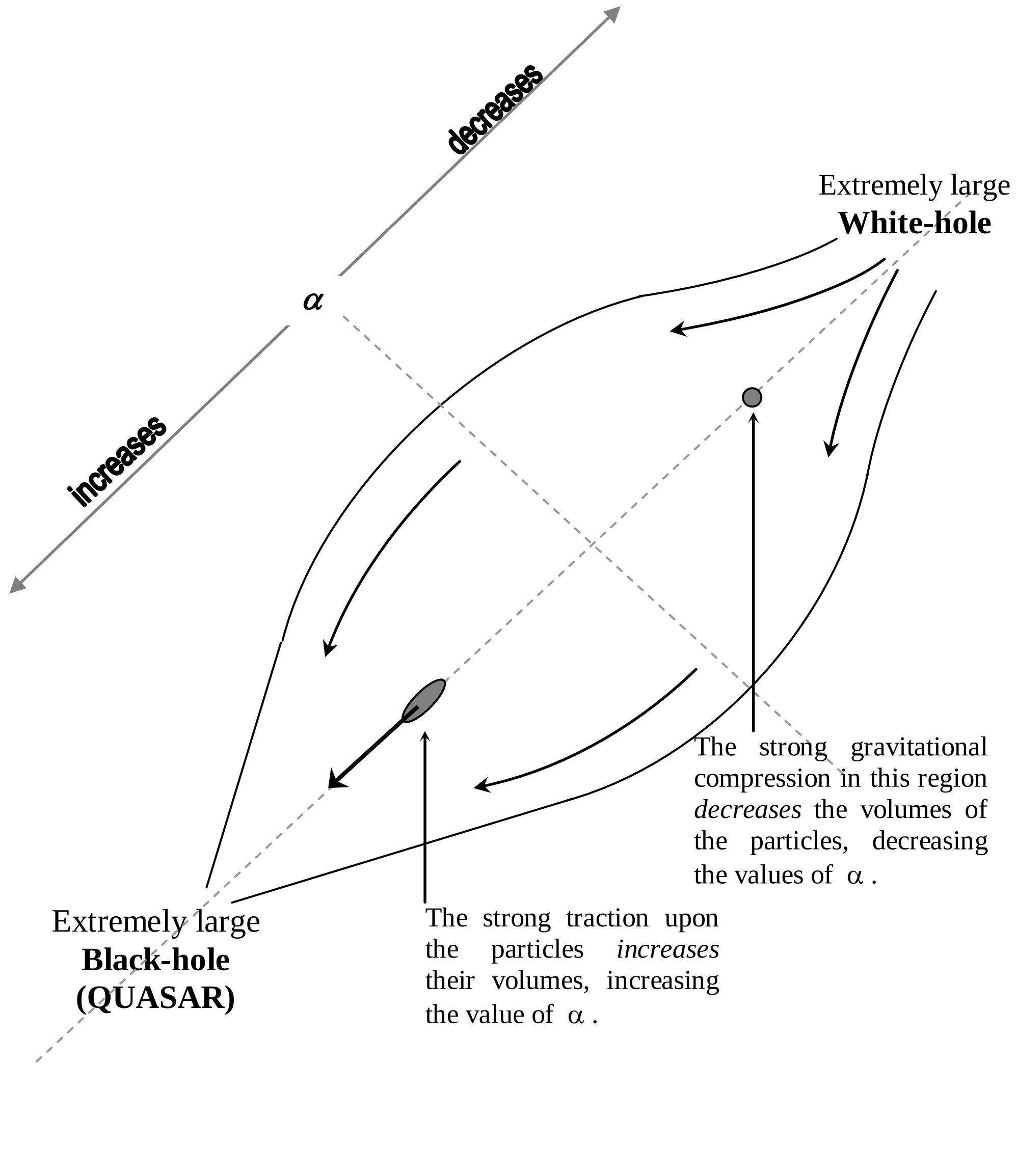

Fig. 1 – A spatial dipole that can explain the dipole variation of α reported by Webb. J.K. *et al*.

# The velocity of neutrinos

**Fran De Aquino**
Maranhao State University, Physics Department, S.Luis/MA, Brazil.

Recently, the OPERA neutrino experiment at the underground Gran Sasso Laboratory has measured the velocity of neutrinos from the CERN CNGS beam over a baseline of about 730 km. The experiment shows that neutrinos can have *superluminal* velocities. This result could, in principle, be taken as a clear violation of the Special Relativity. However, it will be show here that neutrinos can actually travel at velocities faster than light speed, without violating Special Relativity.

**Key words:** Neutrino mass, Neutrino interactions, Special relativity.
PACS: 14.60.Pq, 13.15.+g, 03.30.+p.

The mass of the *electron neutrino* $(\nu_e)$ is usually measured using the beta decay. The continuous spectrum of beta decay electrons terminates at a maximal energy, which depends on the neutrino mass and on the emitting nucleus type. Because of the way that the neutrino mass affects the electron energy spectrum, the measured quantity is the *square* of the neutrino mass. All recent measurements show that the *neutrino mass squared* is *negative* [1]. However, the square root of a negative number is an *imaginary number*. Thus, the measurements suggest that the *electron neutrino has an imaginary mass*. Assuming that the neutrino has no *real mass*, and considering that the *imaginary momentum* has a *real* value, i.e., $L_{(im)} = I_{(im)}\omega_{(im)} \equiv S_{(real)}$ and $p_{(im)} = M_{g(im)}V_{(im)} \equiv p_{(real)}$, we can infer that the neutrino is an *imaginary particle* with a measurable property; *the square of its imaginary mass*.

The OPERA neutrino experiment [2] at the underground Gran Sasso Laboratory (LNGS) was designed to perform the first detection of neutrino oscillations. Recently, it was reported that the OPERA neutrino experiment had discovered neutrinos with velocities greater than the light speed [3]. The neutrinos in question appear to be reaching the detector 60 nanoseconds faster than light would take to cover the same distance. That translates to a speed 0.002% higher than $c = 299{,}792{,}458$ m.s$^{-1}$ (the speed upper limit for *real particles* in the *real* spacetime).

The quantization of velocity shows that there is a speed upper limit, $c_i > c$, for *imaginary particles* in the *real* spacetime (real Universe)*. This means that Einstein's speed limit $(c)$ not applies to imaginary particles propagating in the real spacetime. Theoretical predictions show that $c_i \approx 10^{12}\, m.s^{-1}$ [4]. Consequently, the imaginary particles, such as the neutrinos, can reaches velocities faster than light speed. Therefore, in the case of imaginary particles, we must replace $c$ in the *Lorentz transformation* by $C_{(im)} = c_i i$ in order to generalize the equations of Special relativity. Thus, the *imaginary* kinetic energy of *imaginary particles*, for example, is written in the following form:

$$K_{(im)} = \left(m_{i(im)} - m_{i0(im)}\right)C^2_{(im)} = \left(\frac{1}{\sqrt{1-\frac{V^2_{(im)}}{C^2_{(im)}}}} - 1\right)m_{i0(im)}C^2_{(im)} =$$

$$= \left(\frac{1}{\sqrt{1-\frac{V^2}{c_i^2}}} - 1\right)m_{i0(im)}C^2_{(im)}$$

where $m_{i0(im)}$ is the *imaginary mass* of the particle at rest. The expression above shows

* The speed upper limit for *real particles* in the *imaginary* spacetime is $c$, because the relativistic expression of the mass shows that the velocity of *real* particles cannot be larger than $c$ in *any space-time*.

that the imaginary particle has a *real* velocity $V$. This means that imaginary particles propagating in the real spacetime can be detected. This is the case, for example, of the neutrinos with $V > c$ observed in the OPERA neutrino experiment.

Note that the *imaginary kinetic energy* of the particle is what gives to the neutrino its real velocity $\left(K_{(im)} \rightarrow V\right)$. This solves therefore, the problem of how the neutrino propagates in the space.

In addition, we can conclude that in the neutrino-electron reactions, mediated by the Z particle, the neutrino does not enter as a real mass but as a *real* angular *momentum* (spin ½). The real mass of the neutrino is null, but the *real* angular *momentum* and the *imaginary* angular *momentum* of the neutrino are not null. The *real* angular *momentum* of the neutrino, $S_{(real)}$, derives from its *imaginary* angular *momentum*, according to the following relation: $L_{(im)} = I_{(im)}\omega_{(im)} \equiv S_{(real)} = \sqrt{s(s+1)}\hbar$.

# Proca Equations and the Photon Imaginary Mass

**Fran De Aquino**
Maranhao State University, Physics Department, S.Luis/MA, Brazil.

It has been recently proposed that the photon has *imaginary mass* and null real mass. Proca equations are the unique simplest relativistic generalization of Maxwell equations. They are the theoretical expressions of possible nonzero photon rest mass. The fact that the photon has imaginary mass introduces relevant modifications in Proca equations which point to a deviation from the Coulomb's inverse square law.

**Key words:** Quantum electrodynamics, Specific calculations, Photons
PACS: 12.20.-m, 12.20.Ds, 14.70.Bh.

For quite a long time it has been known that the effects of a *nonzero photon rest mass* can be incorporated into electromagnetism through the *Proca equations* [1-2]. It is also known that particles with imaginary mass can be described by a real Proca field with a negative mass square [3-5]. They could be generated in storage rings, jovian magnetosphere, and supernova remnants. The existence of imaginary mass associated to the neutrino is already well-known. It has been reported by different groups of experimentalists that the mass square of the neutrino is *negative* [6]. Although the imaginary mass is not a measurable amount, its square is [7]. Recently, it was shown that an imaginary mass exist associated to the *electron* and the *photon* too [8]. The photon *imaginary mass* is given by

$$m_\gamma = \tfrac{2}{\sqrt{3}}\left(hf/c^2\right) i \qquad (1)$$

This means that the *photon* has null *real* mass and an *imaginary mass*, $m_\gamma$, expressed by the previous equation.

Proca equations may be found in many textbooks [9-11]. They provide a complete and self-consistent description of electromagnetic phenomena [12]. In the presence of sources $\rho$ and $\vec{j}$, these equations may be written as (in SI units)

$$\nabla\cdot\vec{E} = \frac{\rho}{\varepsilon_0} - \mu_\gamma^2\phi \qquad (2)$$

$$\nabla\cdot\vec{B} = 0 \qquad (3)$$

$$\nabla\times\vec{E} = -\frac{\partial\vec{B}}{\partial t} \qquad (4)$$

$$\nabla\times\vec{B} = \mu_0\vec{j} + \mu_0\varepsilon_0\frac{\partial\vec{E}}{\partial t} - \mu_\gamma^2\vec{A} \qquad (5)$$

where $\mu_\gamma = m_\gamma c/\hbar$, with the *real* variables $\mu_\gamma$ and $m_\gamma$. However, according to Eq. (1) $m_\gamma$ is an *imaginary mass*. Then, $\mu_\gamma$ must be also an imaginary variable. Thus, $\mu_\gamma^2$ is a *negative real* number similarly to $m_\gamma^2$. Consequently, we can write that

$$\mu_\gamma^2 = \frac{m_\gamma^2 c^2}{\hbar^2} = \frac{4}{3}\left(\frac{2\pi}{\lambda}\right)^2 = \frac{4}{3}k_r^2 \qquad (6)$$

whence we recognize $k_r = 2\pi/\lambda$ as the real part of the *propagation vector* $\vec{k}$ ;

$$k = \left|\vec{k}\right| = \left|k_r + ik_i\right| = \sqrt{k_r^2 + k_i^2} \qquad (7)$$

Substitution of Eq. (6) into Proca equations, gives

$$\nabla\cdot\vec{E} = \frac{\rho}{\varepsilon_0} - \frac{4}{3}k_r^2\phi \qquad (8)$$

$$\nabla\cdot\vec{B} = 0 \qquad (9)$$

$$\nabla\times\vec{E} = -\frac{\partial\vec{B}}{\partial t} \qquad (10)$$

$$\nabla\times\vec{B} = \mu_0\vec{j} + \mu_0\varepsilon_0\frac{\partial\vec{E}}{\partial t} - \frac{4}{3}k_r^2\vec{A} \qquad (11)$$

In four-dimensional space these equations can be rewritten as

$$\left(\nabla^2 - \frac{1}{c^2}\frac{\partial^2}{\partial t^2} - \frac{4}{3}k_r^2\right)A_\mu = -\mu_0\vec{j}_\mu \qquad (12)$$

where $A_\mu$ and $\vec{j}_\mu$ are the 4-vector of potential $(A, i\phi/c)$ and the current density $(\vec{j}, ic\rho)$, respectively. In free space the above equation reduces to

$$\left(\nabla^2 - \frac{1}{c^2}\frac{\partial^2}{\partial t^2} - \frac{4}{3}k_r^2\right)A_\mu = 0 \qquad (13)$$

which is essentially the Klein-Gordon equation for the photon.

Therefore, the presence of a photon in a static electric field modifies the wave equation for all potentials (including the Coulomb potential) in the form

$$\left(\nabla^2 - \frac{1}{c^2}\frac{\partial^2}{\partial t^2} - \frac{4}{3}k_r^2\right)\phi = -\frac{\rho}{\varepsilon_0} \qquad (14)$$

For a point charge, we obtain

$$\phi(r) = \frac{1}{4\pi\varepsilon_0}\frac{q}{r}e^{-\frac{2}{\sqrt{3}}(k_r r)} \qquad (15)$$

and the electric field

$$E(r) = \frac{q}{4\pi\varepsilon_0 r^2}\left[1 + \frac{2}{\sqrt{3}}(k_r r)\right]e^{-\frac{2}{\sqrt{3}}(k_r r)} \qquad (16)$$

Note that only in the absence of the photon $(k_r = 0)$ the expression of $E(r)$ reduces to the well-known expression: $E(r) = q/4\pi\varepsilon_0 r^2$. Thus, these results point to an exponential deviation from Coulomb's inverse square law, which, as we know, is expressed by the following equation (in SI units):

$$\vec{F}_{12} = -\vec{F}_{21} = \frac{1}{4\pi\varepsilon_0}\frac{q_1 q_2 \vec{r}_{12}}{|\vec{r}_{12}|^3} \qquad (17)$$

As seen in Eq. (16), the term

$$\frac{2}{\sqrt{3}}(k_r r)$$

only becomes significant if

$$r > \sim 10^{-4}\lambda \qquad (18)$$

This means that the Coulomb's law is a good approximation when $r < \sim 10^{-4}\lambda$. However, if $r > \sim 10^{-4}\lambda$, the expression of the force departs from the prediction of Maxwell's equations.

The lowest-frequency photons of the primordial radiation of 2.7K is about $10^8 Hz$ [13]. Therefore, the wavelength of these photons is $\lambda \approx 1m$. Consider the presence of these photons in a terrestrial experiment designed to measure the force between two electric charges separated by a distance $r$. According to Eq. (18), the deviation from the Coulomb's law only becomes relevant if $r > 10^{-4}m$. Then, if we take $r = 0.1m$, the result is

$$\frac{2}{\sqrt{3}}(k_r r) = \frac{4\pi}{\sqrt{3}}\left(\frac{r}{\lambda}\right) = 0.73$$

and

$$\left[1 + \frac{2}{\sqrt{3}}(k_r r)\right]e^{-\frac{2}{\sqrt{3}}(k_r r)} = 0.83$$

Therefore, a deviation of 17% in respect to the value predicted by the Coulomb's law.

Then, why the above deviation is not experimentally observed? Theoretically because of the presence of *Schumann radiation* $(f_1 = 7.83\,Hz, \lambda_1 = 3.8 \times 10^7 m)$ [14-15]. According to Eq. (18), for $\lambda_1 = 3.8 \times 10^7 m$, the deviation only becomes significant if

$$r > \sim 10^{-4}\lambda_1 = 3.8Km$$

Since the values of $r$ in usual experiments are much smaller than $3.8Km$ the result is that the deviation is negligible. In fact, this is easy to verify. For example, if $r = 0.1m$, we get

$$\frac{2}{\sqrt{3}}(k_r r) = \frac{4\pi}{\sqrt{3}}\left(\frac{r}{\lambda_1}\right) = \frac{4\pi}{\sqrt{3}}\left(\frac{0.1}{3.8\times 10^7}\right) = 1.9\times 10^{-8}$$

and

$$\left[1 + \frac{2}{\sqrt{3}}(k_r r)\right]e^{-\frac{2}{\sqrt{3}}(k_r r)} = 0.999999999$$

Now, if we put the experiment inside an aluminum box whose thickness of the walls are equal to 21cm * the experiment will be shielded for the Schumann radiation. By putting inside the box a photons source of $\lambda \approx 1m$, and making $r = 0.1m$, then it will be possible to observe the deviation previously computed of 17% in respect to the value predicted by the Coulomb's law.

* The thickness $\delta$ necessary to shield the experiment for Schumann radiation can be calculated by means of the well-known expression [16]: $\delta = 5z = 10/\sqrt{2\pi\mu\sigma f}$ where $\mu$ and $\sigma$ are, respectively, the permeability and the electric conductivity of the material; $f$ is the frequency of the radiation to be shielded.

# Gravity Control by means of Modified Electromagnetic Radiation

**Fran De Aquino**
Maranhao State University, Physics Department, S.Luis/MA, Brazil.

Here a new way for gravity control is proposed that uses electromagnetic radiation modified to have a smaller wavelength. It is known that when the velocity of a radiation is reduced its wavelength is also reduced. There are several ways to strongly reduce the velocity of an electromagnetic radiation. Here, it is shown that such a reduction can be done simply by making the radiation cross a conductive foil.

**Key words:** Modified theories of gravity, Experimental studies of gravity, Electromagnetic wave propagation.
PACS: 04.50.Kd , 04.80.-y, 41.20.Jb, 75.70.-i.

It was shown that the *gravitational mass* $m_g$ and *inertial mass* $m_i$ are correlated by means of the following factor [1]:

$$\frac{m_g}{m_{i0}} = \left\{1 - 2\left[\sqrt{1 + \left(\frac{\Delta p}{m_{i0}c}\right)^2} - 1\right]\right\} \qquad (1)$$

where $m_{i0}$ is the *rest* inertial mass of the particle and $\Delta p$ is the variation in the particle's *kinetic momentum*; $c$ is the speed of light.

When $\Delta p$ is produced by the absorption of a photon with wavelength $\lambda$, it is expressed by $\Delta p = h/\lambda$. In this case, Eq. (1) becomes

$$\frac{m_g}{m_{i0}} = \left\{1 - 2\left[\sqrt{1 + \left(\frac{h/m_{i0}c}{\lambda}\right)^2} - 1\right]\right\}$$
$$= \left\{1 - 2\left[\sqrt{1 + \left(\frac{\lambda_0}{\lambda}\right)^2} - 1\right]\right\} \qquad (2)$$

where $\lambda_0 = h/m_{i0}c$ is the *De Broglie wavelength* for the particle with *rest* inertial mass $m_{i0}$.

It is easily seen that $m_g$ cannot be strongly reduced simply by using electromagnetic waves with wavelength $\lambda$ because $\lambda_0$ is very smaller than $10^{-10}m$. However, it is known that the wavelength of a radiation can be strongly reduced simply by strongly reducing its velocity.

There are several ways to reduce the velocity of an electromagnetic radiation. For example, by making light cross an *ultra cold atomic gas,* it is possible to reduce its velocity down to *17m/s* [2-7]. Here, it is shown that the velocity of an electromagnetic radiation can be strongly reduced simply by making the radiation cross a conductive foil.

From Electrodynamics we know that when an electromagnetic wave with frequency $f$ and velocity $c$ incides on a material with relative permittivity $\varepsilon_r$, relative magnetic permeability $\mu_r$ and electrical conductivity $\sigma$, its *velocity is reduced* to $v = c/n_r$ where $n_r$ is the index of refraction of the material, given by [8]

$$n_r = \frac{c}{v} = \sqrt{\frac{\varepsilon_r \mu_r}{2}\left(\sqrt{1 + (\sigma/\omega\varepsilon)^2} + 1\right)} \qquad (3)$$

If $\sigma >> \omega\varepsilon$, $\omega = 2\pi f$, the Eq. (3) reduces to

$$n_r = \sqrt{\frac{\mu_r \sigma}{4\pi\varepsilon_0 f}} \qquad (4)$$

Thus, the wavelength of the incident radiation becomes

$$\lambda_{mod} = \frac{v}{f} = \frac{c/f}{n_r} = \frac{\lambda}{n_r} = \sqrt{\frac{4\pi}{\mu f \sigma}} \qquad (5)$$

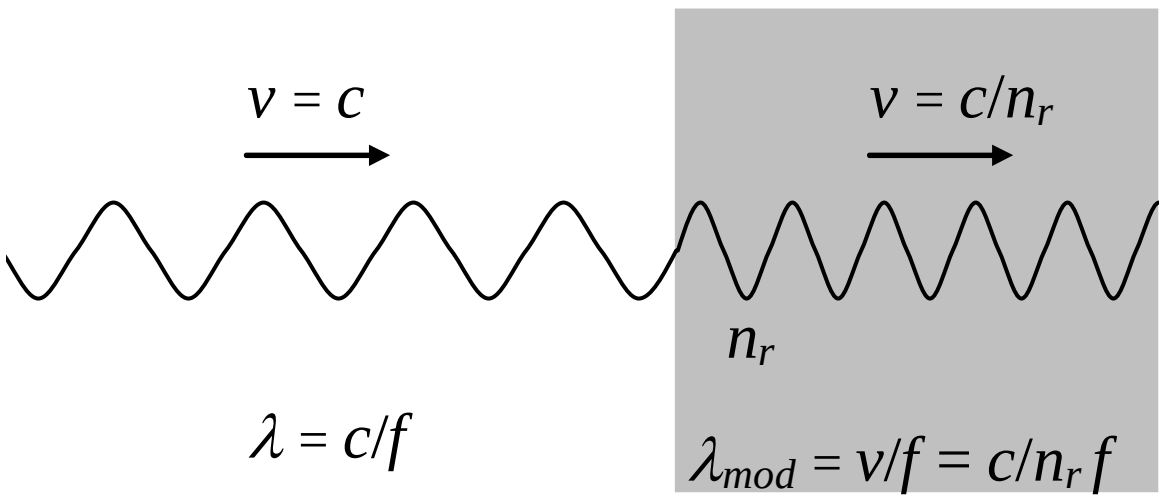

Fig. 1 – *Modified Electromagnetic Wave*. The wavelength of the electromagnetic wave can be strongly reduced, but its frequency remains the same.

Now consider a $1GHz$ ($\lambda \cong 0.3m$) radiation incident on *Aluminum foil* with $\sigma = 3.82\times10^7 S/m$ and thickness $\xi = 10.5\mu m$. According to Eq. (5), the *modified wavelength* is

$$\lambda_{mod} = \sqrt{\frac{4\pi}{\mu f \sigma}} = 1.6 \times 10^{-5} m \qquad (6)$$

Consequently, the wavelength of the $1GHz$ radiation *inside the foil* will be $\lambda_{mod} = 1.6\times10^{-5} m$ and not $\lambda \cong 0.3m$.

It is known that a radiation with frequency *f*, propagating through a material with electromagnetic characteristics $\varepsilon$, $\mu$ and $\sigma$, has the amplitudes of its waves decreased in $e^{-1}$=0.37 (37%), when it passes through a distance *z*, given by

$$z = \frac{1}{\omega\sqrt{\frac{1}{2}\varepsilon\mu\left(\sqrt{1+(\sigma/\omega\varepsilon)^2}-1\right)}} \qquad (7)$$

The radiation is totally absorbed at a distance δ≅5z [8].

In the case of the $1GHz$ radiation propagating through the Aluminum foil Eq. (7), gives

$$z = \frac{1}{\sqrt{\pi\mu\sigma f}} = 2.57\times10^{-6} = 2.57\mu m \qquad (8)$$

Since the thickness of the Aluminum foil is $\xi = 10.5\mu m$ then, we can conclude that, practically all the incident $1GHz$ radiation is absorbed by the foil.

If the foil contains $n$ atoms/m$^3$, then the number of atoms per area unit is $n\xi$. Thus, if the electromagnetic radiation with frequency $f$ incides on an area $S$ of the foil it reaches $nS\xi$ atoms. If it incides on the total area of the foil, $S_f$, then the total number of atoms reached by the radiation is $N = nS_f\xi$. The number of atoms per unit of volume, $n$, is given by

$$n = \frac{N_0 \rho}{A} \qquad (9)$$

where $N_0 = 6.02\times10^{26} atoms/kmole$ is the Avogadro's number ; $\rho$ is the matter density of the foil (in $kg/m^3$) and $A$ is the atomic mass. In the case of the *Aluminum* $\left(\rho = 2700kg/m^3, A = 26.98kmole\right)$ the result is

$$n_{Al} = 6.02\times10^{28} atoms/m^3 \qquad (10)$$

The *total number of photons* inciding on the foil is $n_{total\ photons} = P/hf^2$, where $P$ is the power of the radiation flux incident on the foil.

When an electromagnetic wave incides on the Aluminum foil, it strikes on $N_f$ front atoms, where $N_f \cong (nS_f)\phi_{atom}$. Thus, the wave incides effectively on an area $S = N_f S_a$, where $S_a = \frac{1}{4}\pi\phi^2_{atom}$ is the cross section area of one Aluminum atom. After these collisions, it carries out $n_{collisions}$ with the other atoms of the foil (See Fig.2).

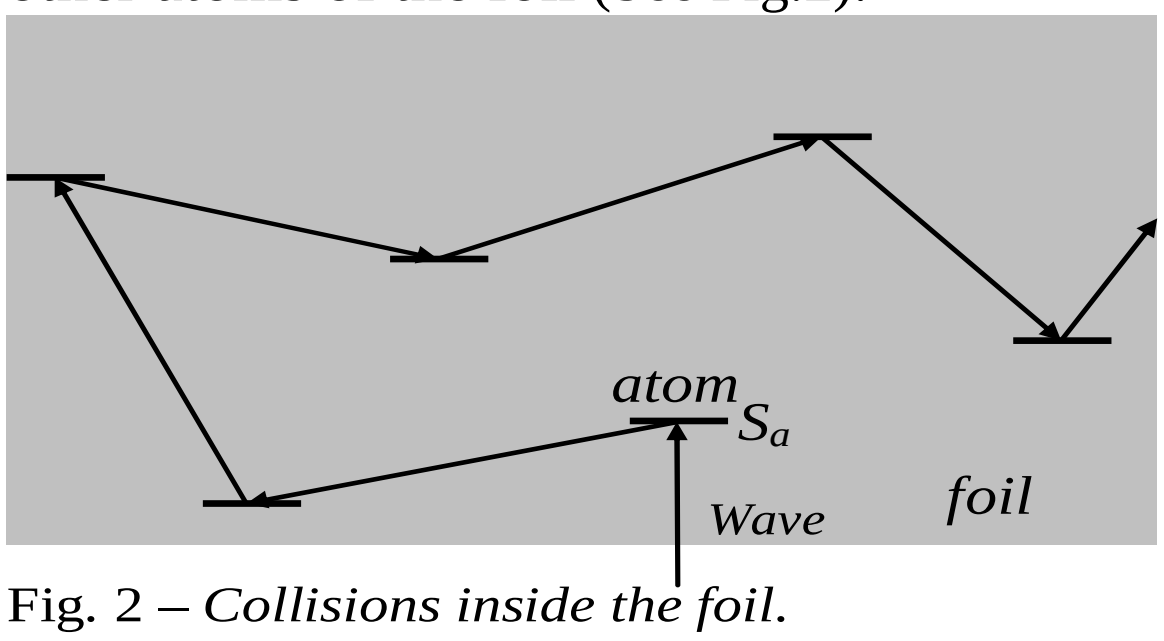

Fig. 2 – *Collisions inside the foil.*

Thus, the total number of collisions in the volume $S\xi$ is

$$N_{collisions} = N_f + n_{collisions} = nS\phi_{atom} + (nS\xi - nS\phi_{atom}) =$$
$$= nS\xi \qquad (11)$$

The power density, $D$, of the radiation on the foil can be expressed by

$$D = \frac{P}{S} = \frac{P}{N_f S_a} \qquad (12)$$

The same power density as a function of the power $P_0$ radiated from the antenna, is given by

$$D = \frac{P_0}{4\pi r^2} \qquad (13)$$

where $r$ is the distance between the antenna and the foil. Comparing equations (12) and (13), we get

$$P = \left(\frac{N_f S_a}{4\pi r^2}\right) P_0 \qquad (14)$$

We can express the *total mean number of collisions in each atom*, $n_1$, by means of the following equation

$$n_1 = \frac{n_{total\ photons} N_{collisions}}{N} \qquad (15)$$

Since in each collision is transferred a *momentum* $h/\lambda$ to the atom, then the *total momentum* transferred to the foil will be $\Delta p = (n_1 N) h/\lambda$. Therefore, in accordance with Eq. (1), we can write that

$$\frac{m_g}{m_{i0}}=\left\{1-2\left[\sqrt{1+\left[(n_1N)\frac{\lambda_0}{\lambda}\right]^2}-1\right]\right\}=$$

$$=\left\{1-2\left[\sqrt{1+\left[n_{total\ photons}N_{collisions}\frac{\lambda_0}{\lambda}\right]^2}-1\right]\right\} \quad (16)$$

Since Eq. (11) gives $N_{collisions}=nS\xi$, we get

$$n_{total\ photons}N_{collisions}=\left(\frac{P}{hf^2}\right)(nS\xi) \quad (17)$$

Substitution of Eq. (17) into Eq. (16) yields

$$\frac{m_g}{m_{i0}}=\left\{1-2\left[\sqrt{1+\left[\left(\frac{P}{hf^2}\right)(nS\xi)\frac{\lambda_0}{\lambda}\right]^2}-1\right]\right\} \quad (18)$$

Substitution of Eq. (14) into Eq. (18) gives

$$\frac{m_g}{m_{i0}}=\left\{1-2\left[\sqrt{1+\left[\left(\frac{N_fS_aP_0}{4\pi r^2f^2}\right)\left(\frac{nS\xi}{m_{i0}c}\right)\frac{1}{\lambda}\right]^2}-1\right]\right\} \quad (19)$$

Substitution of $N_f\cong(nS_f)\phi_{atom}$ and $S=N_fS_a$ into Eq. (19) it reduces to

$$\frac{m_g}{m_{i0}}=\left\{1-2\left[\sqrt{1+\left[\left(\frac{n^3S_f^2S_a^2\phi_{atom}^2P_0\xi}{4\pi r^2m_{i0}cf^2}\right)\frac{1}{\lambda}\right]^2}-1\right]\right\} \quad (20)$$

In the case of a 20cm square Aluminum foil, with thickness $\xi=10.5\mu m$, we get $m_{i0}=1.1\times10^{-3}kg$, $S_f=4\times10^{-2}m^2$, $\phi_{atom}\cong10^{-10}m^2$ $S_a\cong10^{-20}m^2$, $n=n_{Al}=6.02\times10^{28}atoms/m^3$, Substitution of these values into Eq. (20), gives

$$\frac{m_{g(Al)}}{m_{i0(Al)}}=\left\{1-2\left[\sqrt{1+\left[\left(8.84\times10^{11}\frac{P_0}{r^2f^2}\right)\frac{1}{\lambda}\right]^2}-1\right]\right\} \quad (21)$$

Thus, if the Aluminum foil is at a distance $r=1m$ from the antenna, and the power radiated from the antenna is $P_0=32W$, and the frequency of the radiation is $f=1GHz$ then Eq.(21) gives

$$\frac{m_{g(Al)}}{m_{i0(Al)}}=\left\{1-2\left[\sqrt{1+\left[\frac{2.8\times10^{-5}}{\lambda}\right]^2}-1\right]\right\} \quad (22)$$

In the case of the Aluminum foil and $1Ghz$ radiation, Eq. (6) shows that $\lambda_{mod}=1.6\times10^{-5}m$. Thus, by substitution of $\lambda$ by $\lambda_{mod}$ into Eq. (22), we get the following expression

$$\frac{m_{g(Al)}}{m_{i0(Al)}}\cong-1 \quad (23)$$

Since $\vec{P}=m_g\vec{g}$ then the result is

$$\vec{P}_{(Al)}=m_{g(Al)}\vec{g}\cong-m_{i0(Al)}\vec{g} \quad (24)$$

This means that, in the mentioned conditions, *the weight force* of the Aluminum foil *is inverted*.

It was shown [1] that there is an additional effect of *Gravitational Shielding* produced by a substance whose gravitational mass was reduced or made negative. This effect shows that just *above the substance* the gravity acceleration $g_1$ will be reduced at the same ratio $\chi_1=m_g/m_{i0}$, i.e., $g_1=\chi_1 g$, ($g$ is the gravity acceleration *bellow* the substance). This means that above the Aluminum foil the gravity acceleration will be modified according to the following expression

$$g_1=\chi_1 g=\left(\frac{m_{g(Al)}}{m_{i0(Al)}}\right)g \quad (25)$$

where the factor $\chi_1=m_{g(Al)}/m_{i0(Al)}$ will be given Eq. (21).

In order to check the theory presented here, we propose the experimental set-up shown in Fig. 3. The distance between the Aluminum foil and the antenna is $r=1m$. The maximum output power of the $1GHz$ transmitter is 32W CW. A 10g body is placed above Aluminum foil , in order to check the *Gravitational Shielding Effect*. The distance between the Aluminum foil and the 10g body is approximately 10 *cm*. The alternative device to measure the weight variations of the foil and the body (including the *negative* values) uses two balances (200g / 0.01g) as shown in Fig .3.

In order to check the effect of a *second* Gravitational Shielding above the first one(Aluminum foil), we can remove the 10g body, putting in its place a second Aluminum foil, with the same characteristics of the first one. The 10g body can be then placed at a distance of 10cm above of the second Aluminum foil. Obviously, it must be connected to a third balance.

As shown in a previous paper [9] the gravity above the second Gravitational Shielding, in the case of $\chi_2 = \chi_1$, is given by

$$g_2 = \chi_2 g_1 = \chi_1^2 g \qquad (26)$$

If a third Aluminum foil is placed above the second one, then the gravity above this foil is $g_3 = \chi_3 g_2 = \chi_3 \chi_2 \chi_1 g = \chi_1^3 g$, and so on.

In practice, *Multiple Gravitational Shieldings* can be constructed by inserting $N$ several *parallel* Aluminum foils inside the dielectric of a parallel plate capacitor (See Fig. 4). In this case, the resultant capacity of the capacitor becomes $C_r = C/N = \varepsilon_r \varepsilon_0 S_f / Nd$, where $S_f$ is the area of the Aluminum foils and $d$ the distance between them; $\varepsilon_r$ is the relative permeability of the dielectric. By applying a voltage $V_{rms}$ on the plates of the capacitor a current $i_{rms}$ is produced through the Aluminum foils. It is expressed by $i_{rms} = V_{rms}/X_C = 2\pi f C_r V_{rms}$.

Since $j_{rms} = \sigma E_{rms}$ and $j_{rms} = i_{rms}/S_f$ we get $E_{rms} = i_{rms}/S_f\,\sigma$, which is the oscillating electric field through the Aluminum foils. By substituting this expression into Eq. (20), and considering that $\lambda = \lambda_{mod} = (4\pi/\mu f \sigma)^{\frac{1}{2}}$ (Eq.6) and $D = P_0/4\pi r^2 = n_r E_{rms}^2 / 2\mu_r \mu_0 c$, where $n_r = (\mu_r \sigma / 4\pi\varepsilon_0 f)^{\frac{1}{2}}$ (Eq. 4), we obtain:

$$\chi = \left\{1 - 2\left[\sqrt{1 + \frac{n_{Al}^6 S_a^4 \phi_{atom}^4 i_{rms}^4}{64\pi^2 \rho_{Al}^2 c^2 S_f^2 \sigma_{Al}^2 f^4}} - 1\right]\right\} \qquad (27)$$

Since

$$i_{rms} = V_{rms}/X_C = 2\pi f C_r V_{rms} = 2\pi f \left(\varepsilon_r \varepsilon_0 S_f / Nd\right) V_{rms}$$

Then

$$\frac{i_{rms}}{f} = 2\pi \left(\varepsilon_r \varepsilon_0 S_f / Nd\right) V_{rms} \qquad (28)$$

Substitution of this equation into Eq. (27) gives

$$\chi = \left\{1 - 2\left[\sqrt{1 + \frac{\pi^2 n_{Al}^6 S_a^4 \phi_{atom}^4 \varepsilon_r^4 \varepsilon_0^4 S_f^2 V_{rms}^4}{4\rho_{Al}^2 c^2 \sigma_{Al}^2 N^4 d^4}} - 1\right]\right\} \qquad (29)$$

Substitution of the known value of $n_{Al} = 6.02\times10^{28} atoms/m^3$, $\phi_{atom} \cong 1\times10^{-10} m$, $S_a = \frac{1}{2}\left[4\pi(\phi_{atom}/2)^2\right] = \frac{1}{2}\pi\phi_{atom}^2 \cong 1\times10^{-20} m^2$, $\varepsilon_r = 2.1$ (Teflon $24KV/mm$, Short Time, 1.6 mm [10]), $\rho_{Al} = 2700 kg.m^{-3}$, we get

$$\chi = \left\{1 - 2\left[\sqrt{1 + 1.4\times10^{-29}\frac{S_f^2}{N^4}\left(\frac{V_{rms}}{d}\right)^4} - 1\right]\right\} \qquad (30)$$

Note that, based on the equation above, it is possible to create a device for moving very heavy loads such as large monoliths, for example.

Imagine a large monolith on the Earth's surface. If we place below the monolith some sets with *Multiple Gravitational Shieldings* (See Fig.4), the value of the gravity acceleration above each set of Gravitational Shieldings becomes

$$g_R = \chi^\eta g \qquad (31)$$

where $\eta$ is the number of Gravitational Shieldings in each set.

Since we must have $V_{rms}/d < 24KV/mm$ (dielectric strength of Teflon) [10] then, for $d = 1.6mm \rightarrow V_{rms} < 38.4KV$. For $V_{rms} = 37KV$, $d = 1.6mm$, $S_f = 2.7m^2$, $N = 2$ and $\eta = 3$ Eq. (30) gives $\chi = -0.36$ and Eq. (31) shows that $g_R = \chi^3 g \cong -0.46 m/s^2$. The sign (-) shows that *the gravity acceleration above the six sets of Gravitational Shieldings* becomes *repulsive* in respect to the Earth. Thus, by controlling the value of $\chi$ it is possible to make the total mass of the monolith slightly negative in order to the monolith can float and, in this way, it can be displaced and carried to anywhere with ease.

Considering the dielectric strength of known dielectrics, we can write that $(V_{rms}/d)_{max} < 200KV/mm$. Thus, for a single capacitor $(N = 1)$ Eq. (30) gives

$$\chi = \left\{1 - 2\left[\sqrt{1 + \left(<< 2.2\times10^4 S_f^2\right)} - 1\right]\right\} \qquad (31)$$

The Gravitational Shielding effect becomes negligible for $\chi < 0.01$ (variation smaller than 1% in the gravitational mass). Thus, considering Eq. (31), we can conclude that *the Gravitational Shielding effect becomes significant only for* $S_f >> 10^{-2} m^2$. Possibly this is why it was not yet detected.

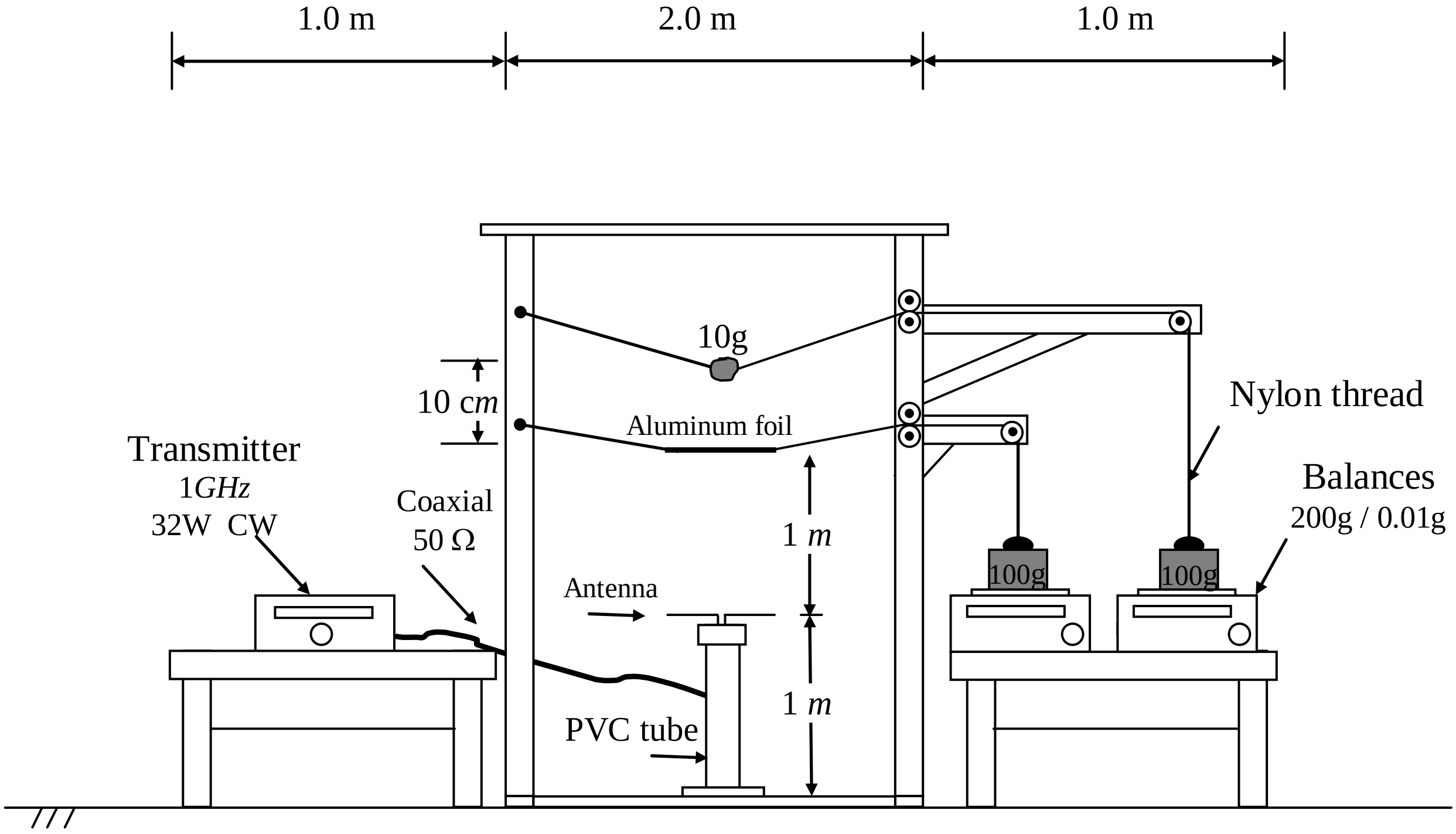

Fig. 3 – Experimental Set-up

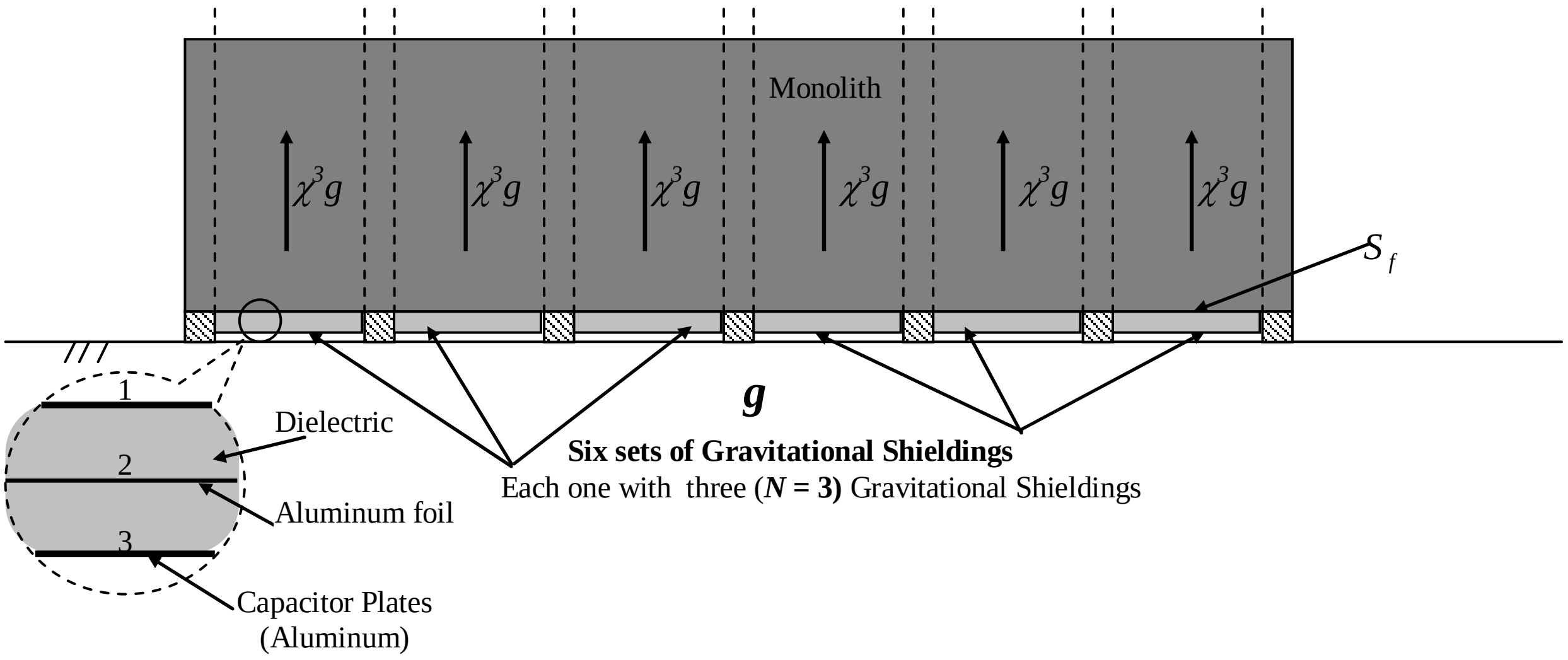

(a)

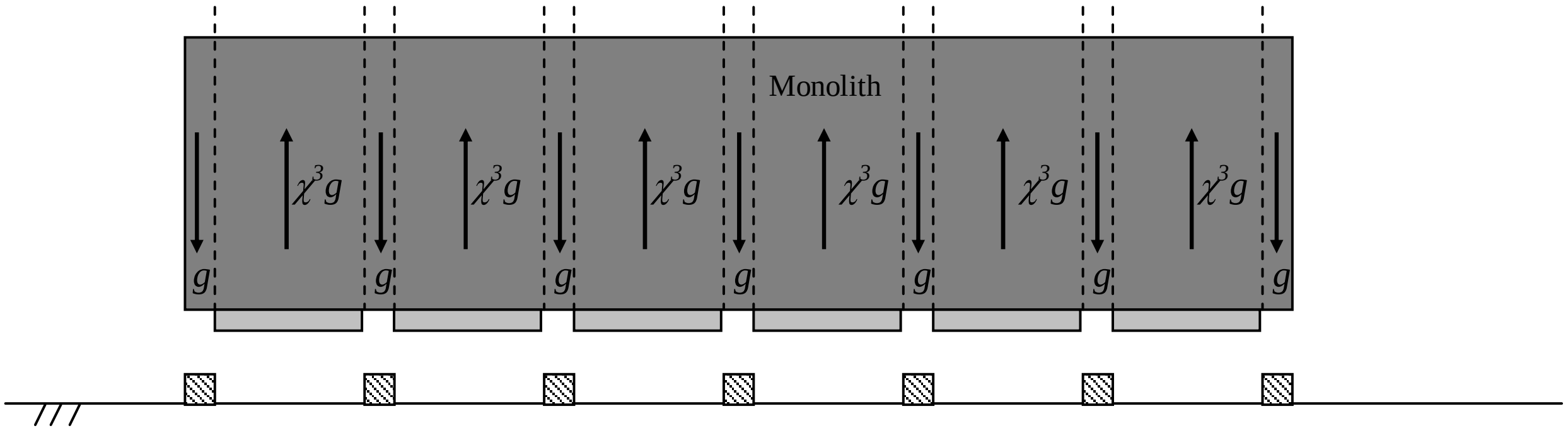

(b)

Fig. 4 – *System with six sets of Gravitational Shieldings for moving very heavy loads.* For $V_{rms} = 37KV$, $d = 1.6mm$, $S_f = 2.7m^2$, $N = 2$ and $\eta = 3$ Eq. (30) gives $\chi = -0.36$ and Eq. (31) shows that $g_R = \chi^3 g \cong -0.46 m/s^2$. The sign (-) shows that *the gravity acceleration above the six sets of Gravitational Shieldings* becomes *repulsive* in respect to the Earth. Thus, by controlling the value of $\chi$ it is possible to make the total mass of the monolith slightly negative in order to the monolith can float and, in this way, it can be displaced and carried to anywhere with ease.

# Transmission of DNA Genetic Information into Water by means of Electromagnetic Fields of Extremely-low Frequencies

**Fran De Aquino**
Maranhao State University, Physics Department, S.Luis/MA, Brazil.

Recently it was experimentally shown that the DNA genetic information can be transmitted into water when the DNA and the water are subjected jointly to an electromagnetic field with 7Hz frequency. As announced, the reported phenomenon could allow developing highly sensitive detection systems for chronic bacterial and viral infections. Here, it is shown a possible explanation for the phenomenon based on the recent framework of Quantum Gravity. It is shown that, if volume of water with a DNA molecule is placed *near* another volume of pure water, and the *gravitational masses* of the two water volumes are simultaneously reduced to values in the range + $0.159m_{i0}$ to – $0.159m_{i0}$, by means of electromagnetic fields of extremely-low frequency (ELF), then the DNA genetic information are transmitted to pure water, imprinting onto it the structure of the DNA molecule. After several hours, as final result, a replication of the DNA can arise in the pure water.

**Key words:** DNA, Modified theories of gravity, Experimental studies of gravity, Memory of Water.
PACS: 87.14.gk , 04.50.Kd , 04.80.-y, 41.20.Jb.

## 1. Introduction

A recent experiment showed that the DNA genetic information can be transmitted into *water* when the DNA and the water are subjected jointly to an electromagnetic field with 7Hz frequency. The main researcher behind the new DNA experiment is a recent Nobel prizewinner, Luc Montagnier. He and his research partners have made a summary of his findings [1]. Montagnier's experiment basically consists in two test tubes, one of which contained a tiny piece of bacterial DNA, the other pure water. The tubes were then placed close to one another inside a horizontally oriented solenoid. Both tubes were jointly subjected to a weak electromagnetic field with 7Hz frequency. Eighteen hours later, after DNA amplification using a polymerase chain reaction, as if by magic, *the DNA was detectable in the test tube containing pure water*, showing that, under certain conditions, DNA can project copies of itself in another place.

As mentioned in a recently published article in the *New Scientist* [2], 'physicists in Montagnier's team suggest that DNA emits low-frequency electromagnetic waves which imprint the structure of the molecule onto the water. This structure, they claim, is *preserved* and amplified through quantum coherence effects, and because it mimics the shape of the original DNA, the enzymes in the PCR process mistake it for DNA itself, and somehow use it as a template to make DNA match that which "sent" the signal'.

Here, based on the framework of a recently proposed theory of Quantum Gravity [3], is presented a consistent explanation showing how an exact copy of the structure of the DNA molecule is imprinted onto the *pure water*.

## 2. Theory

The quantization of gravity showed that the *gravitational mass* $m_g$ and *inertial mass* $m_i$ are correlated by means of the following factor [3]:

$$\frac{m_g}{m_{i0}} = \left\{1 - 2\left[\sqrt{1+\left(\frac{\Delta p}{m_{i0}c}\right)^2} - 1\right]\right\} \qquad (1)$$

where $m_{i0}$ is the *rest* inertial mass of the particle and $\Delta p$ is the variation in the particle's *kinetic momentum*; $c$ is the speed of light.

When $\Delta p$ is produced by the absorption of a photon with wavelength $\lambda$, it is expressed by $\Delta p = h/\lambda$. In this case, Eq. (1) becomes

$$\frac{m_g}{m_{i0}} = \left\{1 - 2\left[\sqrt{1+\left(\frac{h/m_{i0}c}{\lambda}\right)^2} - 1\right]\right\}$$

$$= \left\{1 - 2\left[\sqrt{1+\left(\frac{\lambda_0}{\lambda}\right)^2} - 1\right]\right\} \qquad (2)$$

where $\lambda_0 = h/m_{i0}c$ is the *De Broglie wavelength* for the particle with *rest* inertial mass $m_{i0}$.

It is easily seen that $m_g$ cannot be strongly reduced simply by using electromagnetic waves with wavelength $\lambda$ because $\lambda_0$ is much smaller than $10^{-10}m$. However, it is known that the wavelength of a radiation can be strongly reduced simply by strongly reducing its velocity.

From Electrodynamics we know that when an electromagnetic wave with frequency $f$ and velocity $c$ incides on a material with relative permittivity $\varepsilon_r$, relative magnetic permeability $\mu_r$ and electrical conductivity $\sigma$, its *velocity is reduced* to $v = c/n_r$ where $n_r$ is the index of refraction of the material, given by [3]

$$n_r = \frac{c}{v} = \sqrt{\frac{\varepsilon_r \mu_r}{2}\left(\sqrt{1+(\sigma/\omega\varepsilon)^2}+1\right)} \qquad (3)$$

If $\sigma >> \omega\varepsilon$, $\omega = 2\pi f$, the Eq. (3) reduces to

$$n_r = \sqrt{\frac{\mu_r \sigma}{4\pi\varepsilon_0 f}} \qquad (4)$$

Thus, the wavelength of the incident radiation becomes

$$\lambda_{\text{mod}} = \frac{v}{f} = \frac{c/f}{n_r} = \frac{\lambda}{n_r} = \sqrt{\frac{4\pi}{\mu f \sigma}} \qquad (5)$$

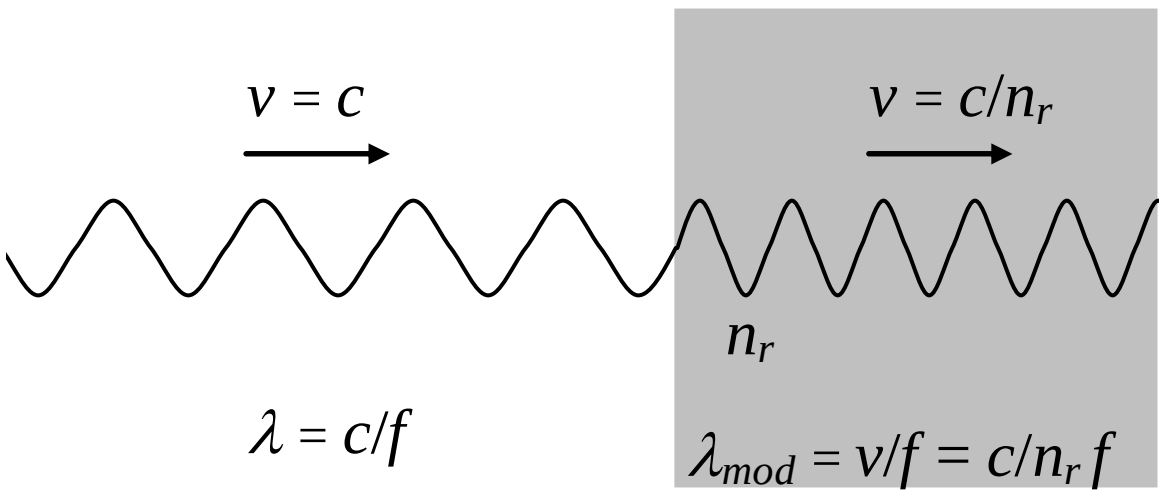

Fig. 1 – *Modified Electromagnetic Wave*. The wavelength of the electromagnetic wave can be strongly reduced, but its frequency remains the same.

Now consider a $7Hz$ ($\lambda \cong 4.3\times10^7 m$) radiation incident on pure water ($\sigma = 2\times10^{-4} S/m$). According to Eq. (5), the *modified wavelength* is

$$\lambda_{\text{mod}} = \sqrt{\frac{4\pi}{\mu f \sigma}} = 8.4\times10^4 m \qquad (6)$$

Consequently, the wavelength of the $7Hz$ radiation *inside the water* will be $\lambda_{\text{mod}} = 8.4\times10^4 m$ and not $\lambda \cong 4.3\times10^7 m$.

If a water lamina with thickness equal to $\xi$ contains $n$ molecules/m$^3$, then the number of molecules per unit area is $n\xi$. Thus, if the electromagnetic radiation with frequency $f$ incides on an area $S$ of the lamina it reaches $nS\xi$ molecules. If it incides on the total area of the lamina, $S_f$, then the total number of molecules reached by the radiation is $N = nS_f\xi$. The number of molecules per unit volume, $n$, is given by

$$n = \frac{N_0 \rho}{A} \qquad (7)$$

where $N_0 = 6.02\times10^{26}\, molecules/kmole$ is the Avogadro's number; $\rho$ is the matter density of the lamina($kg/m^3$)and $A$ is the Molar Mass. In the case of pure $Water\left(\rho = 10^3 kg/m^3, A = 18.01 kg.kmole^{-1}\right)$ the result is

$$n_{water} = 3.34\times10^{28}\, molecules/m^3 \qquad (8)$$

The *total number of photons* inciding on the water is $n_{total\ photons} = P/hf^2$, where $P$ is the power of the radiation flux incident on the water.

When an electromagnetic wave incides on the water, it strikes on $N_f$ front molecules, where $N_f \cong (nS_f)\phi_m$. Thus, the wave incides effectively on an area $S = N_f S_m$, where $S_m = \frac{1}{4}\pi\phi_m^2 \cong 7\times10^{-21} m^2$ is the cross section area of one molecule of the water molecule. After these collisions, it carries out $n_{collisions}$ with the other atoms of the foil (See Fig.2).

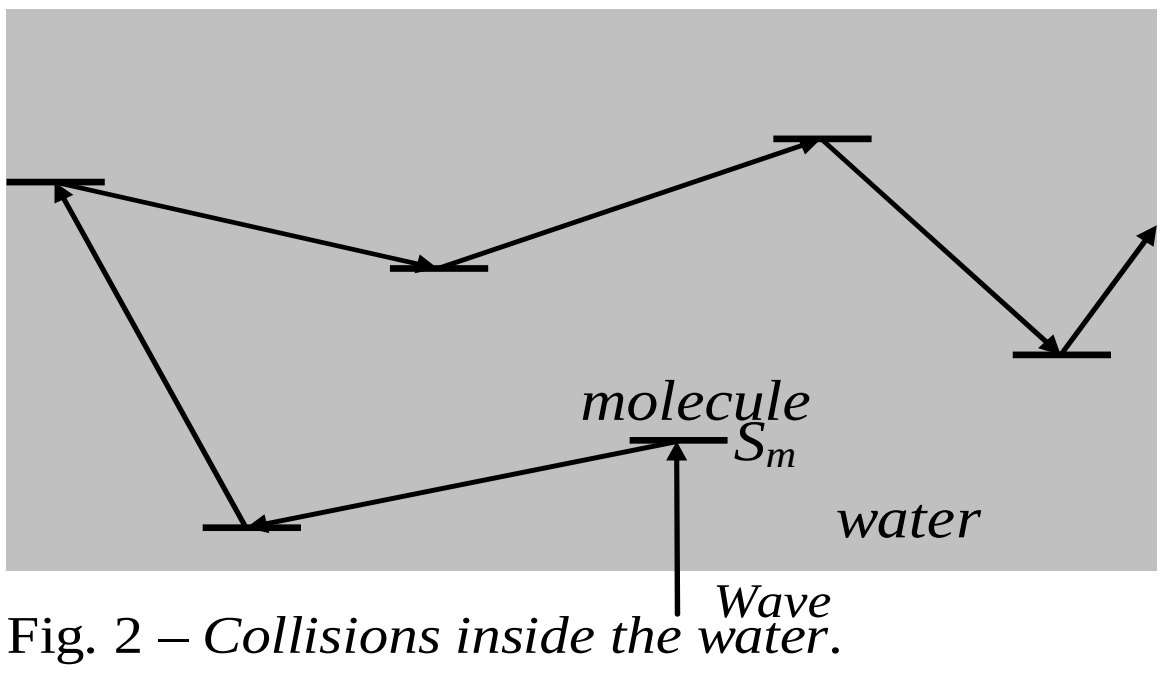

Fig. 2 – *Collisions inside the water.*

Thus, the total number of collisions in the volume $S\xi$ is

$$N_{collisions} = N_f + n_{collisions} = nS_f\delta + \left(nS\xi - nS_f\delta\right) = \\ = nS\xi \qquad (9)$$

The power density, $D$, of the radiation on the water can be expressed by

$$D = \frac{P}{S} = \frac{P}{N_f S_m} \qquad (10)$$

We can express the *total mean number of collisions in each molecule*, $n_1$, by means of the following equation

$$n_1 = \frac{n_{total\ \ photons} N_{collisions}}{N} \qquad (11)$$

Since in each collision a *momentum* $h/\lambda$ is transferred to the molecule, then the *total momentum* transferred to the water will be $\Delta p = \left(n_1 N\right)h/\lambda$. Therefore, in accordance with Eq. (1), we can write that

$$\frac{m_g}{m_{i0}} = \left\{1 - 2\left[\sqrt{1 + \left[\left(n_1 N\right)\frac{\lambda_0}{\lambda}\right]^2} - 1\right]\right\} = \\ = \left\{1 - 2\left[\sqrt{1 + \left[n_{total\ photons} N_{collisions}\frac{\lambda_0}{\lambda}\right]^2} - 1\right]\right\} \qquad (12)$$

Since Eq. (9) gives $N_{collisions} = nS\xi$, we get

$$n_{total\ \ photons} N_{collisions} = \left(\frac{P}{hf^2}\right)\left(nS\xi\right) \qquad (13)$$

Substitution of Eq. (13) into Eq. (12) yields

$$\frac{m_g}{m_{i0}} = \left\{1 - 2\left[\sqrt{1 + \left[\left(\frac{P}{hf^2}\right)\left(nS\xi\right)\frac{\lambda_0}{\lambda}\right]^2} - 1\right]\right\} \qquad (14)$$

Substitution of $P$ given by Eq. (10) into Eq. (14) gives

$$\frac{m_g}{m_{i0}} = \left\{1 - 2\left[\sqrt{1 + \left[\left(\frac{N_f S_m D}{f^2}\right)\left(\frac{nS\xi}{m_{i0}c}\right)\frac{1}{\lambda}\right]^2} - 1\right]\right\} \qquad (15)$$

Substitution of $N_f \cong \left(nS_f\right)\phi_m$ and $S = N_f S_m$ into Eq. (15) the result is

$$\frac{m_g}{m_{i0}} = \left\{1 - 2\left[\sqrt{1 + \left[\left(\frac{n^3 S_f^2 S_m^2 \phi_m^2 D\xi}{m_{i0} c f^2}\right)\frac{1}{\lambda}\right]^2} - 1\right]\right\} \\ = \left\{1 - 2\left[\sqrt{1 + \left[\left(\frac{n^3 S_m^2 \phi_m^2 S_f D}{\rho c f^2}\right)\frac{1}{\lambda}\right]^2} - 1\right]\right\} \qquad (16)$$

In the case of the *water*, we can take the following values: $n = 3.34\times10^{28}\, molecules/m^3$ ; $S_f \cong 1.9\times10^{-5} m^2$ ($S_f$ is the area of the horizontal cross-section of the test tube); $S_m \cong 7\times10^{-21} m^2$; $\phi_m \cong 1\times10^{-10} m$; $\xi$ (height of water inside the test tube). Substitution of these values into Eq. (16), gives

$$\frac{m_{g(water)}}{m_{i0(water)}} = \left\{1 - 2\left[\sqrt{1 + \left[\left(1.1\times10^9 \frac{D}{f^2}\right)\frac{1}{\lambda}\right]^2} - 1\right]\right\} \qquad (17)$$

In the case of a $7Hz$ radiation, Eq. (6) shows that $\lambda_{mod} = 8.4\times10^4 m$. Thus, by substitution of $\lambda$ by $\lambda_{mod}$ into Eqs. (17) , we get the following expression

$$\frac{m_{g(water)}}{m_{i0(water)}} \cong \left\{1 - 2\left[\sqrt{1 + 7.1\times10^4 D^2} - 1\right]\right\} \qquad (18)$$

Now, considering that the water is inside a solenoid, which produces a weak ELF electromagnetic field with $E_m$ and $B_m$, then we can write that [4]

$$D = \frac{E_m^2}{2\mu_0 v_{water}} = \frac{v_{water}^2 B_m^2}{2\mu_0 v_{water}} = \frac{cB_m^2}{2\mu_0 n_{(water)}} \qquad (19)$$

Equation (4) shows that for $f = 7Hz$, $n_{r(water)} = 506.7$ .Substitution of this value into Eq.(19) gives

$$D = 2.3\times10^{11} B_m^2$$

Substitution of this value into Eq. (19) gives

$$\frac{m_{g(water)}}{m_{i0(water)}} \cong \left\{1 - 2\left[\sqrt{1 + 3.7\times10^{27} B_m^4} - 1\right]\right\} \qquad (20)$$

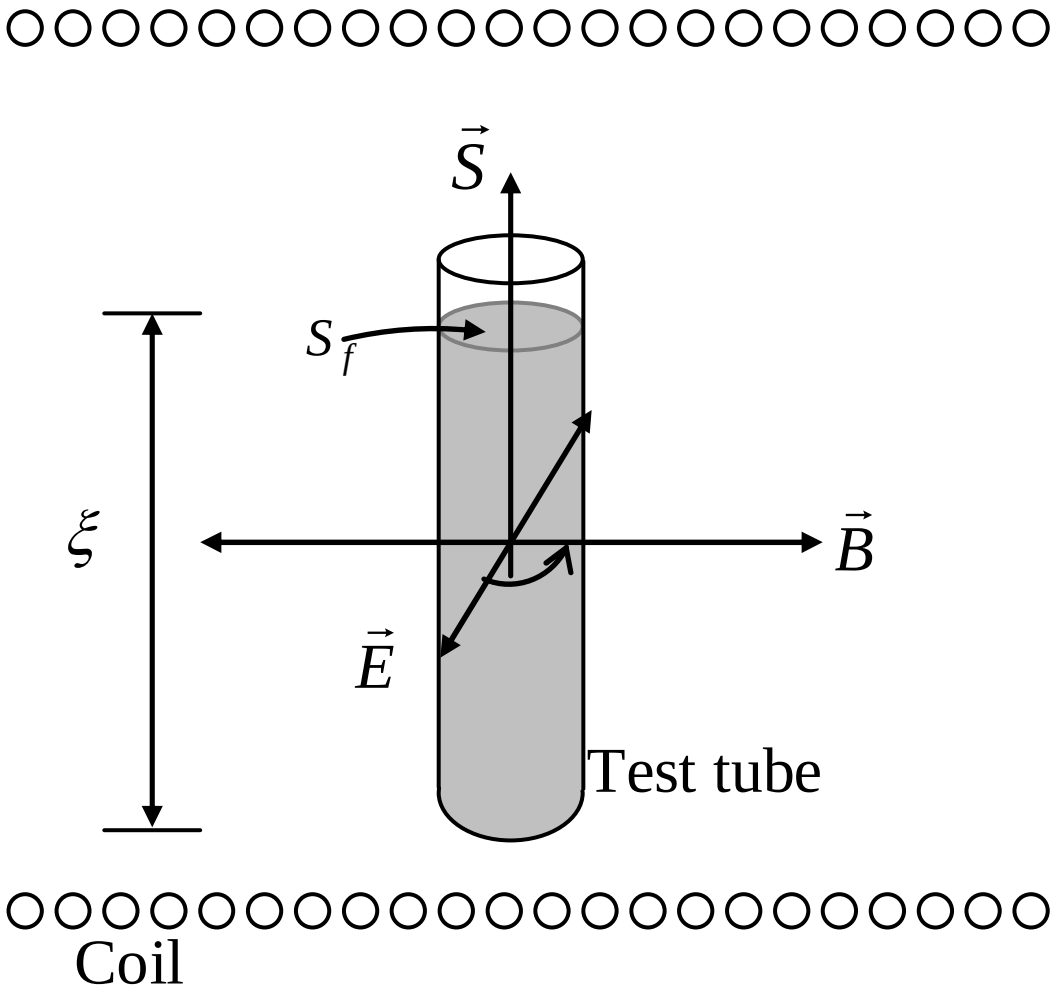

Fig. 3 – The vector of *Pointing* $\vec{S} = \vec{E} \times \vec{B}$ at the test tube. The electromagnetic radiation propagates in the direction of the vector of *Pointing*

In Montagnier's experiment, the set-up was placed in a container shielded by 1 mm thick layer of *mumetal* in order to avoid interference from the earth's natural magnetic field, whose intensity is $B_{\oplus} \cong 6 \times 10^{-5} T$. This is because the intensity of magnetic field in Montagnier's experiment was much smaller than $B_{\oplus}$. Note that, if the intensity of the magnetic field is in the range $1.2 \times 10^{-7} T < B_m < 1.4 \times 10^{-7} T$, then, according to Eq. (20), the gravitational masses of the *water with DNA* and the *water inside the other test tube* are reduced to values in the range $+0.159 m_{i0}$ to $-0.159 m_{i0}$. It was shown in a previous paper [3] that, when this occurs the gravitational masses becomes *imaginaries* and the bodies leave our *Real* Universe, i.e., they perform transitions to the *Imaginary* Universe, which contains our *Real* Universe. The terms real and imaginary are borrowed from mathematics (real and imaginary numbers). It was also shown that in the Imaginary Universe the imaginary bodies are subjected to the *Imaginary Interaction* that is similar to the Gravitational Interaction. If the masses of the bodies have the same sign, then the interaction among them will be attractive.

The masses of the water with DNA and the pure water are decreased at the same ratio,

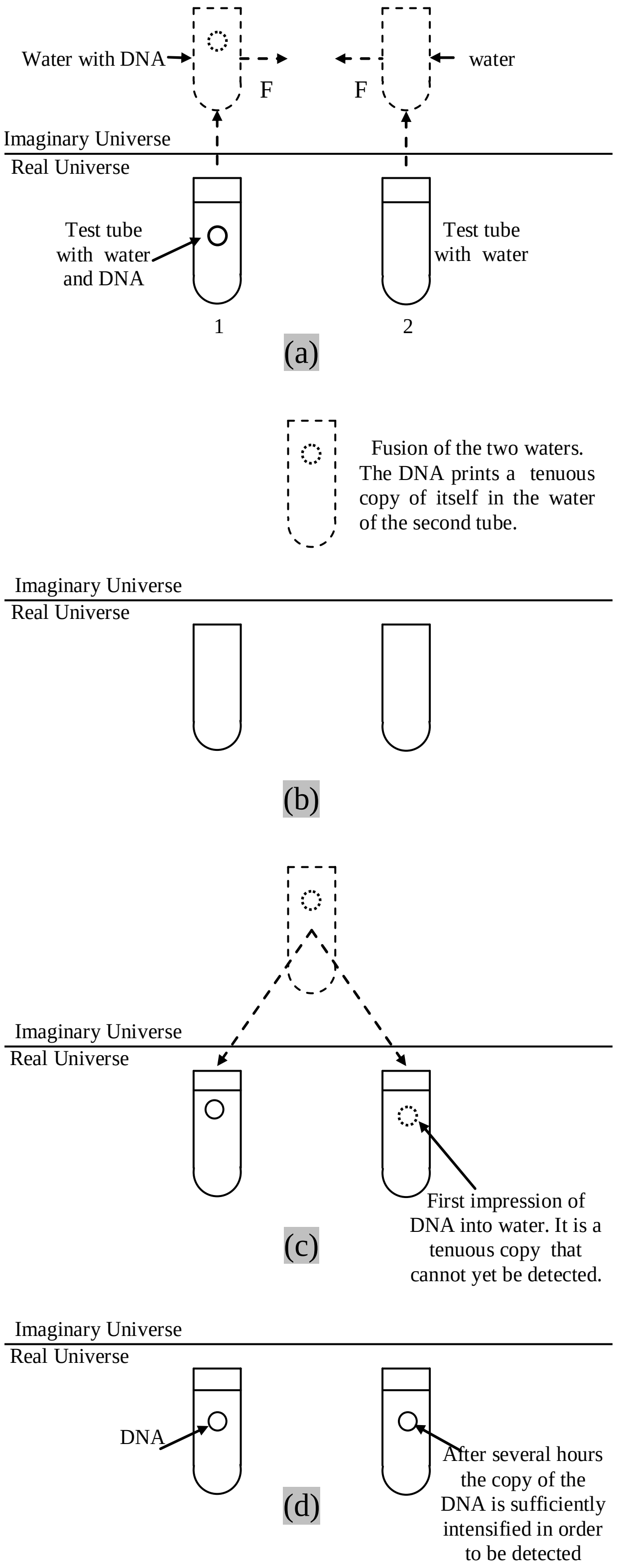

Fig. 4 – (a) Transition to the Imaginary Universe and attraction. (b) Fusion of the two waters. (c) Return to Real Universe. (d) After several comings and goings to the Imaginary Universe a real copy of the DNA can be detected in the tube with pure water.

in such way that they remain with the same sign. Thus, when they arrive the Imaginary Universe the *attractive* imaginary interaction approaches each other. *Due to the small distance between them**, they are subjected to a significative attraction. Consequently, they entered one another (fusion). This imprints in the *pure water* an exactly copy of the DNA molecule. However, the water with DNA and the pure water return immediately to the real universe because the ELF electromagnetic field does not accompany them during the transition. When they get back to the real universe, the effect previously produced by the ELF electromagnetic field sends again the water with DNA and pure water to Imaginary Universe, and again a new imprint of the DNA is produced at the same place of the first one, strengthening the copy of DNA onto the water.

Thus, during the time interval in what the ELF electromagnetic field remains on, the process continue. After some hours (16 to 18 hours in the case of Montagnier's experiment) the copy of the DNA can become sufficiently strong to be detected. Thus, when the ELF electromagnetic field is turned off, the water can contain a real DNA molecule, which is an exactly equal to that one that exists in the other tube.

The physicists in Montagnier's team suggest that the imprints of the DNA are preserved through *quantum coherence effects* [1]. This conclusion is based on the framework of a recently proposed theory of liquid water based on Quantum Field Theory (QFT) [5-10]. Jacques Benveniste [11] has been the first to propose (1988) that water has memory. The fact that the water contains electric dipoles, which can give to it a significant memory capacity, has been also considered by Brian Josephson [12] and, more recently by J. Dunning-Davies [13].

---

* Due to the small distance between the two test tubes. The tubes were then placed near to one another inside a horizontally oriented solenoid.

# A Possible Explanation for Anomalous Heat Production in Ni-H Systems

**Fran De Aquino**
Maranhao State University, Physics Department, S.Luis/MA, Brazil.

Anomalous heat production has been detected in Ni-H Systems. Several evidences point to the occurrence of nuclear fusion reactions. A possible explanation for this phenomenon is shown here based on the recent discovery that electromagnetic fields of extremely-low frequencies (ELF) can increase the intensities of gravitational forces. Under certain circumstances, the intensities of gravitational forces can even overcome the intensity of the electrostatic repulsion forces, and, in this way, produce nuclear fusion reactions, without need high temperatures for these reactions occur.

**Key words:** Modified theories of gravity, Nuclear Fusion**,** Fusion Reactors.
PACS: 04.50.Kd , 89.30.Jj , 28.52.-s.

## 1. Introduction

Since the experiment of Fleischmann, Hawkins and Pons [1] the anomalous production of heat has been searched for in various systems. Recently, a large anomalous production of heat has been reported by Focardi et al., [2] in a nickel rod filled with hydrogen. This phenomenon was posteriorly confirmed by Cerron-Zeballos et al., [3].

The called "cold fusion" was a process of nuclear fusion that was first conceived of by Fleischmann, Hawkins and Pons during their experiment that involved heavy water electrolysis through hydrogen on a palladium electrode surface [1]. They made claims originally that there was heat and energy being created from the reaction taking place at room temperature. This is why it is referred to as cold fusion, because it occurred in an environment that was previously considered too cool for *nuclear fusion to occur*.

Here it is shown that nuclear fusion can be produced at room temperature by *increasing the gravitational forces in order to overcome the electrostatic repulsion forces between the nucle*i. This process became feasible after the Quantization of Gravity [4], with the discovery that the *gravitational mass* $m_g$ can be made negative and strongly intensified by means of electromagnetic fields of extremely-low frequencies.

This effect can provide a consistent and coherent explanation for anomalous heat production detected in Ni-H Systems.

## 2. Theory

The quantization of gravity shown that the *gravitational mass* $m_g$ and *inertial mass* $m_i$ are correlated by means of the following factor [4]:

$$\frac{m_g}{m_{i0}} = \left\{1 - 2\left[\sqrt{1+\left(\frac{\Delta p}{m_{i0}c}\right)^2} - 1\right]\right\} \quad (1)$$

where $m_{i0}$ is the *rest* inertial mass of the particle and $\Delta p$ is the variation in the particle's *kinetic momentum*; $c$ is the speed of light.

When $\Delta p$ is produced by the absorption of a photon with wavelength $\lambda$, it is expressed by $\Delta p = h/\lambda$. In this case, Eq. (1) becomes

$$\frac{m_g}{m_{i0}} = \left\{1 - 2\left[\sqrt{1+\left(\frac{h/m_{i0}c}{\lambda}\right)^2} - 1\right]\right\}$$
$$= \left\{1 - 2\left[\sqrt{1+\left(\frac{\lambda_0}{\lambda}\right)^2} - 1\right]\right\} \quad (2)$$

where $\lambda_0 = h/m_{i0}c$ is the *De Broglie wavelength* for the particle with *rest* inertial mass $m_{i0}$.

From Electrodynamics we know that when an electromagnetic wave with frequency $f$ and velocity $c$ incides on a material with relative permittivity $\varepsilon_r$, relative magnetic permeability $\mu_r$ and electrical conductivity $\sigma$, its *velocity is* *reduced* to $v = c/n_r$ where $n_r$ is the index of refraction of the material, given by [5]

$$n_r = \frac{c}{v} = \sqrt{\frac{\varepsilon_r \mu_r}{2}\left(\sqrt{1+(\sigma/\omega\varepsilon)^2}+1\right)} \qquad (3)$$

If $\sigma >> \omega\varepsilon$, $\omega = 2\pi f$, Eq. (3) reduces to

$$n_r = \sqrt{\frac{\mu_r \sigma}{4\pi\varepsilon_0 f}} \qquad (4)$$

Thus, the wavelength of the incident radiation (See Fig . 1) becomes

$$\lambda_{\text{mod}} = \frac{v}{f} = \frac{c/f}{n_r} = \frac{\lambda}{n_r} = \sqrt{\frac{4\pi}{\mu f \sigma}} \qquad (5)$$

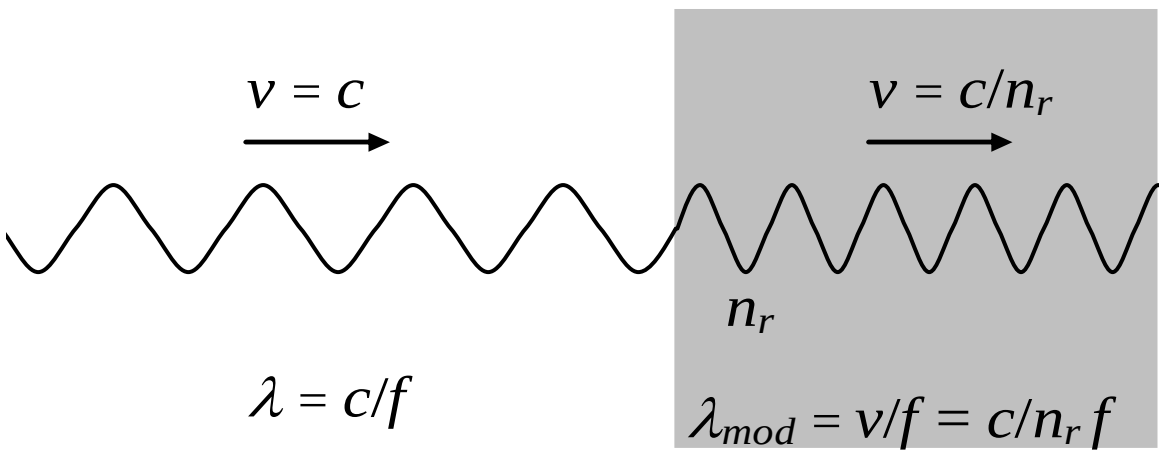

Fig. 1 – *Modified Electromagnetic Wave*. The wavelength of the electromagnetic wave can be strongly reduced, but its frequency remains the same.

It is known that the *Schumann resonances* [6] are global electromagnetic resonances (a set of spectrum peaks in the extremely low frequency ELF), excited by lightning discharges in the *spherical resonant cavity* formed by the Earth's surface and the inner edge of the ionosphere (60km from the Earth's surface). The Earth–ionosphere waveguide behaves like a resonator at ELF frequencies and amplifies the spectral signals from lightning at the resonance frequencies. In the normal mode descriptions of Schumann resonances, the fundamental mode $(n=1)$ is a standing wave in the Earth–ionosphere cavity with a wavelength equal to the circumference of the Earth. This lowest-frequency (and highest-intensity) mode of the Schumann resonance occurs at a frequency $f_1 = 7.83Hz$ [7].

Now consider a $7.83Hz$ $(\lambda \cong 3.8\times10^7 m)$ radiation passing through a *Nickel powder* cylinder $(\sigma = 1.6\times10^7 S/m;$ $\mu_r = 2.17$ [8,9]) as shown in Fig.2. According to Eq. (5), the *modified wavelength* is

$$\lambda_{\text{mod}} = \sqrt{\frac{4\pi}{\mu f \sigma}} = 0.19m \qquad (6)$$

Consequently, the wavelength of the $7.83Hz$ radiation *inside the Nickel powder* will be $\lambda_{\text{mod}} = 0.19m$ and not $\lambda \cong 3.8\times10^7 m$.

If a *Nickel powder*[*] lamina with thickness equal to $\xi$ contains $n$ molecules/m$^3$, then the number of molecules per area unit is $n\xi$. Thus, if the electromagnetic radiation with frequency $f$ incides on an area $S$ of the lamina it reaches $nS\xi$ molecules. If it incides on the total area of the lamina, $S_f$, then the total number of molecules reached by the radiation is $N = nS_f\xi$. The number of molecules per unit of volume, $n$, is given by

$$n = \frac{N_0 \rho}{A} \qquad (7)$$

where $N_0 = 6.02\times10^{26} molecules/kmole$ is the Avogadro's number; $\rho$ is the matter density of the lamina (in $kg/m^3$) and $A$ is the molar mass. In the case of *Nickel powder* $\left(\rho = 8800 kg/m^3, A = 58.71 kg.kmole^{-1}\right)$ the result is

$$n_{(Ni)} = 9.02\times10^{28} molecules/m^3 \qquad (8)$$

The *total number of photons* inciding on the Nickel powder is $n_{total\ photons} = P/hf^2$, where $P$ is the power of the radiation flux incident on the Nickel powder.

[*] Ultra fine nickel powder (e.g. Inco type 210) with particle size of 0.5-1.0μm.

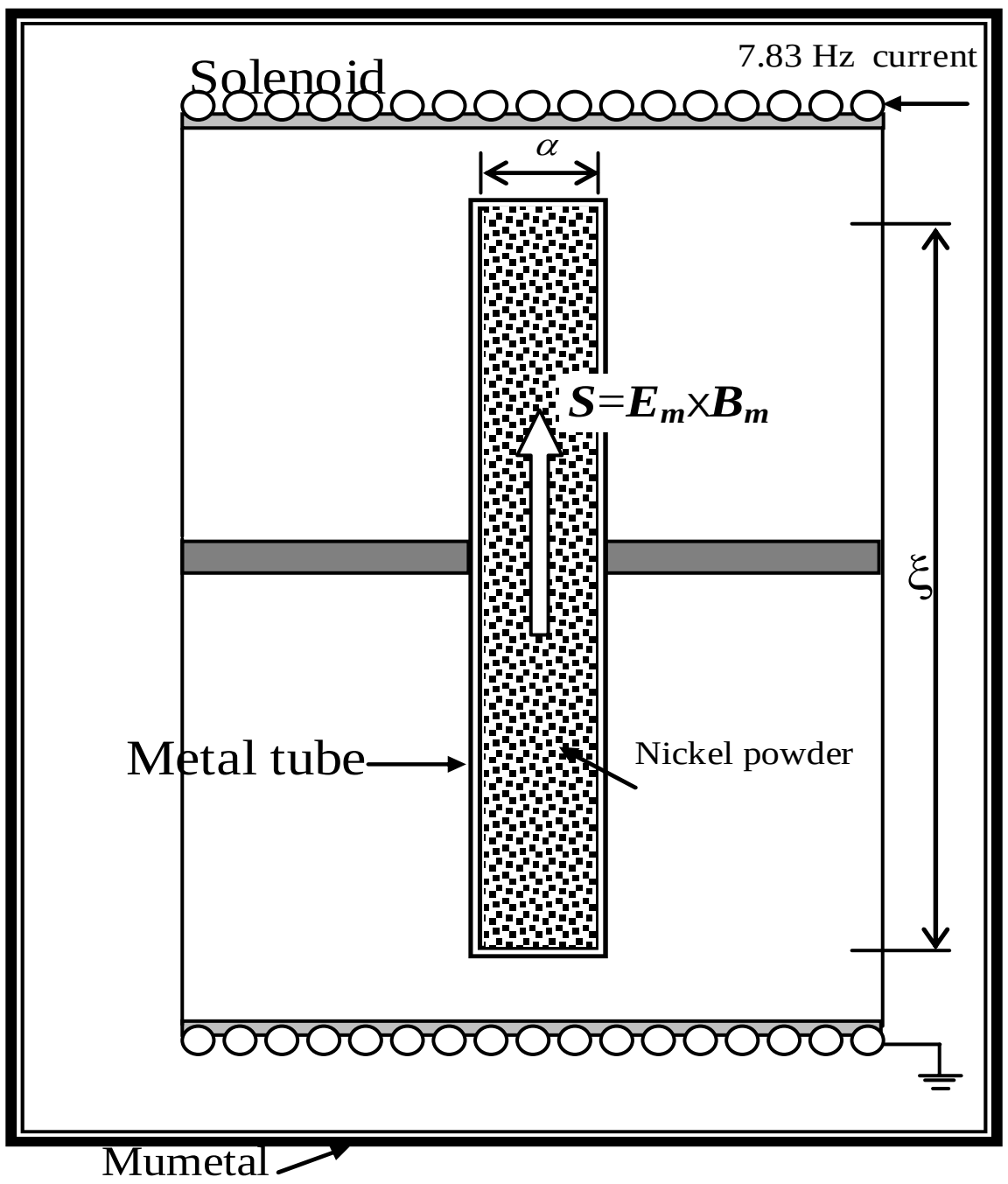

Fig.2 – *Experimental set-up*. The electromagnetic radiation propagates in the direction of the vector of *Pointing* $\vec{S} = \vec{E} \times \vec{B}$. The set-up is placed in a container shielded by 1 mm thick layer of *mumetal* in order to avoid interference from external electromagnetic fields. In practice, the solenoid is not necessary, since the 7.83 Hz electromagnetic field naturally exists inside the *spherical resonant cavity* formed by the Earth's surface and the inner edge of the ionosphere. (Schumann resonance).

When an electromagnetic wave incides on a *solid* lamina of Nickel, it strikes on $N_f$ front molecules, where $N_f \cong \left(n_{(Ni)}S_f\right)\phi_{(Ni)}$. Thus, the electromagnetic wave incides effectively on an area $S = N_f S_{(Ni)}$, where $S_{(Ni)} = \frac{1}{4}\pi\phi_{(Ni)}^2 \cong 1.2 \times 10^{-20} m^2$ is the cross section area of one *Ni* atom. After these collisions, it carries out $n_{collisions}$ with the other atoms of the Nickel powder (See Fig.3).

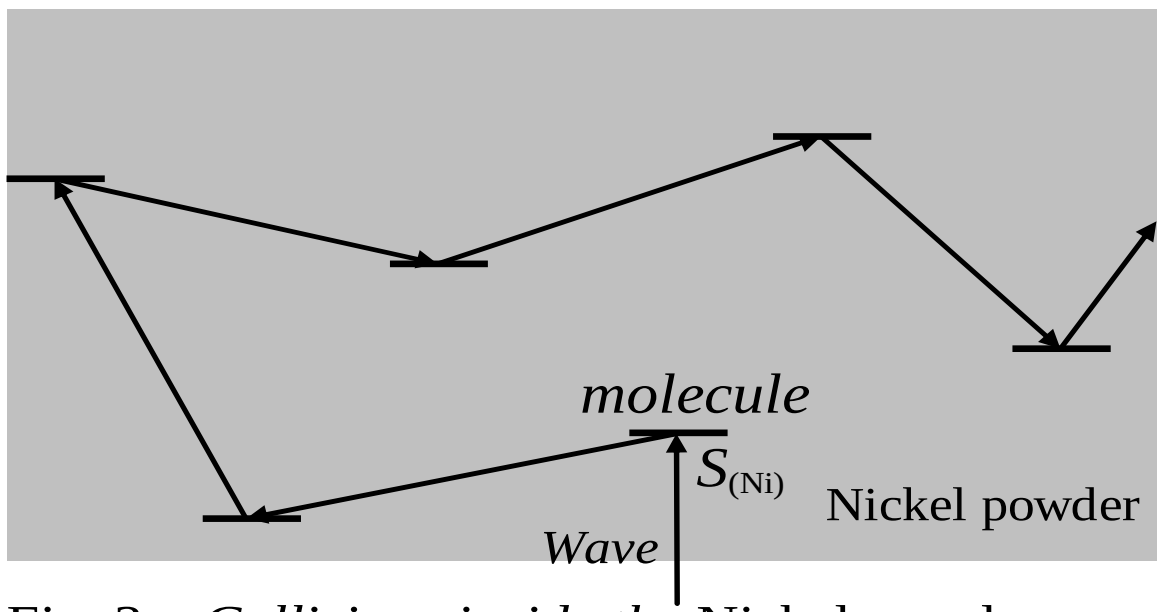

Fig. 3 – *Collisions inside the* Nickel powder.

Thus, the total number of collisions in the volume $S\xi$ is

$$N_{collisions} = N_f + n_{collisions} = n_{(Ni)}S\phi_{(Ni)} + \left(n_{(Ni)}S\xi - n_{(Ni)}S\phi_{(Ni)}\right) = \\ = n_{(Ni)}S\xi \qquad (9)$$

The power density, $D$, of the radiation on the Nickel powder can be expressed by

$$D = \frac{P}{S} = \frac{P}{N_f S_{(Ni)}} \qquad (10)$$

We can express the *total mean number of collisions in each Ni molecule,* $n_1$, by means of the following equation

$$n_1 = \frac{n_{total\ \ photons} N_{collisions}}{N} \qquad (11)$$

Since in each collision a *momentum* $h/\lambda$ is transferred to the molecule, then the *total momentum* transferred to the Nickel will be $\Delta p = (n_1 N)h/\lambda$. Therefore, in accordance with Eq. (1), we can write that

$$\frac{m_{g(w)}}{m_{i0(w)}} = \left\{1 - 2\left[\sqrt{1 + \left[(n_1 N)\frac{\lambda_0}{\lambda}\right]^2} - 1\right]\right\} = \\ = \left\{1 - 2\left[\sqrt{1 + \left[n_{total\ photons} N_{collisions}\frac{\lambda_0}{\lambda}\right]^2} - 1\right]\right\} \qquad (12)$$

Since Eq. (9) gives $N_{collisions} = n_{(Ni)}S\xi$, we get

$$n_{total\ \ photons} N_{collisions} = \left(\frac{P}{hf^2}\right)\left(n_{(Ni)}S\xi\right) \qquad (13)$$

Substitution of Eq. (13) into Eq. (12) yields

$$\frac{m_{g(Ni)}}{m_{i0(Ni)}} = \left\{1 - 2\left[\sqrt{1 + \left[\left(\frac{P}{hf^2}\right)\left(n_{(Ni)}S\xi\right)\frac{\lambda_0}{\lambda}\right]^2} - 1\right]\right\} \qquad (14)$$

Substitution of $P$ given by Eq. (10) into Eq. (14) gives

$$\frac{m_{g(Ni)}}{m_{i0(Ni)}} = \left\{1 - 2\left[\sqrt{1 + \left[\left(\frac{N_f S_{(Ni)} D}{f^2}\right)\left(\frac{n_{(Ni)}S\xi}{m_{i0(Ni)}c}\right)\frac{1}{\lambda}\right]^2} - 1\right]\right\} \qquad (15)$$

Substitution of $N_f \cong \left(n_{(Ni)}S_f\right)\phi_{(Ni)}$ and $S = N_f S_{(Ni)}$ into Eq. (15) results

$$\frac{m_{g(Ni)}}{m_{i0(Ni)}}=\left\{1-2\left[\sqrt{1+\left[\left(\frac{n^3_{(Ni)}S_f^2S^2_{(Ni)}\phi^2_{(Ni)}\xi D}{m_{i0(Ni)}cf^2}\right)\frac{1}{\lambda}\right]^2}-1\right]\right\} \quad (16)$$

where $m_{i0(Ni)} = \rho_{(Ni)}V_{cyl} = \rho_{(Ni)}\left(\pi\alpha^2/4\right)\xi$. Thus, Eq. (16) reduces to

$$\frac{m_{g(Ni)}}{m_{i0(Ni)}}=\left\{1-2\left[\sqrt{1+\left[\left(\frac{n^3_{(Ni)}S_f^2S^2_{(Ni)}\phi^2_{(Ni)}D}{\rho_{(Ni)}\left(\pi\alpha^2/4\right)cf^2}\right)\frac{1}{\lambda}\right]^2}-1\right]\right\} \quad (17)$$

For $\phi = 5cm$ we get $S_\alpha = \pi\alpha^2/4 = 1.9\times10^{-3}m^2$. Note that $S_f$ is not equal to $S_\alpha$ because the area is not continuous, but expressed by $S_f = nS_p$, where $S_p$ is the area of the cross-section of one Ni particle, and $n$ is *the number of particles in the front area*, which is expressed by $n = x\left(n_p\phi_pS_\alpha\right)$, $x << 1$, where $n_p\phi_pS_\alpha$ is the number of particles inside area $S_\alpha$; $n_p$ is the number of Ni particles/$m^3$, given by $n_p = N_p/S_\alpha\xi$ where $N_p = S_\alpha\xi/V_p + V_v$; $V_p$ is the mean volume of one Ni particle and $V_V$ is the void volume, corresponding to that particle. This volume can be calculated considering one sphere with $\phi_p$ - diameter inside a cube whose edge is $\phi_p$. The result is $V_V \cong 0.48\phi_p^3$. The mean size of the particles is $\phi_p = 0.75\mu m$. Thus, $V_p \cong 2.2\times10^{-19}m^3$ and $S_p \cong 4.4\times10^{-13}m^2$. Consequently, $V_p + V_V \cong 4.2\times10^{-19}m^2$. Then, we get $n_p = 2.4\times10^{18}\,particles/m^3$. Now, we can calculate the value of $S_f$:

$$S_f = x\left(n_p\phi_pS_\alpha\right)S_p \cong x\left(1.5\times10^{-3}\right)m^2$$

Substitution of this value jointly with $n_{(Ni)} = 9.02\times10^{28}moles/m^3$; $\phi_{(Ni)} = 1.24\times10^{-10}m$; $S_\alpha = \pi\alpha^2/4 = 1.9\times10^{-3}m^2$; $S_{(Ni)} \cong 1.2\times10^{-20}m^2$; $\rho_{(Ni)} = 8800\ kg.m^{-3}$; $f = 7.83\ Hz$ (Note that, this is lowest-frequency mode of the Schumann resonance. *Therefore, in practice, is not necessary to provide the 7.83 Hz electromagnetic field*) and $\lambda = \lambda_{mod} = 0.19m$ into Eq.(17), gives

$$\frac{m_{g(Ni)}}{m_{i0(Ni)}}=\left\{1-2\left[\sqrt{1+3.9\times10^{21}x^4D^2}-1\right]\right\} \quad (18)$$

Now, considering that the Nickel powder is inside a solenoid, which produces a weak ELF electromagnetic field with $E_m$ and $B_m$, then we can write that [10]

$$D=\frac{E_m^2}{2\mu_0v_{Ni}}=\frac{v_{Ni}^2B_m^2}{2\mu_0v_{Ni}}=\frac{cB_m^2}{2\mu_0n_{r(Ni)}} \quad (19)$$

Equation (4) shows that, for $f = 7.83Hz$, $n_{r(Ni)} = 2\times10^8$. Substitution of this value into Eq.(19) gives

$$D = 5.9\times10^5B_m^2 \quad (21)$$

Substitution of this value into Eq. (18) gives

$$\chi=\frac{m_{g(Ni)}}{m_{i0(Ni)}}\cong\left\{1-2\left[\sqrt{1+1.3\times10^{33}x^4B_m^4}-1\right]\right\} \quad (22)$$

The value of $B_m$ is limited by the *ionization energy* of the atoms, which is, as we known, the energy required to remove electrons from atoms. Since the minimum energy required for the electron to leave the atom is: $U_{min} = -e^2/4\pi\varepsilon_0\phi_{max} = 7.7\times10^{-19}\,joules$ then, for the ionization does not occur, the energy of the wave $(hf)$ must be smaller than $U_{min}$. Thus, it follows that

$$hf^2/S_a < U_{min}f/S_a \Rightarrow D < U_{min}f/S_a \cong f$$

According to Eq. (19), $D_{max} = cB^2_{max}/2\mu_0$. Then, the result is

$$B_{max} < 9\times10^{-8}\sqrt{f}$$

In the case of $f = 7.83Hz$, we conclude that

$$B_{max} < 2\times10^{-7}T \quad (23)$$

Assuming that $B_{max} \cong 2\times10^{-7}T$ then Eq. (22) yields

$$\chi=\frac{m_{g(Ni)}}{m_{i0(Ni)}}\cong\left\{1-2\left[\sqrt{1+2.1\times10^6x^4}-1\right]\right\} \quad (24)$$

Since $x << 1$, we can conclude that *there is no significant variation in the gravitational mass of the Nickel powder.*

However, if the air inside the Nickel powder is evacuated by means of a vacuum pump, and after Hydrogen (or Deuterium,

Tritium, Helium, etc) is injected into the Nickel powder (See Fig.4) then, the area $S_f$ to be considered, in order to calculate the gravitational mass of the Hydrogen, is the *surface area* of the Nickel powder, which can be obtained by multiplying the *specific surface area of the Nickel powder*[†] ($\sim 4\times10^3 m^2/Kg$) *by the total mass of the Nickel powder* ($m_{i0(Ni)} = \rho_{(Ni)}(\pi\alpha^2/4)\xi$). Thus, we get $S_f \cong 4\times10^3 \rho_{(Ni)} S_\alpha \xi$.

The characteristics of the Nickel prevail on those of the Hydrogen, in the Ni-H systems, because the Nickel amount is much larger than the Hydrogen amount. Thus, we must take the values of $\rho$, $\mu_r$, and $\sigma$ equal to $\rho_{(Ni)}$, $\mu_{r(Ni)}$ and $\sigma_{(Ni)}$ respectively, in order to calculate $m_{g(H)}$, in Ni-H systems. In addition, since $n = N_0\rho/A$ and $\lambda_{mod} = \sqrt{4\pi/\mu f\sigma}$ we can conclude that also $n \equiv n_{(Ni)}$ and $\lambda_{mod} = \lambda_{mod(Ni)} = 0.19m$. Therefore, in order to obtain the expression $m_{g(H)}/m_{i0(H)}$ we can take Eq. (17) only substituting $S_f$ for the expression above obtained $(S_f \cong 4\times10^3 \rho_{(Ni)} S_\alpha \xi)$. Thus we get

$$\frac{m_{g(H)}}{m_{i0(H)}} = \left\{1-2\left[\sqrt{1+\left[\left(\frac{n^3_{(Ni)}\rho_{(Ni)}S_\alpha\xi^2S^2_{(Ni)}\phi^2_{(Ni)}D}{18.7f^2}\right)\frac{1}{\lambda}\right]^2}-1\right]\right\} \quad (25)$$

For $\xi = 0.1m$ (length of the Ni-H cylinder in Focardi experiment) Eq.(25) gives

$$\chi = \frac{m_{g(H)}}{m_{i0(H)}} = \left\{1-2\left[\sqrt{1+1.5\times10^{48}D^2}-1\right]\right\} \quad (26)$$

Based on Eq. (19), we can write that $D = cB_m^2/2\mu_0 n_{r(H)}$, where $n_{r(H)} \cong 1$. Thus, we get $D = 1.2\times10^{14} B_m^2$. Substitution of this expression into Eq. (26) yields

$$\chi = \frac{m_{g(H)}}{m_{i0(H)}} = \left\{1-2\left[\sqrt{1+2.1\times10^{76}B_m^4}-1\right]\right\} \quad (27)$$

It is known that, at any time in the *spherical resonant cavity* formed by the Earth's surface and the inner edge of the ionosphere (60km from the Earth's surface) there is a drop voltage of 200KV. This, produces an electric field with intensity $E_m \cong 3V/m$, which gives $B_m \cong 1\times10^{-8}T$. Substitution of this value into Eq. (27), yields

$$\chi \cong -2\times10^{22} \quad (28)$$

Thus, the gravitational forces between two protons (hydrogen nuclei) becomes

$$F = -Gm_{gp}^2/r^2 = -\chi^2 Gm_{ip}^2/r^2 \cong -7\times10^{-20}/r^2$$

Comparing with the electrostatic repulsion forces between the nuclei, which is given by

$$F_e = e^2/4\pi\varepsilon_0 r^2 = 2.3\times10^{-28}/r^2$$

We conclude that the intensities of the gravitational forces *overcome the intensities of the electrostatic repulsion forces between the nucle*i. This is sufficient to produce their fusion.

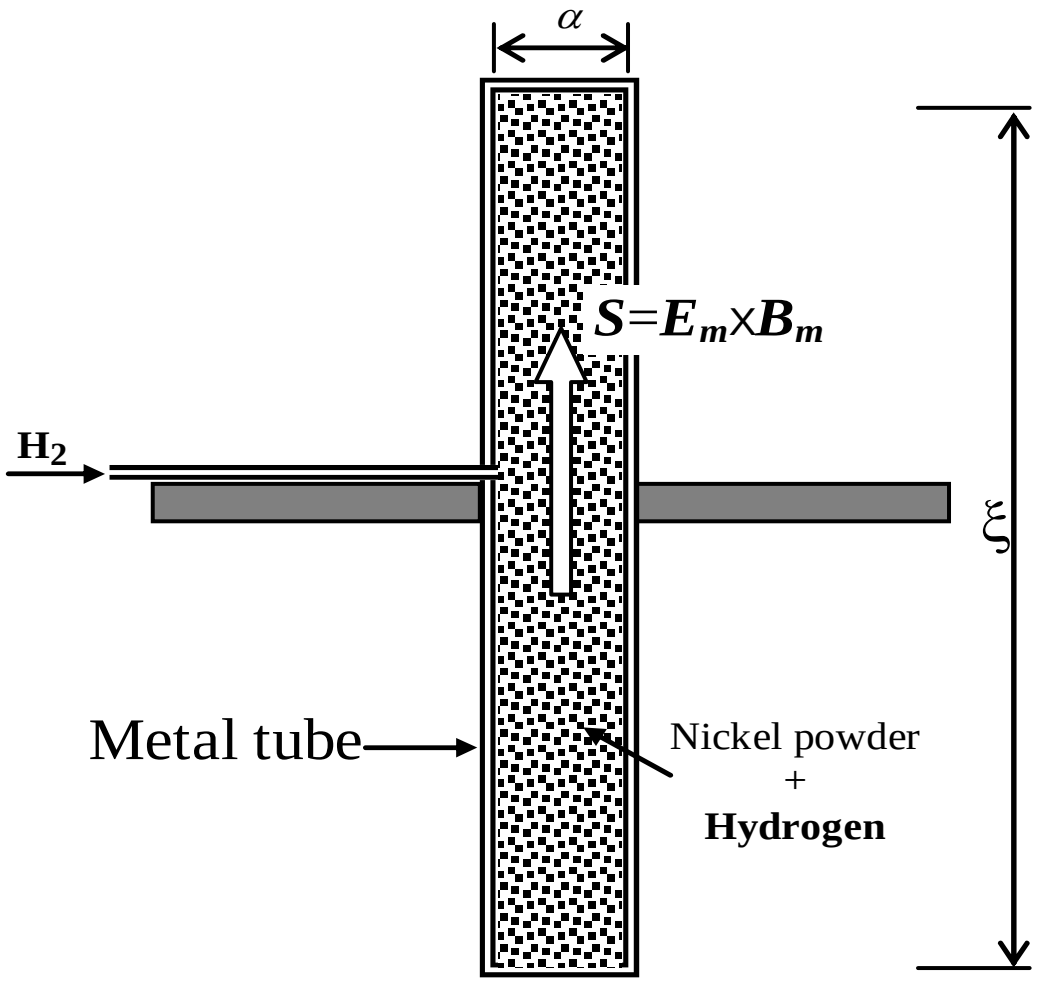

Fig.4 – *Cold Fusion Reactor on Earth*. Note that here the 7.83 Hz electromagnetic field is what naturally exists inside the *spherical resonant cavity* formed by the Earth's surface and the inner edge of the ionosphere. (Schumann resonance).

† Ultra fine nickel powder (e.g. Inco type 210) with particle size of 0.5-1.0μm has specific surface areas range from 1.5 to 6$m^2$/g [11]. Hydrogen production with *nickel powder cathode* points to a value of 4.31$m^2$ /g in the case of new cathodes, and 3.84 $m^2$ /g in the case of used cathodes [12].

The enormous value of $\chi$ (Eq. 28) strongly increases the gravitational masses of the Hydrogen nuclei ($m_{gp} = \chi m_{i0p}$) and their respective electrons ($m_{ge} = \chi m_{i0e}$). Thus, the gravitational force between a nucleus (proton) and the corresponding electron is given by $F_{pe} = -\chi^2 G m_{i0p} m_{i0e} / r^2$ and the gravitational force between two Hydrogen nuclei is $F_{pp} = -\chi^2 G m_{i0p} m_{i0p} / r^2$. Therefore, two well-known types of fusions can occur, i.e.,

$$p + e^- \rightarrow n + \nu_e \qquad (29)$$

$$p + p \rightarrow d + \nu_e + e^+ + 0.42MeV \qquad (30)$$

Due to the strong gravitational attraction, the following fusions occur instantaneously:

$$d + \nu_e \rightarrow n + p + \nu_e$$

and

$$n + e^+ \rightarrow p + \bar{\nu}_e$$

These reactions are widely known because they have been studied extensively due to their importance in astrophysics and neutrino physics [13–16]. Thus, the term $p + \nu_e + e^+$ in Eq. (30) reduces instantaneously to $p + p + \nu_e + \bar{\nu}_e$.

In these fusion reactions, *neutrons* (Eq. (29)), *neutrinos* and *antineutrinos*, and *energy* (0.42MeV at each fusion of two Hydrogen nuclei) are produced. Note that *there is no gamma ray emission* during the process. The evidence of neutron emission during energy production in Ni-H systems has been reported by Battaglia, A. et al., [17].

In order to calculate the number of Hydrogen atoms/m$^3$ inside the Nickel powder we will calculate the density of the Hydrogen. According to *Focardi's* experiments, the pressure of the Hydrogen is $P = 0.051 atm = 5.166\times10^3 N/m^2$ at temperature $T = 400K$. Thus, according to the well-kwon *Equation of State* $\rho = PM_0/ZRT$, we get

$$\rho_H = \frac{\left(5.166\times10^3 N/m^2\right)\left(2\times10^{-3} kg.mol^{-1}\right)}{(\sim 1)\left(8.314\, joule.mol^{-1}.K^{-1}\right)(400K)} = $$
$$= 3.1\times10^{-3} kg/m^3$$

Thus, the number of Hydrogen atoms/m$^3$ inside the Nickel powder is

$$n_H = N_0 \rho_H / A_{H2} = 3.01\times10^{26} \rho_H \quad atoms/m^3$$

Then, the number of H atoms inside the Nickel powder is given by

$$n_H V_H = n_H S_f \delta_H \cong 8.3\times10^{24} \rho_H \alpha^2 \xi$$

where $\delta_H = \Delta_{Ni} - \phi_{Ni} \cong 1nm$; $\phi_{Ni}$ is the diameter of Ni atom; $\Delta_{Ni}$ is the *average molecular separation* in the Ni. Then, we get $n_H V_H = n_H S_f \delta_H \cong 6.4\times10^{18}\, atoms$. Thus, the total energy realized in the p-p fusions is

$$E = \frac{n_H V_H}{2}(0.42MeV) =$$
$$= \frac{6.4\times10^{18}}{2}(0.42MeV) = 1.3\times10^{24} eV \cong$$
$$\cong 2.1\times10^5 J \cong 0.05\ Kwh$$

This energy correspond to a power of $0.05 Kwh/h = 50W$, which is the same value detected in the *Focardi's* experiments.

This explains the anomalous heat production in Ni-H Systems detected in the *Focardi's* experiments.

Since the 7.83 Hz electromagnetic field (Schumann resonances) *does not disappear when the device is switched off, the energy conversion can remain running for long period after it is switched off* because, when the device is switched off, the value of the electrical conductivity of the Ni-H system, which was approximately equal to $\sigma_{Ni}$, slowly decreases, tending to $\sigma_H$, which is much smaller than 1. When the electrical conductivity becomes smaller than $\omega\varepsilon$ the value of $n_r$ becomes approximately equal to 1. Consequently, $\lambda_{mod}$ becomes equal to $c/f = 3.8\times10^7 m$ and, according to Eq.(17), the result is $\chi \cong 1$.

This explain why in the Focardi's experiment the device remained running for twenty four days after being switched off.

It is evident that the discovery of this energy conversion device is highly relevant. However, this system is not an efficient energy source if compared to the *Gravitational Motor* [18], which can provide 219KW/m$^3$ while the Ni-H system only 20Kw/m$^3$ (by increasing $\alpha$ from 5cm up to 100cm). Furthermore, the Gravitational Motor converts gravitational energy into rotational mechanical energy directly from the gravitational field, while the Ni-H system needs to produce vapor in order to convert the energy into rotational mechanical energy.

## 3. Transforming a Ni-H system into a Hydrogen Bomb.

It is easy to see that a Ni-H System can be transformed into a Hydrogen bomb, simply increasing the volume of the Ni-H cylinder and substituting the Hydrogen by *a liquid deuterium* LD (12.5 MeV of *energy* is produced at each fusion of two *deuterium nuclei* ‡ ). For example, if $\alpha = 0.27$m, $\xi = 2$ m, and, if a liquid deuterium ($\rho_H = 67.8\ kg.m^{-3}$ [19]) is injected into the Ni powder, then the total energy realized in the fusions becomes

$$E = \frac{n_H V_H}{2}(12.5MeV) =$$
$$= \frac{8.4\times10^{24}\rho_H\alpha^2\xi}{2}(12.5MeV) \cong \qquad (2)$$
$$\cong 5.2\times10^{31}\rho_H\alpha^2\xi\ eV \cong 8.2\times10^{13}J \cong 20\ kilotons$$

The Hiroshima's atomic bomb had *20 kilotons*.

It is important to note that *this bomb type is much easier to build than the conventional nuclear bombs*. Basically, these bombs are made of *Nickel powder* (99%), *liquid deuterium-tritium mixture* and *Mumetal*. These materials can be easily obtained. Due to the simplicity of its construction *these bombs can be built at the very place of the target* (For example, *inside a house or apartment at the target city*.). This means that, in the most of cases missiles are not necessary to launch them. In addition, they cannot be easily detected during their building because the necessary materials are trivial, and there is no radioactive material.

‡ The $d + d$ fusion reaction has two branches that occur with nearly equal probability: ($T + p + 4.03MeV$ and $^3He + n + 3.27MeV$ ). Then, a *deuteron* $d$ is produced by the fusion of the proton $p$ (produced in the first branch) with the neutron (produced in the second branch). Next, occurs the fusion of this deuteron with the *tritium* $T$ produced in the first branch, i.e., ($d + T \rightarrow {}^3_2He + n + 17.6\ MeV$). Thus, we count the $d + d$ fusion energy as $E_{fus} = (4.03+17.6+3.27)/2 = 12.5\ MeV$.

# Engineering the Ni-H Bomb

**Fran De Aquino**
Maranhao State University, Physics Department, S.Luis/MA, Brazil.

The anomalous heat production detected in Ni-H systems was recently explained based on the fact that electromagnetic fields of extremely-low frequencies (ELF) can increase the intensities of gravitational forces and overcome the intensity of the electrostatic repulsion forces, producing nuclear fusion reactions. This effect can provide a consistent and coherent explanation for anomalous heat production detected in Ni-H Systems, and shows that a Ni-H System can be easily transformed into a Hydrogen bomb. Here, a Ni-H bomb of 20 kilotons is engineered.

## 1. Introduction

Recently, a large anomalous production of heat in a nickel rod filled with hydrogen has been reported by Focardi et al., [1]. This phenomenon was posteriorly confirmed by Cerron-Zeballos et al., [2].

Nuclear fusion can be produced by *increasing the gravitational forces in order to overcome the electrostatic repulsion forces between the nucle*i. This process became feasible after the Quantization of Gravity [3], with the discovery that the *gravitational mass* $m_g$ can be made negative and strongly intensified by means of electromagnetic fields of extremely-low frequencies. This effect can provide a consistent and coherent explanation for anomalous heat production detected in Ni-H Systems, and shows that a Ni-H System can be easily transformed into a Hydrogen bomb [4]. Here, a Ni-H bomb of 20 kilotons is engineered.

## 2. Theory

Consider the Ni-H system showed in Fig. 1. In a previous paper [4] it was showed that, if the air inside the Nickel powder is evacuated by means of a vacuum pump (down to $P = 0.051 atm = 5.166\times10^3 N/m^2$ at temperature $T = 400K$) and after Hydrogen is injected into the Nickel powder, then, the number of Hydrogen atoms/m$^3$ inside the Nickel powder is

$$n_H = N_0\rho_H / A_{H2} = 1.94\times10^{29}\rho_H \quad atoms/m^3$$

where $\rho_H$ is the Hydrogen density; $N_0 = 6.02\times10^{26} molecules/kmole$ is the Avogadro's number and $A$ is the molar mass.

Then, the number of atoms inside the Nickel powder is given by

$$n_H V_H = n_H S_f \delta_H \cong 8.3\times10^{24}\rho_H \alpha^2 \xi$$

where $S_f \cong 4\times10^3 \rho_{(Ni)} S_\alpha \xi$; $\rho_{(Ni)} = 8800 kg.m^{-3}$; $S_\alpha = \pi\alpha^2/4$ and $\delta_H = \Delta_{Ni} - \phi_{Ni} \cong 1nm$; $\phi_{Ni}$ is the diameter of Ni atom; $\Delta_{Ni}$ is the *average molecular separation* in the Ni.

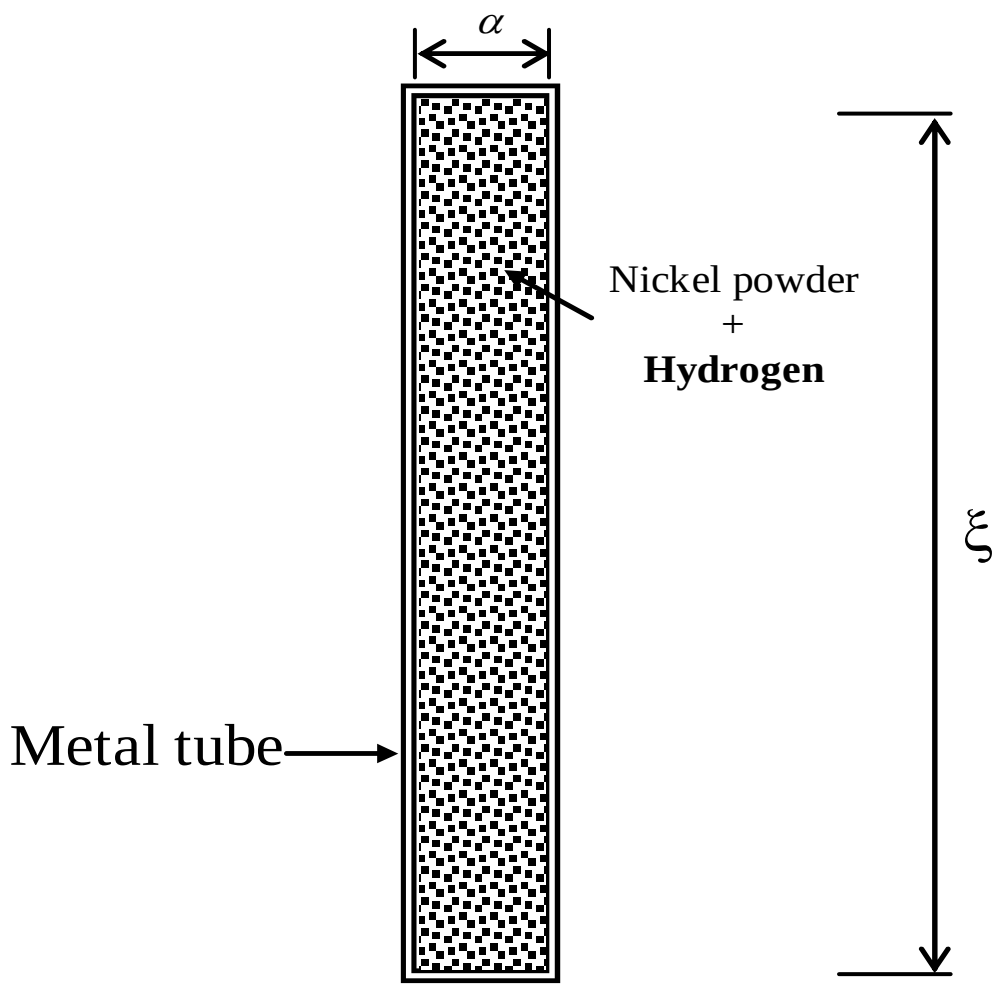

Fig.1 – *Ni-H system*. Note that, on Earth, the system is subjected to a 7.83 Hz electromagnetic field. This field is what naturally exists inside the *spherical resonant cavity* formed by the Earth's surface and the inner edge of the ionosphere. (Schumann resonance).

Thus, the total energy realized in the *protons fusions* is[*]

$$E=\frac{n_H V_H}{2}=\frac{8.3\times10^{24}\rho_H\alpha^2\xi}{2}(0.42MeV)\cong \quad (1)$$

$$\cong 1.7\times10^{30}\rho_H\alpha^2\xi\ \ eV\cong 2.7\times10^{11}\rho_H\alpha^2\xi\ \ Joules$$

It is easy to see that a Ni-H System can be transformed into a Hydrogen bomb, simply increasing the volume of the Ni-H cylinder and substituting the Hydrogen by *a liquid deuterium* LD (12.5 MeV of *energy* is produced at each fusion of two *deuterium nuclei* [†]). For example, if α = 0.27m, ξ = 2 m (See Fig.2), and, if a liquid deuterium ($\rho_H = 67.8\ \ kg.m^{-3}$ [5]) is injected into the Ni powder, then the total energy realized in the fusions becomes

$$E=\frac{8.4\times10^{24}\rho_H\alpha^2\xi}{2}(12.5MeV)\cong \quad (2)$$

$$\cong 5.2\times10^{31}\rho_H\alpha^2\xi\ \ eV\cong 8.3\times10^{13}J\cong 20\ \ kilotons$$

The Hiroshima's atomic bomb had *20 kilotons*.

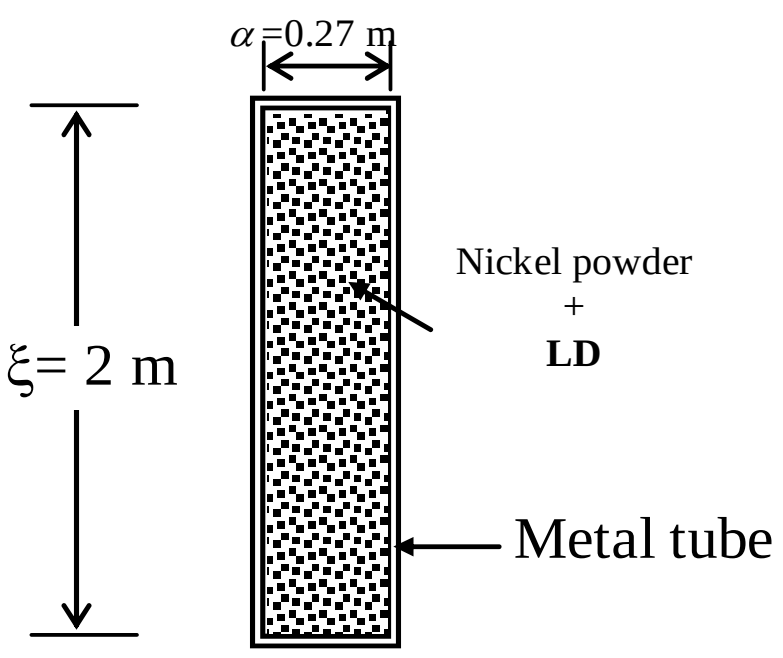

Fig.2 – The core of the Ni-H bomb of *20 kilotons.*

It is important to note that *this bomb type is much easier to build than the conventional nuclear bombs*. Basically, these bombs are made of *Nickel powder* (99%), *liquid deuterium* and *Mumetal*. These materials can be easily obtained. Due to the simplicity of its construction, *these bombs can be built at the very location of the target* (For example, *inside a house or apartment at the target city*.). This means that, in most of cases missiles are not necessary to deliver them, except for launching the Ni-H bomb at the height of explosion (<1Km [‡]) when necessary.

Thus, the Ni-H bomb so far seems to be the simplest atomic bomb ever to be built. *It can be made by every nation*, in such a way that, peace in the World will be reached in the future due to the equilibrium of forces among nations.

Figure 3 shows the Ni-H bomb. It is enveloped by a Mumetal box in order to avoid the action of the 7.83Hz electromagnetic field that naturally exists inside the *spherical resonant cavity* formed by the Earth's surface and the inner edge of the ionosphere. (Schumann resonance [6, 7]). *When the mumetal shielding is exploded the 7.83Hz electromagnetic field acts on the core of the Ni-H bomb and it explodes.*

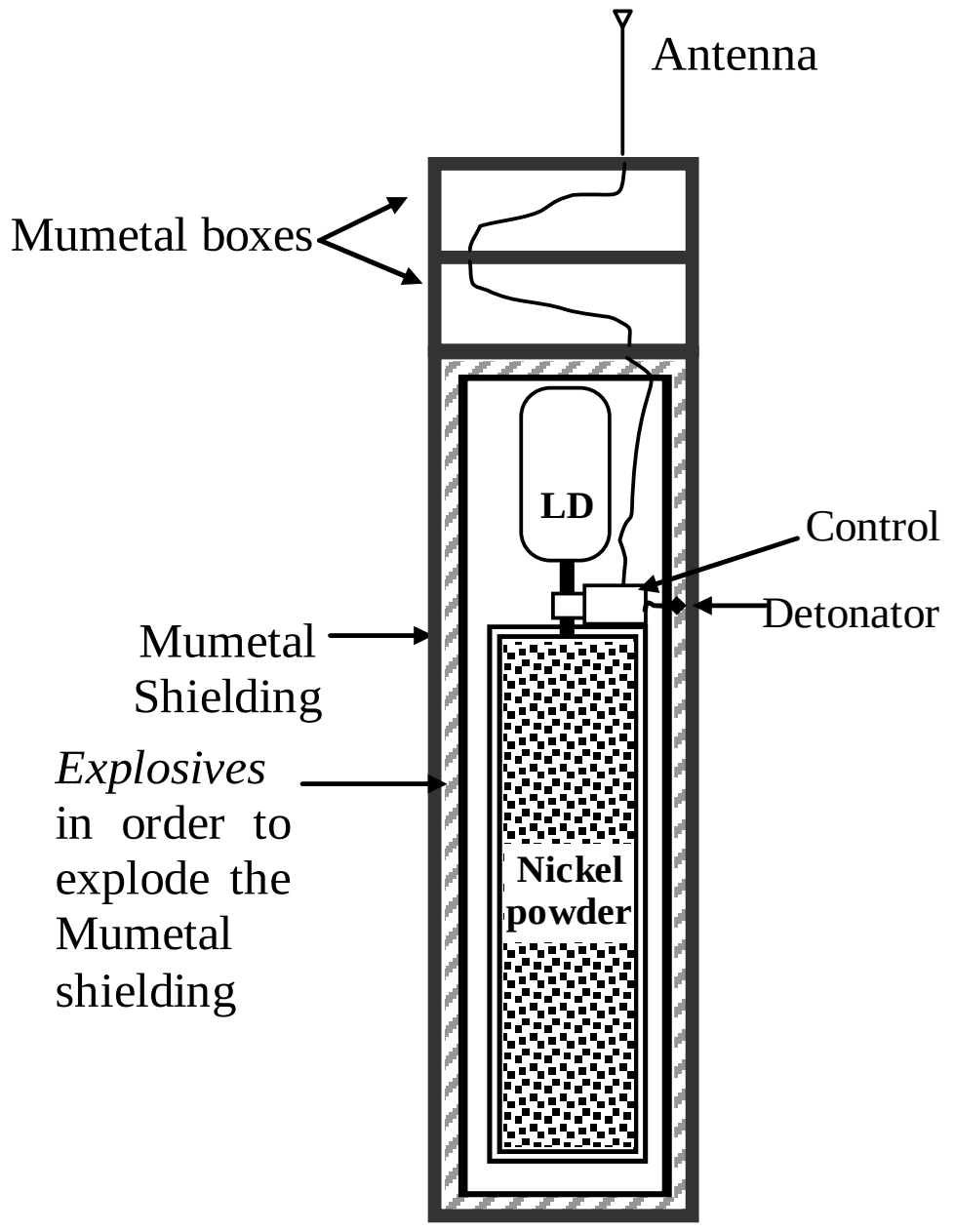

Fig.3 – The Ni-H bomb of *20 kilotons.*

[*] 0.42MeV are realized at each fusion of *two* Hydrogen nuclei.

[†] The *d* + *d* fusion reaction has two branches that occur with nearly equal probability: ($T + p + 4.03MeV$ and $^3He + n + 3.27MeV$). Then, a *deuteron d* is produced by the fusion of the proton *p* (produced in the first branch) with the neutron (produced in the second branch). Next, occurs the fusion of this deuteron with the *tritium T* produced in the first branch, i.e., ($d + T \rightarrow {}^3_2He + n + 17.6\ MeV$). Thus, we count the $d + d$ fusion energy as $E_{fus}$ = (4.03+17.6+3.27)/2 = 12.5 *MeV.*

[‡] Hiroshima 600m above. Nagasaki 500m above.

# Gravitational Shockwave Weapons

**Fran De Aquino**
Maranhao State University, Physics Department, S.Luis/MA, Brazil.

Detonation velocities, greater than that generated by high explosives (~$10^4$m/s), can be achieved by using the gravitational technology recently discovered. This possibility leads to the conception of powerful shockwave weapons. Here, we show the design of a portable gravitational shockwave weapon, which can produce detonation velocities greater than $10^5$m/s, and detonation pressures greater than $10^{10}$N/m$^2$ .

**Key words:** Modified theories of gravity, Detonation waves, Shockwaves, Nozzle flow.
PACS: 04.50.Kd, 47.40.Rs, 47.40.-x, 47.60.Kz.

## 1. Introduction

The most important single property of an explosive is the *detonation velocity*. It is the speed at which the detonation wave travels through the explosive. Typical detonation velocities in solid explosives often range beyond 3,000 m/s to 10,300 m/s [1].

At the front of the detonation zone, an energy pulse or "shockwave" is generated and transmitted to the adjacent region. The shockwave travels outward as a compression wave, moving at or near detonation velocity. When the intensity of the shockwave exceeds the compression strength of the materials they are destroyed. If the mass of the body is too large the wave energy is simply absorbed by the body [2].

The pressure produced in the explosion zone is called *Detonation Pressure*. It expresses the intensity of the generated shockwave. A high detonation pressure is necessary when blasting hard, dense bodies. Detonation pressures of high explosives are in the range from $10^6$N/m$^2$ to over $10^7$ N/m$^2$ [3].

Here, we show the design of a portable shockwave weapon, which uses the *Gravitational Shielding Effect* (BR Patent Number: PI0805046-5, July 31, 2008) *in order to produce detonations velocities greater than 100,000m/s*, and detonation pressures greater than $10^{10}$N/m$^2$. It is important to remember that an aluminum-nitrate truck bomb has a relatively low detonation velocity of 3,500 m/s (sound speed is 343.2m/s)*. High explosives such as TNT has a detonation velocity of 6,900m/s; Military explosives used to destroy strong concrete and steel structures have a detonation velocity of 7,000 to 8,000 m/s [3].

## 2. Theory

The contemporary greatest challenge of the Theoretical Physics was to prove that, Gravity is a *quantum* phenomenon. The quantization of gravity showed that the *gravitational mass* $m_g$ and *inertial mass* $m_i$ are correlated by means of the following factor [4]:

$$\chi = \frac{m_g}{m_{i0}} = \left\{1 - 2\left[\sqrt{1 + \left(\frac{\Delta p}{m_{i0}c}\right)^2} - 1\right]\right\} \qquad (1)$$

where $m_{i0}$ is the *rest* inertial mass of the particle and $\Delta p$ is the variation in the particle's *kinetic momentum*; $c$ is the speed of light.

When $\Delta p$ is produced by the absorption of a photon with wavelength $\lambda$, it is expressed by $\Delta p = h/\lambda$. In this case, Eq. (1) becomes

$$\frac{m_g}{m_{i0}} = \left\{1 - 2\left[\sqrt{1 + \left(\frac{h/m_{i0}c}{\lambda}\right)^2} - 1\right]\right\}$$
$$= \left\{1 - 2\left[\sqrt{1 + \left(\frac{\lambda_0}{\lambda}\right)^2} - 1\right]\right\} \qquad (2)$$

* When a shockwave is created by high explosives it will always travel at high supersonic velocity from its point of origin.

where $\lambda_0 = h/m_{i0}c$ is the *De Broglie wavelength* for the particle with *rest* inertial mass $m_{i0}$.

It was shown that there is an additional effect - *Gravitational Shielding* effect - produced by a substance whose gravitational mass was reduced or made negative [5]. The effect extends beyond substance (gravitational shielding) , up to a certain distance from it (along the central axis of gravitational shielding). This effect shows that in this region the gravity acceleration, $g_1$, is reduced at the same proportion, i.e., $g_1 = \chi_1 g$ where $\chi_1 = m_g/m_{i0}$ and $g$ is the gravity acceleration *before* the gravitational shielding). Consequently, *after a second gravitational shielding*, the gravity will be given by $g_2 = \chi_2 g_1 = \chi_1 \chi_2 g$, where $\chi_2$ is the value of the ratio $m_g/m_{i0}$ for the *second* gravitational shielding. In a generalized way, we can write that after the *nth* gravitational shielding the gravity, $g_n$, will be given by

$$g_n = \chi_1 \chi_2 \chi_3 ... \chi_n g \tag{3}$$

This possibility shows that, by means of a battery of gravitational shieldings, we can make particles acquire enormous accelerations. In practice, this can lead to the conception of powerful particles accelerators, kinetic weapons or weapons of shockwaves.

From Electrodynamics we know that when an electromagnetic wave with frequency $f$ and velocity $c$ incides on a material with relative permittivity $\varepsilon_r$, relative magnetic permeability $\mu_r$ and electrical conductivity $\sigma$, its *velocity is reduced* to $v = c/n_r$ where $n_r$ is the index of refraction of the material, given by [6]

$$n_r = \frac{c}{v} = \sqrt{\frac{\varepsilon_r \mu_r}{2}\left(\sqrt{1 + (\sigma/\omega\varepsilon)^2} + 1\right)} \tag{4}$$

If $\sigma >> \omega\varepsilon$, $\omega = 2\pi f$ , Eq. (4) reduces to

$$n_r = \sqrt{\frac{\mu_r \sigma}{4\pi\varepsilon_0 f}} \tag{5}$$

Thus, the wavelength of the incident radiation (See Fig. 1) becomes

$$\lambda_{\text{mod}} = \frac{v}{f} = \frac{c/f}{n_r} = \frac{\lambda}{n_r} = \sqrt{\frac{4\pi}{\mu f \sigma}} \tag{6}$$

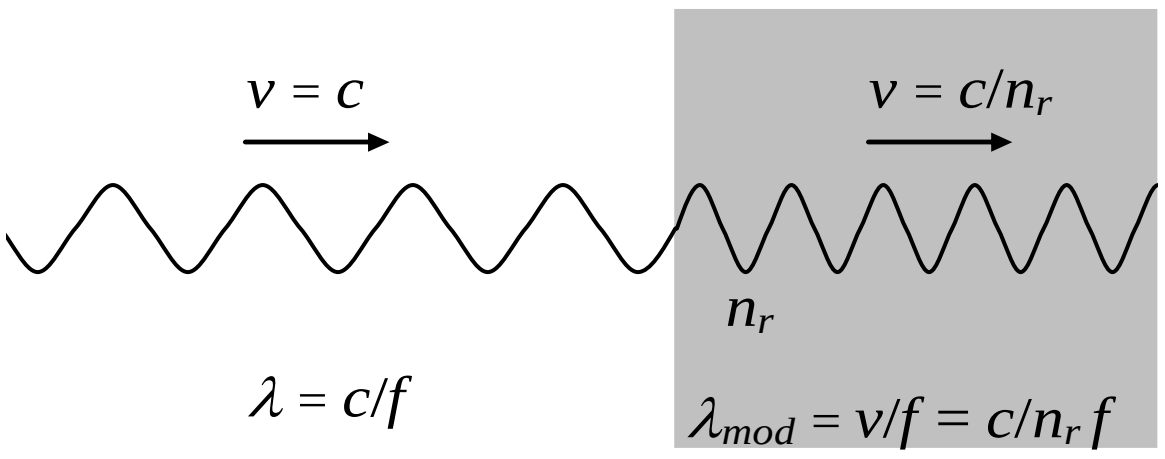

Fig. 1 – *Modified Electromagnetic Wave*. The wavelength of the electromagnetic wave can be strongly reduced, but its frequency remains the same.

If a lamina with thickness equal to $\xi$ contains $n$ molecules/m$^3$, then the number of molecules per area unit is $n\xi$. Thus, if the electromagnetic radiation with frequency $f$ incides on an area $S$ of the lamina it reaches $nS\xi$ molecules. If it incides on the *total area of the lamina*, $S_f$, then the total number of molecules reached by the radiation is $N = nS_f\xi$. The number of molecules per unit of volume, $n$, is given by

$$n = \frac{N_0 \rho}{A} \tag{7}$$

where $N_0 = 6.02 \times 10^{26} \, molecules/kmole$ is the Avogadro's number; $\rho$ is the matter density of the lamina (in *kg/m*$^3$) and $A$ is the molar mass.

When an electromagnetic wave incides on the lamina, it strikes on $N_f$ front molecules, where $N_f \cong (n S_f)\phi_m$ , $\phi_m$ is the "diameter" of the molecule. Thus, the electromagnetic wave incides effectively on an area $S = N_f S_m$ , where $S_m = \frac{1}{4}\pi\phi_m^2$ is the cross section area of one molecule. After these collisions, it carries out $n_{collisions}$ with the other molecules (See Fig.2).

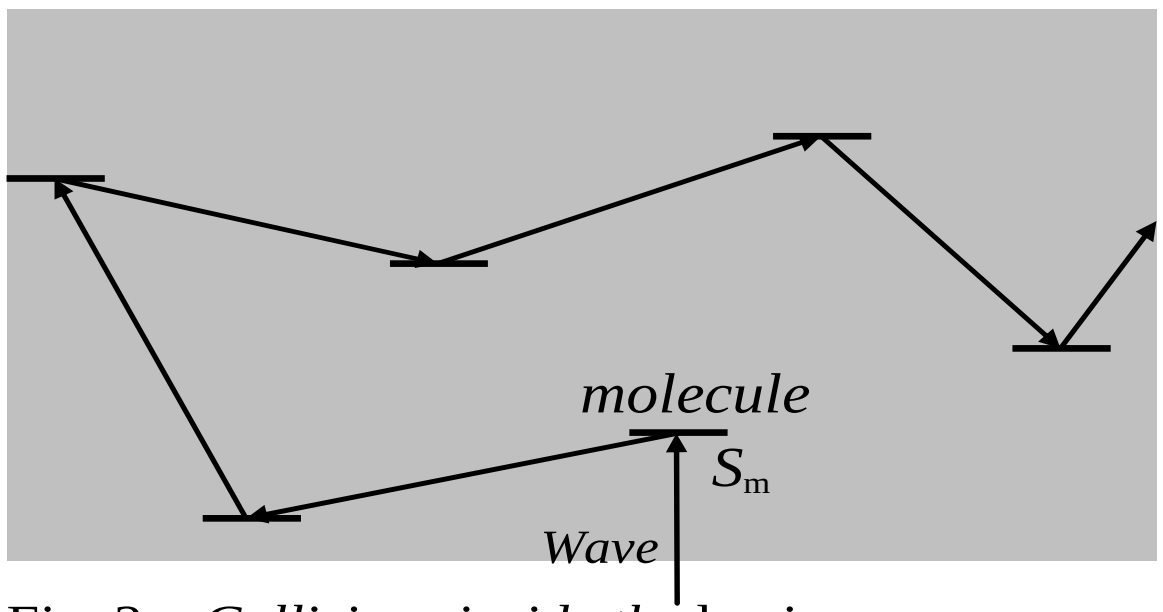

Fig. 2 – *Collisions inside the* lamina.

Thus, the total number of collisions in the volume $S\xi$ is

$$N_{collisions} = N_f + n_{collisions} = n S\phi_m + (n S\xi - n_m S\phi_m) = = n_m S\xi \qquad (8)$$

The power density, $D$, of the radiation on the lamina can be expressed by

$$D = \frac{P}{S} = \frac{P}{N_f S_m} \qquad (9)$$

We can express the *total mean number of collisions in each molecule*, $n_1$, by means of the following equation

$$n_1 = \frac{n_{total\ \ photons} N_{collisions}}{N} \qquad (10)$$

Since in each collision a *momentum* $h/\lambda$ is transferred to the molecule, then the *total momentum* transferred to the lamina will be $\Delta p = (n_1 N) h/\lambda$. Therefore, in accordance with Eq. (1), we can write that

$$\frac{m_{g(l)}}{m_{i0(l)}} = \left\{1 - 2\left[\sqrt{1 + \left[(n_1 N)\frac{\lambda_0}{\lambda}\right]^2} - 1\right]\right\} = = \left\{1 - 2\left[\sqrt{1 + \left[n_{total\ photons} N_{collisions}\frac{\lambda_0}{\lambda}\right]^2} - 1\right]\right\} \qquad (11)$$

Since Eq. (8) gives $N_{collisions} = n\ S\xi$, we get

$$n_{total\ \ photons} N_{collisions} = \left(\frac{P}{hf^2}\right)(n\ S\xi) \qquad (12)$$

Substitution of Eq. (12) into Eq. (11) yields

$$\frac{m_{g(l)}}{m_{i0(l)}} = \left\{1 - 2\left[\sqrt{1 + \left[\left(\frac{P}{hf^2}\right)(n\ S\xi)\frac{\lambda_0}{\lambda}\right]^2} - 1\right]\right\} \qquad (13)$$

Substitution of $P$ given by Eq. (9) into Eq. (13) gives

$$\frac{m_{g(l)}}{m_{i0(l)}} = \left\{1 - 2\left[\sqrt{1 + \left[\left(\frac{N_f S_m D}{f^2}\right)\left(\frac{n\ S\xi}{m_{i0(l)} c}\right)\frac{1}{\lambda}\right]^2} - 1\right]\right\} \qquad (14)$$

Substitution of $N_f \cong (n\ S_f)\phi_m$ and $S = N_f S_m$ into Eq. (14) results

$$\frac{m_{g(l)}}{m_{i0(l)}} = \left\{1 - 2\left[\sqrt{1 + \left[\left(\frac{n^3 S_f^2 S_m^2 \phi_m^2 \xi D}{m_{i0(l)} c f^2}\right)\frac{1}{\lambda}\right]^2} - 1\right]\right\} \qquad (15)$$

where $m_{i0(l)} = \rho_{(l)} V_{(l)}$.

Now, considering that the lamina is inside a ELF electromagnetic field with $E$ and $B$, then we can write that [7]

$$D = \frac{n_{r(l)} E^2}{2\mu_0 c} \qquad (16)$$

Substitution of Eq. (16) into Eq. (15) gives

$$\frac{m_{g(l)}}{m_{i0(l)}} = \left\{1 - 2\left[\sqrt{1 + \left[\left(\frac{n_{r(l)} n^3 S_f^2 S_m^2 \phi_m^2 \xi E^2}{2\mu_0 m_{i0(l)} c^2 f^2}\right)\frac{1}{\lambda}\right]^2} - 1\right]\right\} \qquad (17)$$

Now assuming that the lamina is a *cylindrical air lamina* (diameter = $\alpha$ ; thickness = $\xi$) where $n_{r(l)} \cong 1$; $n = N_0 \rho_{(l)}/A = 2.6 \times 10^{25}\ molecules\ /m^3$; $\phi_m = 1.55 \times 10^{-10} m$; $S_m = \pi\phi_m^2/4 = 1.88 \times 10^{-20} m^2$, then, Eq. (17) reduces to

$$\frac{m_{g(l)}}{m_{i0(l)}}=\left\{1-2\left[\sqrt{1+\left[6.6\times10^{5}\left(\frac{S_f^2\xi}{m_{i0(l)}f^2}\right)\frac{E^2}{\lambda}\right]^2}-1\right]\right\} \quad (18)$$

An *atomized water spray* is created by forcing the water through an orifice. The energy required to overcome the pressure drop is supplied by the *spraying pressure* at each detonation. Spraying pressure depends on feed characteristics and desired *particle size*. If *atomizing water* is injected into the air lamina, then the area $S_f$ to be considered is the *surface area* of the *atomizing water*, which can be obtained by multiplying the *specific surface area*(SSA) *of the atomizing water* (which is given by $SSA = A/\rho_w V = 3/\rho_w r_d$ ) *by the total mass of the atomizing water* ($m_{i0(w)} = \rho_w V_{water\ droplets} N_d$). Assuming that the *atomizing water* is composed of monodisperse particles with $10\mu m$ radius $\left(r_d = 1\times10^{-5} m\right)$, and that the *atomizing water* has $N_p \approx 10^8 droplets/m^3$ [8] then we obtain $SSA = 3/\rho_w r_d = 300 m^2/kg$ and $m_{i0(w)} = \rho_w V_{water\ droplets} N_d \approx 10^{-5} kg$. Thus, we get

$$S_f = (SSA) m_{i0(w)} \approx 10^{-3} m^2 \quad (18)$$

Substitution of $S_f \approx 10^{-3} m^2$ and $m_{i0(l)} = \rho_{air} S_\alpha \xi = 1.2 S_\alpha \xi$ into Eq. (18) gives

$$\frac{m_{g(l)}}{m_{i0(l)}}=\left\{1-2\left[\sqrt{1+\sim\frac{E^4}{S_\alpha^2 f^4 \lambda^2}}-1\right]\right\} \quad (19)$$

The injection of atomized water increases the electrical conductivity of the mean, making it greater than the conductivity of water $(\sigma >> 0.005 S/m)$. Under these conditions, the value of $\lambda$, given by Eq. (6), becomes

$$\lambda = \lambda_{mod} = \sqrt{\frac{4\pi}{\mu_0 f \sigma}} \quad (20)$$

where $f$ is the frequency of the ELF electromagnetic field.

Substitution of Eq. (20) into Eq. (19) yields

$$\chi = \frac{m_{g(l)}}{m_{i0(l)}}=\left\{1-2\left[\sqrt{1+\sim10^{-7}\frac{\sigma E^4}{S_\alpha^2 f^3}}-1\right]\right\} \quad (21)$$

Note that $E = E_m \sin\omega t$.The average value for $E^2$ is equal to $\frac{1}{2}E_m^2$ because $E$ varies sinusoidaly ($E_m$ is the maximum value for $E$). On the other hand, $E_{rms} = E_m/\sqrt{2}$. Consequently we can change $E^4$ by $E_{rms}^4$, and the equation above can be rewritten as follows

$$\chi = \frac{m_{g(l)}}{m_{i0(l)}}=\left\{1-2\left[\sqrt{1+\sim10^{-7}\frac{\sigma E_{rms}^4}{S_\alpha^2 f^3}}-1\right]\right\} \quad (22)$$

Now consider the weapon showed in Fig. 3 $(\alpha = 12.7mm)$. When an ELF electromagnetic field with frequency $f = 10Hz$ is activated, an electric field $E_{rms}$ passes through the 7 cylindrical air laminas. Then, according to Eq. (22) the value of $\chi$ (for $\sigma >> 0.005 S/m$) at each lamina is

$$\chi >> \left\{1-2\left[\sqrt{1+\sim10^{-9}E_{rms}^4}-1\right]\right\} \quad (23)$$

For example, if $E_{rms} \cong 10^4 V/m$ we get

$$\chi >> -10^3 \quad (24)$$

Therefore, according to Eq. (3) the gravitational acceleration produced by the gravitational mass $M_g = 4.23kg$, just after the $7^{th}$ cylindrical air lamina $(r_7 = 150mm)$, will be given by

$$g_7 = \chi^7 g = -\chi^7 \frac{GM_g}{r_7^2} >> +10^{13} m/s^2 \qquad (25)$$

This is the acceleration acquired by the air molecules that are just after the $7^{th}$ cylindrical air lamina. Obviously, this produces enormous pressure in the air after the $7^{th}$ cylindrical air lamina, in a similar way that pressure produced by a detonation. The detonation velocity after the $7^{th}$ cylindrical air lamina is

$$v_d = \sqrt{2g_7(\Delta r)} >> 10^5 m/s \qquad (26)$$

Consequently, the detonation pressure is

$$p = 2\rho_{air} v_d^2 >> 10^{10} N/m^2 \qquad (27)$$

These values show how powerful can be the gravitational shockwaves weapons. The maxima resistance of the most resistant steels is of the order of $10^{11}$N/m$^2$ (*Graphene* ~$10^{12}$N/m$^2$). Since the gravitational shockwave weapons can be designed to produce detonation pressures of these magnitudes, we can conclude that it can destroy anything.

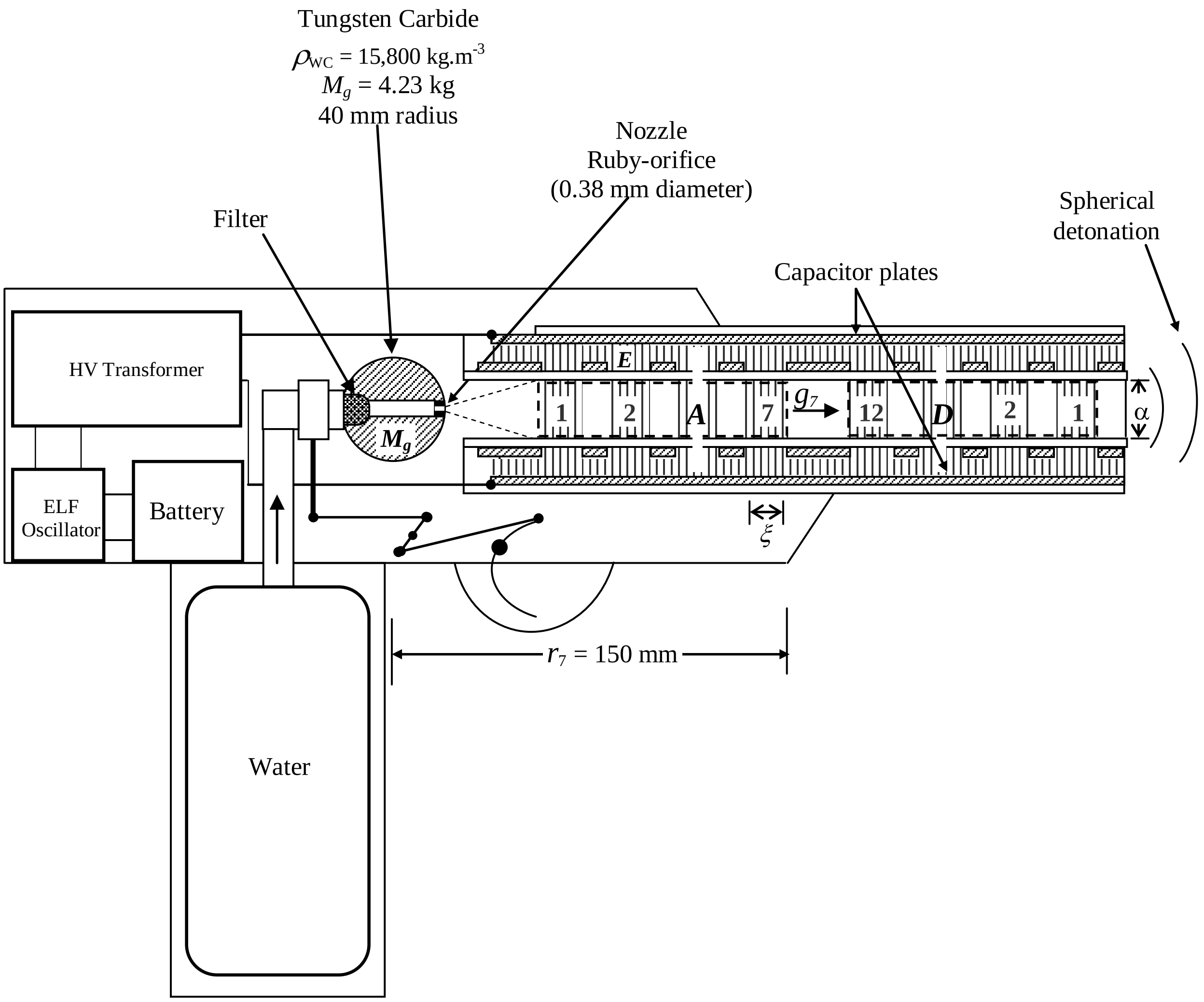

Fig. 3 – *Portable Weapon of Gravitational Shockwaves*. Note that there are two sets of Gravitational Shieldings (GS): the set ***A*** (accelerator) with 7 GS and the set ***D*** (decelerator) with 12 GS. The objective of the set ***D***, with 12 GCC, is to reduce strongly the value of the *external* gravity along the axis of the tube (in the opposite direction of the acceleration $g_7$). In this case, the value of the external gravity, $g_{ext}$, is reduced by the factor $\chi_d^{12} g_{ext}$, where $\chi_d = 10^{-2}$. For example, if the opening of the tube $(\alpha)$ of the weapon is positioned on the Earth surface then $g_{ext} = 9.81 m/s^2$ is reduced to $\chi_d^{12} g_{ext}$ and, after in the set ***A***, it is increased by $\chi^7$. Without the set ***D***, the back of the weapon can explode.

# A System to convert Gravitational Energy directly into Electrical Energy

**Fran De Aquino**
Maranhao State University, Physics Department, S.Luis/MA, Brazil.

We show that it is possible to produce *strong gravitational accelerations on the free electrons of a conductor* in order to obtain electrical current. This allows the conversion of gravitational energy directly into electrical energy. Here, we propose a system that can produce several tens of kilowatts of electrical energy converted from the gravitational energy.

**Key words:** Modified theories of gravity, Electric fields effects on material flows, Electron tubes, Electrical instruments.
PACS: 04.50.Kd, 83.60.Np , 84.47.+w, 07.50.-e.

## 1. Introduction

In a previous paper [1], we have proposed a system to convert gravitational energy into rotational kinetic energy (*Gravitational Motor*), which can be converted into electrical energy by means of a conventional electrical generator. Now, we propose a novel system to convert gravitational energy *directly* into electrical energy.

It is known that, in some materials, called *conductors*, the free electrons are so loosely held by the atom and so close to the neighboring atoms that they tend to drift randomly from one atom to its neighboring atoms. This means that the electrons move in all directions by the same amount. However, if some outside force acts upon the free electrons their movement becomes not random, and they move from atom to atom at the same direction of the applied force. This flow of electrons (their electric charge) through the conductor produces the *electrical current*, which is defined as a flow of electric charge through a medium [2]. This charge is typically carried by moving electrons in a conductor, but it can also be carried by ions in an electrolyte, or by both ions and electrons in a plasma [3].

Thus, the electrical current arises in a conductor when an outside force acts upon the free electrons. This force is called, in a generic way, of *electromotive force* (EMF). Usually, it is of *electrical* nature $(F = eE)$. Here, it is shown that the electrical flow can also be achieved by means of *gravitational* forces $(F = m_e g)$. The *Gravitational Shielding Effect* (BR Patent Number: PI0805046-5, July 31, 2008 [4]), shows that a battery of Gravitational Shieldings can strongly intensify the gravitational acceleration in any direction and, in this way, it is possible *to produce strong gravitational accelerations on the free electrons of a conductor* in order to obtain electrical current.

## 2. Theory

From the quantization of gravity it follows that the *gravitational mass* $m_g$ and the *inertial mass* $m_i$ are correlated by means of the following factor [1]:

$$\chi = \frac{m_g}{m_{i0}} = \left\{1 - 2\left[\sqrt{1 + \left(\frac{\Delta p}{m_{i0}c}\right)^2} - 1\right]\right\} \qquad (1)$$

where $m_{i0}$ is the *rest* inertial mass of the particle and $\Delta p$ is the variation in the particle's *kinetic momentum*; $c$ is the speed of light.

When $\Delta p$ is produced by the absorption of a photon with wavelength $\lambda$, it is expressed by $\Delta p = h/\lambda$. In this case, Eq. (1) becomes

$$\frac{m_g}{m_{i0}} = \left\{1 - 2\left[\sqrt{1 + \left(\frac{h/m_{i0}c}{\lambda}\right)^2} - 1\right]\right\}$$

$$= \left\{1 - 2\left[\sqrt{1 + \left(\frac{\lambda_0}{\lambda}\right)^2} - 1\right]\right\} \quad (2)$$

where $\lambda_0 = h/m_{i0}c$ is the *De Broglie wavelength* for the particle with *rest* inertial mass $m_{i0}$.

It has been shown that there is an additional effect - *Gravitational Shielding* effect - produced by a substance whose gravitational mass was reduced or made negative [5]. The effect extends beyond substance (gravitational shielding) , up to a certain distance from it (along the central axis of gravitational shielding). This effect shows that in this region the gravity acceleration, $g_1$, is reduced at the same proportion, i.e., $g_1 = \chi_1 g$ where $\chi_1 = m_g/m_{i0}$ and $g$ is the gravity acceleration *before* the gravitational shielding). Consequently, *after a second gravitational shielding*, the gravity will be given by $g_2 = \chi_2 g_1 = \chi_1\chi_2 g$, where $\chi_2$ is the value of the ratio $m_g/m_{i0}$ for the *second* gravitational shielding. In a generalized way, we can write that after the *nth* gravitational shielding the gravity, $g_n$, will be given by

$$g_n = \chi_1\chi_2\chi_3 \cdots \chi_n g \quad (3)$$

This possibility shows that, by means of a battery of gravitational shieldings, we can make particles acquire enormous accelerations. In practice, this can lead to the conception of powerful particles accelerators, kinetic weapons or weapons of shockwaves.

From Electrodynamics we know that when an electromagnetic wave with frequency $f$ and velocity $c$ incides on a material with relative permittivity $\varepsilon_r$, relative magnetic permeability $\mu_r$ and electrical conductivity $\sigma$, its *velocity is reduced* to $v = c/n_r$ where $n_r$ is the index of refraction of the material, given by [6]

$$n_r = \frac{c}{v} = \sqrt{\frac{\varepsilon_r\mu_r}{2}\left(\sqrt{1 + (\sigma/\omega\varepsilon)^2} + 1\right)} \quad (4)$$

If $\sigma >> \omega\varepsilon$, $\omega = 2\pi f$, Eq. (4) reduces to

$$n_r = \sqrt{\frac{\mu_r\sigma}{4\pi\varepsilon_0 f}} \quad (5)$$

Thus, the wavelength of the incident radiation (See Fig. 1) becomes

$$\lambda_{mod} = \frac{v}{f} = \frac{c/f}{n_r} = \frac{\lambda}{n_r} = \sqrt{\frac{4\pi}{\mu f\sigma}} \quad (6)$$

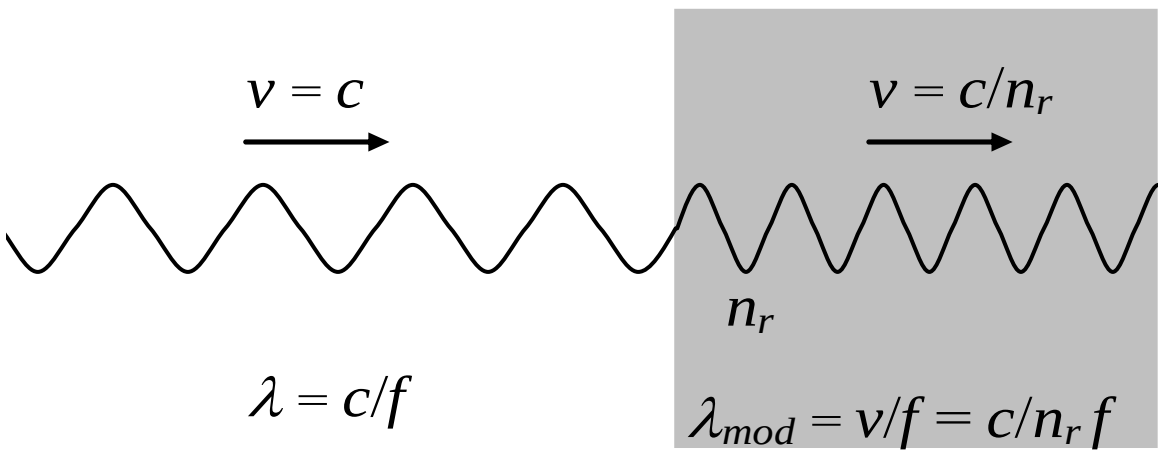

Fig. 1 – *Modified Electromagnetic Wave*. The wavelength of the electromagnetic wave can be strongly reduced, but its frequency remains the same.

If a lamina with thickness equal to $\xi$ contains $n$ atoms/m$^3$, then the number of atoms per area unit is $n\xi$. Thus, if the electromagnetic radiation with frequency $f$ incides on an area $S$ of the lamina it reaches $nS\xi$ atoms. If it incides on the *total area of the lamina*, $S_f$, then the total number of atoms reached by the radiation is $N = nS_f\xi$. The number of atoms per unit of volume, $n$, is given by

$$n = \frac{N_0\rho}{A} \quad (7)$$

where $N_0 = 6.02\times10^{26}\, atoms/kmole$ is the Avogadro's number; $\rho$ is the matter density of the lamina (in *kg/m³*) and $A$ is the molar mass(*kg/kmole*).

When an electromagnetic wave incides on the lamina, it strikes $N_f$ front atoms, where $N_f \cong (nS_f)\phi_m$, $\phi_m$ is the "diameter" of the atom. Thus, the electromagnetic wave incides effectively on an area $S = N_f S_m$, where $S_m = \frac{1}{4}\pi\phi_m^2$ is the cross section area of one atom.

After these collisions, it carries out $n_{collisions}$ with the other atoms (See Fig.2).

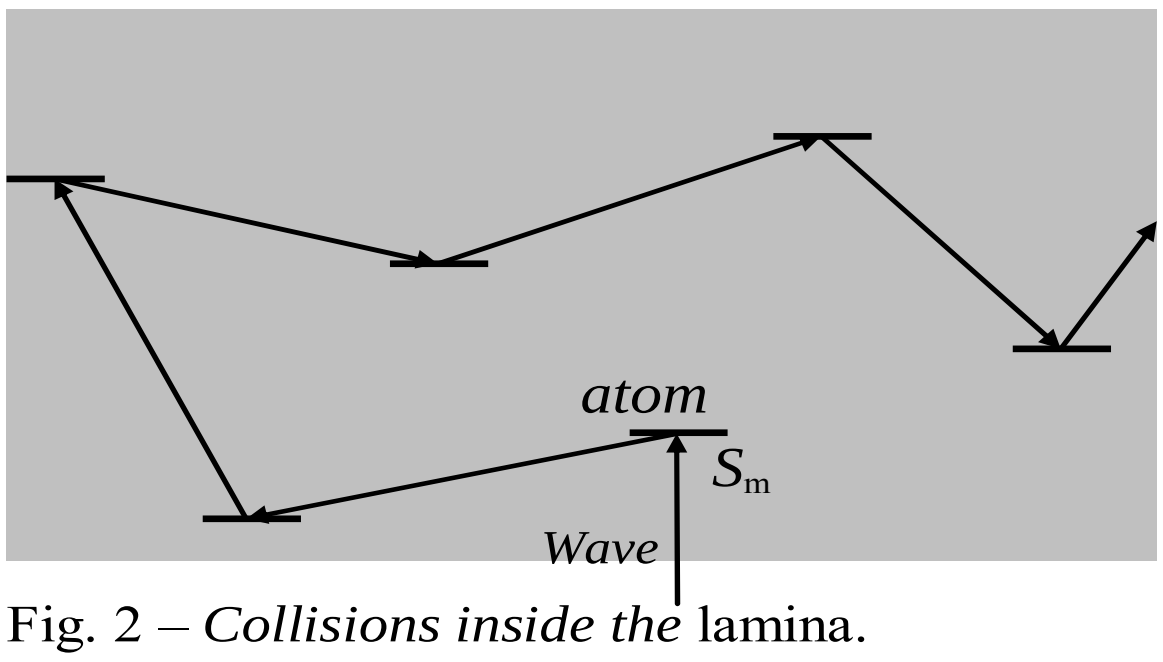

Fig. 2 – *Collisions inside the* lamina.

Thus, the total number of collisions in the volume $S\xi$ is

$$N_{collisions} = N_f + n_{collisions} = n_l S\phi_m + \left(n_l S\xi - n_m S\phi_m\right) = \\ = n_l S\xi \qquad (8)$$

The power density, $D$, of the radiation on the lamina can be expressed by

$$D = \frac{P}{S} = \frac{P}{N_f S_m} \qquad (9)$$

We can express the *total mean number of collisions in each atom*, $n_1$, by means of the following equation

$$n_1 = \frac{n_{total\ photons} N_{collisions}}{N} \qquad (10)$$

Since in each collision a *momentum* $h/\lambda$ is transferred to the atom, then the *total momentum* transferred to the lamina will be $\Delta p = (n_1 N) h/\lambda$. Therefore, in accordance with Eq. (1), we can write that

$$\frac{m_{g(l)}}{m_{i0(l)}} = \left\{1 - 2\left[\sqrt{1 + \left[(n_1 N)\frac{\lambda_0}{\lambda}\right]^2} - 1\right]\right\} = \\ = \left\{1 - 2\left[\sqrt{1 + \left[n_{total\ photons} N_{collisions}\frac{\lambda_0}{\lambda}\right]^2} - 1\right]\right\} \qquad (11)$$

Since Eq. (8) gives $N_{collisions} = n_l S\xi$, we get

$$n_{total\ photons} N_{collisions} = \left(\frac{P}{hf^2}\right)(n_l S\xi) \qquad (12)$$

Substitution of Eq. (12) into Eq. (11) yields

$$\frac{m_{g(l)}}{m_{i0(l)}} = \left\{1 - 2\left[\sqrt{1 + \left[\left(\frac{P}{hf^2}\right)(n_l S\xi)\frac{\lambda_0}{\lambda}\right]^2} - 1\right]\right\} \qquad (13)$$

Substitution of $P$ given by Eq. (9) into Eq. (13) gives

$$\frac{m_{g(l)}}{m_{i0(l)}} = \left\{1 - 2\left[\sqrt{1 + \left[\left(\frac{N_f S_m D}{f^2}\right)\left(\frac{n_l S\xi}{m_{i0(l)} c}\right)\frac{1}{\lambda}\right]^2} - 1\right]\right\} \qquad (14)$$

Substitution of $N_f \cong (n_l S_f)\phi_m$ and $S = N_f S_m$ into Eq. (14) results

$$\frac{m_{g(l)}}{m_{i0(l)}} = \left\{1 - 2\left[\sqrt{1 + \left[\left(\frac{n_l^3 S_f^2 S_m^2 \phi_m^2 \xi D}{m_{i0(l)} c f^2}\right)\frac{1}{\lambda}\right]^2} - 1\right]\right\} \qquad (15)$$

where $m_{i0(l)} = \rho_{(l)} V_{(l)}$.

Now, considering that the lamina is inside an ELF electromagnetic field with $E$ and $B$, then we can write that [7]

$$D = \frac{n_{r(l)} E^2}{2\mu_0 c} \qquad (16)$$

Substitution of Eq. (16) into Eq. (15) gives

$$\frac{m_{g(l)}}{m_{i0(l)}} = \left\{1 - 2\left[\sqrt{1 + \left[\left(\frac{n_{r(l)} n_l^3 S_f^2 S_m^2 \phi_m^2 \xi E^2}{2\mu_0 m_{i0(l)} c^2 f^2}\right)\frac{1}{\lambda}\right]^2} - 1\right]\right\} \qquad (17)$$

In the case in which the area $S_f$ is just the *area of the cross-section of the lamina* $(S_\alpha)$, we obtain from Eq. (17), considering that $m_{i0(l)} = \rho_{(l)} S_\alpha \xi$, the following expression

$$\frac{m_{g(l)}}{m_{i0(l)}}=\left\{1-2\left[\sqrt{1+\left[\left(\frac{n_{r(l)}n_l^3 S_\alpha S_m^2\phi_m^2 E^2}{2\mu_0\rho_{(l)}c^2 f^2}\right)\frac{1}{\lambda}\right]^2}-1\right]\right\} \quad (18)$$

If the electrical conductivity of the lamina, $\sigma_{(l)}$, is such that $\sigma_{(l)} >> \omega\varepsilon$, then the value of $\lambda$ is given by Eq. (6), i.e.,

$$\lambda=\lambda_{\text{mod}}=\sqrt{\frac{4\pi}{\mu f\sigma}} \quad (19)$$

Substitution of Eq. (19) into Eq. (18) gives

$$\frac{m_{g(l)}}{m_{i0(l)}}=\left\{1-2\left[\sqrt{1+\frac{n_{r(l)}^2 n_l^6 S_\alpha^2 S_m^4\phi_m^4\sigma_{(l)}E^4}{16\pi\mu_0\rho_{(l)}^2c^4 f^3}}-1\right]\right\} \quad (20)$$

Note that $E=E_m\sin\omega t$. The average value for $E^2$ is equal to $\frac{1}{2}E_m^2$ because $E$ varies sinusoidaly ($E_m$ is the maximum value for $E$). On the other hand, $E_{rms}=E_m/\sqrt{2}$. Consequently we can change $E^4$ by $E_{rms}^4$, and the equation above can be rewritten as follows

$$\chi=\frac{m_{g(l)}}{m_{i0(l)}}=$$

$$=\left\{1-2\left[\sqrt{1+\frac{n_{r(l)}^2 n_l^6 S_\alpha^2 S_m^4\phi_m^4\sigma_{(l)}E_{rms}^4}{16\pi\mu_0\rho_{(l)}^2c^4 f^3}}-1\right]\right\} \quad (21)$$

Now consider the system shown in Fig.3. It was designed to convert *Gravitational Energy* directly into *Electrical Energy*. Thus, we can say that it is a *Gravitational EMF Source*.

Inside the system there is a *dielectric tube* ($\varepsilon_r \cong 1$) with the following characteristics: $\alpha = 8mm$(diameter), $S_\alpha=\pi\alpha^2/4=5.03\times10^{-5}m^2$. Inside the tube there is a *Lead sphere* ($\rho_s = 11340 Kg/m^3$) with 4mm radius and mass $M_{gs}=3.04\times10^{-3}kg$. The tube is filled with *air* at ambient temperature and 1atm. Thus, inside the tube, the air density is

$$\rho_{air}=1.2 \ kg.m^{-3} \quad (22)$$

The number of atoms of air (Nitrogen) per unit of volume, $n_{air}$, according to Eq.(7), is given by

$$n_{air}=\frac{N_0\rho_{air}}{A_N}=5.16\times10^{25}atoms/m^3 \quad (23)$$

The *parallel metallic plates* (p), shown in Fig.3 are subjected to different drop voltages. The two sets of plates (*D*), placed on the extremes of the tube, are subjected to $V_{(D)rms}=3.847 \ kV$ at $f=60Hz$, while the central set of plates (*A*) is subjected to $V_{(A)rms}=7.827 \ kV$ at $f=60Hz$. Since $d=14mm$, then the intensity of the electric field, which passes through the 36 *cylindrical air laminas* (each one with 5mm thickness) of the *two* sets (*D*), is

$$E_{(D)rms}=V_{(D)rms}/d=2.748\times10^5 V/m$$

and the intensity of the electric field, which passes through the 19 *cylindrical air laminas* of the central set (*A*), is given by

$$E_{(A)rms}=V_{(A)rms}/d=5.591\times10^5 V/m$$

Note that the *metallic rings* (5*mm* thickness) are positioned in such way to block the electric field out of the cylindrical air laminas. The objective is to turn each one of these laminas into a *Gravity Control Cells* (GCC) [5]. Thus, the system shown in Fig. 3 has 3 sets of GCC. Two with 18 GCC each, and one with 19 GCC. The two sets with 18 GCC each are positioned at the extremes of the tube (*D*). They work as gravitational *decelerator* while the other set with 19 GCC (*A*) works as a gravitational *accelerator*, intensifying the gravity acceleration produced by the mass $M_{gs}$ of the Lead sphere. According to Eq. (3), this gravity, after the $19^{th}$ GCC becomes $g_{19}=\chi^{19}GM_{gs}/r_0^2$, where $\chi=m_{g(l)}/m_{i(l)}$ given by Eq. (21) and $r_0=9mm$ is the distance between the center of the Lead sphere and the surface of the first GCC of the set (A).

The objective of the sets (*D*), with 18 GCC each, is to reduce strongly the value of the external gravity along the axis of the tube. In this case, the value of the external gravity, $g_{ext}$, is reduced by the factor $\chi_d^{18}g_{ext}$, where $\chi_d=10^{-2}$. For example, if the base BS of the system is positioned on the Earth surface, then $g_{ext}=9.81m/s^2$ is reduced to $\chi_d^{18}g_{ext}$ and, after the set A, it is increased by $\chi^{19}$. Since the system is designed for $\chi=-6.4138$, then the gravity acceleration on the sphere becomes $\chi^{19}\chi_d^{18}g_{ext}=2.1\times10^{-20}m/s^2$, this value is much smaller than $g_{sphere}=GM_{gs}/r_s^2=1.27\times10^{-8}m/s^2$.

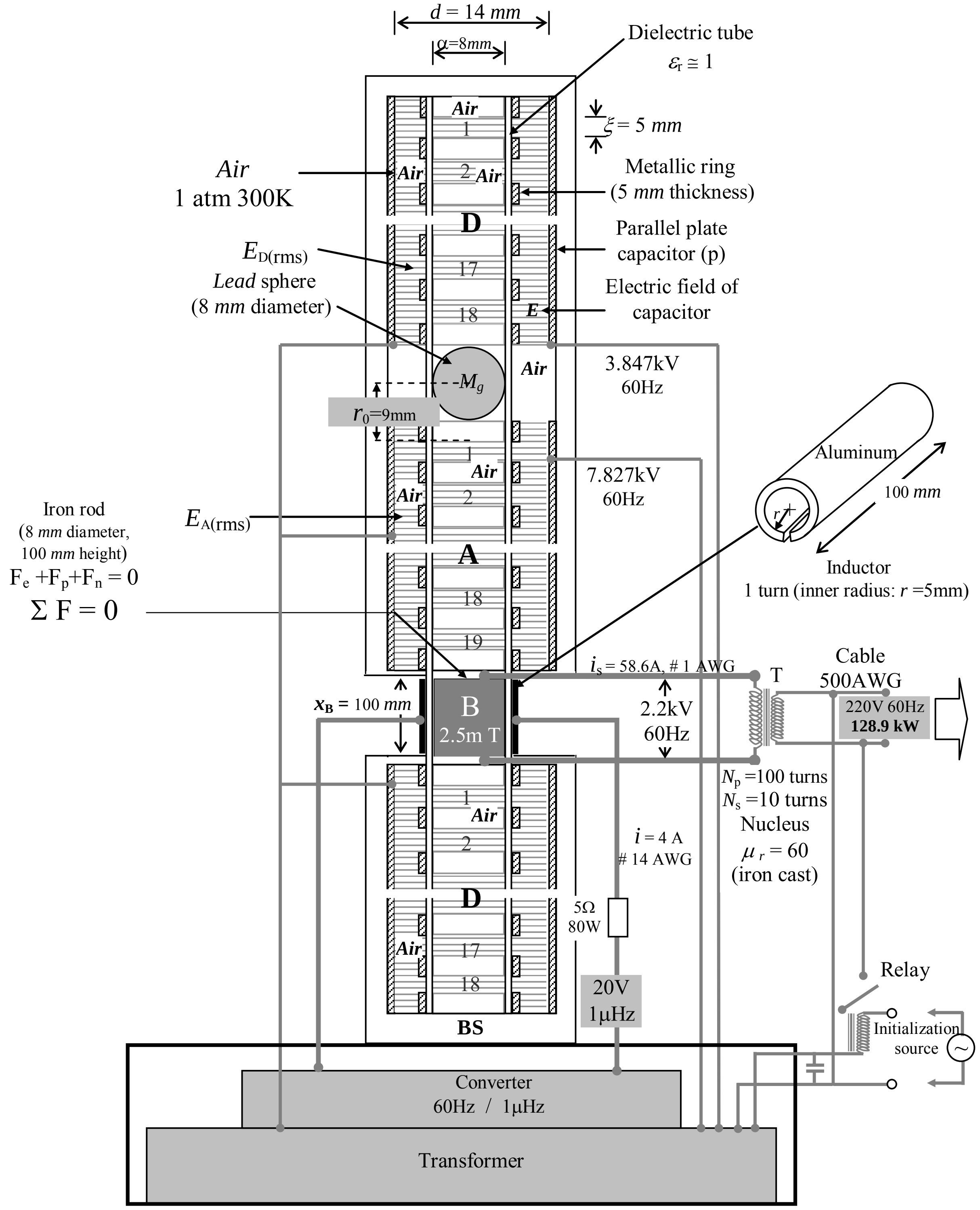

Fig. 3 – A *Gravitational EMF Source* (Developed from a process *patented* in July, 31 2008, PI0805046-5)

The electrical conductivity of air, *inside the dielectric tube*, is equal to the electrical conductivity of Earth's atmosphere near the land, whose average value is $\sigma_{air}\cong1\times10^{-14}S/m$ [8]. This value is of fundamental importance in order to obtain the convenient values of the electrical current $i$ and the value of $\chi$ and $\chi_d$, which are given by Eq. (21), i.e.,

$$\chi=\left\{1-2\left[\sqrt{1+\frac{n_{r(air)}^2 n_{air}^6 S_\alpha^2 S_m^4 \phi_m^4 \sigma_{air} E_{(A)rms}^4}{16\pi\mu_0\rho_{air}^2 c^4 f^3}}-1\right]\right\}=$$
$$=\left\{1-2\left[\sqrt{1+2.165\times10^{-22}E_{(A)rms}^4}-1\right]\right\} \quad (24)$$

$$\chi_d=\left\{1-2\left[\sqrt{1+\frac{n_{r(air)}^2 n_{air}^6 S_\alpha^2 S_m^4 \phi_m^4 \sigma_{air} E_{(D)rms}^4}{16\pi\mu_0\rho_{air}^2 c^4 f^3}}-1\right]\right\}=$$
$$=\left\{1-2\left[\sqrt{1+2.165\times10^{-22}E_{(D)rms}^4}-1\right]\right\} \quad (25)$$

where $n_{r(air)}=\sqrt{\varepsilon_r\mu_r}\cong1$, since $(\sigma<<\omega\varepsilon)$; $n_{air}=5.16\times10^{25}atoms/m^3$, $\phi_m=1.55\times10^{-10}m$, $S_m=\pi\phi_m^2/4=1.88\times10^{-20}m^2$ and $f=60Hz$. Since $E_{(A)rms}=5.591\times10^5V/m$ and $E_{(D)rms}=2.748\times10^5V/m$, we get

$$\chi=-6.4138 \quad (26)$$

and

$$\chi_d\cong10^{-2} \quad (27)$$

Note that there is a *uniform magnetic field*, $B$, through the *Iron rod*. Then, the *gravitational forces* due to the gravitational mass of the sphere $(M_{gs})$ acting on *electrons* $(F_e)$, *protons* $(F_p)$ and *neutrons* $(F_p)$ of the Iron rod, are respectively expressed by the following relations

$$F_e=m_{ge}a_e=\chi_{Be}m_e\left(\chi^7G\frac{M_{gs}}{r_0^2}\right) \quad (28)$$

$$F_p=m_{gp}a_p=\chi_{Bp}m_p\left(\chi^7G\frac{M_{gs}}{r_0^2}\right) \quad (29)$$

$$F_n=m_{gn}a_n=\chi_{Bn}m_n\left(\chi^7G\frac{M_{gs}}{r_0^2}\right) \quad (30)$$

The factors $\chi_B$ are due to the electrons, protons and neutrons are inside the *magnetic field B*.

*In order to make null the resultant of these forces in the Iron* (and *also in the sphere*) we must have $F_e=F_p+F_n$, i.e.,

$$m_e\chi_{Be}=m_p\chi_{Bp}+m_n\chi_{Bn} \quad (31)$$

It is important to note that *the set with* 19 GCC (*A*) *cannot be turned on before the magnetic field B is on.* Because the gravitational accelerations on the *Iron* rod and Lead sphere will be enormous $\left(\chi^{19}GM_{gs}/r_0^2\cong5.4\times10^6m/s^2\right)$, and will explode the device.

The force $F_e$ is the electromotive force (EMF), which produces the electrical current. Here, this force has *gravitational* nature. The corresponding force of *electrical* nature is $F_e=eE$. Thus, we can write that

$$m_{ge}a_e=eE \quad (32)$$

The electrons inside the *Iron* rod (See Fig. 3) are subjected to the gravity acceleration produced by the sphere, and increased by the 19 GCC in the region (A). The result is

$$a_e=\chi^{19}g_s=\chi^{19}G\frac{M_{gs}}{r_0^2} \quad (33)$$

Comparing Eq. (32) with Eq.(33), we obtain

$$E=\left(\frac{m_{ge}}{e}\right)\chi^{19}G\frac{M_{gs}}{r_0^2} \quad (34)$$

The electron mobility, $\mu_e$, considering various scattering mechanisms can be obtained by solving the Boltzmann equation in the relaxation time approximation. The result is [9]

$$\mu_e = \frac{e\langle\tau\rangle}{m_{ge}} \quad (35)$$

where $\langle\tau\rangle$ is the average relaxation time over the electron energies and $m_{ge}$ is the gravitational mass of electron, which is the effective mass of electron.

Since $\langle\tau\rangle$ can be expressed by $\langle\tau\rangle = m_{ge}\sigma/ne^2$ [10], then Eq. (35) can be written as follows

$$\mu_e = \frac{\sigma}{ne} \quad (36)$$

Thus, the *drift velocity* will be expressed by

$$v_d = \mu_e E = \frac{\sigma}{ne}\left(\frac{m_{ge}}{e}\right)\chi^{19} G\frac{M_{gs}}{r_0^2} \quad (37)$$

and the *electrical current density* expressed by

$$j_e = \rho_{qe} v_d = \sigma_{iron}\left(\frac{m_{ge}}{e}\right)\chi^{19} G\frac{M_{gs}}{r_0^2} \quad (38)$$

where $\rho_{qe} = ne$, and $m_{ge} = \chi_{Be} m_e$. Therefore, Eq. (38) reduces to

$$j_e = \sigma_{iron}\left(\frac{m_e}{e}\right)\chi_{Be}\chi^{19} G\frac{M_{gs}}{r_0^2} \quad (39)$$

In order to calculate the expressions of $\chi_{Be}$, $\chi_{Bp}$ and $\chi_{Bn}$ we start from Eq. (17), for the particular case of *single electron* in the region subjected to the magnetic field *B* (*Iron* rod). In this case, we must substitute $n_{r(l)}$ by $n_{riron} = \left(\mu_{r(iron)}\sigma_{iron}/4\pi\varepsilon_0 f\right)^{\frac{1}{2}}$; $n_l$ by $1/V_e = 1/\frac{4}{3}\pi r_e^3$ ($r_e$ is the electrons radius), $S_f$ by $(SSA_e)\rho_e V_e$ ($SSA_e$ is the *specific surface area* for electrons in this case: $SSA_e = \frac{1}{2}A_e/m_e = \frac{1}{2}A_e/\rho_e V_e = 2\pi r_e^2/\rho_e V_e$), $S_m$ by $S_e = \pi r_e^2$, $\xi$ by $\phi_m = 2r_e$ and $m_{i0(l)}$ by $m_e$. The result is

$$\chi_{Be} = \left\{1 - 2\left[\sqrt{1 + \frac{45.56\pi^2 r_e^4 n_{riron}^2 E^4}{\mu_0^2 m_e^2 c^4 f^4 \lambda^2}} - 1\right]\right\} \quad (40)$$

Electrodynamics tells us that $E_{rms} = vB_{rms} = \left(c/n_{r(iron)}\right)B_{rms}$, and Eq. (19) gives $\lambda = \lambda_{mod} = \left(4\pi/\mu_{iron}\sigma_{iron} f\right)$. Substitution of these expressions into Eq. (40) yields

$$\chi_{Be} = \left\{1 - 2\left[\sqrt{1 + \frac{45.56\pi^2 r_e^4 B_{rms}^4}{\mu_0^2 m_e^2 c^2 f^2}} - 1\right]\right\} \quad (41)$$

Similarly, in the case of *proton* and *neutron* we can write that

$$\chi_{Bp} = \left\{1 - 2\left[\sqrt{1 + \frac{45.56\pi^2 r_p^4 B_{rms}^4}{\mu_0^2 m_p^2 c^2 f^2}} - 1\right]\right\} \quad (42)$$

$$\chi_{Bn} = \left\{1 - 2\left[\sqrt{1 + \frac{45.56\pi^2 r_n^4 B_{rms}^4}{\mu_0^2 m_n^2 c^2 f^2}} - 1\right]\right\} \quad (43)$$

The radius of *free* electron is $r_e = 6.87 \times 10^{-14} m$ (See *Appendix* A) and the radius of *protons inside the atoms* (nuclei) is $r_p = 1.2 \times 10^{-15} m$, $r_n \cong r_p$, then we obtain from Eqs. (41) (42) and (43) the following expressions:

$$\chi_{Be} = \left\{1 - 2\left[\sqrt{1 + 8.49 \times 10^4 \frac{B_{rms}^4}{f^2}} - 1\right]\right\} \quad (44)$$

$$\chi_{Bn} \cong \chi_{Bp} = \left\{1 - 2\left[\sqrt{1 + 2.35 \times 10^{-9} \frac{B_{rms}^4}{f^2}} - 1\right]\right\} \quad (45)$$

Then, from Eq. (31) it follows that

$$m_e \chi_{Be} \cong 2m_p \chi_{Bp} \quad (46)$$

Substitution of Eqs. (44) and (45) into Eq. (46) gives

$$\frac{\left\{1 - 2\left[\sqrt{1 + 8.49 \times 10^4 \frac{B_{rms}^4}{f^2}} - 1\right]\right\}}{\left\{1 - 2\left[\sqrt{1 + 2.35 \times 10^{-9} \frac{B_{rms}^4}{f^2}} - 1\right]\right\}} = 3666.3 \quad (47)$$

For $f = 1\mu Hz$, we get

$$\frac{\left\{1 - 2\left[\sqrt{1 + 8.49 \times 10^6 B_{rms}^4} - 1\right]\right\}}{\left\{1 - 2\left[\sqrt{1 + 2.35 \times 10^{-7} B_{rms}^4} - 1\right]\right\}} = 3666.3 \quad (48)$$

whence we obtain

$$B_{rms} = 2.5m\ T \quad (49)$$

Consequently, Eq. (44) and (45) yields

$$\chi_{Be} = -3666.3 \quad (50)$$

and

$$\chi_{Bn} \cong \chi_{Bp} \cong 0.999 \qquad (51)$$

In order to the forces $F_e$ and $F_p$ have *contrary direction* (such as occurs in the case, in which the nature of the electromotive force is electrical) we must have $\chi_{Be} < 0$ and $\chi_{Bn} \cong \chi_{Bp} > 0$ (See equations (28) (29) and (30)), i.e.,

$$\left\{1-2\left[\sqrt{1+8.49\times10^{4}\frac{B_{rms}^4}{f^2}}-1\right]\right\}<0 \qquad (52)$$

and

$$\left\{1-2\left[\sqrt{1+2.35\times10^{9}\frac{B_{rms}^4}{f^2}}-1\right]\right\}>0 \qquad (53)$$

This means that we must have

$$0.06\sqrt{f} < B_{rms} < 151.86\sqrt{f} \qquad (54)$$

In the case of $f = 1\mu Hz = 10^{-6}\,Hz$ the result is

$$6.5\times10^{-5}T < B_{rms} < 0.151T \qquad (55)$$

Note the cylindrical format (1turn, $r = 5mm$) of the inductor (Figs. 3 and 6). By using only 1 turn it is possible to eliminate the *capacitive effect* between the turns. This is highly relevant in this case because the extremely-low frequency $f = 1\mu Hz$ would strongly increase the capacitive reactance ($X_C$) associated to the inductor. When a current $i$ passes through this inductor, the value of $B$ inside the *Iron rod* is given by $B = \mu_r\,\mu_0 i/x_B$ where $x_B = 100mm$ is inductor's length and $\mu_r = 4000$ (very pure Iron). However, the effective permeability is defined as $\mu_{r(eff)} = \mu_r / 1 + (\mu_r - 1)N_m$, where $N_m$ is the average *demagnetizing factor* [11]. Since the iron rod has 5mm diameter and 100mm height, then we obtain the factor $\gamma = 100mm/5mm = 20$ which gives $N_m = 0.02$ (See table V[12]). Therefore, we obtain $\mu_{r(eff)} = 49.4$. Thus, for $B_{rms} = 2.5mT$ (See Eq. (49)) ,i.e., $B = \mu_{r(eff)}\,\mu_0 i/x_B = 2.5mT$, the value of $i$ must be $i = 4A$. Then, the resistor in Fig.3. must have $20V/4A = 5\Omega$. The dissipated power is 80W.

Let us now calculate the *current density through the Iron* rod (Fig. 3). According to Eq. (39) we have

$$j_e = \sigma_{iron}\left(\frac{m_e}{e}\right)\chi_{Be}\chi^{19}G\frac{M_{gs}}{r_0^2}$$

Since $\sigma_{iron} = 1.03\times10^{7}\,S/m$, $\chi = -6.4138$, $\chi_{Be} = -3666.3$, $M_{gs} = 3.04\times10^{-3}\,kg$ and $r_0 = 9mm$, we obtain

$$j_e = 1.164\times10^{6}\,A/m^2$$

Given that $S_\alpha = \pi\alpha^2/4 = 5.03\times10^{-5}\,m^2$ we get

$$i_{source} = j_e S_\alpha \cong 58.6\ A$$

The resistance of the Iron rod is

$$R_{source} = \left(\frac{x_B}{\sigma_{iron} S_\alpha}\right) = 1.93\times10^{-4}\,\Omega$$

Thus, the *dissipated power* by the Iron rod is

$$P_d = R_{source} i_{source}^2 \cong 0.66W \qquad (56)$$

Note that this *Gravitational EMF source* is a *Current Source*. As we known, *a Current Source is a device that keeps invariable the electric current between its terminals*. So, if the source is connected to an external load, and the resistance of the load varies, then the own source will increase or decrease its output voltage in order to maintain invariable the value of the current in the circuit.

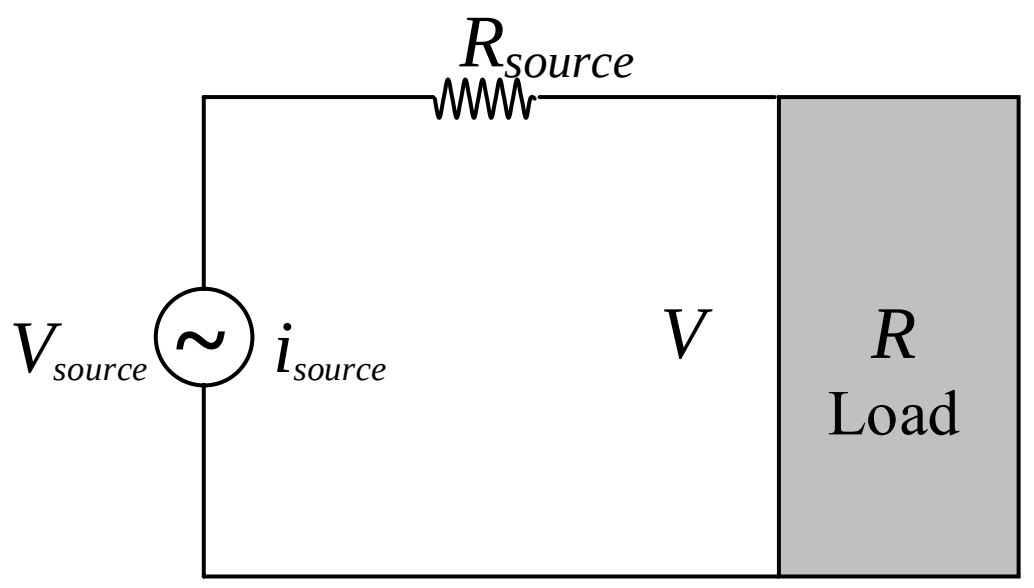

Fig. 4 – Current Source

Based on Kirchhoff's laws we can express the electric voltage between the terminals of the Current Source, $V_s$, by means of the following relation (See Fig.4):

$$V_{source} = R_{source}\, i_{source} + V$$

where $V$ is the voltage applied on the charge.

The transformer $T$ connected to Gravitational EMF Source (See Fig. 3) is designed* to make the voltage $V = 2.2kV@60Hz$. Since $R_{source} i_{source} << V$, then we can write that $V_{source} \cong V$ . Thus, in the primary circuit, the voltage is $V_p = V_{source} \cong V = 2.2kV$ and the current is $i_p = i_{source} = 58.6A$; the winding *turns ratio* is $N_p / N_s = 10$; thus, in the secondary circuit the output voltage is $V_s = 220V@60Hz$ and the current is $i_s = 586A$. Consequently, the source output power is

$$P = V_s i_s = 128.9kW$$

Note that, in order to *initializing* the Gravitational EMF Source, is used an external source, which is removed after the initialization of the Gravitational EMF Source.

Now it will be shown that this Gravitational EMF source can be *miniaturized*.

We start making $x_B = 10mm$ and $\xi = 0.5mm; \alpha = 2mm$, $d_A = 8mm$ , $d_D = 16mm$ and $r_0 = 1.5mm$ (See Fig. 5). The sphere with 2mm diameter is now of *Tungsten carbide* (W+Cobalt) with $15{,}630 kg/m^3$ density. Then $M_{gs} = 6.54 \times 10^{-5} kg$ and $S_\alpha = 3.14 \times 10^{-6} m^2$ . Thus, for $f = 1\mu Hz$ Eq.(24) gives

$$\chi = \left\{1 - 2\left[\sqrt{1 + 0.1822 E^4_{(A)rms}} - 1\right]\right\} \qquad (57)$$

For $V_{A(rms)} = 26mV$ and $d_A = 8mm$ we get $E_{(A)rms} = 3.250V/m$, and Eq. (57) yields

$$\chi = -6.236$$

For $V_{D(rms)} = 26mV$ and $d_D = 16mm$ we get $E_{(D)rms} = 1.625V/m$ and $\chi_d = -0.01$.

Since $\sigma_{iron} = 1.03 \times 10^7 S/m$ and $\chi_{Be} = -3666.3$, then the value of $j_e$ is

$$j_e = \sigma_{iron}\left(\frac{m_e}{e}\right)\chi_{Be}\chi^{19} G \frac{M_{gs}}{r_0^2} =$$

$$= 5.29 \times 10^5 A/m^2$$

and

$$i_{source} = j_e S_\alpha \cong 1.66 \ A$$

The resistance of the iron rod is given by

$$R_{source} = \left(\frac{x_B}{\sigma_{iron} S_\alpha}\right) = 3.1 \times 10^{-4} \Omega$$

Thus, the *dissipated power* by the Iron rod is

$$P_d = R_{source} i^2_{source} \cong 0.9mW$$

In the case of the *miniaturized source*, the iron rod has 2mm diameter and 10mm height, then we obtain the factor $\gamma = 10mm/2mm = 5$ which gives $N_m = 0.06$ (See table V[12]). Therefore, we obtain $\mu_{r(eff)} = 16.6$.

Since $V_s = V_{A(rms)} = V_{D(rms)} = 26mV$ and the resistance of the resistor $R_1$ is $21.6m\Omega / 31mW$ (See Fig.5), then the current from the *first* source is $i = 1.2A$. Thus, we get $B = \mu_{r(eff)} \mu_0 i / x_B = 2.5mT$.

Since the current through the *second source* is $i_{source} = 1.66A$, and, if the voltage required by the charge, is $V = 3.7V$ (usual lithium batteries' voltage), then the source voltage is given by

$$V_{source} = R_{source}\, i_{source} + V \cong 3.7V$$

Consequently, the miniaturized source can provide the power:

$$P = V_{source}\, i_{sorce} = (3.7V)1.66A \cong 6.1W$$

This is the magnitude of the power of lithium batteries used in mobiles. Note that the *miniaturized source of Gravitational EMF* does *not need to be recharged* and it occupies a volume (8mm x 70mm x 80mm. See Fig.6) similar to the volume of the mobile batteries. In addition, note that the dimensions of this miniaturized source can be further reduced (possibly down to a few *millimeters* or less).

---

* The impedances are respectively,

$$Z_p \cong 2\pi f L_p = 2\pi f\left(\mu_r \mu_0 N_p^2 A_p / l_p\right) = 14.36\Omega$$

$$Z_s \cong 2\pi f L_s = 2\pi f\left(\mu_r \mu_0 N_s^2 A_s / l_s\right) = 6.154 \times 10^{-4}\Omega$$

$$Z_{p(total)} = Z_p + Z_{reflected} = Z_p + (N_p/N_s) Z_s = 37.56\Omega$$

where $\mu_r = 60$ (iron cast), $N_p = 100, N_s = 10$, $l_p = l_s = 0.18m, \phi_p = 107.6mm, \phi_s = 136.8mm$, $A_p = 0.0091 m^2$, $A_s = 0.0147\ m^2$.

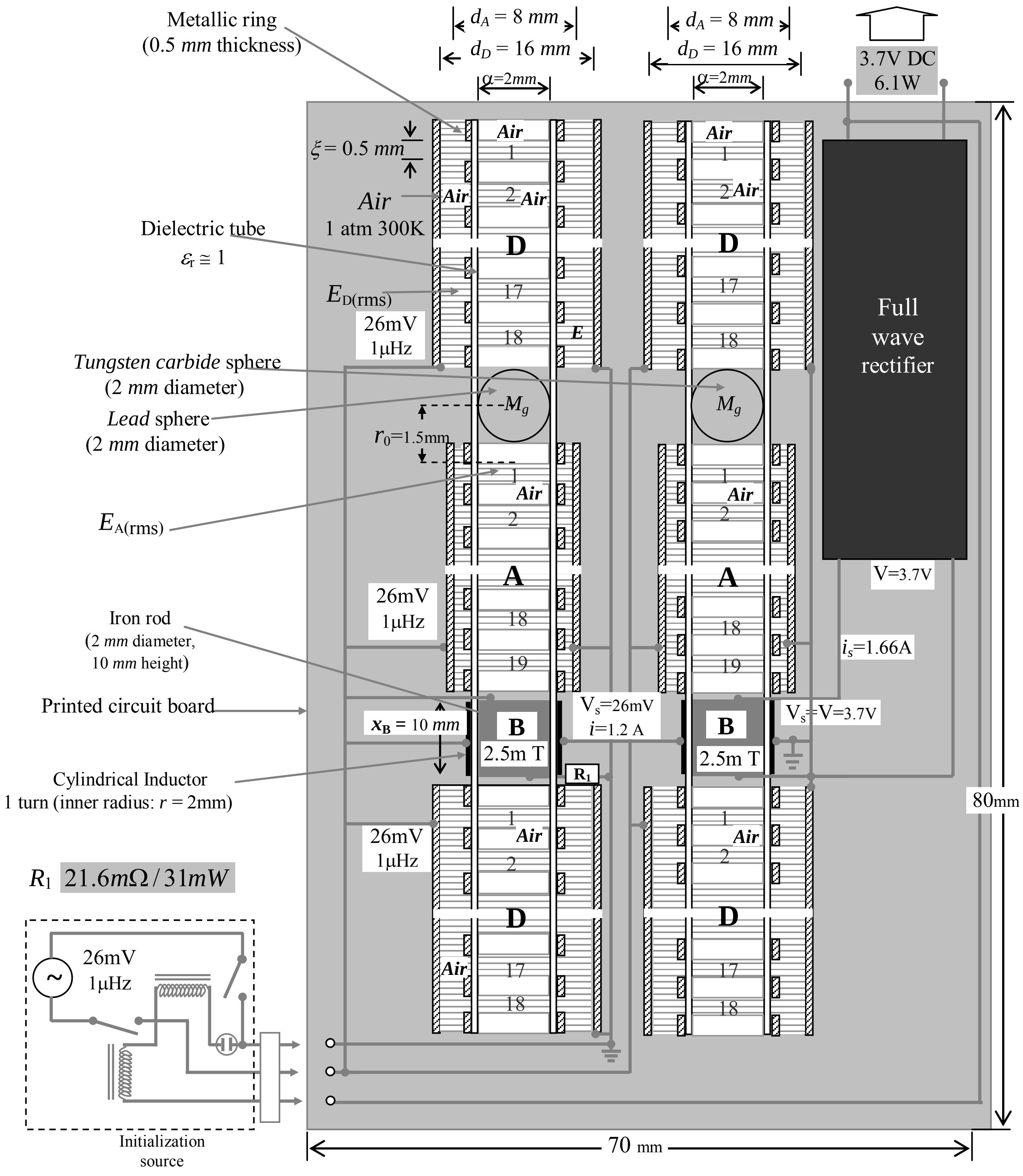

Fig. 5 – A *Miniaturized Source* of *Gravitational EMF*

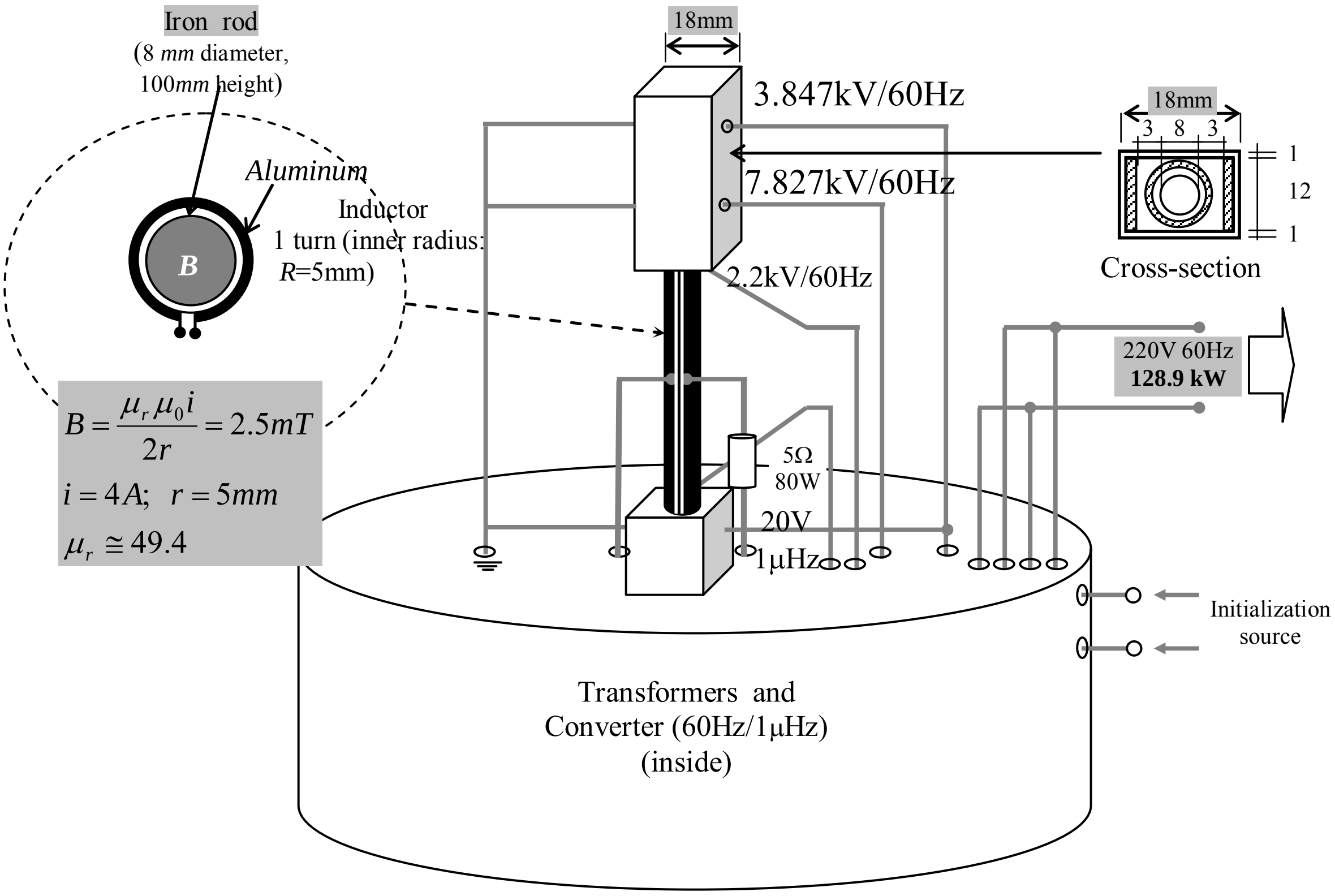

**High-power Source**

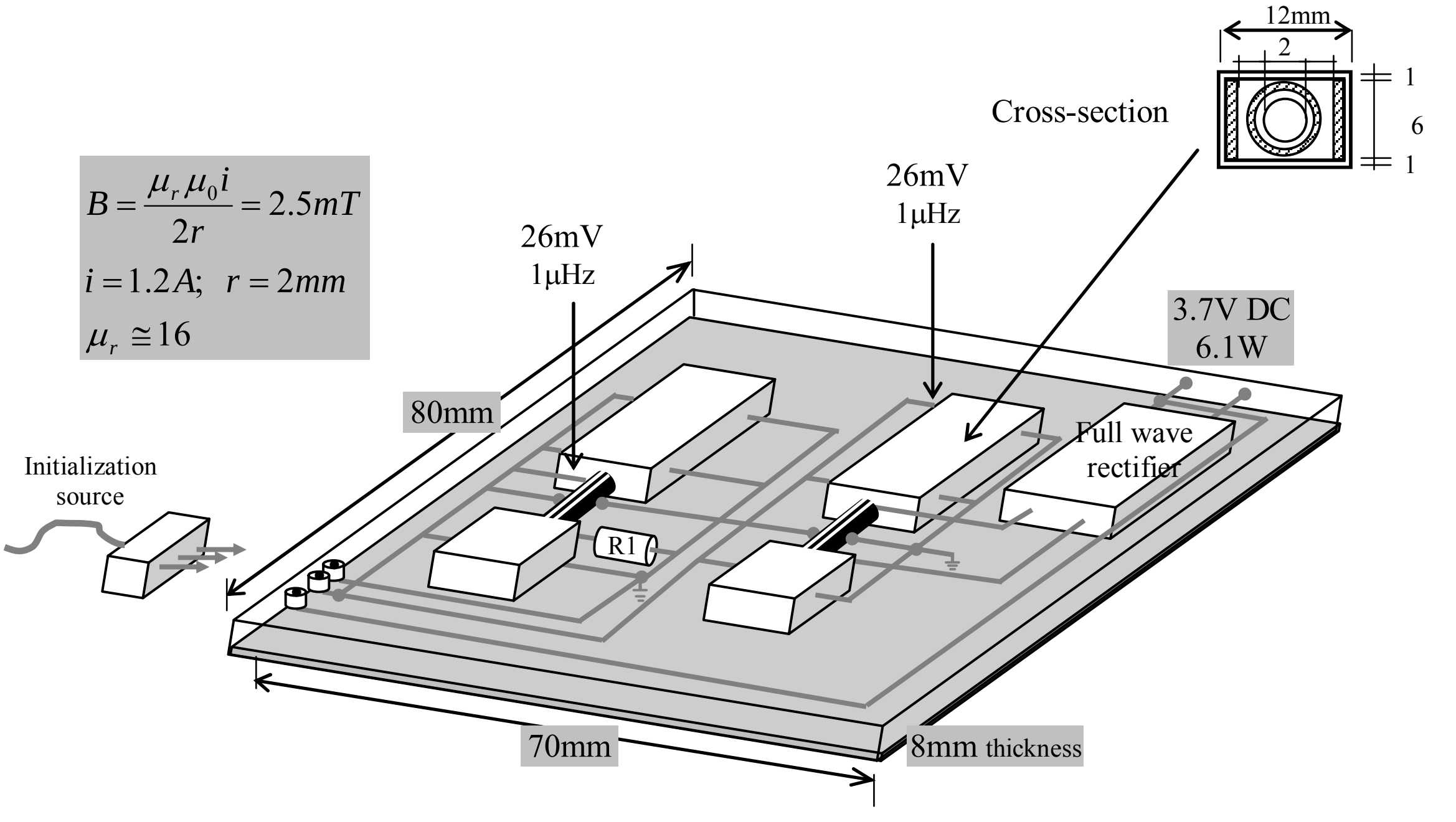

**Low-power Source**

Fig. 6 – Schematic Diagram in 3D of the *Gravitational EMF Sources*

## **Appendix A**: *The "Geometrical Radii" of Electron and Proton*

It is known that the frequency of oscillation of a simple spring oscillator is

$$f = \frac{1}{2\pi}\sqrt{\frac{K}{m}} \qquad (A1)$$

where $m$ is the inertial mass attached to the spring and $K$ is the spring constant (in N·m$^{-1}$). In this case, the restoring *force* exerted by the spring *is linear* and given by

$$F = -Kx \qquad (A2)$$

where $x$ is the displacement from the equilibrium position.

Now, consider the gravitational force: For example, above the surface of the Earth, the force follows the familiar Newtonian function, i.e., $F = -GM_{g\oplus}m_g/r^2$, where $M_{g\oplus}$ the mass of Earth is, $m_g$ is the gravitational mass of a particle and $r$ is the distance between the centers. *Below* Earth's surface the *force is linear* and given by

$$F = -\frac{GM_{g\oplus}m_g}{R_\oplus^3}r \qquad (A3)$$

where $R_\oplus$ is the radius of Earth.

By comparing (A3) with (A2) we obtain

$$\frac{K}{m_g} = \frac{K}{\chi\ m} = \frac{GM_{g\oplus}}{R_\oplus^3}\left(\frac{r}{x}\right) \qquad (A4)$$

Making $x = r = R_\oplus$, and substituting (A4) into (A1) gives

$$f = \frac{1}{2\pi}\sqrt{\frac{GM_{g\oplus}\chi}{R_\oplus^3}} \qquad (A5)$$

In the case of an *electron* and a *positron*, we substitute $M_{g\oplus}$ by $m_{ge}$, $\chi$ by $\chi_e$ and $R_\oplus$ by $R_e$, where $R_e$ is the radius of electron (or positron). Thus, Eq. (A5) becomes

$$f = \frac{1}{2\pi}\sqrt{\frac{Gm_{ge}\chi_e}{R_e^3}} \qquad (A6)$$

The value of $\chi_e$ varies with the density of energy [1]. When the electron and the positron are distant from each other and the local density of energy is small, the value of $\chi_e$ becomes very close to 1. However, *when the electron and the positron are penetrating one another*, the energy densities in each particle become very strong due to the proximity of their electrical charges $e$ and, consequently, the value of $\chi_e$ strongly increases. In order to calculate the value of $\chi_e$ under these conditions ($x = r = R_e$), we start from the expression of correlation between *electric charge* $q$ and *gravitational mass*, obtained in a previous work [1]:

$$q = \sqrt{4\pi\varepsilon_0 G}\ m_{g(imaginary)}\ i \qquad (A7)$$

where $m_{g(imaginary)}$ is the *imaginary* gravitational mass, and $i = \sqrt{-1}$.

In the case of *electron*, Eq. (A7) gives

$$\begin{aligned} q_e &= \sqrt{4\pi\varepsilon_0 G}\ m_{ge(imaginary)}\ i = \\ &= \sqrt{4\pi\varepsilon_0 G}\left(\chi_e m_{i0e(imaginary)} i\right) = \\ &= \sqrt{4\pi\varepsilon_0 G}\left(-\chi_e \tfrac{2}{\sqrt{3}} m_{i0e(real)} i^2\right) = \\ &= \sqrt{4\pi\varepsilon_0 G}\left(\tfrac{2}{\sqrt{3}}\chi_e m_{i0e(real)}\right) = -1.6\times10^{-19}C \end{aligned} \qquad (A8)$$

where we obtain

$$\chi_e = -1.8\times10^{21} \qquad (A9)$$

This is therefore, the value of $\chi_e$ increased by the strong density of energy produced by the electrical charges $e$ of the two particles, under previously mentioned conditions.

Given that $m_{ge} = \chi_e m_{i0e}$, Eq. (A6) yields

$$f = \frac{1}{2\pi}\sqrt{\frac{G\chi_e^2 m_{i0e}}{R_e^3}} \qquad (A10)$$

From Quantum Mechanics, we know that

$$hf = m_{i0}c^2 \qquad (A11)$$

where $h$ is the Planck's constant. Thus, in the case of $m_{i0} = m_{i0e}$ we get

$$f = \frac{m_{i0e}c^2}{h} \qquad (A12)$$

By comparing (A10) and (A12) we conclude that

$$\frac{m_{i0e}c^2}{h} = \frac{1}{2\pi}\sqrt{\frac{G\chi_e^2 m_{i0e}}{R_e^3}} \qquad (A13)$$

Isolating the radius $R_e$, we get:

$$R_e = \left(\frac{G}{m_{i0e}}\right)^{\frac{1}{3}}\left(\frac{\chi_e h}{2\pi\ c^2}\right)^{\frac{2}{3}} = 6.87\times10^{-14}m \quad (A14)$$

Compare this value with the *Compton sized electron*, which predicts $R_e = 3.86\times10^{-13}m$ and also with standardized result recently obtained of $R_e = 4-7\times10^{-13}m$ [13].

In the case of *proton*, we have

$$q_p = \sqrt{4\pi\varepsilon_0 G}\ m_{gp(imaginary)}\, i =$$
$$= \sqrt{4\pi\varepsilon_0 G}\left(\chi_p m_{i0p(imaginary)} i\right) =$$
$$= \sqrt{4\pi\varepsilon_0 G}\left(-\chi_p \tfrac{2}{\sqrt{3}}\, m_{i0p(real)} i^2\right) =$$
$$= \sqrt{4\pi\varepsilon_0 G}\left(\tfrac{2}{\sqrt{3}}\,\chi_p m_{i0p(real)}\right) = -1.6\times10^{-19}C \quad (A15)$$

where we obtain

$$\chi_p = -9.7\times10^{17} \qquad (A16)$$

Thus, the result is

$$R_p = \left(\frac{G}{m_{i0p}}\right)^{\frac{1}{3}}\left(\frac{\chi_p h}{2\pi\ c^2}\right)^{\frac{2}{3}} = 3.72\times10^{-17}m \quad (A17)$$

Note that these radii, given by Equations $(A14)$ and $(A17)$, are the radii of *free* electrons and *free* protons (when the particle and antiparticle (in isolation) penetrate themselves mutually).

Inside the atoms (nuclei) the radius of protons is well-known. For example, protons, as the hydrogen nuclei, have a radius given by $R_p \cong 1.2\times10^{-15}m$ [14, 15]. The strong increase in respect to the value given by Eq. (A17) is due to the interaction with the electron of the atom.

# Superconducting State generated by Cooper Pairs bound by Intensified Gravitational Interaction

**Fran De Aquino**
Maranhao State University, Physics Department, S.Luis/MA, Brazil.

We show that by *intensifying the gravitational interaction* between electron pairs it is possible to produce pair binding energies on the order of $10^{-1}$eV, enough to keep electron's pairs (*Cooper Pairs*) at *ambient temperatures*. By means of this method, metals can be transformed into superconductors at ambient temperature.

**Key words:** Modified theories of gravity, Theories and models of superconducting state, superconducting materials, Nonconventional mechanisms.
PACS: 04.50.Kd, 74.20.-z, 74.70.-b, 74.20.Mn.

## 1. Introduction

A pair of weakly bound electrons in a superconductor is called *Cooper pair*; it was first described in 1956 by Leon Cooper [1]. As showed by Cooper, an attraction between electrons in a metal can cause a paired state of electrons to have a lower energy than the Fermi energy, which implies that the pair is *bound*. In conventional superconductors, this attraction is due to the electron–phonon interaction. *The Cooper pair state is responsible for superconductivity, as described in* the BCS theory developed by John Bardeen, John Schrieffer and Leon Cooper for which they shared the 1972 Nobel Prize [2].

In spite of Cooper pairing to be a quantum effect the reason for the pairing can be seen from a simplified classical explanation [3]. In order to understand how an attraction between two electrons can occur, it is necessary to consider the interaction with the positive ions lattice of the metal. Usually an electron in a metal behaves as a free particle. Its negative charge causes attraction between *the positive ions that make up the rigid lattice of the metal.* This attraction distorts the ion lattice, moving the ions slightly toward the electron, *increasing the positive charge density of the lattice in the local* (See gray glow in Fig.1 (a)). Then, another electron is attracted to the positive charge density (gray glow) created by the first electron distorting the lattice around itself. This attraction can overcome the electrons' repulsion due to their negative charge and create a binding between the two electrons (See Fig.1 (b)). *The electrons can then travel through the lattice as a single entity*, known as a *Cooper Pair* (See Fig.1 (c)). While conventional conduction is resisted by thermal vibrations within the lattice, Cooper Pairs carry the supercurrent relatively unresisted by thermal vibrations.

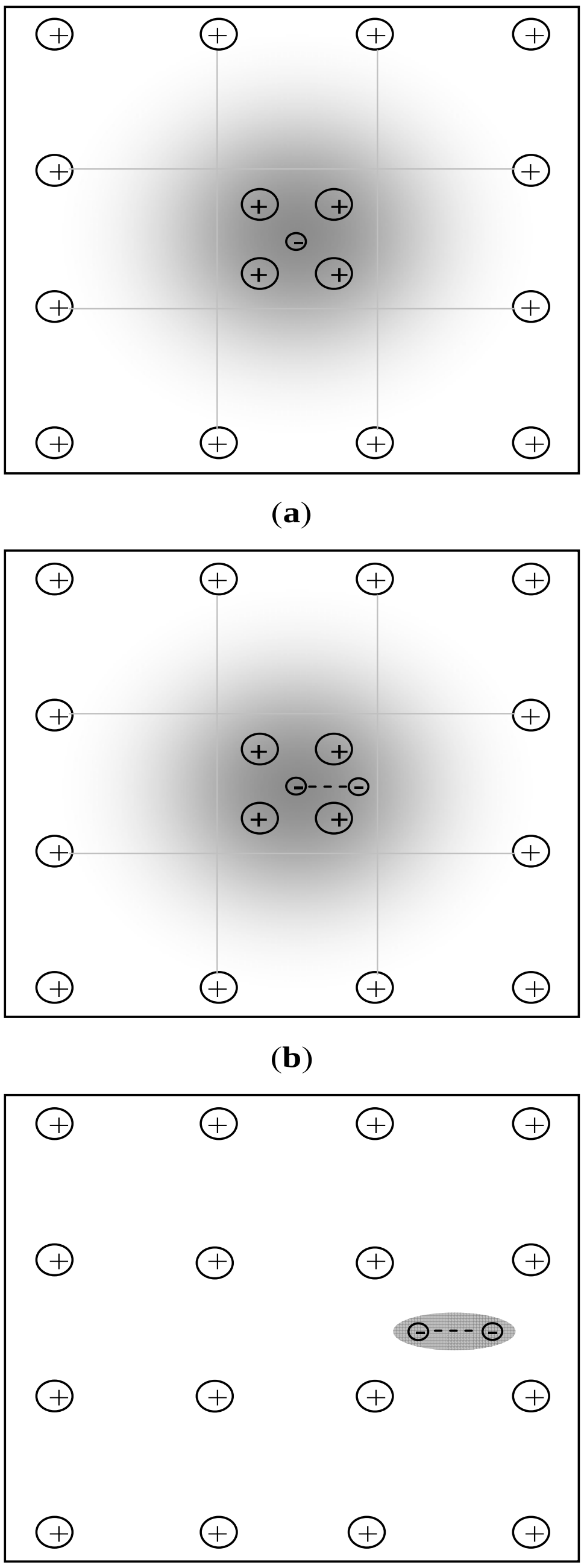

(a)

(b)

(c)

Fig. 1 – Cooper Pair Formation

The energy of the pairing interaction is quite weak, of the order of $10^{-3}$eV, and thermal energy can easily break the pairs. *So only at low temperatures, are a significant number of the electrons in a metal in Cooper pairs*.

Here is showed that, by *intensifying the gravitational interaction* * [4] between electrons pairs, it is possible to produce pair binding energies on the order of $10^{-1}$eV, enough to keep them paired at *ambient temperatures*. Thus, by this way, metals at ambient temperature can have a significant number of the electrons in Cooper pairs, transforming such metals in superconductors at ambient temperature.

## 2. Theory

The quantization of gravity showed that the *gravitational mass* $m_g$ and the *inertial mass* $m_i$ are correlated by means of the following factor [4]:

$$\chi = \frac{m_g}{m_{i0}} = \left\{1 - 2\left[\sqrt{1 + \left(\frac{\Delta p}{m_{i0}c}\right)^2} - 1\right]\right\} \quad (1)$$

where $m_{i0}$ is the *rest* inertial mass of the particle and $\Delta p$ is the variation in the particle's *kinetic momentum*; $c$ is the speed of light.

When $\Delta p$ is produced by the absorption of a photon with wavelength $\lambda$, it is expressed by $\Delta p = h/\lambda$. In this case, Eq. (1) becomes

$$\frac{m_g}{m_{i0}} = \left\{1 - 2\left[\sqrt{1 + \left(\frac{h/m_{i0}c}{\lambda}\right)^2} - 1\right]\right\} = \left\{1 - 2\left[\sqrt{1 + \left(\frac{\lambda_0}{\lambda}\right)^2} - 1\right]\right\} \quad (2)$$

where $\lambda_0 = h/m_{i0}c$ is the *DeBroglie wavelength* for the particle with *rest* inertial mass $m_{i0}$.

In general, the *momentum* variation $\Delta p$ is expressed by $\Delta p = F\Delta t$ where $F$ is the applied force during a time interval $\Delta t$. Note that there is no restriction concerning the *nature* of the force, i.e., it can be mechanical, electromagnetic, etc. For example, we can look on the *momentum* variation $\Delta p$ as due to absorption or emission of *electromagnetic energy* by the particle.

This means that, by means of electromagnetic fields, the *gravitational mass* can be decreased down to become negative and *increased* (*independently* of the inertial mass $m_i$). In this way, *the gravitational forces can be intensified*. Consequently, we can use, for example, oscillating magnetic fields in order to *intensify the gravitational interaction* between electrons pairs, in order to produce pair binding energies enough to keep them paired at *ambient temperatures*. We will show that the magnetic field used in this case must have extremely-low frequency (ELF).

From Electrodynamics we know that when an electromagnetic wave with frequency $f$ and velocity $c$ incides on a material with relative permittivity $\varepsilon_r$, relative magnetic permeability $\mu_r$ and electrical conductivity $\sigma$, its *velocity is reduced* to $v = c/n_r$ where $n_r$ is the index of refraction of the material, given by [5]

$$n_r = \frac{c}{v} = \sqrt{\frac{\varepsilon_r \mu_r}{2}\left(\sqrt{1 + (\sigma/\omega\varepsilon)^2} + 1\right)} \quad (3)$$

---

* De Aquino, F. (2008) *Process and Device for* Controlling the Locally the Gravitational Mass and the Gravity Acceleration, BR Patent Number: PI0805046-5, July 31, 2008.

If $\sigma >> \omega\varepsilon$, $\omega = 2\pi f$, Eq. (3) reduces to

$$n_r = \sqrt{\frac{\mu_r \sigma}{4\pi\varepsilon_0 f}} \qquad (4)$$

Thus, the wavelength of the incident radiation (See Fig. 2) becomes

$$\lambda_{\text{mod}} = \frac{v}{f} = \frac{c/f}{n_r} = \frac{\lambda}{n_r} = \sqrt{\frac{4\pi}{\mu f \sigma}} \qquad (5)$$

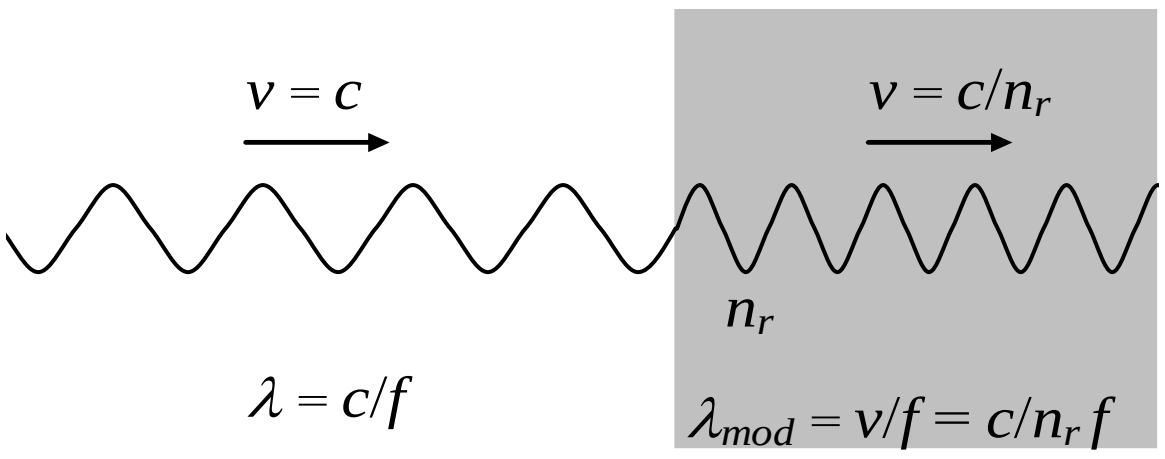

Fig. 2 – *Modified Electromagnetic Wave*. The wavelength of the electromagnetic wave can be strongly reduced, but its frequency remains the same.

If a lamina with thickness equal to $\xi$ contains $n$ atoms/m$^3$, then the number of atoms per area unit is $n\xi$. Thus, if the electromagnetic radiation with frequency $f$ incides on an area $S$ of the lamina it reaches $nS\xi$ atoms. If it incides on the *total area of the lamina*, $S_f$, then the total number of atoms reached by the radiation is $N = nS_f\xi$. The number of atoms per unit of volume, $n$, is given by

$$n = \frac{N_0 \rho}{A} \qquad (6)$$

where $N_0 = 6.02 \times 10^{26}\, atoms/kmole$ is the Avogadro's number; $\rho$ is the matter density of the lamina (in *kg/m*$^3$) and $A$ is the molar mass(*kg/kmole*).

When an electromagnetic wave incides on the lamina, it strikes $N_f$ front atoms, where $N_f \cong (n S_f)\phi_m$, $\phi_m$ is the "diameter" of the atom. Thus, the electromagnetic wave incides effectively on an area $S = N_f S_m$, where $S_m = \frac{1}{4}\pi\phi_m^2$ is the cross section area of one atom. After these collisions, it carries out $n_{collisions}$ with the other atoms (See Fig.3).

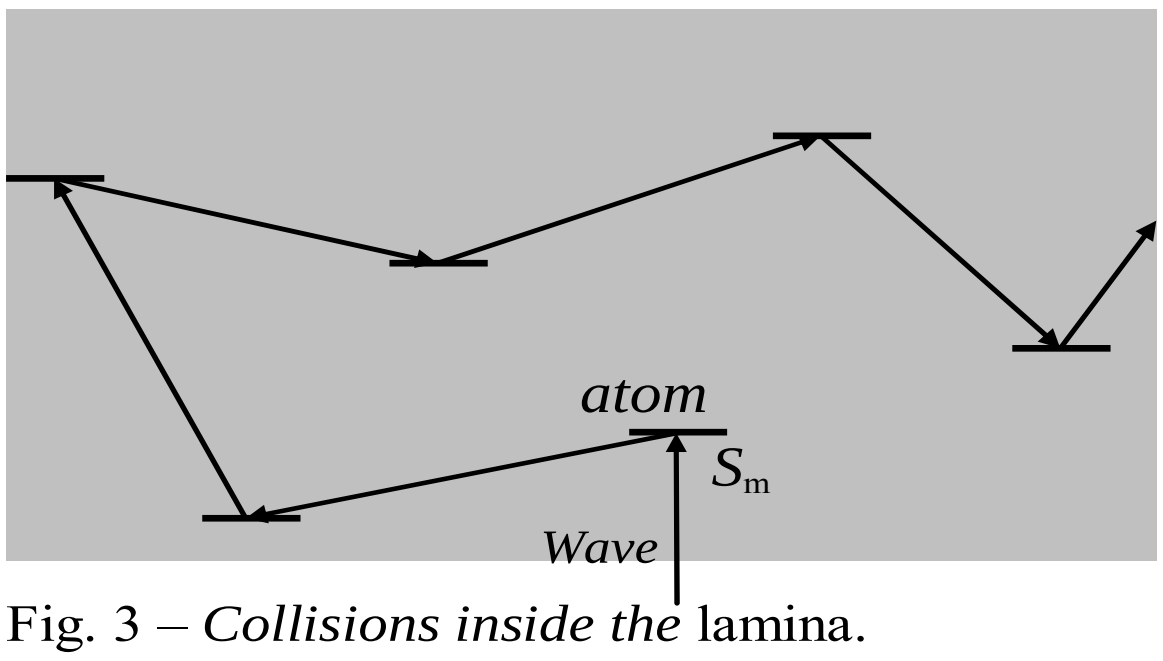

Fig. 3 – *Collisions inside the* lamina.

Thus, the total number of collisions in the volume $S\xi$ is

$$N_{collisions} = N_f + n_{collisions} = n_l S\phi_m + (n_l S\xi - n_m S\phi_m) = \\ = n_m S\xi \qquad (7)$$

The power density, $D$, of the radiation on the lamina can be expressed by

$$D = \frac{P}{S} = \frac{P}{N_f S_m} \qquad (8)$$

We can express the *total mean number of collisions in each atom*, $n_1$, by means of the following equation

$$n_1 = \frac{n_{total\ photons} N_{collisions}}{N} \qquad (9)$$

Since in each collision a *momentum* $h/\lambda$ is transferred to the atom, then the *total momentum* transferred to the lamina will be $\Delta p = (n_1 N) h/\lambda$. Therefore, in accordance with Eq. (1), we can write that

$$\frac{m_{g(l)}}{m_{i0(l)}} = \left\{1 - 2\left[\sqrt{1 + \left[(n_1 N)\frac{\lambda_0}{\lambda}\right]^2} - 1\right]\right\} = \\ = \left\{1 - 2\left[\sqrt{1 + \left[n_{total\ photons} N_{collisions} \frac{\lambda_0}{\lambda}\right]^2} - 1\right]\right\} \qquad (10)$$

Since Eq. (7) gives $N_{collisions} = n_l S\xi$ , we get

$$n_{total\ photons} N_{collisions} = \left(\frac{P}{hf^2}\right)(n_l S\xi) \qquad (11)$$

Substitution of Eq. (11) into Eq. (10) yields

$$\frac{m_{g(l)}}{m_{i0(l)}} = \left\{1 - 2\left[\sqrt{1 + \left[\left(\frac{P}{hf^2}\right)(n_l S\xi)\frac{\lambda_0}{\lambda}\right]^2} - 1\right]\right\} \qquad (12)$$

Substitution of $P$ given by Eq. (8) into Eq. (12) gives

$$\frac{m_{g(l)}}{m_{i0(l)}} = \left\{1 - 2\left[\sqrt{1 + \left[\left(\frac{N_f S_m D}{f^2}\right)\left(\frac{n_l S\xi}{m_{i0(l)}c}\right)\frac{1}{\lambda}\right]^2} - 1\right]\right\} \qquad (13)$$

Substitution of $N_f \cong (n_l S_f)\phi_m$ and $S = N_f S_m$ into Eq. (13) results

$$\frac{m_{g(l)}}{m_{i0(l)}} = \left\{1 - 2\left[\sqrt{1 + \left[\left(\frac{n_l^3 S_f^2 S_m^2 \phi_m^2 \xi D}{m_{i0(l)}cf^2}\right)\frac{1}{\lambda}\right]^2} - 1\right]\right\} \qquad (14)$$

where $m_{i0(l)} = \rho_{(l)} V_{(l)}$.

Now, considering that the lamina is inside an ELF electromagnetic field with $E$ and $B$, then we can write that [6]

$$D = \frac{n_{r(l)} E^2}{2\mu_0 c} \qquad (15)$$

Substitution of Eq. (15) into Eq. (14) gives

$$\frac{m_{g(l)}}{m_{i0(l)}} = \left\{1 - 2\left[\sqrt{1 + \left[\left(\frac{n_{r(l)} n_l^3 S_f^2 S_m^2 \phi_m^2 \xi E^2}{2\mu_0 m_{i0(l)} c^2 f^2}\right)\frac{1}{\lambda}\right]^2} - 1\right]\right\} \qquad (16)$$

Note that $E = E_m \sin \omega t$ .The average value for $E^2$ is equal to $\frac{1}{2}E_m^2$ because $E$ varies sinusoidaly ( $E_m$ is the maximum value for $E$ ). On the other hand, $E_{rms} = E_m/\sqrt{2}$ . Consequently we can replace $E^4$ for $E_{rms}^4$ . Thus, for $\lambda = \lambda_{mod}$ , the equation above can be rewritten as follows

$$\frac{m_{g(l)}}{m_{i0(l)}} = \left\{1 - 2\left[\sqrt{1 + \left[\left(\frac{n_{r(l)} n_l^3 S_f^2 S_m^2 \phi_m^2 \xi E_{rms}^2}{2\mu_0 m_{i0(l)} c^2 f^2}\right)\frac{1}{\lambda_{mod}}\right]^2} - 1\right]\right\} \qquad (17)$$

Electrodynamics tells us that $E_{rms} = vB_{rms} = (c/n_{r(l)})B_{rms}$ . Substitution of this expression into Eq. (17) gives

$$\chi = \frac{m_{g(l)}}{m_{i0(l)}} = \left\{1 - 2\left[\sqrt{1 + \frac{n_l^6 S_f^4 S_m^4 \phi_m^4 \xi^2 B_{rms}^4}{4\mu_0^2 m_{i0(l)}^2 f^4 \lambda_{mod}^2 n_{r(l)}^2}} - 1\right]\right\} \qquad (18)$$

Since $\lambda_{mod} = \lambda/n_{r(l)}$ then Eq. (18) can be rewritten in the following form

$$\chi = \frac{m_{g(l)}}{m_{i0(l)}} = \left\{1 - 2\left[\sqrt{1 + \frac{n_l^6 S_f^4 S_m^4 \phi_m^4 \xi^2 B_{rms}^4}{4\mu_0^2 m_{i0(l)}^2 c^2 f^2}} - 1\right]\right\} \qquad (19)$$

In order to calculate the expressions of $\chi_{Be}$ for the particular case of a *free electron* inside a conductor, subjected to an external magnetic field $B_{rms}$ with frequency $f$ , we must consider the interaction with the *positive ions that make up the rigid lattice of the metal*.

The negative charge of the free electron causes attraction between *the positive ions lattice of the metal*. This attraction distorts the ion lattice, moving the ions slightly toward the electron, *increasing the positive charge density of the lattice in the local* (See gray glow in Fig.1 (a)). Then, another electron is attracted to the positive charge density (gray glow) created by the first electron distorting the lattice around itself, *which produces a strong attraction upon the electron* deforming its surface as showed in Fig. 4. Under these circumstances, *the volume of the electron does not vary*, but its external surface is strongly increased, becomes equivalent to the external area of a sphere with radius $r_{xe} >> r_e$ ( $r_e$ is the radius of the *free* electron out of the ions “gage” showed in Fig. 1 (a)). Based on such conclusions, we substitute in Eq.(19) $n_l$ by $1/V_e = 1/\frac{4}{3}\pi r_e^3$, $S_f$ by $(SSA_e)\rho_e V_e$ ($SSA_e$ is the *specific surface area* for electrons in this case: $SSA_e = \frac{1}{2}A_e/m_e = \frac{1}{2}A_e/\rho_e V_e = 2\pi r_{xe}^2/\rho_e V_e$), $S_m$ by $S_e = \pi r_{xe}^2$, $\xi$ by $\phi_m = 2r_{xe}$ and $m_{i0(l)}$ by $m_e$. The result is

$$\chi_{Be} = \left\{1 - 2\left[\sqrt{1 + \frac{45.56\pi^2 r_{xe}^{22} B_{rms}^4}{c^2 \mu_0^2 m_e^2 r_e^{18} f^2}} - 1\right]\right\} \qquad (20)$$

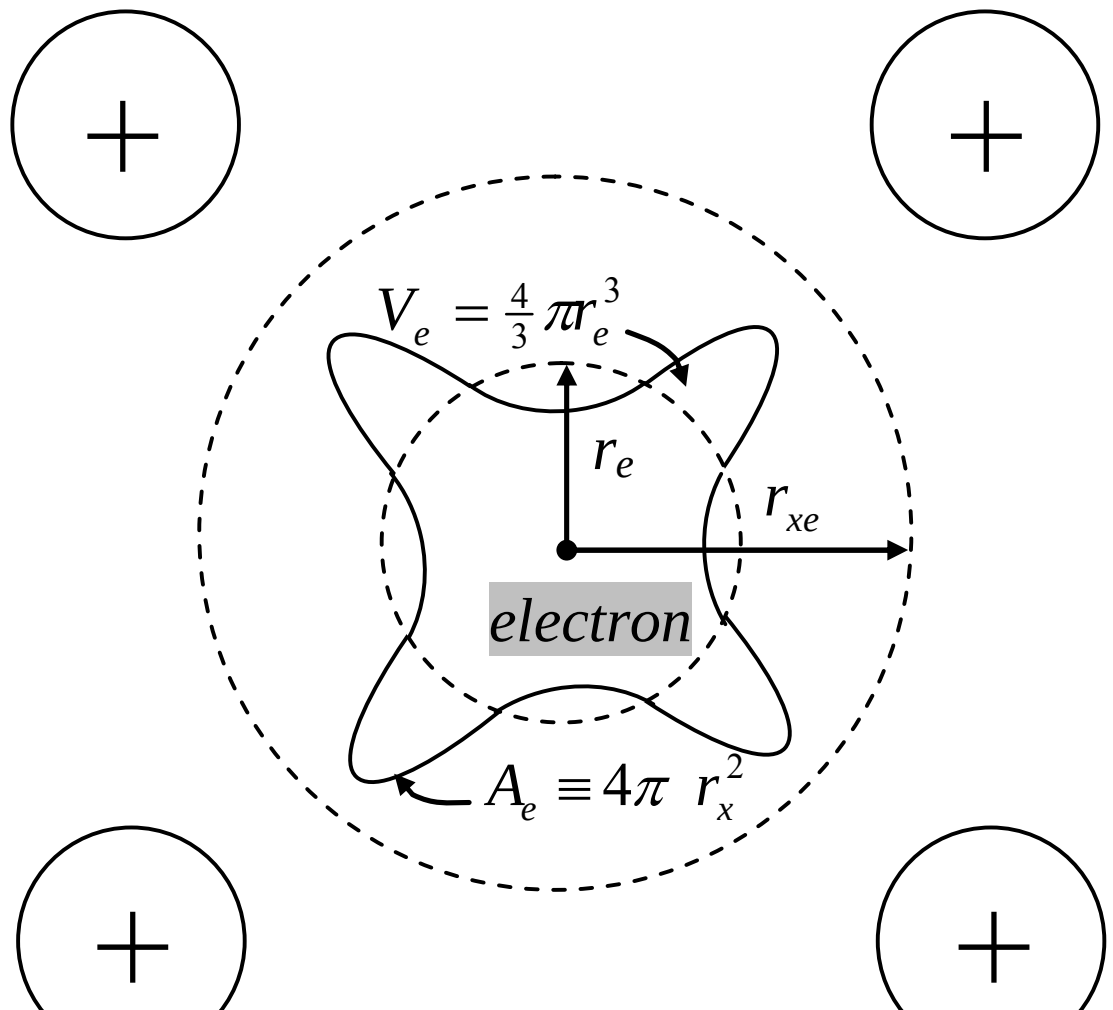

Fig. 4 – *Schematic diagram of Electrons' structure inside the ion lattice.* The positive ions lattice around the electron produces a strong attraction upon the electron deforming its surface. The volume of the electron does not vary, but its external surface is increased and becomes equivalent to the area of a sphere with radius $r_{xe} >> r_e$.

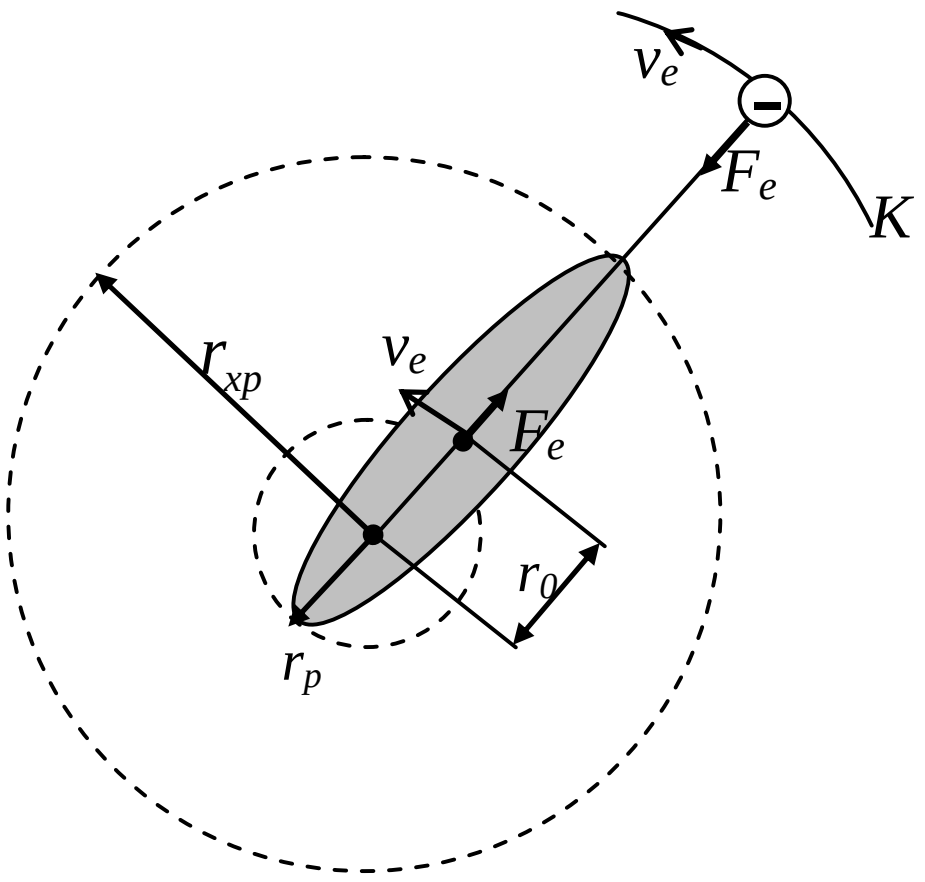

Fig. 5 – *The deformation of the proton.*

In order to calculate the value of $r_{xe}$ we start considering a hydrogen atom, where the electron spins around the proton with a velocity $v_e = 3\times10^6 m.s^{-1}$. The electrical force acting on the proton is $F_e = e^2/4\pi\varepsilon_0 r_1^2$, which is equal to the centrifuge force $F_c = m_p\omega_e^2 r_0$ where $\omega_e$ is the angular velocity of the electron and $r_0$ is the distance between the inertial center of the proton and the center of the moving proton (See Fig. 5, where we conclude that $2(r_0 + r_p) = r_{xp} + r_p$; $r_{xp}$ is the radius of the sphere whose external area is equivalent to the increased area of the proton). Thus, we get $r_0 = \frac{1}{2}(r_{xp} - r_p)$. Substitution of this value into expression of $F_c = F_e$ gives

$$r_{xp} = \frac{e^2}{4\pi\varepsilon_0 m_p v_e^2} + r_p = 3.2\times10^{-14} m$$

Therefore, we can write that $r_{xp} = k_{xp} r_p$, where

$$k_{xp} = \frac{r_{xp}}{r_p} = 25.6$$

The *electron is similarly deformed by the relative movement of the proton in respect to the electron*. In this case, by analogy, we can write that

$$r_{xe} = \frac{e^2}{4\pi\varepsilon_0 m_e v_e^2} + r_e = 6.4\times10^{-11} m$$

and $r_{xe} = k_{xe} r_e$, where $r_{xe}$ is the radius of the sphere whose external area is equivalent to the increased area of the electron. The radius of *free* electron is $r_e = 6.87\times10^{-14} m$ (See *Appendix* A). However, for electrons in the atomic eletrosphere the value of $r_e$ can be calculated starting from Quantum Mechanics.

The wave packet that describes the electron satisfies an *uncertainty principle* $(\Delta p \Delta x \geq \frac{1}{2}\hbar)$, where $\Delta p = \hbar \Delta k$ and $\Delta k$ is the approximate extension of the wave packet. Thus, we can write that $(\Delta k \Delta x \geq \frac{1}{2})$. For the ``square" packet the full width in $k$ is $\Delta k = 2\pi/\lambda_0$ ($\lambda_0 = h/m_e c$ is the average wavelength). The width in $x$ is a little harder to define, but, lets use the first node in the probability found at $(2\pi/\lambda_0)x/2 = \pi$ or $x = \lambda_0$. So, the width of the wave packet is twice this or $\Delta x = 2\lambda_0$. Obviously, $2r_e$ cannot be greater than $\Delta x$, i.e., $r_e$ *must be smaller and close to* $\lambda_0 = h/m_e c = 2.43\times10^{-12} m$. Then, assuming that $r_e \cong 2.4\times10^{-12} m$, we get

$$k_{xe} = \frac{r_{xe}}{r_e} = 26.6$$

Note that $k_{xe} \cong k_{xp}$. In the case *of electrons inside the ion lattice* (See Fig. 4), we can note that, in spite of the electron speed $v_e$ be null, the deformations are similar, in such way that, in this case, we can take the values above.

Substitution of these values into Eq. (20) gives

$$\chi_{Be} = \left\{1 - 2\left[\sqrt{1 + 3.8\times10^{57}\frac{k_{xe}^{22} r_e^4 B_{rms}^4}{f^2}} - 1\right]\right\} =$$

$$= \left\{1 - 2\left[\sqrt{1 + 2.8\times10^{42}\frac{B_{rms}^4}{f^2}} - 1\right]\right\} \qquad (21)$$

Similarly, in the case of *proton* and *neutron* we can write that

$$\chi_{Bp} = \left\{1 - 2\left[\sqrt{1 + \frac{45.56\pi^2 k_{xp}^{22} r_p^4 B_{rms}^4}{c^2 \mu_0^2 m_p^2 f^2}} - 1\right]\right\} \qquad (22)$$

$$\chi_{Bn} = \left\{1 - 2\left[\sqrt{1 + \frac{45.56\pi^2 r_n^4 B_{rms}^4}{\mu_0^2 m_n^2 c^2 f^2}} - 1\right]\right\} \qquad (23)$$

In the case of the neutron, $k_{xn} = 1$ due to its electric charge be null. The radius of *protons inside the atoms* (nuclei) is $r_p = 1.2\times10^{-15} m$ [7,8], $r_n \cong r_p$, then we obtain from Eqs. (22) and (23) following expressions:

$$\chi_{Bp} = \left\{1 - 2\left[\sqrt{1 + 2.2\times10^{22}\frac{B_{rms}^4}{f^2}} - 1\right]\right\} \qquad (24)$$

$$\chi_{Bn} = \left\{1 - 2\left[\sqrt{1 + 2.35\times10^{9}\frac{B_{rms}^4}{f^2}} - 1\right]\right\} \qquad (24a)$$

Since $m_{ge} = \chi_{Be} m_e$, $m_{gp} = \chi_{Bp} m_p$ and $m_{gn} = \chi_{Bn} m_n$, it easy to see, by means of Eqs. (21), (24) and (24a), that $m_{ge}$ is much greater than $m_{gp}$ and $m_{gp}$. This means that, in the calculation of the gravitational force $F_g$ (between the positive ions + electron and the external electron), we can disregard the effects of the gravitational masses of the ions. Thus, the expression of $F_g$ reduces to the expression of the gravitational forces between the two electrons, i.e.,

$$F_g = -G\frac{m_{ge}^2}{r^2} = -\chi_{Be}^2 G \frac{m_e^2}{r^2} \qquad (25)$$

*For the creation* of the Cooper Pairs $F_g$ must overcome the electrons' repulsion due to their negative charge $(e^2/4\pi\varepsilon_0 r^2)$. Thus, we must have $\chi_{Be}^2 G m_e^2 > e^2/4\pi\varepsilon_0$ or

$$\chi_{Be} > \frac{(e/m_e)}{\sqrt{4\pi\varepsilon_0 G}} = -2\times10^{21} \qquad (26)$$

For the Cooper Pairs *not be destructed by the thermal vibrations* due to the temperature $T$, we must have $\chi_{Be}^2 G m_e^2/r > kT$ whence we conclude that $T < \chi_{Be}^2 G m_e^2/r$. Consequently, the *transition temperature*, $T_c$, can be expressed by the following expression

$$T_c = \frac{\chi_{Be}^2 G m_e^2}{k\xi} \qquad (27)$$

where $\xi$ is the size of the Cooper pair, which is given by the *coherence length* of the Cooper-pair wavefunction. It is known that the coherence length is typically 1000 Å (though it can be as small as 30Å in the copper oxides). The coherence length of the Cooper-pair in Aluminum superconductor is quite large ($\xi \cong 1\ micron$ [9]). Substitution of this value into Eq. (27) gives

$$T_c = 4\times10^{-42}\chi_{Be}^2 \qquad (28)$$

For $T_c = 400K\ (\sim 127°C)$ we obtain

$$\chi_{Be} = -1\times10^{22} \qquad (29)$$

By comparing (29) with (26), we can conclude that this value of $\chi_e$ is sufficient *for the creation* of the Cooper Pairs, and also in order that they do not *be destructed by the thermal vibrations* due to the temperature up to $T_c = 400K\ (\sim 127°C)$.

In order to calculate the intensity of the magnetic field $B_m$ with frequency $f$, necessary to produce the value given by Eq.(29), it is necessary the substitution of Eq. (29) into Eq. (21). Thus, we get

$$\left\{1-2\left[\sqrt{1+2.8\times10^{42}\frac{B_{rms}^4}{f^2}}-1\right]\right\} \cong -1\times10^{22} \qquad (30)$$

For $f = 2Hz$ the value of $B_{rms}$ is

$$B_{rms} > 3T$$

Therefore, if a magnetic field with frequency $f = 2Hz$ and intensity $B_{rms} > 3T$ † is applied upon an Aluminum wire it becomes superconductor at ambient temperature ($T_c = 400K\ (\sim 127°C)$). Note that the magnetic field is used only during a time interval sufficient to transform the Aluminum into a superconductor. This means that the process is a some sort of "magnetization" that transforms a conductor into a "permanent" superconductor. After the "magnetization" the magnetic field can be turned off, similarly to the case of "magnetization" that transforms an iron rod into a "permanent" magnet.

† Modern magnetic resonance imaging systems work with magnetic fields up to $8T$ [10, 11].

## **Appendix A**: *The "Geometrical Radii" of Electron and Proton*

It is known that the frequency of oscillation of a simple spring oscillator is

$$f = \frac{1}{2\pi}\sqrt{\frac{K}{m}} \qquad (A1)$$

where $m$ is the inertial mass attached to the spring and $K$ is the spring constant (in $N \cdot m^{-1}$). In this case, the restoring *force* exerted by the spring *is linear* and given by

$$F = -Kx \qquad (A2)$$

where $x$ is the displacement from the equilibrium position.

Now, consider the gravitational force: For example, above the surface of the Earth, the force follows the familiar Newtonian function, i.e., $F = -GM_{g\oplus}m_g/r^2$, where $M_{g\oplus}$ is the mass of Earth, $m_g$ is the gravitational mass of a particle and $r$ is the distance between the centers. *Below* Earth's surface the *force is linear* and given by

$$F = -\frac{GM_{g\oplus}m_g}{R_\oplus^3} r \qquad (A3)$$

where $R_\oplus$ is the radius of Earth.

By comparing (A3) with (A2) we obtain

$$\frac{K}{m_g} = \frac{K}{\chi\, m} = \frac{GM_{g\oplus}}{R_\oplus^3}\left(\frac{r}{x}\right) \qquad (A4)$$

Making $x = r = R_\oplus$, and substituting (A4) into (A1) gives

$$f = \frac{1}{2\pi}\sqrt{\frac{GM_{g\oplus}\chi}{R_\oplus^3}} \qquad (A5)$$

In the case of an *electron* and a *positron*, we substitute $M_{g\oplus}$ by $m_{ge}$, $\chi$ by $\chi_e$ and $R_\oplus$ by $R_e$, where $R_e$ is the radius of electron (or positron). Thus, Eq. (A5) becomes

$$f = \frac{1}{2\pi}\sqrt{\frac{Gm_{ge}\chi_e}{R_e^3}} \qquad (A6)$$

The value of $\chi_e$ varies with the density of energy [4]. When the electron and the positron are distant from each other and the local density of energy is small, the value of $\chi_e$ becomes very close to 1. However, *when the electron and the positron are penetrating one another*, the energy densities in each particle become very strong due to the proximity of their electrical charges $e$ and, consequently, the value of $\chi_e$ strongly increases. In order to calculate the value of $\chi_e$ under these conditions ($x = r = R_e$), we start from the expression of correlation between *electric charge* $q$ and *gravitational mass*, obtained in a previous work [4]:

$$q = \sqrt{4\pi\varepsilon_0 G}\; m_{g(imaginary)}\, i \qquad (A7)$$

where $m_{g(imaginary)}$ is the *imaginary* gravitational mass, and $i = \sqrt{-1}$.

In the case of *electron*, Eq. (A7) gives

$$\begin{aligned} q_e &= \sqrt{4\pi\varepsilon_0 G}\; m_{ge(imaginary)}\, i = \\ &= \sqrt{4\pi\varepsilon_0 G}\left(\chi_e m_{i0e(imaginary)} i\right) = \\ &= \sqrt{4\pi\varepsilon_0 G}\left(-\chi_e \tfrac{2}{\sqrt{3}} m_{i0e(real)} i^2\right) = \\ &= \sqrt{4\pi\varepsilon_0 G}\left(\tfrac{2}{\sqrt{3}} \chi_e m_{i0e(real)}\right) = -1.6\times 10^{-19} C \end{aligned} \qquad (A8)$$

where we obtain

$$\chi_e = -1.8\times 10^{21} \qquad (A9)$$

This is therefore, the value of $\chi_e$ increased by the strong density of energy produced by the electrical charges $e$ of the two particles, under previously mentioned conditions.

Given that $m_{ge} = \chi_e m_{i0e}$, Eq. (A6) yields

$$f = \frac{1}{2\pi}\sqrt{\frac{G\chi_e^2 m_{i0e}}{R_e^3}} \qquad (A10)$$

From Quantum Mechanics, we know that

$$hf = m_{i0}c^2 \qquad (A11)$$

where $h$ is the Planck's constant. Thus, in the case of $m_{i0} = m_{i0e}$ we get

$$f = \frac{m_{i0e}c^2}{h} \qquad (A12)$$

By comparing (A10) and (A12) we conclude that

$$\frac{m_{i0e}c^2}{h} = \frac{1}{2\pi}\sqrt{\frac{G\chi_e^2 m_{i0e}}{R_e^3}} \qquad (A13)$$

Isolating the radius $R_e$, we get:

$$R_e = \left(\frac{G}{m_{i0e}}\right)^{\frac{1}{3}}\left(\frac{\chi_e h}{2\pi\ c^2}\right)^{\frac{2}{3}} = 6.87\times10^{-14}m \quad (A14)$$

Compare this value with the *Compton sized electron*, which predicts $R_e = 3.86\times10^{-13}m$ and also with standardized result recently obtained of $R_e = 4-7\times10^{-13}m$ [12].

In the case of *proton*, we have

$$\begin{aligned} q_p &= \sqrt{4\pi\varepsilon_0 G}\ m_{gp(imaginary)}\, i = \\ &= \sqrt{4\pi\varepsilon_0 G}\left(\chi_p m_{i0p(imaginary)} i\right) = \\ &= \sqrt{4\pi\varepsilon_0 G}\left(-\chi_p \tfrac{2}{\sqrt{3}} m_{i0p(real)} i^2\right) = \\ &= \sqrt{4\pi\varepsilon_0 G}\left(\tfrac{2}{\sqrt{3}}\chi_p m_{i0p(real)}\right) = -1.6\times10^{-19}C \quad (A15) \end{aligned}$$

where we obtain

$$\chi_p = -9.7\times10^{17} \qquad (A16)$$

Thus, the result is

$$R_p = \left(\frac{G}{m_{i0p}}\right)^{\frac{1}{3}}\left(\frac{\chi_p h}{2\pi\ c^2}\right)^{\frac{2}{3}} = 3.72\times10^{-17}m \quad (A17)$$

Note that these radii, given by Equations $(A14)$ and $(A17)$, are the radii of *free* electrons and *free* protons (when the particle and antiparticle (in isolation) penetrate themselves mutually).

Inside the atoms (nuclei) the radius of protons is well-known. For example, protons, as the hydrogen nuclei, have a radius given by $R_p \cong 1.2\times10^{-15}m$ [7, 8]. The strong increase in respect to the value given by Eq. (A17) is due to the interaction with the electron of the atom.

# Gravitational Separator of Isotopes

**Fran De Aquino**
Maranhao State University, Physics Department, S.Luis/MA, Brazil.

In this work we show a *gravitational* separator of isotopes which can be much more effective than those used in the conventional processes of isotopes separation. It is based on intensification of the gravitational acceleration, and can generate accelerations tens of times more intense than those generated in the most powerful centrifuges used for *Uranium enrichment.*

**Key words:** Modified theories of gravity, Isotope separation and enrichment, Nonconventional mechanisms.
PACS: 04.50.Kd, 28.60.+ s, 74.20.Mn.

## 1. Introduction

A conventional gas centrifuge is basically a cylinder that spins around its central axis with ultra-high angular speed while a gas is injected inside it. Under these conditions, the heavier molecules of the gas move towards the cylinder wall and the lighter ones remain close to the center. In addition, if one creates a thermal gradient in a perpendicular direction by keeping the top of the rotating column cool and the bottom hot, the resulting convection current *carries the lighter molecules to the top* while the *heavier ones settle at the bottom*, from which they can be continuously withdrawn.

An important use of gas centrifuges is for the separation of uranium-235 from uranium-238. As a first step, the uranium metal is turned into a gas (uranium hexafluoride, $UF_6$). Next, the $UF_6$ is injected inside a gas centrifuge, which spins at about 100.000 rpm in order to produce a strong centrifugal force upon the $UF_6$ molecules. Thus, the $UF_6$ is separated by the difference in molecular weight between $^{235}UF_6$ and $^{238}UF_6$ [1]. The heavier molecules of the gas ($^{238}UF_6$) move towards the cylinder wall and the lighter ones ($^{235}UF_6$) remain close to the center. The convection current *carries the lighter molecules* ($^{235}UF_6$) *to the top* while the *heavier ones* ($^{238}UF_6$) *settle at the bottom*. However, the gas at the top is not composed totally by $^{235}UF_6$ it contains also $^{238}UF_6$, in such way that we can say that the gas at the top is only *a gas rich in* $^{235}U$. In practice, several of such centrifuges are connected in series. A cascade of identical stages produces successively higher concentrations of $^{235}U$. This process is called Uranium enrichment.

Uranium occurs naturally as two isotopes: 99.3% is Uranium-238 and 0.7% is Uranium-235. Their atoms are identical except for the number of neutrons in the nucleus: Uranium-238 has three more and this makes it less able to fission. Uranium enrichment is used to increase the percentage of the fissile U-235. Nuclear reactors typically require uranium fuel enriched to about 3% to 5% U-235. Nuclear bombs typically use 'Highly Enriched Uranium', enriched to 90% U-235 [2].

In order to extract the $^{235}U$ from the $^{235}UF_6$ it is necessary to add Calcium. The Calcium reacts with the gas producing a salt and pure $^{235}U$.

The conventional gas centrifuges used in the Uranium enrichment are very expensive, and they consume much energy during the process. Here, it is proposed a new type of separator of isotopes based on the *intensification the gravitational acceleration* * [3]. It can generate accelerations tens times more intense than those generated in the most powerful centrifuges used for *Uranium enrichment*.

---

* De Aquino, F. (2008) *Process and Device for* Controlling the Locally the Gravitational Mass and the Gravity Acceleration, BR Patent Number: PI0805046-5, July 31, 2008.

## 2. Theory

From the quantization of gravity it follows that the *gravitational mass* $m_g$ and the *inertial mass* $m_i$ are correlated by means of the following factor [3]:

$$\chi = \frac{m_g}{m_{i0}} = \left\{1 - 2\left[\sqrt{1+\left(\frac{\Delta p}{m_{i0}c}\right)^2} - 1\right]\right\} \tag{1}$$

where $m_{i0}$ is the *rest* inertial mass of the particle and $\Delta p$ is the variation in the particle's *kinetic momentum*; $c$ is the speed of light.

When $\Delta p$ is produced by the absorption of a photon with wavelength $\lambda$, it is expressed by $\Delta p = h/\lambda$. In this case, Eq. (1) becomes

$$\frac{m_g}{m_{i0}} = \left\{1 - 2\left[\sqrt{1+\left(\frac{h/m_{i0}c}{\lambda}\right)^2} - 1\right]\right\} = \left\{1 - 2\left[\sqrt{1+\left(\frac{\lambda_0}{\lambda}\right)^2} - 1\right]\right\} \tag{2}$$

where $\lambda_0 = h/m_{i0}c$ is the *De Broglie wavelength* for the particle with *rest* inertial mass $m_{i0}$.

It has been shown that there is an additional effect - *Gravitational Shielding* effect - produced by a substance whose gravitational mass was reduced or made negative [4]. The effect extends beyond substance (gravitational shielding) , up to a certain distance from it (along the central axis of gravitational shielding). This effect shows that in this region the gravity acceleration, $g_1$, is reduced at the same proportion, i.e., $g_1 = \chi_1 g$ where $\chi_1 = m_g/m_{i0}$ and $g$ is the gravity acceleration *before* the gravitational shielding). Consequently, *after a second gravitational shielding*, the gravity will be given by $g_2 = \chi_2 g_1 = \chi_1\chi_2 g$, where $\chi_2$ is the value of the ratio $m_g/m_{i0}$ for the *second* gravitational shielding. In a generalized way, we can write that after the *nth* gravitational shielding the gravity, $g_n$, will be given by

$$g_n = \chi_1\chi_2\chi_3 \cdots \chi_n g \tag{3}$$

This possibility shows that, by means of a battery of gravitational shieldings, we can make particles acquire enormous accelerations.

From Electrodynamics we know that when an electromagnetic wave with frequency $f$ and velocity $c$ incides on a material with relative permittivity $\varepsilon_r$, relative magnetic permeability $\mu_r$ and electrical conductivity $\sigma$, its *velocity is reduced* to $v = c/n_r$ where $n_r$ is the index of refraction of the material, given by [5]

$$n_r = \frac{c}{v} = \sqrt{\frac{\varepsilon_r\mu_r}{2}\left(\sqrt{1+(\sigma/\omega\varepsilon)^2}+1\right)} \tag{4}$$

If $\sigma >> \omega\varepsilon$, $\omega = 2\pi f$, Eq. (4) reduces to

$$n_r = \sqrt{\frac{\mu_r\sigma}{4\pi\varepsilon_0 f}} \tag{5}$$

Thus, the wavelength of the incident radiation (See Fig. 1) becomes

$$\lambda_{mod} = \frac{v}{f} = \frac{c/f}{n_r} = \frac{\lambda}{n_r} = \sqrt{\frac{4\pi}{\mu f\sigma}} \tag{6}$$

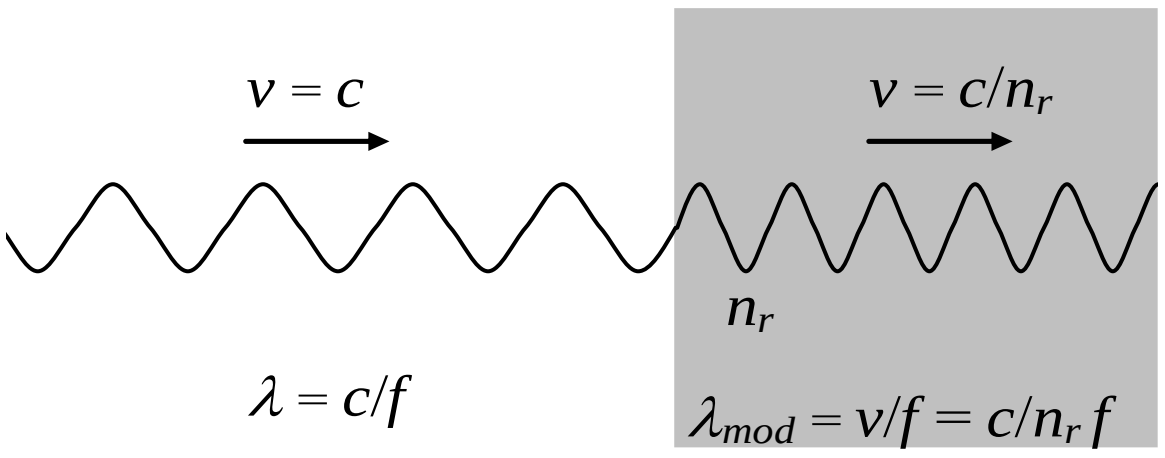

Fig. 1 – *Modified Electromagnetic Wave*. The wavelength of the electromagnetic wave can be strongly reduced, but its frequency remains the same.

If a lamina with thickness equal to $\xi$ contains $n$ atoms/m³, then the number of atoms per area unit is $n\xi$. Thus, if the electromagnetic radiation with frequency $f$ incides on an area $S$ of the lamina it reaches $nS\xi$ atoms. If it incides on the *total area of the lamina*, $S_f$, then the total number of atoms reached by the radiation is $N = nS_f\xi$. The number of atoms per unit of volume, $n$, is given by

$$n = \frac{N_0 \rho}{A} \qquad (7)$$

where $N_0 = 6.02 \times 10^{26}\, atoms/kmole$ is the Avogadro's number; $\rho$ is the matter density of the lamina (in *kg/m*$^3$) and $A$ is the molar mass(*kg/kmole*).

When an electromagnetic wave incides on the lamina, it strikes $N_f$ front atoms, where $N_f \cong (n\, S_f)\phi_m$ , $\phi_m$ is the "diameter" of the atom. Thus, the electromagnetic wave incides effectively on an area $S = N_f S_m$ , where $S_m = \frac{1}{4}\pi\phi_m^2$ is the cross section area of one atom. After these collisions, it carries out $n_{collisions}$ with the other atoms (See Fig.2).

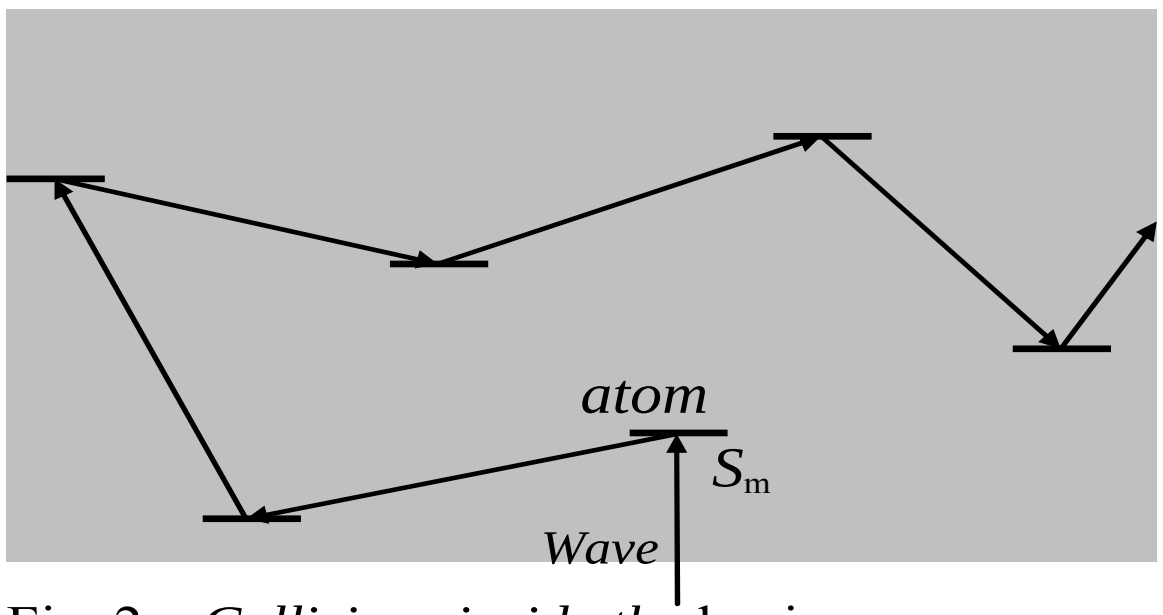

Fig. 2 – *Collisions inside the* lamina.

Thus, the total number of collisions in the volume $S\xi$ is

$$N_{collisions} = N_f + n_{collisions} = n_l S\phi_m + (n_l S\xi - n_m S\phi_m) = \\ = n_m S\xi \qquad (8)$$

The power density, $D$, of the radiation on the lamina can be expressed by

$$D = \frac{P}{S} = \frac{P}{N_f S_m} \qquad (9)$$

We can express the *total mean number of collisions in each atom*, $n_1$, by means of the following equation

$$n_1 = \frac{n_{total\ \ photons} N_{collisions}}{N} \qquad (10)$$

Since in each collision a *momentum* $h/\lambda$ is transferred to the atom, then the *total momentum* transferred to the lamina will be $\Delta p = (n_1 N) h/\lambda$ . Therefore, in accordance with Eq. (1), we can write that

$$\frac{m_{g(l)}}{m_{i0(l)}} = \left\{1 - 2\left[\sqrt{1 + \left[(n_1 N)\frac{\lambda_0}{\lambda}\right]^2} - 1\right]\right\} = \\ = \left\{1 - 2\left[\sqrt{1 + \left[n_{total\ photons} N_{collisions}\frac{\lambda_0}{\lambda}\right]^2} - 1\right]\right\} \qquad (11)$$

Since Eq. (8) gives $N_{collisions} = n_l S\xi$ , we get

$$n_{total\ \ photons} N_{collisions} = \left(\frac{P}{hf^2}\right)(n_l S\xi) \qquad (12)$$

Substitution of Eq. (12) into Eq. (11) yields

$$\frac{m_{g(l)}}{m_{i0(l)}} = \left\{1 - 2\left[\sqrt{1 + \left[\left(\frac{P}{hf^2}\right)(n_l S\xi)\frac{\lambda_0}{\lambda}\right]^2} - 1\right]\right\} \qquad (13)$$

Substitution of $P$ given by Eq. (9) into Eq. (13) gives

$$\frac{m_{g(l)}}{m_{i0(l)}} = \left\{1 - 2\left[\sqrt{1 + \left[\left(\frac{N_f S_m D}{f^2}\right)\left(\frac{n_l S\xi}{m_{i0(l)} c}\right)\frac{1}{\lambda}\right]^2} - 1\right]\right\} \qquad (14)$$

Substitution of $N_f \cong (n_l S_f)\phi_m$ and $S = N_f S_m$ into Eq. (14) results

$$\frac{m_{g(l)}}{m_{i0(l)}} = \left\{1 - 2\left[\sqrt{1 + \left[\left(\frac{n_l^3 S_f^2 S_m^2 \phi_m^2 \xi D}{m_{i0(l)} c f^2}\right)\frac{1}{\lambda}\right]^2} - 1\right]\right\} \qquad (15)$$

where $m_{i0(l)} = \rho_{(l)} V_{(l)}$.

Now, considering that the lamina is inside an ELF electromagnetic field with $E$ and $B$ , then we can write that [6]

$$D = \frac{n_{r(l)} E^2}{2\mu_0 c} \qquad (16)$$

Substitution of Eq. (16) into Eq. (15) gives

$$\frac{m_{g(l)}}{m_{i0(l)}}=\left\{1-2\left[\sqrt{1+\left[\left(\frac{n_{r(l)}n_l^3S_f^2S_m^2\phi_m^2E^2}{2\mu_0 m_{i0(l)}c^2f^2}\right)\frac{1}{\lambda}\right]^2}-1\right]\right\} \quad (17)$$

In the case in which the area $S_f$ is just the *area of the cross-section of the lamina* $(S_\alpha)$, we obtain from Eq. (17), considering that $m_{i0(l)}=\rho_{(l)}S_\alpha\xi$, the following expression

$$\frac{m_{g(l)}}{m_{i0(l)}}=\left\{1-2\left[\sqrt{1+\left[\left(\frac{n_{r(l)}n_l^3S_\alpha S_m^2\phi_m^2E^2}{2\mu_0\rho_{(l)}c^2f^2}\right)\frac{1}{\lambda}\right]^2}-1\right]\right\} \quad (18)$$

If the electrical conductivity of the lamina, $\sigma_{(l)}$, is such that $\sigma_{(l)} >> \omega\varepsilon$ , then the value of $\lambda$ is given by Eq. (6), i.e.,

$$\lambda=\lambda_{mod}=\sqrt{\frac{4\pi}{\mu f\sigma}} \quad (19)$$

Substitution of Eq. (19) into Eq. (18) gives

$$\frac{m_{g(l)}}{m_{i0(l)}}=\left\{1-2\left[\sqrt{1+\frac{n_{r(l)}^2n_l^6S_\alpha^2S_m^4\phi_m^4\sigma_{(l)}E^4}{16\pi\mu_0\rho_{(l)}^2c^4f^3}}-1\right]\right\} \quad (20)$$

Note that $E=E_m\sin\omega t$ .The average value for $E^2$ is equal to $\frac{1}{2}E_m^2$ because $E$ varies sinusoidaly ($E_m$ is the maximum value for $E$). On the other hand, $E_{rms}=E_m/\sqrt{2}$ . Consequently we can replace $E^4$ for $E_{rms}^4$, and the equation above can be rewritten as follows

$$\chi=\frac{m_{g(l)}}{m_{i0(l)}}=$$

$$=\left\{1-2\left[\sqrt{1+\frac{n_{r(l)}^2n_l^6S_\alpha^2S_m^4\phi_m^4\sigma_{(l)}E_{rms}^4}{16\pi\mu_0\rho_{(l)}^2c^4f^3}}-1\right]\right\} \quad (21)$$

Now consider the system shown in Fig.3. It was originally designed to convert *Gravitational Energy* directly into *Electrical Energy* [7]. Here, it works as *Separator of Isotopes*. These systems are basically similar, except in the core. The core of the original system has been replaced by the one shown in Fig.3 and Fig.4 (detailed).

Inside the *Gravitational Separator of Isotopes* there is a *dielectric tube* ($\varepsilon_r \cong 1$) with the following characteristics: $\alpha = 60mm$, $S_\alpha = \pi\alpha^2/4 = 2.83\times10^{-3}m^2$. Inside the tube there is an *Aluminum sphere* with 30mm radius and mass $M_{gs}=0.30536kg$. The tube is filled with *air* at ambient temperature and 1atm. Thus, inside the tube, the air density is

$$\rho_{air}=1.2 \ kg.m^{-3} \quad (22)$$

The number of atoms of air (Nitrogen) per unit of volume, $n_{air}$, according to Eq.(7), is given by

$$n_{air}=\frac{N_0\rho_{air}}{A_N}=5.16\times10^{25}atoms/m^3 \quad (23)$$

The *parallel metallic plates* (p), shown in Fig.3 are subjected to different drop voltages. The two sets of plates (*D*), placed on the extremes of the tube, are subjected to $V_{(D)rms}=1.64kV$ at $f=1Hz$, while the central set of plates (*A*) is subjected to $V_{(A)rms}=19.7kV$ at $f=1Hz$. Since $d=98mm$, then the intensity of the electric field, which passes through the 36 *cylindrical air laminas* (each one with 5mm thickness) of the *two* sets (*D*), is

$$E_{(D)rms}=V_{(D)rms}/d=1.67\times10^4V/m$$

and the intensity of the electric field, which passes through the 7 *cylindrical air laminas* of the central set (*A*), is given by

$$E_{(A)rms}=V_{(A)rms}/d=2.012\times10^5V/m$$

Note that the *metallic rings* (5*mm* thickness) are positioned in such way to block the electric field out of the cylindrical air laminas. The objective is to turn each one of these laminas into a *Gravity Control Cells* (GCC) [4]. Thus, the system shown in Fig. 3 has 3 sets of GCC. Two with 18 GCC each, and one with 7 GCC. The two sets with 18 GCC each are positioned at the extremes of the tube (*D*). They work as gravitational *decelerator* while the other set with 7 GCC ($A$) works as a gravitational *accelerator*, intensifying the gravity acceleration produced by the mass $M_{gs}$ of the Aluminum sphere. According to Eq. (3), this gravity, after the $7^{th}$ GCC becomes $g_7 = \chi^7 GM_{gs}/r_0^2$, where $\chi = m_{g(l)}/m_{i(l)}$ given by Eq. (21) and $r_0 = 35mm$ is the distance between the center of the Aluminum sphere and the surface of the first GCC of the set (A).

The objective of the sets ($D$), with 18 GCC each, is to reduce strongly the value of the external gravity along the axis of the tube. In this case, the value of the external gravity, $g_{ext}$, is reduced by the factor $\chi_d^{18} g_{ext}$, where $\chi_d = 10^{-2}$. For example, if the base BS of the system is positioned on the Earth surface, then $g_{ext} = 9.81 m/s^2$ is reduced to $\chi_d^{18} g_{ext}$ and, after the set A, it is increased by $\chi^7$. Since the system is designed for $\chi = -308.5$ (See Eq. (26)), then the gravity acceleration on the sphere becomes $\chi^7 \chi_d^{18} g_{ext} = 2.6\times10^{-18} m/s^2$. This value is much smaller than $g_{sphere} = GM_{gs}/r_s^2 = 2.26\times10^{-8} m/s^2$.

Note that there is a *uniform magnetic field*, $B$, through the *core* of the Gravitational Separator of Isotopes (a cylindrical *Dielectric Chamber* with 60mm external diameter; 50mm internal diameter and 100mm height). The electrical conductivity of air, *inside the dielectric tube*, is equal to the electrical conductivity of Earth's atmosphere near the land, whose average value is $\sigma_{air} \cong 1\times10^{-14} S/m$ [8]. This value is of fundamental importance in order to obtain the convenient values of the electrical current $i$ and the value of $\chi$ and $\chi_d$, which are given by Eq. (21), i.e.,

$$\chi = \left\{1 - 2\left[\sqrt{1 + \frac{n_{r(air)}^2 n_{air}^6 S_\alpha^2 S_m^4 \phi_m^4 \sigma_{air} E_{(A)rms}^4}{16\pi\mu_0 \rho_{air}^2 c^4 f^3}} - 1\right]\right\} = \left\{1 - 2\left[\sqrt{1 + 1.480\times10^{-17} E_{(A)rms}^4} - 1\right]\right\} \qquad (24)$$

$$\chi_d = \left\{1 - 2\left[\sqrt{1 + \frac{n_{r(air)}^2 n_{air}^6 S_\alpha^2 S_m^4 \phi_m^4 \sigma_{air} E_{(D)rms}^4}{16\pi\mu_0 \rho_{air}^2 c^4 f^3}} - 1\right]\right\} = \left\{1 - 2\left[\sqrt{1 + 1.48\times10^{-17} E_{(D)rms}^4} - 1\right]\right\} \qquad (25)$$

where $n_{r(air)} = \sqrt{\varepsilon_r \mu_r} \cong 1$, since $(\sigma << \omega\varepsilon)$; $n_{air} = 5.16\times10^{25} atoms/m^3$, $\phi_m = 1.55\times10^{-10} m$, $S_m = \pi\phi_m^2/4 = 1.88\times10^{-20} m^2$ and $f = 60Hz$. Since $E_{(A)rms} = 2.012\times10^5 V/m$ and $E_{(D)rms} = 1.67\times10^4 V/m$, we get

$$\chi = -308.5 \qquad (26)$$

and

$$\chi_d \cong 10^{-2} \qquad (27)$$

It is important to note that *the set with* 7 GCC ($A$) *cannot be turned on before the magnetic field* $B$ *is on.* Because the gravitational accelerations on the dielectric chamber and *Al* sphere will be enormous $\left(\chi^7 GM_{gs}/r_0^2 \cong 4.4\times10^9 m/s^2\right)$, and will explode the device.

The *isotopes* inside the *Dielectric Chamber* are subjected to the gravity acceleration produced by the sphere, and increased by the 7 GCC in the region (A). Its value is

$$a_i = \chi^7 g_s = \chi^7 G\frac{M_{gs}}{r_s^2} \cong 6.0\times10^9 m/s^2 \qquad (28)$$

Comparing this value with the produced in the most powerful centrifuges (at 100,000 rpm), which is of the order of $10^7 m/s^2$, we conclude that the accelerations in the *Gravitational Separator of Isotopes* is about 600 times greater than the values of the centrifuges.

In the case of Uranium enrichment, the gas $UF_6$ is injected inside this core where it is strongly accelerated. Thus, the $UF_6$ is separated by the difference in molecular weight between $^{235}UF_6$ and $^{238}UF_6$ (See Fig.4). The heavier molecules of the gas ($^{238}UF_6$) move towards the cylinder bottom and the lighter ones ($^{235}UF_6$) remain close to the center. The convection current, produced by a thermal gradient of about 300°C between the bottom and the top of the cylinder, *carries the lighter molecules* ($^{235}UF_6$) *to the top* while the *heavier ones* ($^{238}UF_6$) *settle at the bottom*, from which they can be continuously withdrawn. The gas withdrawn at the top of the cylinder is *a gas rich in* $^{235}U$.

The *gravitational forces* due to the gravitational mass of the sphere $(M_{gs})$ acting on *electrons* $(F_e)$, *protons* $(F_p)$ and *neutrons* $(F_p)$ of the *dielectric* of the *Dielectric Chamber*, are respectively expressed by the following relations

$$F_e = m_{ge}a_e = \chi_{Be}m_e\left(\chi^7 G\frac{M_{gs}}{r_0^2}\right) \qquad (29)$$

$$F_p = m_{gp}a_p = \chi_{Bp}m_p\left(\chi^7 G\frac{M_{gs}}{r_0^2}\right) \qquad (30)$$

$$F_n = m_{gn}a_n = \chi_{Bn}m_n\left(\chi^7 G\frac{M_{gs}}{r_0^2}\right) \qquad (31)$$

*In order to make null the resultant of these forces in the Dielectric Chamber* (and *also in the sphere*) we must have $F_e = F_p + F_n$, i.e.,

$$m_e\chi_{Be} = m_p\chi_{Bp} + m_n\chi_{Bn} \qquad (32)$$

In order to calculate the expressions of $\chi_{Be}$, $\chi_{Bp}$ and $\chi_{Bn}$ we start from Eq. (17), for the particular case of *single electron* in the region subjected to the magnetic field *B*. In this case, we must substitute $n_{r(l)}$ by $n_r = (\mu_r\sigma/4\pi\varepsilon_0 f)^{\frac{1}{2}}$; $n_l$ by $1/V_e = 1/\frac{4}{3}\pi r_e^3$ ($r_e$ is the electrons radius), $S_f$ by $(SSA_e)\rho_e V_e$ ($SSA_e$ is the *specific surface area* for electrons in this case: $SSA_e = \frac{1}{2}A_e/m_e = \frac{1}{2}A_e/\rho_e V_e = 2\pi r_e^2/\rho_e V_e$), $S_m$ by $S_e = \pi r_e^2$, $\xi$ by $\phi_m = 2r_e$ and $m_{i0(l)}$ by $m_e$. The result is

$$\chi_{Be} = \left\{1 - 2\left[\sqrt{1 + \frac{45.56\pi^2 r_e^4 n_r^2 E^4}{\mu_0^2 m_e^2 c^4 f^4 \lambda^2}} - 1\right]\right\} \qquad (33)$$

Electrodynamics tells us that $E_{rms} = vB_{rms} = (c/n_r)B_{rms}$, and Eq. (19) gives $\lambda = \lambda_{mod} = (4\pi/\mu_r\sigma f)$. Substitution of these expressions into Eq. (33) yields

$$\chi_{Be} = \left\{1 - 2\left[\sqrt{1 + \frac{45.56\pi^2 r_e^4 B_{rms}^4}{\mu_0^2 m_e^2 c^2 f^2}} - 1\right]\right\} \qquad (34)$$

Similarly, in the case of *proton* and *neutron* we can write that

$$\chi_{Bp} = \left\{1 - 2\left[\sqrt{1 + \frac{45.56\pi^2 r_p^4 B_{rms}^4}{\mu_0^2 m_p^2 c^2 f^2}} - 1\right]\right\} \qquad (35)$$

$$\chi_{Bn} = \left\{1 - 2\left[\sqrt{1 + \frac{45.56\pi^2 r_n^4 B_{rms}^4}{\mu_0^2 m_n^2 c^2 f^2}} - 1\right]\right\} \qquad (36)$$

The radius of *free* electron is $r_e = 6.87\times10^{-14}m$ (See *Appendix* A) and the radius of *protons inside the atoms* (nuclei) is $r_p = 1.2\times10^{-15}m$, $r_n \cong r_p$ [9, 10], then we obtain from Eqs. (34) (35) and (36) the following expressions:

$$\chi_{Be} = \left\{1 - 2\left[\sqrt{1 + 8.49\times10^4\frac{B_{rms}^4}{f^2}} - 1\right]\right\} \qquad (37)$$

$$\chi_{Bn} \cong \chi_{Bp} = \left\{1 - 2\left[\sqrt{1 + 2.35\times10^{-9}\frac{B_{rms}^4}{f^2}} - 1\right]\right\} \qquad (38)$$

Then, from Eq. (32) it follows that

$$m_e\chi_{Be} \cong 2m_p\chi_{Bp} \qquad (39)$$

Substitution of Eqs. (37) and (38) into Eq. (39) gives

$$\frac{\left\{1-2\left[\sqrt{1+8.49\times10^{4}\frac{B_{rms}^{4}}{f^{2}}}-1\right]\right\}}{\left\{1-2\left[\sqrt{1+2.35\times10^{-9}\frac{B_{rms}^{4}}{f^{2}}}-1\right]\right\}}=3666.3 \quad (40)$$

For $f = 0.1Hz$, we get

$$\frac{\left\{1-2\left[\sqrt{1+8.49\times10^{6}B_{rms}^{4}}-1\right]\right\}}{\left\{1-2\left[\sqrt{1+2.35\times10^{-7}B_{rms}^{4}}-1\right]\right\}}=3666.3 \quad (41)$$

whence we obtain

$$B_{rms} = 0.793\ T \quad (42)$$

Consequently, Eq. (37) and (38) yields

$$\chi_{Be} = -3666.3 \quad (43)$$

and

$$\chi_{Bn} \cong \chi_{Bp} \cong 0.999 \quad (44)$$

In order for the forces $F_e$ and $F_p$ have *contrary direction* (such as it occurs in the case, in which the nature of the electromotive force is electrical) we must have $\chi_{Be} < 0$ and $\chi_{Bn} \cong \chi_{Bp} > 0$ (See equations (29) (30) and (31)), i.e.,

$$\left\{1-2\left[\sqrt{1+8.49\times10^{4}\frac{B_{rms}^{4}}{f^{2}}}-1\right]\right\}<0 \quad (45)$$

and

$$\left\{1-2\left[\sqrt{1+2.35\times10^{-9}\frac{B_{rms}^{4}}{f^{2}}}-1\right]\right\}>0 \quad (46)$$

This means that we must have

$$0.06\sqrt{f} < B_{rms} < 151.86\sqrt{f} \quad (47)$$

In the case of $f = 0.1Hz$ the result is

$$0.01T < B_{rms} < 48.02T \quad (48)$$

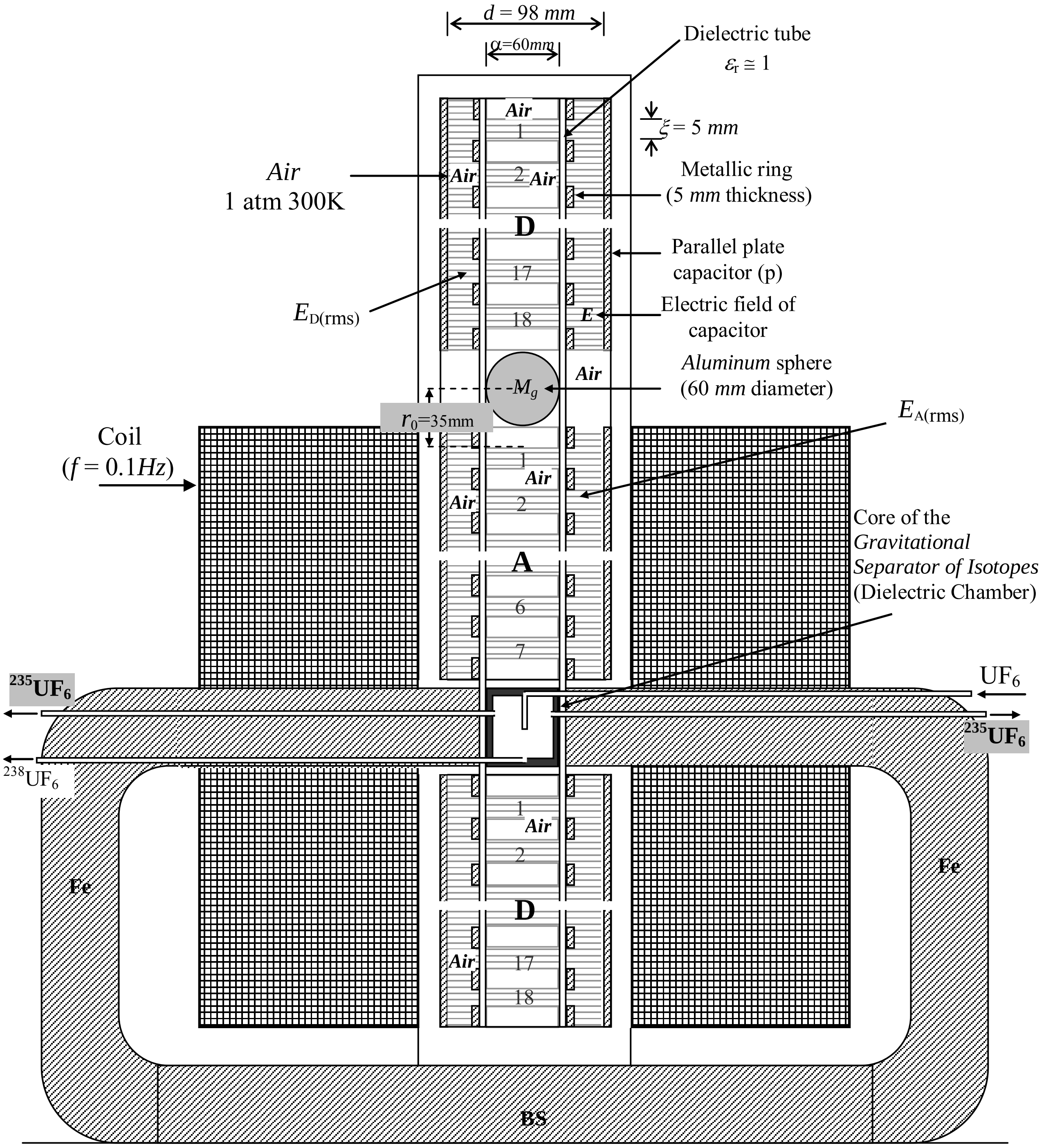

Fig. 3 – Schematic Diagram of a *Gravitational Separator of Isotopes* (Based on a process of gravity control *patented* in July, 31 2008, PI0805046-5). In the case of Uranium enrichment, the gas $UF_6$ is injected inside the core of the *Gravitational Separator of Isotopes* where it is strongly accelerated. Thus, the $UF_6$ is separated by the difference in molecular weight between $^{235}UF_6$ and $^{238}UF_6$. The heavier molecules of the gas ($^{238}UF_6$) move towards the cylinder bottom and the lighter ones ($^{235}UF_6$) remain close to the center. The convection current, produced by a thermal gradient of about 300°C between the bottom and the top of the cylinder, *carries the lighter molecules* ($^{235}UF_6$) *to the top* while the *heavier ones* ($^{238}UF_6$) *settle at the bottom*, from which they can be continuously withdrawn. The gas withdrawn at the top of the cylinder is *a gas rich in* $^{235}U$.

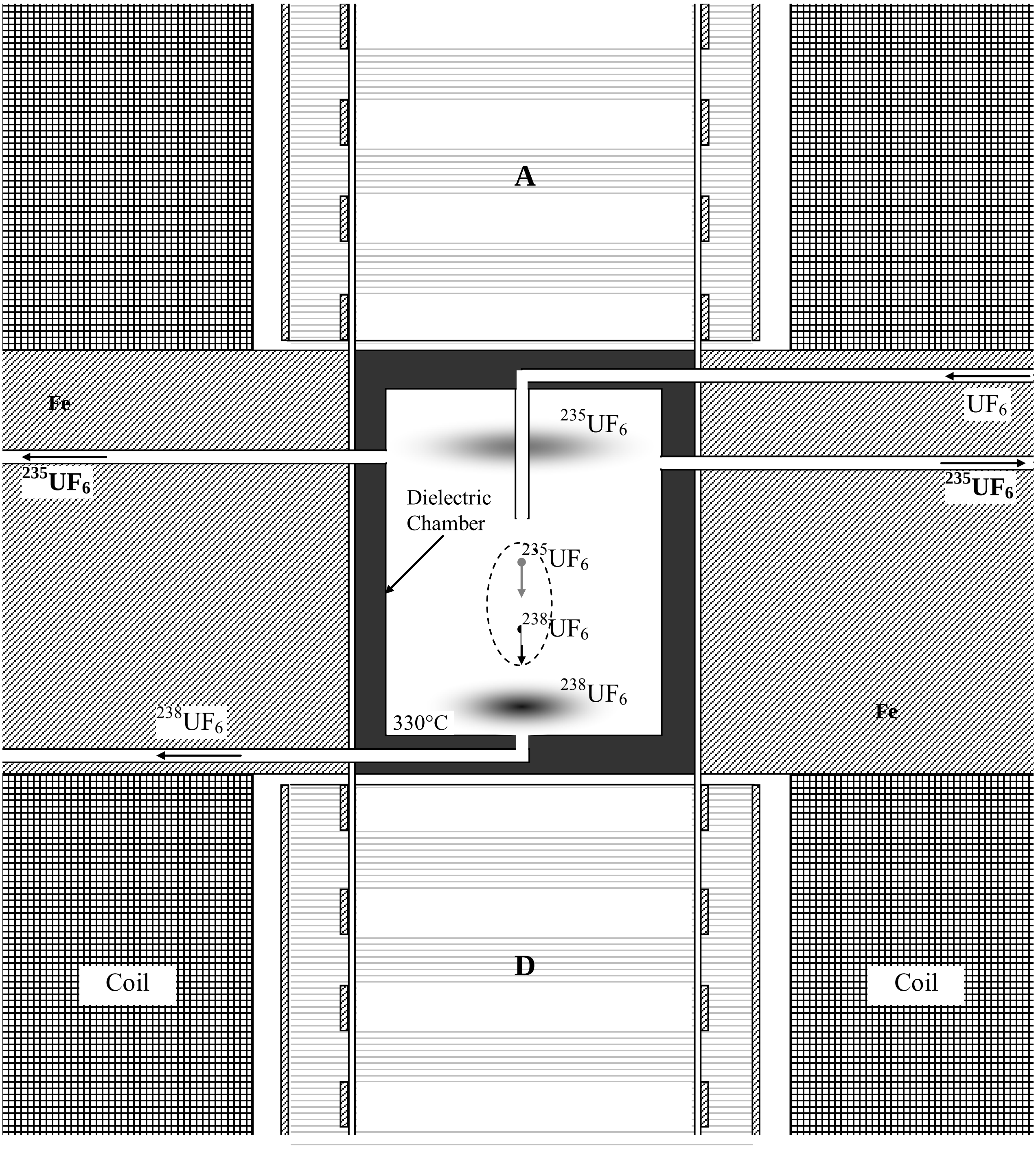

Fig. 4 – *Details of the core of the Gravitational Separator of Isotopes*. In the case of Uranium enrichment, the heavier molecules of the gas ($^{238}UF_6$) move towards the cylinder bottom and the lighter ones ($^{235}UF_6$) remain close to the center. The convection current, produced by a thermal gradient of about 300°C between the bottom and the top of the *Dielectric Chamber*, *carries the lighter molecules* ($^{235}UF_6$) *to the top* while the *heavier ones* ($^{238}UF_6$) *settle at the bottom*, from which they can be continuously withdrawn. The gas withdrawn at the top of the chamber is *a gas rich in* $^{235}U$.

## **Appendix A**: *The "Geometrical Radii" of Electron and Proton*

It is known that the frequency of oscillation of a simple spring oscillator is

$$f = \frac{1}{2\pi}\sqrt{\frac{K}{m}} \qquad (A1)$$

where $m$ is the inertial mass attached to the spring and $K$ is the spring constant (in $N{\cdot}m^{-1}$). In this case, the restoring *force* exerted by the spring *is linear* and given by

$$F = -Kx \qquad (A2)$$

where $x$ is the displacement from the equilibrium position.

Now, consider the gravitational force: For example, above the surface of the Earth, the force follows the familiar Newtonian function, i.e., $F = -GM_{g\oplus}m_g/r^2$, where $M_{g\oplus}$ is the mass of Earth, $m_g$ is the gravitational mass of a particle and $r$ is the distance between the centers. *Below* Earth's surface the *force is linear* and given by

$$F = -\frac{GM_{g\oplus}m_g}{R_\oplus^3}r \qquad (A3)$$

where $R_\oplus$ is the radius of Earth.

By comparing (A3) with (A2) we obtain

$$\frac{K}{m_g} = \frac{K}{\chi\ m} = \frac{GM_{g\oplus}}{R_\oplus^3}\left(\frac{r}{x}\right) \qquad (A4)$$

Making $x = r = R_\oplus$, and substituting (A4) into (A1) gives

$$f = \frac{1}{2\pi}\sqrt{\frac{GM_{g\oplus}\chi}{R_\oplus^3}} \qquad (A5)$$

In the case of an *electron* and a *positron*, we substitute $M_{g\oplus}$ by $m_{ge}$, $\chi$ by $\chi_e$ and $R_\oplus$ by $R_e$, where $R_e$ is the radius of electron (or positron). Thus, Eq. (A5) becomes

$$f = \frac{1}{2\pi}\sqrt{\frac{Gm_{ge}\chi_e}{R_e^3}} \qquad (A6)$$

The value of $\chi_e$ varies with the density of energy [3]. When the electron and the positron are distant from each other and the local density of energy is small, the value of $\chi_e$ becomes very close to 1. However, *when the electron and the positron are penetrating one another*, the energy densities in each particle become very strong due to the proximity of their electrical charges $e$ and, consequently, the value of $\chi_e$ strongly increases. In order to calculate the value of $\chi_e$ under these conditions ($x = r = R_e$), we start from the expression of correlation between *electric charge* $q$ and *gravitational mass*, obtained in a previous work [3]:

$$q = \sqrt{4\pi\varepsilon_0 G}\ m_{g(imaginary)}\ i \qquad (A7)$$

where $m_{g(imaginary)}$ is the *imaginary* gravitational mass, and $i = \sqrt{-1}$.

In the case of *electron*, Eq. (A7) gives

$$q_e = \sqrt{4\pi\varepsilon_0 G}\ m_{ge(imaginary)}\ i =$$
$$= \sqrt{4\pi\varepsilon_0 G}\left(\chi_e m_{i0e(imaginary)} i\right) =$$
$$= \sqrt{4\pi\varepsilon_0 G}\left(-\chi_e \tfrac{2}{\sqrt{3}} m_{i0e(real)} i^2\right) =$$
$$= \sqrt{4\pi\varepsilon_0 G}\left(\tfrac{2}{\sqrt{3}}\chi_e m_{i0e(real)}\right) = -1.6\times10^{-19}C \qquad (A8)$$

where we obtain

$$\chi_e = -1.8\times10^{21} \qquad (A9)$$

This is therefore, the value of $\chi_e$ increased by the strong density of energy produced by the electrical charges $e$ of the two particles, under previously mentioned conditions.

Given that $m_{ge} = \chi_e m_{i0e}$, Eq. (A6) yields

$$f = \frac{1}{2\pi}\sqrt{\frac{G\chi_e^2 m_{i0e}}{R_e^3}} \qquad (A10)$$

From Quantum Mechanics, we know that

$$hf = m_{i0}c^2 \qquad (A11)$$

where $h$ is the Planck's constant. Thus, in the case of $m_{i0} = m_{i0e}$ we get

$$f = \frac{m_{i0e}c^2}{h} \qquad (A12)$$

By comparing (A10) and (A12) we conclude that

$$\frac{m_{i0e}c^2}{h} = \frac{1}{2\pi}\sqrt{\frac{G\chi_e^2 m_{i0e}}{R_e^3}} \qquad (A13)$$

Isolating the radius $R_e$, we get:

$$R_e = \left(\frac{G}{m_{i0e}}\right)^{\frac{1}{3}}\left(\frac{\chi_e h}{2\pi\ c^2}\right)^{\frac{2}{3}} = 6.87\times10^{-14}m \quad (A14)$$

Compare this value with the *Compton sized electron*, which predicts $R_e = 3.86\times10^{-13}m$ and also with standardized result recently obtained of $R_e = 4-7\times10^{-13}m$ [11].

In the case of *proton*, we have

$$\begin{aligned} q_p &= \sqrt{4\pi\varepsilon_0 G}\ m_{gp(imaginary)}\, i = \\ &= \sqrt{4\pi\varepsilon_0 G}\left(\chi_p m_{i0p(imaginary)}\, i\right) = \\ &= \sqrt{4\pi\varepsilon_0 G}\left(-\chi_p \tfrac{2}{\sqrt{3}} m_{i0p(real)}\, i^2\right) = \\ &= \sqrt{4\pi\varepsilon_0 G}\left(\tfrac{2}{\sqrt{3}}\chi_p m_{i0p(real)}\right) = -1.6\times10^{-19}C \quad (A15) \end{aligned}$$

where we obtain

$$\chi_p = -9.7\times10^{17} \qquad (A16)$$

Thus, the result is

$$R_p = \left(\frac{G}{m_{i0p}}\right)^{\frac{1}{3}}\left(\frac{\chi_p h}{2\pi\ c^2}\right)^{\frac{2}{3}} = 3.72\times10^{-17}m \quad (A17)$$

Note that these radii, given by Equations $(A14)$ and $(A17)$, are the radii of *free* electrons and *free* protons (when the particle and antiparticle (in isolation) penetrate themselves mutually).

Inside the atoms (nuclei) the radius of protons is well-known. For example, protons, as the hydrogen nuclei, have a radius given by $R_p \cong 1.2\times10^{-15}m$ [9, 10]. The strong increase in respect to the value given by Eq. (A17) is due to the interaction with the electron of the atom.

# Gravitational Atomic Synthesis at Room Temperature

**Fran De Aquino**
Maranhao State University, Physics Department, S.Luis/MA, Brazil.

It is described a process for creating new atoms starting from pre-existing atoms. We show that all the elements of the periodic table can be synthesized, at room temperature, by a gravitational process based on the intensification of the gravitational interaction by means of electromagnetic fields.

**Key words:** Modified theories of gravity, Atom manipulation, Atomic forces.
PACS: 04.50.Kd, 37.90.+j, 34.20.-b.

## 1. Introduction

Rutherford [1] was the first to observe the transmutation of the atoms, and also the first to perform transmutation of the atoms. That gave him a double justification for being labeled an alchemist.

It is currently believed that the synthesis of precious metals, a symbolic goal long sought by alchemists, is only possible with methods involving either nuclear reactors or particle accelerators. However, it will be shown here that *all the elements* of the periodic table can be synthesized, at room temperature, by a gravitational process based on the intensification of the gravitational interaction by means of electromagnetic fields. The process is very simple, but it requires extremely-low frequency (ELF) magnetic field with very strong intensity ($B_{rms} > 2{,}500T$).

The strongest continuous magnetic field yet produced in a laboratory had *45 T* (Florida State University's National High Magnetic Field Laboratory in Tallahassee, USA) [2]. The strongest (pulsed) magnetic field yet obtained non-destructively in a laboratory had about *100T*. (National High Magnetic Field Laboratory, Los Alamos National Laboratory, USA) [3]. The strongest pulsed magnetic field yet obtained in a laboratory, destroying the used equipment, but not the laboratory itself (Institute for Solid State Physics, Tokyo) reached *730 T*. The strongest (pulsed) magnetic field ever obtained (with explosives) in a laboratory (VNIIEF in Sarov, Russia, 1998) reached *2,800T* [4].

## 2. Theory

The quantization of gravity showed that the *gravitational mass* $m_g$ and the *inertial mass* $m_i$ are correlated by means of the following factor [5]:

$$\chi = \frac{m_g}{m_{i0}} = \left\{1 - 2\left[\sqrt{1 + \left(\frac{\Delta p}{m_{i0}c}\right)^2} - 1\right]\right\} \qquad (1)$$

where $m_{i0}$ is the *rest* inertial mass of the particle and $\Delta p$ is the variation in the particle's *kinetic momentum*; $c$ is the speed of light.

When $\Delta p$ is produced by the absorption of a photon with wavelength $\lambda$, it is expressed by $\Delta p = h/\lambda$. In this case, Eq. (1) becomes

$$\frac{m_g}{m_{i0}} = \left\{1 - 2\left[\sqrt{1 + \left(\frac{h/m_{i0}c}{\lambda}\right)^2} - 1\right]\right\}$$
$$= \left\{1 - 2\left[\sqrt{1 + \left(\frac{\lambda_0}{\lambda}\right)^2} - 1\right]\right\} \qquad (2)$$

where $\lambda_0 = h/m_{i0}c$ is the *DeBroglie wavelength* for the particle with *rest* inertial mass $m_{i0}$.

In general, the *momentum* variation $\Delta p$ is expressed by $\Delta p = F\Delta t$ where $F$ is the applied force during a time interval $\Delta t$. Note that there is no restriction concerning the *nature* of the force, i.e., it can be mechanical, electromagnetic, etc. For example, we can look on the *momentum* variation $\Delta p$ as due to absorption or emission of *electromagnetic energy* by the particle.

This means that, by means of electromagnetic fields, the *gravitational mass* can be decreased down to become negative and *increased* (*independently* of the inertial mass $m_i$). In this way, *the gravitational forces can be intensified*. Consequently, we can use, for example, oscillating magnetic fields in order to *intensify the gravitational interaction* between electrons and protons.

From Electrodynamics we know that when an electromagnetic wave with frequency $f$ and velocity $c$ incides on a material with relative permittivity $\varepsilon_r$, relative magnetic permeability $\mu_r$ and electrical conductivity $\sigma$, its *velocity is reduced* to $v = c/n_r$ where $n_r$ is the index of refraction of the material, given by [6]

$$n_r = \frac{c}{v} = \sqrt{\frac{\varepsilon_r \mu_r}{2}\left(\sqrt{1+(\sigma/\omega\varepsilon)^2}+1\right)} \qquad (3)$$

If $\sigma >> \omega\varepsilon$, $\omega = 2\pi f$, Eq. (3) reduces to

$$n_r = \sqrt{\frac{\mu_r \sigma}{4\pi\varepsilon_0 f}} \qquad (4)$$

Thus, the wavelength of the incident radiation (See Fig. 2) becomes

$$\lambda_{mod} = \frac{v}{f} = \frac{c/f}{n_r} = \frac{\lambda}{n_r} = \sqrt{\frac{4\pi}{\mu f \sigma}} \qquad (5)$$

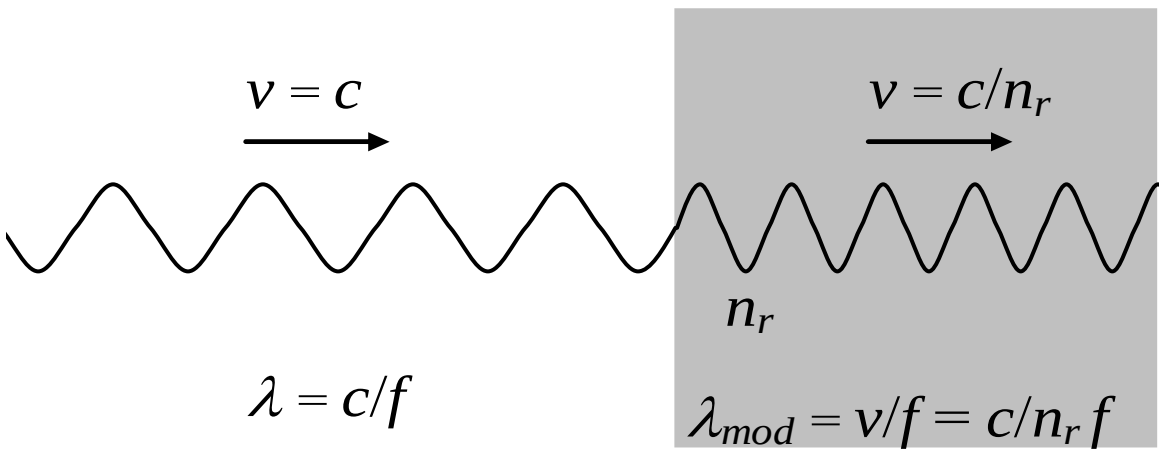

Fig. 2 – *Modified Electromagnetic Wave*. The wavelength of the electromagnetic wave can be strongly reduced, but its frequency remains the same.

If a lamina with thickness equal to $\xi$ contains $n$ atoms/m$^3$, then the number of atoms per area unit is $n\xi$. Thus, if the electromagnetic radiation with frequency $f$ incides on an area $S$ of the lamina it reaches $nS\xi$ atoms. If it incides on the *total area of the lamina*, $S_f$, then the total number of atoms reached by the radiation is $N = nS_f\xi$. The number of atoms per unit of volume, $n$, is given by

$$n = \frac{N_0 \rho}{A} \qquad (6)$$

where $N_0 = 6.02\times10^{26}\, atoms/kmole$ is the Avogadro's number; $\rho$ is the matter density of the lamina (in *kg/m*$^3$) and $A$ is the molar mass(*kg/kmole*).

When an electromagnetic wave incides on the lamina, it strikes $N_f$ front atoms, where $N_f \cong (n\,S_f)\phi_m$, $\phi_m$ is the "diameter" of the atom. Thus, the electromagnetic wave incides effectively on an area $S = N_f S_m$, where $S_m = \frac{1}{4}\pi\phi_m^2$ is the cross section area of one atom. After these collisions, it carries out $n_{collisions}$ with the other atoms (See Fig.3).

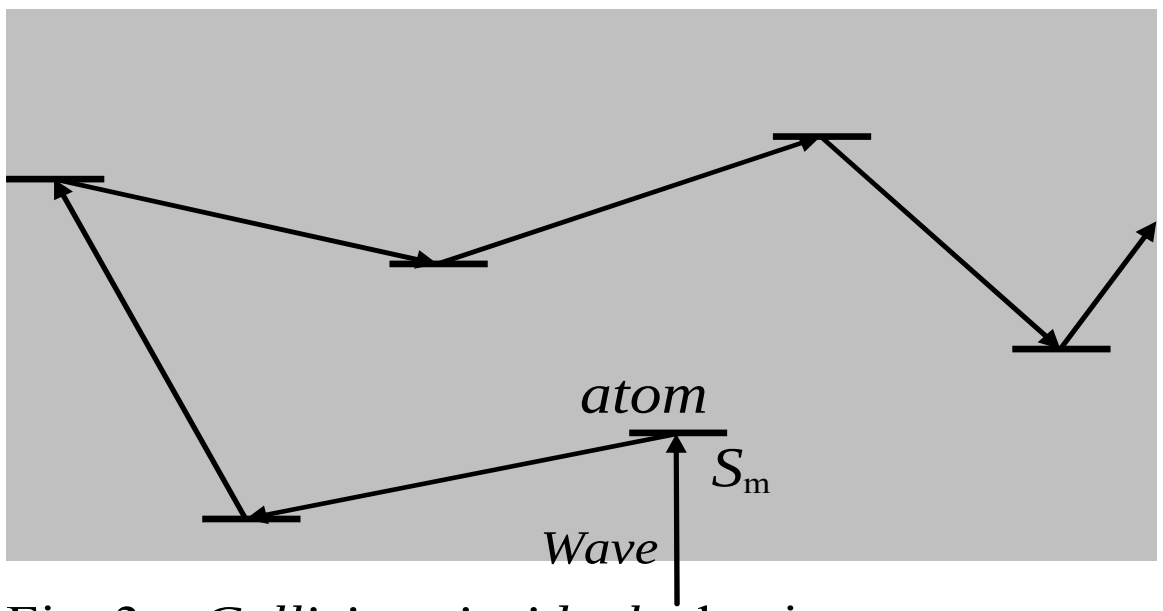

Fig. 3 – *Collisions inside the* lamina.

Thus, the total number of collisions in the volume $S\xi$ is

$$N_{collisions} = N_f + n_{collisions} = n_l S\phi_m + (n_l S\xi - n_m S\phi_m) = \qquad$$
$$= n_m S\xi \qquad (7)$$

The power density, $D$, of the radiation on the lamina can be expressed by

$$D = \frac{P}{S} = \frac{P}{N_f S_m} \qquad (8)$$

We can express the *total mean number of collisions in each atom*, $n_1$, by means of the following equation

$$n_1 = \frac{n_{total\ \ photons} N_{collisions}}{N} \qquad (9)$$

Since in each collision a *momentum* $h/\lambda$ is transferred to the atom, then the *total momentum* transferred to the lamina will be $\Delta p = (n_1 N)h/\lambda$. Therefore, in accordance with Eq. (1), we can write that

$$\frac{m_{g(l)}}{m_{i0(l)}} = \left\{1 - 2\left[\sqrt{1+\left[(n_1 N)\frac{\lambda_0}{\lambda}\right]^2} - 1\right]\right\} =$$

$$= \left\{1 - 2\left[\sqrt{1+\left[n_{total\ photons} N_{collisions}\frac{\lambda_0}{\lambda}\right]^2} - 1\right]\right\} \qquad (10)$$

Since Eq. (7) gives $N_{collisions} = n_l S \xi$, we get

$$n_{total\ \ photons} N_{collisions} = \left(\frac{P}{hf^2}\right)(n_l S \xi) \qquad (11)$$

Substitution of Eq. (11) into Eq. (10) yields

$$\frac{m_{g(l)}}{m_{i0(l)}} = \left\{1 - 2\left[\sqrt{1+\left[\left(\frac{P}{hf^2}\right)(n_l S\xi)\frac{\lambda_0}{\lambda}\right]^2} - 1\right]\right\} \qquad (12)$$

Substitution of $P$ given by Eq. (8) into Eq. (12) gives

$$\frac{m_{g(l)}}{m_{i0(l)}} = \left\{1 - 2\left[\sqrt{1+\left[\left(\frac{N_f S_m D}{f^2}\right)\left(\frac{n_l S\xi}{m_{i0(l)}c}\right)\frac{1}{\lambda}\right]^2} - 1\right]\right\} \qquad (13)$$

Substitution of $N_f \cong (n_l S_f)\phi_m$ and $S = N_f S_m$ into Eq. (13) results

$$\frac{m_{g(l)}}{m_{i0(l)}} = \left\{1 - 2\left[\sqrt{1+\left[\left(\frac{n_l^3 S_f^2 S_m^2 \phi_m^2 \xi D}{m_{i0(l)}cf^2}\right)\frac{1}{\lambda}\right]^2} - 1\right]\right\} \qquad (14)$$

where $m_{i0(l)} = \rho_{(l)} V_{(l)}$.

Now, considering that the lamina is inside an ELF electromagnetic field with $E$ and $B$, then we can write that [7]

$$D = \frac{n_{r(l)} E^2}{2\mu_0 c} \qquad (15)$$

Substitution of Eq. (15) into Eq. (14) gives

$$\frac{m_{g(l)}}{m_{i0(l)}} = \left\{1 - 2\left[\sqrt{1+\left[\left(\frac{n_{r(l)} n_l^3 S_f^2 S_m^2 \phi_m^2 \xi E^2}{2\mu_0 m_{i0(l)} c^2 f^2}\right)\frac{1}{\lambda}\right]^2} - 1\right]\right\} \qquad (16)$$

Note that $E = E_m \sin \omega t$. The average value for $E^2$ is equal to $\frac{1}{2}E_m^2$ because $E$ varies sinusoidaly ($E_m$ is the maximum value for $E$). On the other hand, $E_{rms} = E_m/\sqrt{2}$. Consequently we can replace $E^4$ for $E_{rms}^4$. Thus, for $\lambda = \lambda_{\text{mod}}$, the equation above can be rewritten as follows

$$\frac{m_{g(l)}}{m_{i0(l)}} = \left\{1 - 2\left[\sqrt{1+\left[\left(\frac{n_{r(l)} n_l^3 S_f^2 S_m^2 \phi_m^2 \xi E_{rms}^2}{2\mu_0 m_{i0(l)} c^2 f^2}\right)\frac{1}{\lambda_{\text{mod}}}\right]^2} - 1\right]\right\} \qquad (17)$$

Electrodynamics tells us that $E_{rms} = vB_{rms} = (c/n_{r(l)})B_{rms}$. Substitution of this expression into Eq. (17) gives

$$\chi = \frac{m_{g(l)}}{m_{i0(l)}} = \left\{1 - 2\left[\sqrt{1+\frac{n_l^6 S_f^4 S_m^4 \phi_m^4 \xi^2 B_{rms}^4}{4\mu_0^2 m_{i0(l)}^2 f^4 \lambda_{\text{mod}}^2 n_{r(l)}^2}} - 1\right]\right\} \qquad (18)$$

Since $\lambda_{\text{mod}} = \lambda/n_{r(l)}$ then Eq. (18) can be rewritten in the following form

$$\chi = \frac{m_{g(l)}}{m_{i0(l)}} = \left\{1 - 2\left[\sqrt{1+\frac{n_l^6 S_f^4 S_m^4 \phi_m^4 \xi^2 B_{rms}^4}{4\mu_0^2 m_{i0(l)}^2 c^2 f^2}} - 1\right]\right\} \qquad (19)$$

In order to calculate the expressions of $\chi_{Be}$ for the particular case of a *electron of the electrophere of a atom*, subjected to an external magnetic field $B_{rms}$ with frequency $f$, we must substitute in Eq. (19) $n_l$ for $1/V_e = 1/\frac{4}{3}\pi r_e^3$, $S_f$ for $(SSA_e)\rho_e V_e$ ($SSA_e$ is the *specific surface area* for electrons in this case: $SSA_e = \frac{1}{2}A_e/m_e = \frac{1}{2}A_e/\rho_e V_e = 2\pi r_{xe}^2/\rho_e V_e$), $S_m$ by $S_e = \pi r_{xe}^2$, $\xi$ by $\phi_m = 2r_{xe}$ and $m_{i0(l)}$ by $m_e$. The result is

$$\chi_{Be} = \left\{1 - 2\left[\sqrt{1 + \frac{45.56\pi^2 r_{xe}^{22} B_{rms}^4}{c^2 \mu_0^2 m_e^2 r_e^{18} f^2}} - 1\right]\right\} \qquad (20)$$

In order to calculate the value of $r_{xe}$ we start considering a hydrogen atom, where the electron spins around the proton with a velocity $v_e = 3 \times 10^6 m.s^{-1}$. The electrical force acting on the proton is $F_e = e^2/4\pi\varepsilon_0 r_1^2$, which is equal to the centrifuge force $F_c = m_p \omega_e^2 r_0$ where $\omega_e$ is the angular velocity of the electron and $r_0$ is the distance between the inertial center of the proton and the center of the moving proton (See Fig. 5, where we conclude that $2(r_0 + r_p) = r_{xp} + r_p$, where $r_{xp}$ is the radius of the sphere whose external area is equivalent to the increased area of the proton). Thus, we get $r_0 = \frac{1}{2}(r_{xp} - r_p)$.

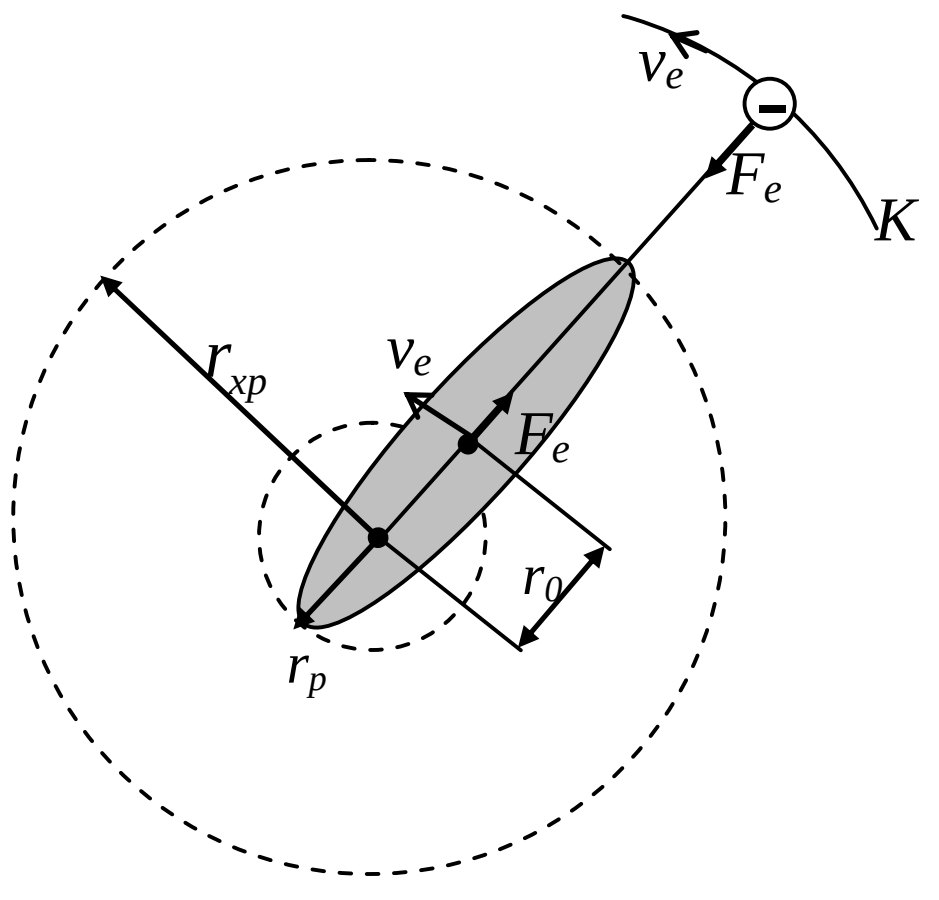

Fig. 5 – *The deformation of the proton.*

Substitution of this value into expression of $F_c = F_e$ gives

$$r_{xp} = \frac{e^2}{4\pi\varepsilon_0 m_p v_e^2} + r_p = 3.2 \times 10^{-14} m$$

Therefore, we can write that $r_{xp} = k_{xp} r_p$, where

$$k_{xp} = \frac{r_{xp}}{r_p} = 25.6$$

The electron is similarly deformed by the relative movement of the proton in respect to it. In this case, by analogy, we can write that

$$r_{xe} = \frac{e^2}{4\pi\varepsilon_0 m_e v_e^2} + r_e = 6.4 \times 10^{-11} m$$

and $r_{xe} = k_{xe} r_e$, where $r_{xe}$ is the radius of the sphere whose external area is equivalent to the increased area of the electron. The radius of *free* electron is $r_e = 6.87 \times 10^{-14} m$ (See *Appendix* A). However, for *electrons in the atomic eletrosphere of atoms* the value of $r_e$ must be calculated starting from Quantum Mechanics. The wave packet that describes the electron satisfies an *uncertainty principle* $(\Delta p \Delta x \geq \frac{1}{2}\hbar)$, where $\Delta p = \hbar \Delta k$ and $\Delta k$ is the approximate extension of the wave packet. Thus, we can write that $(\Delta k \Delta x \geq \frac{1}{2})$. For the ``square" packet the full width in $k$ is $\Delta k = 2\pi/\lambda_0$ ($\lambda_0 = h/m_e c$ is the average wavelength). The width in $x$ is a little harder to define, but, lets use the first node in the probability found at $(2\pi/\lambda_0)x/2 = \pi$ or $x = \lambda_0$. So, the width of the wave packet is twice this or $\Delta x = 2\lambda_0$. Obviously, $2r_e$ cannot be greater than $\Delta x$, i.e., $r_e$ *must be smaller and close to* $\lambda_0 = h/m_e c = 2.43 \times 10^{-12} m$. Then, assuming that $r_e \cong 2.4 \times 10^{-12} m$, we get

$$k_{xe} = \frac{r_{xe}}{r_e} = 26.6$$

Note that $k_{xe} \cong k_{xp}$.

Substitution of these values into Eq. (20) gives

$$\chi_{Be} = \left\{1 - 2\left[\sqrt{1 + 3.8\times10^{57}\frac{k_{xe}^{22} r_e^4 B_{rms}^4}{f^2}} - 1\right]\right\} =$$

$$= \left\{1 - 2\left[\sqrt{1 + 2.8\times10^{42}\frac{B_{rms}^4}{f^2}} - 1\right]\right\} \qquad (21)$$

Similarly, in the case of *proton* and *neutron* we can write that

$$\chi_{Bp} = \left\{1 - 2\left[\sqrt{1 + \frac{45.56\pi^2 k_{xp}^{22} r_p^4 B_{rms}^4}{c^2 \mu_0^2 m_p^2 f^2}} - 1\right]\right\} \qquad (22)$$

$$\chi_{Bn} = \left\{1 - 2\left[\sqrt{1 + \frac{45.56\pi^2 r_n^4 B_{rms}^4}{\mu_0^2 m_n^2 c^2 f^2}} - 1\right]\right\} \qquad (23)$$

In the case of the neutron, $k_{xn} = 1$ due to its electric charge be null. The radius of *protons inside the atoms* (nuclei) is $r_p = 1.2\times10^{-15} m$ [8,9], $r_n \cong r_p$, then we obtain from Eqs. (22) and (23) the following expressions:

$$\chi_{Bp} = \left\{1 - 2\left[\sqrt{1 + 2.2\times10^{22}\frac{B_{rms}^4}{f^2}} - 1\right]\right\} \qquad (24)$$

$$\chi_{Bn} = \left\{1 - 2\left[\sqrt{1 + 2.35\times10^{9}\frac{B_{rms}^4}{f^2}} - 1\right]\right\} \qquad (25)$$

When a strong magnetic field $B_{rms}$ is applied on the atom, the enormous value of $\chi_{Be}$ (See Eq. 21) makes the gravitational force between the electrons greater than the electric force due to its charges, and consequently, the electrons are joined in pairs (Cooper pairs)[10]. However, due to the vales of $\chi_{Be}$ and $\chi_{Bp}$, the gravitational attraction between the electrons of the K shell and the protons of the nucleus becomes greater than the nuclear force, i.e.,

$$G\frac{m_{gp} m_{ge}}{r_1^2} = \chi_{Bp}\chi_{Be} G\frac{m_p m_e}{r_1^2} > F_N \qquad (26)$$

Then, *the proton more weakly bound to the nucleus is ejected towards the nearest electron*. When they collide, there occurs the formation of *one neutron* and one neutrino, according the well-known reaction $p + e \rightarrow n + \nu_e$. Since the neutron is beyond the reach of the nuclear force, it is not attracted to the nucleus and leaves the atom. *The final result is that the atom loses a proton and an electron and is transformed in a new atom*. But the transmutation is not completed until the magnetic field is turned off. When this occurs the Cooper pairs are broken, and the *new atom* leaves the transitory state, and passes to the normal state.

In order to satisfy the condition expressed by Eq. (26), we must have

$$\chi_{Bp}\chi_{Be} > \frac{F_N r_1^2}{G m_p m_e} \approx 3\times10^{48} \qquad (27)$$

By substitution of Eq. (21) and (22) into Eq. (27), we obtain

$$\frac{B_{rms}^2}{f} > 6\times10^7$$

Thus, for $f = 0.1Hz$ we conclude that the required value of $B_{rms}$ is

$$B_{rms} > 2{,}500T$$

This means that, if we subject, for example, an amount of $^{198}$Hg (80 electrons, 80 protons, 118 neutrons) to a magnetic field with *$B_{rms}$>2,500 T* and frequency 0.1Hz the $^{198}$Hg loses 1 proton and 1 electron and consequently will be transmuted to $^{197}$Au (79 electrons, 79 protons, 118 neutrons) when the magnetic field is turned off. Besides the transformation of mercury into gold, we can make several transmutations. For example, if the $^{197}$Au is after subjected to the same magnetic field it will be transmuted to $^{196}$Pt (78 electrons, 78 protons, 118 neutrons). Similarly, if $^{110}$Cd (48 electrons, 48 protons, 62 neutrons) is subjected to the mentioned field it will be transmuted to $^{109}$Ag (47 electrons, 47 protons, 62 neutrons). Also $^{235}$U can be easily obtained by this process, i.e., if we subject an amount of $^{236}$Np (93 electrons, 93 protons, 143 neutrons) to the magnetic field with *>2,500 T* and frequency 0.1Hz, the $^{236}$Np loses 1 proton and 1 electron and consequently will be transmuted to $^{235}$U (92 electrons, 92 protons, 143 neutrons).

## **Appendix A**: *The "Geometrical Radii" of Electron and Proton*

It is known that the frequency of oscillation of a simple spring oscillator is

$$f = \frac{1}{2\pi}\sqrt{\frac{K}{m}} \qquad (A1)$$

where $m$ is the inertial mass attached to the spring and $K$ is the spring constant (in $N \cdot m^{-1}$). In this case, the restoring *force* exerted by the spring *is linear* and given by

$$F = -Kx \qquad (A2)$$

where $x$ is the displacement from the equilibrium position.

Now, consider the gravitational force: For example, above the surface of the Earth, the force follows the familiar Newtonian function, i.e., $F = -GM_{g\oplus}m_g/r^2$, where $M_{g\oplus}$ is the mass of Earth, $m_g$ is the gravitational mass of a particle and $r$ is the distance between the centers. *Below* Earth's surface the *force is linear* and given by

$$F = -\frac{GM_{g\oplus}m_g}{R_\oplus^3}r \qquad (A3)$$

where $R_\oplus$ is the radius of Earth.

By comparing (A3) with (A2) we obtain

$$\frac{K}{m_g} = \frac{K}{\chi\, m} = \frac{GM_{g\oplus}}{R_\oplus^3}\left(\frac{r}{x}\right) \qquad (A4)$$

Making $x = r = R_\oplus$, and substituting (A4) into (A1) gives

$$f = \frac{1}{2\pi}\sqrt{\frac{GM_{g\oplus}\chi}{R_\oplus^3}} \qquad (A5)$$

In the case of an *electron* and a *positron*, we substitute $M_{g\oplus}$ by $m_{ge}$, $\chi$ by $\chi_e$ and $R_\oplus$ by $R_e$, where $R_e$ is the radius of electron (or positron). Thus, Eq. (A5) becomes

$$f = \frac{1}{2\pi}\sqrt{\frac{Gm_{ge}\chi_e}{R_e^3}} \qquad (A6)$$

The value of $\chi_e$ varies with the density of energy [5]. When the electron and the positron are distant from each other and the local density of energy is small, the value of $\chi_e$ becomes very close to 1. However, *when the electron and the positron are penetrating one another*, the energy densities in each particle become very strong due to the proximity of their electrical charges $e$ and, consequently, the value of $\chi_e$ strongly increases. In order to calculate the value of $\chi_e$ under these conditions ($x = r = R_e$), we start from the expression of correlation between *electric charge* $q$ and *gravitational mass*, obtained in a previous work [5]:

$$q = \sqrt{4\pi\varepsilon_0 G}\; m_{g(imaginary)}\, i \qquad (A7)$$

where $m_{g(imaginary)}$ is the *imaginary* gravitational mass, and $i = \sqrt{-1}$.

In the case of *electron*, Eq. (A7) gives

$$\begin{aligned} q_e &= \sqrt{4\pi\varepsilon_0 G}\; m_{ge(imaginary)}\, i = \\ &= \sqrt{4\pi\varepsilon_0 G}\left(\chi_e m_{i0e(imaginary)} i\right) = \\ &= \sqrt{4\pi\varepsilon_0 G}\left(-\chi_e \tfrac{2}{\sqrt{3}} m_{i0e(real)} i^2\right) = \\ &= \sqrt{4\pi\varepsilon_0 G}\left(\tfrac{2}{\sqrt{3}}\chi_e m_{i0e(real)}\right) = -1.6\times10^{-19}C \qquad (A8) \end{aligned}$$

where we obtain

$$\chi_e = -1.8\times10^{21} \qquad (A9)$$

This is therefore, the value of $\chi_e$ increased by the strong density of energy produced by the electrical charges $e$ of the two particles, under previously mentioned conditions.

Given that $m_{ge} = \chi_e m_{i0e}$, Eq. (A6) yields

$$f = \frac{1}{2\pi}\sqrt{\frac{G\chi_e^2 m_{i0e}}{R_e^3}} \qquad (A10)$$

From Quantum Mechanics, we know that

$$hf = m_{i0}c^2 \qquad (A11)$$

where $h$ is the Planck's constant. Thus, in the case of $m_{i0} = m_{i0e}$ we get

$$f = \frac{m_{i0e}c^2}{h} \qquad (A12)$$

By comparing (A10) and (A12) we conclude that

$$\frac{m_{i0e}c^2}{h} = \frac{1}{2\pi}\sqrt{\frac{G\chi_e^2 m_{i0e}}{R_e^3}} \qquad (A13)$$

Isolating the radius $R_e$, we get:

$$R_e = \left(\frac{G}{m_{i0e}}\right)^{\frac{1}{3}}\left(\frac{\chi_e h}{2\pi\ c^2}\right)^{\frac{2}{3}} = 6.87\times10^{-14}m \quad (A14)$$

Compare this value with the *Compton sized electron*, which predicts $R_e = 3.86\times10^{-13}m$ and also with standardized result recently obtained of $R_e = 4-7\times10^{-13}m$ [11].

In the case of *proton*, we have

$$\begin{aligned} q_p &= \sqrt{4\pi\varepsilon_0 G}\ m_{gp(imaginary)}\, i = \\ &= \sqrt{4\pi\varepsilon_0 G}\left(\chi_p m_{i0p(imaginary)} i\right) = \\ &= \sqrt{4\pi\varepsilon_0 G}\left(-\chi_p \tfrac{2}{\sqrt{3}} m_{i0p(real)} i^2\right) = \\ &= \sqrt{4\pi\varepsilon_0 G}\left(\tfrac{2}{\sqrt{3}}\chi_p m_{i0p(real)}\right) = -1.6\times10^{-19}C \quad (A15) \end{aligned}$$

where we obtain

$$\chi_p = -9.7\times10^{17} \qquad (A16)$$

Thus, the result is

$$R_p = \left(\frac{G}{m_{i0p}}\right)^{\frac{1}{3}}\left(\frac{\chi_p h}{2\pi\ c^2}\right)^{\frac{2}{3}} = 3.72\times10^{-17}m \quad (A17)$$

Note that these radii, given by Equations $(A14)$ and $(A17)$, are the radii of *free* electrons and *free* protons (when the particle and antiparticle (in isolation) penetrate themselves mutually).

Inside the atoms (nuclei) the radius of protons is well-known. For example, protons, as the hydrogen nuclei, have a radius given by $R_p \cong 1.2\times10^{-15}m$ [8,9]. The strong increase in respect to the value given by Eq. (A17) is due to the interaction with the electron of the atom.

# Ultrafast Conversion of Graphite to Diamond in Gravitational Pressure Apparatus

**Fran De Aquino**
Maranhao State University, Physics Department, S.Luis/MA, Brazil.

Currently the artificial production of diamond is very expensive because it consumes large amounts of energy in order to produce a single diamond. Here, we propose a new type of press based on the *intensification of the gravitational acceleration*. This Gravitational Press can generate pressures several times more intense than the *80GPa* required for *ultrafast transformation of graphite into diamond*. In addition, due to the enormous pressure that the Gravitational Press can produce, the "synthesis capsule" may be very large (up to about 1000 $cm^3$ in size). This is sufficient to produce diamonds with up to 100 carats (20g) or more. On the other hand, besides the ultrafast conversion, the energy required for the Gravitational Presses is very low, in such a way that the production cost of the diamonds becomes very low, what means that they could be produced on a large scale.

**Key words:** Modified theories of gravity, High-pressure apparatus, Graphite, Diamond.
PACS: 04.50.Kd, 07.35.+k, 81.05.uf , 81.05.ug

## 1. Introduction

After the discovery that diamond was pure carbon, many attempts were made to convert various carbon forms into diamond. Converting graphite into diamond has been a long held dream of alchemists. The artificial production of diamond was first achieved by H.T Hall in 1955. He used a press capable of producing pressures above 10 GPa and temperatures above 2,000 °C [1].

Today, there are several methods to produce synthetic diamond. The more widely utilized method uses high pressure and high temperature (HPHT) of the order of 10 GPa and 2500°C *during many hours* in order to produce a single diamond. The fact that this process requires high pressure and high temperatures *during a long time* means that it consumes large amounts of energy, and this is the reason why the production cost of artificial diamond is so expensive. The second method, using chemical vapor deposition (CVD), creates a carbon plasma over a substrate onto which the carbon atoms deposit to form diamond. Other methods include explosive formation and sonication of graphite solutions [2,3,4].

In the HPHT method, there are three main press designs used to supply the pressure and temperature necessary to produce synthetic diamond: the *Belt press*, the *Cubic press* and the split-sphere (*BARS*) press. Typical pressures and temperatures achievable are of the order of 10 GPa and 2500°C [5].

Diamonds may be formed in the Earth's mantle mainly by direct transition, graphite to diamond or by systems involving carbon dissolved in molten metals. The classic high-pressure, high-temperature synthesis of diamond utilizes molten transition metals as solvent/catalysts. *Converting diamond from graphite in the absence of a catalyst requires pressures that are significantly higher than those at equilibrium coexistence* [6-12]. At lower temperatures, the formation of the metastable hexagonal polymorph of diamond is favored instead of the more stable cubic diamond [7, 10-12]. These phenomena cannot be explained by the concerted mechanism suggested in previous theoretical studies [13-17]. However, recently Michele Parrinello, Professor of Computational Science at ETH Zurich, and his team have developed a method by which they have successfully simulated this phase transition accurately and adequately using computer models [18]. Instead of happening concerted, all at once, the conversion evidently takes place in a step by step process involving the formation of a diamond seed in the graphite, *which is then transformed completely at high pressure*. In quantitative agreement with the *ab initio* calculations of Tateyama at al. [19], the stability of diamond relative to graphite increases with pressure whereas the barrier separating two phases decreases. Parrinello's work shows that *at a pressure of 80 GPa and temperature between* 0 *and* 1,000*K graphite* reaches a lattice instability point and *undergoes an ultrafast transformation to diamond* as was previously observed in *ab initio* simulations by Scandolo et al [20].

Here, we propose a new type of press based on the *intensification of the gravitational acceleration*[*]. This press can generate pressures several times more intense[†] than the *80GPa* required for the *ultrafast transformation of graphite to diamond*. In addition, due to the enormous pressure that the Gravitational Press can produce (*>>80GPa*), the ceramic cube ("synthesis capsule") can be very large (up to about 1000 $cm^3$ in size). This is sufficient to produce diamonds up to 100 carats (20g) or more. On the other hand, besides the ultrafast conversion, the energy required for the Gravitational Presses is very low, in such a way that the production cost of the diamonds becomes very low, what means that they could be produced on a large scale.

## 2. Theory

From the quantization of gravity it follows that the *gravitational mass* $m_g$ and the *inertial mass* $m_i$ are correlated by means of the following factor [21]:

$$\chi = \frac{m_g}{m_{i0}} = \left\{1 - 2\left[\sqrt{1 + \left(\frac{\Delta p}{m_{i0}c}\right)^2} - 1\right]\right\} \tag{1}$$

where $m_{i0}$ is the *rest* inertial mass of the particle and $\Delta p$ is the variation in the particle's *kinetic momentum*; $c$ is the speed of light.

When $\Delta p$ is produced by the absorption of a photon with wavelength $\lambda$, it is expressed by $\Delta p = h/\lambda$. In this case, Eq. (1) becomes

$$\frac{m_g}{m_{i0}} = \left\{1 - 2\left[\sqrt{1 + \left(\frac{h/m_{i0}c}{\lambda}\right)^2} - 1\right]\right\} = \left\{1 - 2\left[\sqrt{1 + \left(\frac{\lambda_0}{\lambda}\right)^2} - 1\right]\right\} \tag{2}$$

where $\lambda_0 = h/m_{i0}c$ is the *De Broglie wavelength* for the particle with *rest* inertial mass $m_{i0}$.

It has been shown that there is an additional effect - *Gravitational Shielding* effect - produced by a substance whose gravitational mass was reduced or made negative [22]. The effect extends beyond substance (gravitational shielding) , up to a certain distance from it (along the central axis of gravitational shielding). This effect shows that in this region the gravity acceleration, $g_1$, is reduced at the same proportion, i.e., $g_1 = \chi_1 g$ where $\chi_1 = m_g/m_{i0}$ and $g$ is the gravity acceleration *before* the gravitational shielding). Consequently, *after a second gravitational shielding*, the gravity will be given by $g_2 = \chi_2 g_1 = \chi_1 \chi_2 g$, where $\chi_2$ is the value of the ratio $m_g/m_{i0}$ for the *second* gravitational shielding. In a generalized way, we can write that after the *nth* gravitational shielding the gravity, $g_n$, will be given by

$$g_n = \chi_1 \chi_2 \chi_3 \cdots \chi_n g \tag{3}$$

This possibility shows that, by means of a battery of gravitational shieldings, we can make particles acquire enormous accelerations. In practice, this is the basis to the conception of the *Gravitational Press*.

---

[*] De Aquino, F. (2008) *Process and Device for* Controlling the Locally the Gravitational Mass and the Gravity Acceleration, BR Patent Number: PI0805046-5, July 31, 2008.

[†] The limit is basically determined by the compression resistance of the material of the piston and anvils of the Gravitational Press since it can produce pressures far beyond 1000 GPa.

From Electrodynamics we know that when an electromagnetic wave with frequency $f$ and velocity $c$ incides on a material with relative permittivity $\varepsilon_r$, relative magnetic permeability $\mu_r$ and electrical conductivity $\sigma$, its *velocity is reduced* to $v = c/n_r$ where $n_r$ is the index of refraction of the material, given by [23]

$$n_r = \frac{c}{v} = \sqrt{\frac{\varepsilon_r \mu_r}{2}\left(\sqrt{1+(\sigma/\omega\varepsilon)^2}+1\right)} \qquad (4)$$

If $\sigma >> \omega\varepsilon$, $\omega = 2\pi f$, Eq. (4) reduces to

$$n_r = \sqrt{\frac{\mu_r \sigma}{4\pi\varepsilon_0 f}} \qquad (5)$$

Thus, the wavelength of the incident radiation (See Fig. 1) becomes

$$\lambda_{mod} = \frac{v}{f} = \frac{c/f}{n_r} = \frac{\lambda}{n_r} = \sqrt{\frac{4\pi}{\mu f \sigma}} \qquad (6)$$

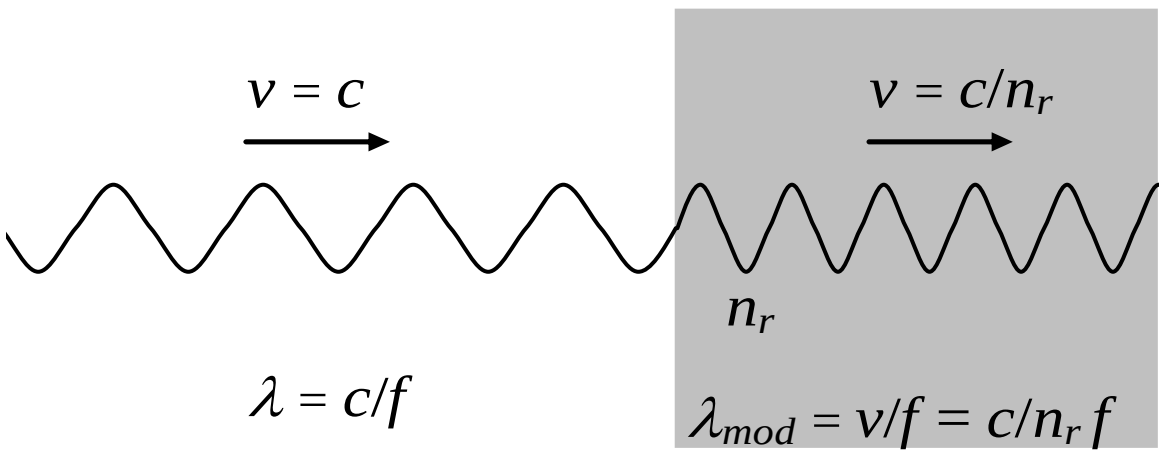

Fig. 1 – *Modified Electromagnetic Wave.* The wavelength of the electromagnetic wave can be strongly reduced, but its frequency remains the same.

If a lamina with thickness equal to $\xi$ contains $n$ atoms/m$^3$, then the number of atoms per area unit is $n\xi$. Thus, if the electromagnetic radiation with frequency $f$ incides on an area $S$ of the lamina it reaches $nS\xi$ atoms. If it incides on the *total area of the lamina*, $S_f$, then the total number of atoms reached by the radiation is $N = nS_f\xi$. The number of atoms per unit of volume, $n$, is given by

$$n = \frac{N_0 \rho}{A} \qquad (7)$$

where $N_0 = 6.02\times10^{26}\, atoms/kmole$ is the Avogadro's number; $\rho$ is the matter density of the lamina (in *kg/m*$^3$) and $A$ is the molar mass(*kg/kmole*).

When an electromagnetic wave incides on the lamina, it strikes $N_f$ front atoms, where $N_f \cong (n S_f)\phi_m$, $\phi_m$ is the "diameter" of the atom. Thus, the electromagnetic wave incides effectively on an area $S = N_f S_m$, where $S_m = \frac{1}{4}\pi\phi_m^2$ is the cross section area of one atom. After these collisions, it carries out $n_{collisions}$ with the other atoms (See Fig.2).

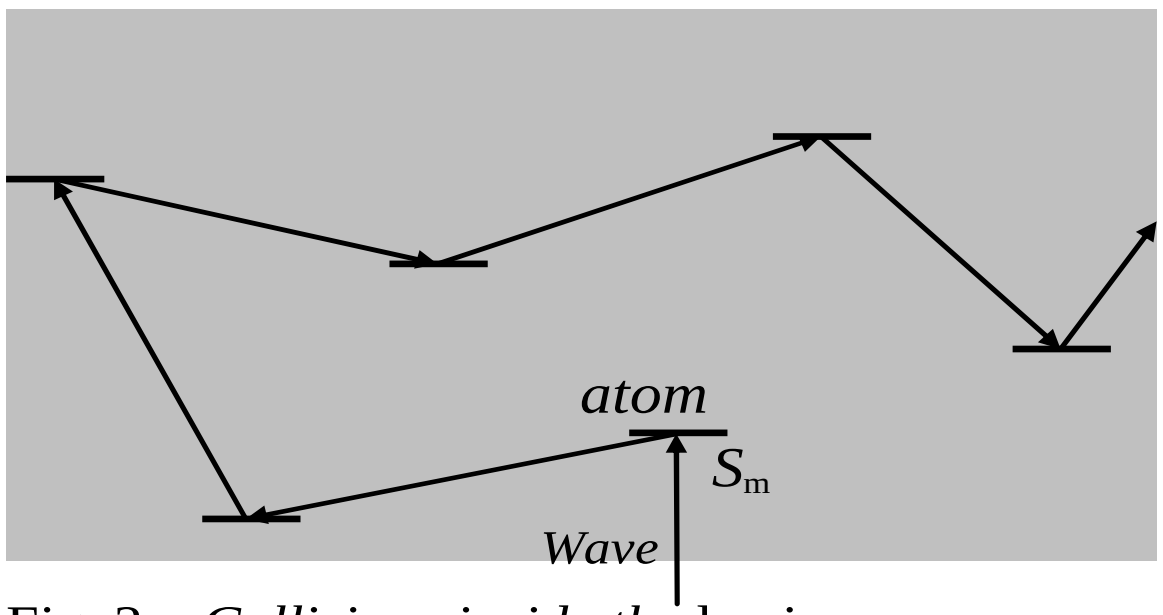

Fig. 2 – *Collisions inside the* lamina.

Thus, the total number of collisions in the volume $S\xi$ is

$$N_{collisions} = N_f + n_{collisions} = n_l S\phi_m + (n_l S\xi - n_m S\phi_m) = \\ = n_m S\xi \qquad (8)$$

The power density, $D$, of the radiation on the lamina can be expressed by

$$D = \frac{P}{S} = \frac{P}{N_f S_m} \qquad (9)$$

We can express the *total mean number of collisions in each atom*, $n_1$, by means of the following equation

$$n_1 = \frac{n_{total\ photons} N_{collisions}}{N} \qquad (10)$$

Since in each collision a *momentum* $h/\lambda$ is transferred to the atom, then the *total momentum* transferred to the lamina will be

$\Delta p = (n_1 N)h/\lambda$. Therefore, in accordance with Eq. (1), we can write that

$$\frac{m_{g(l)}}{m_{i0(l)}} = \left\{1 - 2\left[\sqrt{1 + \left[(n_1 N)\frac{\lambda_0}{\lambda}\right]^2} - 1\right]\right\} =$$

$$= \left\{1 - 2\left[\sqrt{1 + \left[n_{total\ photons} N_{collisions}\frac{\lambda_0}{\lambda}\right]^2} - 1\right]\right\} \quad (11)$$

Since Eq. (8) gives $N_{collisions} = n_l S\xi$, we get

$$n_{total\ \ photons} N_{collisions} = \left(\frac{P}{hf^2}\right)(n_l S\xi) \qquad (12)$$

Substitution of Eq. (12) into Eq. (11) yields

$$\frac{m_{g(l)}}{m_{i0(l)}} = \left\{1 - 2\left[\sqrt{1 + \left[\left(\frac{P}{hf^2}\right)(n_l S\xi)\frac{\lambda_0}{\lambda}\right]^2} - 1\right]\right\} \quad (13)$$

Substitution of $P$ given by Eq. (9) into Eq. (13) gives

$$\frac{m_{g(l)}}{m_{i0(l)}} = \left\{1 - 2\left[\sqrt{1 + \left[\left(\frac{N_f S_m D}{f^2}\right)\left(\frac{n_l S\xi}{m_{i0(l)} c}\right)\frac{1}{\lambda}\right]^2} - 1\right]\right\} \quad (14)$$

Substitution of $N_f \cong (n_l S_f)\phi_m$ and $S = N_f S_m$ into Eq. (14) results

$$\frac{m_{g(l)}}{m_{i0(l)}} = \left\{1 - 2\left[\sqrt{1 + \left[\left(\frac{n_l^3 S_f^2 S_m^2 \phi_m^2 \xi D}{m_{i0(l)} c f^2}\right)\frac{1}{\lambda}\right]^2} - 1\right]\right\} \quad (15)$$

where $m_{i0(l)} = \rho_{(l)} V_{(l)}$.

Now, considering that the lamina is inside an ELF electromagnetic field with $E$ and $B$, then we can write that [24]

$$D = \frac{n_{r(l)} E^2}{2\mu_0 c} \qquad (16)$$

Substitution of Eq. (16) into Eq. (15) gives

$$\frac{m_{g(l)}}{m_{i0(l)}} = \left\{1 - 2\left[\sqrt{1 + \left[\left(\frac{n_{r(l)} n_l^3 S_f^2 S_m^2 \phi_m^2 \xi E^2}{2\mu_0 m_{i0(l)} c^2 f^2}\right)\frac{1}{\lambda}\right]^2} - 1\right]\right\} \quad (17)$$

In the case in which the area $S_f$ is just the *area of the cross-section of the lamina* $(S_\alpha)$, we obtain from Eq. (17), considering that $m_{i0(l)} = \rho_{(l)} S_\alpha \xi$, the following expression

$$\frac{m_{g(l)}}{m_{i0(l)}} = \left\{1 - 2\left[\sqrt{1 + \left[\left(\frac{n_{r(l)} n_l^3 S_\alpha S_m^2 \phi_m^2 E^2}{2\mu_0 \rho_{(l)} c^2 f^2}\right)\frac{1}{\lambda}\right]^2} - 1\right]\right\} \quad (18)$$

If the electrical conductivity of the lamina, $\sigma_{(l)}$, is such that $\sigma_{(l)} >> \omega\varepsilon$, then the value of $\lambda$ is given by Eq. (6), i.e.,

$$\lambda = \lambda_{\text{mod}} = \sqrt{\frac{4\pi}{\mu f \sigma}} \qquad (19)$$

Substitution of Eq. (19) into Eq. (18) gives

$$\frac{m_{g(l)}}{m_{i0(l)}} = \left\{1 - 2\left[\sqrt{1 + \frac{n_{r(l)}^2 n_l^6 S_\alpha^2 S_m^4 \phi_m^4 \sigma_{(l)} E^4}{16\pi\mu_0 \rho_{(l)}^2 c^4 f^3}} - 1\right]\right\} \quad (20)$$

Note that $E = E_m \sin\omega t$. The average value for $E^2$ is equal to $\frac{1}{2}E_m^2$ because $E$ varies sinusoidaly ($E_m$ is the maximum value for $E$). On the other hand, $E_{rms} = E_m/\sqrt{2}$. Consequently we can change $E^4$ by $E_{rms}^4$, and the equation above can be rewritten as follows

$$\chi = \frac{m_{g(l)}}{m_{i0(l)}} =$$

$$= \left\{1 - 2\left[\sqrt{1 + \frac{n_{r(l)}^2 n_l^6 S_\alpha^2 S_m^4 \phi_m^4 \sigma_{(l)} E_{rms}^4}{16\pi\mu_0 \rho_{(l)}^2 c^4 f^3}} - 1\right]\right\} \quad (21)$$

Now consider the system (*Gravitational Press*) shown in Fig.3.

Inside the system there is a *dielectric tube* ($\varepsilon_r \cong 1$) with the following characteristics: $\alpha = 60mm$, $S_\alpha = \pi\alpha^2/4 = 2.83\times10^{-3}m^2$. Inside the tube there is an *Aluminum sphere* with 30mm radius and mass $M_{gs} = 0.30536kg$. The tube is filled with *air* at ambient temperature and 1atm. Thus, inside the tube, the air density is

$$\rho_{air} = 1.2 \ \ kg.m^{-3} \qquad (22)$$

The number of atoms of air (Nitrogen) per unit of volume, $n_{air}$, according to Eq.(7), is given by

$$n_{air} = \frac{N_0\rho_{air}}{A_N} = 5.16\times10^{25}\,atoms/m^3 \qquad (23)$$

The *parallel metallic plates* (p), shown in Fig.3 are subjected to different drop voltages. The two sets of plates (*D*), placed on the extremes of the tube, are subjected to $V_{(D)rms} = 1.64kV$ at $f = 1Hz$, while the central set of plates (*A*) is subjected to $V_{(A)rms} = 19.7kV$ at $f = 1Hz$. Since $d = 98mm$, then the intensity of the electric field, which passes through the 36 *cylindrical air laminas* (each one with 5mm thickness) of the *two* sets (*D*), is

$$E_{(D)rms} = V_{(D)rms}/d = 1.67\times10^4 V/m$$

and the intensity of the electric field, which passes through the 7 *cylindrical air laminas* of the central set (*A*), is given by

$$E_{(A)rms} = V_{(A)rms}/d = 2.012\times10^5 V/m$$

Note that the *metallic rings* (5*mm* thickness) are positioned in such way to block the electric field out of the cylindrical air laminas. The objective is to turn each one of these laminas into a *Gravity Control Cell* (GCC) [22]. Thus, the system shown in Fig. 3 has 3 sets of GCC. Two with 18 GCC each and one with 7 GCC. The two sets with 18 GCC each are positioned at the extremes of the tube (*D*). They work as gravitational *decelerator* while the other set with 7 GCC (*A*) works as a gravitational *accelerator*, intensifying the gravity acceleration produced by the mass $M_{gs}$ of the Aluminum sphere. According to Eq. (3), this gravity, after the $7^{th}$ GCC becomes $g_7 = \chi^7 GM_{gs}/r_0^2$, where $\chi = m_{g(l)}/m_{i(l)}$ given by Eq. (21) and $r_0 = 35mm$ is the distance between the center of the Aluminum sphere and the surface of the first GCC of the set (A).

The objective of the sets (*D*), with 18 GCC each, is to reduce strongly the value of the external gravity along the axis of the tube. In this case, the value of the external gravity, $g_{ext}$, is reduced by the factor $\chi_d^{18} g_{ext}$, where $\chi_d = 10^{-2}$. For example, if the base BS of the system is positioned on the Earth surface, then $g_{ext} = 9.81m/s^2$ is reduced to $\chi_d^{18} g_{ext}$ and, after the set A, it is increased by $\chi^7$. Since the system is designed for $\chi = -308.5$, then the gravity acceleration on the sphere becomes $\chi^7\chi_d^{18} g_{ext} = 2.6\times10^{-18} m/s^2$, this value is much smaller than $g_{sphere} = GM_{gs}/r_s^2 = 2.26\times10^{-8} m/s^2$.

The electrical conductivity of air, *inside the dielectric tube*, is equal to the electrical conductivity of Earth's atmosphere near the land, whose average value is $\sigma_{air} \cong 1\times10^{-14} S/m$ [25]. This value is of fundamental importance in order to obtain the convenient values of $\chi$ and $\chi_d$, which are given by Eq. (21), i.e.,

$$\chi = \left\{1 - 2\left[\sqrt{1 + \frac{n_{r(air)}^2 n_{air}^6 S_\alpha^2 S_m^4 \phi_m^4 \sigma_{air} E_{(A)rms}^4}{16\pi\mu_0\rho_{air}^2 c^4 f^3}} - 1\right]\right\} =$$

$$= \left\{1 - 2\left[\sqrt{1 + 1.480\times10^{-17} E_{(A)rms}^4} - 1\right]\right\} \qquad (24)$$

$$\chi_d = \left\{1 - 2\left[\sqrt{1 + \frac{n^2_{r(air)} n^6_{air} S^2_\alpha S^4_m \phi^4_m \sigma_{air} E^4_{(D)rms}}{16\pi\mu_0\rho^2_{air} c^4 f^3}} - 1\right]\right\} =$$

$$= \left\{1 - 2\left[\sqrt{1 + 1.48\times10^{-17} E^4_{(D)rms}} - 1\right]\right\} \qquad (25)$$

where $n_{r(air)} = \sqrt{\varepsilon_r \mu_r} \cong 1$, since $(\sigma << \omega\varepsilon)$; $n_{air} = 5.16\times10^{25}\,atoms/m^3$, $\phi_m = 1.55\times10^{-10}m$, $S_m = \pi\phi_m^2/4 = 1.88\times10^{-20}m^2$ and $f = 1Hz$. Since $E_{(A)rms} = 2.012\times10^5\,V/m$ and $E_{(D)rms} = 1.67\times10^4\,V/m$, we get

$$\chi = -308.5 \qquad (26)$$

and

$$\chi_d \cong 10^{-2} \qquad (27)$$

Then, the gravitational acceleration upon the piston of the Gravitational Press shown in Fig. 3 is equal to the value of the gravitational acceleration *after the 7th gravitational shielding*, i.e.,

$$g_7 = \chi^7 g = -\chi^7 \frac{GM_{gs}}{r_0^2} \cong 4.4\times10^9\,m/s^2 \qquad (28)$$

If the mass of the piston is $m_{piston} = 15kg$ with 20cm diameter then the *pressure* upon the *cubic-anvil apparatus* (Fig. 4) is

$$p = \frac{F}{S} = \frac{m_{piston}g_7}{\pi(0.1)^2} = 2\times10^{12}\,N/m^2 = 2000GPa$$

It is important to note that the pressure can be easily increased by increasing the value of $\chi$. However, the pressure limit is basically determined by the compression resistance of the material of the piston and anvils of the Gravitational Press since it can produce pressures far beyond 2000 GPa.

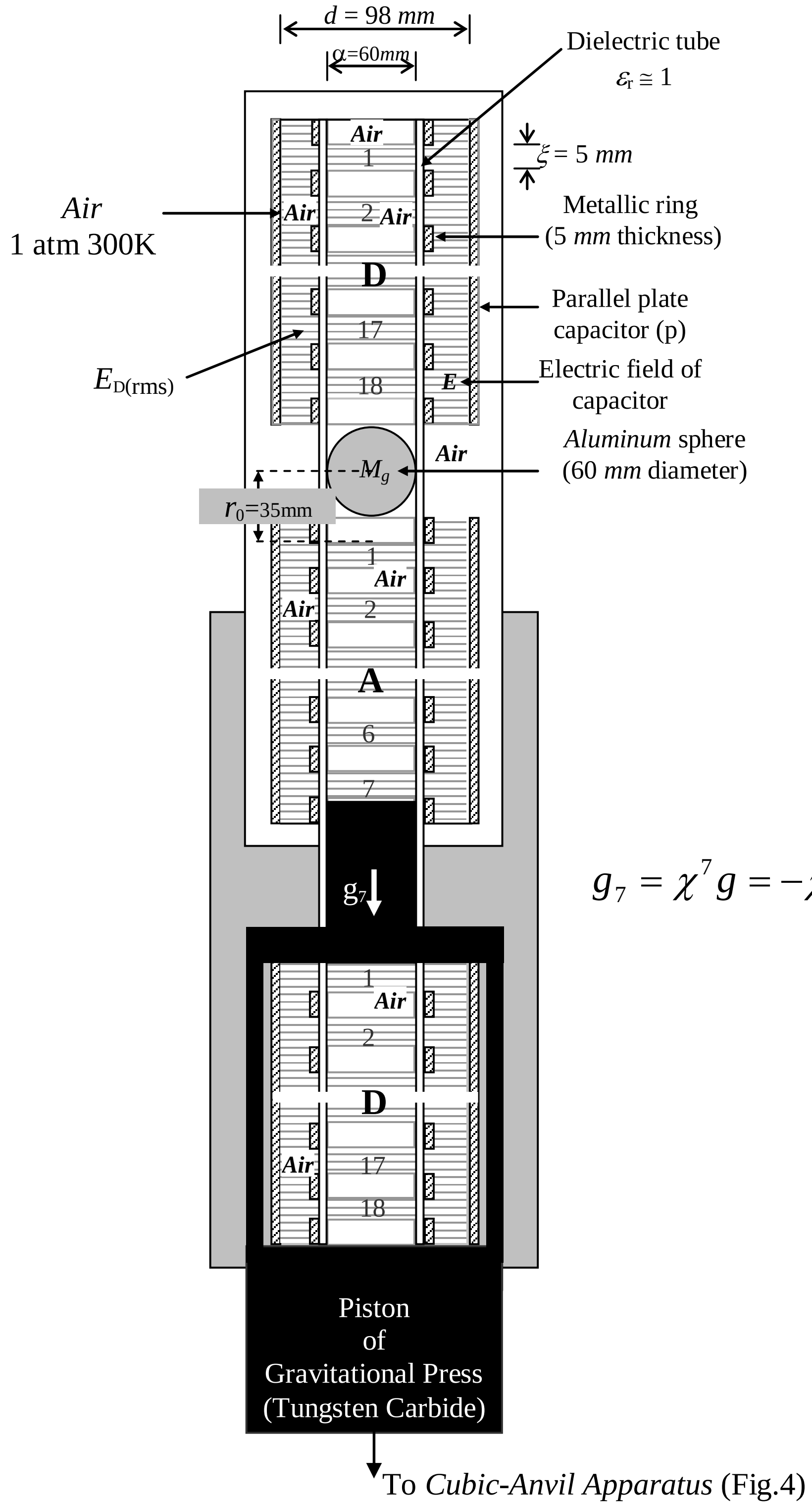

$$g_7 = \chi^7 g = -\chi^7 \frac{GM_{gs}}{r_0^2}$$

Fig. 3 – *Gravitational Press* (Developed from a process *patented* in July, 31 2008, PI0805046-5)

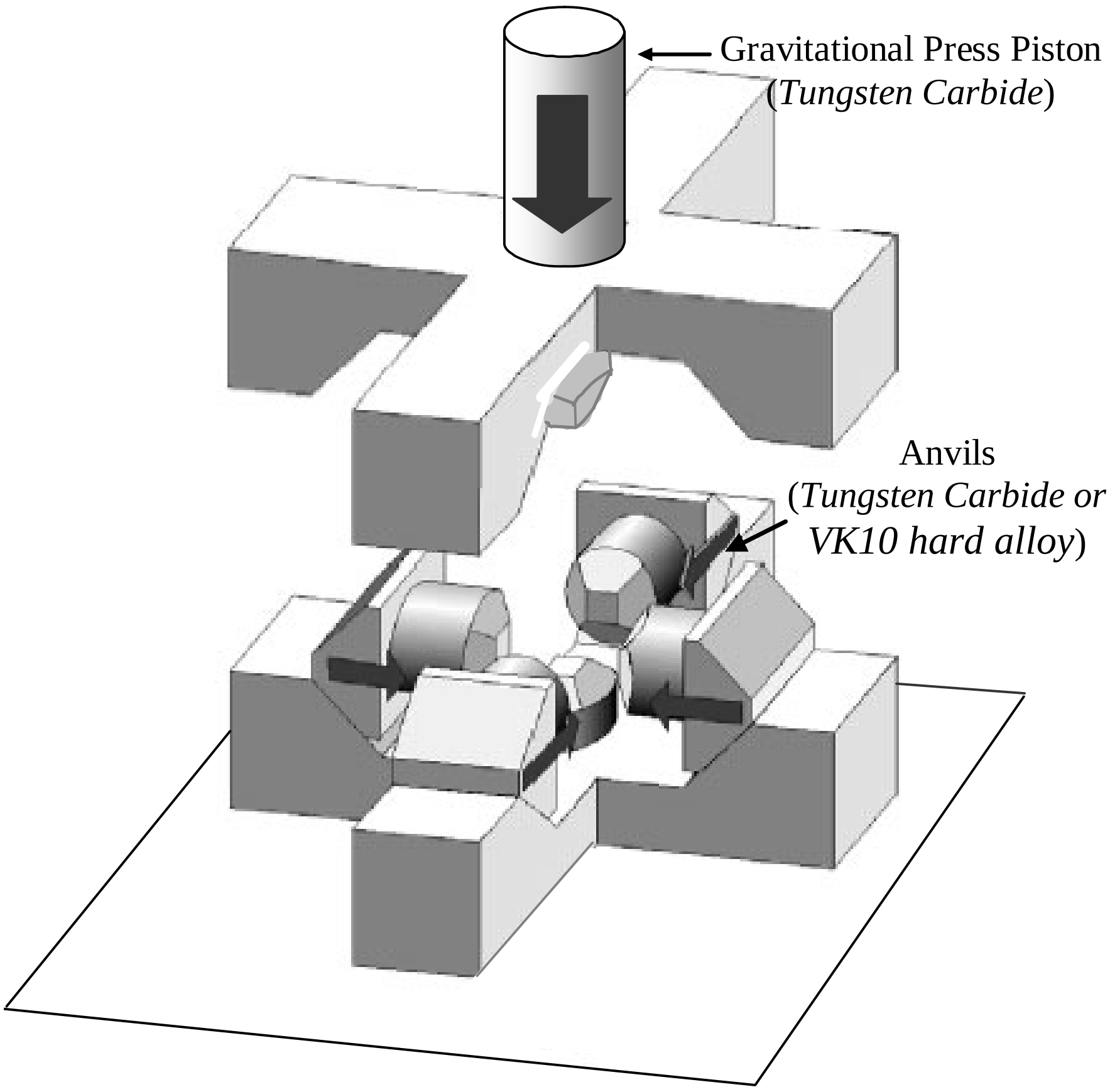

Fig.4 - *Diagram of Cubic-Anvil Apparatus for the Gravitational Press*. – In the center of the apparatus, is placed a ceramic cube ("synthesis capsule") of pyrophyllite ceramics, which contains graphite and is pressed by the anvils made from cemented carbide (e.g., tungsten carbide or VK10 hard alloy). Note that, due to the enormous pressure that the Gravitational Press can produce (>>*80GPa*), the "synthesis capsule" can be very large (up to about 1000 $cm^3$ in size). This is sufficient to produce diamonds with up to 100 carats (20g) or more.

# Artificial Gravitational Lenses

**Fran De Aquino**
Maranhao State University, Physics Department, S.Luis/MA, Brazil.

We show that it is possible to produce gravitational lenses at laboratory scale by means of a toroidal device which strongly intensifies the radial gravitational acceleration at its nucleus, and can make the acceleration repulsive besides attractive. This means that a light flux through the toroid can become convergent or divergent from its central axis. These lenses are similar to optical lenses and can be very useful for telescopes, microscopes, and for the concentration of solar light in order to convert solar energy into thermal energy.

**Key words:** Modified theories of gravity, Gravitational lenses, Solar instruments.
PACS: 04.50.Kd, 98.62.Sb, 95.55.Ev.

## 1. Introduction

It is known that Gravitational fields can bend light. This effect was confirmed in 1919 during a solar eclipse, when Arthur Eddington observed the light from stars passing close to the sun was slightly bent, so that stars appeared slightly out of position [1]. Einstein realized that a massive astronomical object can bend light making what is called a *gravitational lens*. The *gravitational lensing* is one of the predictions of Einstein's general theory of relativity. Although this phenomenon was first mentioned in 1924 by Orest Chwolson [2], the effect is more commonly associated with Einstein, who published a more famous article on the subject in 1936 [3, 4].

Here we show that it is possible to produce gravitational lenses at laboratory scale, by means of a toroidal device which strongly intensifies the radial *gravitational acceleration* at its nucleus, and can make the acceleration repulsive besides attractive * [5]. This means that a light flux through the toroid can becomes convergent or divergent from its central axis. These lenses are similar to optical lenses and can be very useful for telescopes, microscopes, and for the concentration of solar light in order to convert solar energy into thermal energy.

## 2. Theory

From the quantization of gravity it follows that the *gravitational mass* $m_g$ and the *inertial mass* $m_i$ are correlated by means of the following factor [5]:

$$\chi = \frac{m_g}{m_{i0}} = \left\{1 - 2\left[\sqrt{1+\left(\frac{\Delta p}{m_{i0}c}\right)^2} - 1\right]\right\} \qquad (1)$$

where $m_{i0}$ is the *rest* inertial mass of the particle and $\Delta p$ is the variation in the particle's *kinetic momentum*; $c$ is the speed of light.

When $\Delta p$ is produced by the absorption of a photon with wavelength $\lambda$, it is expressed by $\Delta p = h/\lambda$. In this case, Eq. (1) becomes

$$\frac{m_g}{m_{i0}} = \left\{1 - 2\left[\sqrt{1+\left(\frac{h/m_{i0}c}{\lambda}\right)^2} - 1\right]\right\} = \left\{1 - 2\left[\sqrt{1+\left(\frac{\lambda_0}{\lambda}\right)^2} - 1\right]\right\} \qquad (2)$$

where $\lambda_0 = h/m_{i0}c$ is the *De Broglie wavelength* for the particle with *rest* inertial mass $m_{i0}$.

It has been shown that there is an additional effect - *Gravitational Shielding* effect - produced by a substance whose gravitational mass was reduced or made negative [6]. The effect extends beyond substance (gravitational shielding) , up to a certain distance from it (along the central axis of gravitational shielding). This effect shows that in this region the gravity acceleration, $g_1$, is reduced at the same proportion, i.e., $g_1 = \chi_1 g$ where $\chi_1 = m_g / m_{i0}$ and $g$ is the gravity acceleration *before* the gravitational shielding). Consequently, *after a second gravitational shielding*, the gravity will be given by $g_2 = \chi_2 g_1 = \chi_1 \chi_2 g$, where $\chi_2$ is the value of the ratio $m_g / m_{i0}$ for the *second* gravitational shielding. In a generalized way, we can write that after the *nth* gravitational shielding the gravity, $g_n$, will be given by

$$g_n = \chi_1 \chi_2 \chi_3 ... \chi_n g \tag{3}$$

---

* De Aquino, F. (2008) *Process and Device for* Controlling the Locally the Gravitational Mass and the Gravity Acceleration, BR Patent Number: PI0805046-5, July 31, 2008.

This possibility shows that, by means of a battery of gravitational shieldings, we can strongly intensify the gravitational acceleration.

From Electrodynamics we know that when an electromagnetic wave with frequency $f$ and velocity $c$ incides on a material with relative permittivity $\varepsilon_r$, relative magnetic permeability $\mu_r$ and electrical conductivity $\sigma$, its *velocity is reduced* to $v = c/n_r$ where $n_r$ is the index of refraction of the material, given by [7]

$$n_r = \frac{c}{v} = \sqrt{\frac{\varepsilon_r \mu_r}{2}\left(\sqrt{1+(\sigma/\omega\varepsilon)^2}+1\right)} \tag{4}$$

If $\sigma >> \omega\varepsilon$, $\omega = 2\pi f$ , Eq. (4) reduces to

$$n_r = \sqrt{\frac{\mu_r \sigma}{4\pi\varepsilon_0 f}} \tag{5}$$

Thus, the wavelength of the incident radiation (See Fig. 1) becomes

$$\lambda_{\text{mod}} = \frac{v}{f} = \frac{c/f}{n_r} = \frac{\lambda}{n_r} = \sqrt{\frac{4\pi}{\mu f \sigma}} \tag{6}$$

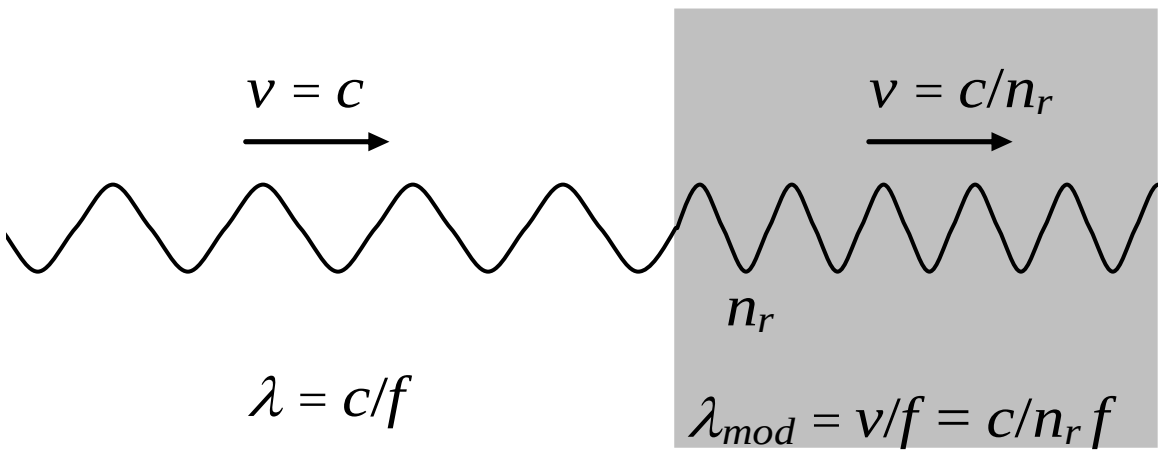

Fig. 1 – *Modified Electromagnetic Wave*. The wavelength of the electromagnetic wave can be strongly reduced, but its frequency remains the same.

If a lamina with thickness equal to $\xi$ contains $n$ atoms/m$^3$, then the number of atoms per area unit is $n\xi$. Thus, if the electromagnetic radiation with frequency $f$ incides on an area $S$ of the lamina it reaches $nS\xi$ atoms. If it incides on the *total area of the lamina*, $S_f$, then the total number of atoms reached by the radiation is $N = nS_f\xi$. The number of atoms per unit of volume, $n$, is given by

$$n = \frac{N_0 \rho}{A} \tag{7}$$

where $N_0 = 6.02 \times 10^{26}\, atoms/kmole$ is the Avogadro's number; $\rho$ is the matter density of the lamina (in *kg/m*$^3$) and $A$ is the molar mass(*kg/kmole*).

When an electromagnetic wave incides on the lamina, it strikes $N_f$ front atoms, where $N_f \cong (n S_f)\phi_m$ , $\phi_m$ is the "diameter" of the atom. Thus, the electromagnetic wave incides effectively on an area $S = N_f S_m$ , where $S_m = \frac{1}{4}\pi\phi_m^2$ is the cross section area of one atom. After these collisions, it carries out $n_{collisions}$ with the other atoms (See Fig.2).

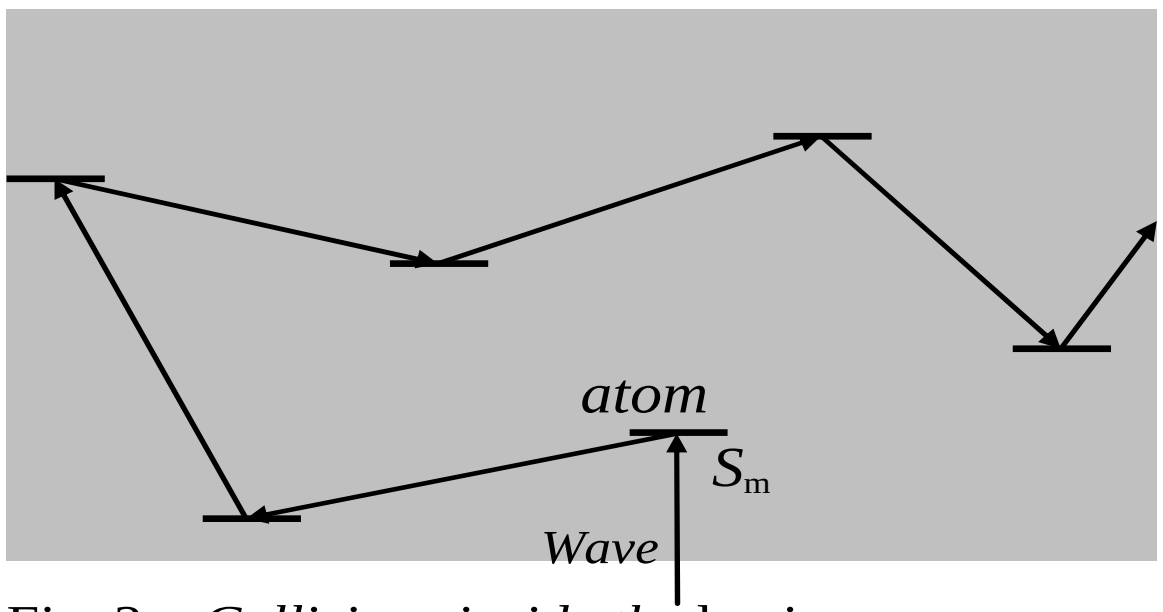

Fig. 2 – *Collisions inside the* lamina.

Thus, the total number of collisions in the volume $S\xi$ is

$$N_{collisions} = N_f + n_{collisions} = n_l S\phi_m + \left(n_l S\xi - n_m S\phi_m\right) =$$
$$= n_m S\xi \qquad (8)$$

The power density, $D$, of the radiation on the lamina can be expressed by

$$D = \frac{P}{S} = \frac{P}{N_f S_m} \qquad (9)$$

We can express the *total mean number of collisions in each atom*, $n_1$, by means of the following equation

$$n_1 = \frac{n_{total\ \ photons} N_{collisions}}{N} \qquad (10)$$

Since in each collision a *momentum* $h/\lambda$ is transferred to the atom, then the *total momentum* transferred to the lamina will be $\Delta p = \left(n_1 N\right)h/\lambda$. Therefore, in accordance with Eq. (1), we can write that

$$\frac{m_{g(l)}}{m_{i0(l)}} = \left\{1 - 2\left[\sqrt{1 + \left[\left(n_1 N\right)\frac{\lambda_0}{\lambda}\right]^2} - 1\right]\right\} =$$
$$= \left\{1 - 2\left[\sqrt{1 + \left[n_{total\ photons} N_{collisions}\frac{\lambda_0}{\lambda}\right]^2} - 1\right]\right\} \qquad (11)$$

Since Eq. (8) gives $N_{collisions} = n_l S\xi$, we get

$$n_{total\ \ photons} N_{collisions} = \left(\frac{P}{hf^2}\right)\left(n_l S\xi\right) \qquad (12)$$

Substitution of Eq. (12) into Eq. (11) yields

$$\frac{m_{g(l)}}{m_{i0(l)}} = \left\{1 - 2\left[\sqrt{1 + \left[\left(\frac{P}{hf^2}\right)\left(n_l S\xi\right)\frac{\lambda_0}{\lambda}\right]^2} - 1\right]\right\} \qquad (13)$$

Substitution of $P$ given by Eq. (9) into Eq. (13) gives

$$\frac{m_{g(l)}}{m_{i0(l)}} = \left\{1 - 2\left[\sqrt{1 + \left[\left(\frac{N_f S_m D}{f^2}\right)\left(\frac{n_l S\xi}{m_{i0(l)}c}\right)\frac{1}{\lambda}\right]^2} - 1\right]\right\} \qquad (14)$$

Substitution of $N_f \cong \left(n_l S_f\right)\phi_m$ and $S = N_f S_m$ into Eq. (14) results

$$\frac{m_{g(l)}}{m_{i0(l)}} = \left\{1 - 2\left[\sqrt{1 + \left[\left(\frac{n_l^3 S_f^2 S_m^2 \phi_m^2 \xi D}{m_{i0(l)} c f^2}\right)\frac{1}{\lambda}\right]^2} - 1\right]\right\} \qquad (15)$$

where $m_{i0(l)} = \rho_{(l)} V_{(l)}$.

Now, considering that the lamina is inside an ELF electromagnetic field with $E$ and $B$, then we can write that [8]

$$D = \frac{n_{r(l)} E^2}{2\mu_0 c} \qquad (16)$$

Substitution of Eq. (16) into Eq. (15) gives

$$\frac{m_{g(l)}}{m_{i0(l)}} = \left\{1 - 2\left[\sqrt{1 + \left[\left(\frac{n_{r(l)} n_l^3 S_f^2 S_m^2 \phi_m^2 \xi E^2}{2\mu_0 m_{i0(l)} c^2 f^2}\right)\frac{1}{\lambda}\right]^2} - 1\right]\right\} \qquad (17)$$

In the case in which the area $S_f$ is just the *area of the cross-section of the lamina* $\left(S_\alpha\right)$, we obtain from Eq. (17), considering that $m_{i0(l)} = \rho_{(l)} S_\alpha \xi$, the following expression

$$\frac{m_{g(l)}}{m_{i0(l)}} = \left\{1 - 2\left[\sqrt{1 + \left[\left(\frac{n_{r(l)} n_l^3 S_\alpha S_m^2 \phi_m^2 E^2}{2\mu_0 \rho_{(l)} c^2 f^2}\right)\frac{1}{\lambda}\right]^2} - 1\right]\right\} \qquad (18)$$

If the electrical conductivity of the lamina, $\sigma_{(l)}$, is such that $\sigma_{(l)} >> \omega\varepsilon$ , then the value of $\lambda$ is given by Eq. (6), i.e.,

$$\lambda = \lambda_{\text{mod}} = \sqrt{\frac{4\pi}{\mu f \sigma}} \qquad (19)$$

Substitution of Eq. (19) into Eq. (18) gives

$$\frac{m_{g(l)}}{m_{i0(l)}} = \left\{1 - 2\left[\sqrt{1 + \frac{n_{r(l)}^2 n_l^6 S_\alpha^2 S_m^4 \phi_m^4 \sigma_{(l)} E^4}{16\pi\mu_0 \rho_{(l)}^2 c^4 f^3}} - 1\right]\right\} \qquad (20)$$

Note that $E = E_m \sin \omega t$ .The average value for $E^2$ is equal to $\frac{1}{2}E_m^2$ because $E$ varies sinusoidaly ( $E_m$ is the maximum value for $E$ ). On the other hand, $E_{rms} = E_m/\sqrt{2}$ . Consequently we can change $E^4$ by $E_{rms}^4$, and the equation above can be rewritten as follows

$$\chi = \frac{m_{g(l)}}{m_{i0(l)}} =$$

$$= \left\{1 - 2\left[\sqrt{1 + \frac{n_{r(l)}^2 n_l^6 S_\alpha^2 S_m^4 \phi_m^4 \sigma_{(l)} E_{rms}^4}{16\pi\mu_0 \rho_{(l)}^2 c^4 f^3}} - 1\right]\right\} \qquad (21)$$

Now consider the *Artificial Gravitational Lenses* shown in Fig.3.

Basically they are rectangular toroids. Inside them there are two *dielectric rings* with $\varepsilon_r \cong 1$ and an *Aluminum ring* with *mass density* $\rho = 2700 kg.m^{-3}$ (See Fig.3). The rectangular toroid is filled with *air* at ambient temperature and 1atm. Thus, inside the tube, the air density is

$$\rho_{air} \cong 1.2 \ \ kg.m^{-3} \qquad (22)$$

The number of atoms of air (Nitrogen) per unit of volume, $n_{air}$, according to Eq.(7), is given by

$$n_{air} = \frac{N_0 \rho_{air}}{A_N} = 5.16 \times 10^{25} atoms/m^3 \qquad (23)$$

Here, the area $S_\alpha$ refers to the area of the ring inside the air toroid, with average radius $\bar{r} = r_e + r_i/2$ and height $\alpha$ , i.e.,

$$S_\alpha = 2\pi\alpha\,(r_i + r_e)/2 = \pi\alpha(r_i + r_e)$$

where $r_i$ is the inner radius and $r_e$ the outer radius of the rectangular toroid. For $r_i = 400mm$, $r_e = 650mm$ and $\alpha = 60mm$, we get

$$S_\alpha = \pi\alpha(r_i + r_e) = 0.198\,m^2 \qquad (24)$$

The *parallel metallic plates* (p), shown in Fig.3 are subjected to different drop voltages. The *two* sets of plates (*D*), placed on the extremes of the toroid, are subjected to $V_{(D)rms} = 576.2V$ at $f = 2.5Hz$, while the central set of plates (*A*) is subjected to $V_{(A)rms} = 19.7kV$ at $f = 2.5Hz$. Since $d = 98mm$, then the intensity of the electric field, which passes through the 36 *cylindrical air laminas* (each one with 5mm thickness) of the *two* sets (*D*), is .

$$E_{(D)rms} = V_{(D)rms}/d = 5.88 \times 10^3 V/m$$

and the intensity of the electric field, which passes through the 9 *cylindrical air laminas* of the two sets (*A*), is given by

$$E_{(A)rms} = V_{(A)rms}/d = 1.97 \times 10^5 V/m$$

Note that the *metallic rings* (5*mm* thickness) are positioned in such way to block the electric field out of the cylindrical air laminas (also 5mm thickness). The objective is to turn each one of these laminas into a *Gravity Control Cell* (GCC) [6]. Thus, the system shown in Fig. 3 has 4 sets of GCC. Two with 18 GCC each and two with 9 GCC each. The two sets with 18 GCC each are positioned at the extremes of the tube (*D*). They work as gravitational *decelerator* while the other two set with 9 GCC (*A*) each works as a gravitational *accelerator*, intensifying the gravity acceleration produced by the *Aluminum ring*. According to Eq. (3), this gravity after the $9^{th}$ GCC becomes $g_9 = \chi^9 g_0$, where $\chi = m_{g(l)}/m_{i(l)}$ given by Eq. (21), and $g_0$ can be calculated starting from the expression of the gravitational mass of the *half-toroid* of Aluminum, $M_{g\left(\frac{1}{2}toroid\right)}$, which is given by

$$\int_0^{M_g} dM_{g\left(\frac{1}{2}toroid\right)} = \rho\alpha(r_e - r_i)\int_0^{\pi r_i} dz$$

*whence*

$$M_{g\left(\frac{1}{2}toroid\right)} = \pi\rho\alpha(r_e - r_i)r_i \qquad (25)$$

On the other hand, we have that

$$\int_0^g dg = -\frac{G}{r^2}\int_0^{M_g} dM_g = -\frac{G\alpha\rho(r_e - r_i)}{r^2}\int_0^{\pi r_i} dz$$

whence we get

$$g = -\frac{G\pi\alpha\rho(r_e - r_i)r_i}{r^2} \qquad (26)$$

which gives the value of $g$ produced by the half-toroid at a point inside the nucleus of the toroid, distant $r$ from the center of the cross-section of the rectangular toroid. Thus, the value of $g_0'$ $(r = r_0)$, due to the first half-toroid is

$$g_0' \cong -G\pi\rho\alpha\left(\frac{r_e - r_i}{r_0^2}\right)r_i$$

The value of $g_0''$, due to the opposite half-toroid is

$$g_0'' \cong -G\pi\rho\alpha\left\{\frac{(r_e - r_i)r_i}{\left[2r_e - \left(\frac{r_e - r_i}{2} + r_0\right) - \left(\frac{r_e - r_i}{2}\right)\right]^2}\right\}$$

Consequently, the resultant is

$$g_0 \cong -G\pi\rho\alpha\left(\frac{r_e - r_i}{r_0^2}\right)r_i - G\pi\rho\alpha\left\{\frac{(r_e - r_i)r_i}{\left[2r_e - \left(\frac{r_e - r_i}{2} + r_0\right) - \left(\frac{r_e - r_i}{2}\right)\right]^2}\right\} = -G\pi\rho\alpha\left[\frac{(r_e - r_i)r_i}{(r_i + r_0)^2}\right]$$

In the case of $r_i >> r_0$, the equation above reduces to

$$g_0 \cong -G\pi\rho\alpha\left(\frac{r_e - r_i}{r_0^2}\right)r_i \qquad (27)$$

where $r_i$ is the inner radius of the toroid; $r_0$ is the distance between the center of the cross-section of the Aluminum ring and the surface of the first GCC of the set (A); $\alpha$ is the thickness of the Aluminum ring. Here, $r_0 = 35mm$ and $\alpha = 60mm$ (See Fig. 3 (a)).

The objective of the sets (*D*), with 18 GCC each, is to reduce strongly the value of the external gravity along the *rectangular toroid of air* in *D* region. In this case, the value of the external gravity, $g_{ext}$, is reduced by the factor $\chi_d^{18} g_{ext}$, where $\chi_d = 10^{-2}$. For example, if $g_{ext} = 9.81 m/s^2$ then this value is reduced to $\chi_d^{18} g_{ext}$ and, after the set A, it is increased by $\chi^9$. Since the system is designed for $\chi = -627.1$, then the gravity acceleration on the Aluminum ring becomes $\chi^9 \chi_d^{18} g_{ext} = 1.47 \times 10^{-10} m/s^2$, this value is smaller than $g_0 \cong -G\pi\rho\alpha\left[(r_e - r_i)r_i/r_0^2\right] = 9.9 \times 10^{-8} m.s^{-2}$.

The electrical conductivity of air, *inside the dielectric tube*, is equal to the electrical conductivity of Earth's atmosphere near the land, whose average value is $\sigma_{air} \cong 1 \times 10^{-14} S/m$ [9]. This value is of fundamental importance in order to obtain the convenient values of $\chi$ and $\chi_d$, which are given by Eq. (21), i.e.,

$$\chi = \left\{1 - 2\left[\sqrt{1 + \frac{n_{r(air)}^2 n_{air}^6 S_\alpha^2 S_m^4 \phi_m^4 \sigma_{air} E_{(A)rms}^4}{16\pi\mu_0 \rho_{air}^2 c^4 f^3}} - 1\right]\right\} =$$
$$= \left\{1 - 2\left[\sqrt{1 + 6.59\times10^{-17} E_{(A)rms}^4} - 1\right]\right\} \quad (28)$$

$$\chi_d = \left\{1 - 2\left[\sqrt{1 + \frac{n_{r(air)}^2 n_{air}^6 S_\alpha^2 S_m^4 \phi_m^4 \sigma_{air} E_{(D)rms}^4}{16\pi\mu_0 \rho_{air}^2 c^4 f^3}} - 1\right]\right\} =$$
$$= \left\{1 - 2\left[\sqrt{1 + 6.59\times10^{-17} E_{(D)rms}^4} - 1\right]\right\} \quad (29)$$

where $n_{r(air)} = \sqrt{\varepsilon_r \mu_r} \cong 1$, since $(\sigma << \omega\varepsilon)$; $n_{air} = 5.16\times10^{25} atoms/m^3$, $\phi_m = 1.55\times10^{-10} m$, $S_m = \pi\phi_m^2/4 = 1.88\times10^{-20} m^2$ and $f = 2.5Hz$. Since $E_{(A)rms} = 1.97\times10^5 V/m$, $E_{(D)rms} = 5.88\times10^3 V/m$, we get

$$\chi = -627.1 \quad (30)$$

and

$$\chi_d \cong 10^{-2} \quad (31)$$

Then the gravitational acceleration *after the 9th gravitational shielding* is

$$g_9 = \chi^9 g_0 = -\chi^9 G\pi\rho\alpha\left[\frac{(r_e - r_i)r_i}{r_0^2}\right] \quad (32)$$

It is known that gravitational fields can bend light, and that due to this effect, a light ray that passes very close to a body with gravitational mass $M_g$ is deviated of an angle $\delta$ (deflection angle) given by [3]

$$\delta = -\frac{4GM_g}{c^2 d} \quad (33)$$

where $d$ is the distance of closest approach.

Here, we can obtain the expression of $\delta$ as follows: by comparing Eq. (26) with Eq. (25) we obtain $GM_g = gr^2$. Substitution of this expression into Eq. (33) leads to the following equation

$$\delta = \frac{4gr^2}{c^2 d} \quad (34)$$

For $r = r_0$ we have $g = g_0$ and equation above can be rewritten as follows

$$\delta = \frac{4g_0 r_0^2}{c^2 d} \quad (35)$$

However, considering the symmetry of the gravitational lenses shown in Fig.3, it is easy to see that Eq. (35) must be rewritten as follows

$$\delta = \frac{4g_0 r_0^2}{c^2 d'} - \frac{4g_0 r_0^2}{c^2 d''} \quad (36)$$

where $d'$ and $d''$ are respectively, the distances of closest approach of the light ray with respect to the two sides of the Aluminum ring (See Fig 3 (b)).

When the gravitational lenses are activated the value of $g_0$ is amplified to $\chi^9 g_0$, then Eq. (36) becomes

$$\delta = \frac{4\chi^9 g_0 r_0^2}{c^2 d'} - \frac{4\chi^9 g_0 r_0^2}{c^2 d''} \quad (37)$$

Note that, for $d' = d''$ (light ray *at the center* of the Gravitational lens) Eq. (37) gives $\delta = 0$ ( *null deflection*). On the other hand, if $d' < d''$ we have $\delta > 0$ (the light ray is *gravitationally attracted* to the inner edge of rectangular toroid). Under these conditions, when a light flux crosses the gravitational lens (nucleus of the rectangular toroid), it becomes *divergent* in respect to the central axis of the toroid (See Fig. 3(c)). If $d' > d''$ then Eq. (37) shows that $\delta < 0$ (the light ray is *gravitationally repelled* from the inner edge of rectangular toroid). In this case, when a light flux crosses the gravitational lens, it becomes *convergent* in respect to the central axis of the toroid (See Fig. 3(b)).

Substitution of the known values into Eq. (37) yields

$$\delta \cong 0.1\left(\frac{1}{d'} - \frac{1}{d''}\right) \qquad (38)$$

Note that the values of $\delta$ can be easily controlled simply by controlling of the value of $\chi$. Also note that the curvatures of the light rays are proportional to the distances $d'$ and $d''$, similarly to the curvature of the light rays in the *optical lenses*. Then it is easy to see that these gravitational lenses can be very useful in building of telescopes, microscopes, and in concentrating solar light in order to convert solar energy into thermal energy.

I would like to thank Physicist *André Luis Martins* (RJ, Brazil) who came up with the original idea to build Artificial Gravitational Lens using sets of Gravitational Shieldings, as shown in my previous papers.

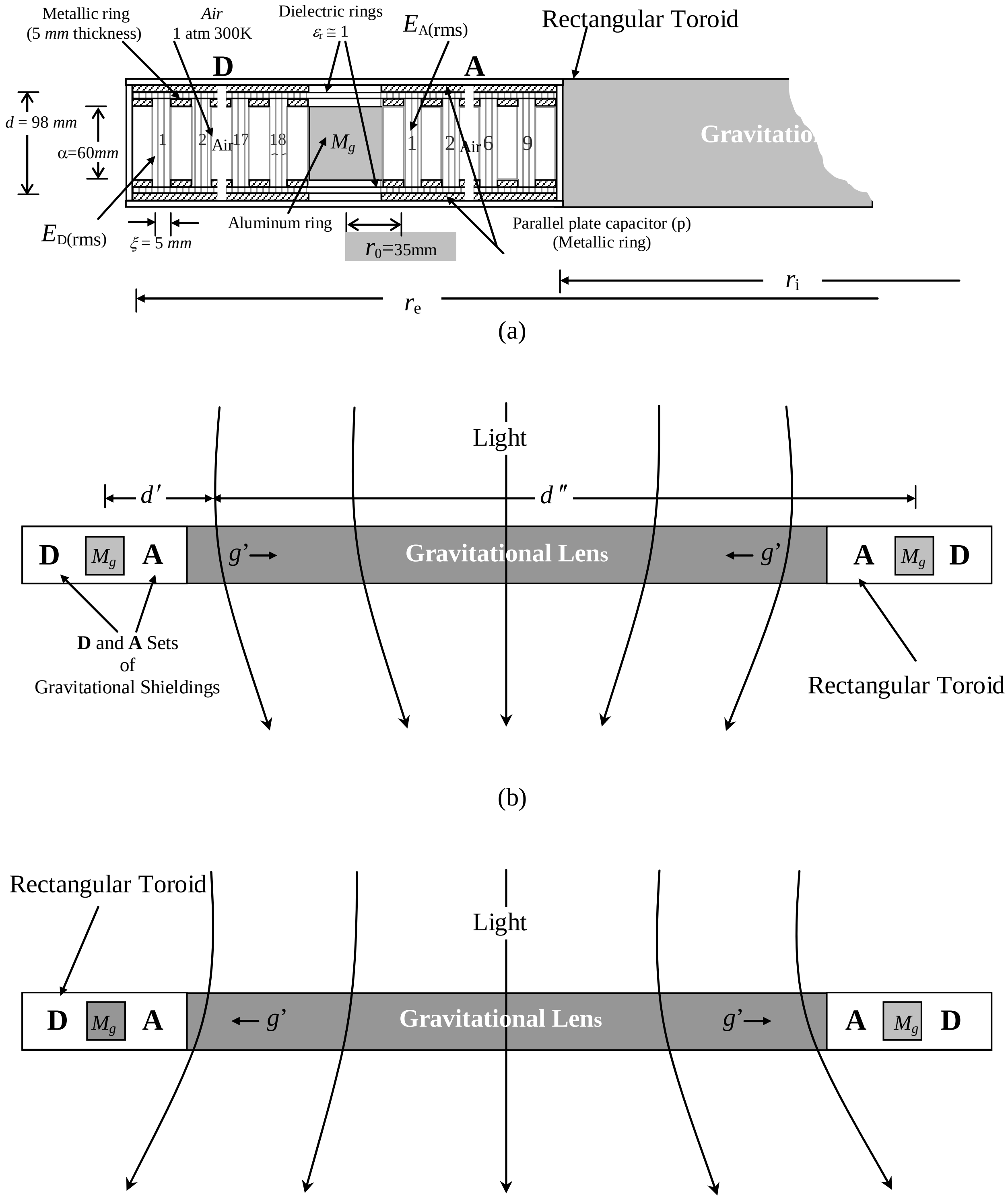

Fig. 3 – *Artificial Gravitational Lens*. (a) Cross-section of the Artificial Gravitational Lens. (b) Cross-section of a *Convergent* Gravitational Lens. The light rays are *gravitationally repelled* from the inner edge of toroid (c) Cross-section of a *Divergent* Gravitational Lens. The light rays are *gravitationally attracted* to the inner edge of toroid.

# Gravitational Blueshift and Redshift generated at Laboratory Scale

**Fran De Aquino**
Maranhao State University, Physics Department, S.Luis/MA, Brazil.

In this paper we show that it is possible to produce gravitational *blueshift* and *redshift* at laboratory scale by means of a device that can strongly intensify the local gravitational potential. Thus, by using this device, it is possible to generate electromagnetic radiation of *any frequency*, from ELF radiation ($f$ < 10*Hz*) up to *high energy gamma-rays*. In this case, several uses, such as medical imaging, radiotherapy and radioisotope production for PET (positron emission tomography) scanning, could be realized. The device is smaller and less costly than conventional sources of gamma rays.

**Key words:** Modified theories of gravity, Relativity and Gravitation, Gravitational Redshift and Blueshift.
PACS: 04.50.Kd, 95.30.Sf, 98.62.Py.

## 1. Introduction

It is known that electromagnetic radiation is *blueshifted* when propagating from a region of weaker gravitational field to a region of stronger gravitational field. In this case the radiation is blueshifted because it gains energy during propagation. In the contrary case, the radiation is *redshifted*. This effect was predicted by Einstein's Relativity Theory [1, 2] and was widely confirmed by several experiments [3, 4]. It was first confirmed in 1959 in the Pound and Rebka experiment [3].

Here we show that it is possible to produce gravitational *blueshift* and *redshift* at laboratory scale by means of a device that can strongly intensify the local gravitational potential* [5]. Thus, by using this device, it is possible to generate electromagnetic radiation of *any frequency*, from ELF radiation ($f$ < 10*Hz*) up to *high energy gamma-rays*. In this case, several uses, such as in medical imaging, radiotherapy and radioisotope production for PET (positron emission tomography) scanning and others, could be devised. The device is smaller and less costly than conventional sources of gamma rays.

* De Aquino, F. (2008) *Process and Device for* Controlling the Locally the Gravitational Mass and the Gravity Acceleration, BR Patent Number: PI0805046-5, July 31, 2008.

## 2. Theory

From the quantization of gravity it follows that the *gravitational mass* $m_g$ and the *inertial mass* $m_i$ are correlated by means of the following factor [5]:

$$\chi = \frac{m_g}{m_{i0}} = \left\{1 - 2\left[\sqrt{1+\left(\frac{\Delta p}{m_{i0}c}\right)^2} - 1\right]\right\} \quad (1)$$

where $m_{i0}$ is the *rest* inertial mass of the particle and $\Delta p$ is the variation in the particle's *kinetic momentum*; $c$ is the speed of light.

When $\Delta p$ is produced by the absorption of a photon with wavelength $\lambda$, it is expressed by $\Delta p = h/\lambda$. In this case, Eq. (1) becomes

$$\frac{m_g}{m_{i0}} = \left\{1 - 2\left[\sqrt{1+\left(\frac{h/m_{i0}c}{\lambda}\right)^2} - 1\right]\right\}$$
$$= \left\{1 - 2\left[\sqrt{1+\left(\frac{\lambda_0}{\lambda}\right)^2} - 1\right]\right\} \quad (2)$$

where $\lambda_0 = h/m_{i0}c$ is the *De Broglie wavelength* for the particle with *rest* inertial mass $m_{i0}$.

It has been shown that there is an additional effect - *Gravitational Shielding*

effect - produced by a substance whose gravitational mass was reduced or made negative [6]. The effect extends beyond substance (gravitational shielding) , up to a certain distance from it (along the central axis of gravitational shielding). This effect shows that in this region the gravity acceleration, $g_1$, is reduced at the same proportion, i.e., $g_1 = \chi_1 g$ where $\chi_1 = m_g / m_{i0}$ and $g$ is the gravity acceleration *before* the gravitational shielding). Consequently, *after a second gravitational shielding*, the gravity will be given by $g_2 = \chi_2 g_1 = \chi_1 \chi_2 g$, where $\chi_2$ is the value of the ratio $m_g / m_{i0}$ for the *second* gravitational shielding. In a generalized way, we can write that after the *nth* gravitational shielding the gravity, $g_n$, will be given by

$$g_n = \chi_1 \chi_2 \chi_3 \cdots \chi_n g \qquad (3)$$

This possibility shows that, by means of a battery of gravitational shieldings, we can strongly intensify the gravitational acceleration.

In order to measure the extension of the shielding effect, samples were placed above a superconducting disk with radius $r_D = 0.1375m$, which was producing a gravitational shielding. The effect has been detected up to *a distance of about 3m* from the disk (along the central axis of disk) [7]. This means that *the gravitational shielding effect extends, beyond the disk by approximately 20 times* the disk *radius*.

From Electrodynamics we know that when an electromagnetic wave with frequency $f$ and velocity $c$ incides on a material with relative permittivity $\varepsilon_r$, relative magnetic permeability $\mu_r$ and electrical conductivity $\sigma$, its *velocity is reduced* to $v = c/n_r$ where $n_r$ is the index of refraction of the material, given by [8]

$$n_r = \frac{c}{v} = \sqrt{\frac{\varepsilon_r \mu_r}{2}\left(\sqrt{1 + (\sigma/\omega\varepsilon)^2} + 1\right)} \qquad (4)$$

If $\sigma >> \omega\varepsilon$, $\omega = 2\pi f$, Eq. (4) reduces to

$$n_r = \sqrt{\frac{\mu_r \sigma}{4\pi\varepsilon_0 f}} \qquad (5)$$

Thus, the wavelength of the incident radiation (See Fig. 1) becomes

$$\lambda_{\text{mod}} = \frac{v}{f} = \frac{c/f}{n_r} = \frac{\lambda}{n_r} = \sqrt{\frac{4\pi}{\mu f \sigma}} \qquad (6)$$

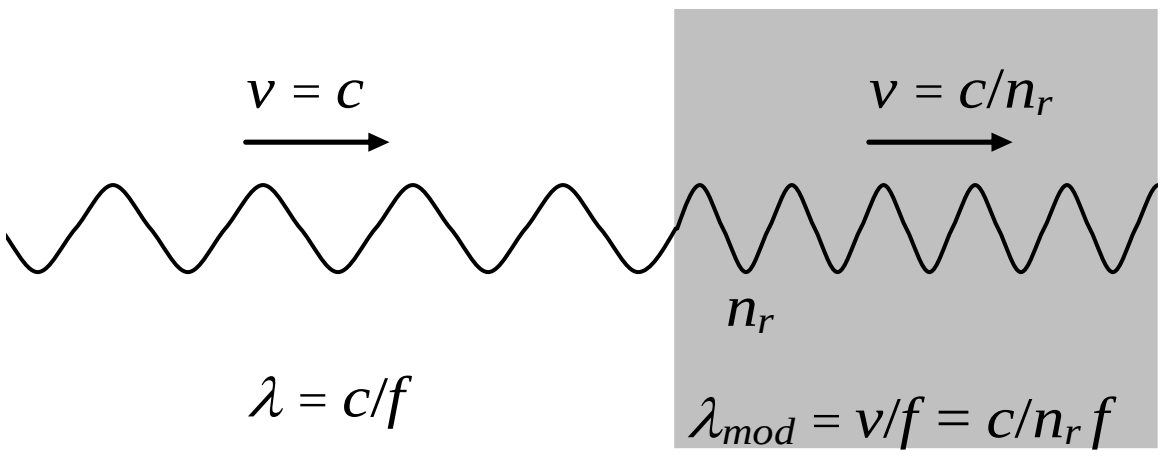

Fig. 1 – *Modified Electromagnetic Wave*. The wavelength of the electromagnetic wave can be strongly reduced, but its frequency remains the same.

If a lamina with thickness equal to $\xi$ contains $n$ atoms/m$^3$, then the number of atoms per area unit is $n\xi$. Thus, if the electromagnetic radiation with frequency $f$ incides on an area $S$ of the lamina it reaches $nS\xi$ atoms. If it incides on the *total area of the lamina*, $S_f$, then the total number of atoms reached by the radiation is $N = nS_f\xi$. The number of atoms per unit of volume, $n$, is given by

$$n = \frac{N_0 \rho}{A} \qquad (7)$$

where $N_0 = 6.02 \times 10^{26}\, atoms/kmole$ is the Avogadro's number; $\rho$ is the matter density of the lamina (in *kg/m*$^3$) and $A$ is the molar mass(*kg/kmole*).

When an electromagnetic wave incides on the lamina, it strikes $N_f$ front atoms, where $N_f \cong (n S_f)\phi_m$, $\phi_m$ is the "diameter" of the atom. Thus, the electromagnetic wave incides effectively on an area $S = N_f S_m$, where $S_m = \frac{1}{4}\pi\phi_m^2$ is the cross section area of one atom. After these collisions, it carries out $n_{collisions}$ with the other atoms (See Fig.2).

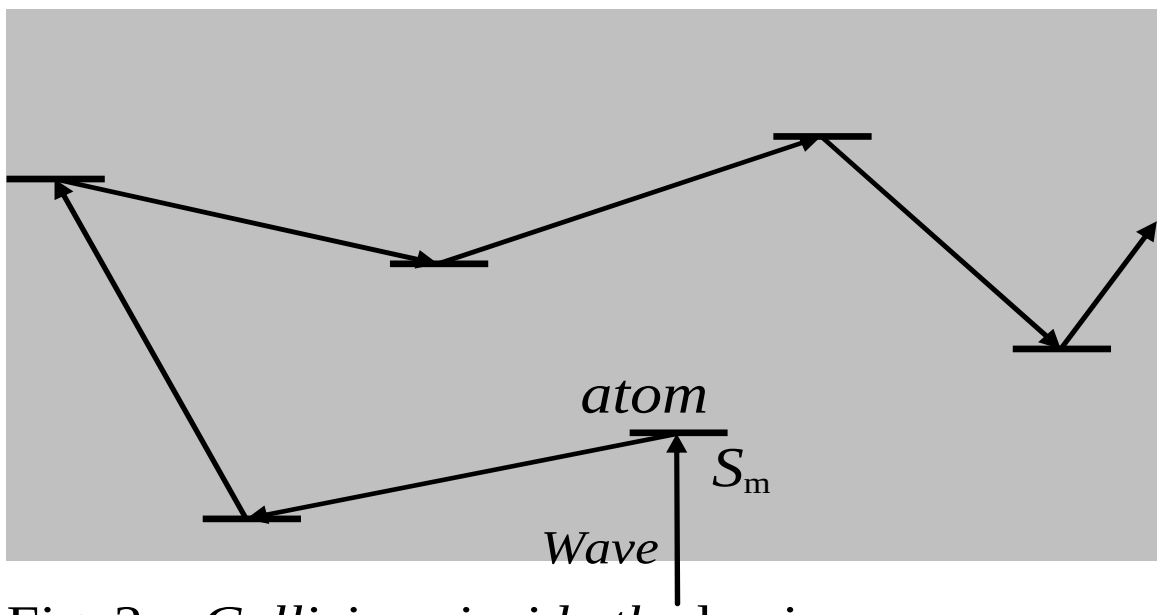

Fig. 2 – *Collisions inside the* lamina.

Thus, the total number of collisions in the volume $S\xi$ is

$$N_{collisions} = N_f + n_{collisions} = n_l S\phi_m + \left(n_l S\xi - n_m S\phi_m\right) = \\ = n_m S\xi \qquad (8)$$

The power density, $D$, of the radiation on the lamina can be expressed by

$$D = \frac{P}{S} = \frac{P}{N_f S_m} \qquad (9)$$

We can express the *total mean number of collisions in each atom,* $n_1$, by means of the following equation

$$n_1 = \frac{n_{total\ \ photons} N_{collisions}}{N} \qquad (10)$$

Since in each collision a *momentum* $h/\lambda$ is transferred to the atom, then the *total momentum* transferred to the lamina will be $\Delta p = \left(n_1 N\right)h/\lambda$. Therefore, in accordance with Eq. (1), we can write that

$$\frac{m_{g(l)}}{m_{i0(l)}} = \left\{1 - 2\left[\sqrt{1 + \left[\left(n_1 N\right)\frac{\lambda_0}{\lambda}\right]^2} - 1\right]\right\} = \\ = \left\{1 - 2\left[\sqrt{1 + \left[n_{total\ photons} N_{collisions}\frac{\lambda_0}{\lambda}\right]^2} - 1\right]\right\} \qquad (11)$$

Since Eq. (8) gives $N_{collisions} = n_l S\xi$, we get

$$n_{total\ \ photons} N_{collisions} = \left(\frac{P}{hf^2}\right)\left(n_l S\xi\right) \qquad (12)$$

Substitution of Eq. (12) into Eq. (11) yields

$$\frac{m_{g(l)}}{m_{i0(l)}} = \left\{1 - 2\left[\sqrt{1 + \left[\left(\frac{P}{hf^2}\right)\left(n_l S\xi\right)\frac{\lambda_0}{\lambda}\right]^2} - 1\right]\right\} \qquad (13)$$

Substitution of $P$ given by Eq. (9) into Eq. (13) gives

$$\frac{m_{g(l)}}{m_{i0(l)}} = \left\{1 - 2\left[\sqrt{1 + \left[\left(\frac{N_f S_m D}{f^2}\right)\left(\frac{n_l S\xi}{m_{i0(l)}c}\right)\frac{1}{\lambda}\right]^2} - 1\right]\right\} \qquad (14)$$

Substitution of $N_f \cong \left(n_l S_f\right)\phi_m$ and $S = N_f S_m$ into Eq. (14) results

$$\frac{m_{g(l)}}{m_{i0(l)}} = \left\{1 - 2\left[\sqrt{1 + \left[\left(\frac{n_l^3 S_f^2 S_m^2 \phi_m^2 \xi D}{m_{i0(l)} c f^2}\right)\frac{1}{\lambda}\right]^2} - 1\right]\right\} \qquad (15)$$

where $m_{i0(l)} = \rho_{(l)} V_{(l)}$.

Now, considering that the lamina is inside an ELF electromagnetic field with $E$ and $B$, then we can write that [9]

$$D = \frac{n_{r(l)} E^2}{2\mu_0 c} \qquad (16)$$

Substitution of Eq. (16) into Eq. (15) gives

$$\frac{m_{g(l)}}{m_{i0(l)}} = \left\{1 - 2\left[\sqrt{1 + \left[\left(\frac{n_{r(l)} n_l^3 S_f^2 S_m^2 \phi_m^2 \xi E^2}{2\mu_0 m_{i0(l)} c^2 f^2}\right)\frac{1}{\lambda}\right]^2} - 1\right]\right\} \qquad (17)$$

In the case in which the area $S_f$ is just the *area of the cross-section of the lamina* $\left(S_\alpha\right)$, we obtain from Eq. (17), considering that $m_{i0(l)} = \rho_{(l)} S_\alpha \xi$, the following expression

$$\frac{m_{g(l)}}{m_{i0(l)}} = \left\{1 - 2\left[\sqrt{1 + \left[\left(\frac{n_{r(l)} n_l^3 S_\alpha S_m^2 \phi_m^2 E^2}{2\mu_0 \rho_{(l)} c^2 f^2}\right)\frac{1}{\lambda}\right]^2} - 1\right]\right\} \quad (18)$$

If the electrical conductivity of the lamina, $\sigma_{(l)}$, is such that $\sigma_{(l)} >> \omega\varepsilon$, then the value of $\lambda$ is given by Eq. (6), i.e.,

$$\lambda = \lambda_{mod} = \sqrt{\frac{4\pi}{\mu f \sigma}} \quad (19)$$

Substitution of Eq. (19) into Eq. (18) gives

$$\frac{m_{g(l)}}{m_{i0(l)}} = \left\{1 - 2\left[\sqrt{1 + \frac{n_{r(l)}^2 n_l^6 S_\alpha^2 S_m^4 \phi_m^4 \sigma_{(l)} E^4}{16\pi\mu_0 \rho_{(l)}^2 c^4 f^3}} - 1\right]\right\} \quad (20)$$

Note that $E = E_m \sin\omega t$ .The average value for $E^2$ is equal to $\frac{1}{2}E_m^2$ because $E$ varies sinusoidaly ($E_m$ is the maximum value for $E$). On the other hand, $E_{rms} = E_m/\sqrt{2}$. Consequently we can change $E^4$ by $E_{rms}^4$, and the equation above can be rewritten as follows

$$\chi = \frac{m_{g(l)}}{m_{i0(l)}} =$$

$$= \left\{1 - 2\left[\sqrt{1 + \frac{n_{r(l)}^2 n_l^6 S_\alpha^2 S_m^4 \phi_m^4 \sigma_{(l)} E_{rms}^4}{16\pi\mu_0 \rho_{(l)}^2 c^4 f^3}} - 1\right]\right\} \quad (21)$$

Now consider the *Gravitational Shift Device* shown in Fig.3.

Inside the device there is a *dielectric tube* ($\varepsilon_r \cong 1$) with the following characteristics: $\alpha = 60mm$, $S_\alpha = \pi\alpha^2/4 = 2.83\times10^{-3}m^2$. Inside the tube there is an *Aluminum sphere* with 30mm radius and mass $M_{gs} = 0.30536kg$. The tube is filled with *air* at ambient temperature and 1atm. Thus, inside the tube, the air density is

$$\rho_{air} = 1.2 \ kg.m^{-3} \quad (22)$$

The number of atoms of air (Nitrogen) per unit of volume, $n_{air}$, according to Eq.(7), is given by

$$n_{air} = \frac{N_0 \rho_{air}}{A_N} = 5.16\times10^{25} atoms/m^3 \quad (23)$$

The *parallel metallic plates* (p), shown in Fig.3 are subjected to different drop voltages. The two sets of plates (*D*), placed on the extremes of the tube, are subjected to $V_{(D)rms} = 1.64kV$ at $f = 1Hz$, while the central set of plates (*A*) is subjected to $V_{(A)rms} = 19.7kV$ at $f = 1Hz$. Since $d = 98mm$, then the intensity of the electric field, which passes through the 36 *cylindrical air laminas* (each one with 5mm thickness) of the *two* sets (*D*), is

$$E_{(D)rms} = V_{(D)rms}/d = 1.67\times10^4 V/m$$

and the intensity of the electric field, which passes through the 7 *cylindrical air laminas* of the central set (*A*), is given by

$$E_{(A)rms} = V_{(A)rms}/d = 2.012\times10^5 V/m$$

Note that the *metallic rings* (5*mm* thickness) are positioned in such way to block the electric field out of the cylindrical air laminas. The objective is to turn each one of these laminas into a *Gravity Control Cells* (GCC) [10]. Thus, the system shown in Fig. 3 has 3 sets of GCC. Two with 18 GCC each, and one with 19 GCC. The two sets with 18 GCC each are positioned at the extremes of the tube (*D*). They work as gravitational *decelerator* while the other set with 19 GCC (*A*) works as a gravitational *accelerator*, intensifying the gravity acceleration and the *gravitational potential* produced by the mass $M_{gs}$ of the Aluminum sphere. According to Eq. (3) the gravity, after the $19^{th}$ GCC becomes $g_{19} = \chi^{19} GM_{gs}/r_1^2$, and the gravitational potential $\varphi = \chi^{19} GM_{gs}/r_1$ where $\chi = m_{g(l)}/m_{i(l)}$ is given by Eq. (21) and $r_1 = 35mm$ is the distance between the center of the Aluminum sphere and the surface of the first GCC of the set (A).

The objective of the sets (*D*), with 18 GCC each, is to reduce strongly the value of the external gravity along the axis of the tube. In this case, the value of the external gravity, $g_{ext}$, is reduced by the factor $\chi_d^{18} g_{ext}$, where $\chi_d = 10^{-2}$. For example, if the base BS of the system is positioned on the Earth surface, then $g_{ext} = 9.81 m/s^2$ is reduced to $\chi_d^{18} g_{ext}$ and, after the set A, it is increased by $\chi^{19}$. Since the system is designed for $\chi = -308.5$, then the gravity acceleration on the sphere becomes $\chi^{19}\chi_d^{18} g_{ext} = 2.4 \times 10^{-12} m/s^2$, this value is much smaller than $g_{sphere} = GM_{gs}/r_s^2 = 2.26 \times 10^{-8} m/s^2$.

The electrical conductivity of air, *inside the dielectric tube*, is equal to the electrical conductivity of Earth's atmosphere near the land, whose average value is $\sigma_{air} \cong 1 \times 10^{-14} S/m$ [11]. This value is of fundamental importance in order to obtain the convenient values of the electrical current $i$ and the value of $\chi$ and $\chi_d$, which are given by Eq. (21), i.e.,

$$\chi = \left\{1 - 2\left[\sqrt{1 + \frac{n_{r(air)}^2 n_{air}^6 S_\alpha^2 S_m^4 \phi_m^4 \sigma_{air} E_{(A)rms}^4}{16\pi\mu_0 \rho_{air}^2 c^4 f^3}} - 1\right]\right\} = \\ = \left\{1 - 2\left[\sqrt{1 + 6.81 \times 10^{-11} E_{(A)rms}^4} - 1\right]\right\} \qquad (24)$$

$$\chi_d = \left\{1 - 2\left[\sqrt{1 + \frac{n_{r(air)}^2 n_{air}^6 S_\alpha^2 S_m^4 \phi_m^4 \sigma_{air} E_{(D)rms}^4}{16\pi\mu_0 \rho_{air}^2 c^4 f^3}} - 1\right]\right\} = \\ = \left\{1 - 2\left[\sqrt{1 + 6.81 \times 10^{-11} E_{(D)rms}^4} - 1\right]\right\} \qquad (25)$$

where $n_{r(air)} = \sqrt{\varepsilon_r \mu_r} \cong 1$, since $(\sigma << \omega\varepsilon)$; $n_{air} = 5.16 \times 10^{25} atoms/m^3$, $\phi_m = 1.55 \times 10^{-10} m$, $S_m = \pi\phi_m^2/4 = 1.88 \times 10^{-20} m^2$ and $f = 60Hz$.
Since $E_{(A)rms} = 2.012 \times 10^5 V/m$ and $E_{(D)rms} = 1.67 \times 10^4 V/m$, we get

$$\chi = -308.5 \qquad (26)$$

and

$$\chi_d \cong 10^{-2} \qquad (27)$$

Then the gravitational acceleration *after the 19*$^{th}$ *gravitational shielding* of the set A (See Fig.3) † is

$$g_{19} = \chi^{19} g_1 = \chi^{19} GM_{gs}/r_1^2 \qquad (28)$$

and the *gravitational potential* is

$$\varphi = \chi^{19}\varphi_1 = \chi^{19} GM_{gs}/r_1 \qquad (29)$$

Thus, if photons with frequency $f_0$ are emitted from a point 0 near the Earth's surface, where the gravitational potential is $\varphi_0 \cong -GM_\oplus/r_\oplus$ (See photons source in Fig.3), and these photons pass through the region in front of the 19$^{th}$ gravitational shielding, where the gravitational potential is increased to the value expressed by Eq. (29) then the frequency of the photons in this region, according to Einstein's relativity theory, becomes $f = f_0 + \Delta f$, where $\Delta f$ is given by

$$\Delta f = \frac{\varphi - \varphi_0}{c^2} f_0 = \frac{-\chi^{19} GM_{gs}/r_1 + GM_\oplus/r_\oplus}{c^2} \qquad (30)$$

If $\chi < 0$, then $\chi^{19} < 0$ and $\Delta f > 0$ (*blueshift*). Note that, if the number $n$ of Gravitational Shieldings in the set **A** is *odd* ($n = 1,3,5,7,...$) then the result is $\Delta f > 0$ (*blueshift*). But, if $n$ is *even* ($n = 2,4,6,8,...$) and $\left|\chi^n M_{gs}/r_1\right| > \left|M_\oplus/r_\oplus\right|$ then the result is $\Delta f < 0$ (*redshift*). Note that to reduce $f_0 = 10^{14} Hz$ down to $f \cong 10^{11} Hz$ it is

---

† *The gravitational shielding effect extends beyond the gravitational shielding by approximately 20 times* its *radius* (along the central axis of *the gravitational shielding*). [7] Here, this means that, *in absence of the set D* (*bottom of the device*), *the gravitational shielding effect extends, beyond the 19*$^{th}$*gravitational shielding, by approximately 20* ($\alpha/2$) ≈ 600mm.

necessary that $\Delta f = -0.999 \times 10^{14} Hz$. This precision is not easy to be obtained in practice. On the other hand, if for example, $f_0 = 10^{14} Hz$ and $\Delta f = -10^{10} Hz$ then $f = f_0 + \Delta f \cong 10^{14} Hz$ i.e., the redshift is negligible. However, the device can be useful to generate *ELF radiation* by redshift. For example, if $f_0 = 1GHz$, $n = 18$ and $\chi = 95.15278521$, then we obtain ELF radiation with frequency $f \cong 1Hz$. Radiation of any frequency can be generated by gravitational blueshift. For example, if $f_0 = 10^{14} Hz$ and $\Delta f = +10^{18} Hz$ then $f = f_0 + \Delta f \cong 10^{18} Hz$. What means that a light beam with frequency $10^{14} Hz$ was converted into a gamma-ray beam with frequency $10^{18} Hz$. Similarly, if $f_0 = 1MHz$ and $\Delta f = +9MHz$, then $f = f_0 + \Delta f \cong 10MHz$, and so on.

Now, consider the device shown in Fig. 3, where $\chi = -308.5$, $M_{gs} = 0.30536kg$, $r_1 = 35mm$. According to Eq. (30), it can produce a $\Delta f$ given by

$$\Delta f \cong \frac{-\chi^{19} GM_{gs} / r_1}{c^2} \cong 3.6 \times 10^{22} Hz \qquad (31)$$

Thus, we get

$$f = f_0 + \Delta f \cong 3.6 \times 10^{22} Hz \qquad (32)$$

What means that the device is able to convert any type of electromagnetic radiation (frequency $f_0$) into a gamma-ray beam with frequency $3.6 \times 10^{22} Hz$. Thus, by controlling the value of $\chi$ and $f_0$, it is possible to generate radiation of any frequency.

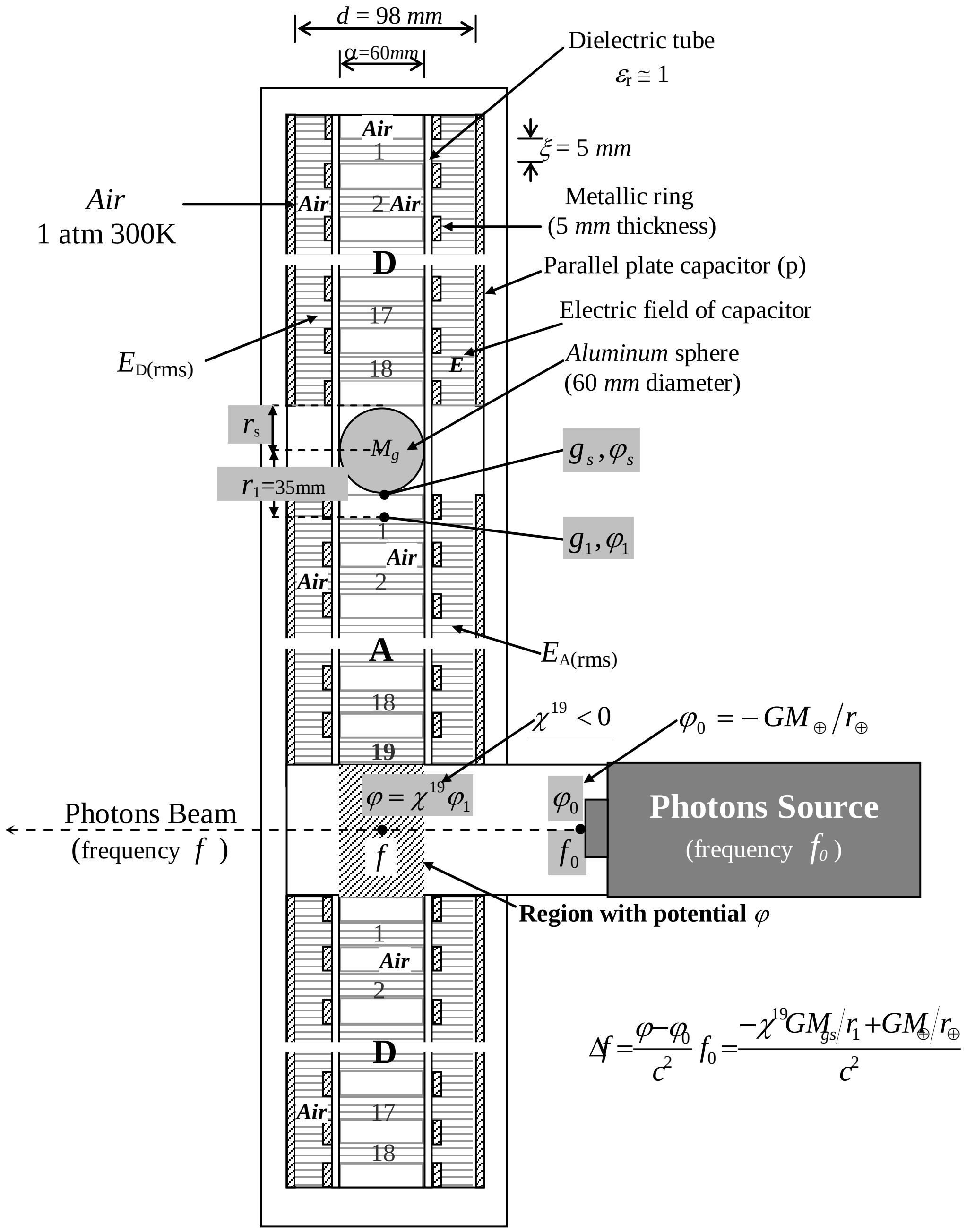

Fig. 3 – *Schematic diagram of the Gravitational Shift Device* (*Blueshift and Redshift*) – The device can generate electromagnetic radiation of *any frequency*, since ELF radiation ($f < 10Hz$) up to *high energy gamma-rays*.

# Quantum Reversal of Soul Energy

**Fran De Aquino**
Maranhao State University, Physics Department, S.Luis/MA, Brazil.

In the last decades, the existence of the Soul has been seriously considered by Quantum Physics. It has been frequently described as a body of unknown *energy* coupled to human body by means of a mutual interaction. The Quantum Physics shows that energy is *quantized*, i.e., it has discrete values that are defined as energy levels. Thus, along the life of a person, the energy of its soul is characterized by several quantum levels of energy. Here, we show by means of application of specific electromagnetic radiations on the human body, that it is possible to revert the energy of the soul to previous energy levels. This process can have several therapeutic applications.

**Key words:** Modified theories of gravity, Microwave fields effects, Therapeutic applications, Quantum information.
PACS: 04.50.Kd, 87.50.S-,87.50.st, 03.67.-a.

## 1. Introduction

Since long the Soul has remained an element of strongly consideration by Religion. Some authors claim that Religion is the science of the Soul [1]. Others claim that Soul and Religion are related to evolution. Sir Julian Huxley, a leading evolutionary biologist, the first Director-General of UNESCO and signatory to the Humanist Manifesto II, wrote: "*Human Soul and Religion are just the product of evolution*" [2]. This show how important the Soul is for the Religion. Philosophy also realizes the importance of the Soul. Plato, drawing on the words of his teacher Socrates, considered the Soul the essence of a person, being that which decides how we behave. As bodies die, the Soul is continually reborn in subsequent bodies.

Nowadays, Quantum Physics and other branches of Science are seriously considering the existence of the Soul.

It has been frequently described as a body of unknown *energy* coupled to human body by means of a mutual interaction. This type of energy from the viewpoint of Physics has been considered as *Imaginary* Energy. The term imaginary are borrowed from Mathematics (real and imaginary numbers) [3].

Quantum Physics shows that *energy* is *quantized*, i.e., that it has discrete values that are defined as discrete energy levels that correspond to all positive integer values of the *quantum number* $n,(n=1,2,3,...)$ [4]. Thus, along the life of a person, the energy of its Soul is characterized by several quantum levels of energy. Here, we show that, by means of application of specific electromagnetic radiations on the human body (its Soul), it is possible to revert the energy of the Soul to previous energy levels. This process can have several therapeutic applications.

## 2. Theory

From the quantization of gravity it follows that the *imaginary* gravitational mass $m_{g\,(im)}$ and the *imaginary* inertial mass $m_{i0\,(im)}$ are correlated by means of the following factor [5]:

$$\chi=\frac{m_{g(im)}}{m_{i0(im)}}=\left\{1-2\left[\sqrt{1+\left(\frac{\Delta p_{(im)}}{m_{i0(im)}c}\right)^2}-1\right]\right\} \qquad (1)$$

where $m_{i0(im)}=-\frac{2}{\sqrt{3}}m_{i0}i$ is the *imaginary* inertial mass at rest of the particle and $\Delta p_{(im)}=U_{(im)}n_r/c=(Ui)n_r/c$ is the variation in the particle's *imaginary* kinetic momentum; $c$ is the speed of light. Thus, Eq. (1) can be rewritten as follows

$$\chi=\frac{m_{g(im)}}{m_{i0(im)}}=\left\{1-2\left[\sqrt{1+\frac{3}{4}\left(\frac{Un_r}{m_{i0}c^2}\right)^2}-1\right]\right\} \quad (2)$$

When $\Delta p$ is produced by the absorption of a photon with wavelength $\lambda$, i.e., $U=hf$, Eq. (2) becomes

$$\chi=\frac{m_{g(im)}}{m_{i0(im)}}=\left\{1-2\left[\sqrt{1+\frac{3}{4}\left(\frac{\lambda_0}{\lambda_{\text{mod}}}\right)^2}-1\right]\right\} \quad (3)$$

where $\lambda_0=h/m_{i0}c$ is the *De Broglie wavelength* for the particle with *rest* inertial mass (real) $m_{i0}$ and $\lambda_{\text{mod}}=\lambda/n_r$.

From Electrodynamics we know that when an electromagnetic wave with frequency $f$ and velocity $c$ incides on a material with relative permittivity $\varepsilon_r$, relative magnetic permeability $\mu_r$ and electrical conductivity $\sigma$, its *velocity is reduced* to $v=c/n_r$ where $n_r$ is the index of refraction of the material, given by [6]

$$n_r=\frac{c}{v}=\sqrt{\frac{\varepsilon_r\mu_r}{2}\left(\sqrt{1+(\sigma/\omega\varepsilon)^2}+1\right)} \quad (4)$$

If $\sigma >> \omega\varepsilon$, $\omega=2\pi f$, Eq. (4) reduces to

$$n_r=\sqrt{\frac{\mu_r\sigma}{4\pi\varepsilon_0 f}} \quad (5)$$

Thus, the wavelength of the incident radiation (See Fig. 1) becomes

$$\lambda_{\text{mod}}=\frac{v}{f}=\frac{c/f}{n_r}=\frac{\lambda}{n_r}=\sqrt{\frac{4\pi}{\mu f\sigma}} \quad (6)$$

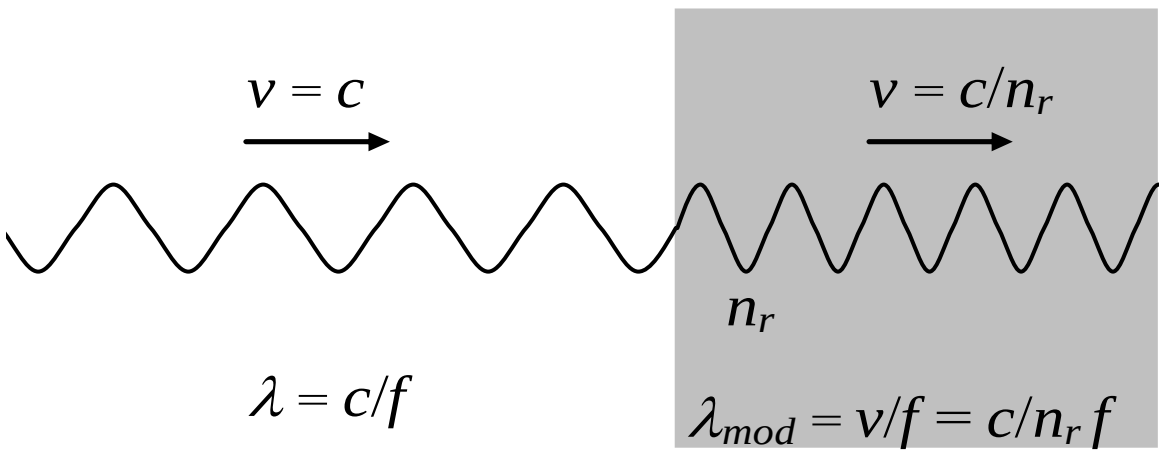

Fig. 1 – *Modified Electromagnetic Wave*. The wavelength of the electromagnetic wave can be strongly reduced, but its frequency remains the same.

If an *imaginary* lamina with thickness equal to $\xi$ contains $n$ *imaginary molecules*/m$^3$, then the number of molecules per unit area is $n\xi$. Thus, if the electromagnetic radiation with frequency $f$ incides on an area $S$ of the lamina it reaches $nS\xi$ molecules. If it incides on the *total area of the lamina*, $S_f$, then the total number of molecules reached by the radiation is $N=nS_f\xi$. The number of molecules per unit volume, $n$, is given by

$$n=\frac{N_0\rho}{A} \quad (7)$$

where $N_0=6.02\times10^{26}\,moleculess/kmole$ is the Avogadro's number; $\rho$ is the matter density of the lamina (in *kg/m*$^3$) and $A$ is the molar mass(*kg/kmole*).

When an electromagnetic wave incides on the lamina, it strikes $N_f$ front molecules, where $N_f\cong(nS_f)\phi_m$, $\phi_m$ is the "diameter" of the molecule. Thus, the electromagnetic wave incides effectively on an area $S=N_fS_m$, where $S_m=\frac{1}{4}\pi\phi_m^2$ is the cross section area of one molecule. After these collisions, it carries out $n_{collisions}$ with the other molecules (See Fig.2).

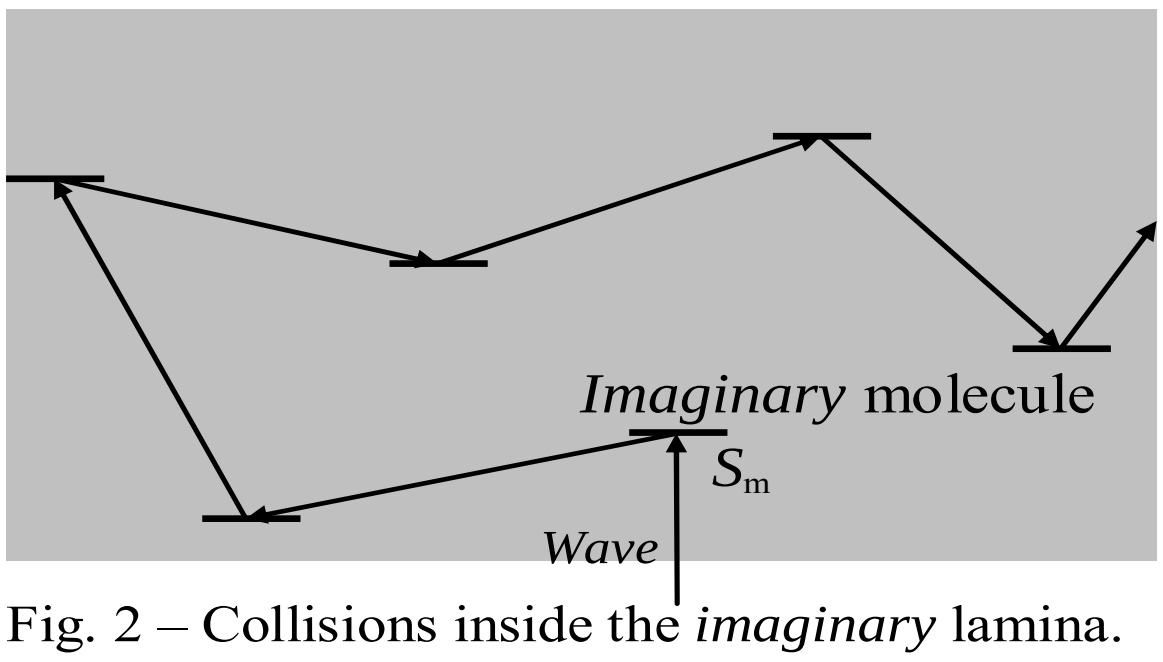

Fig. 2 – Collisions inside the *imaginary* lamina.

Thus, the total number of collisions in the volume $S\xi$ is

$$N_{collisions}=N_f+n_{collisions}=n_lS\phi_m+(n_lS\xi-n_mS\phi_m)=$$
$$=n_mS\xi \quad (8)$$

The power density, $D$, of the radiation on the lamina can be expressed by

$$D = \frac{P}{S} = \frac{P}{N_f S_m} \qquad (9)$$

We can express the *total mean number of collisions in each molecule*, $n_1$, by means of the following equation

$$n_1 = \frac{n_{total\ \ photons} N_{collisions}}{N} \qquad (10)$$

Since in each collision a *momentum* $h/\lambda$ is transferred to the molecule, then the *total momentum* transferred to the lamina will be $\Delta p = (n_1 N)h/\lambda$, i.e., $U n_r/c = (n_1 N) n_r h/\lambda = (n_1 N) h/\lambda_{mod}$ Therefore, in accordance with Eq. (2), we can write that

$$\frac{m_{g(l)im}}{m_{i0(l)im}} = \left\{1 - 2\left[\sqrt{1 + \frac{3}{4}\left[(n_1 N)\frac{\lambda_0}{\lambda_{mod}}\right]^2} - 1\right]\right\} =$$

$$= \left\{1 - 2\left[\sqrt{1 + \frac{3}{4}\left[n_{total\ photons} N_{collisions}\frac{\lambda_0}{\lambda_{mod}}\right]^2} - 1\right]\right\} \qquad (11)$$

Since Eq. (8) gives $N_{collisions} = n_l S\xi$, we get

$$n_{total\ \ photons} N_{collisions} = \left(\frac{P}{hf^2}\right)(n_l S\xi) \qquad (12)$$

Substitution of Eq. (12) into Eq. (11) yields

$$\frac{m_{g(l)im}}{m_{i0(l)im}} = \left\{1 - 2\left[\sqrt{1 + \frac{3}{4}\left[\left(\frac{P}{hf^2}\right)(n_l S\xi)\frac{\lambda_0}{\lambda_{mod}}\right]^2} - 1\right]\right\} \qquad (13)$$

Substitution of $P$ given by Eq. (9) into Eq. (13) gives

$$\frac{m_{g(l)im}}{m_{i0(l)im}} = \left\{1 - 2\left[\sqrt{1 + \frac{3}{4}\left[\left(\frac{N_f S_m D}{f^2}\right)\left(\frac{n_l S\xi}{m_{i0(l)}c}\right)\frac{1}{\lambda_{mod}}\right]^2} - 1\right]\right\} \qquad (14)$$

Substitution of $N_f \cong (n_l S_f)\phi_m$ and $S = N_f S_m$ into Eq. (14) results

$$\frac{m_{g(l)im}}{m_{i0(l)im}} = \left\{1 - 2\left[\sqrt{1 + \frac{3}{4}\left[\left(\frac{n_l^3 S_f^2 S_m^2 \phi_m^2 \xi D}{m_{i0(l)} c f^2}\right)\frac{1}{\lambda_{mod}}\right]^2} - 1\right]\right\} \qquad (15)$$

where $m_{i0(l)} = \rho_{(l)} V_{(l)}$.

The case in which the area $S_f$ is just the *area of the cross-section of the lamina* $(S_\alpha)$, we obtain from Eq. (15), considering that $m_{i0(l)} = \rho_{(l)} S_\alpha \xi$, the following expression

$$\frac{m_{g(l)im}}{m_{i0(l)im}} = \left\{1 - 2\left[\sqrt{1 + \frac{3}{4}\left[\left(\frac{n_l^3 S_\alpha S_m^2 \phi_m^2 D}{\rho_{(l)} c f^2}\right)\frac{1}{\lambda_{mod}}\right]^2} - 1\right]\right\} \qquad (16)$$

If the electrical conductivity of the lamina, $\sigma_{(l)}$, is such that $\sigma_{(l)} >> \omega\varepsilon$, then the value of $\lambda$ is given by Eq. (6), i.e.,

$$\lambda = \lambda_{mod} = \sqrt{\frac{4\pi}{\mu f \sigma}} \qquad (17)$$

Substitution of Eq. (17) into Eq. (16) gives

$$\chi = \frac{m_{g(l)im}}{m_{i0(l)im}} = \left\{1 - 2\left[\sqrt{1 + \frac{3 n_l^6 S_\alpha^2 S_m^4 \phi_m^4 \mu\sigma D^2}{16\pi\rho^2 c^2 f^3}} - 1\right]\right\} \qquad (18)$$

The Soul has been frequently described as a body of *unknown energy* coupled to human body by means of a mutual interaction. This type of energy from the viewpoint of Physics, has been considered as *Imaginary* Energy. The term imaginary is borrowed from Mathematics (real and imaginary numbers) [3]. As imaginary energy, the Soul can be now defined as an *imaginary body*, made of *imaginary particles* each one them described by *imaginaries wavefunctions* $\psi_{im}$, by similarity to the real bodies, which are made of real particles described in *Quantum* Mechanics by its real wavefunction $\psi$. The extension of the *imaginary wavefunction* to the *relativistic form* can be then made in a consistent way with the Lorentz transformations equations of the special theory of relativity [7, 8], similarly to the real wavefunction [4]. In addition, the Soul's energy can be now expressed by the well-known Einstein's energy expression $\left(E = mc^2\right)$ extended to the imaginary form, i.e., $E_{g(S)im} = m_{g(S)im}c^2$. Therefore, we can say that the Soul has an *imaginary* energy $E_{g(S)im} = m_{g(S)im}c^2$ where $m_{g(S)im}$ is the *imaginary* gravitational mass of Soul, which according to Eq. (18), is correlated to *imaginary* inertial mass of Soul at rest, $m_{i0(S)im}$, by means of the following expression: $\chi_S = m_{g(S)im}/m_{i0(S)im}$. This means that the value of $E_{g(S)im}$ can be decreased and also made negative by means of absorption of energy of radiation incident upon the Soul (See Eq.18).

As widely mentioned in the literature of Spiritualistic Philosophy, the Soul has 2 parts: *Perispirit* and *Spirit* (See Fig.1). The Spirit is inside the human body (HB); *the Perispirit is an involucre of the spirit*, its boundaries coincide with the boundaries of the human body. The *perispirit* density $(\rho_{pe} = m_{i0(pe)im}/V_{(pe)im} = real)$ is equal to the density of the mean where the Soul is [9]. This occurs by the imaginary mass decrease or by the imaginary mass increase, resultant, respectively, from the emission or absorption of imaginary energy from the Universe. Thus, in the *human body* the *perispirit* density is

$$\rho_{pe} = \rho_{HB} \cong 1000 kg.m^{-3}$$

Therefore, according to Eq. (7), we can write that the density of molecules in the perispirit is given by

$$n_{pe} = (N_0 \rho_{pe}/A) \cong 3.3 \times 10^{28} molecules.m^{-3}$$

where $A = A_{H2O} = 18$kg/kmole. Out of the Earth's atmosphere (outer space) the density of the perispirit is equal to the density of the spirit $(\rho_s = m_{i0(s)im}/V_{(s)im} = real)$. In the outer space, the Earth's atmospheric pressure drops to about 3.2 × $10^{-7}$ atm [10]. Thus using the well-known Equation of State $(\rho = PM_0/ZT)$, we can write the following correlation expression:

$$\frac{\rho_{air(1atm)}}{\rho_{air(outer\ space)}} = \frac{1atm}{3.2 \times 10^{-7} atm}$$

This means that the *density of spirit* is given by

$$\rho_s = \rho_{pe(outer\ space)} = \rho_{air(outer\ space)} = $$
$$= 3.8 \times 10^{-7} kg.m^{-3}$$

Thus, inside the human body the perispirit density is $\rho_{pe} = \rho_{HB} \cong 1000 kg.m^{-3}$ and the spirit density is $\rho_s = 3.8 \times 10^{-7} kg.m^{-3}$. Since the Perispirit is just an involucre of the spirit, we can assume that $\rho_S \cong \rho_s$.

The gravitational mass of the Soul, $m_{g(S)im}$, is given by the sum of the spirit's gravitational mass with the perispirit's gravitational mass, i.e.,

$$m_{g(S)im} = m_{g(s)im} + m_{g(pe)im}$$

As the *perispirit* is the unique part of the Soul with sufficient density to absorb measurable amounts of electromagnetic energy, we can neglect the contribution of the energy absorbed by the spirit in the calculation of the total energy absorbed by the Soul making $m_{g(s)im} = 0$. Under these conditions, we can write that the *gravitational mass of the Soul*, $m_{g(S)im}$, is given by

$$m_{g(S)im} = \underbrace{m_{g(s)im}}_{0} + m_{g(pe)im} = m_{g(pe)im} \qquad (19)$$

By analogy, the expression of the inertial mass of the Soul, $m_{i0(s)im}$ can be written as follows

$$m_{i0(S)im} = \underbrace{m_{i0(s)im}}_{0} + m_{i0(pe)im} \cong m_{i0(pe)im} \qquad (20)$$

Based on Eq. (19) we can write that $\rho_S V_{(S)im} = \rho_{pe} V_{(pe)im}$, where $V_{(pe)im} = S_{(\alpha)im} \Delta x_{pe}$; $\Delta x_{pe}$ is the thickness of perispirit. Since $\rho_S \cong \rho_s$ and $V_{(S)im} \cong V_{(s)im}$, we can write that $\rho_S V_{(S)im} \cong \rho_s V_{(s)im} = \rho_{pe} V_{(pe)im}$. In addition, we have $V_{(S)im}/S_{(\alpha)im} \cong V_{(s)im}/S_{(\alpha)im} = V_{HB}/S_{HB} \cong 0.4m^3/1.1m^2 \cong 0.4m$. Thus, we can conclude that

$$\Delta x_{pe} = \left(\frac{\rho_s}{\rho_{pe}}\right)\left(\frac{V_{(s)im}}{S_{(\alpha)im}}\right) \cong 2 \times 10^{-10} m \qquad (21)$$

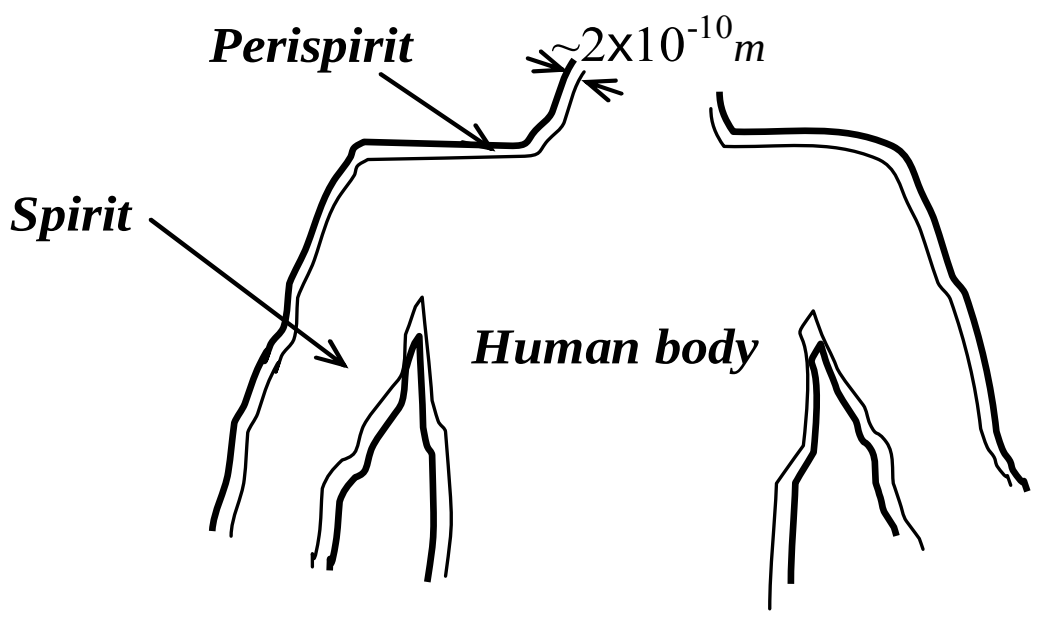

Fig. 3 – *Perispirit and Spirit*. The Spirit is inside the human body; the Perispirit is an involucre of the spirit, its boundaries coincide with the boundaries of the human body.

The power density of radiation, $D_{pe}$, absorbed by the perispirit can be expressed by $D_{pe} = (\Delta x_{pe}/\delta_{pe})D$ where $\delta_{pe}$ is length scale for *total* absorption of the radiation with frequency $\omega = 2\pi f$. As we know, if the electrical conductivity of the mean, $\sigma$, is such that $\sigma >> \omega\varepsilon$, where $\varepsilon$ is the permittivity of the mean, then $\delta$ is given by [11]:

$$\delta = 2.5\times10^{3}\Big/\sqrt{f\sigma} \qquad (22)$$

Therefore, we can write that

$$D_{pe} = (\Delta x_{pe}/\delta_{pe})D \cong 8\times10^{-14}\sqrt{f\sigma_{pe}}D \qquad (23)$$

where $D$ is the total power density of the incident radiation on the perispirit.

By dividing Eq. (19) by Eq. (20) we obtain

$$\chi_S = \frac{m_{g(S)im}}{m_{i0(S)im}} = \frac{m_{g(pe)im}}{m_{i0(pe)im}} = \chi_{pe}$$

Thus, based on Eq. (18), we can write that

$$\chi_S = \chi_{pe} = \frac{m_{g(pe)im}}{m_{i0(pe)im}} =$$

$$= \left\{1 - 2\left[\sqrt{1 + \frac{3n_{pe}^6 S_\alpha^2 S_m^4 \phi_m^4 \mu_0 \sigma_{pe} D_{pe}^2}{16\pi\rho_{pe}^2 c^2 f^3}} - 1\right]\right\} \qquad (24)$$

Further on, we will show that the electrical conductivity of perispirit is enormous (10 trillion times greater than that of the metals), what shows that it contains a *plasma*. For the population of excited states for the elements in the plasma to be predominately caused by collisions with electrons and not by radioactive processes, it requires a minimal electron density to ensure these collisions. This *minimal electron density* is known as the McWhirter criterion and is defined as [12]:

$$N_e \geq 1.6\times10^{18} T^{\frac{1}{2}} (\Delta E)^3 \qquad (25)$$

where $\Delta E$ (in *eV*) is the largest gap between 2 adjacent energy levels; $T$ (in *K*) is the plasma temperature, and $N_e$ is in electrons/m$^3$.

This condition is deduced for hydrogen and hydrogen-like atoms in an optically thin, stationary and homogenous plasma [13]. The largest gap for hydrogen is indeed between the ground state and the first excited energy state and corresponds to 4 *eV*. This is not always the case for other elements. The largest energy gap for oxygen does not include the ground state and is 10 *eV*. In order to calculate the value of $n_e$ for the perispirit at the human body, we must take these values: $\Delta E = 10eV$, $T \cong 300K$. The result is

$$N_e \geq 3\times10^{22}\, electrons.m^{-3} \qquad (26)$$

As we have already shown $n_{pe} \cong 3.3\times10^{28}\, molecules.m^{-3}$. Thus, we can assume that

$$N_{pe} \approx 10^{28}\, ions.m^{-3}$$

It is known that the electrical conductivity is proportional to both the concentration and the mobility of the *ions* and the *free electrons*, and is expressed by

$$\sigma = n_e\mu_e + n_i\mu_i$$

where $n_e$ and $n_i$ express respectively the concentrations $(C/m^3)$ of *electrons* and atom-*ions*; $\mu_e$ and $\mu_i$ are respectively the mobilities of the electrons and the ions.

In order to calculate the electrical conductivity of the perispirit, we first need to calculate the concentrations $n_e$ and $n_i$. We start by calculating the value of $n_i$, which is given by

$$n_i = n_e = N_{pe}e \approx 10^9 C/m^3$$

This corresponds to the concentration level in the case of *conducting materials*. For these materials, at temperature of 300K, the mobilities $\mu_e$ and $\mu_i$ are of the order of $10^{-1}m^2V^{-1}s^{-1}$ [14]. Very high mobility has been found in several low-dimensional systems, such as two-dimensional electron gases (2DEG) ($300m^2V^{-1}s^{-1}$), [15] carbon nanotubes ($10m^2V^{-1}s^{-1}$) [16] and more recently, graphene ($20 m^2V^{-1}s^{-1}$)[17]. It is known that the mobility $\mu_d$ is related to the drift velocity $v_d$ by means of the following equation:

$$v_d = \mu_d E$$

where $E$ is the electric field. Thus, based on this equation, we can relate the mobility of free electrons of the Soul, $\mu_e$, with the typical mobility of conductors, $\mu_{cond} \approx 10^{-1} m^2V^{-1}s^{-1}$, by means of the following expression:

$$\mu_e = \frac{v_{d(pe)}}{v_{d(cond)}} \mu_{cond} \qquad (27)$$

where the typical drift velocity in conductors is $v_{d(cond)} \approx 10^{-4} m/s$ [18], and the drift velocity in perispirit is $v_{d(pe)} \approx c$ (since there are no collisions among the imaginary electrons). Thus, we get

$$\mu_e \approx 10^{12} m^2V^{-1}s^{-1}$$

Consequently, we can write that the electrical conductivity of perispirit is given by

$$\sigma_{pe} = n_e\mu_e + n_i\mu_i \approx 10^{21} S.m^{-1} \qquad (28)$$

By substitution of this value into Eq. (22), we get

$$\delta_{pe} \cong 10^{-7}/\sqrt{f} \qquad (29)$$

By substitution of $\sigma_{HB} \approx 0.1 S/m$ (conductivity of human body) into Eq. (22), we obtain

$$\delta_{HB} \cong 10^{4}/\sqrt{f} \qquad (30)$$

Substitution of the values of $n_{pe}, \sigma_{pe}$, $\rho_{pe}$, $D_{pe}$, $\phi_m = 1.55\times10^{-10} m$, (average "diameter" of the *molecules*), $S_m = \pi\phi_m^2/4 = 1.88\times10^{-20} m^2$ and $S_\alpha \cong 1.1 m^2$ into Eq. (24), gives

$$\chi_S = \left\{1 - 2\left[\sqrt{1 + \frac{\sim 10^{39} D^2}{f^2}} - 1\right]\right\} \qquad (31)$$

In this expression, the *minimum value* of $D$ is limited by the *uncertainty principle*, i.e., by the amount of energy $\Delta E$ that can be detectable by our instruments. According to the uncertainty principle, $\Delta E \Delta t \geq \hbar$. Thus, $\Delta E \geq \hbar/\Delta t \rightarrow \Delta E_{min} = \hbar/\Delta t_{max}$. Since we can write that $D = \Delta E/S_\alpha \Delta t \geq \hbar/S_\alpha \Delta t^2 \rightarrow D_{min} = \hbar/S_\alpha \Delta t^2_{max}$, we obtain $D_{min} = \Delta E^2_{min}/\hbar S_\alpha$. Here, $\Delta E_{min} = kT$, because if $< kT$, then the action of the incident radiation will be hidden by the action of the thermal radiation $(kT)$. Consequently, we can write that $D_{min} = k^2T^2/\hbar S_\alpha$. Therefore, for $T = 300K$ and $S_\alpha \cong 1.1 m^2$, we get

$$D_{min} \cong 1.5\times10^{-7} W/m^2$$

.

Based on Eq.(31), we can write that the *Soul imaginary* energy $E_{g(S)im} = m_{g(S)im}c^2$ can be expressed by

$$E_{g(S)im} = m_{g(S)im}c^2 = \chi_S m_{i0(S)im}c^2 \cong$$
$$\cong \left\{1 - 2\left[\sqrt{1 + \frac{10^{39} D^2}{f^2}} - 1\right]\right\} m_{i0(S)im}c^2 \qquad (32)$$

This energy varies along the time, having a minimum value at the beginning of life and a maximum value, $m^{max}_{g(S)im}c^2$, in a specific instant of the life of the person. After this maximum value, the energy decreases progressively down to the instant of the death of the person. This means that the average variation of this energy along the time can be expressed by the well-known *bell curve* (probability curve [19]), in the following form

$$m_{g(S)im}c^2 = m^{max}_{g(S)im}c^2 e^{-4\pi^2 b^2 t^2} \qquad (33)$$

where $b$ is a time-constant to be defined. Since $m_{g(S)im} = \chi m_{i0(S)im}$ and $m_{g(S)}^{max} = \chi m_{i0(S)im}^{max}$ Eq. (33) can be rewritten as follows

$$m_{i0(S)} = m_{i0(S)}^{max} e^{-4\pi^2 b^2 t^2} \tag{34}$$

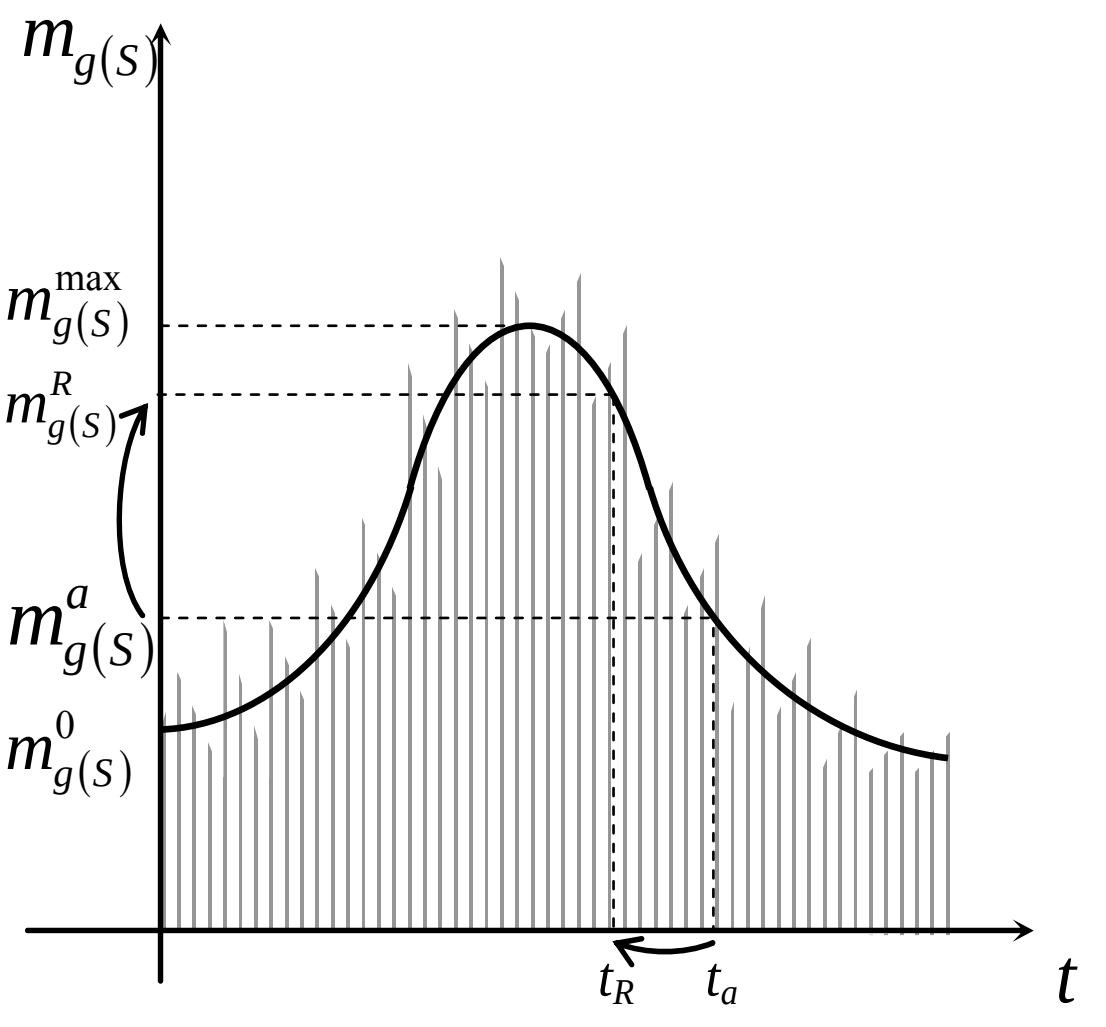

Fig.4 - The variation of the *gravitational mass of the Soul* (Soul energy $m_{g(S)}c^2$) along the life of a person, and Quantum Reversal from the current energy level to a previous energy level.

Making $t = t_a$ (current time in the life of the person), and after $t = t_R$ (*reversal time**, see Fig.4) into Eq. (34), we obtain the following expressions

$$m_{i0(S)}^{a} = m_{i0(S)}^{max} e^{-4\pi^2 b^2 t_a^2} \tag{35}$$

and

$$m_{i0(S)}^{R} = m_{i0(S)}^{max} e^{-4\pi^2 b^2 t_R^2} \tag{36}$$

By dividing Eq.(35) by Eq. (36), we get

$$m_{i0(S)}^{a} = m_{i0(S)}^{R} e^{-4\pi^2 b^2 \left(t_a^2 - t_R^2\right)} \tag{37}$$

Positive values of the Soul energy express the progress of the energy levels and obviously, a given positive value cannot to cause reversion. Thus, in order to occur the reversion it is necessary that $\chi_S < 0$, When this occurs, the energy of the Soul ($m_{g(S)im}^{a} c^2 = \chi_S m_{i0(S)im}^{a} c^2$) becomes *positive and returns to the corresponding* value ($m_{i0(S)im}^{R} c^2 = |\chi_S| m_{i0(S)im}^{a} c^2$). Thus, we can write that $m_{i0(S)}^{R} = |\chi_S| m_{i0(S)}^{a}$. Substitution of this expression into Eq. (37) gives

$$m_{i0(S)}^{a} = |\chi_S| m_{i0(S)}^{a} e^{-4\pi^2 b^2 \left(t_a^2 - t_R^2\right)} \tag{38}$$

or

$$|\chi_S| = e^{4\pi^2 b^2 \left(t_a^2 - t_R^2\right)} \tag{39}$$

Note that the time-constant $b^{-2}$ must be a very big number, because the values of $e^{\left(t_a^2 - t_R^2\right)}$ are enormous in the case of $0 << t_a < 3.1 \times 10^9$s (100years). Thus, if the value of $b^{-2}$ is not very big then the values of $\chi_S$ lose their meaning. A very big number related to the time is, certainly, the age of the Universe. Thus, we will define the time-constant $b^{-2}$ as follows

$$b^{-2} = 4.26 \times 10^{17} s = current\ age\ of\ Universe$$

For example, if a person is exactly 62 years old ($t_a = 1.928 \times 10^9 s$), and wants to revert its Soul energy to the energy that it had at *5 years* ago (57 years old, $t_R = 1.774 \times 10^9 s$), then the value of $\chi_S$, according Eq.(39), must be given by

$$|\chi_S| = e^{4\pi^2 b^2 \left(t_a^2 - t_R^2\right)} \cong e^{51.32} \cong 1.9 \times 10^{22} \tag{40}$$

Equation (31) shows that, in order to obtain $\chi_S = -1.9 \times 10^{22}$, is necessary to apply on the *Soul* (body) an electromagnetic radiation with frequency $f$ and power density $D$, given by

$$D \cong 300 f \tag{41}$$

Maximum Permissible Exposure (MPE) levels have been established by ANSI Z136.1 [20] for various laser wavelengths and exposure durations. The MPE is the level of laser radiation to which a person may be exposed without hazardous effect or adverse biological changes in the eye or skin. This limit is ~1000 W/m$^2$. Here, considering this limit, we can conclude that, according to Eq.

* This is not a return in time. It is only return to *energy level* that the Soul had at a specific time of its life.

(41), the maximum value for the frequency is about 3.3Hz.

Now, we can verify the effect of the ELF radiation upon the *gravitational mass* of the *human body*. By substitution of $n_{HB} = n_{pe}$, $\rho_{HB} = \rho_{pe}$, $\sigma_{HB} \sim 0.1 S/m$, $\phi_m = 1.55\times10^{-10} m$, $S_m = \pi\phi_m^2/4 = 1.88\times10^{-20} m^2$ and $S_\alpha \cong 1.1 m^2$ into Eq. (18) we obtain

$$\chi_{HB} = \left\{1 - 2\left[\sqrt{1 + \frac{\sim 10^{23}\sigma_{HB} D_{HB}^2}{f^3}} - 1\right]\right\} \quad (42)$$

The expression of $D_{HB}$ can be obtained from the following relation

$$\frac{D_{HB}}{D_{pe}} = \frac{\delta_{pe}}{\delta_{HB}} \quad (43)$$

where $D_{pe} \cong 8\times10^{-14}\sqrt{f\sigma_{pe}}D$ (Eq. 23). Thus, Eq. (43) can be rewritten as follows

$$D_{HB} \cong 10^{-14}\sqrt{f}\,D \quad (44)$$

Substitution of this equation into Eq. (42), gives

$$\chi_{HB} = \left\{1 - 2\left[\sqrt{1 + \frac{\sim 10^{-6} D^2}{f^2}} - 1\right]\right\} \quad (45)$$

Substitution of Eq. (41) into (45) yields $\chi_{HB} > 0.9$. This corresponds to a weight decrease of less than 10%, which shows that, here, in the case of $D \cong 300 f$, the effect of the ELF radiation upon the *gravitational mass* of the *human body* is negligible.

Quantum Physics shows that the *energy* is *quantized*, i.e., it has discrete values that are defined as discrete energy levels that correspond to all positive integer values of the *quantum number* $n, (n = 1,2,3,...)$. Thus, along the life of a person, the energy of its Soul is characterized by several quantum levels of energy. Then, we can say that, when occurs a reversal of soul energy, it carries out a *quantum reversal* to a previous level of energy.

Any action once performed leaves an impression on the Soul (its energy). Thus, each energy level of a Soul expresses, at the corresponding moment, the human being in its totality. This means, for example, that our current human shape results from the current energy level (spectrum) of our Soul. Imagine that a person break a leg when it is 50 years old. If he is subjected to an electromagnetic radiation flux with $D \approx 300 W/m^2 at\ 1Hz$, then its Soul carries out a quantum reversal to the energy level that he had 5 years ago. At this energy level the leg was not broken in the human body. Consequently, we can expect that the broken part disappears, and the leg returns to the form that it had in this energy level. By means of this process it seems possible *the immediate cure of any wound*, *any kind of disease*, and also the *resuscitation* of persons who have died some seconds ago (before the spirit leaves the human body).

It is known that the brain is able to generate electromagnetic waves with frequencies smaller than 100HZ. The brainwaves of lowest frequencies are the *Delta waves*. Delta waves are defined as having a frequency between 0.5 and 2 hertz. They are the highest amplitude brainwaves. In adults they are radiated from their forehead [21]. Also, it is known that the total electromagnetic power (all the frequencies) generated by the brain can reach up to 25Wor more [22]. This means that at a distance of 1m from the brain a maximum power density is about 2W/m$^2$. Substitution of this value and $f = 1Hz$ into Eq. (31) gives

$$\chi_S = -1.265\times10^{20} \quad (46)$$

Comparing with Eq. (40), yields

$$|\chi_S| = e^{4\pi^2 b^2\left(t_a^2 - t_R^2\right)} \cong e^{46.286} \cong 1.265\times10^{20} \quad (47)$$

whence we obtain

$$t_R = \sqrt{t_a^2 - 4.994\times10^{17}} \quad (48)$$

For $t_a = 1.928\times10^9 s = 62 years$, we obtain

$$t_R \cong 1.794\times10^9 s \cong 58 years \quad (49)$$

This means a return of approximately *4 years* in the Soul energy level. Note that, while it is necessary 2W/m$^2$ at 1Hz to return 4 years, it is necessary 300W/m$^2$ at 1Hz to return 5 years.

However, only a small part of the 25W is due to the delta waves. This means that the return is yet smaller. It obvious that the power densities of the delta waves radiated from the brains vary of persons for persons. Possibly, for most the persons the power densities of delta waves radiated from its brains are negligible (smaller than the critical value $D_{\min} \cong 1.5\times10^{-7} W/m^2$). Since we can relate the radiation density, $D$, with the intensity of the electric field, $E$, by means of the following expression $D = n_r E^2 / 2\mu_0 c$ [23], then, considering that the value of electric field in the forehead of a person (when emitting delta waves) is $E = V/r$, where $V$ is the local electric potential (for ordinary persons $V \approx 150\mu V$ [24]), and $r$ is the radius of the sphere with the same volume of the brain $(r \approx 0.1m)$, then we can write that

$$D = \frac{E^2}{2\mu_0 c} = \frac{V^2}{2\mu_0 c r^2} \approx 10^{-9} W/m^2 << D_{\min}$$

This shows why the ordinary persons cannot to produce immediate cures. Equation above shows that to produce $D > D_{\min}$ is necessary that $V > 1mV$ (approximately 10 times greater than that of ordinary persons).

Note that, if $t_a = 1.866\times10^8 s = 6 years$, and we want to return 1 year, then for *1Hz* the necessary value of $\chi_S$, according to Eq. (39) is $|\chi_S| = e^{0.986} \rightarrow \chi_S = -2.68$. Thus, according to Eq. (31) the value of $D$ is $8.4\times10^{-20} W/m^2$. However, as we have already seen, the value of $D$ is limited to $D_{\min} \cong 1.5\times10^{-7} W/m^2$. This means that the return of 1 year, in the case what $t_a = 1.866\times10^8 s = 6 years$, is impossible. Also, note that for $D \geq D_{\min}$ the values of $t_R$ become imaginaries. What means that it is impossible to return the soul energy of a child with 6 years old. In general, it is impossible to return the soul energy of any person with $t_a < \sqrt{\ln|\chi_S| / 4\pi^2 b^2}$.

Since $D_{\max} \cong 10^3 W/m^2$ and $f_{\min} \cong 0.1Hz$, we obtain from Eqs. (31) and (39) the following expression:

$$t_R^{\max} = \sqrt{t_a^2 - 5.914 \times 10^{17}} \qquad (50)$$

For $t_a = 1.928\times10^9 s = 62 years$, we get $t_R^{\max} = 1.768\times10^9 s \cong 56.8 years$. This means a maximum return of ~ *5.2 years* in the soul energy level. For $t_a = 9.330\times10^8 s = 30 years$, Eq. (50) gives $t_R^{\max} = 5.282\times10^8 s \cong 17 years$. Therefore, a return of approximately *13 years,* in the soul energy level. Note that, the maximum return possible, $\sim 13.8 years$, occurs for $t_a \cong 29 years$.

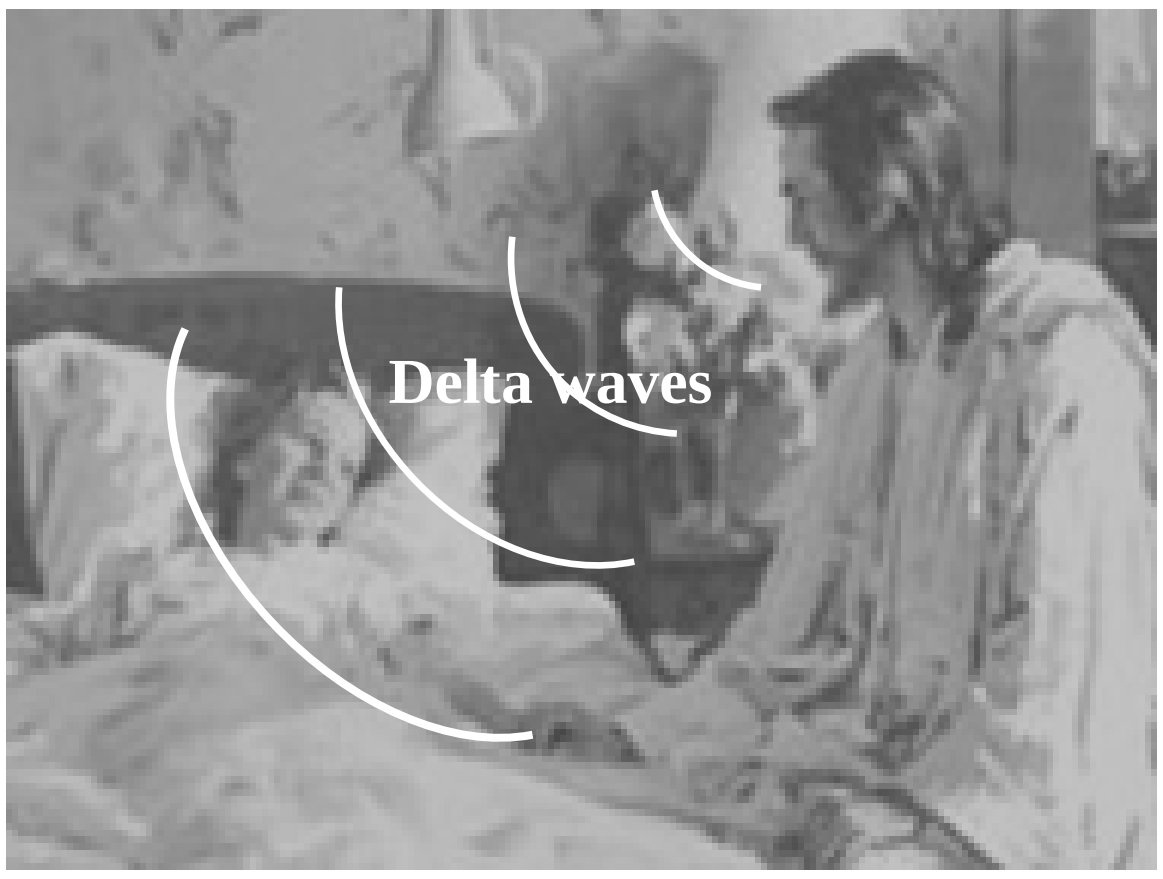

Fig. 5 – *The immediate Cure or Resuscitation.* Delta waves are defined as having a frequency between 0.5 and 2 hertz. They are the highest amplitude brainwaves. In adults they are radiated from their forehead.

The building of ELF transmitters is very difficult because the length of the antenna is enormous. In the case of 1Hz the antenna length must be of the order of 100.000km. However, by using the process of gravitational *redshift* at laboratory scale, shown in a previous paper [25] it is possible for example, to reduce frequencies $f \cong 1GHz$ down to ~1Hz. In order to produce a power density $D \cong 10^{-6} W/m^2$ at ~1Hz, by the mentioned redshift process, it is necessary an initial flux with $D \cong 10^3 W/m^2$ at ~1GHz, what corresponds to the minimum frequency band of MASERS. These devices were invented before the laser, but have languished in obscurity because they required high magnetic fields and difficult cooling schemes. Hydrogen masers oscillate at a frequency at around 1.42GHZ and have a typical power of ~ $10^{-13}$W [26]. They are very complex and expensive devices.

Recently, it was discovered a *room-temperature solid-state maser*, which oscillates at frequency of 1.45GHZ. Basically, it is a simple crystal called *p-terphenyl*. This new device is very simple to build and operate, and removes totally the masers' complexity. When configured as an oscillator, this solid-state maser's measured output power density of around 1*m*W/mm$^2$ ≅ 1000W/m$^2$ (approximately 100 million times greater than that of an atomic hydrogen maser) [27].

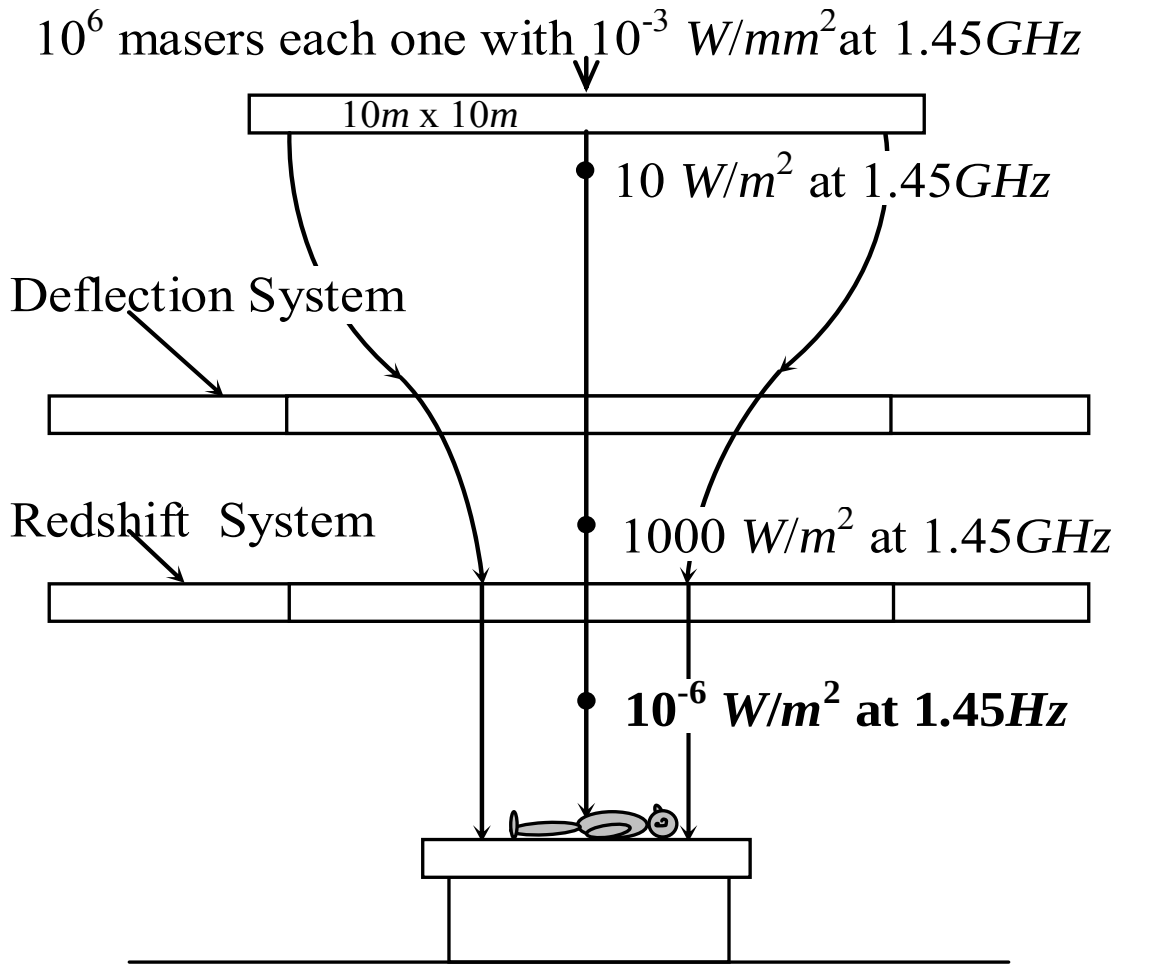

Fig. 6 – Schematic diagram of a system, using solid-state masers, to produce an ELF radiation flux with **$10^{-6}$ $W/m^2$ at 1.45*Hz***.

Considering that each one of these masers radiates 1*m*W/mm$^2$, then it is necessary a set of $10^6$ masers placed inside an area of 10mx10m (See Fig.6), and *after concentrated* into an area of *1m$^2$* (by means of the process of deflection of electromagnetic waves at laboratory scale [28]), in order to obtain a flux of $10^3$W/m$^2$ at 1.45 GHz, which is posteriorly redshifted to a flux of $10^{-6}$W/m$^2$ at 1.45Hz $\left(D_{min} \cong 1.5\times10^{-7} W/m^2\right)$. Substitution of these values into Eq. (31) gives $\chi_S \cong -4\times10^{13}$. By substitution of this value into Eq. (39), we get

$$t_R = \sqrt{t_a^2 - 3\times10^{17}} \qquad (51)$$

This equation shows that the system will only be useful to produce *short returns* in the soul energy of persons with $t_a > 18 years$. Similar systems with higher power densities can provide *higher returns*, for persons with $t_a >> 18 years$.

Probably all human brains are able to generate delta waves. But, only few brains can generate fluxes of this kind of radiation with the necessary power density to return the energy of the Soul to a previous energy level, sufficient to carry out the immediate cure of any wound, any kind of disease, or the resuscitation of persons. The history shows the existence of several persons who have realized immediate cures, and someone that has carried out even resuscitations. Among them, the most known is Jesus of Nazareth.

As we have already shown, the ordinary persons usually are not able to produce fluxes of delta waves with power densities sufficient to carry out immediate cures ($D > D_{min} \cong 1.5\times10^{-7} W/m^2$). Also, we have shown that, in order to carry out these cures is necessary power densities about 100 times more intense then those produced by the ordinary persons $\left(D \approx 10^{-9} W/m^2\right)$. In addition, it is very rare to remain conscious during the emission of delta waves. Thus, the persons who, at conscious state, are able to radiate fluxes of delta waves with power densities 100 times more high than those produced by the ordinary persons - which just radiate delta waves at sleep state - are really *extraordinary persons*.

What is necessary for the brain of a person acquire this capacity? A special diet? Specific physical exercises? Or the persons only acquire this capacity by means of the evolution? That is, all persons have this capacity on a latent stage, but it is only awakened at a specific evolution level.

Recently, it was proved that the state of mature cells, with a specific disease, can be reverted to a previous state (pluripotent stem cell state), where the cells become healthy† [29,30]. This is in agreement with the process of *quantum reversal of the soul energy*, proposed in this work, which shows that it is possible to revert the current state of a human body to a healthy previous state.

This matter is unprecedented in the literature. It is necessary more than one paper to deepen it. We will return to this matter in a future work.

† This work, initiated by Gurdon (1962) and concluded by Yamanaka (2006), has been awarded with the 2012 Nobel Prize in Medicine.